\documentclass[fleqn,usenatbib]{mnras}
\pdfoutput=1 
\usepackage[T1]{fontenc}
\usepackage{ae,aecompl}
\usepackage{hyperref}
\usepackage{graphicx}	
\usepackage{amsmath}	
\usepackage{amssymb}	

\usepackage{txfonts}
\usepackage{rotating}
\usepackage{accents}
\usepackage{float}
\usepackage{epsfig}
\usepackage{textcomp}
\usepackage{upgreek}
\usepackage{epstopdf}
\usepackage{color}
\usepackage{pdflscape}
\usepackage{appendix}

\begin{document}
\title[MEGASTAR  First Release]{MEGARA-GTC Stellar Spectral Library (II). MEGASTAR  First Release}

\author[E. Carrasco et al. ]
{E. Carrasco$^{1}$\thanks{E-mail:}, M. Moll\'{a}$^{2}$, M.L. Garc{\'\i}a-Vargas$^{3}$, A. Gil de Paz$^{4, 5}$,  N. Cardiel$^{4, 5}$, P. \newauthor
G\'{o}mez-Alvarez$^{3}$ and S. R. Berlanas$^{6}$
\\
$^{1}$ Instituto Nacional de Astrof{\'\i}sica, {\'O}ptica y Electr{\'o}nica, INAOE, Calle Luis Enrique Erro 1, C.P. 72840 Santa Mar{\'\i}a Tonantzintla, Puebla, Mexico\\
$^{2}$ Dpto. de Investigaci\'{o}n B\'{a}sica, CIEMAT, Avda. Complutense 40, E-28040 Madrid, Spain\\
$^{3}$ FRACTAL S.L.N.E., Calle Tulip{\'a}n 2, portal 13, 1A, E-28231 Las Rozas de Madrid, Spain \\
$^{4}$ Dpto. de F{\'\i}sica de la Tierra y Astrof{\'\i}sica, Fac. CC. F{\'\i}sicas, Universidad Complutense de Madrid, Plaza de las Ciencias, 1, E-28040 Madrid, Spain \\
$^{5}$ Instituto de F{\'\i}sica de Part{\'\i}culas y del Cosmos, IPARCOS, Fac. CC. F{\'\i}sicas, Universidad Complutense de Madrid, Plaza de las Ciencias 1, E-28040 Madrid, Spain\\
$^{6}$ Departamento de Física Aplicada, Universidad de Alicante, 03690 San Vicente del Raspeig, Alicante, Spain}

\date{Accepted Received ; in original form }
\pagerange{\pageref{firstpage}--\pageref{lastpage}} \pubyear{2019}

\maketitle
\label{firstpage}

\begin{abstract}
MEGARA  is an optical integral field and  multi-object   fibre-based  spectrograph for the  10.4m Gran Telescopio CANARIAS that offers medium to high spectral resolutions (FWHM) of  R~$\simeq$~6000, 12000, 20000.  Commissioned at the telescope in 2017, it started operation as a common-user instrument in  2018. We are creating an instrument-oriented empirical spectral library from MEGARA-GTC stars observations, MEGASTAR, crucial  for the  correct interpretation of MEGARA  data. This piece of work describes the content of the first release of MEGASTAR, formed by the spectra of 414 stars observed with R~$\simeq$~20000 in the spectral intervals 6420\ -- 6790\,\AA\ and  8370\ -- 8885\,\AA, and obtained with a continuum average signal to noise ratio around 260. We describe the release sample, the observations, the data reduction procedure and the MEGASTAR database. Additionally, we include in Appendix A, an atlas with the complete set of 838 spectra of this first release of the MEGASTAR catalogue.

\end{abstract}


\section{Introduction}
\label{Introduction}
MEGARA, acronymous of Multi Espectr{\'o}grafo en GTC de Alta Resoluci{\'o}n para Astronom{\'\i}a,   is an optical  integral field and multi-object spectrograph (MOS) for the Gran Telescopio CANARIAS (GTC).  The MEGARA project was carried out by a consortium formed by the Universidad Complutense de Madrid, (Spain), as the leading institution, the Instituto Nacional de Astrofísica, Óptica y Electrónica, (México), the Instituto de Astrofísica de Andaluc\'ia, (Spain) and the Universidad Polit\'ecnica de Madrid, (Spain), with the participation of European, Mexican and USA companies. In particular, the Spanish company FRACTAL S.L.N.E. played a key role as responsible of the project  management and the system engineering, among other work packages. MEGARA accomplished its mission as a case of success: it was finished fulfilling all the requirements within budget and schedule. The instrument  combines  versatility and performance offering  both bi-dimensional and MOS high efficiency spectroscopy with three spectral resolutions. MEGARA commissioning at GTC  concluded on  August 31, 2017 and was offered to the community in the second observing semester of 2018.    
\begin{table}
\caption{Main characteristics of MEGARA at GTC.}
\label{mainchar}
\begin{tabular}{llrr}
\hline
 IFU (LCB) FoV & & 12.5$\,\arcsec\times 11.3\,\arcsec$ \\
LCB  multiplexing  &   & 623 + 56 sky fibres &  \\
MOS FoV &  & 3.5\,$\arcmin \times 3.5\,\arcmin$\\
MOS multiplexing  &  &   644\\
Spaxel (LCB/MOS)&  &  0.62\,$\arcsec$ \\
Resolving power & LR &  6000  &\\
 & MR &12000  & \\ 
& HR & 20000 & \\
& \\
Spectral configurations& {6 LR, 10 MR and 2 HR} \\
& \\
Wavelength intervals& LR &   {3650\ -- 9750\,\AA}\\
 & MR &   {3650\ -- 9750\,\AA}\\
   &  HR &{6420\ -- 6790\,\AA} & \\
  &   &{8370\ -- 8885\,\AA} &\\
\hline
\end{tabular}
\end{table}

\begin{figure*}
\centering
\includegraphics[width=0.39\textwidth,angle=0]{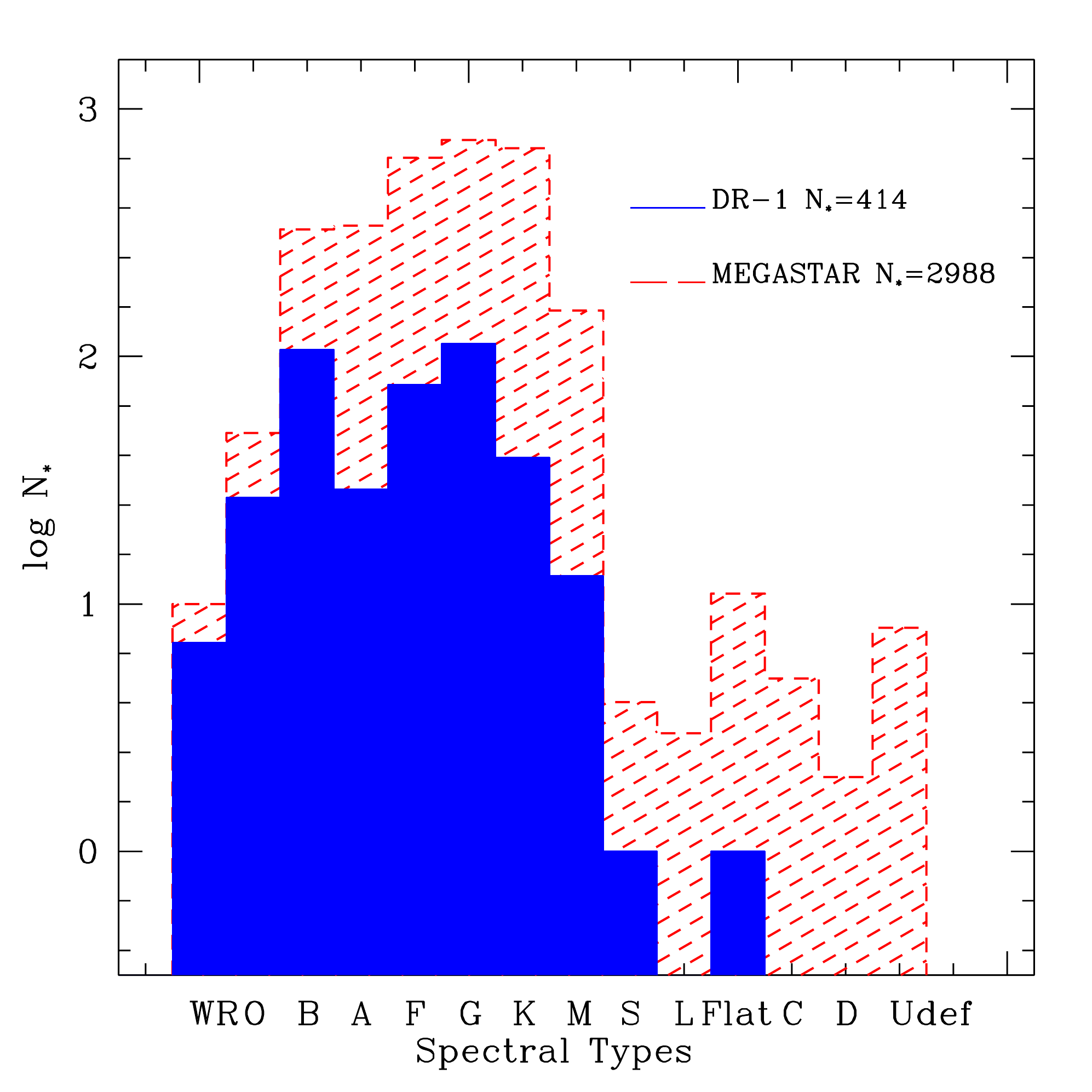}
\includegraphics[width=0.46\textwidth,angle=0]{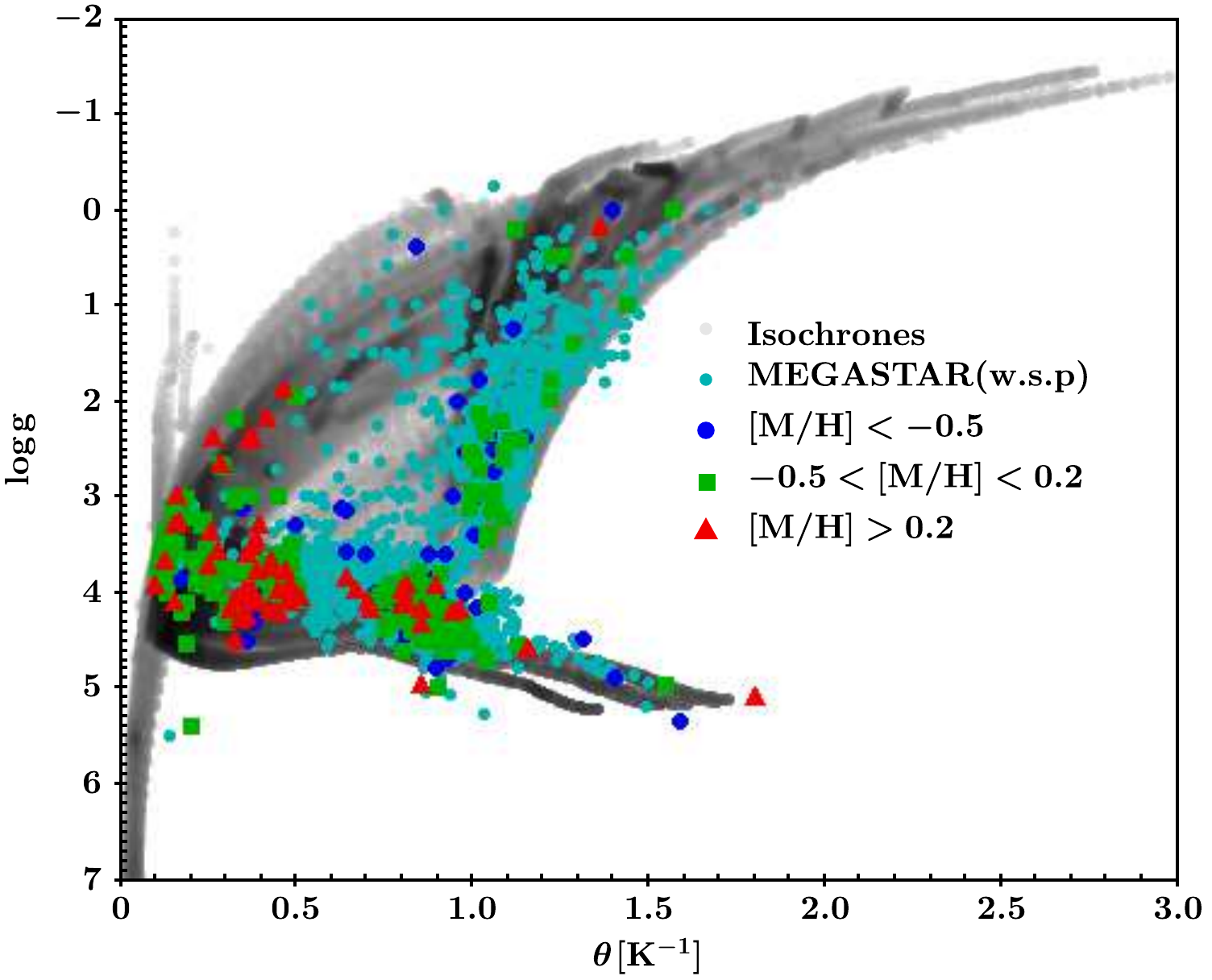}
\caption{Properties of the release stars compared with the whole MEGASTAR library. Left) Distribution of spectral types retrieved from SIMBAD. In red the complete library MEGASTAR, in blue the corresponding to this data release (1.0)  DR-1. The y-axis is in logarithmic scale. Right) Surface gravity  $\rm \log{g}$ {\sl vs.} $\Theta$=5040/$\rm T_{eff}$ diagram. There we show: 1) The values given by the Padova isochrones which we would need for a synthesis code as {\sc PopStar}
as grey scale (more intense in the regions where there are more points); 2) The stars from the MEGASTAR library, with stellar parameters (w.s.p.)  available in the literature, as cyan dots, and 3) the stars of this release  shown with three metallicity intervals, indicated by different colors as labelled.}
\label{theta-logg}
\end{figure*}

For a detailed description of the instrument  and its scientific validation see \citet{carspie18},  \citet{gilspie18}, \citet{dull19} and  \citet{gilAAS20}. Here,  we present a summary for completeness. Its main characteristics are shown in Table~\ref{mainchar}. In the bidimensional mode, an integral field unit (IFU) named as Large Compact Bundle (LCB),  provides a Field Of View (FoV) of 12.5$\arcsec~\times$~11.3$\arcsec$, plus eight additional 7-fibres minibundles for  sky subtraction, located in the external part  of the  MOS field.  In the MOS  mode,   92  robotic positioners, each with a  7-fibres minibundle, cover an area on sky of 3.5$\arcmin$~$\times$~3.5$\arcmin$. The spatial sampling in both modes is 0.62$\arcsec$ per fibre. Each spaxel size is the combination of a 100~$\upmu$m core fibre coupled to a microlens that converts the f/17 entrance telescope beam to f/3  to minimise  focal ratio degradation.

\begin{figure*}
\includegraphics[width=0.22\textwidth,angle=-90]{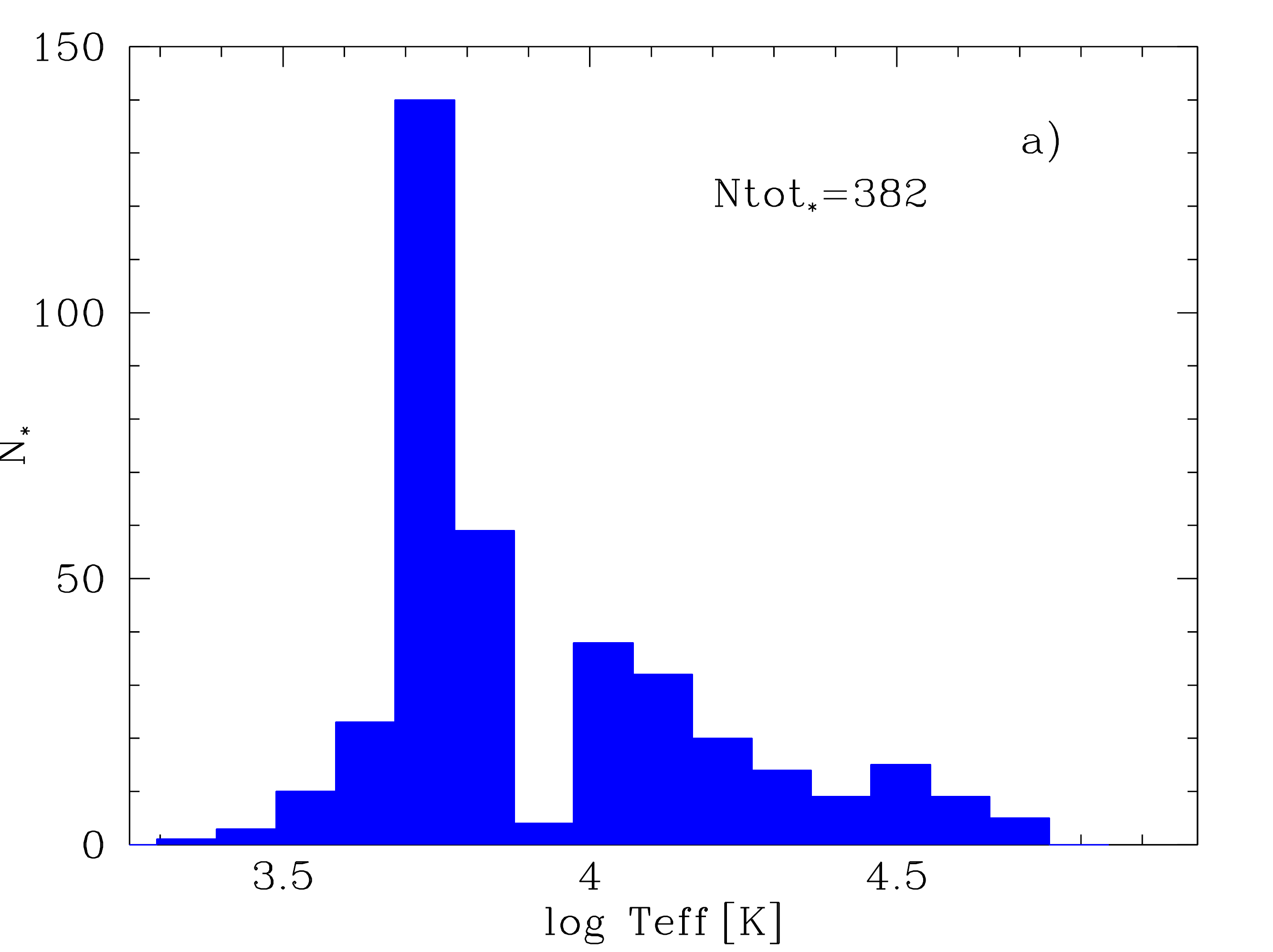}
\includegraphics[width=0.22\textwidth,angle=-90]{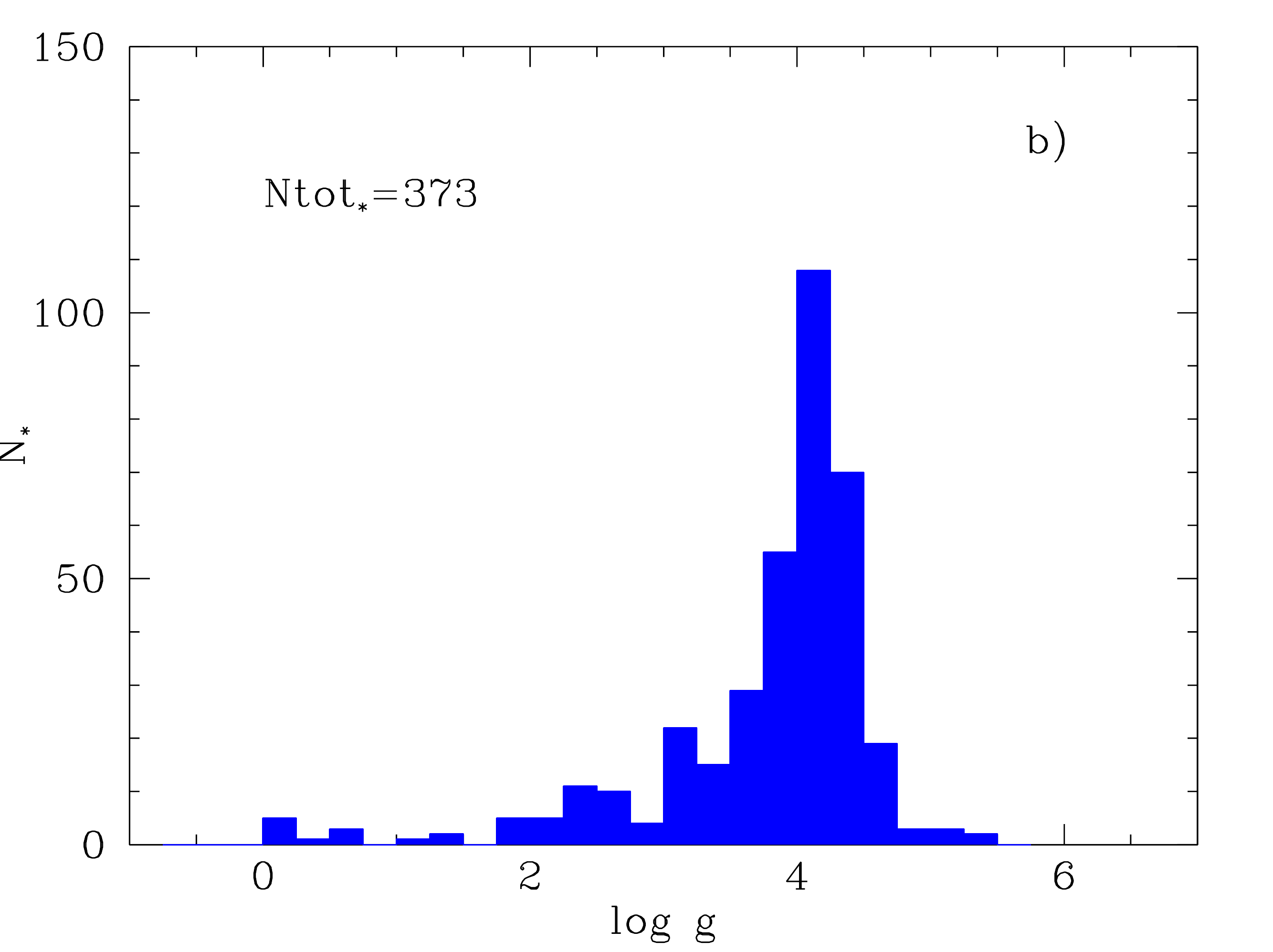}
\includegraphics[width=0.22\textwidth,angle=-90]{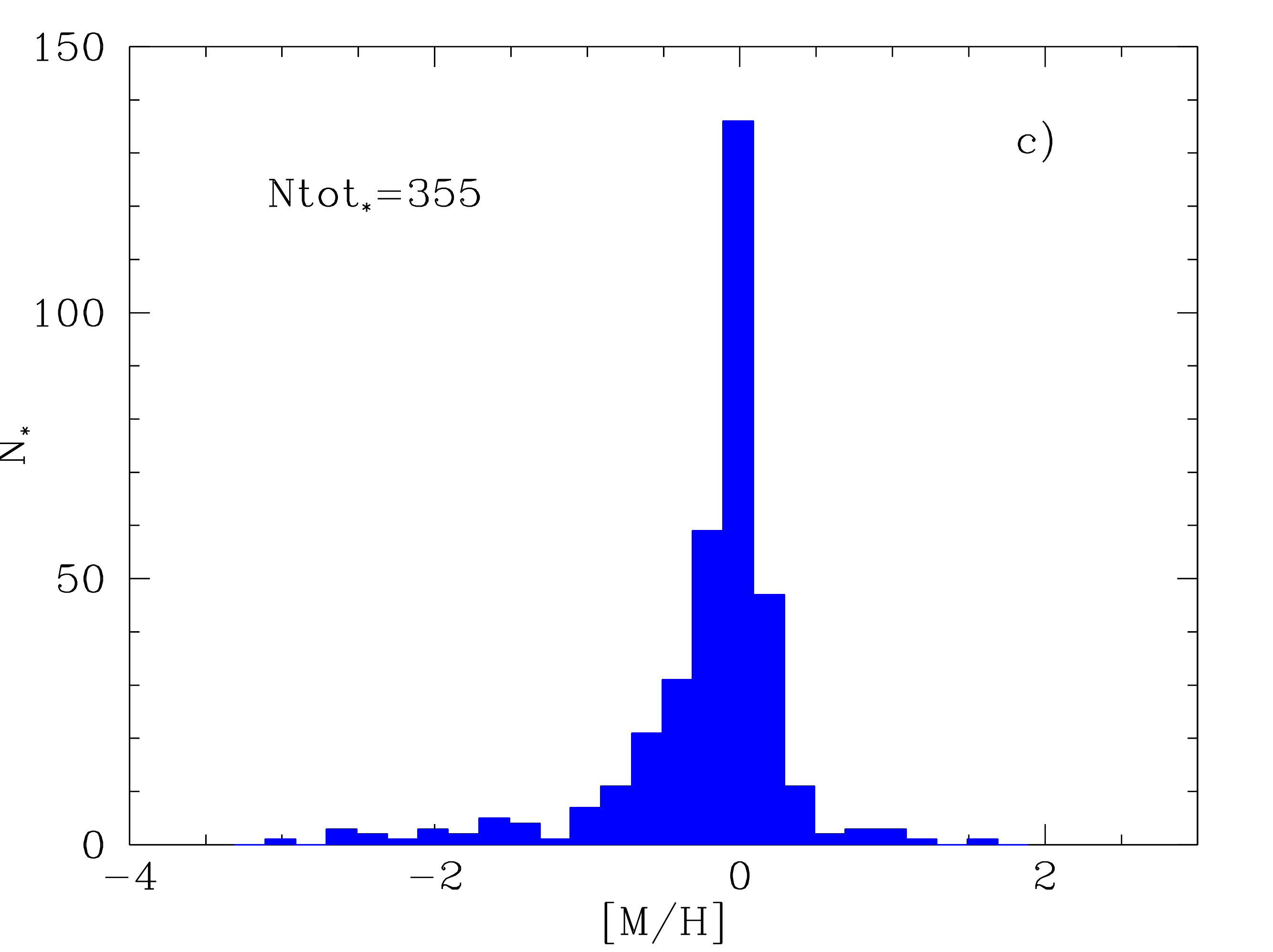}
 \caption{Histogram of the number of stars in this release  as a function of   $\rm T_{eff}$ (a); $\rm \log{g}$ (b) and  $\rm [M/H]$ (c).  {\em Ntot} indicates the number of points in each graph.}
\label{histograms}
\end{figure*}

MEGARA provides Low Resolution, LR, R(FWHM)~$=$~6000; Medium Resolution, MR, R(FWHM)~$=$~12000; and High Resolution, HR, R(FWHM)~$=$~20000. In LR and MR with six and ten volume phase holographic  (VPH) gratings, respectively, the wavelength interval coverage is 3650 -- 9750\,\AA. In HR, using two gratings, the wavelength range for HR-R is 6405 -- 6797\,\AA\ and for HR-I is 8360 -- 8890\,\AA.

MEGASTAR library is an ambitious long-term project with the goal of having stellar spectra in as many as spectral configuration shown in Table \ref{mainchar}, although the priority is now to complete a wide enough library in the HR setups.  This is the second of a series of papers related to MEGARA-GTC library. In the first one, \citep[][hereafter Paper-I]{gvargas20},  the authors comprehensively described this MEGARA-GTC spectral library and the rationale behind the 2988 stars catalogue. It  was created from  libraries whose spectral resolutions were similar  to that of MEGARA at LR, MR  and HR covering a wide interval in $\rm T_{eff}$, $\rm \log{g}$ and abundance $\rm [M/H]$, generally as $\rm [Fe/H]$,  and that could be observed from the Observatorio del Roque de los Muchachos with geographical coordinates  28$^{\circ}$ 45$\arcmin$ 25$\arcsec$ N Latitude, 17$^{\circ}$ 53$\arcmin$ 
 33$\arcsec$ W Longitude.   

A discussion of the fundamental role of spectral libraries in Single Stellar Population (SSP) models and the advantages and constrains introduced by theoretical and empirical libraries in these models is presented in Paper-I. The main motivation of the MEGASTAR library  is to produce a spectral atlas to be used as input spectra for {\sc PopStar} models \citep[see  e.g.][]{mol09, mman10,gvargas13} to create the synthetic templates required to interpret the observations taken with the same instrument setup.  To delimit the goal we have concentrated  on HR-R  and HR-I spectral configuration  with R~$\simeq$~20000, centred in rest-frame at H~${\alpha}$  and the brightest line of \ion{Ca}{ii} triplet, respectively, as there are not any published theoretical or empirical catalogues with these resolution and spectral intervals. Moreover, such resolution  with the combination of efficiency and telescope collecting area has not been  offered in any other integral field  instrument.

In this paper we present the first release of MEGASTAR library formed by 838 spectra  obtained in the two high resolution spectral configurations: HR-R and HR-I. This is the result of 152.25 hours of {\it filler}-type observing time of 414 stars, covering  different spectral types, effective temperature, surface gravity and abundances. The complete atlas with fully reduced and calibrated spectra will be available to the community at the time of the acceptance of this publication (\url{https://www.fractal-es.com/megaragtc-stellarlibrary/private/home}).   In section \ref{sample}, we describe the main characteristic of the stars of this release. The observations and the  data reduction pipeline are summarised in sections \ref{observations} and \ref{megaradrp}, respectively. The database produced  to manage  the star catalogue, the observing proposals preparation, the resulting observations  and the release itself is presented in section \ref{database}. Some   examples of our spectra  are given in section \ref{library-stars} and in  section \ref{summary}  a  summary and final remarks are included. We present in Appendix~A an atlas with the 838 spectra, corresponding to the 414 stars of this first release. Appendix~B is the release summary table and Appendix~C contains a table with Gaia DR2 data for the 388 stars of the release for which GAIA data exist. These three appendices will be published in electronic version only.

\section{Sample} 
\label{sample}

The distribution of the stellar types, retrieved from SIMBAD astronomical database - CDS\footnote{\url{http://simbad.u-strasbg.fr/}} or from published papers in the literature is shown on the left panel of Fig.~\ref{theta-logg}. The red dashed-line region represents the parameters for the whole MEGASTAR library while the solid blue bars indicate the present data release (1.0), DR-1 stars. The dominant spectral types of this release are G (113) and B (107) followed by F (77), K (39), A (29), O (27), M (13), W (7), S (1) and Flat (1). On the right panel we present the $\rm \log{g}$ {\sl vs.} $\Theta$=5040/$\rm T_{eff}$ diagram over which we have plotted as cyan dots the values of the stars of the whole MEGARA library with estimated stellar parameters from the literature and the ones from DR-1 stars plotted as blue circles, green squares and red triangles according to the three metallicity ranges as indicated  in the plot. Finally, we have overprinted the Padova isochrones \citep[]{ber94, gir00, mar08} since these are the ones used in the evolutionary synthesis {\sc PopStar} models \citep[see ][]{mol09} which will be used in combination with MEGASTAR to provide a MEGARA-oriented set of spectra. The continuous update of this plot allows us to identify which areas of the physical parameter space must be completed to prioritize the future observations of the corresponding library stars, within the limitations of the {\it filler}-type program.

These stars cover the values of $\rm T_{eff}$, $\rm \log{g}$ and abundance $\rm [M/H]$  presented in Fig.~\ref{histograms}, where {\em Ntot} indicates the number of points used for each histogram, since for some stars one or more stellar parameters were not reported in the literature when we elaborated the MEGASTAR catalogue. One of the goals of our project is to elaborate a method to homogeneously derive the stellar parameters for all the stars in the library. In particular, in Paper-I  we proposed a technique to estimate  these parameters by best fitting theoretical models to the combined spectrum of HR-R and HR-I of 97 stars. We plan to apply this method to the 414 stars of this release in a forthcoming paper.

\begin{figure}
\centering
\includegraphics[trim=5cm 0cm 10cm 4cm, clip=true,width=0.95\hsize,angle=0]{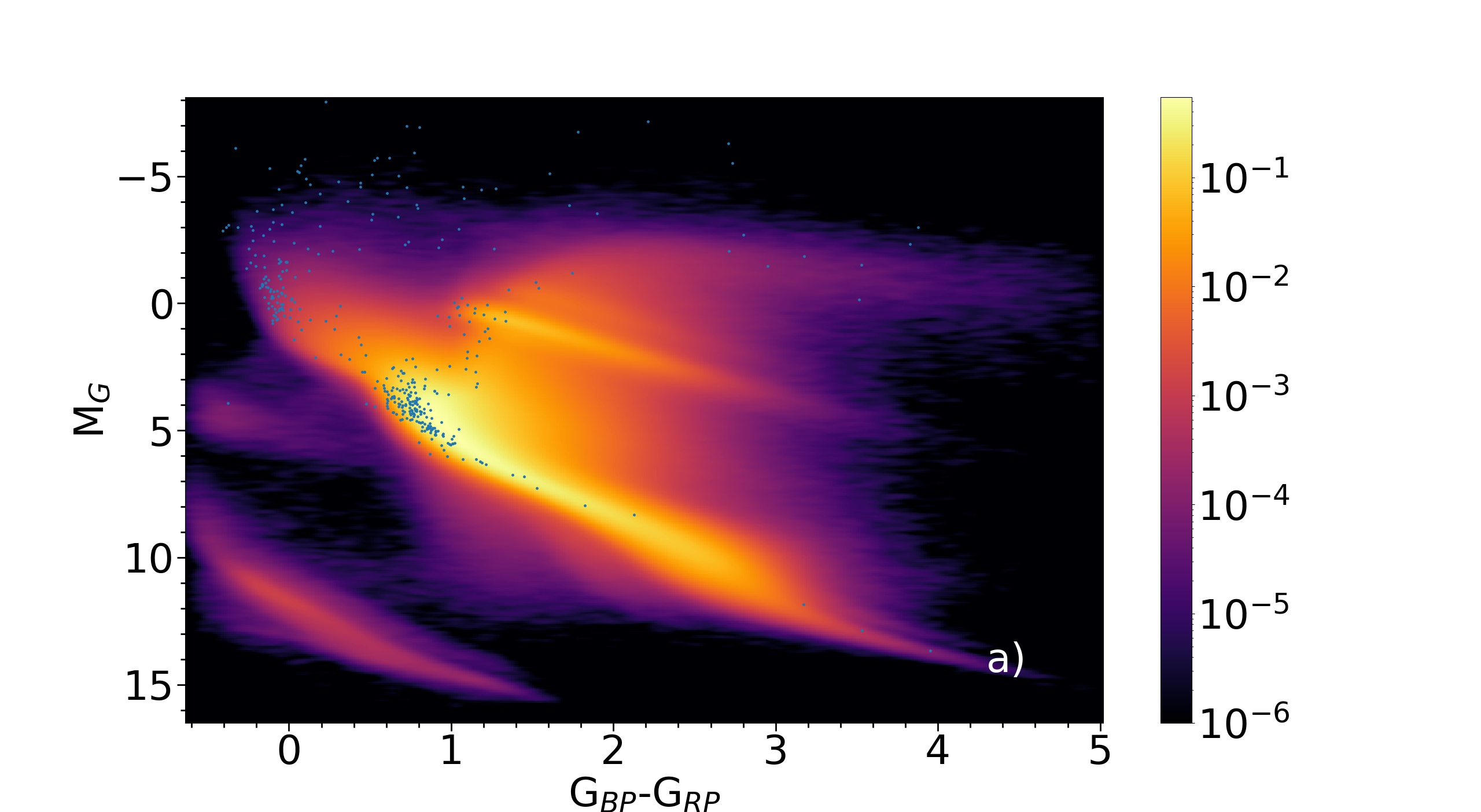}
\includegraphics[trim=5cm 0cm 10cm 4cm, clip=true, width=0.95\hsize,angle=0]{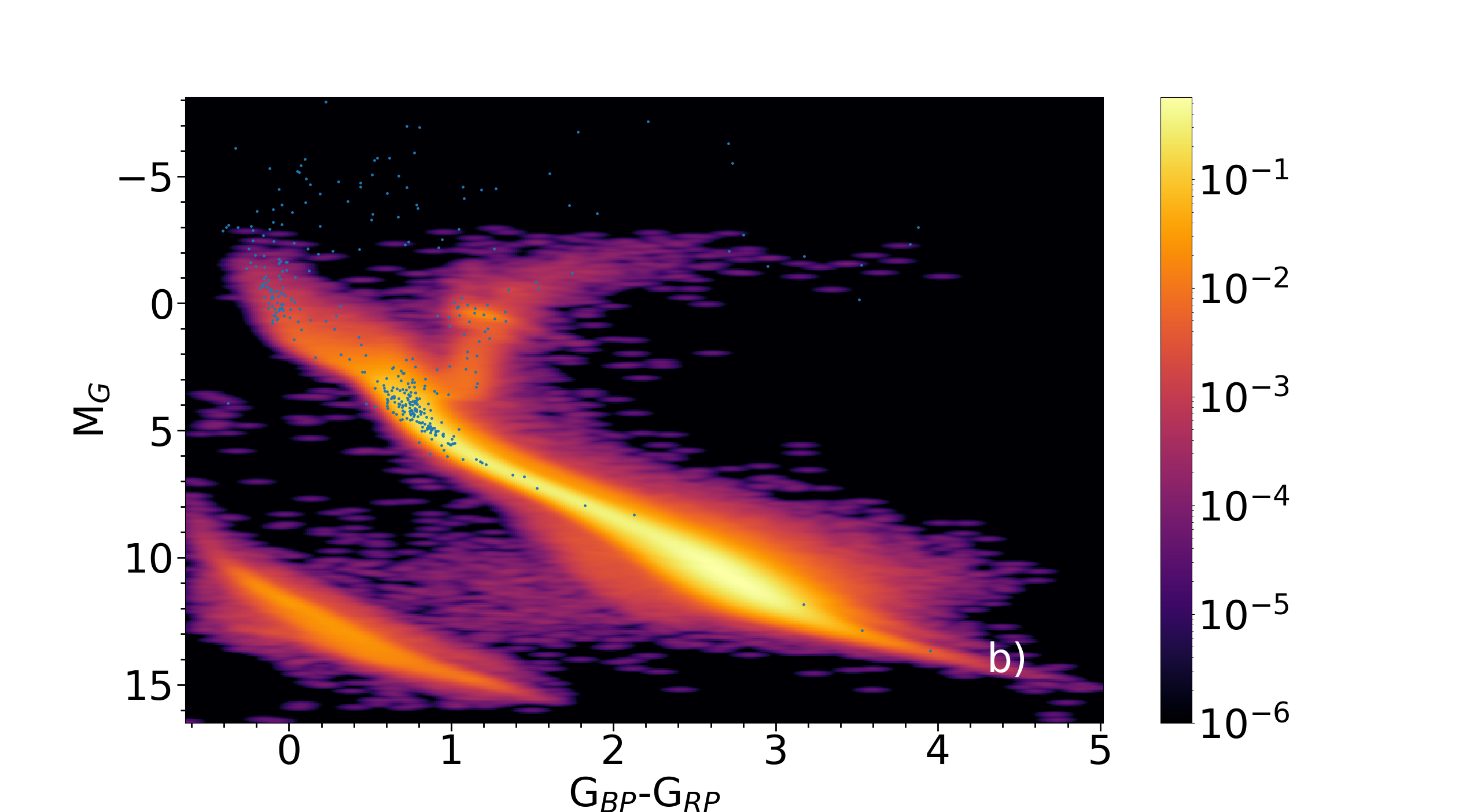}
\caption{a) HRD showing the density of 65,921,112 Gaia stars in DR2; b) HRD of Gaia DR2 stars with identical maximum parallax and photometry
uncertainties as panel (a) but selecting only those 1,322,033 Solar Neighborhood stars. The 388 MEGARA-GTC spectral library with Gaia DR2 measurements are shown as light-blue filled circles.}
\label{gaia}
\end{figure}

To juxtapose our sample with Gaia\footnote{\url{https://gea.esac.esa.int/archive/documentation/GDR/}} data, we found DR2 measurements for 388 MEGARA library stars. On the top panel of Fig.~\ref{gaia} we present the Hertzsprung-Russell Diagram (HRD) showing the density of 65,921,112 Gaia stars in DR2 with good quality parallax and photometry measurements, i.e.  $\Delta\varpi/\varpi$ $<$ 0.1, $\sigma_G$ $<$ 0.022\,mag and $\sigma_{G_{XP}}$ $<$ 0.054\,mag  \citep[see Figure~1 in][ from the Gaia collaboration]{Babusiaux18}. A Gaussian kernel-density estimate  has been applied to the stellar density map. The units in this map are normalised to the maximum star density and the color map is shown in logarithmic scale; a lower threshold of 10$^{-6}$ for the  stellar density has been set.  On the bottom  panel of the same figure we show the HRD of Gaia DR2 stars with identical maximum parallax and photometry uncertainties as panel (a) but selecting only those 1,322,033 Solar Neighborhood stars with parallaxes $\varpi >$ 5\,mas, that is, heliocentric distances below 200\,pc.  The MEGARA library stars with Gaia DR2 measurements are shown as light-blue filled circles. There is a  lack of late-type stars  due to  the low fraction of main-sequence K and M stars in this release sample, around 13\%, specially those with reliable Gaia DR2 measurements. Note that the Gaia DR2 saturation limit is in G$=$3 but at G$<$6 the quality of the astrometry starts to worsen \citep{Lindegren18}.  
In Appendix B we include an extract of the Gaia DR2  measurements for  388  out of  414 stars of MEGARA release 1.0. The details are described in section \ref{Gaia-stars}.  

\section{Observations}
\label{observations}

The observations of the library are in progress and has been awarded GTC open time in four consecutive semesters. The observations included in this release 1.0 of MEGASTAR are from programs GTC22/18B, GTC37/19A and GTC33/19B, with a total of 152.24\,h of observed {\it filler}-type GTC Open Time. Table~\ref{observing time} shows the time requested, awarded and observed through the three semesters whose observations are included in this release. The stars observed were 414 but five of them were observed twice, therefore we are reporting 419 observations carried out in the HR-R and HR-I spectral configurations, producing  a total of 838 spectra. Additionally, the project was awarded another 75\,h in 2020A. 

\begin{figure}
\includegraphics[width=0.35\textwidth,angle=-90]{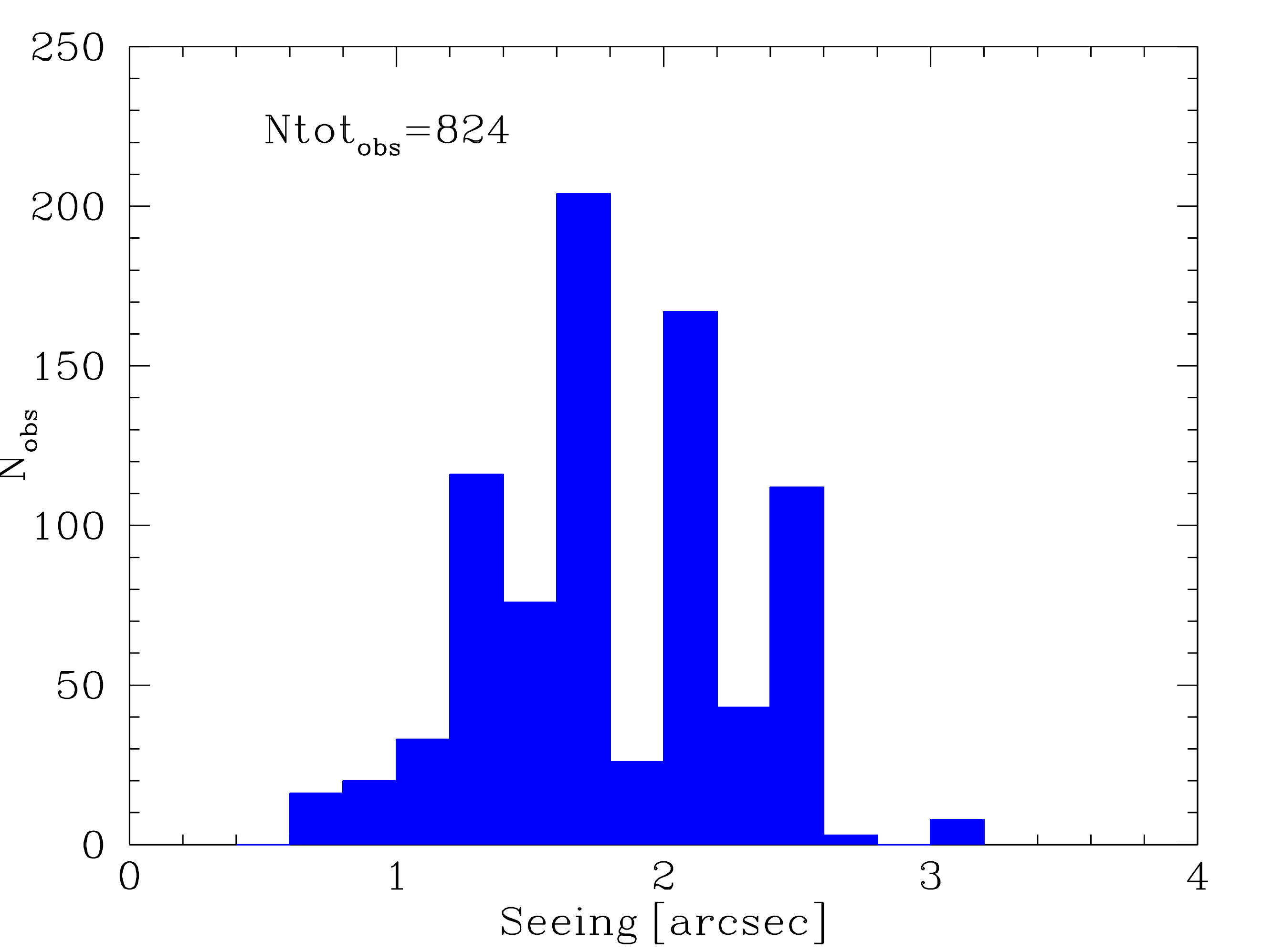}
\caption{Histogram of the seeing conditions during the observations of the MEGASTAR first release. Ntot$_{obs}$ is the number of observations with a seeing measurement reported. The values of the first quartile, the median and the third quartile of the seeing distribution are 1.5, 1.8 and 2.0$\arcsec$ respectively.}
\label{seeing}
\end{figure}

MEGASTAR library observations have been carried out as a {\it filler} type  program that, according to GTC rules, must accept the following values: seeing generally larger than 1.5$\arcsec$, any night-type, especially bright and any kind of sky quality, in particular spectroscopic (non-photometric) time,  being the possibility of having bad seeing the most important criterion. The seeing distribution, as reported in GTC log files, is shown in Fig.~\ref{seeing} with a first quartile,  median  and third quartile values of 1, 1.8 and 2$\arcsec$, respectively. Nevertheless, the  great advantage of creating a stellar library with an IFU instrument is that the spectral resolution is conserved as the slit width is constant as far as the stop is at the microlens + fiber. In fact, to assure that the flux was recovered in all the cases, we added always the individual spectra from 37 spaxels centered in the highest flux fiber in the reconstructed LCB images. Most stars were observed in bright conditions. The telescope delivers a standard star per Observing Block (OB) for flux calibration and instrument response correction.

Thanks to the optimization of the observational strategy, the priority given to bright stars, and overall, the maneuver and GTC overhead time saving to observe HR-R and HR-I in the same OB, in the three already fully observed semesters, the charged time per OB ranges from 850\,s to 1600\,s, for  V magnitude stars brighter than 12.5, with an average value of 1100\,s/star, with the two setups, so that we have increased the originally expected efficiency by a factor of 1.6. 

The  limiting V magnitude of the observed stars  is  12.4.  Up to now we have observed 80 stars with Teff $>$ 20000\,K, being 7 of them WR stars. This will allow us to produce an initial version of our R = 20000 SSP models for young population.  The targets proposed for the complete library are available through MEGARA-GTC-library database described in section~\ref{database}. The database supports the working team for preparing and uploading the OBs to the GTC-Phase 2 tool. To prepare a new OB set, we search unobserved stars in the MEGARA-GTC library filtered by a certain magnitude range in both R and I bands and/or by spectral type, or any stellar parameter, considering that all stars within that group, and for a given setup, will have a similar Signal to Noise (S/N) ratio  when choosing the appropriate exposure time in each setup. The exposure times were estimated using the  MEGARA Exposure Time  calculator (ETC) tool\footnote{\url{http://gtc-phase2.gtc.iac.es/mect/etc/form}} to obtain S/N ratio values between 20 and 300.  When the exposure time resulting from the ETC was longer than 30\,s, we  preferred to divide  it in three exposures to be able to calculate the median of the three images  for  eliminating the  cosmic rays.  Calibration images of halogen and ThNe lamps  were obtained during day time. 

\begin{table}
\caption{MEGARA-GTC spectral library observing time.}
\label{observing time}
\begin{tabular}{cccrr}
\hline
Semester	 &	Requested	&	Granted	&	Observed & Stars \\
 & h & h & h & observed \\
\hline
2018B	&	50	&	50	&	63.85  & 176  \\
2019A	&	50	&	50	&	11.66  &  32  \\
2019B	&	75	&	75	&	76.73  & 206  \\
 \hline
Total	&	175	&	175	&	152.24 & 414  \\
\hline 
\end{tabular}
\end{table}

\section{Data Reduction}
\label{megaradrp}

The data reduction procedure was carried out using MEGARA Data Reduction Pipeline (DRP)\footnote{\url{ https://github.com/guaix-ucm/megaradrp}},  publicly available and open source under GPLv3+ (GNU Public License, version 3 or later). It is a custom-made user-friendly tool formed by a set of processing recipes developed in Python \citep{carandpas18, pasetal18, pasetal19}.  The recipes used for obtaining all the calibration images were  {\it MegaraBiasImage}, {\it MegaraTraceMap}, {\it MegaraModelMap}, {\it MegaraArcCalibration},  {\it MegaraFiberFlatImage} and  {\it MegaraLcbStdStar}.  Each recipe generates a product: a master bias image for bias subtraction; a trace map for fiber tracing; a model map for spectra extraction - as a result of a simultaneous fit to all the fiber profiles; a wavelength map for wavelength calibration; a fiber flat field  map for pixel to pixel relative sensitivity  and a  master sensitivity function for flux calibration and instrument  response correction. Each product has to be copied to a specific directory,  previously defined by a calibration file tree structure, \citep[see][]{casetal18}.  
Additionally, each routine generates a quality control file allowing a full tracking of the process. Halogen and ThNe lamps exposures are required for tracing flat field and wavelength calibration. MEGARA DRP uses a general configuration file  with the information necessary for data reduction like  the data directories, the polynomial degree and number of spectral lines required for wavelength calibration and the observatory  extinction curve, among others. It operates in a sequential mode, i.e. each recipe requires products generated in the previous ones. An individual routine has its own input \textit{.yaml} file that includes the images to be processed  and the specific parameters required  for that particular recipe. For example, when the temperature of the halogen lamp observations differs more than one degree from that of the source a global offset must be applied  for fiber tracing, what is done by given a value to the {\em extraction\_offset} parameter.

\begin{figure}
\includegraphics[width=0.45\textwidth,angle=0]{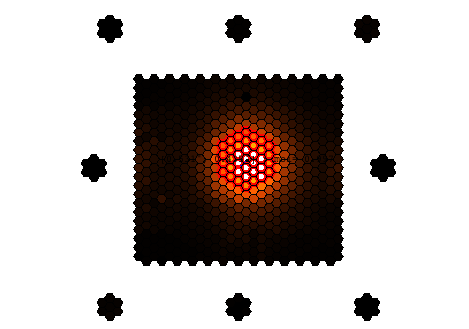}
\includegraphics[width=0.99\hsize,angle=0]{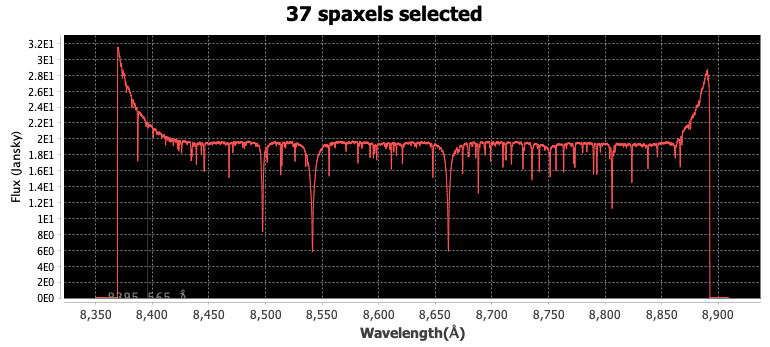}
\caption {Layout of the  QLA tool. Top: reconstructed image of the  star HD048682 on the LCB; the 8 minibundles in the external part, collapsed for visualization, are used for sky subtraction. 
Bottom: final HR-I spectrum corresponding to the 37 selected spaxels marked in red in the LCB. The seeing during this observation was reported to be 2.5\arcsec.}
 \label{QLA}
\end{figure}

\begin{figure}
\includegraphics[width=0.45\textwidth,angle=0]{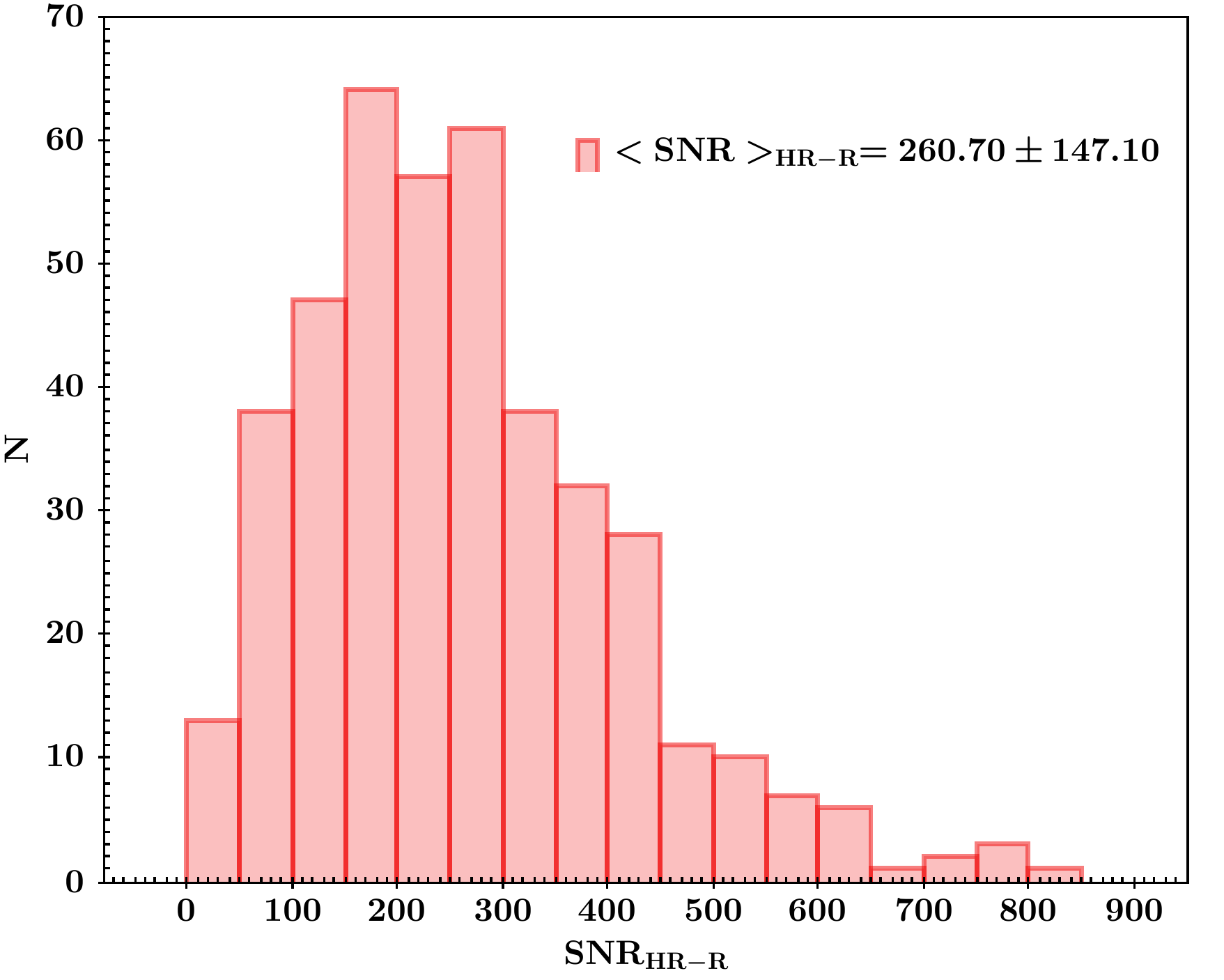}
\includegraphics[width=0.45\textwidth,angle=0]{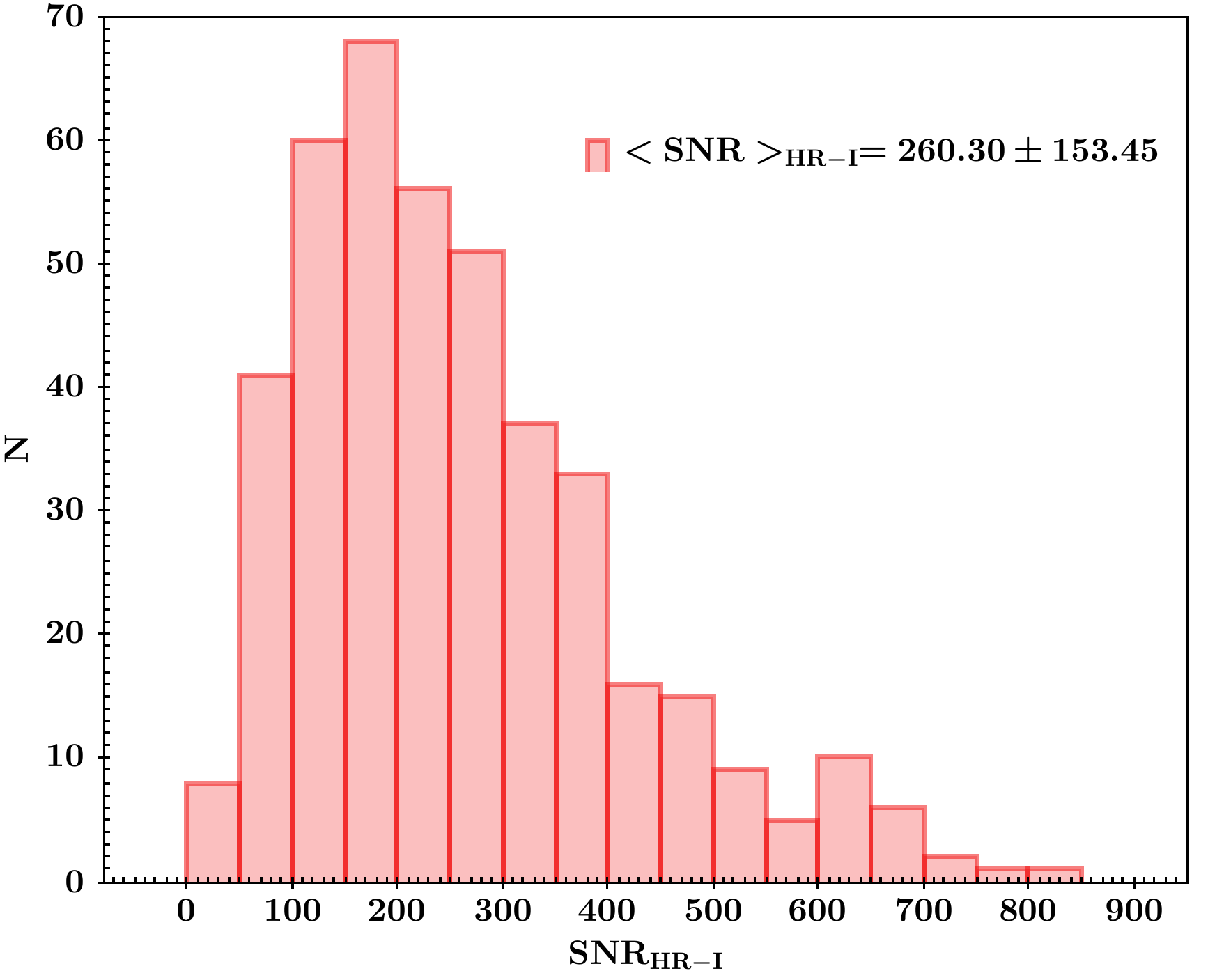}
\caption{Distribution of the  S/N ratio obtained according  to  \citet{stoehr08}  for the HR-R (top) and HR-I (bottom) spectra of MEGASTAR release 1.0.} 
\label{SNR}
\end{figure}

\begin{figure}
\includegraphics[width=0.52\textwidth,angle=0]{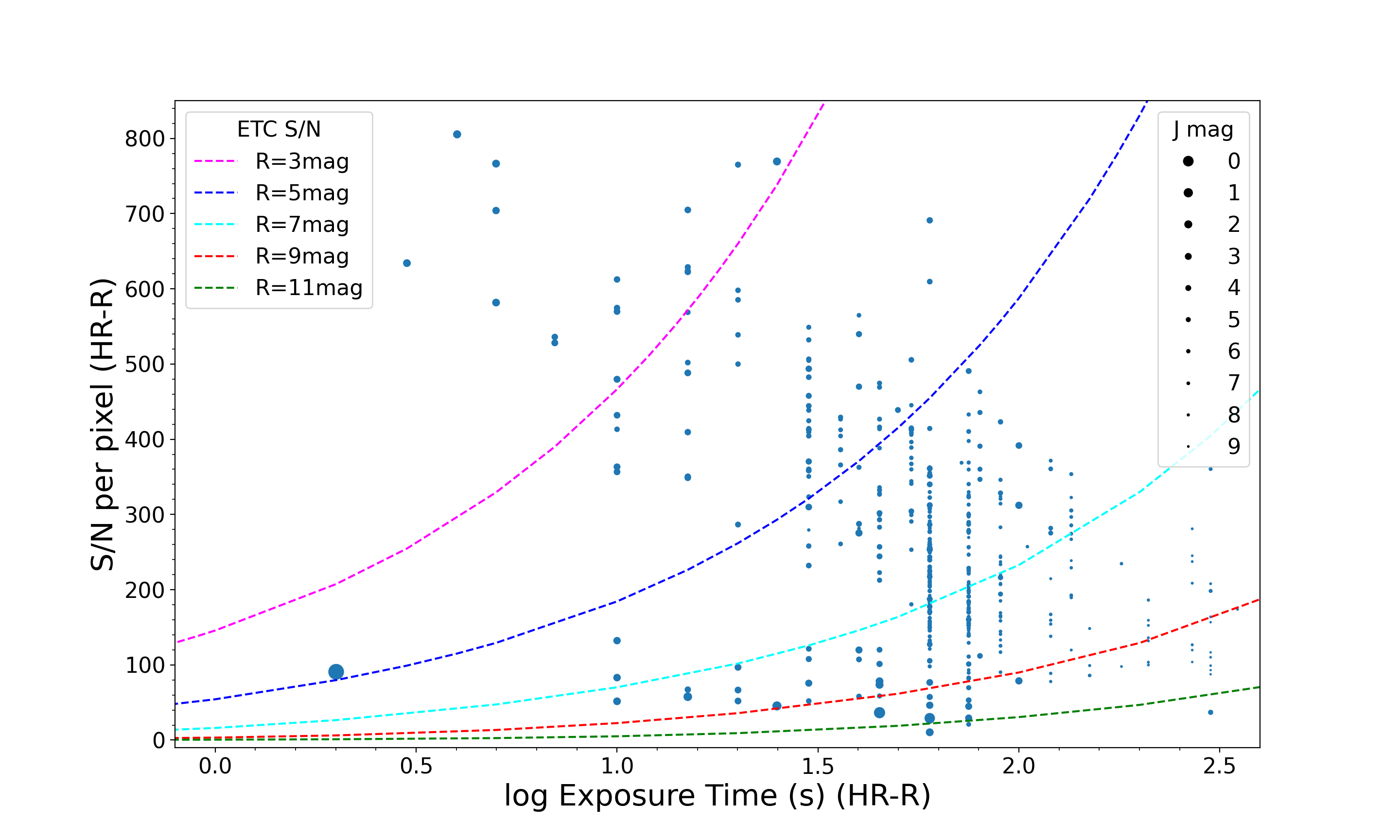}
\includegraphics[width=0.52\textwidth,angle=0]{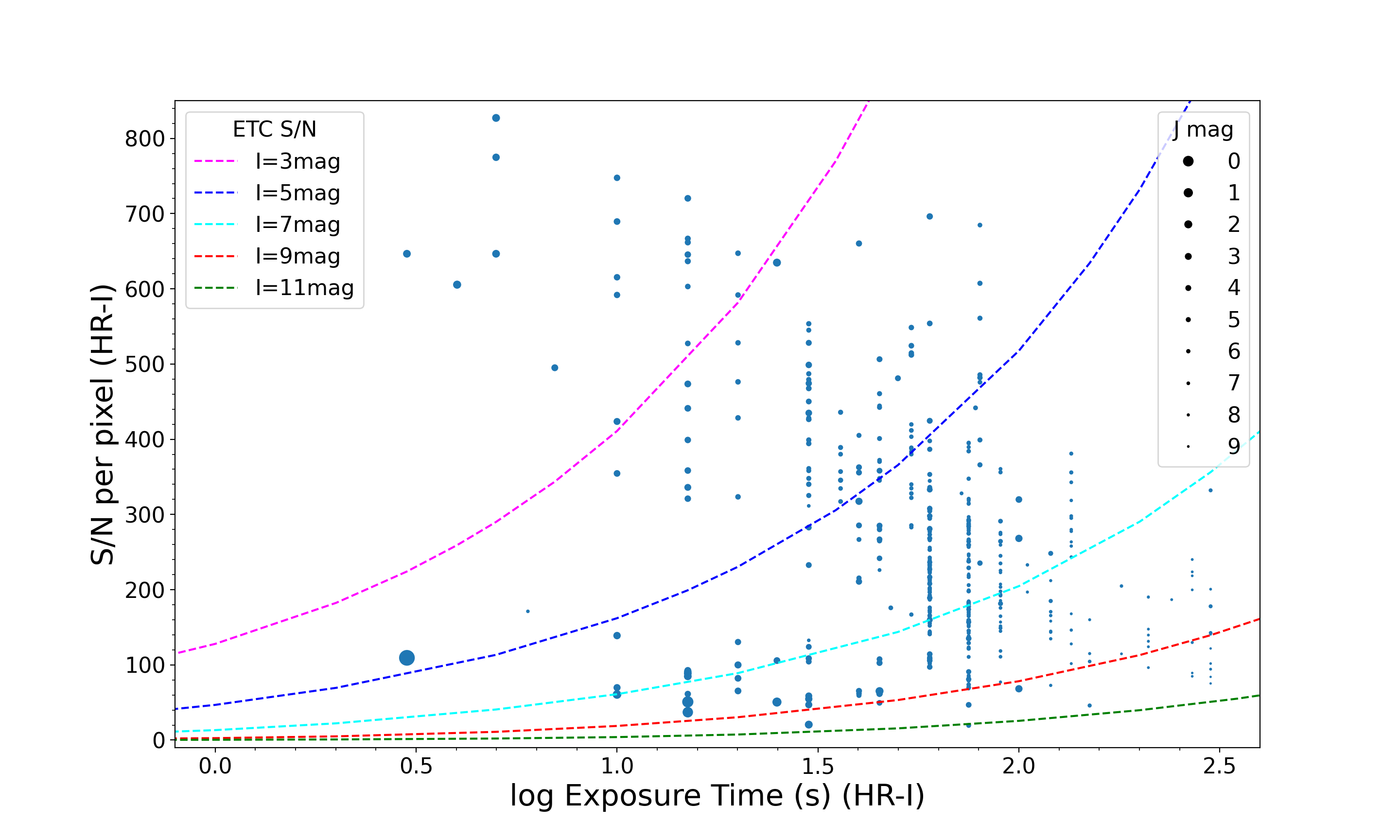}
\caption {S/N ratio  as  a function of the total exposure time. The size of each circle is proportional to the apparent brightness of the star in the $J$-band, for the HR-R (top) and the HR-I (botom) spectra.   As a reference, we include the S/N ratio  predictions of the MEGARA ETC  for different $R$-band  and $I$-band  magnitudes as indicated,  considering average  observing conditions.   More details in section \ref{megaradrp}.}
\label{SNR-mag}
\end{figure}

\begin{figure*}
\centering
\includegraphics[width=1\textwidth,  angle=0]{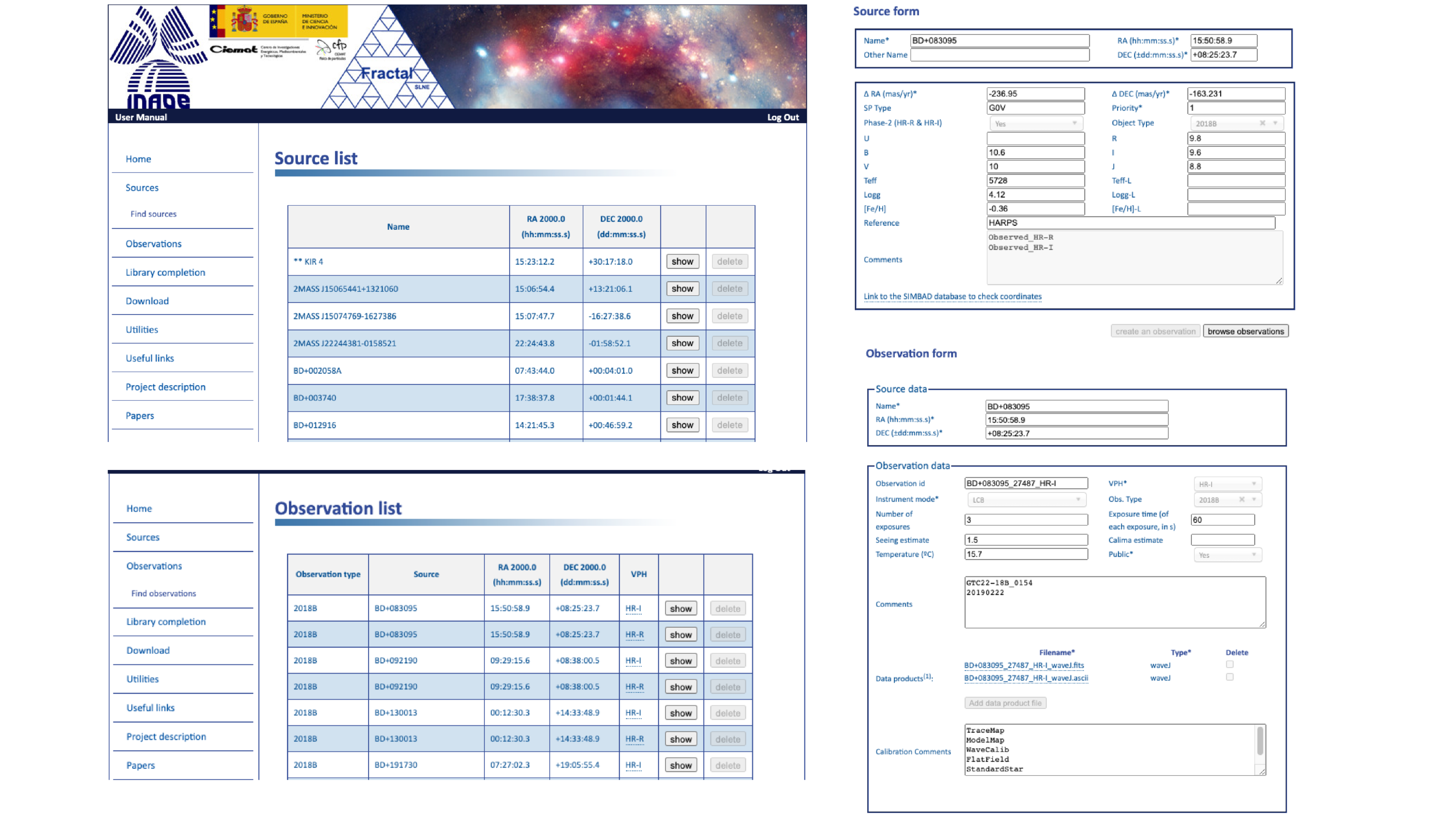}
\caption{MEGARA-GTC library database  user interface. Left: {\em Source list} and {\em Observation list}  options.  Right:  {\em Source form} for BD+083095 with the complete information of the star and  {\em Observation form} with the details of  HR-I observation of the same star.}
\label{source-observation}
\end{figure*}

The last step is to apply the recipe {\it MegaraLcbImage} to  the  star image. It produces a row-stacked-spectra file with the individual fibre spectra corrected for atmospheric extinction, instrument response and flux calibrated. The sky subtraction is carried out automatically by calculating the median of the signal of the 8 sky-minibundles located outside the LCB FoV. The top panel of Fig.~\ref{QLA} shows the reconstructed image of the star HD048682, generated by the Quick Look Analysis (QLA) tool \citep{gomezalv18}. The rectangular shape is the reconstruction of the LCB such as the dimension are equivalent to
$12.5\,\arcsec \times 11.3\,\arcsec$ and the 8 sky minibundles projections are shown in the external part of the LCB. The view of these sky minibundles have been collapsed for visualization purpose but  are located at  between 1.75\,\arcmin\ to 2.5\,\arcmin\ from the center of the IFU.  To extract the 1D-spectrum we integrated three rings formed by 37 spaxels, shown in red, each equivalent to 0.62\,\arcsec\ on the sky. In the bottom panel  the final  HR-I spectrum corresponding to these 37 spaxels,  after flux calibration and instrument response correction, is displayed.  All the spectra were extracted in the same wavelength range: 6420 – 6790\,\AA\ in HR-R and 8370 – 8885\,\AA\ in HR-I. The flux is given in Jy.

We note that at the extreme wavelengths of the spectrum, shown in Fig.~\ref{QLA}, the flux tends to rise.  This effect, present in most of  the spectra, is  due to the coarse reciprocal dispersion and spectral resolution of the tabulated spectra of most spectrophotometric standard stars. During the determination of the system sensitivity function we had to degrade the MEGARA HR-R and HR-I observations of our standard stars. This coarse resolution implies that the tabulated fluxes at the edges of the wavelength range covered by these VPHs are affected by fluxes that correspond to wavelengths that are not explored by our data. Therefore, we ought to degrade the MEGARA observations of these standards using (null) fluxes that would come from wavelengths beyond each specific observed interval. This leads the degraded reference spectrum and the corresponding master sensitivity curve to drop faster at these extreme wavelengths, $\sim$50\,{\AA} at each size, than the actual system sensitivity. As a consequence of this edge effect, the flux-calibrated spectra tend to rise at the very extremes of each specific wavelength range.

In order to obtain a first-order correction to this effect  we have divided every spectrum by a normalised continuum obtained from the fitting to the rather flat HR-R and HR-I spectra of BD+083095. The corresponding fits were performed following the same approach described in Paper-I. Fig.~\ref{spectra-jpg} shows the result for BD+083095.  All the stars of the release have been corrected with this normalised continuum. It is important to remark that this is not producing spectra normalised to the continuum. In Paper-III a specific continuum fitting for each star will be carry out as part of the procedure for  determining  the physical stellar parameters. Accordingly, in the next release we will also include all spectra normalised.

To assess the quality of the spectra, we report in Fig.~\ref{SNR} the distribution of the measured continuum S/N ratio averaged over the whole spectrum, following the recipe  by \citet[]{stoehr08}, for the HR-R and HR-I spectral configurations in the top and bottom panel, respectively. In both cases the average S/N ratio is around 260. The plots are similar since most of the star observations included the two spectroscopy setups in the same OB to be executed in sequence. The  S/N ratio distribution spreading  over a wide range of values is a consequence of the nature of the GTC {\it filler}-type program itself that does not guarantee any specific observing conditions. On the contrary, these observations are conducted  aleatory and out of the standard scheduling, normally done whenever other programmed observations cannot be executed. This is due mostly to bad observing conditions but filler observations can be carried out just to complete the nighttime or during twilight, and sometimes the conditions can be good.

Additionally, any single observation has to be executed with the exposure time pre-defined in the GTC Phase-2 tool. In other standard programs the observing conditions are guaranteed and the exposure time is estimated accordingly. In MEGASTAR observations, the integration time has to be pre-defined no matter   the observing conditions, increasing the  uncertainty to the observation quality. Our decision of extracting  37  spaxels  for all spectra, regardless the seeing value, guarantees that all the flux is integrated, and generally improves the S/N ratio for bad seeing observations, but might degrade it  for very good seeing conditions, since in these cases we are adding mostly noise when extracting all the 37 spaxels.

Fig.~\ref{SNR-mag} shows now the S/N ratio  as a  function of the total exposure time for  the HR-R and HR-I spectra in the top and bottom panel, respectively.   The size of each filled circle is proportional to the apparent brightness of the star in the $J$-band as  this magnitude  is the one available for the vast majority (all but four) of our stars while only 290 (284) stars have $R$-band ($I$-band) magnitudes published.    We also include in this plot, as a reference, the predictions of the MEGARA   ETC of the  S/N ratio per spectral pixel at the central wavelength of the VPH,   for different $R$-band magnitudes  in  the case of HR-R  and $I$-band magnitudes for  HR-I, assuming spectroscopic conditions, full moon, airmass of 1.15, a 1.5\,\arcsec\ seeing and a flat input spectrum, after adding up only 19 spaxels centered on the source, considering that in these ETC simulations the seeing is relatively  good.  Note that all magnitudes used here are in the Vega system.

\section{MEGARA-GTC Library Database}
\label{database}

Our goal in this project has been twofold, to assemble input spectra for {\sc PopStar} models for the interpretation of the stellar populations in a broad range of observations with MEGARA, and to generate a public database of reduced and calibrated spectra for others MEGARA users. We developed a database in MySQL and a web-based tool to manage the stellar data and the observed spectra that will be available to the community as part of this MEGASTAR release 1.0. Via a scheme of  permission levels, different actions are allowed. The public user will be able to retrieve the information compiled of each star and observations and to download the individual reduced spectrum or the full release as is described in this section. 
 
The database resides on \url{https://www.fractal-es.com/megaragtc-stellarlibrary/} and offers different menus: {\em Source}, {\em Observations}, {\em Library completion}, {\em Download}, {\em Utilities}, {\em Project description} and {\em Papers}. The left panel of Fig.~\ref{source-observation} shows the menus available. The {\em Source} functionality allows the user to list the sources of the complete library; to search sources when filtering by different parameters such as source name, RA, DEC and spectral type, among others,  and to browse observations of a source.  
The filtering  parameters are those created in the {\em Source form} where the complete star information resides. In the top right panel of Fig. ~\ref{source-observation} the layout of the  {\em Source form} is shown for the  star BD+083095: its RA and DEC coordinates, the corresponding  $\Delta$ RA and $\Delta$ DEC proper motions, the spectral type and luminosity class and the U, B, V, R, I \& J Johnson-Cousins  magnitudes retrieved from SIMBAD database, are displayed. This star was observed in the GTC observing semester 2018B. In this example, the stellar parameters $\rm T_{eff}$, $\rm \log{g}$ and $\rm [Fe/H]$ were extracted from the HARPS catalogue. In the  $\rm T_{eff}-L$ and  $\rm \log{g}-L$ and  $\rm [Fe/H]-L$  windows the ``L" letter  stands for our own library determination of  the stellar physical parameters that will be provide in the near future. The {\em Comments} section shows that the star was observed in both HR-R and HR-I setups. In {\em Other comments} is the reference to the HARPS project publication from which the star and the values of the physical parameters $\rm T_{eff}$, $\rm \log{g}$ and $\rm [Fe/H]$ were compiled. 
  
\begin{figure}
\centering
\includegraphics[width=0.49\textwidth,angle=0]{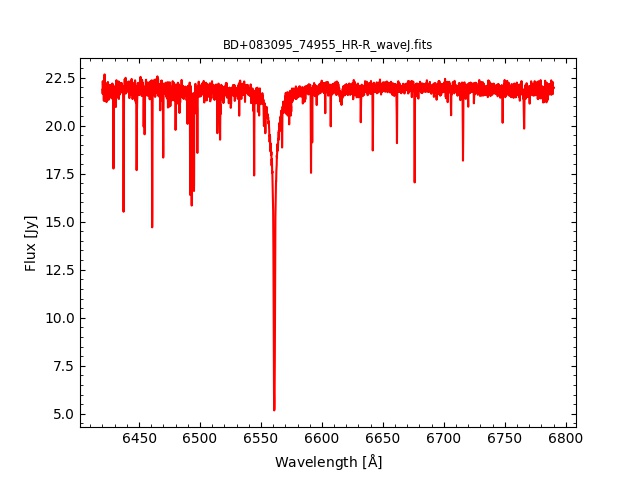}
\includegraphics[width=0.49\textwidth,angle=0]{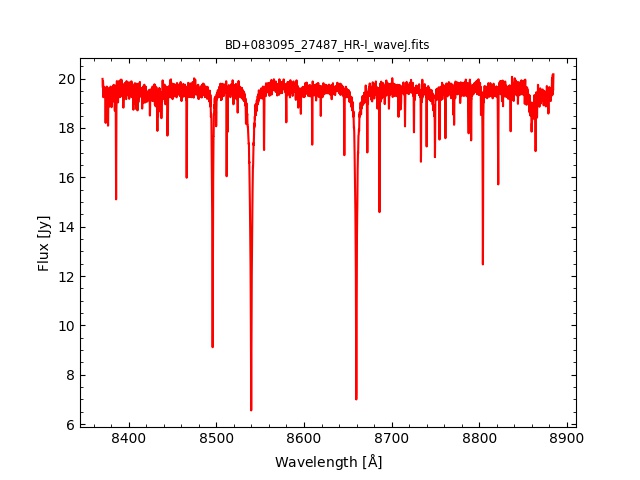}
\caption{Example of the final spectra in JPG of the Star BD+083095 obtained in  HR-R (top) and HR-I (bottom)  setups.  All the spectra of the release are provided in JPG, FITS and ASCII.}
\label{spectra-jpg}
\end{figure}

The {\em Observations} menu allows listing and searching observations filtering them by source name, RA, DEC, VPH and instrument mode, among other parameters, as seen in the bottom-left panel of Fig.~\ref{source-observation}. An {\em Observation  form} is created  for  each  spectral setup observation, as shown in the bottom right panel of the same figure for a HR-I observation. The first window has the name and coordinates of the star retrieved from the {\em Source form}. In the {\em Observation data} window the following information is provided: the BD+083095\_27487\_HR-I id is a unique code used internally to identify the observation; the star was observed in the HR-I setup with 3 exposures of 60\,s\ each, under 1.5\,\arcsec\ seeing, while the temperature at MEGARA spectrograph bench was 15.7\,$^{\circ}$C; the HR-I spectrum is public. In the {\em Comments window} GTC22-18B\_0154 indicates that MEGARA library observing program has been assigned the code GTC22, the OB number is 0154 and the observation was carried out on February 22, 2019. The {\em Data Products} available are the  HR-I spectra in two formats being  BD+083095\_27487\_HR-I\_waveJ.fits and BD+083095\_27487\_HR-I\_waveJ.ascii their filenames,  the  "J" letter in the filenames indicates that the flux calibrated spectra are provided in Jy. The {\em Calibration comments} window shows the DRP recipes used in the data reduction process described in section \ref{megaradrp}. In summary, that the spectrum was obtained after bias subtraction, fiber tracing, wavelength calibration, flat-field, extraction, extinction and spectral response correction and flux calibration. The individual spectrum can be visualised by clicking the HR-I or HR-R buttons in the {\em Observation list} menu. In this case, the spectra that would be displayed for the star BD+083095 are shown in 
Fig.~\ref{spectra-jpg}. All the lists created in the {\em Source} and {\em Observations} menus can be exported to a pdf, MSExcel or ASCII files.

In the {\em Library completion}  menu the {\em Full library} button paints a graph of all the stars of the library  grouped by  three ranges of metallicity: $\rm [Fe/H]$ $< $ $-$0.7, $-$0.7 $\leq$ $\rm [Fe/H]$ $\leq$ $-$0.2 and $\rm [Fe/H]$ $>$ $-$0.2 while in the {\em Observed stars} button paints a plot with all the stars observed in the same defined $\rm [Fe/H]$ intervals. Furthermore, in case the user wishes to upload her/his own stars, she/he can choose to overlay the graph with one of these two graphics: full library or observed stars. In the {\em Utilities} menu the user can download the {\em Spectra Plot} application to plot the data product files. The {\em Useful links} menu includes those to ING Object Visibility-STARALT, to SIMBAD Astronomical Database and to MEGARA Exposure Time Calculator. 

The {\em Download} menu leads to the link {\em Click here to download the lastest release}. The current release 1.0 presented in this paper contains observed and calibrated spectra from the MEGASTAR Library stars after a basic quality control check. All spectra were reduced with MEGARA DRP passing through all standard reduction steps described in section \ref{megaradrp}. The release download includes: i) a README.txt file with the description and the content of the release; ii) a list of the stars  with the description of the observations (release\_summary\_1.0) in MSExcel and ASCII formats; iii) a main directory and three sub-directories  with the release spectra in  ASCII, FITS and JPG formats and iv) a Java application for the display of the observations.
 
The stars information in the release files is mainly that displayed in the {\em Sources} and {\em Observations} forms. The description of the column headers of  the release\_summary\_1.0 (.xls  and .txt) files,  shown in Table~\ref{release-header}, is the following: {\em Name}, main star name; {\em Right Ascension (RA)} and {\em Declination (DEC)} in equatorial coordinates J2000.0, {\em Spectral type and luminosity class}, referenced  available Johnson-Cousins magnitudes {\em U, B, V, R, I} and {\em J}, all these parameters obtained from SIMBAD database,  updated to July 8, 2020. The {\em Other name}  column is for  an alternative  name of the star. The next one shows the {\em VPH} grating used in the observation; the values for  effective temperature, surface gravity (log) and iron abundance (log) $\rm T_{eff}$, $\rm \log{g}$ and $\rm [Fe/H]$ are displayed in the next three columns, respectively, for those stars whose values have been extracted from other libraries; under the {\em Reference} column is the name of the  original catalogue from which the star was selected; {\em Other comments} are those related  to the star. The  next two columns indicate the name of the {\em ASCII/FITS} spectrum file provided by the release and the GTC observing semester {\em Obs. Period}. The observation parameters of each star are  presented in the next three columns: the number of exposures {\em No. Exp.} and  the exposure time {\em Exp. Time} (s) from the images headers and the {\em Seeing} ($\arcsec$) as reported by GTC in the observation log file. The last one, {\em Obs-GTC}, displays comments related to the observations including the OB number and the observation date. The  release\_summary\_1.0 files (.xls and .txt) are included in Appendix~B.
 
The main directory observations\_release\_1.0 has the ASCII, FITS and JPG  sub-directories with the release spectra in the corresponding formats and three files with the  spectra lists. An example of the JPG  spectra is shown in Fig.~\ref{spectra-jpg} for the star BD+083095 in HR-R and HR-I; the header of the  plot includes the  spectrum filename. The Java application spectraplot.jar allows further visualization and basic analysis of  the  ASCII or FITS  spectra. The Java plots includes the star name, the spectrograph setup, the spectral type, and the values of $\rm T_{eff}$, $\rm \log{g}$ and $\rm [Fe/H]$ from literature whenever available.

\begin{table}
\caption{Columns description of the release\_summary\_1.0 file, included in Appendix~B, available in electronic format only.}
\label{release-header}
\begin{tabular}{ll} 
\hline
Column  & Description\\
\hline
Name 	&	Star name (*) \\			
RA	&	Right Ascension (2000.0) (hh:mm:ss.s) (*) \\
DEC	&	Declination (2000.0) (dd:dd:ss.s) (*) \\		
Sp Type  	&	 Spectral Type  (*) \\				
U	&	Jonhson-Cousins U magnitude (*) \\		
B   &	Jonhson-Cousins B magnitude (*) \\			
V   &	Jonhson-Cousins V magnitude (*) \\			
R   &	Jonhson-Cousins R magnitude (*) \\			
I	&	Jonhson-Cousins I magnitude (*)\\			
J	&	Jonhson-Cousins J magnitude (*) \\			
Other name	&	Alternative name of the star \\		
VPH &	 Grating of the observed spectrum	\\
Teff	&	 Effective temperature from literature \\
Logg	&	Surface gravity (log) from literature \\
$[$Fe/H$]$	&	Iron abundance (log) from Literature \\
Reference	& Original catalogue from which it was inherited\\	
Other Comments & Comments related to the star	\\		
ASCII/FITS file &	Name of the ASCII/FITS spectrum file\\	
Obs. Period &	 GTC Observing semester \\			
No. Exp  	&	 Number of exposures	\\				
Exp. Time   &	Time  of  the individual exposures (s)\\
Seeing 	&	Value of the seeing   as reported by GTC (\arcsec)	\\
Obs-GTC 	&	 Comments related to the observations\\	
\hline
(*) Source: SIMBAD
\end{tabular}
\end {table}

\begin{figure*}
\includegraphics[width=0.18\textwidth,angle=-90]{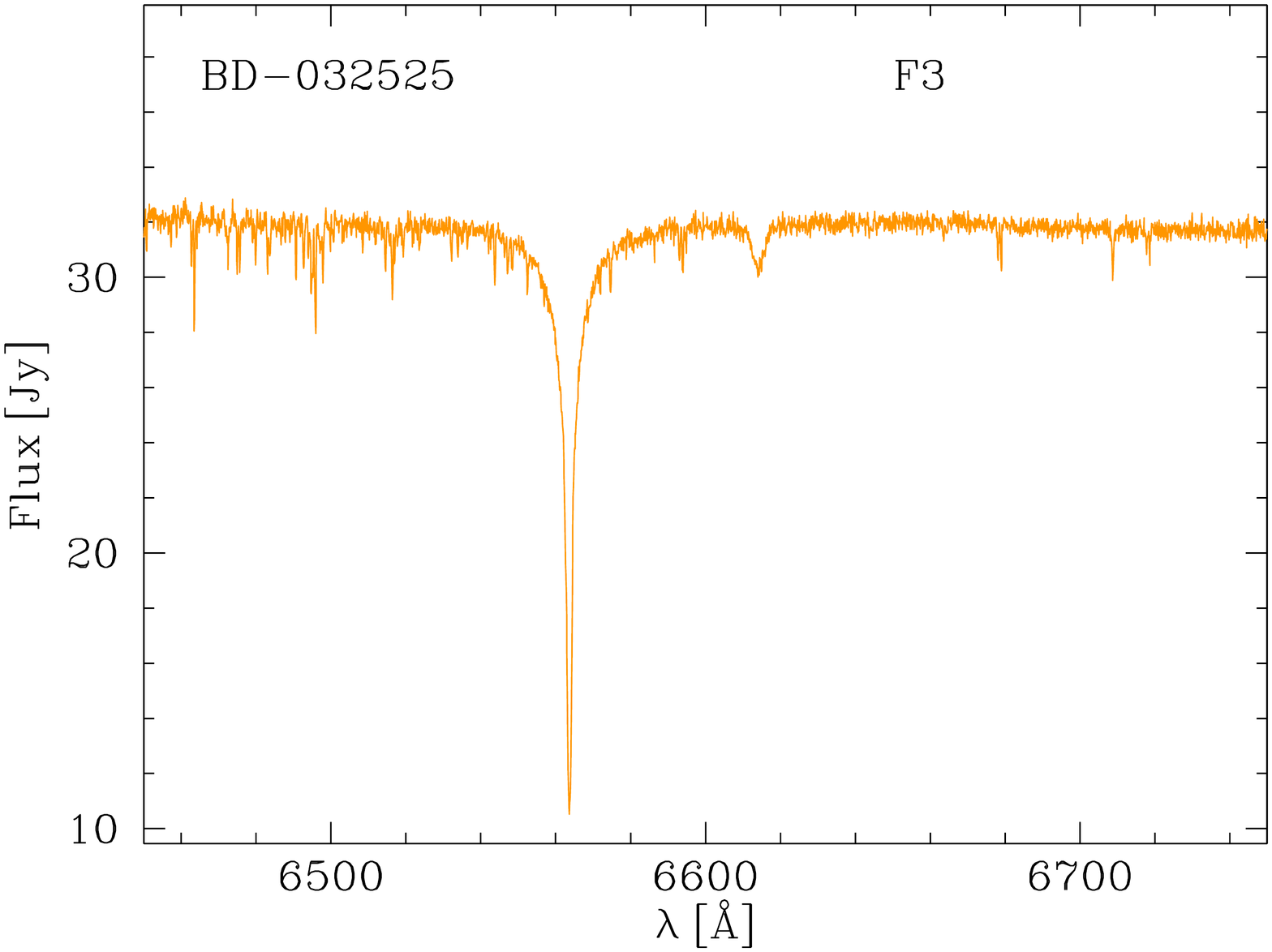}
\includegraphics[width=0.18\textwidth,angle=-90]{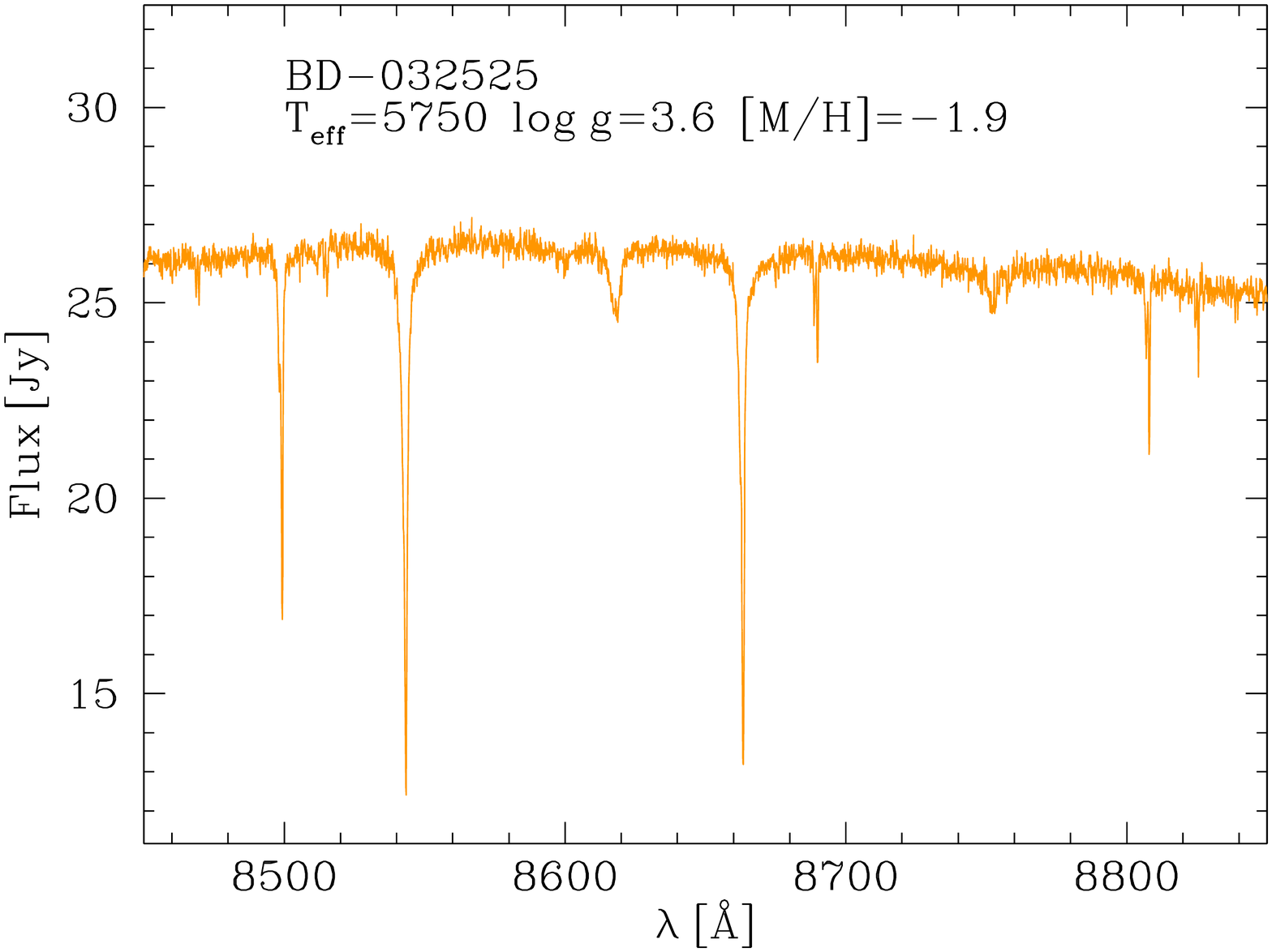}
\includegraphics[width=0.18\textwidth,angle=-90]{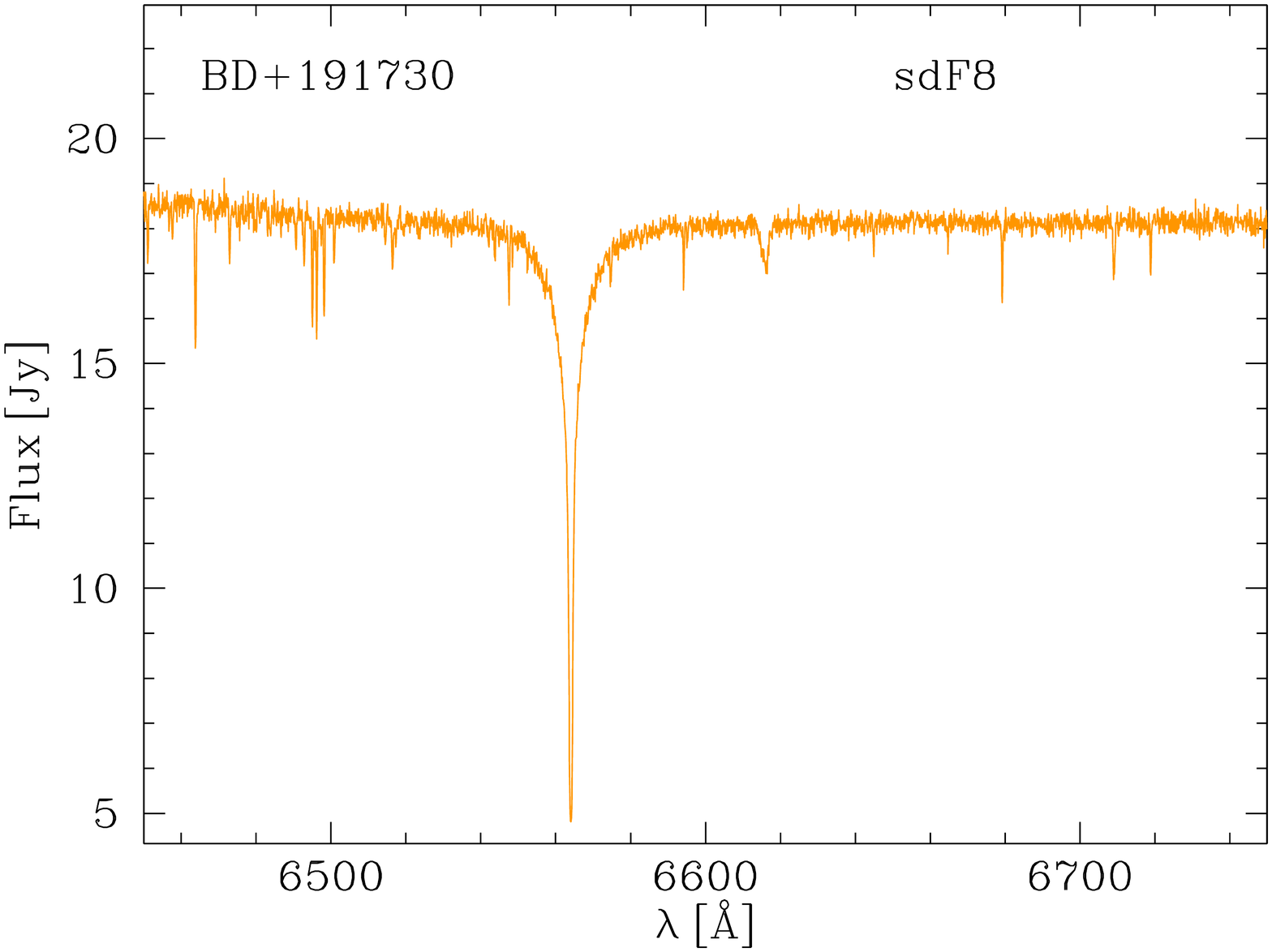}
\includegraphics[width=0.18\textwidth,angle=-90]{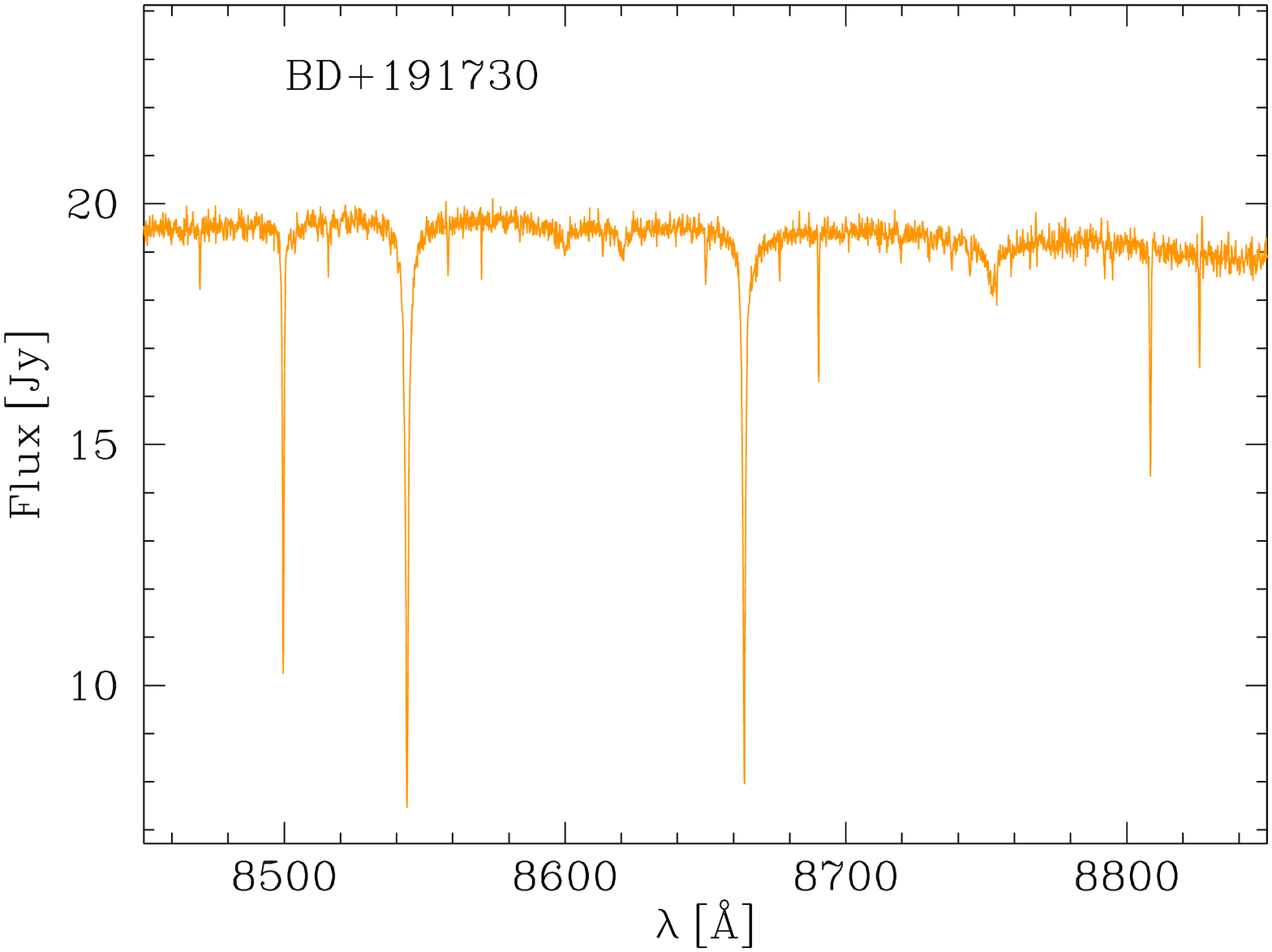}
\includegraphics[width=0.18\textwidth,angle=-90]{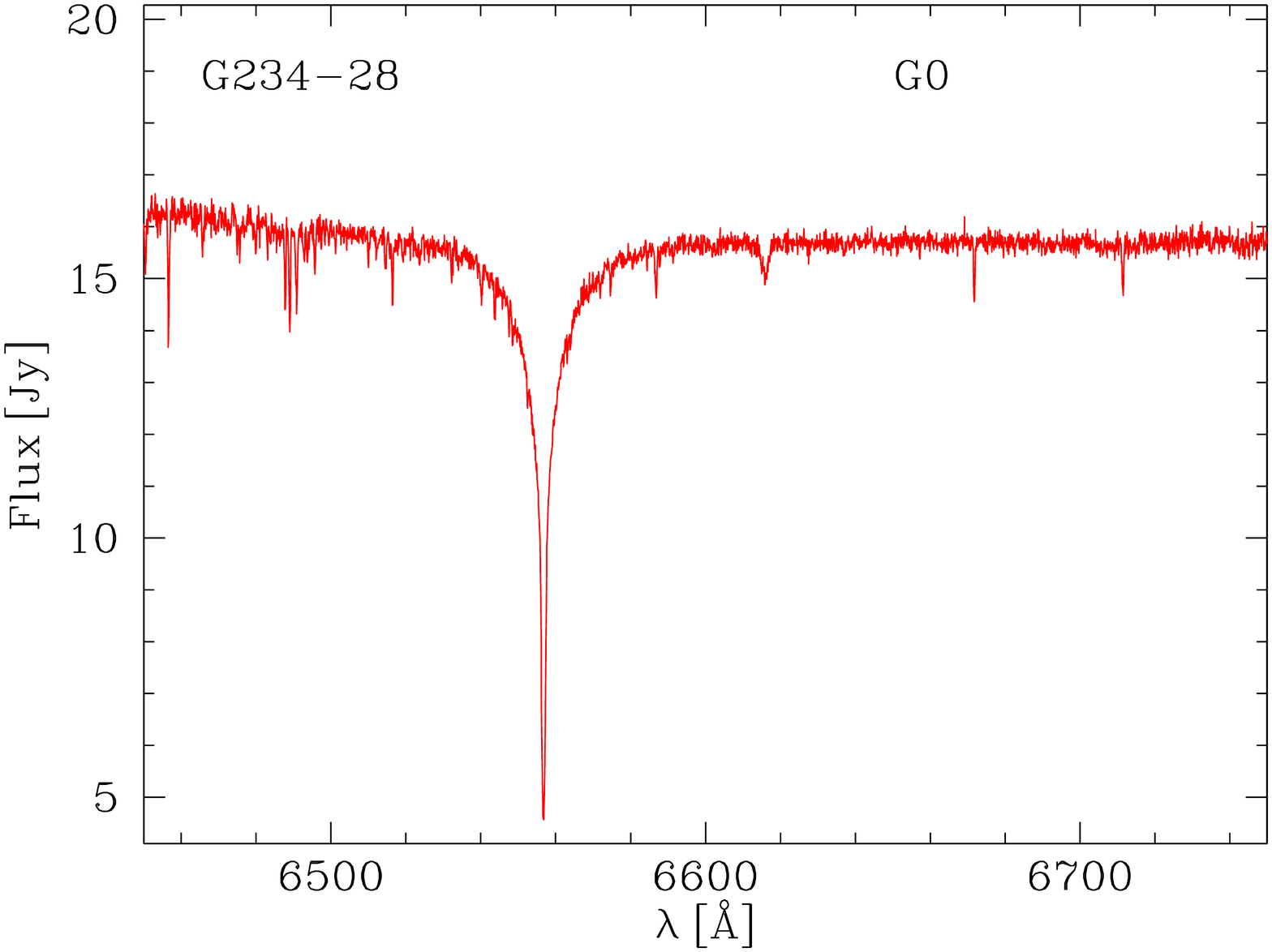}
\includegraphics[width=0.18\textwidth,angle=-90]{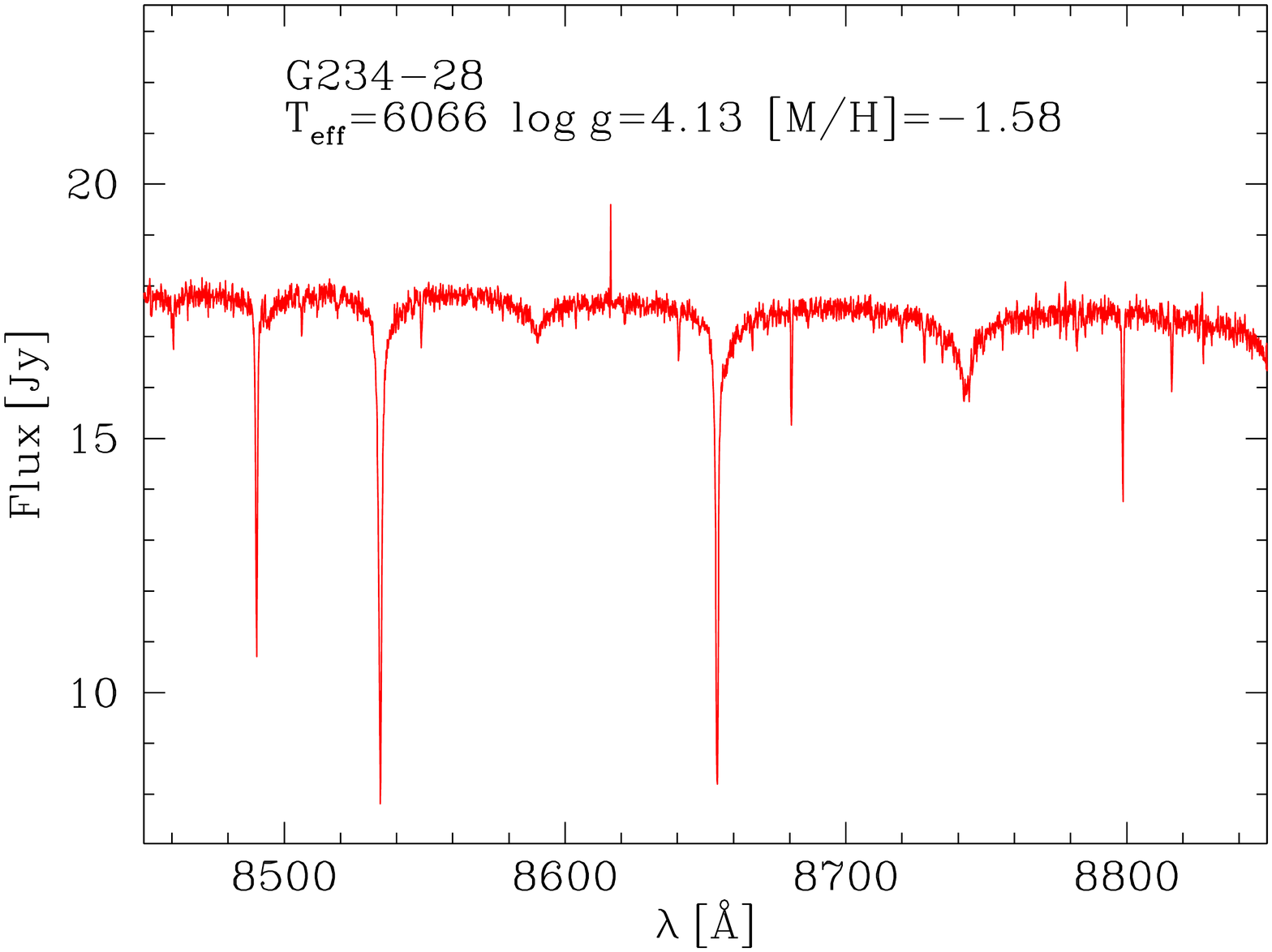}
\includegraphics[width=0.18\textwidth,angle=-90]{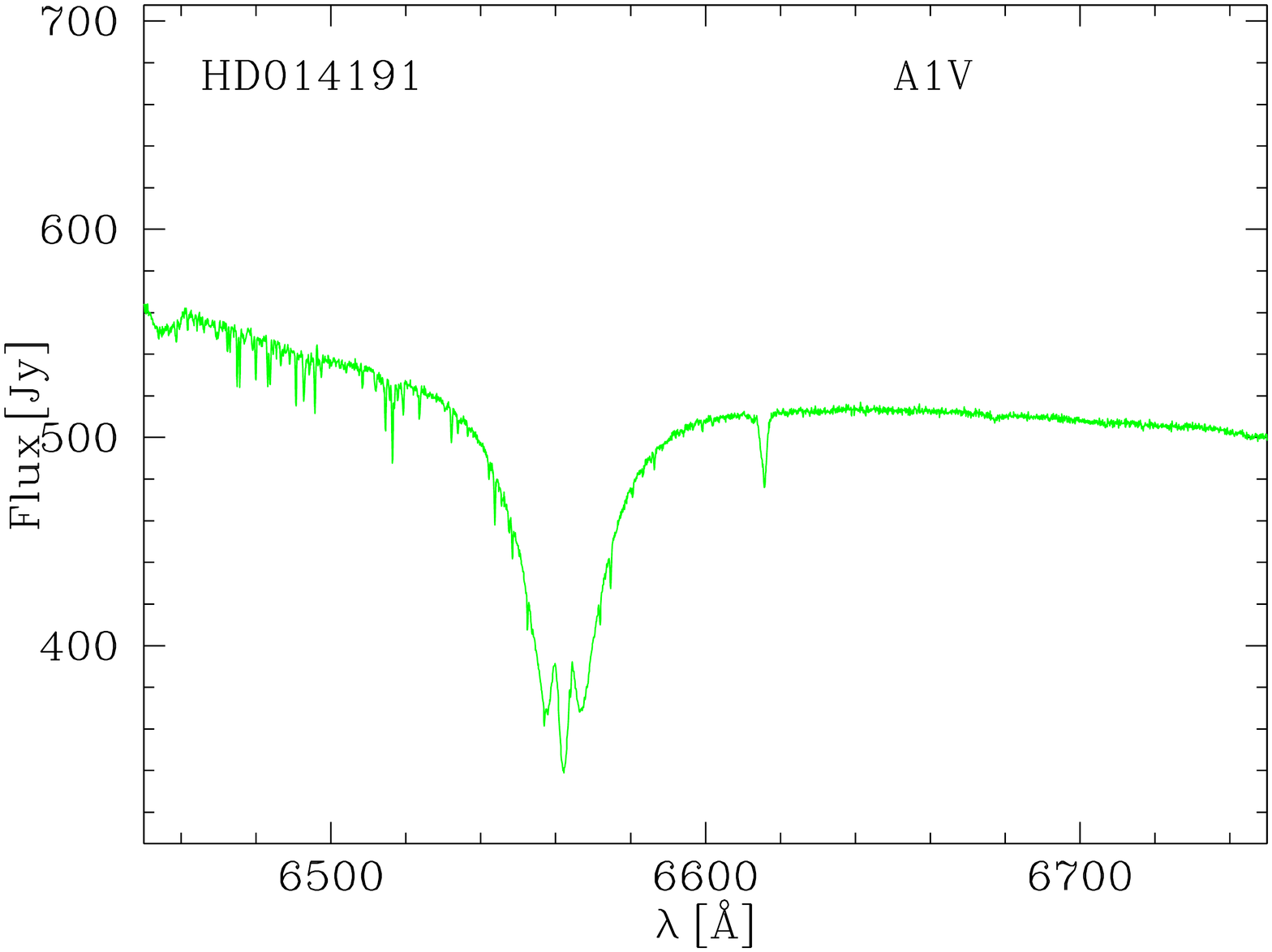}
\includegraphics[width=0.18\textwidth,angle=-90]{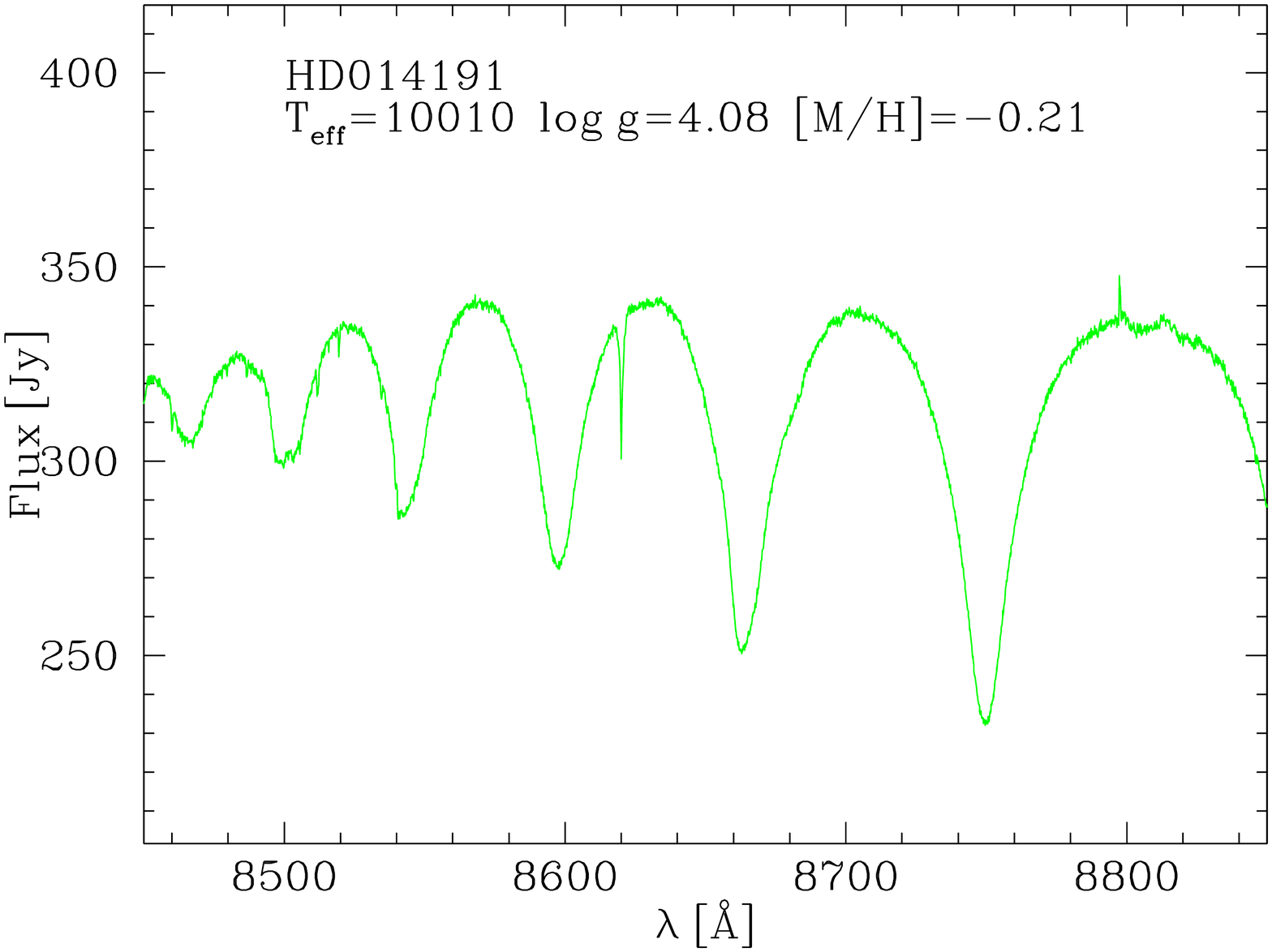}
\includegraphics[width=0.18\textwidth,angle=-90]{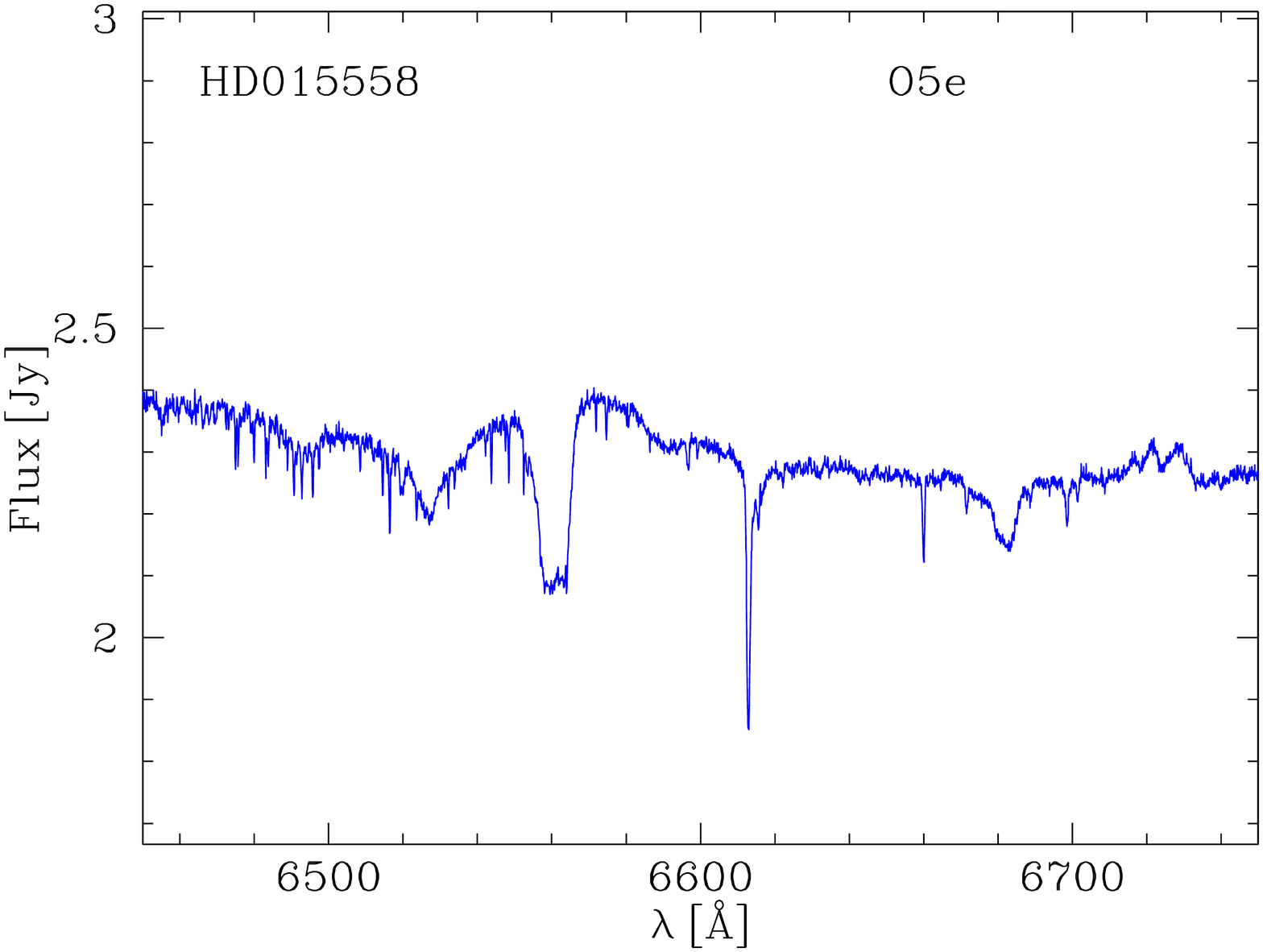}
\includegraphics[width=0.18\textwidth,angle=-90]{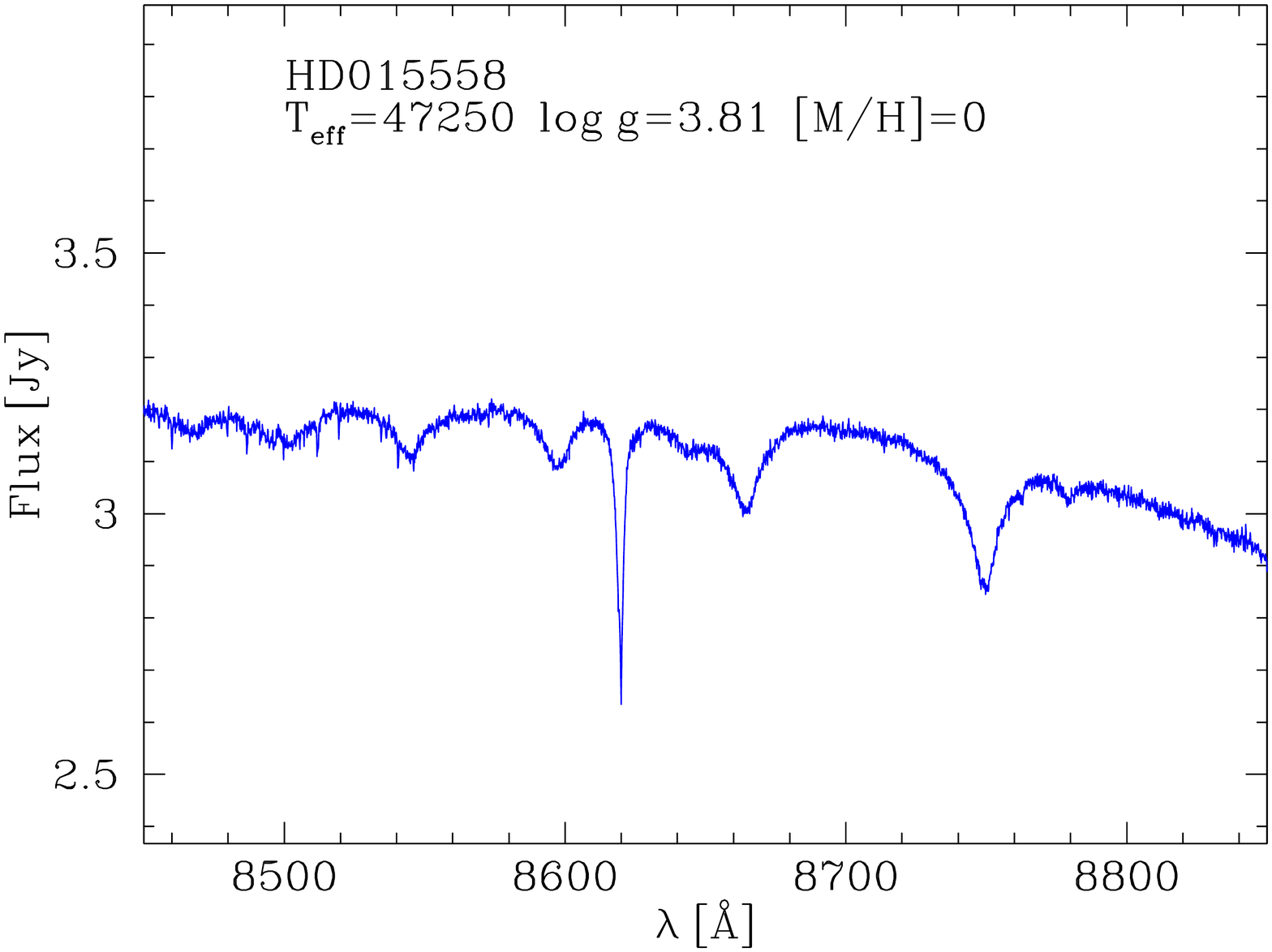}
\includegraphics[width=0.18\textwidth,angle=-90]{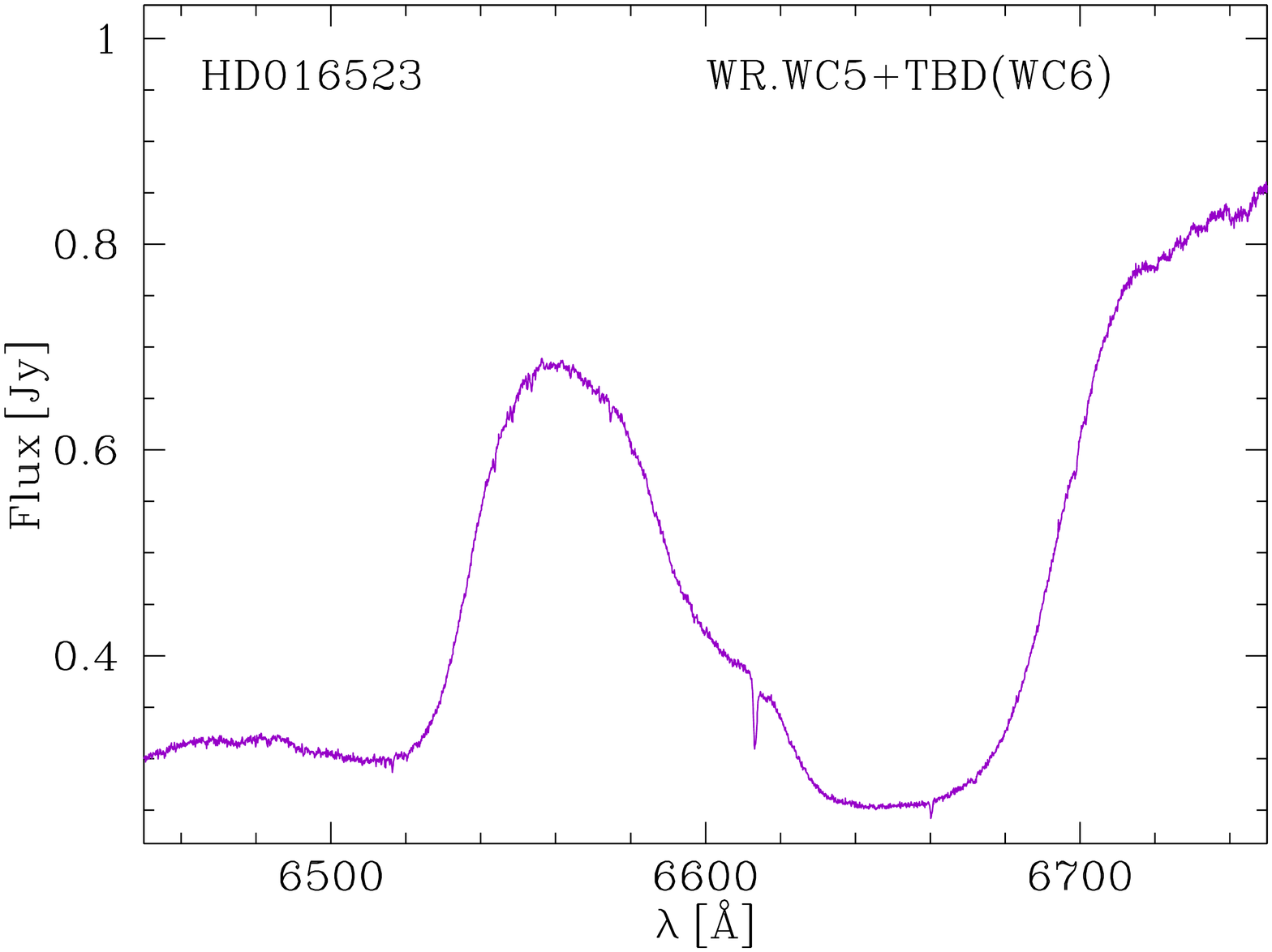}
\includegraphics[width=0.18\textwidth,angle=-90]{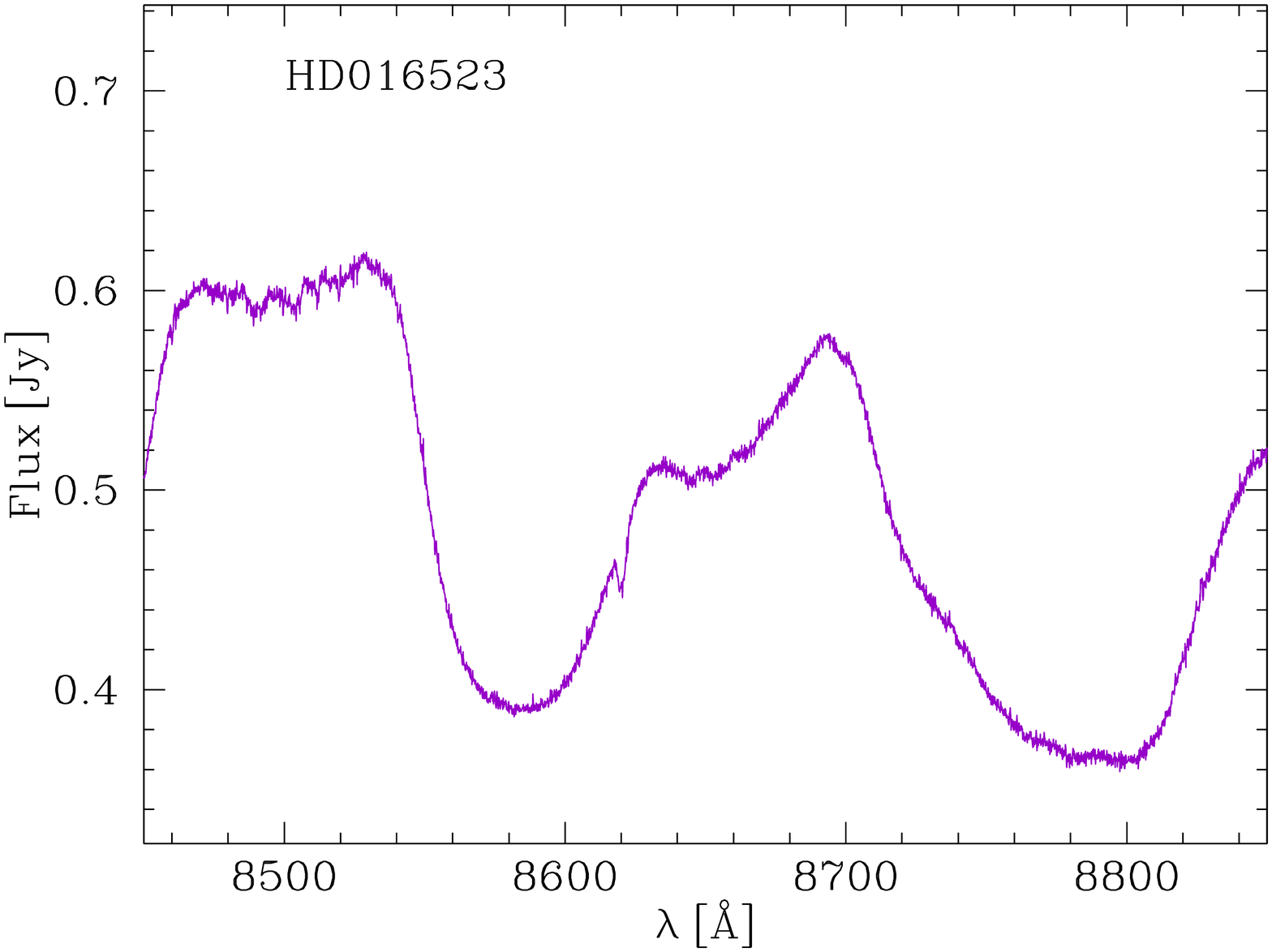}
\includegraphics[width=0.18\textwidth,angle=-90]{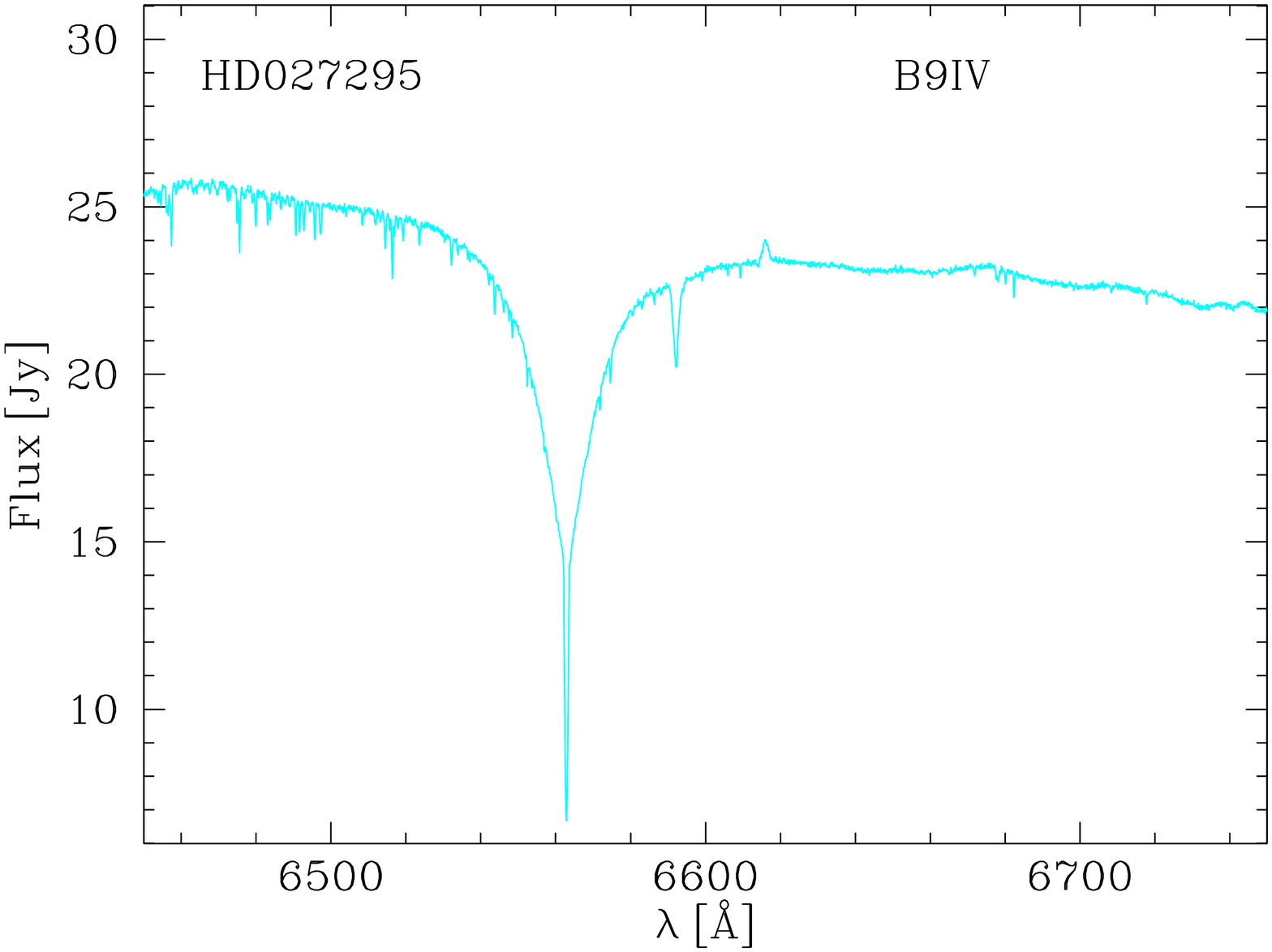}
\includegraphics[width=0.18\textwidth,angle=-90]{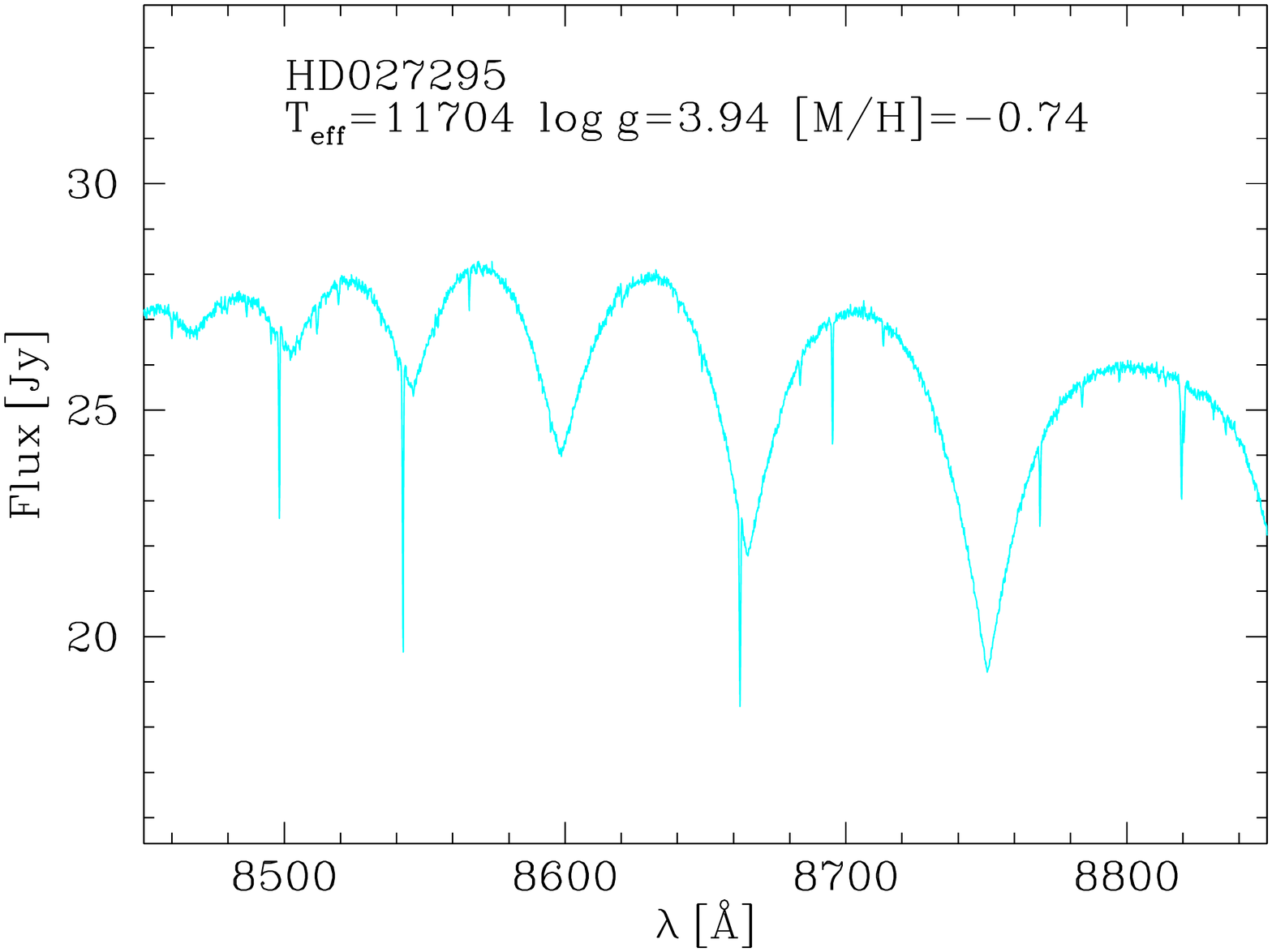}
\includegraphics[width=0.18\textwidth,angle=-90]{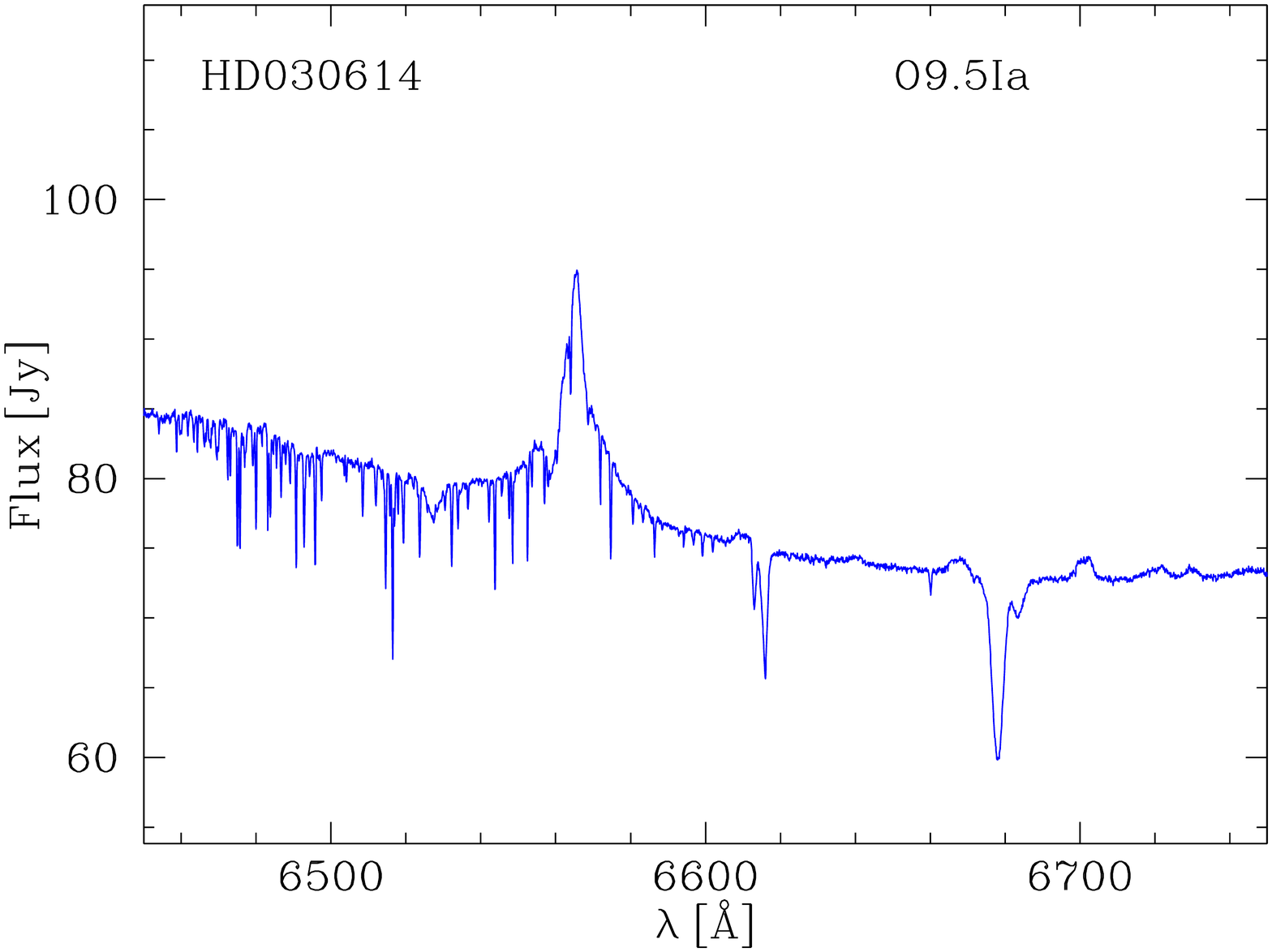}
\includegraphics[width=0.18\textwidth,angle=-90]{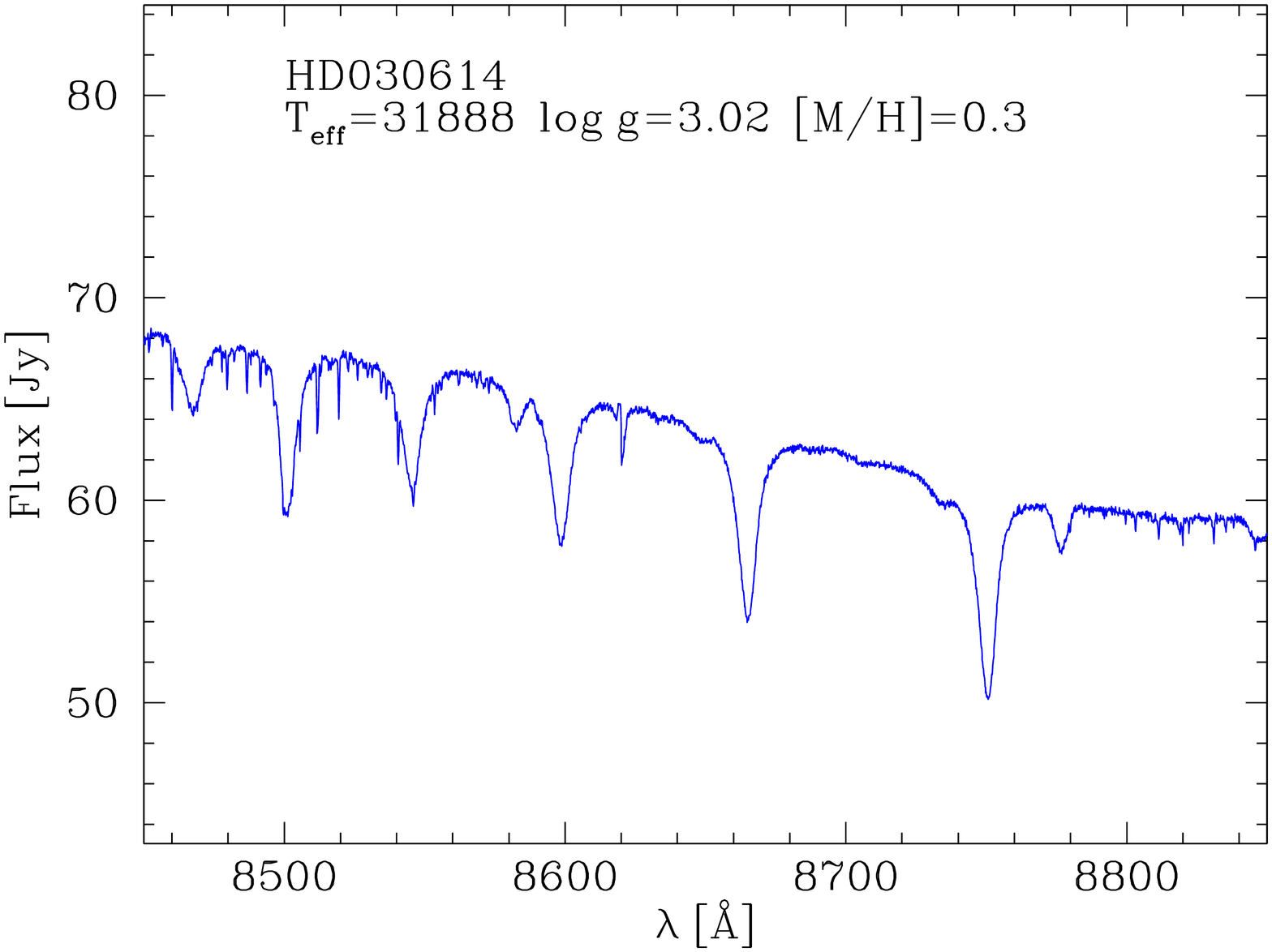}
\includegraphics[width=0.18\textwidth,angle=-90]{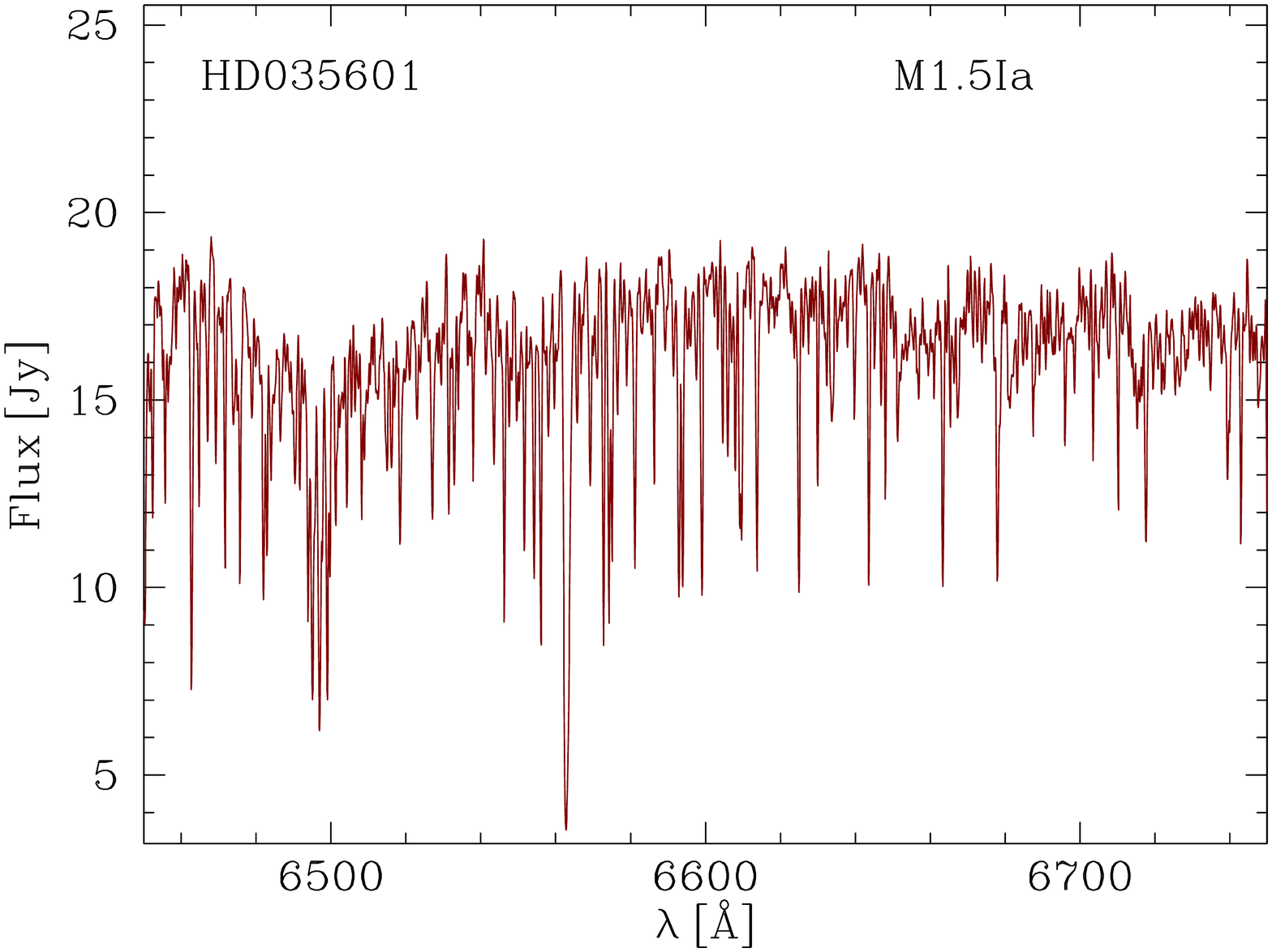}
\includegraphics[width=0.18\textwidth,angle=-90]{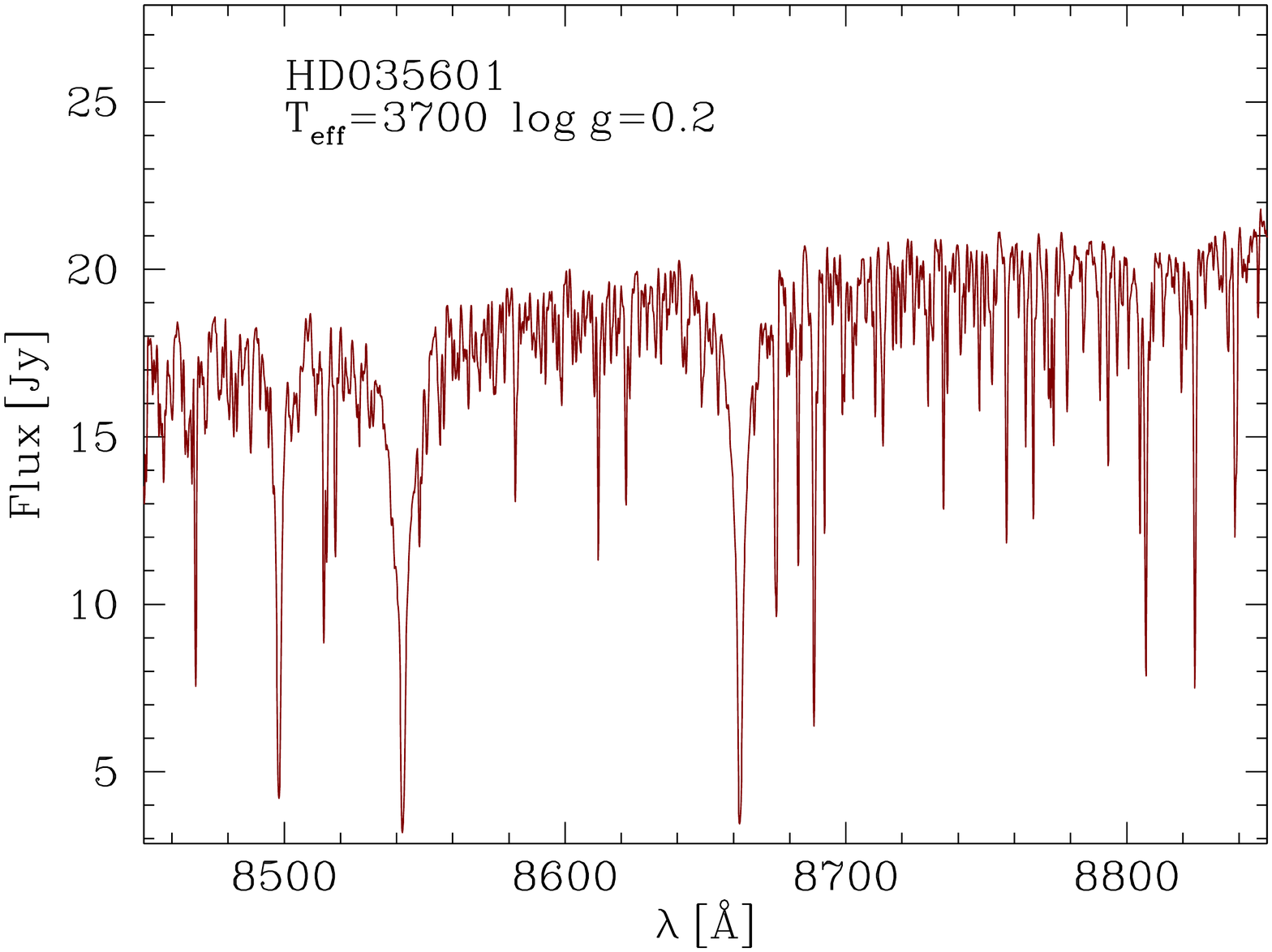}
\includegraphics[width=0.18\textwidth,angle=-90]{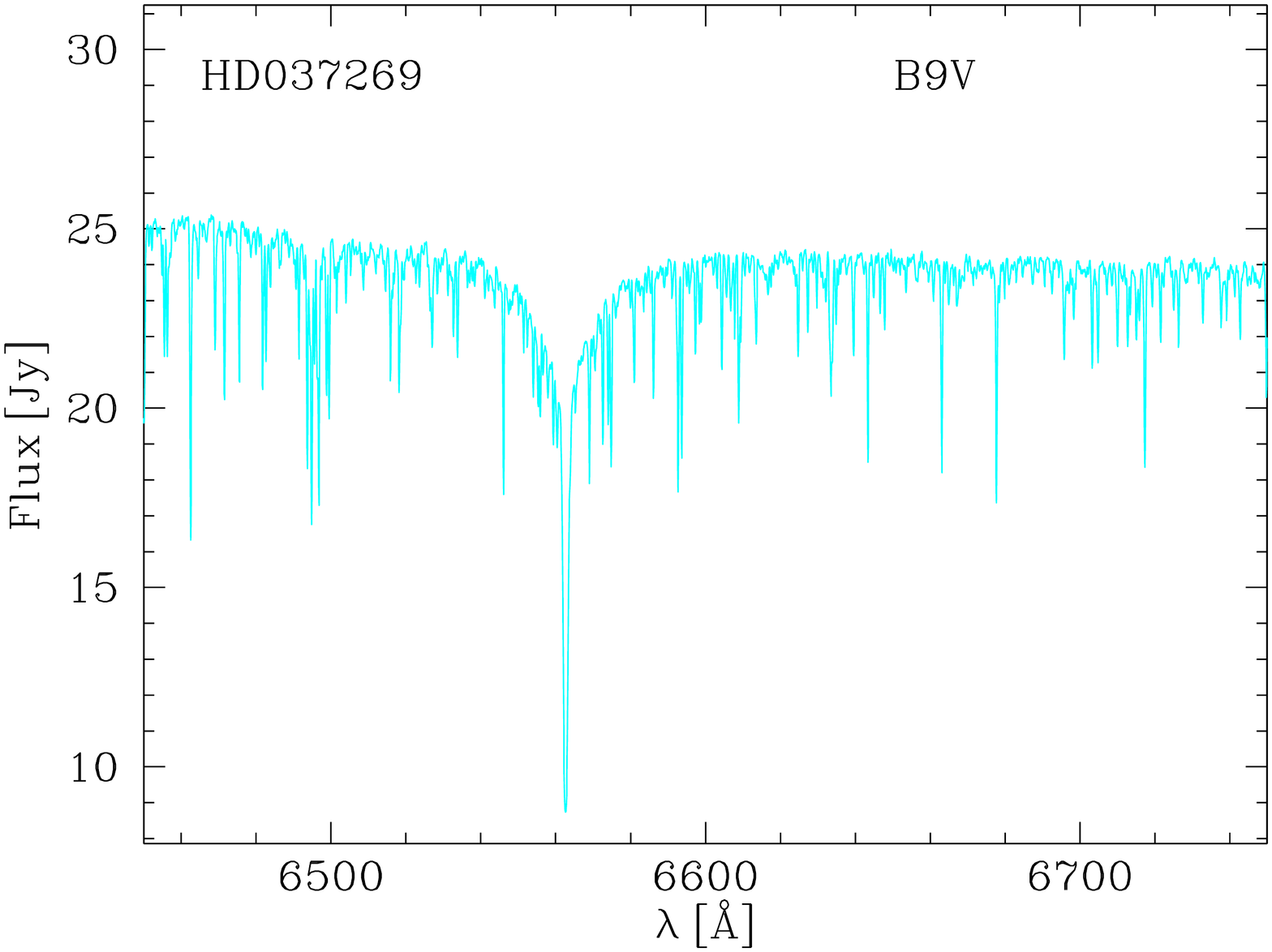}
\includegraphics[width=0.18\textwidth,angle=-90]{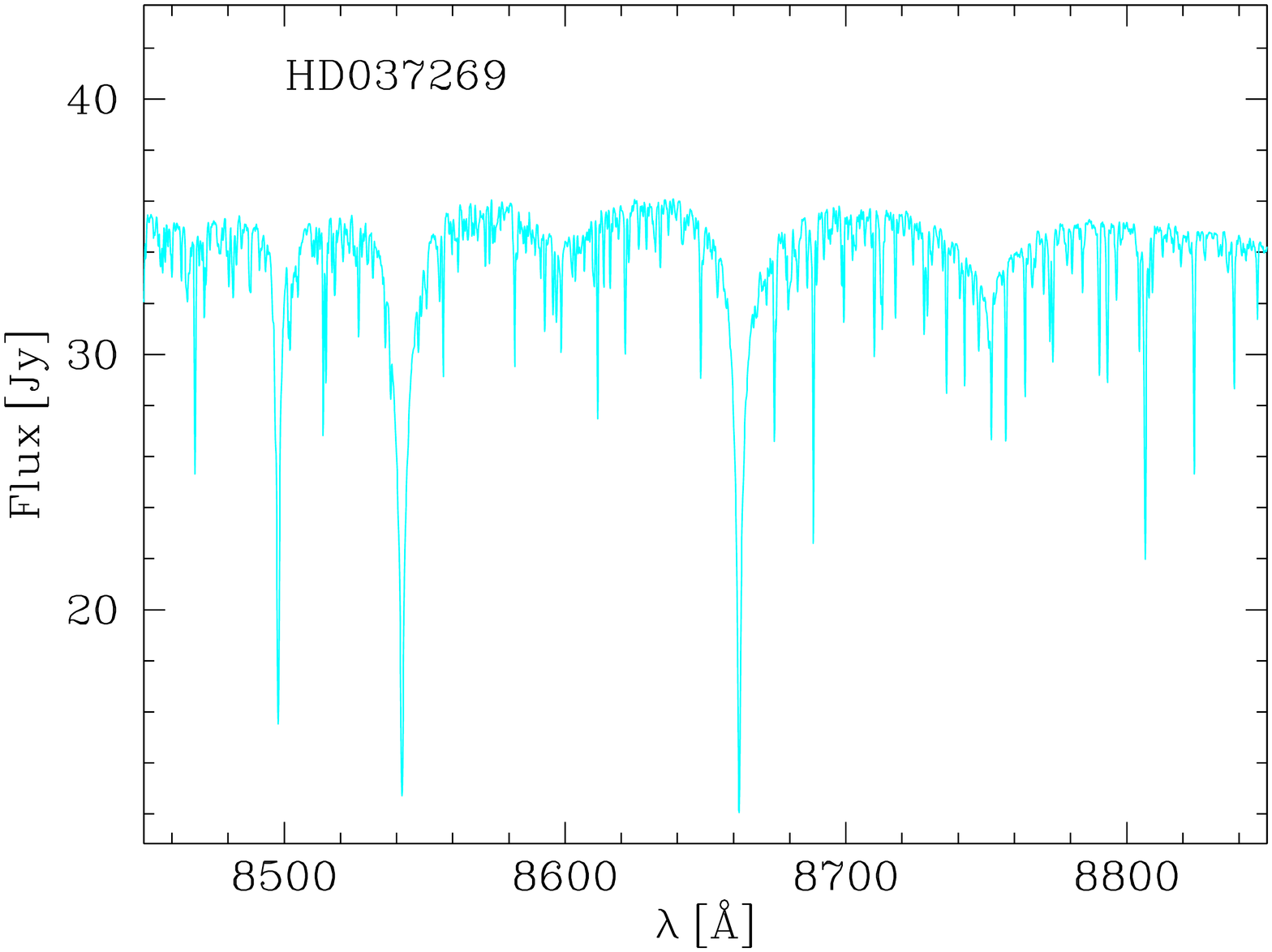}
\includegraphics[width=0.18\textwidth,angle=-90]{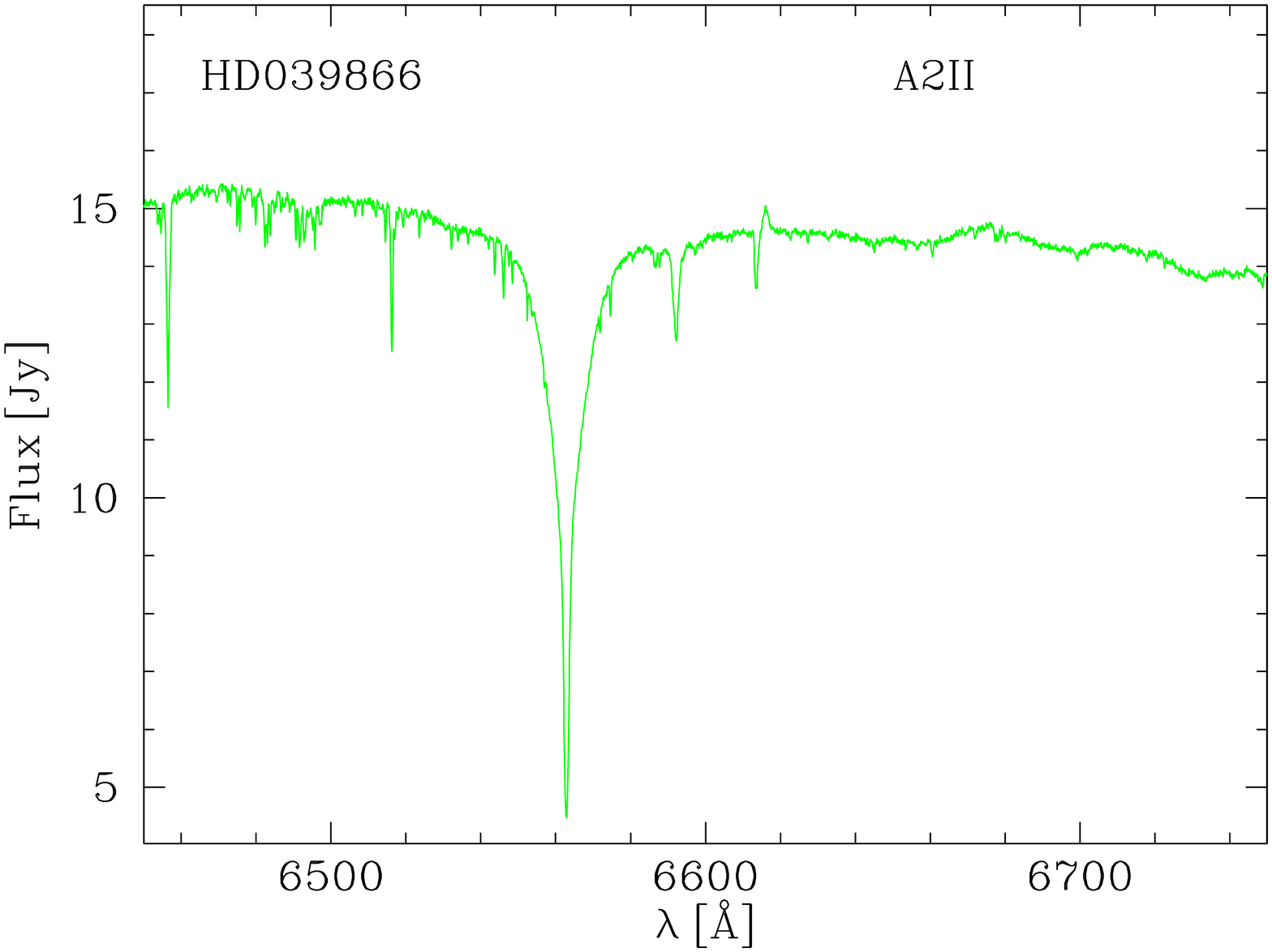}
\includegraphics[width=0.18\textwidth,angle=-90]{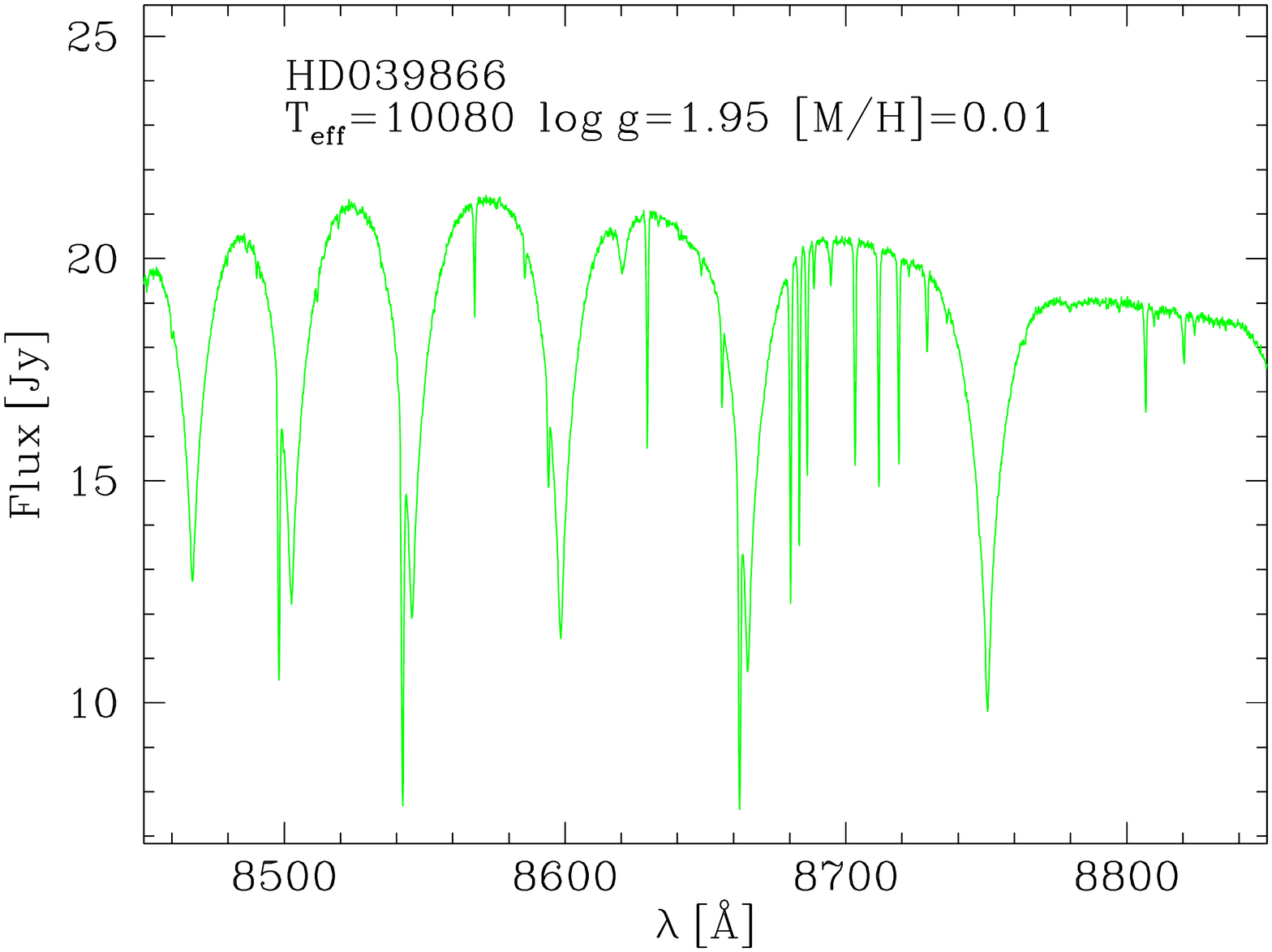}
\includegraphics[width=0.18\textwidth,angle=-90]{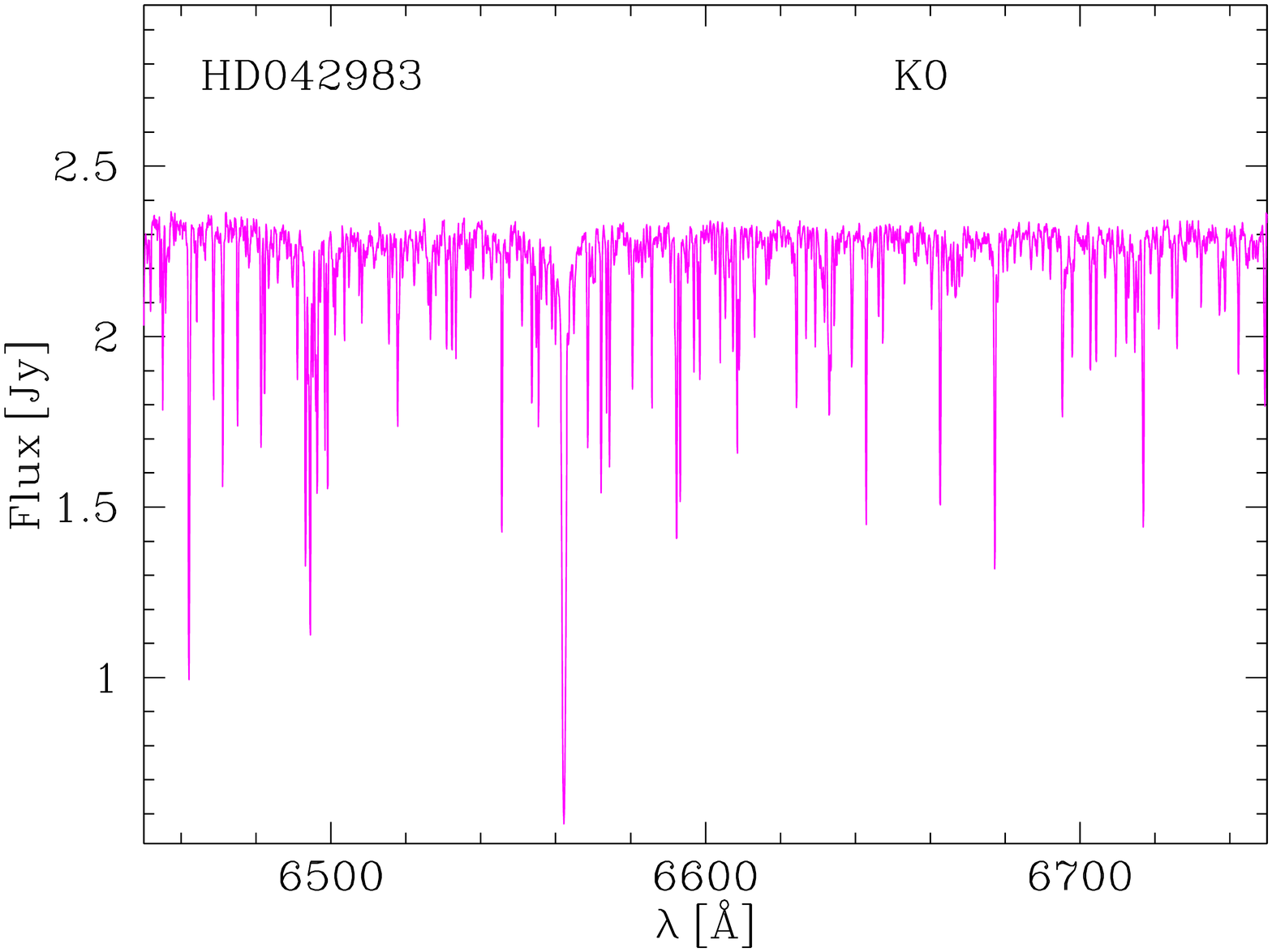}
\includegraphics[width=0.18\textwidth,angle=-90]{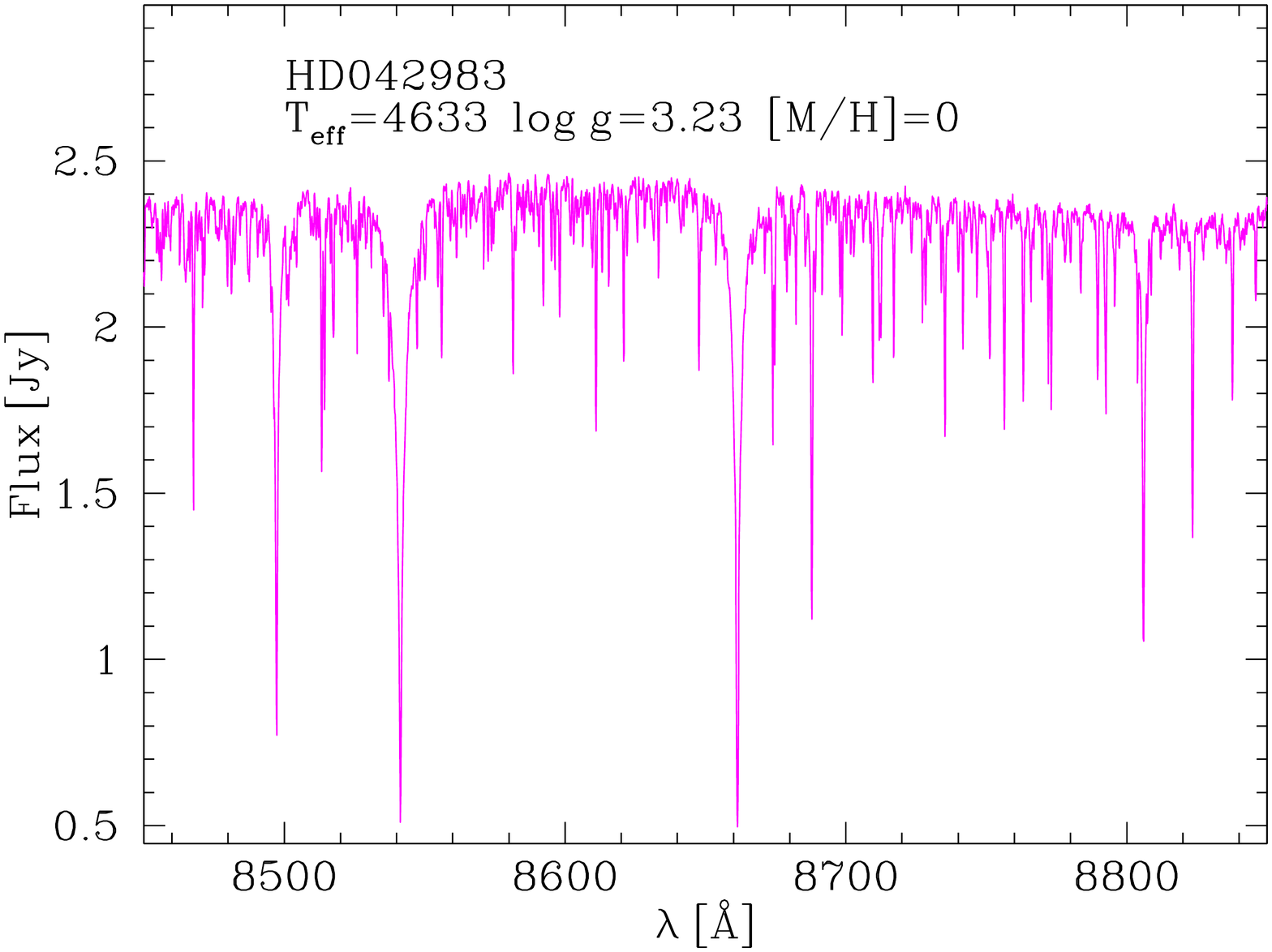}
\includegraphics[width=0.18\textwidth,angle=-90]{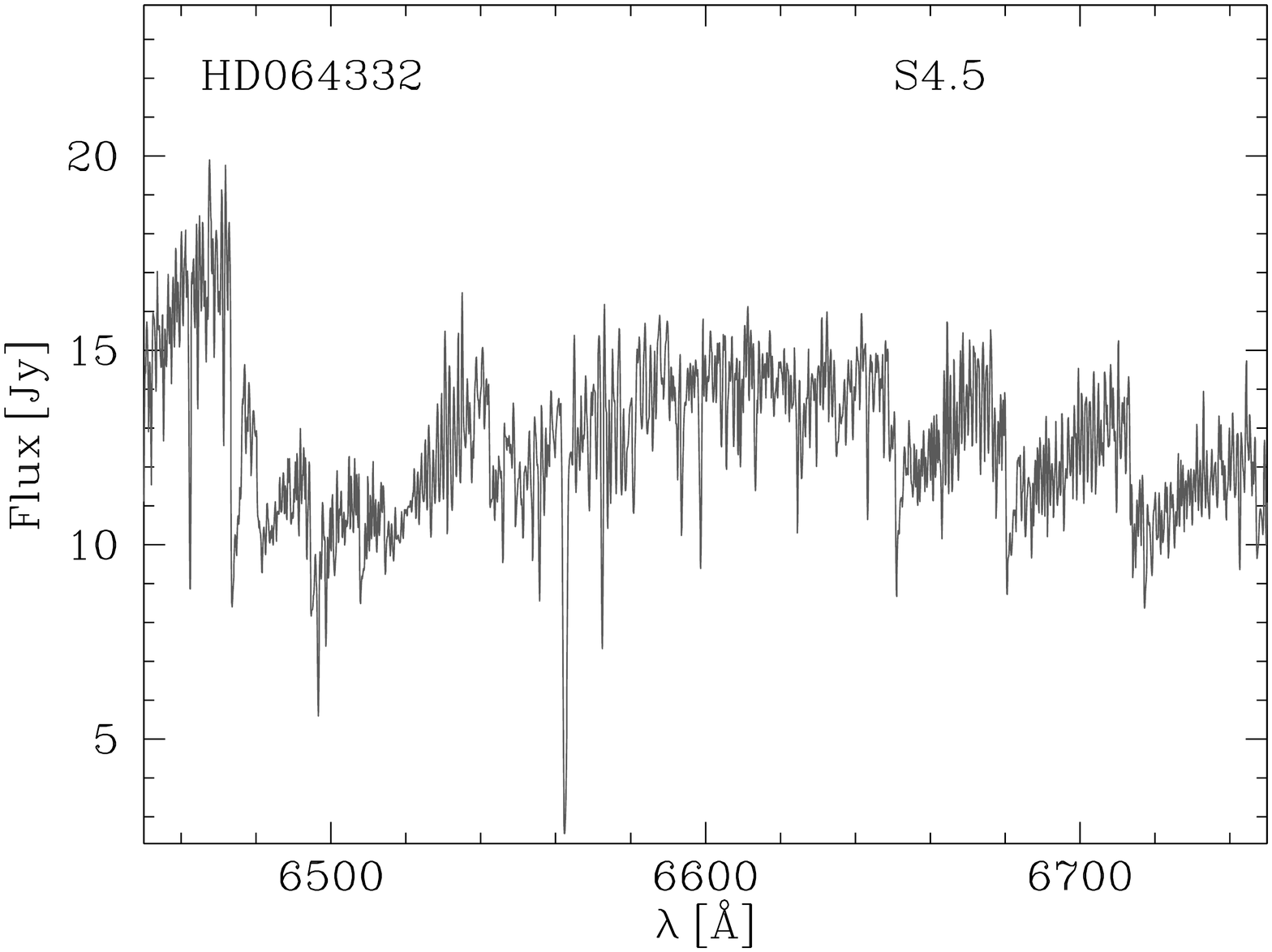}
\includegraphics[width=0.18\textwidth,angle=-90]{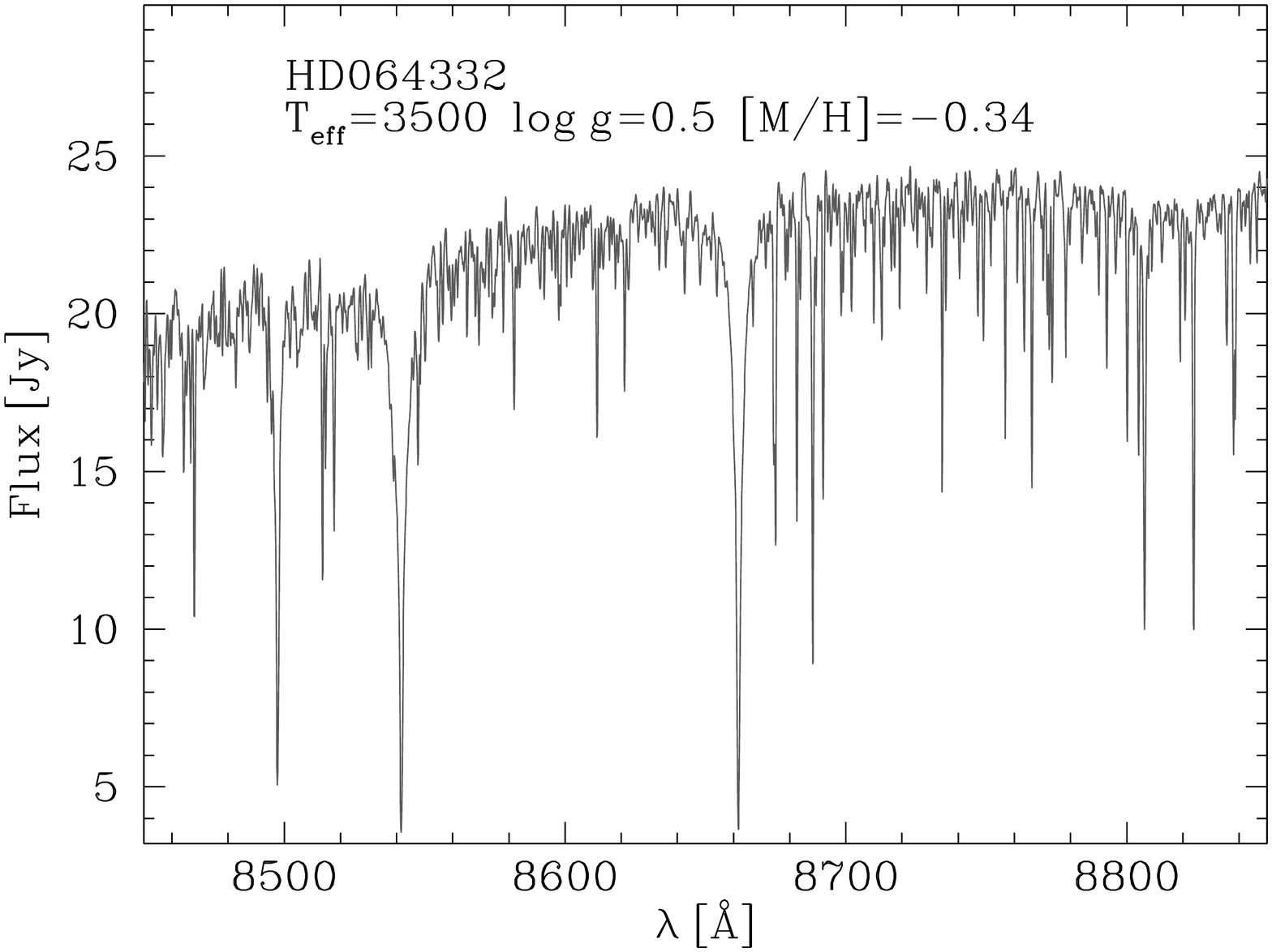}
\includegraphics[width=0.18\textwidth,angle=-90]{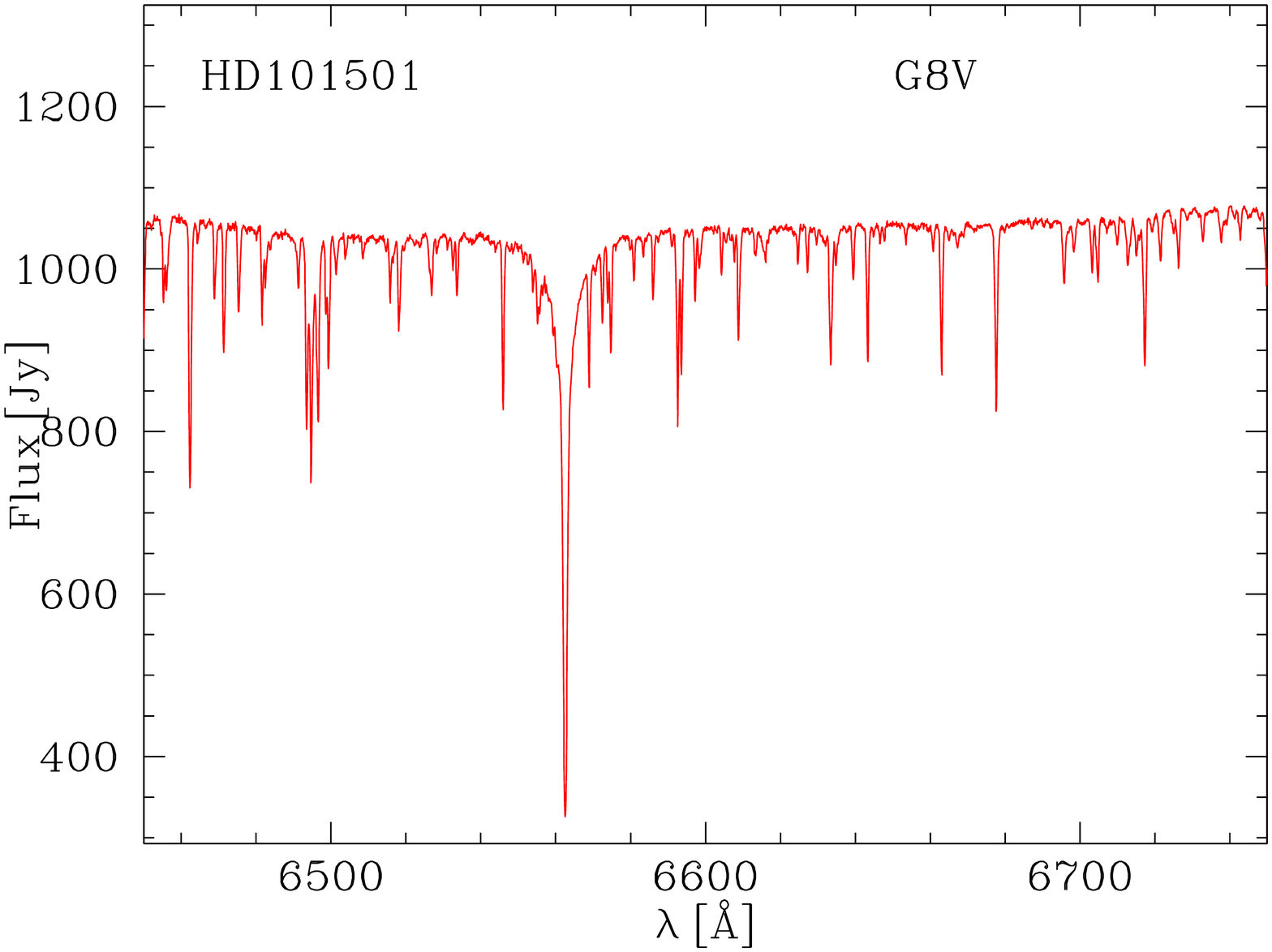}
\includegraphics[width=0.18\textwidth,angle=-90]{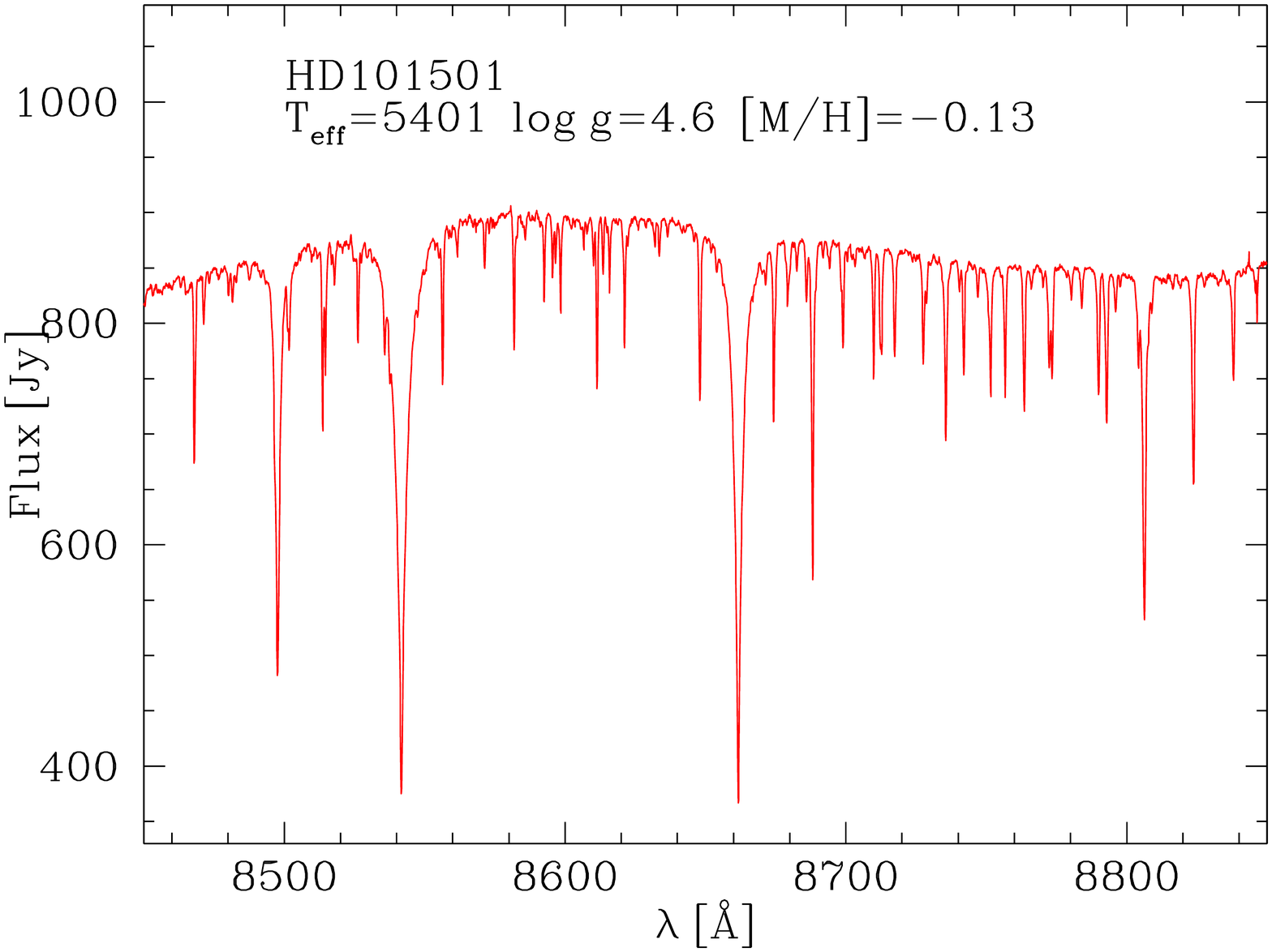}
\caption{Extract of MEGASTAR first release atlas. Each row shows the HR-R (left) and HR-I (right) spectra for two observed stars. Thus, columns 1 and 3 are HR-R spectra while columns 2 and 4 contains HR-I spectra. Columns 1 and 2, or columns 3 and 4, refer to the same observation. The color  indicates the published spectral type of the star: purple, WR; blue, O; cyan, B; green, A; orange, F; red, G; magenta, K; maroon, M; and grey, S. The complete atlas is presented in Appendix A, available in  electronic format only.}
\label{atlas}
\end{figure*}

\begin{figure*}
\includegraphics[width=1.\textwidth,height=0.9\textheight,angle=0]{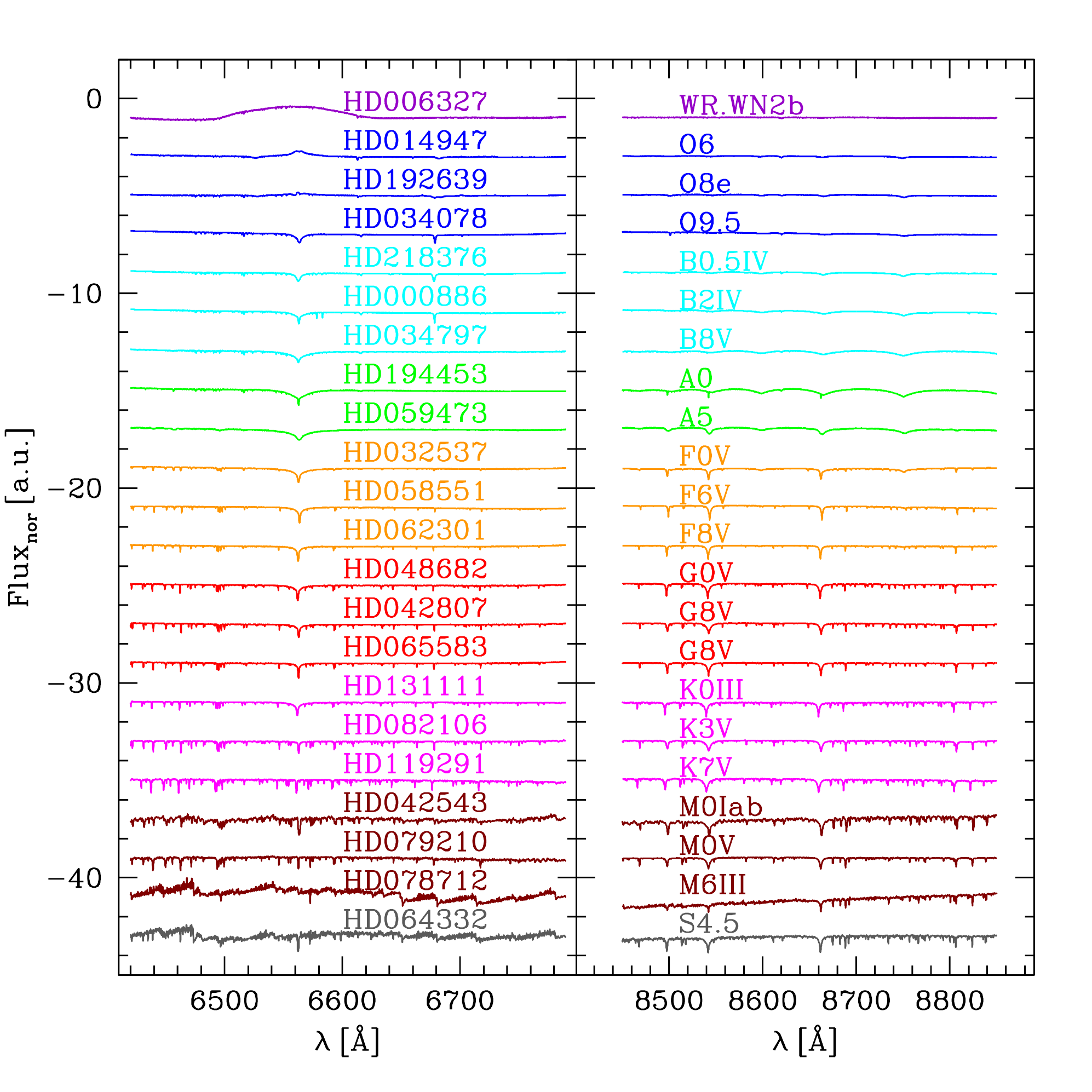}
\caption{Sequence of some spectra from stars of the first release with different spectral type. The left panels show the HR-R setup spectra while the right ones display the corresponding HR-I setup spectra. Each spectrum has the name of the star at the left panel, while the corresponding spectral type has been labelled inside the right one.} 
\label{seq}
\end{figure*}

\begin{figure*}
\centering
\includegraphics[height=0.5\textheight,width=0.49\textwidth,angle=0]{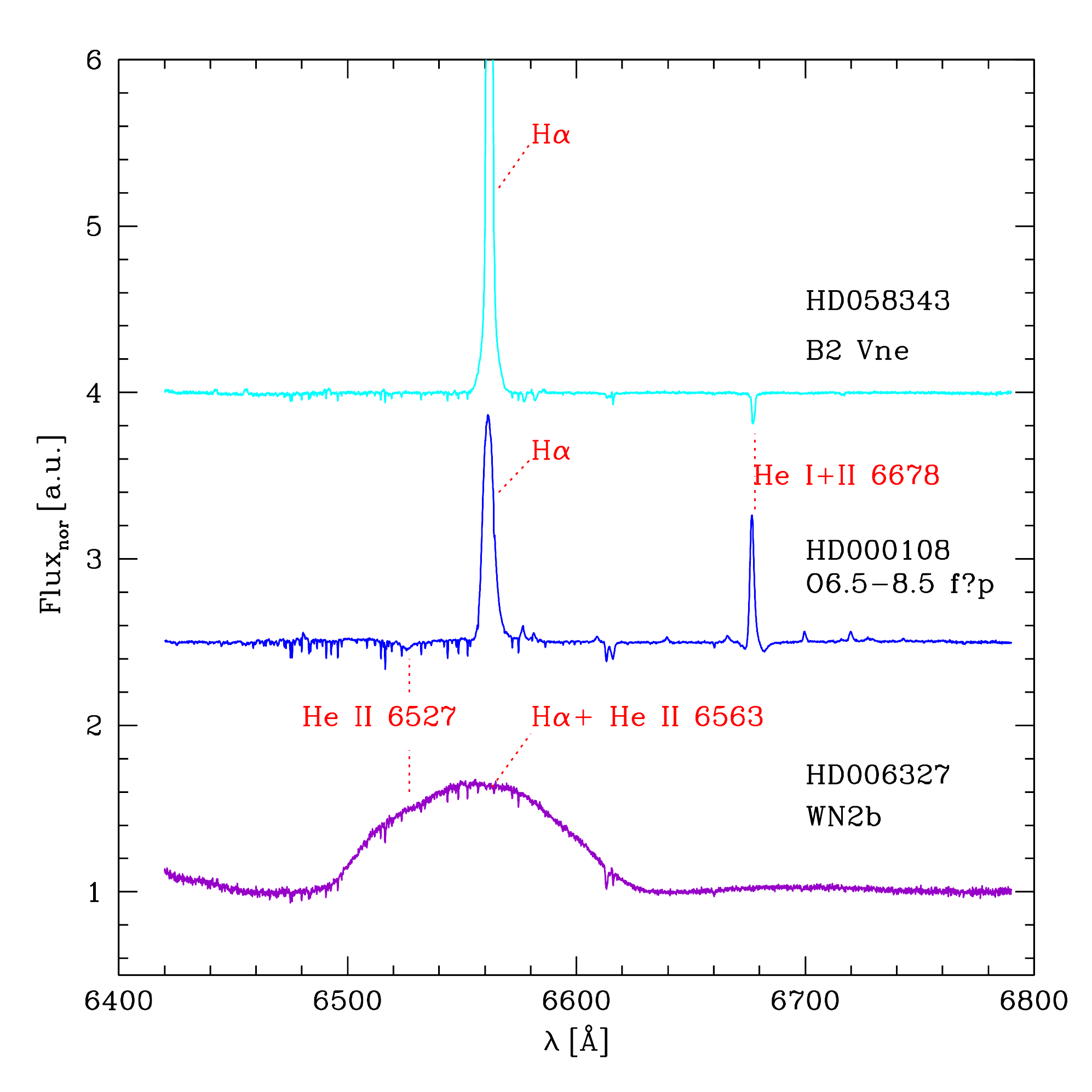}
\includegraphics[height=0.5\textheight, width=0.49\textwidth,angle=0]{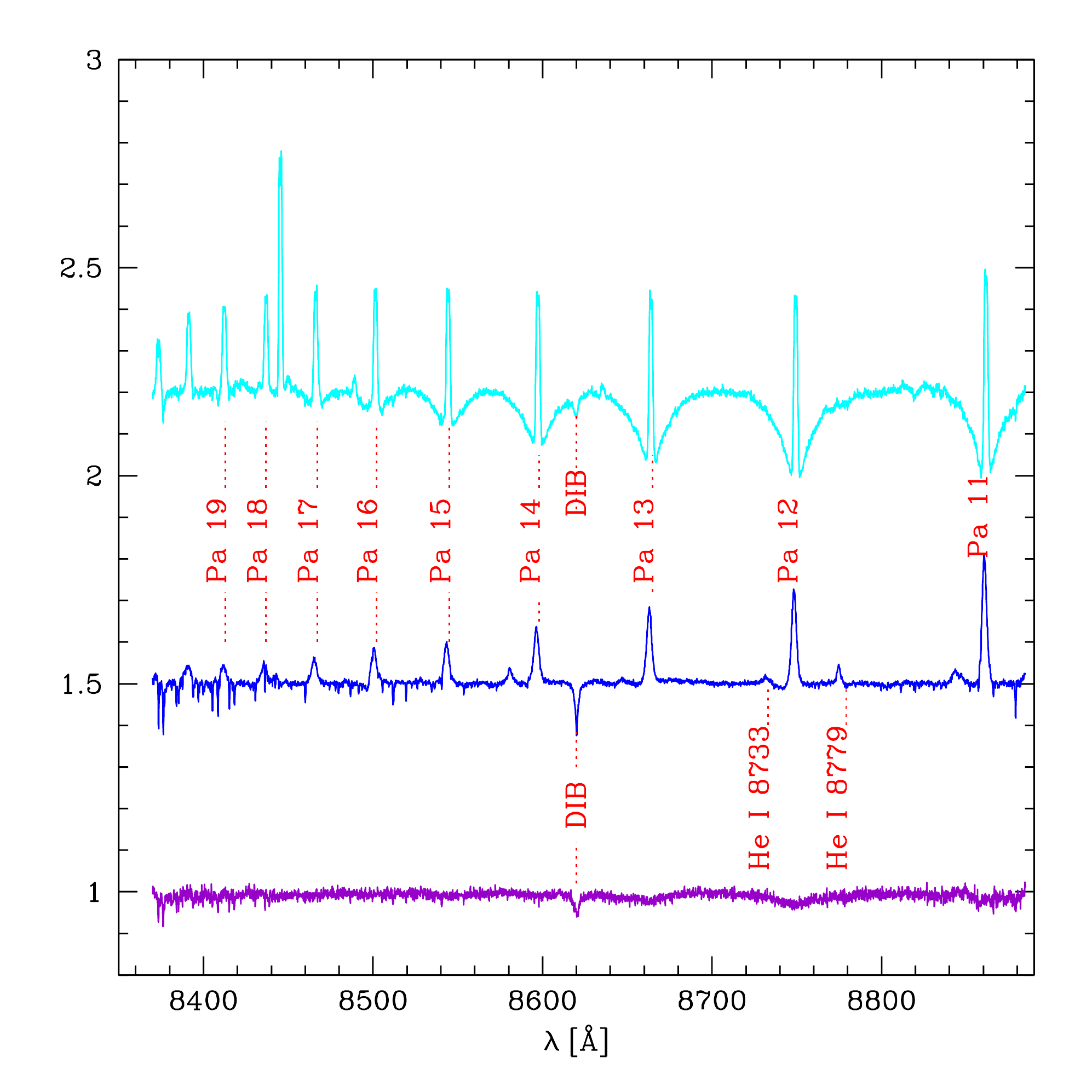}
\caption{Normalised spectra of three hot stars of this first release: HD058343 (cyan), HD000108 (blue) and HD006327 (purple) in the two spectral configurations: HR-R (left) and  HR-I (right). The star names and spectral types are indicated in black while the main hydrogen and helium lines are shown with the identification labelled in red.}
\label{hot-stars}
\end{figure*}
\section{MEGASTAR Library First Release stars}
\label{library-stars}

\subsection{MEGARA stellar atlas}
In Appendix~A, only in the electronic version, we present the atlas of the 838 spectra of this first MEGASTAR release, ordered alphabetically by star name. In Table \ref{appendix-index} we list the name of star in the same way as their spectra are displayed in the corresponding page of that appendix. To illustrate the content of the atlas, Fig.~\ref{atlas} shows an extract of 28 spectra of 14 stars. For each star the atlas displays the HR-R, from 6400 to 6800\,\AA, at the left, followed by the HR-I, from 8450 to 8850\,\AA, at the right. The star name, the spectral type and the stellar parameters compiled from the literature (if they exist) are labelled inside each HR-I plot. 
The color of the spectrum has been associated to the star spectral type in the literature as follows: purple, WR; blue, O; cyan, B; green, A; orange, F;  red, G; magenta, K; maroon, M; grey, S; and black, flat. Spectra shown in Fig.~\ref{atlas} are therefore from stars of different spectral type and luminosity class, in particular from left to right and from top to bottom: F3 (BD-0322525), sdF8 (BD+191730), G0 (G234-28), A1V (HD014191), O5e (HD015558), WC6 (HD016523), B9IV (HD027295), O9.5Ia (HD030614), M1.5Ia (HD035601), B9V (HD037269), A2II (HD039866), K0 (HD042983), S4.5 (HD064332) and G8V (HD101501). 

Fig.~\ref{seq} shows a spectral type sequence of HR-R and HR-I observations of stars of this release, highlighting the diversity of spectral types of the MEGASTAR release. For this figure the spectra were normalized in arbitrary units by dividing for the flux at $\lambda=6650$\,\AA\ in HR-R and at $\lambda=8780$\,\AA\ in HR-I, and shifted vertically. On the left side of the figure the HR-R setup spectra are shown, while on the the right side the corresponding HR-I setup spectra are presented. Each stellar spectrum has the name of the star in the HR-R panel  and the corresponding spectral type in the HR-I panel. 
 
\subsection{Examples of hot stars}

The first release of the MEGARA-GTC library also includes a representative sample of hot stars, i.e. there are 51 OB stars earlier than B3 and 7 Wolf-Rayet stars. As an example, in Fig.~\ref{hot-stars} we show the spectra obtained for three of them: HD006327 (or WR$\#$2), HD000108 and HD058343.  

HD006327 is one  of the hottest stars of the Milky Way, located in the Cas OB1 association. Moreover, it is the only known WN2 star in our Galaxy with an estimated temperature of $\sim$140\,kK \citep*[see][]{hamann06, hamann19}. HD006327 is a firm gamma-ray burst candidate \citep[][]{sander12} that has singular characteristics, such as unusually rounded emission line profiles \cite[][]{hamann06, chene19} that are not present in other early-type WN stars \citep[see e.g.][]{hilter66, conti90}. This feature can be clearly seen in our HR-R spectrum. Although it has been proposed to be a fast rotator or even a visual binary \citep[][]{crowther93}, in a recent deep spectroscopic study by  \cite{chene19}, they found no evidence for any of these scenarios.

HD000108 is an Of?p star \citep[][]{walborn10} that shows spectroscopic variations caused by magnetic effects \citep[][]{naze08}. Recently, \cite{maiz19} identified its two extreme states and provided a combined classification of O6.5--8.5 f?p var. It is an example of extreme rotational braking, with a rotational period of between 50 and 60 years \citep{naze10}. The comparison of the H$\alpha$ line profiles in our HR-R spectrum to those reported by \cite{shultz17} (for the 2015 period) and \cite{martins10} (for the period 2007-2009) suggests that HD000108 might be moving towards a high emission state, since it is much stronger than in both observations.

Finally, HD058343 is a runaway Be star \citep{tetzlff11, hoog01} for which the SIMBAD database provides a spectral classification of B2 Vne, although in the literature we can find spectral types that vary between B2 and B4 \citep[e.g., INDO-US library,][]{cochetti20,ahmed17, silaj10}. Stellar parameters and H$\alpha$ line profile variability has been investigated by \citet{arcos18} and \citet{ahmed17} within the BeSOS and MiMeS surveys, respectively. Our spectrum clearly shows strong emission in the H$\alpha$ line profile. We highlight that all Paschen lines appear in emission with absorption wings.

These examples show the variety of line profiles that can be found in the spectra of massive hot stars. They are highly affected by different factors that play an important role in their evolutionary behaviour, such as rotation, binarity, magnetic fields or strong stellar winds \citep{langer12}. Therefore, the inclusion of high-quality OB star spectra in the MEGASTAR database, which will increase in future releases,  can help  to study their physics, providing empirical data to constrain the complex theoretical evolutionary models of young stellar populations.

\subsection{Additional data:  Gaia-DR2}
\label{Gaia-stars}

As a complementary information to the  first release, in Appendix~C we include Gaia DR2 measurements for 388 stars of MEGASTAR library and additional useful information. In Table \ref{Gaia-table} the  columns headers of Appendix~C are described. 
The included Gaia~DR2 data  were retrieved using the Python interface for querying the VizieR web service provided by Astroquery\footnote{\url{https://astroquery.readthedocs.io/en/latest/vizier/vizier.html}},  and an affiliated package of Astropy\footnote{\url{http://www.astropy.org}, \citep{Astropy18}}. 
The table incorporates the default columns that are pre-selected when accessing the catalogue \texttt{I/345/gaia2} \citep{Gaia18}. For completeness, the table includes, when available, the Hipparcos \citep{Hip00} and Tycho \citep{Tyc07} identifications, as well as the the specific designation when the objects are recognised as known variable stars in the SIMBAD database.

\begin{table}
\centering
\caption{Columns description of the Appendix~C file,  available in electronic format only. Includes data  from Gaia DR2 for 388 stars of MEGASTAR library.}
\small
\begin{tabular}{ll}
\hline
   Column    & Description \\
\hline
 Name	&	Star name  (*) 					 				\\
RA	&	Right Ascension (2000.0)			hh:mm:ss.s (*)	\\
DEC	&	Declination (2000.0)			dd:dd:ss.s (*)	\\
Sp Type	&	Spectral Type 	(*)							\\
RV\_VALUE &   radial velocity (km/s) (*) (n)  \\
U	&	Jonhson-Cousins U magnitude (*) (n)			\\
B	&	Jonhson-Cousins B magnitude (*) (n)			\\
V	&	Jonhson-Cousins V magnitude (*) (n)			\\
R	&	Jonhson-Cousins R magnitude (*) (n)			\\
I	&	Jonhson-Cousins I magnitude (*) (n)			\\
J	&	Jonhson-Cousins J magnitude (*) (n)			\\
H &   Jonhson-Cousins H magnitude (*) (n)\\
K &   Jonhson-Cousins K magnitude (*)  (n) \\
Other name	&	Alternative  name of the star (n)		\\
Teff	&	Effective temperature from literature 	(K)	  	(n)          \\
Logg	&	Surface gravity (log) from literature	(n)					\\
$[$Fe/H$]$	&	Iron abundance (log) from Literature (n)					\\
Reference	&	Original catalogue from which it was inherited (n)	\\
Other Comments	&	Comments related to the star (n)					\\
MAIN\_ID	&	Default name in SIMBAD database					     	\\
id\_variable	&	ID if the star is identified as a known variable (n)	\\
id\_hipparcos	&	Star name in Hipparcus catalogue (n)			        \\
id\_tycho	&	Star name in Tycho catalogue	(n)				   		\\
id\_gaiadr2	&	Star name in  Gaia DR2 catalogue						\\
RA\_ICRS	&	Barycentric right ascension (ICRS) at Ep=2015.5	($^\circ$)   \\
e\_RA\_ICRS	&	Standard error of right ascension           	(mas)   \\
DE\_ICRS	&	 Barycentric declination (ICRS) at Ep=2015.5	($^\circ$)   \\
e\_DE\_ICRS	&	Standard error of declination          			(mas)   \\
Source	& Unique source identifier (unique within 			\\
        & a particular Data Release)                          \\
Plx	&	Absolute stellar parallax  (mas) (n)					         	\\
e\_Plx	&	Standard error of parallax  (mas) (n)					    	\\
pmRA	&	 Proper motion in RA direction  (mas y$^{-1}$)  (n) 		\\
e\_pmRA	&	 Standard error of proper motion in RA (mas y$^{-1}$) (n)  \\
pmDE	&	  Proper motion in DEC direction   (mas y$^{-1}$)   (n)	  	\\
e\_pmDE	&	 Standard error of proper motion in DEC (mas y$^{-1}$)  (n) 	\\
Dup	&	$[$0/1$]$ Source with duplicate sources  	 			\\
FG	&	G-band mean flux  				                     (e- s$^{-1}$)     \\
e\_FG	&	Error on G-band mean flux  (e- s$^{-1}$)  					 		\\
Gmag	&	 G-band mean magnitude (Vega)					   	     	\\
e\_Gmag	&	Standard error of G-band mean magnitude (Vega)		 		\\
FBP	&	 Mean flux in the integrated BP band (e- s$^{-1}$) (n)		\\
e\_FBP	&	 Error on the integrated BP mean flux (e- s$^{-1}$) (n) 	\\
BPmag	&	 Integrated BP mean magnitude (Vega) (n)				 		\\
e\_BPmag	&	 Standard error of BP mean magnitude (Vega) (n)		    \\
FRP	&	Mean flux in the integrated RP band (e- s$^{-1}$) (n)				\\
e\_FRP	&	 Error on the integrated RP mean flux  (e- s$^{-1}$)	(n) 	\\
RPmag	&	 Integrated RP mean magnitude (Vega) (n)			 		\\
e\_RPmag	&	Standard error of RP mean magnitude (Vega)(n)			\\
BP-RP	&	 BP-RP colour 			 		                            \\
RV	&Spectroscopic radial velocity in the solar barycentric             \\
 &reference frame 	(km s$^{-1}$) 	(n)  		                                   	\\
e\_RV	&	 Radial velocity error  	(km s$^{-1}$)		        \\
Teff\_2	&	Stellar effective temperature from A-P	 (K)	     \\
AG	&	 Estimate of extinction in the G band from (n)		\\
E(BP-RP)	&	 Estimate of redenning  from A-P (n)				\\
Rad	&	  Estimate of radius from A-FLAME 		(solRad) (n)		\\
Lum	&	Estimate of luminosity from A-FLAME 		(solLum) (n)		\\
\hline
(*) &  Source: SIMBAD \\
(n)& Indicates possible blank or NULL column \\
A-P & Apsis-Priam \\
A-FLAME & Apsis-FLAME\\

\end{tabular}
\label{Gaia-table}
\end{table}

\begin{table*}
\caption{Summary of the stars of the MEGASTAR first release whose spectra are shown in Appendix A, available in electronic format only. The first column is the number of the Appendix A page where the spectra can be found. The stars are listed in the same order that  they are  displayed in each page of this atlas.}
\label{appendix-index}
 \begin{tabular}{l l l l l l ll}
\hline
Page & Stars& & & & & & \\
\hline 
  A3 & BD-032525& BD-122669& BD+083095& BD+092190& BD+130013& BD+191730& BD+195116B\\
     & BD+203603& BD+241676& BD+262606& BD+351484& BD+381670& BD+511696& BD+541399 \\
 
  A4 & BD+800245& G171-010& G197-45& G202-65& G234-28& HD000108& HD000358\\
     & HD000560  & HD000886& HD003360& HD003369& HD003628& HD003644& HD004004\\
 
  A5 & HD004539 & HD006327& HD006815& HD007374& HD009974& HD009996& HD013267\\
     & HD013268& HD014191& HD014633& HD014947& HD015318& HD015558& HD015570\\
   
  A6 & HD015629& HD016429& HD016523& HD016523& HD016581& HD017081& HD017145\\
     & HD017506& HD017638& HD017638& HD018144& HD018296& HD018409& HD019308\\
 
  A7 & HD020084& HD020512& HD021742& HD022484& HD023862& HD024341& HD024451\\
     & HD024534& HD024912& HD025173& HD025825& HD026756& HD027126& HD027282\\
 
  A8 &  HD027295& HD027371& HD027524& HD027685& HD028005& HD029645& HD030614\\
     & HD030649& HD030676& HD031219& HD031293& HD031374& HD031996& HD032537\\

  A9 & HD033632& HD033904& HD034078& HD034255& HD034797& HD034816& HD035468\\
     & HD035497& HD035601& HD035961& HD036066& HD036130& HD036165& HD036395\\

 A10 & HD036512& HD036960& HD037202& HD037269& HD037272& HD037394& HD037526\\
     & HD037742& HD037958& HD038230& HD038529& HD038650& HD038856& HD038899\\

 A11 & HD039587& HD039773& HD039801& HD039866& HD040801& HD040964& HD041117\\
     & HD041330& HD041357& HD041501& HD041692& HD041808& HD042035& HD042250\\

 A12 & HD042353& HD042543& HD042597& HD042807& HD042983& HD043042& HD043153\\
     & HD043264& HD043285& HD043286& HD043526& HD044109& HD044274& HD044537\\

A13  & HD044614& HD045321& HD045391& HD045410& HD045829& HD045910& HD046223\\
     & HD046317& HD046380& HD046480& HD046588& HD046703& HD047127& HD047309\\
     
A14  & HD047839& HD048279& HD048682& HD049330& HD049409& HD049732& HD050522\\
     & HD050696& HD051219& HD051309& HD051530& HD052711& HD053929& HD054371\\
    
A15  & HD054717& HD055280& HD055575& HD055606& HD056925& HD056925& HD058343\\
     & HD058551& HD058946& HD059473& HD060179& HD060501& HD061606& HD062301\\
     
A16  & HD062613& HD063302& HD063778& HD064090& HD064332& HD064412& HD064606\\
     & HD065123& HD065583& HD066573& HD067767& HD068017& HD068638& HD069897\\
     
A17  & HD069897& HD070298& HD071148& HD071310& HD071881& HD072184& HD072905\\
     & HD072946& HD072968& HD073344& HD073668& HD074000& HD074156& HD074280\\
    
A18  & HD074377& HD075302& HD075318& HD075333& HD075732& HD075782& HD076813\\
     & HD076943& HD078175& HD078209& HD078249& HD078362& HD078418& HD078712\\
    
A19  & HD079028& HD079210& HD079452& HD079765& HD080081& HD080218& HD080536\\
     & HD082106& HD083425& HD084737& HD086133& HD086560& HD086728& HD086986\\
    
A20 & HD088446& HD088609& HD088725& HD088737& HD089010& HD089125& HD089269\\
    & HD089307& HD089744& HD089995& HD090537& HD090839& HD094028& HD094835\\

A21 & HD095128& HD095241& HD096094& HD096436& HD097560& HD097855& HD097916\\
    & HD099028& HD099747& HD100030& HD100446& HD100563& HD100696& HD101107\\
    
A22 & HD101177& HD101177B& HD101227& HD101501& HD101606& HD101690& HD102870\\
    & HD104556& HD104979& HD104985& HD105087& HD106038& HD106156& HD107213\\
     
A23 & HD107328& HD107582& HD108177& HD109358& HD109995& HD110897& HD112735\\
    & HD113002& HD114606& HD114710& HD114762& HD115136& HD115383& HD116316\\
     
A24 & HD117176& HD117243& HD118244& HD119291& HD120136& HD123299& HD124570\\
    & HD125560& HD126271& HD126511& HD126512& HD126660& HD128167& HD128987\\
    
A25 & HD129174& HD129336& HD131111& HD131156& HD131156B& HD131507& HD132756\\
    & HD134083& HD134113& HD135101& HD137391& HD138573& HD138749& HD138764\\
    
A26 & HD139457& HD141004& HD141272& HD142091& HD142860& HD142926& HD143807\\
    & HD144206& HD144284& HD145148& HD145389& HD147394& HD147677& HD148816\\
    
A27 & HD149121& HD149161& HD155358& HD155763& HD160762& HD164353& HD165029\\
    & HD165358& HD165670& HD166046& HD169822& HD173524& HD174912& HD175535\\
    
A28 & HD176437& HD180554& HD183144& HD185936& HD187123& HD187879& HD188001\\
    & HD188209& HD189087& HD190229& HD190603& HD192639& HD192907& HD193432\\
    
A29 & HD193793& HD194453& HD195198& HD195592& HD196426& HD196610& HD198183\\
    & HD198478& HD199478& HD199579& HD200580& HD206165& HD206374& HD208501\\
    
A30 & HD209459& HD209975& HD210809& HD211472& HD212076& HD212442& HD212454\\
    & HD212593& HD213420& HD214080& HD214167& HD214168& HD214680& HD215512\\
     
A31 & HD215704& HD216831& HD216916& HD217086& HD217833& HD217891& HD218045\\
    & HD218059& HD218376& HD220182& HD220787& HD220825& HD220933& HD221585\\
    
A32 & HD221830& HD224544& HD224559& HD224801& HD224926& HD224926& HD225160\\
    & HD233345& HD233511& HD237846& HD241253& LHS10& Ross-889\\
\hline
\end{tabular}
\end{table*}

\section{Summary}
\label{summary}
In this paper we present the first release  (1.0) of MEGARA-GTC stellar spectral (MEGASTAR)
library composed by 838 spectra from 414 different stars of a wide range of spectral types so that physical parameters. 

All the spectra are given reduced and calibrated with MEGARA DRP, which has been proved to be robust and reliable. The data have passed through all standard reduction steps: bias subtraction, fiber tracing and extraction, flat-field and wavelength calibration, instrument response correction and  flux calibration (Jy). The spectra were all taken as {\it filler}-type GTC OpenTime observations and were observed at different seeing conditions. Nevertheless, as we extracted 37 spaxels around the star's centroid position, most of the flux is guaranteed for seeing better than 2.5\arcsec. The spectra were extracted in the same wavelength intervals: 6420 - 6790\,\AA\ in HR-R and 8370 - 8885\,\AA\ in HR-I. 

We have compiled an  atlas with all the 838 spectra (Appendix A), the main characteristics of the stars in the release sample (Appendix B) and the Gaia RD2 data for 388 stars, from the 414 observed,  shared with our release 1.0 sample (Appendix C). We have described in this   work the atlas and tables content but   the three appendices will be published in electronic format only.

We have developed a database that has allowed us to handle the complete stellar library (2988 stars), to select and update the stars to be observed in the different semesters and to generate complex files for the GTC Phase2 tool to prepare the selected observations while guaranteeing a large enough number of stars ready to be observed any night due to the nature of the {\it filler}-type program.

All the 838 reduced and calibrated spectra of this release, MEGASTAR 1.0, will be immediately public after the acceptance of this article and  available to the whole scientific community. To facilitate the use of the released data, we have developed a web page with a web-based tool and a graphic interface to access and visualize the spectra (\url{https://www.fractal-es.com/megaragtc-stellarlibrary}). The users can access the data of 2988 in MEGASTAR catalogue, to retrieve  or visualize the individual spectra obtained in both spectral configurations, HR-R and HR-I, and to download the complete release.

We are confident that the 838 spectra of this MEGASTAR 1.0 release will be a precious tool to generate composed populations for the interpretation of MEGARA data but also they will be an attractive resource for stellar astronomers interested in the study of  the individual stars at this high resolution and wavelength intervals, specifically the HR-I range  with very few available data.

\section{Acknowledgements}
This work has been supported by MINECO-FEDER grants AYA2016-75808-R, AYA2016-79724-C4-3-P, RTI2018-096188-B-I00, AYA2017-90589-REDT and has been partially funded by FRACTAL, INAOE and CIEMAT. This work is based on observations made with the Gran Telescopio Canarias (GTC), installed in the Spanish Observatorio del Roque de los Muchachos, in the island of La Palma. This work is based on data obtained with MEGARA instrument, funded by European Regional Development Funds (ERDF), through {\it Programa Operativo Canarias FEDER 2014-2020}. The authors thank the support given by Dr. Antonio Cabrera and Dr. Daniel Reverte, GTC Operations Group staff, during the preparation and execution of the observations at the GTC. 
This research made use of Astropy, a community-developed core Python package for Astronomy. This research has made use of the SIMBAD database and the VizieR catalogue access tool, CDS, Strasbourg, France \mbox{(DOI: 10.26093/cds/vizier)}. The original description of the VizieR service was published in \mbox{A\&AS 143, 23}. This work has made use of data from the European Space Agency (ESA) mission {\it Gaia} (\url{https://www.cosmos.esa.int/gaia}), processed by the {\it Gaia} Data Processing and Analysis Consortium (DPAC, \url{https://www.cosmos.esa.int/web/gaia/dpac/consortium}). Funding for the DPAC has been provided by national institutions, in particular the institutions participating in the {\it Gaia} Multilateral Agreement.

We are very grateful to the reviewer as her/his comments and suggestions improved the manuscript. 

\section{Data Availability}
The fully reduced and calibrated spectra of MEGASTAR first release are  available at \url{https://www.fractal-es.com/megaragtc-stellarlibrary/private/home}. The access username is {\it{public}} and the password is  {\it{Q50ybAZm}}.

The supplementary material described as Appendices A, B and C is available in electronic format  exclusively. The summary of their content is the  following.  

Appendix A is the atlas of the  838 spectra of the  414 stars of this data release, i.e. the plots of all the spectra are shown. The information in the appendix is described inside including a table to find the page where the spectra for each star are located. This table is in the main manuscript as Table \ref{appendix-index}. The atlas is available as a pdf format file.

Appendix B is a table with the information associated to this data release. The columns description is given in Table \ref{release-header} of the main manuscript. This appendix is provided in ASCII and MSExcel formats. 

Appendix C  is a table with the available information in GAIA DR2 for the 388 stars from MEGASTAR library in common. The columns description is given as Table \ref{Gaia-table} in the main manuscript. This appendix  is provided in ASCII and MSExcel formats.

%
\bsp	
\label{lastpage}
\end{document}


\title[MEGASTAR. First Release. Appendix A]
{MEGARA-GTC Stellar Spectral Library (II). MEGASTAR First Release. Appendix A}

\author[E. Carrasco et al. ]
{E. Carrasco$^{1}$\thanks{E-mail:}, M. Moll\'{a}$^{2}$, M.L. Garc{\'\i}a-Vargas$^{3}$, A. Gil de Paz$^{4, 5}$,  N. Cardiel$^{4, 5}$, P. \newauthor
G\'{o}mez-Alvarez$^{3}$ and S. R. Berlanas$^{6}$
\\
$^{1}$ Instituto Nacional de Astrof{\'\i}sica, {\'O}ptica y Electr{\'o}nica, INAOE, Calle Luis Enrique Erro 1, C.P. 72840 Santa Mar{\'\i}a Tonantzintla, Puebla, Mexico\\
$^{2}$ Dpto. de Investigaci\'{o}n B\'{a}sica, CIEMAT, Avda. Complutense 40, E-28040 Madrid, Spain\\
$^{3}$ FRACTAL S.L.N.E., Calle Tulip{\'a}n 2, portal 13, 1A, E-28231 Las Rozas de Madrid, Spain \\
$^{4}$ Dpto. de F{\'\i}sica de la Tierra y Astrof{\'\i}sica, Fac. CC. F{\'\i}sicas, Universidad Complutense de Madrid, Plaza de las Ciencias, 1, E-28040 Madrid, Spain \\
$^{5}$ Instituto de F{\'\i}sica de Part{\'\i}culas y del Cosmos, IPARCOS, Fac. CC. F{\'\i}sicas, Universidad Complutense de Madrid, Plaza de las Ciencias 1, E-28040 Madrid, Spain\\
$^{6}$ Universidad de Alicante, 03690 San Vicente del Raspeig, Alicante, Spain}

\date{Accepted Received ; in original form }
\pagerange{\pageref{firstpage}--\pageref{lastpage}} \pubyear{2019}

\maketitle
\label{firstpage}

\begin{abstract}

MEGARA  is an optical integral field and  multi-object   fibre-based  spectrograph for the  10.4m Gran Telescopio CANARIAS that offers medium to high spectral resolutions (FWHM) of  R~$\simeq$~6000, 12000, 20000.  Commissioned at the telescope in 2017, it started operation as a common-user instrument in  2018. We are creating an instrument-oriented empirical spectral library from MEGARA-GTC star observations, MEGASTAR, crucial  for the  correct interpretation of MEGARA  data. This piece of work describes the content of the first release of MEGASTAR, formed by the spectra of 414 stars observed with R~$\simeq$~20000 in the spectral intervals 6420\ -- 6790\,\AA\ and  8370\ -- 8885\,{\AA}, and obtained with a continuum average signal to noise ratio around 260. We describe the release sample, the observations, the data reduction procedure and the MEGASTAR database, available to the community with the acceptance of this paper. In this Appendix A we show the 838 spectra of the first release of the MEGASTAR catalogue.
\end{abstract}

\begin{keywords} Astronomical data bases: atlases -- Astronomical data bases:catalogues
stars: abundance -- stars: fundamental parameters (Galaxy:) globular clusters: individual: M15 
\end{keywords}

\section{MEGASTAR First Release: stellar spectra}

In this appendix we show all spectra from the first release of MEGASTAR catalogue. The spectra are flux-calibrated (Jansky, {\it Jy}) using the standard star provided by GTC together with the observing block. However, the observing conditions were not homogeneous because the nature of the {\it filler} GTC-program itself (see main paper), so that the flux calibration error might be large and cannot be determined. 
Figure~\ref{sp-release-838} displays the 838 spectra. The stars are ordered by name. To facilitate the search of the star spectra, Table \ref{appendix-index} lists the name of the stars shown in each page of the atlas. For each star, we display the spectra taken with the HR-R (at the left) and the HR-I (at the right) gratings. The star name and the stellar parameters compiled from the literature have been included as labels in the plots. The color of the spectrum has been associated to the star spectral type in the literature as follows: purple, WR; blue, O; cyan, B; green, A; orange, F;  red, G; magenta, K; maroon, M; grey, S; black, flat.  

\begin{table*}
\caption{Summary of the stars of the MEGASTAR first release whose spectra are shown in this appendix. The first column is the number of the atlas page where the spectra can be found. The stars are listed in the same order than they are  displaying in each page of this atlas.}
\label{appendix-index}
 \begin{tabular}{l l l l l l ll}
\hline
Page & Stars& & & & & & \\
\hline 
  A3 & BD-032525& BD-122669& BD+083095& BD+092190& BD+130013& BD+191730& BD+195116B\\
     & BD+203603& BD+241676& BD+262606& BD+351484& BD+381670& BD+511696& BD+541399 \\
 
  A4 & BD+800245& G171-010& G197-45& G202-65& G234-28& HD000108& HD000358\\
     & HD000560  & HD000886& HD003360& HD003369& HD003628& HD003644& HD004004\\
 
  A5 & HD004539 & HD006327& HD006815& HD007374& HD009974& HD009996& HD013267\\
     & HD013268& HD014191& HD014633& HD014947& HD015318& HD015558& HD015570\\
   
  A6 & HD015629& HD016429& HD016523& HD016523& HD016581& HD017081& HD017145\\
     & HD017506& HD017638& HD017638& HD018144& HD018296& HD018409& HD019308\\
 
  A7 & HD020084& HD020512& HD021742& HD022484& HD023862& HD024341& HD024451\\
     & HD024534& HD024912& HD025173& HD025825& HD026756& HD027126& HD027282\\
 
  A8 &  HD027295& HD027371& HD027524& HD027685& HD028005& HD029645& HD030614\\
     & HD030649& HD030676& HD031219& HD031293& HD031374& HD031996& HD032537\\

  A9 & HD033632& HD033904& HD034078& HD034255& HD034797& HD034816& HD035468\\
     & HD035497& HD035601& HD035961& HD036066& HD036130& HD036165& HD036395\\

 A10 & HD036512& HD036960& HD037202& HD037269& HD037272& HD037394& HD037526\\
     & HD037742& HD037958& HD038230& HD038529& HD038650& HD038856& HD038899\\

 A11 & HD039587& HD039773& HD039801& HD039866& HD040801& HD040964& HD041117\\
     & HD041330& HD041357& HD041501& HD041692& HD041808& HD042035& HD042250\\

 A12 & HD042353& HD042543& HD042597& HD042807& HD042983& HD043042& HD043153\\
     & HD043264& HD043285& HD043286& HD043526& HD044109& HD044274& HD044537\\

A13  & HD044614& HD045321& HD045391& HD045410& HD045829& HD045910& HD046223\\
     & HD046317& HD046380& HD046480& HD046588& HD046703& HD047127& HD047309\\
     
A14  & HD047839& HD048279& HD048682& HD049330& HD049409& HD049732& HD050522\\
     & HD050696& HD051219& HD051309& HD051530& HD052711& HD053929& HD054371\\
    
A15  & HD054717& HD055280& HD055575& HD055606& HD056925& HD056925& HD058343\\
     & HD058551& HD058946& HD059473& HD060179& HD060501& HD061606& HD062301\\
     
A16  & HD062613& HD063302& HD063778& HD064090& HD064332& HD064412& HD064606\\
     & HD065123& HD065583& HD066573& HD067767& HD068017& HD068638& HD069897\\
     
A17  & HD069897& HD070298& HD071148& HD071310& HD071881& HD072184& HD072905\\
     & HD072946& HD072968& HD073344& HD073668& HD074000& HD074156& HD074280\\
    
A18  & HD074377& HD075302& HD075318& HD075333& HD075732& HD075782& HD076813\\
     & HD076943& HD078175& HD078209& HD078249& HD078362& HD078418& HD078712\\
    
A19  & HD079028& HD079210& HD079452& HD079765& HD080081& HD080218& HD080536\\
     & HD082106& HD083425& HD084737& HD086133& HD086560& HD086728& HD086986\\
    
A20 & HD088446& HD088609& HD088725& HD088737& HD089010& HD089125& HD089269\\
    & HD089307& HD089744& HD089995& HD090537& HD090839& HD094028& HD094835\\

A21 & HD095128& HD095241& HD096094& HD096436& HD097560& HD097855& HD097916\\
    & HD099028& HD099747& HD100030& HD100446& HD100563& HD100696& HD101107\\
    
A22 & HD101177& HD101177B& HD101227& HD101501& HD101606& HD101690& HD102870\\
    & HD104556& HD104979& HD104985& HD105087& HD106038& HD106156& HD107213\\
     
A23 & HD107328& HD107582& HD108177& HD109358& HD109995& HD110897& HD112735\\
    & HD113002& HD114606& HD114710& HD114762& HD115136& HD115383& HD116316\\
     
A24 & HD117176& HD117243& HD118244& HD119291& HD120136& HD123299& HD124570\\
    & HD125560& HD126271& HD126511& HD126512& HD126660& HD128167& HD128987\\
    
A25 & HD129174& HD129336& HD131111& HD131156& HD131156B& HD131507& HD132756\\
    & HD134083& HD134113& HD135101& HD137391& HD138573& HD138749& HD138764\\
    
A26 & HD139457& HD141004& HD141272& HD142091& HD142860& HD142926& HD143807\\
    & HD144206& HD144284& HD145148& HD145389& HD147394& HD147677& HD148816\\
    
A27 & HD149121& HD149161& HD155358& HD155763& HD160762& HD164353& HD165029\\
    & HD165358& HD165670& HD166046& HD169822& HD173524& HD174912& HD175535\\
    
A28 & HD176437& HD180554& HD183144& HD185936& HD187123& HD187879& HD188001\\
    & HD188209& HD189087& HD190229& HD190603& HD192639& HD192907& HD193432\\
    
A29 & HD193793& HD194453& HD195198& HD195592& HD196426& HD196610& HD198183\\
    & HD198478& HD199478& HD199579& HD200580& HD206165& HD206374& HD208501\\
    
A30 & HD209459& HD209975& HD210809& HD211472& HD212076& HD212442& HD212454\\
    & HD212593& HD213420& HD214080& HD214167& HD214168& HD214680& HD215512\\
     
A31 & HD215704& HD216831& HD216916& HD217086& HD217833& HD217891& HD218045\\
    & HD218059& HD218376& HD220182& HD220787& HD220825& HD220933& HD221585\\
    
A32 & HD221830& HD224544& HD224559& HD224801& HD224926& HD224926& HD225160\\
    & HD233345& HD233511& HD237846& HD241253& LHS10& Ross-889\\
\hline
\end{tabular}
\end{table*}

\begin{figure*}
\includegraphics[width=0.18\textwidth,angle=-90]{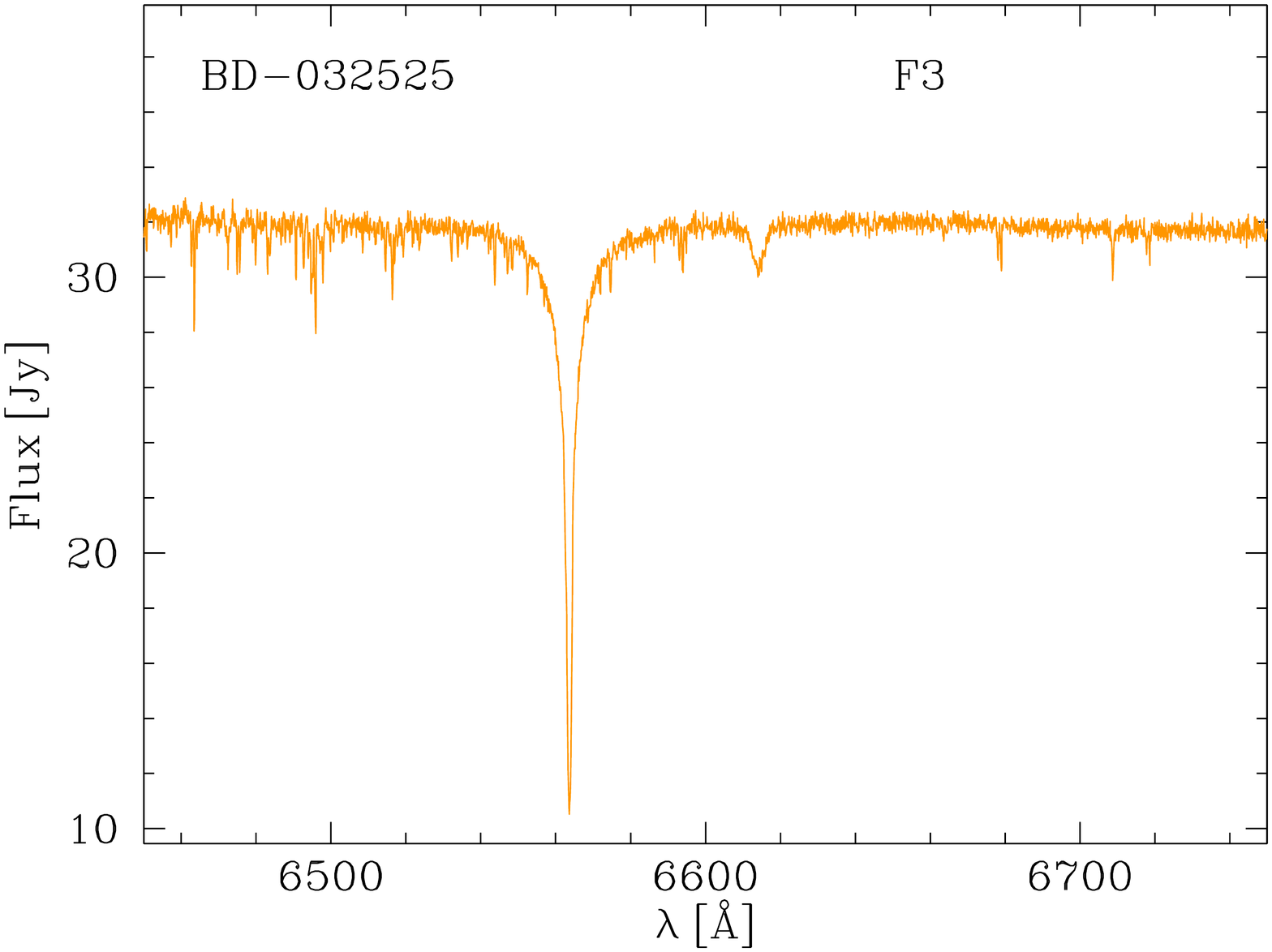}
\includegraphics[width=0.18\textwidth,angle=-90]{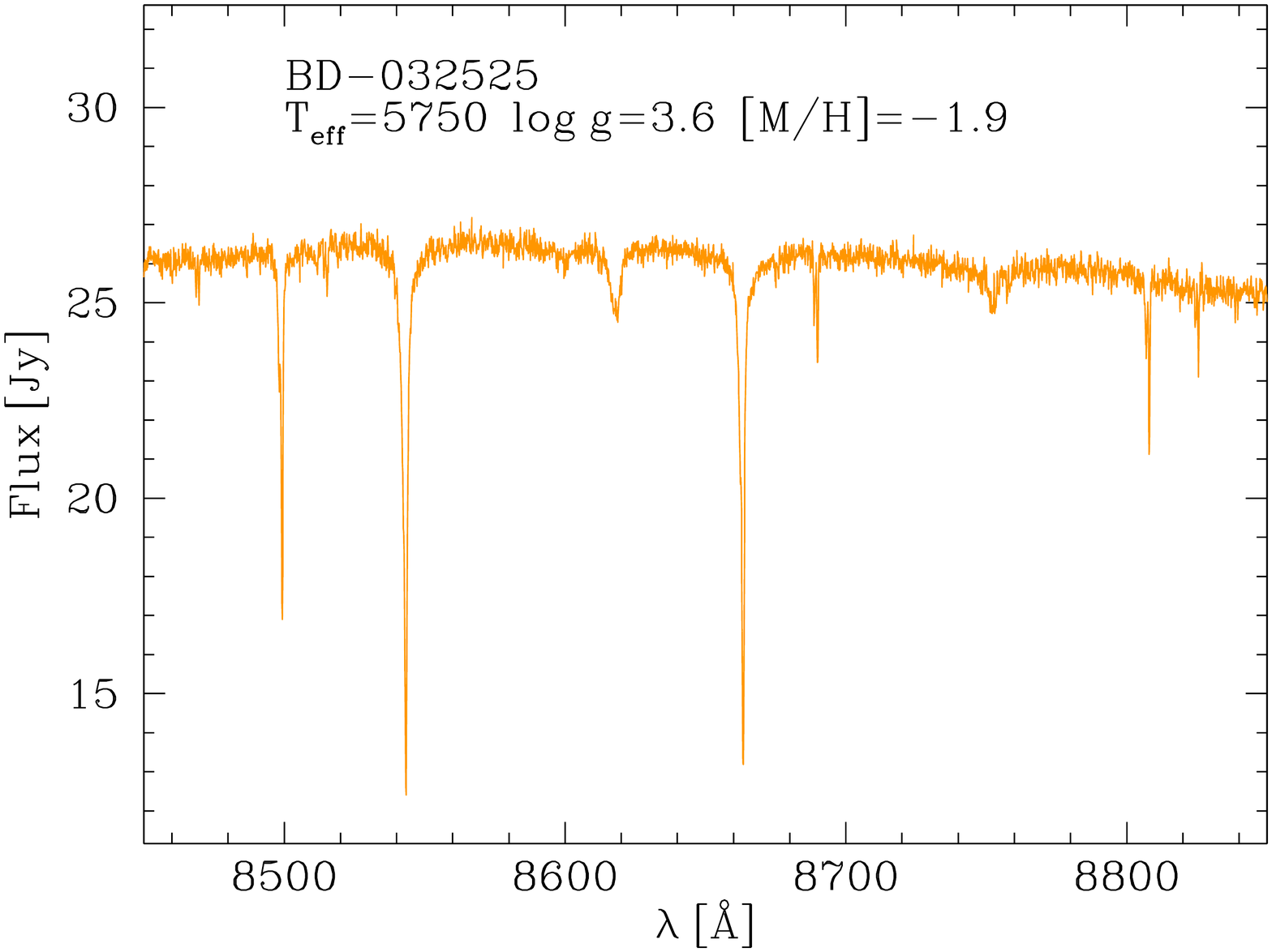}
\includegraphics[width=0.18\textwidth,angle=-90]{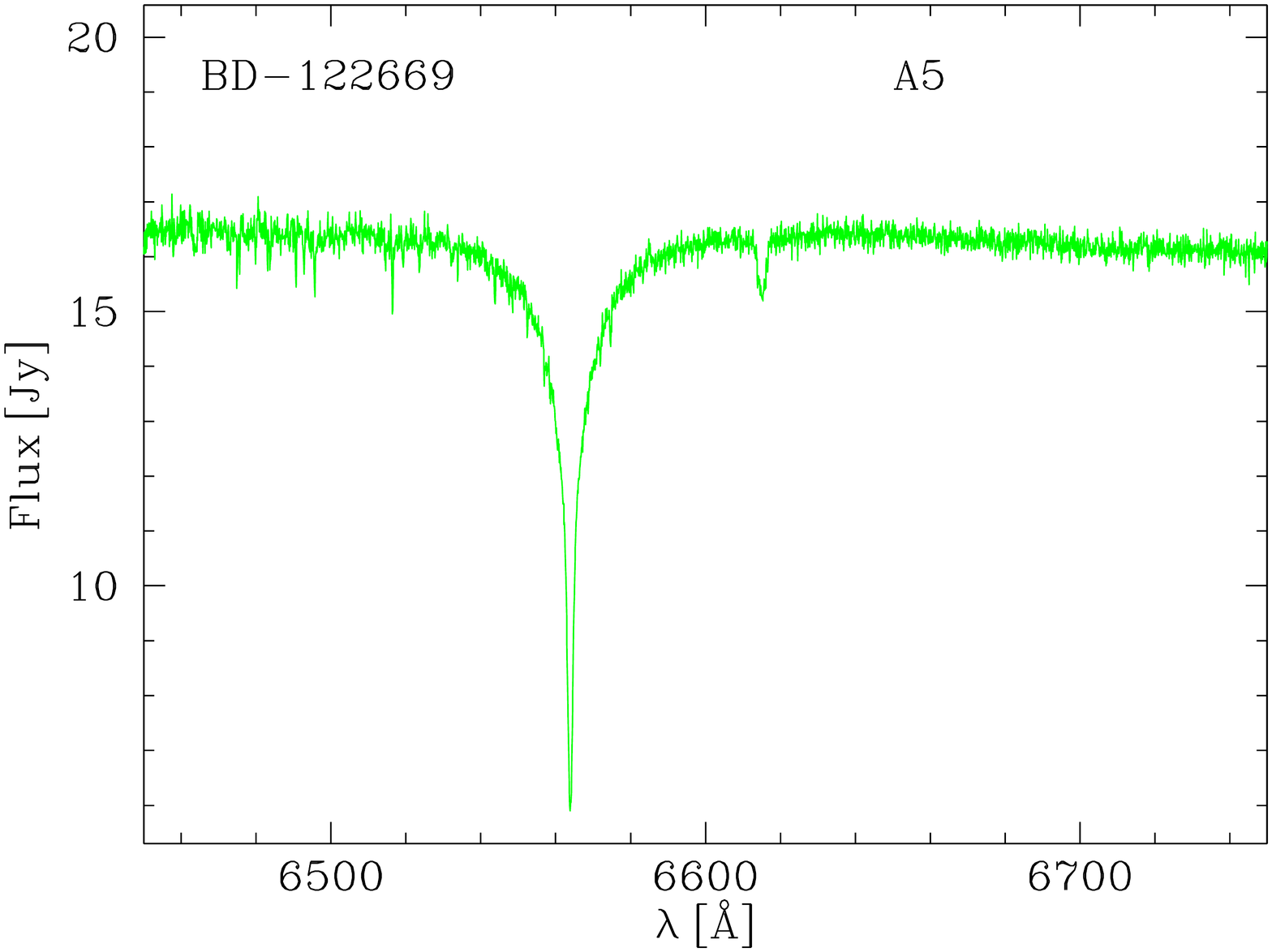}
\includegraphics[width=0.18\textwidth,angle=-90]{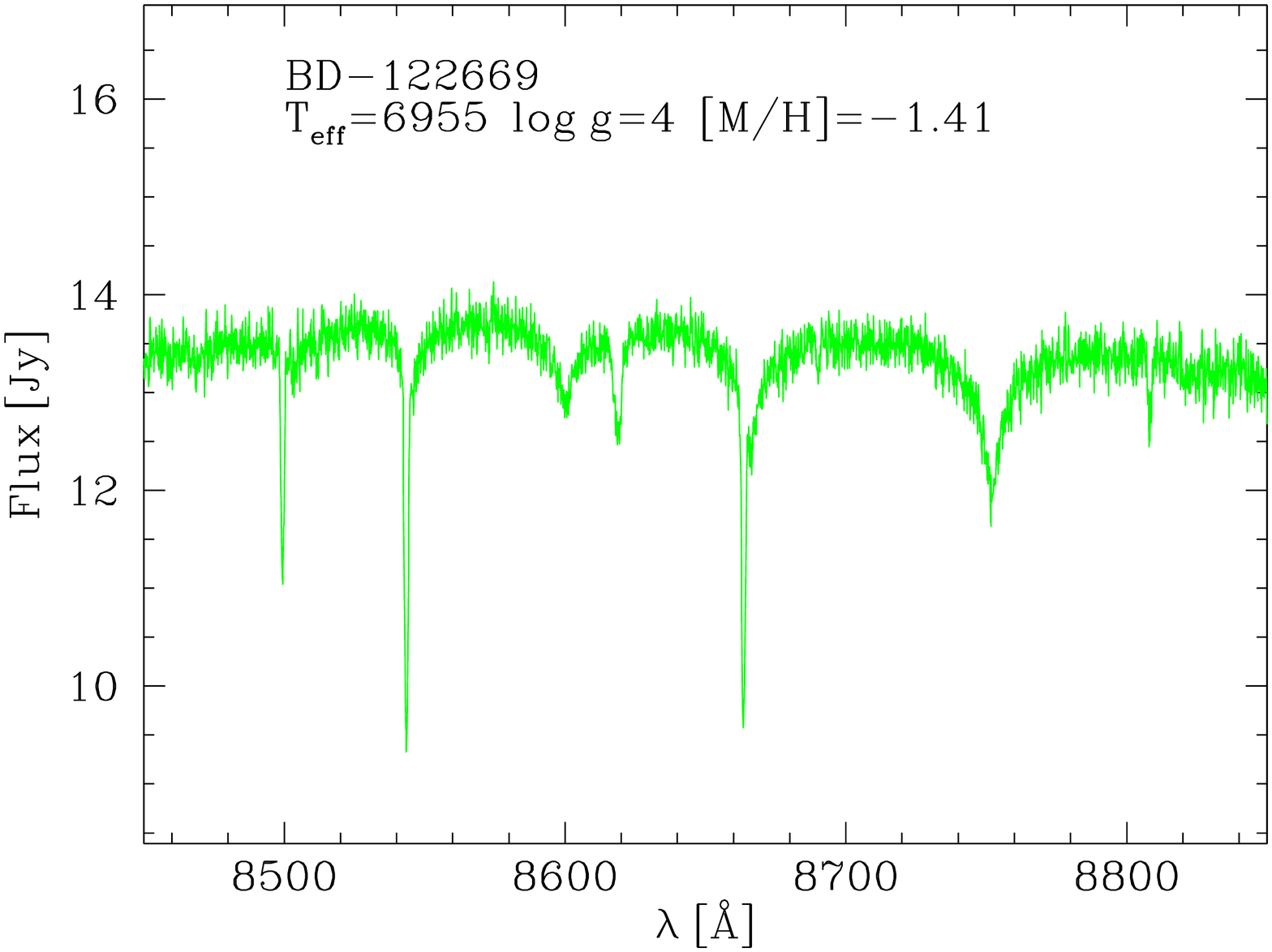}
\includegraphics[width=0.18\textwidth,angle=-90]{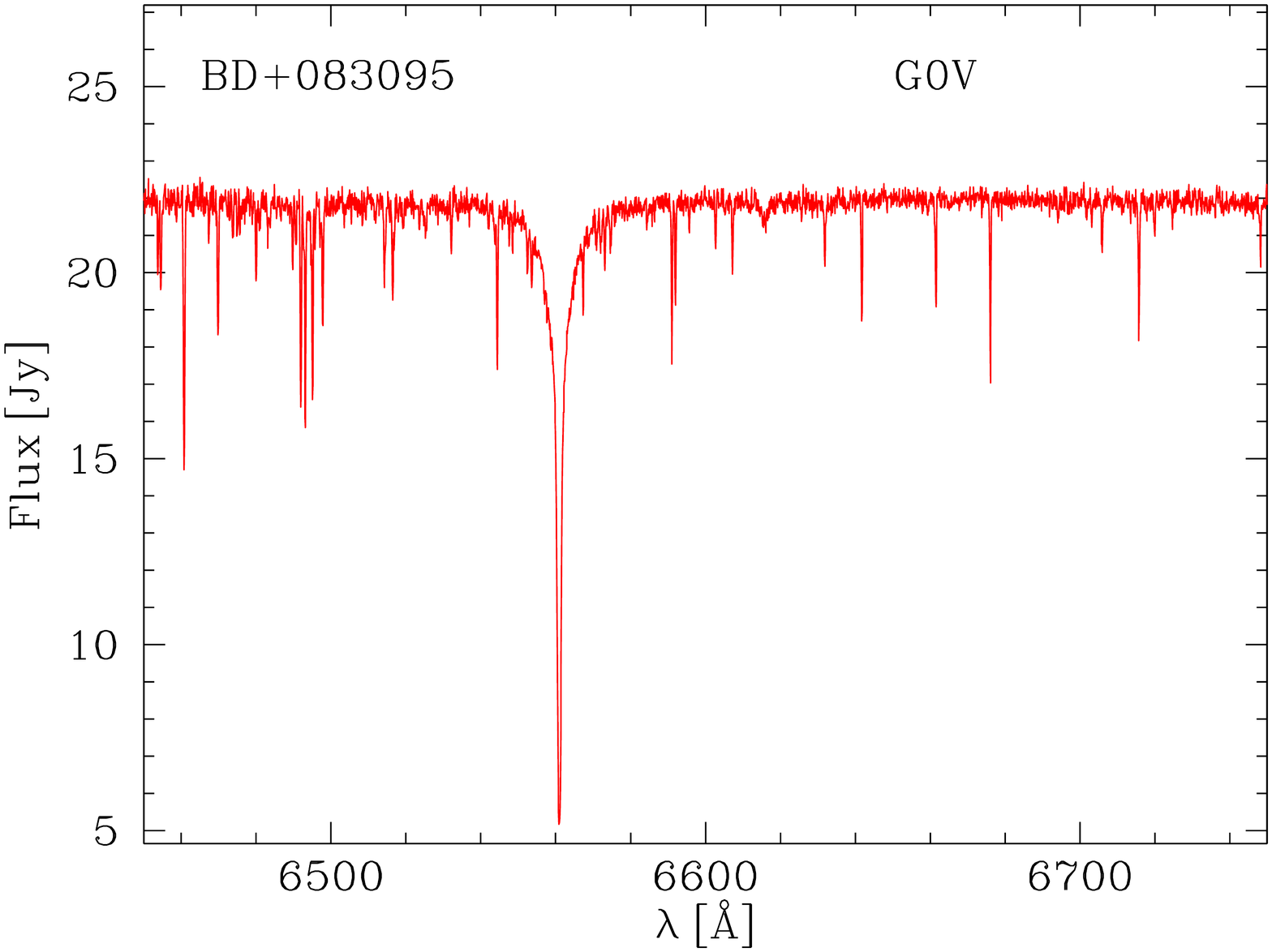}
\includegraphics[width=0.18\textwidth,angle=-90]{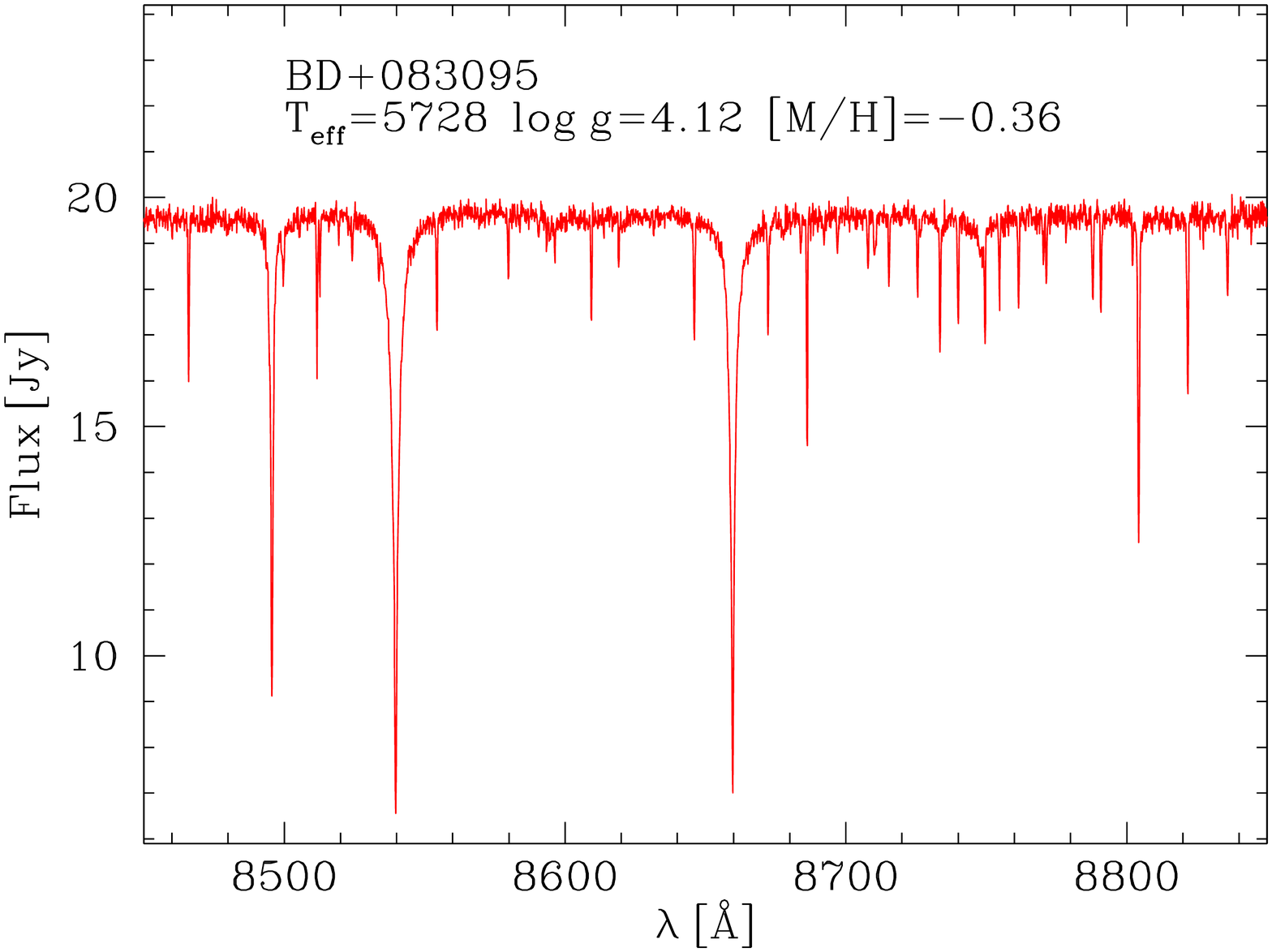}
\includegraphics[width=0.18\textwidth,angle=-90]{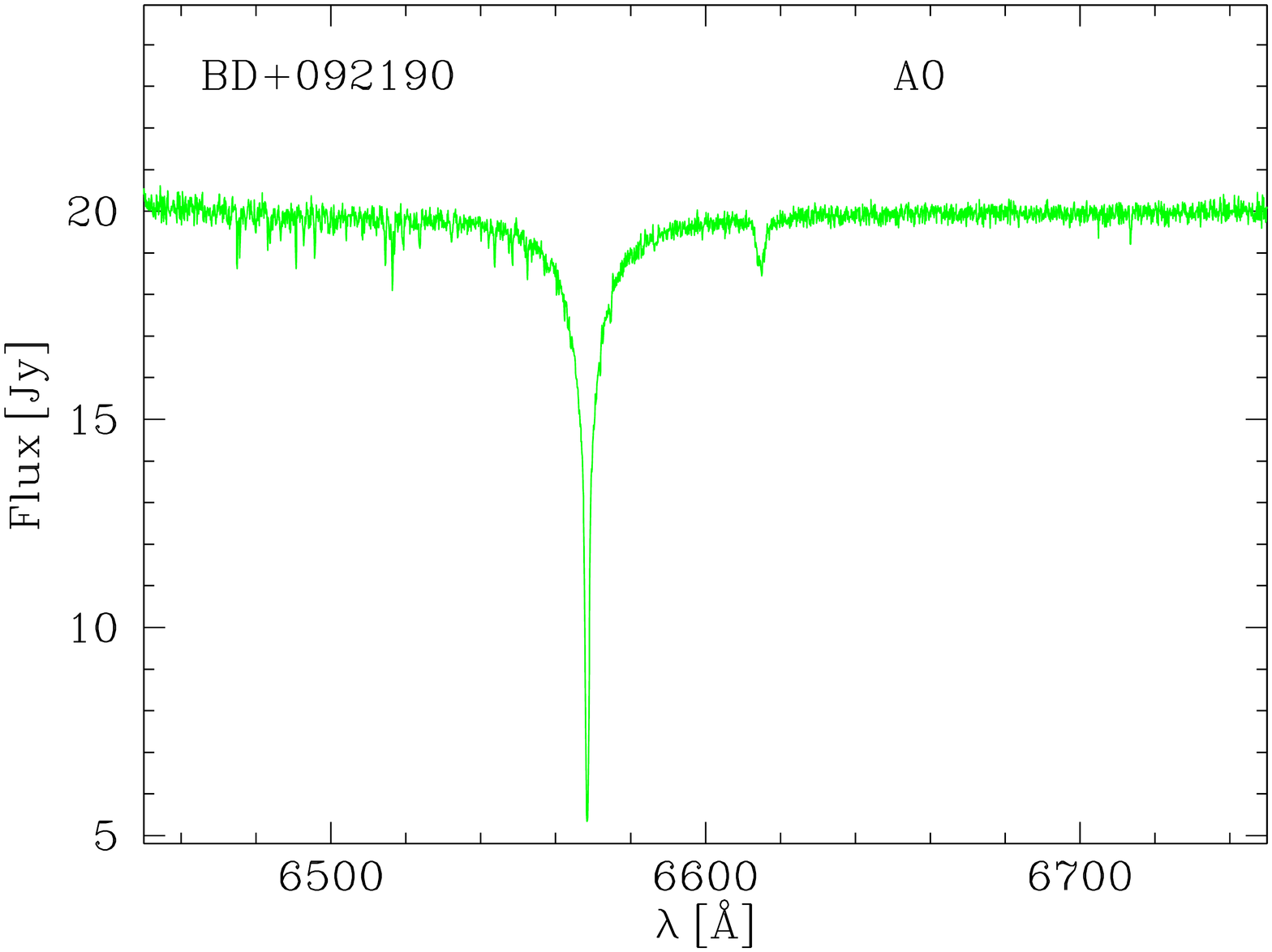}
\includegraphics[width=0.18\textwidth,angle=-90]{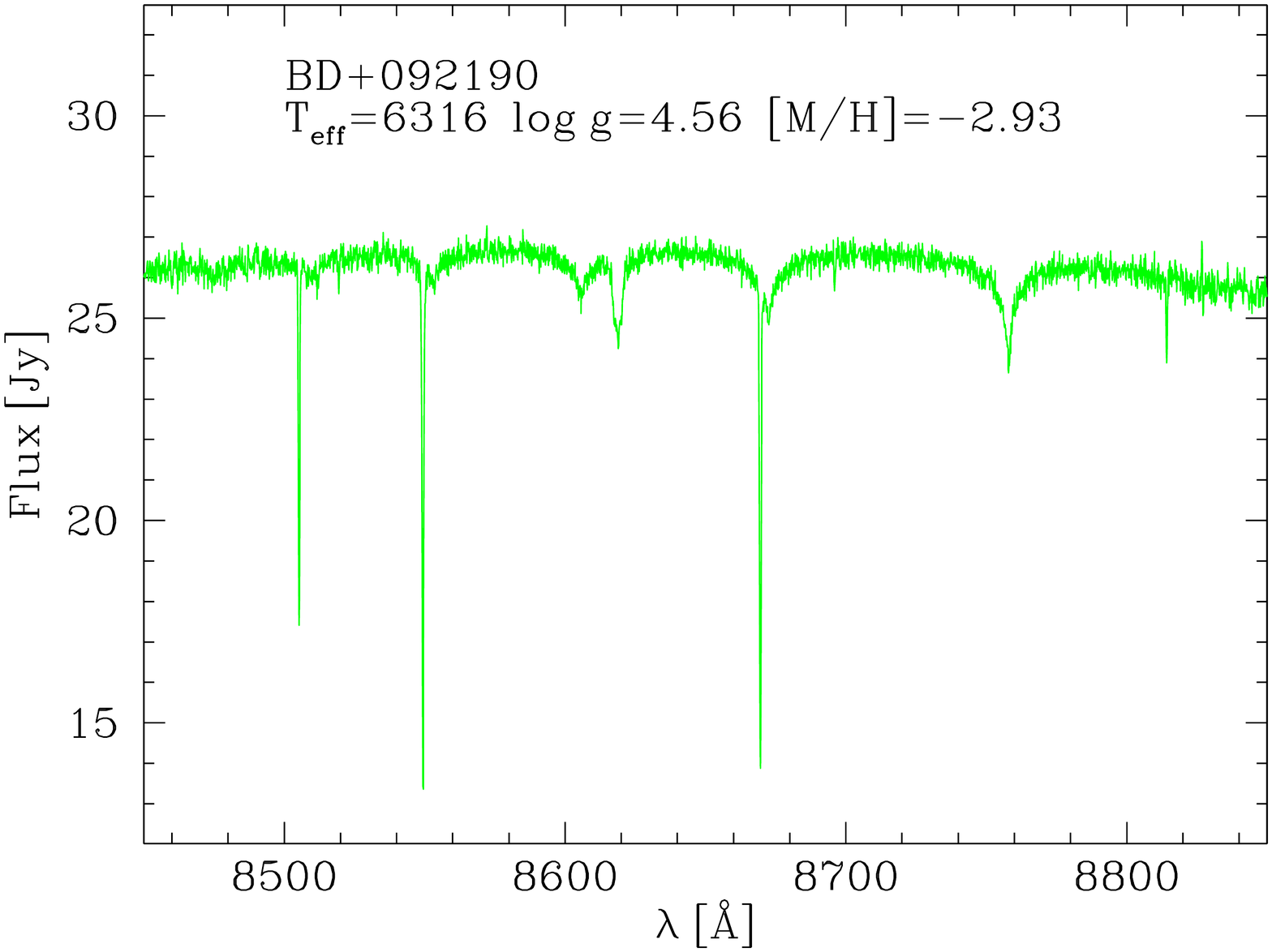}
\includegraphics[width=0.18\textwidth,angle=-90]{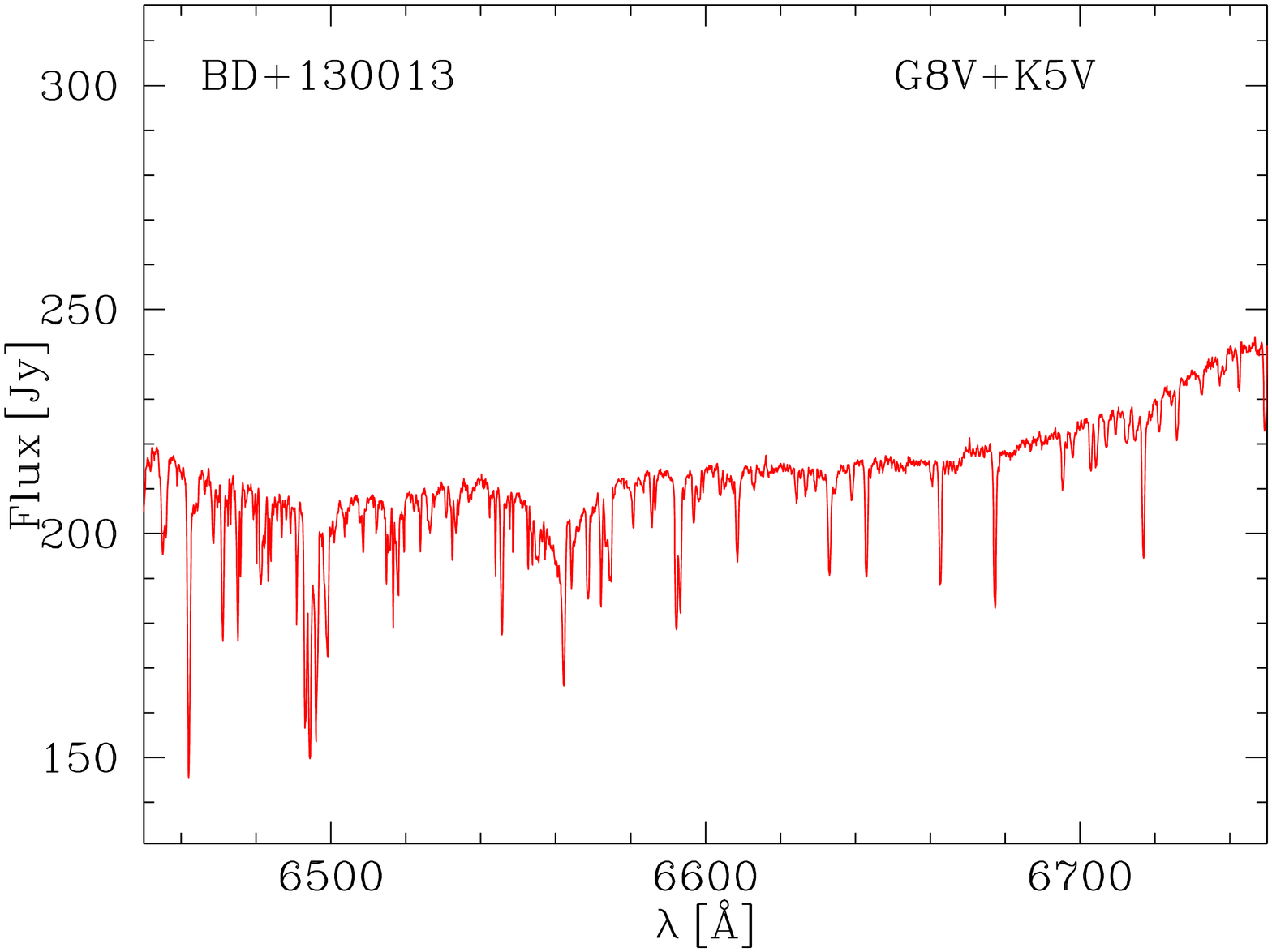}
\includegraphics[width=0.18\textwidth,angle=-90]{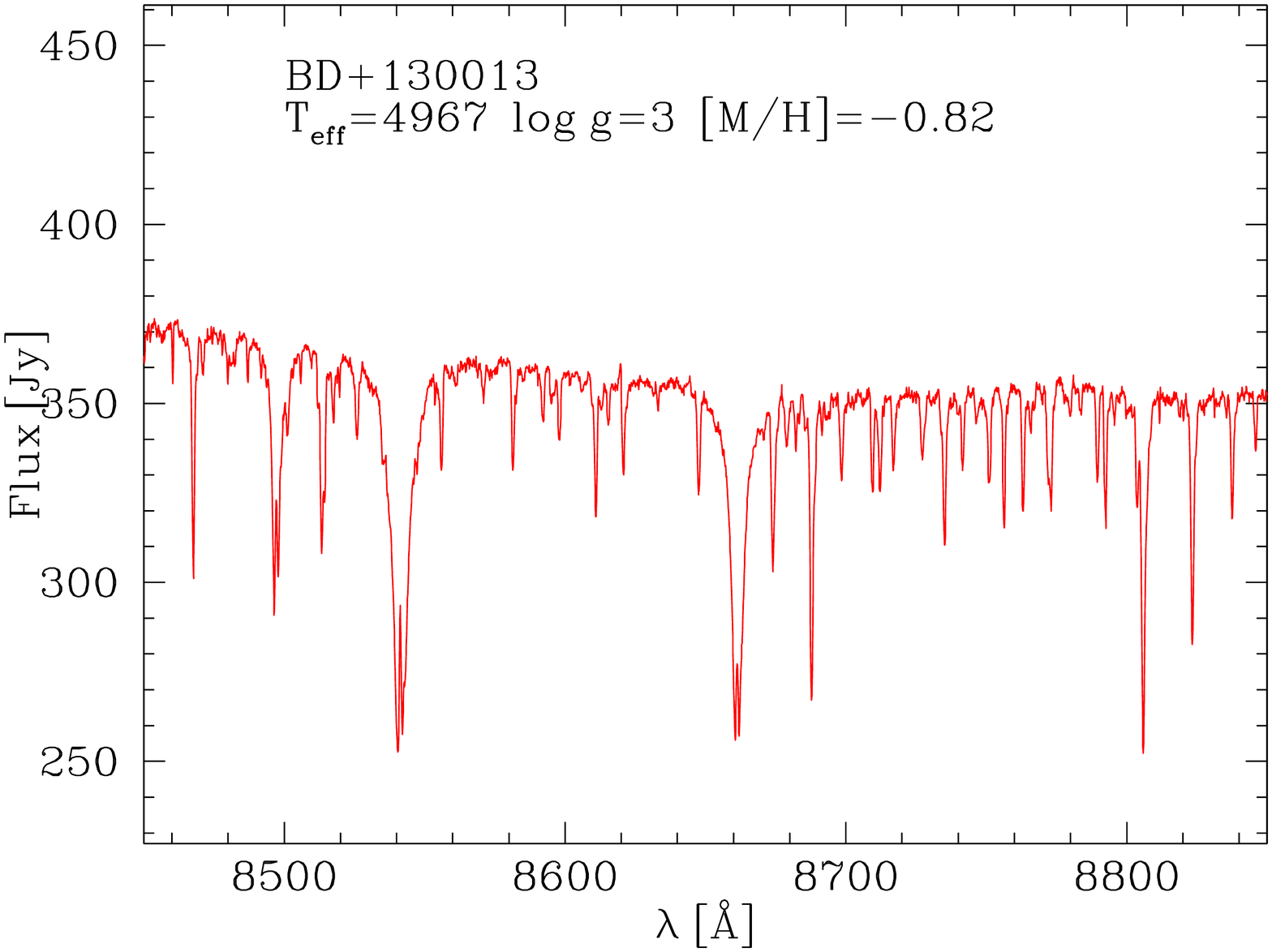}
\includegraphics[width=0.18\textwidth,angle=-90]{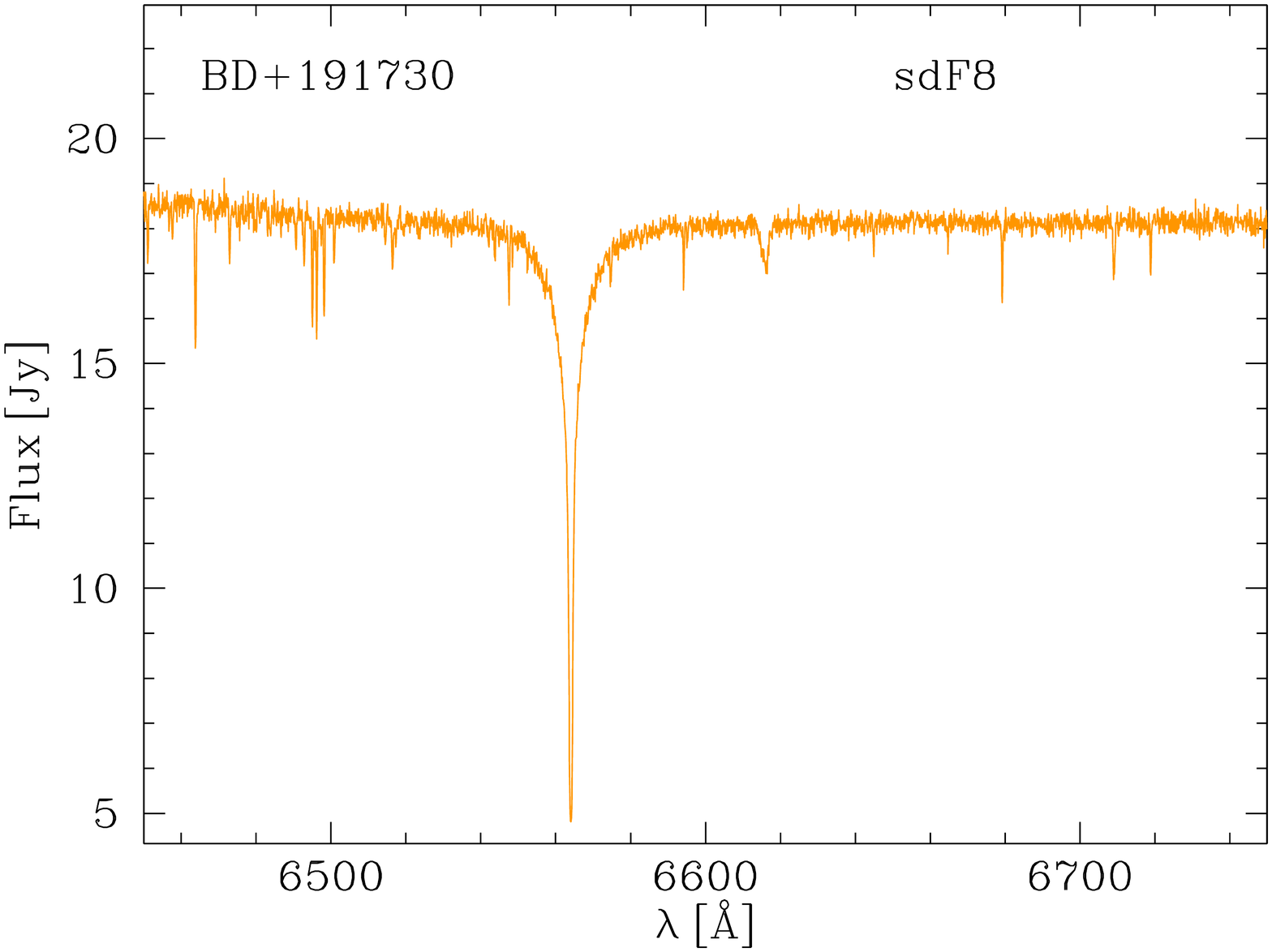}
\includegraphics[width=0.18\textwidth,angle=-90]{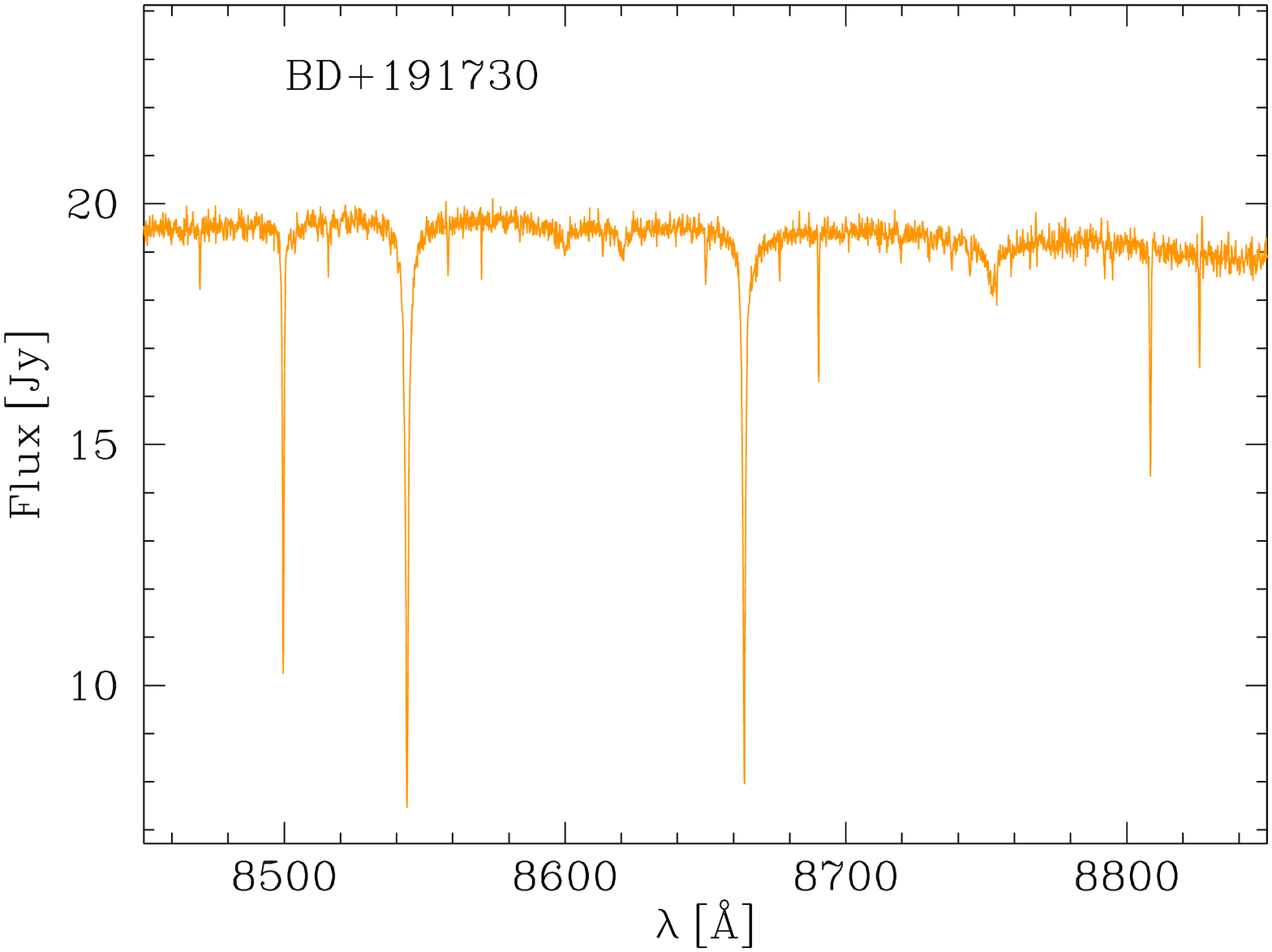}
\includegraphics[width=0.18\textwidth,angle=-90]{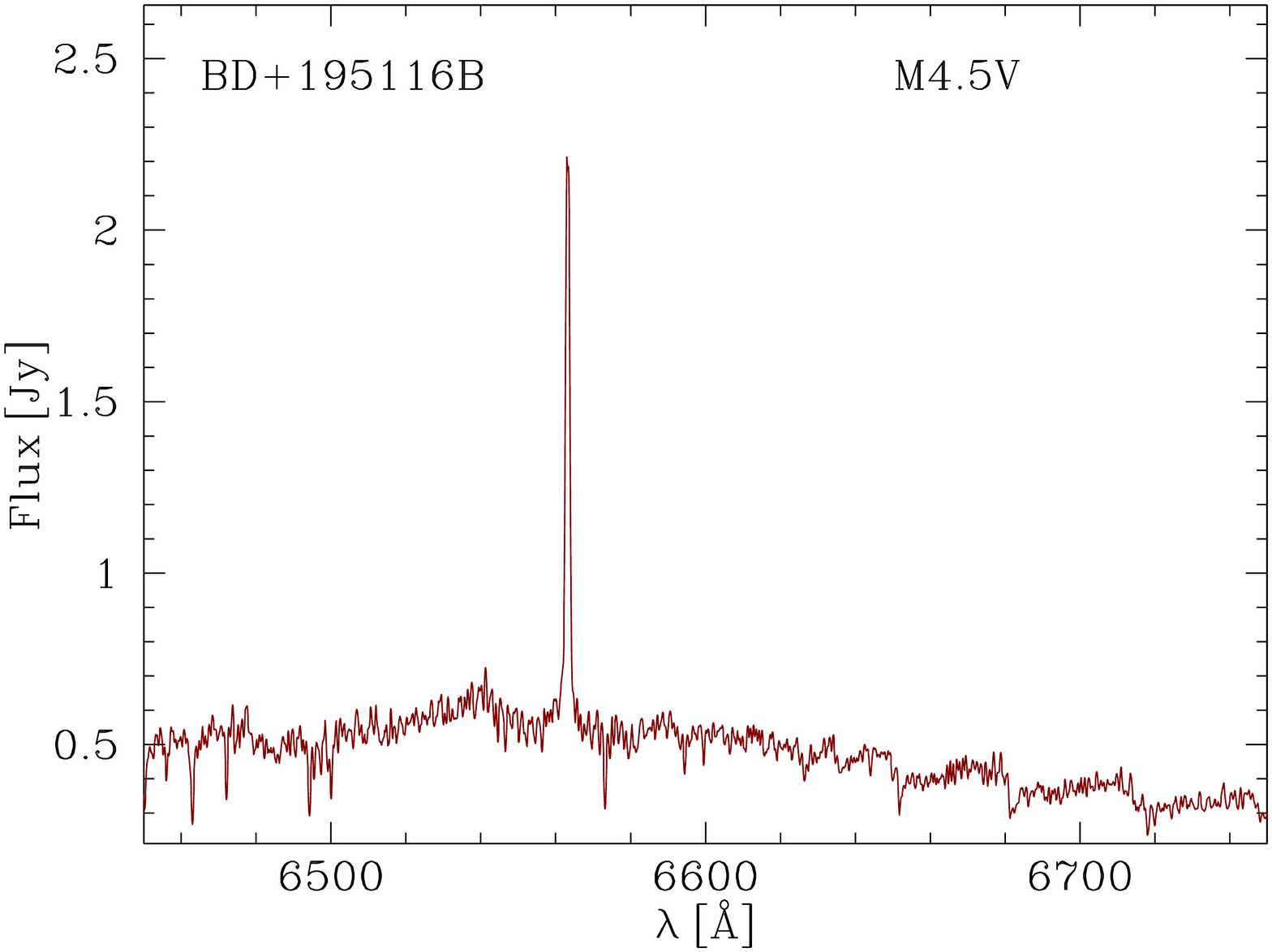}
\includegraphics[width=0.18\textwidth,angle=-90]{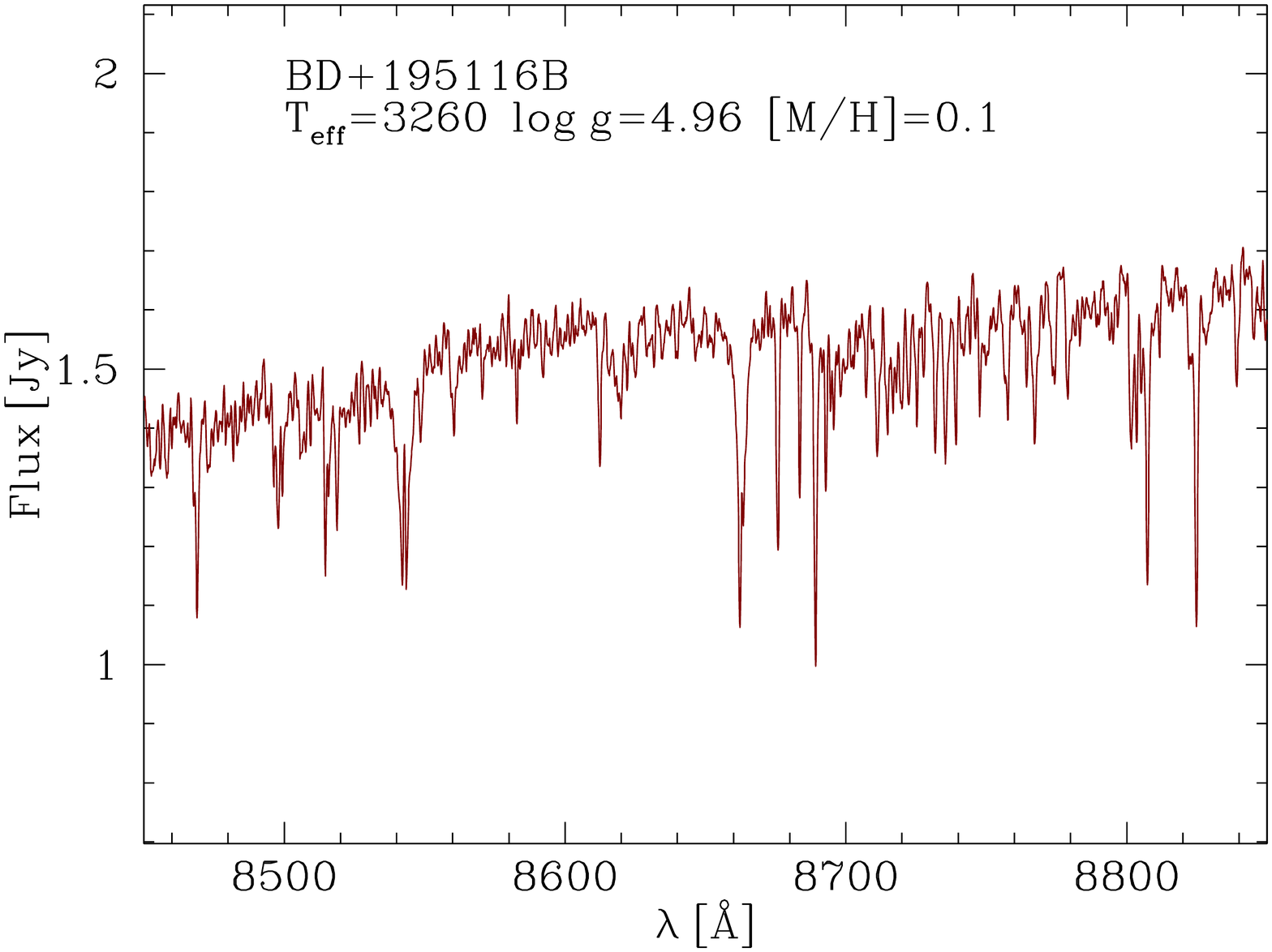}
\includegraphics[width=0.18\textwidth,angle=-90]{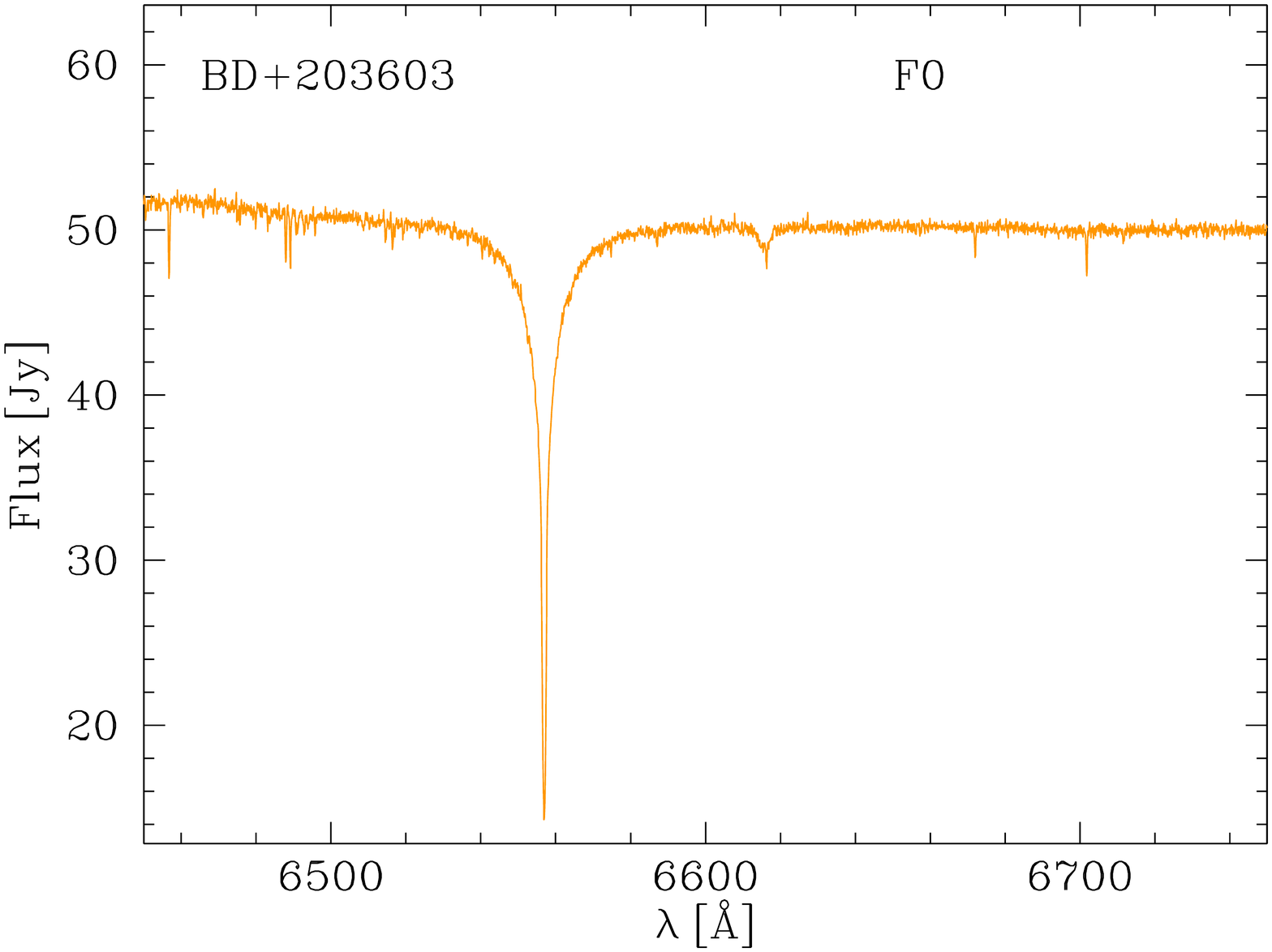}
\includegraphics[width=0.18\textwidth,angle=-90]{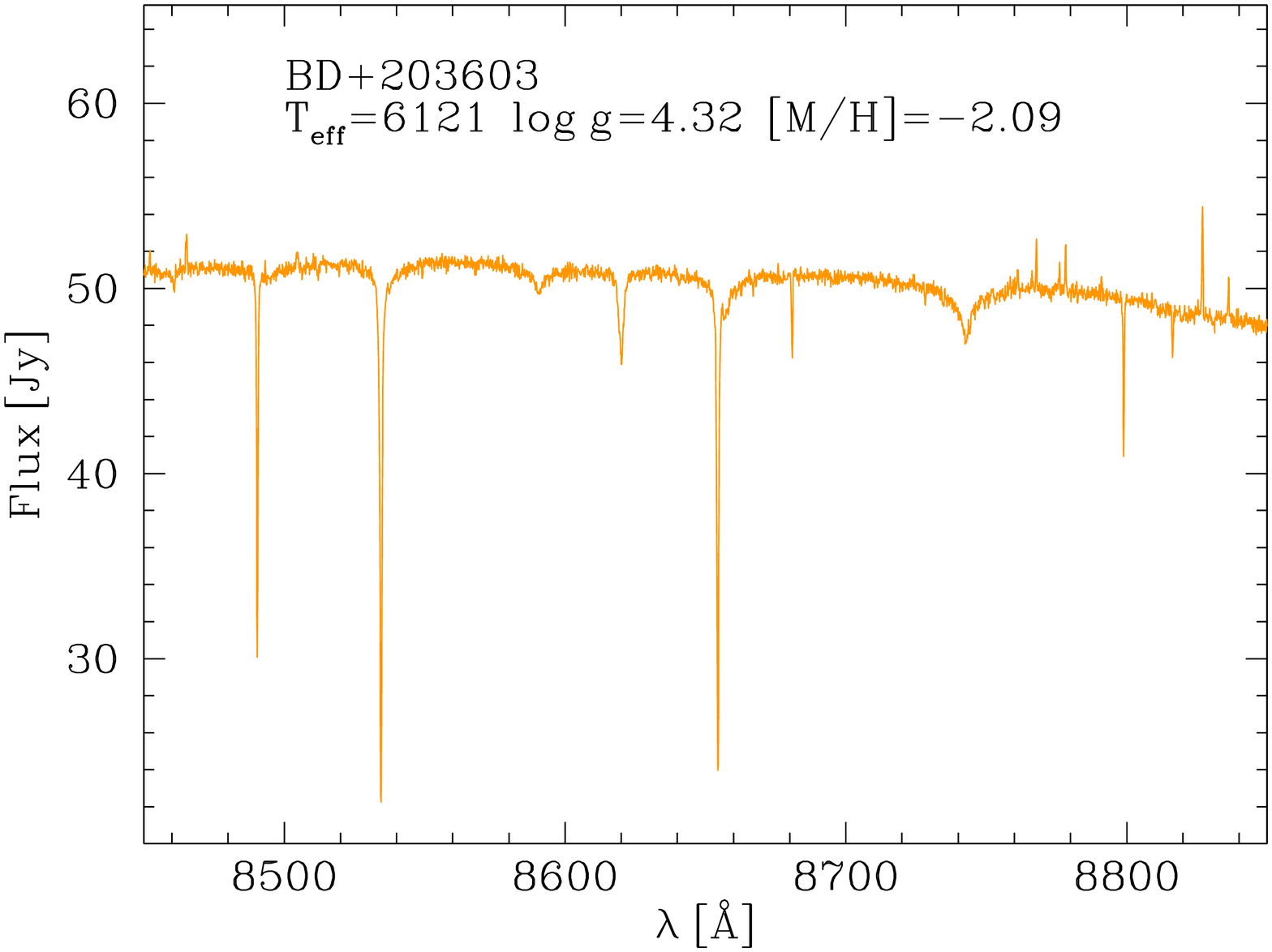}
\includegraphics[width=0.18\textwidth,angle=-90]{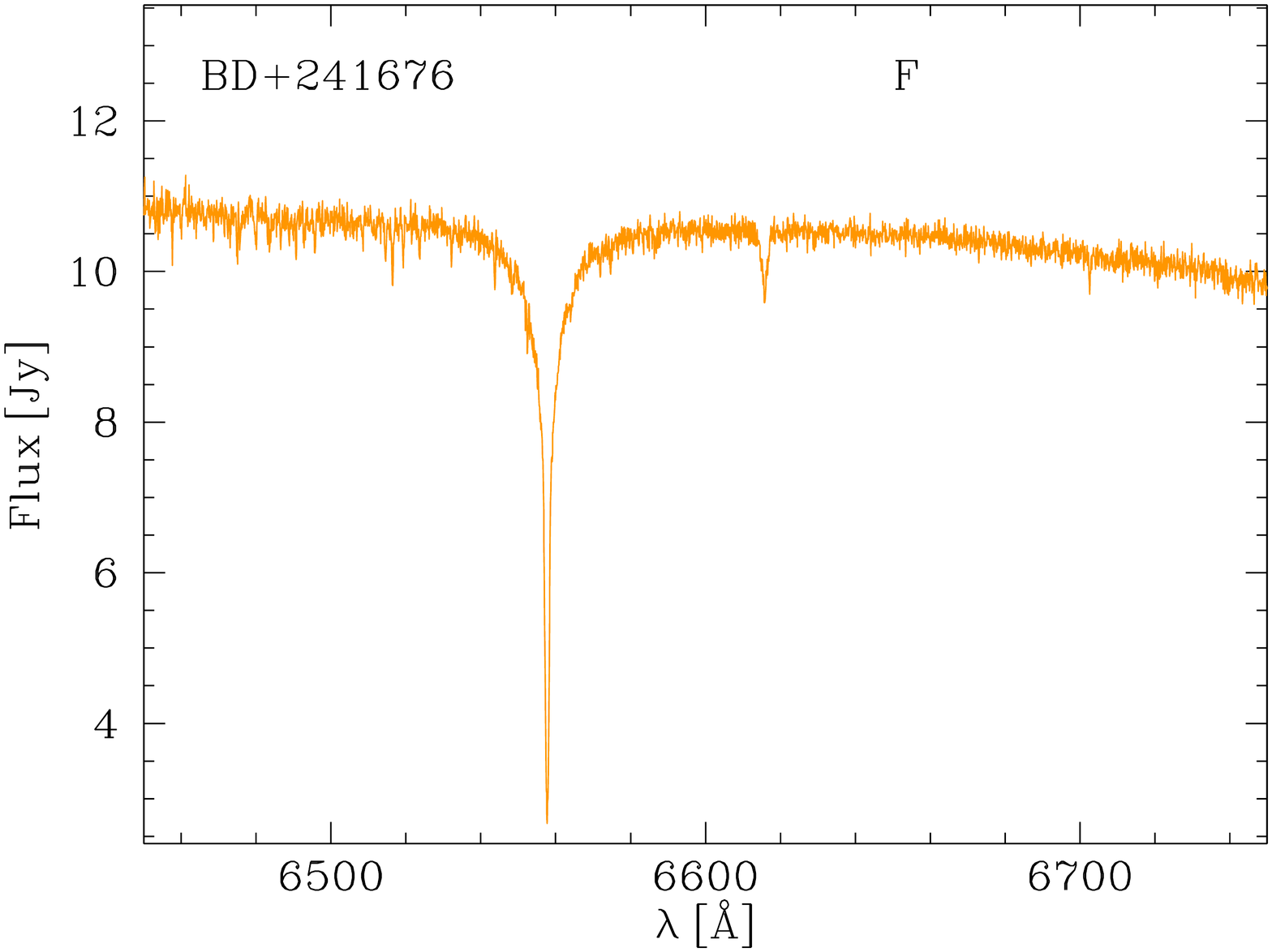}
\includegraphics[width=0.18\textwidth,angle=-90]{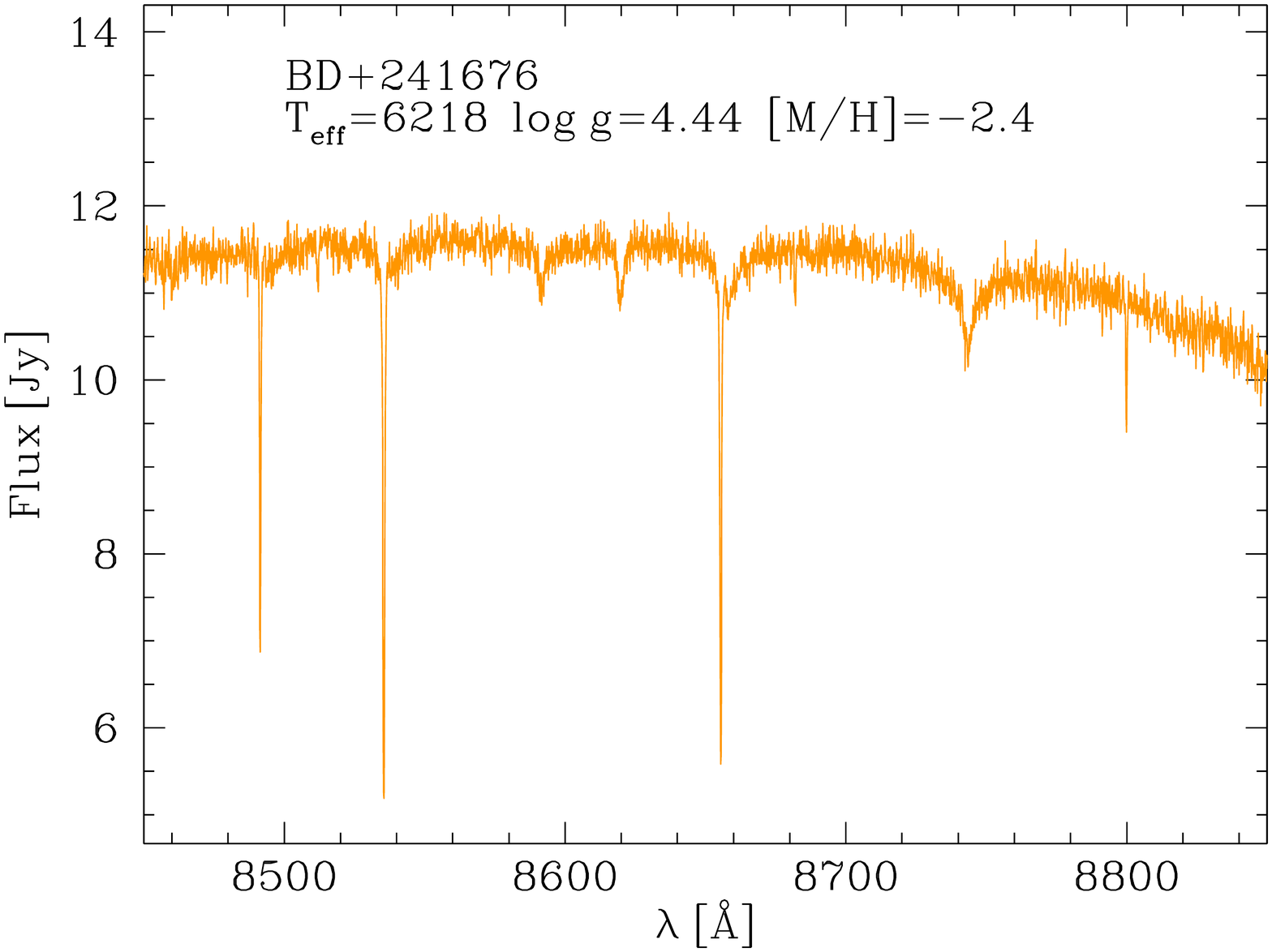}
\includegraphics[width=0.18\textwidth,angle=-90]{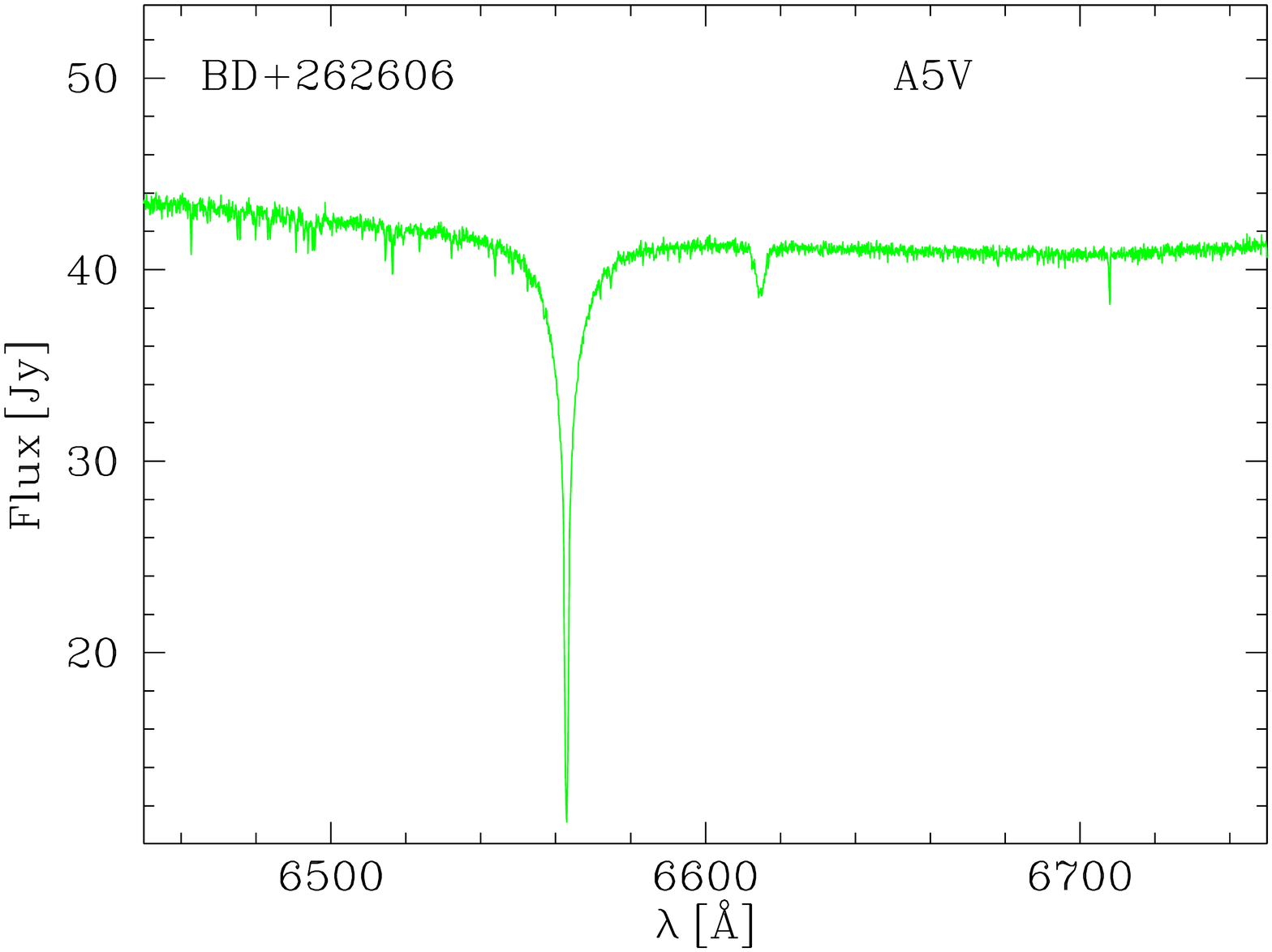}
\includegraphics[width=0.18\textwidth,angle=-90]{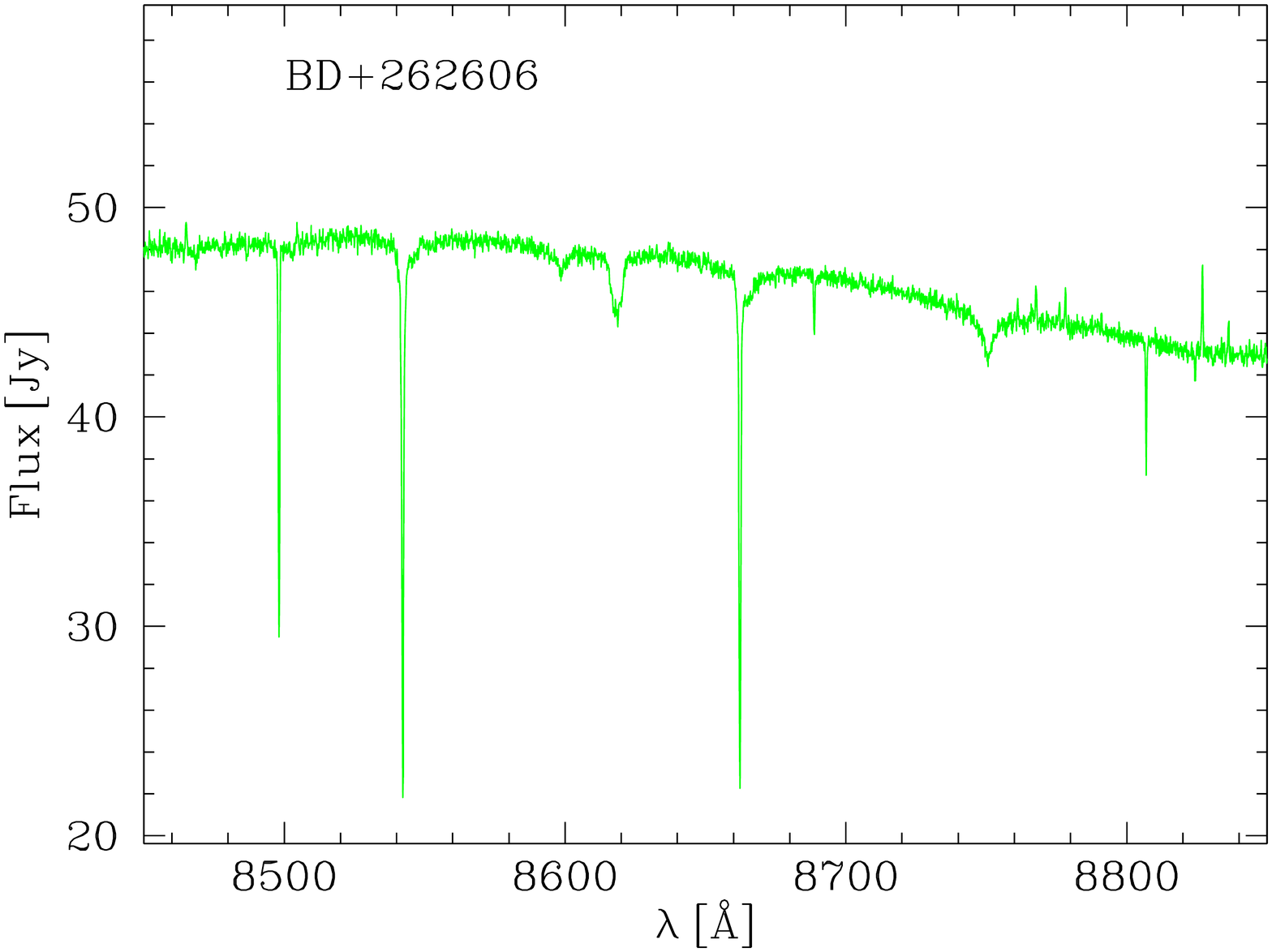}
\includegraphics[width=0.18\textwidth,angle=-90]{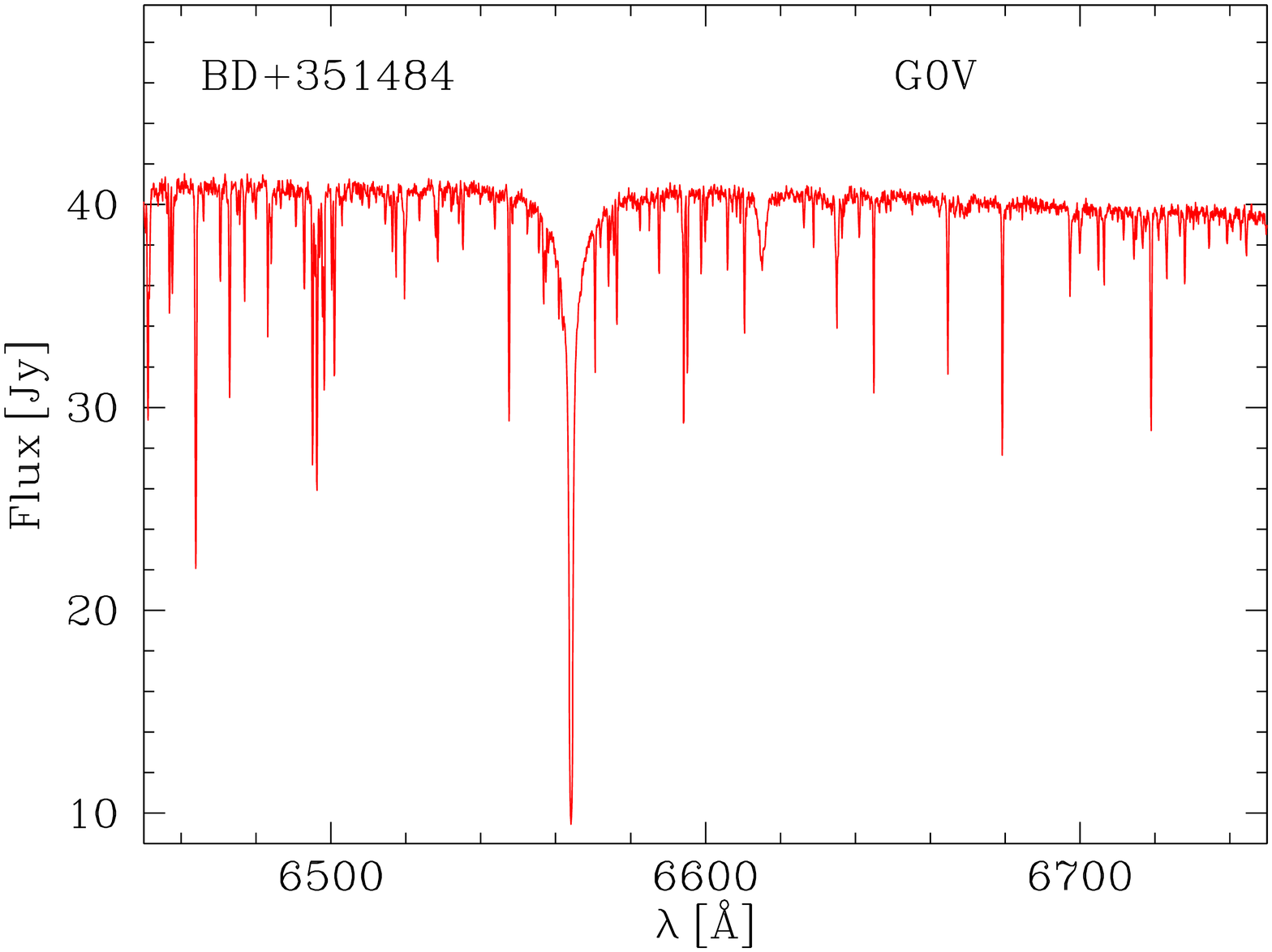}
\includegraphics[width=0.18\textwidth,angle=-90]{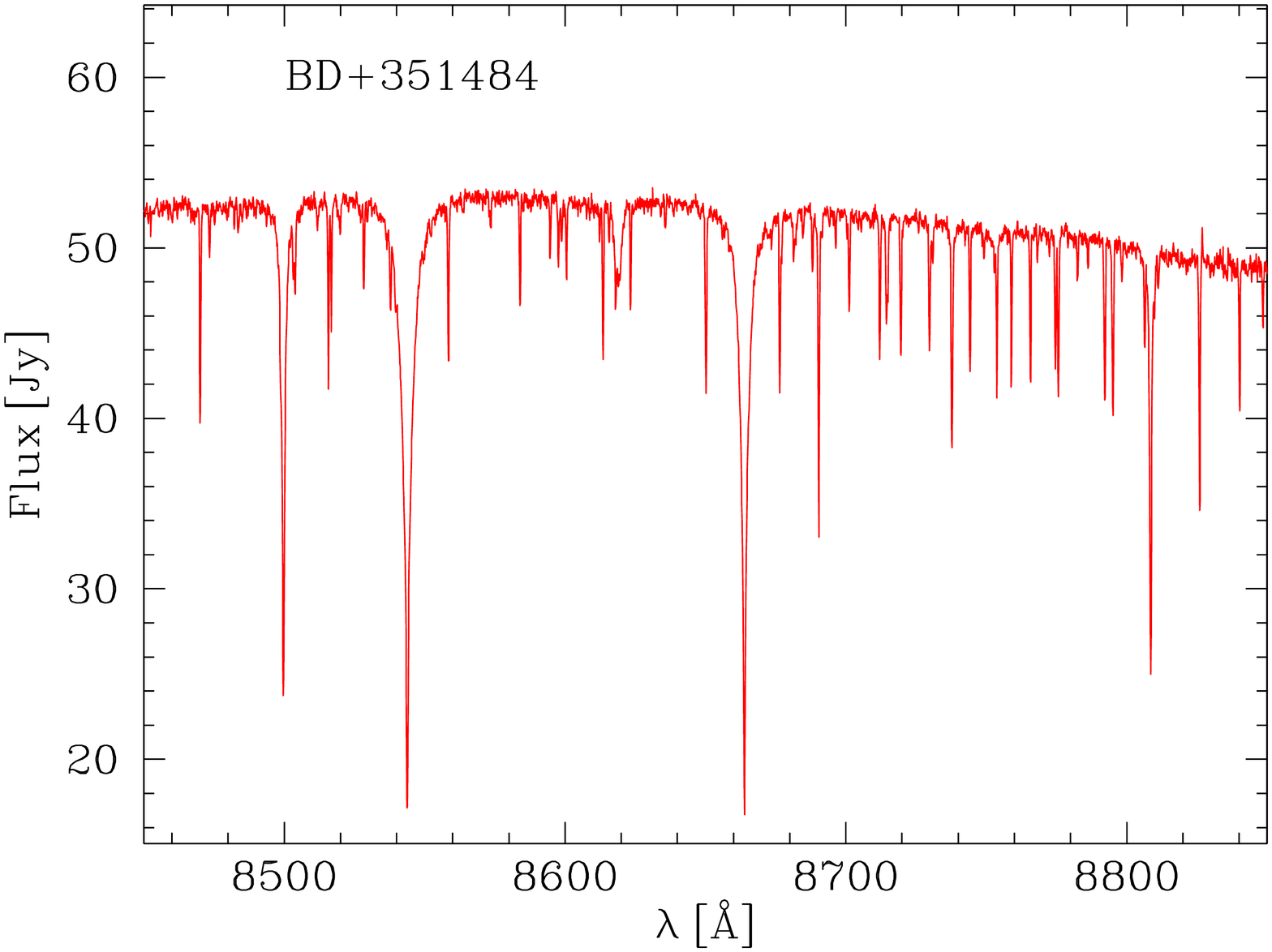}
\includegraphics[width=0.18\textwidth,angle=-90]{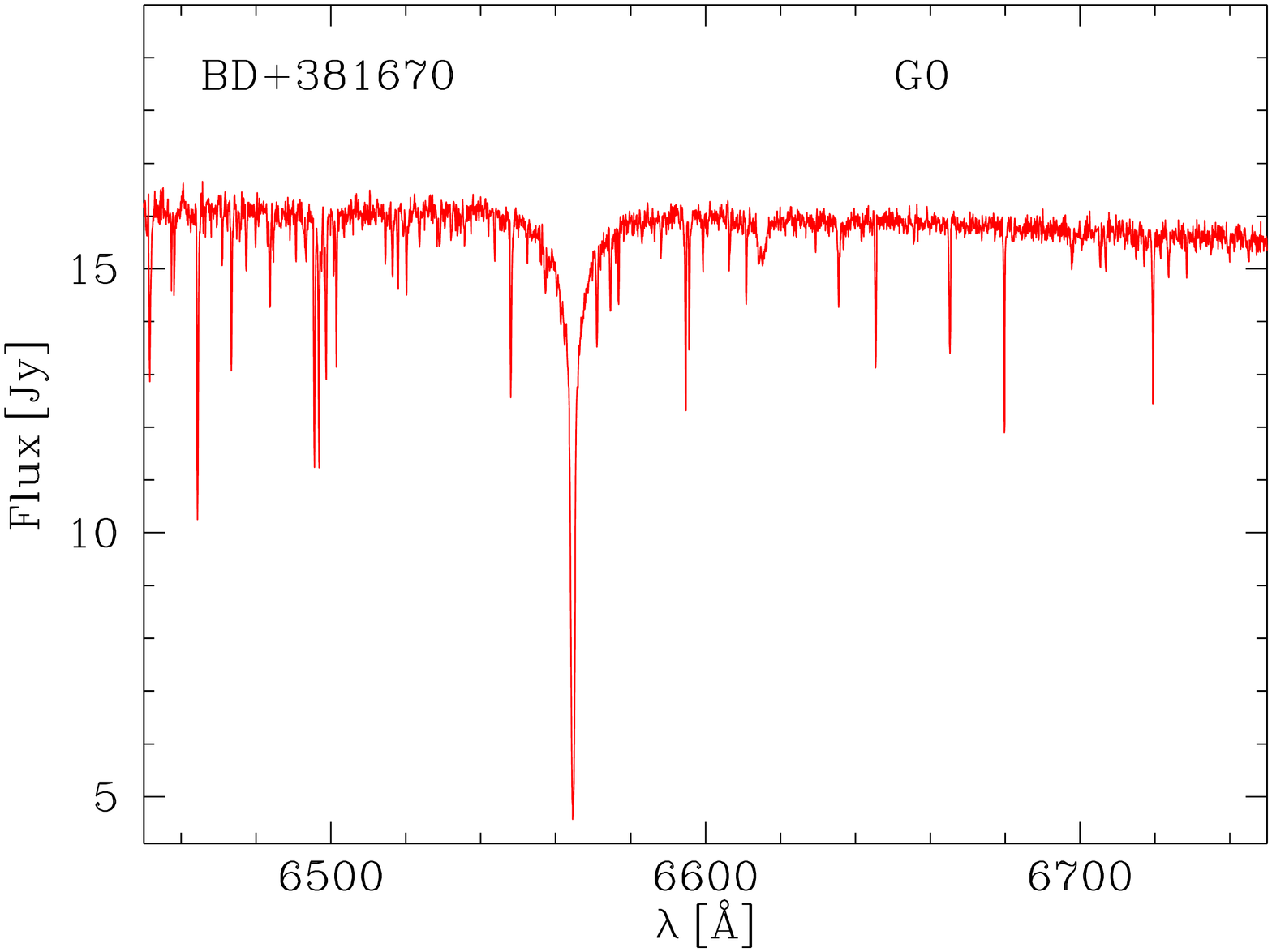}
\includegraphics[width=0.18\textwidth,angle=-90]{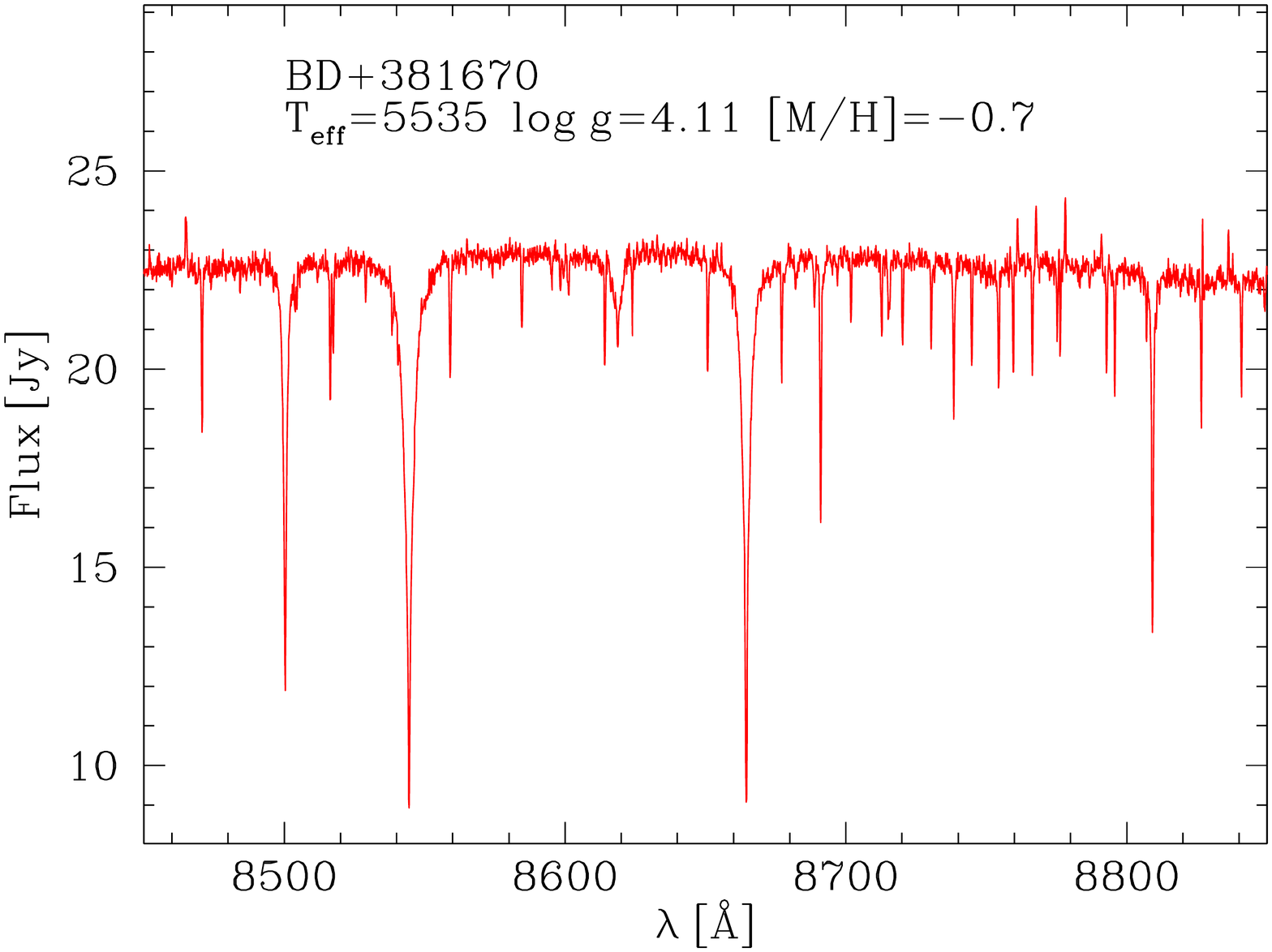}
\includegraphics[width=0.18\textwidth,angle=-90]{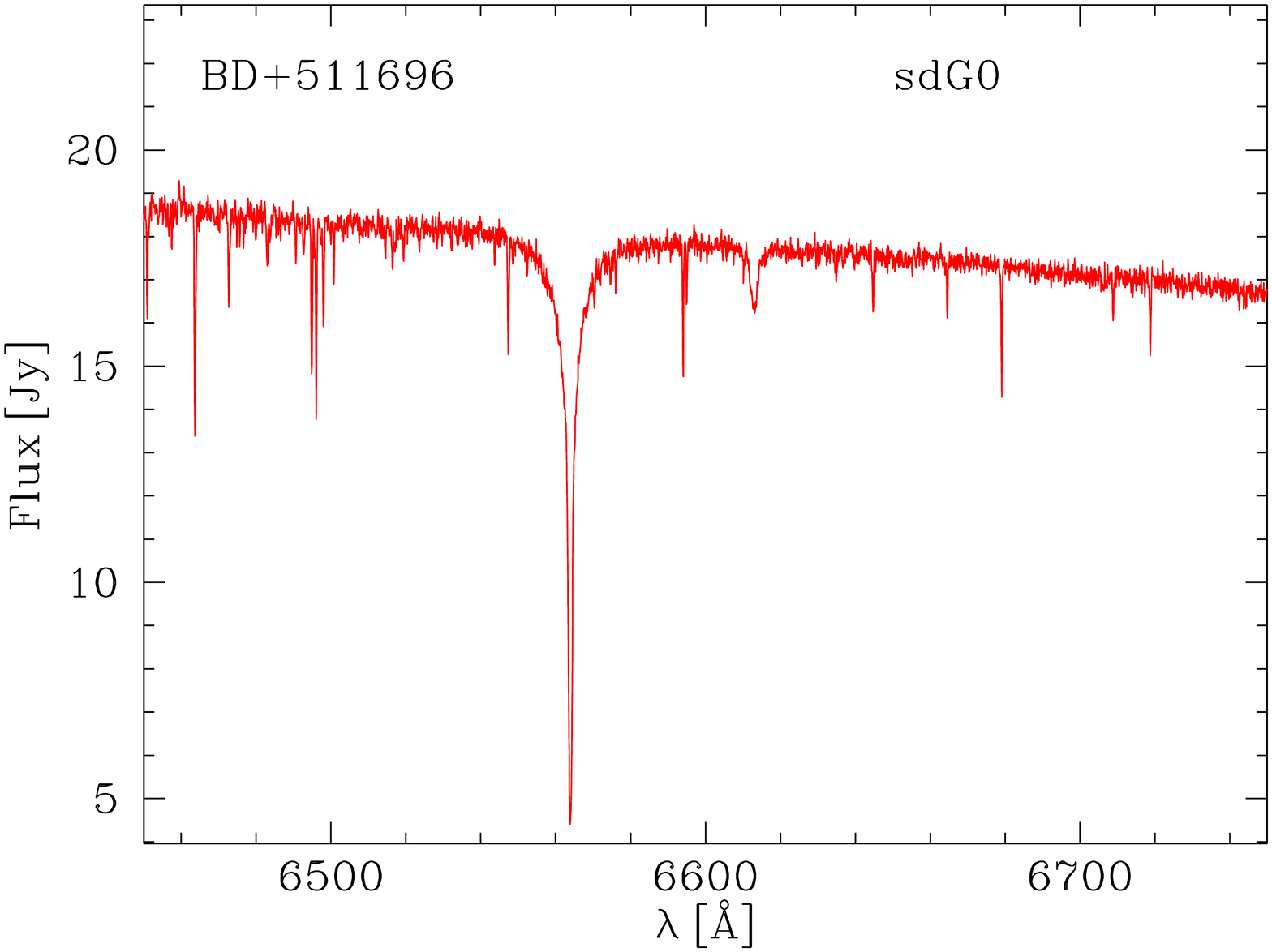}
\includegraphics[width=0.18\textwidth,angle=-90]{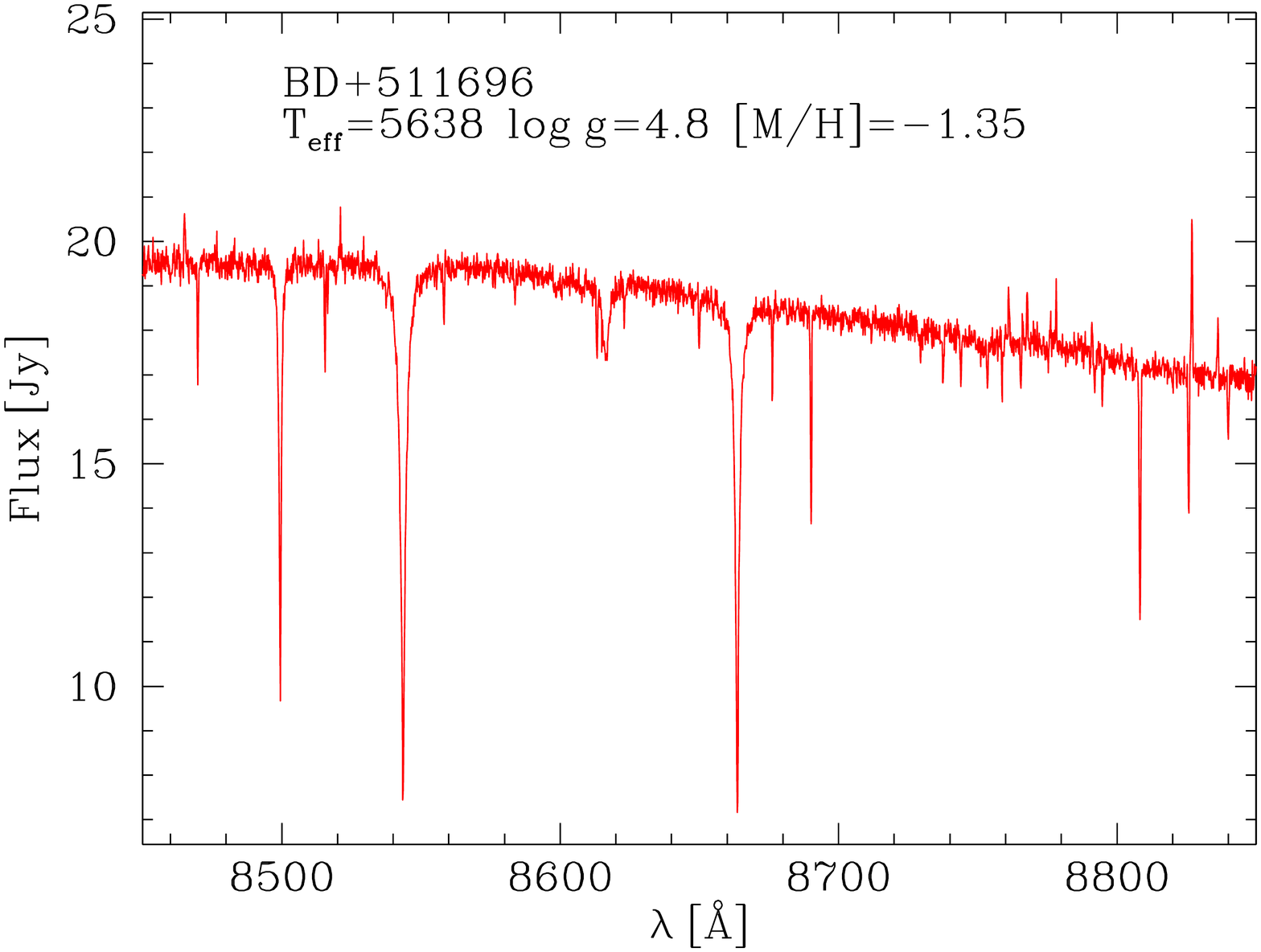}
\includegraphics[width=0.18\textwidth,angle=-90]{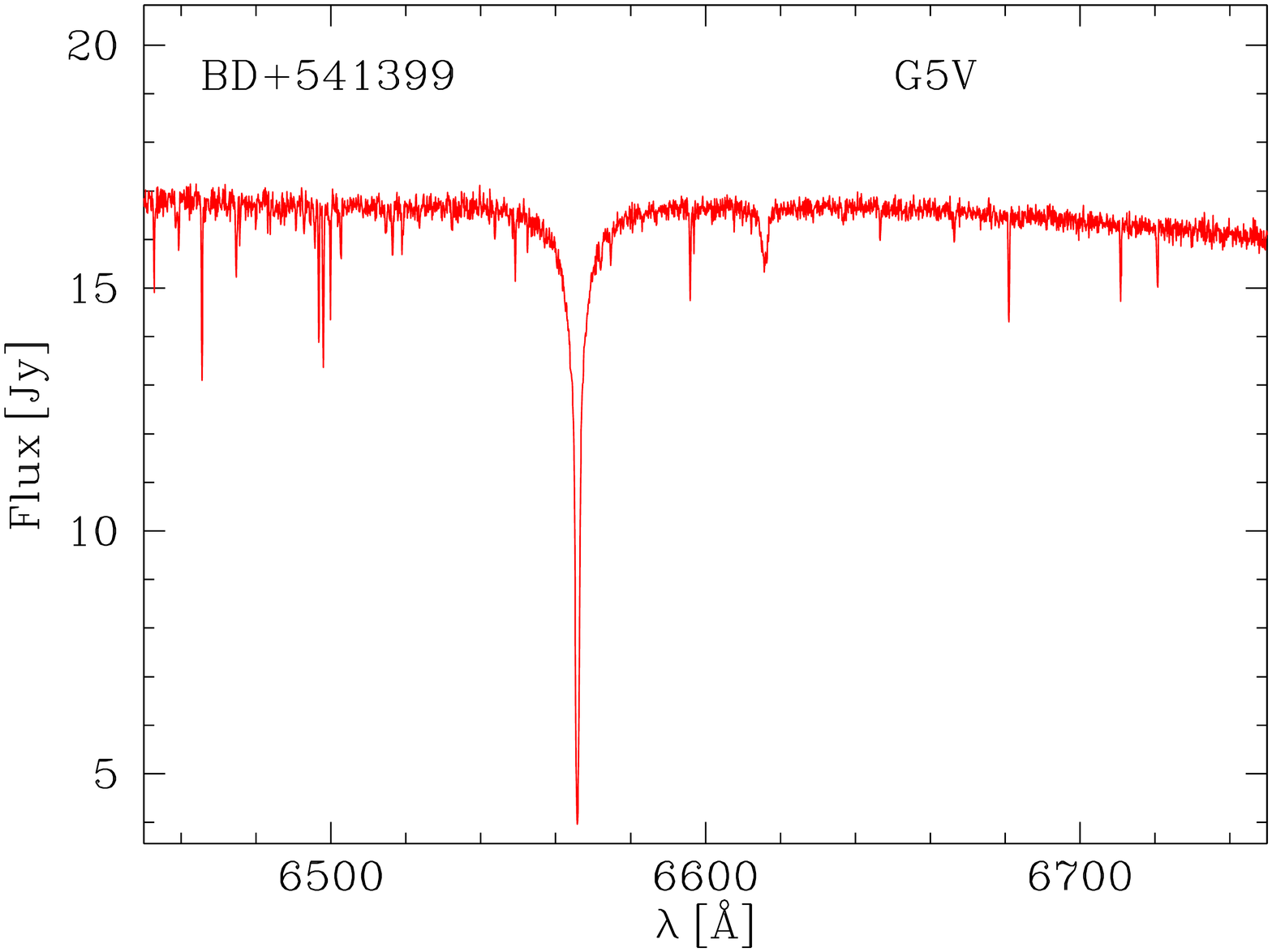}
\includegraphics[width=0.18\textwidth,angle=-90]{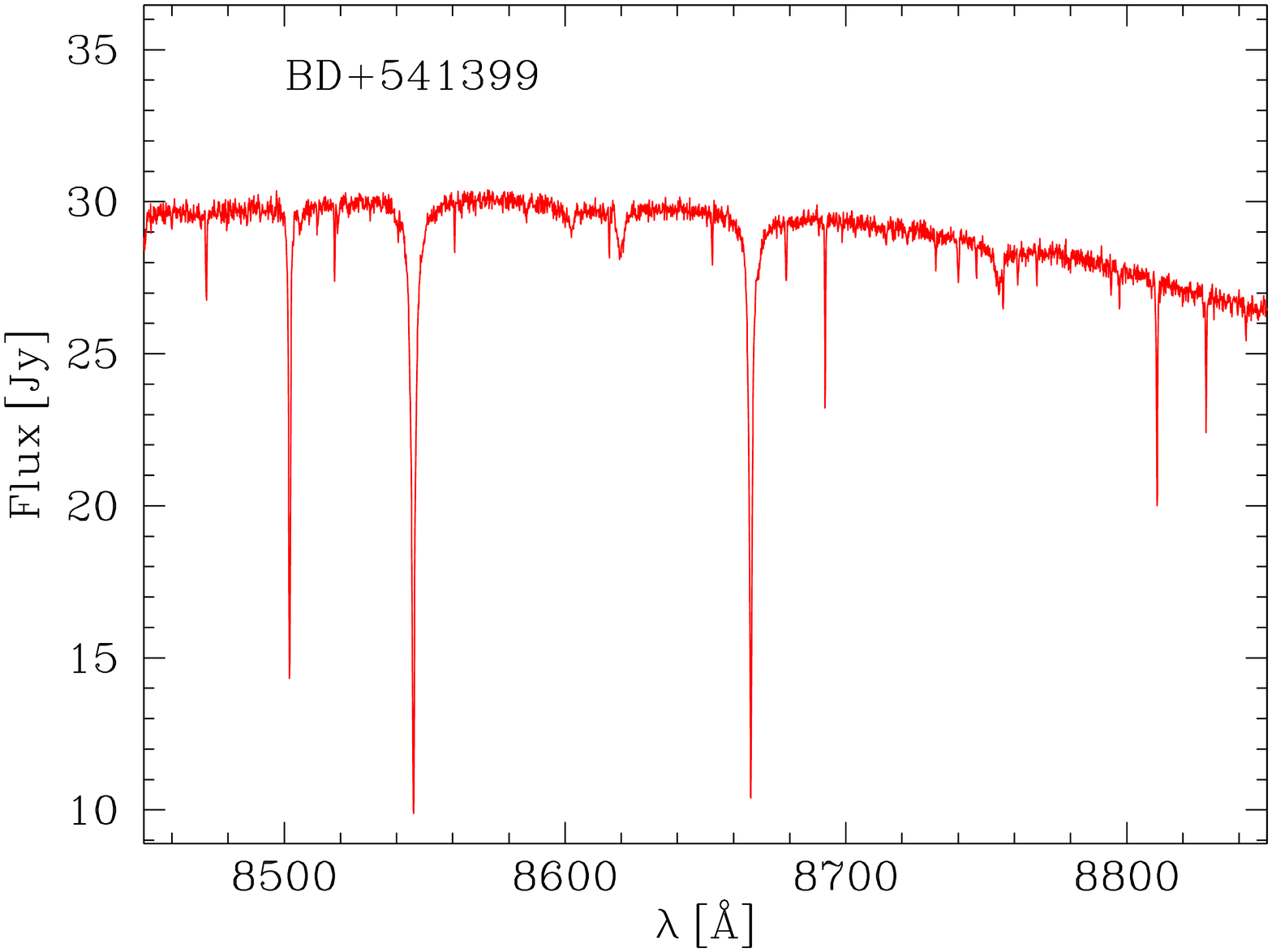}
\caption{Stellar spectra ordered by name, given in each
plot, for this release I catalogue. Each row shows the HR-R (left) and
HR-I (right) spectra for two observed stars. Thus,
columns 1 and 3 are HR-R spectra, while columns 2 and 4 contain HR-I spectra. Columns 1 and 2, or columns 3 and 4, refer to the same observation. We have 5 stars with duplicated observations, which we show separately due to their different observing conditions. Stars shown in this page are: BD-032525, BD-122669, BD+083095, BD+092190, BD+130013, BD+191730, BD+195116B, BD+203603, BD+241676, BD+262606, BD+351484, BD+381670, BD+511696 and BD+541399.}
\label{sp-release-838}
\end{figure*}

\begin{figure*}
\includegraphics[width=0.18\textwidth,angle=-90]{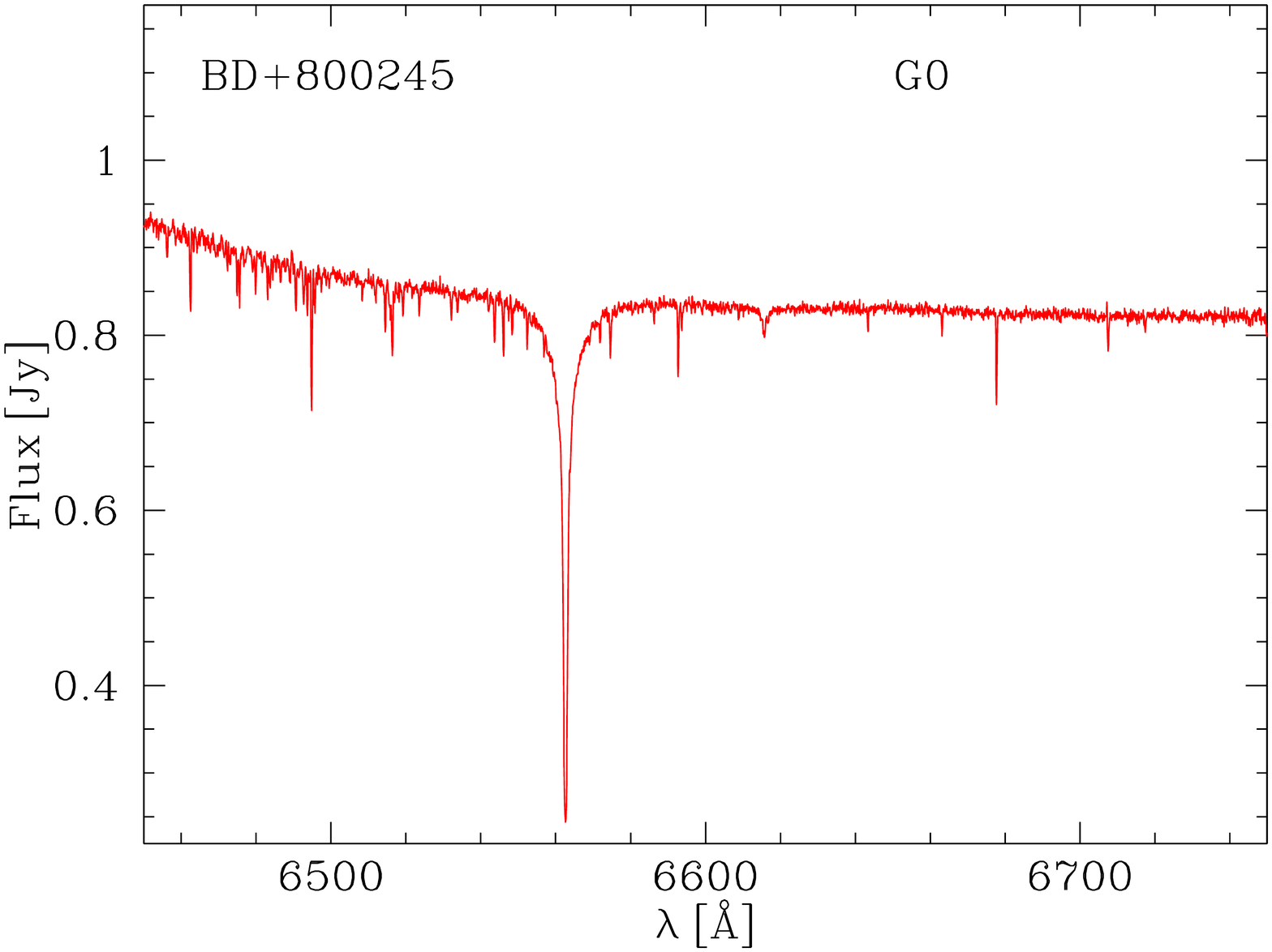}
\includegraphics[width=0.18\textwidth,angle=-90]{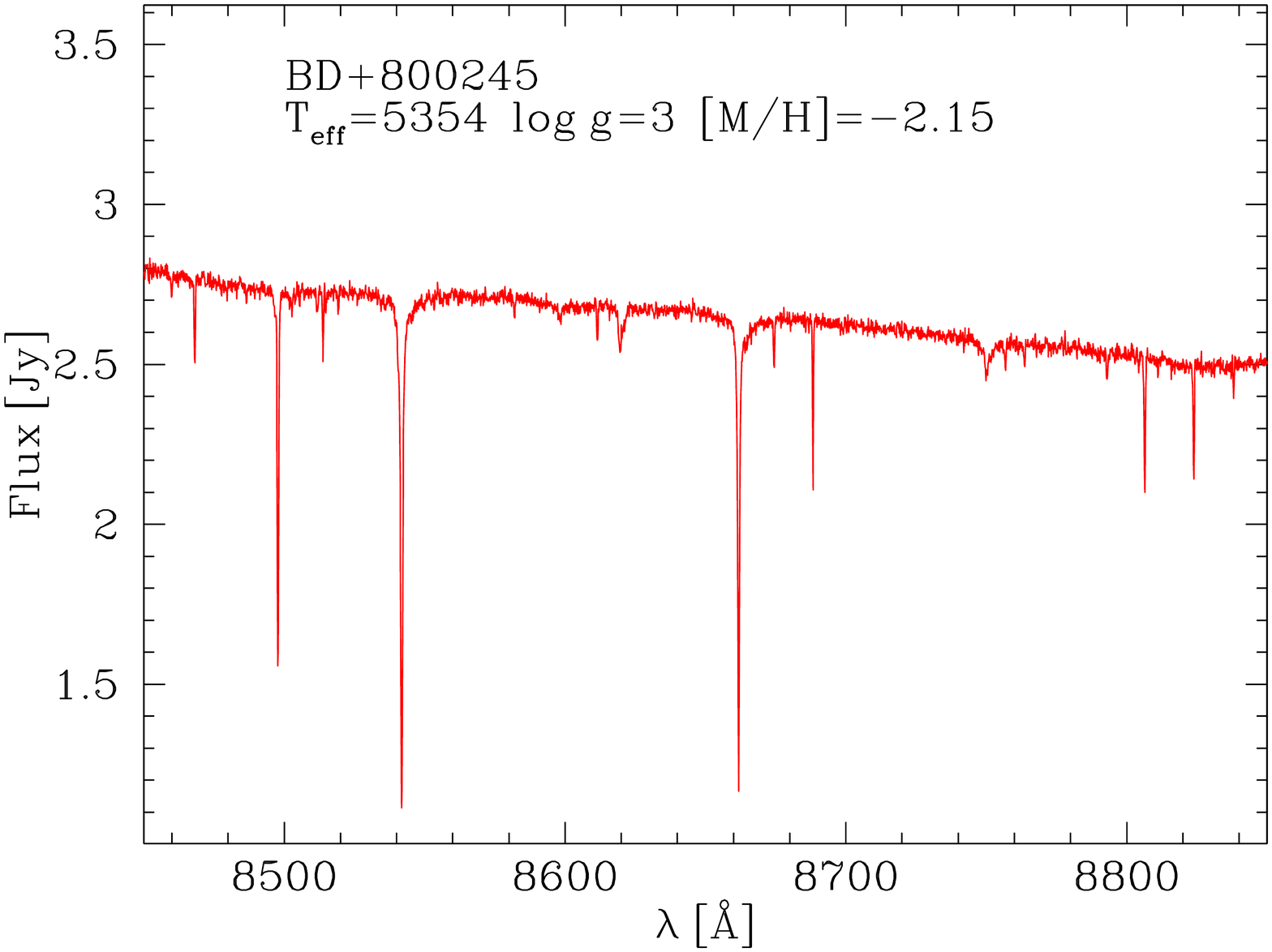}
\includegraphics[width=0.18\textwidth,angle=-90]{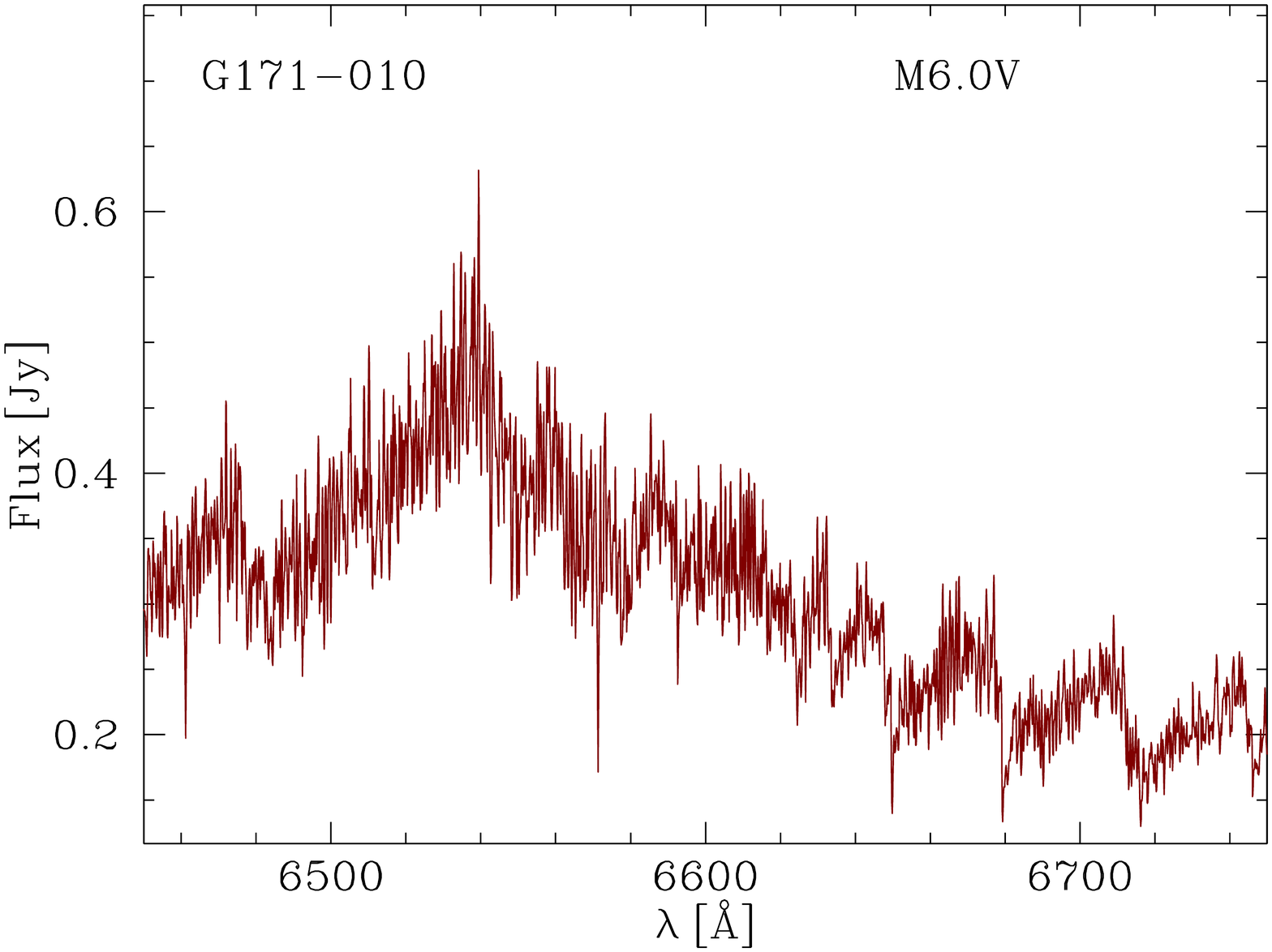}
\includegraphics[width=0.18\textwidth,angle=-90]{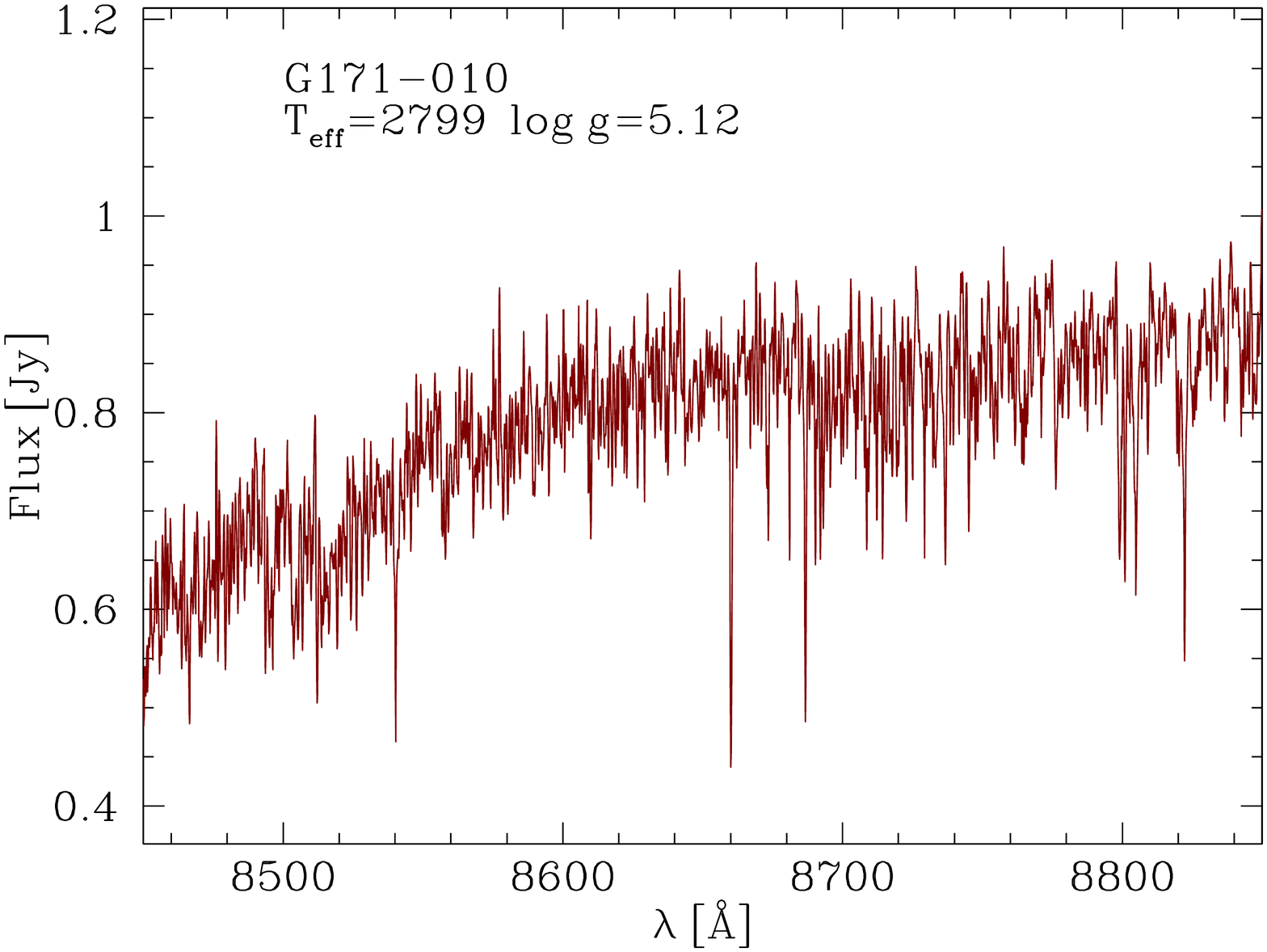}
\includegraphics[width=0.18\textwidth,angle=-90]{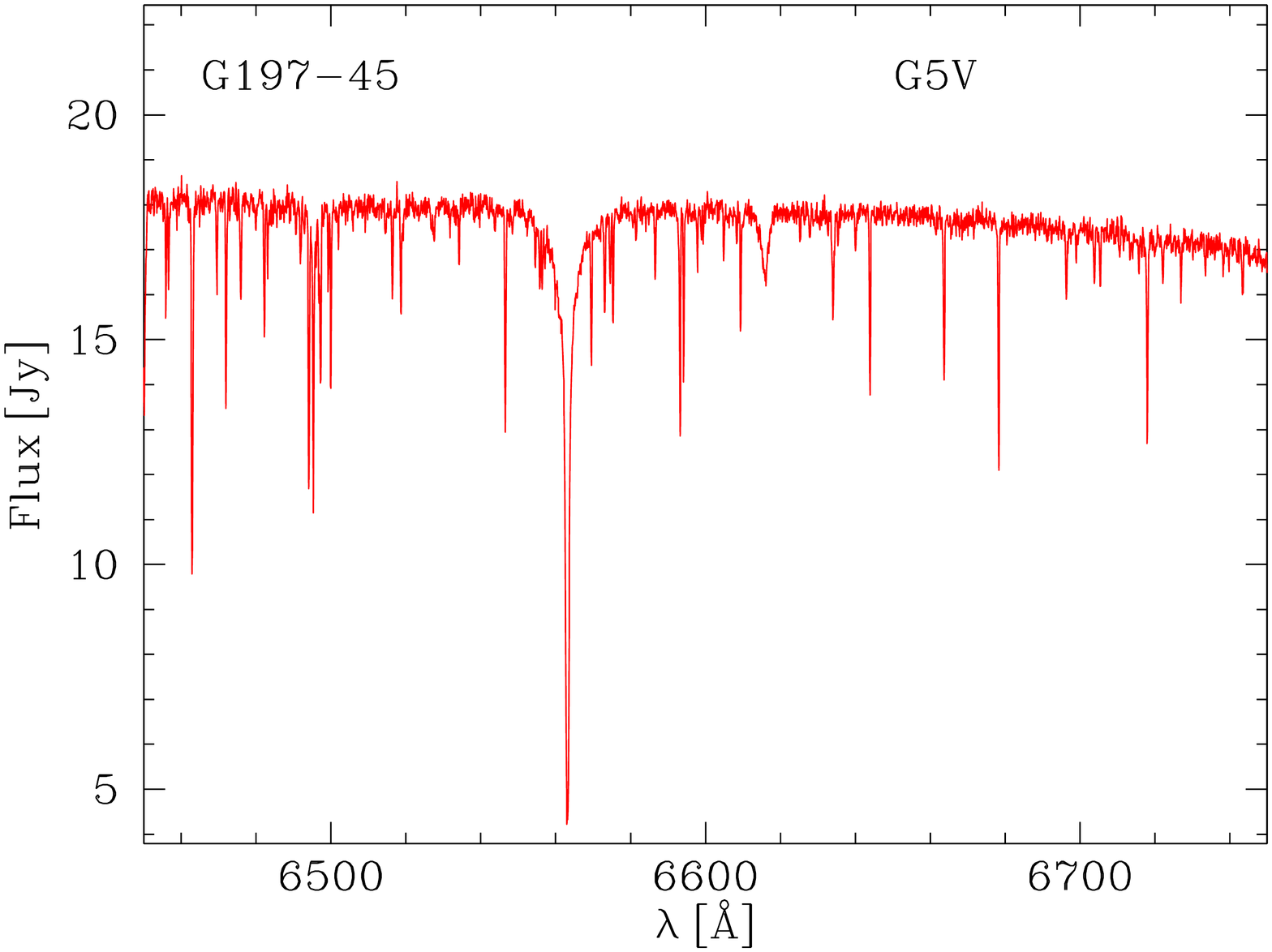}
\includegraphics[width=0.18\textwidth,angle=-90]{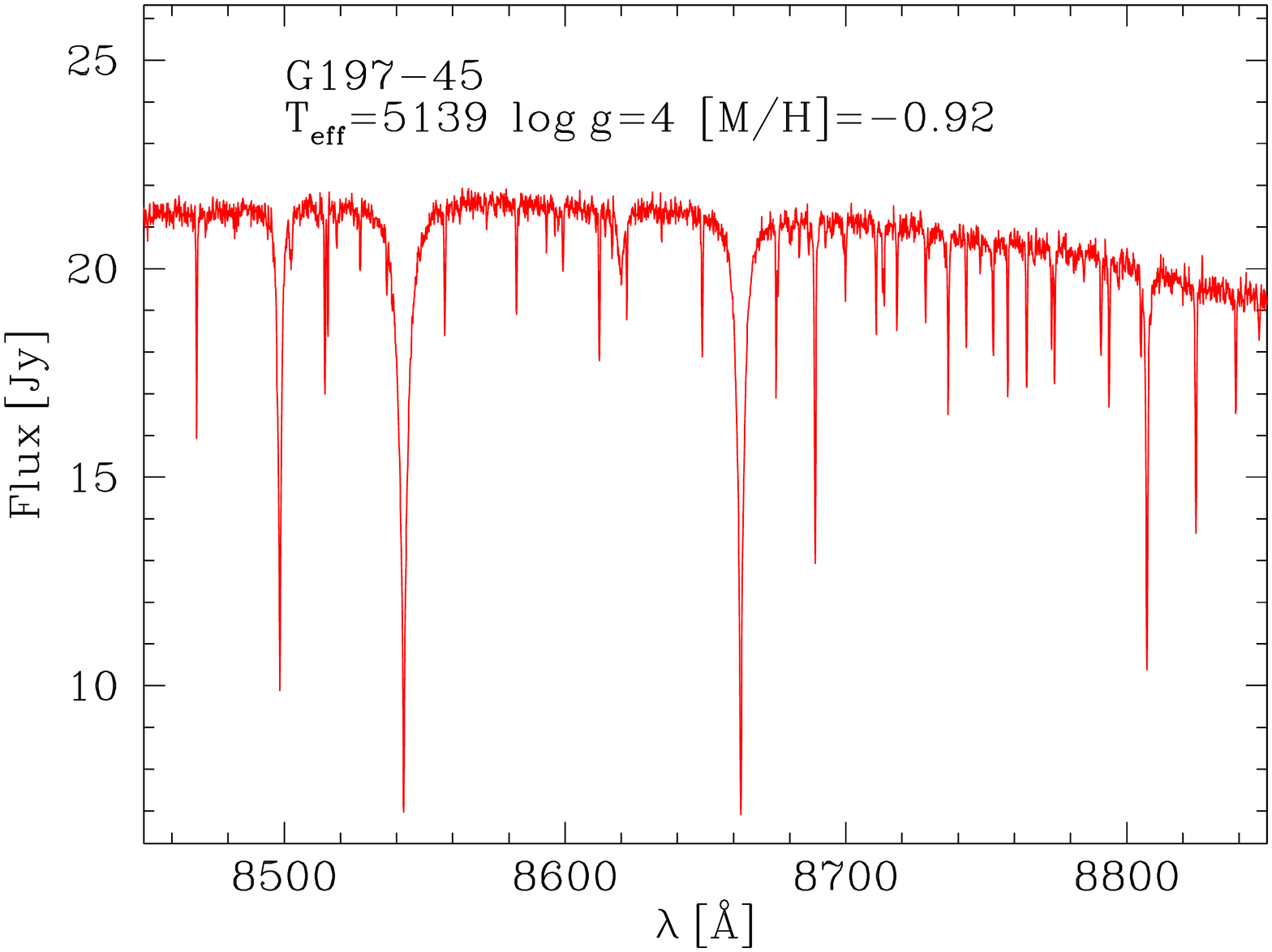}
\includegraphics[width=0.18\textwidth,angle=-90]{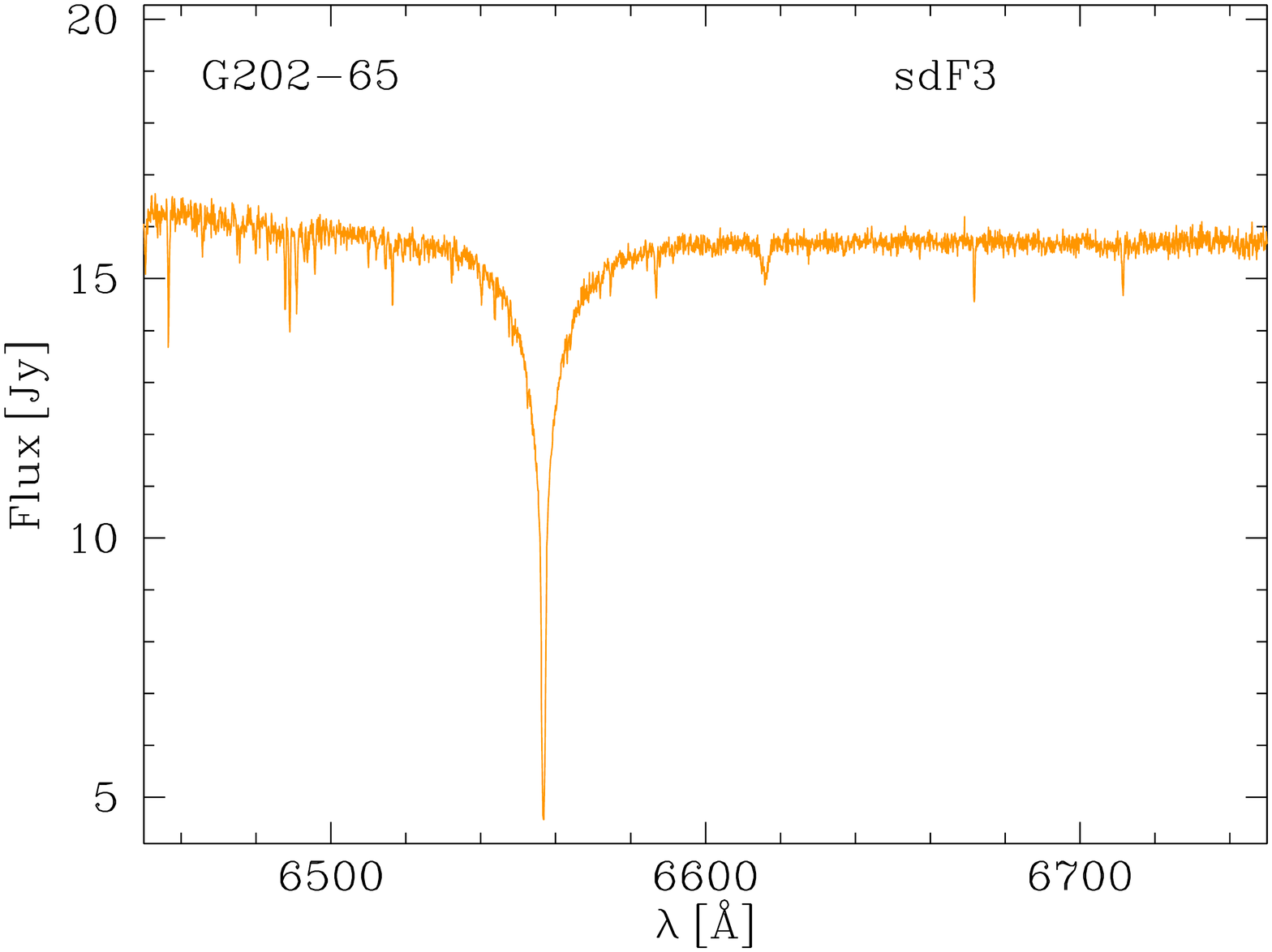}
\includegraphics[width=0.18\textwidth,angle=-90]{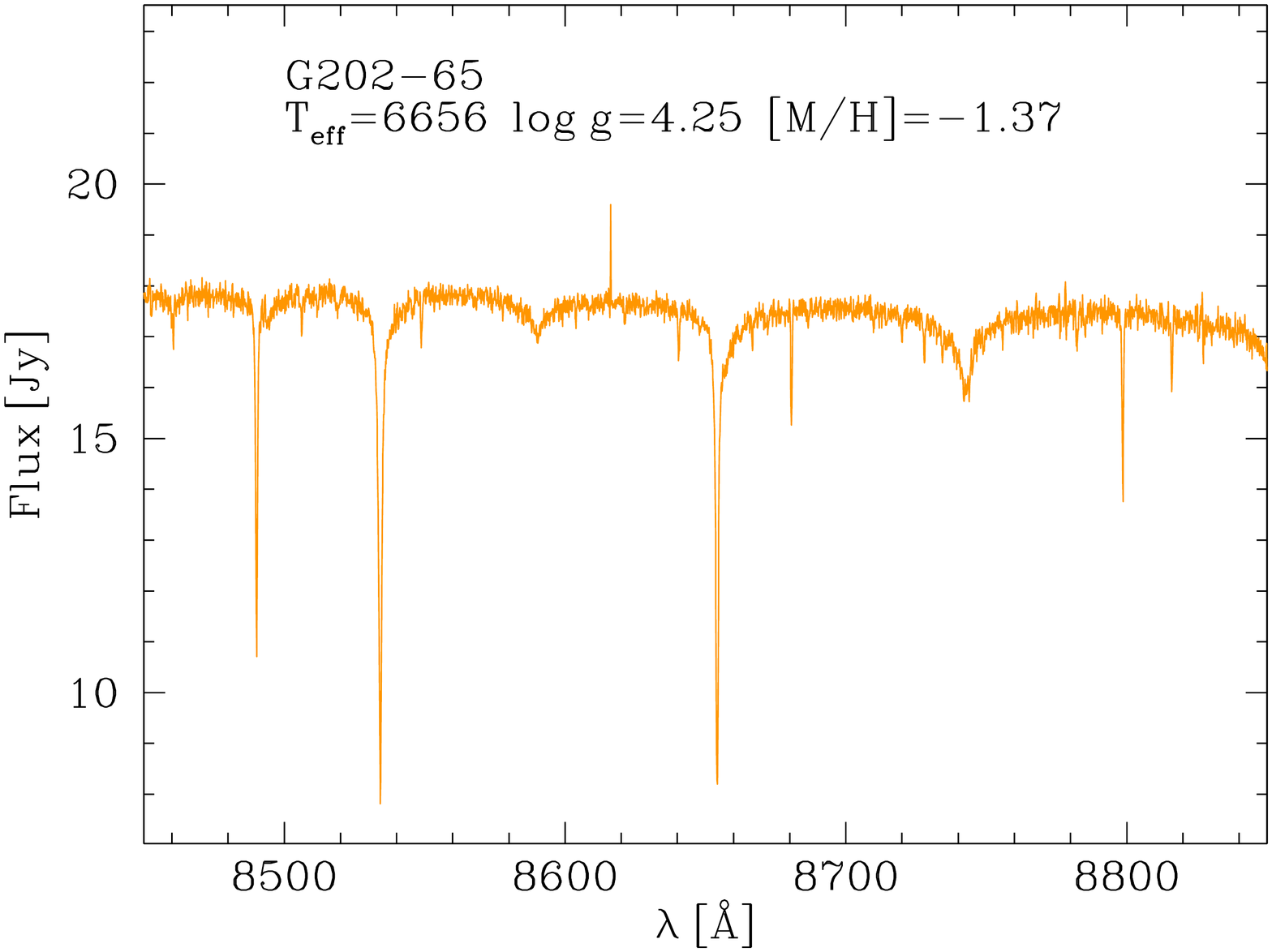}
\includegraphics[width=0.18\textwidth,angle=-90]{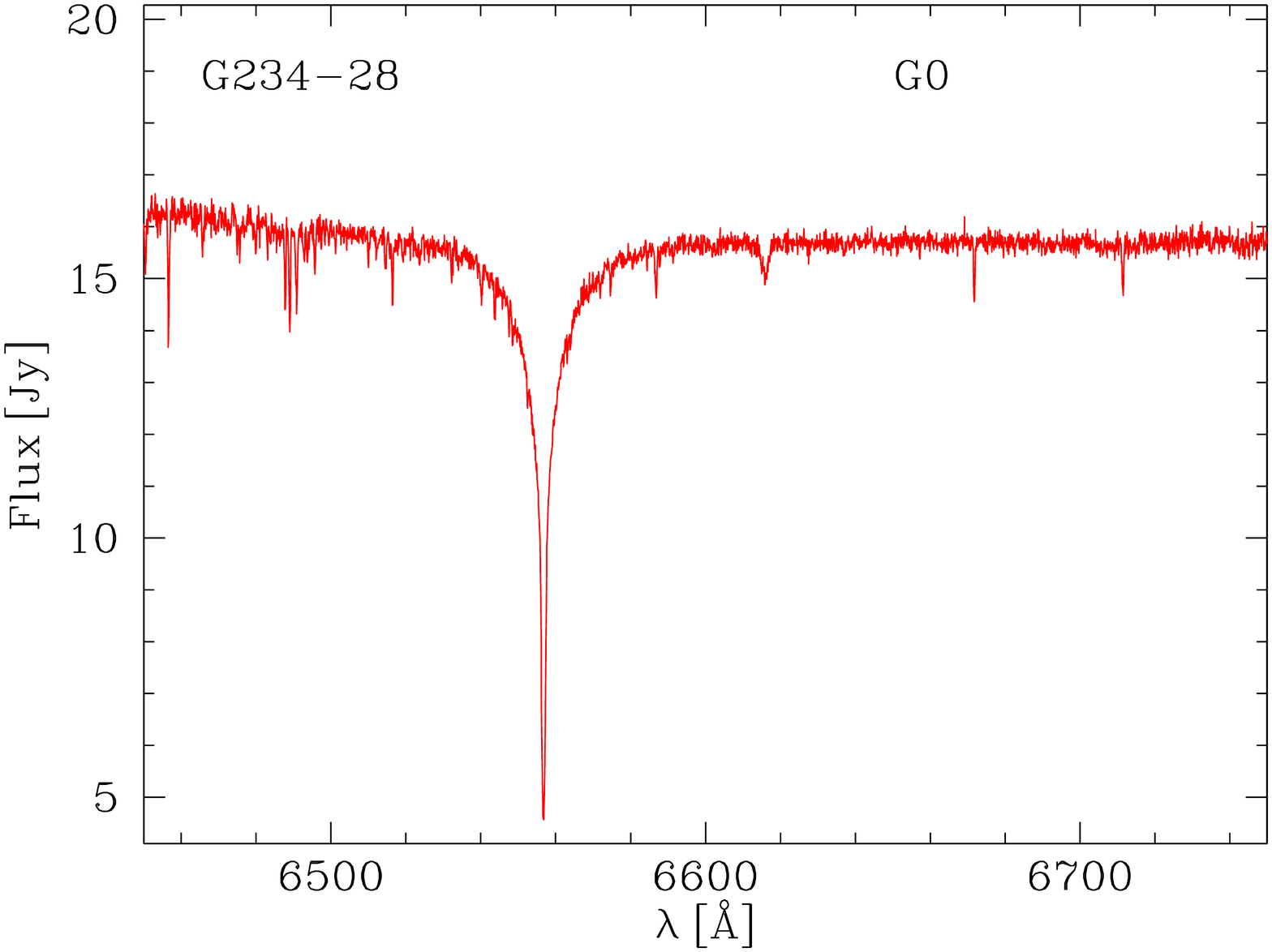}
\includegraphics[width=0.18\textwidth,angle=-90]{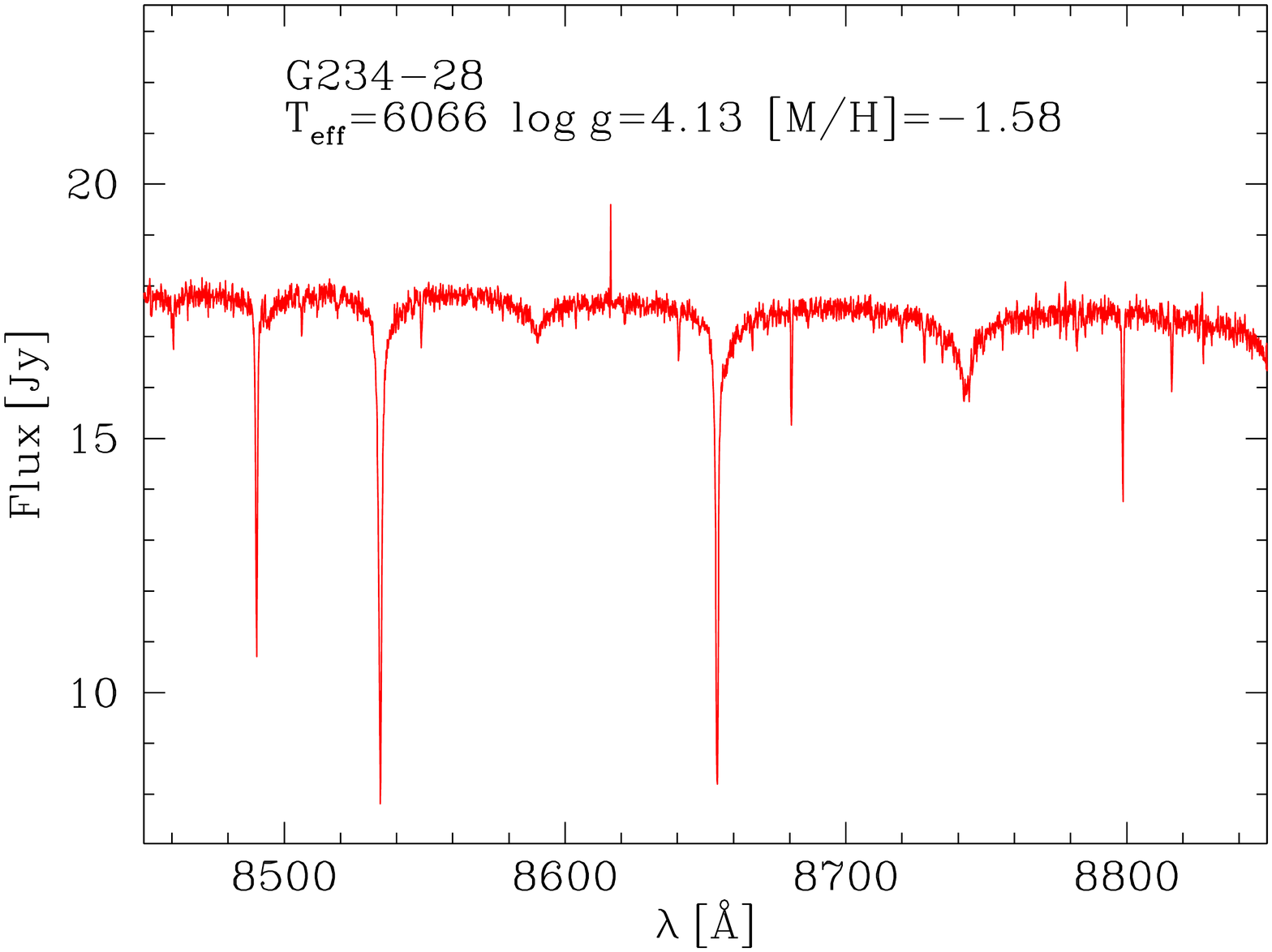}
\includegraphics[width=0.18\textwidth,angle=-90]{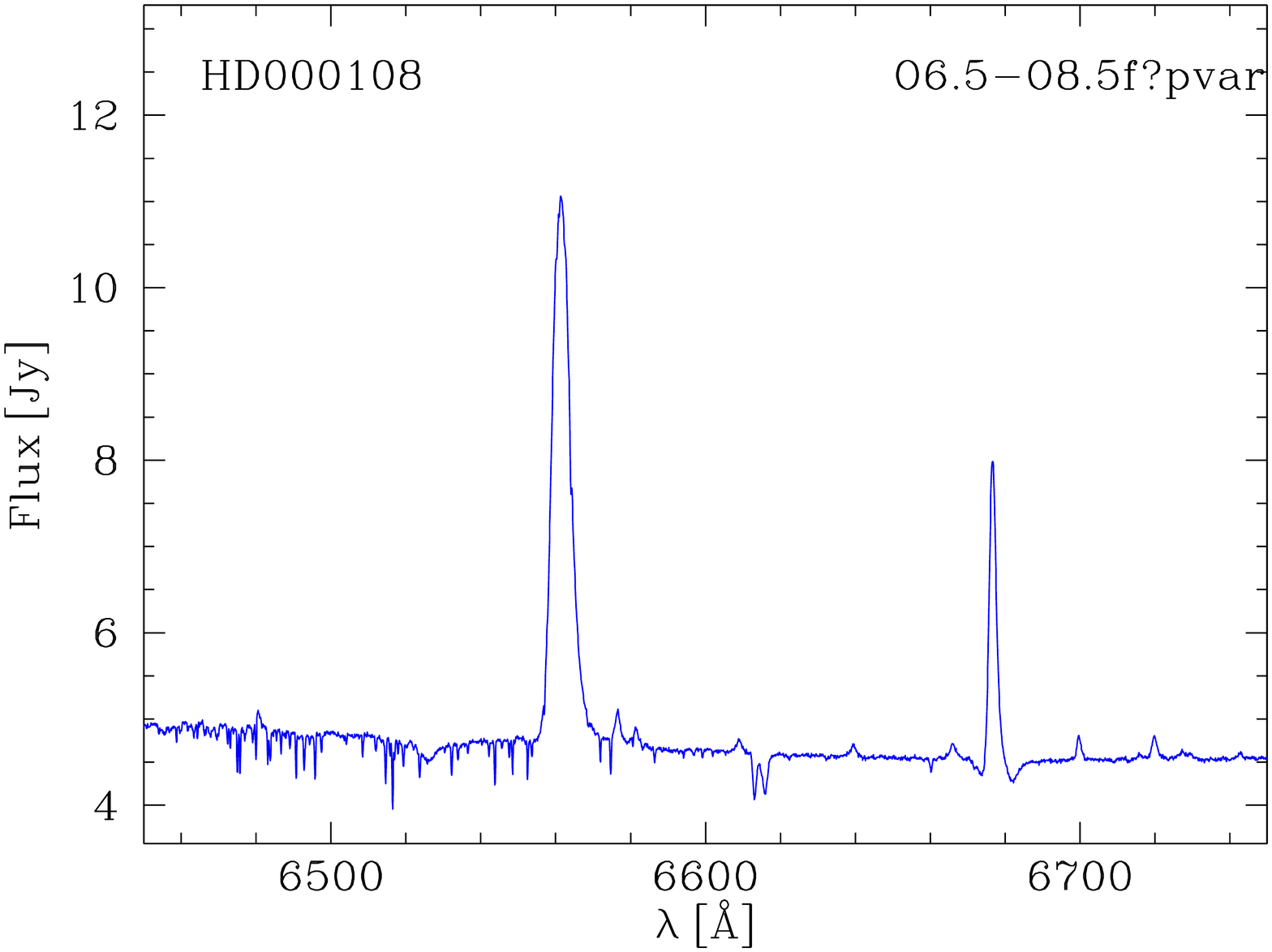}
\includegraphics[width=0.18\textwidth,angle=-90]{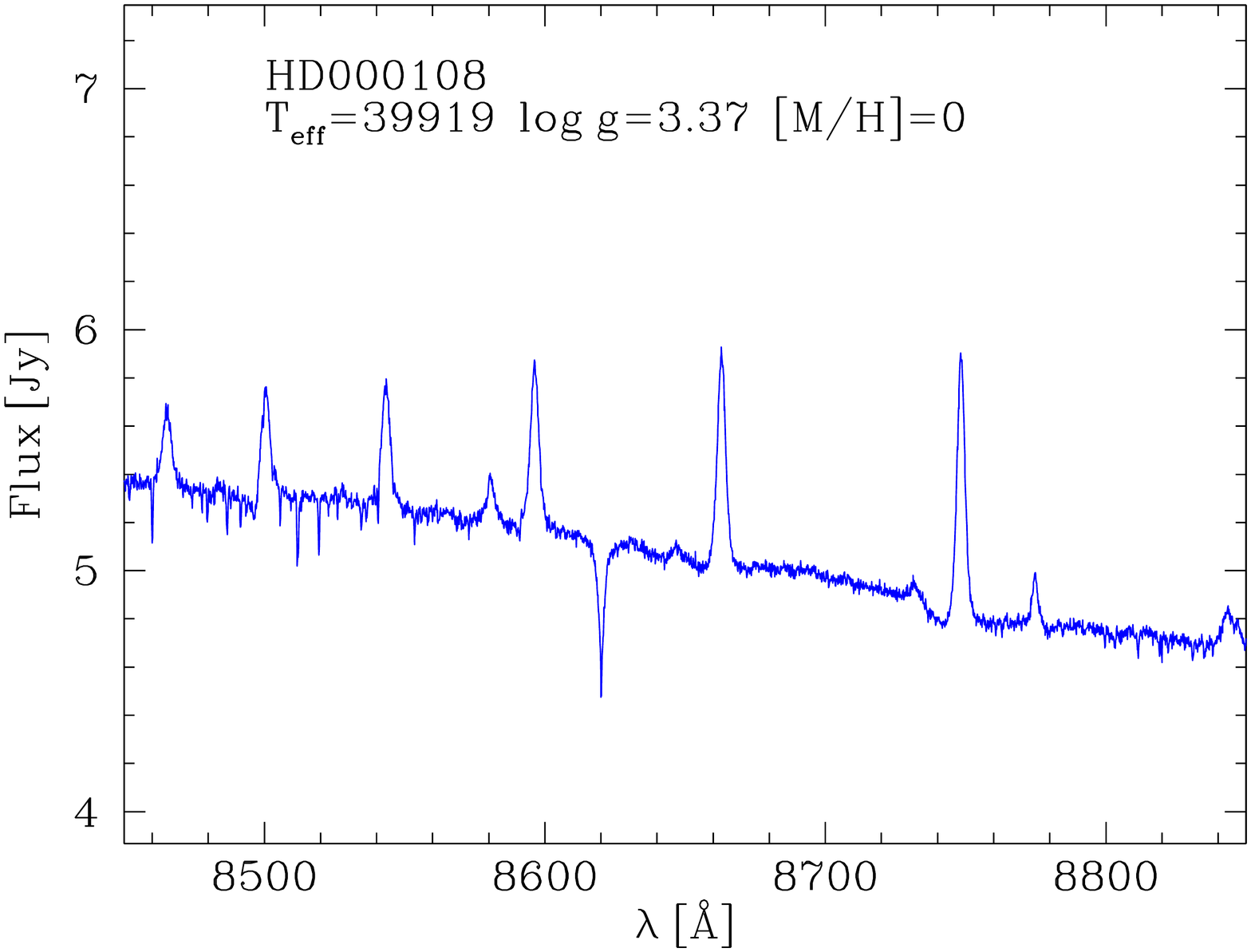}
\includegraphics[width=0.18\textwidth,angle=-90]{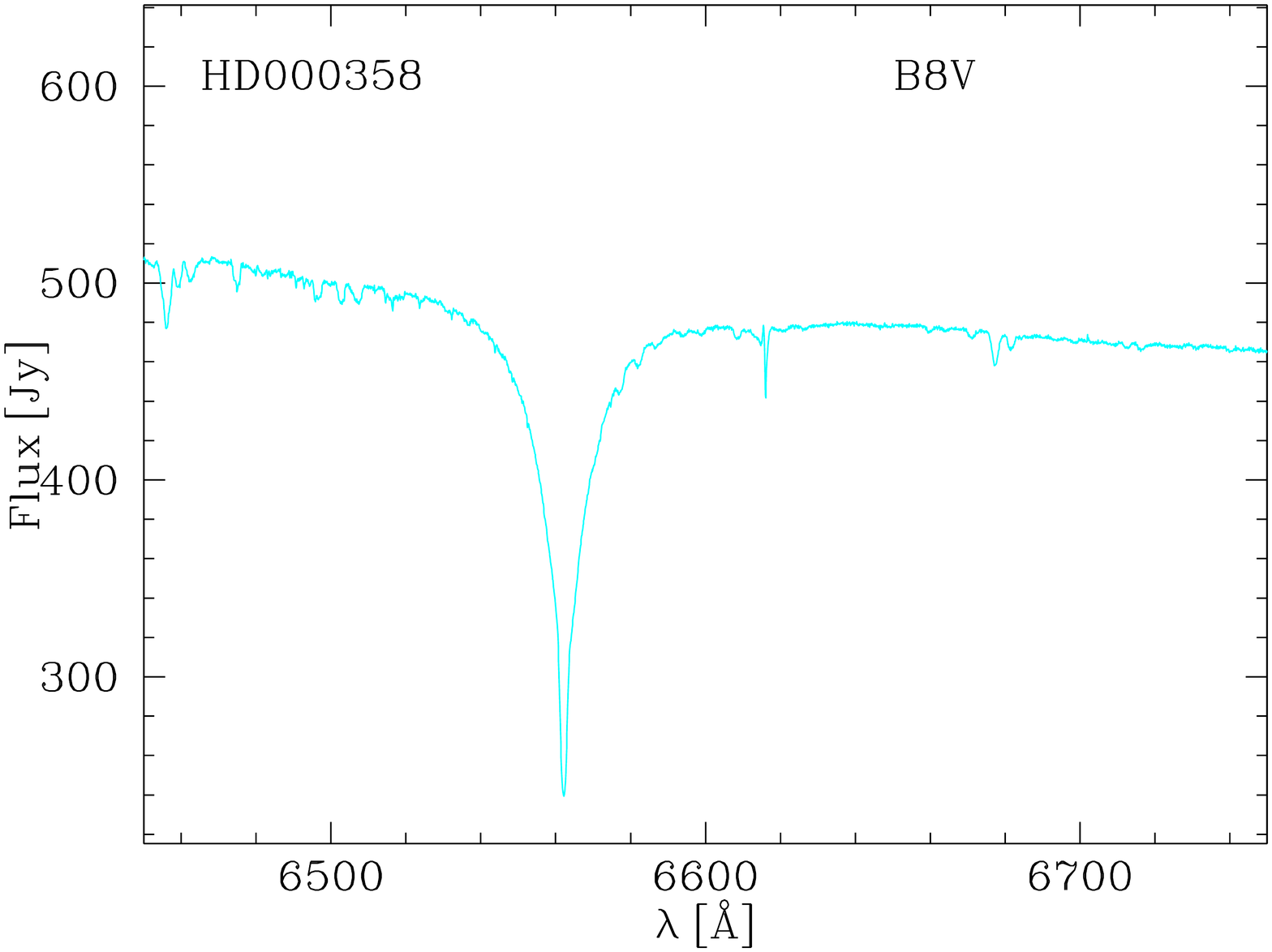}
\includegraphics[width=0.18\textwidth,angle=-90]{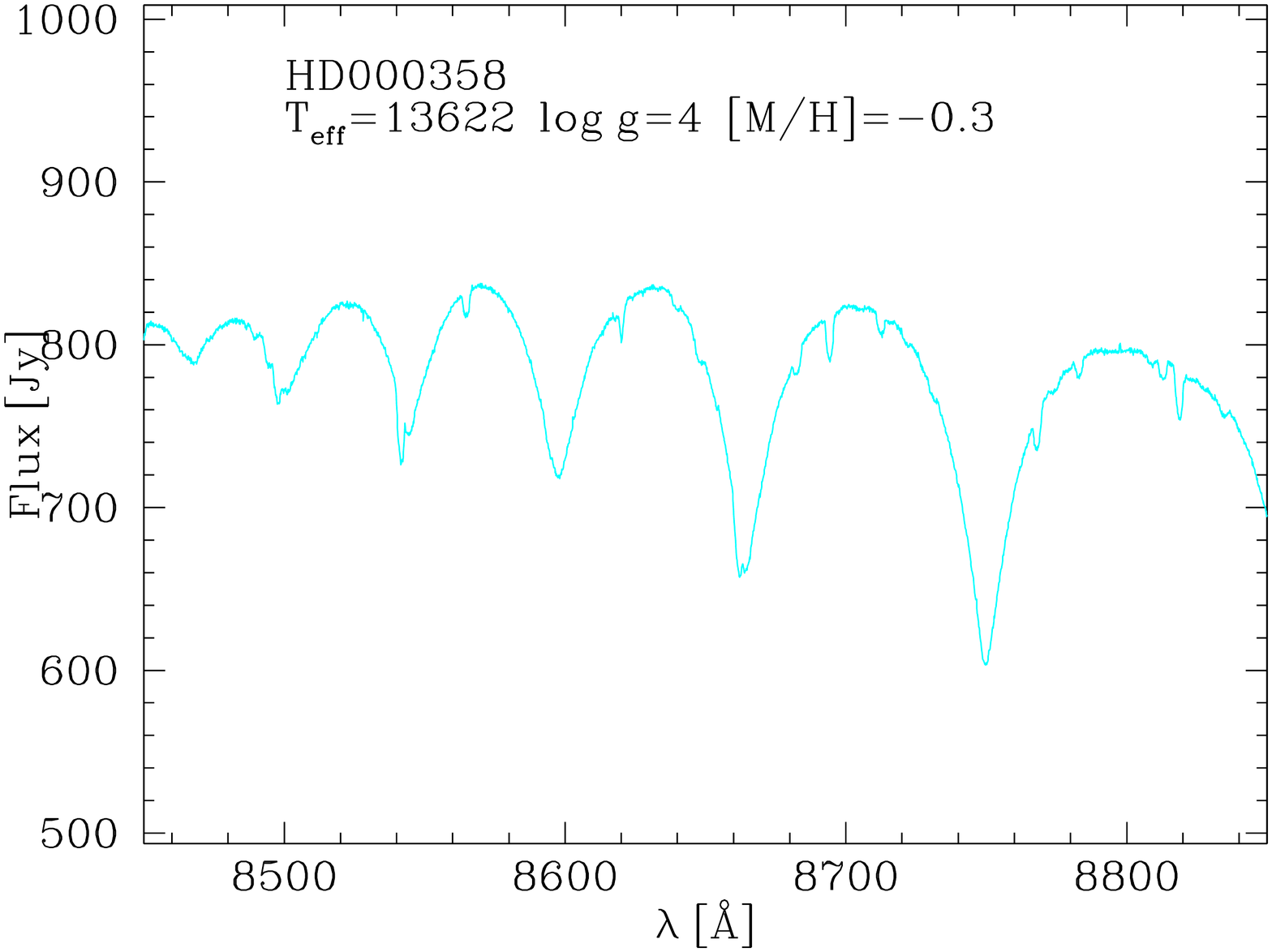}
\includegraphics[width=0.18\textwidth,angle=-90]{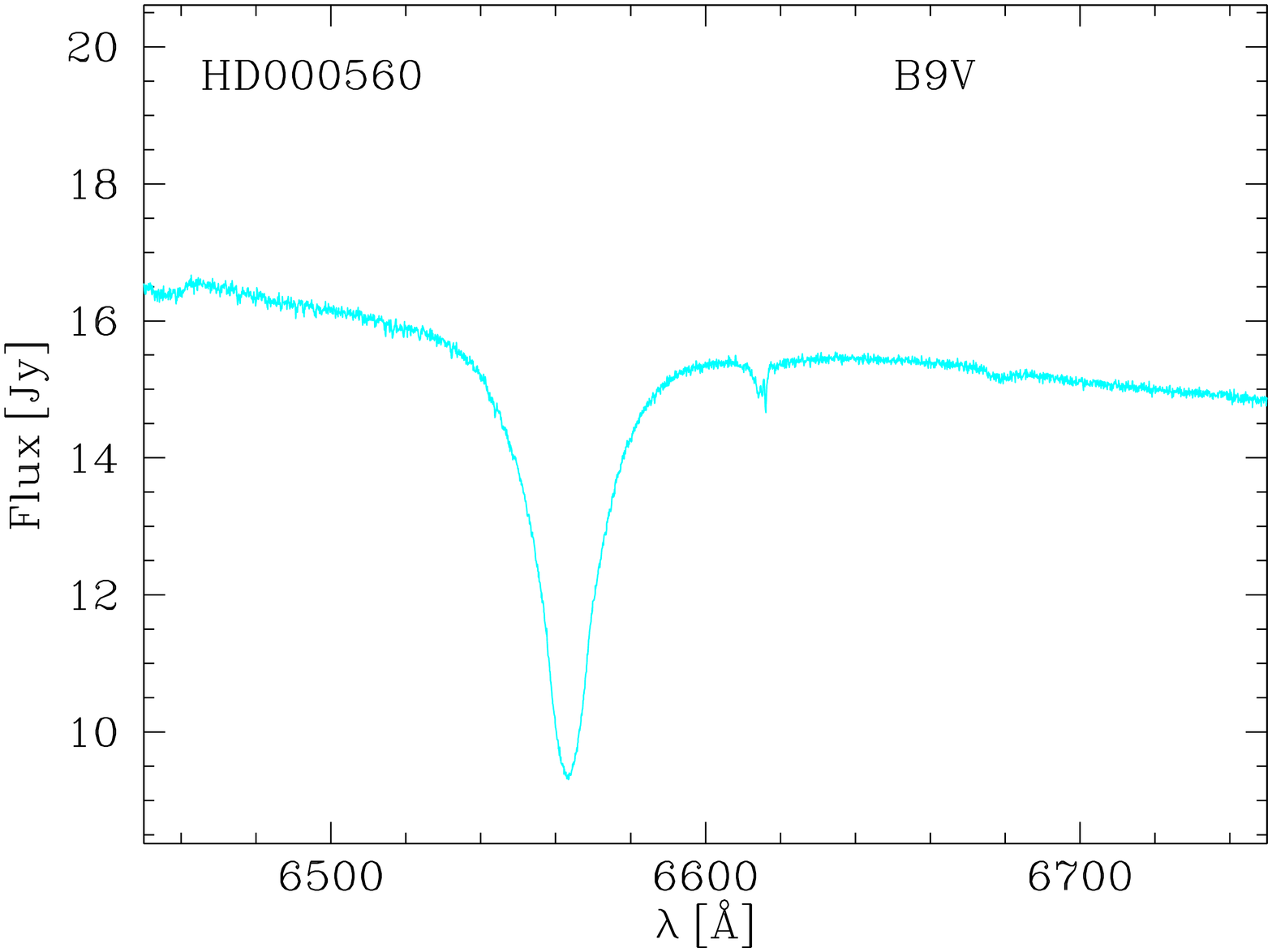}
\includegraphics[width=0.18\textwidth,angle=-90]{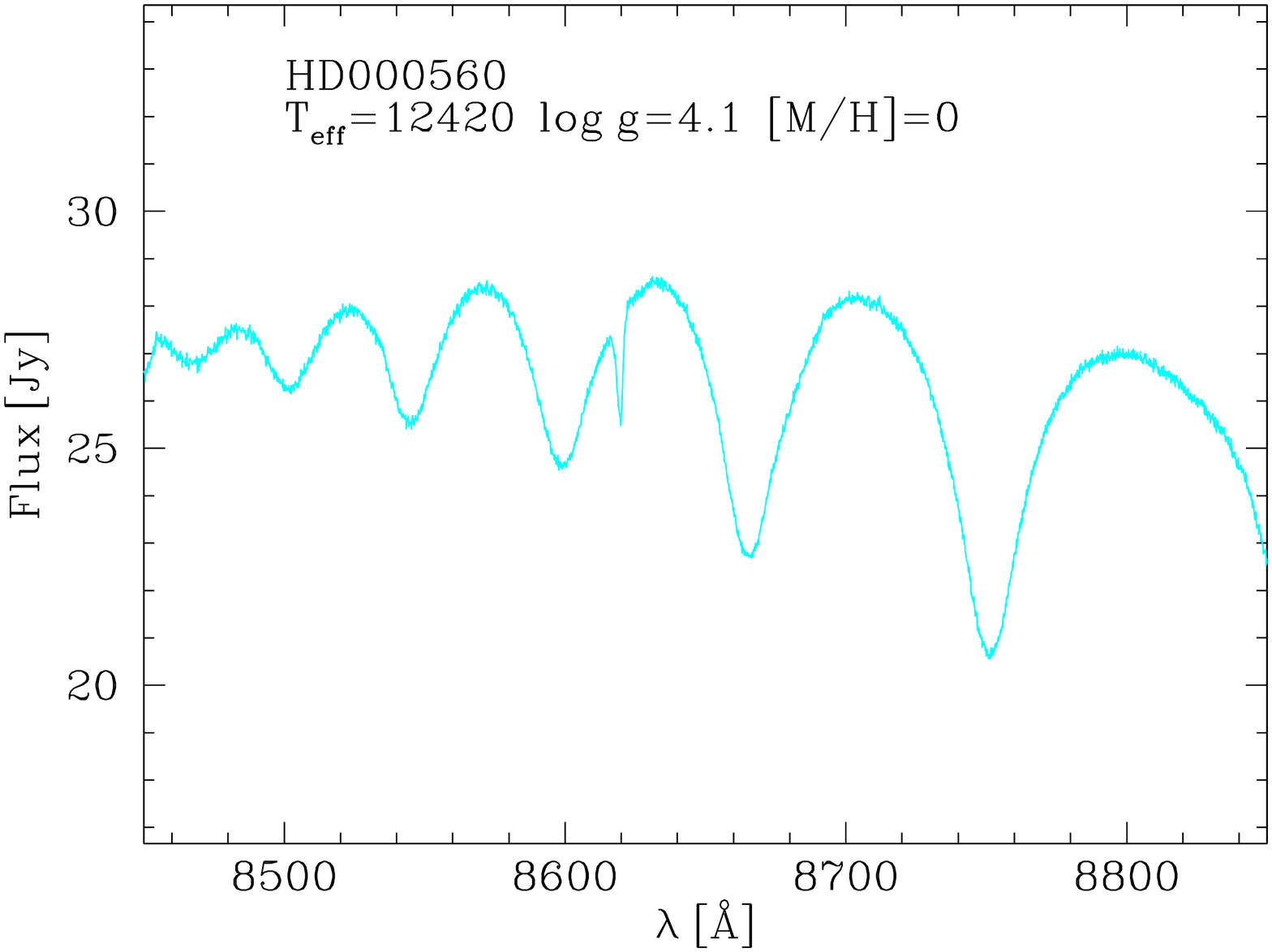}
\includegraphics[width=0.18\textwidth,angle=-90]{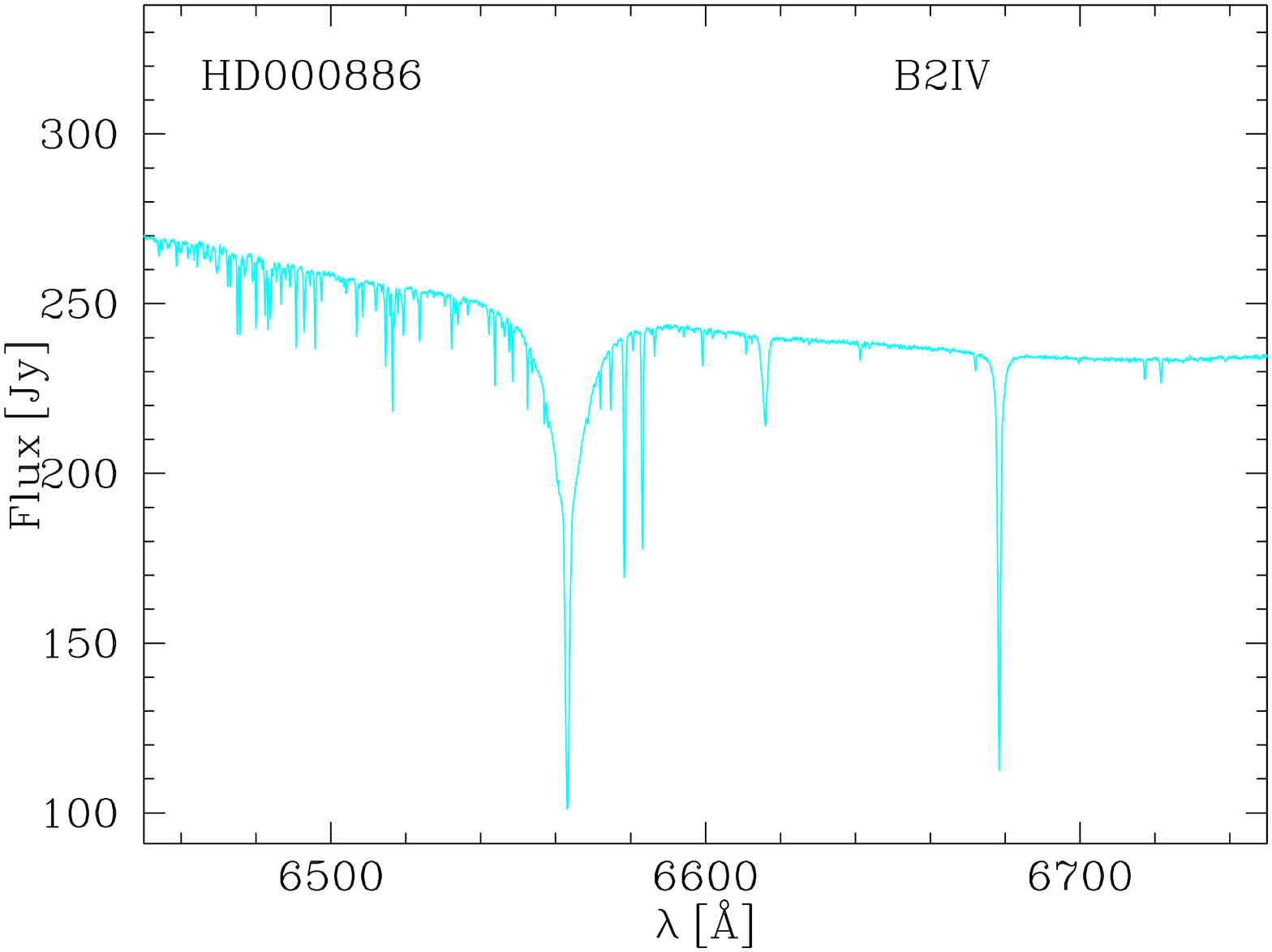}
\includegraphics[width=0.18\textwidth,angle=-90]{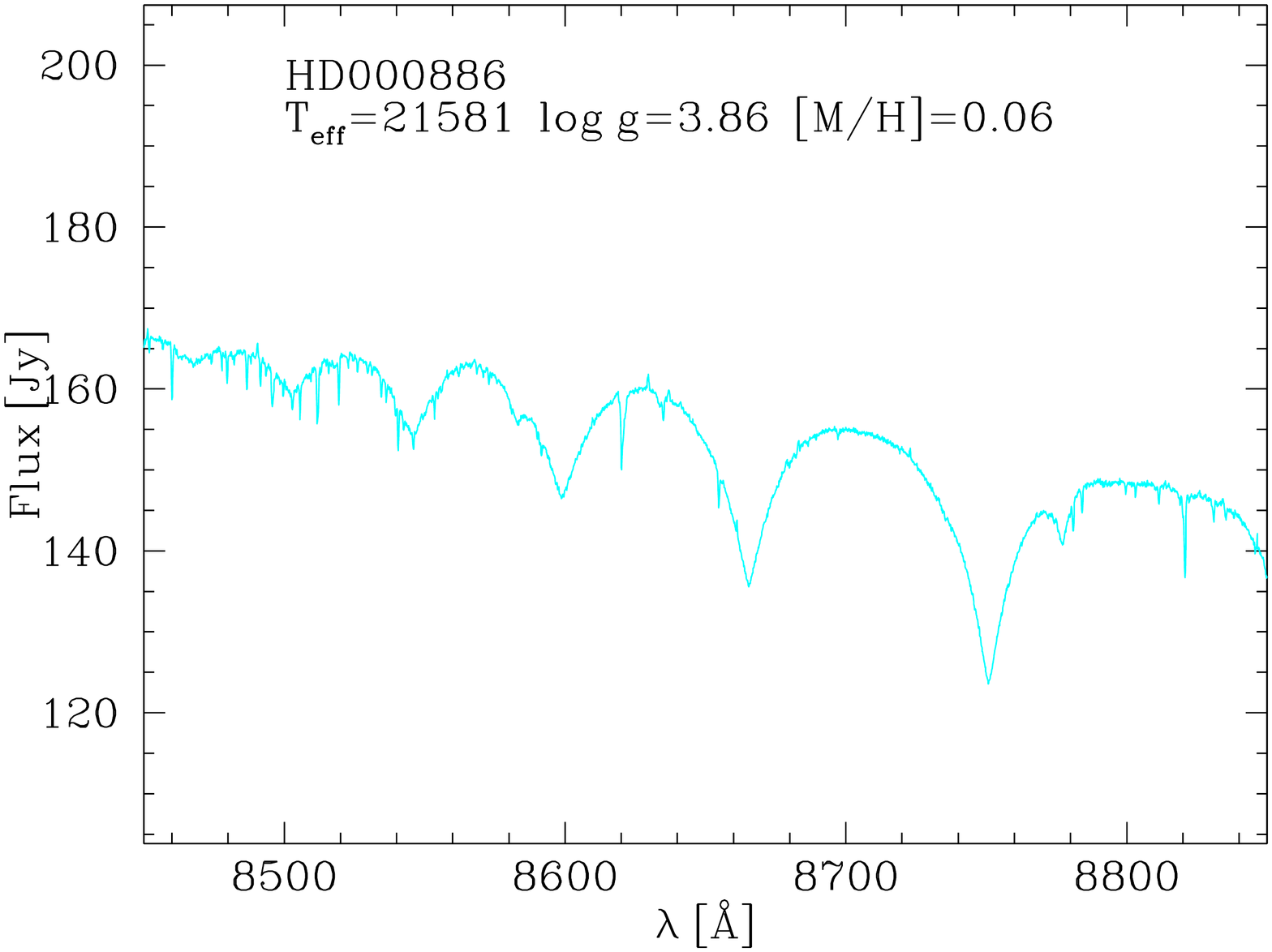}
\includegraphics[width=0.18\textwidth,angle=-90]{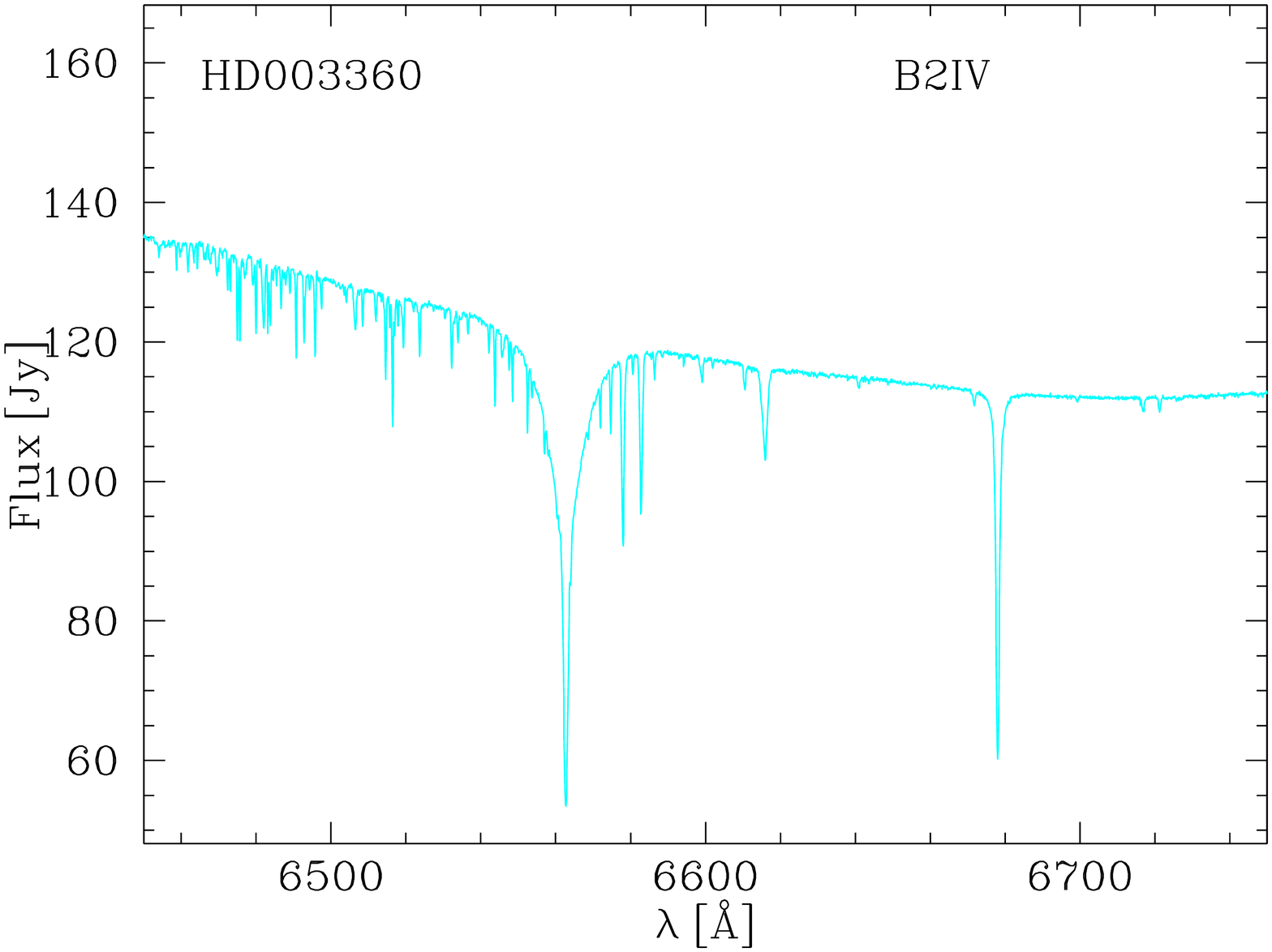}
\includegraphics[width=0.18\textwidth,angle=-90]{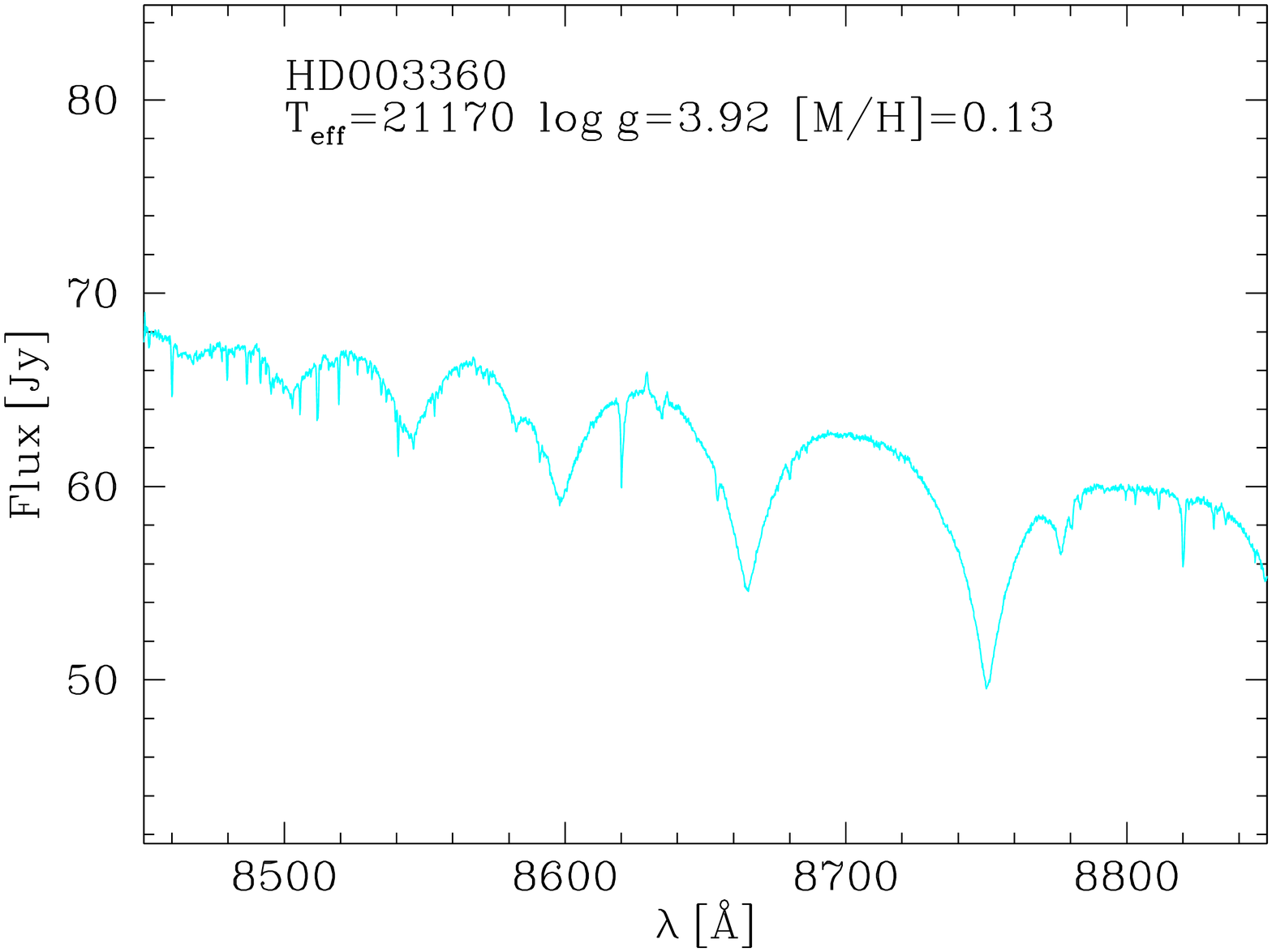}
\includegraphics[width=0.18\textwidth,angle=-90]{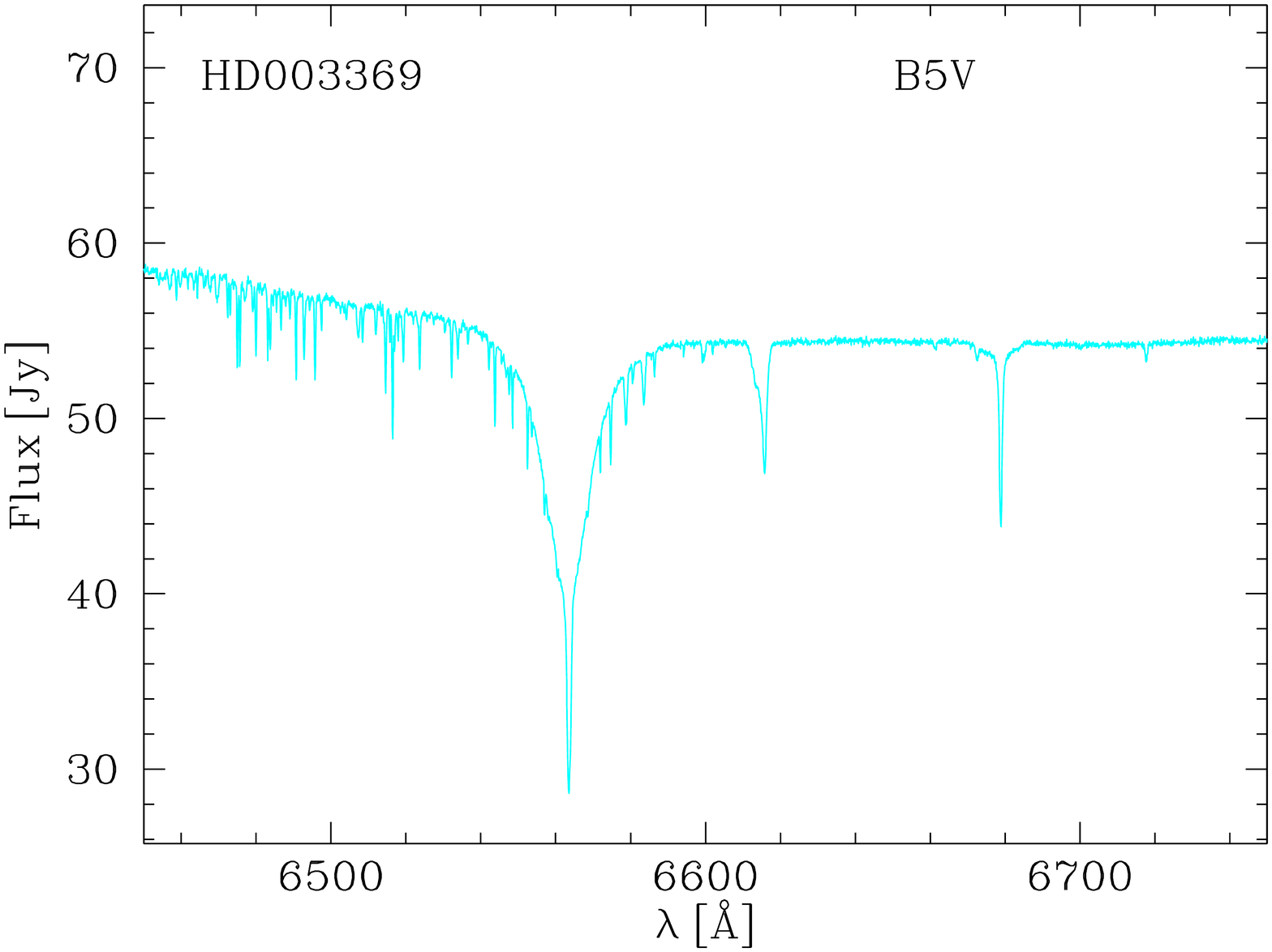}
\includegraphics[width=0.18\textwidth,angle=-90]{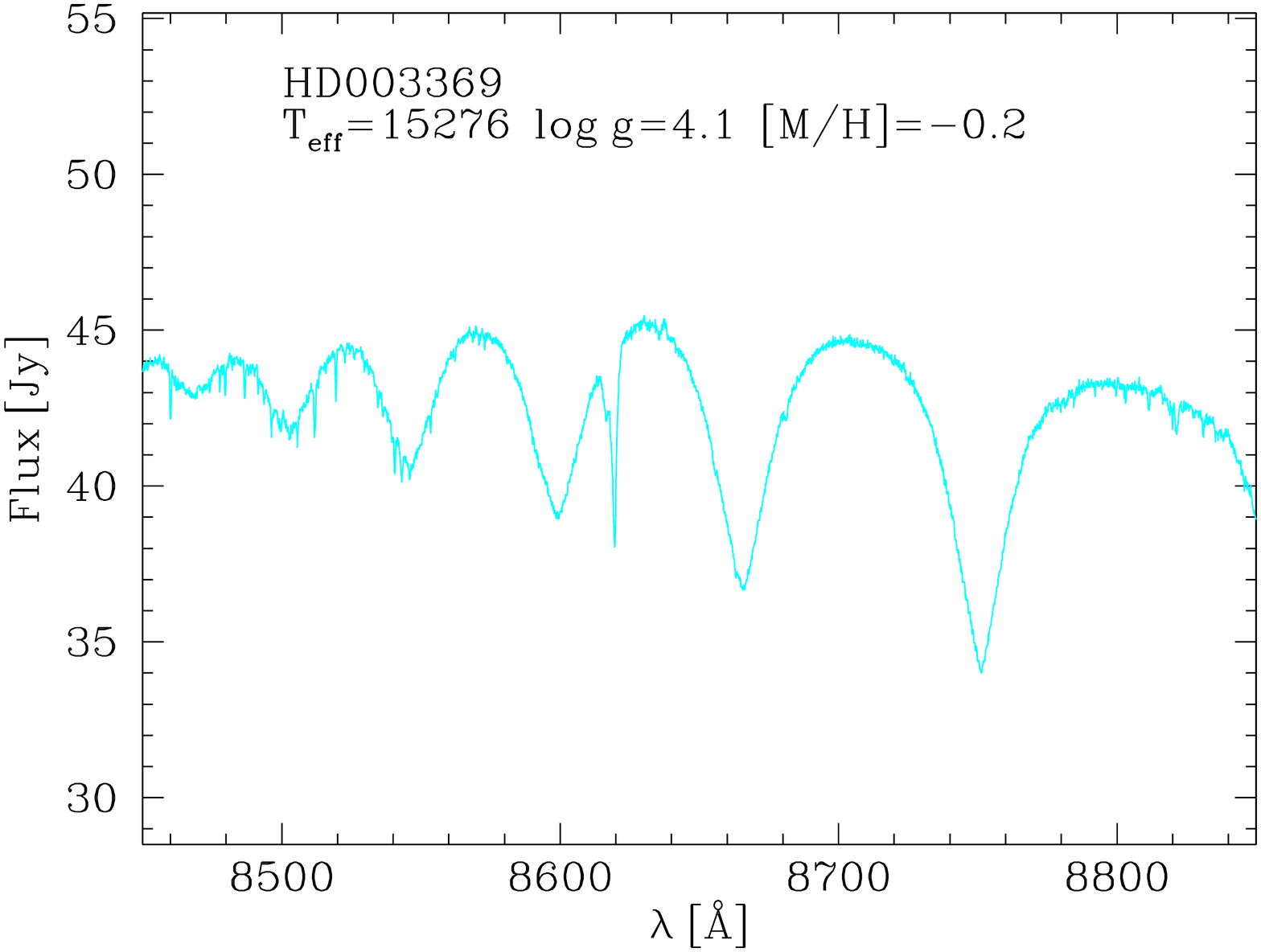}
\includegraphics[width=0.18\textwidth,angle=-90]{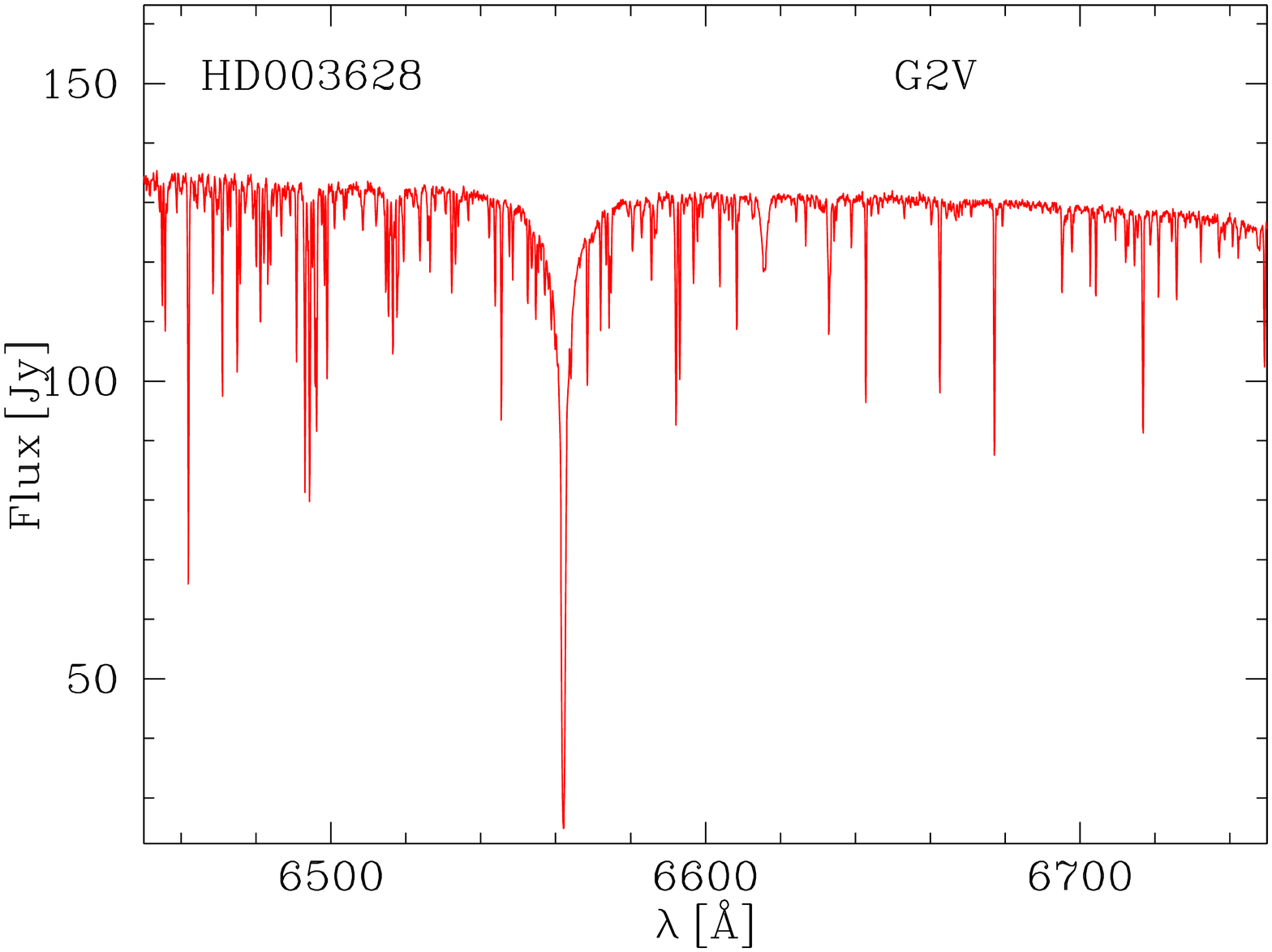}
\includegraphics[width=0.18\textwidth,angle=-90]{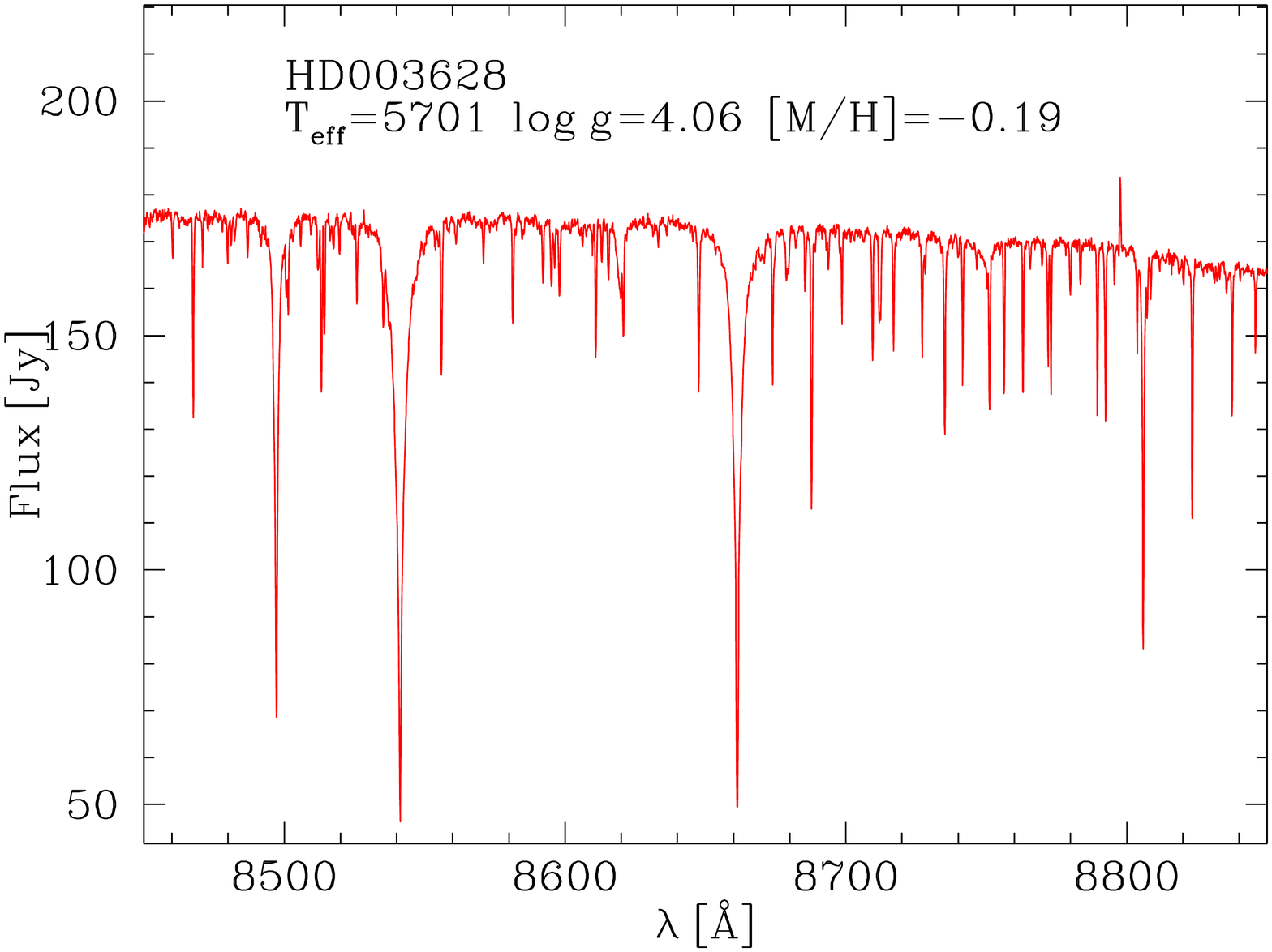}
\includegraphics[width=0.18\textwidth,angle=-90]{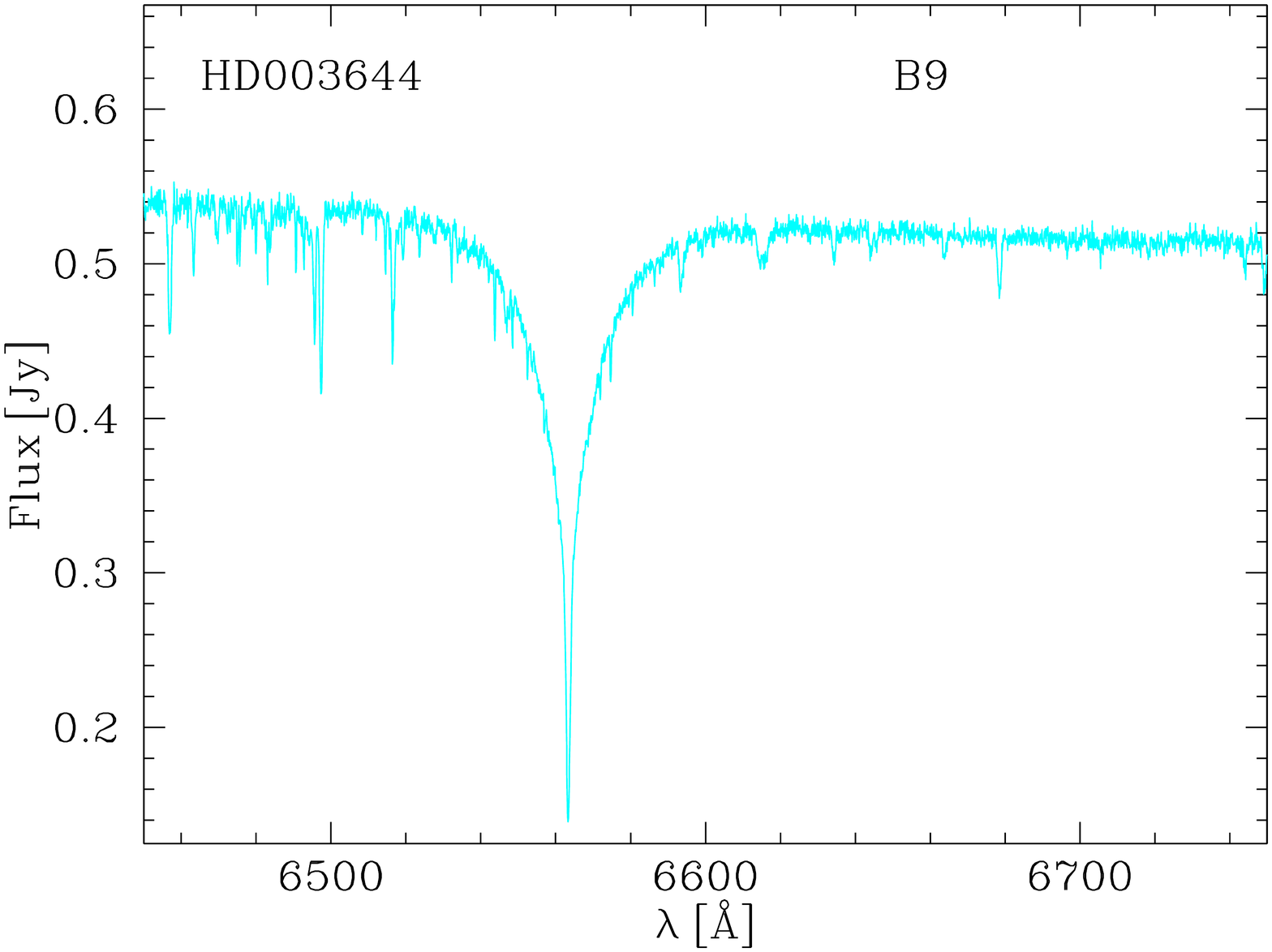}
\includegraphics[width=0.18\textwidth,angle=-90]{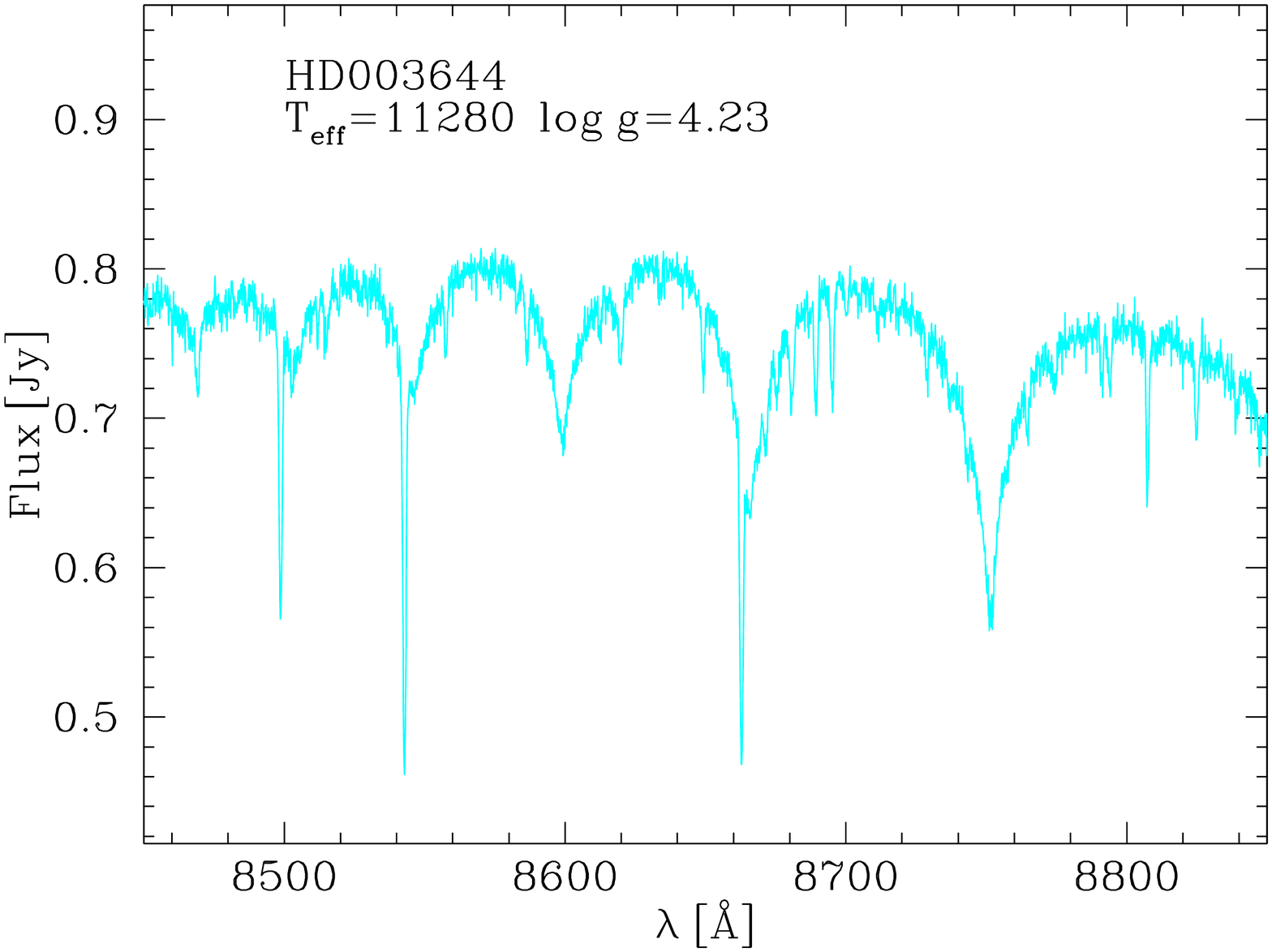}
\includegraphics[width=0.18\textwidth,angle=-90]{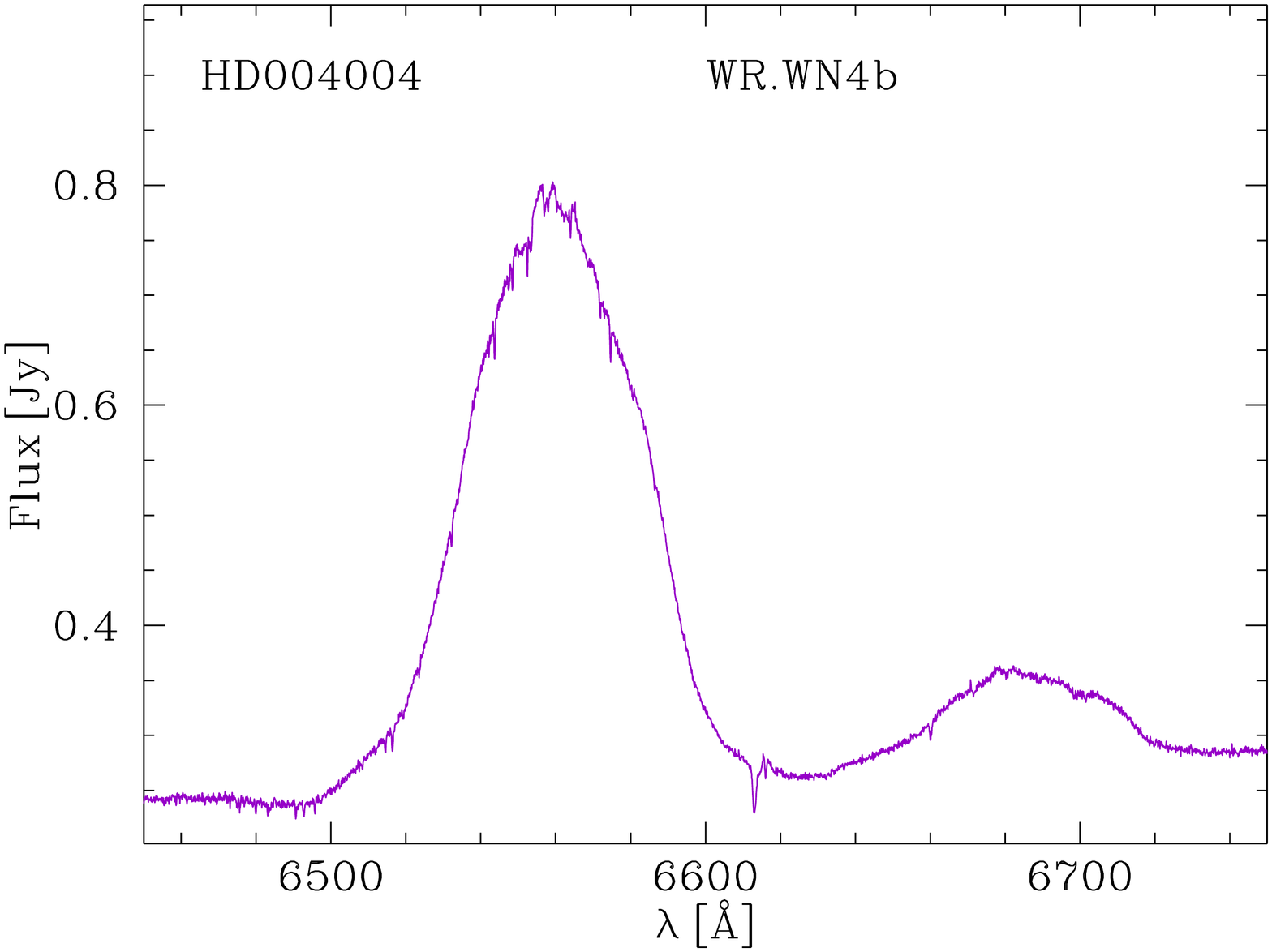}
\includegraphics[width=0.18\textwidth,angle=-90]{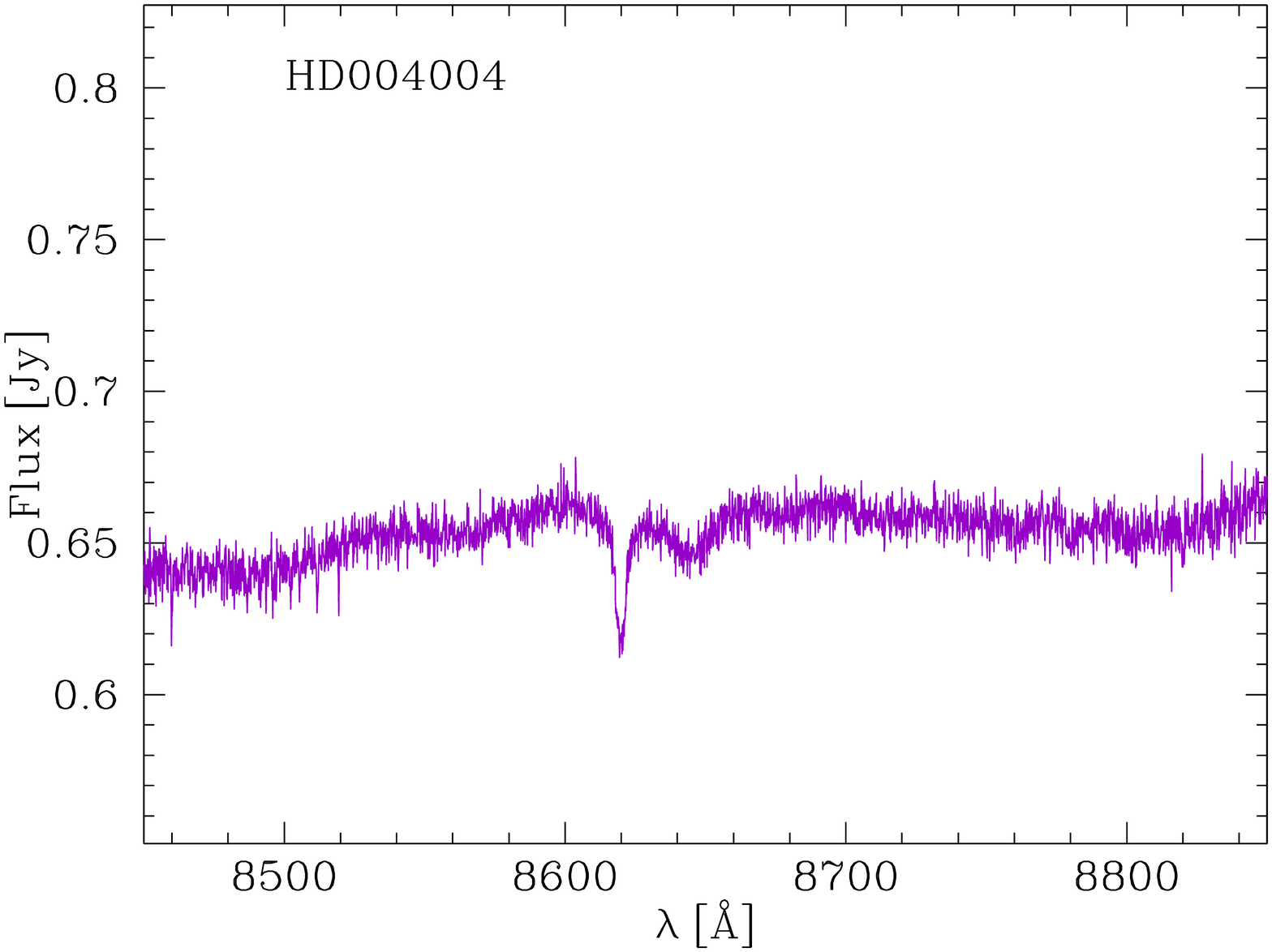}

\contcaption{2. Stars shown in this page are:  BD+800245, G171-010, G197-45, G202-65, G234-28, HD000108, HD000358, HD000560, HD000886, HD003360, HD003369, HD003628, HD003644 and HD004004.}
\end{figure*}

\begin{figure*}
\includegraphics[width=0.18\textwidth,angle=-90]{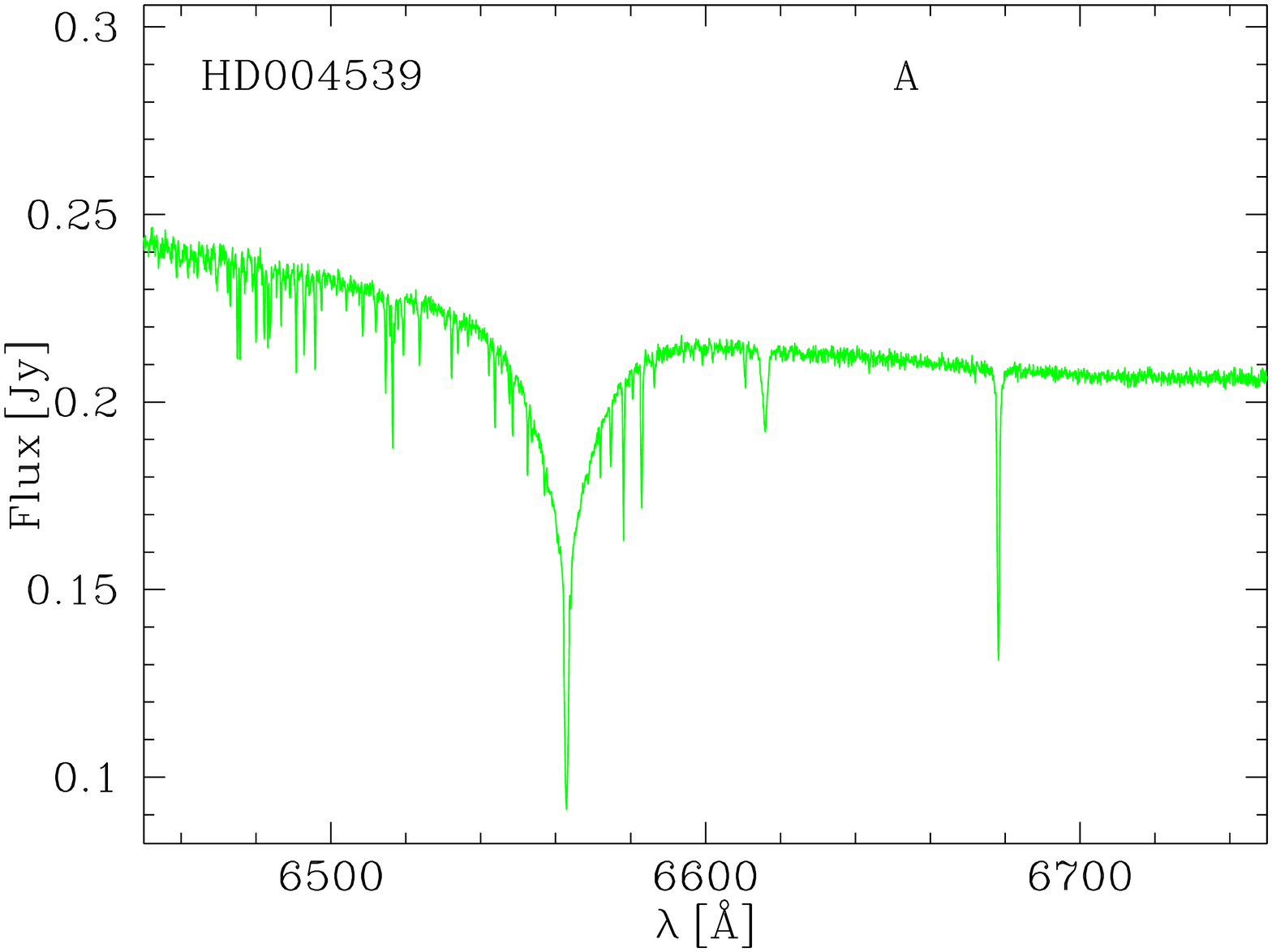}
\includegraphics[width=0.18\textwidth,angle=-90]{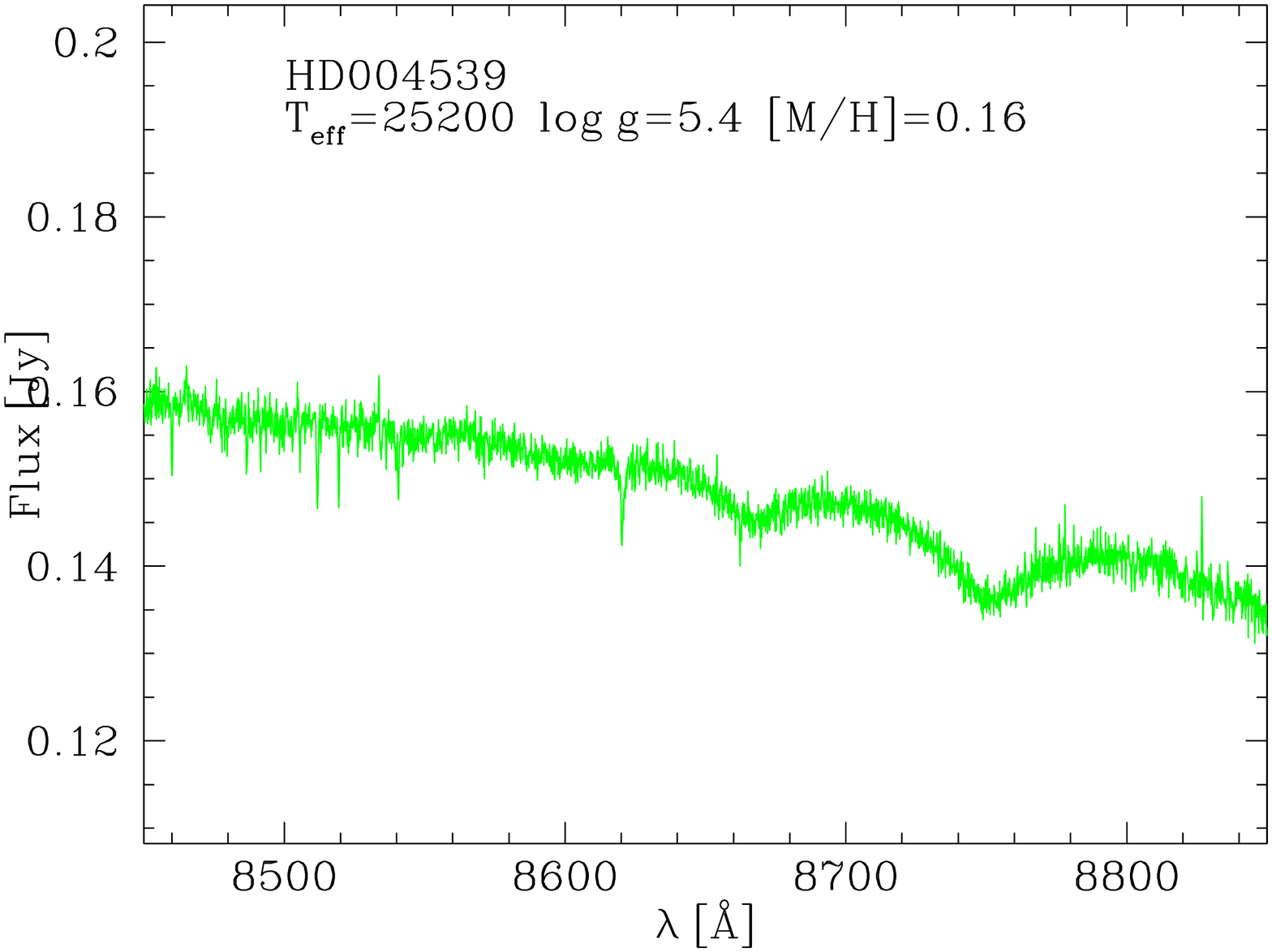}
\includegraphics[width=0.18\textwidth,angle=-90]{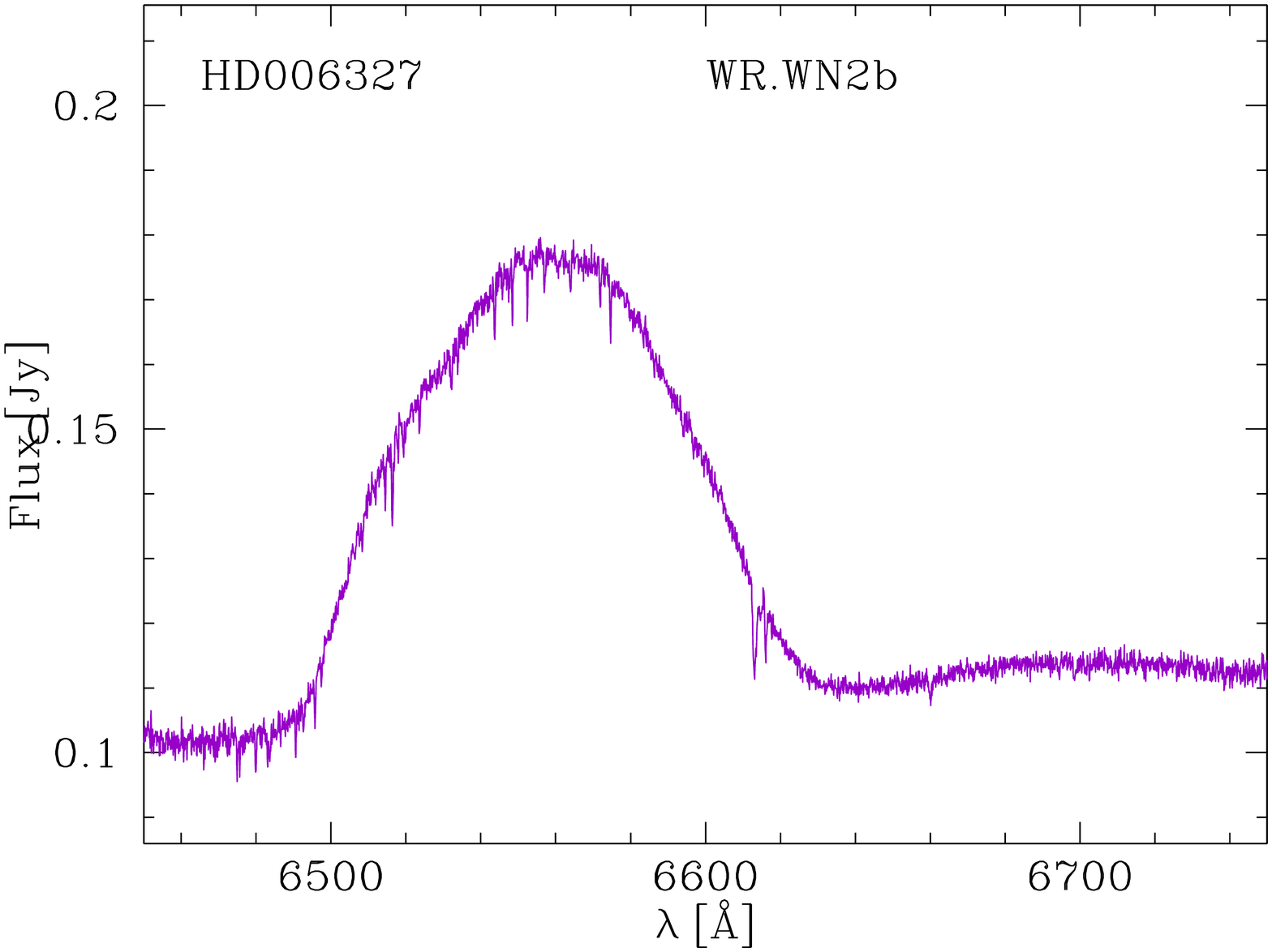}
\includegraphics[width=0.18\textwidth,angle=-90]{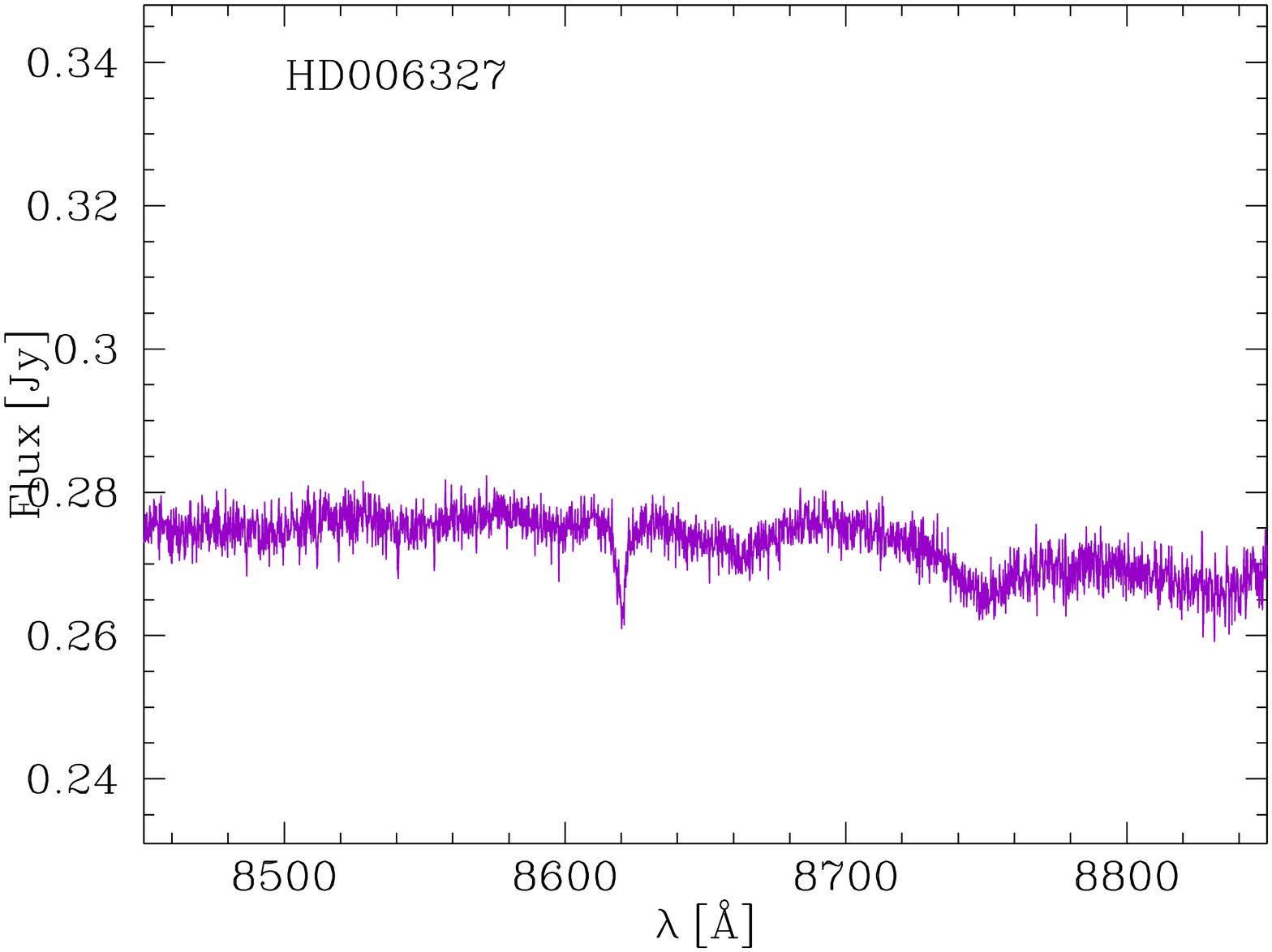}
\includegraphics[width=0.18\textwidth,angle=-90]{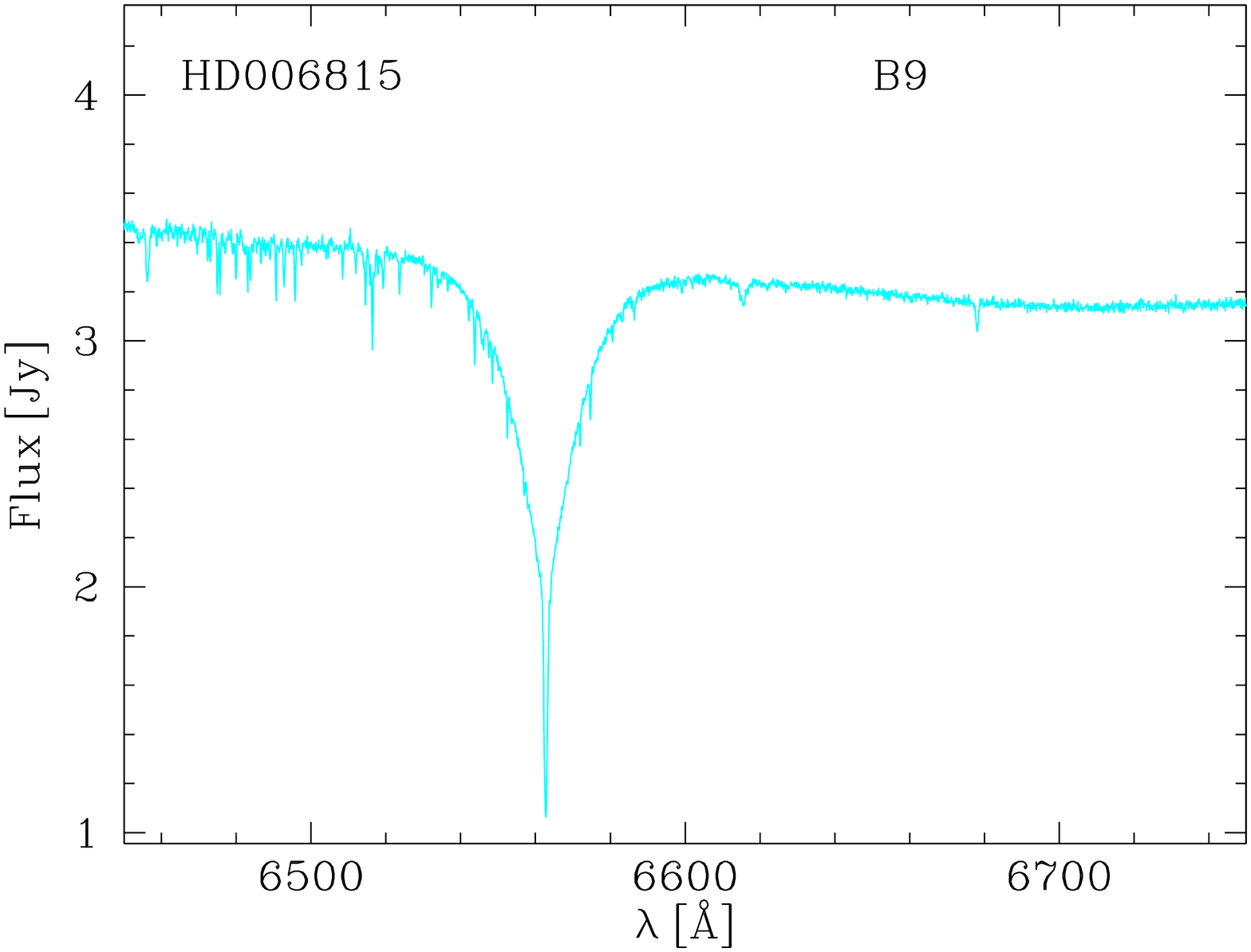}
\includegraphics[width=0.18\textwidth,angle=-90]{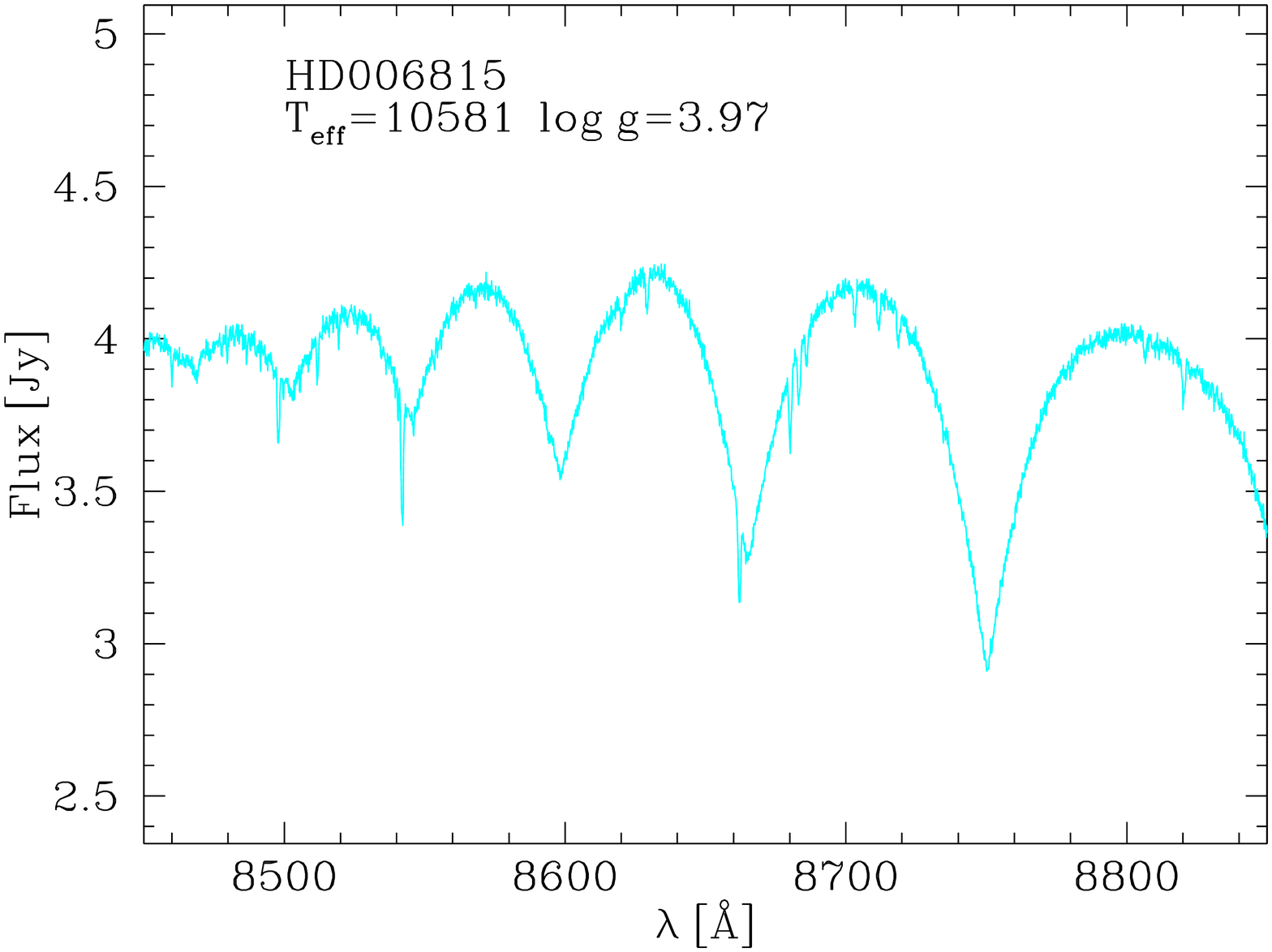}
\includegraphics[width=0.18\textwidth,angle=-90]{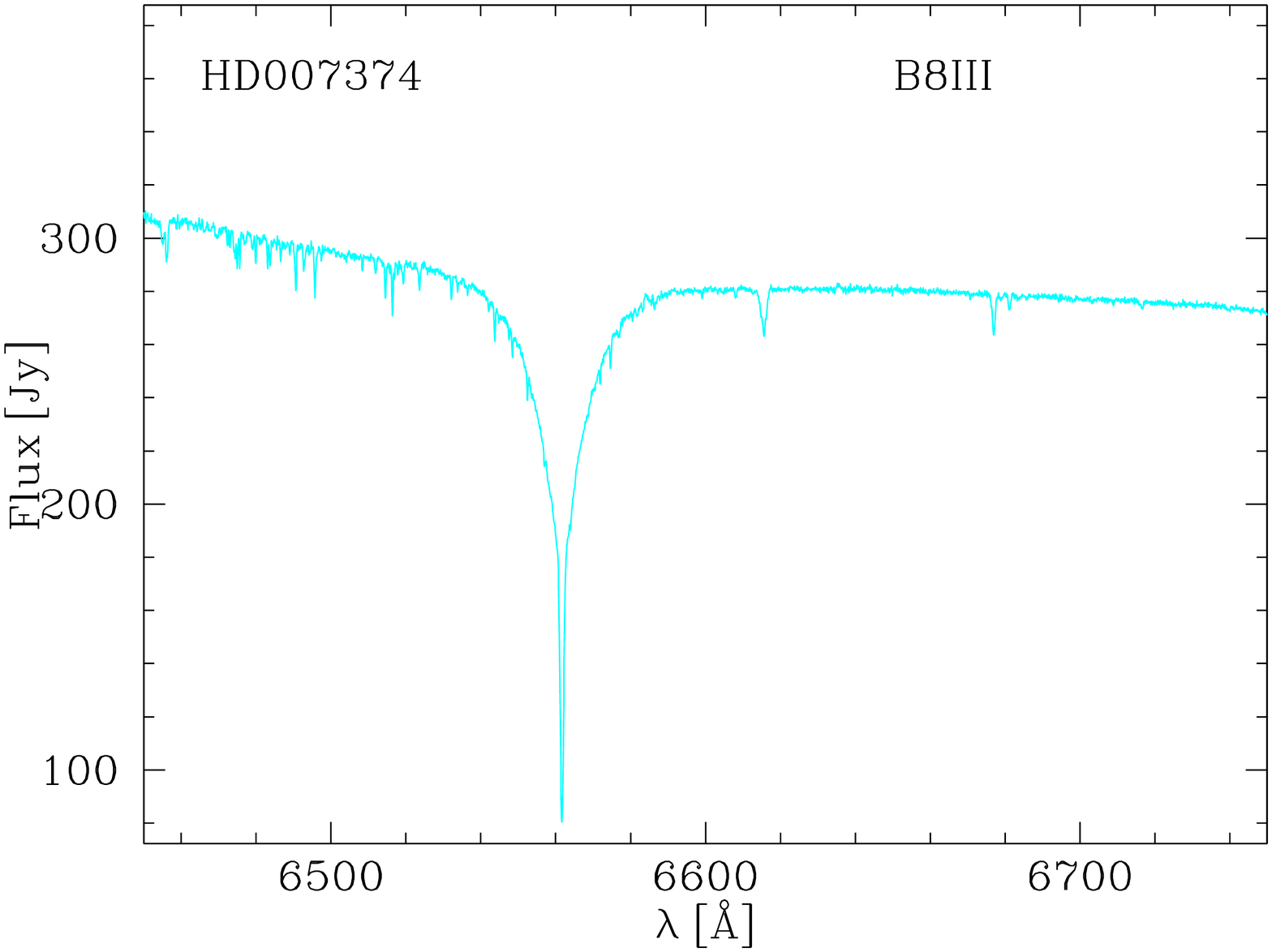}
\includegraphics[width=0.18\textwidth,angle=-90]{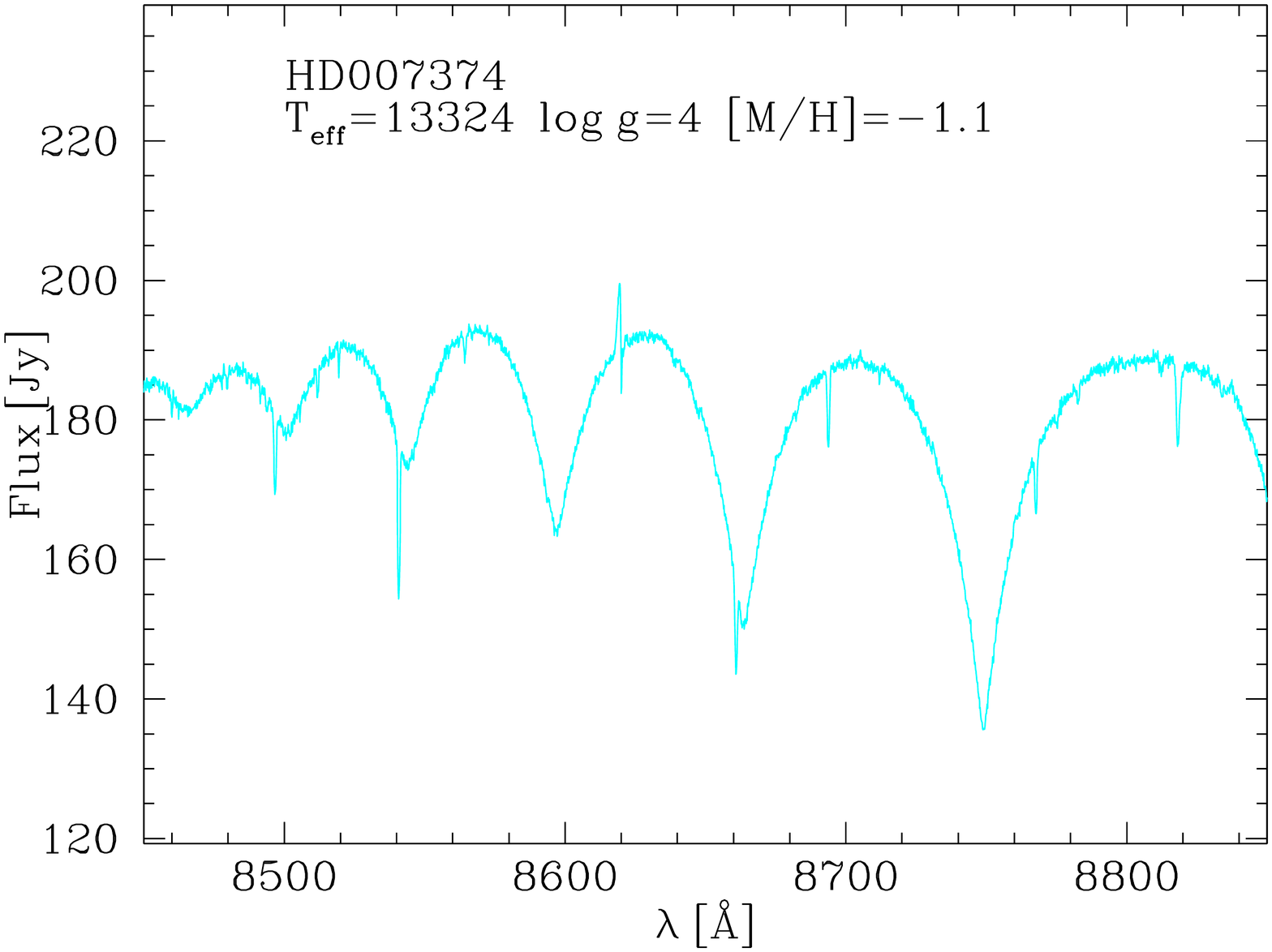}
\includegraphics[width=0.18\textwidth,angle=-90]{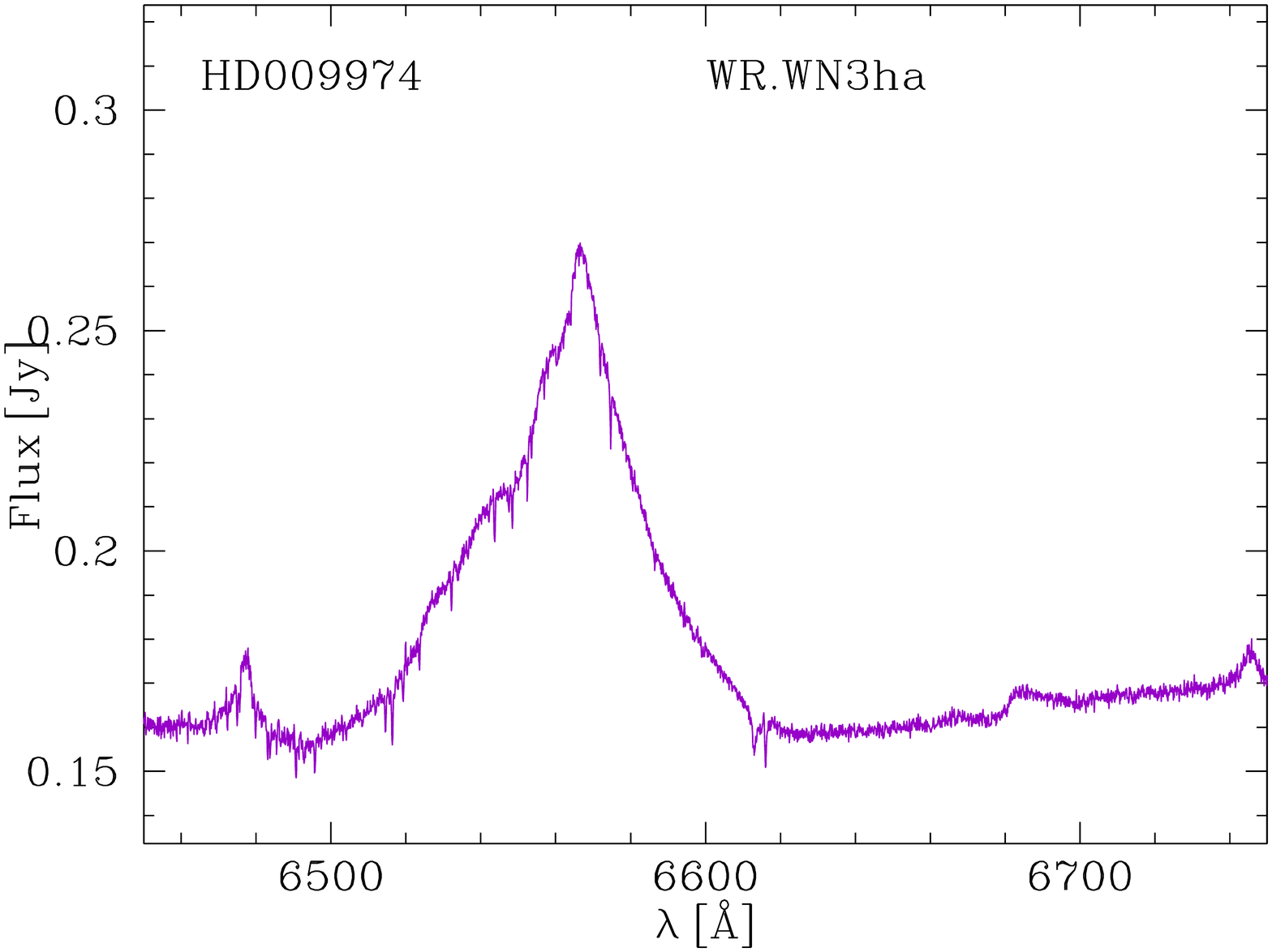}
\includegraphics[width=0.18\textwidth,angle=-90]{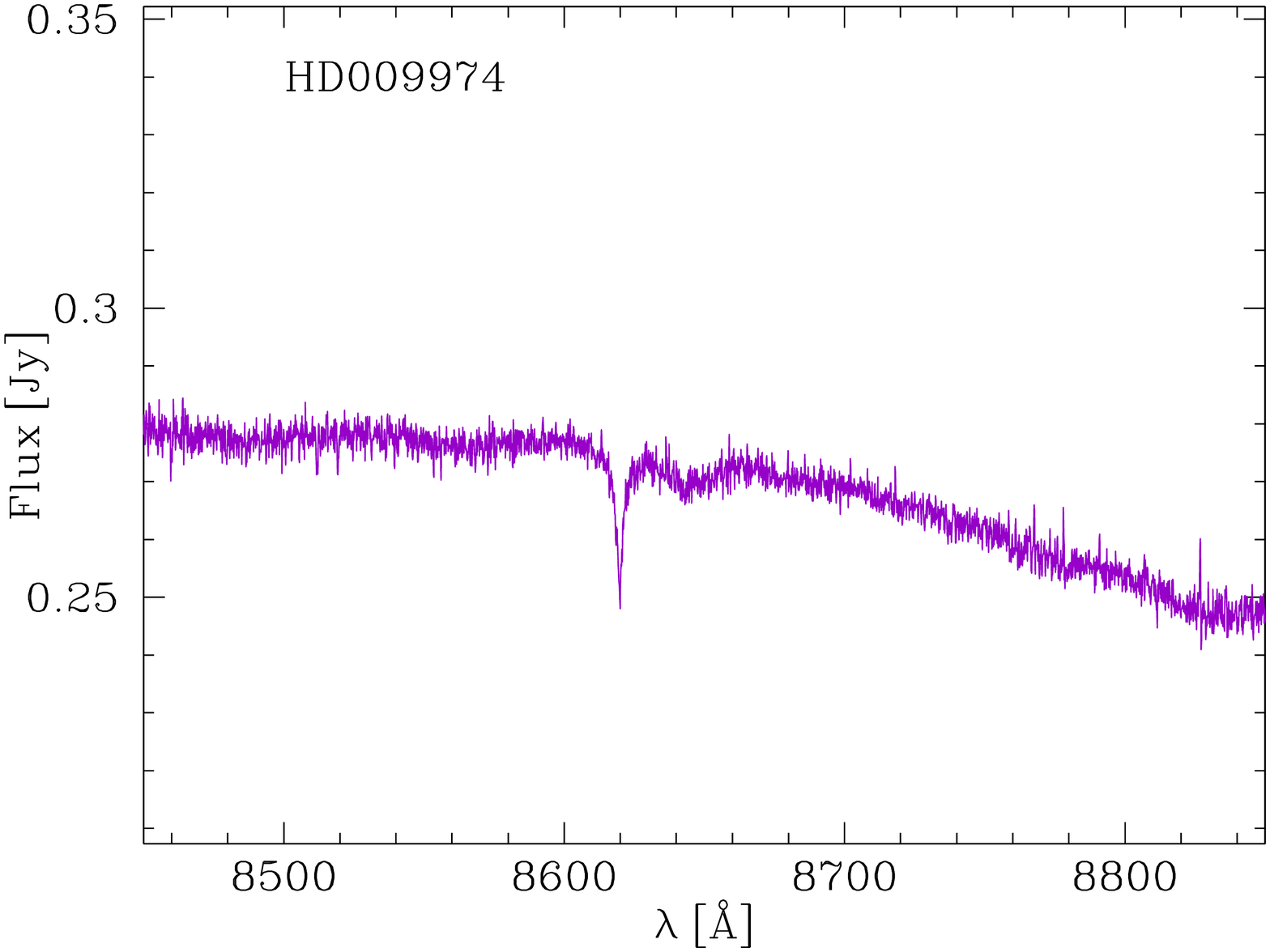}
\includegraphics[width=0.18\textwidth,angle=-90]{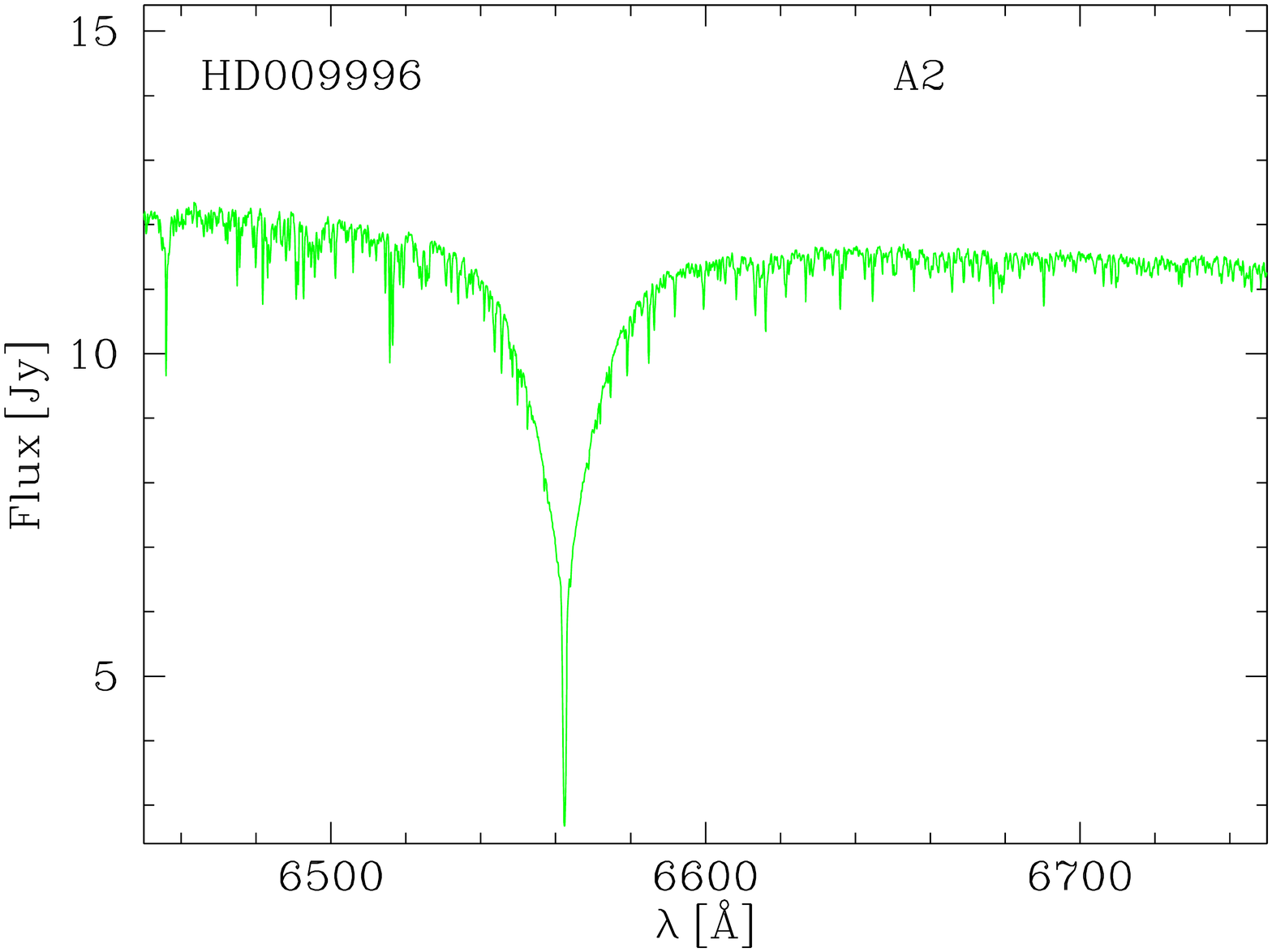}
\includegraphics[width=0.18\textwidth,angle=-90]{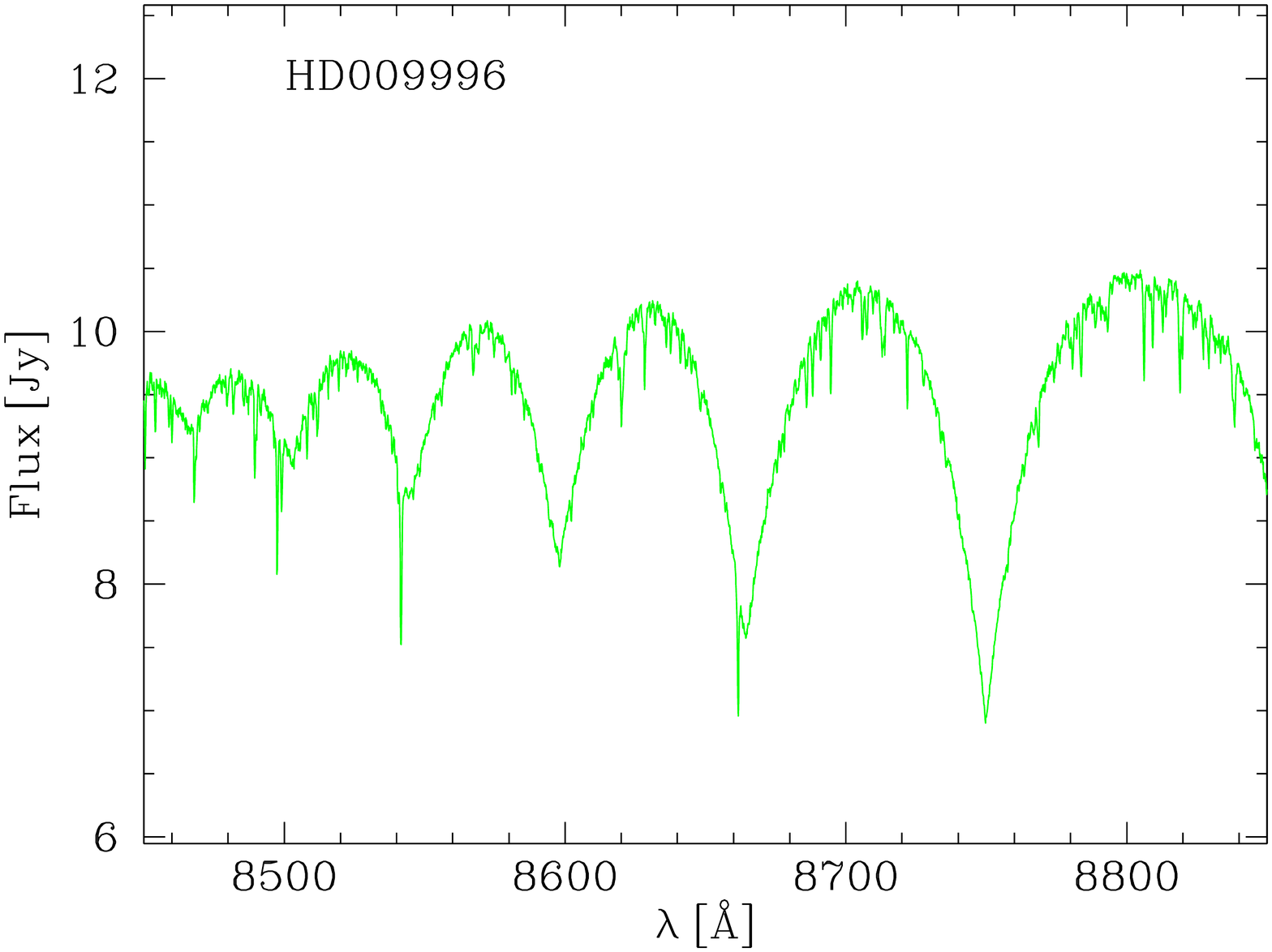}
\includegraphics[width=0.18\textwidth,angle=-90]{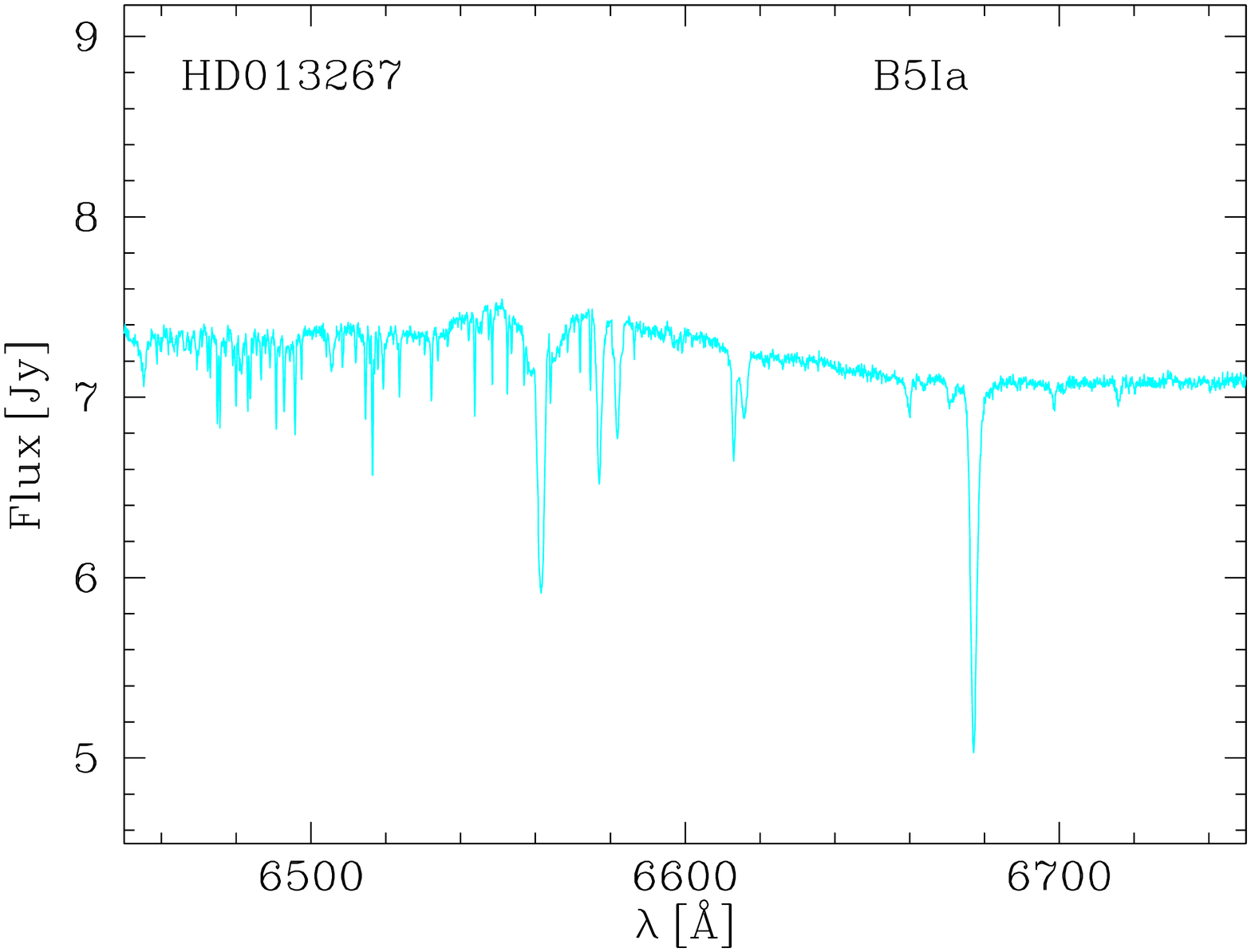}
\includegraphics[width=0.18\textwidth,angle=-90]{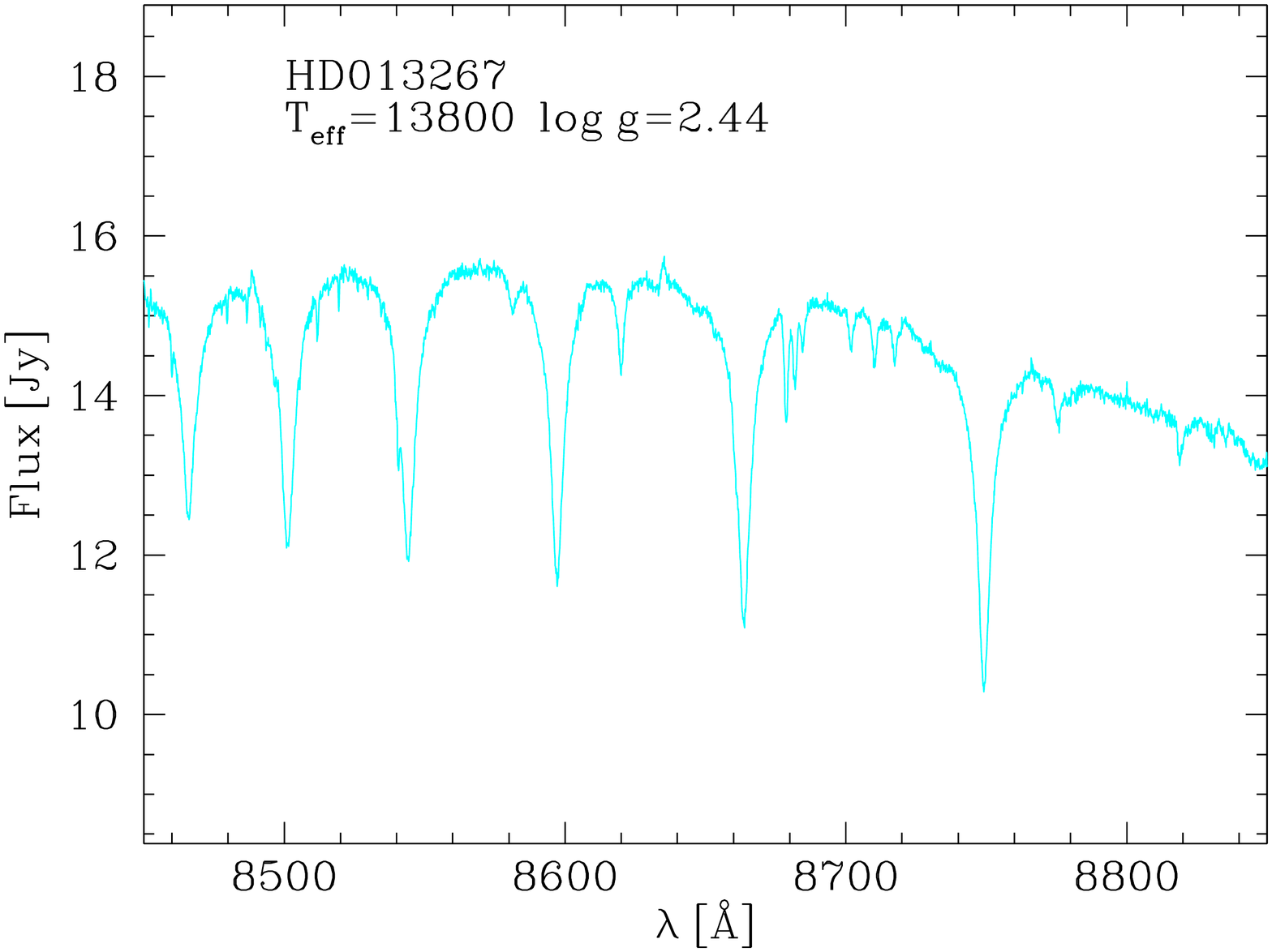}
\includegraphics[width=0.18\textwidth,angle=-90]{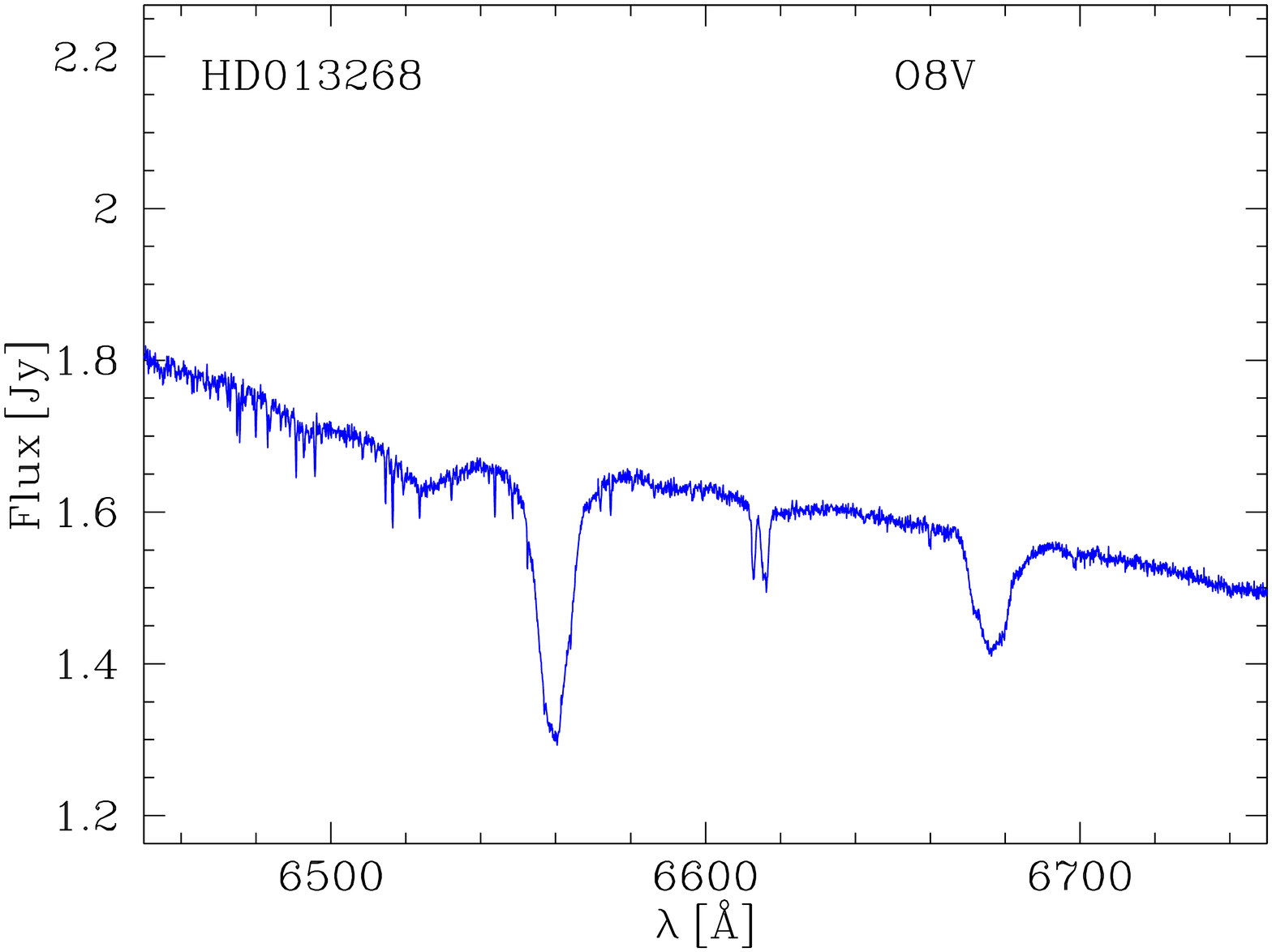}
\includegraphics[width=0.18\textwidth,angle=-90]{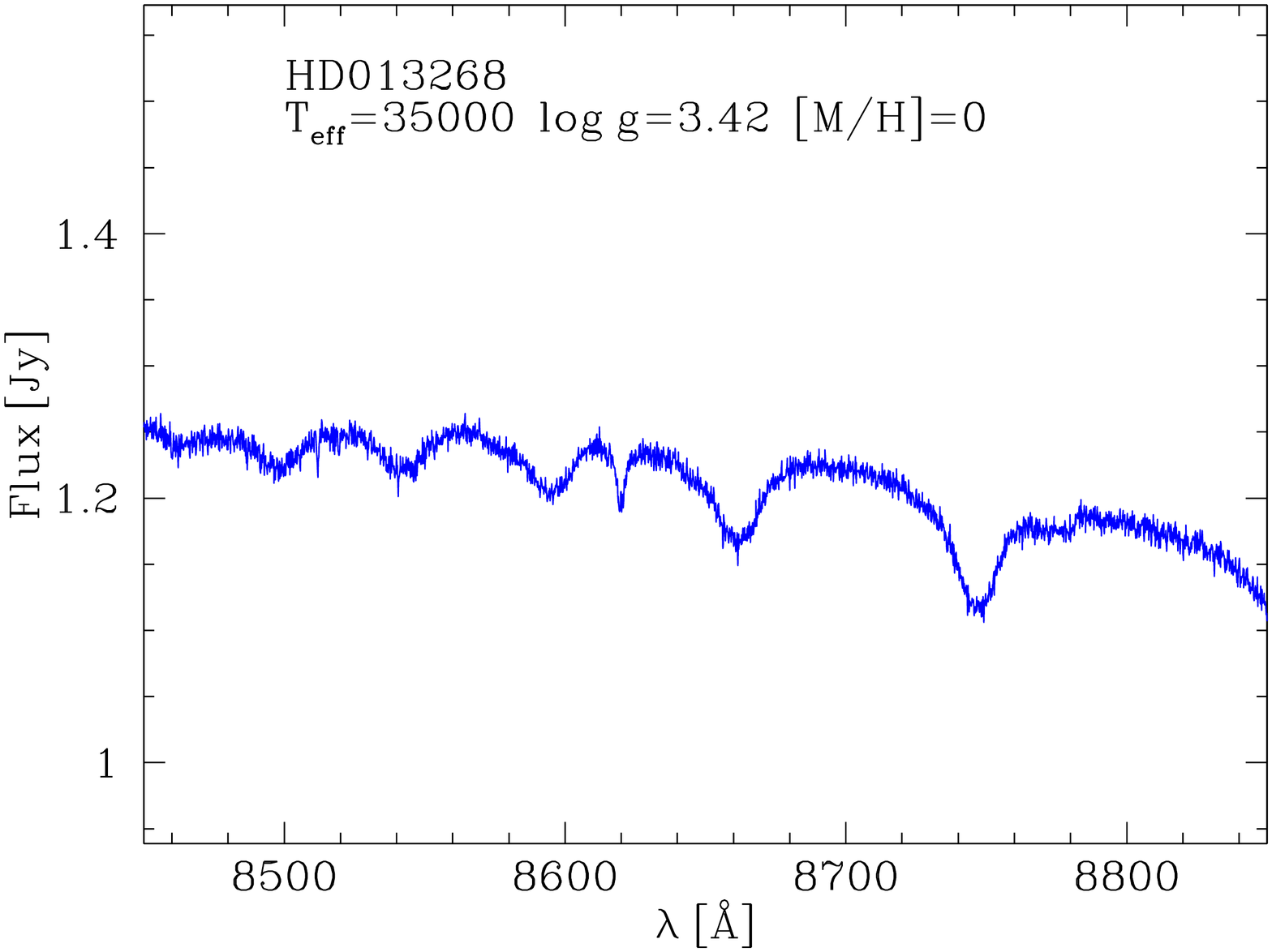}
\includegraphics[width=0.18\textwidth,angle=-90]{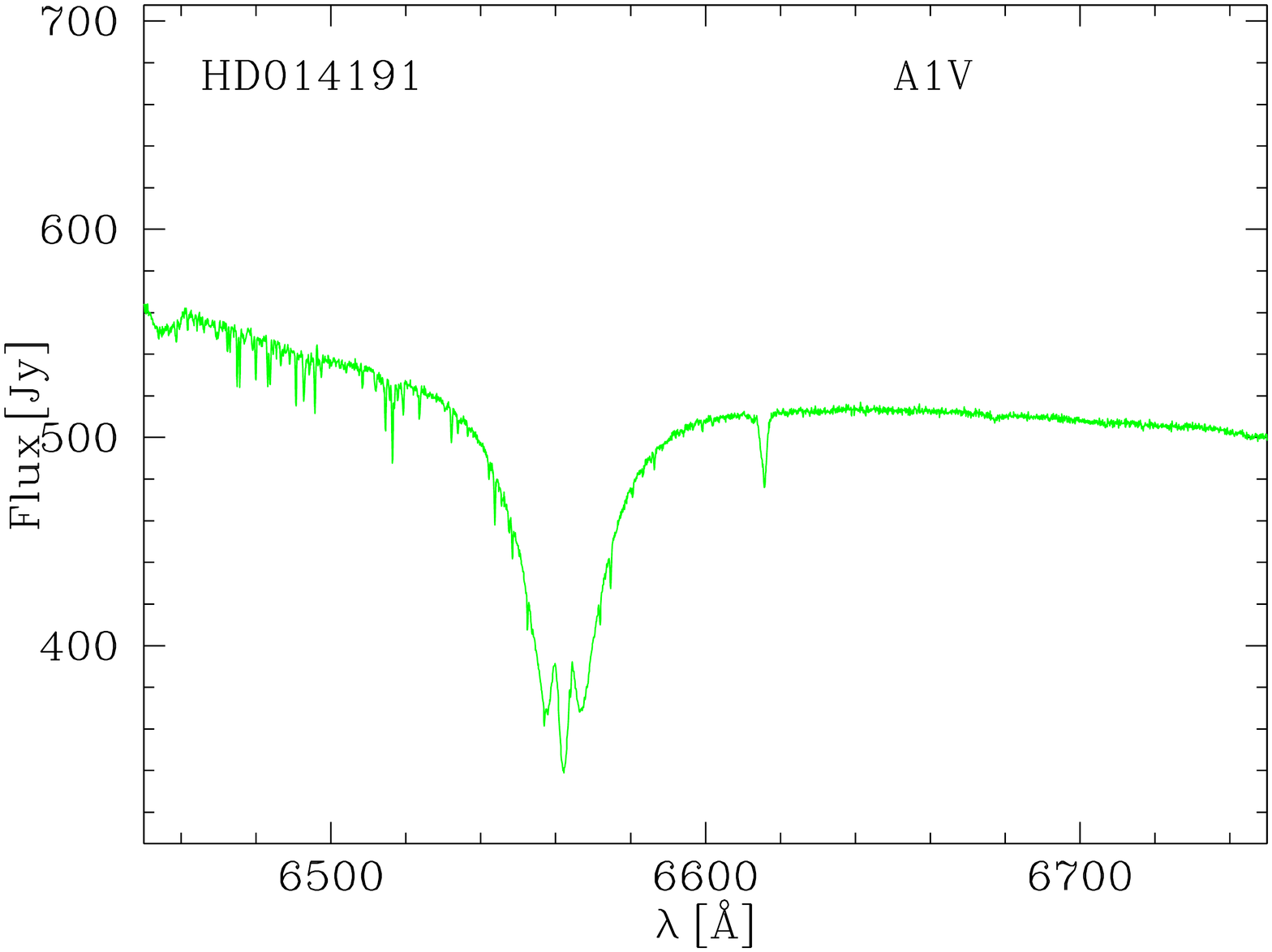}
\includegraphics[width=0.18\textwidth,angle=-90]{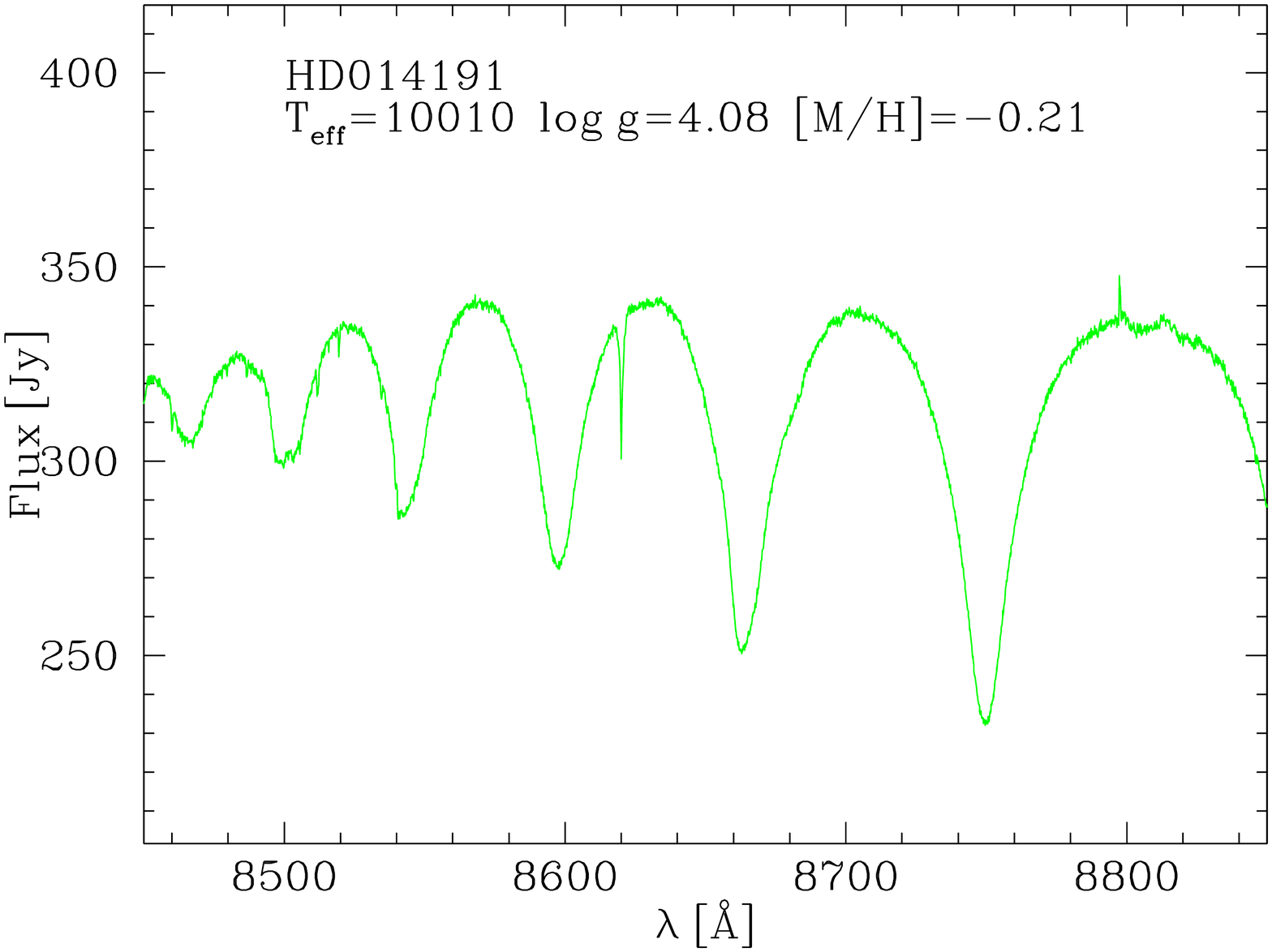}
\includegraphics[width=0.18\textwidth,angle=-90]{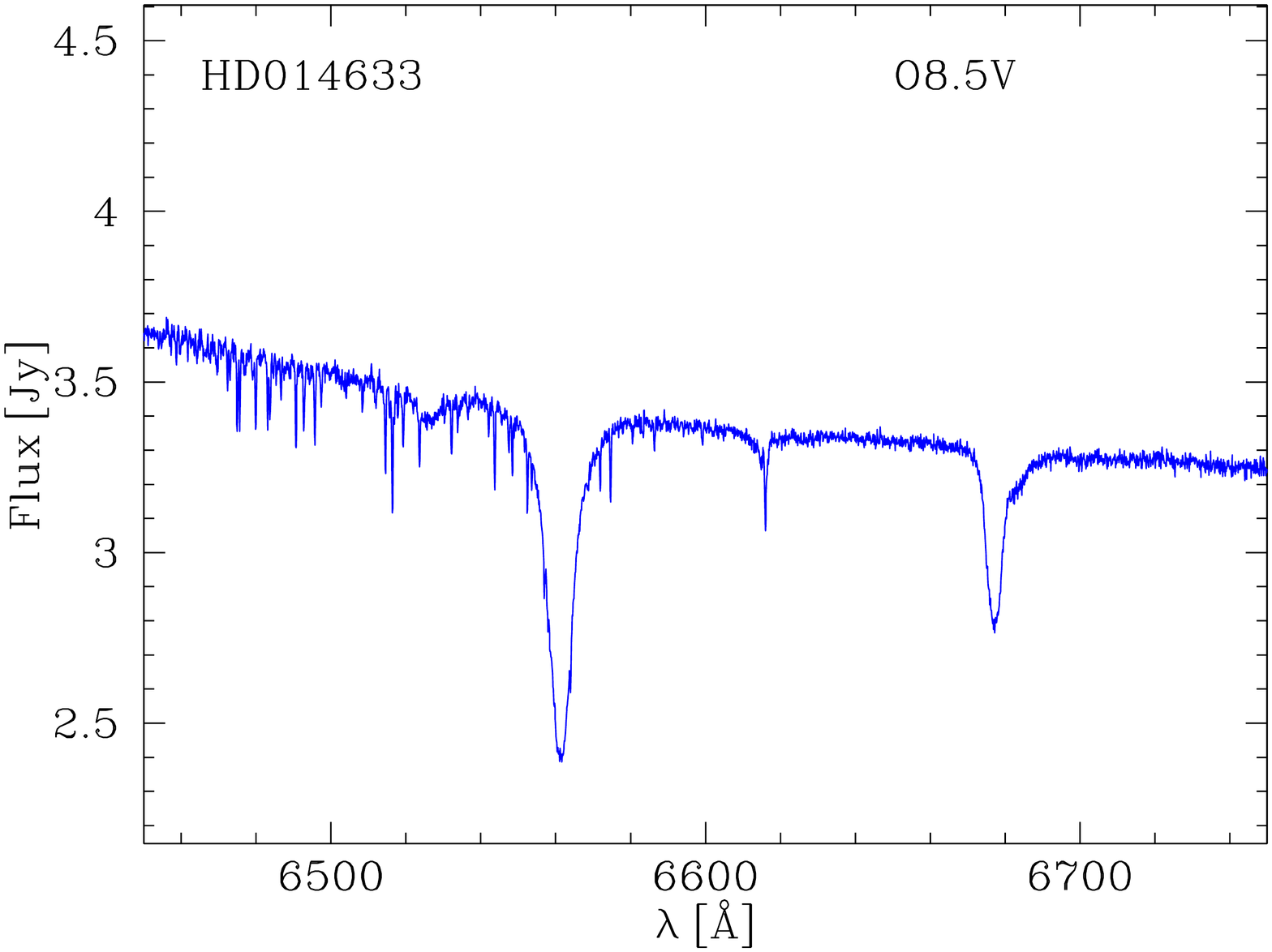}
\includegraphics[width=0.18\textwidth,angle=-90]{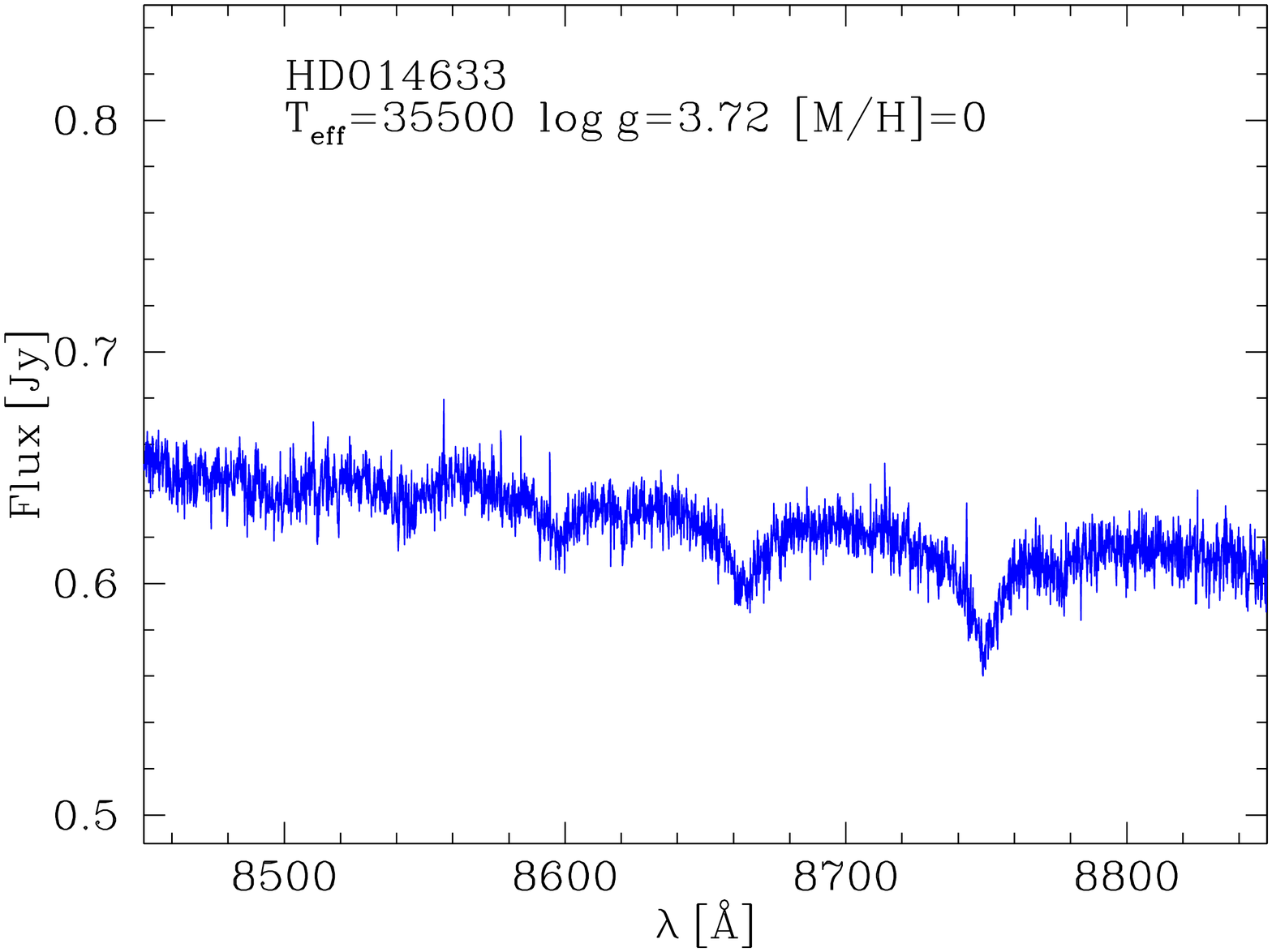}
\includegraphics[width=0.18\textwidth,angle=-90]{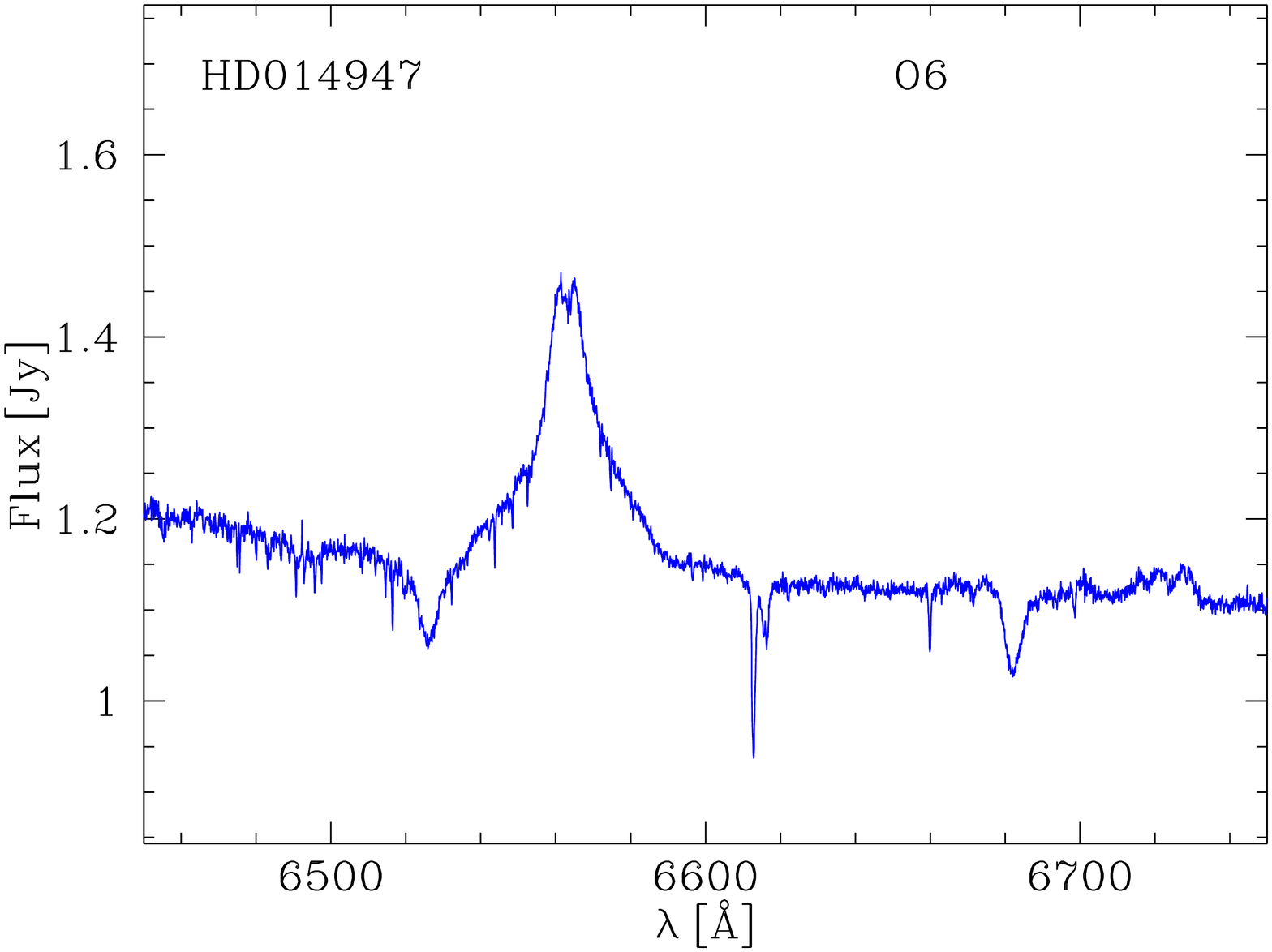}
\includegraphics[width=0.18\textwidth,angle=-90]{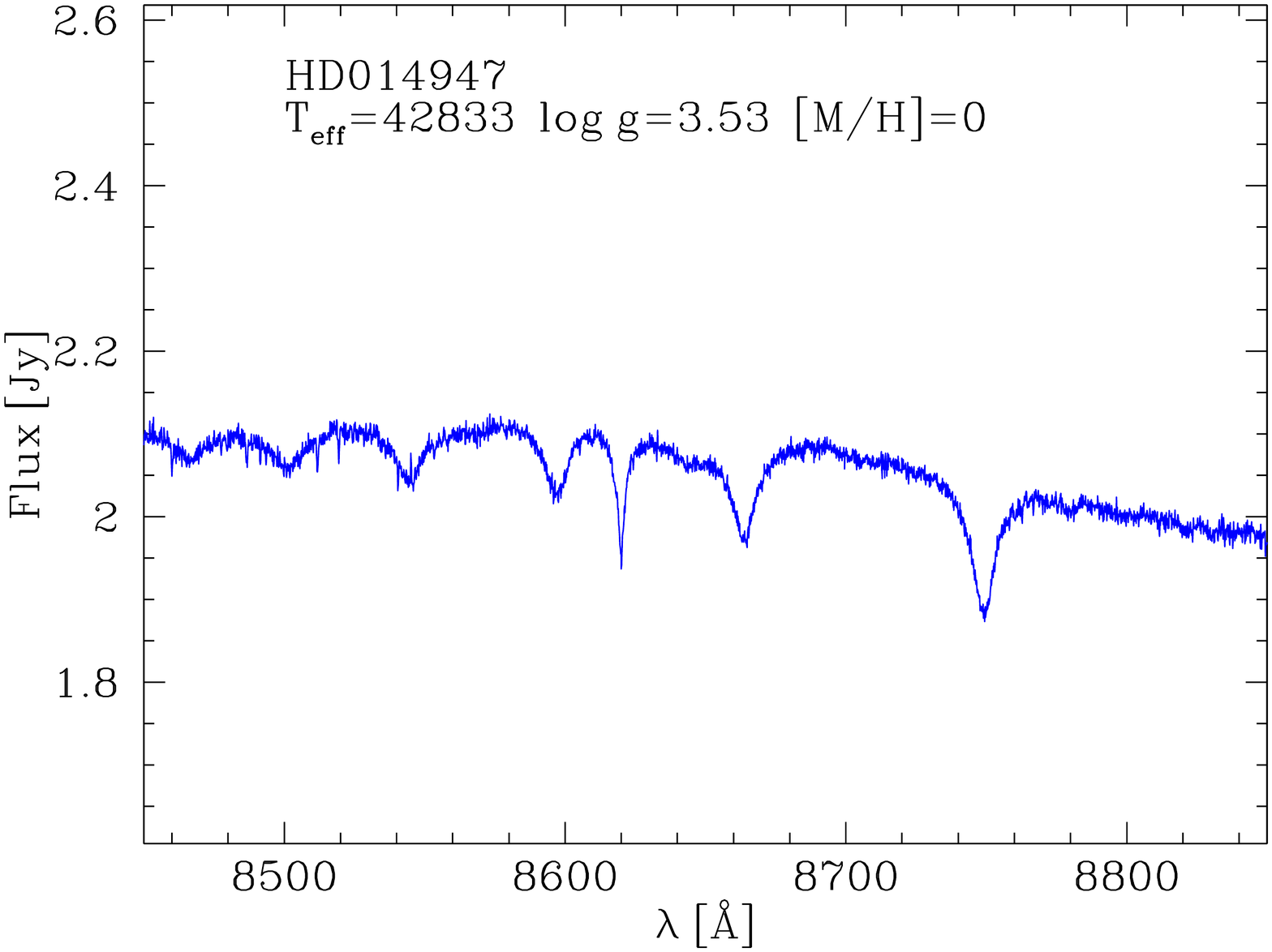}
\includegraphics[width=0.18\textwidth,angle=-90]{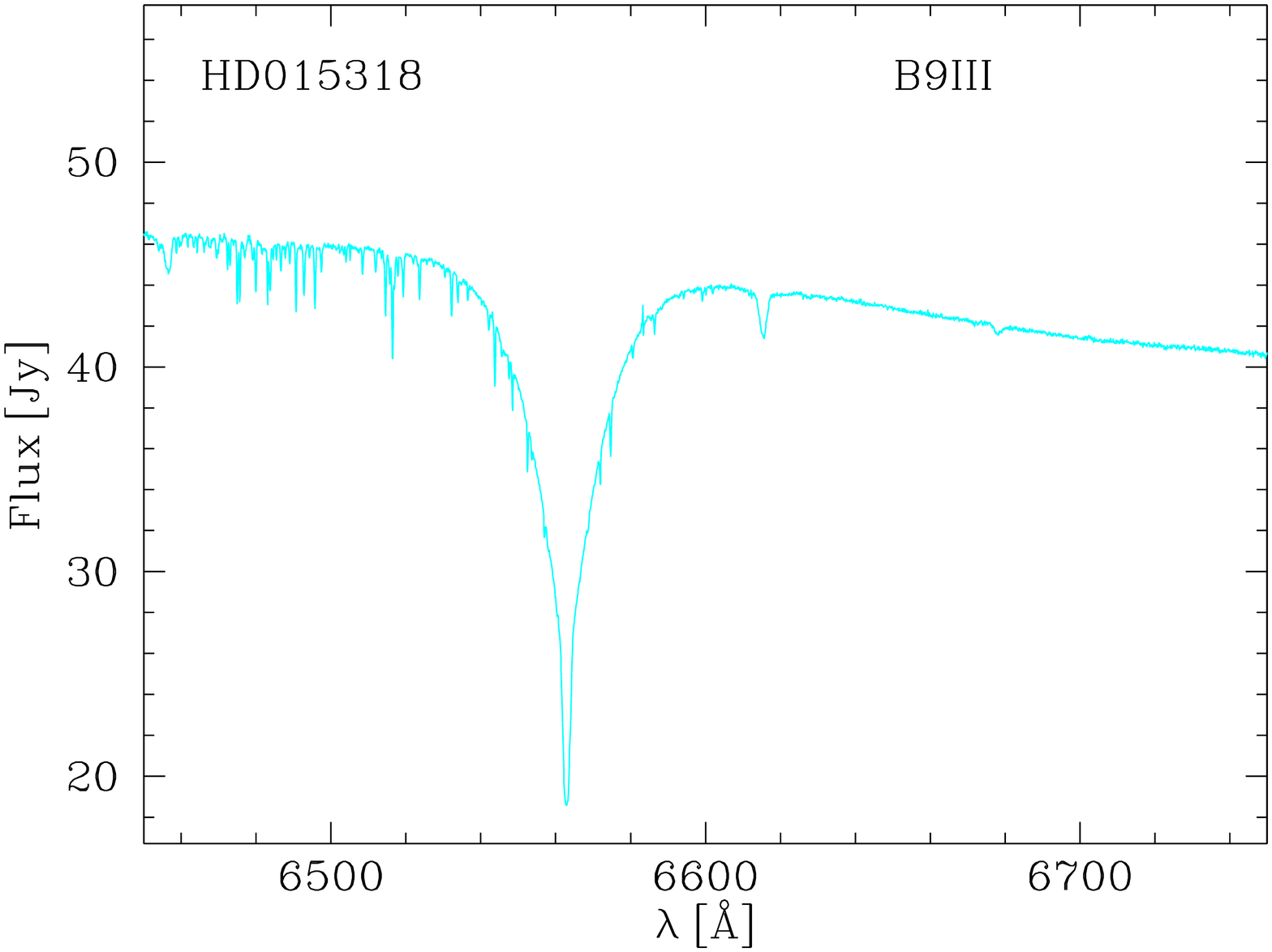}
\includegraphics[width=0.18\textwidth,angle=-90]{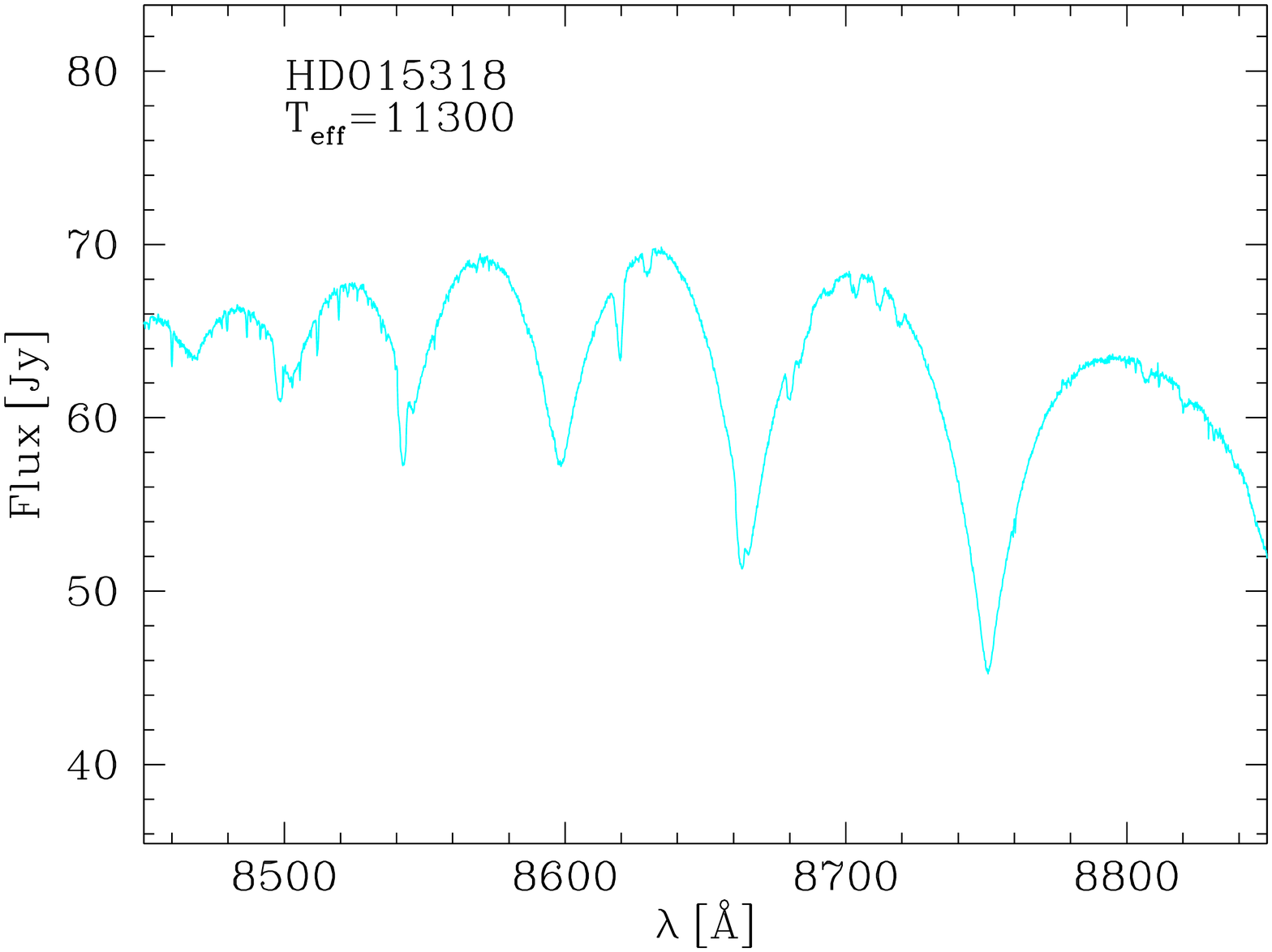}
\includegraphics[width=0.18\textwidth,angle=-90]{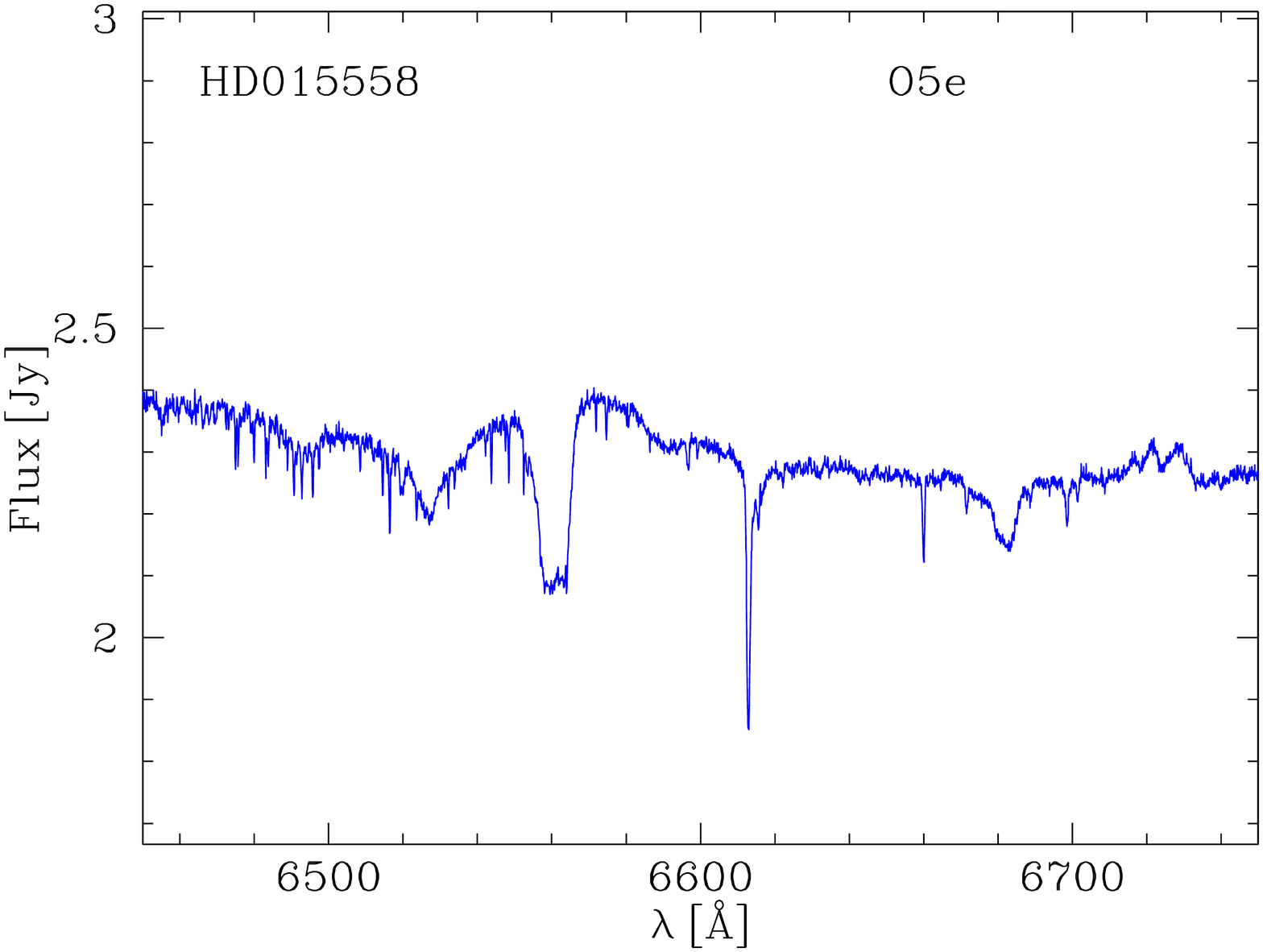}
\includegraphics[width=0.18\textwidth,angle=-90]{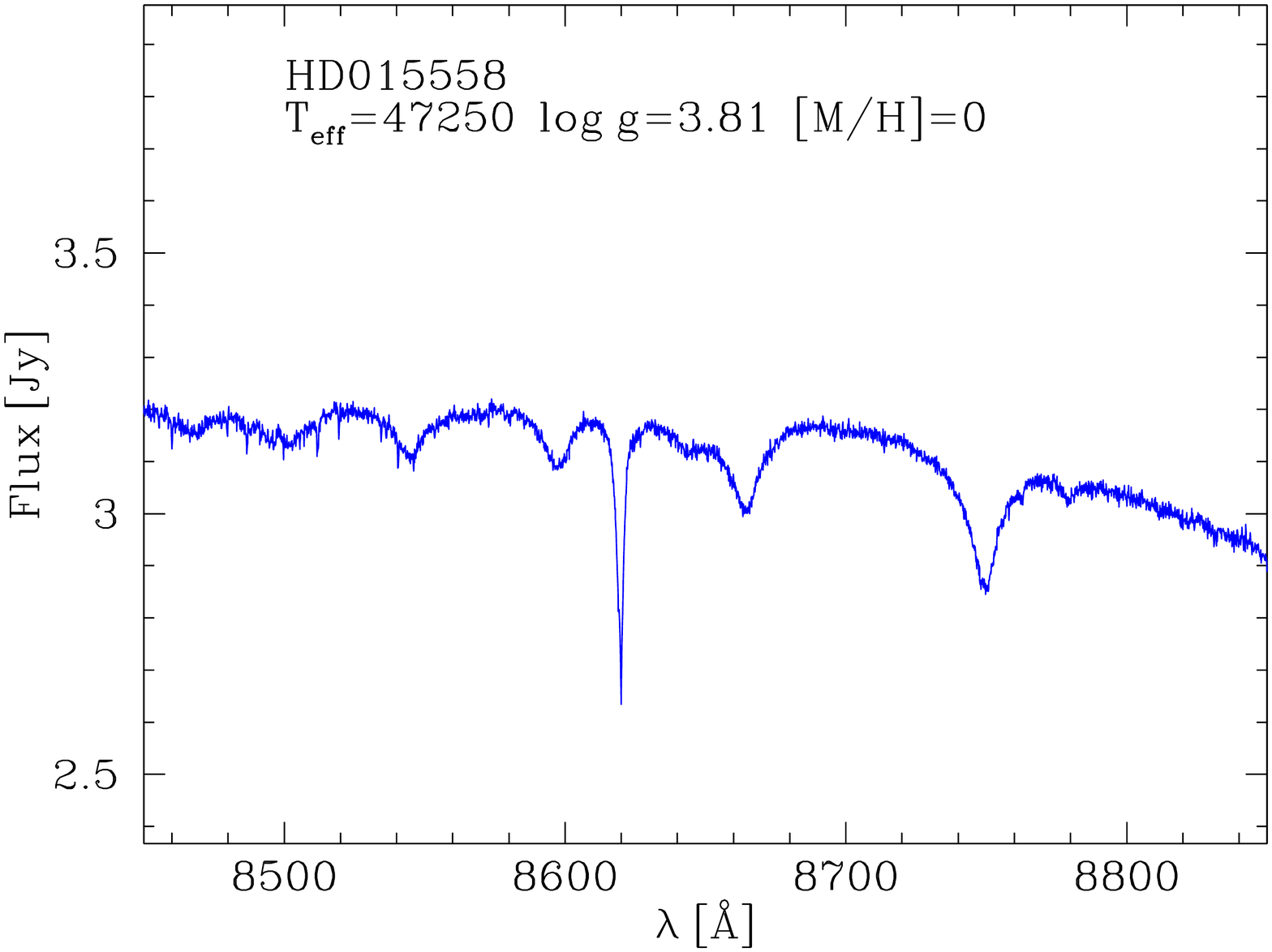}
\includegraphics[width=0.18\textwidth,angle=-90]{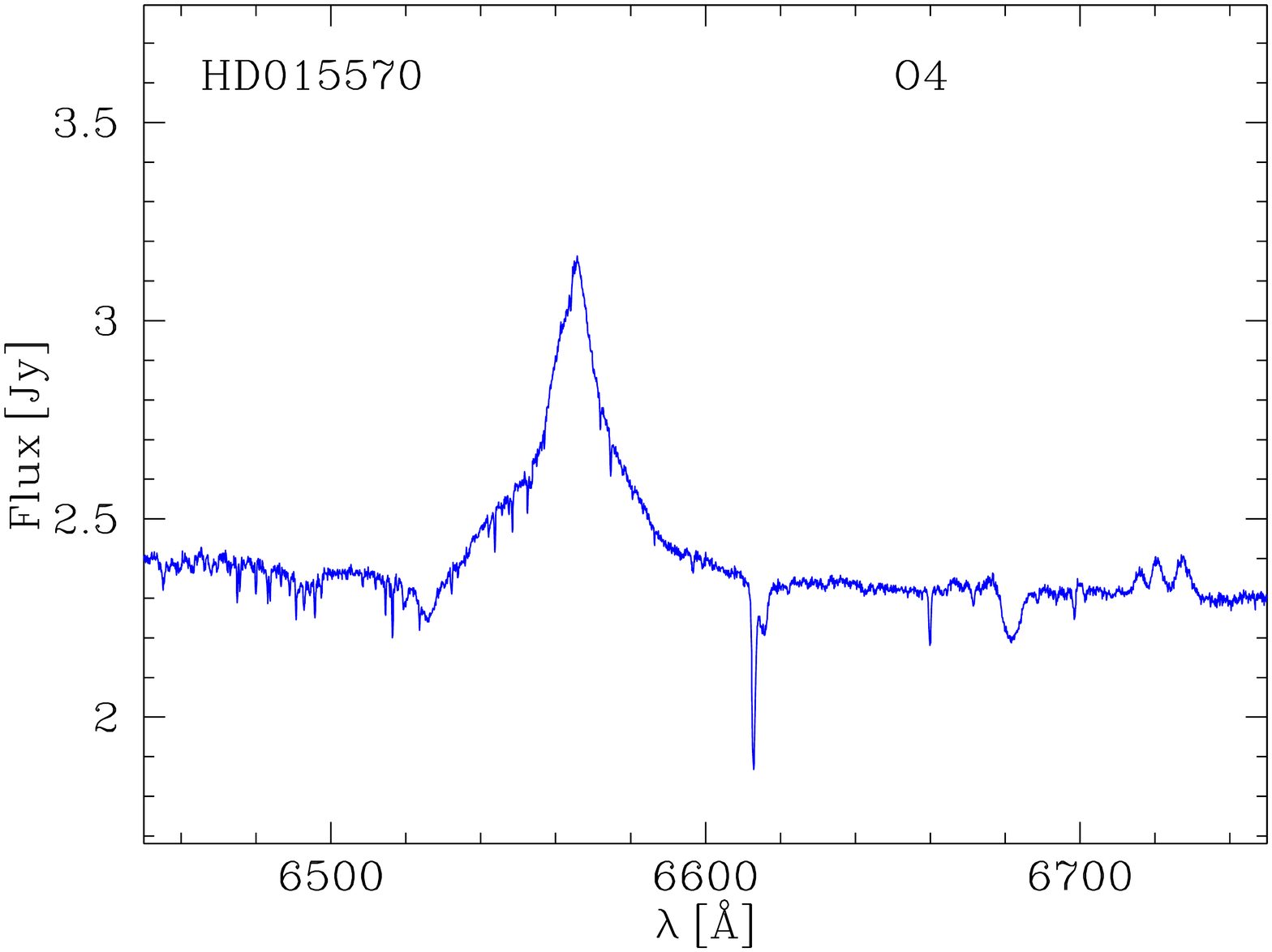}
\includegraphics[width=0.18\textwidth,angle=-90]{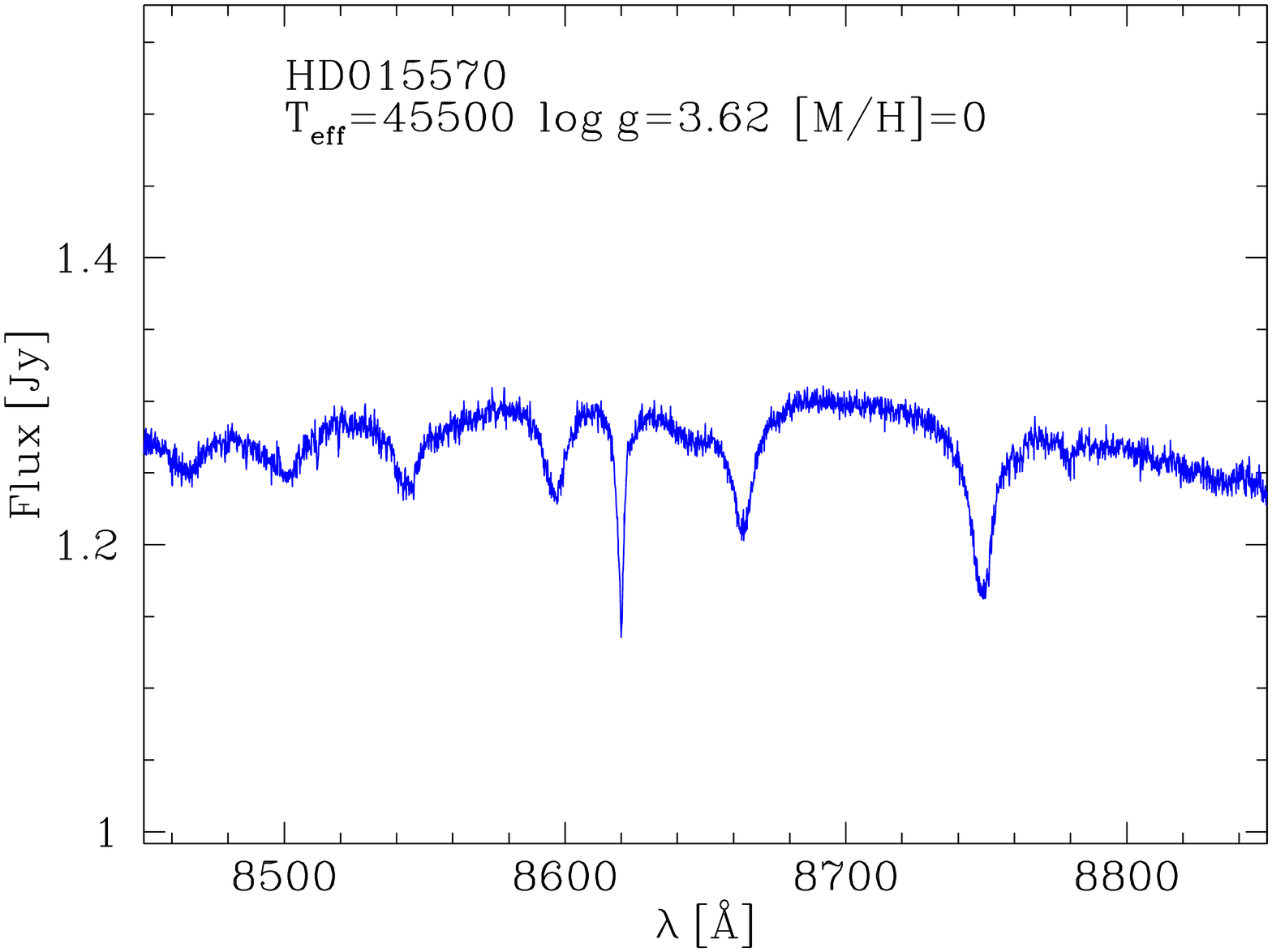}

\contcaption{3. Stars shown in this page are: HD004539, HD006327, HD006815, HD007374, HD009974, HD009996, HD013267, HD013268, HD014191, HD014633, HD014947, HD015318, HD015558 and HD015570.}
\end{figure*}

\begin{figure*}
\includegraphics[width=0.18\textwidth,angle=-90]{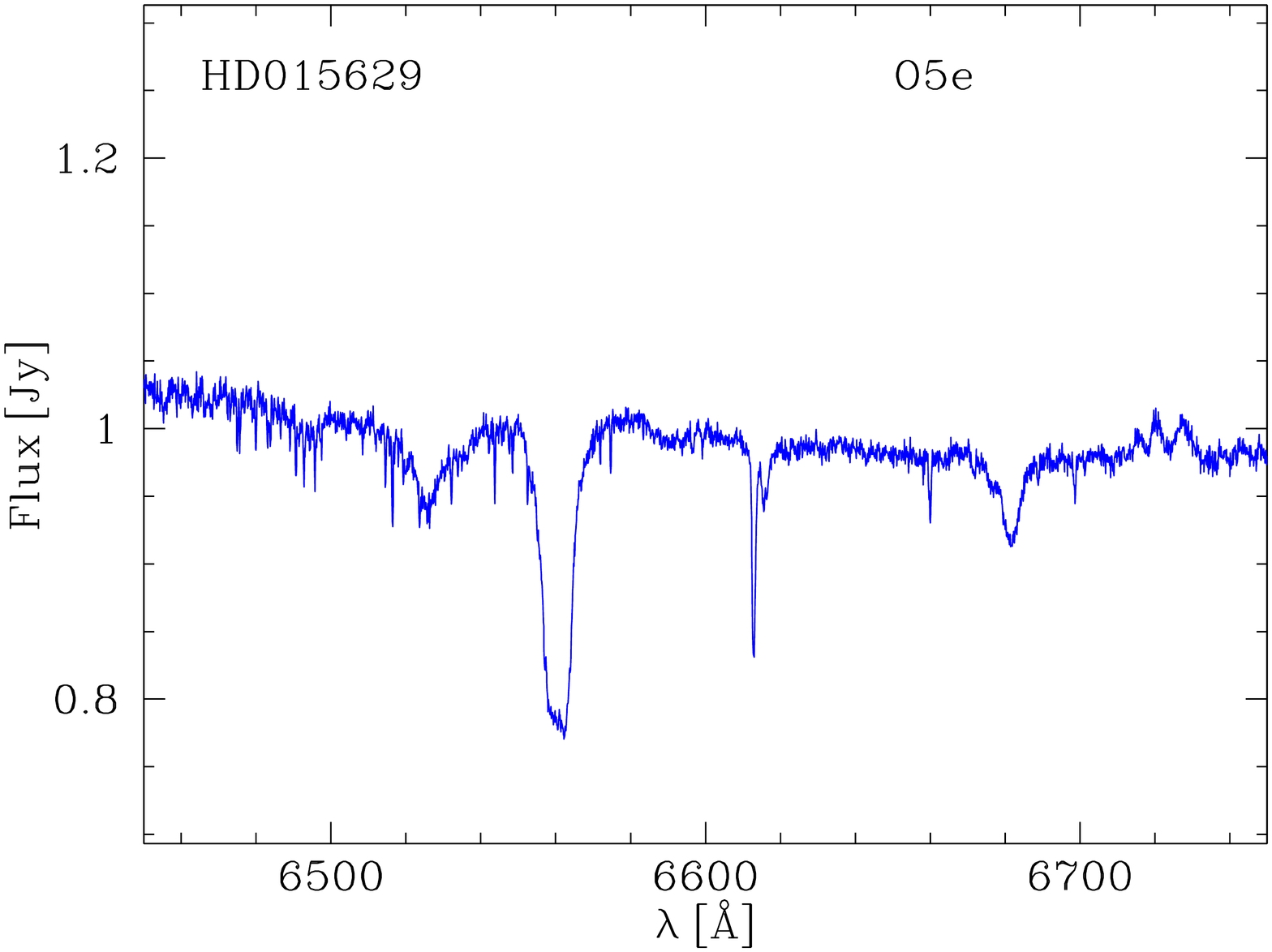}
\includegraphics[width=0.18\textwidth,angle=-90]{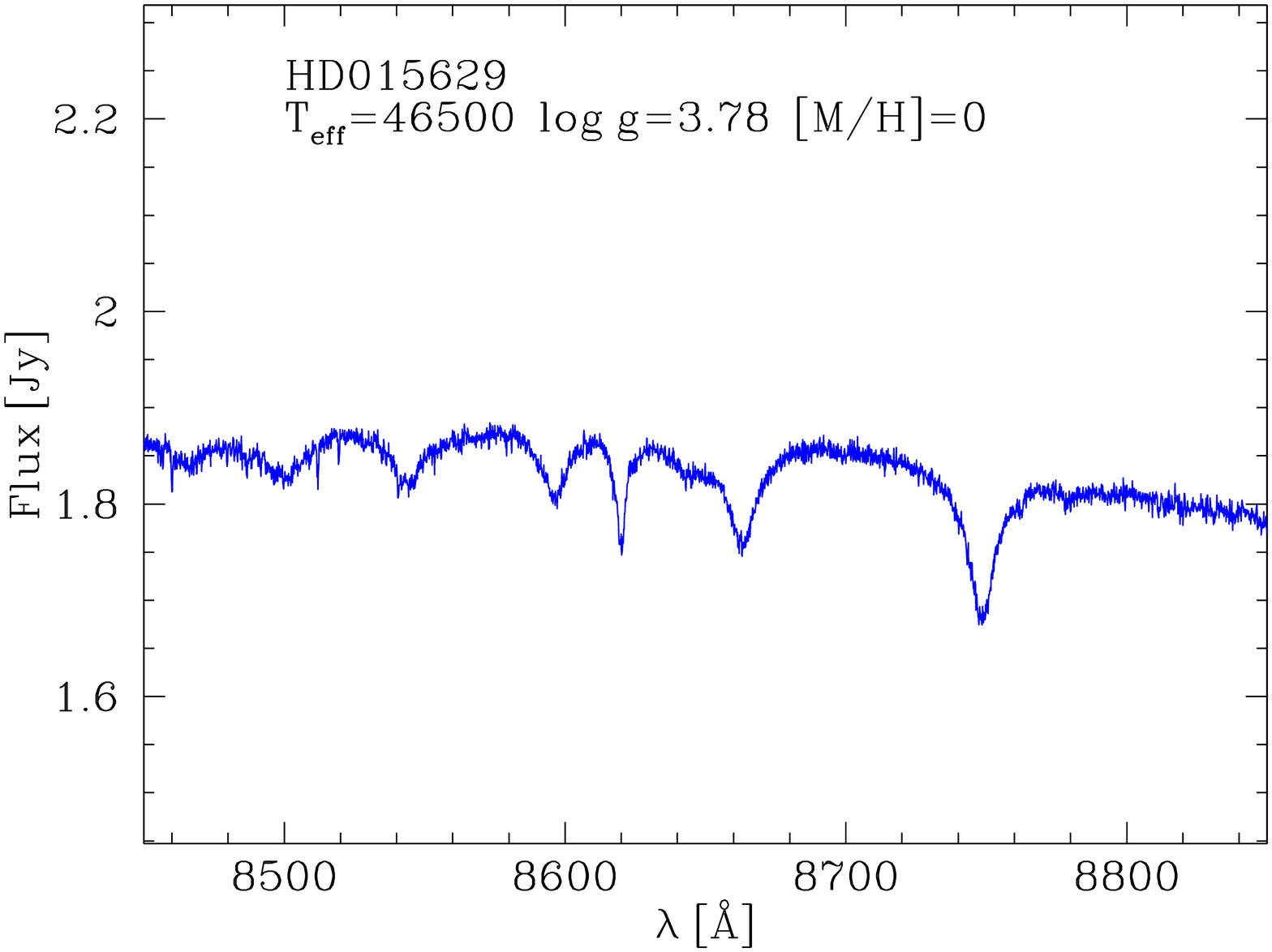}
\includegraphics[width=0.18\textwidth,angle=-90]{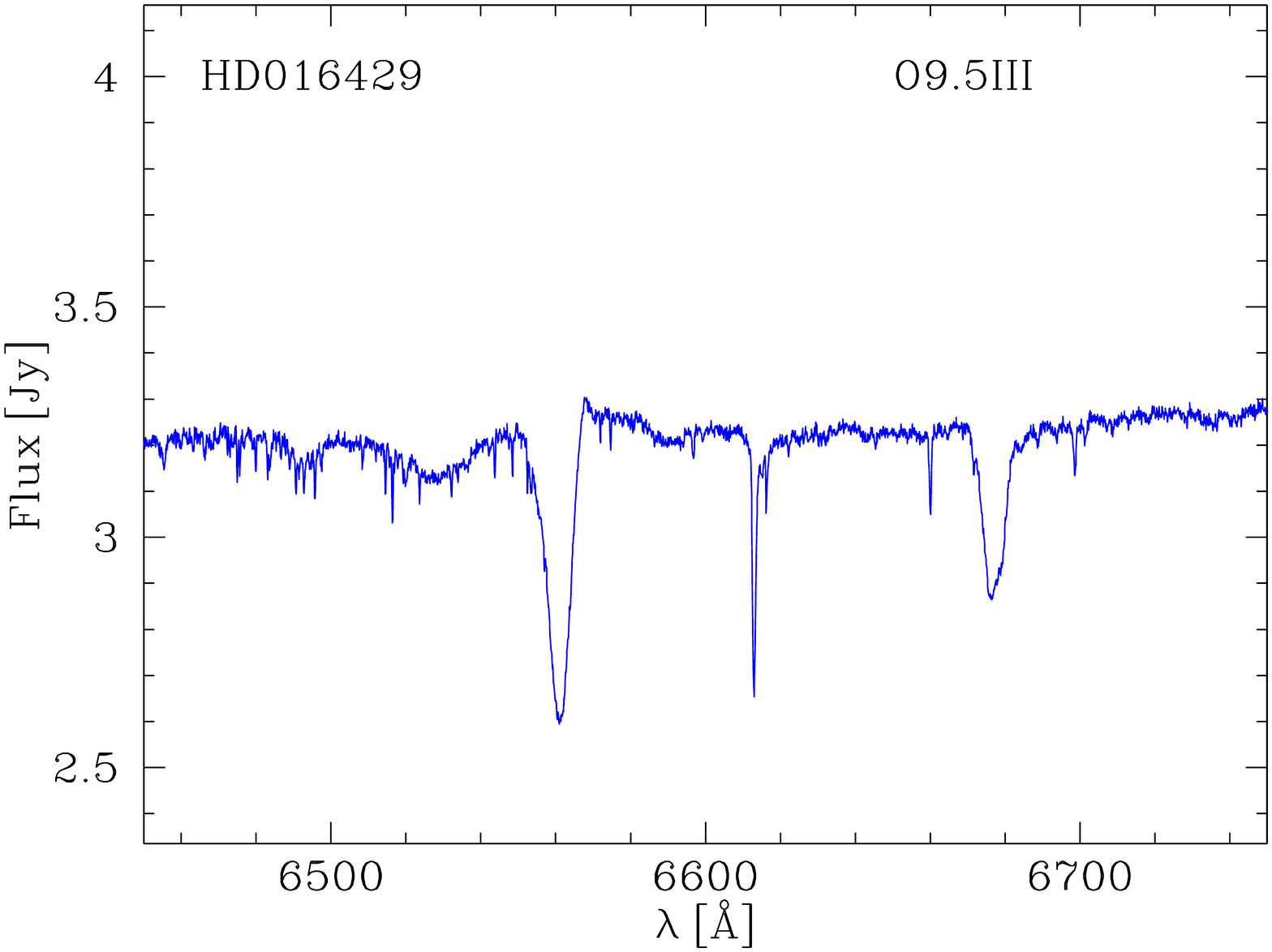} 
\includegraphics[width=0.18\textwidth,angle=-90]{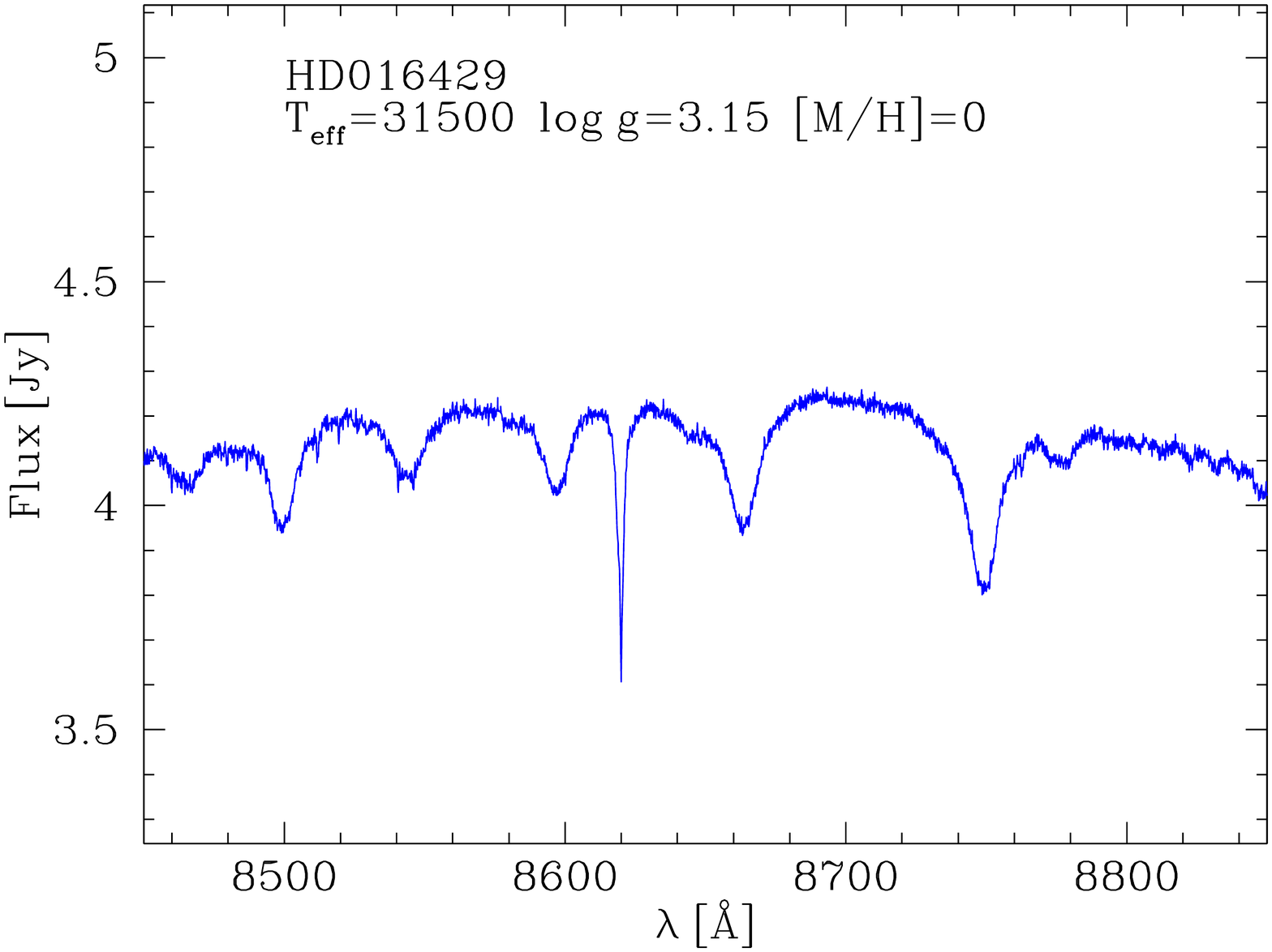}
\includegraphics[width=0.18\textwidth,angle=-90]{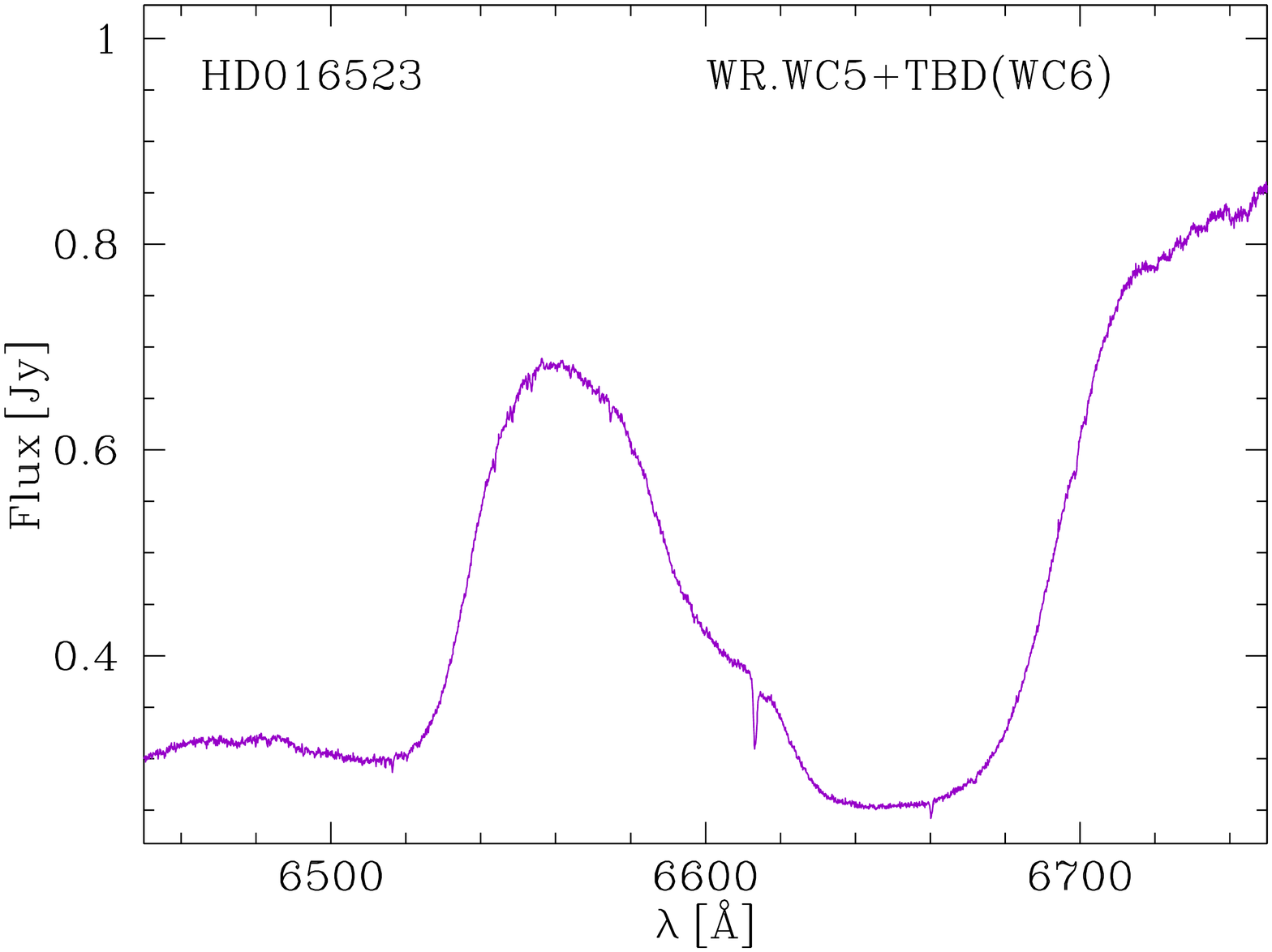}
\includegraphics[width=0.18\textwidth,angle=-90]{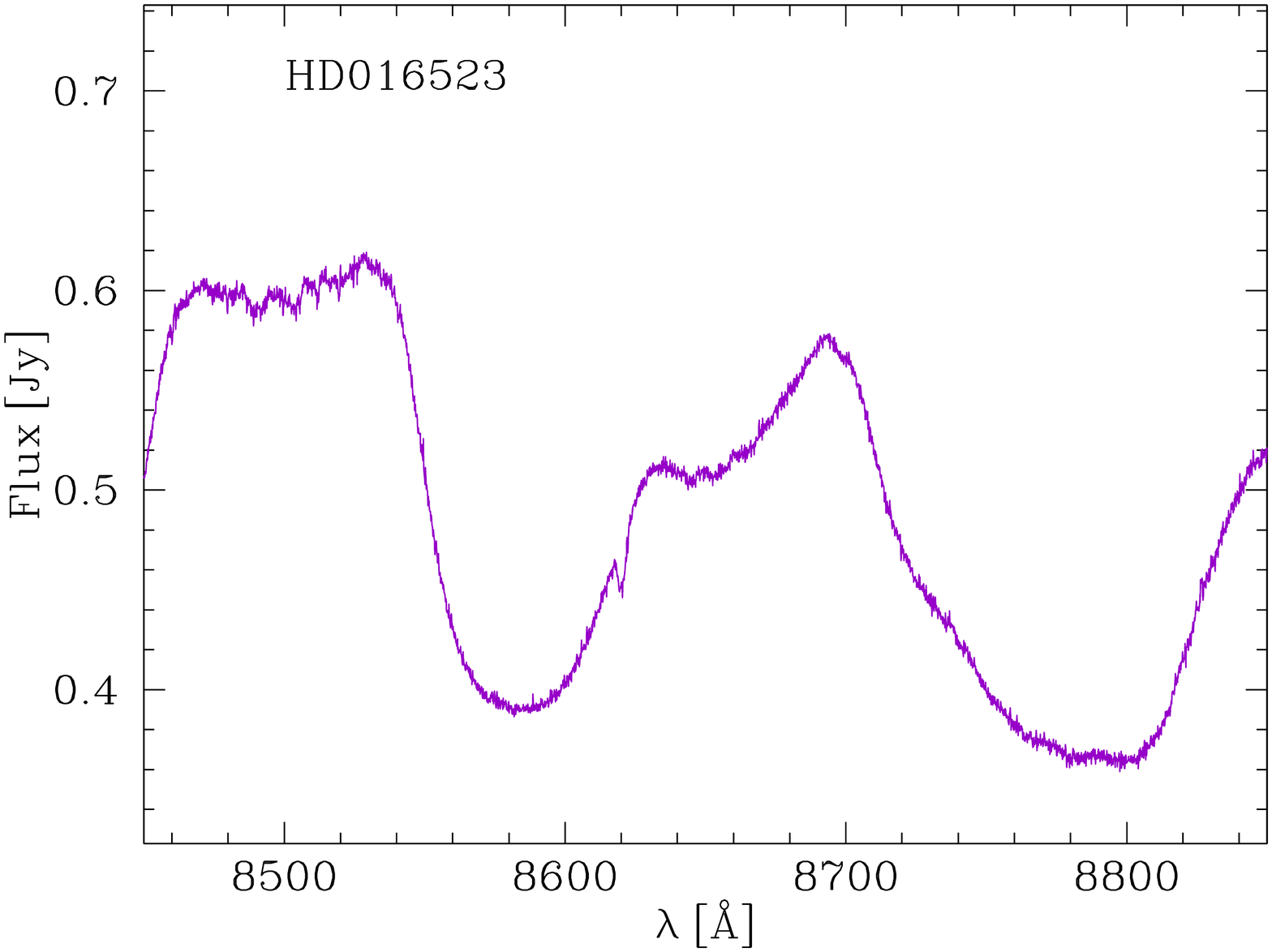}
\includegraphics[width=0.18\textwidth,angle=-90]{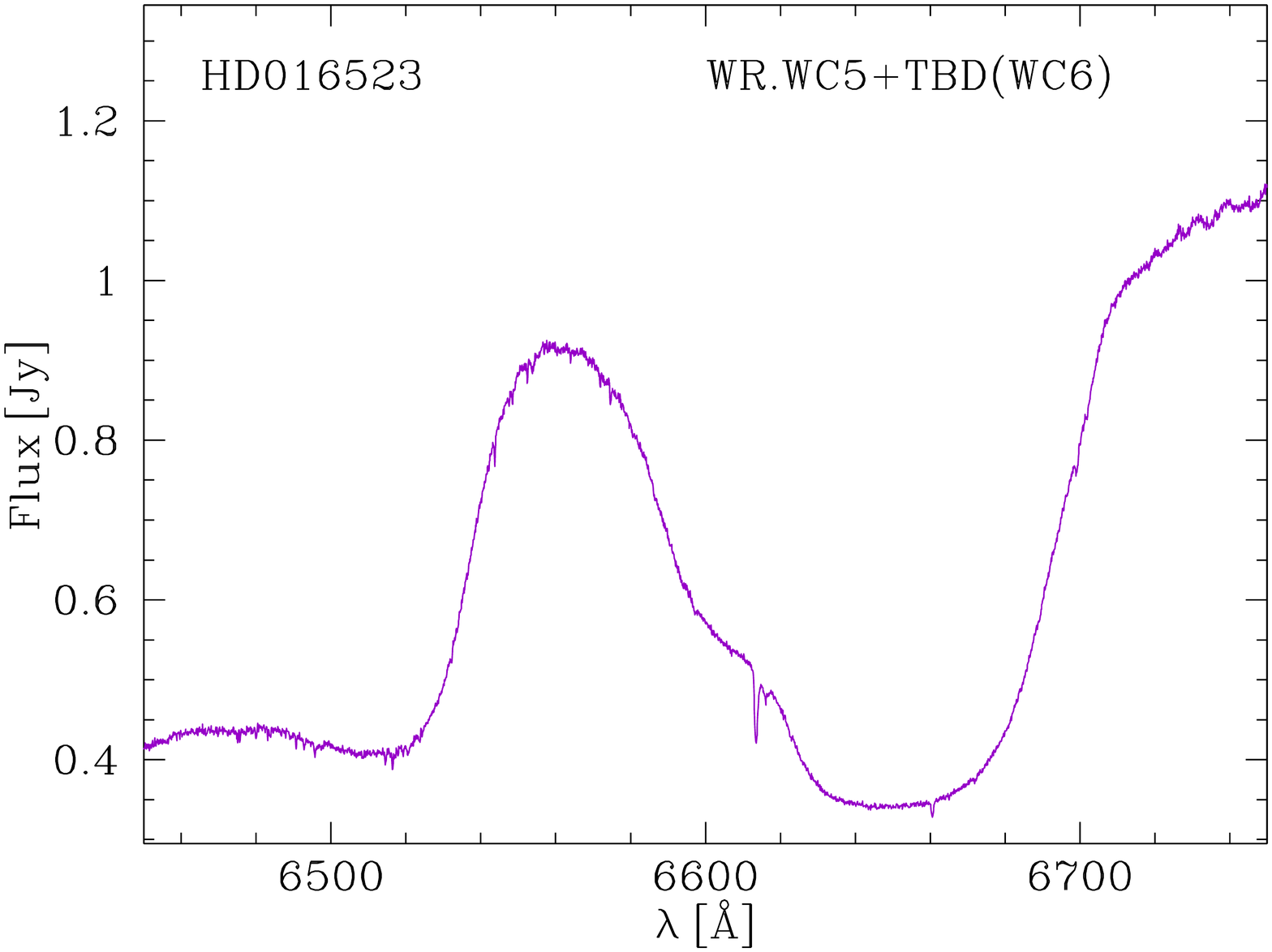}
\includegraphics[width=0.18\textwidth,angle=-90]{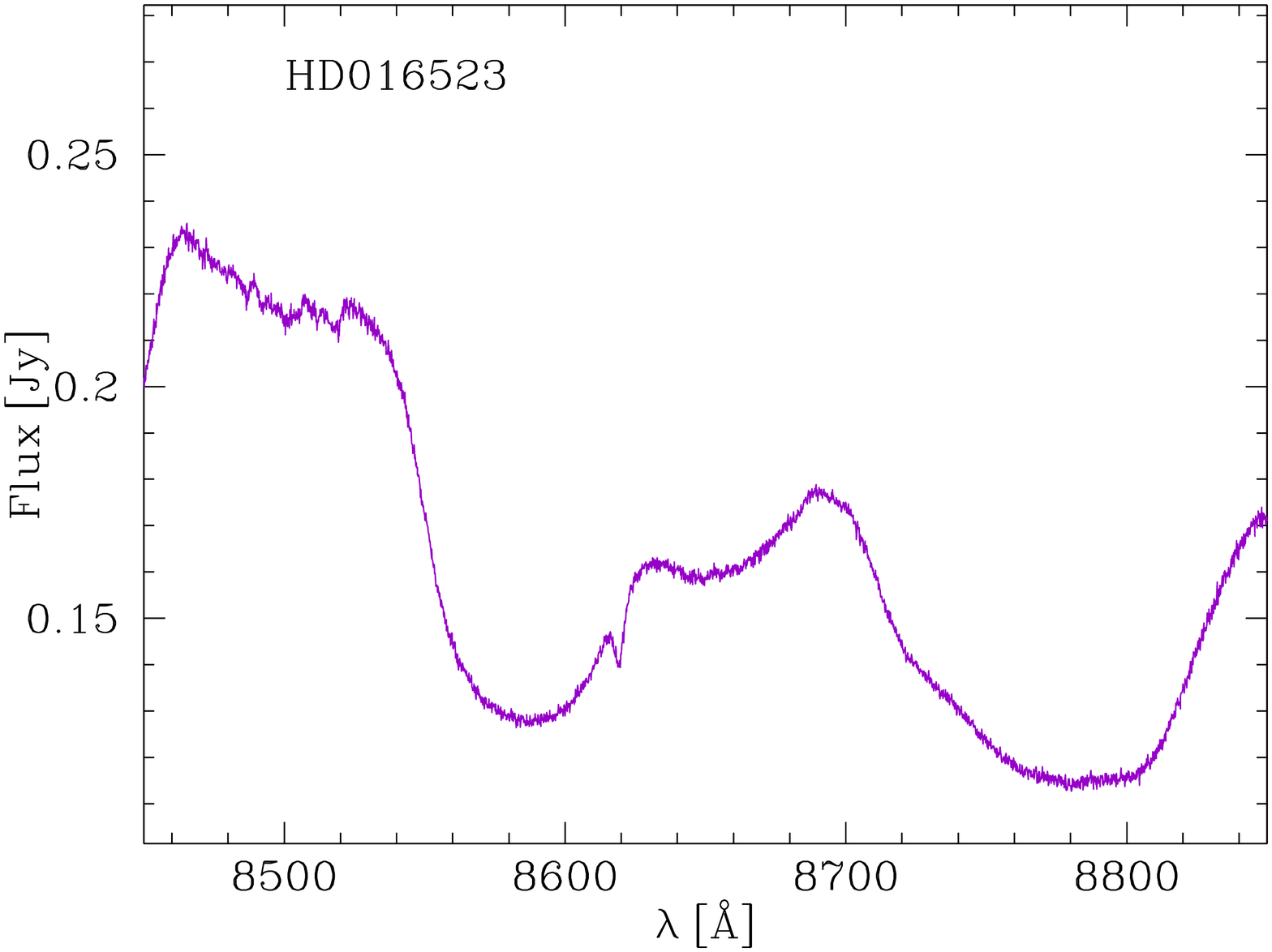}
\includegraphics[width=0.18\textwidth,angle=-90]{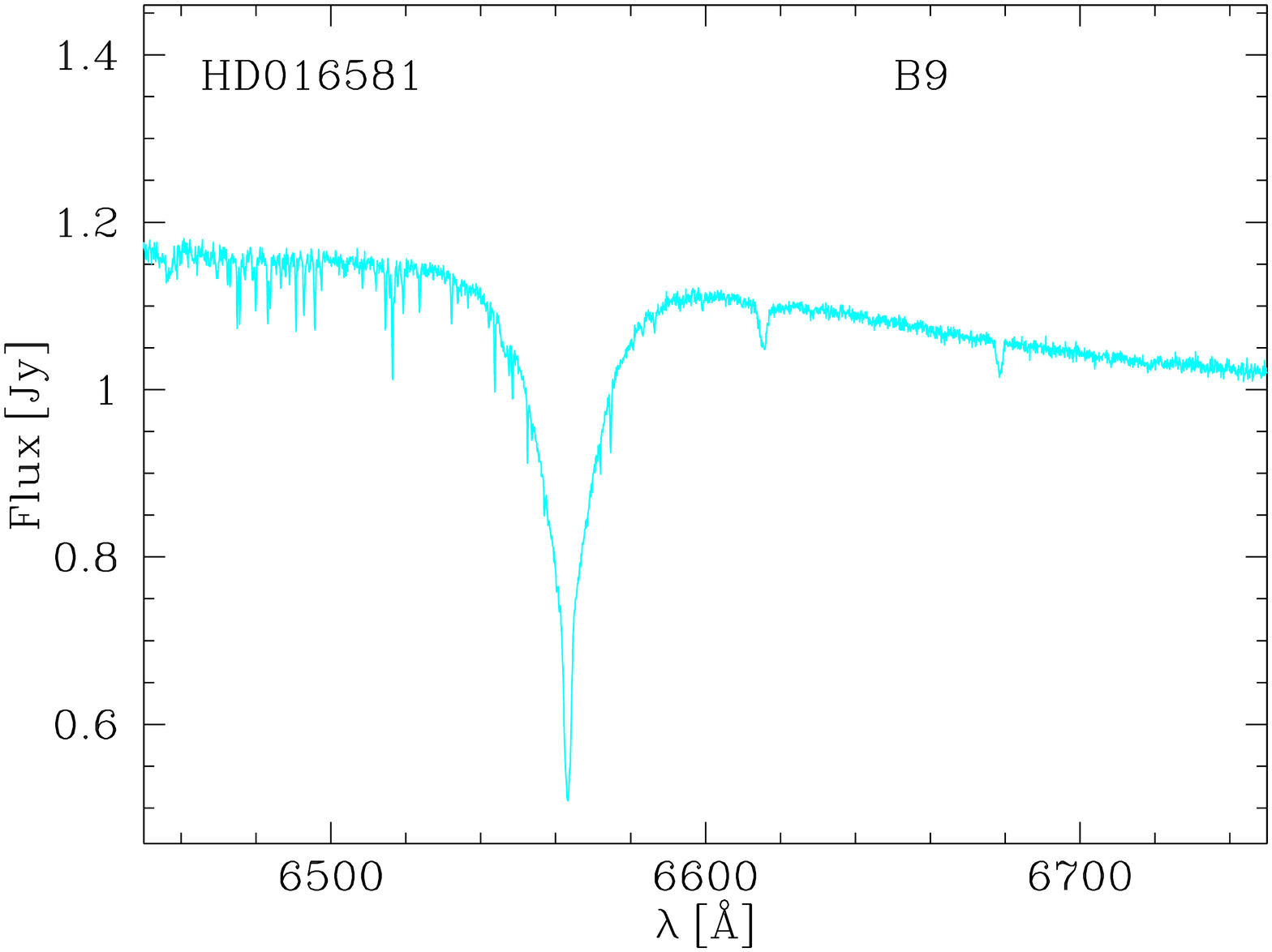}
\includegraphics[width=0.18\textwidth,angle=-90]{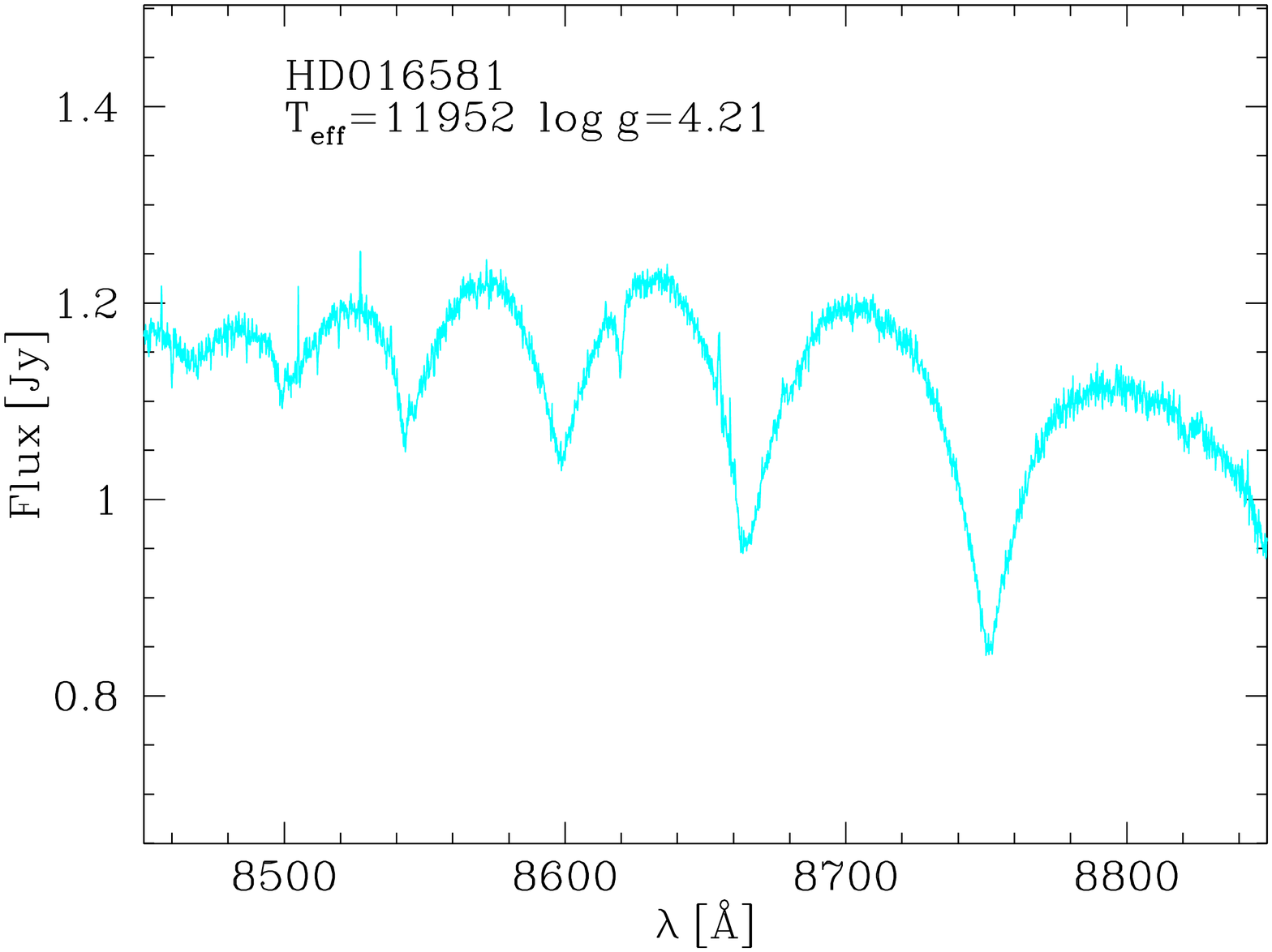}
\includegraphics[width=0.18\textwidth,angle=-90]{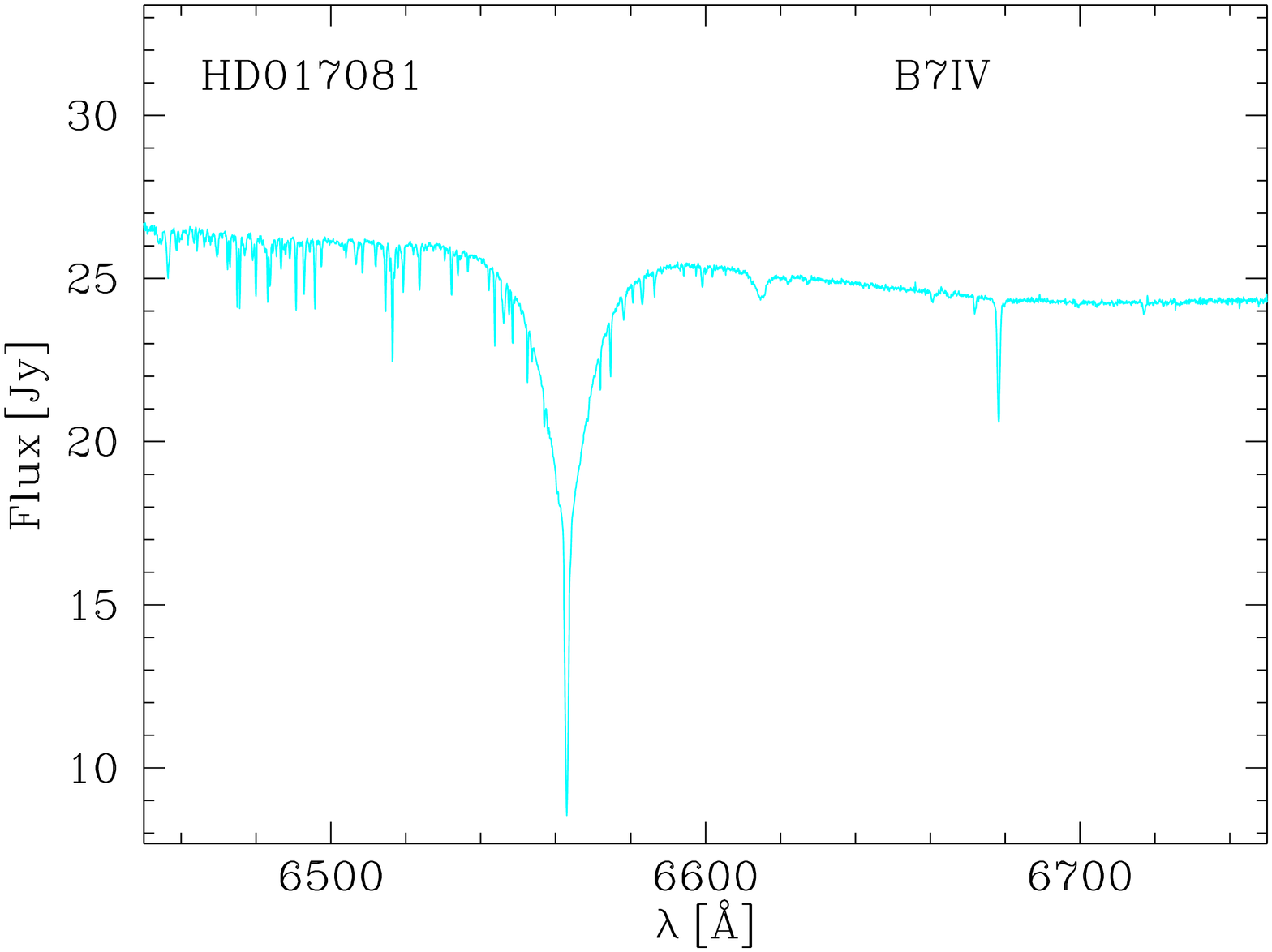} 
\includegraphics[width=0.18\textwidth,angle=-90]{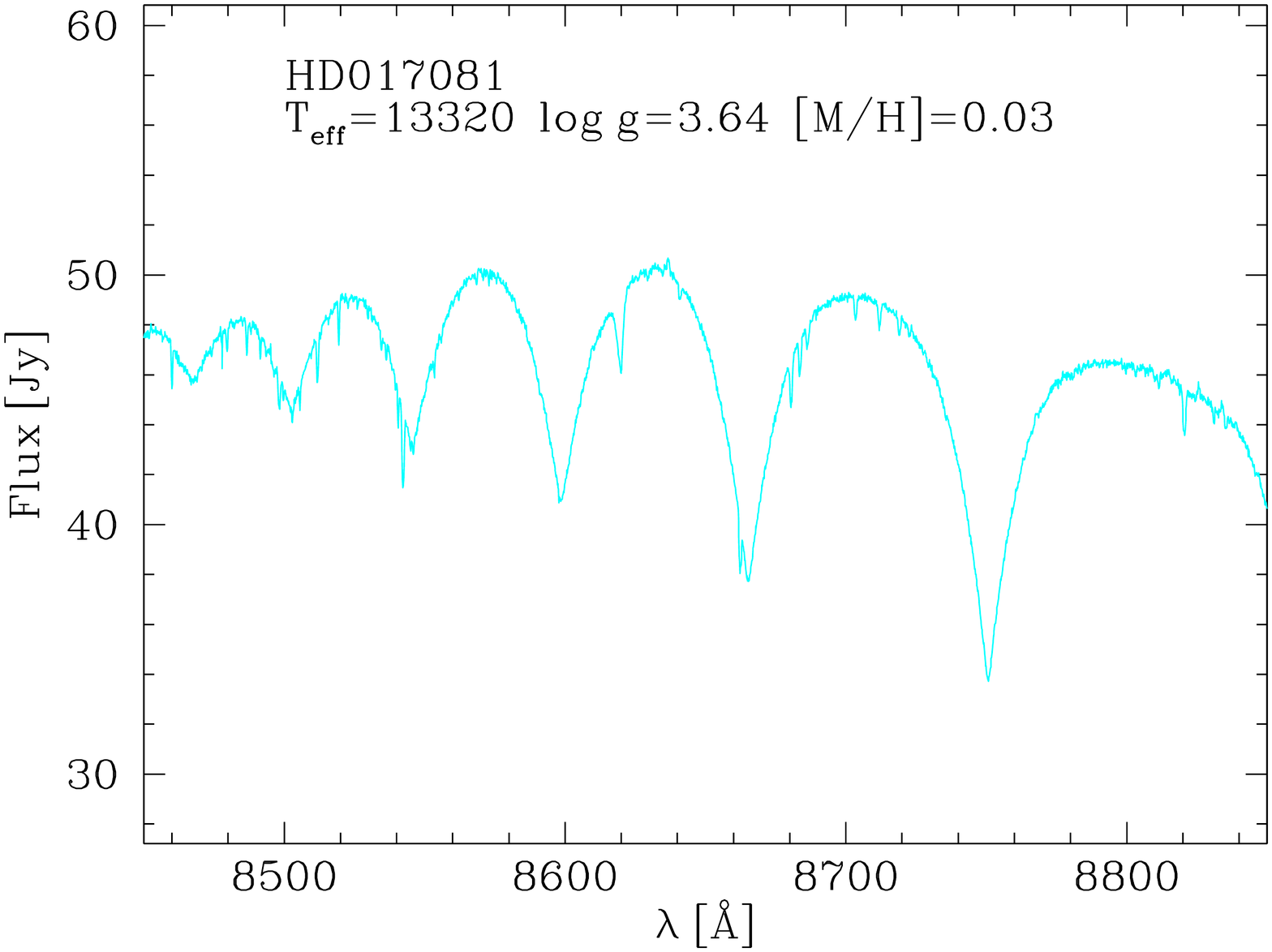}
\includegraphics[width=0.18\textwidth,angle=-90]{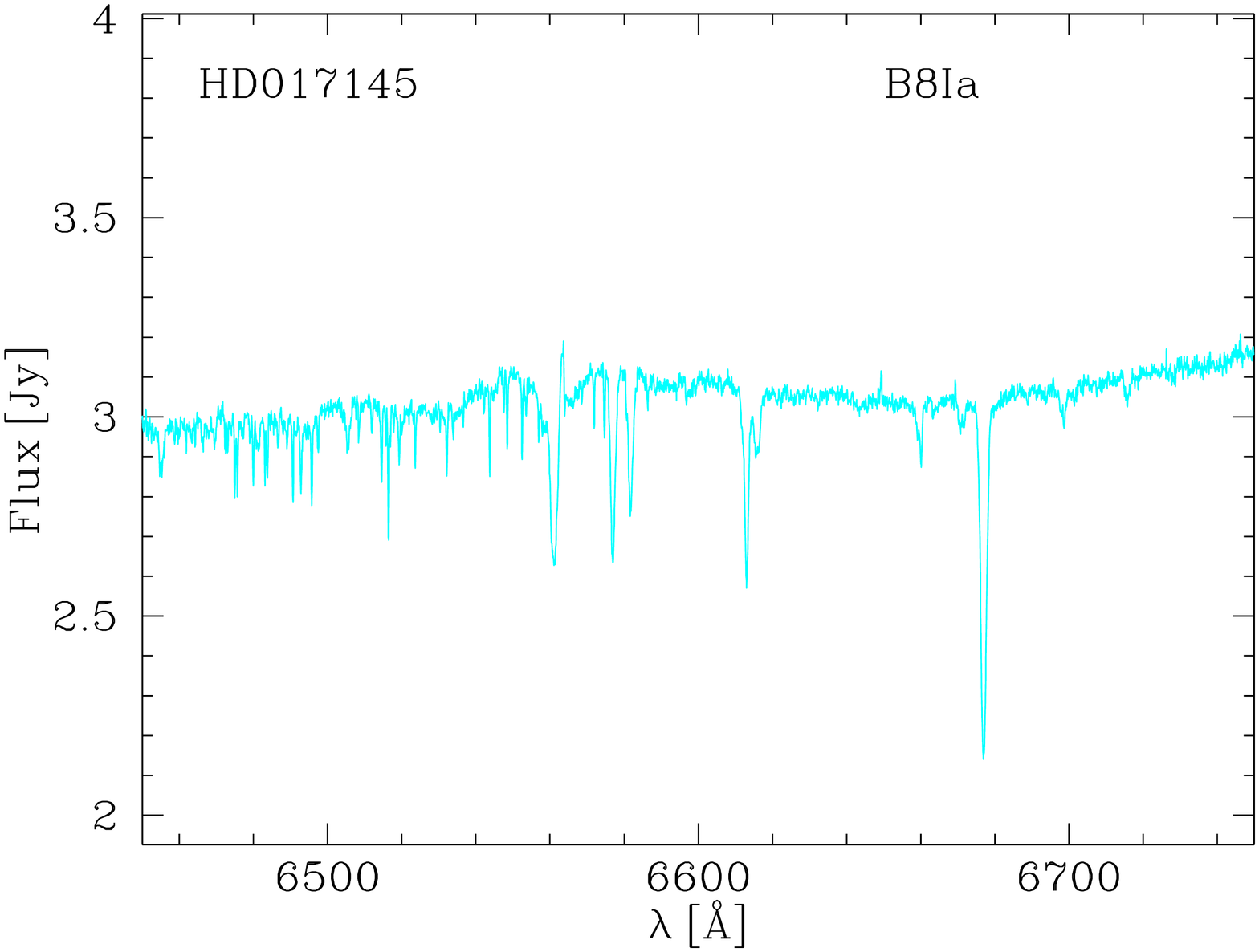}
\includegraphics[width=0.18\textwidth,angle=-90]{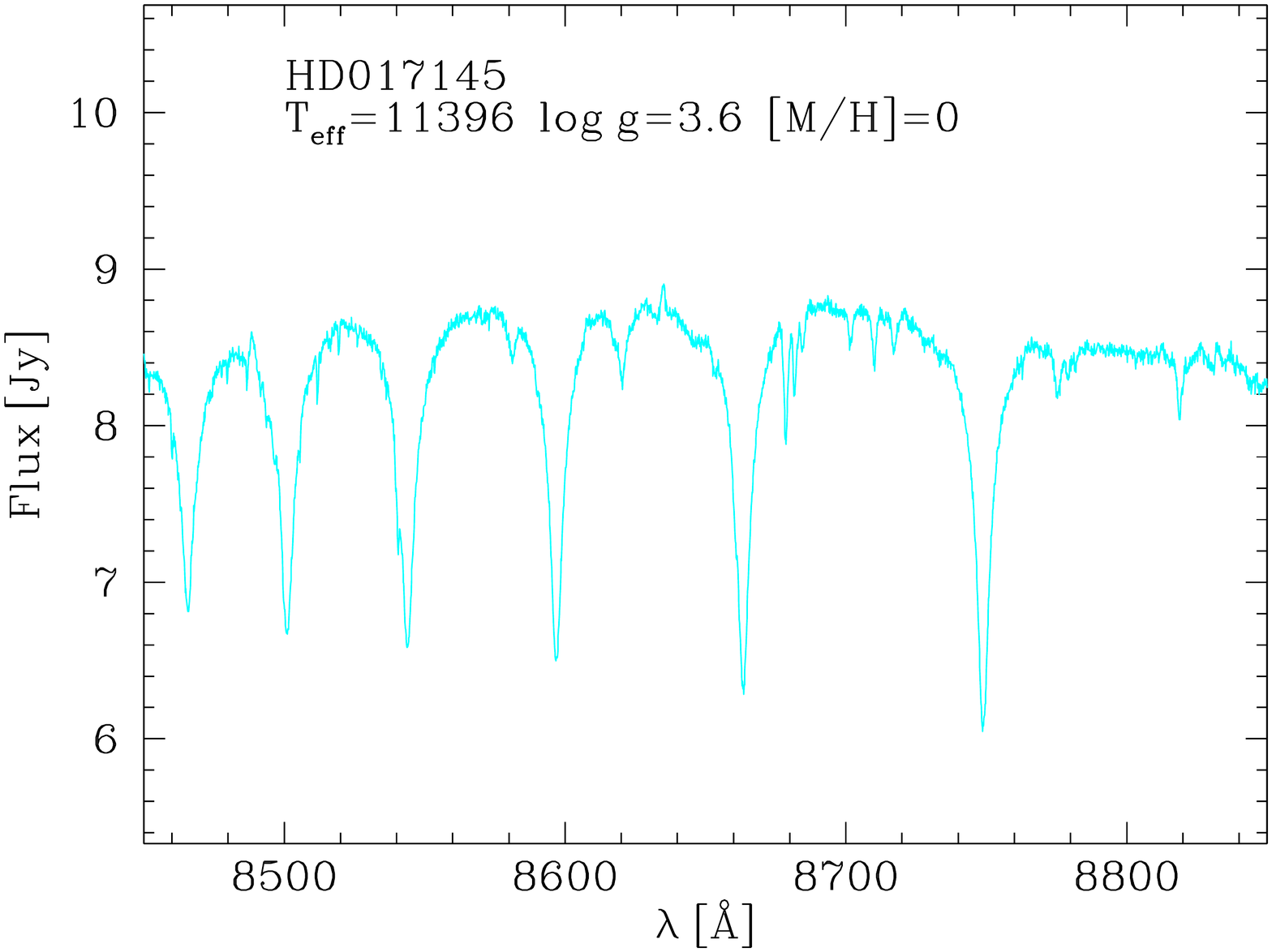}
\includegraphics[width=0.18\textwidth,angle=-90]{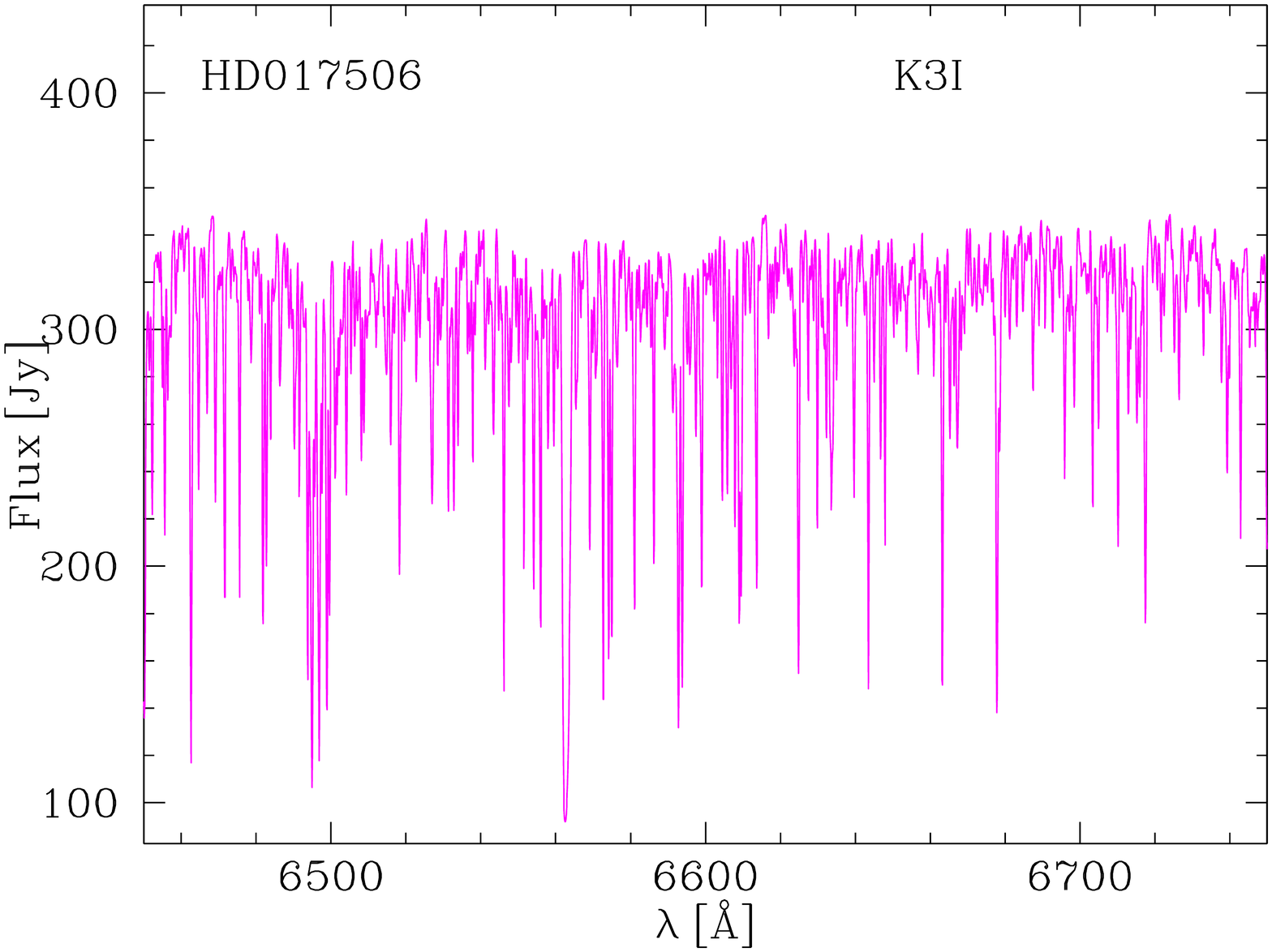}
\includegraphics[width=0.18\textwidth,angle=-90]{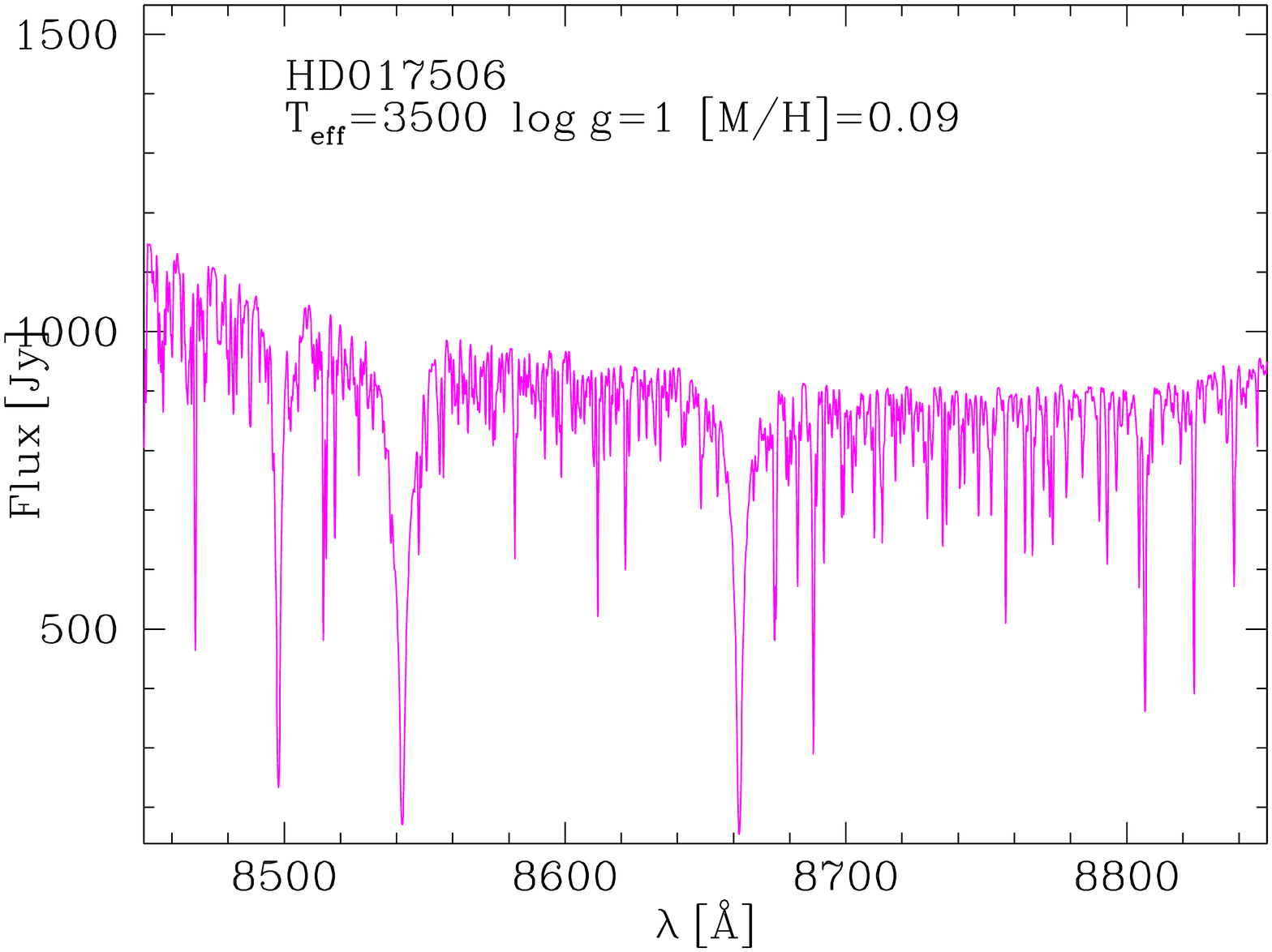}
\includegraphics[width=0.18\textwidth,angle=-90]{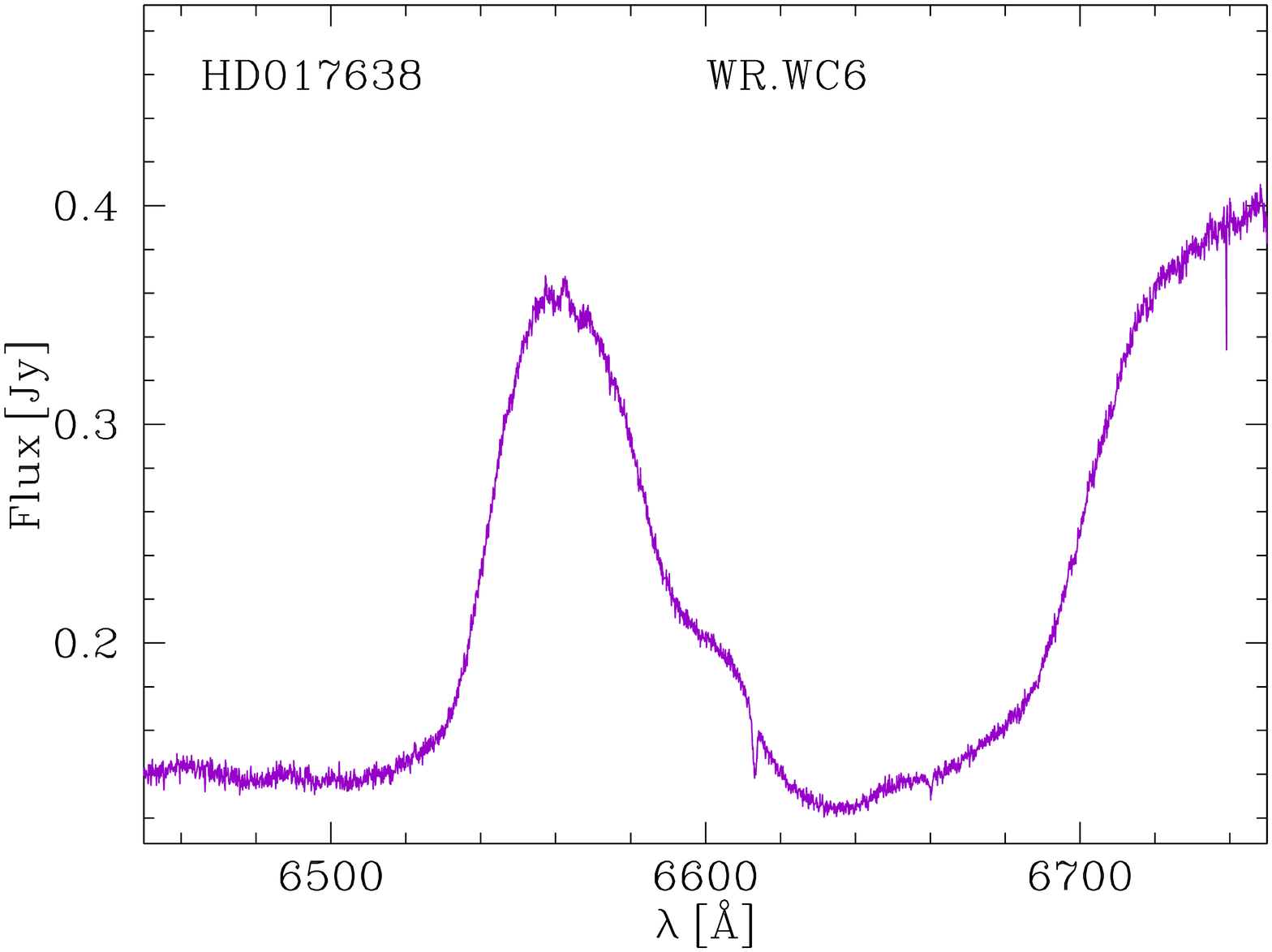}
\includegraphics[width=0.18\textwidth,angle=-90]{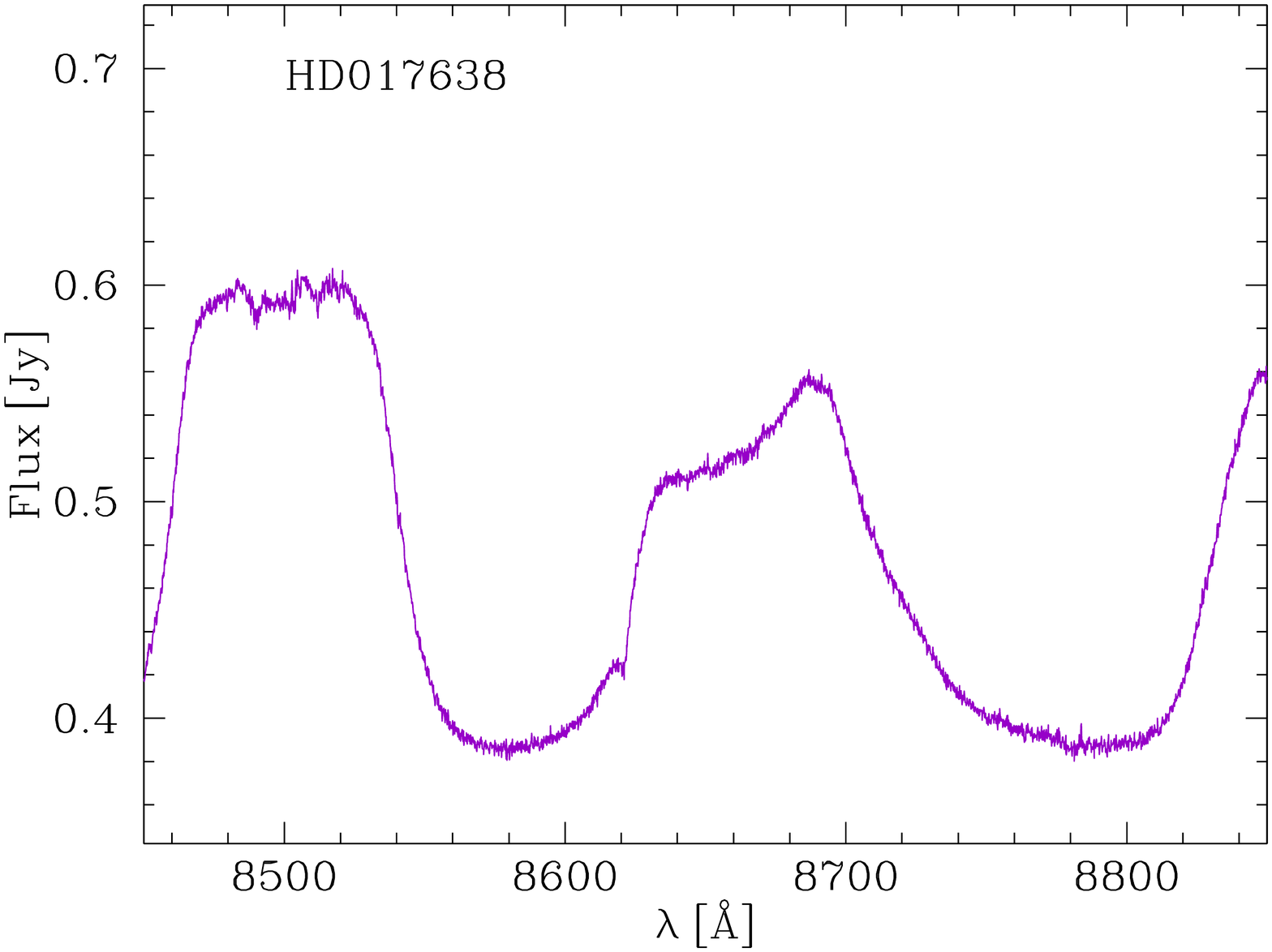}
\includegraphics[width=0.18\textwidth,angle=-90]{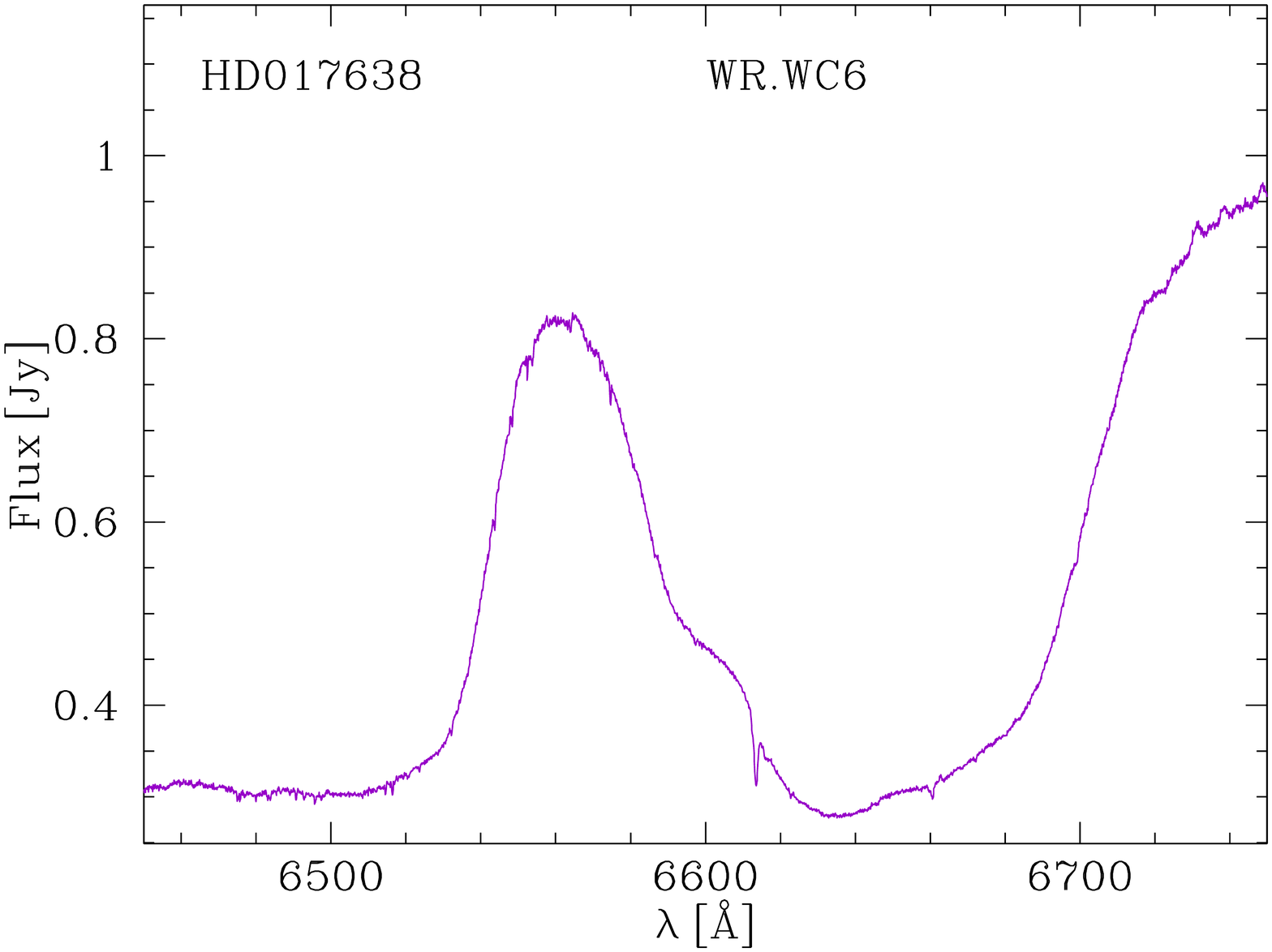}
\includegraphics[width=0.18\textwidth,angle=-90]{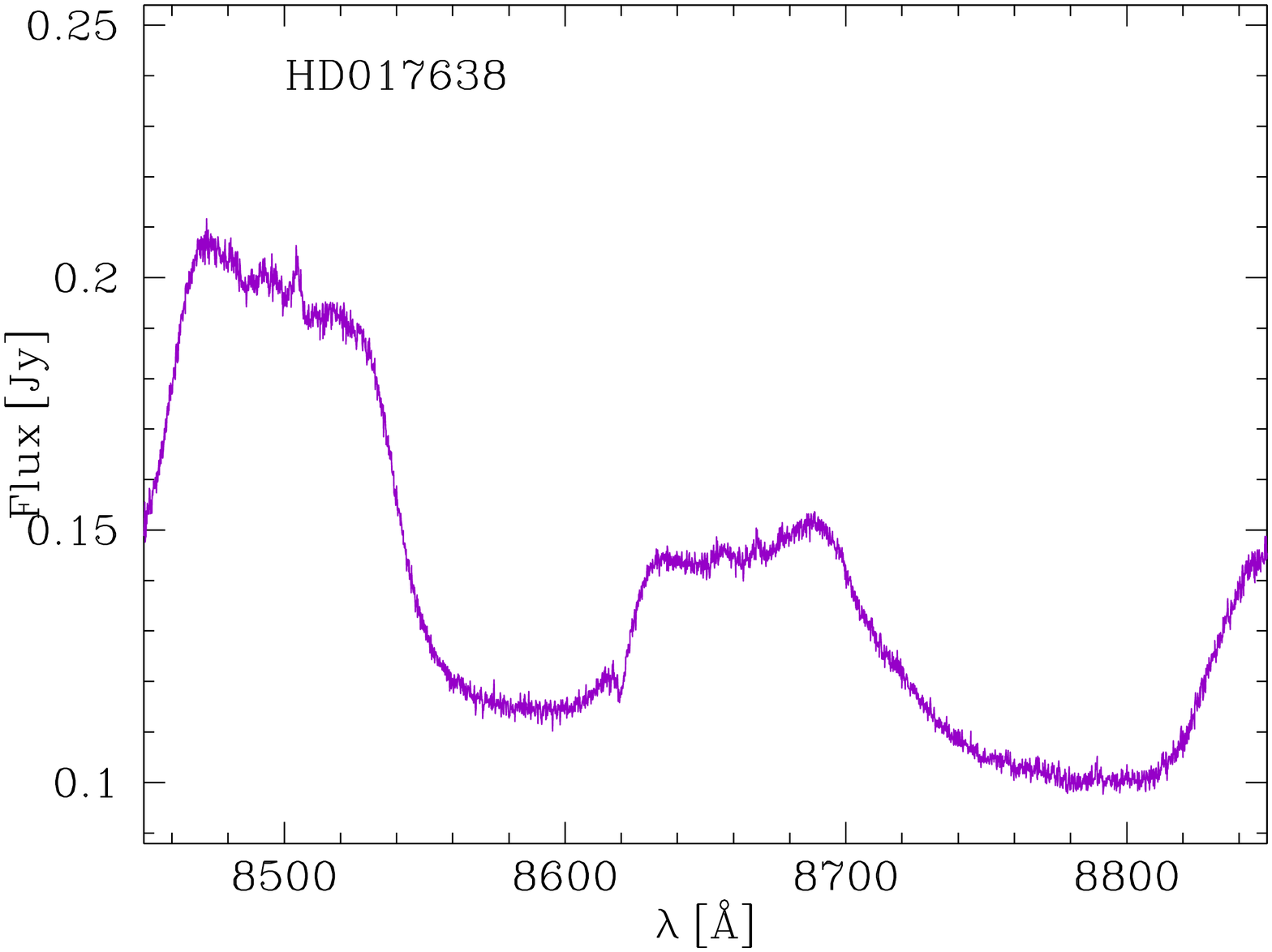}
\includegraphics[width=0.18\textwidth,angle=-90]{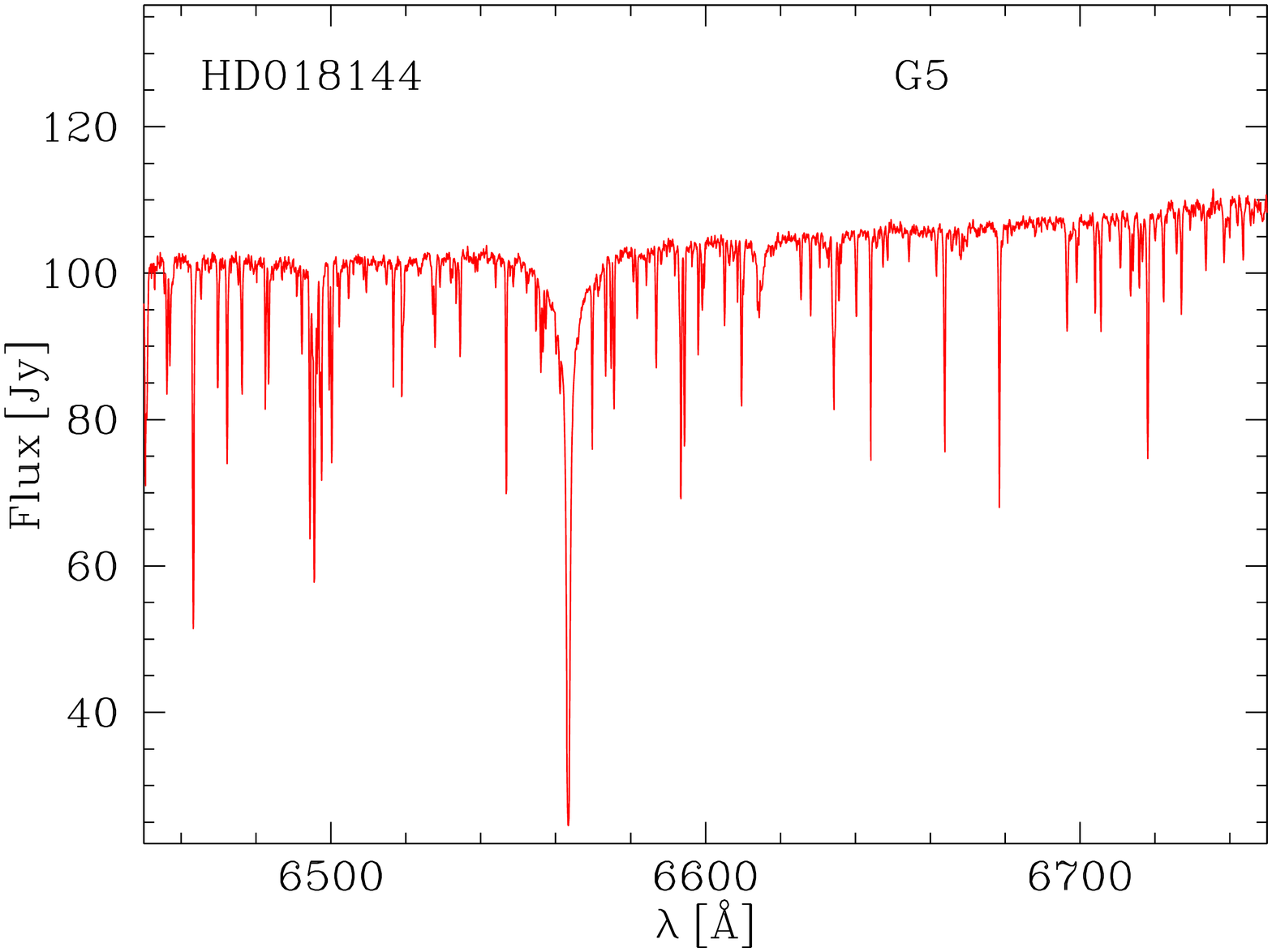}
\includegraphics[width=0.18\textwidth,angle=-90]{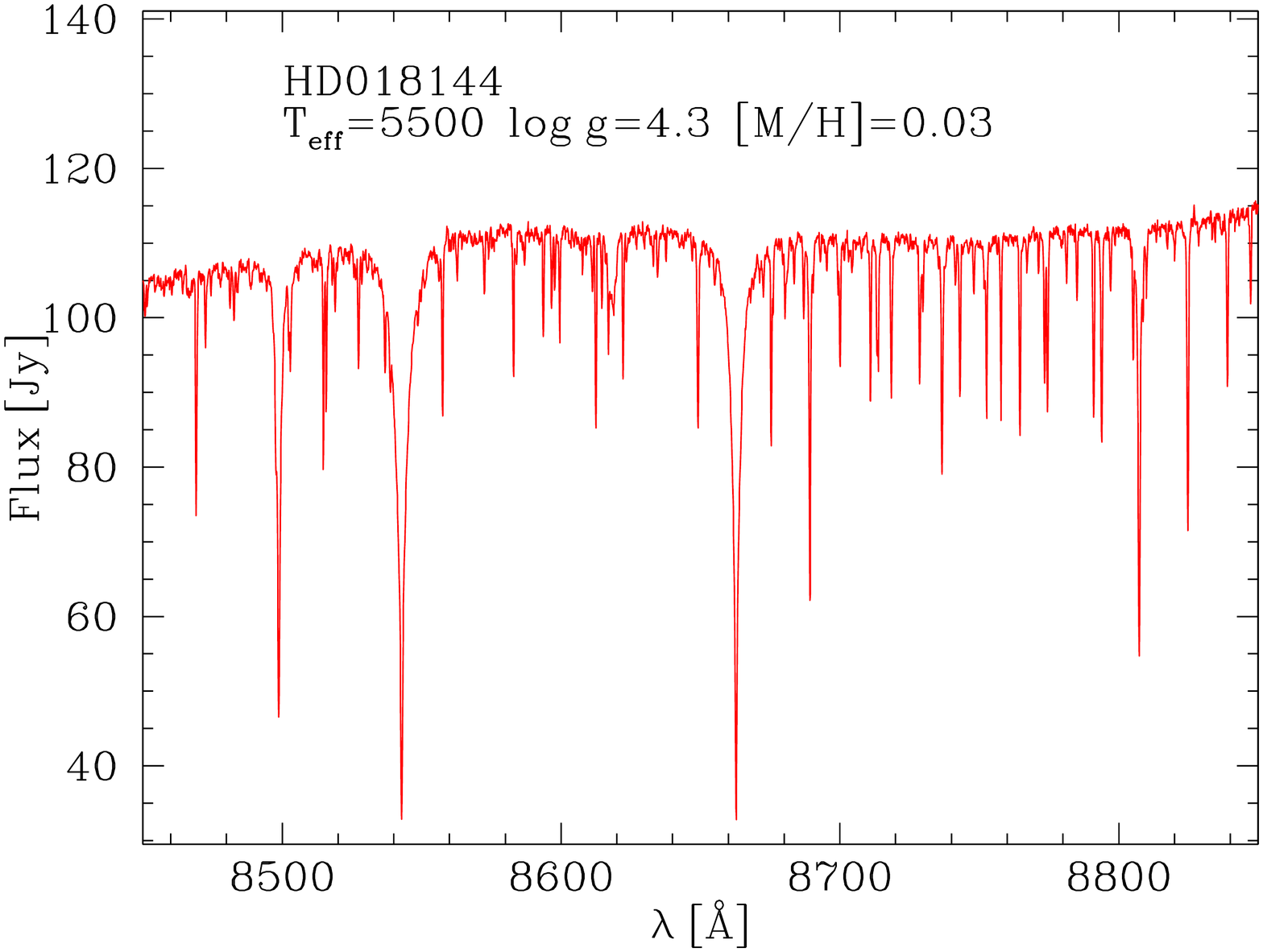}
\includegraphics[width=0.18\textwidth,angle=-90]{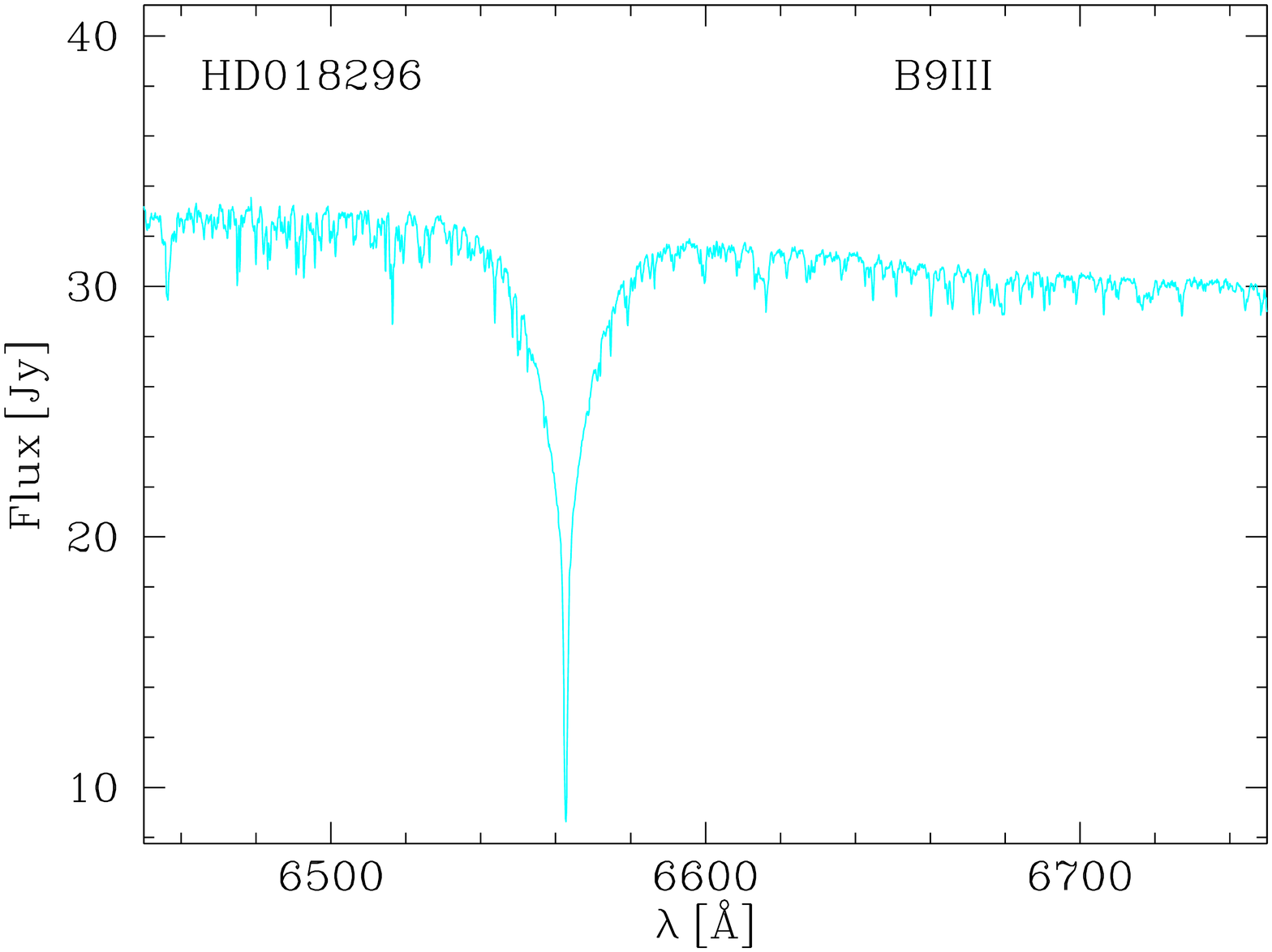}
\includegraphics[width=0.18\textwidth,angle=-90]{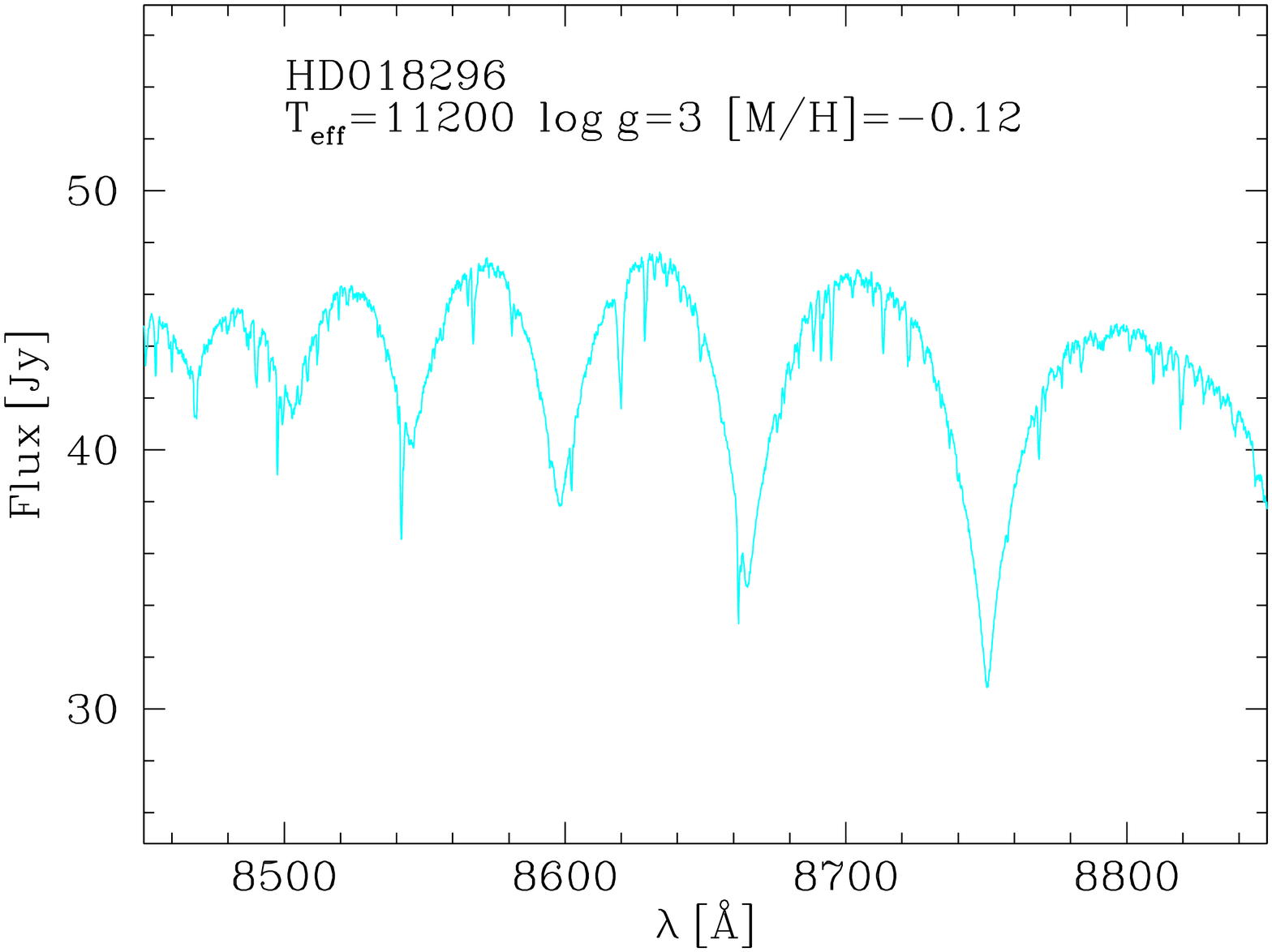}
\includegraphics[width=0.18\textwidth,angle=-90]{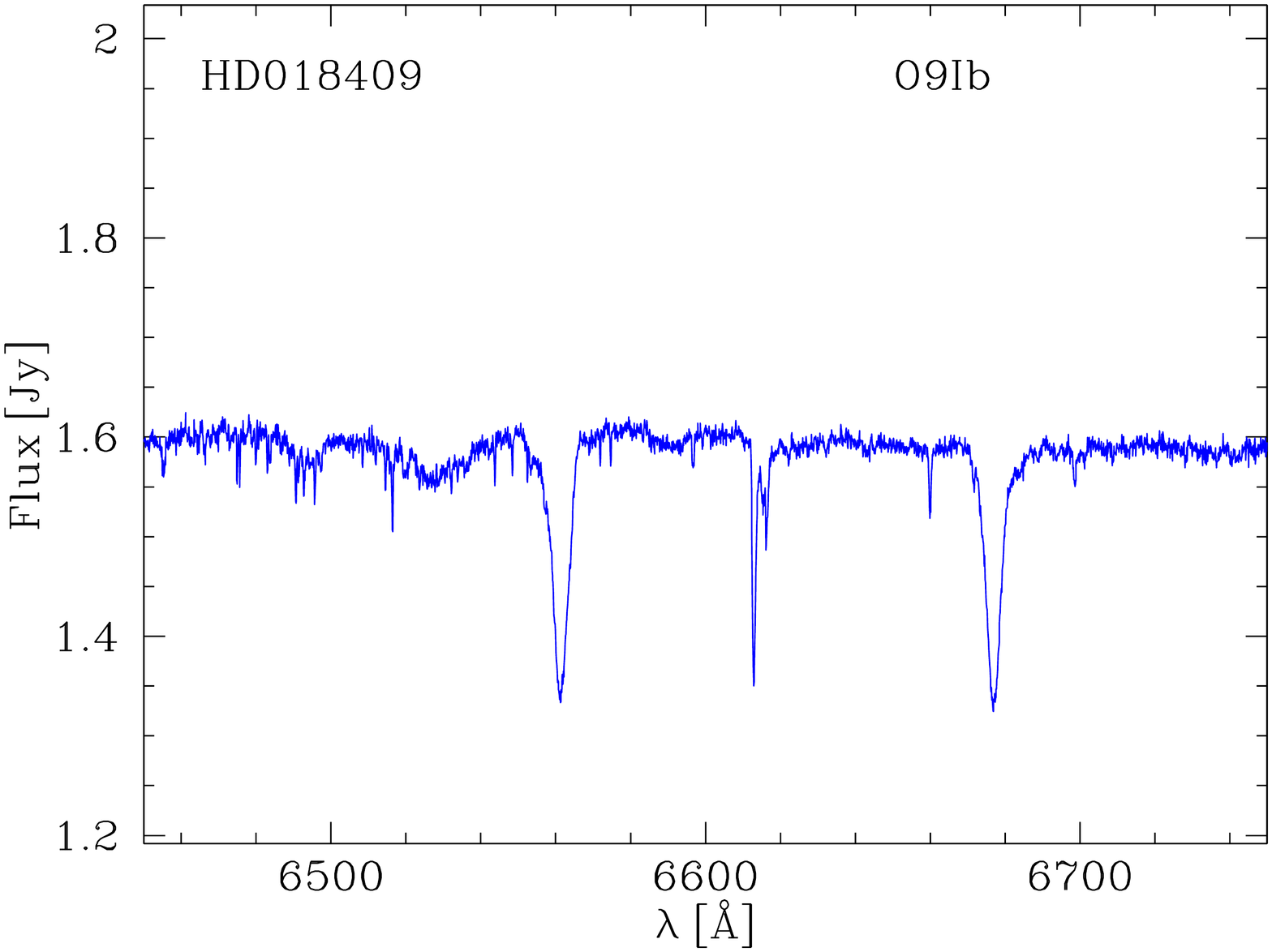}
\includegraphics[width=0.18\textwidth,angle=-90]{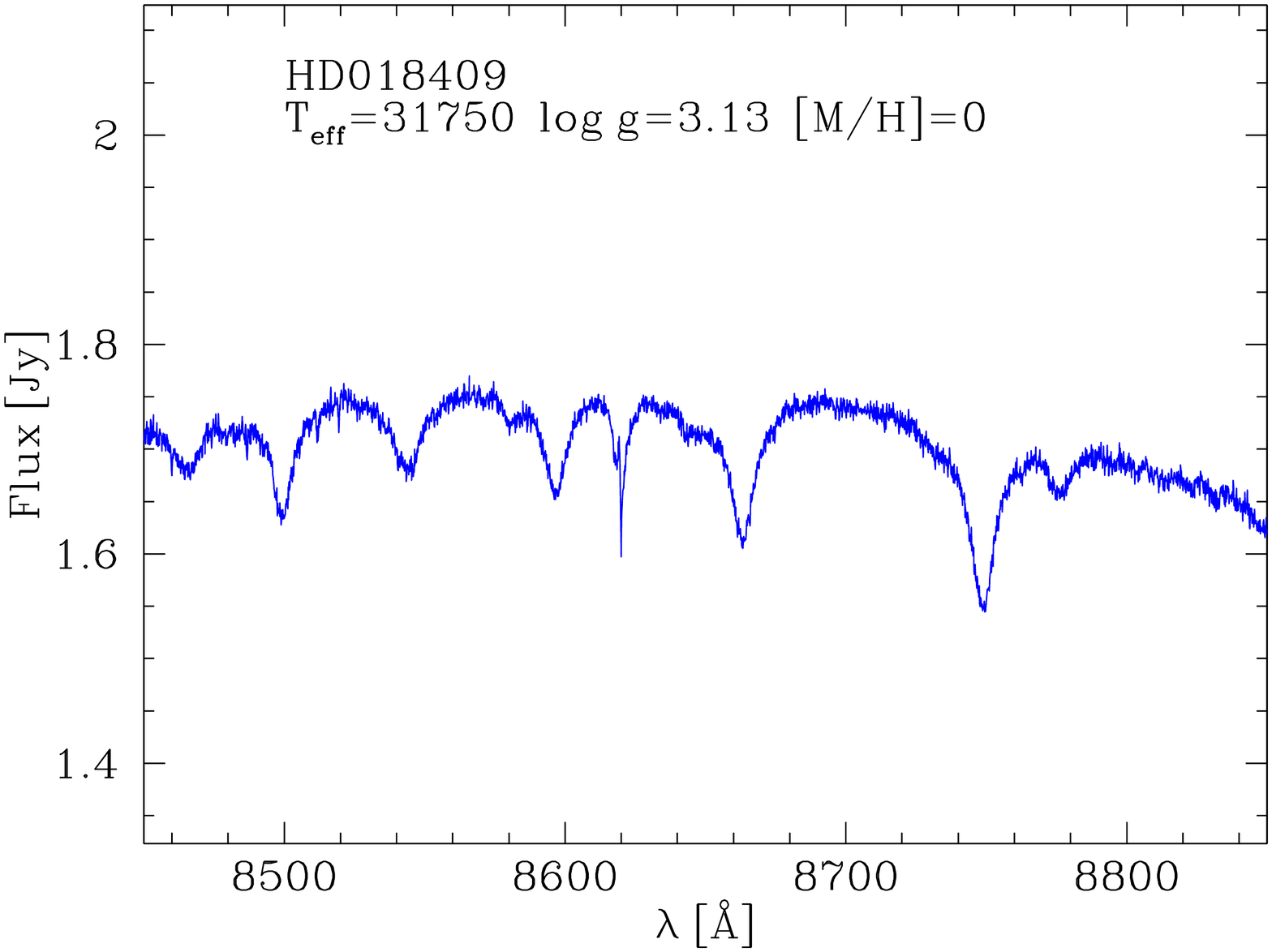}
\includegraphics[width=0.18\textwidth,angle=-90]{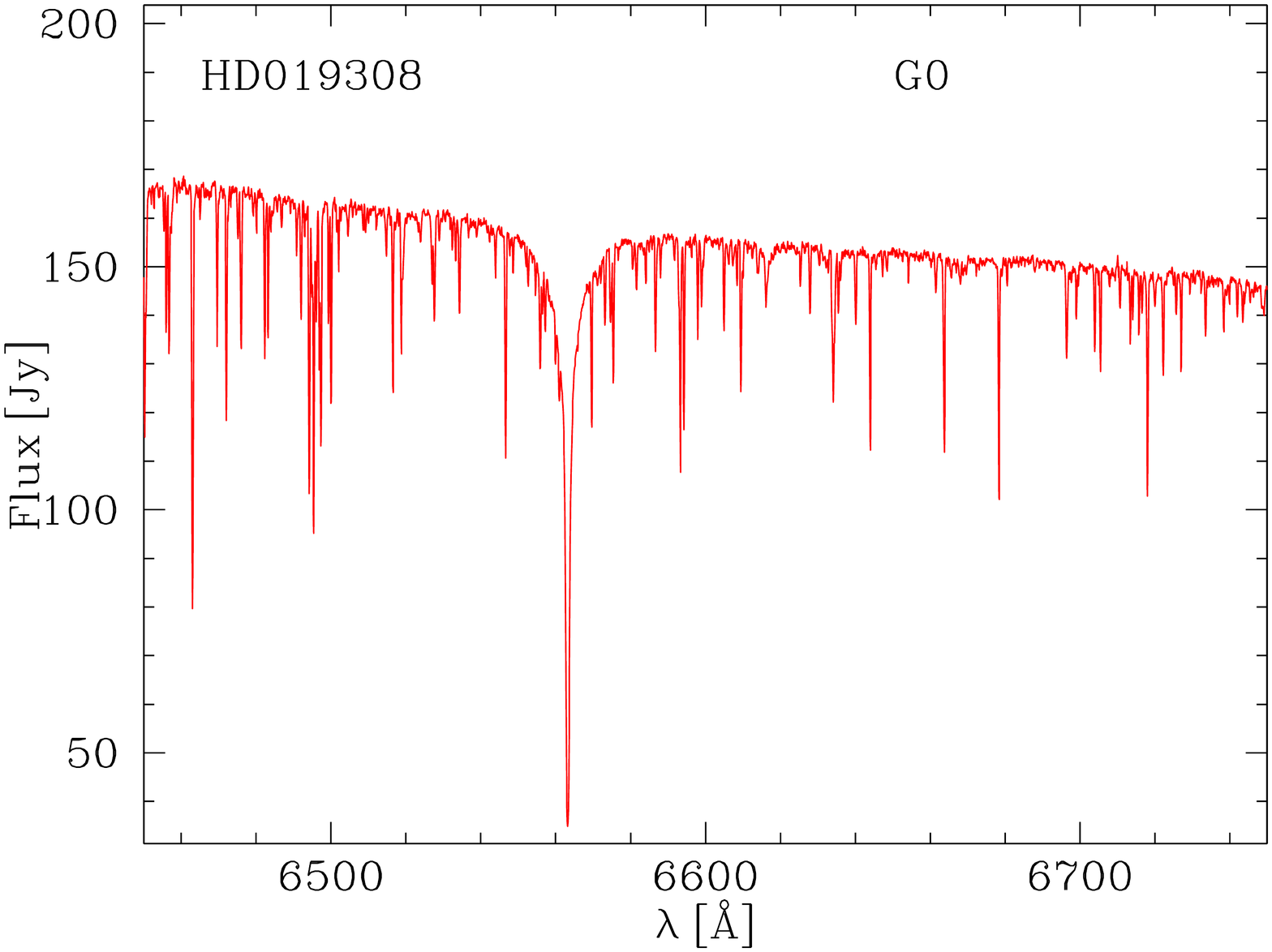}
\includegraphics[width=0.18\textwidth,angle=-90]{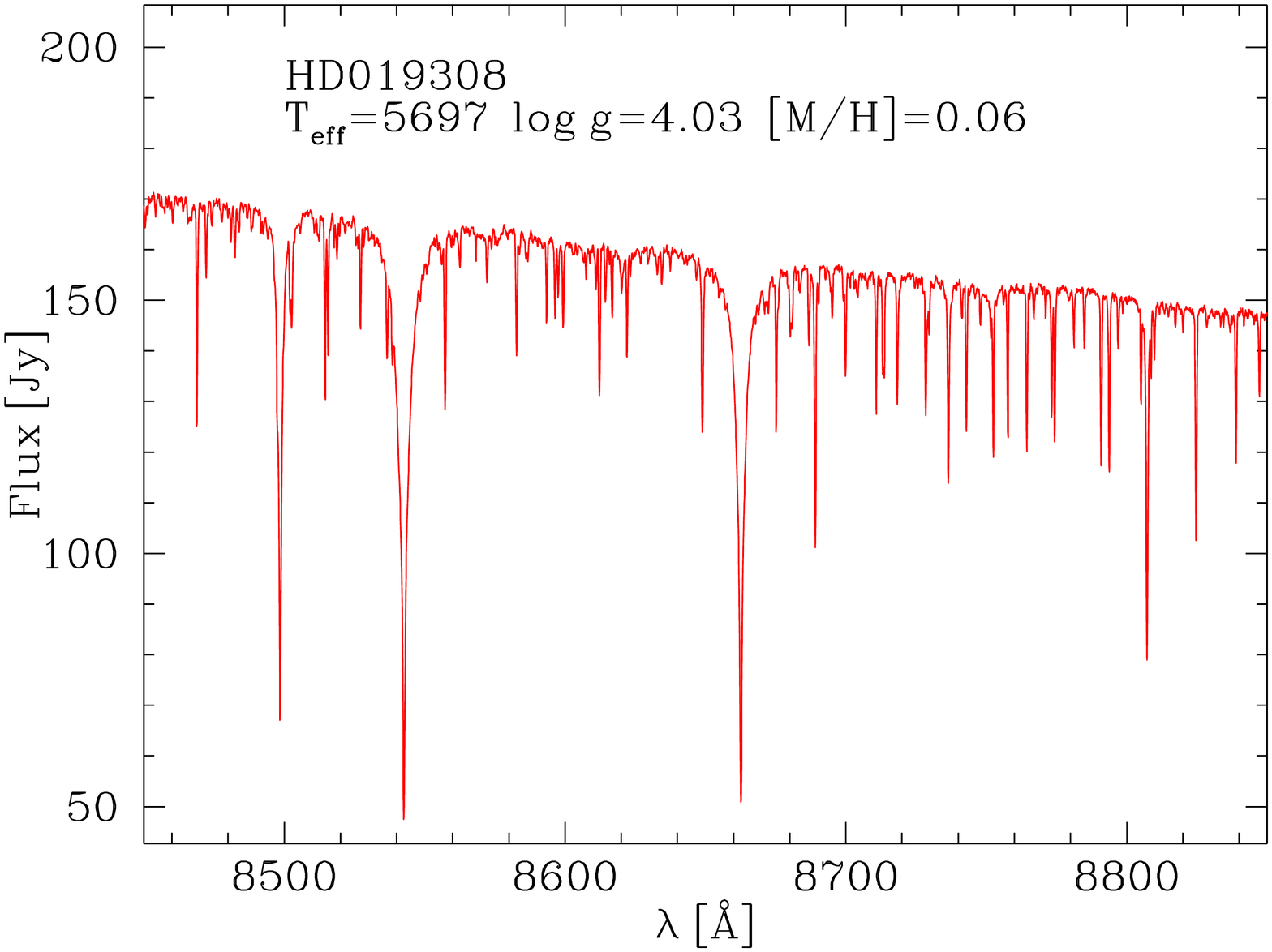}

\contcaption{4. Stars shown in this page are: HD015629, HD016429, HD016523, HD016523, HD016581, HD017081, HD017145, HD017506, HD017638, HD017638, HD018144, HD018296, HD018409 and HD019308.}
\end{figure*}

\begin{figure*}
\includegraphics[width=0.18\textwidth,angle=-90]{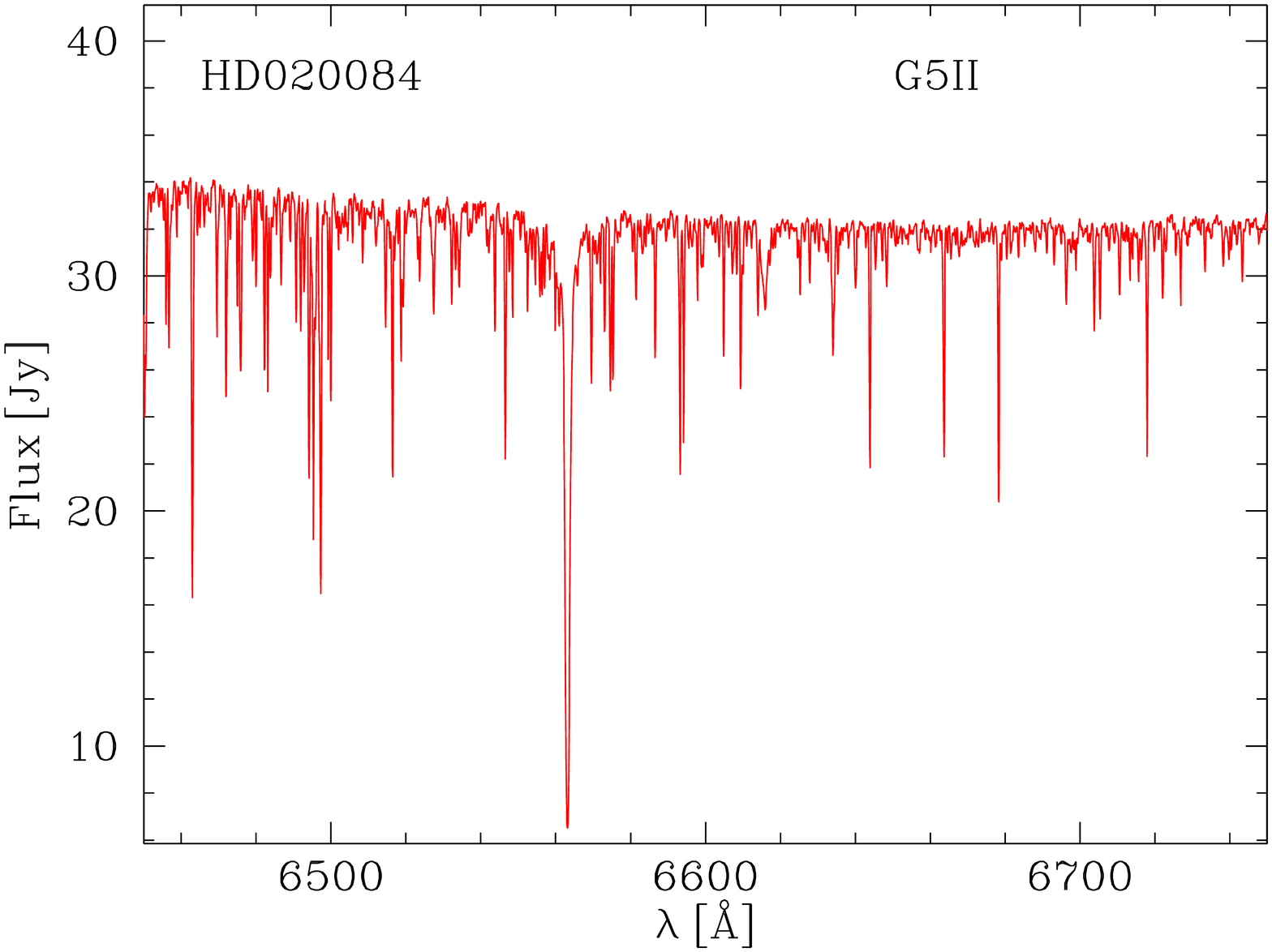}
\includegraphics[width=0.18\textwidth,angle=-90]{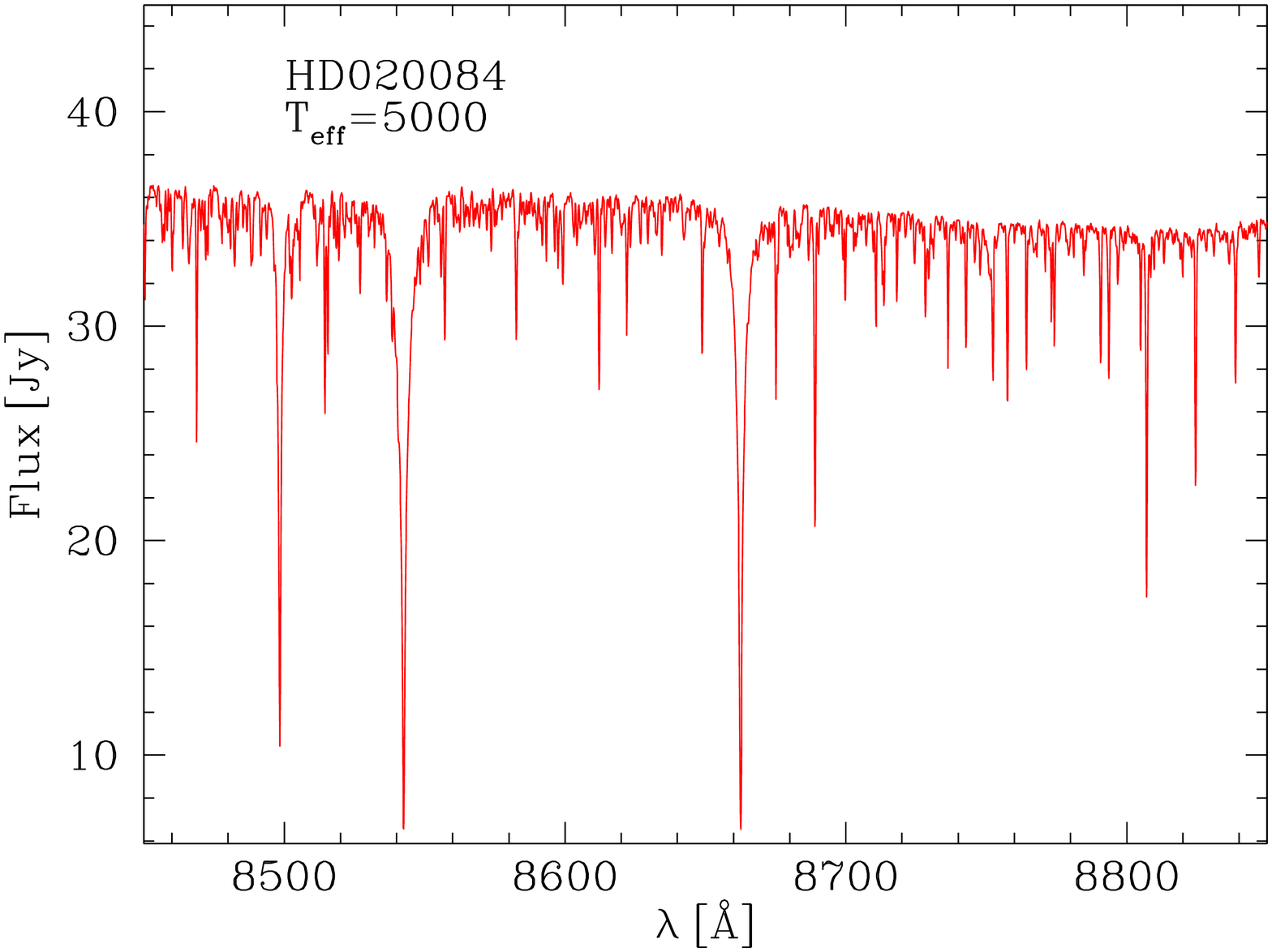}
\includegraphics[width=0.18\textwidth,angle=-90]{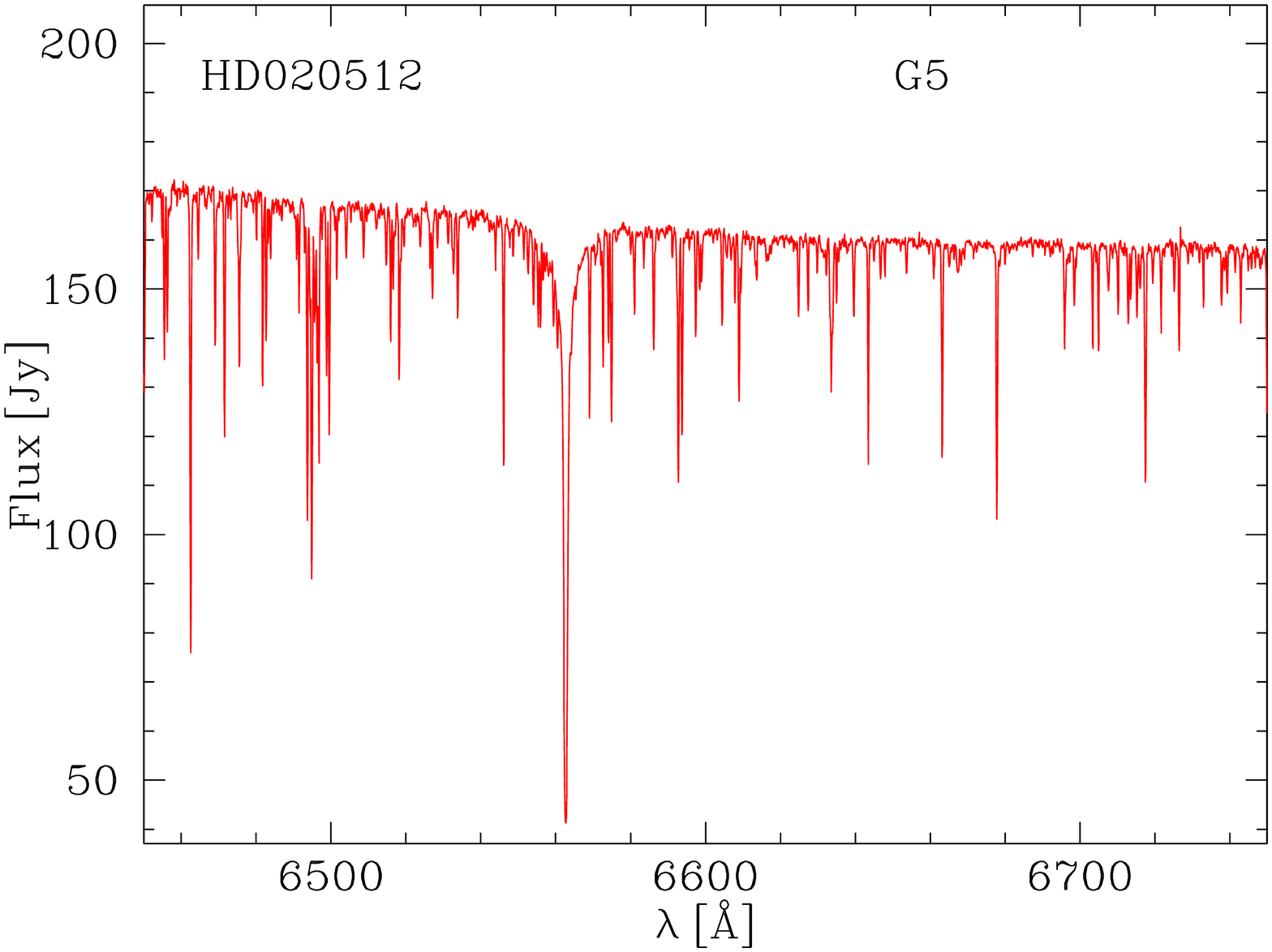}
\includegraphics[width=0.18\textwidth,angle=-90]{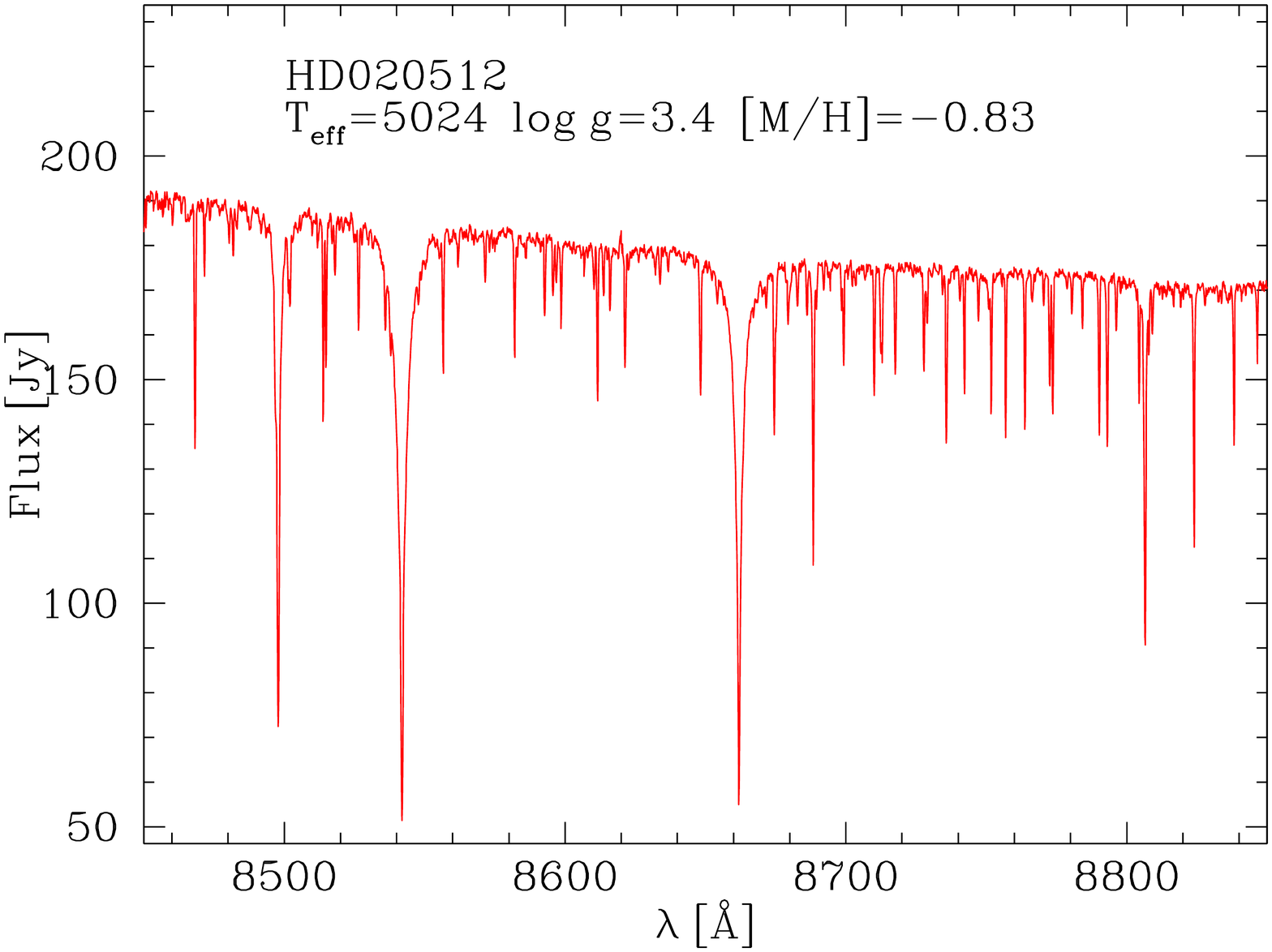}
\includegraphics[width=0.18\textwidth,angle=-90]{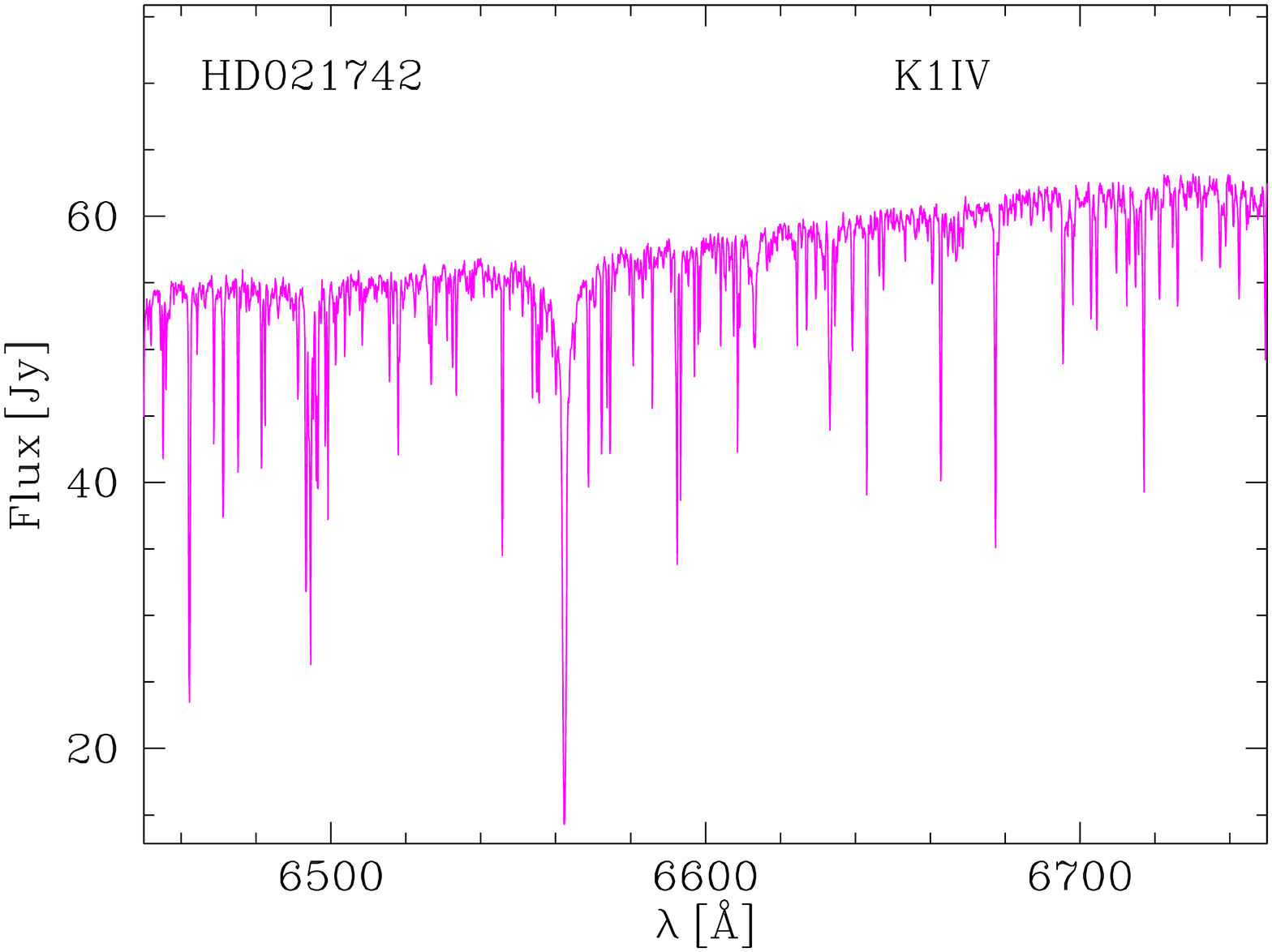}
\includegraphics[width=0.18\textwidth,angle=-90]{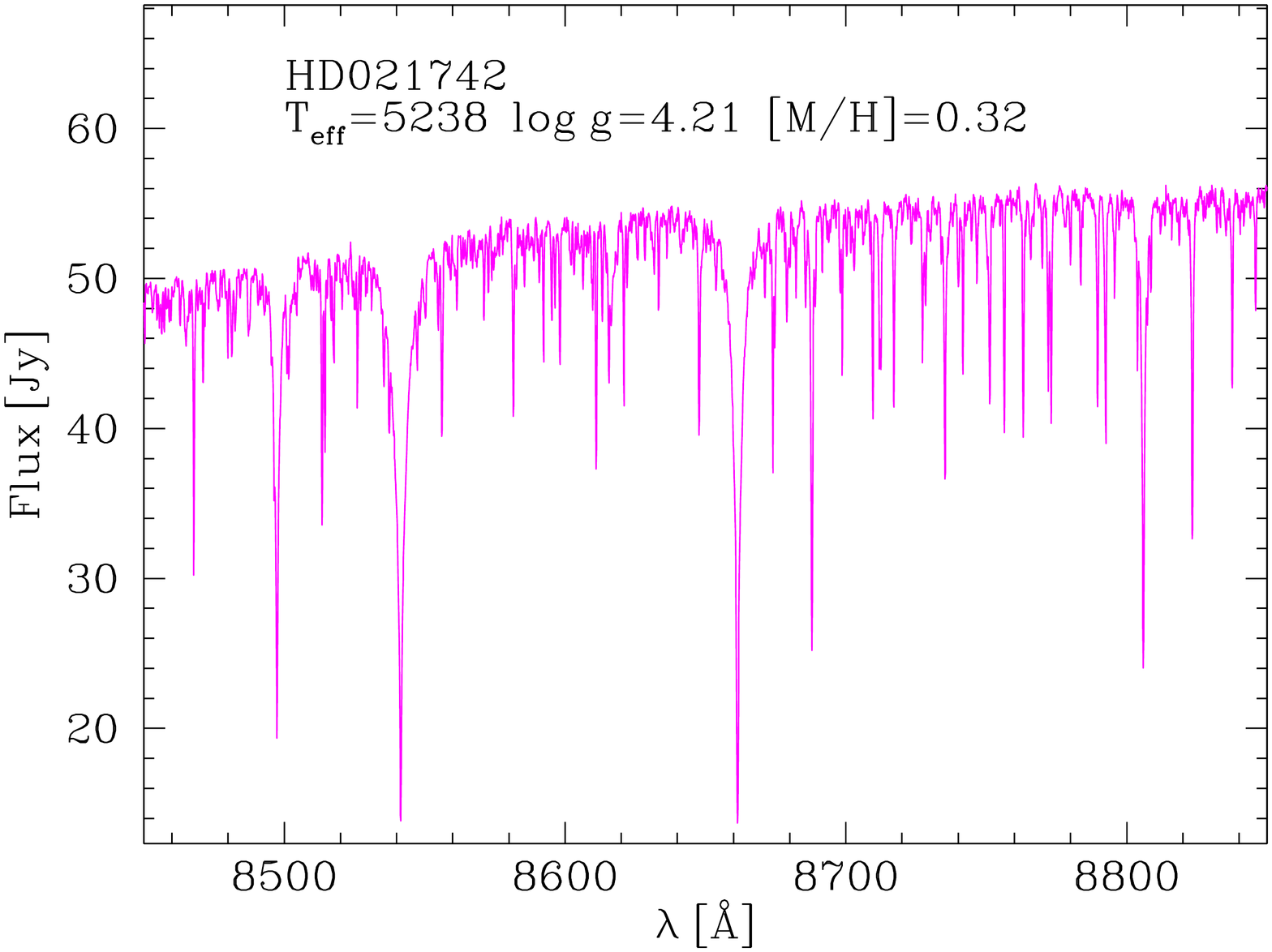}
\includegraphics[width=0.18\textwidth,angle=-90]{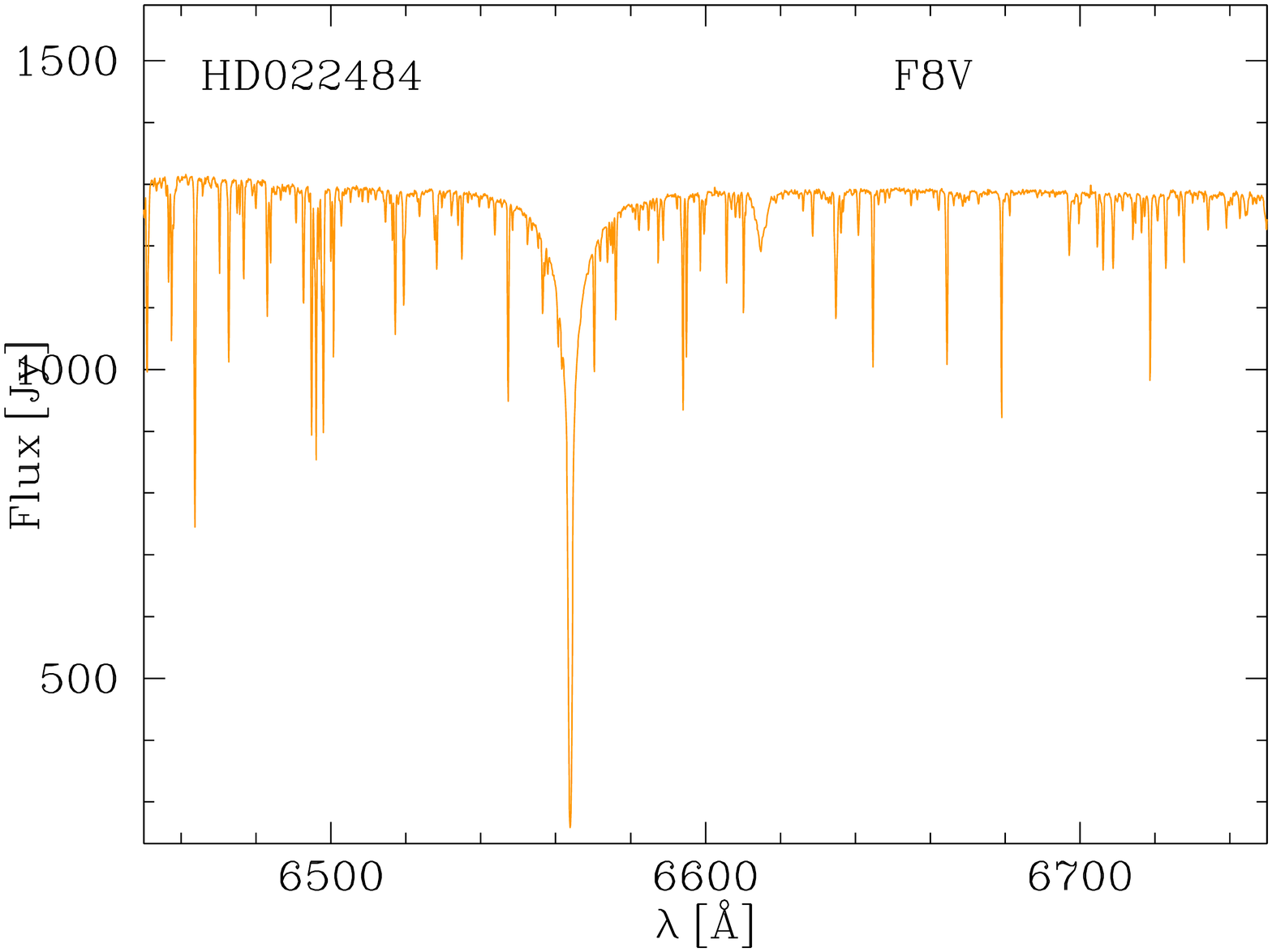}
\includegraphics[width=0.18\textwidth,angle=-90]{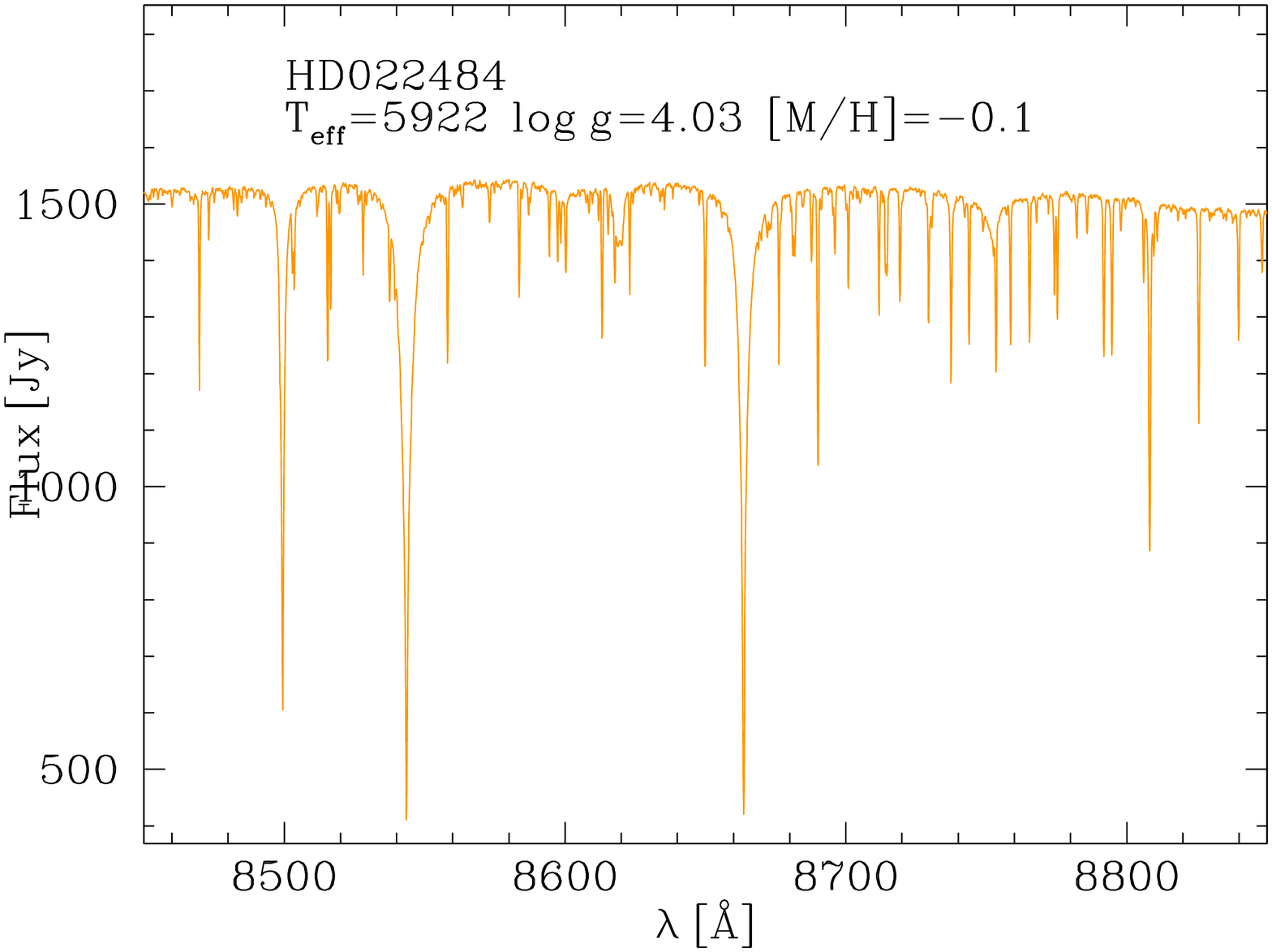}
\includegraphics[width=0.18\textwidth,angle=-90]{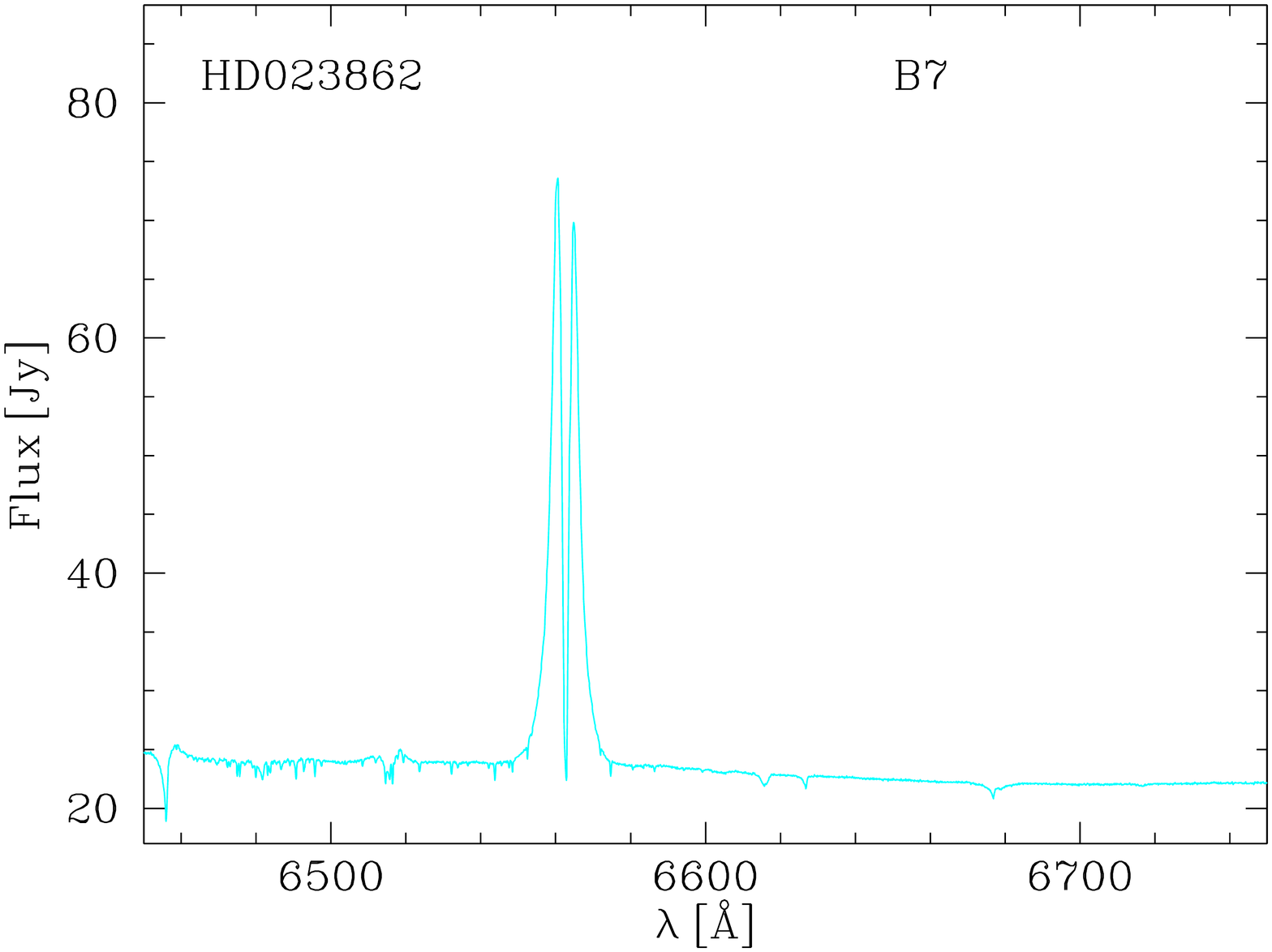}
\includegraphics[width=0.18\textwidth,angle=-90]{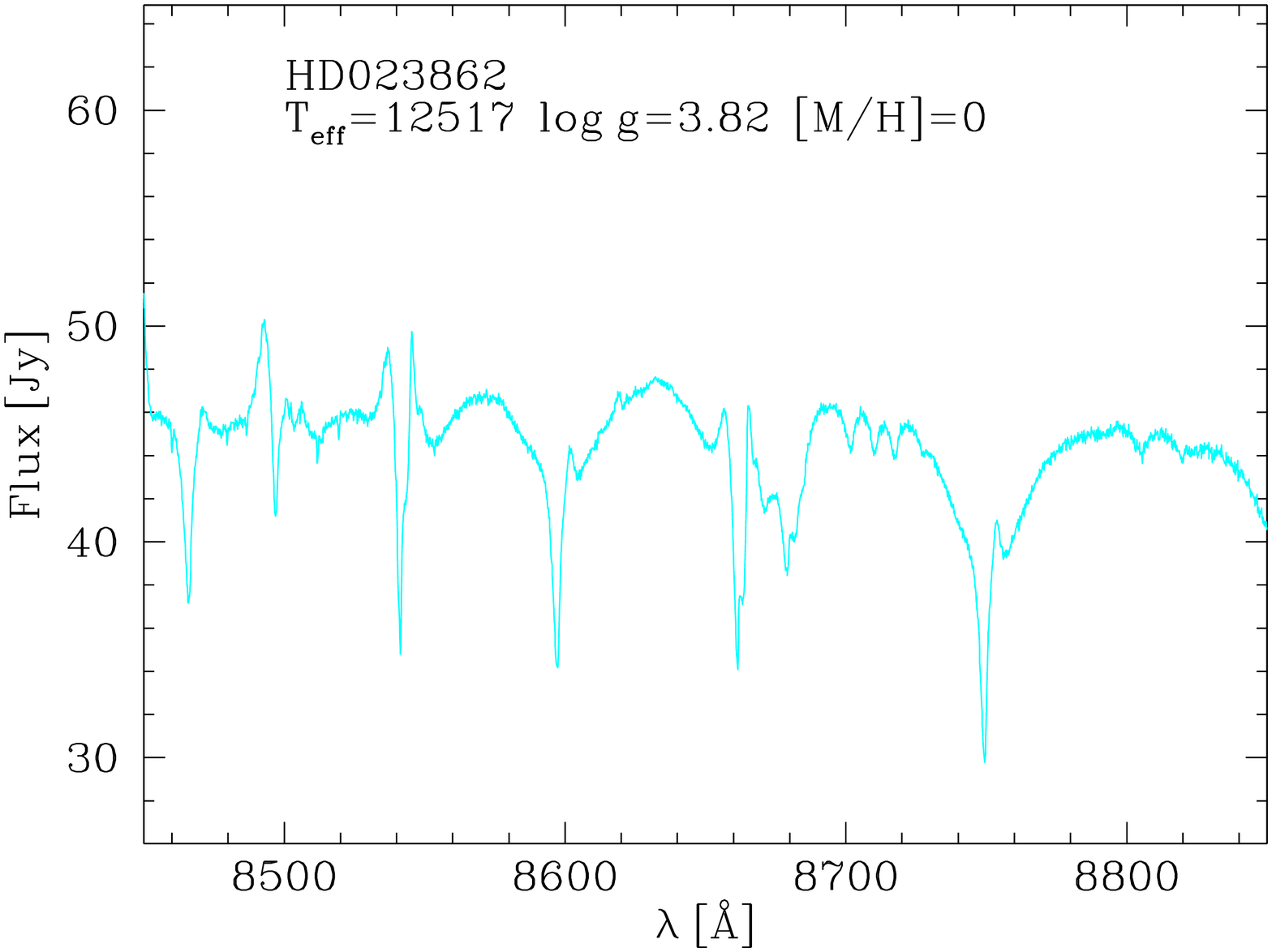}
\includegraphics[width=0.18\textwidth,angle=-90]{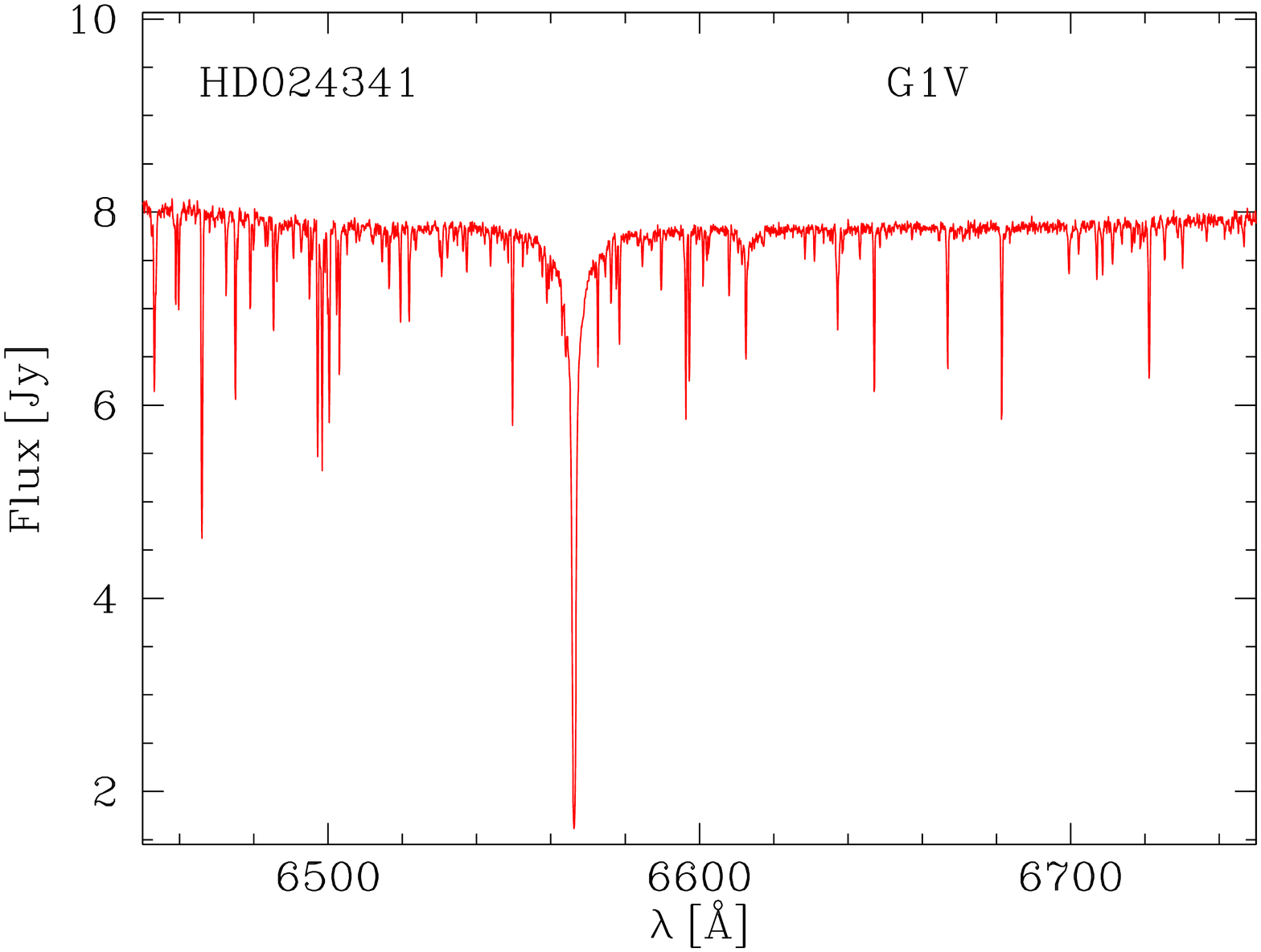}
\includegraphics[width=0.18\textwidth,angle=-90]{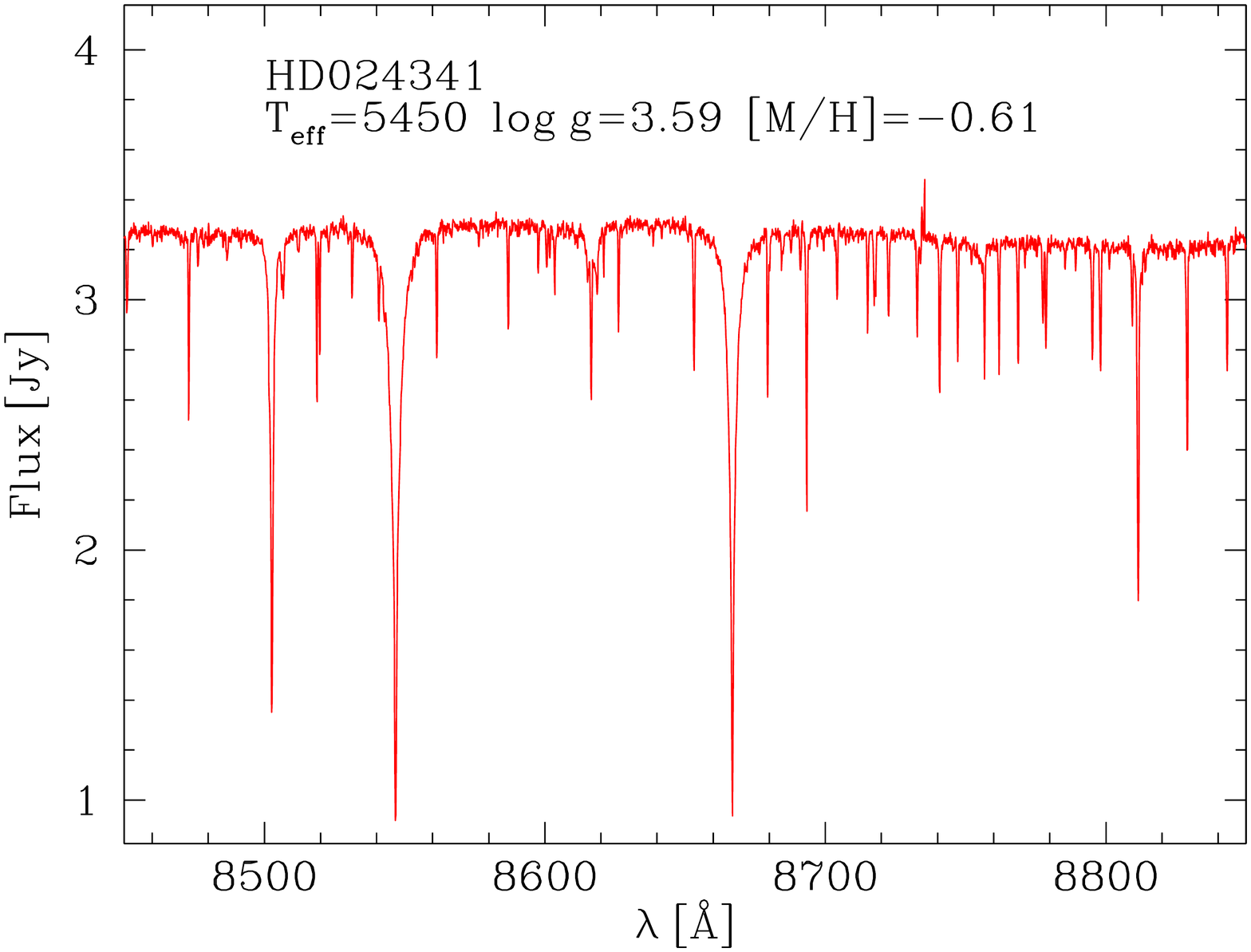}
\includegraphics[width=0.18\textwidth,angle=-90]{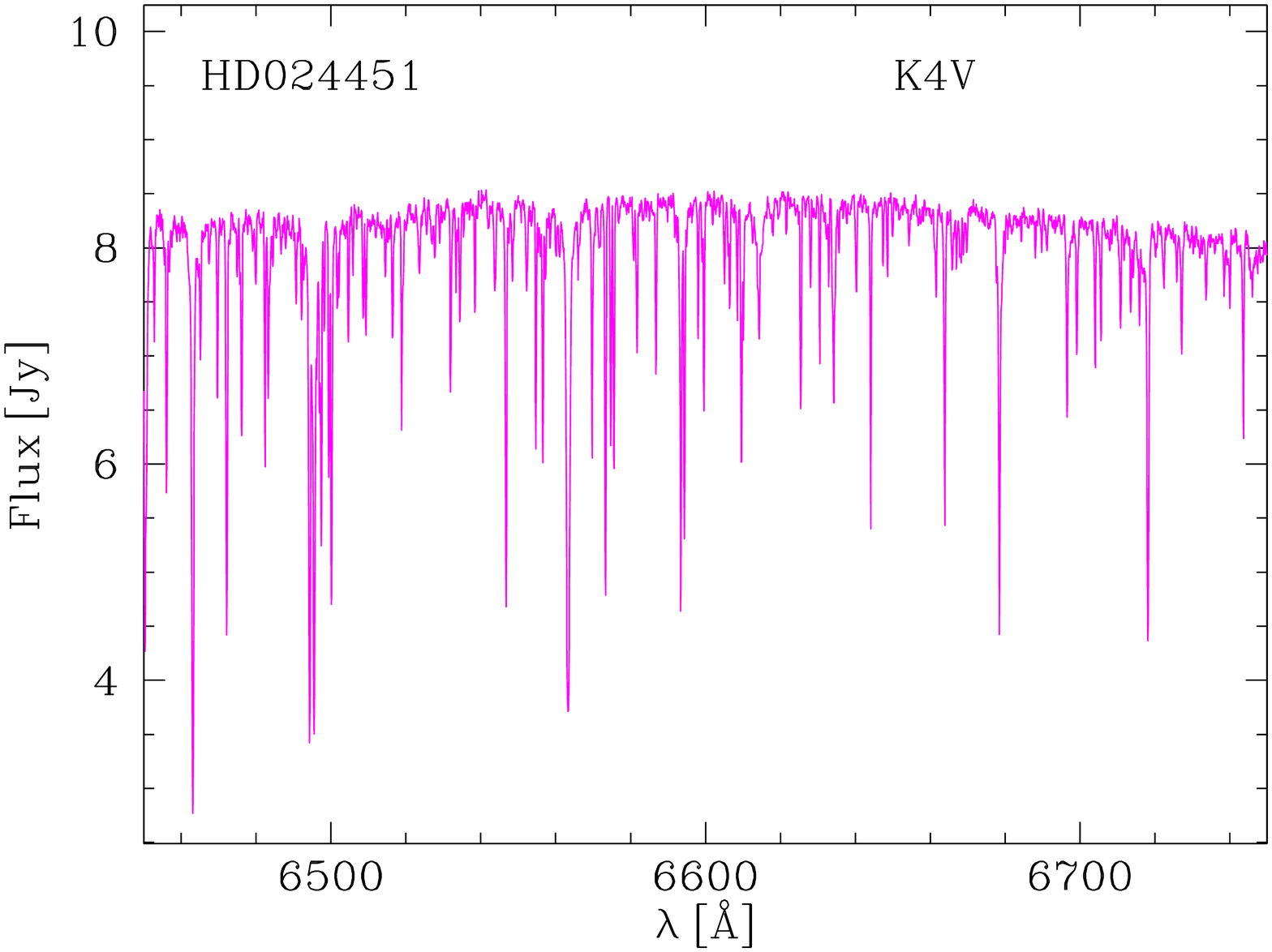}
\includegraphics[width=0.18\textwidth,angle=-90]{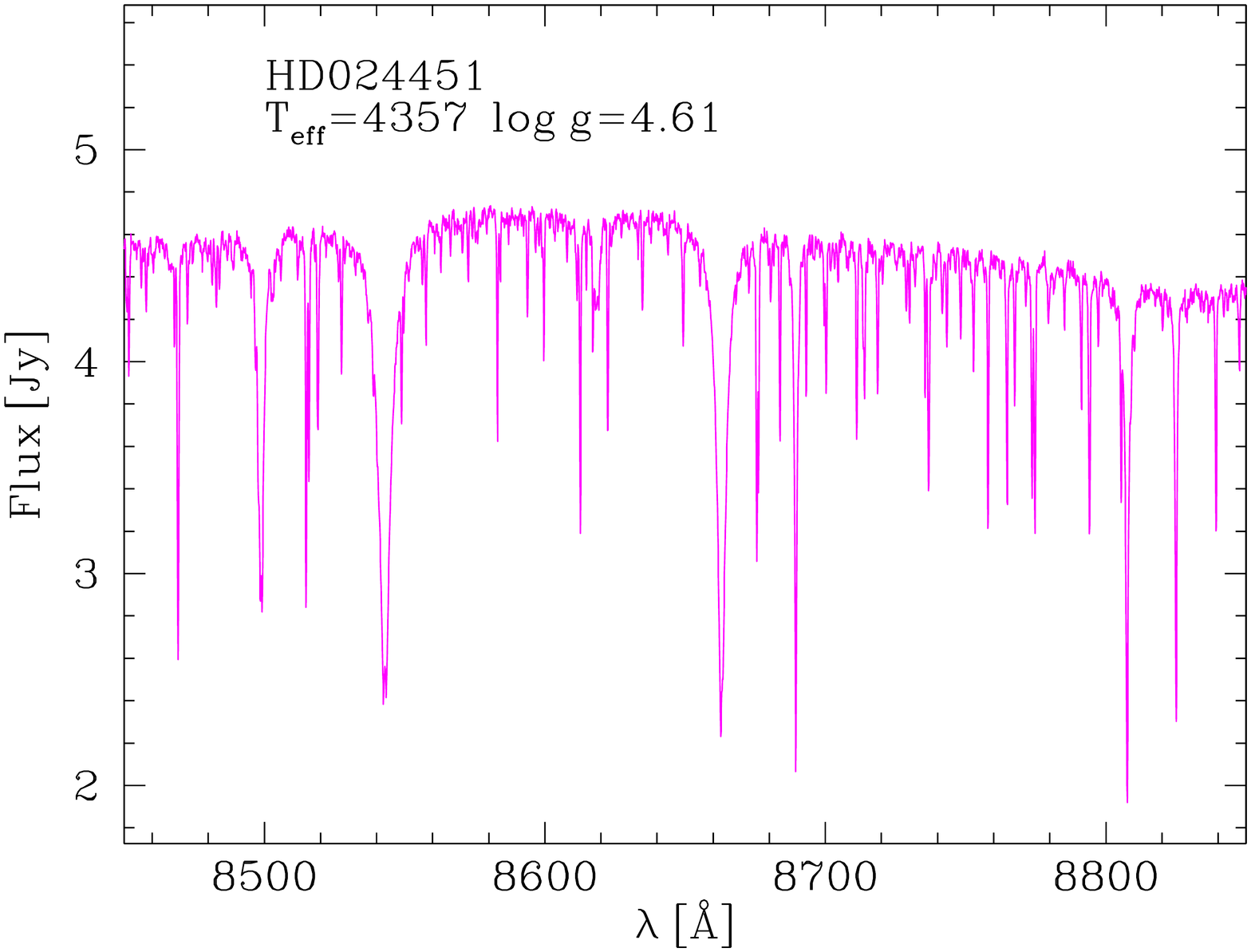}
\includegraphics[width=0.18\textwidth,angle=-90]{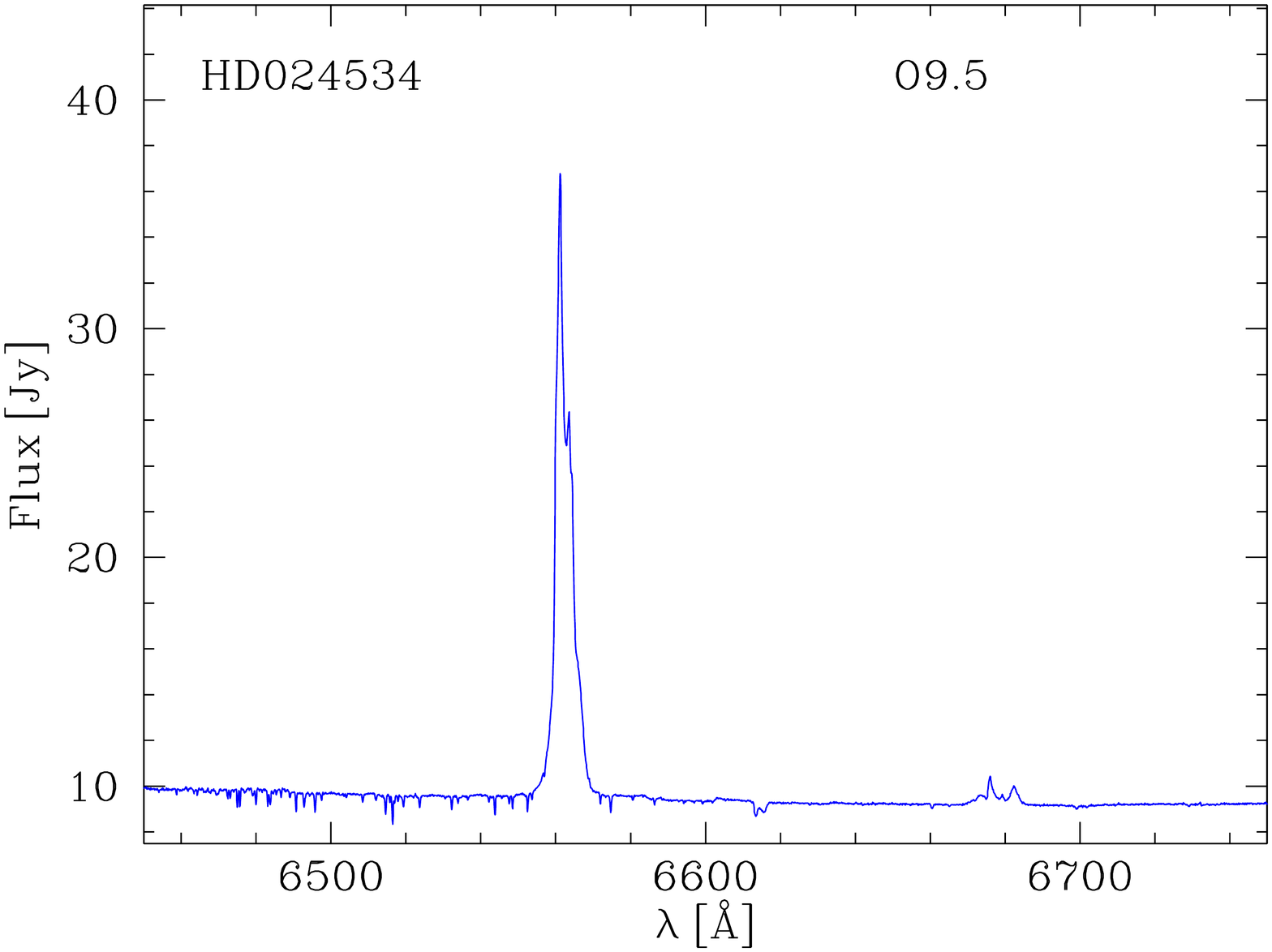}
\includegraphics[width=0.18\textwidth,angle=-90]{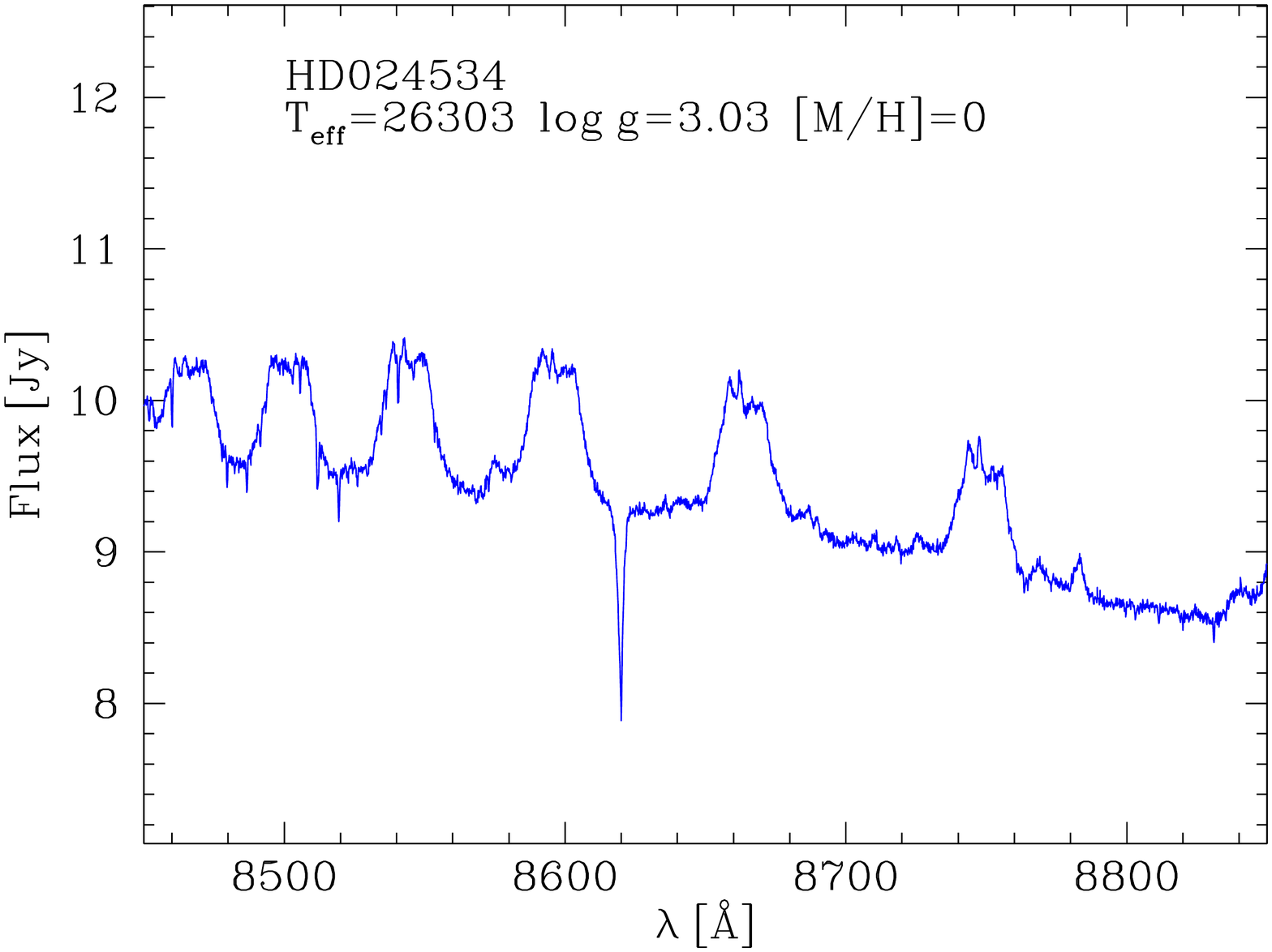}
\includegraphics[width=0.18\textwidth,angle=-90]{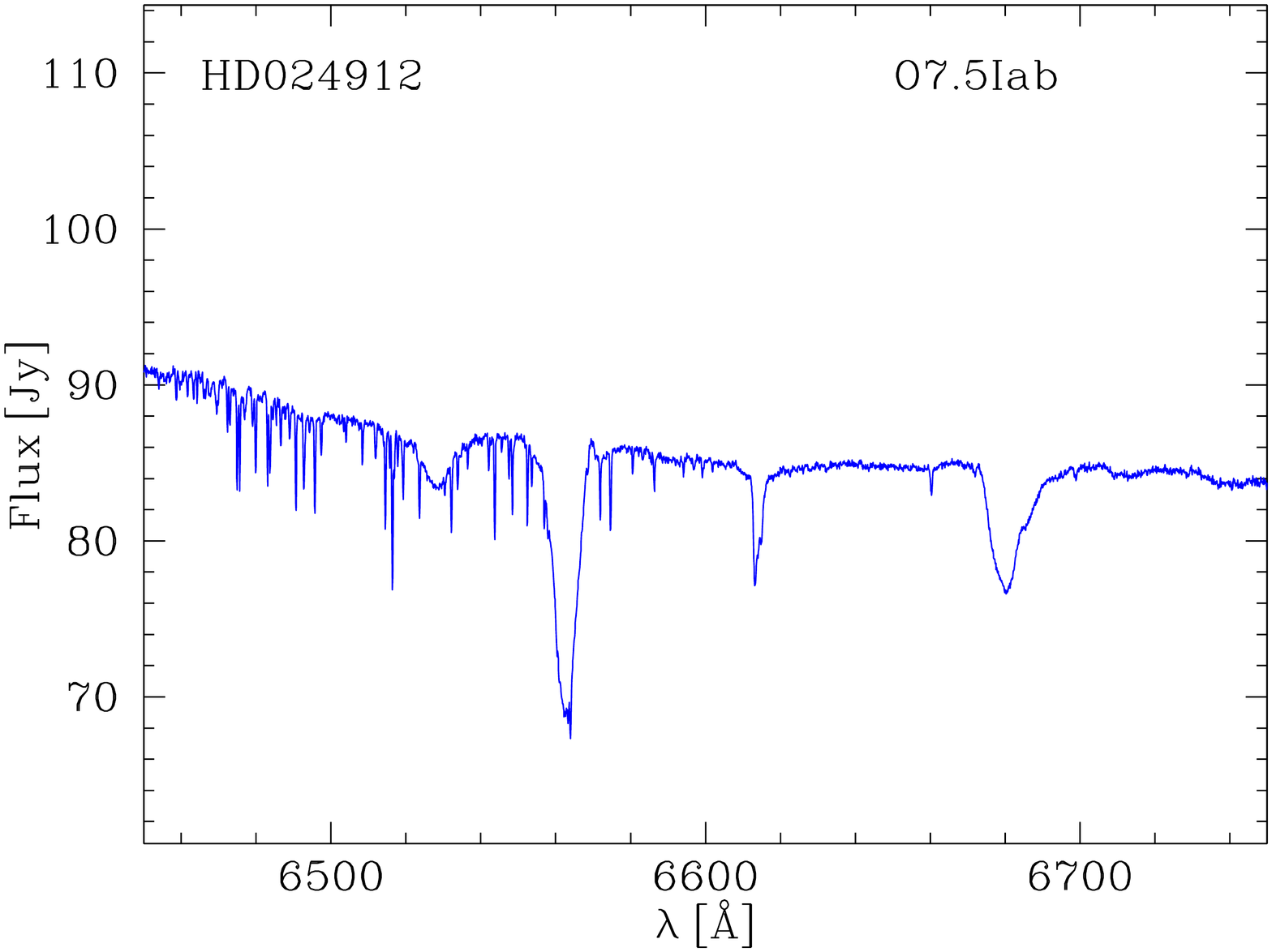} 
\includegraphics[width=0.18\textwidth,angle=-90]{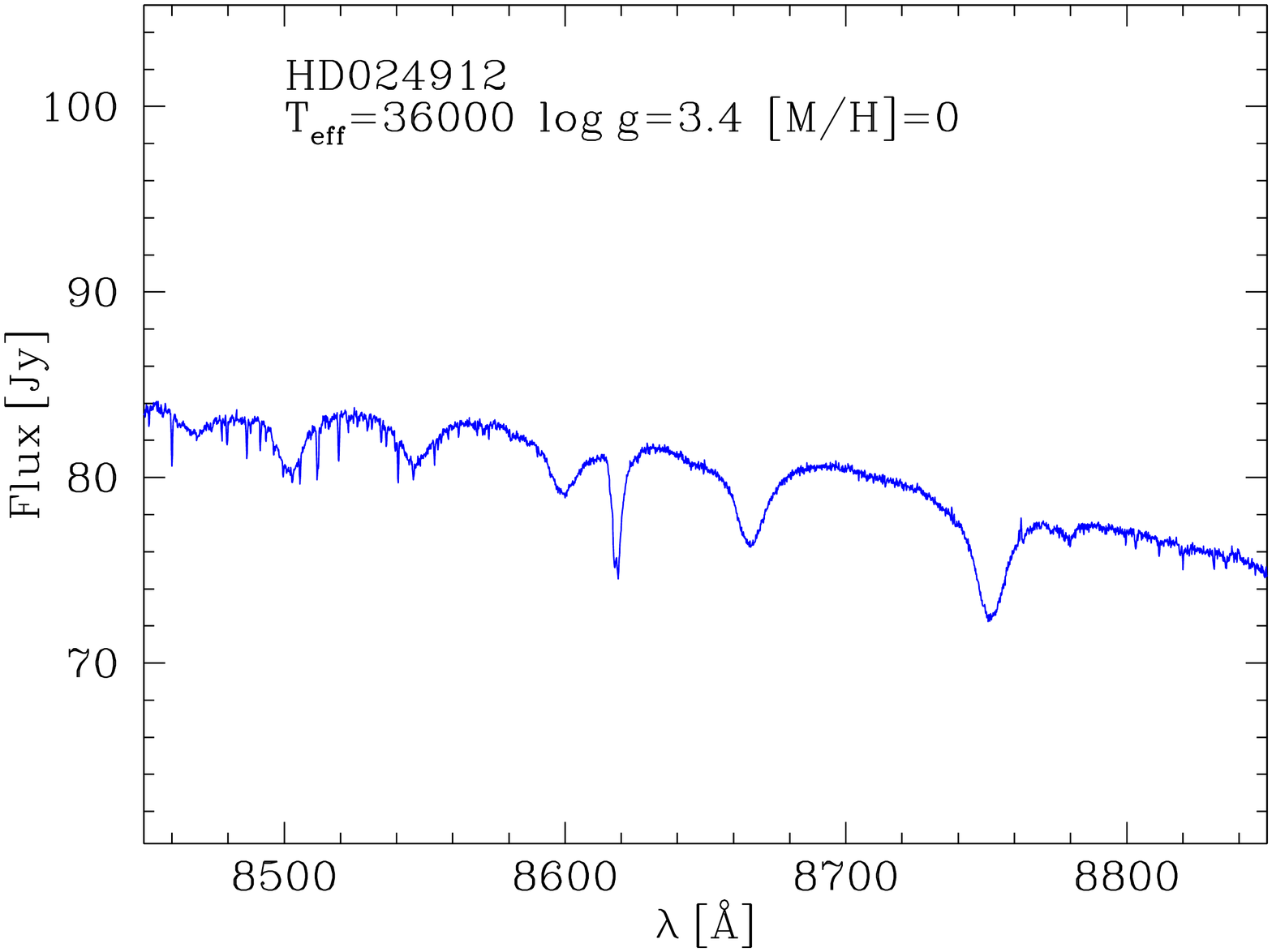}
\includegraphics[width=0.18\textwidth,angle=-90]{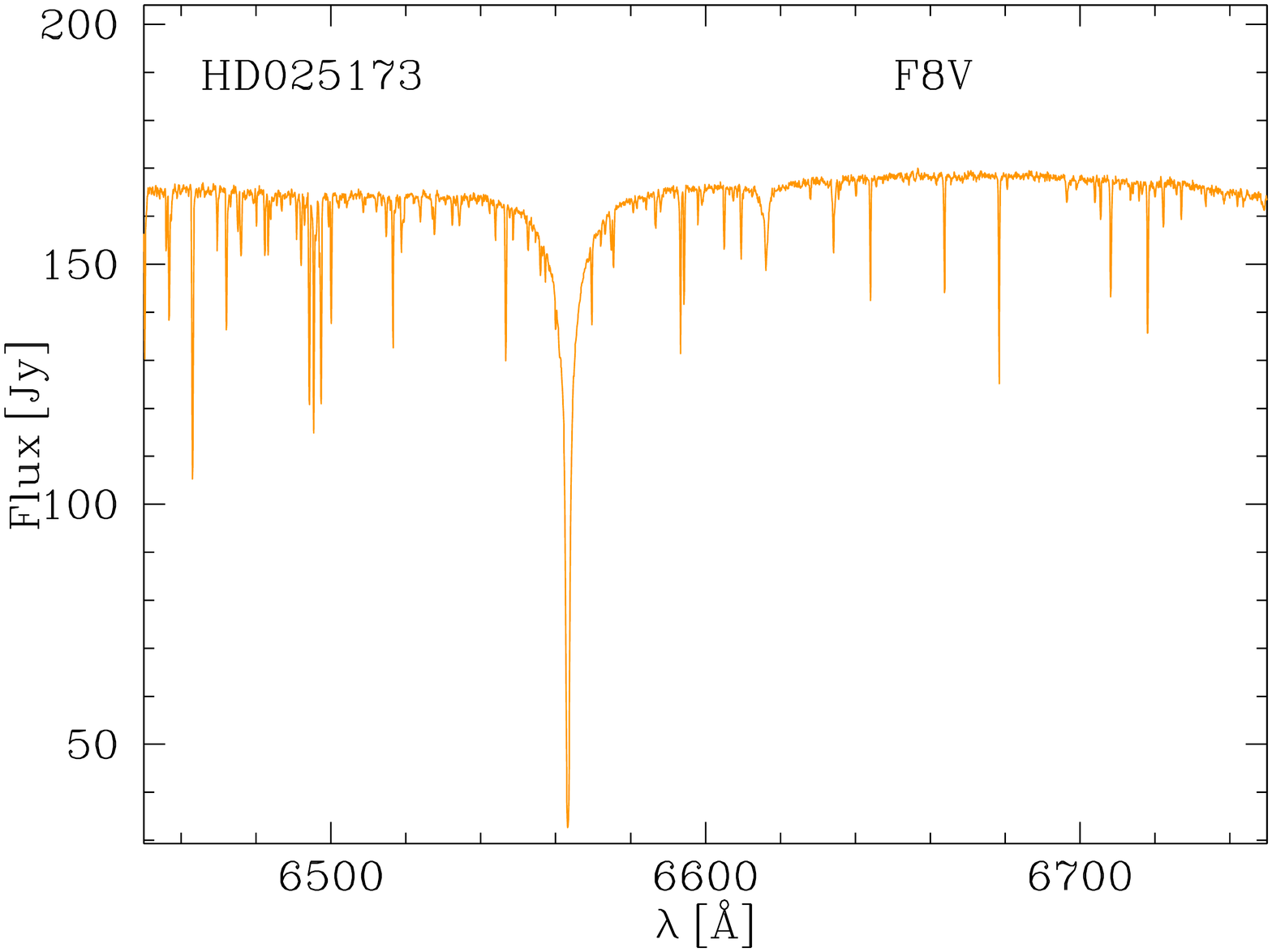}
\includegraphics[width=0.18\textwidth,angle=-90]{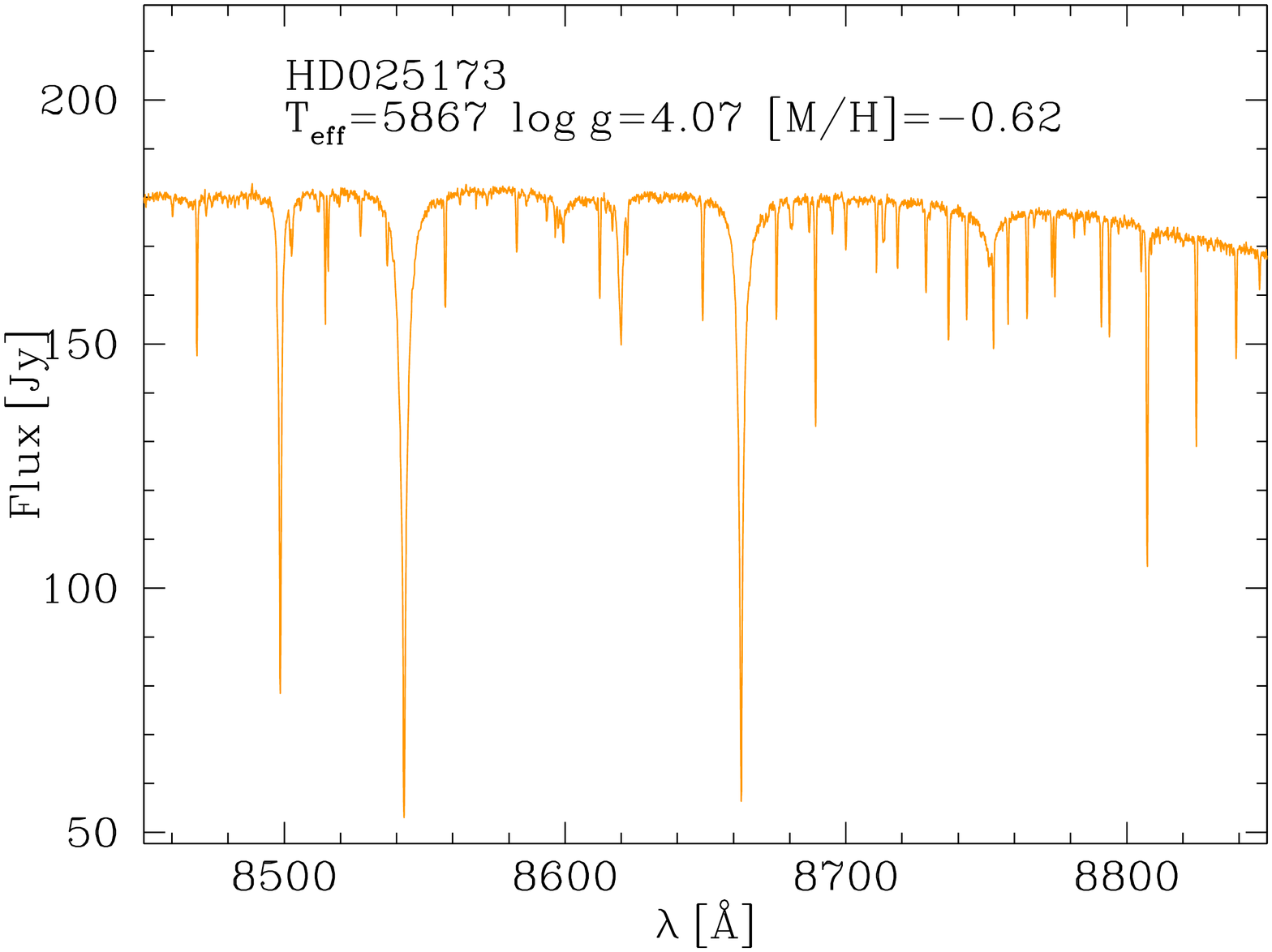}
\includegraphics[width=0.18\textwidth,angle=-90]{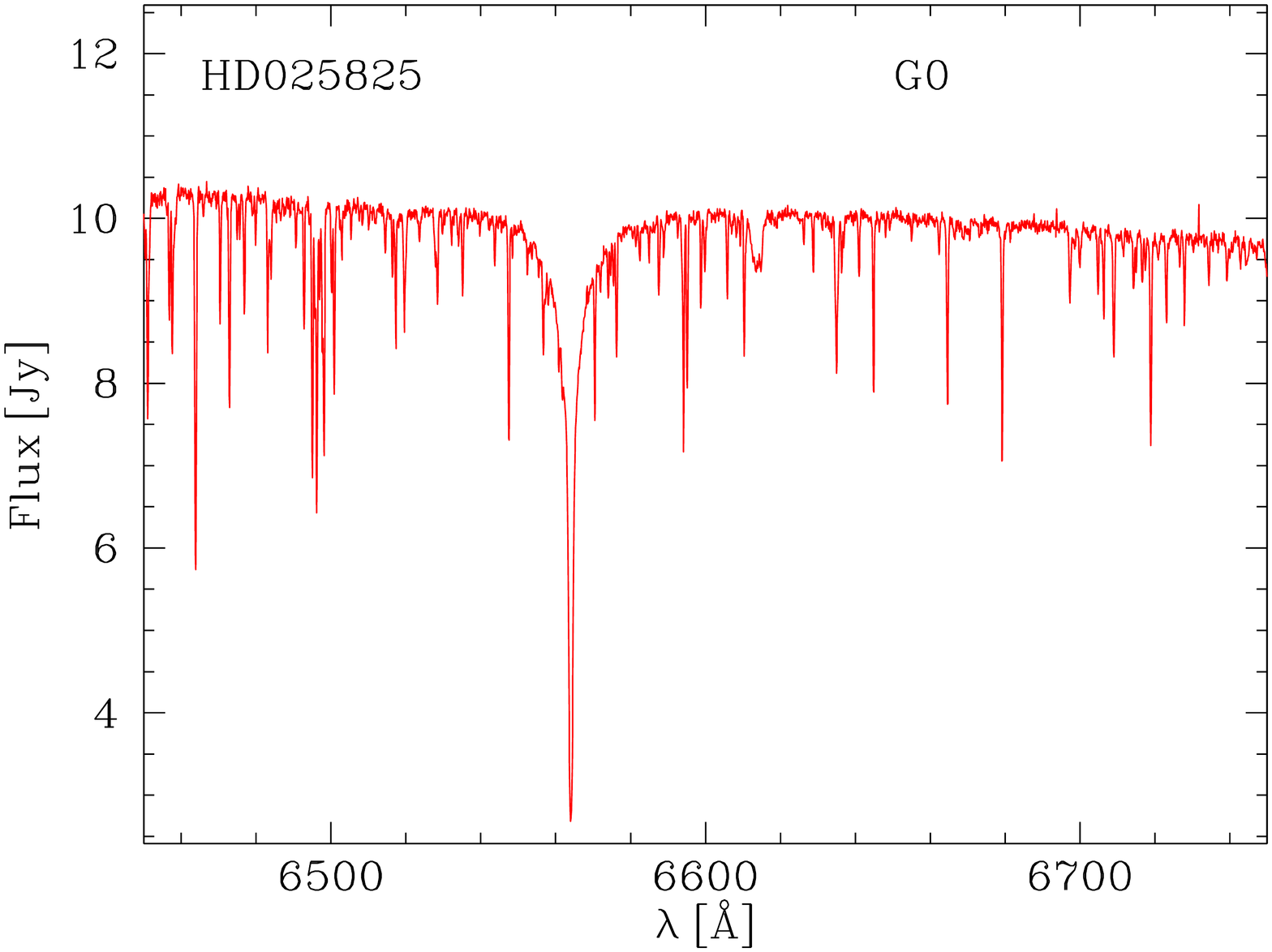}
\includegraphics[width=0.18\textwidth,angle=-90]{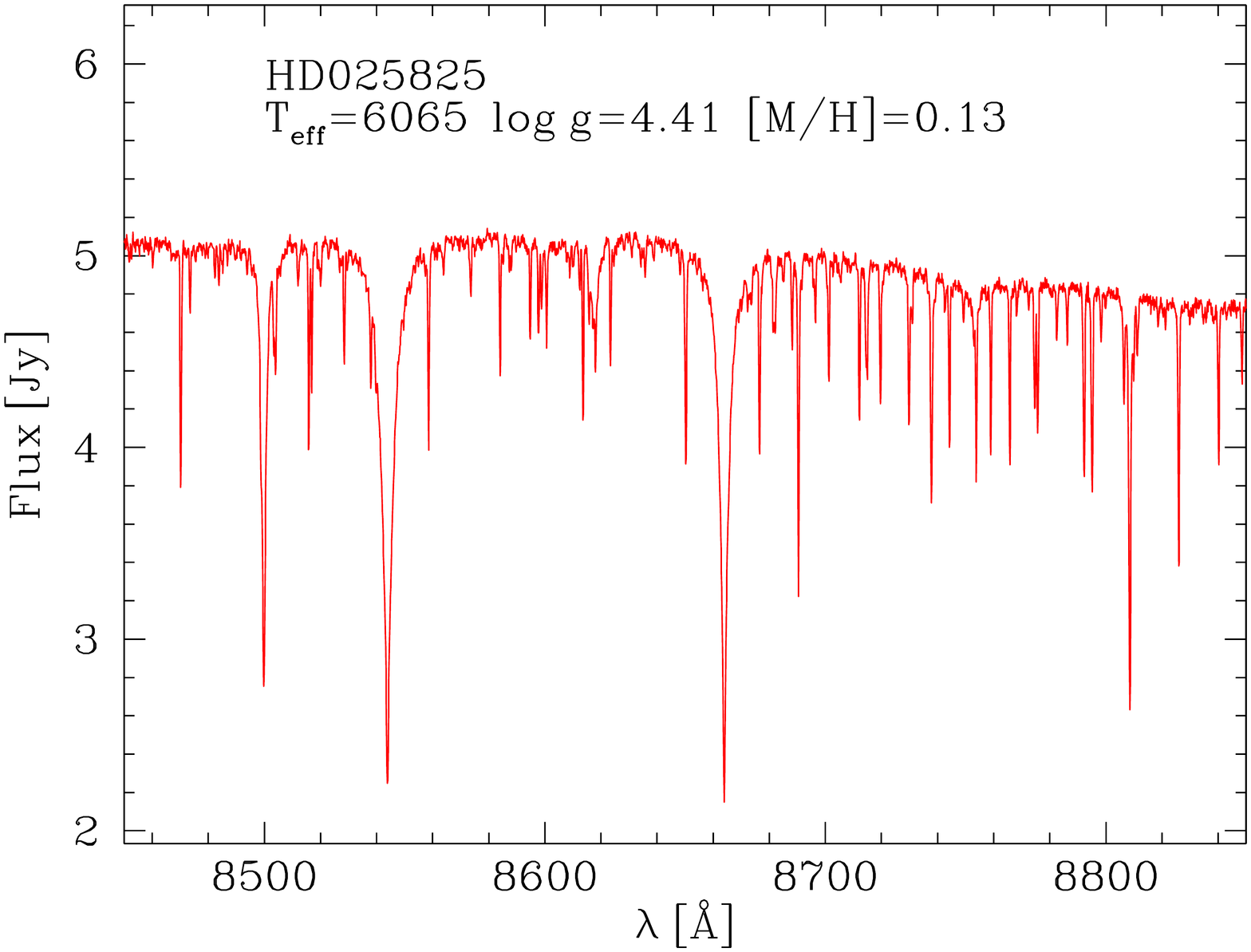}
\includegraphics[width=0.18\textwidth,angle=-90]{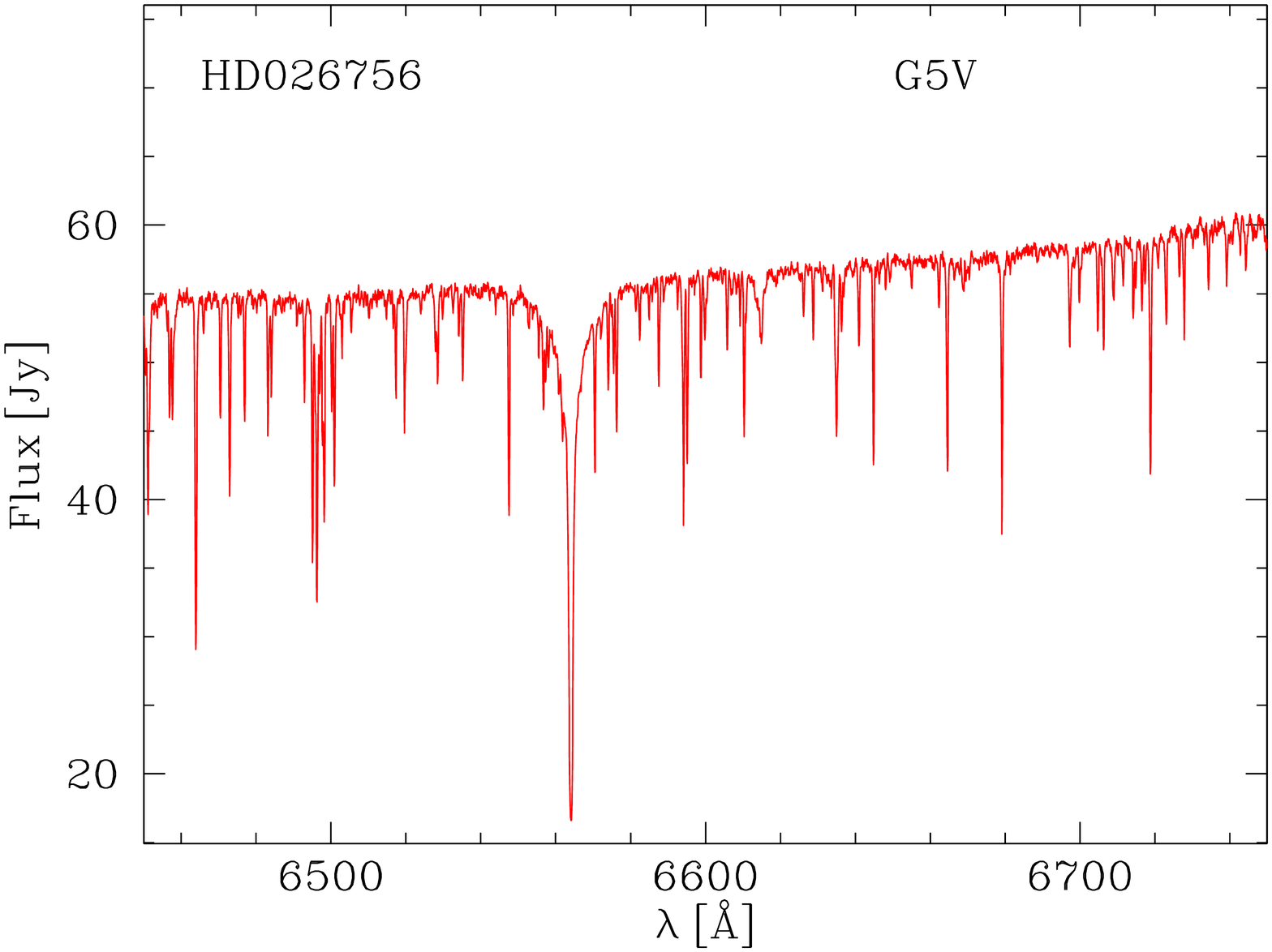}
\includegraphics[width=0.18\textwidth,angle=-90]{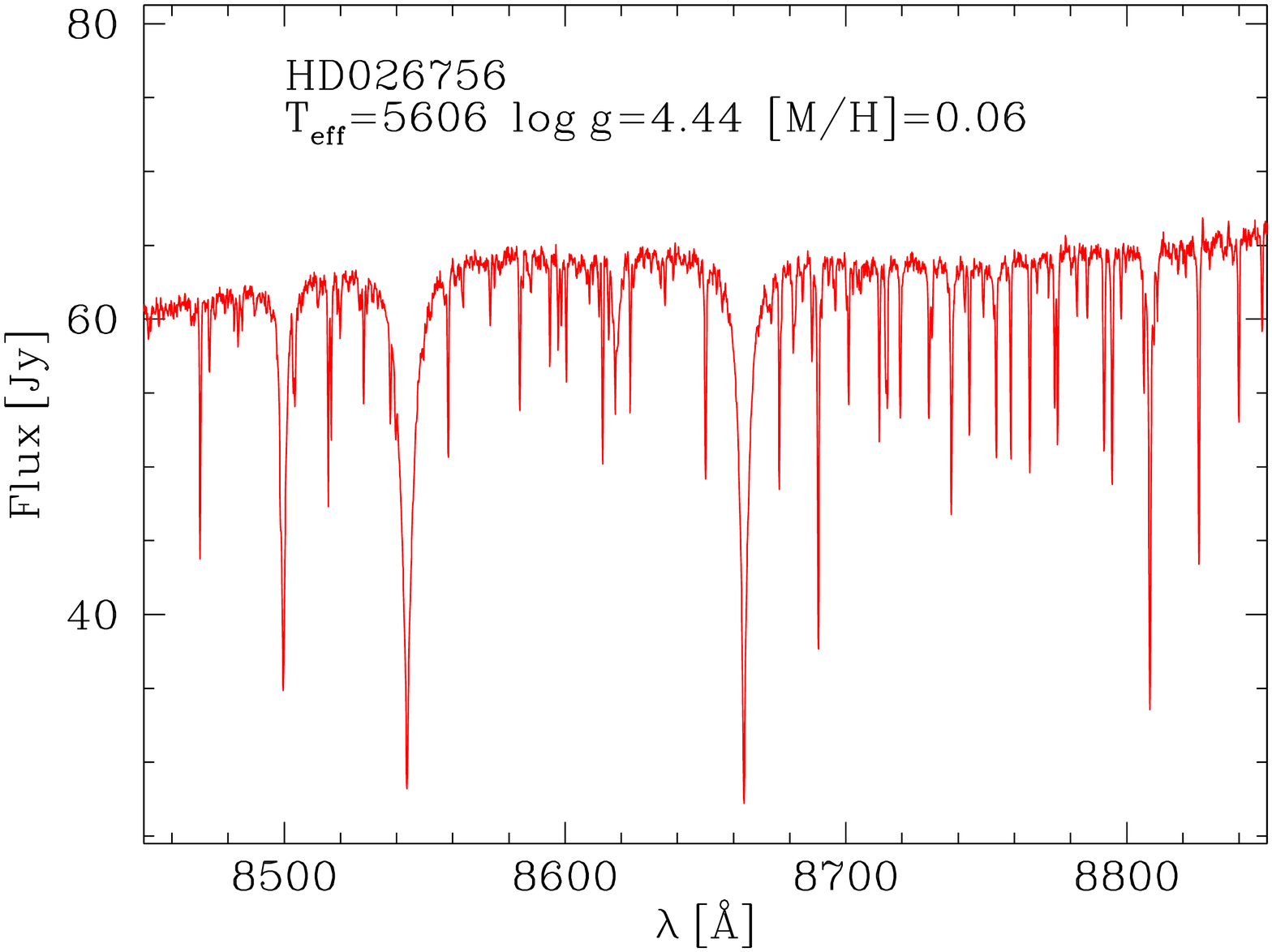}
\includegraphics[width=0.18\textwidth,angle=-90]{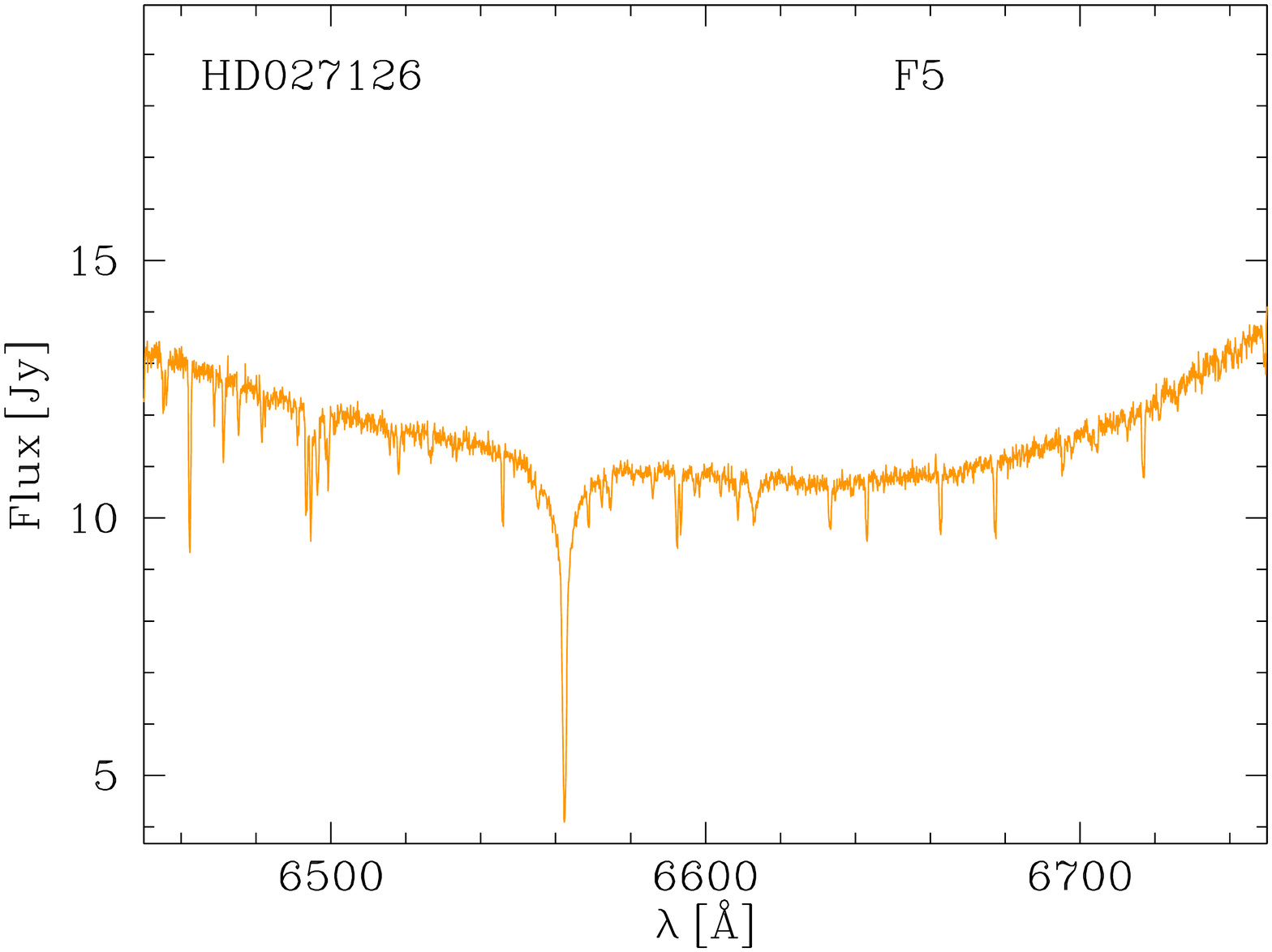}
\includegraphics[width=0.18\textwidth,angle=-90]{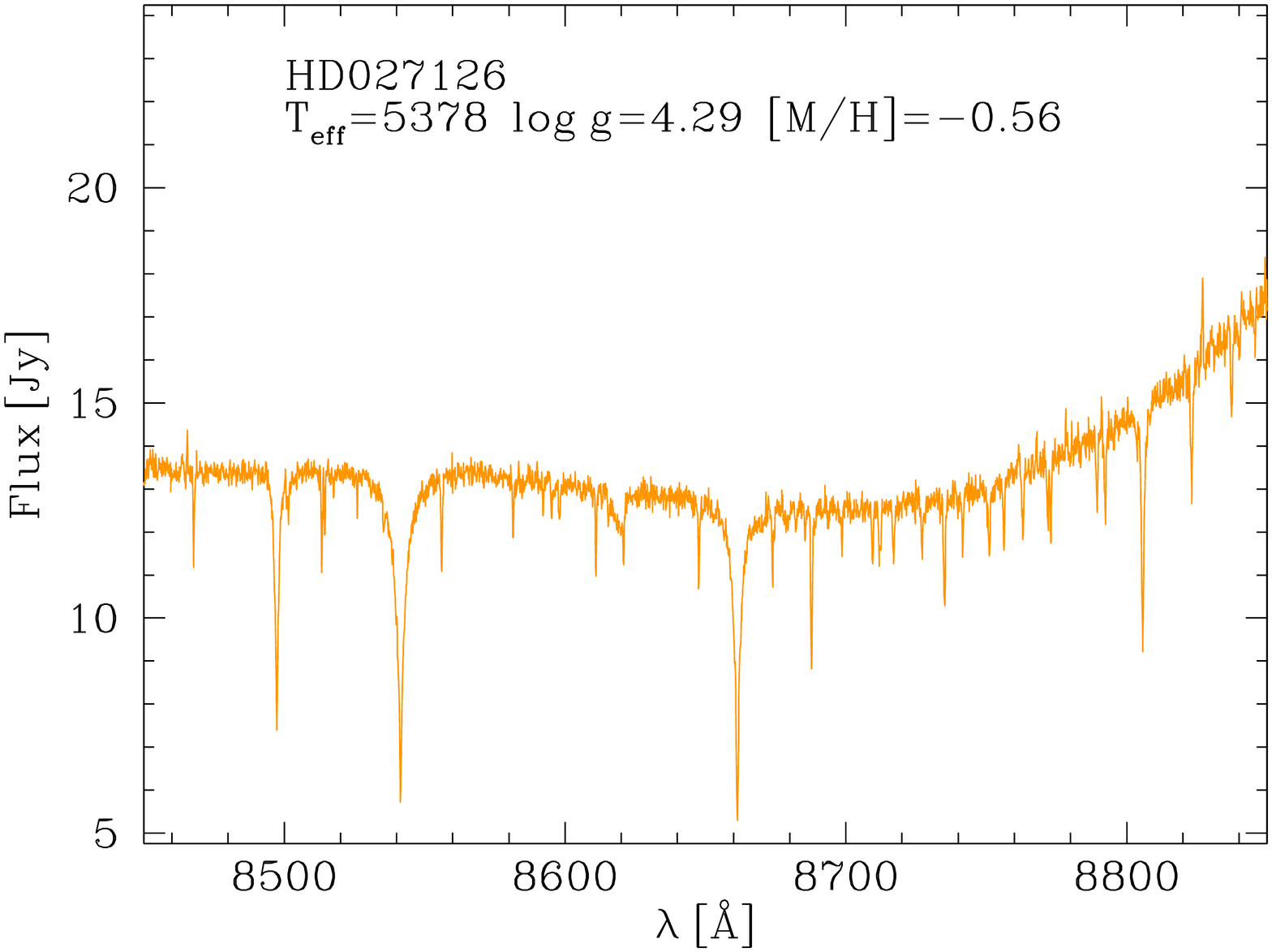}
\includegraphics[width=0.18\textwidth,angle=-90]{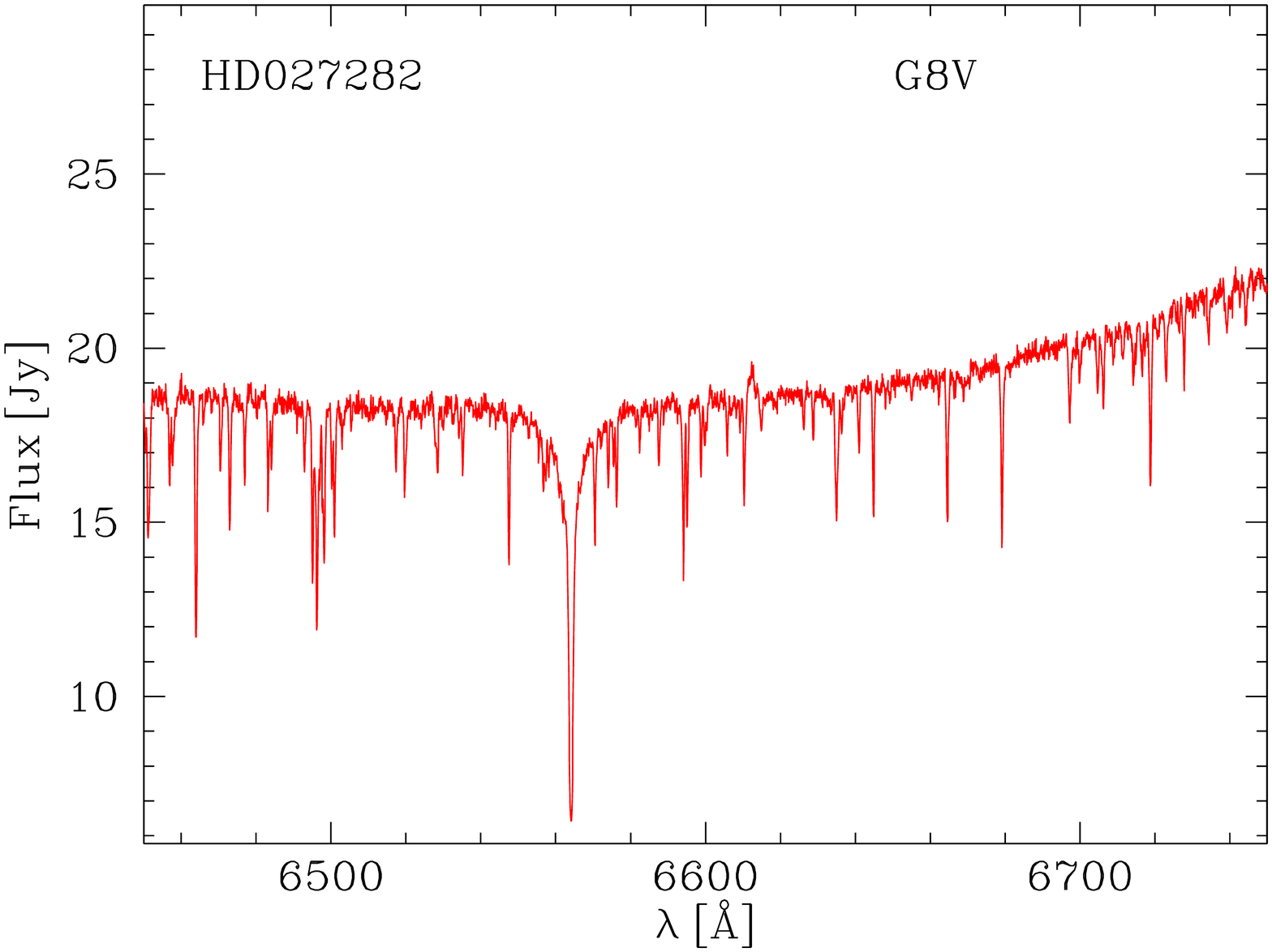}
\includegraphics[width=0.18\textwidth,angle=-90]{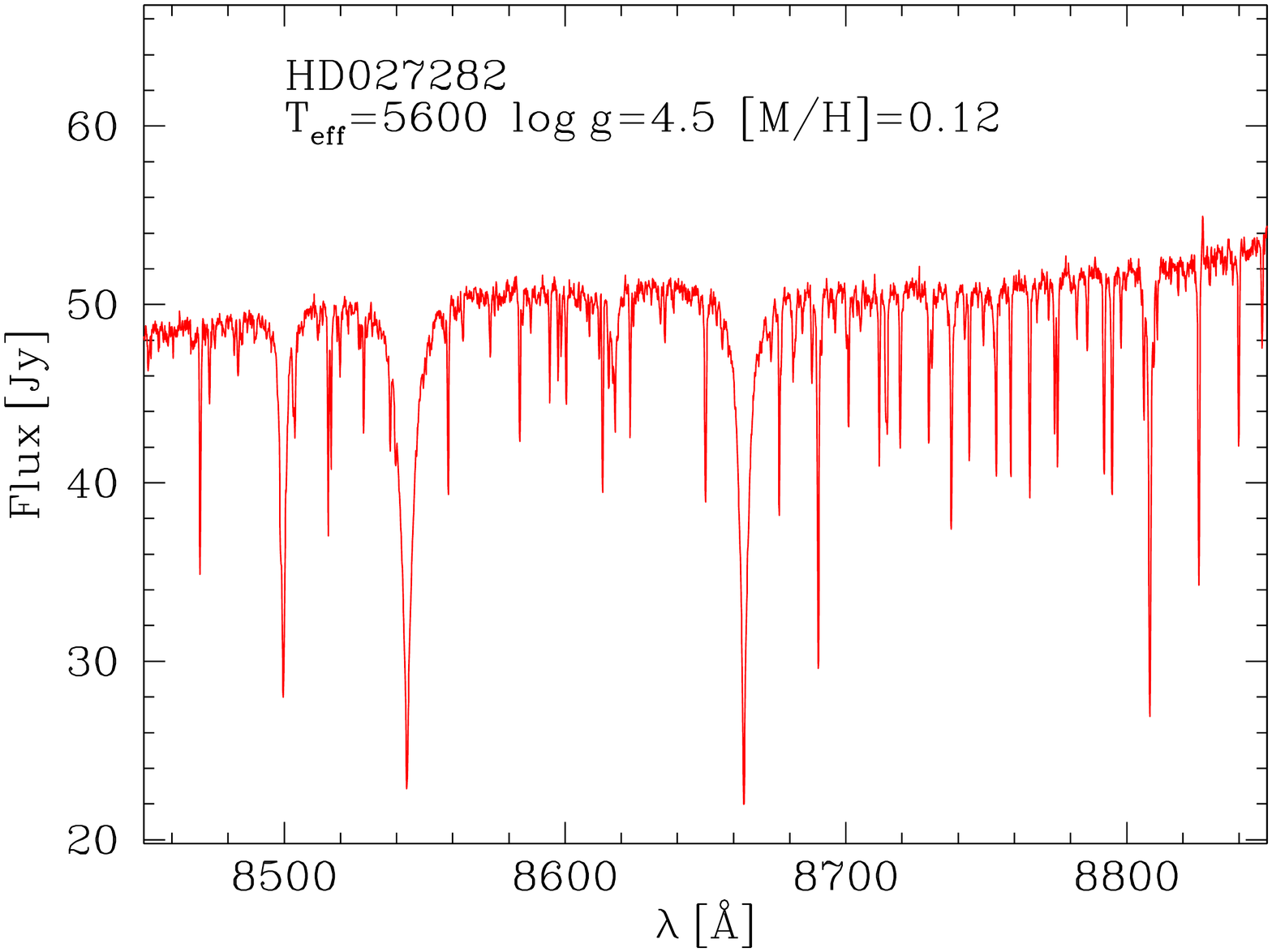}

\contcaption{5. Stars shown in this page are: HD020084, HD020512, HD021742, HD022484, HD023862, HD024341, HD024451, HD024534, HD024912, HD025173, HD025825, HD026756, HD027126 and HD027282.}
\end{figure*}

\begin{figure*}
\includegraphics[width=0.18\textwidth,angle=-90]{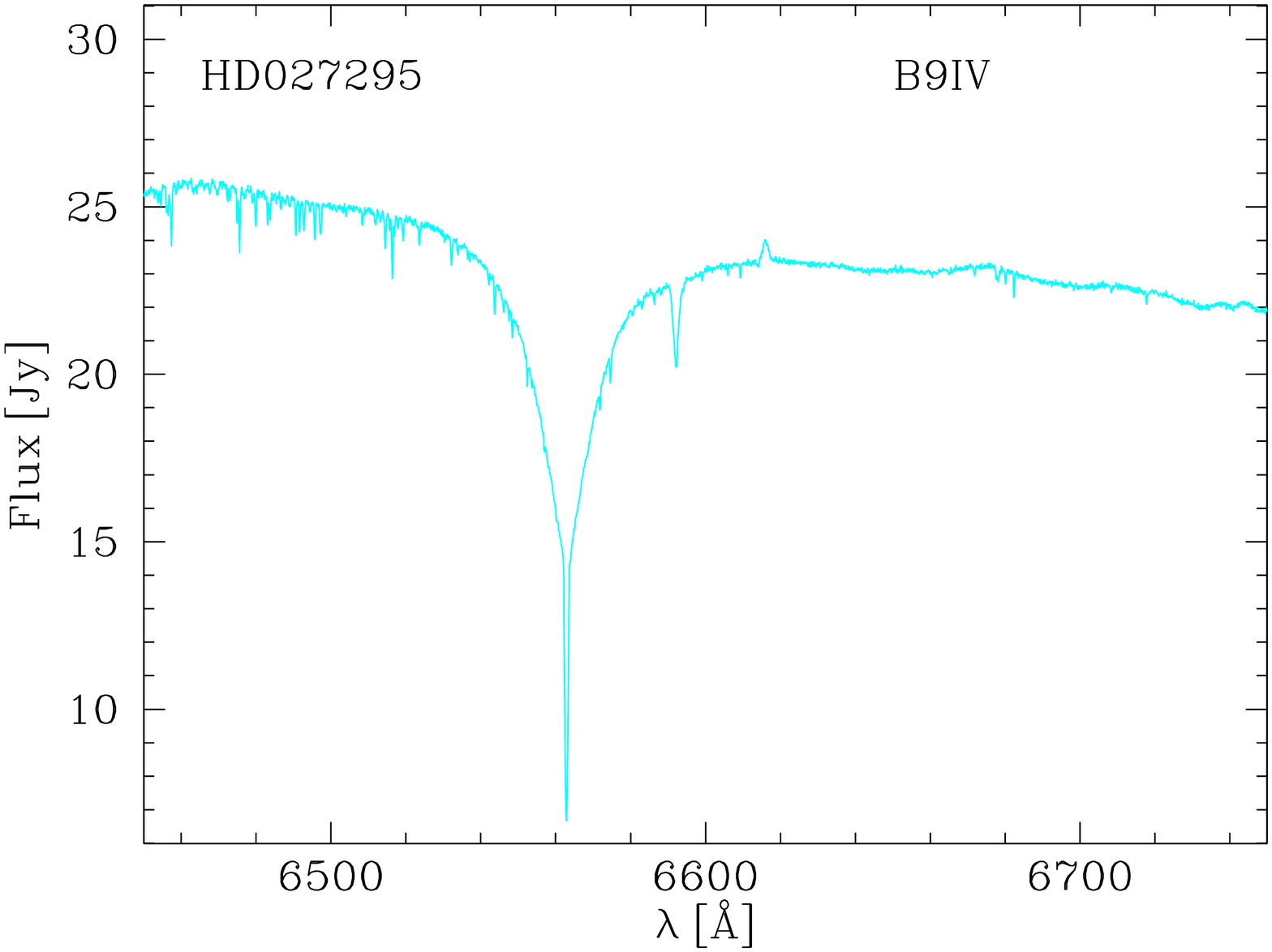}
\includegraphics[width=0.18\textwidth,angle=-90]{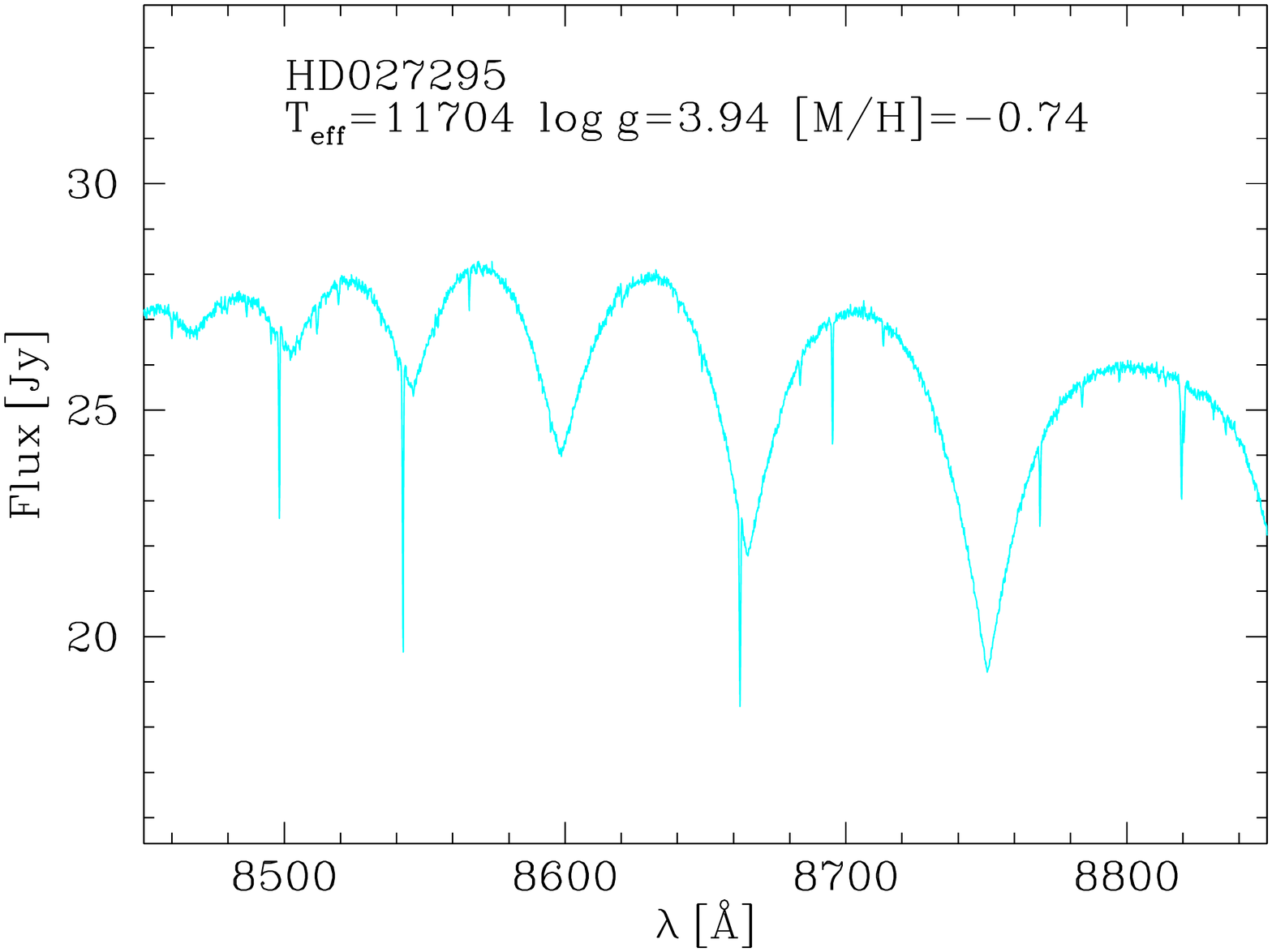}
\includegraphics[width=0.18\textwidth,angle=-90]{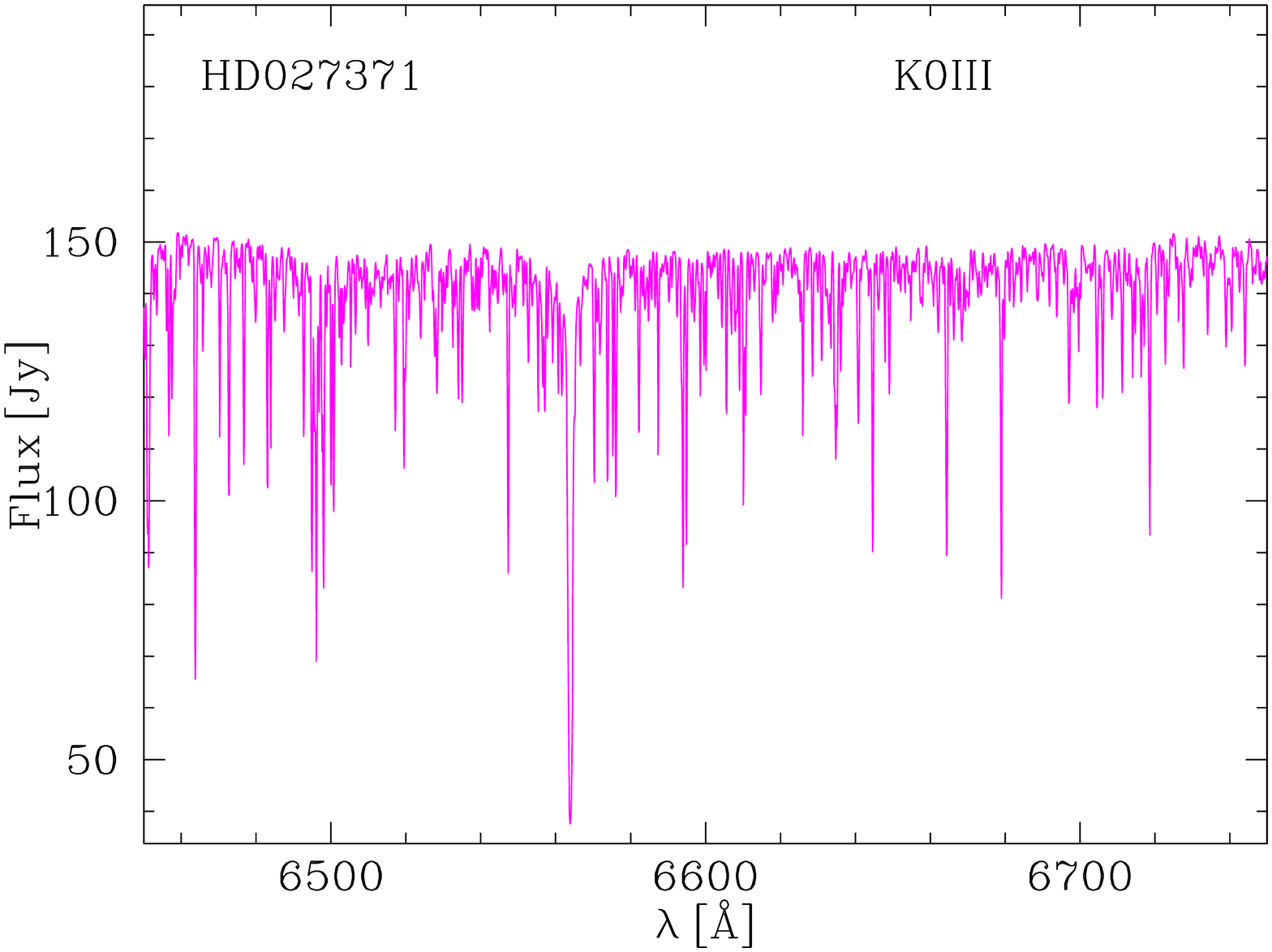}
\includegraphics[width=0.18\textwidth,angle=-90]{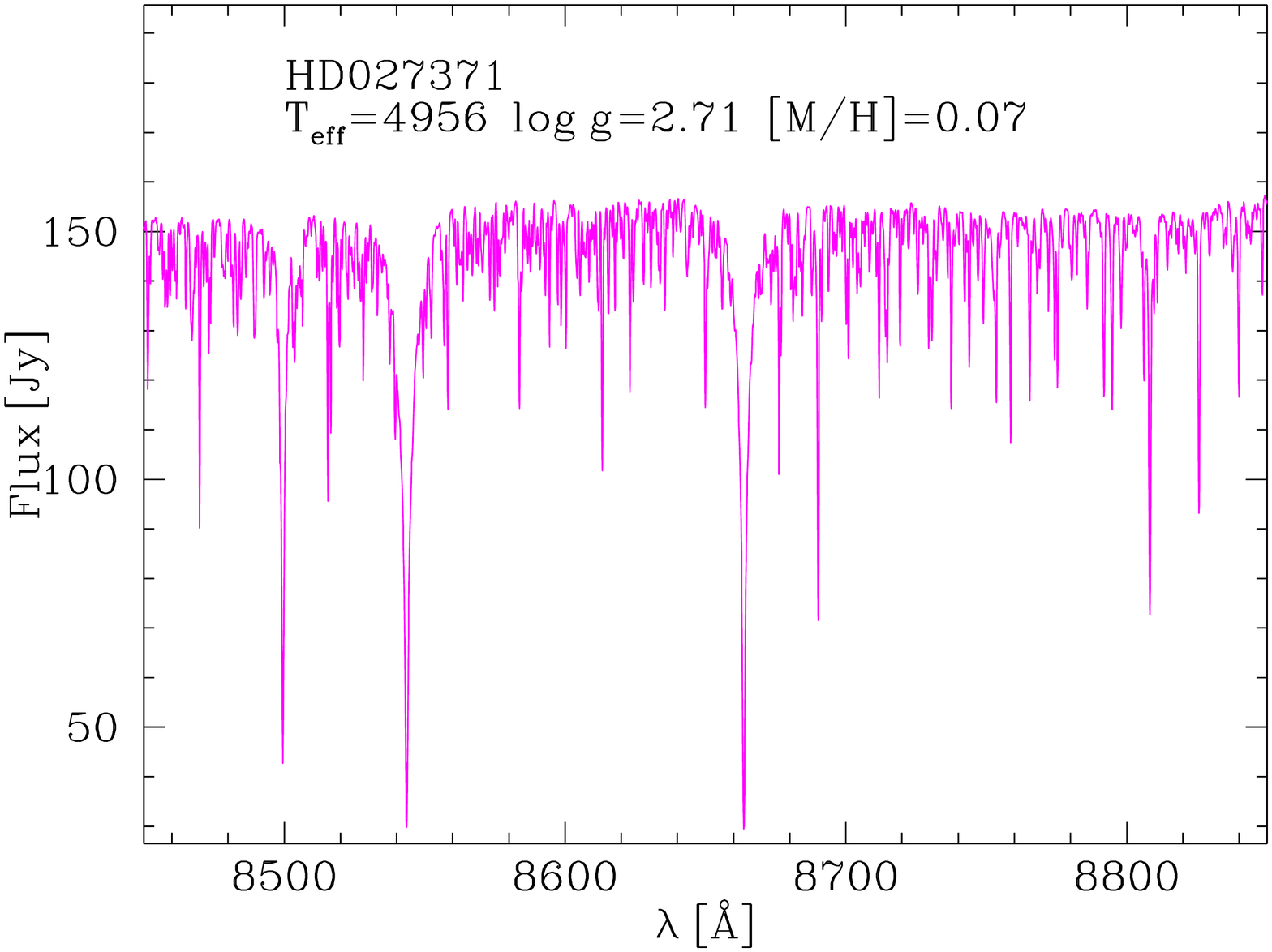}
\includegraphics[width=0.18\textwidth,angle=-90]{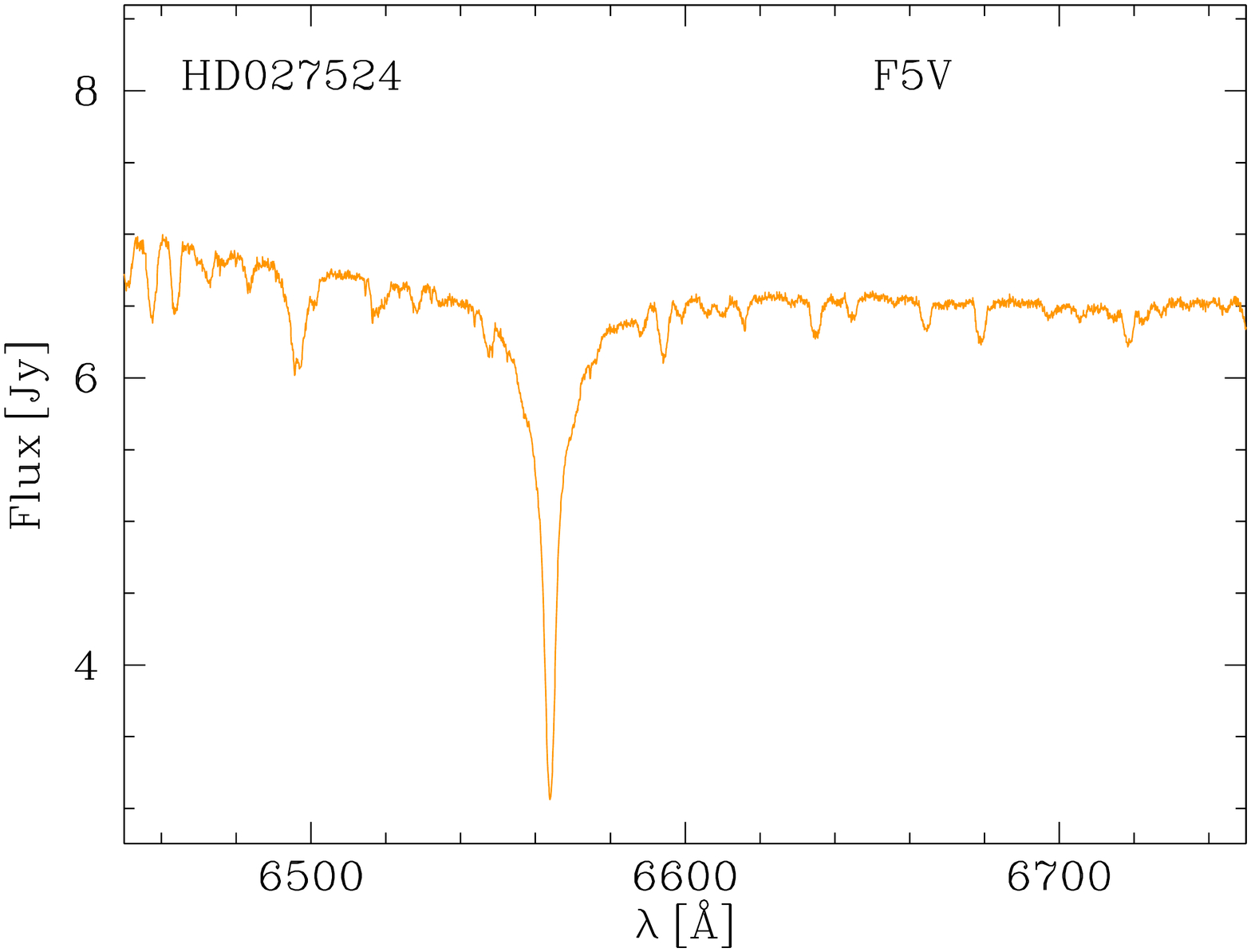}
\includegraphics[width=0.18\textwidth,angle=-90]{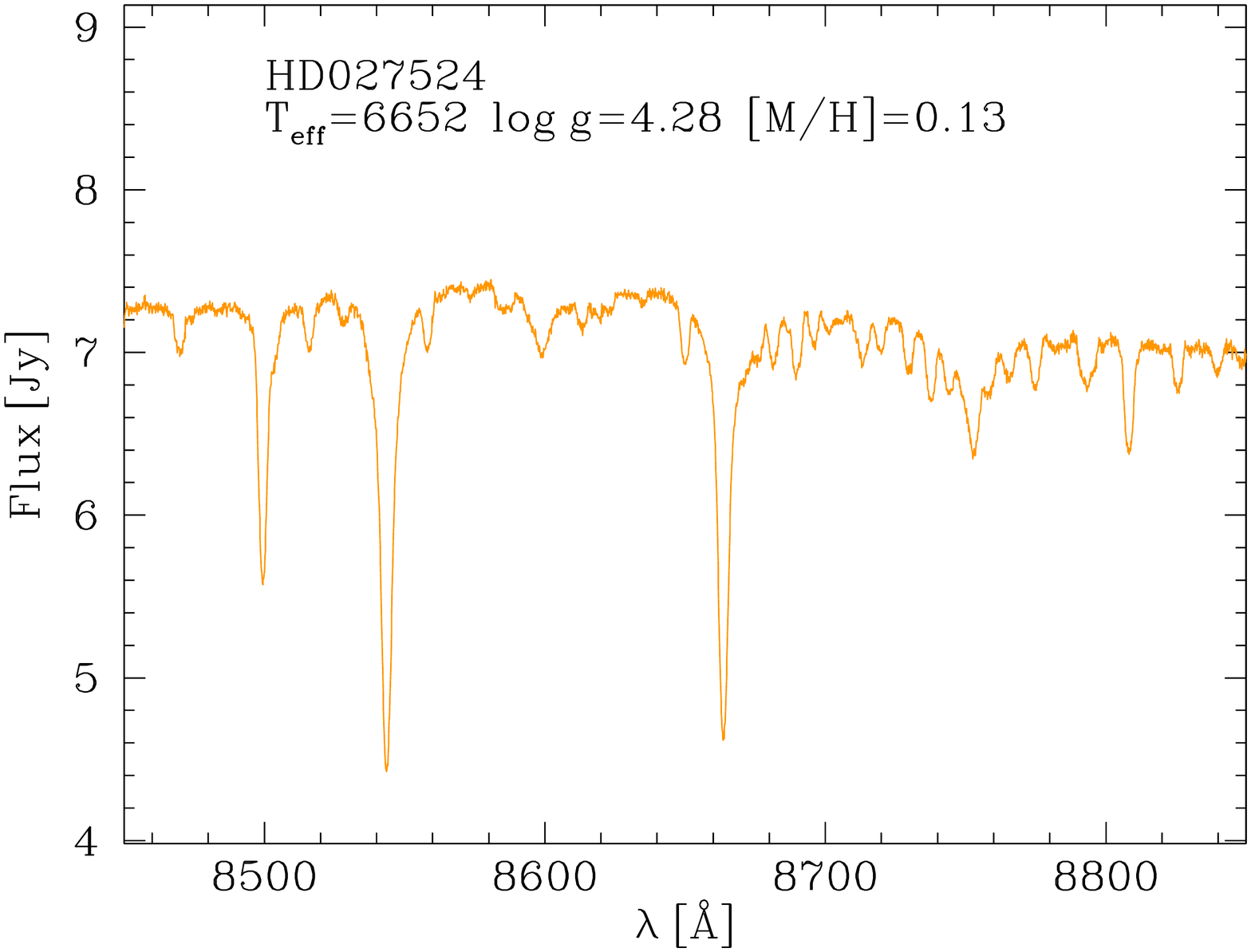}
\includegraphics[width=0.18\textwidth,angle=-90]{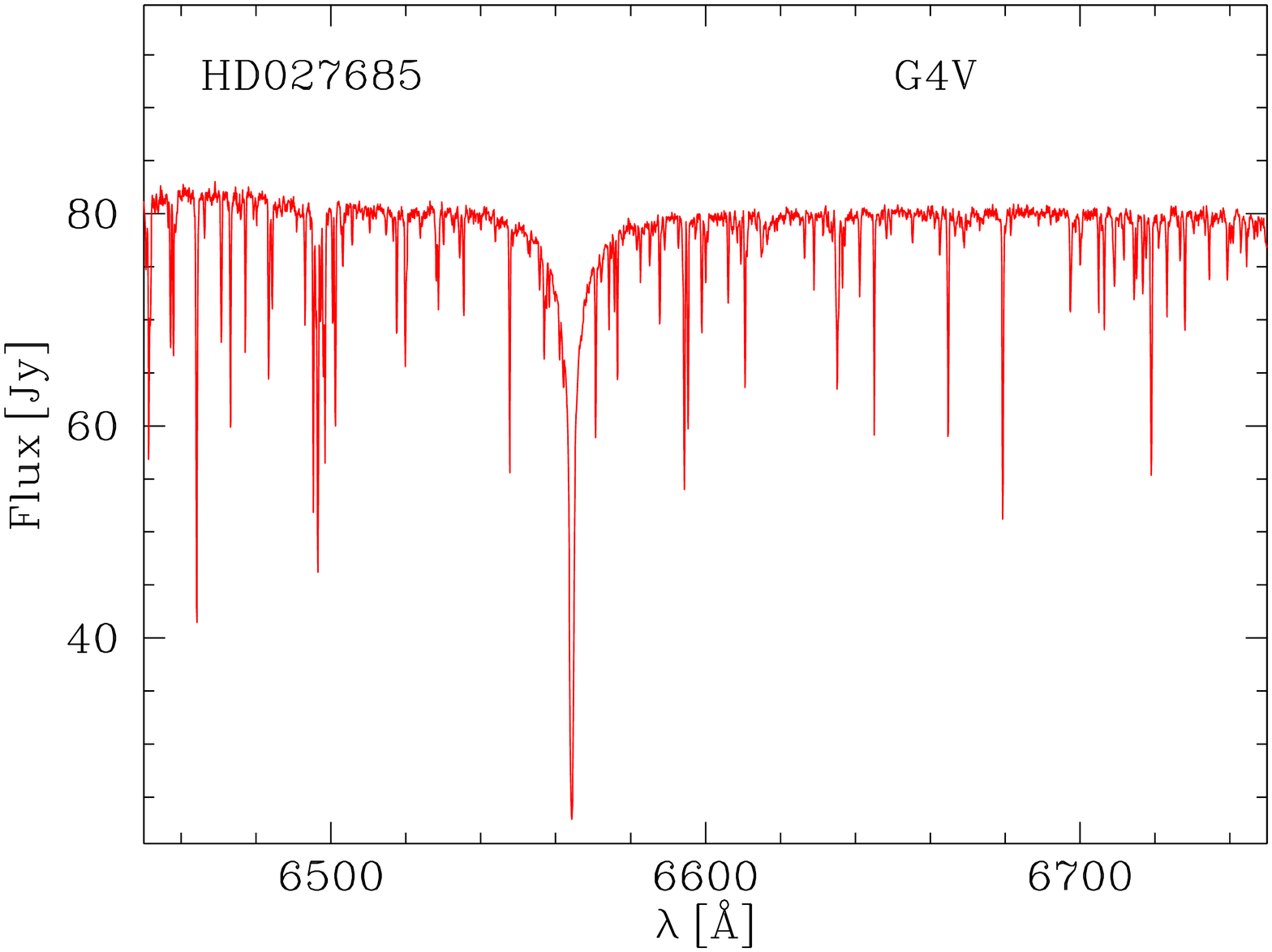}
\includegraphics[width=0.18\textwidth,angle=-90]{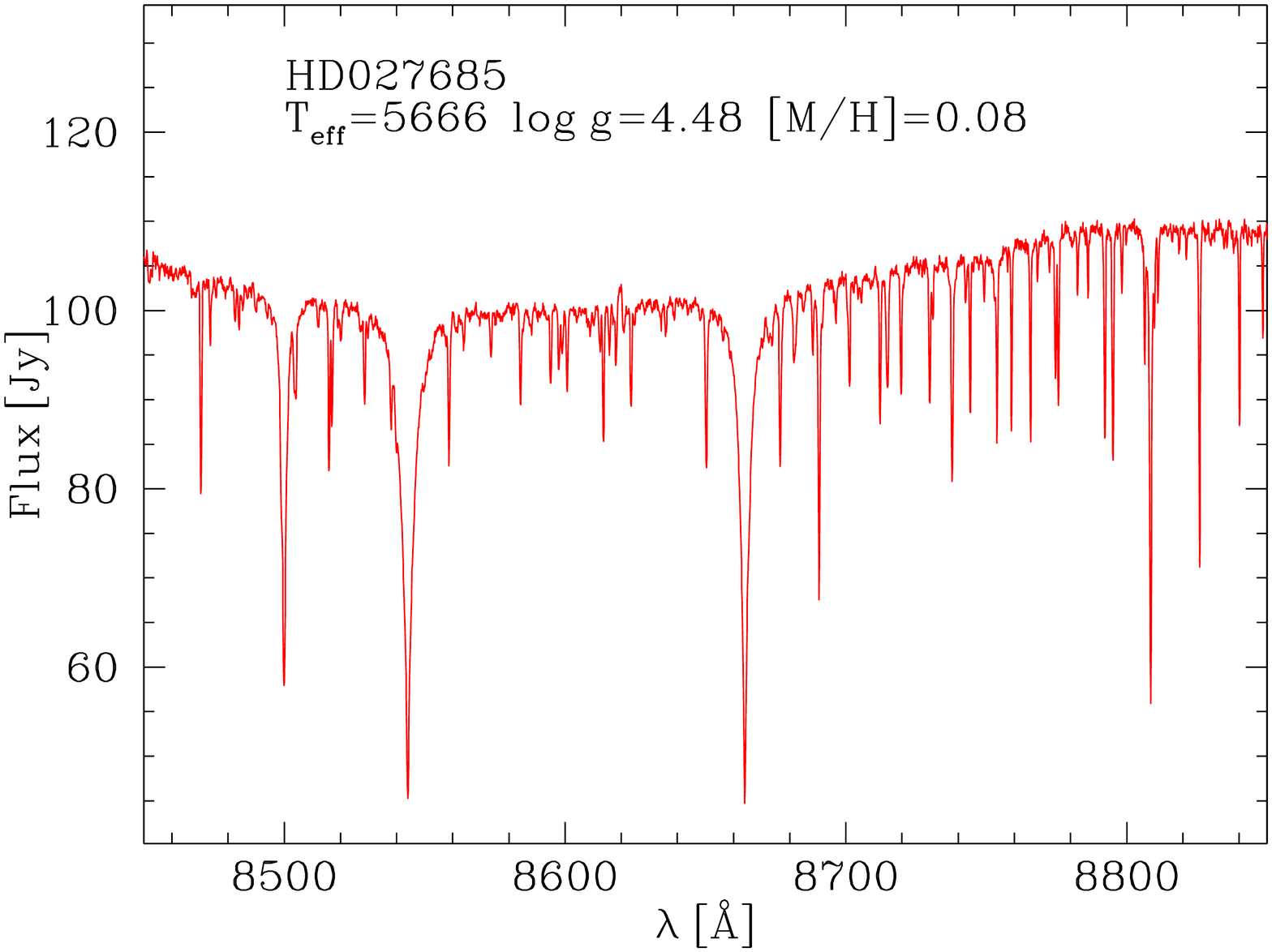}
\includegraphics[width=0.18\textwidth,angle=-90]{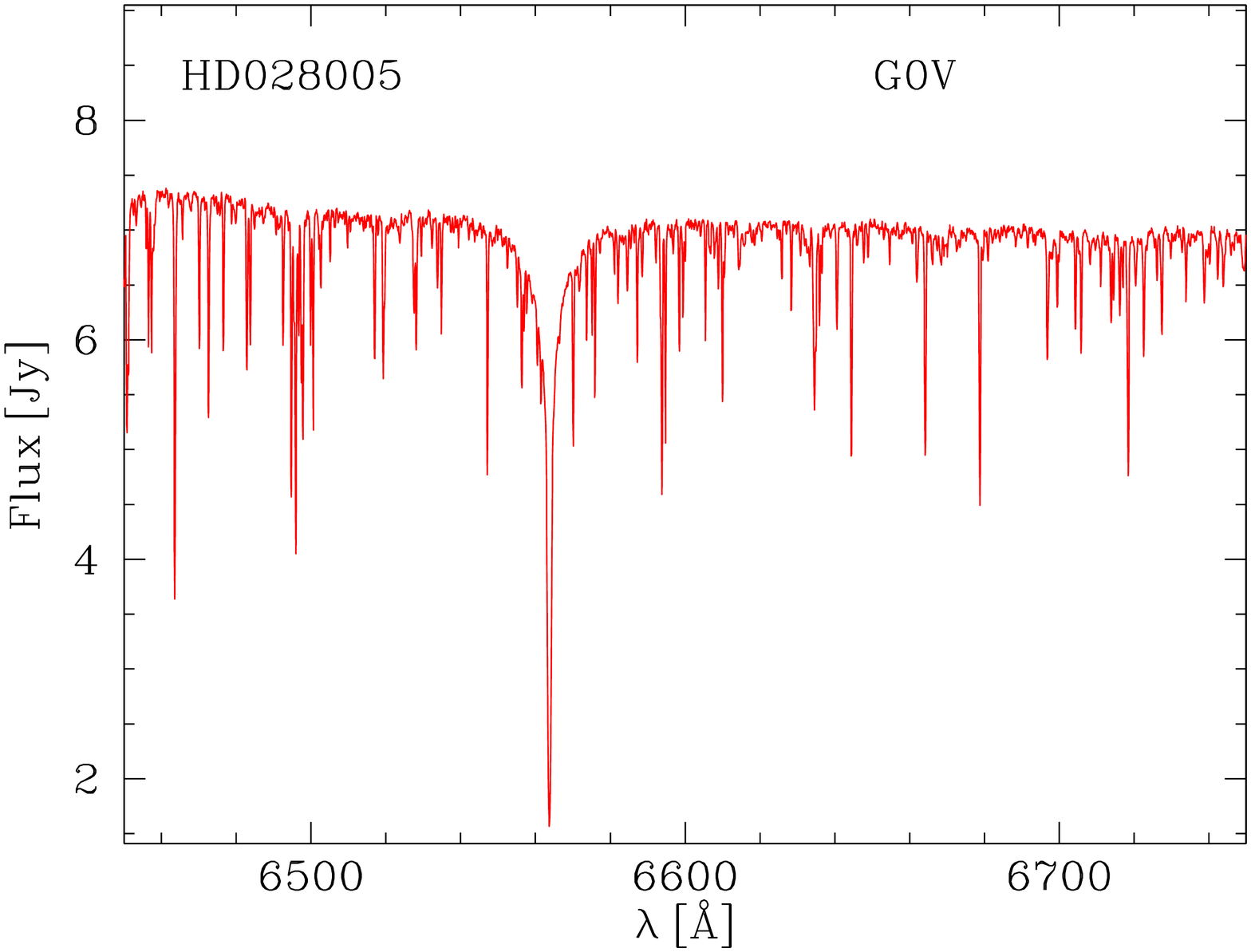}
\includegraphics[width=0.18\textwidth,angle=-90]{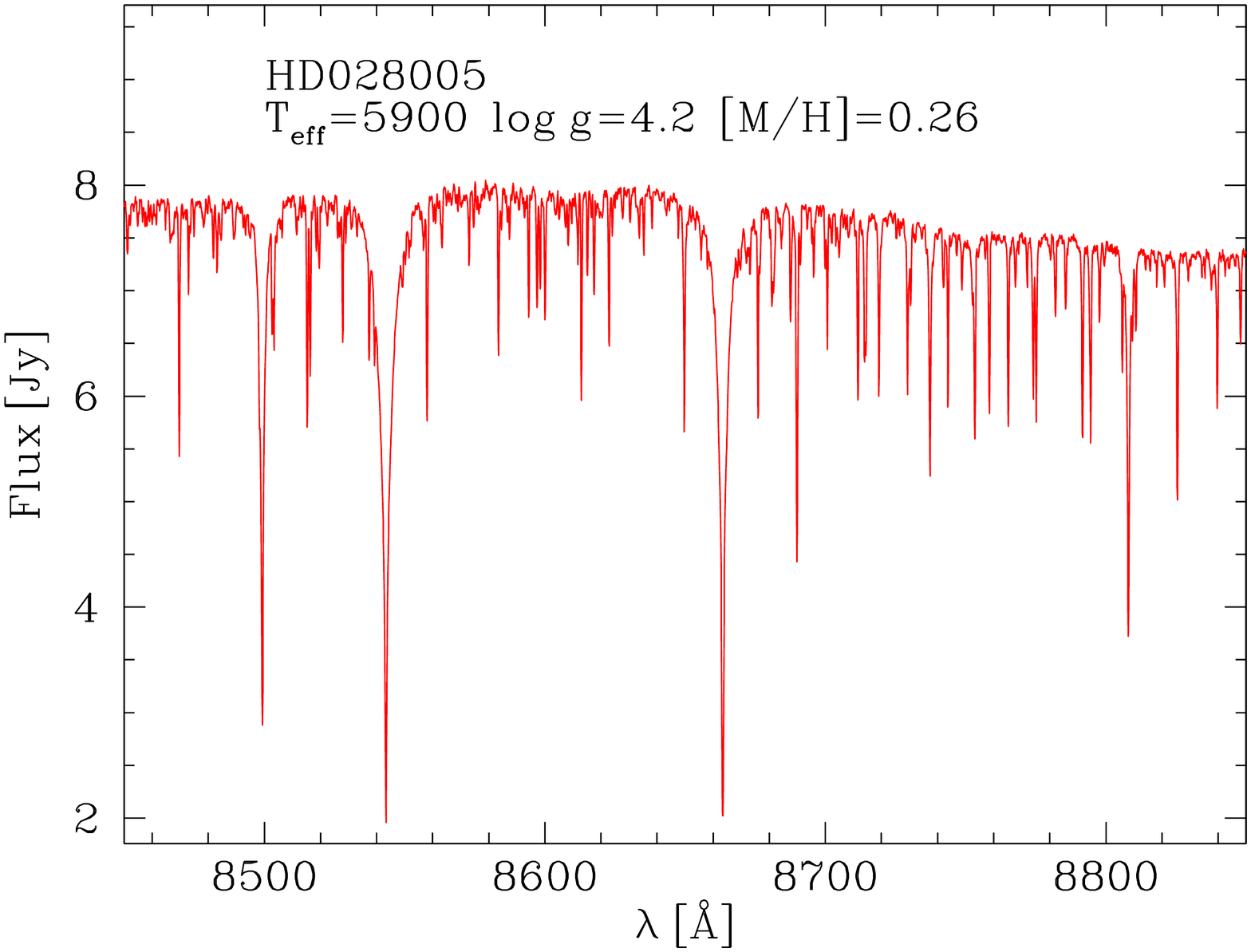}
\includegraphics[width=0.18\textwidth,angle=-90]{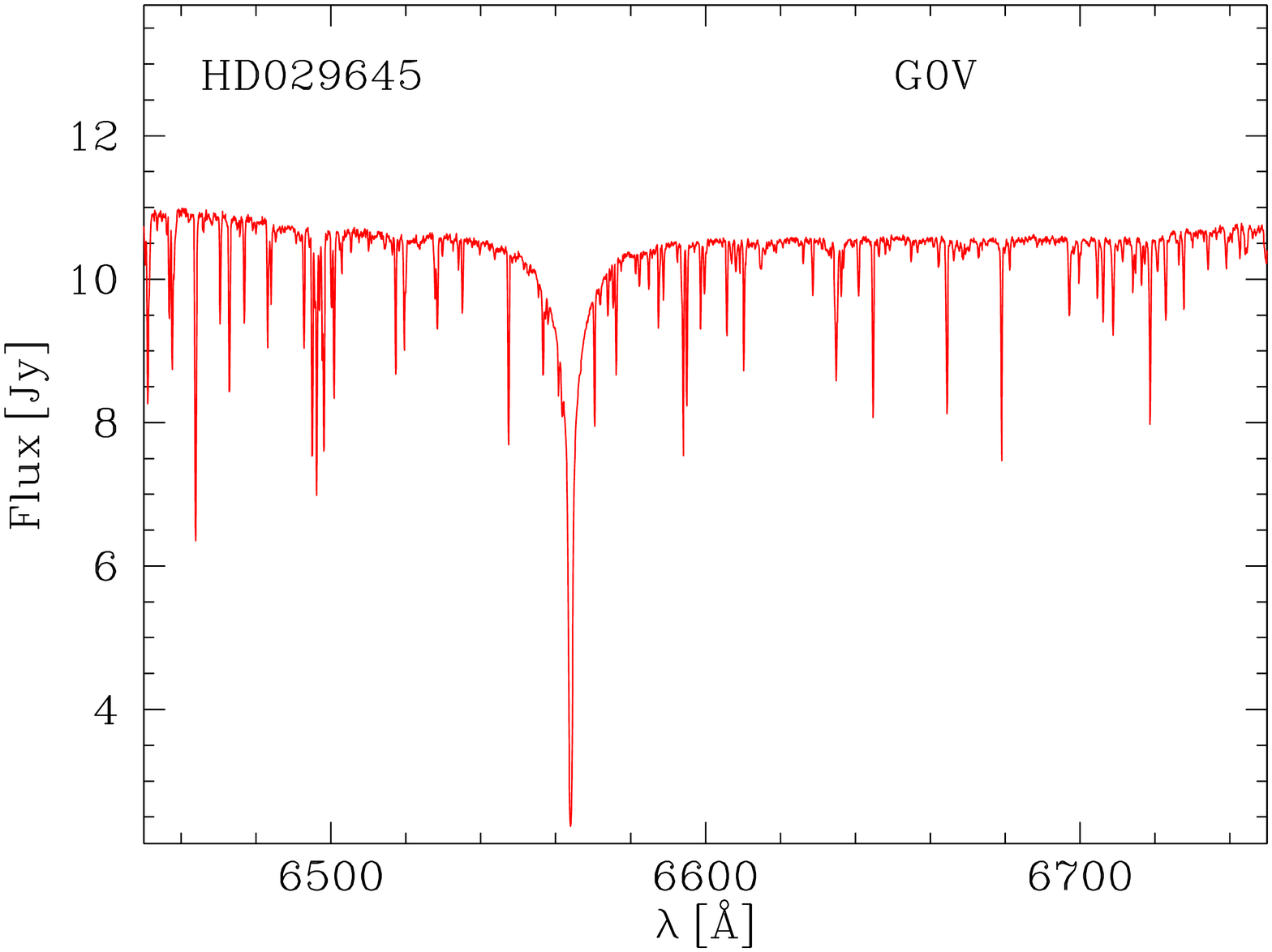}
\includegraphics[width=0.18\textwidth,angle=-90]{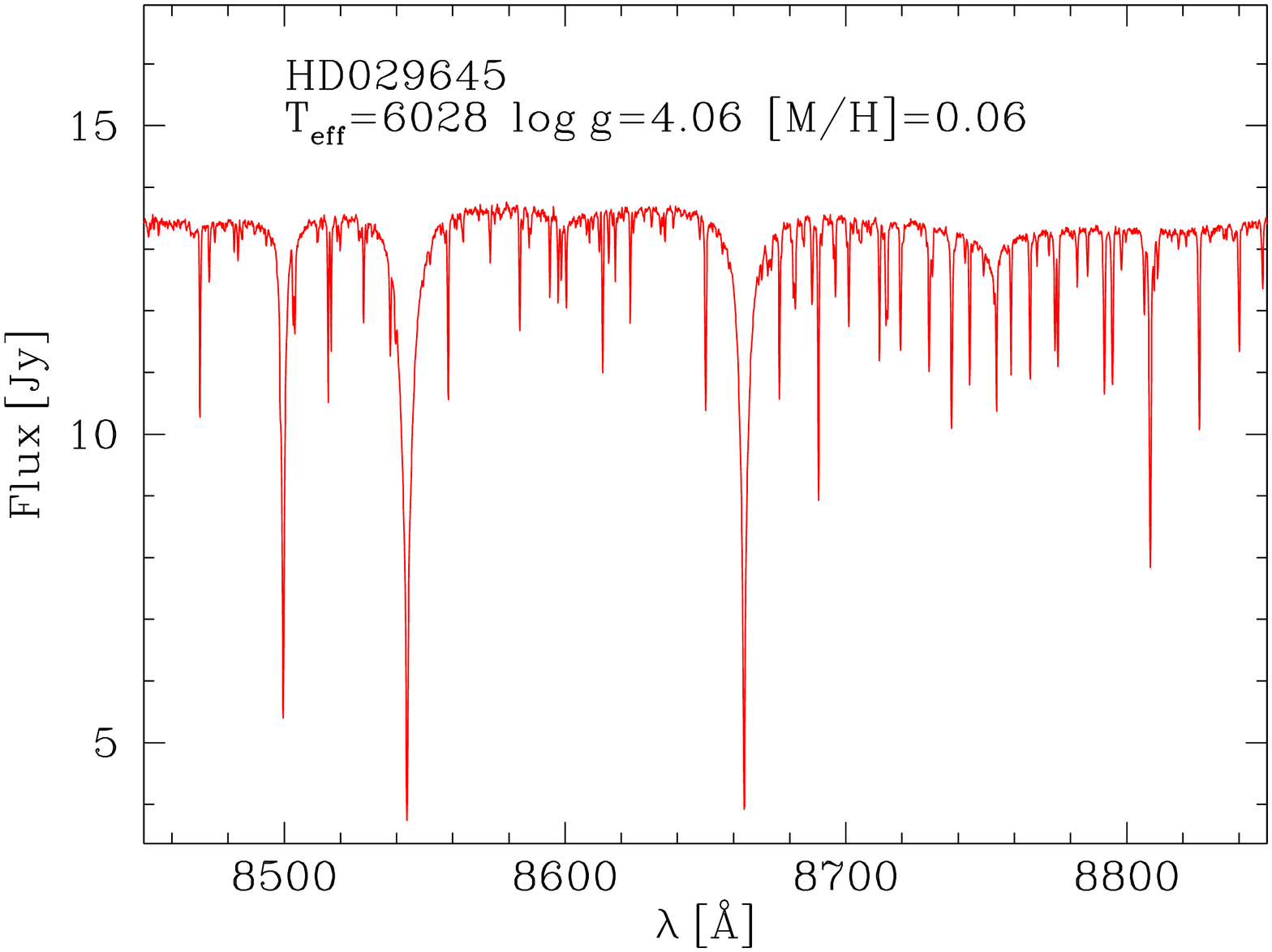}
\includegraphics[width=0.18\textwidth,angle=-90]{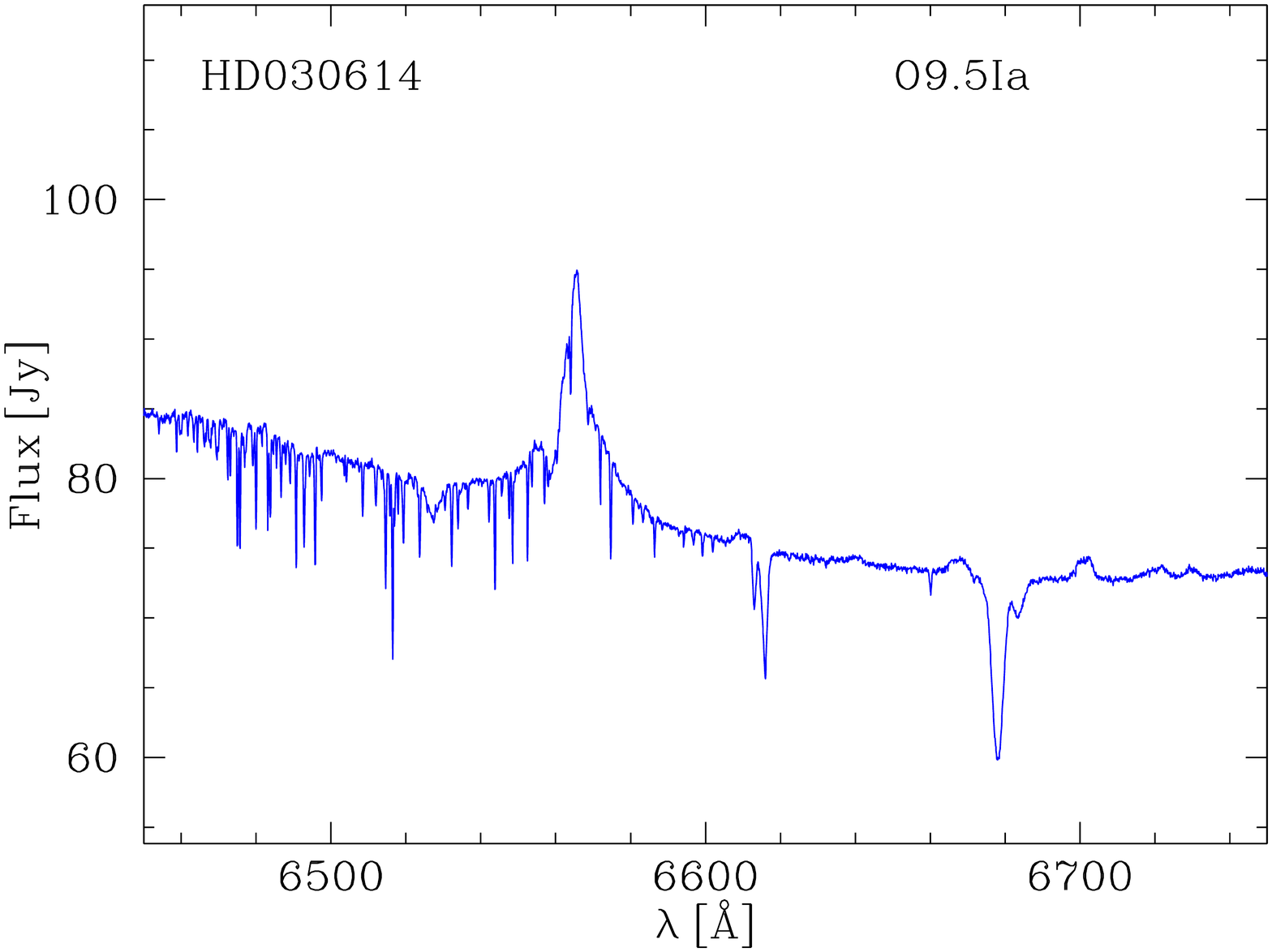}
\includegraphics[width=0.18\textwidth,angle=-90]{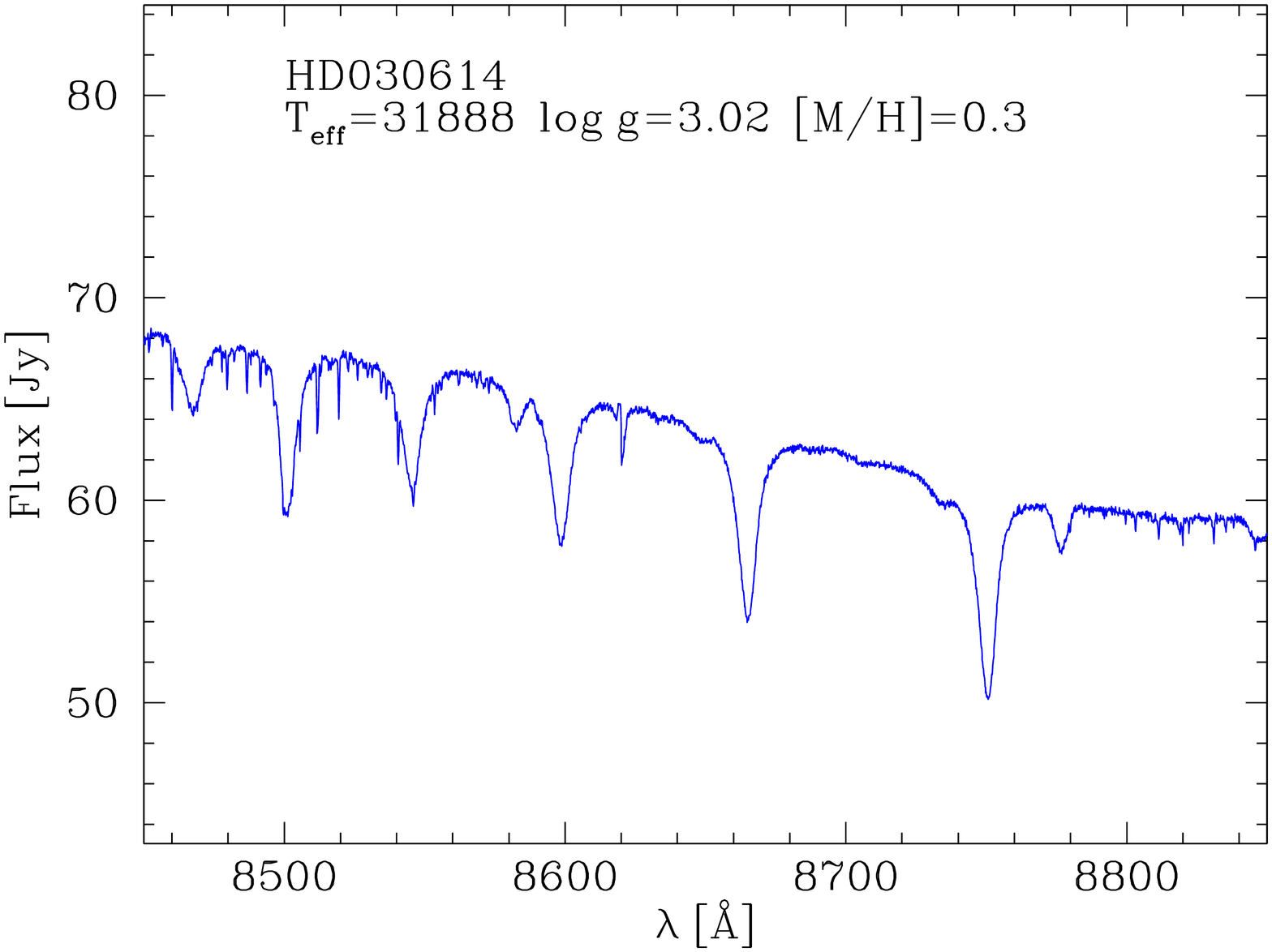}
\includegraphics[width=0.18\textwidth,angle=-90]{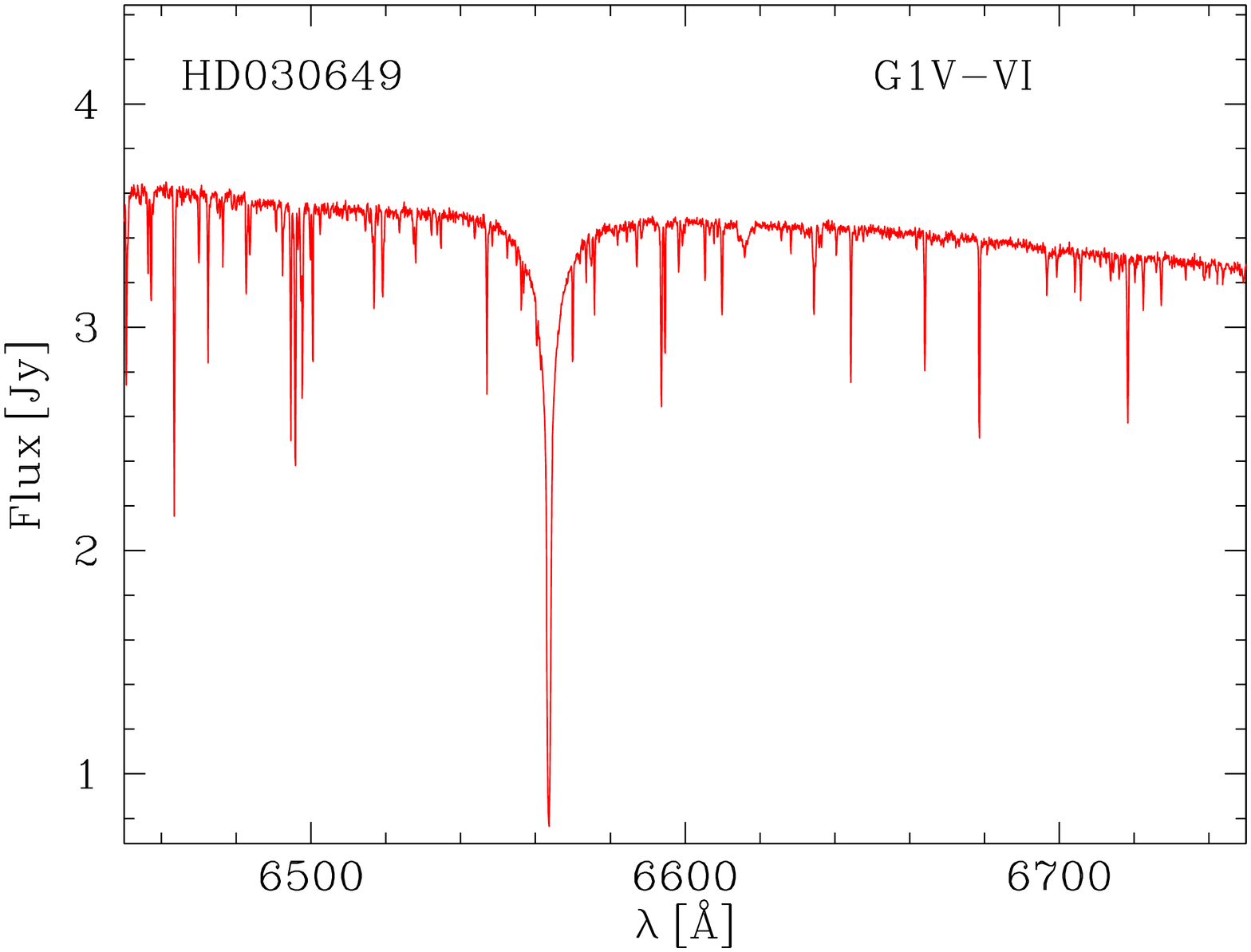}
\includegraphics[width=0.18\textwidth,angle=-90]{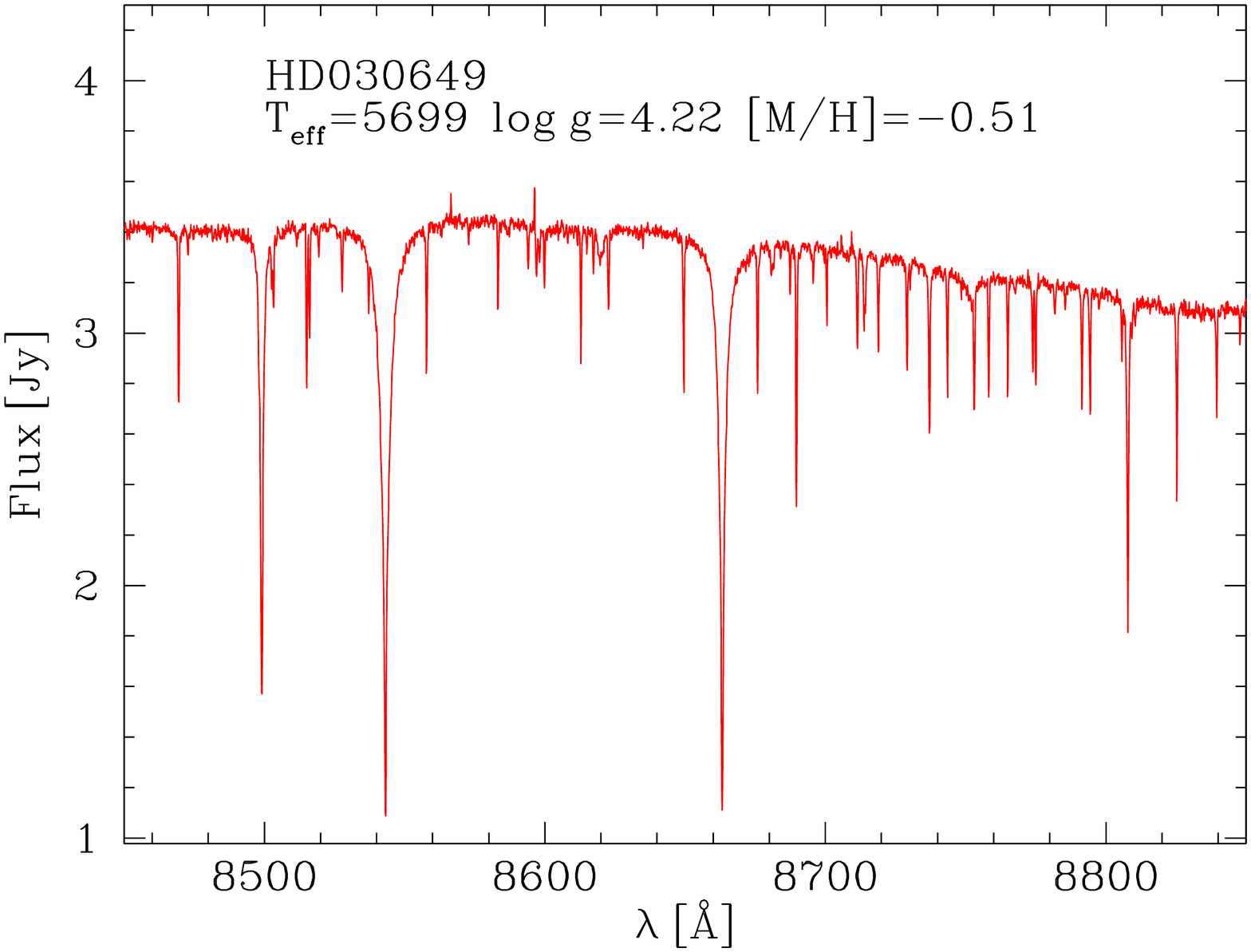}
\includegraphics[width=0.18\textwidth,angle=-90]{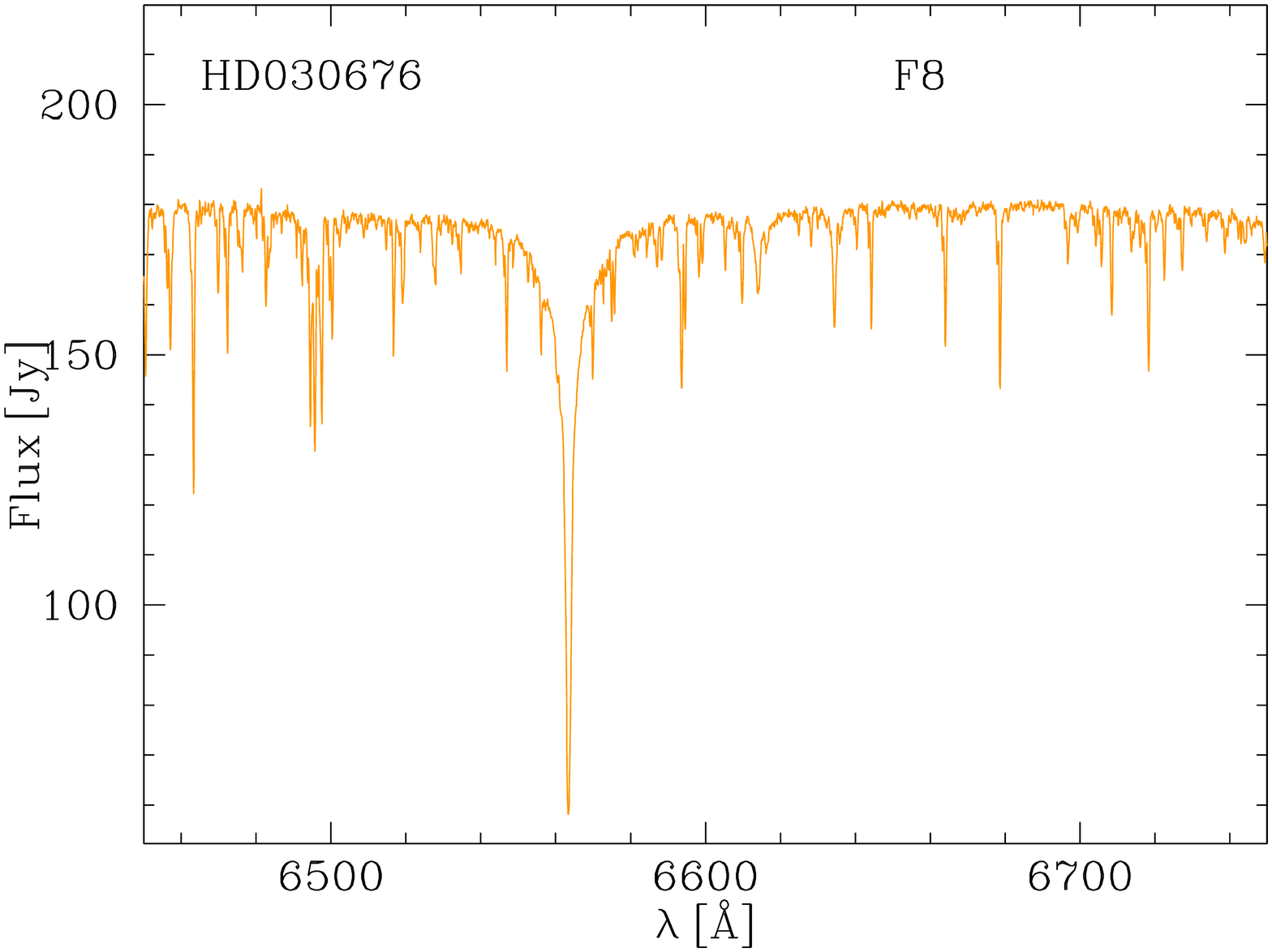}
\includegraphics[width=0.18\textwidth,angle=-90]{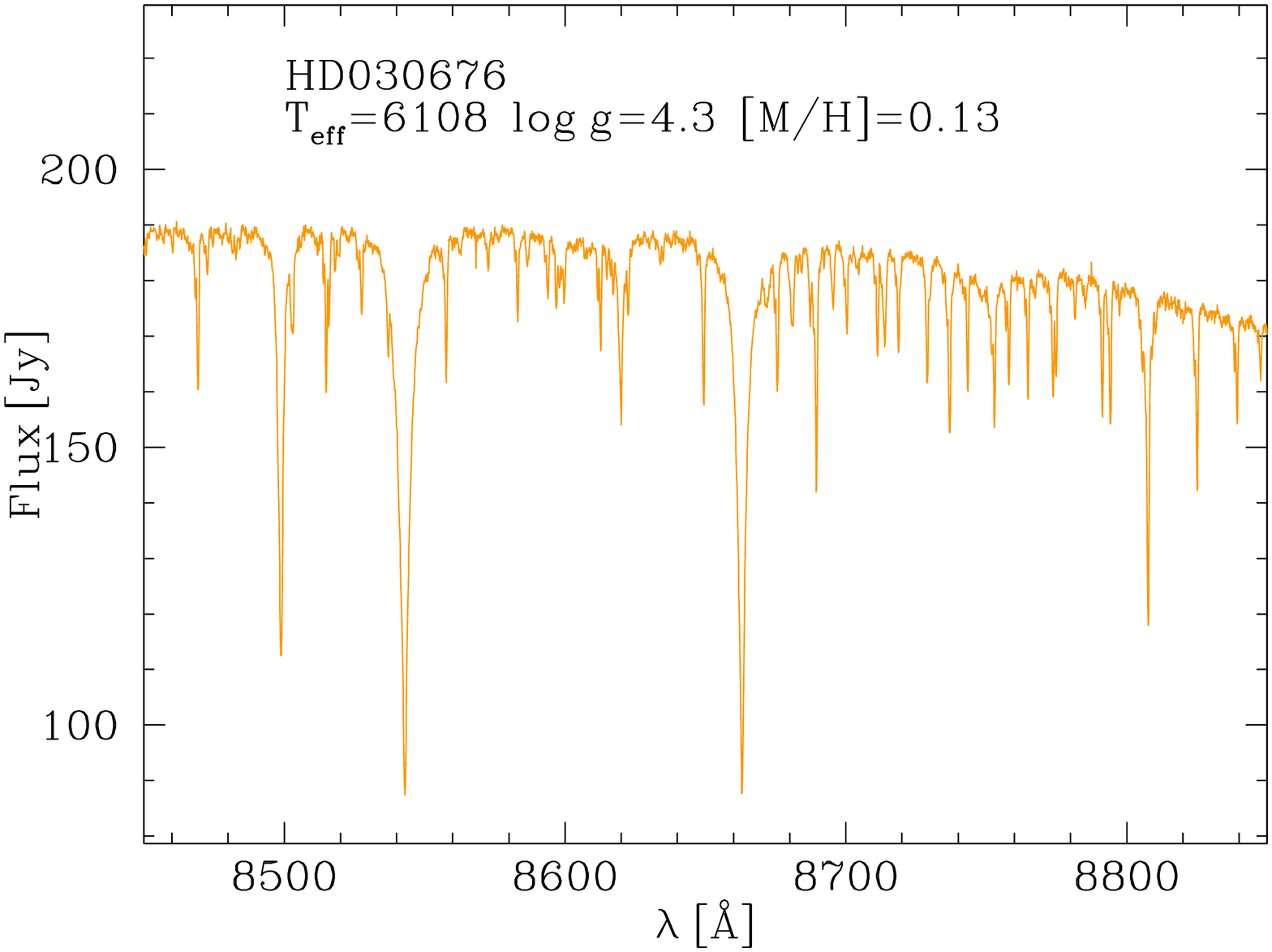}
\includegraphics[width=0.18\textwidth,angle=-90]{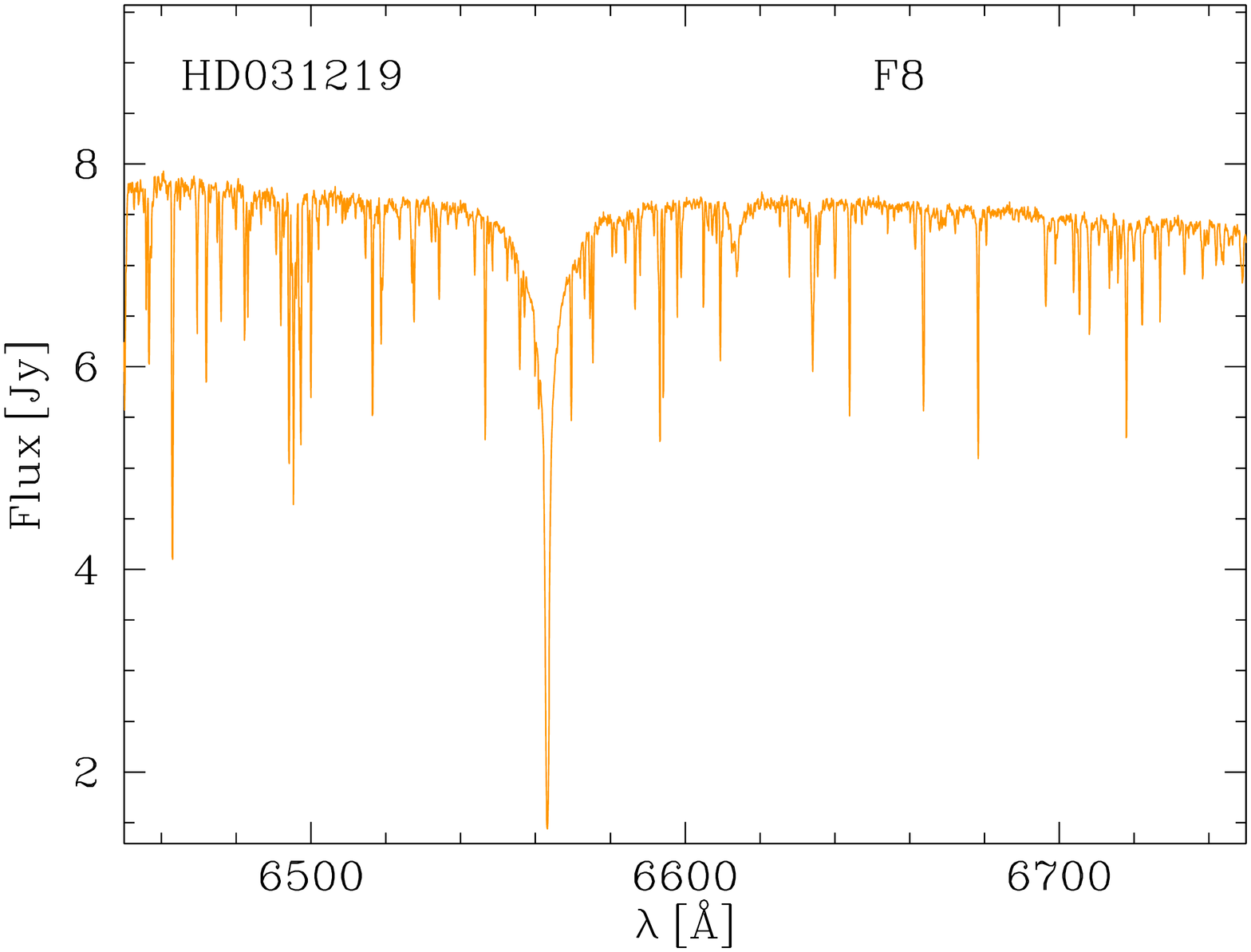}
\includegraphics[width=0.18\textwidth,angle=-90]{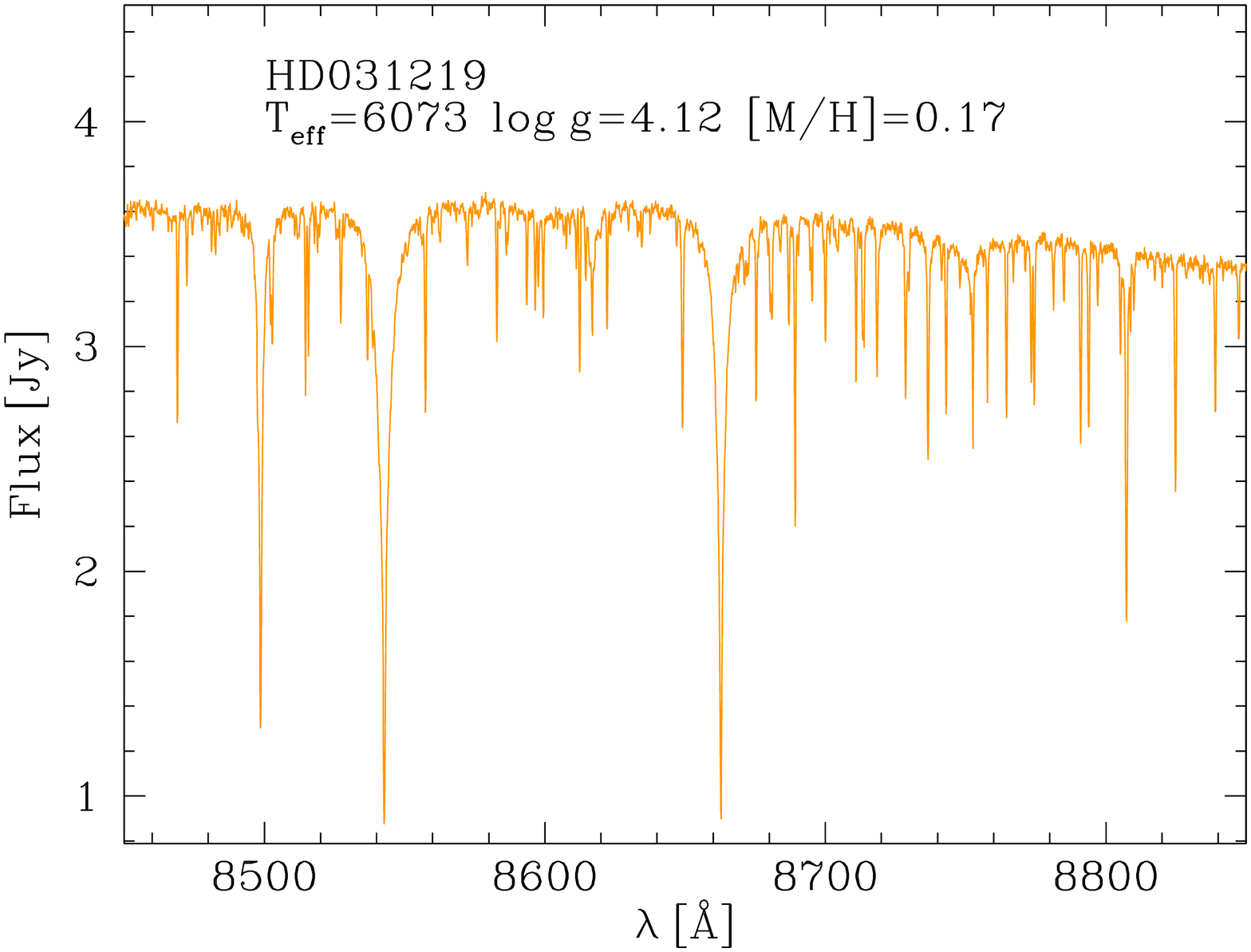}
\includegraphics[width=0.18\textwidth,angle=-90]{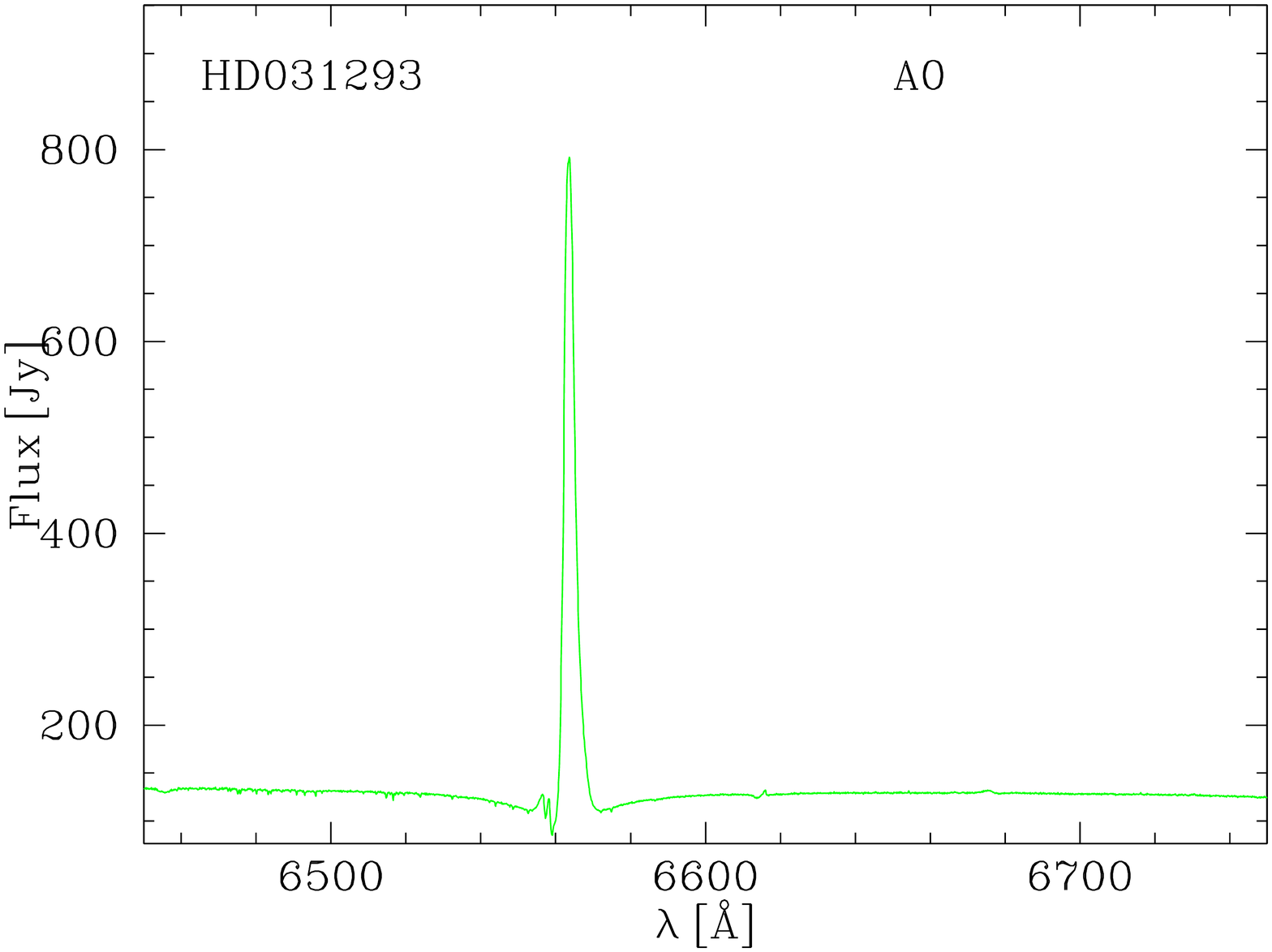}
\includegraphics[width=0.18\textwidth,angle=-90]{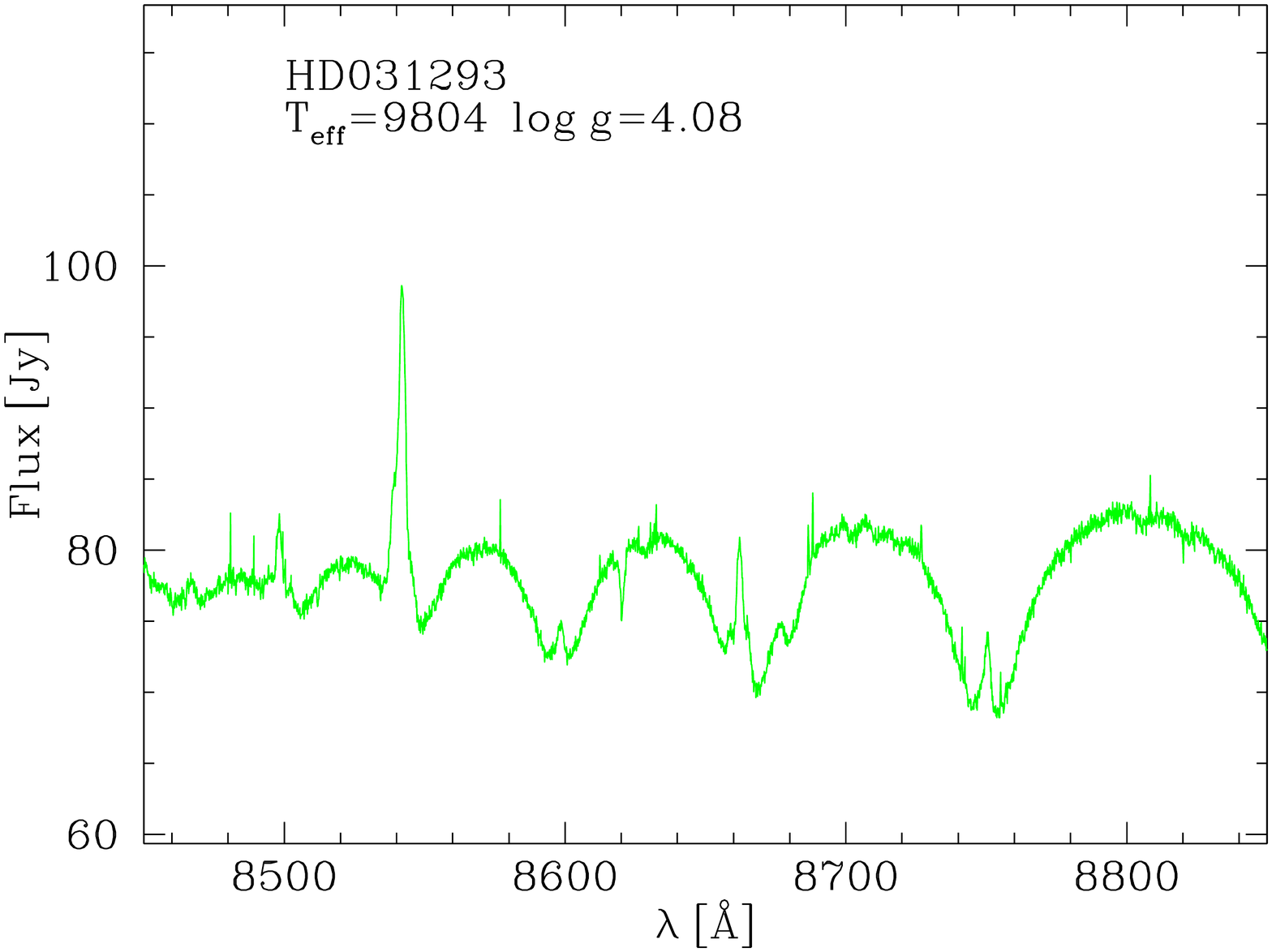}
\includegraphics[width=0.18\textwidth,angle=-90]{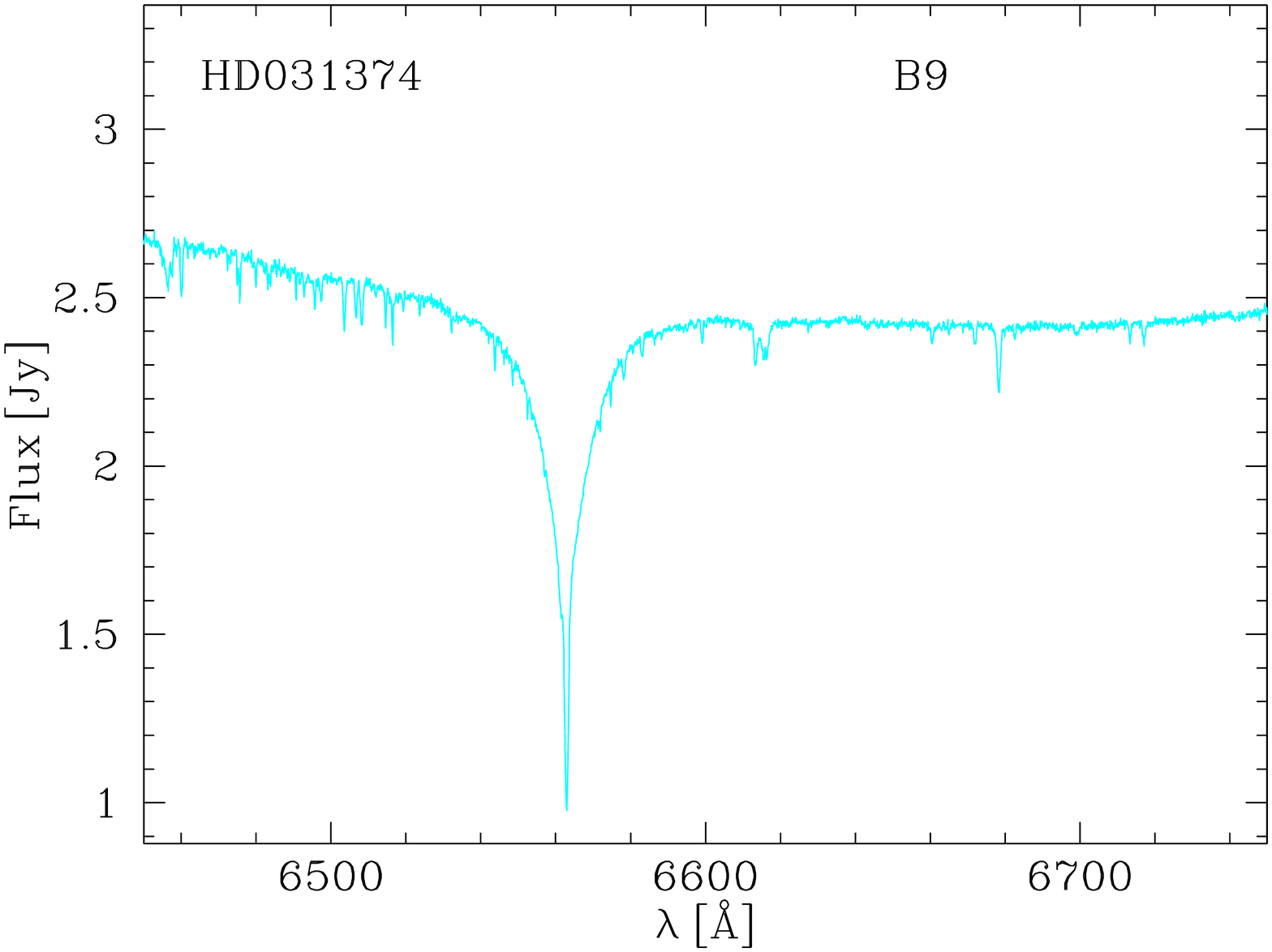}
\includegraphics[width=0.18\textwidth,angle=-90]{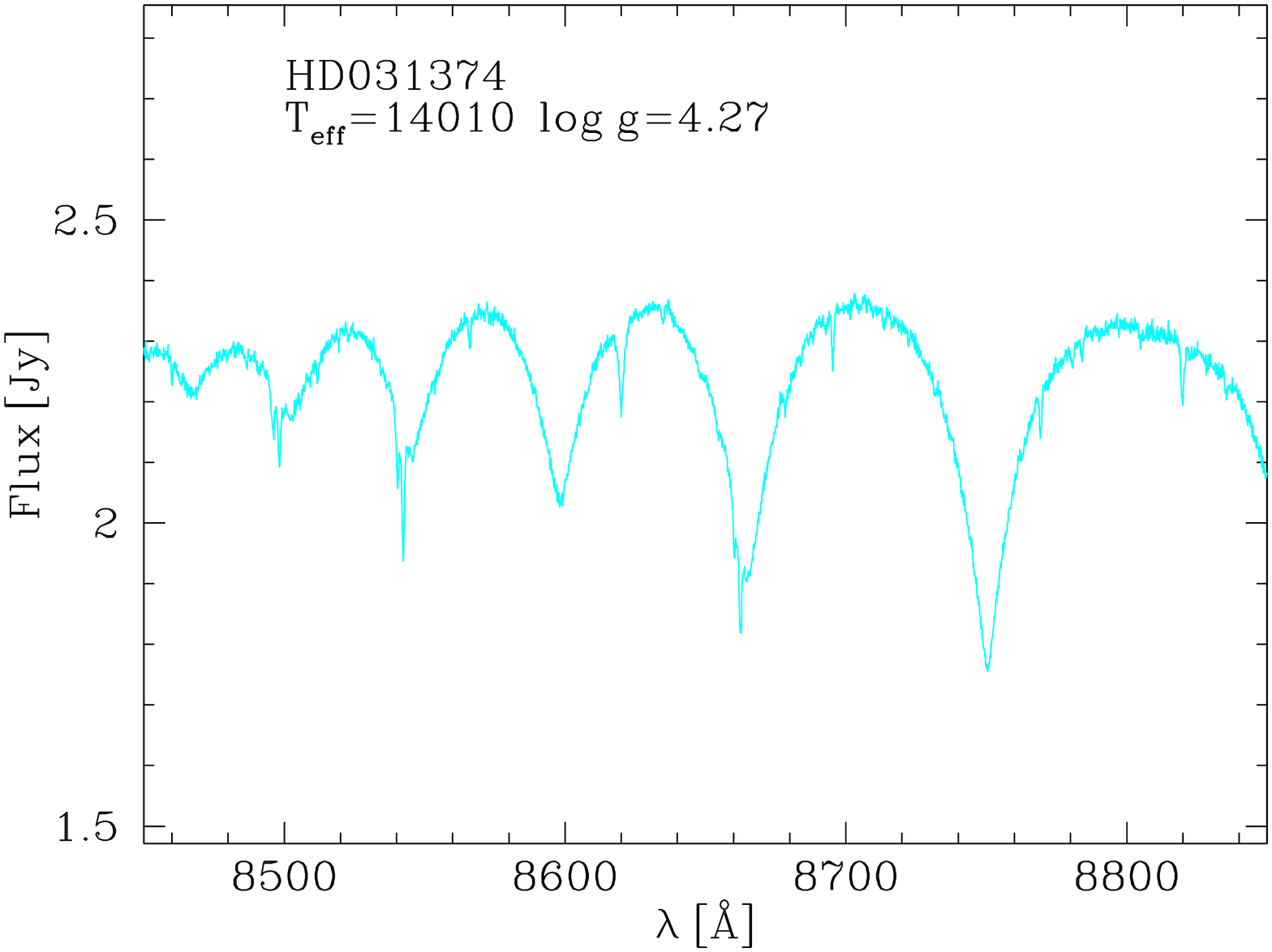}
\includegraphics[width=0.18\textwidth,angle=-90]{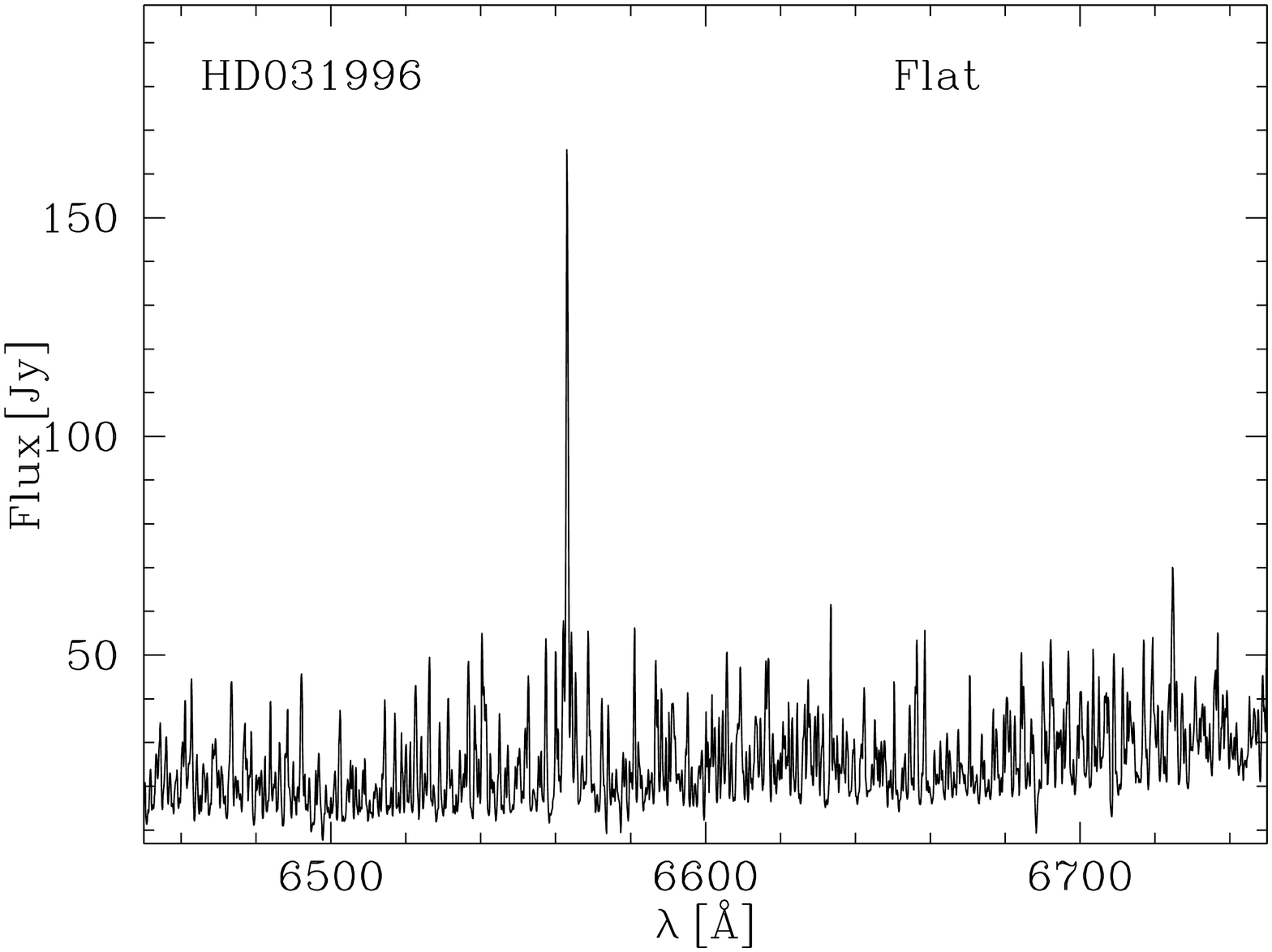}
\includegraphics[width=0.18\textwidth,angle=-90]{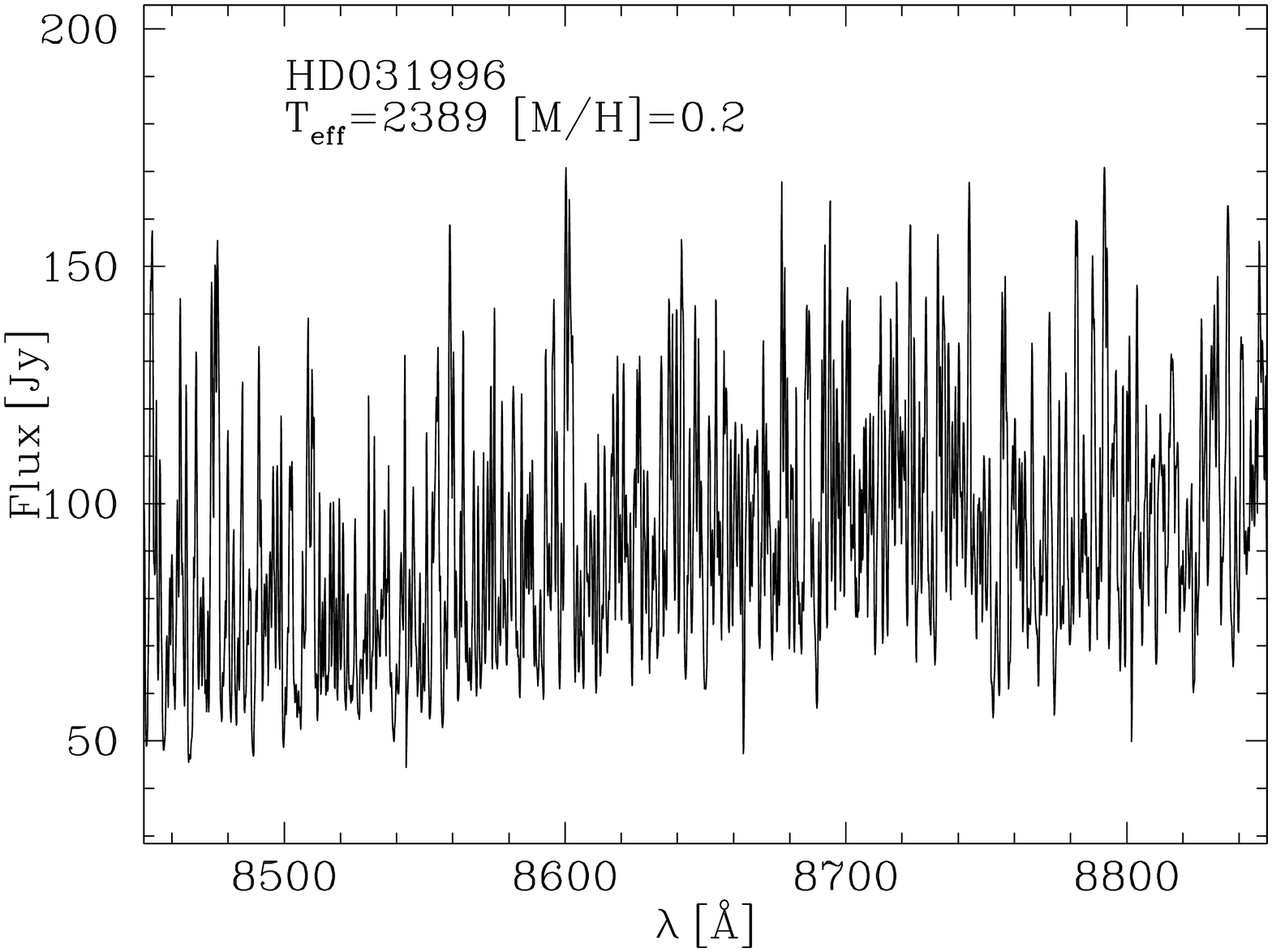}
\includegraphics[width=0.18\textwidth,angle=-90]{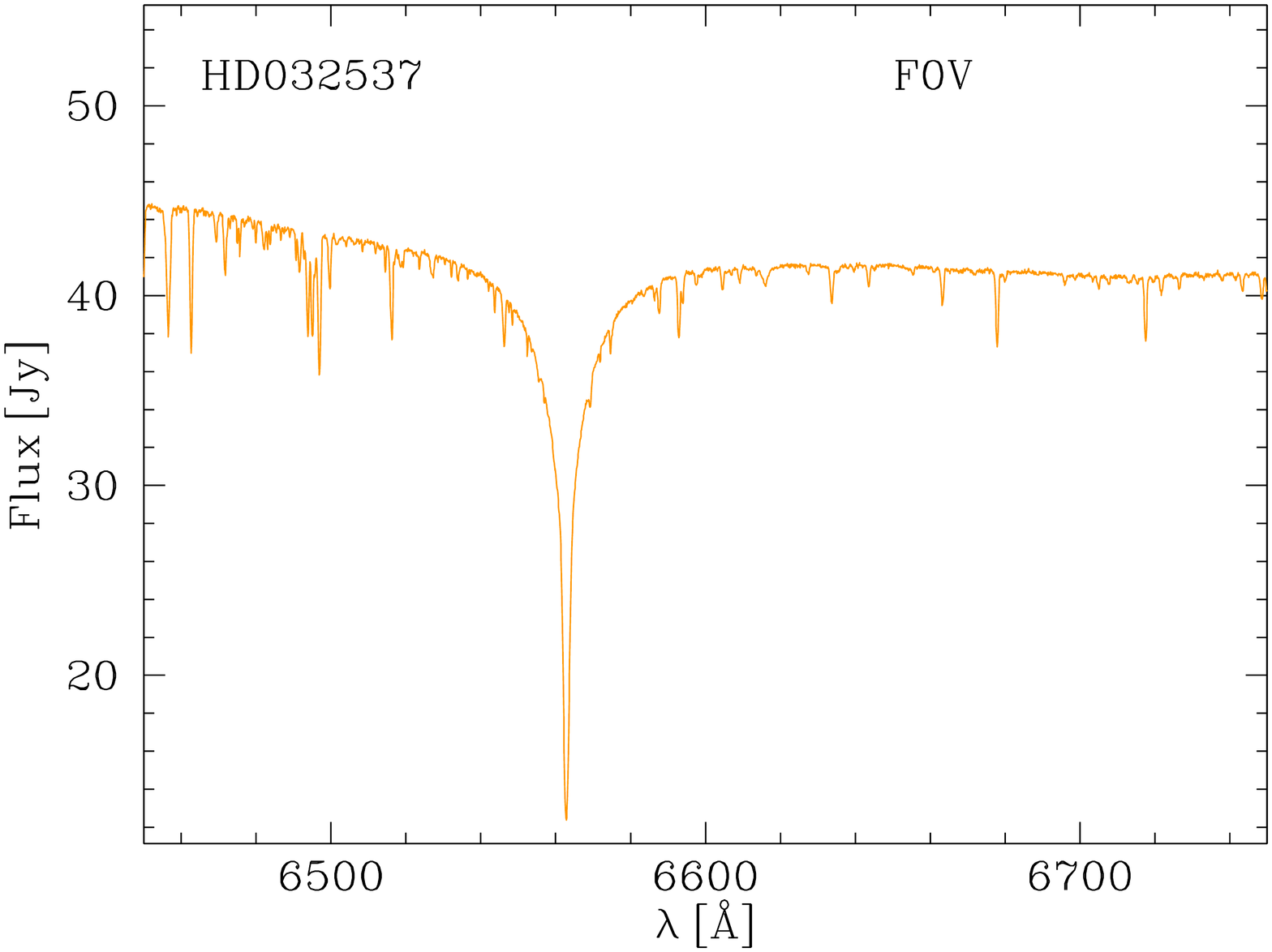}
\includegraphics[width=0.18\textwidth,angle=-90]{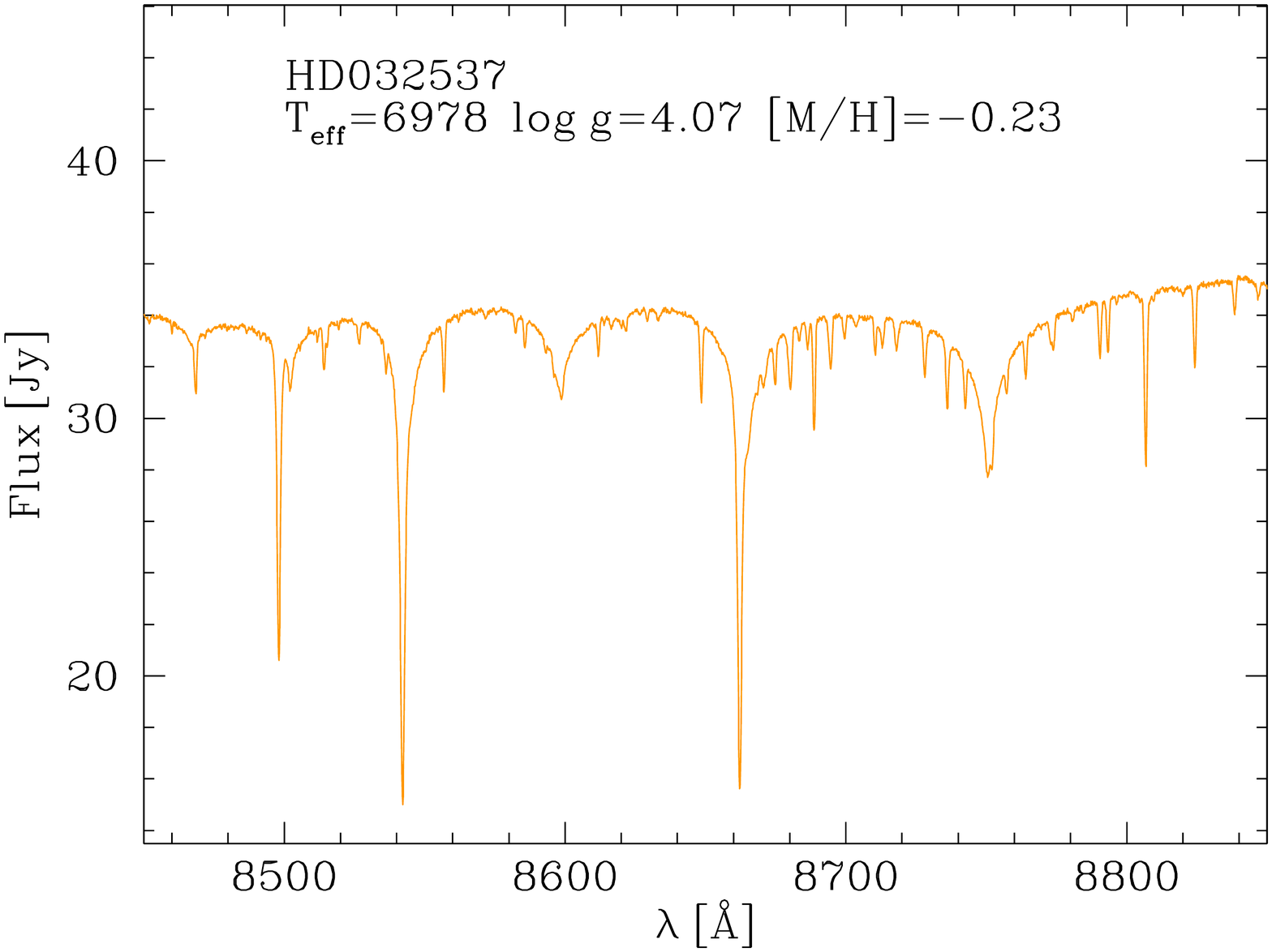}

\contcaption{6. Stars shown in this page are: HD027295, HD027371, HD027524, HD027685, HD028005, HD029645, HD030614, HD030649, HD030676, HD031219, HD031293, HD031374, HD031996 and HD032537.}
\end{figure*}

\begin{figure*}
\includegraphics[width=0.18\textwidth,angle=-90]{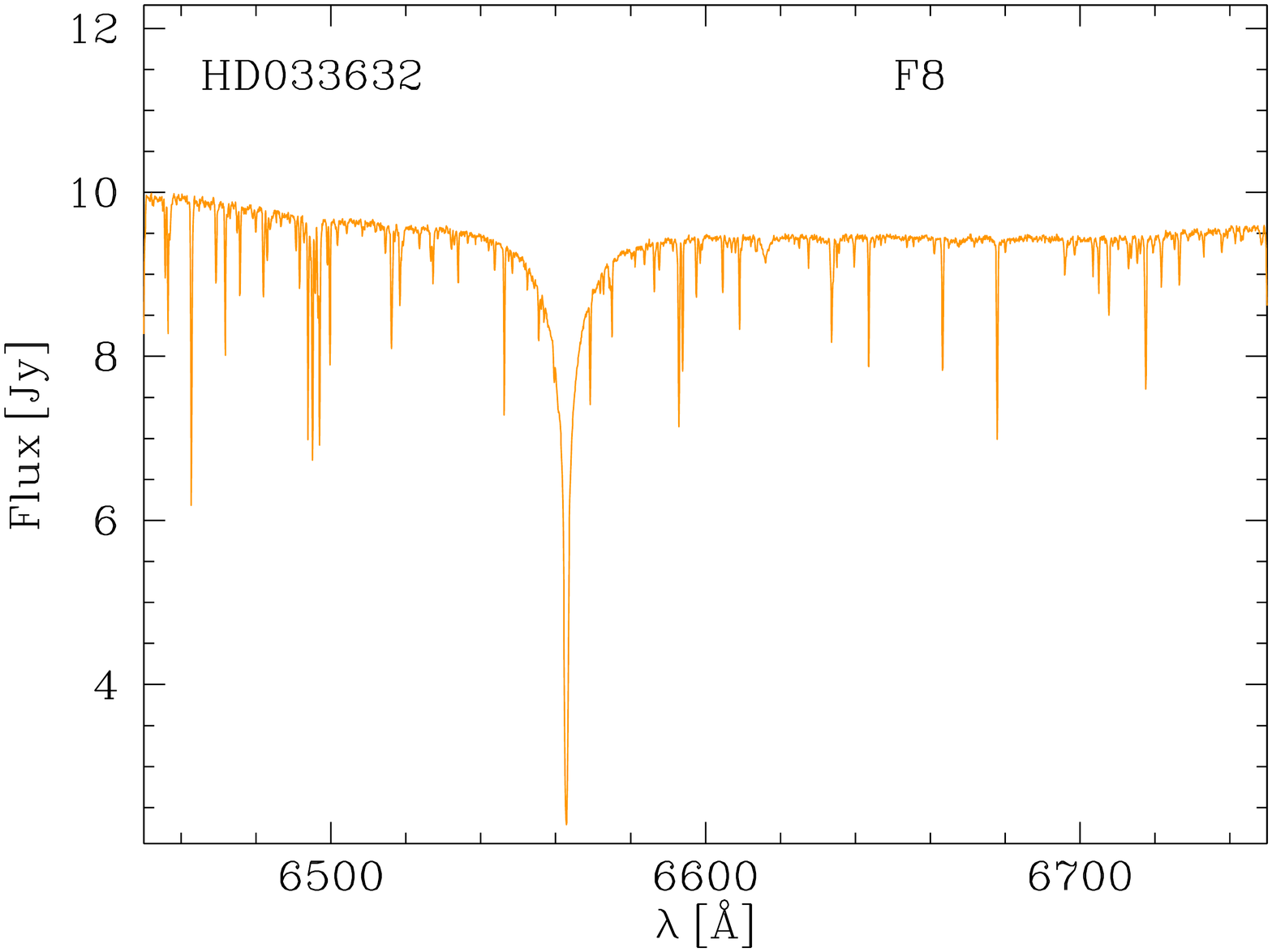}
\includegraphics[width=0.18\textwidth,angle=-90]{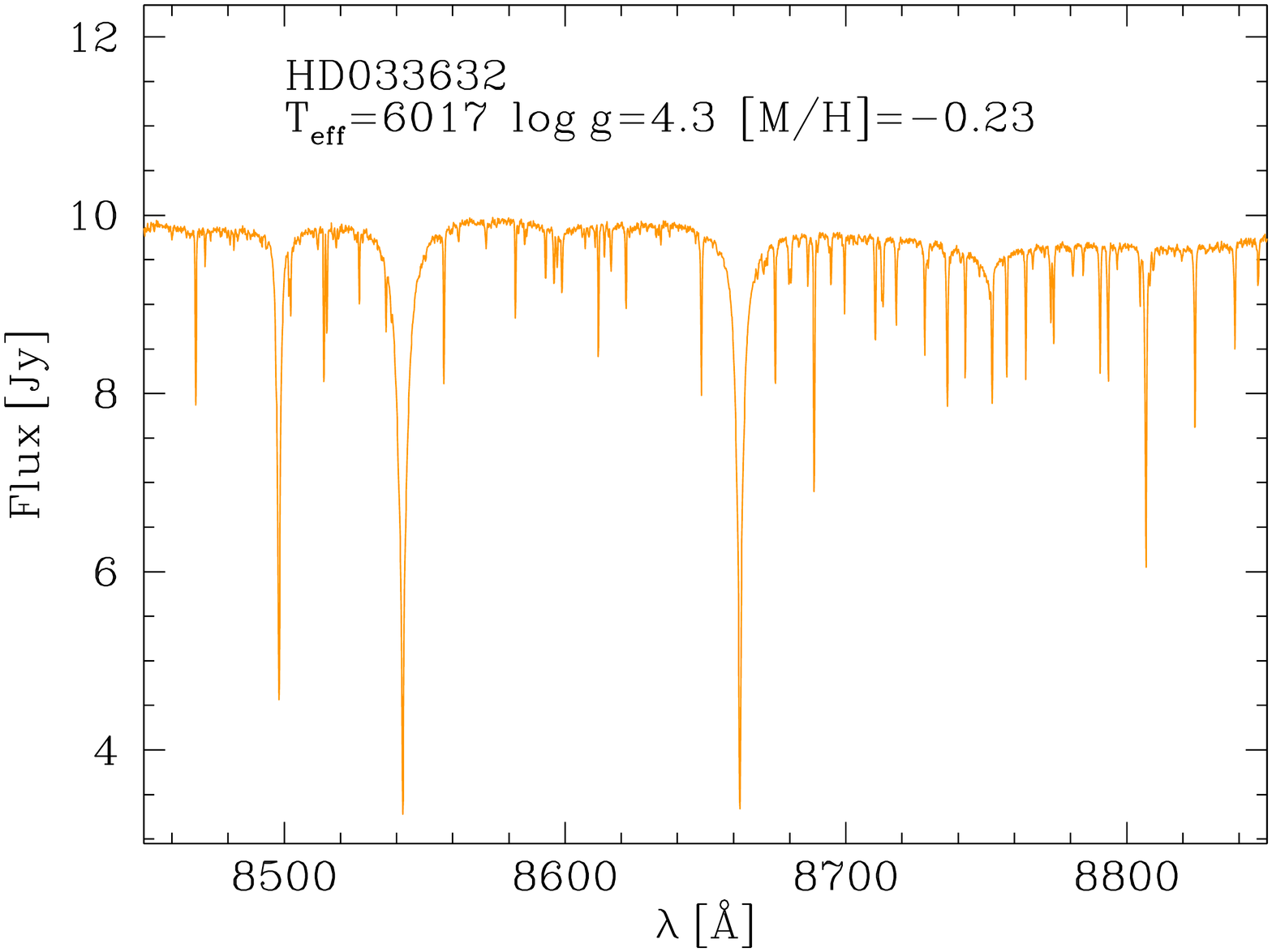}
\includegraphics[width=0.18\textwidth,angle=-90]{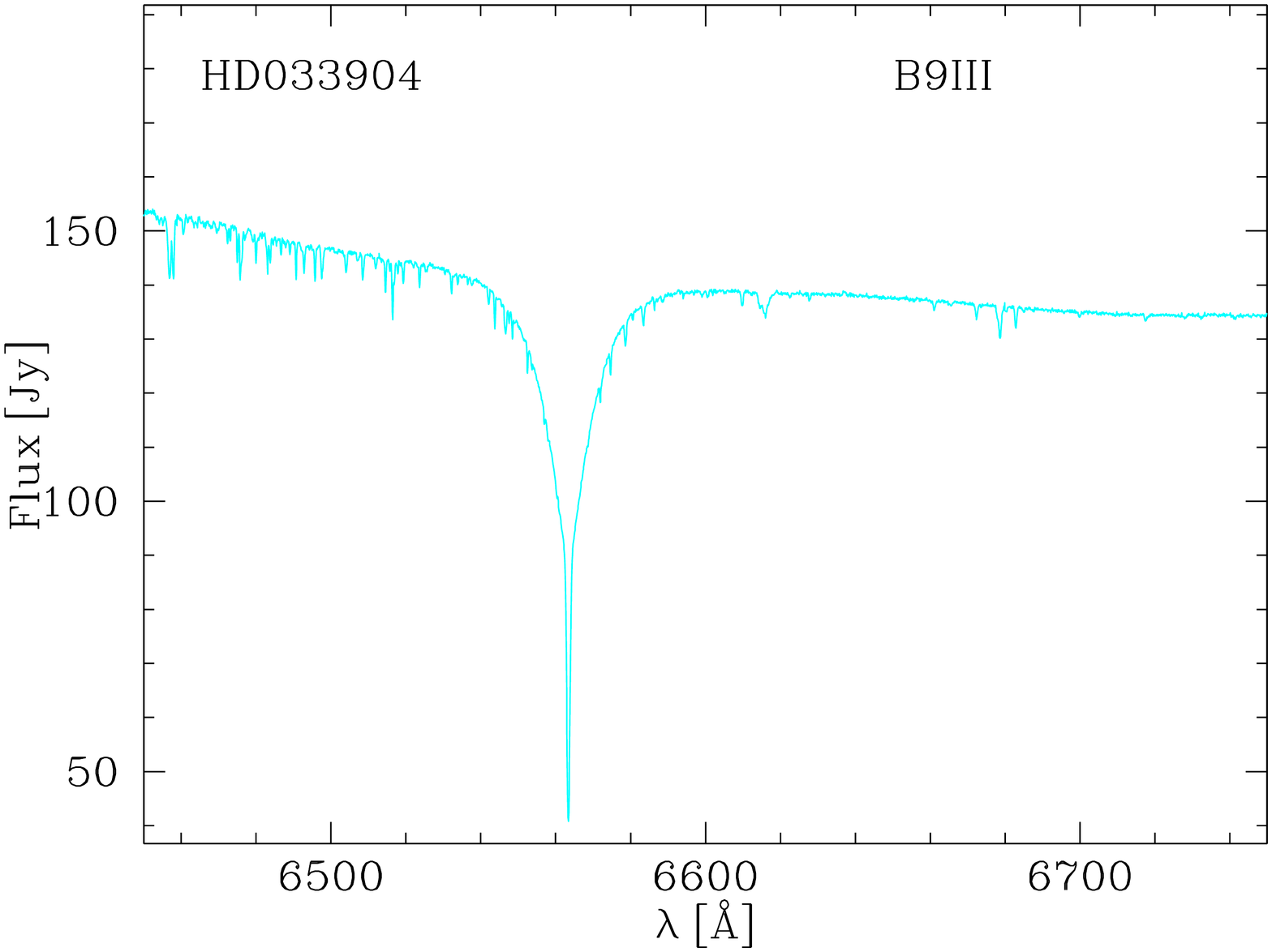}
\includegraphics[width=0.18\textwidth,angle=-90]{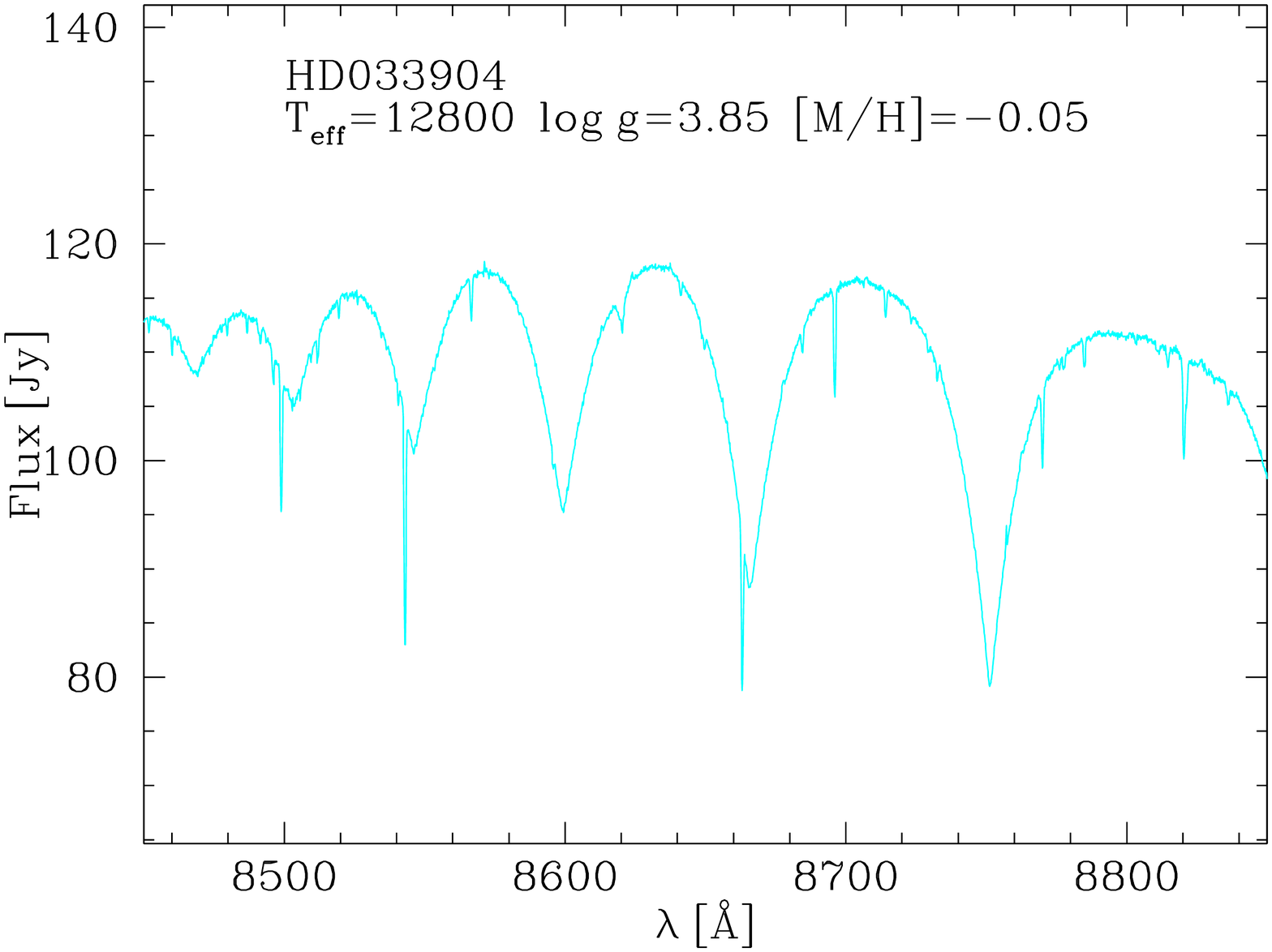}
\includegraphics[width=0.18\textwidth,angle=-90]{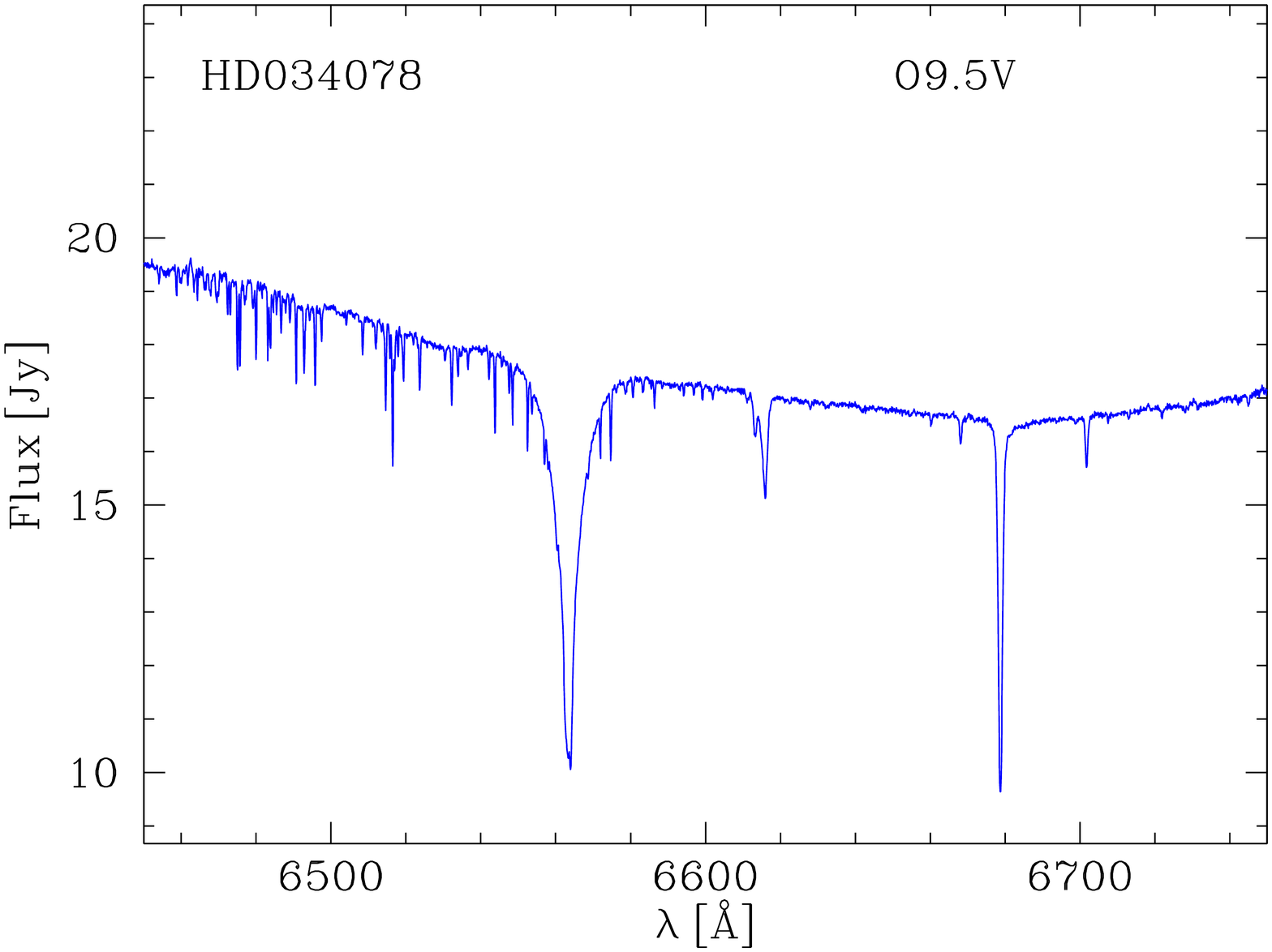}
\includegraphics[width=0.18\textwidth,angle=-90]{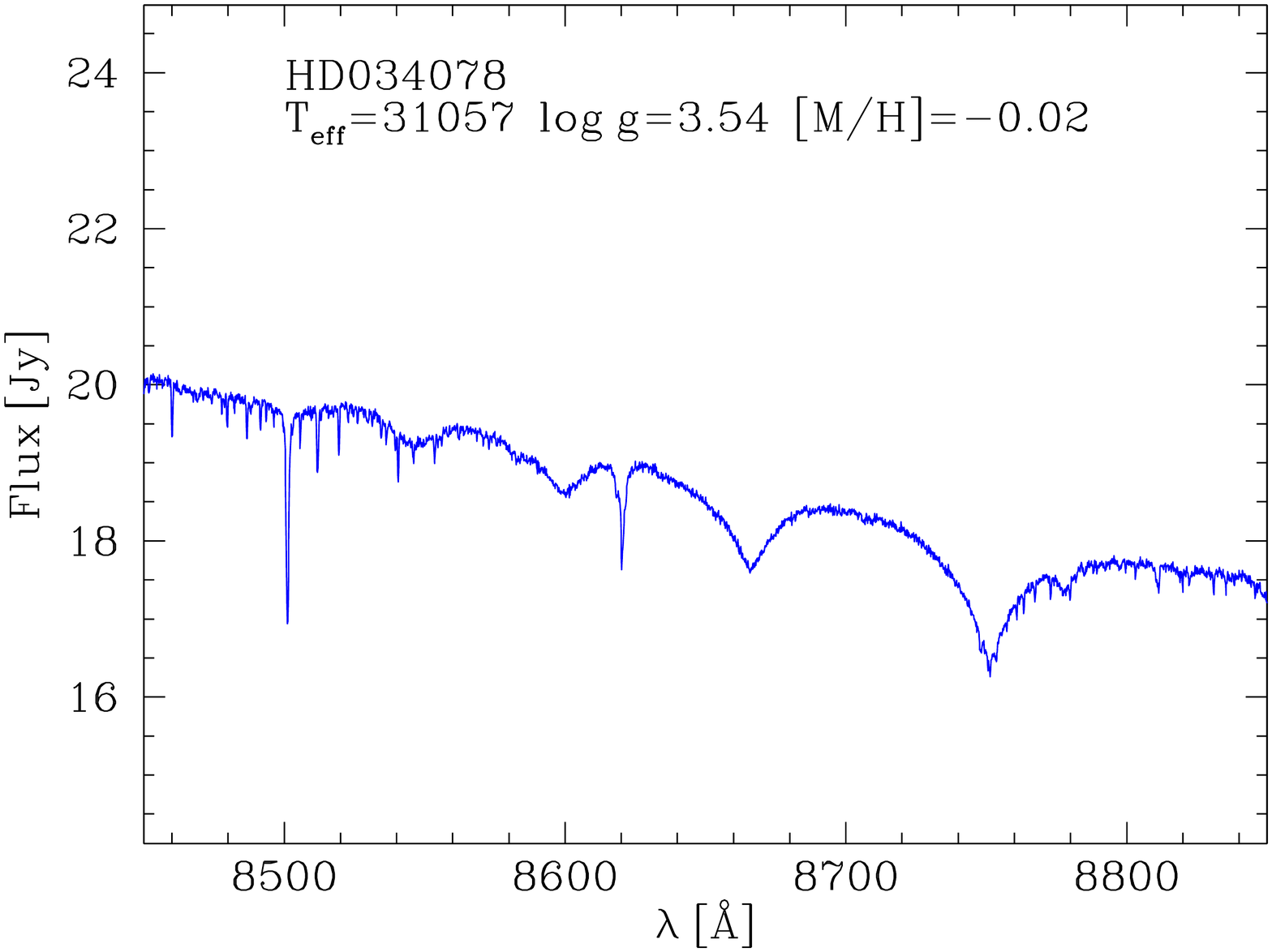}
\includegraphics[width=0.18\textwidth,angle=-90]{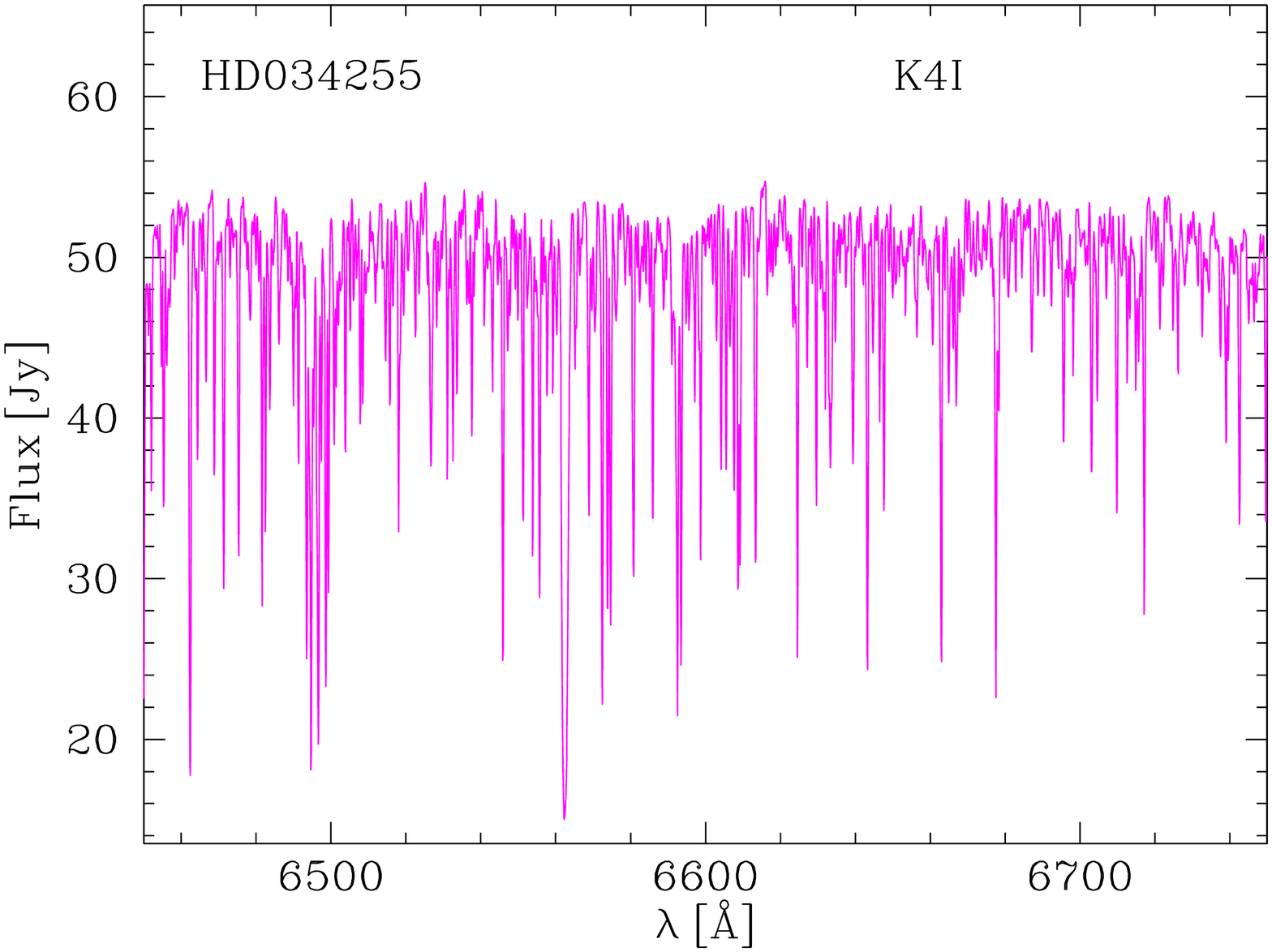}
\includegraphics[width=0.18\textwidth,angle=-90]{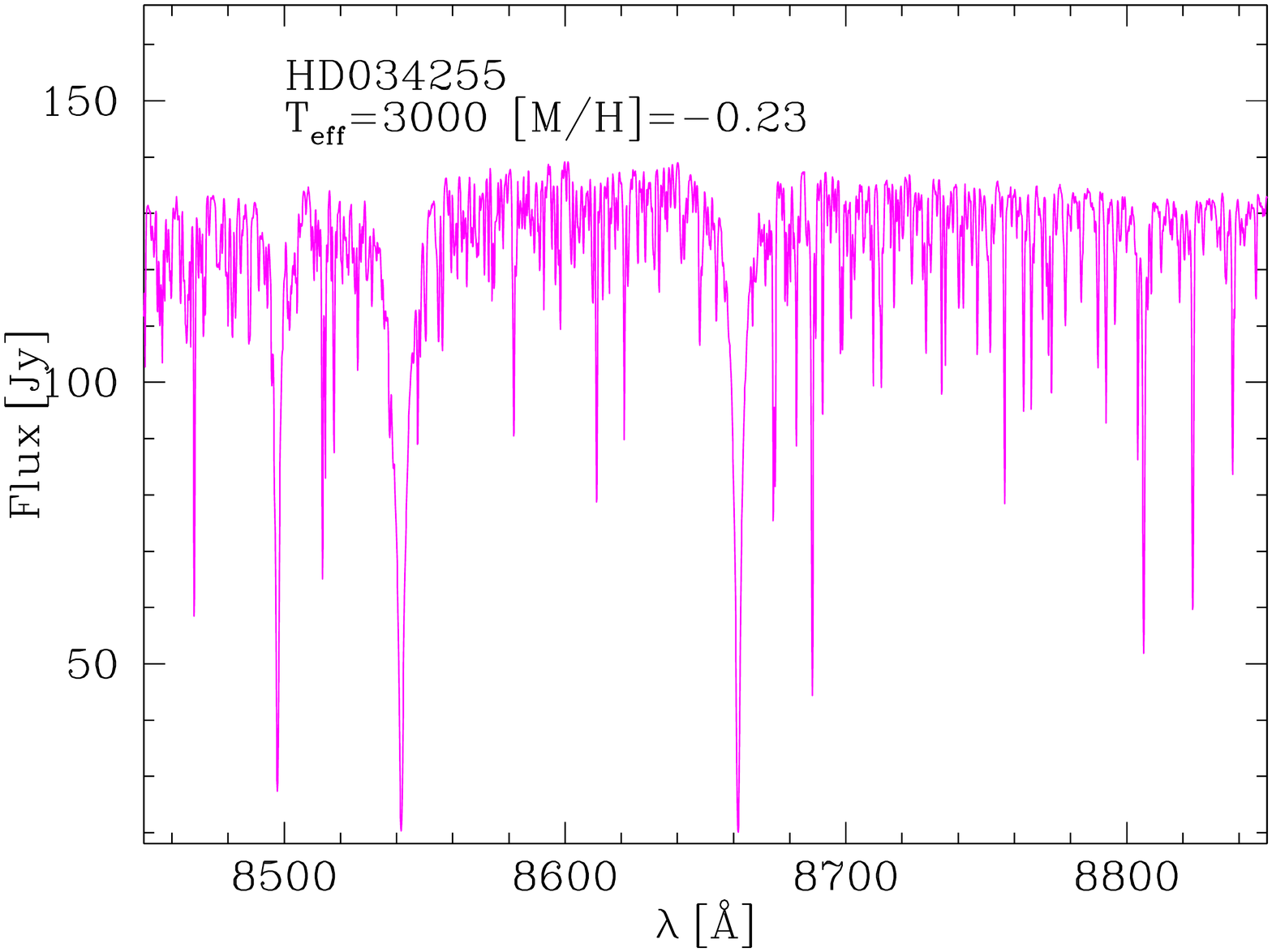}
\includegraphics[width=0.18\textwidth,angle=-90]{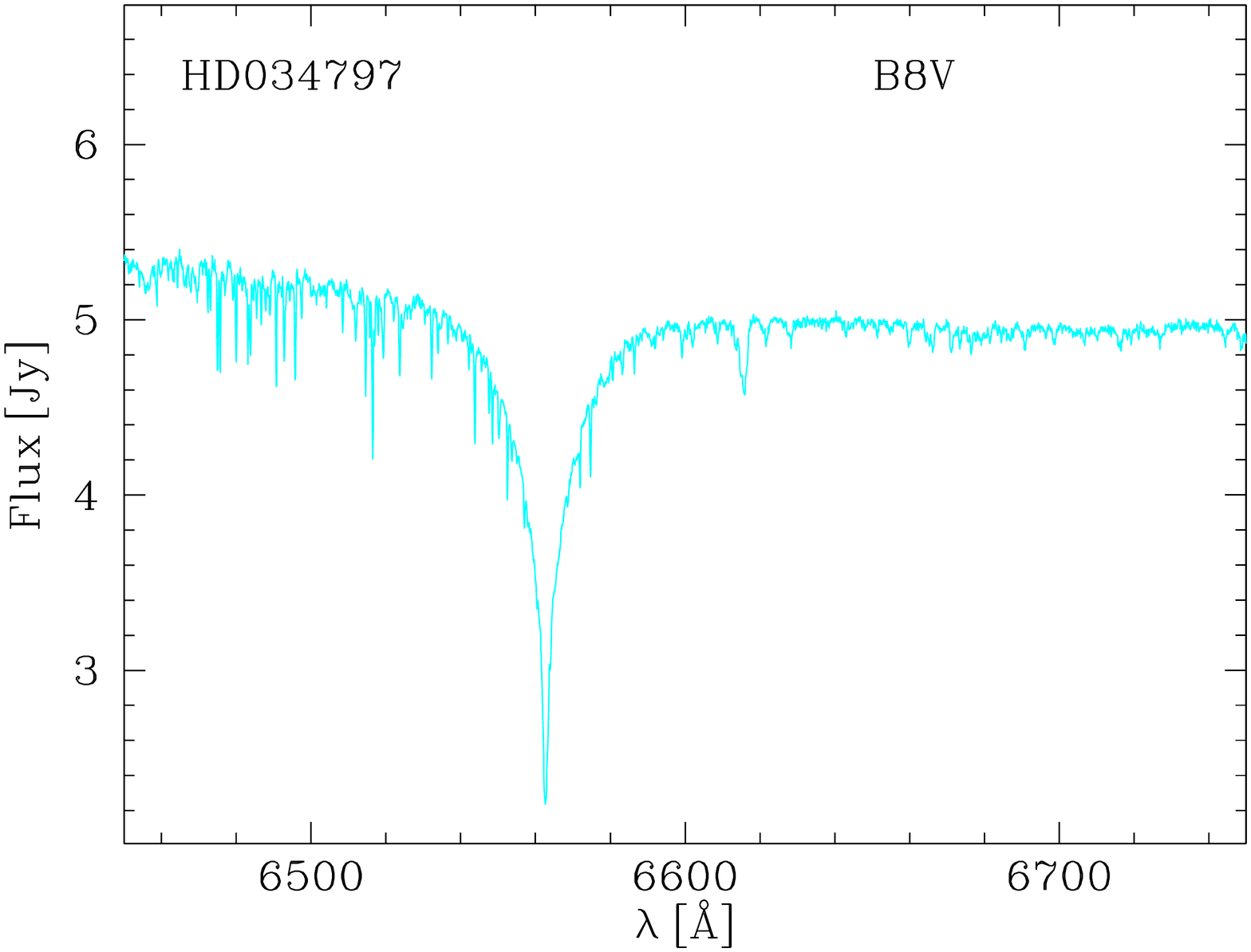}
\includegraphics[width=0.18\textwidth,angle=-90]{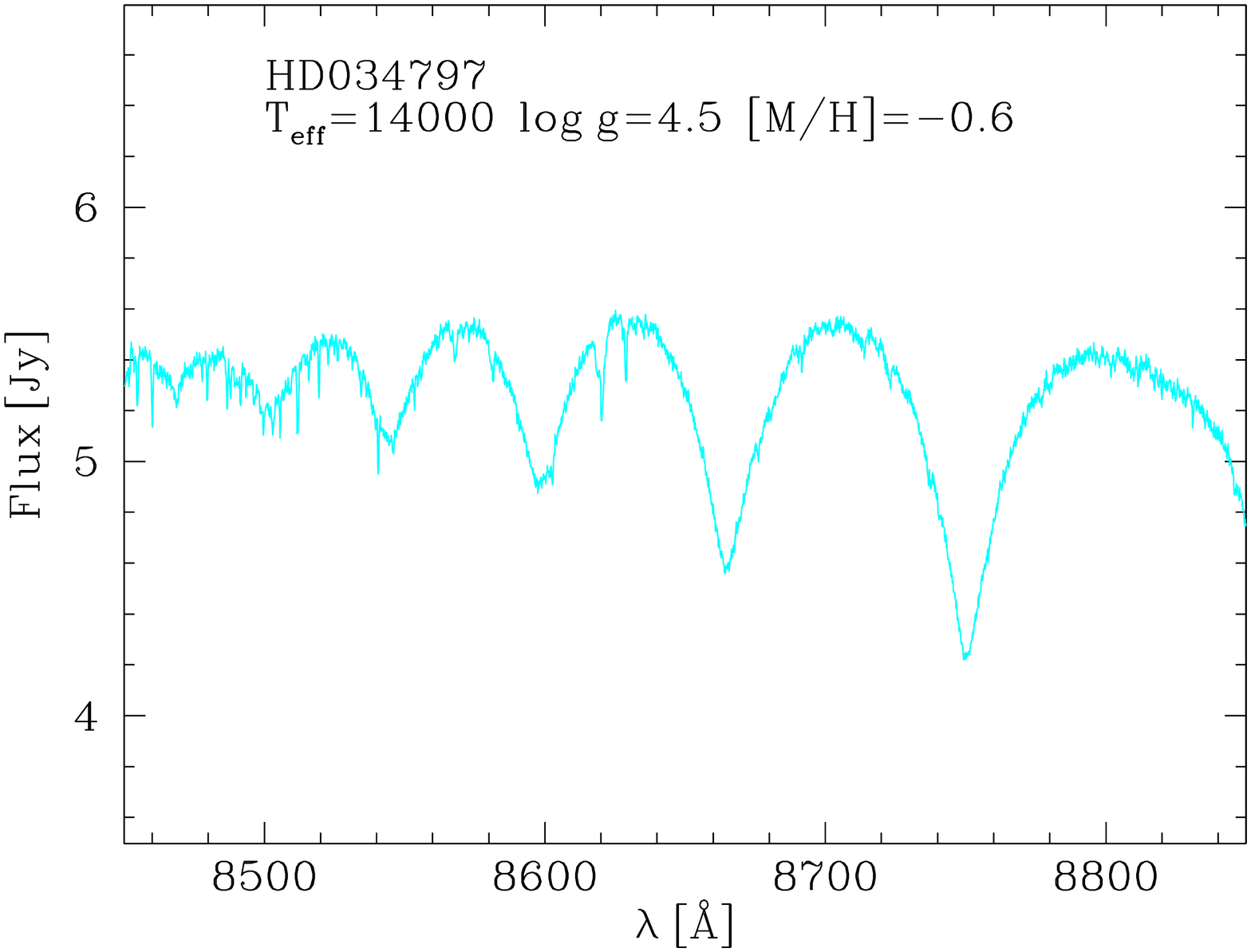}
\includegraphics[width=0.18\textwidth,angle=-90]{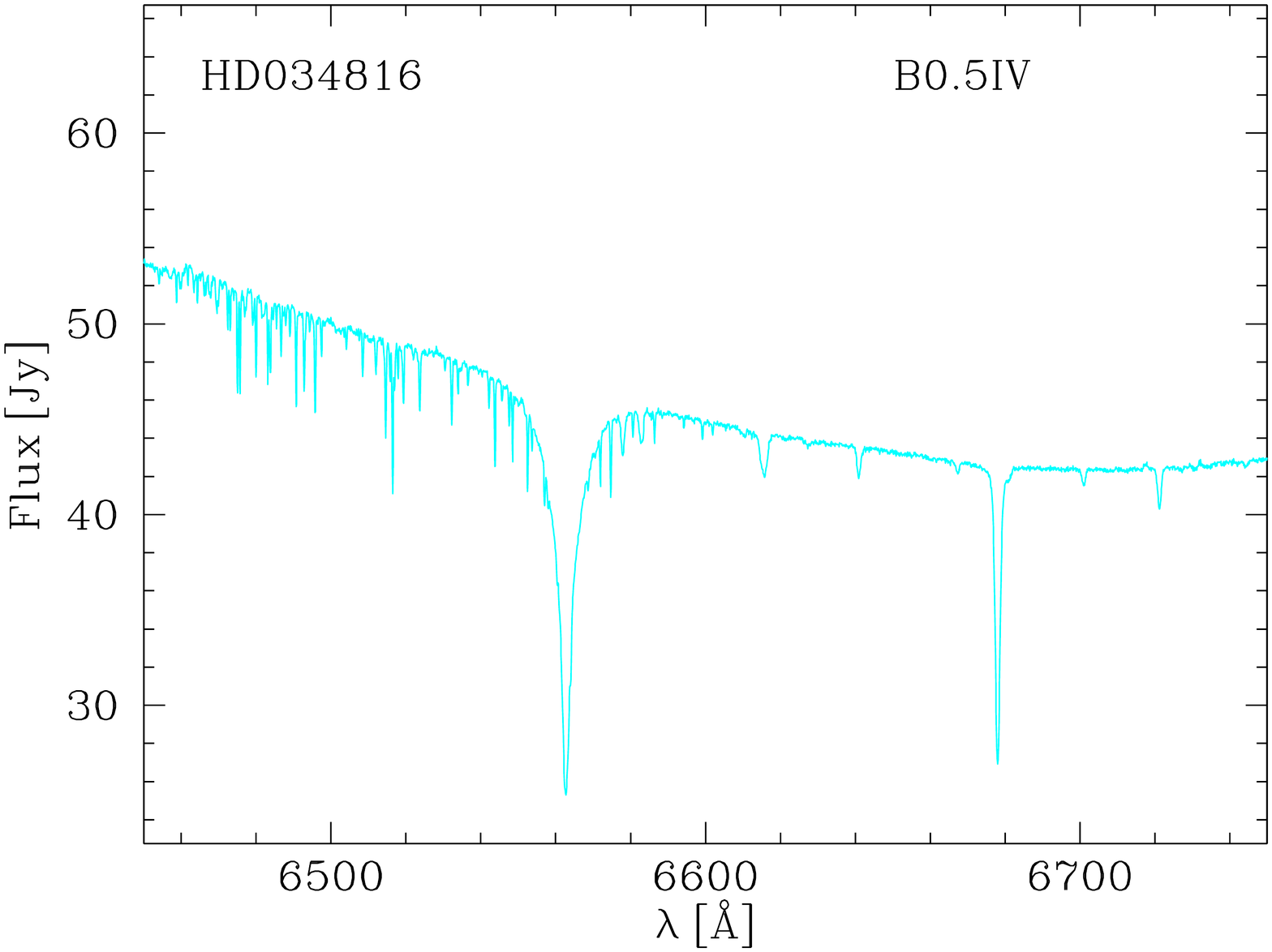}
\includegraphics[width=0.18\textwidth,angle=-90]{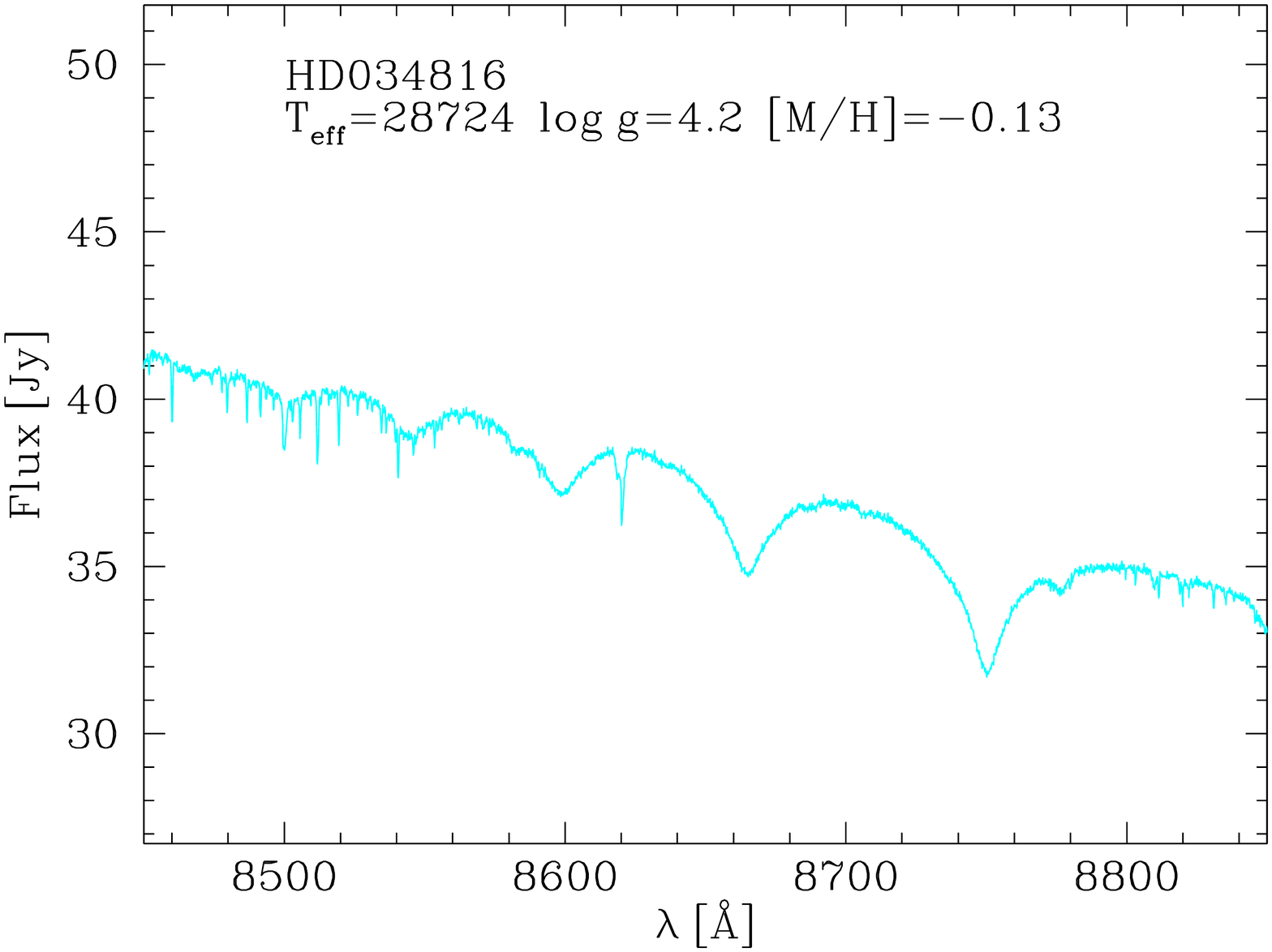}
\includegraphics[width=0.18\textwidth,angle=-90]{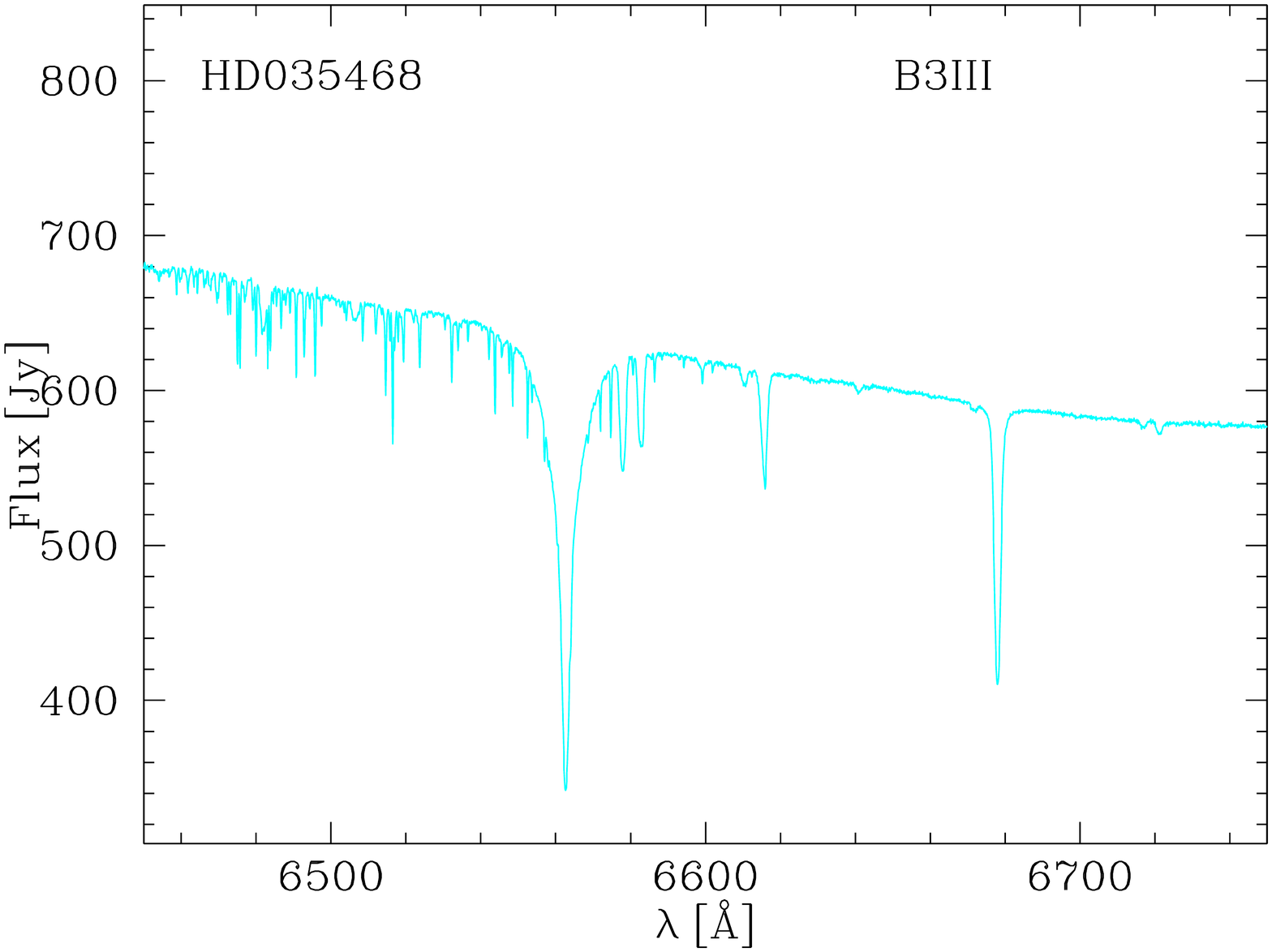}
\includegraphics[width=0.18\textwidth,angle=-90]{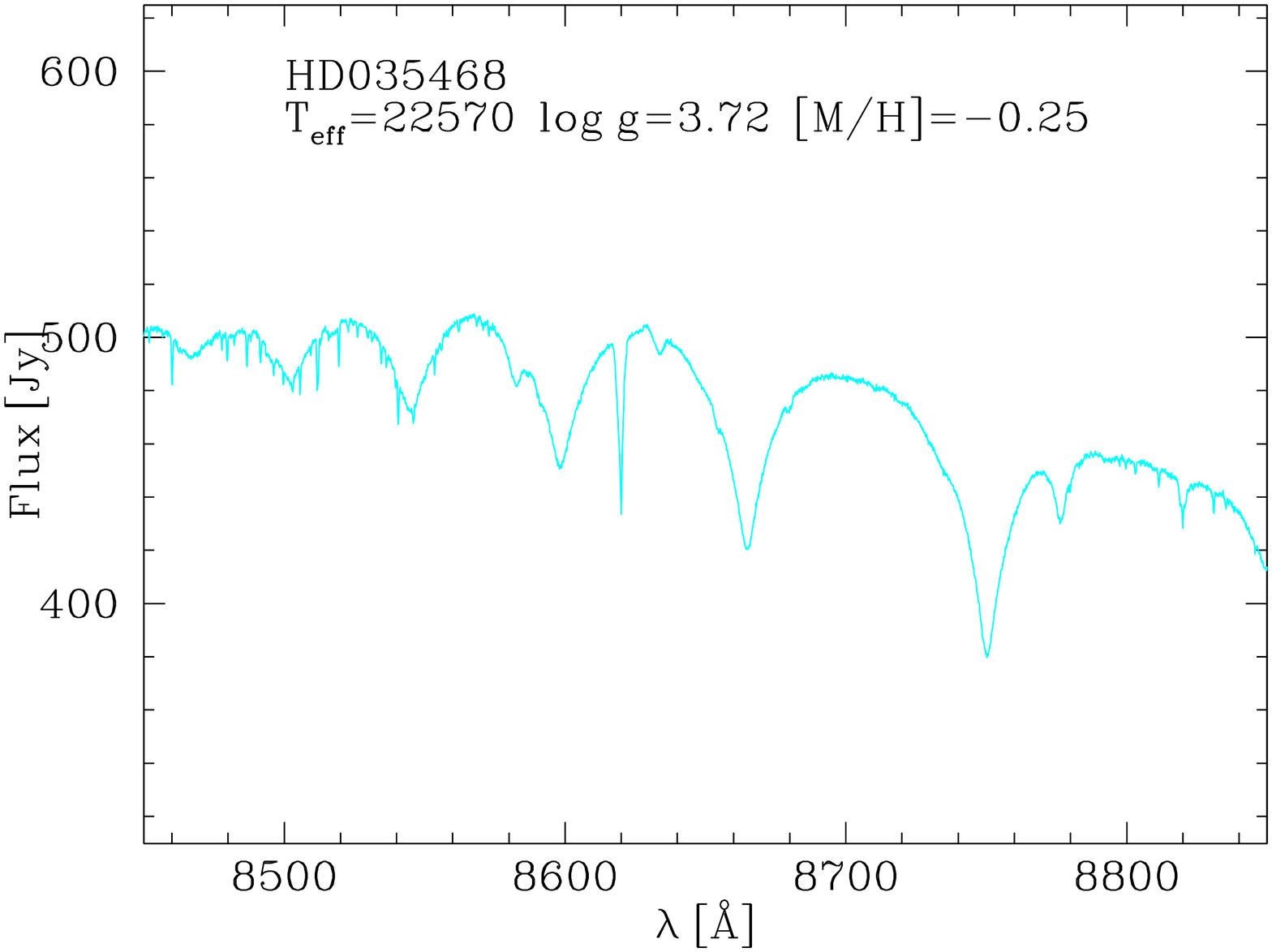}
\includegraphics[width=0.18\textwidth,angle=-90]{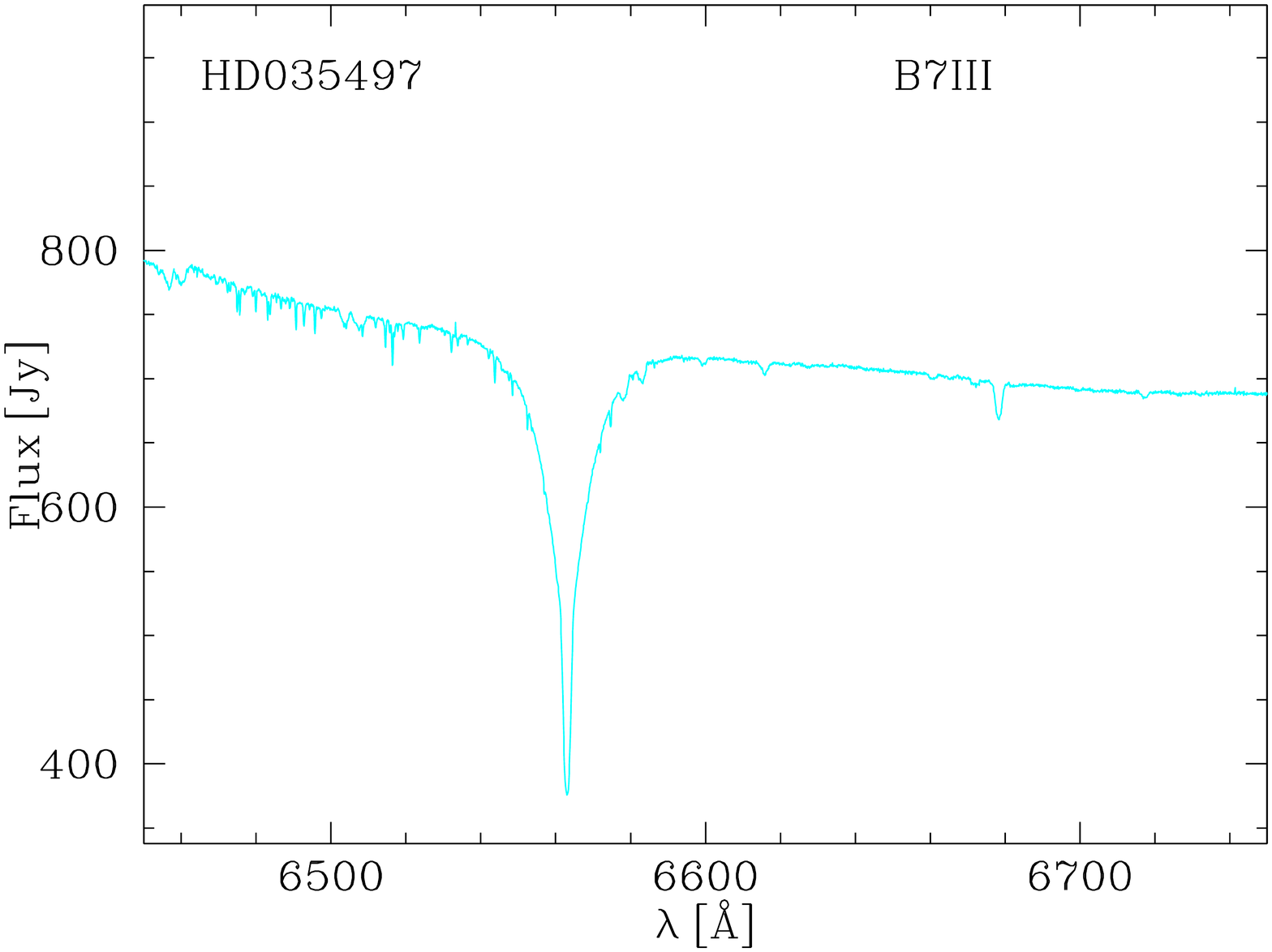}
\includegraphics[width=0.18\textwidth,angle=-90]{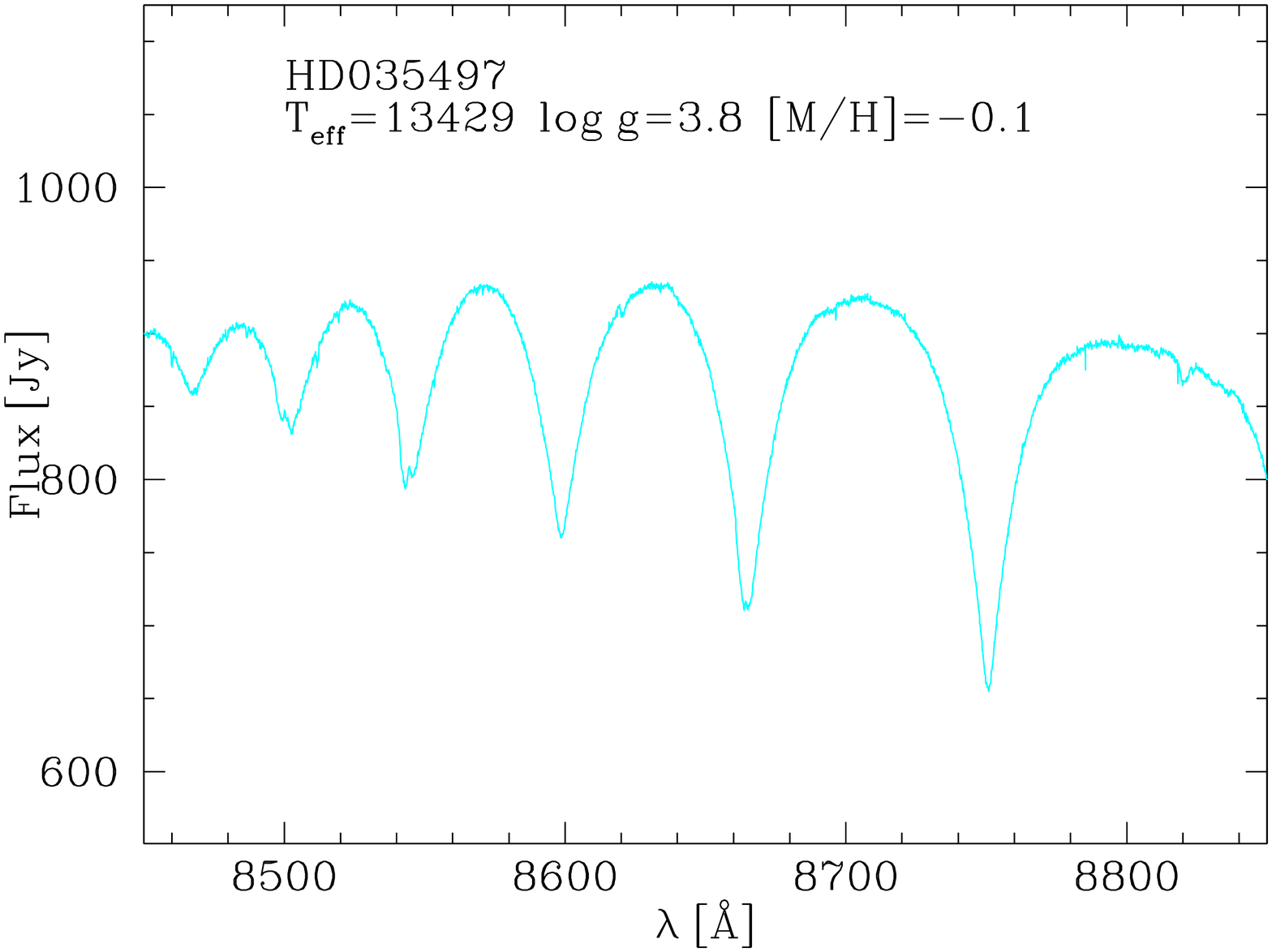}
\includegraphics[width=0.18\textwidth,angle=-90]{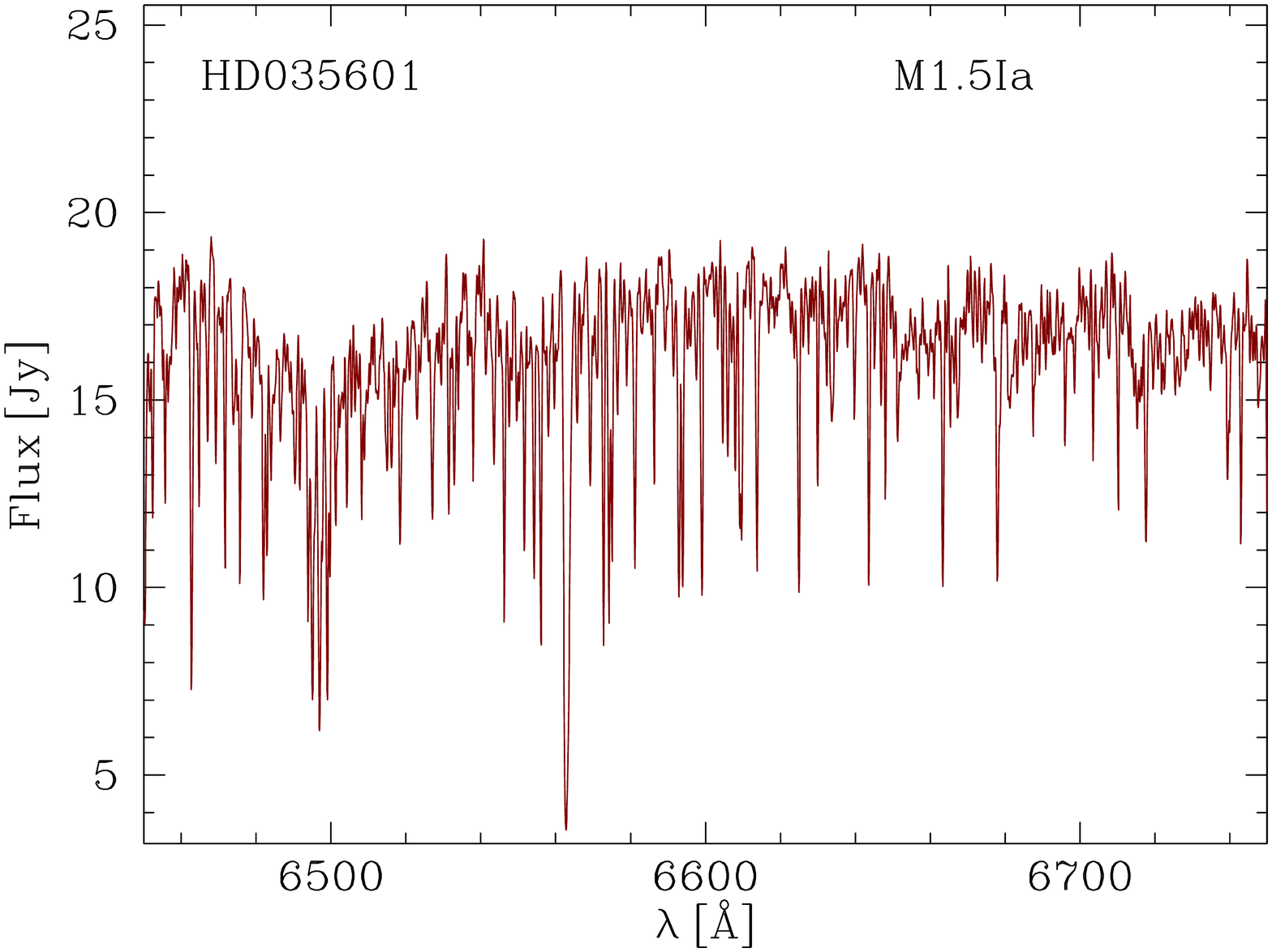}
\includegraphics[width=0.18\textwidth,angle=-90]{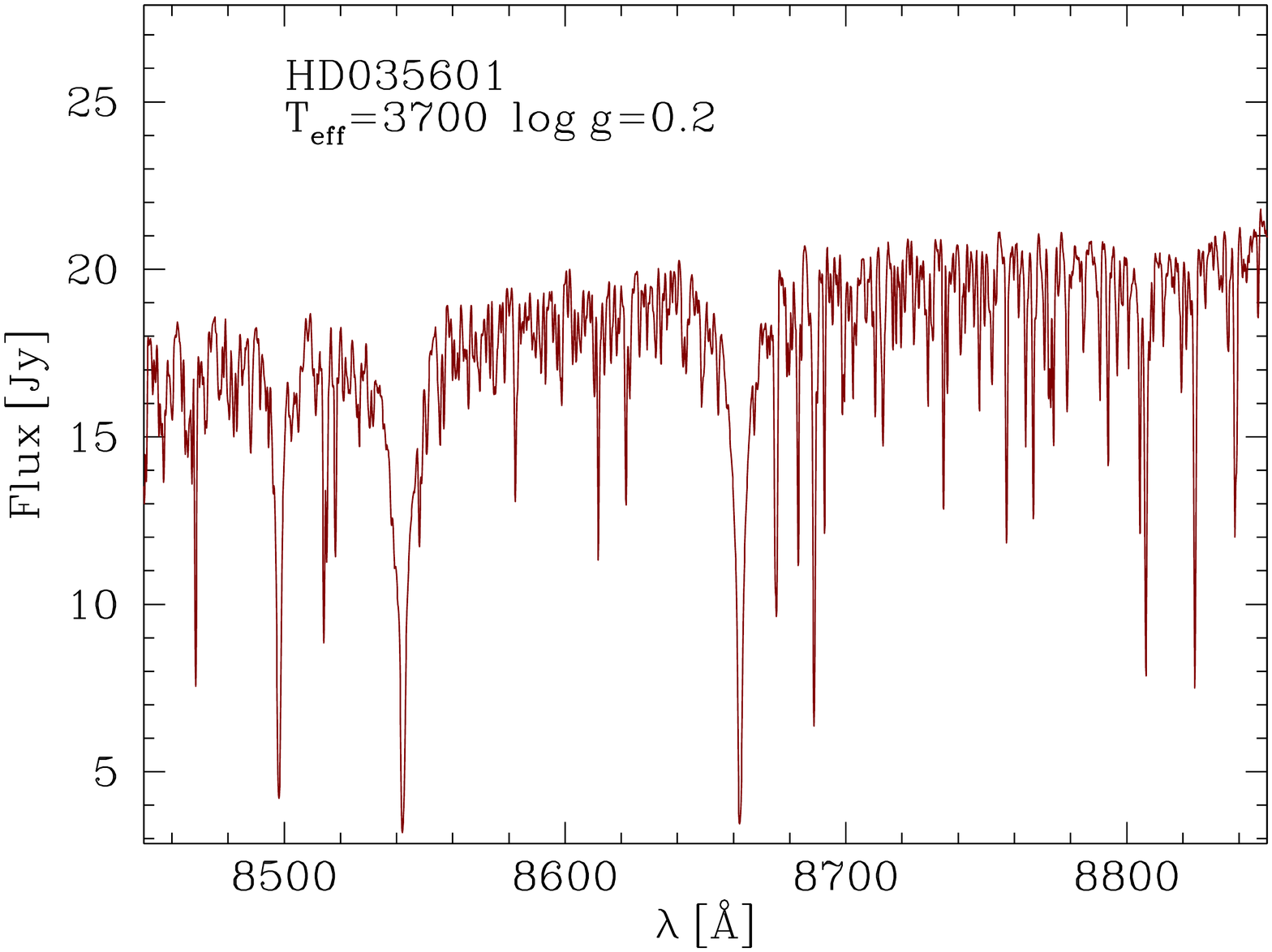}
\includegraphics[width=0.18\textwidth,angle=-90]{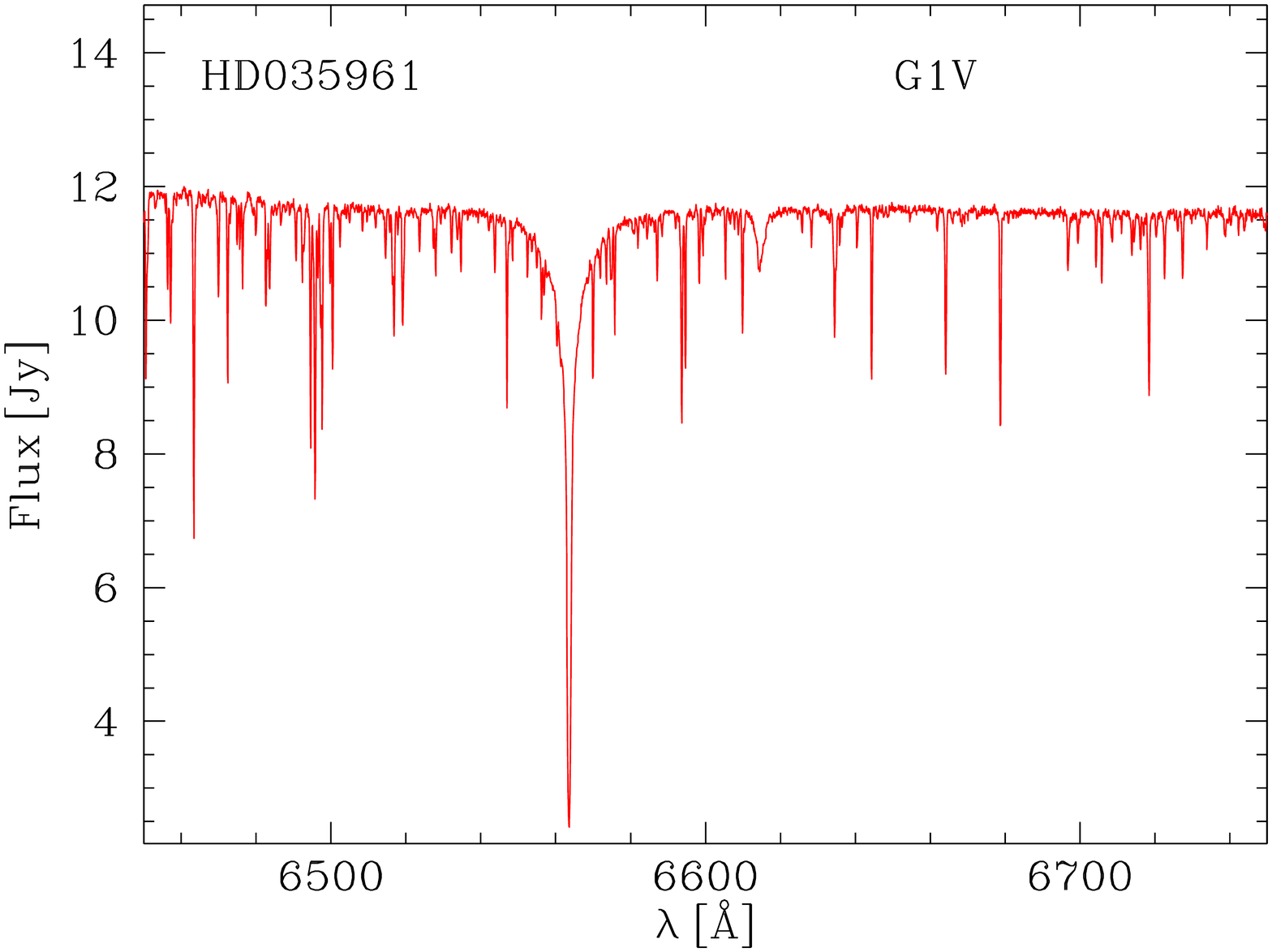}
\includegraphics[width=0.18\textwidth,angle=-90]{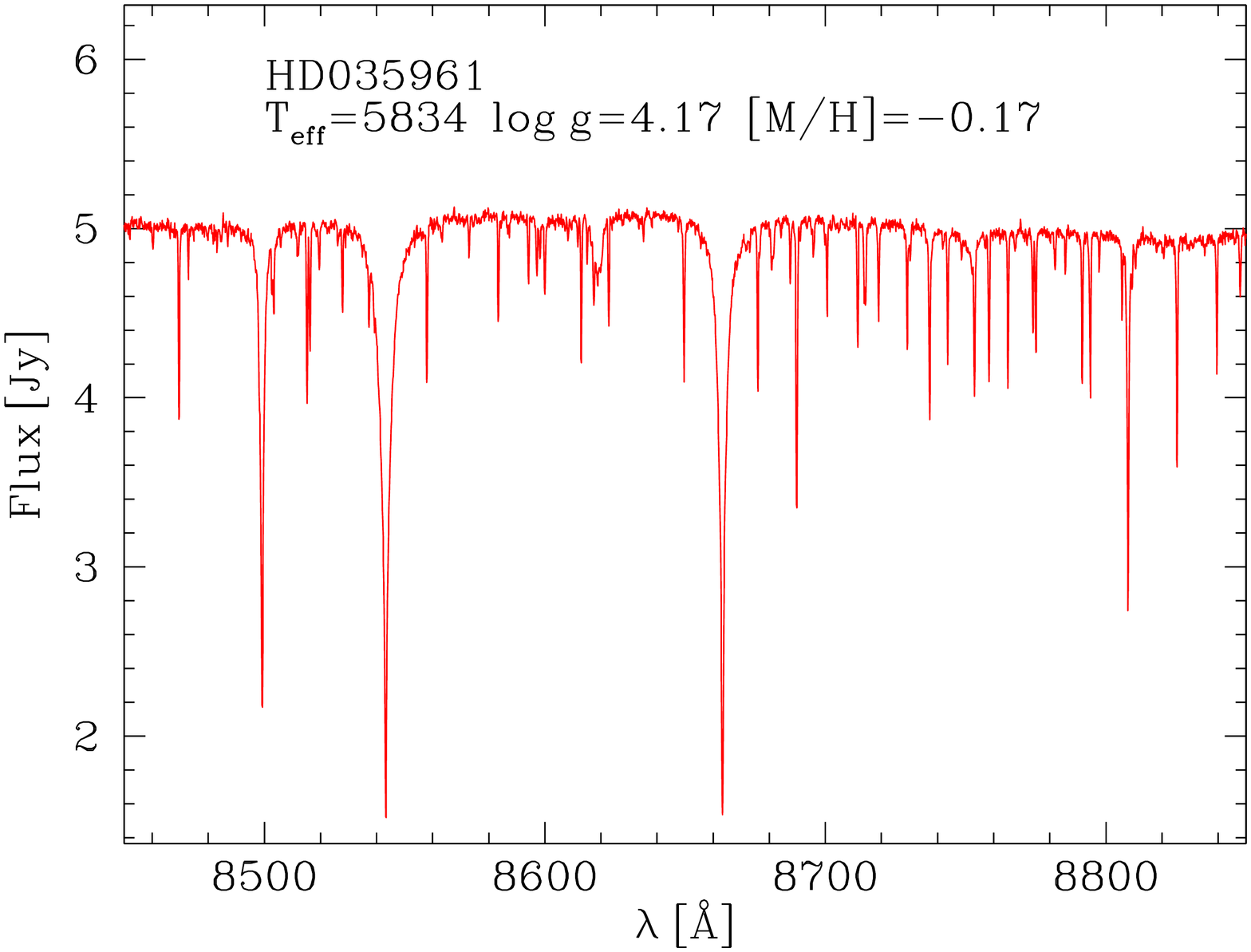}
\includegraphics[width=0.18\textwidth,angle=-90]{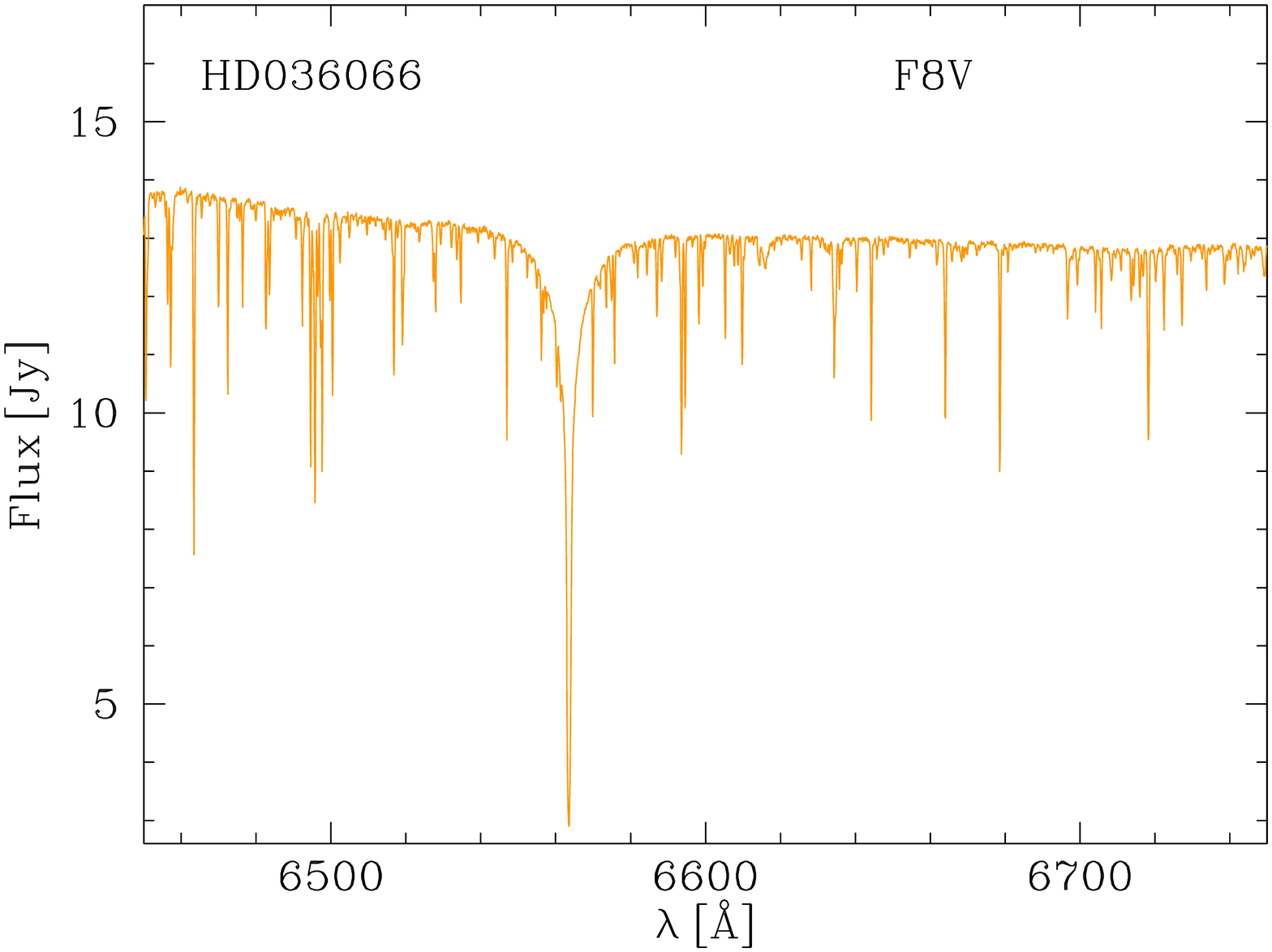}
\includegraphics[width=0.18\textwidth,angle=-90]{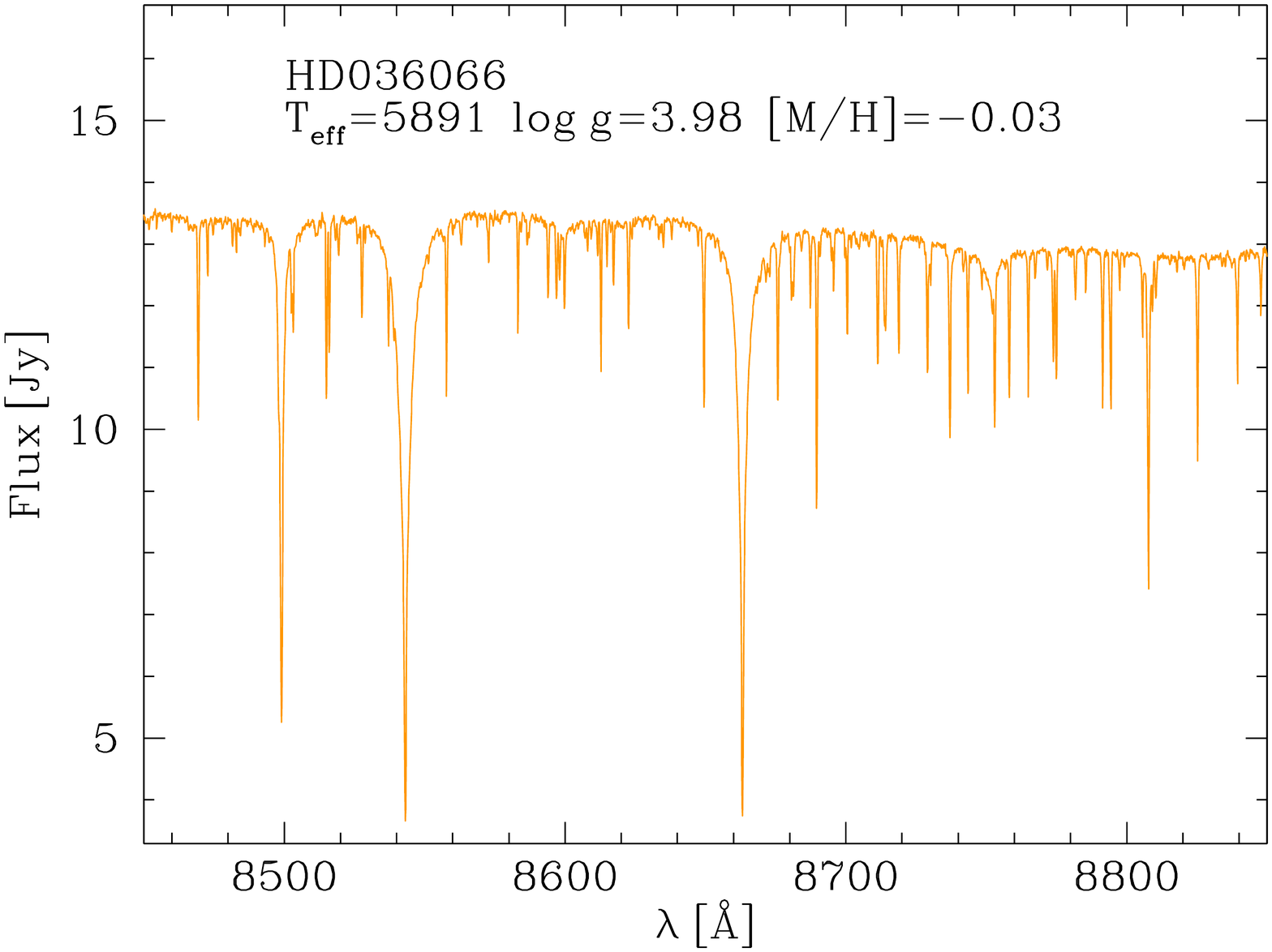}
\includegraphics[width=0.18\textwidth,angle=-90]{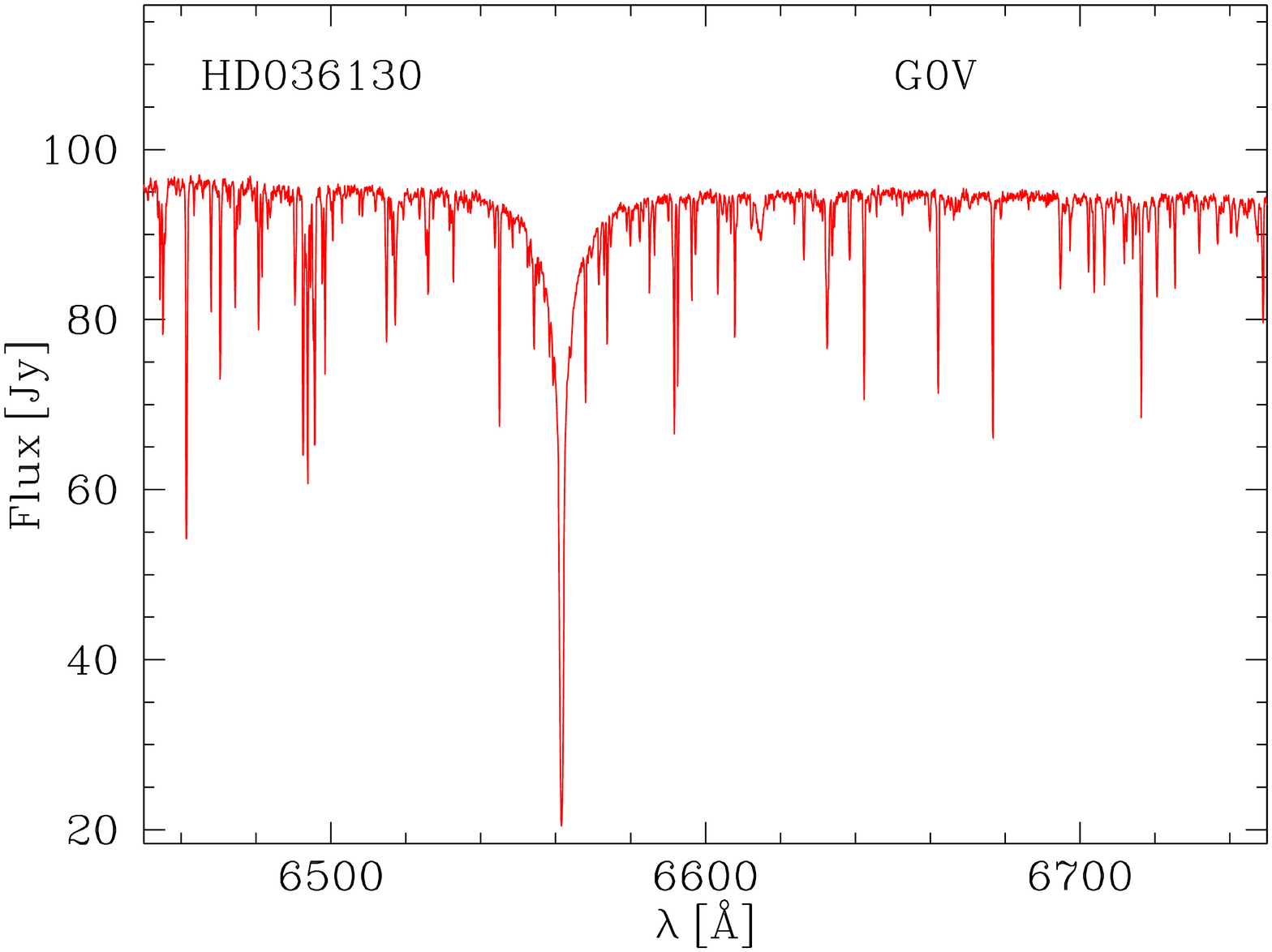}
\includegraphics[width=0.18\textwidth,angle=-90]{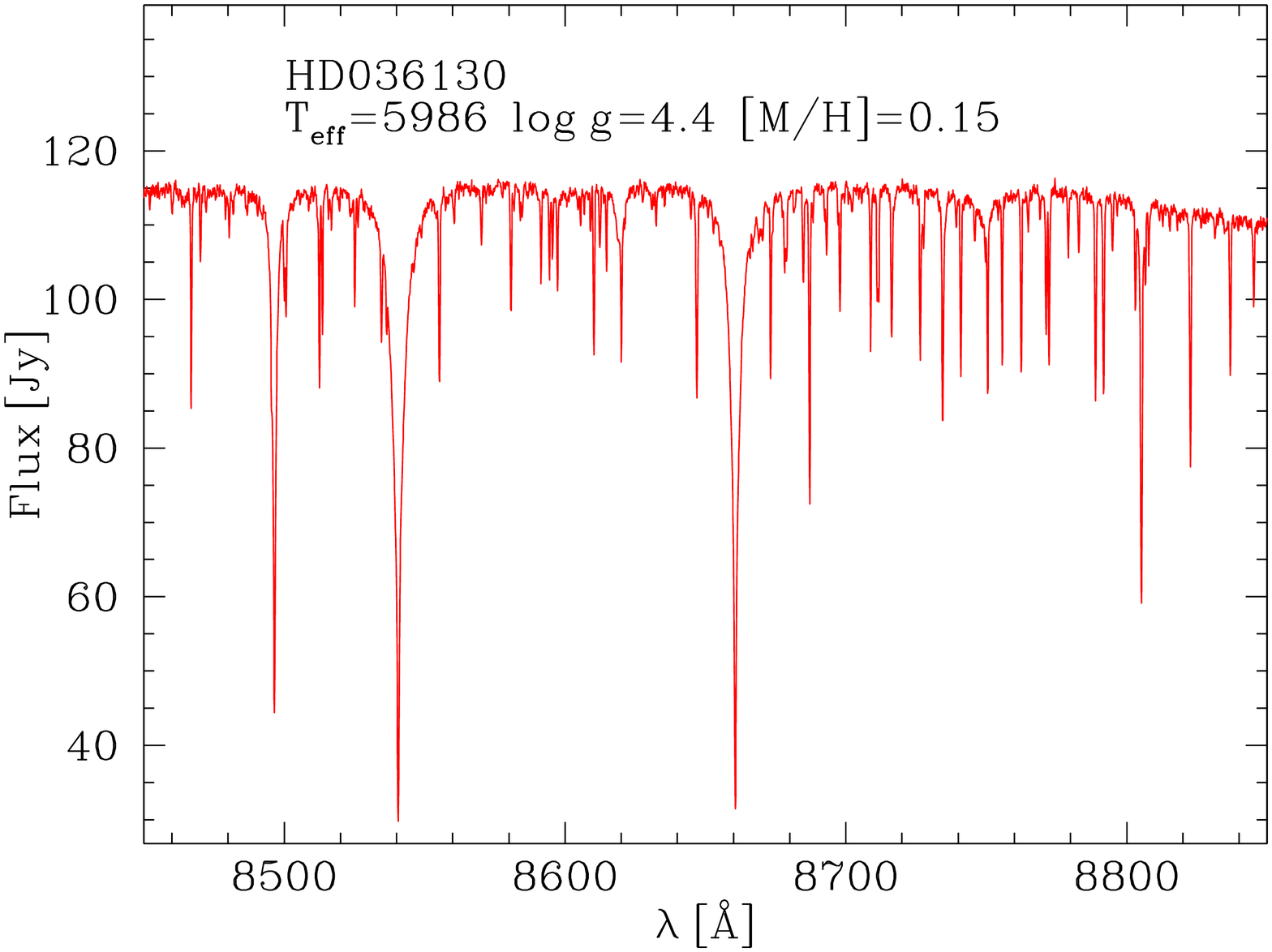}
\includegraphics[width=0.18\textwidth,angle=-90]{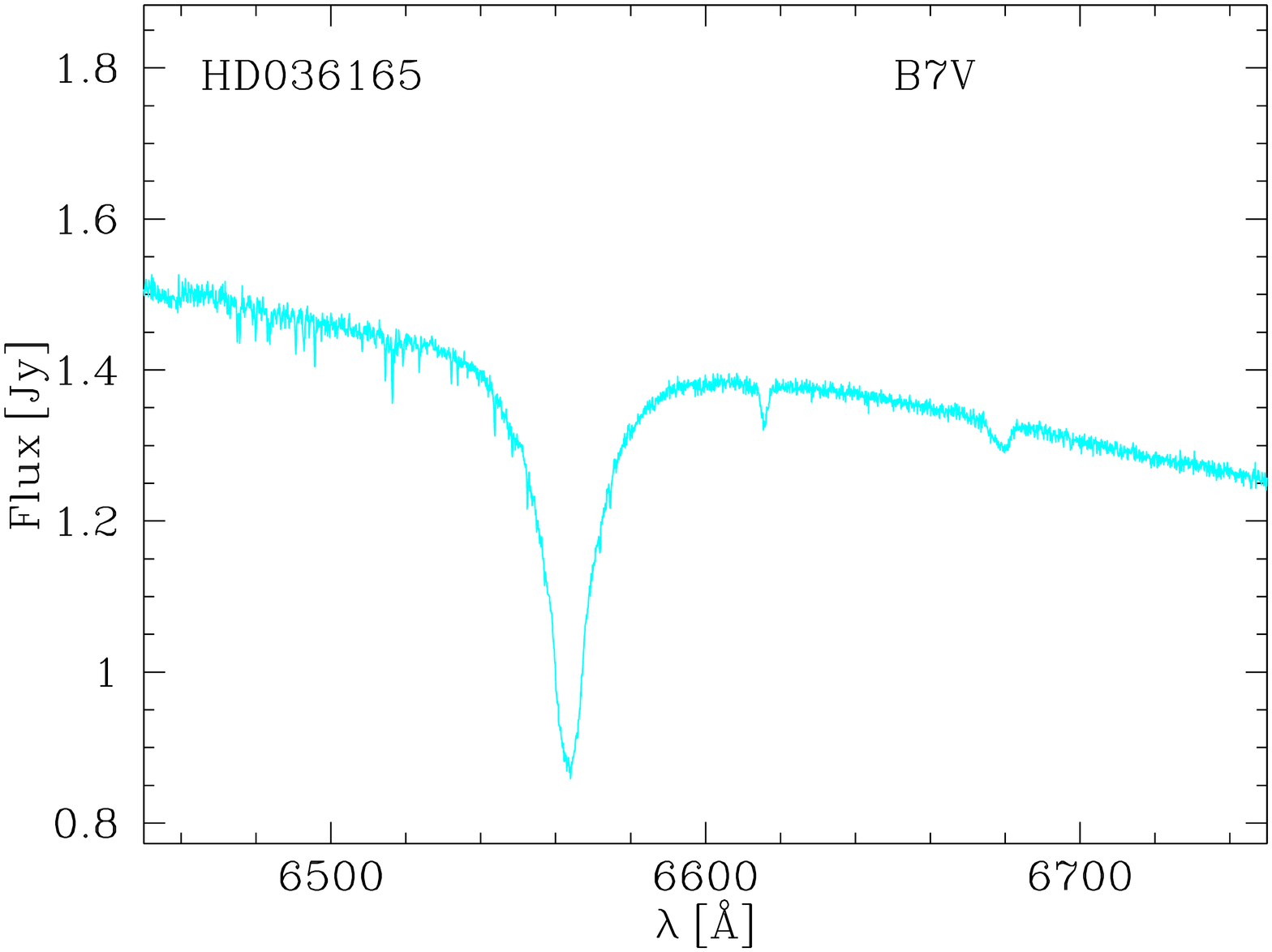}
\includegraphics[width=0.18\textwidth,angle=-90]{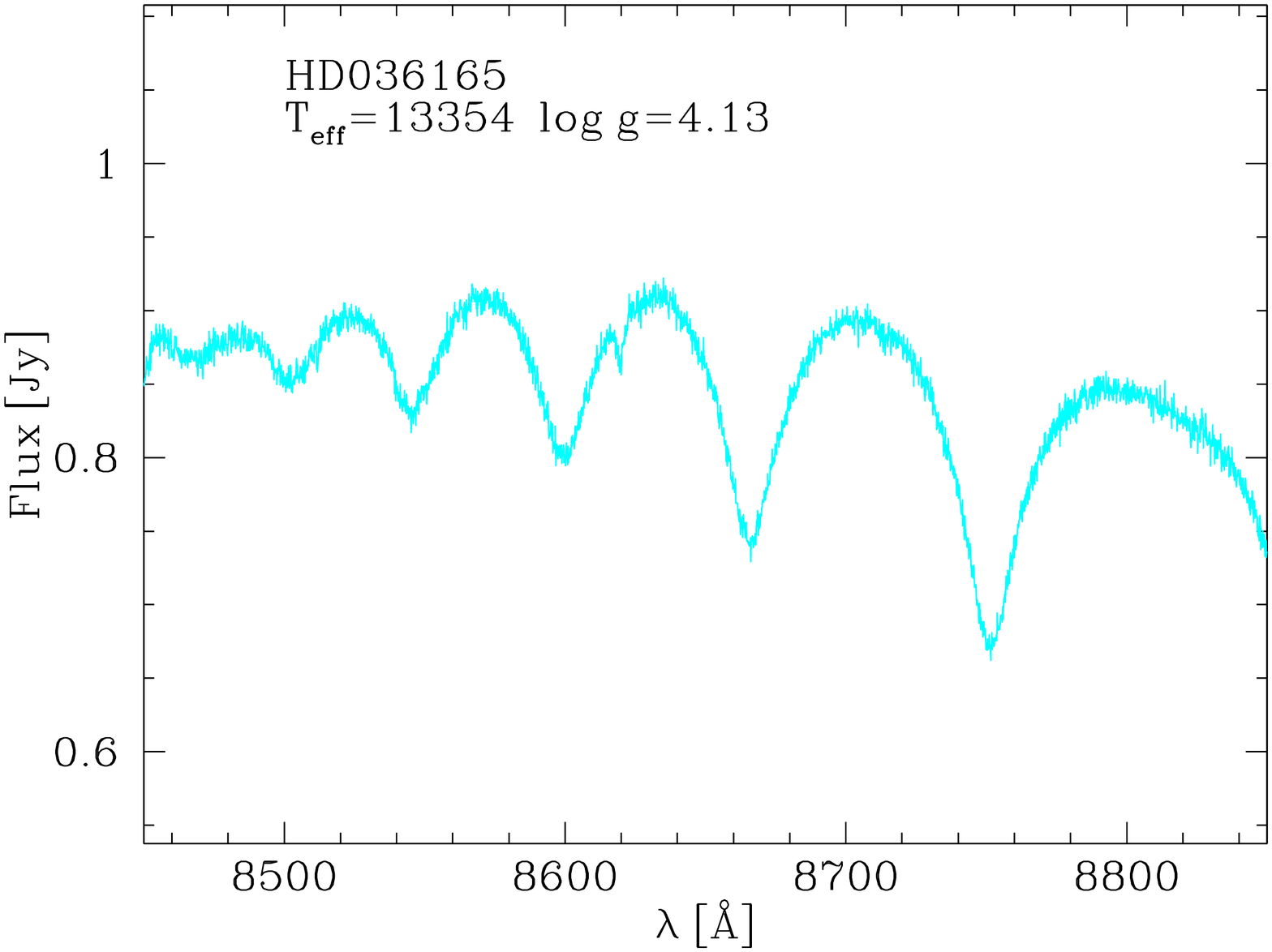}
\includegraphics[width=0.18\textwidth,angle=-90]{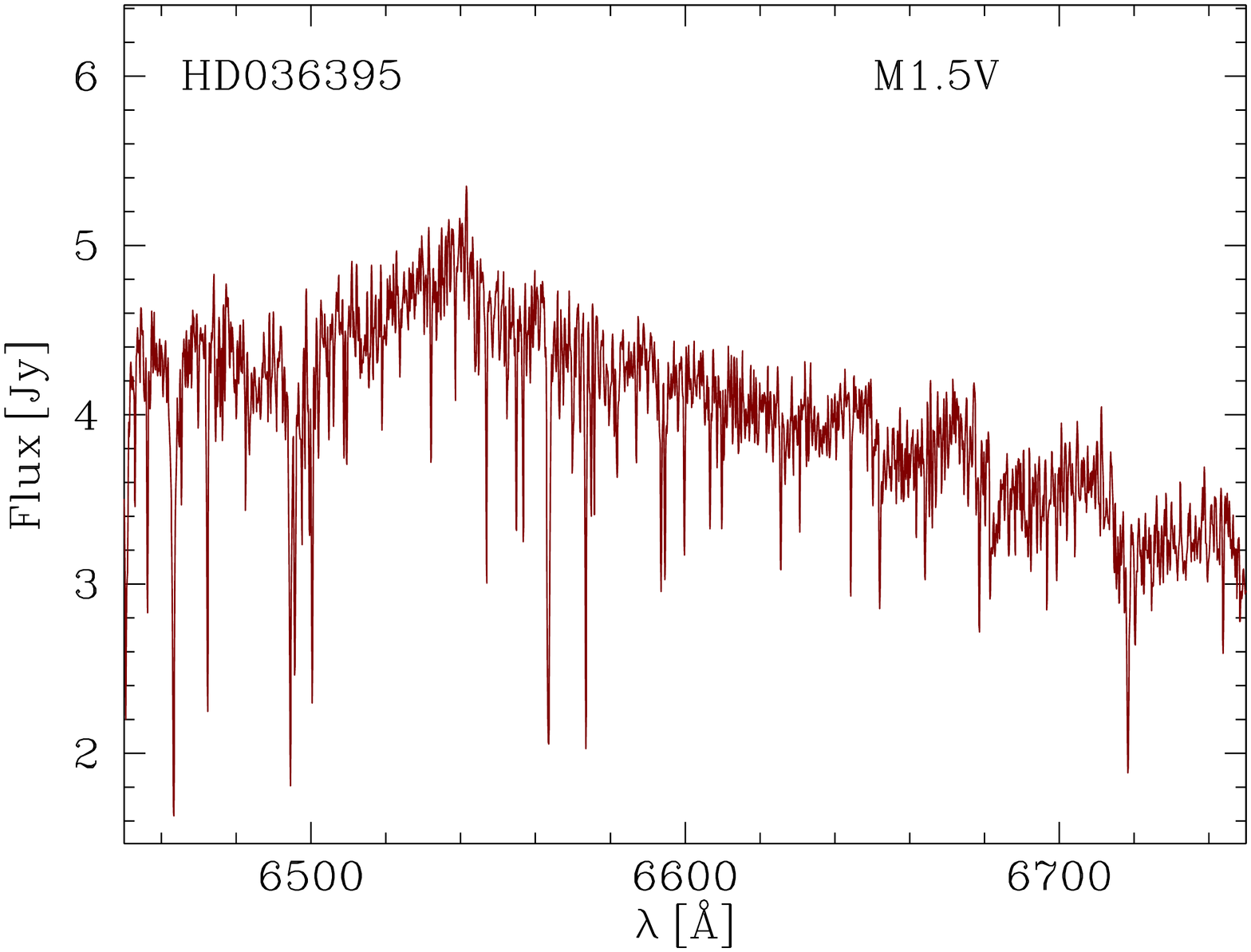}
\includegraphics[width=0.18\textwidth,angle=-90]{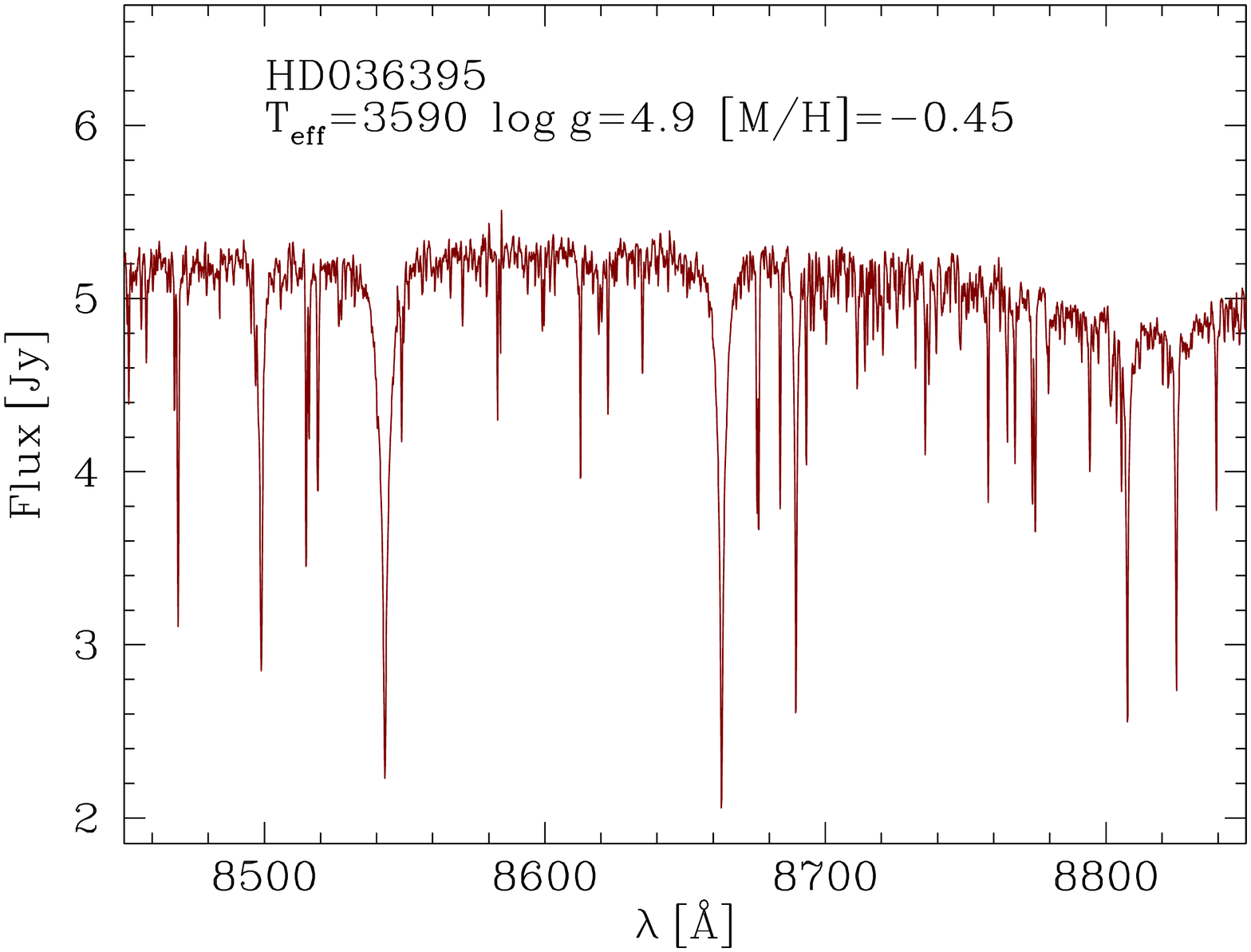}

\contcaption{7. Stars shown in this page are: HD033632, HD033904, HD034078, HD034255, HD034797, HD034816, HD035468, HD035497, HD035601, HD035961, HD036066, HD036130, HD036165 and HD036395.}
\end{figure*}

\begin{figure*}
\includegraphics[width=0.18\textwidth,angle=-90]{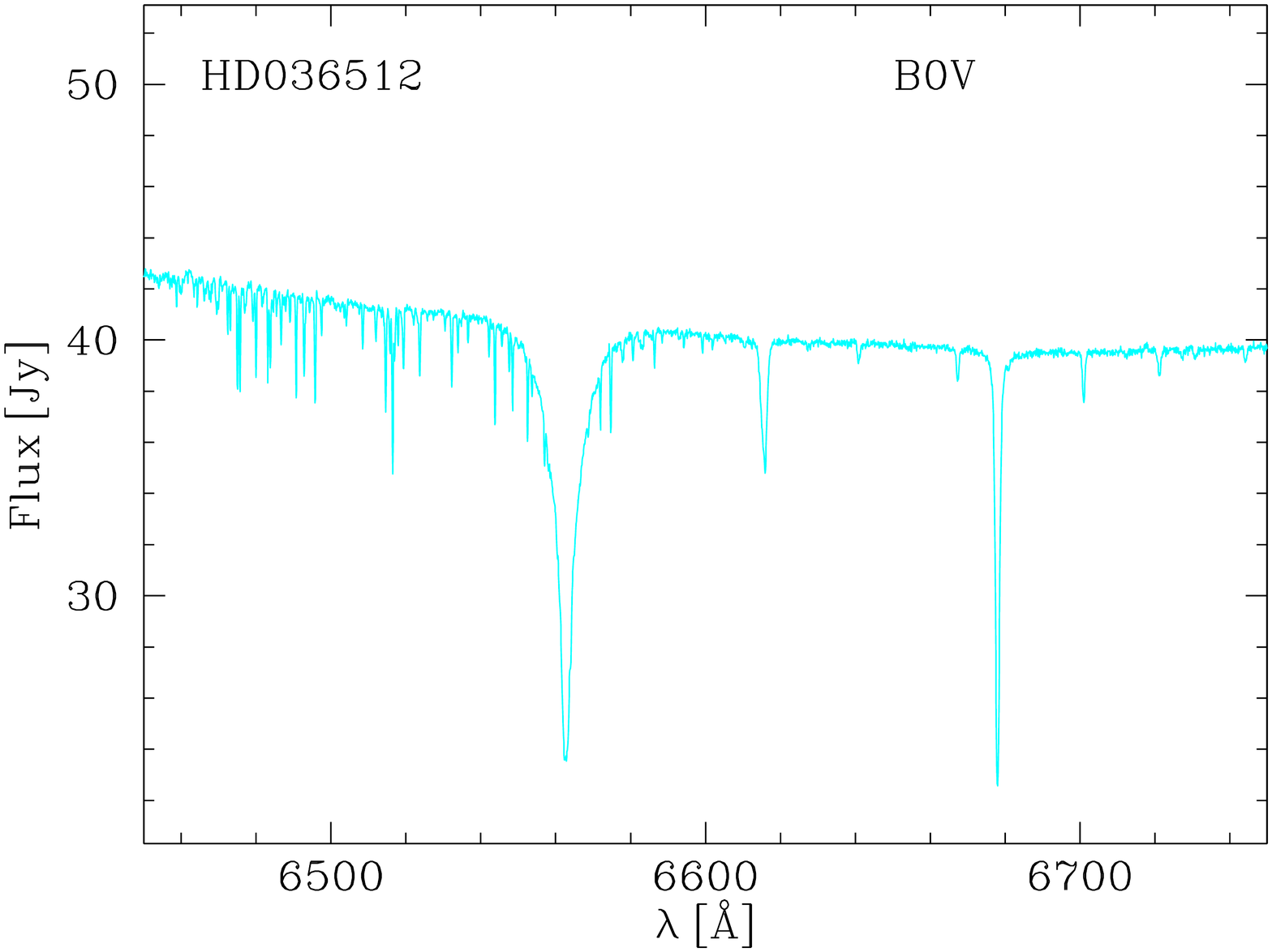}
\includegraphics[width=0.18\textwidth,angle=-90]{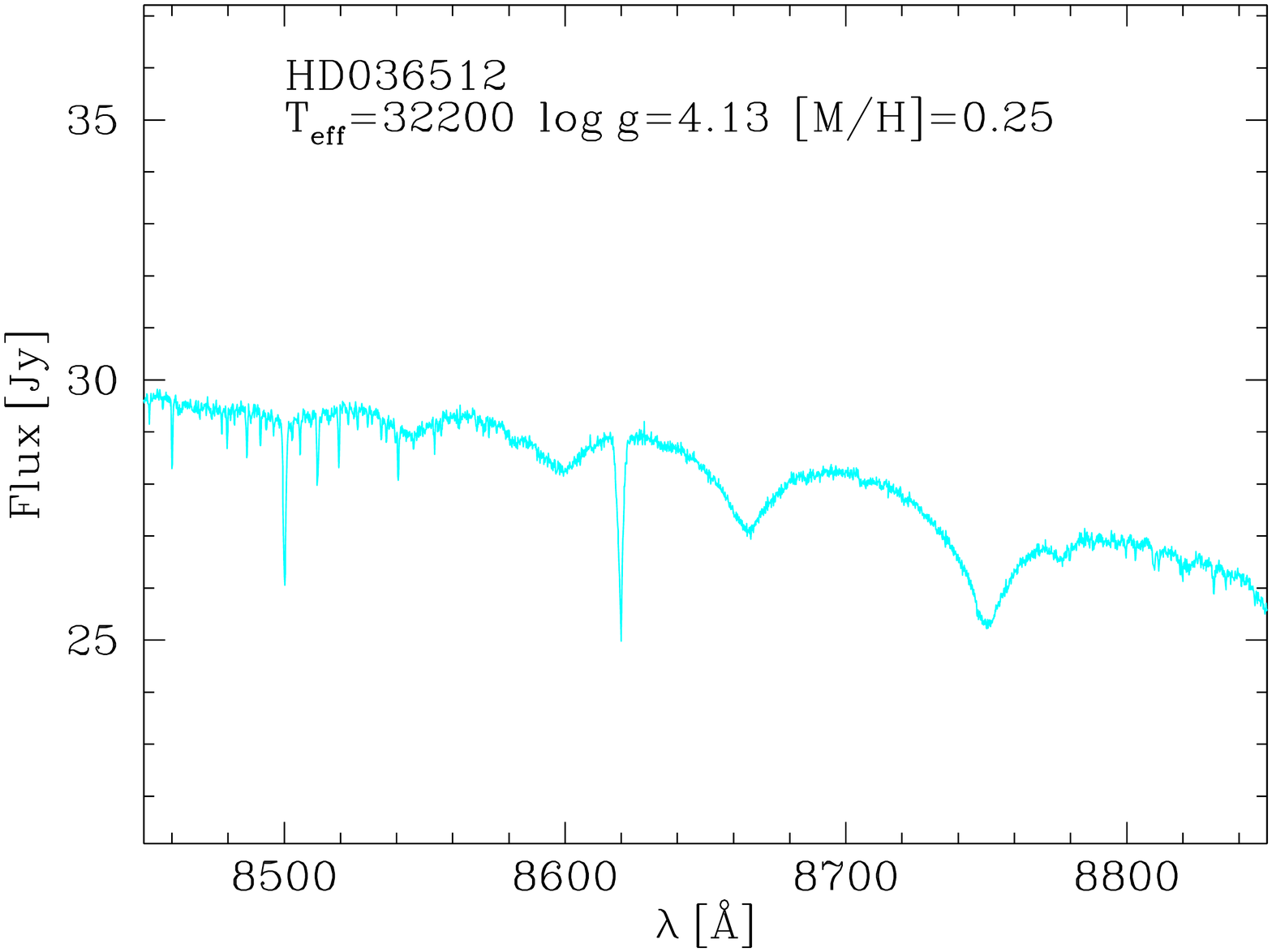}
\includegraphics[width=0.18\textwidth,angle=-90]{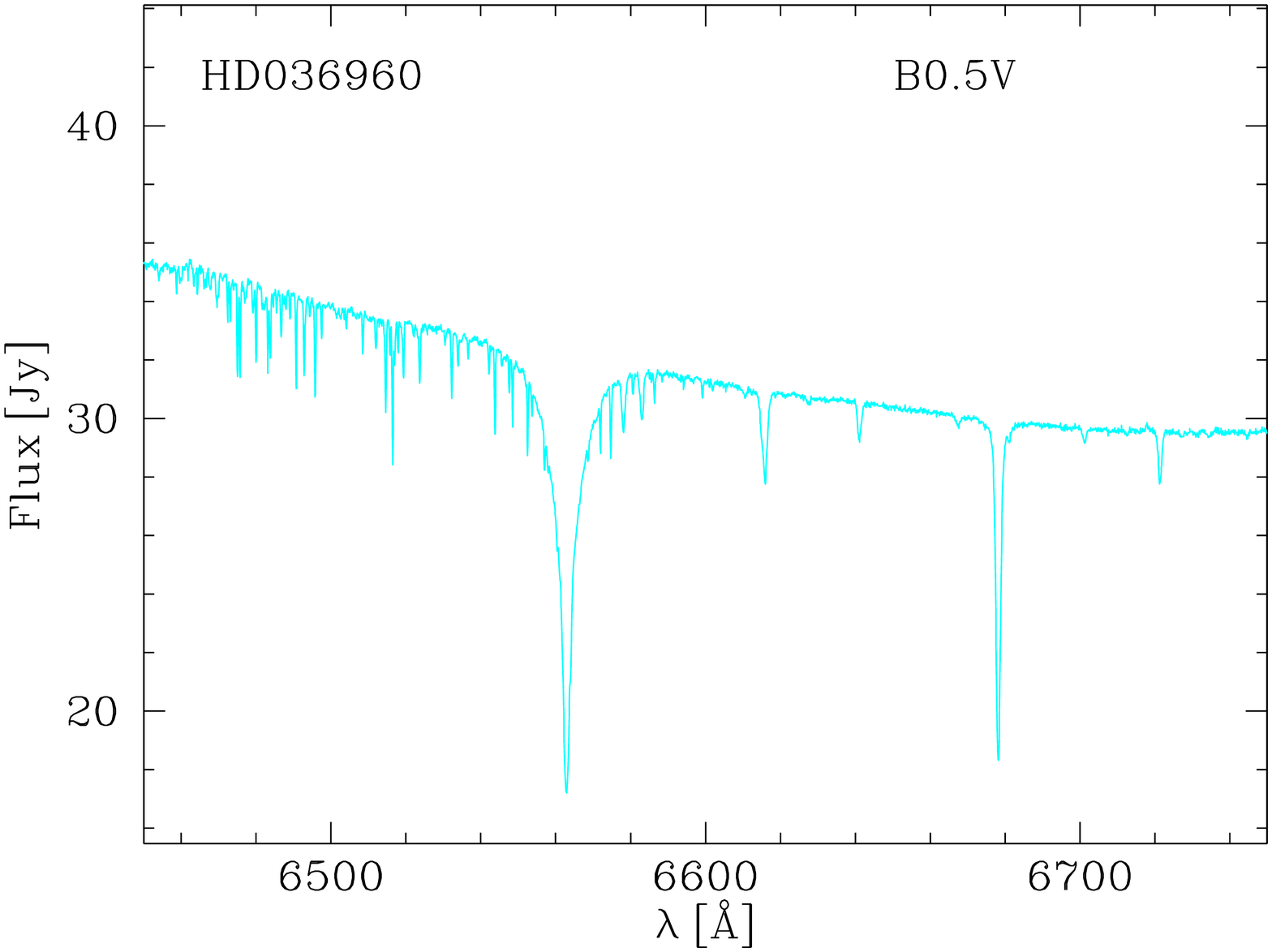}
\includegraphics[width=0.18\textwidth,angle=-90]{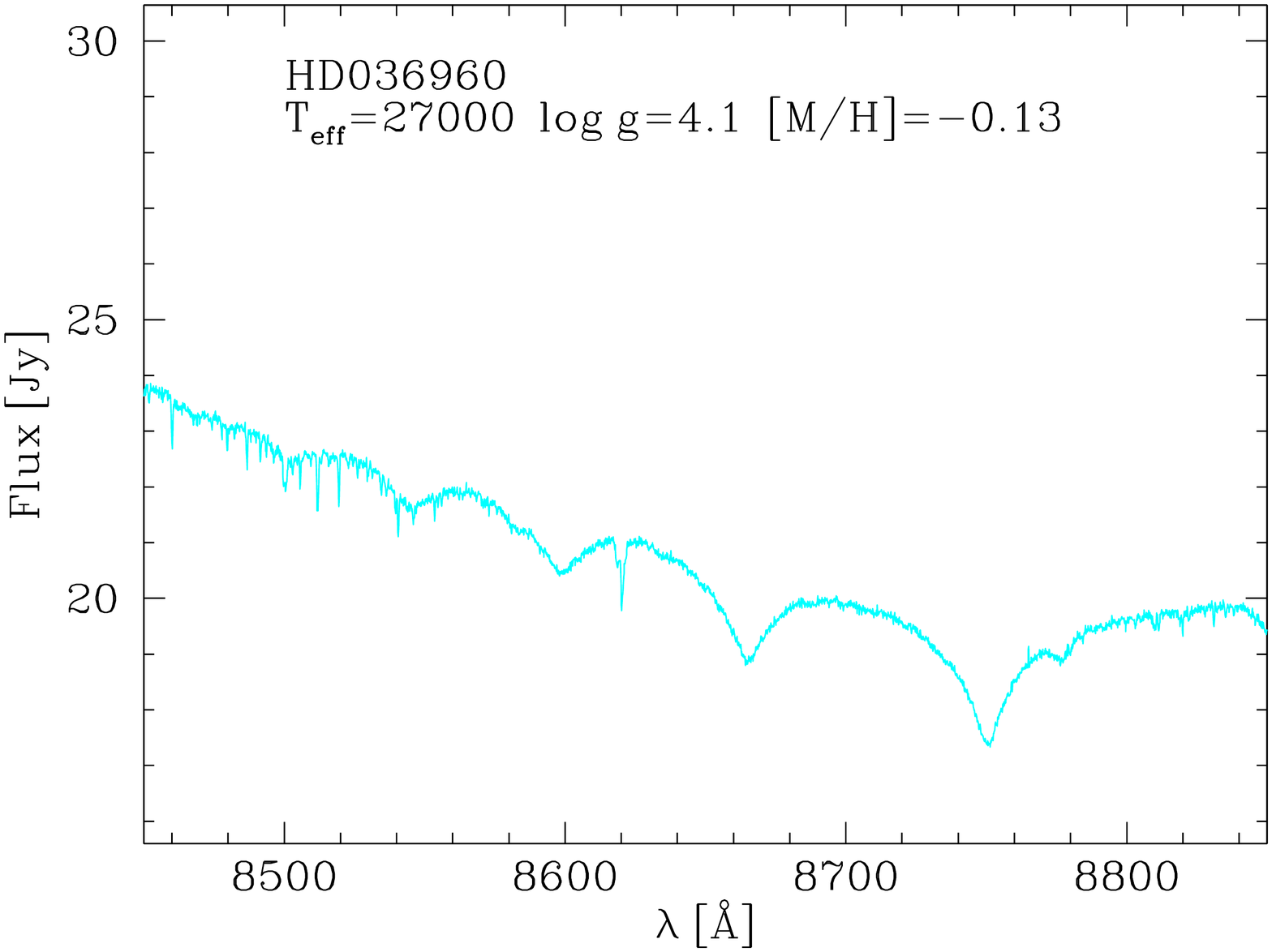}
\includegraphics[width=0.18\textwidth,angle=-90]{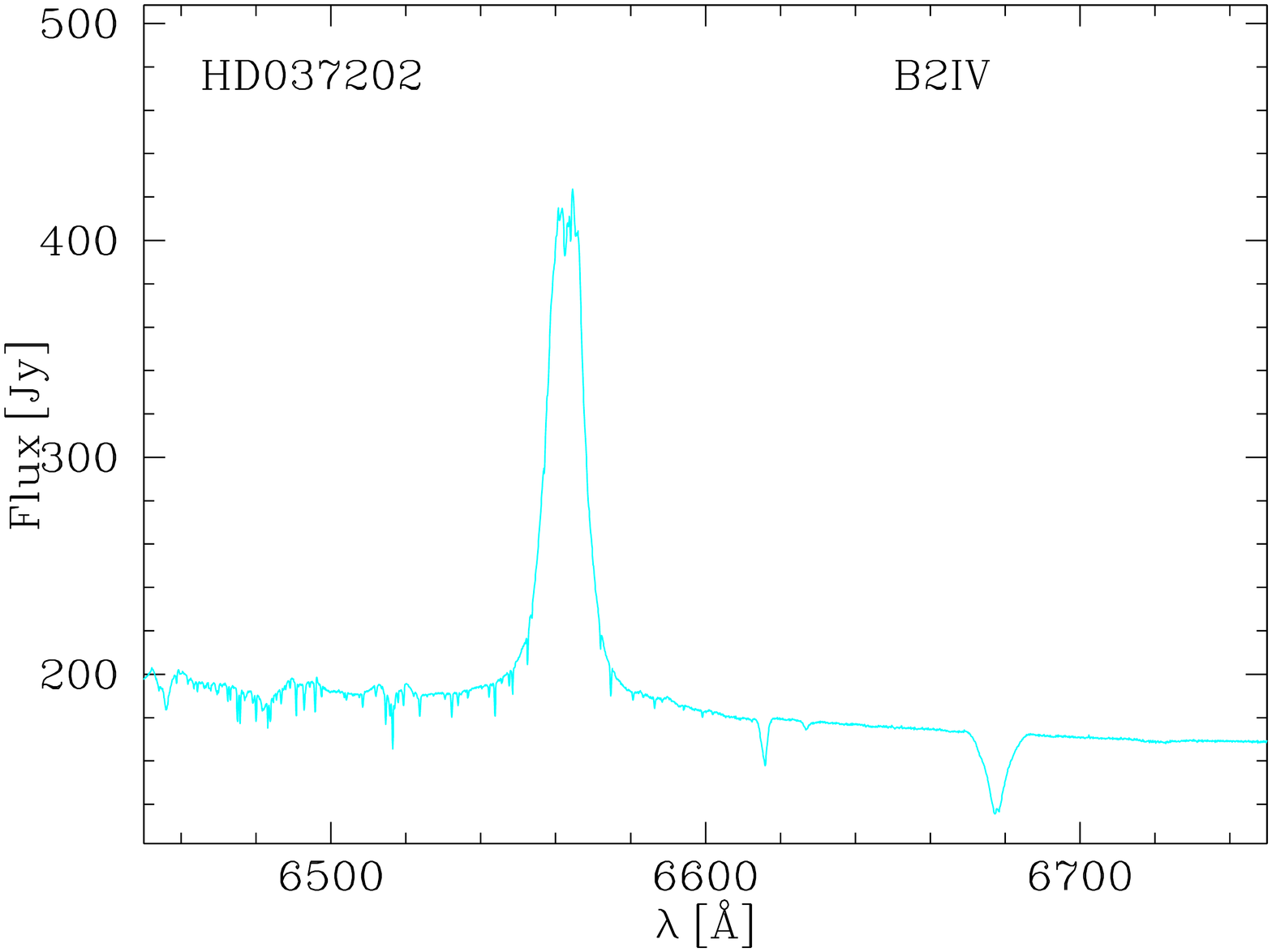}
\includegraphics[width=0.18\textwidth,angle=-90]{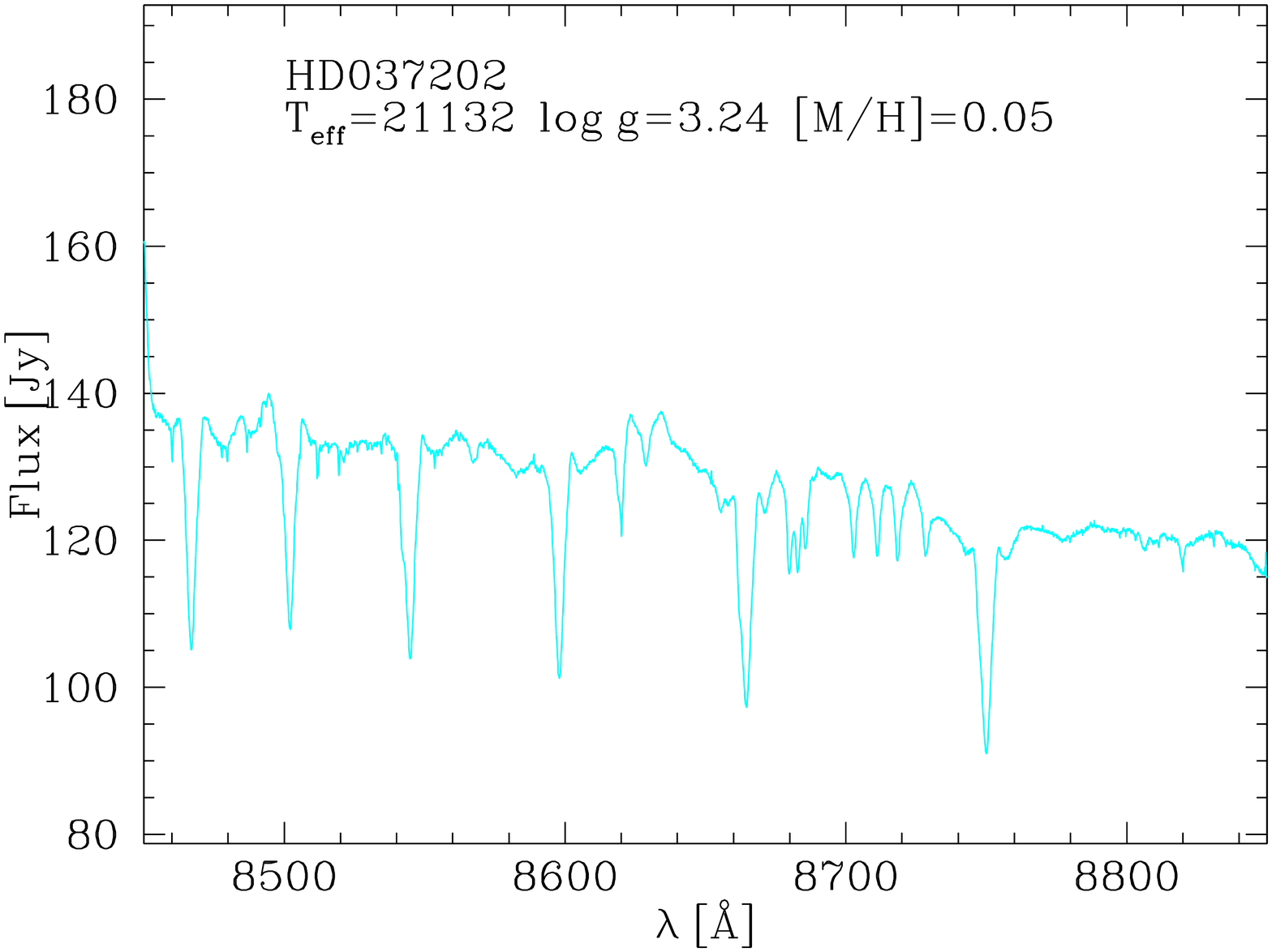}
\includegraphics[width=0.18\textwidth,angle=-90]{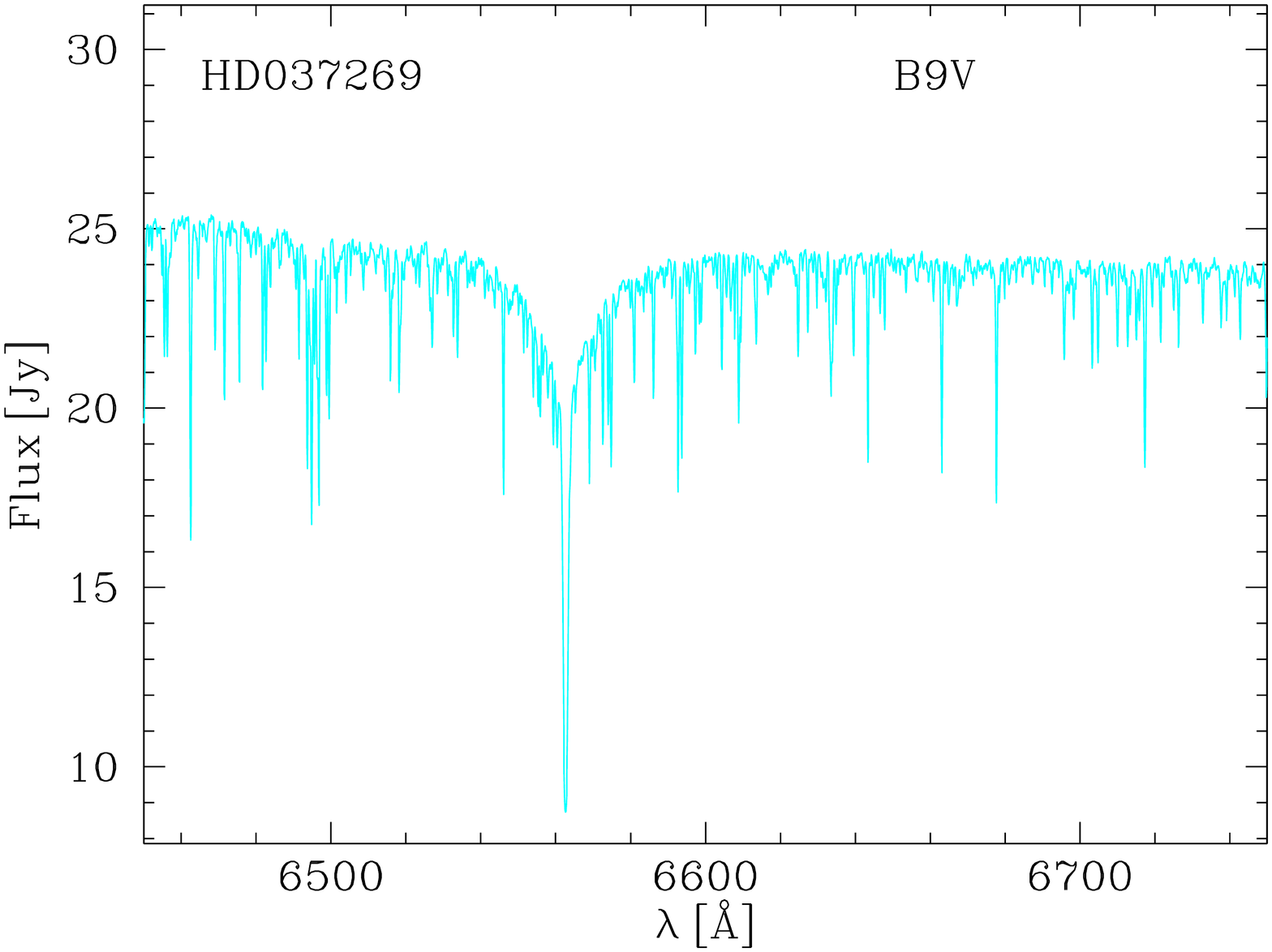}
\includegraphics[width=0.18\textwidth,angle=-90]{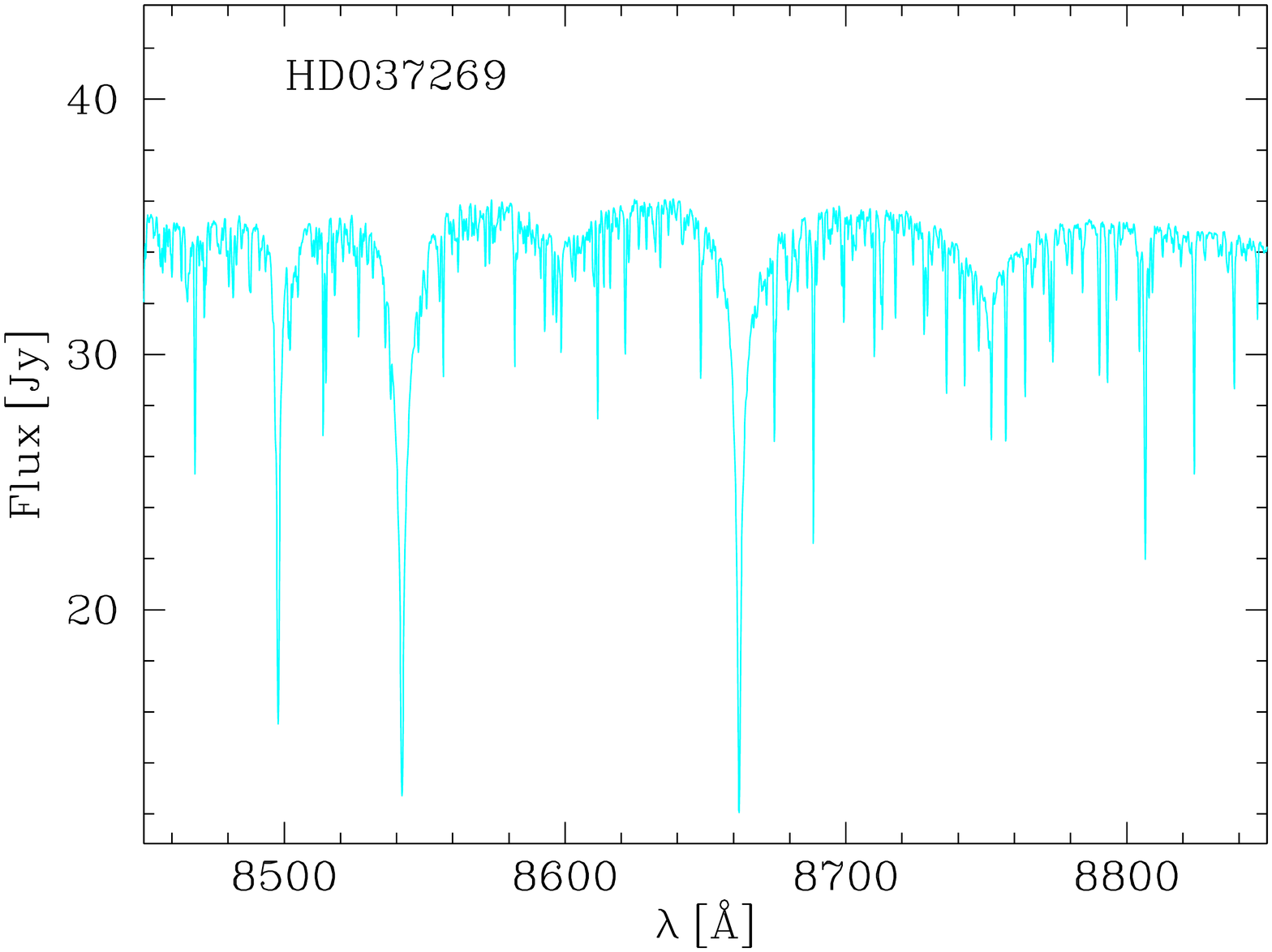}
\includegraphics[width=0.18\textwidth,angle=-90]{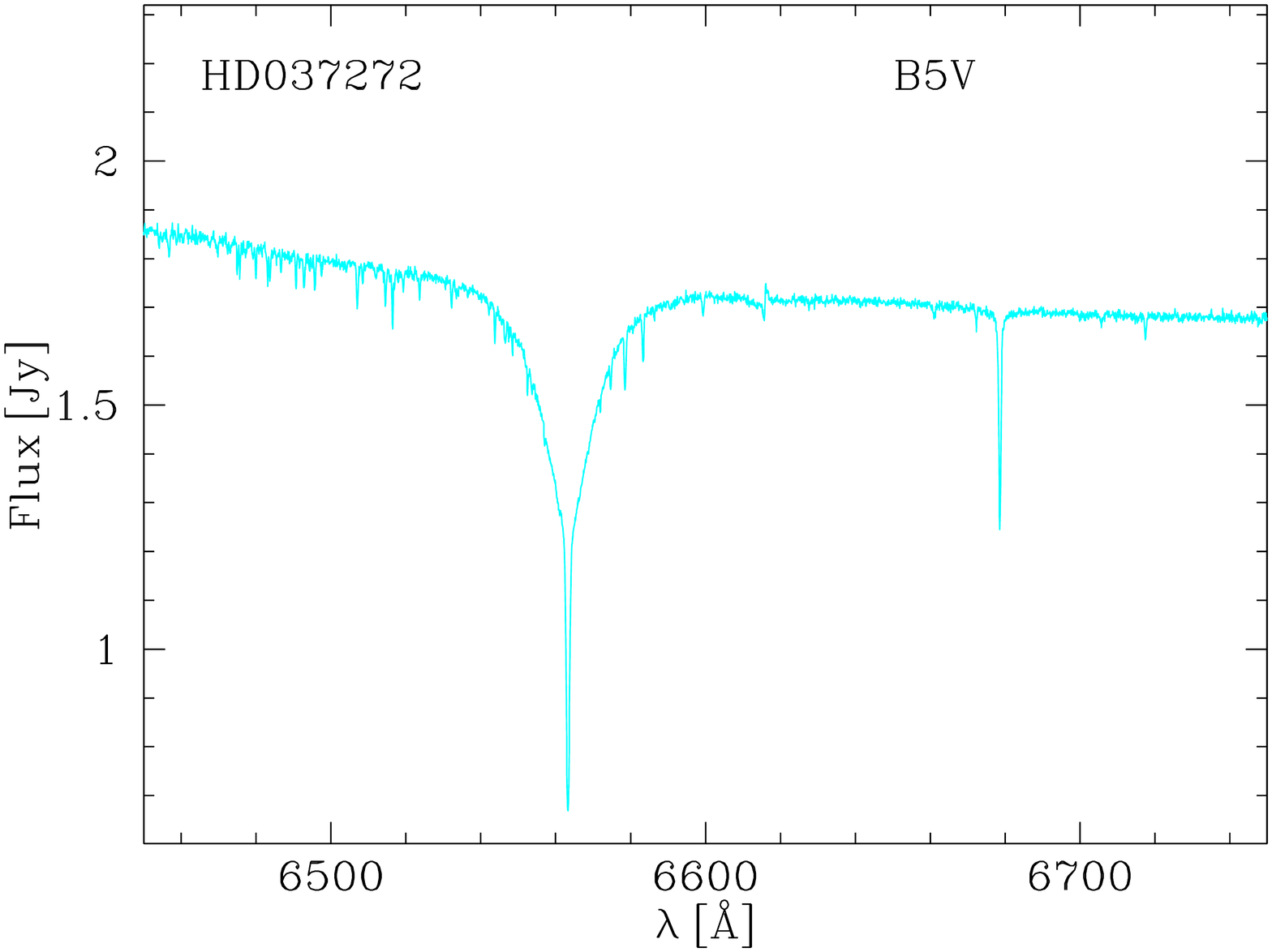}
\includegraphics[width=0.18\textwidth,angle=-90]{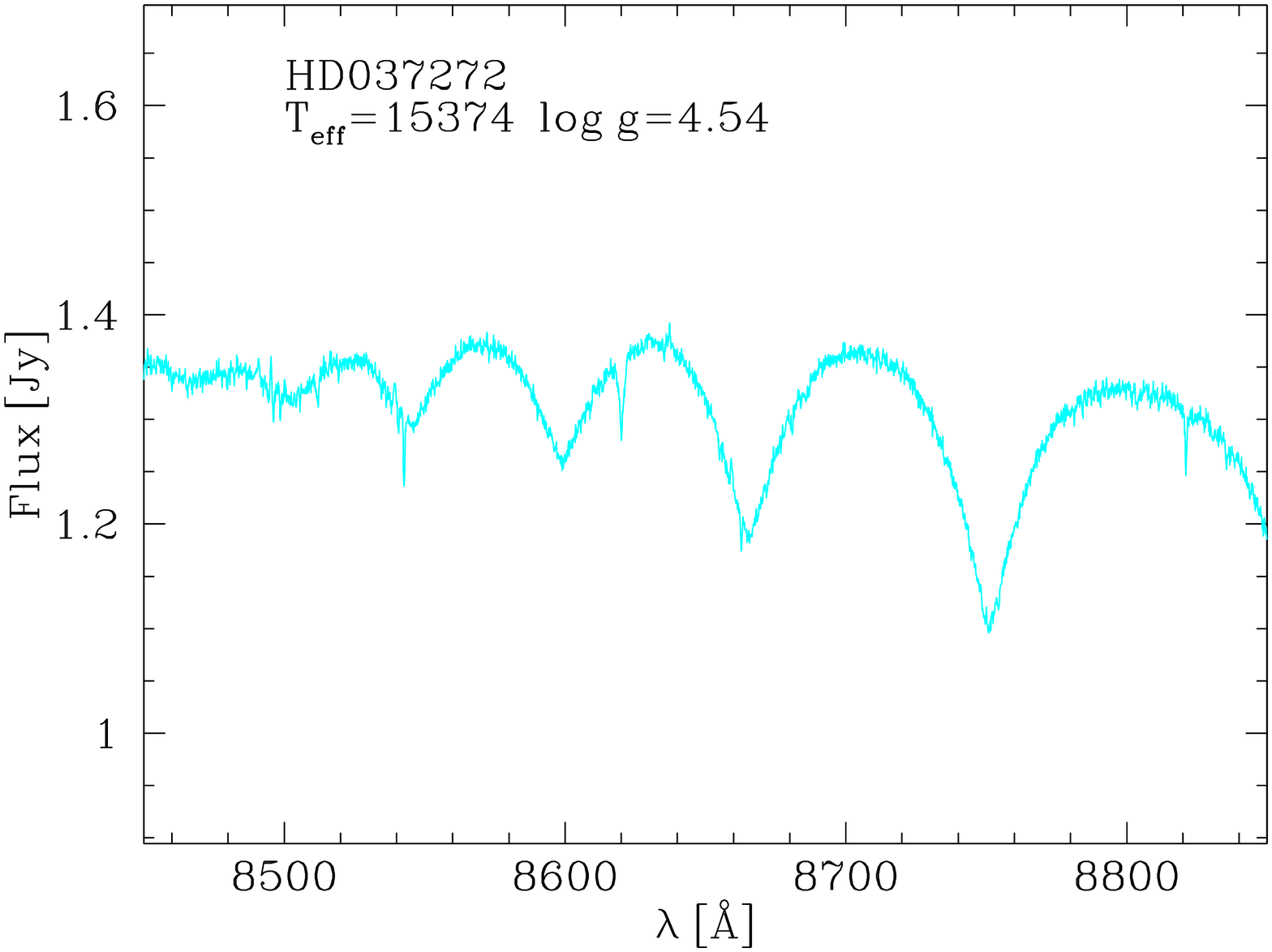}
\includegraphics[width=0.18\textwidth,angle=-90]{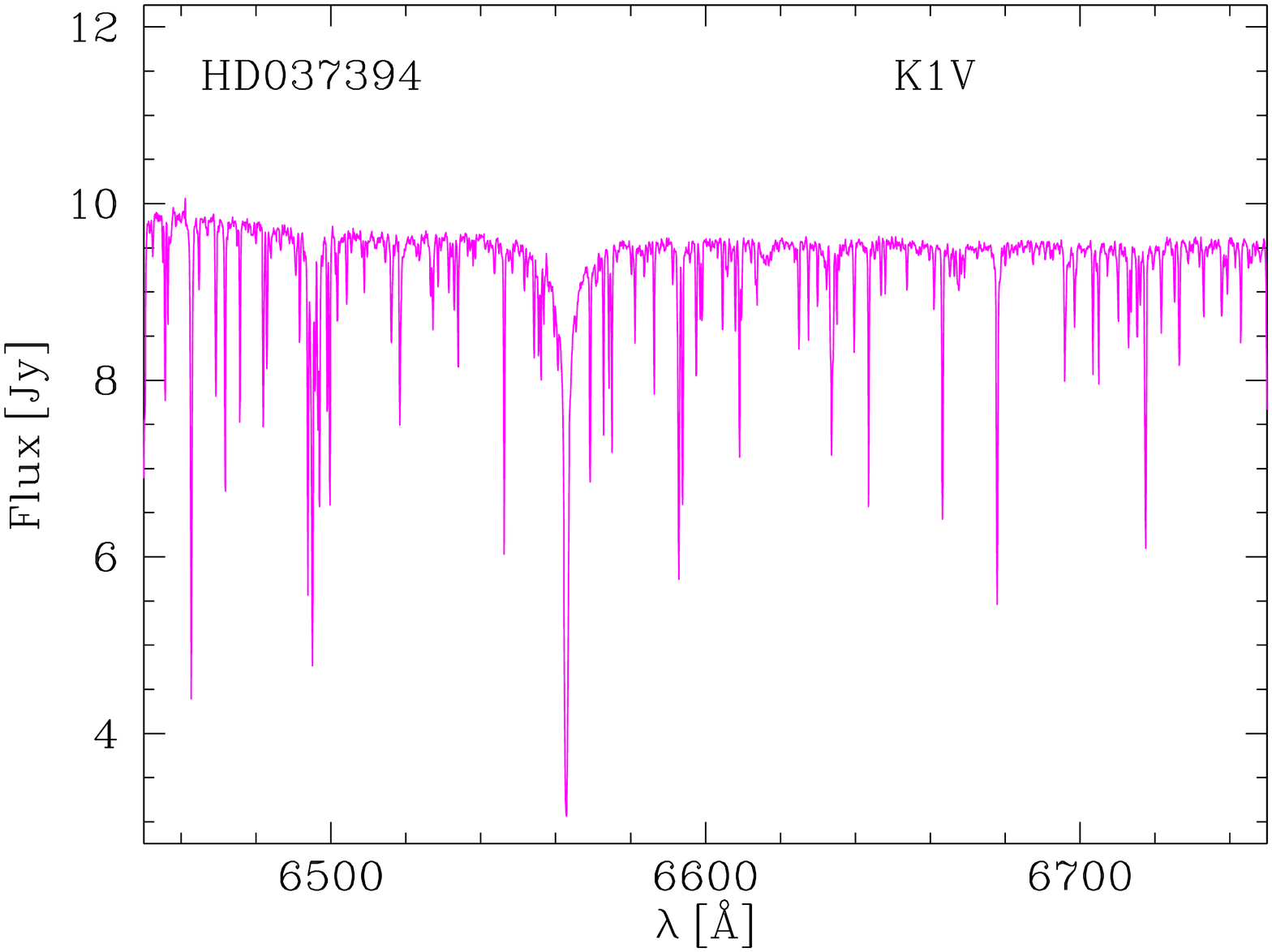}
\includegraphics[width=0.18\textwidth,angle=-90]{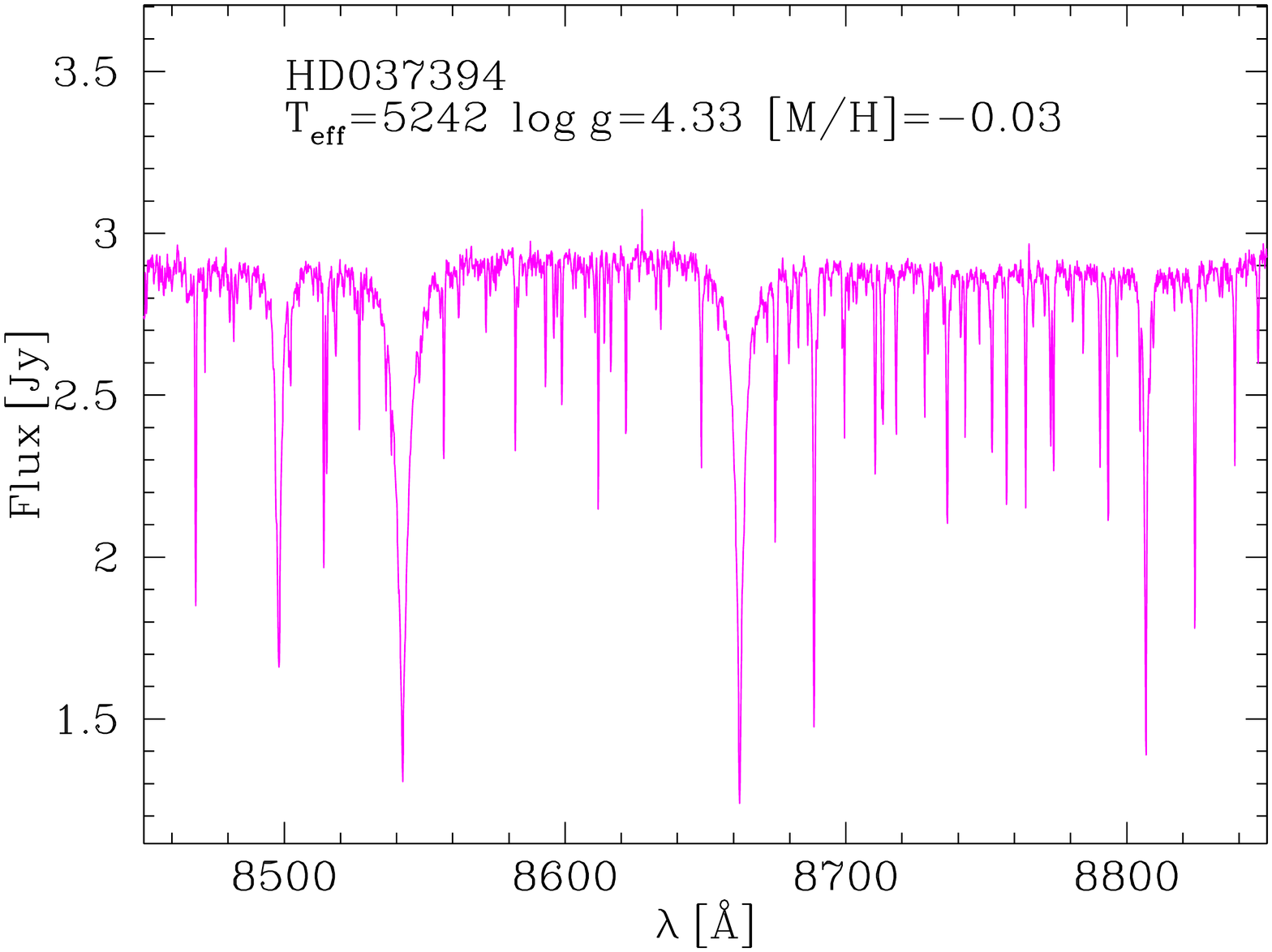}
\includegraphics[width=0.18\textwidth,angle=-90]{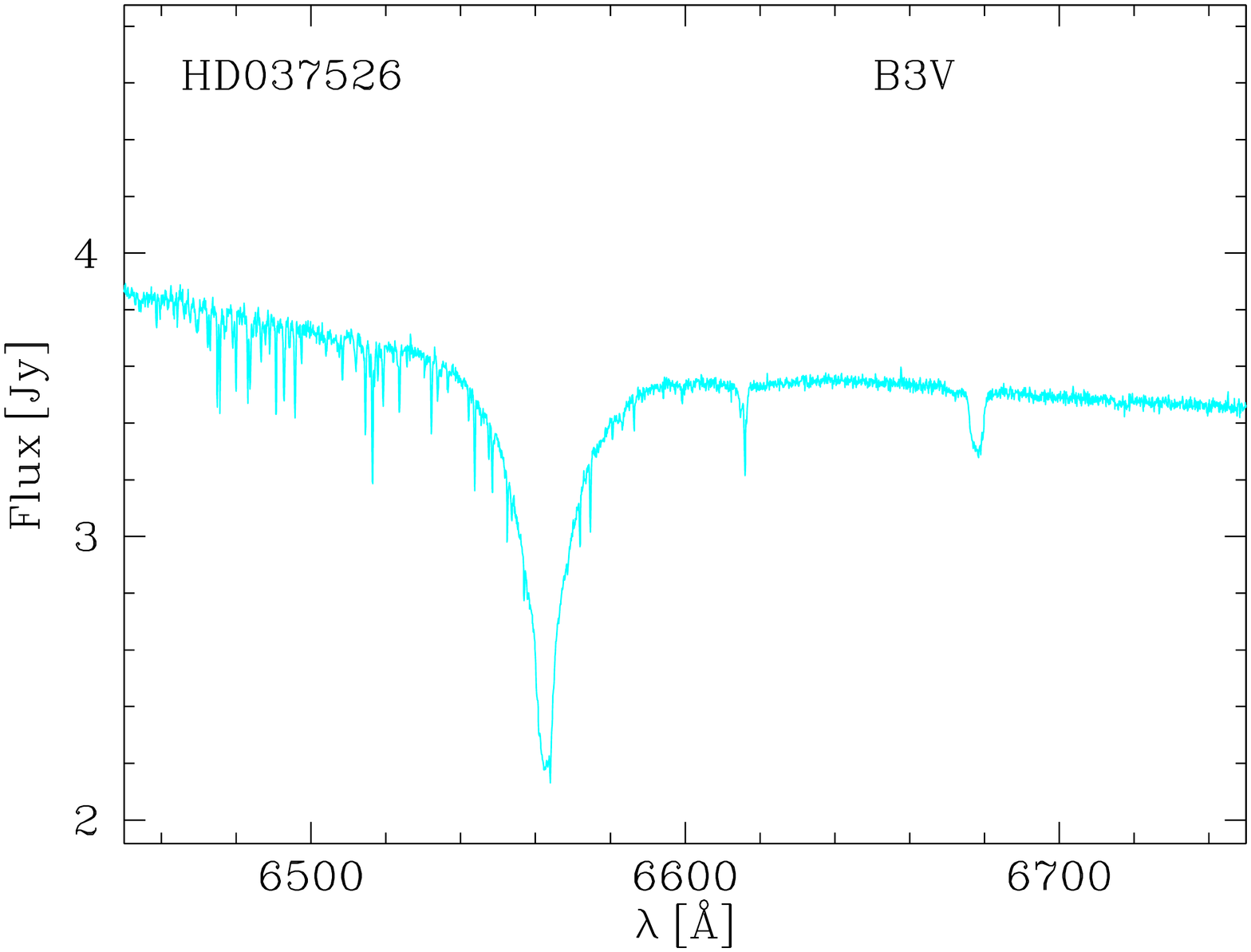}
\includegraphics[width=0.18\textwidth,angle=-90]{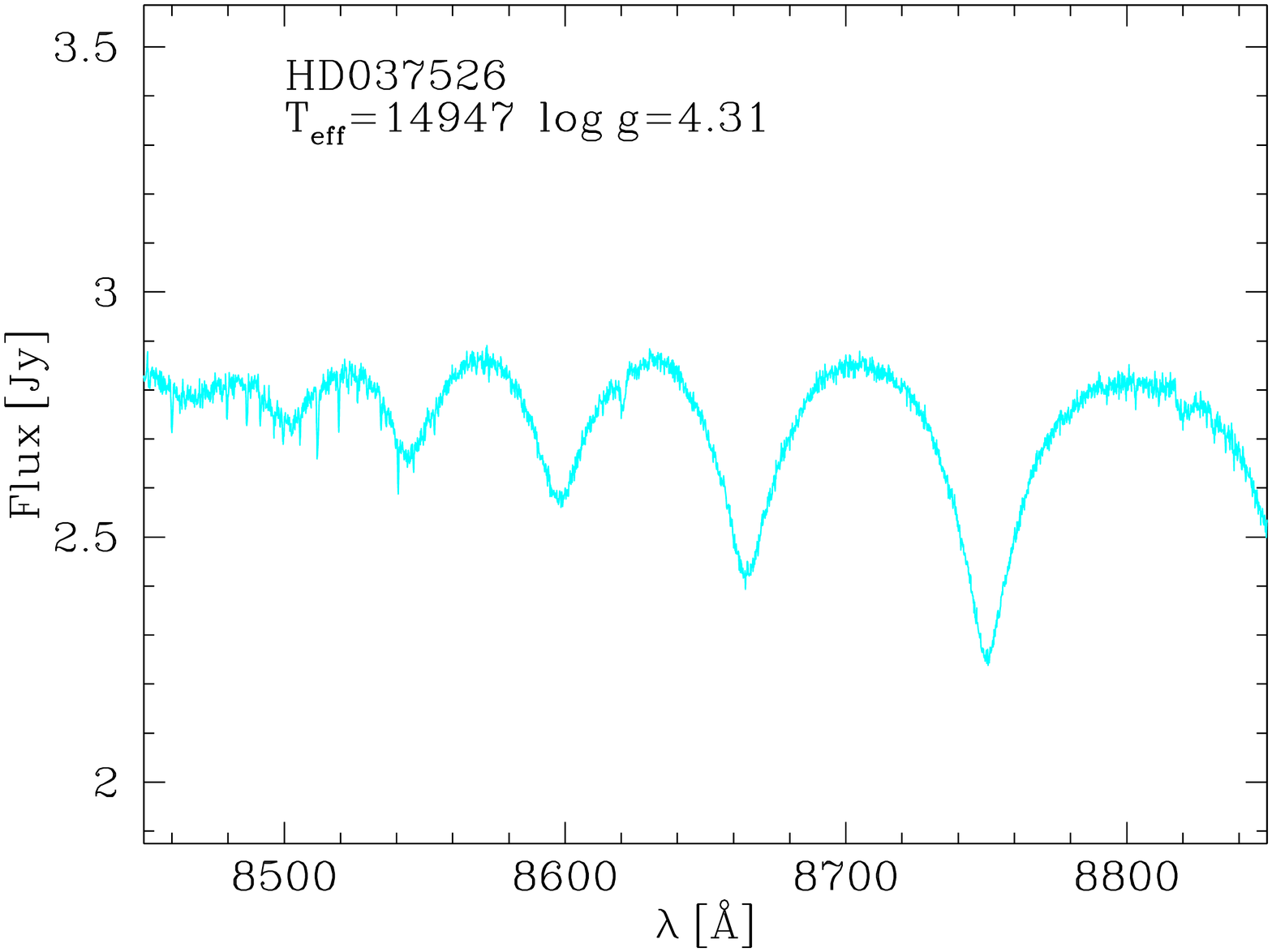}
\includegraphics[width=0.18\textwidth,angle=-90]{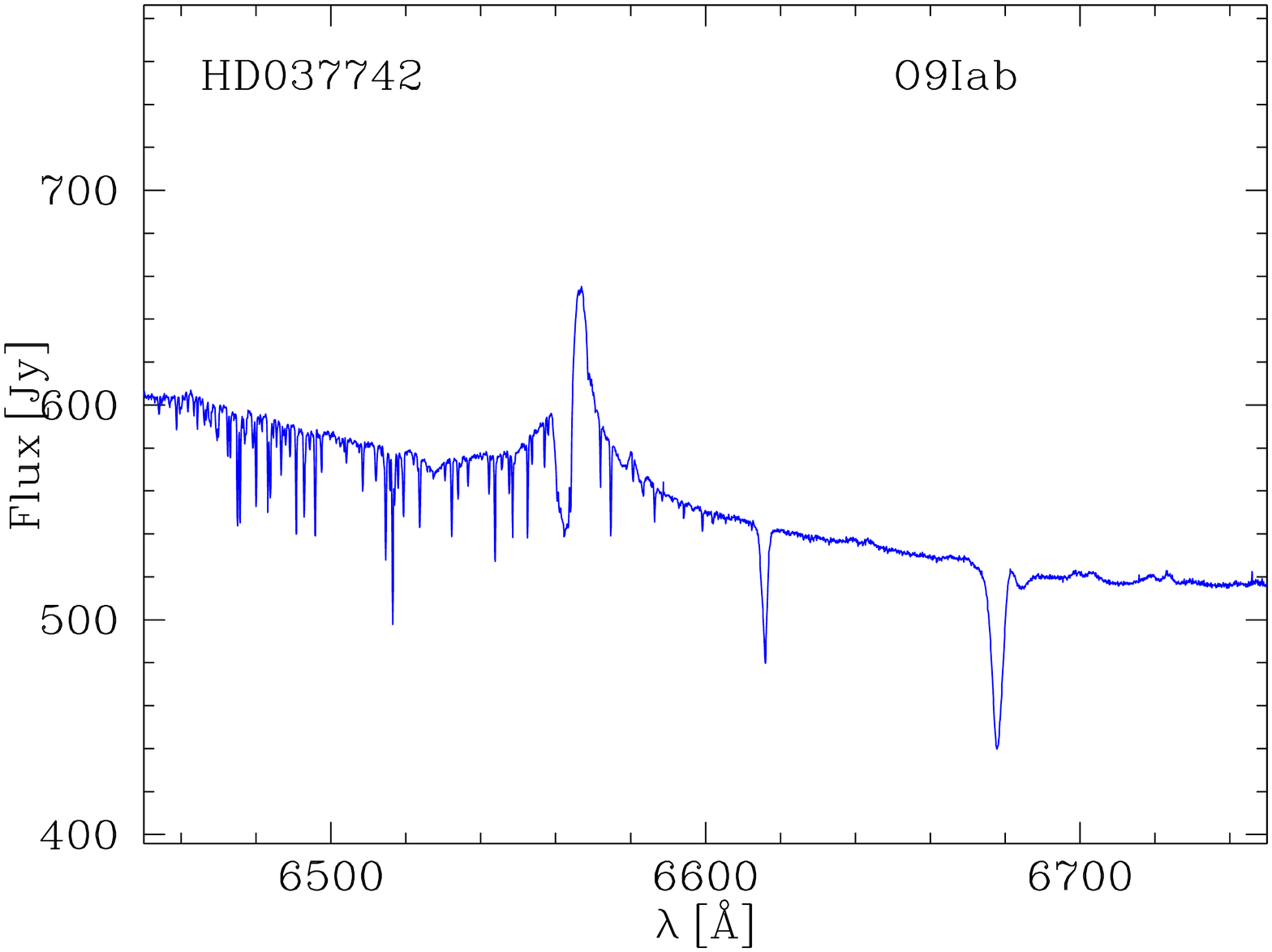}
\includegraphics[width=0.18\textwidth,angle=-90]{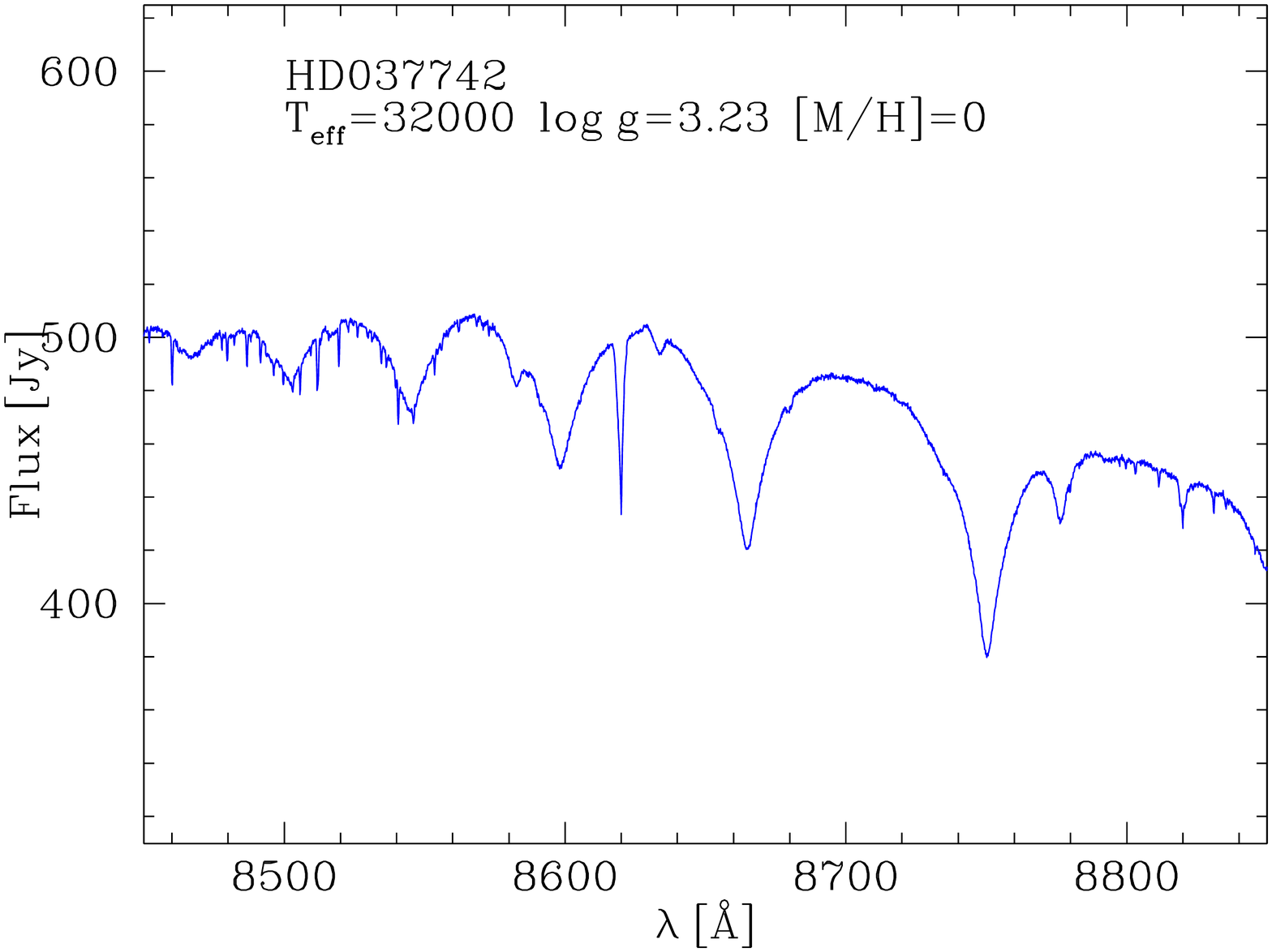}
\includegraphics[width=0.18\textwidth,angle=-90]{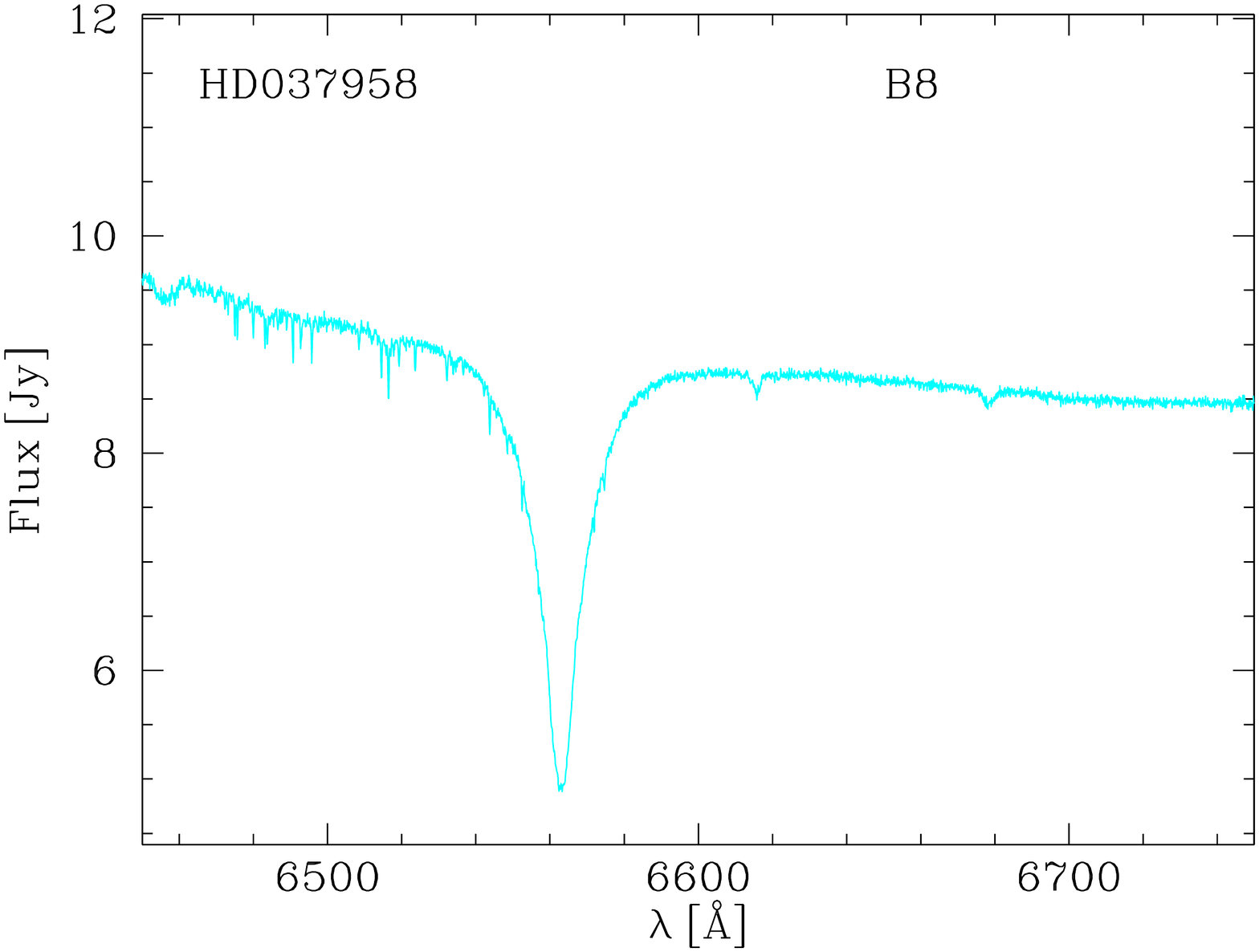}
\includegraphics[width=0.18\textwidth,angle=-90]{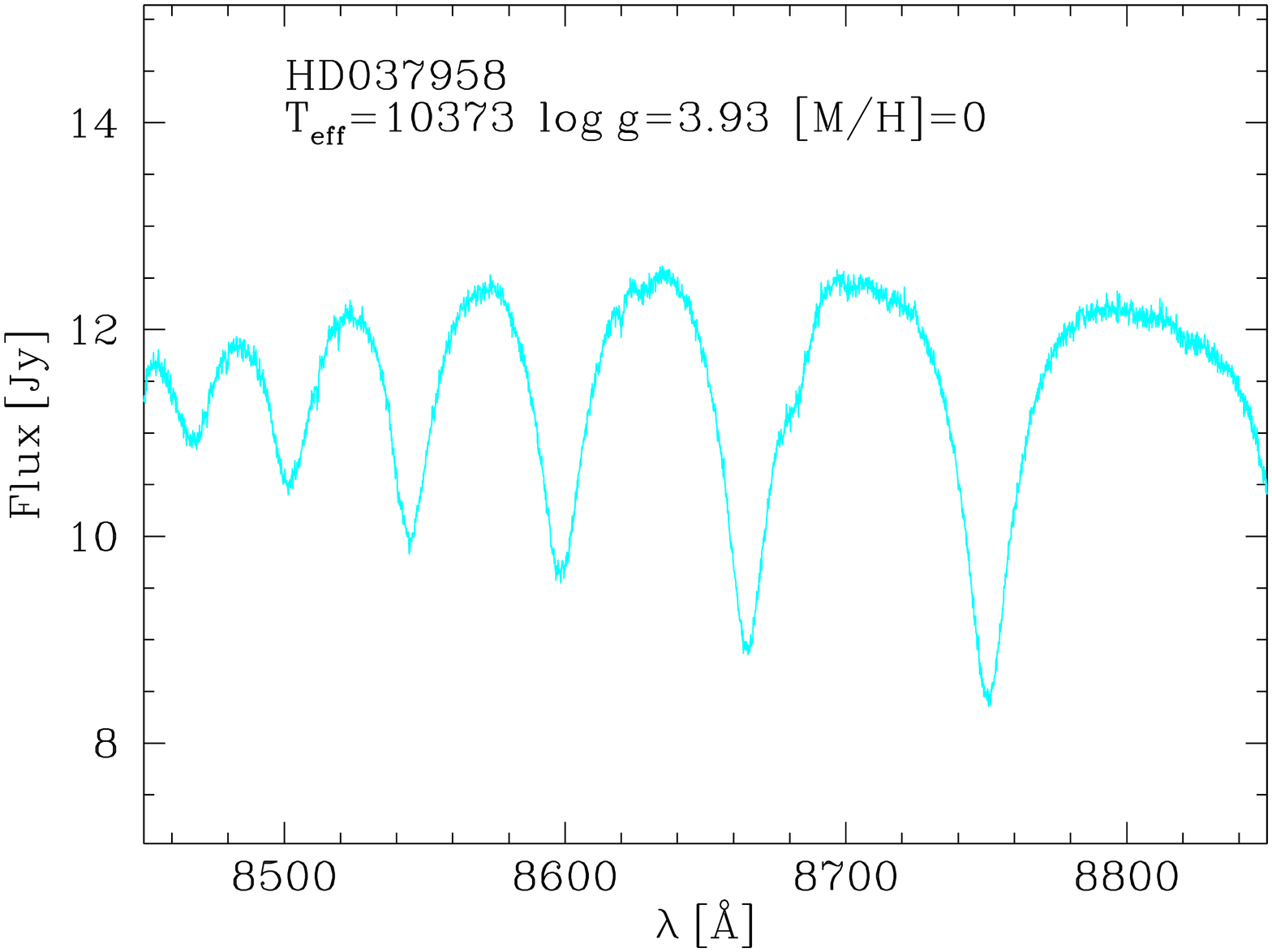}
\includegraphics[width=0.18\textwidth,angle=-90]{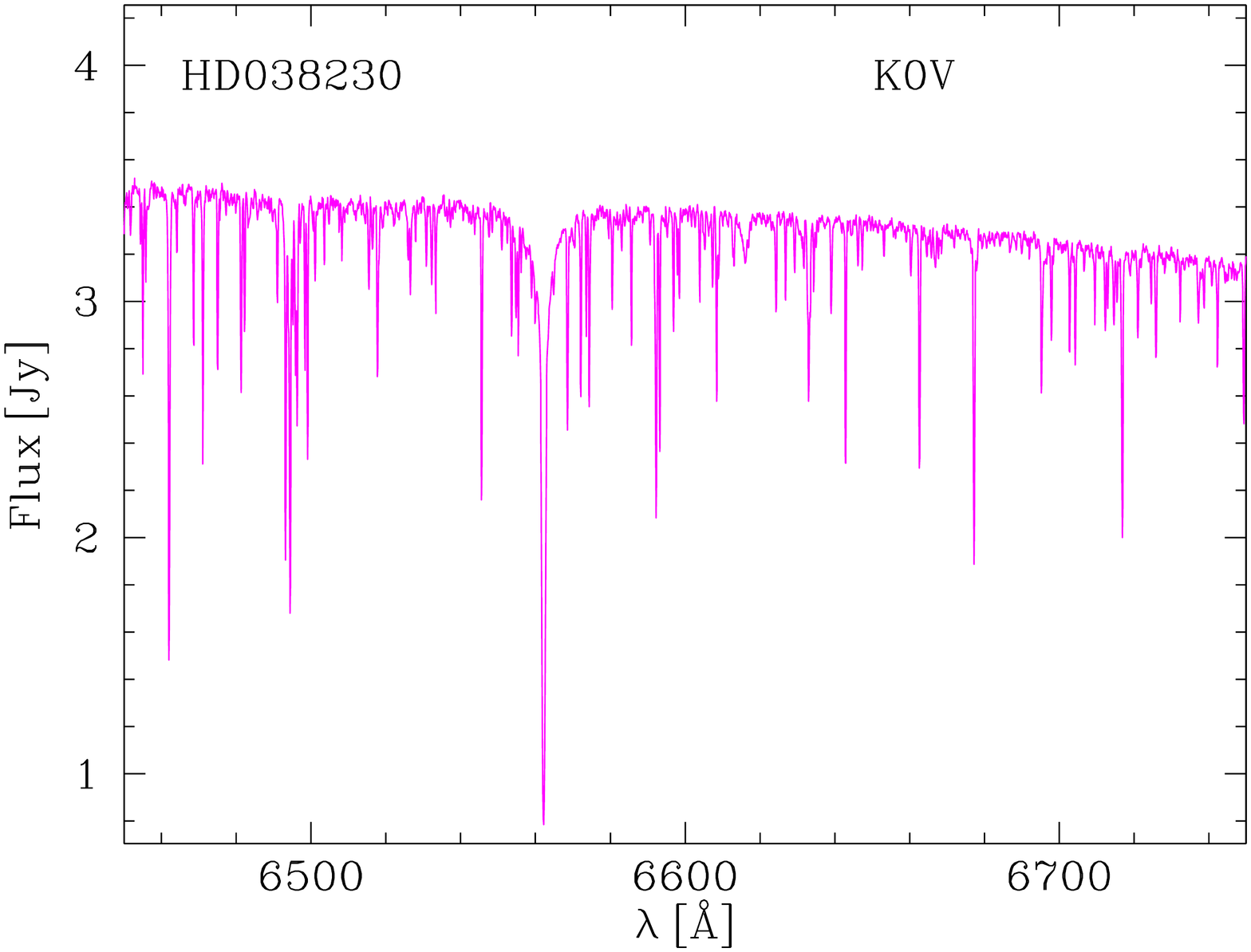}
\includegraphics[width=0.18\textwidth,angle=-90]{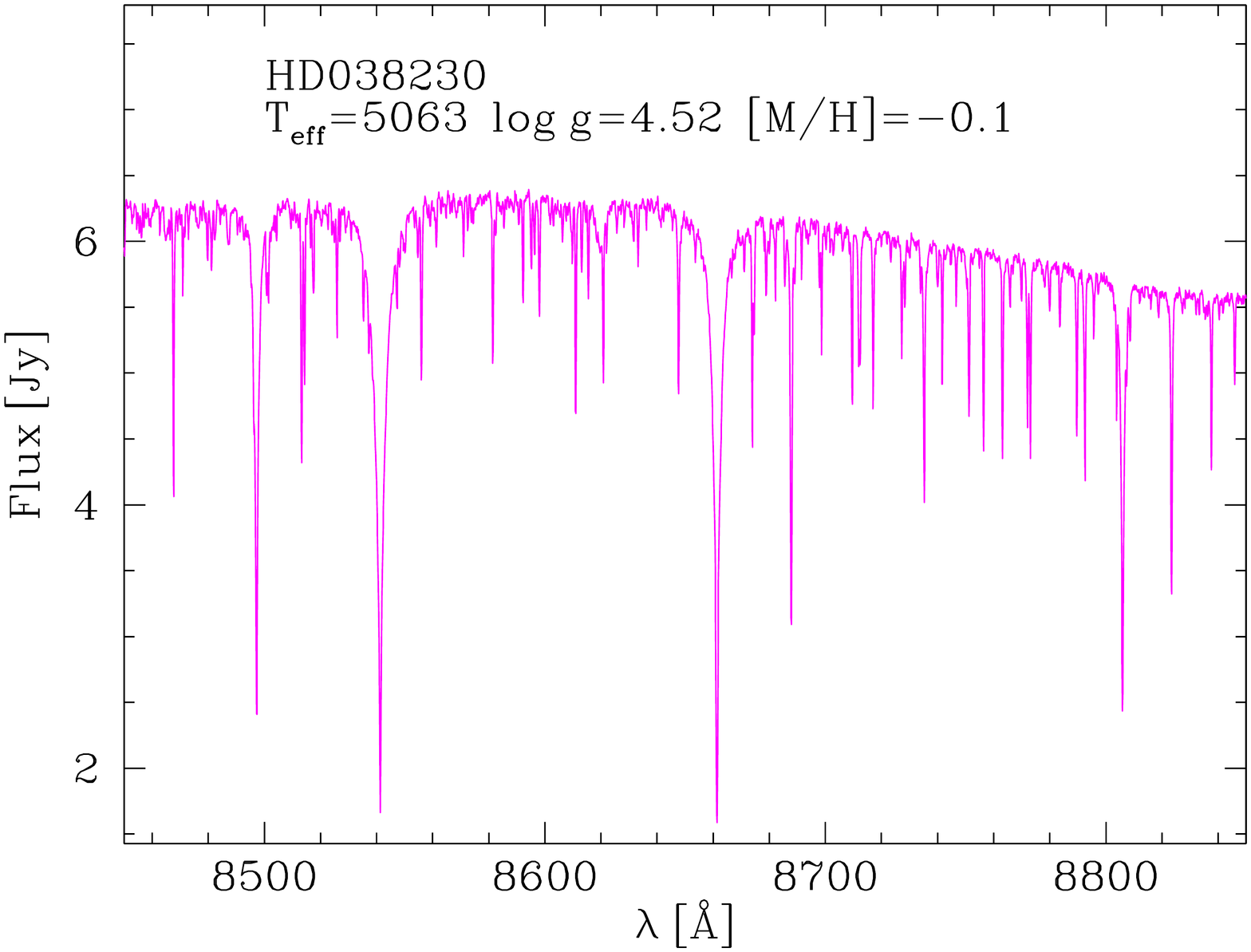}
\includegraphics[width=0.18\textwidth,angle=-90]{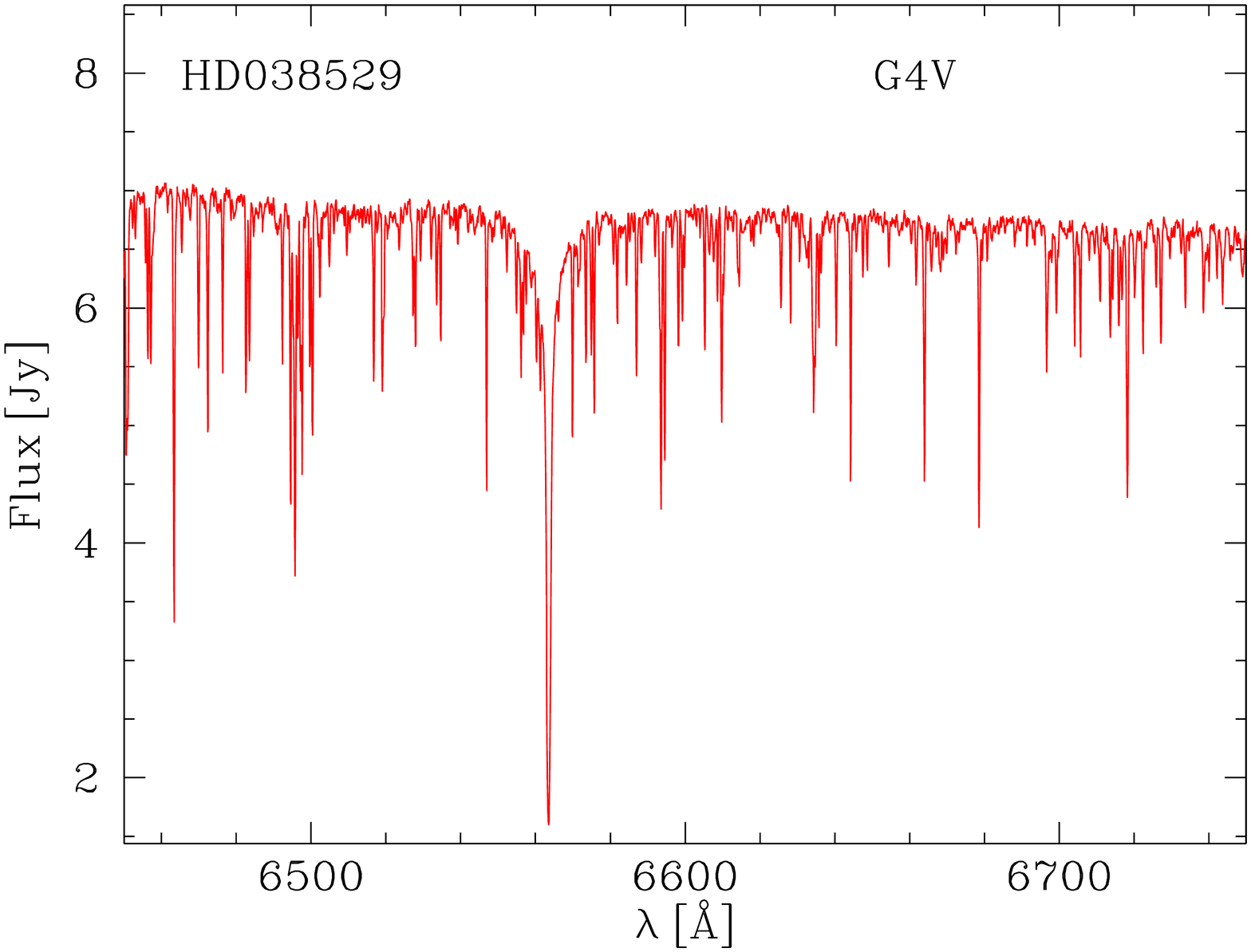}
\includegraphics[width=0.18\textwidth,angle=-90]{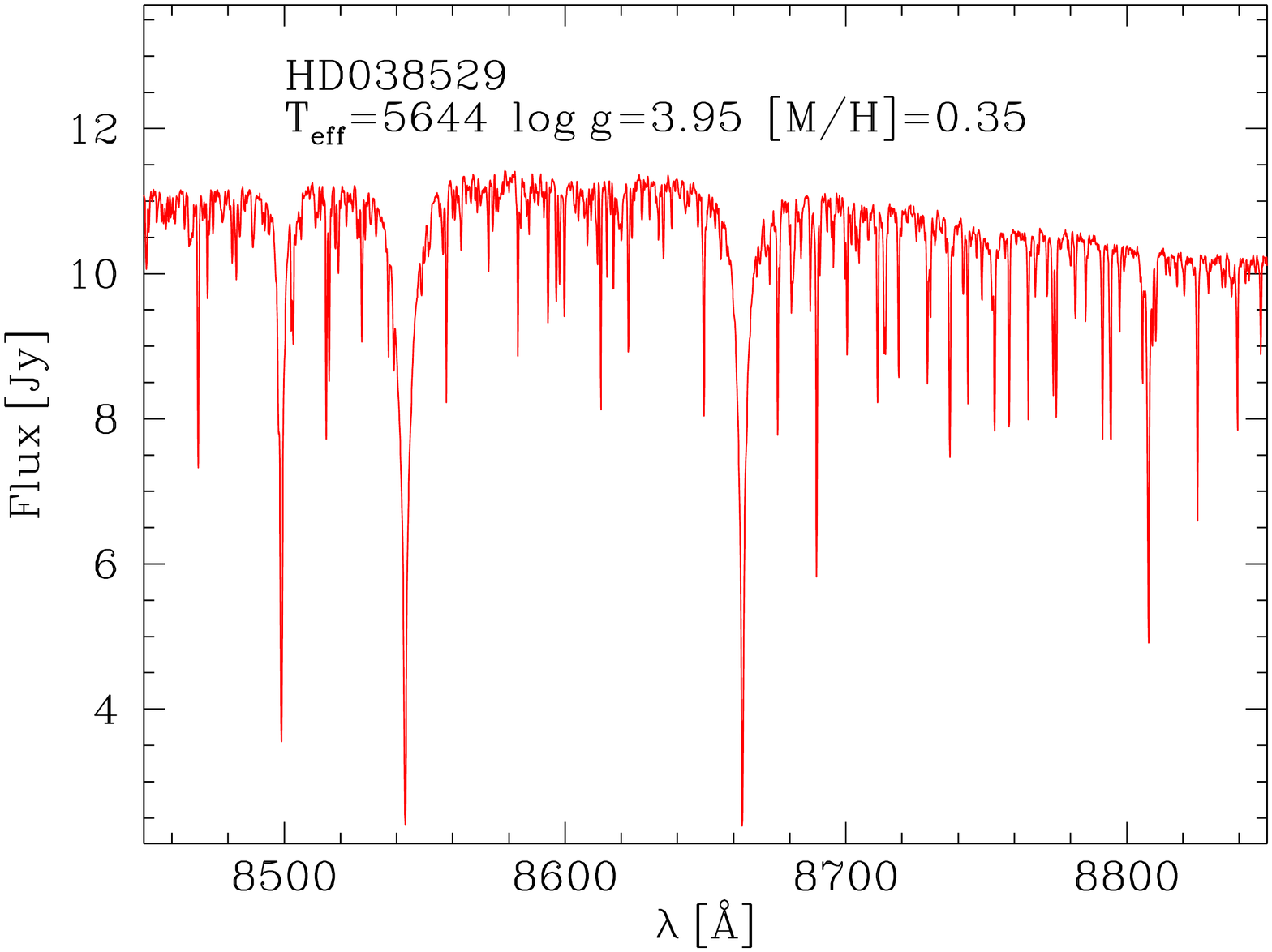}
\includegraphics[width=0.18\textwidth,angle=-90]{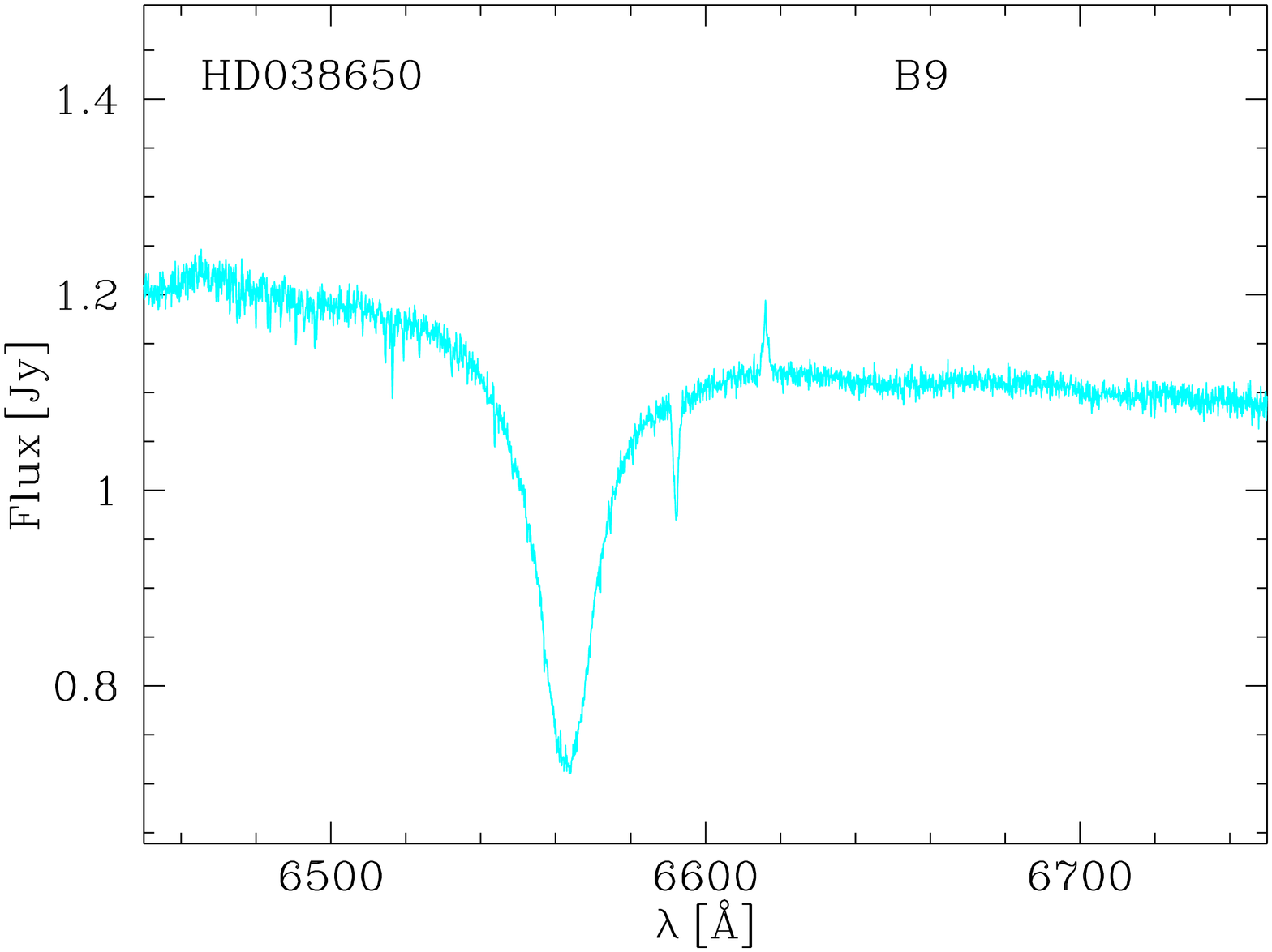}
\includegraphics[width=0.18\textwidth,angle=-90]{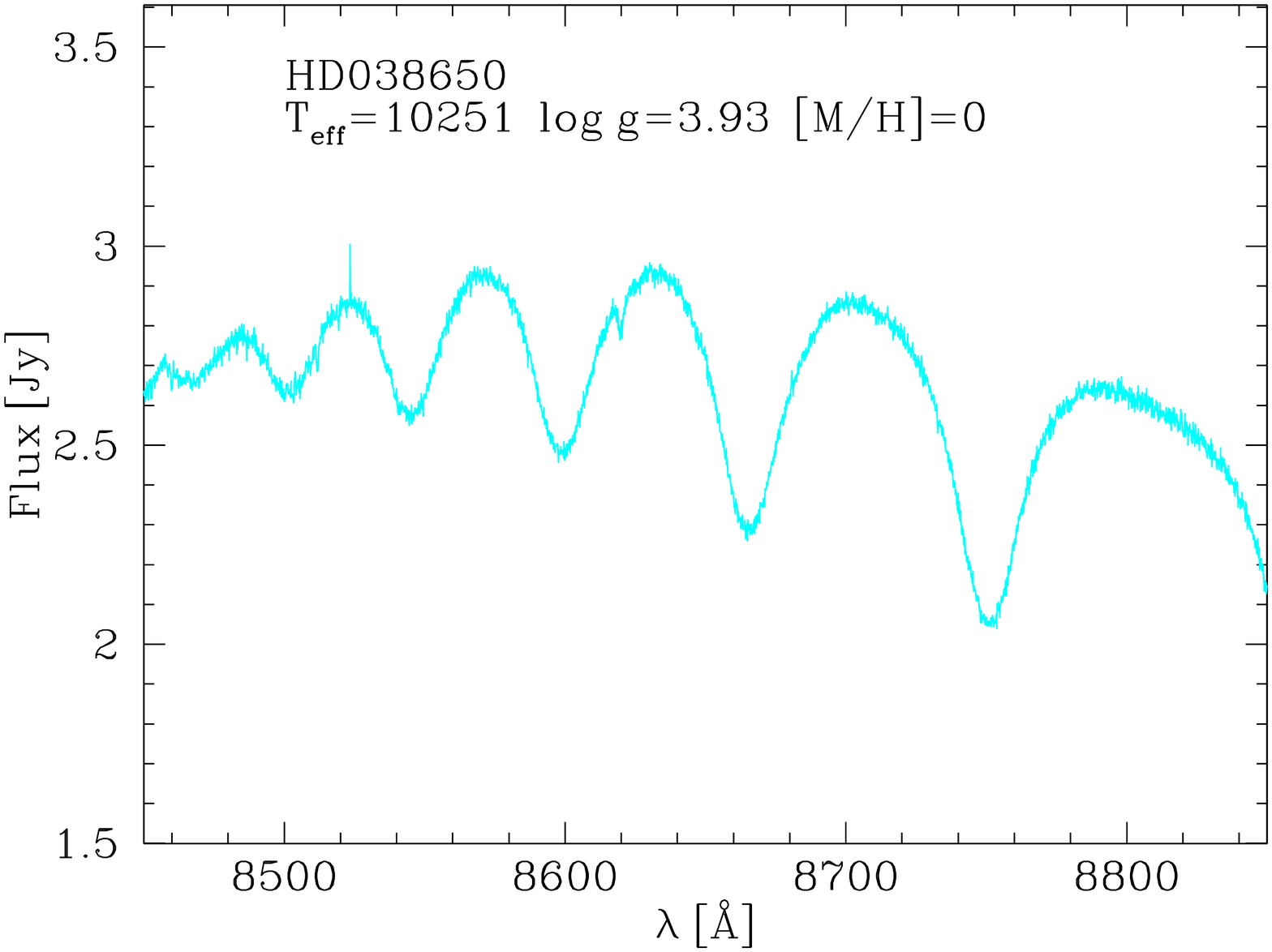}
\includegraphics[width=0.18\textwidth,angle=-90]{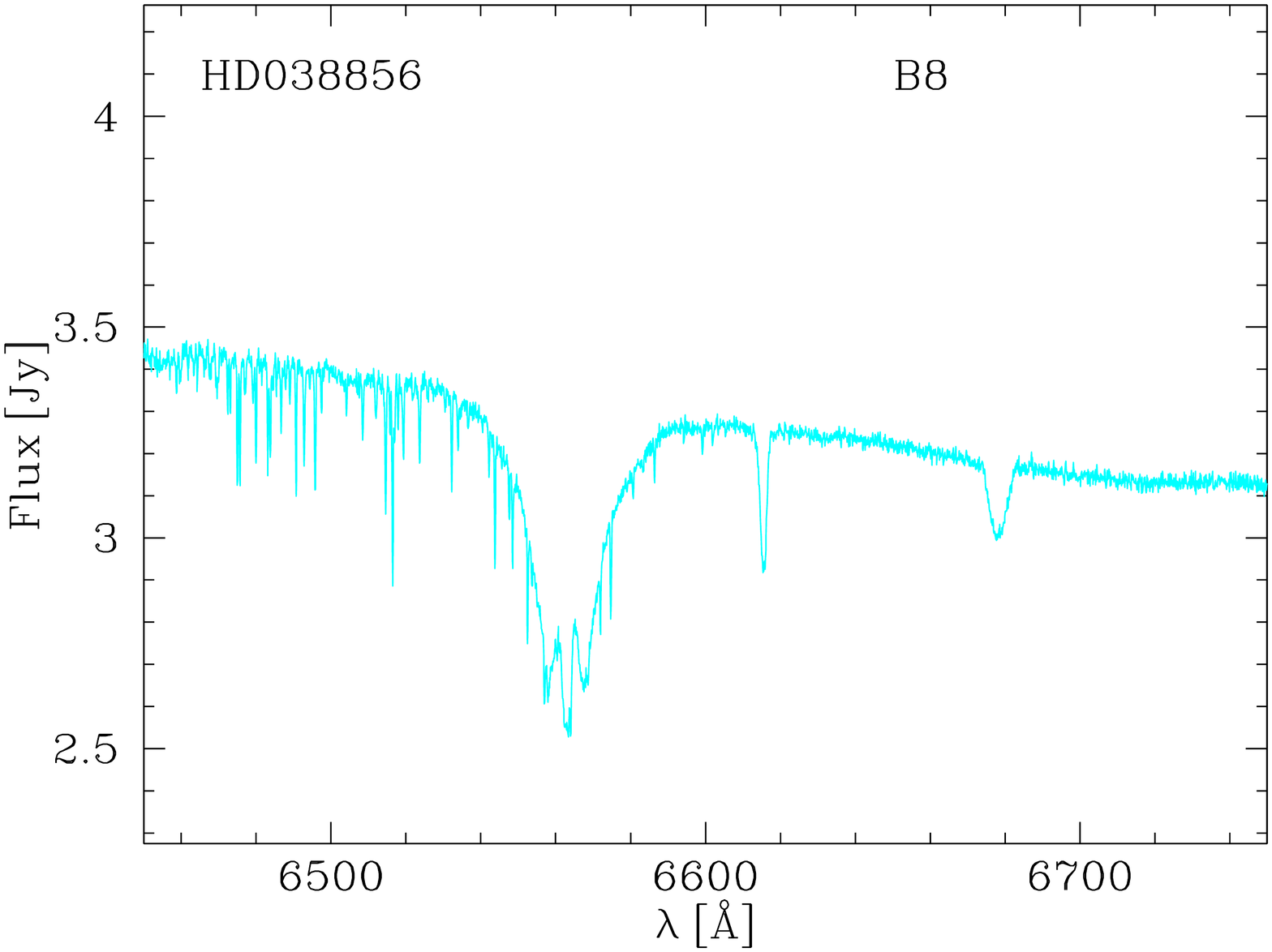}
\includegraphics[width=0.18\textwidth,angle=-90]{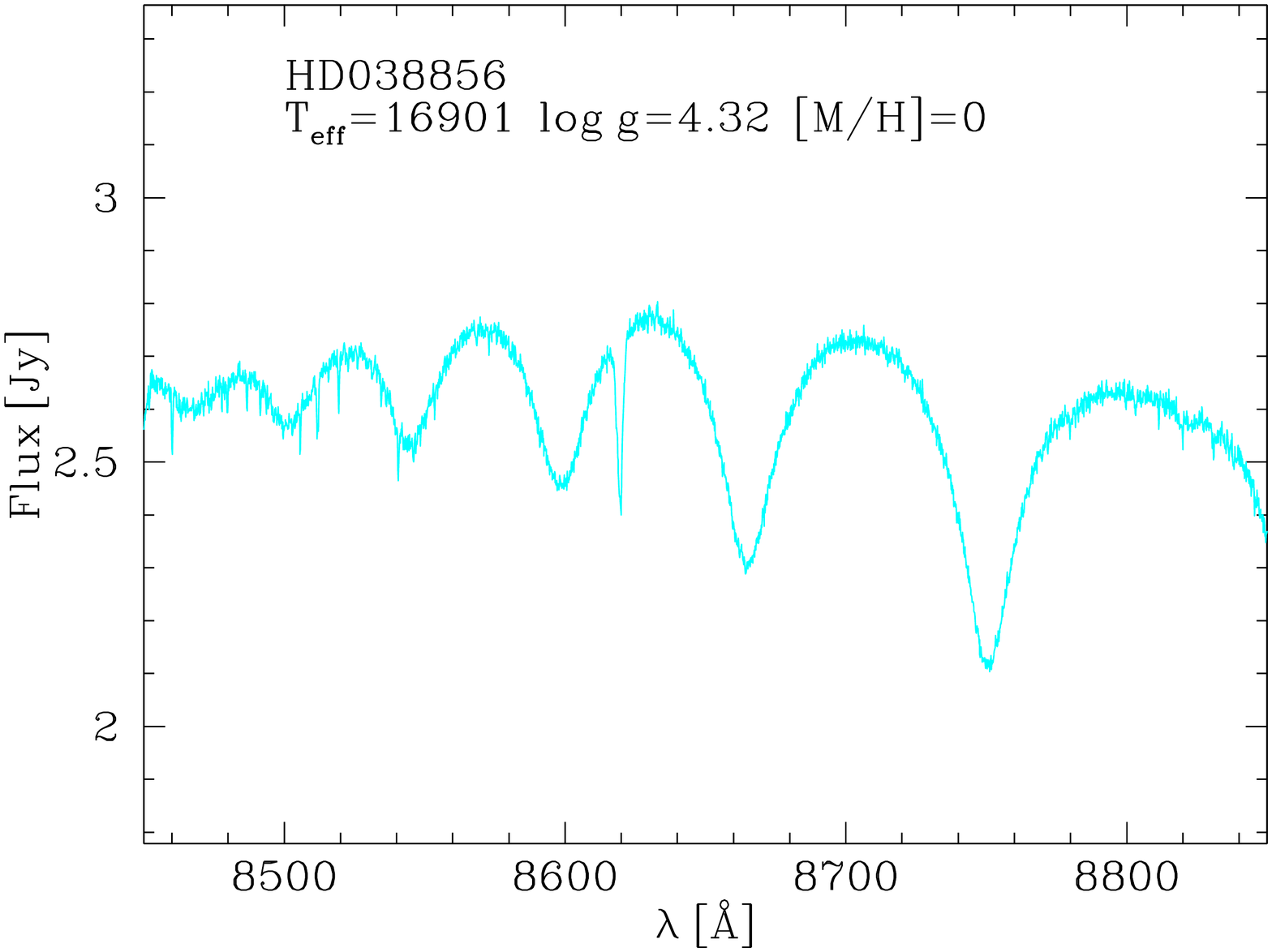}
\includegraphics[width=0.18\textwidth,angle=-90]{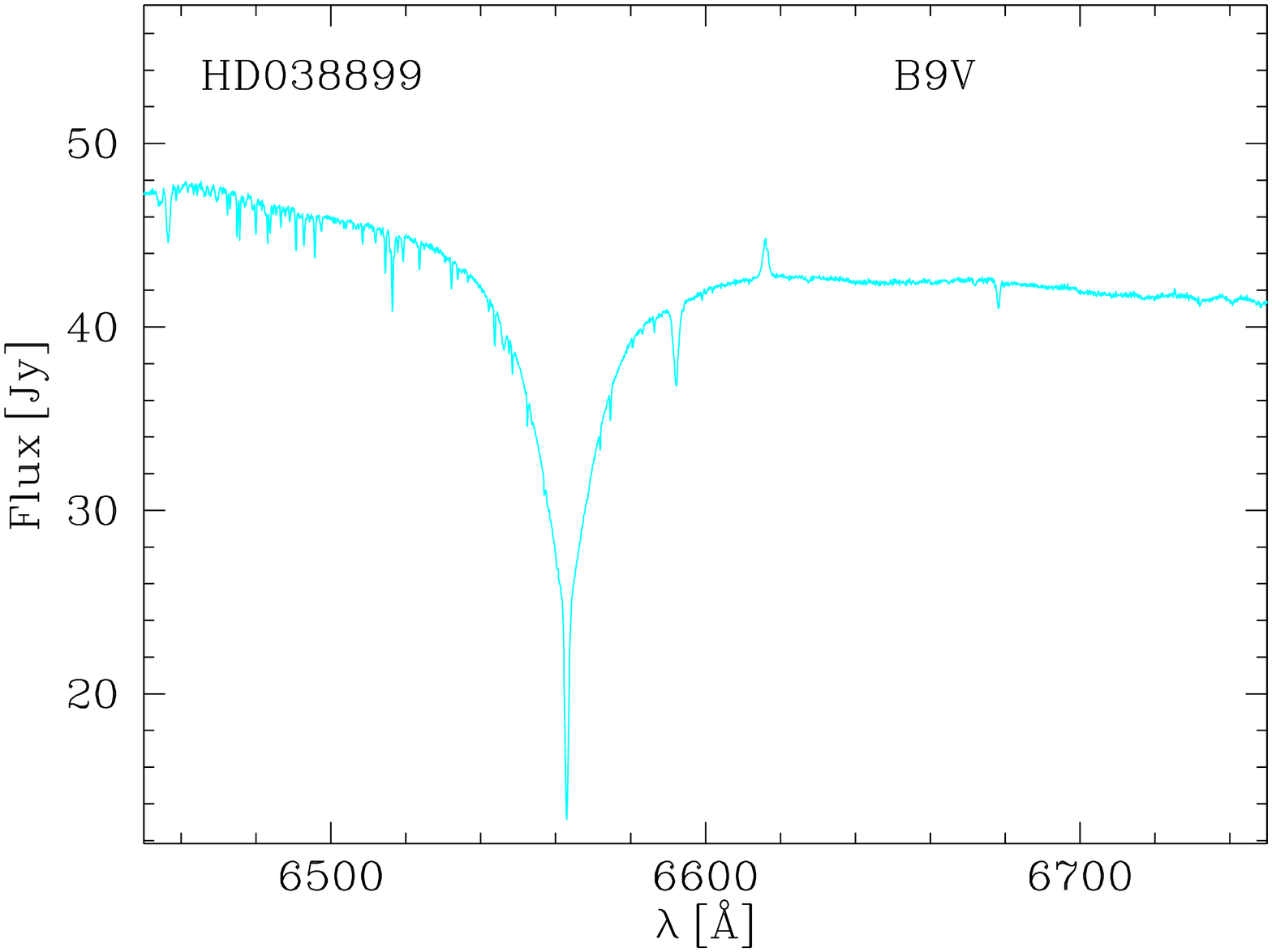}
\includegraphics[width=0.18\textwidth,angle=-90]{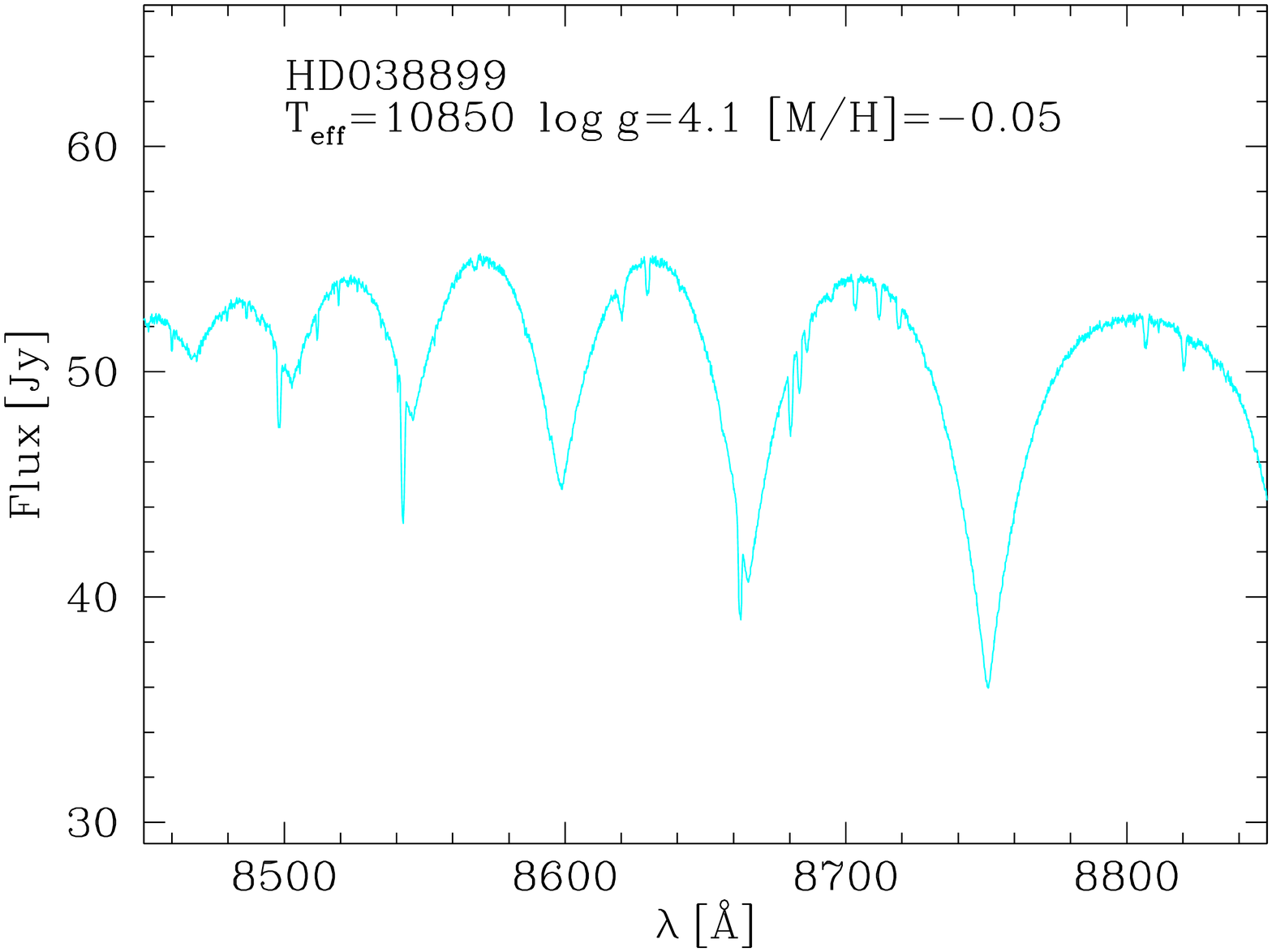}

\contcaption{8. Stars shown in this page are: HD036512, HD036960, HD037202, HD037269, HD037272, HD037394, HD037526, HD037742, HD037958, HD038230, HD038529, HD038650, HD038856 and HD038899.}
\end{figure*}

\begin{figure*}
\includegraphics[width=0.18\textwidth,angle=-90]{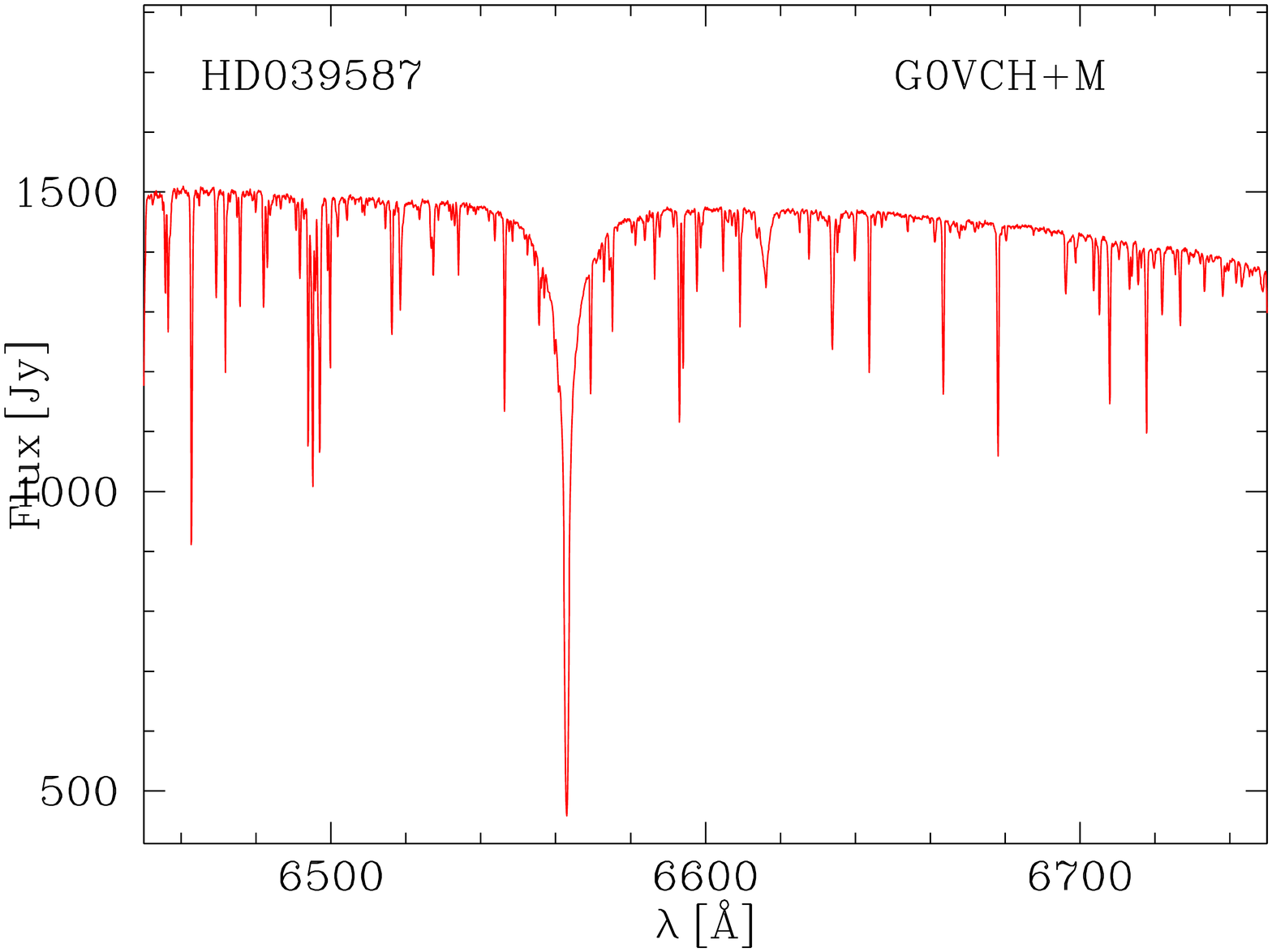}
\includegraphics[width=0.18\textwidth,angle=-90]{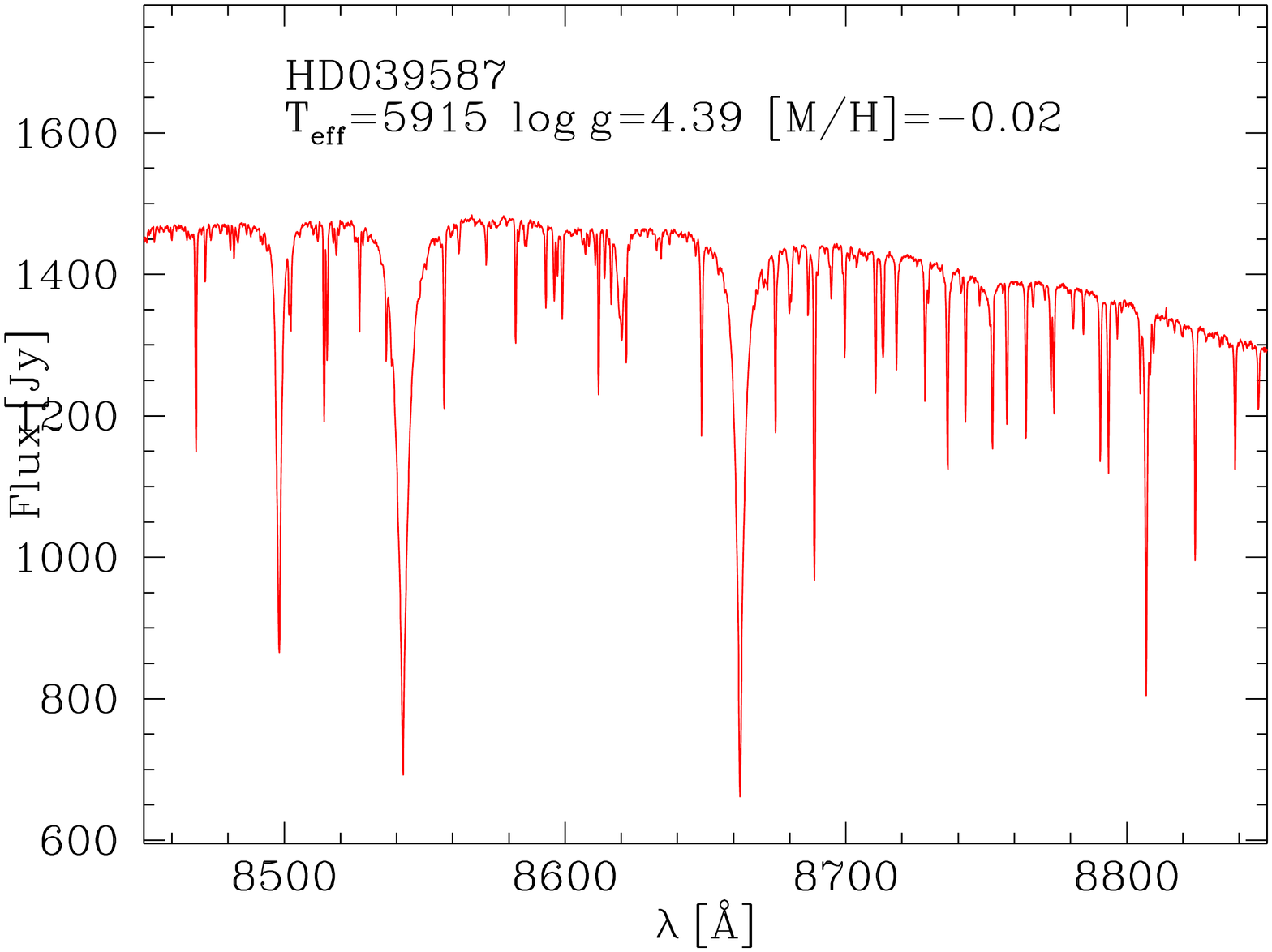}
\includegraphics[width=0.18\textwidth,angle=-90]{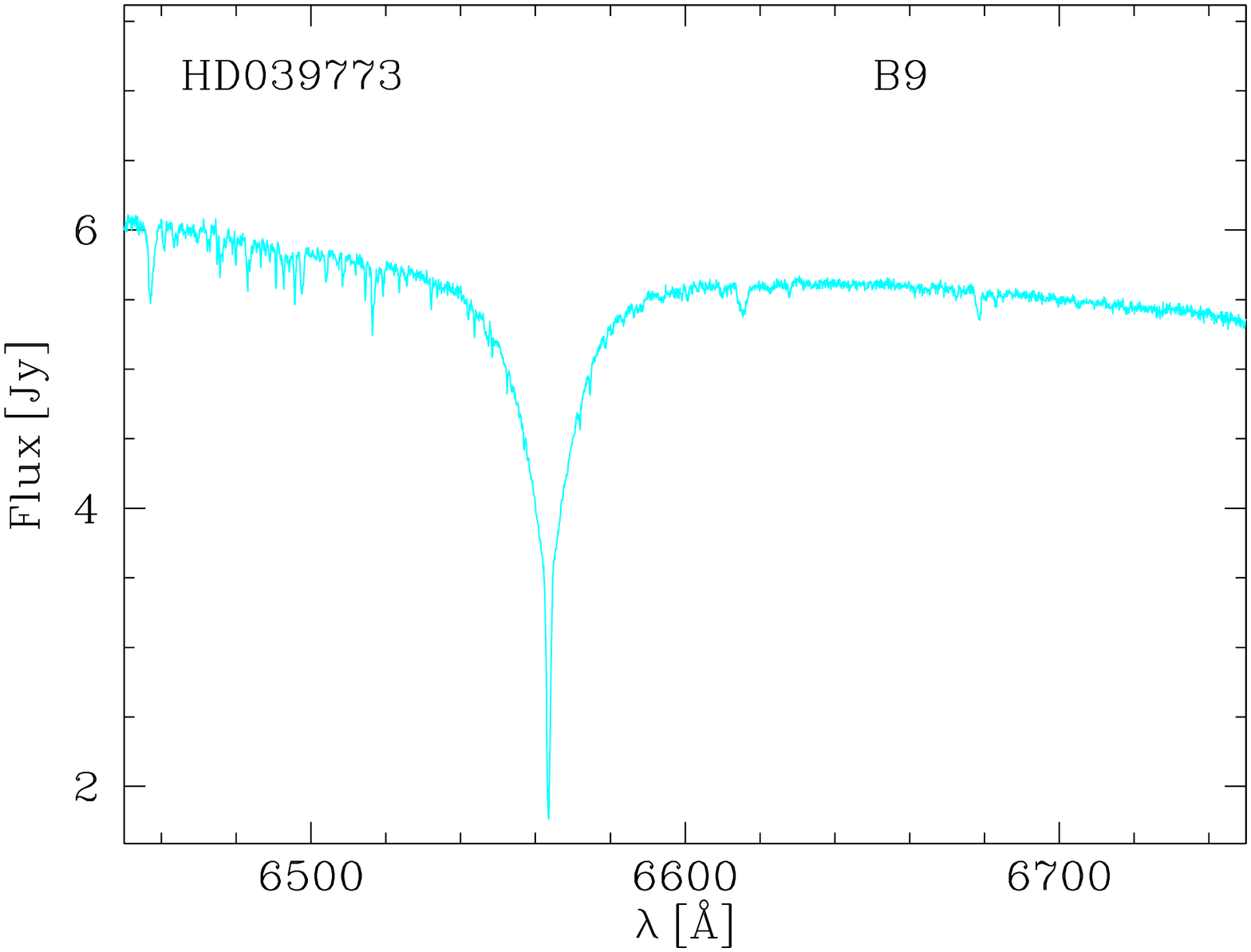}
\includegraphics[width=0.18\textwidth,angle=-90]{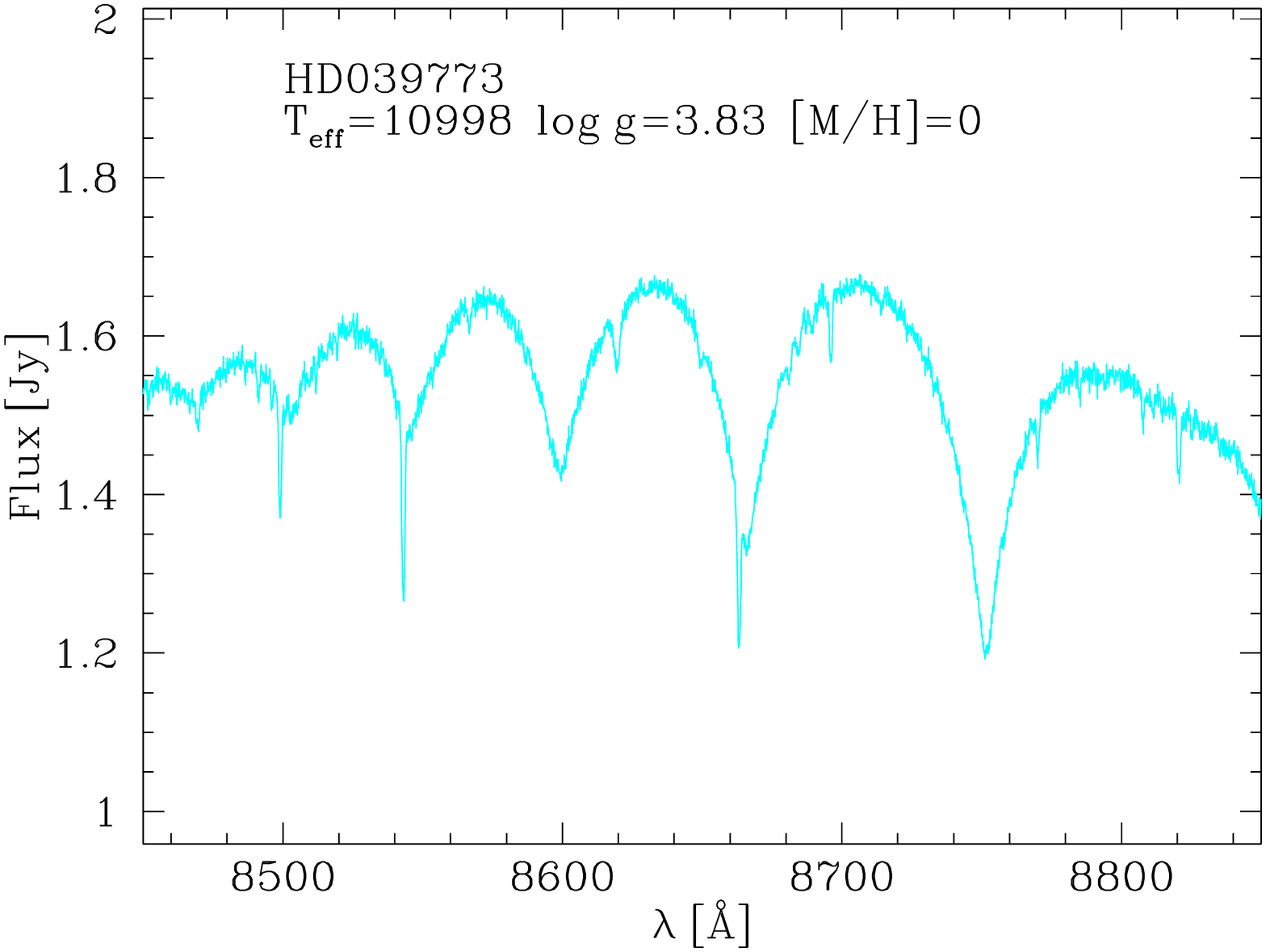}
\includegraphics[width=0.18\textwidth,angle=-90]{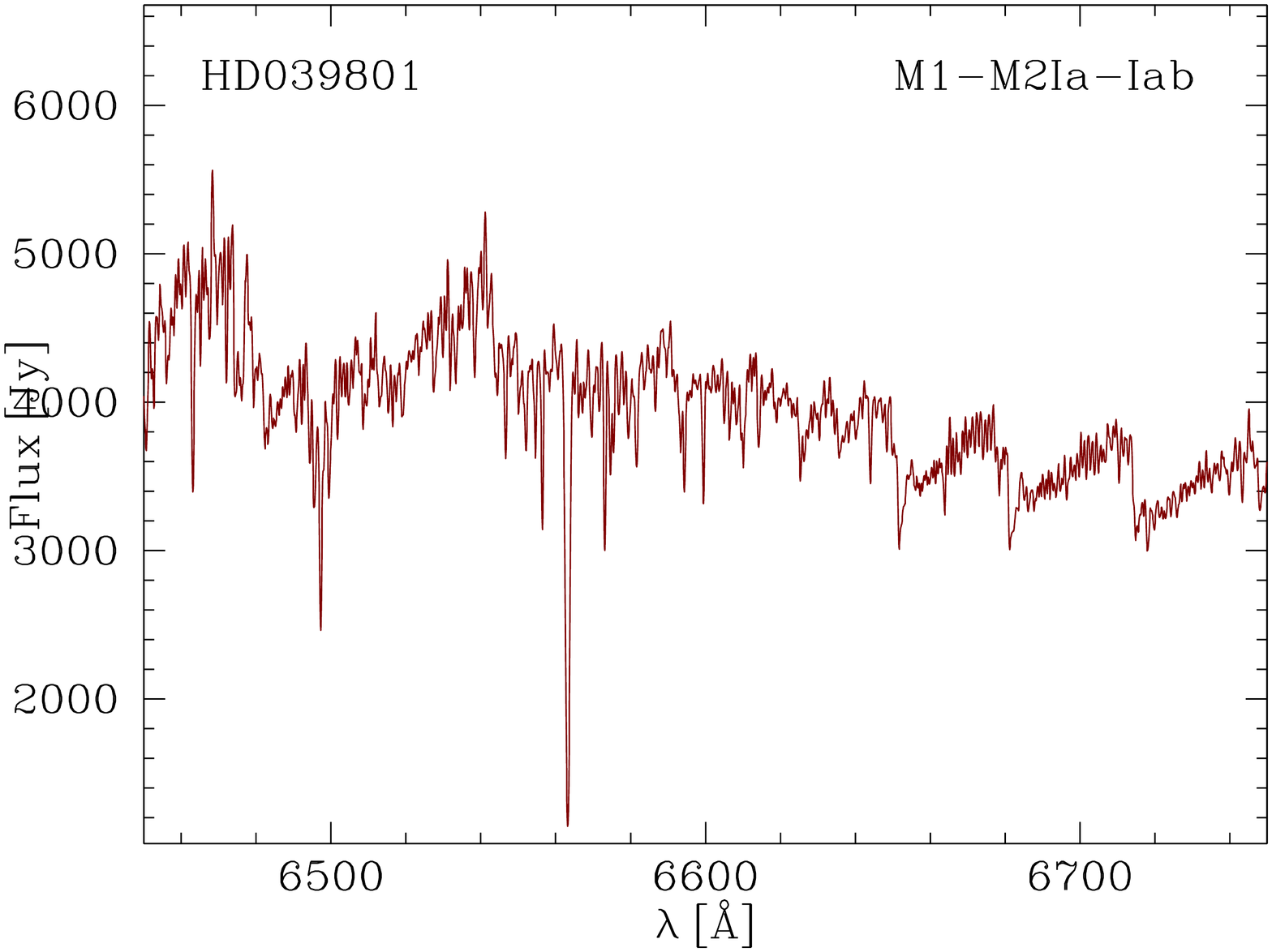}
\includegraphics[width=0.18\textwidth,angle=-90]{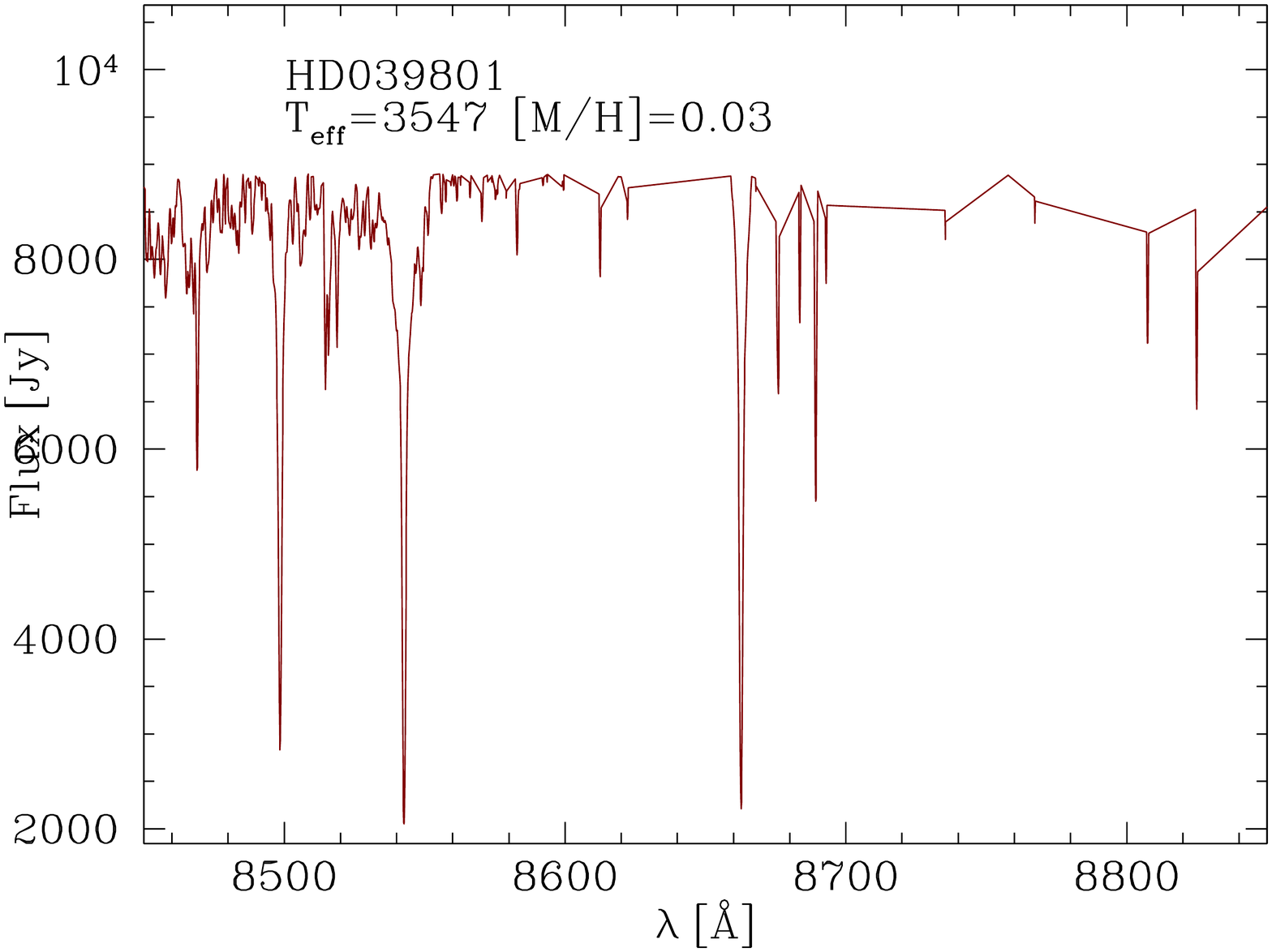}
\includegraphics[width=0.18\textwidth,angle=-90]{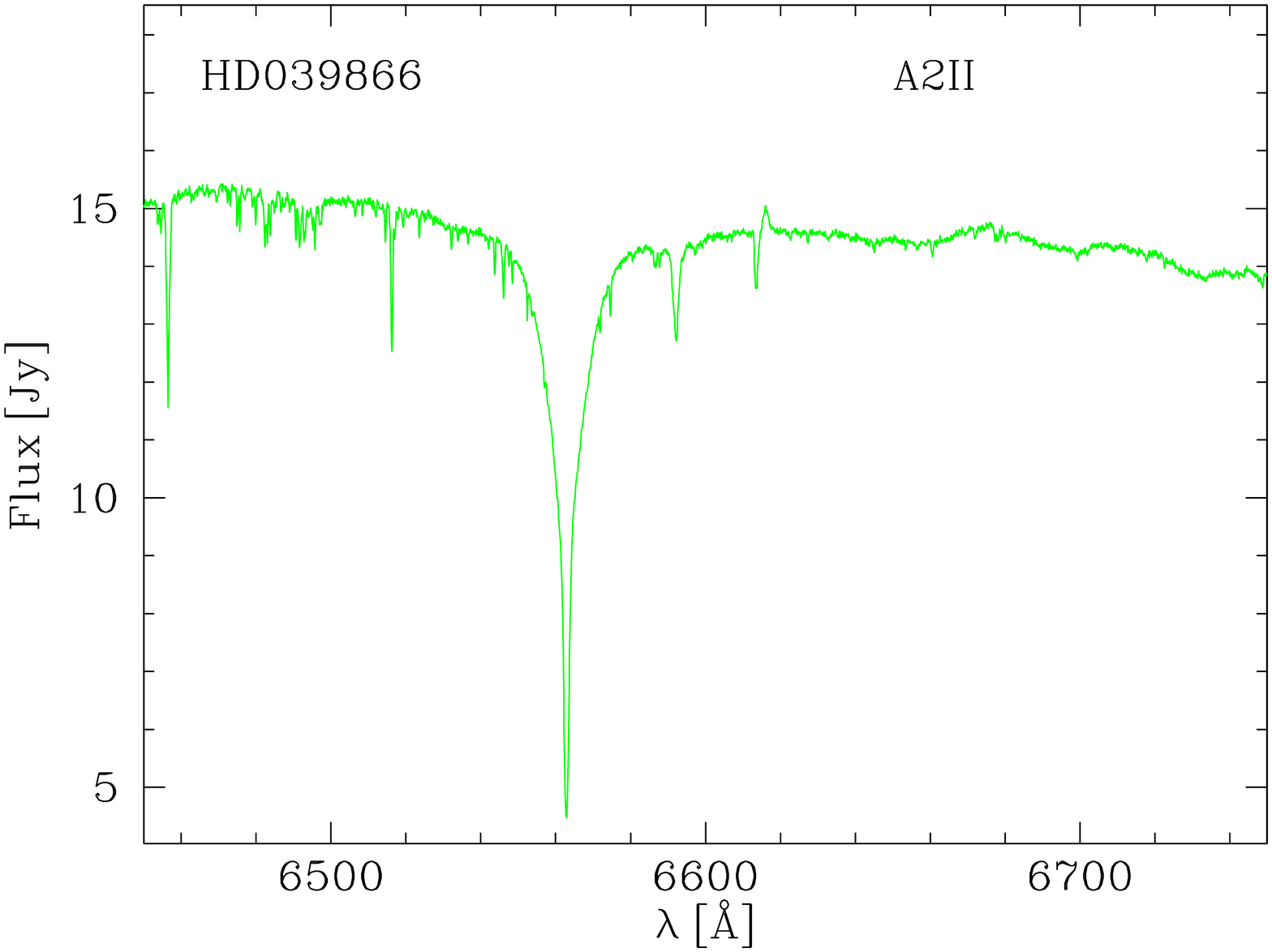}
\includegraphics[width=0.18\textwidth,angle=-90]{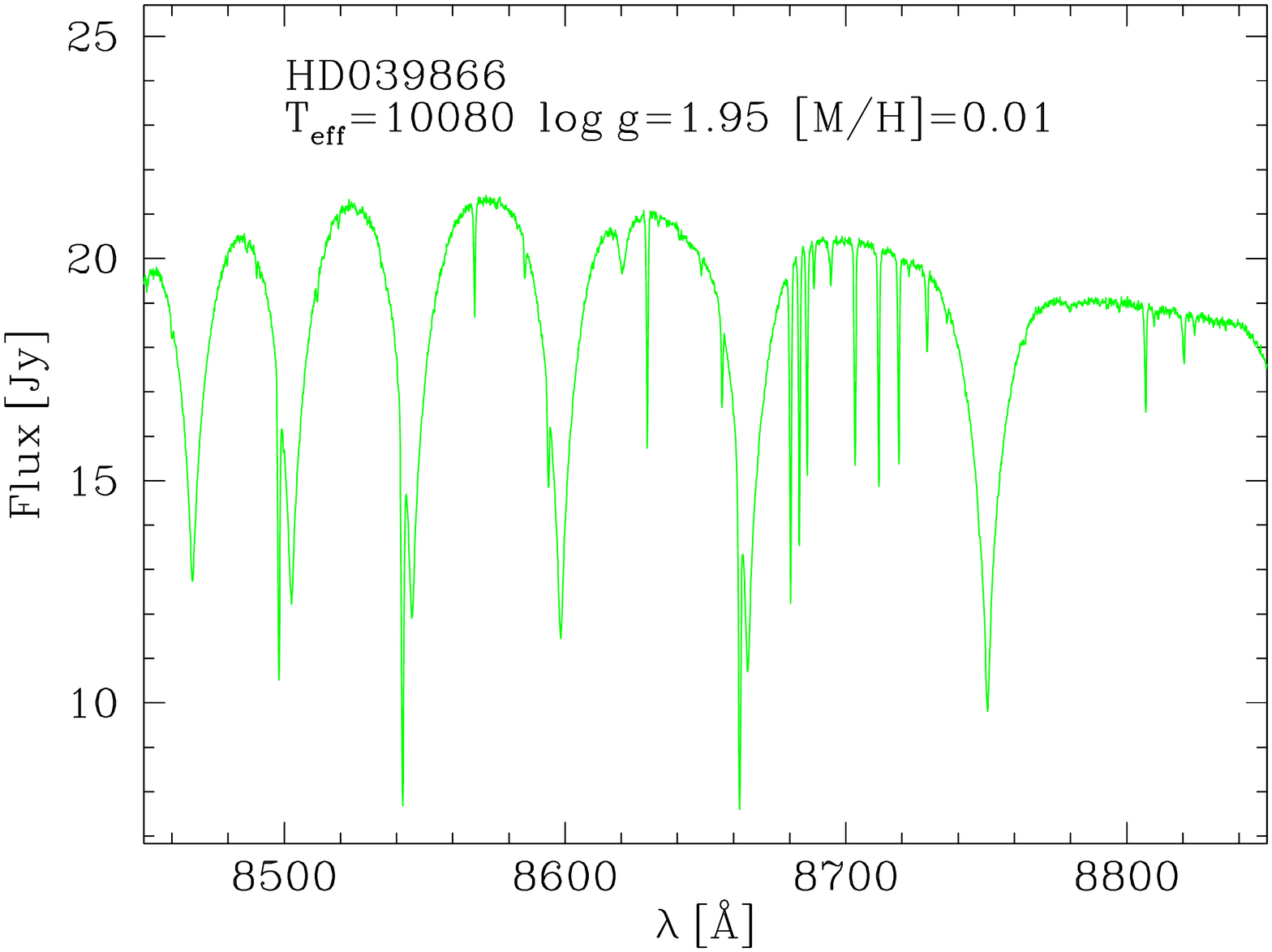}
\includegraphics[width=0.18\textwidth,angle=-90]{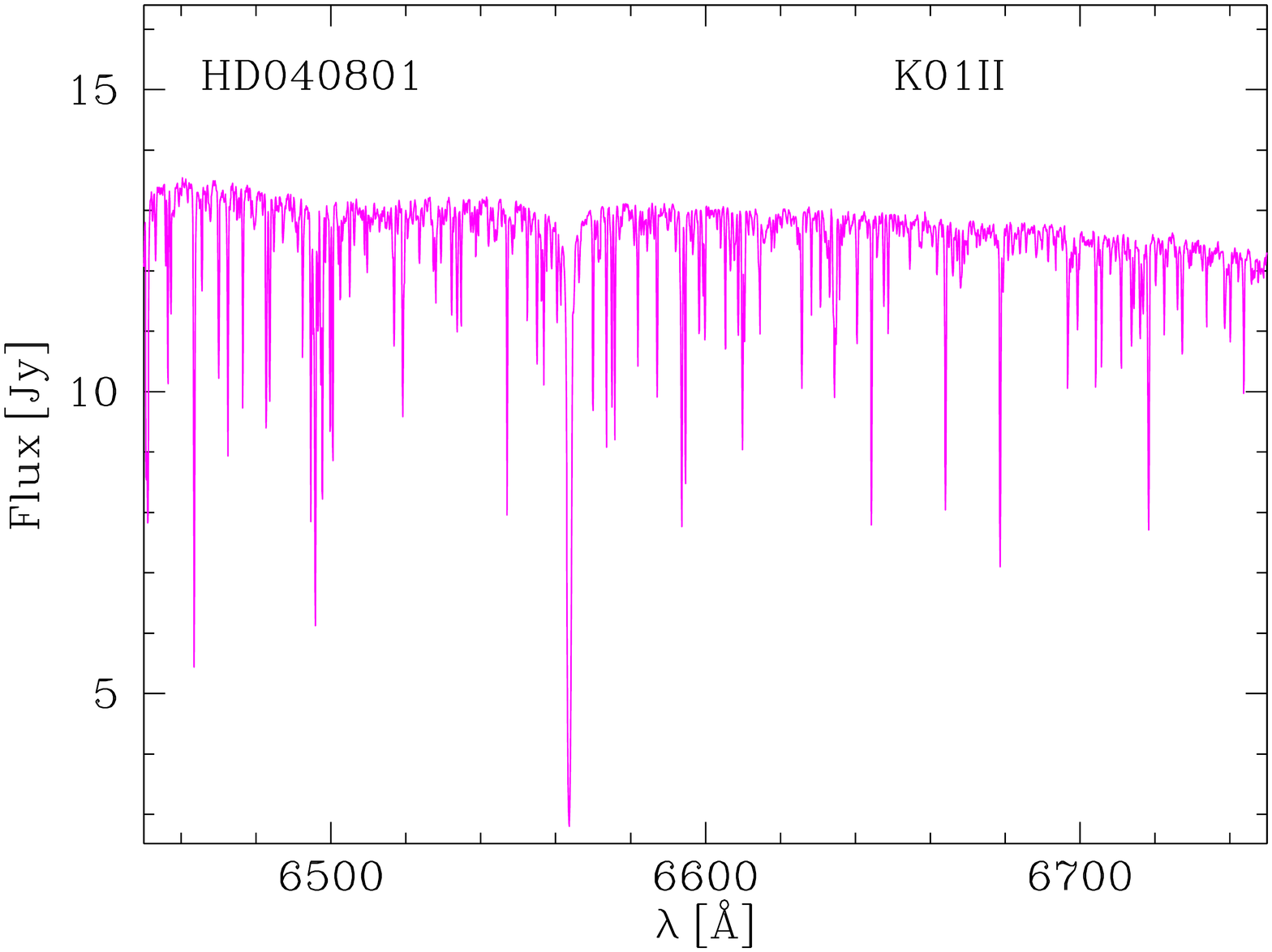}
\includegraphics[width=0.18\textwidth,angle=-90]{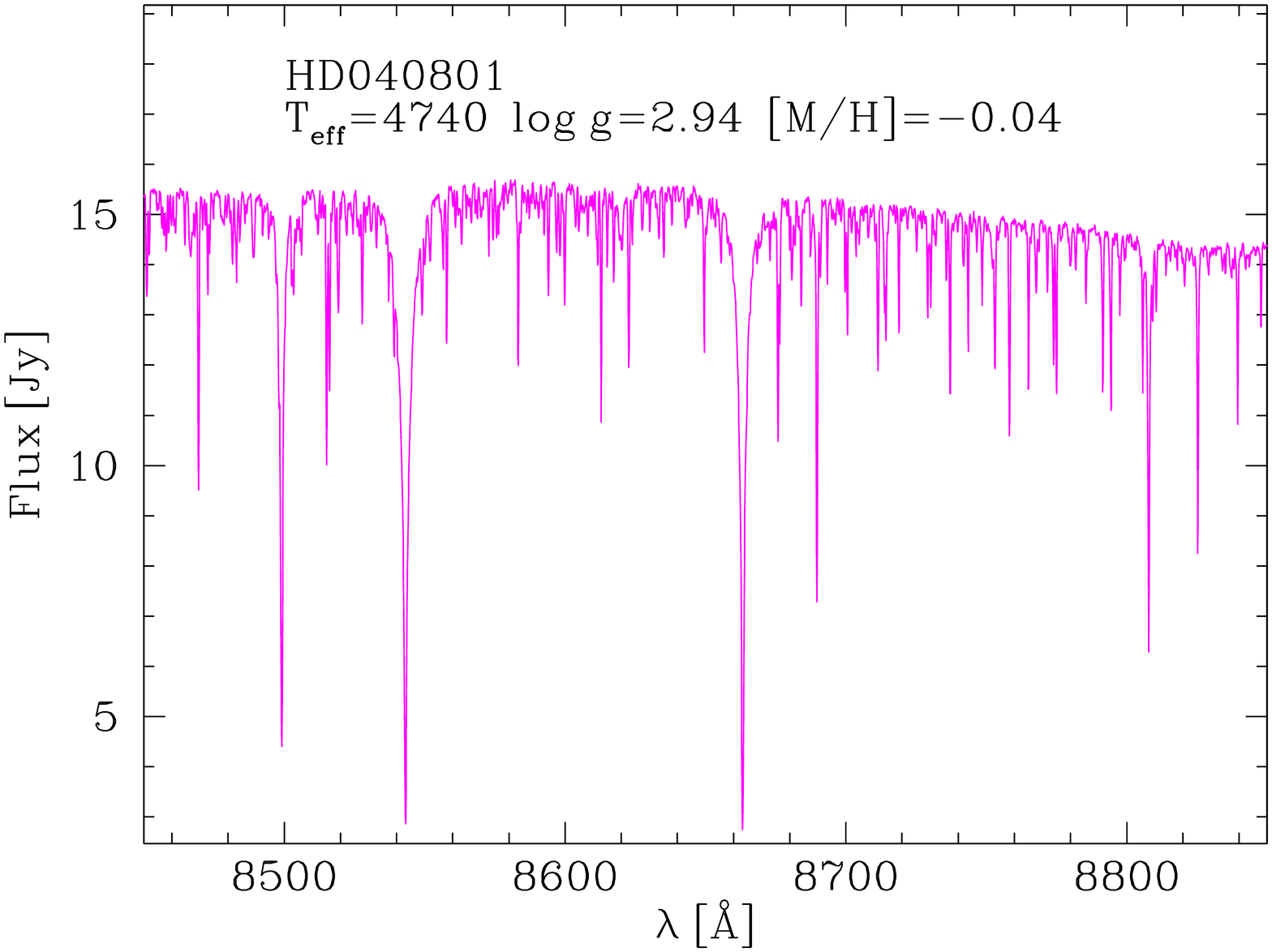}
\includegraphics[width=0.18\textwidth,angle=-90]{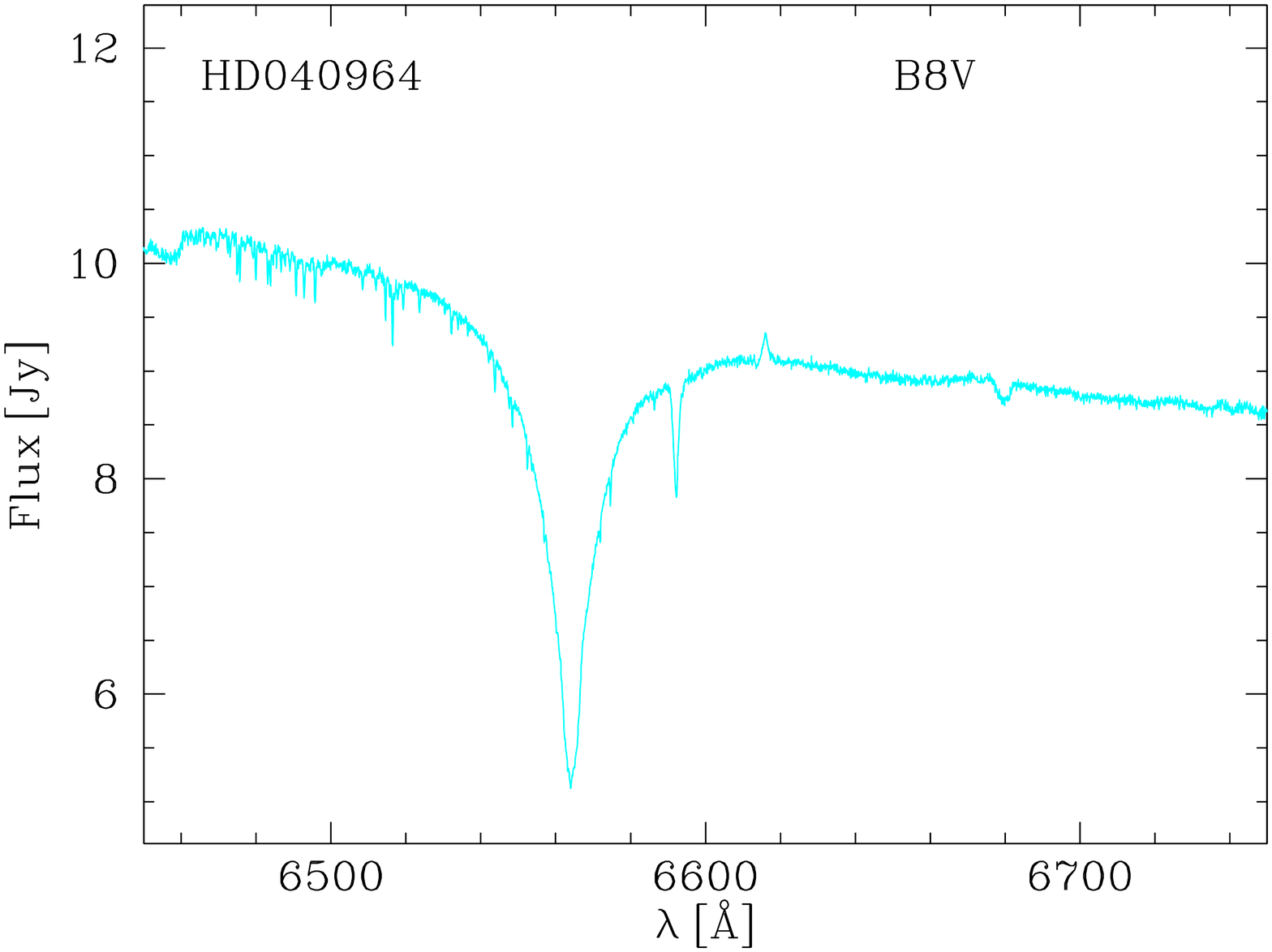}
\includegraphics[width=0.18\textwidth,angle=-90]{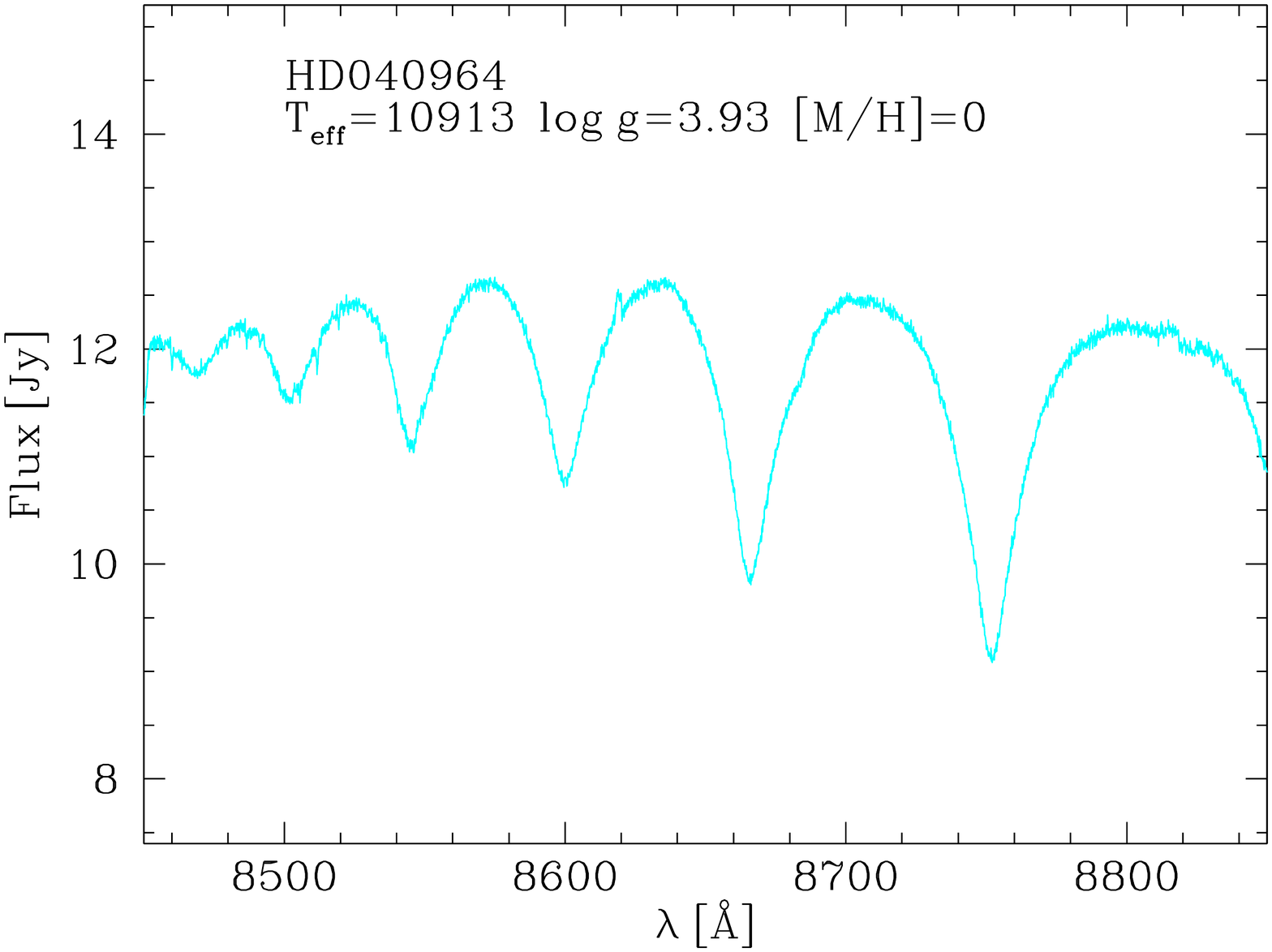}
\includegraphics[width=0.18\textwidth,angle=-90]{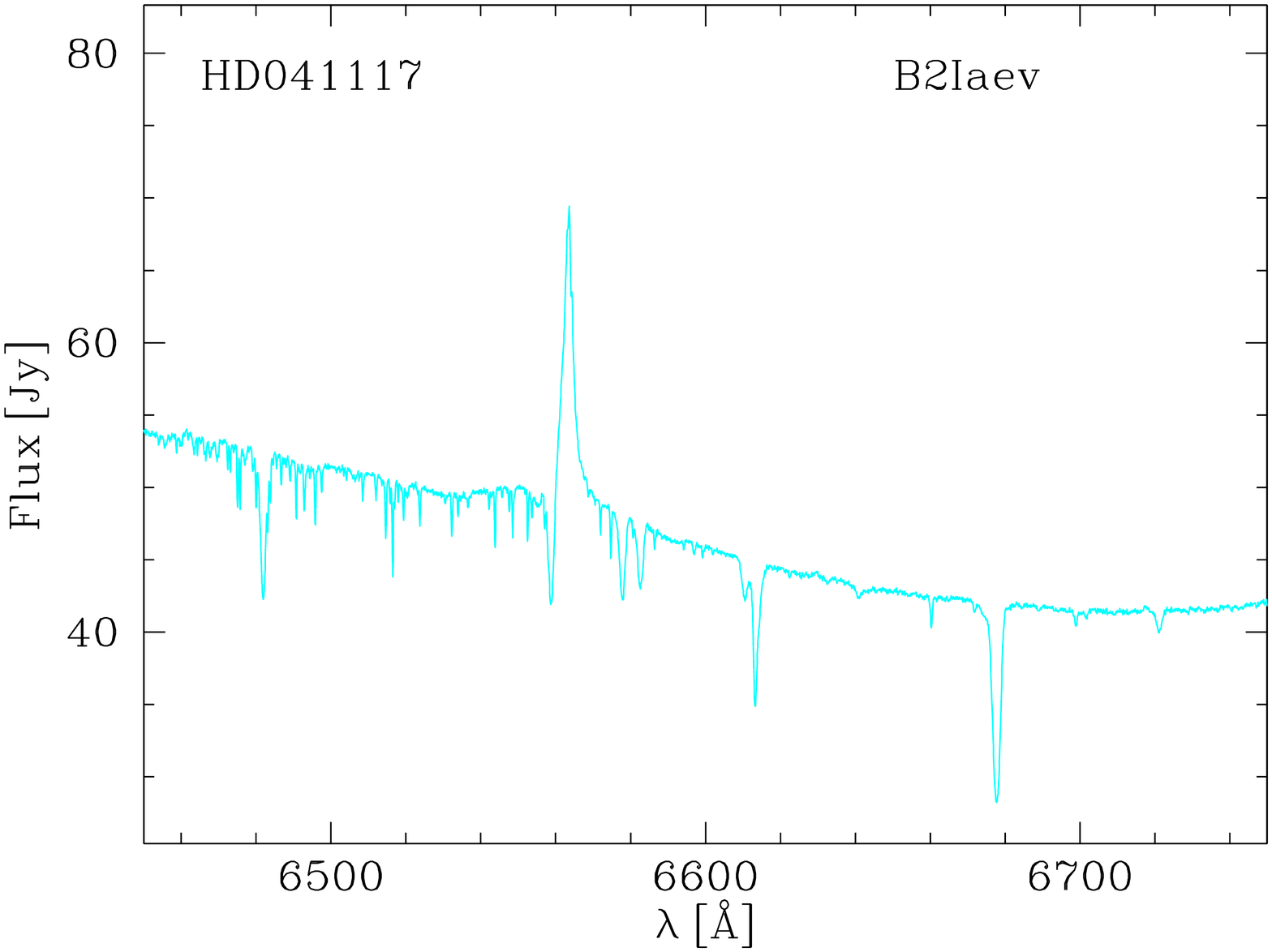}
\includegraphics[width=0.18\textwidth,angle=-90]{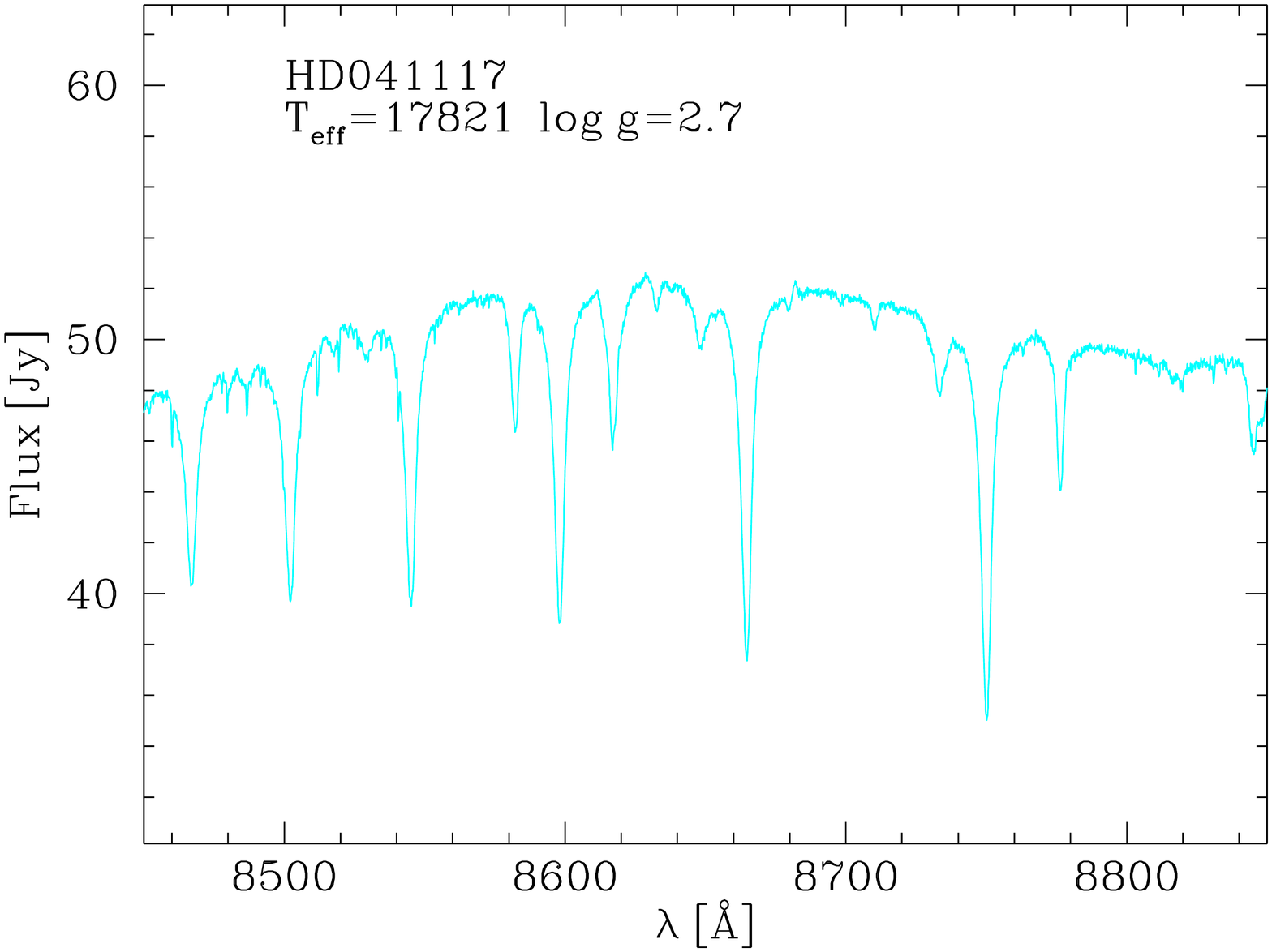}
\includegraphics[width=0.18\textwidth,angle=-90]{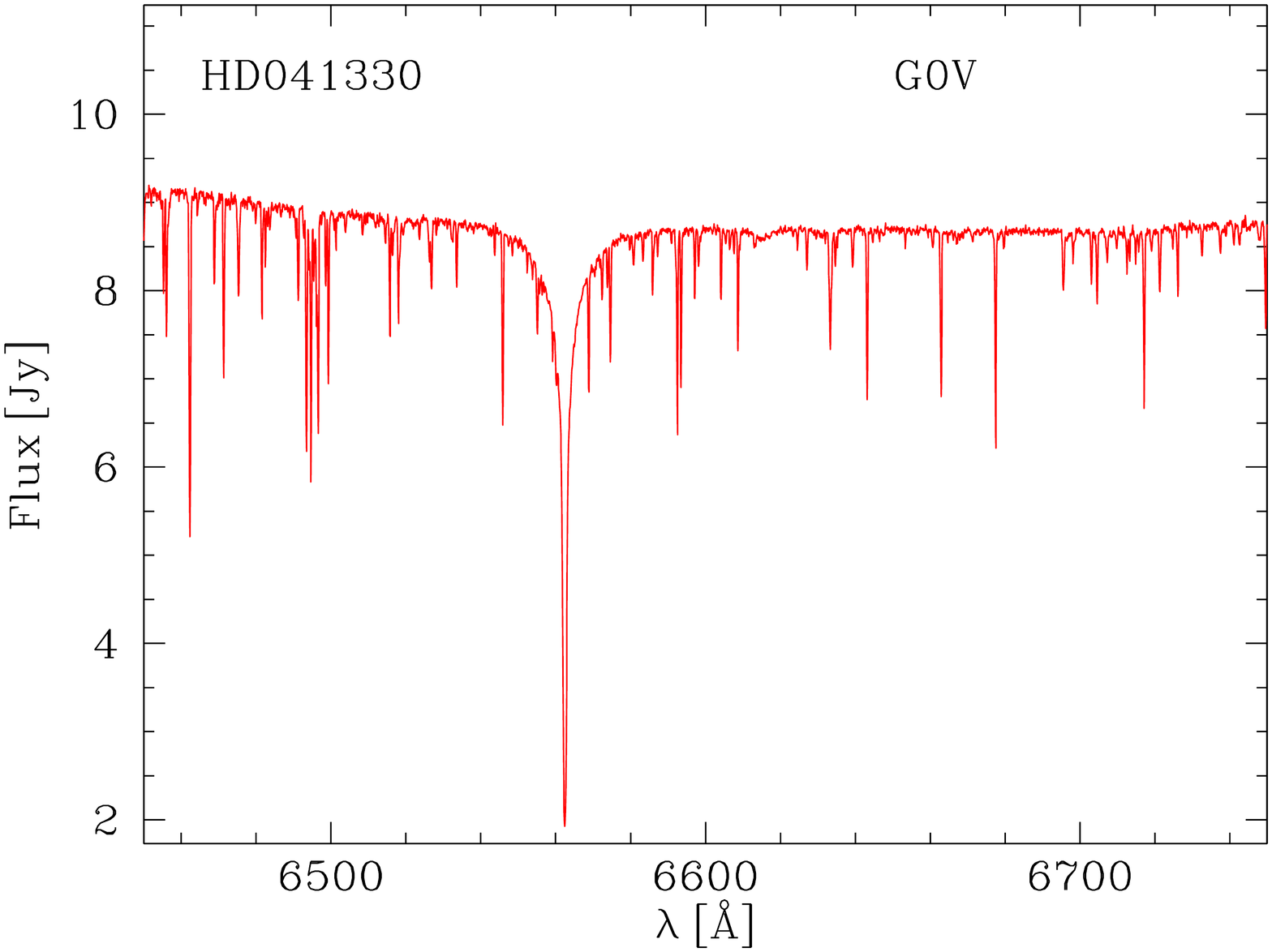}
\includegraphics[width=0.18\textwidth,angle=-90]{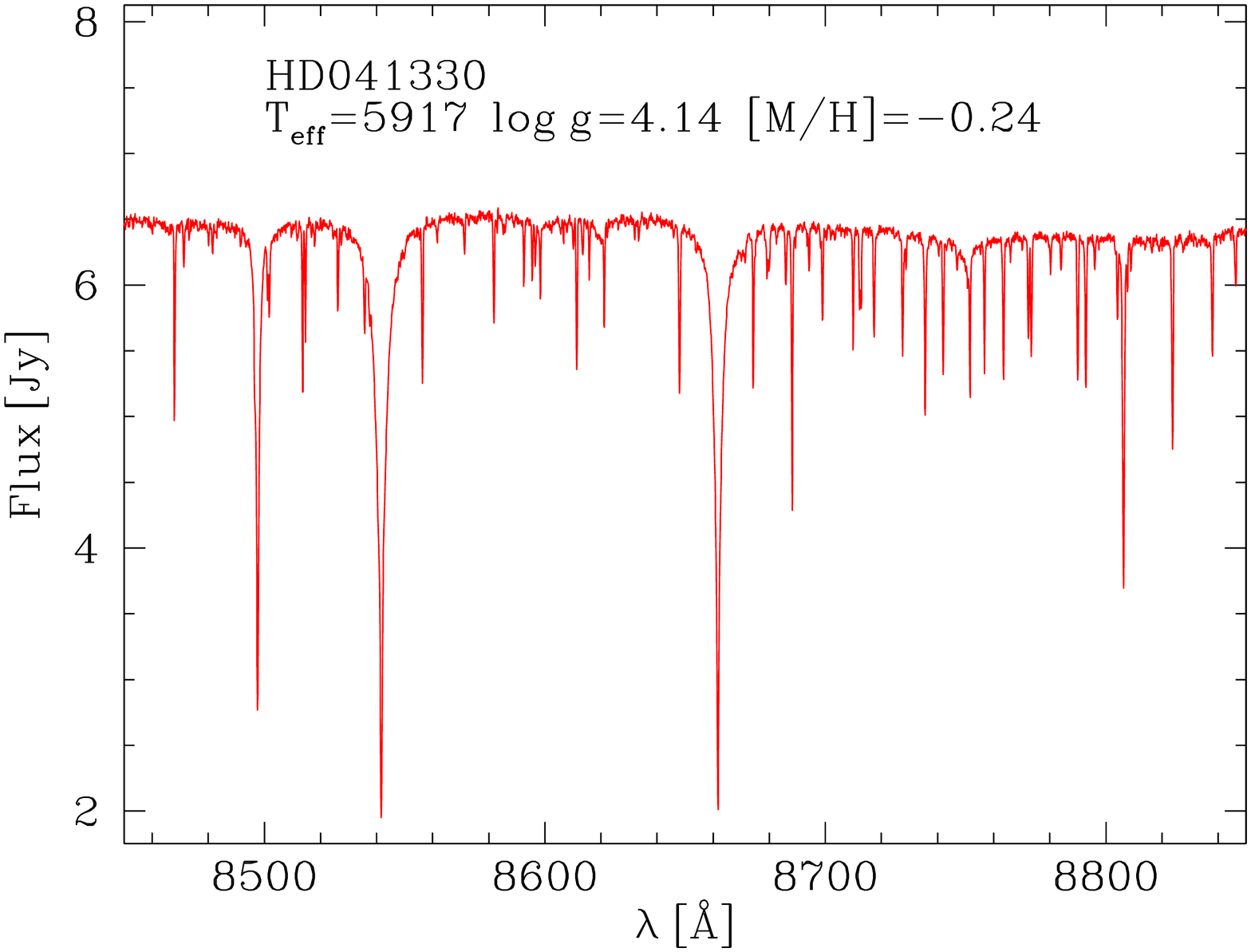}
\includegraphics[width=0.18\textwidth,angle=-90]{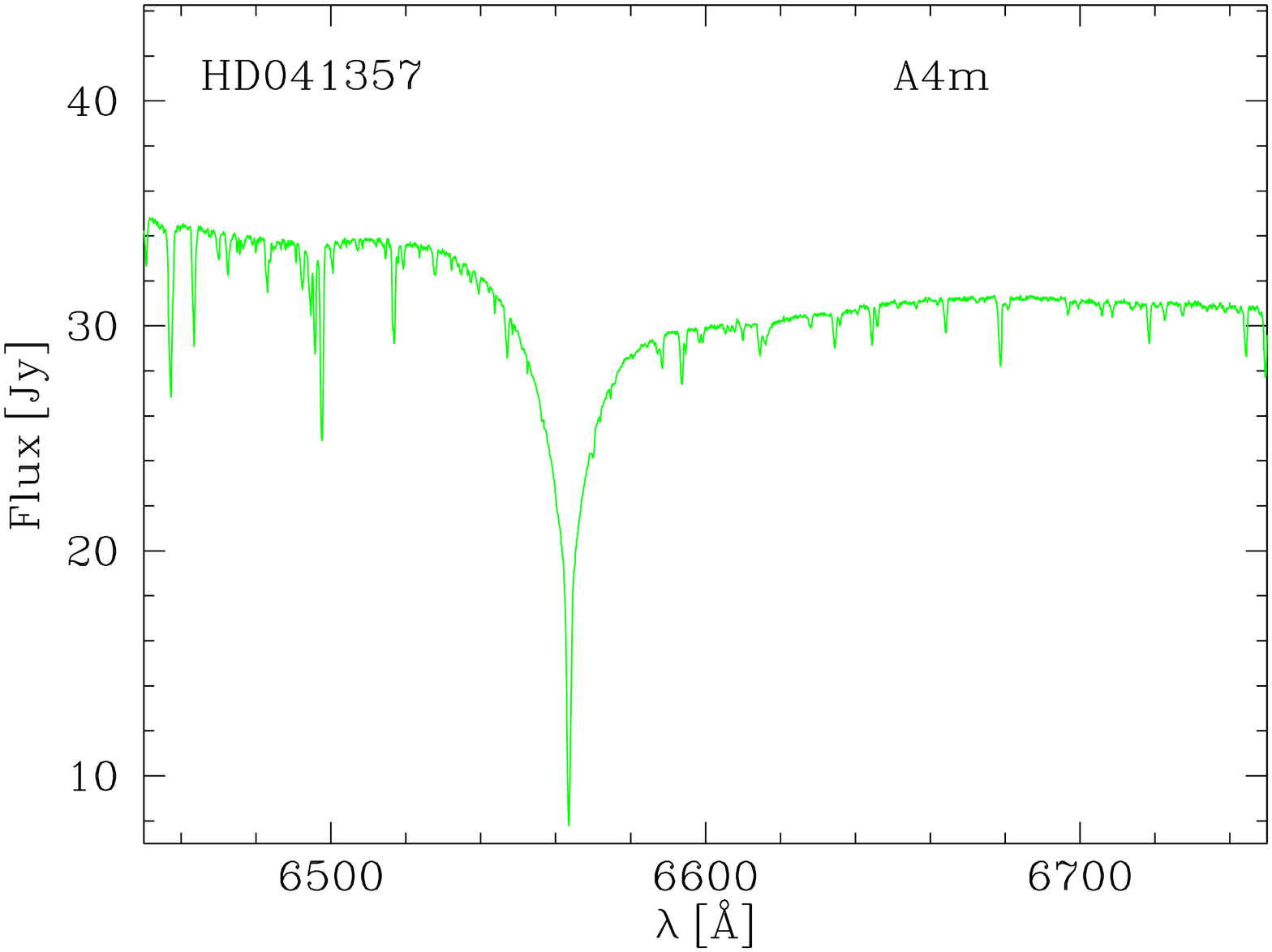}
\includegraphics[width=0.18\textwidth,angle=-90]{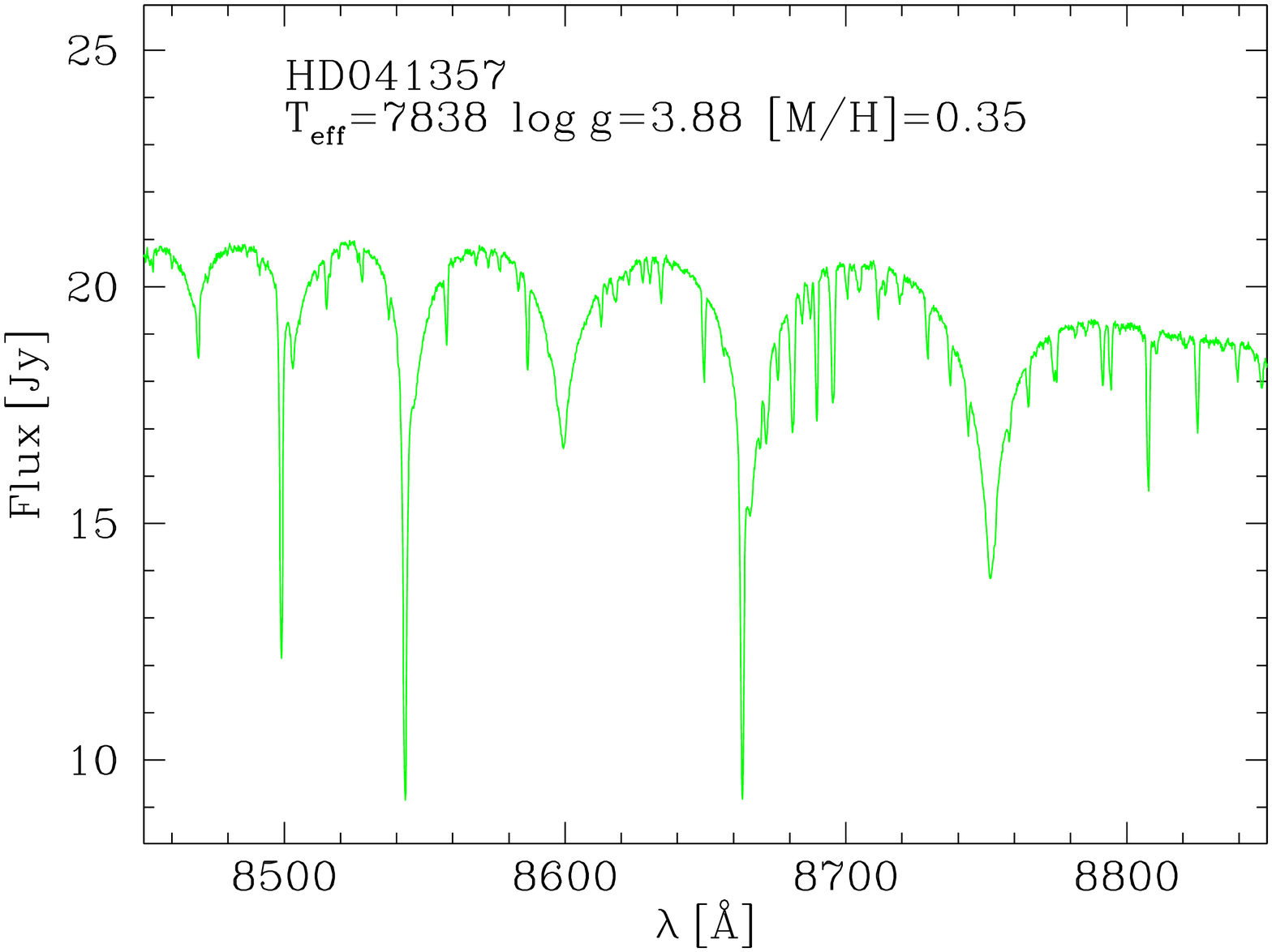}
\includegraphics[width=0.18\textwidth,angle=-90]{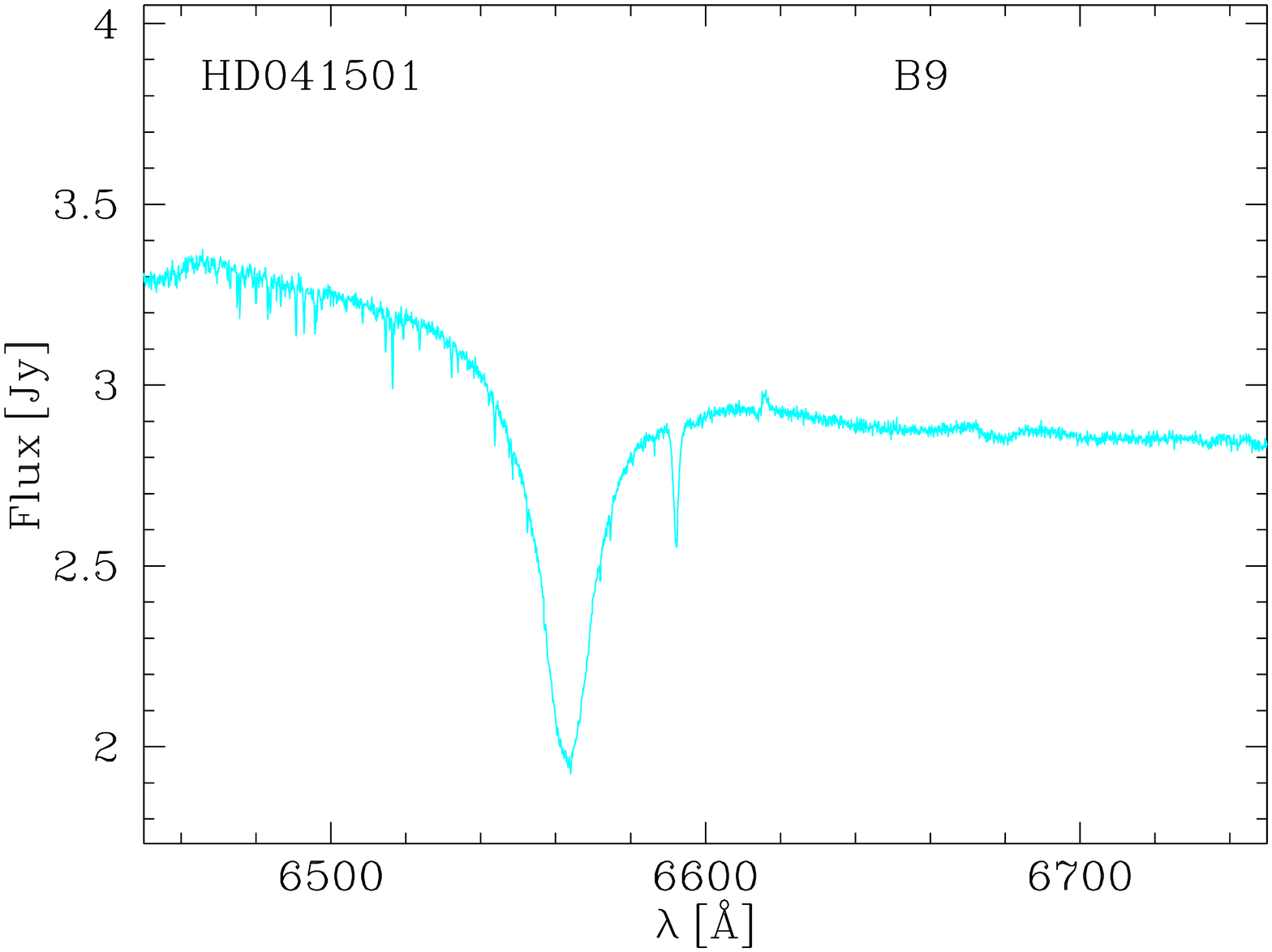}
\includegraphics[width=0.18\textwidth,angle=-90]{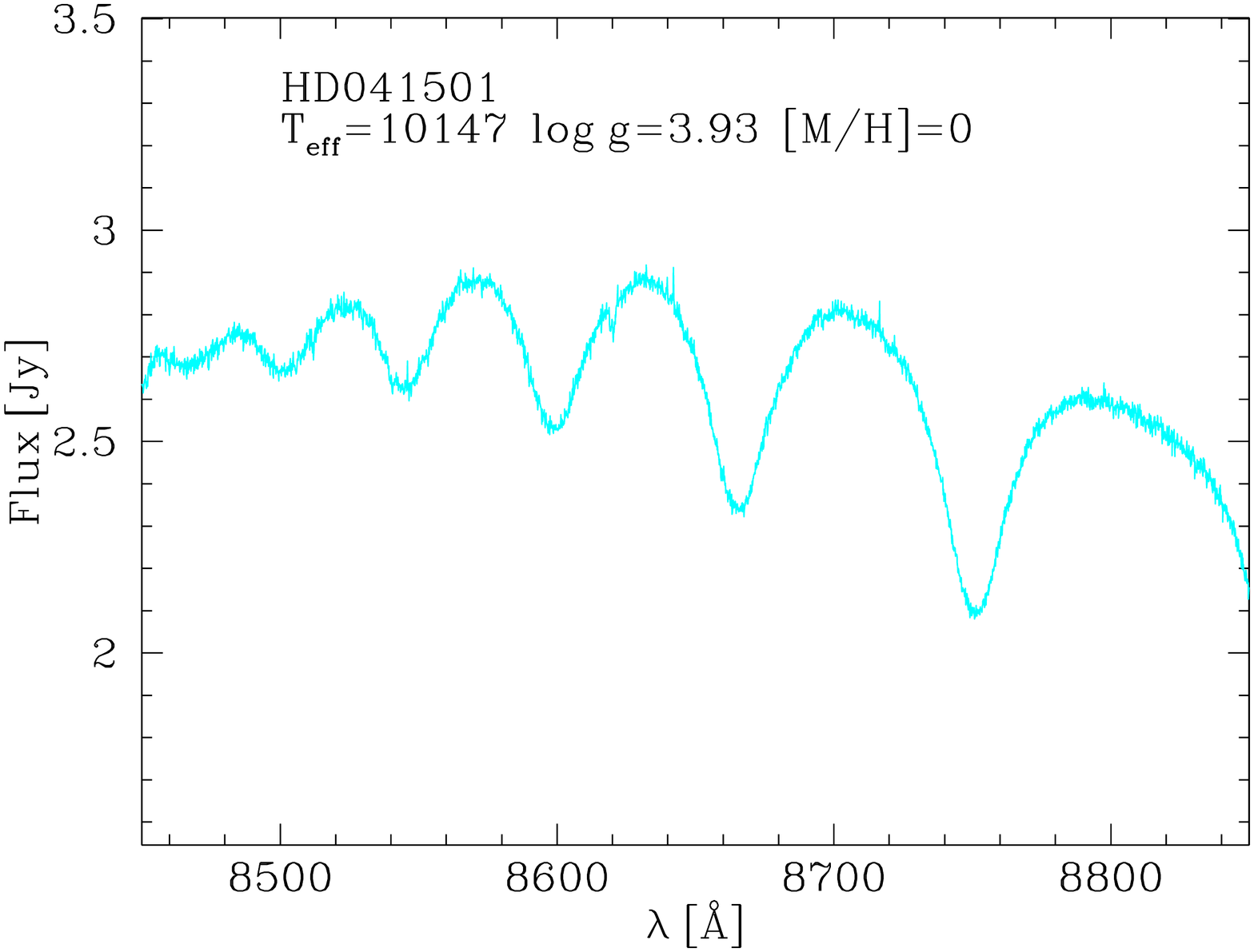}
\includegraphics[width=0.18\textwidth,angle=-90]{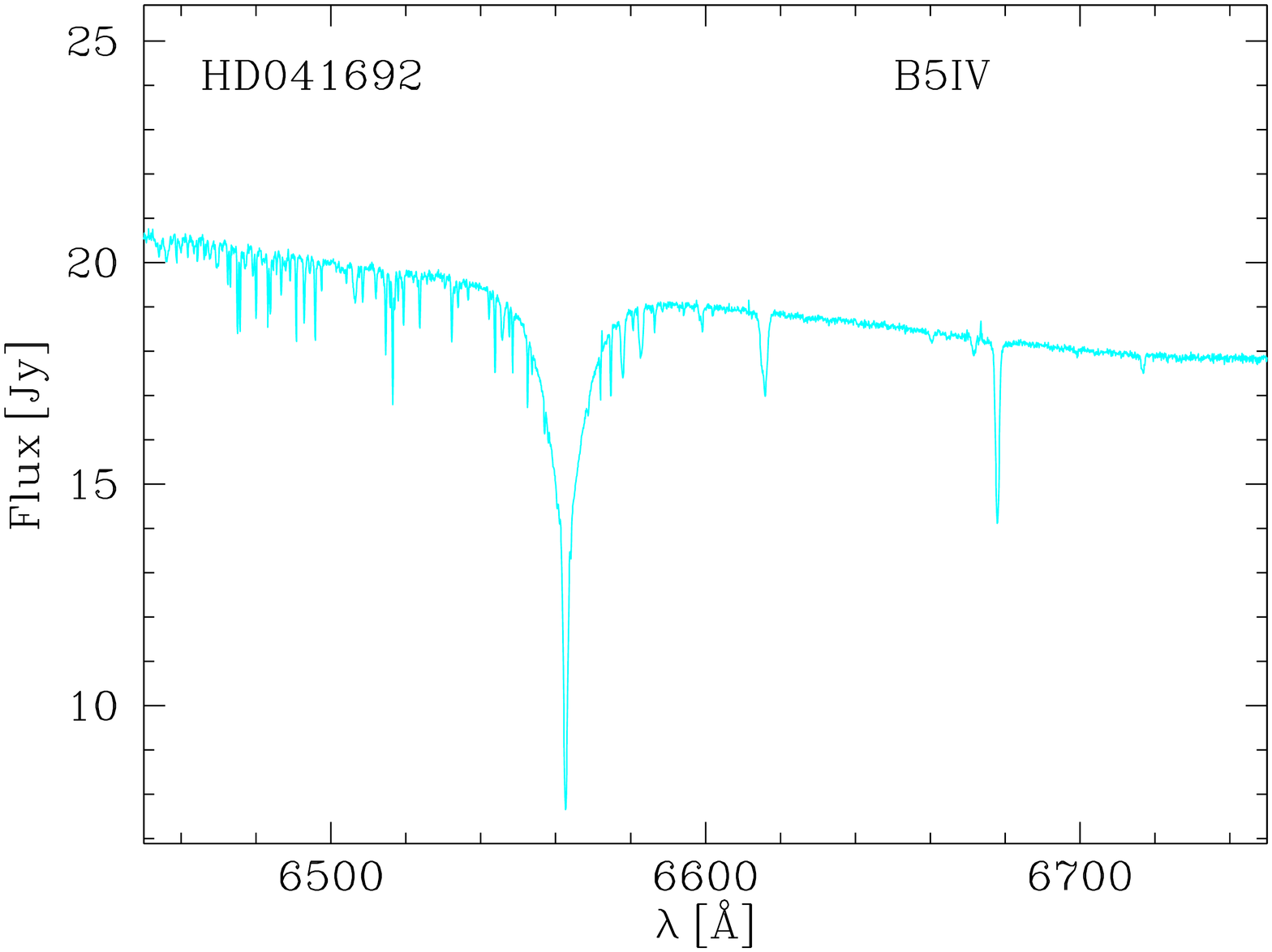}
\includegraphics[width=0.18\textwidth,angle=-90]{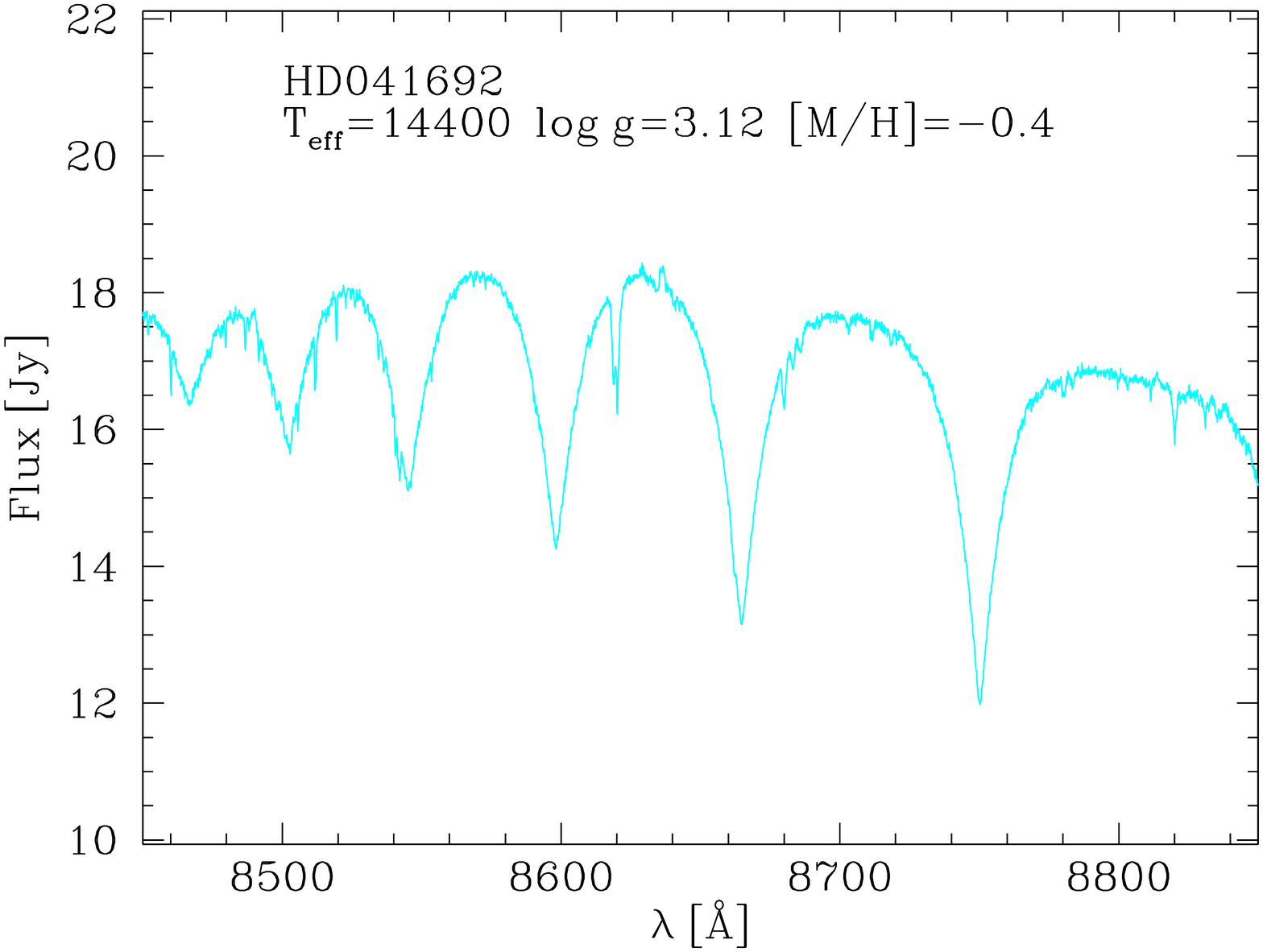}
\includegraphics[width=0.18\textwidth,angle=-90]{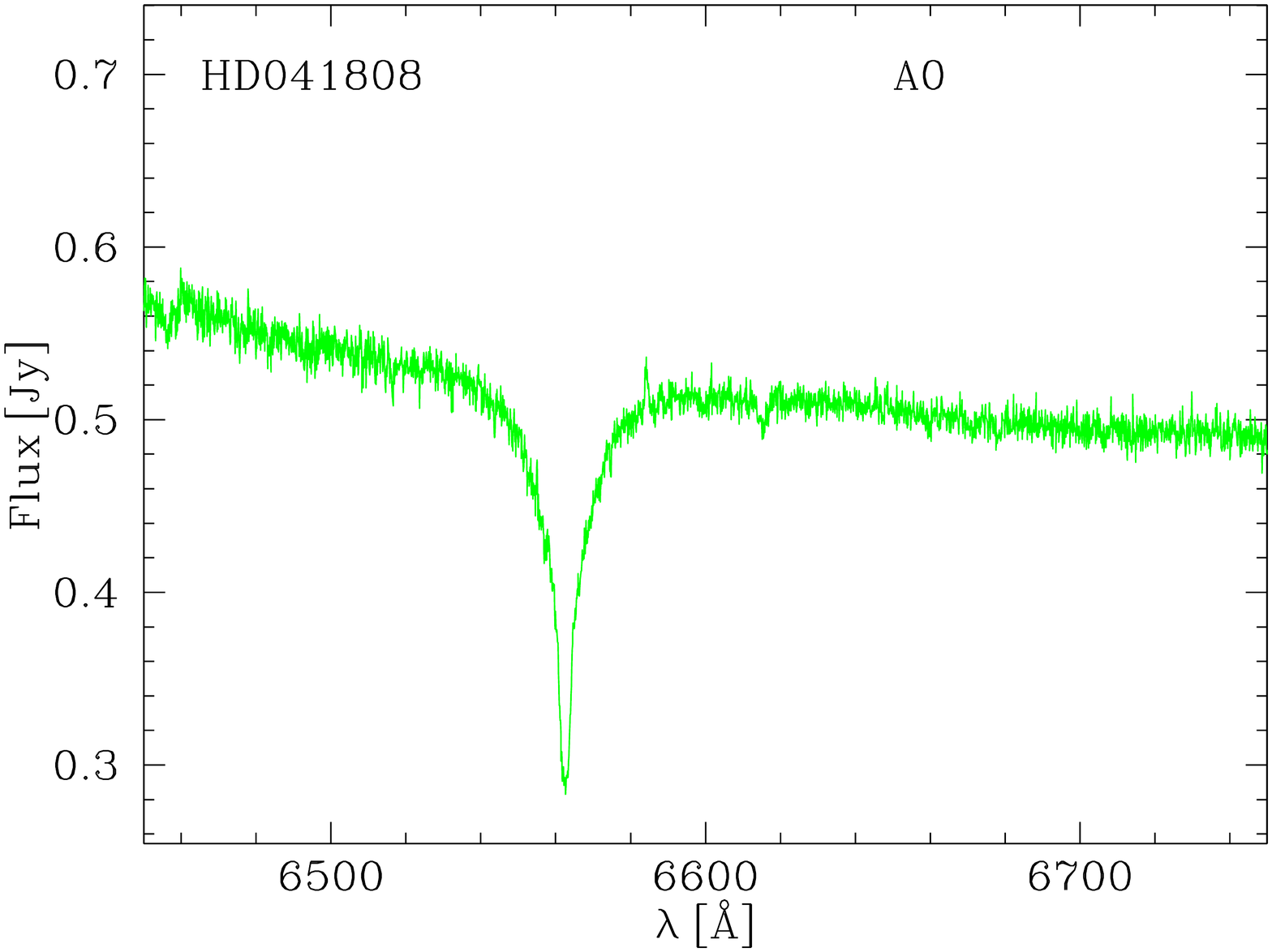}
\includegraphics[width=0.18\textwidth,angle=-90]{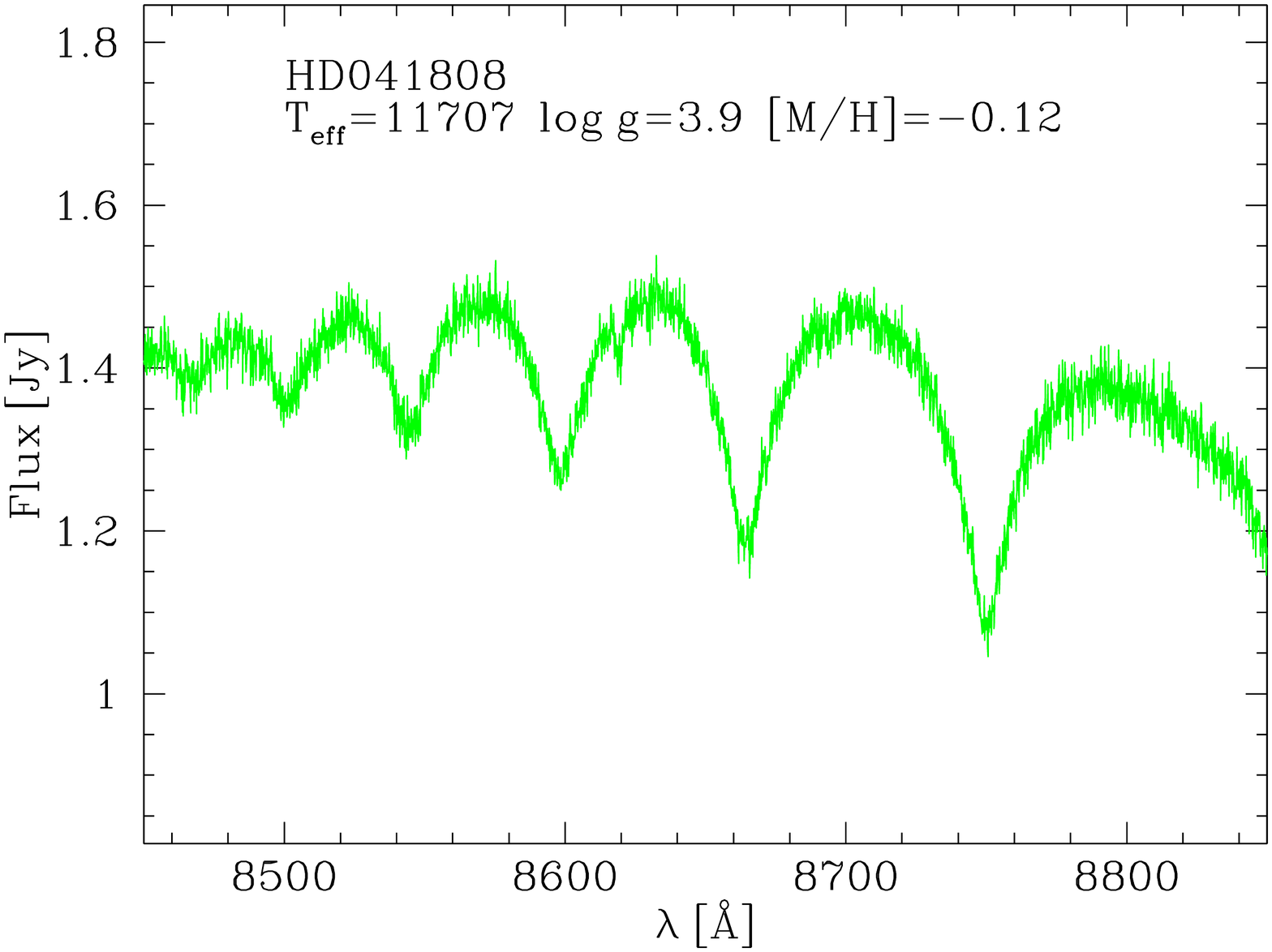}
\includegraphics[width=0.18\textwidth,angle=-90]{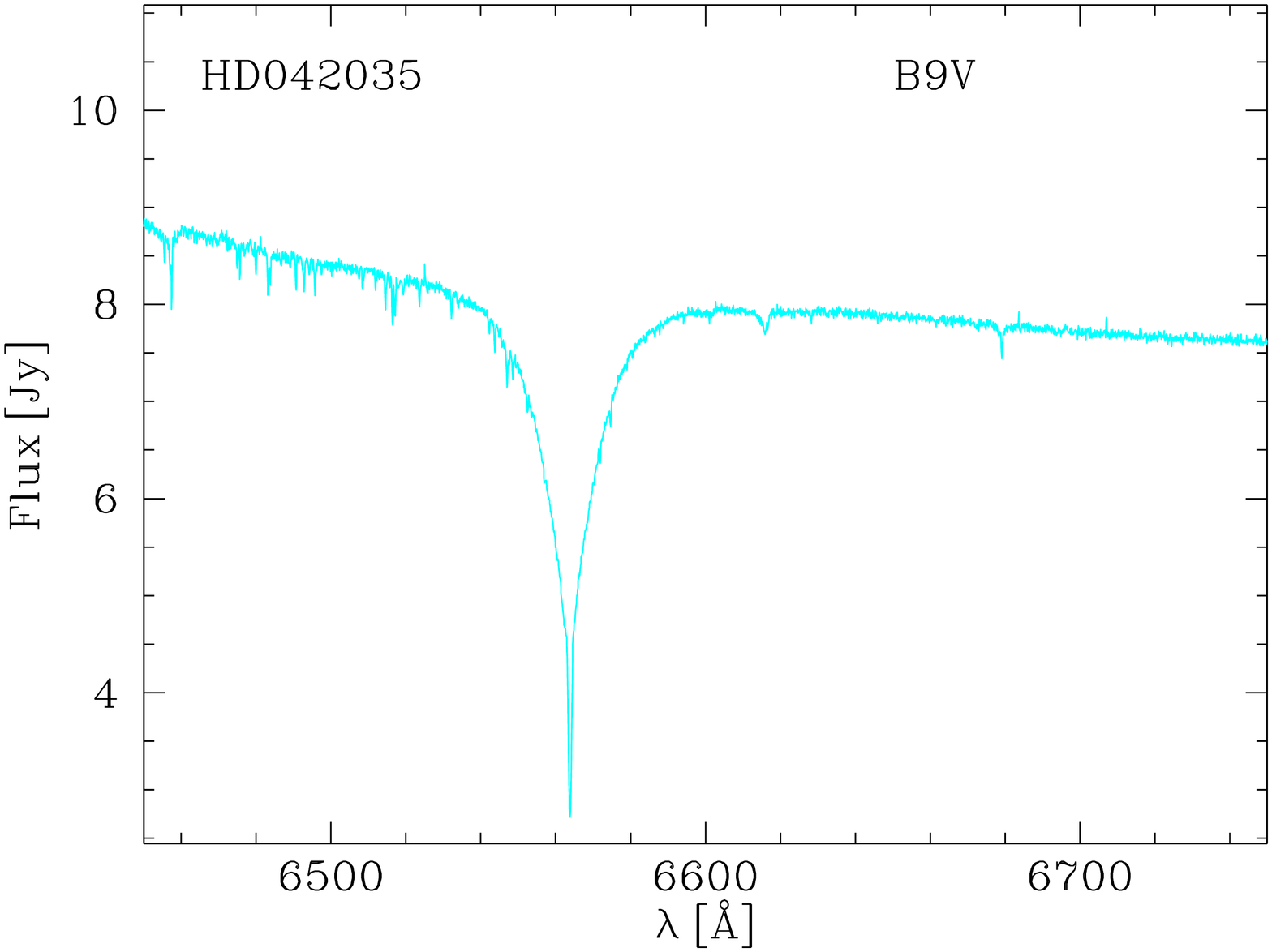} 
\includegraphics[width=0.18\textwidth,angle=-90]{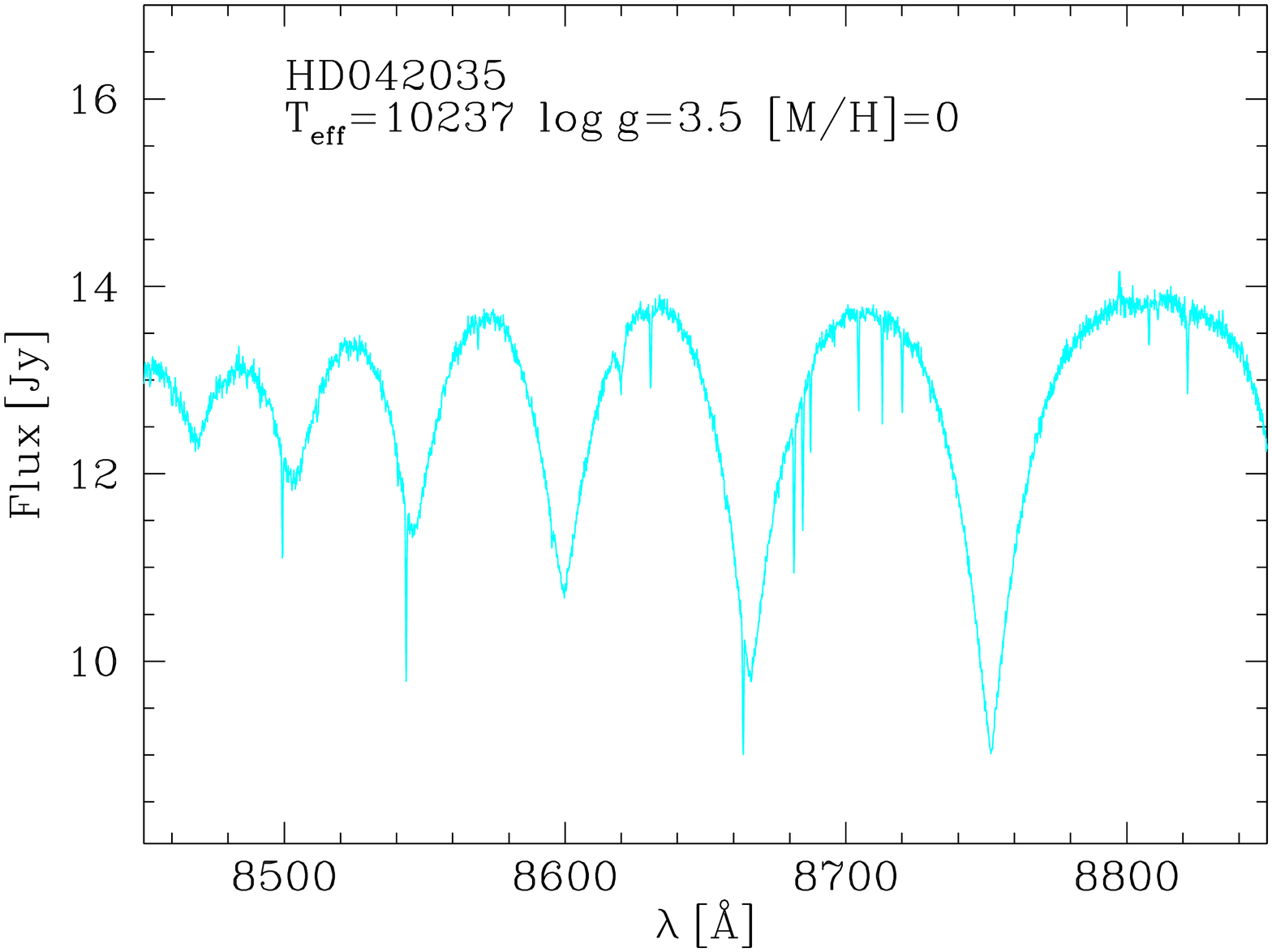}
\includegraphics[width=0.18\textwidth,angle=-90]{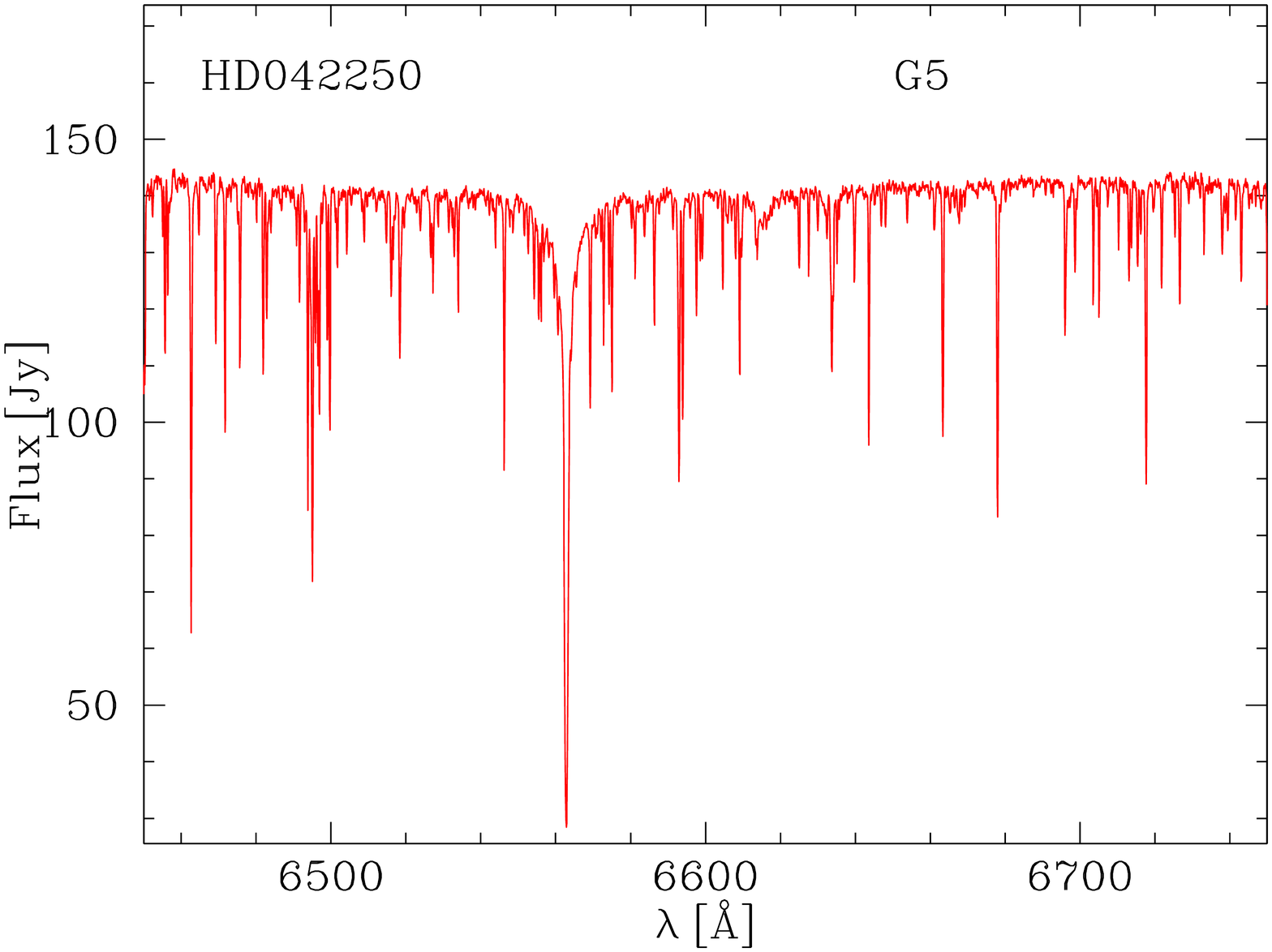}
\includegraphics[width=0.18\textwidth,angle=-90]{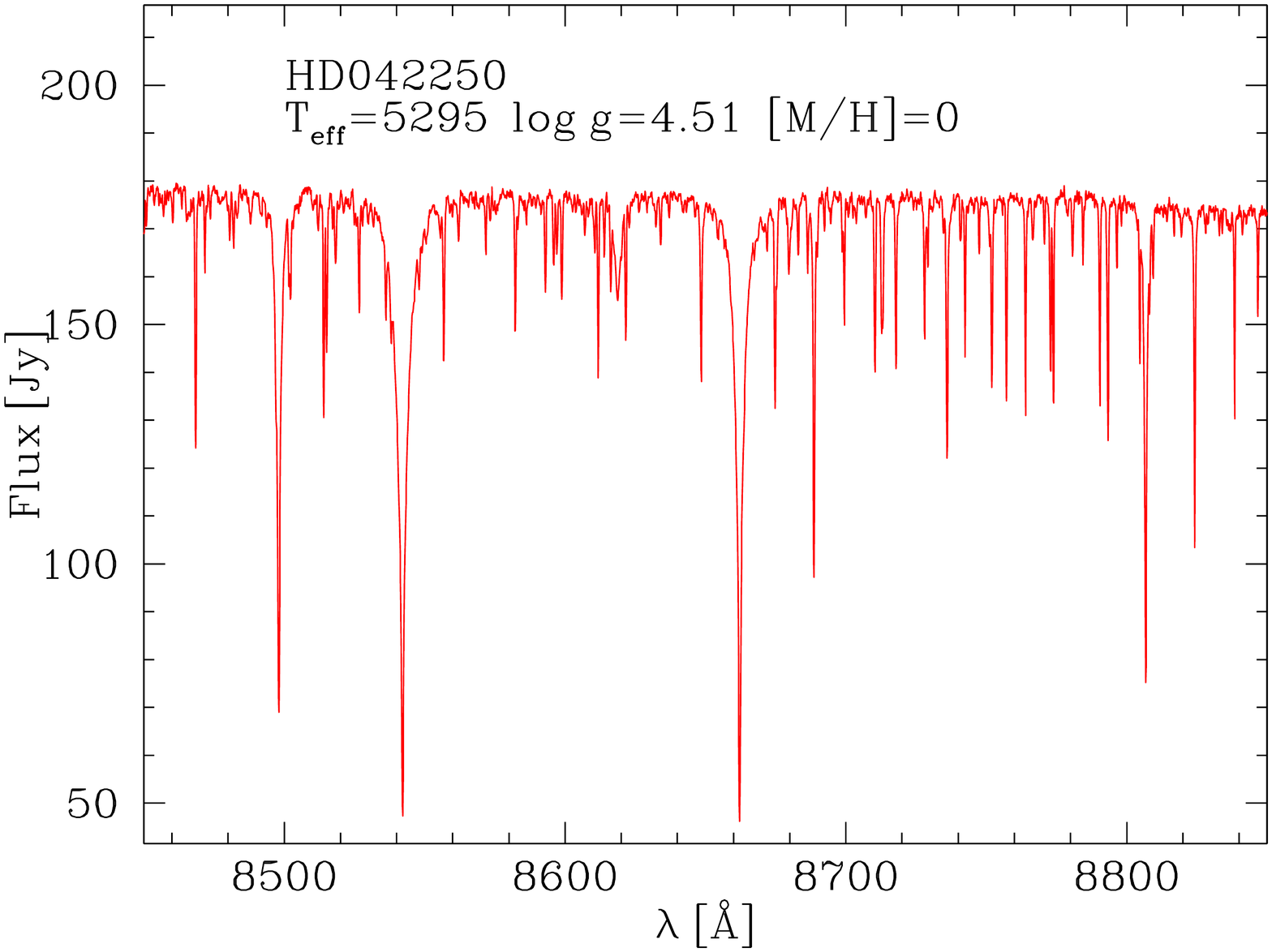}

\contcaption{9. Stars shown in this page are: HD039587, HD039773, HD039801, HD039866, HD040801, HD040964, HD041117, HD041330, HD041357, HD041501, HD041692, HD041808, HD042035 and HD042250.}
\end{figure*}

\begin{figure*}
\includegraphics[width=0.18\textwidth,angle=-90]{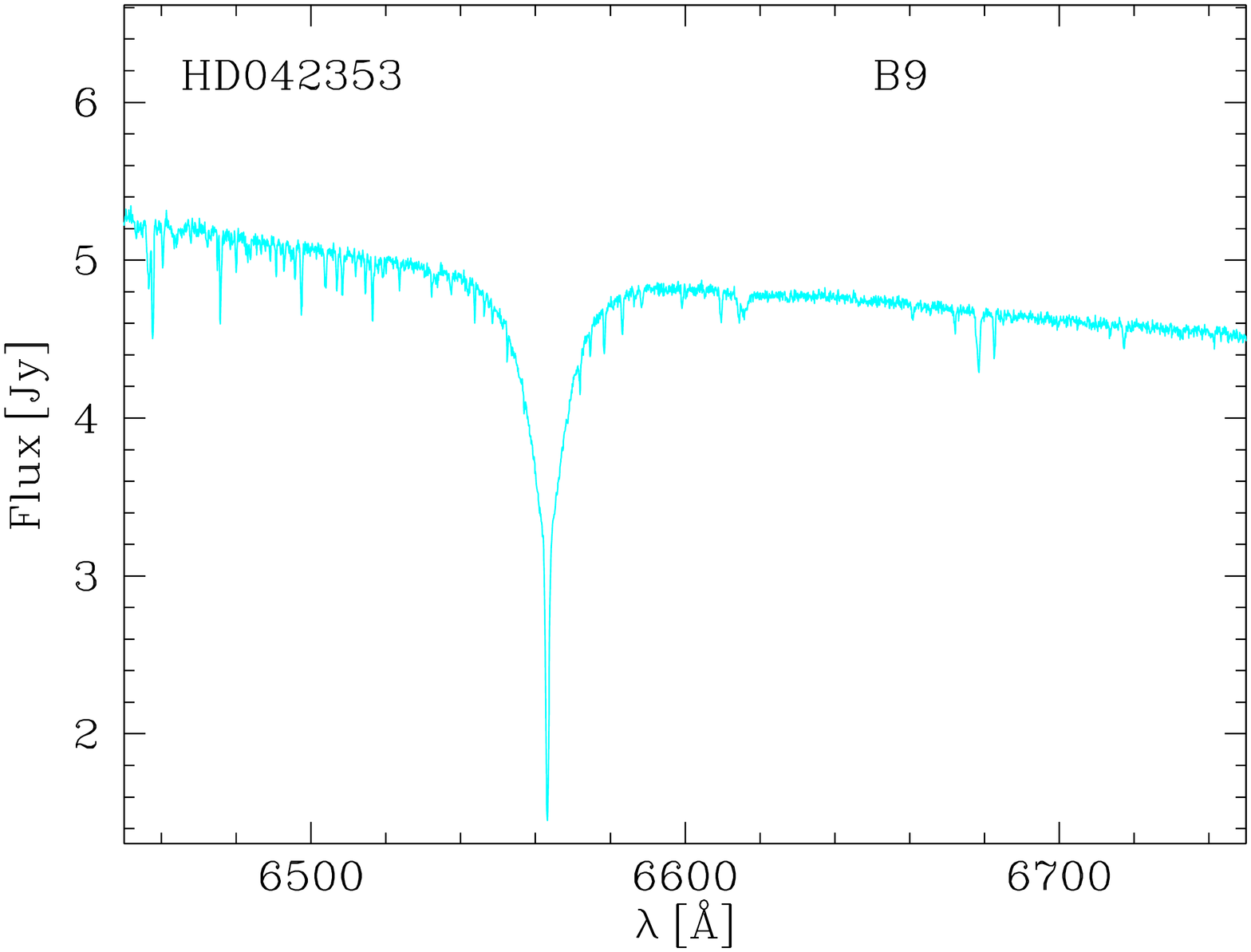}
\includegraphics[width=0.18\textwidth,angle=-90]{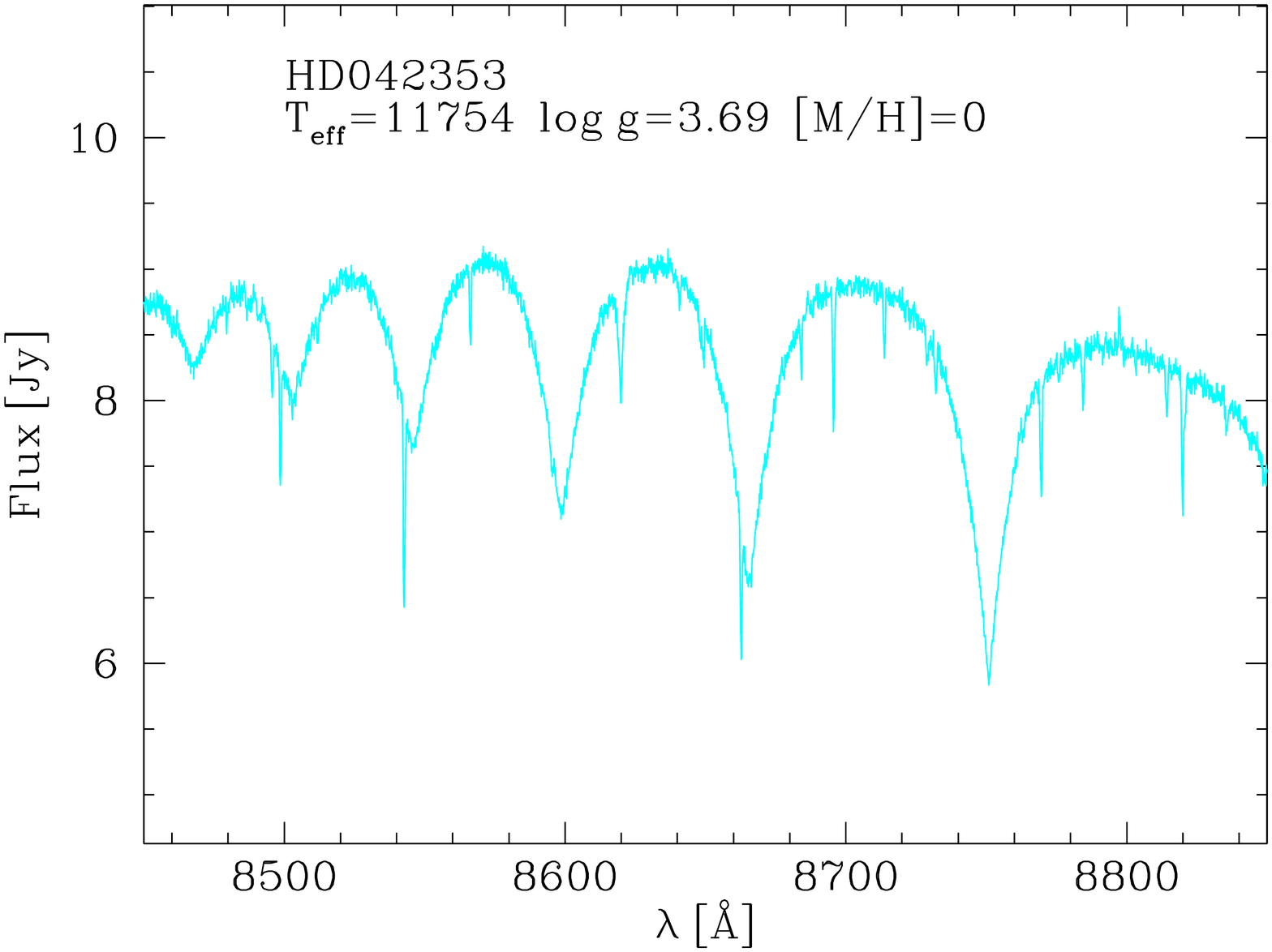}
\includegraphics[width=0.18\textwidth,angle=-90]{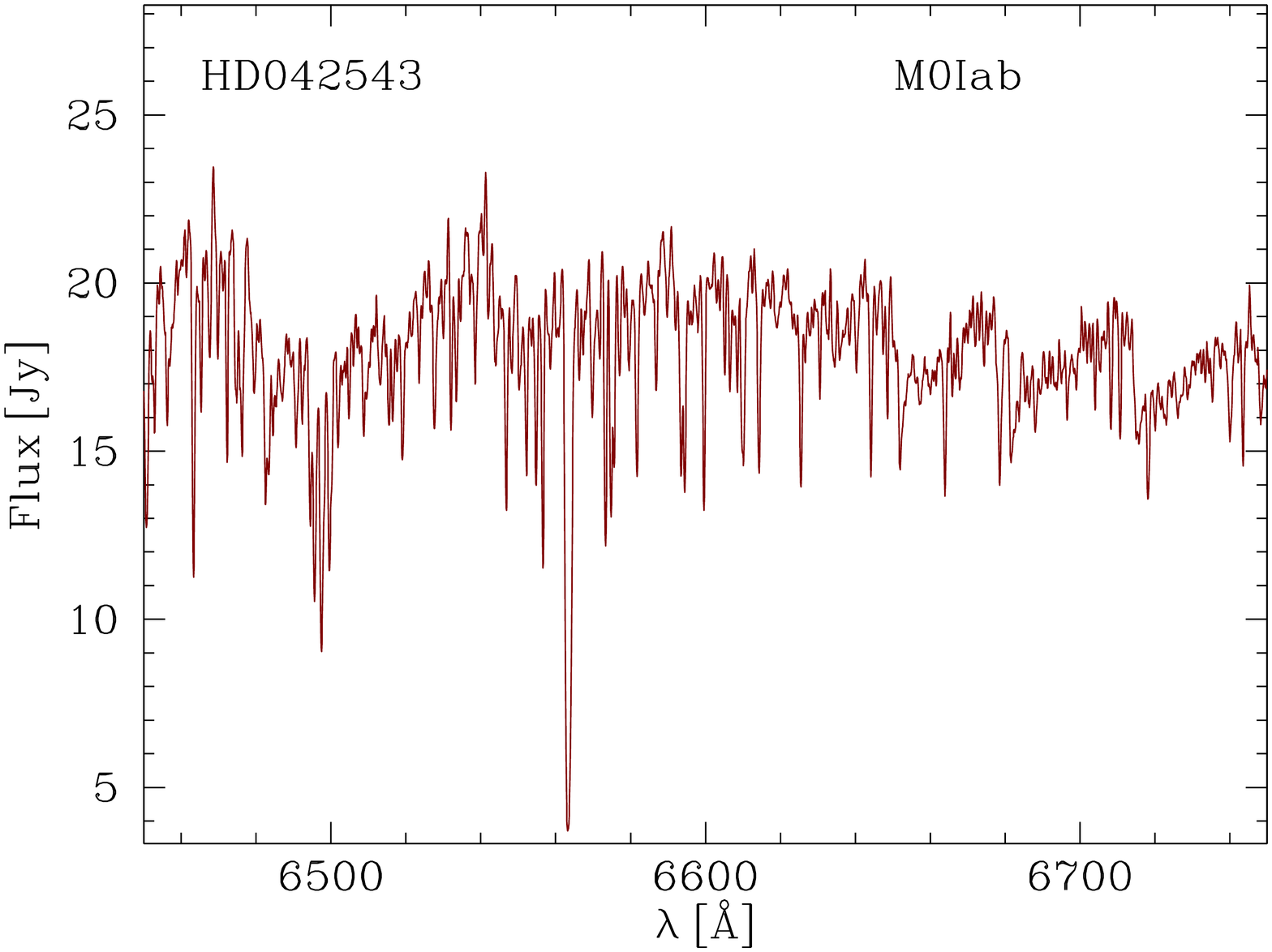}
\includegraphics[width=0.18\textwidth,angle=-90]{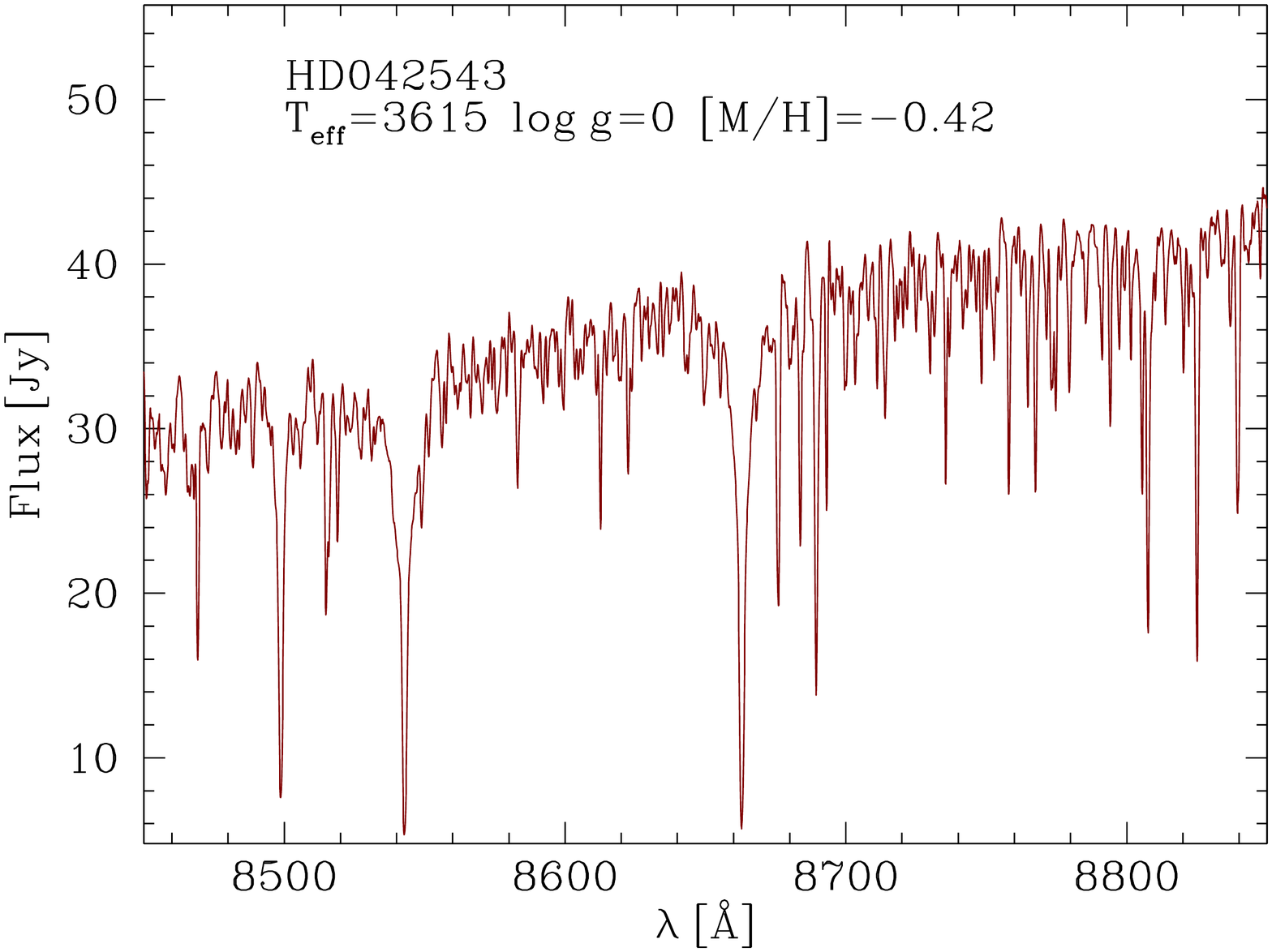}
\includegraphics[width=0.18\textwidth,angle=-90]{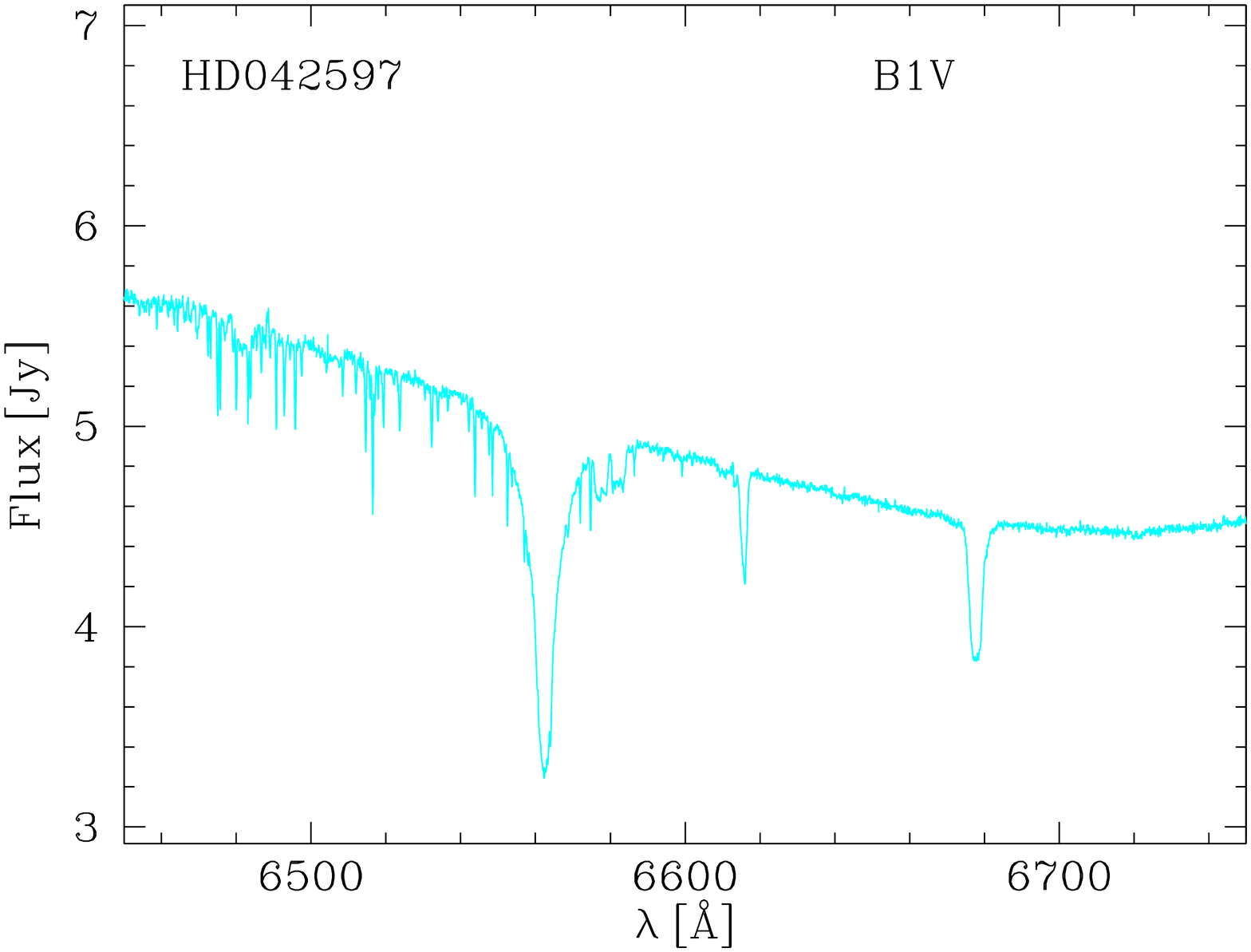}
\includegraphics[width=0.18\textwidth,angle=-90]{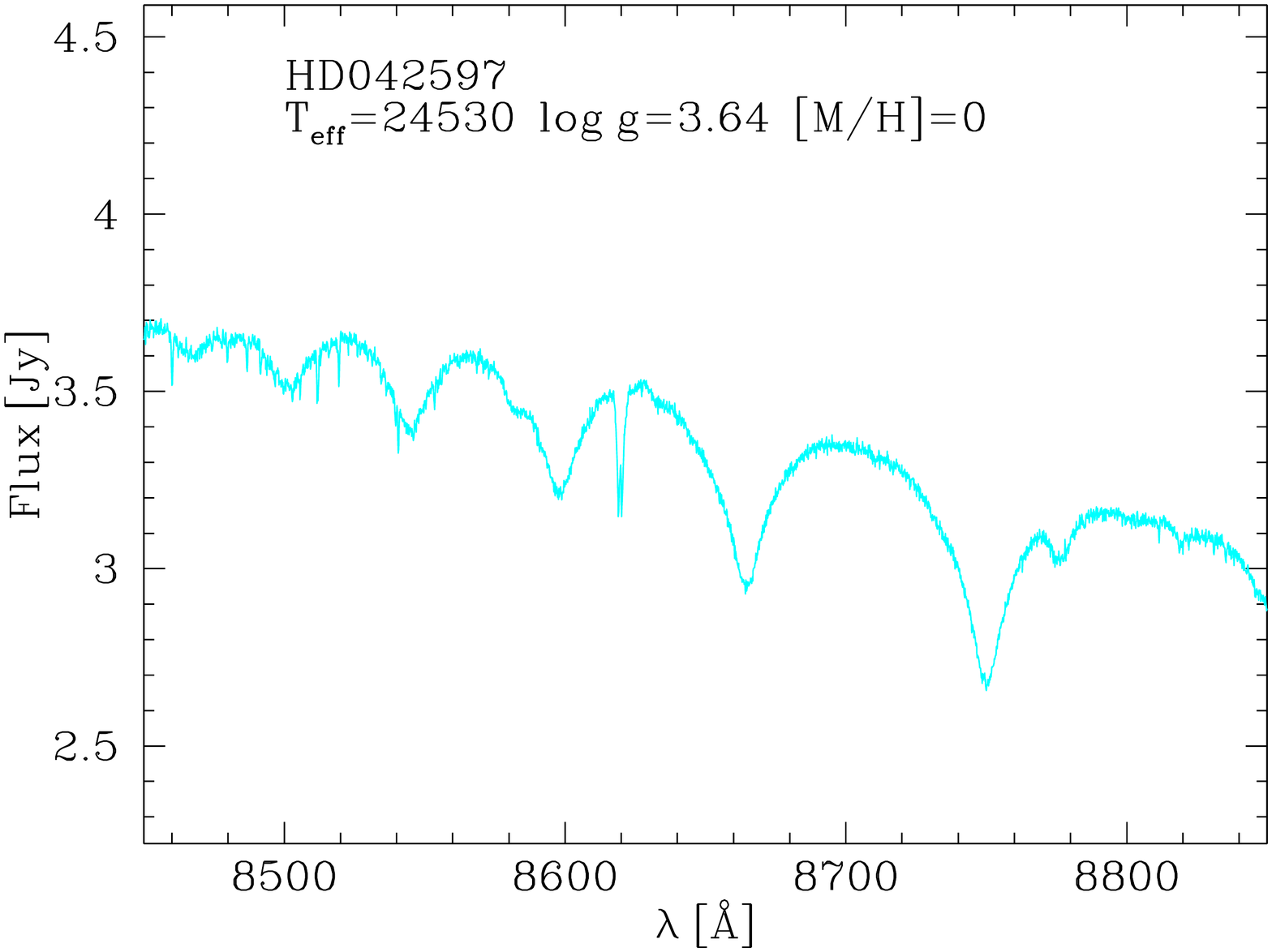}
\includegraphics[width=0.18\textwidth,angle=-90]{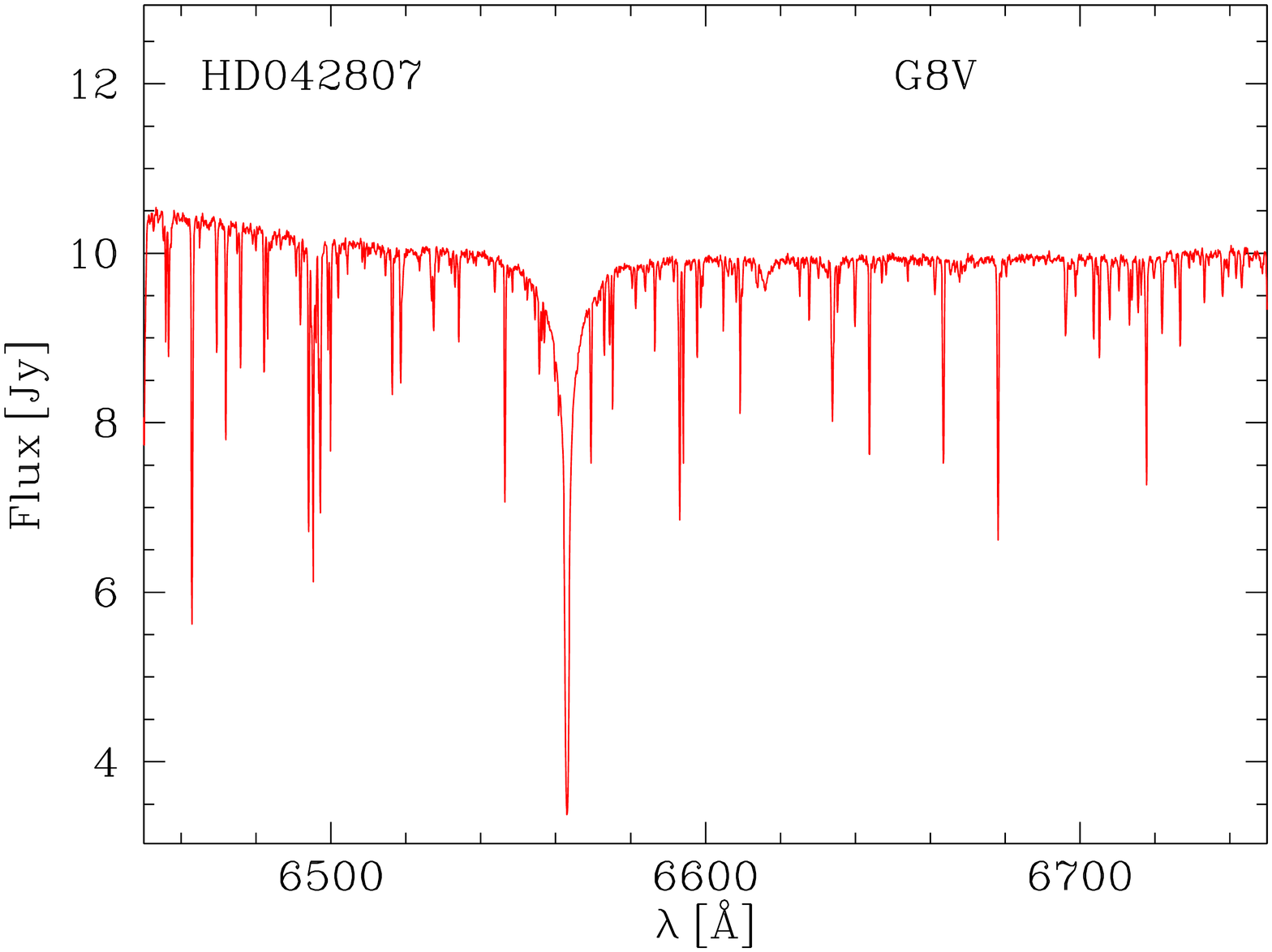}
\includegraphics[width=0.18\textwidth,angle=-90]{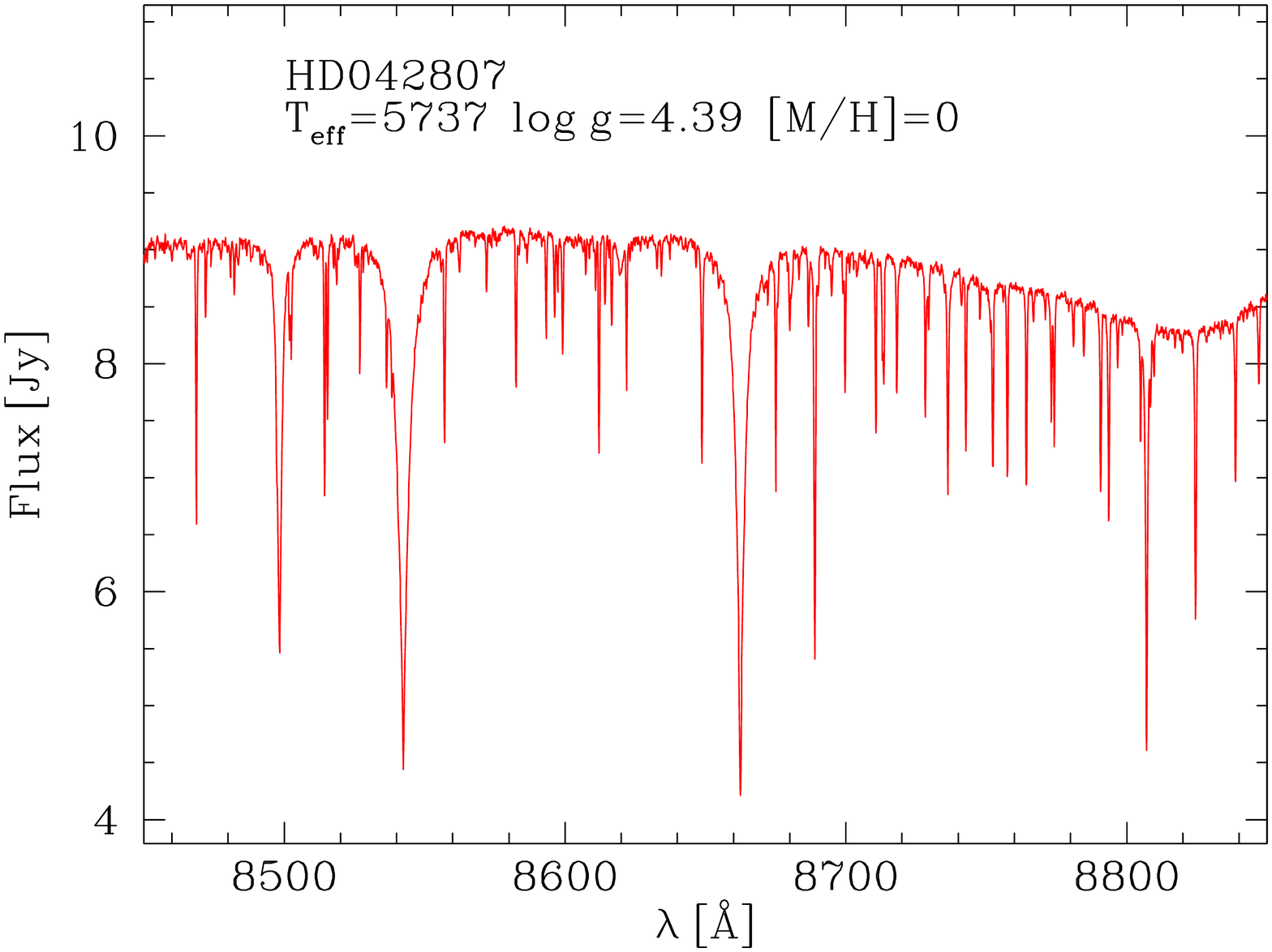}
\includegraphics[width=0.18\textwidth,angle=-90]{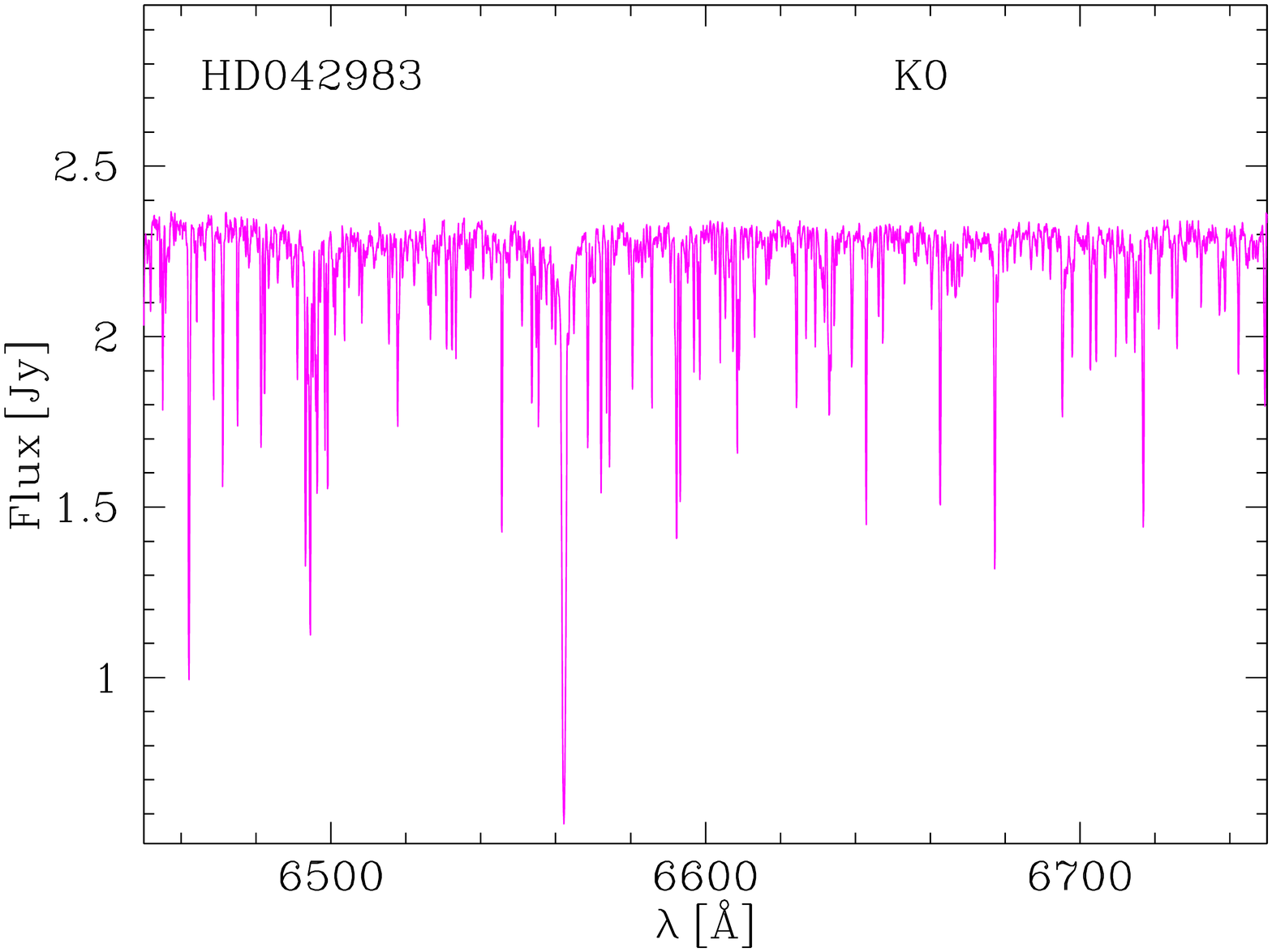}
\includegraphics[width=0.18\textwidth,angle=-90]{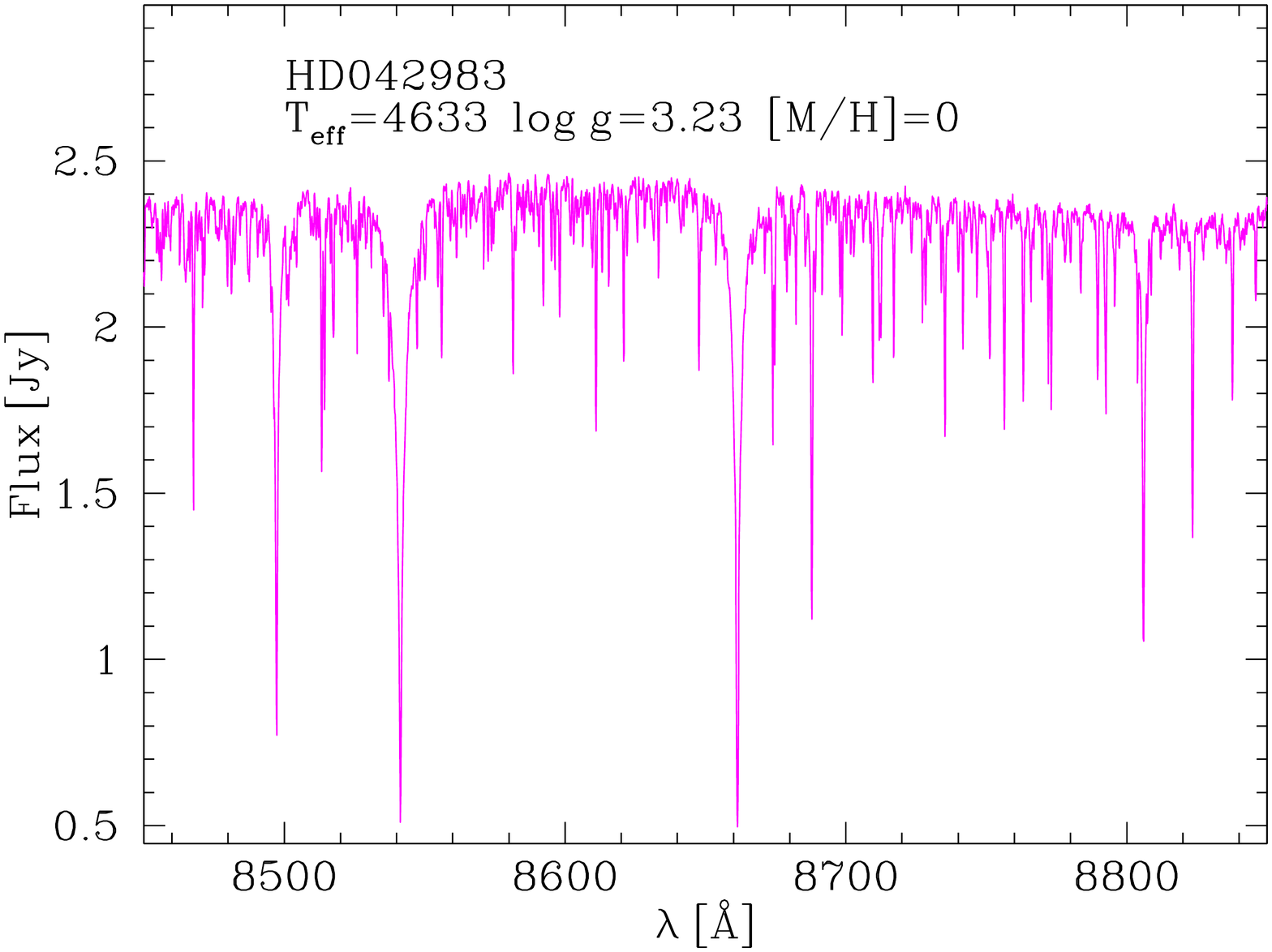}
\includegraphics[width=0.18\textwidth,angle=-90]{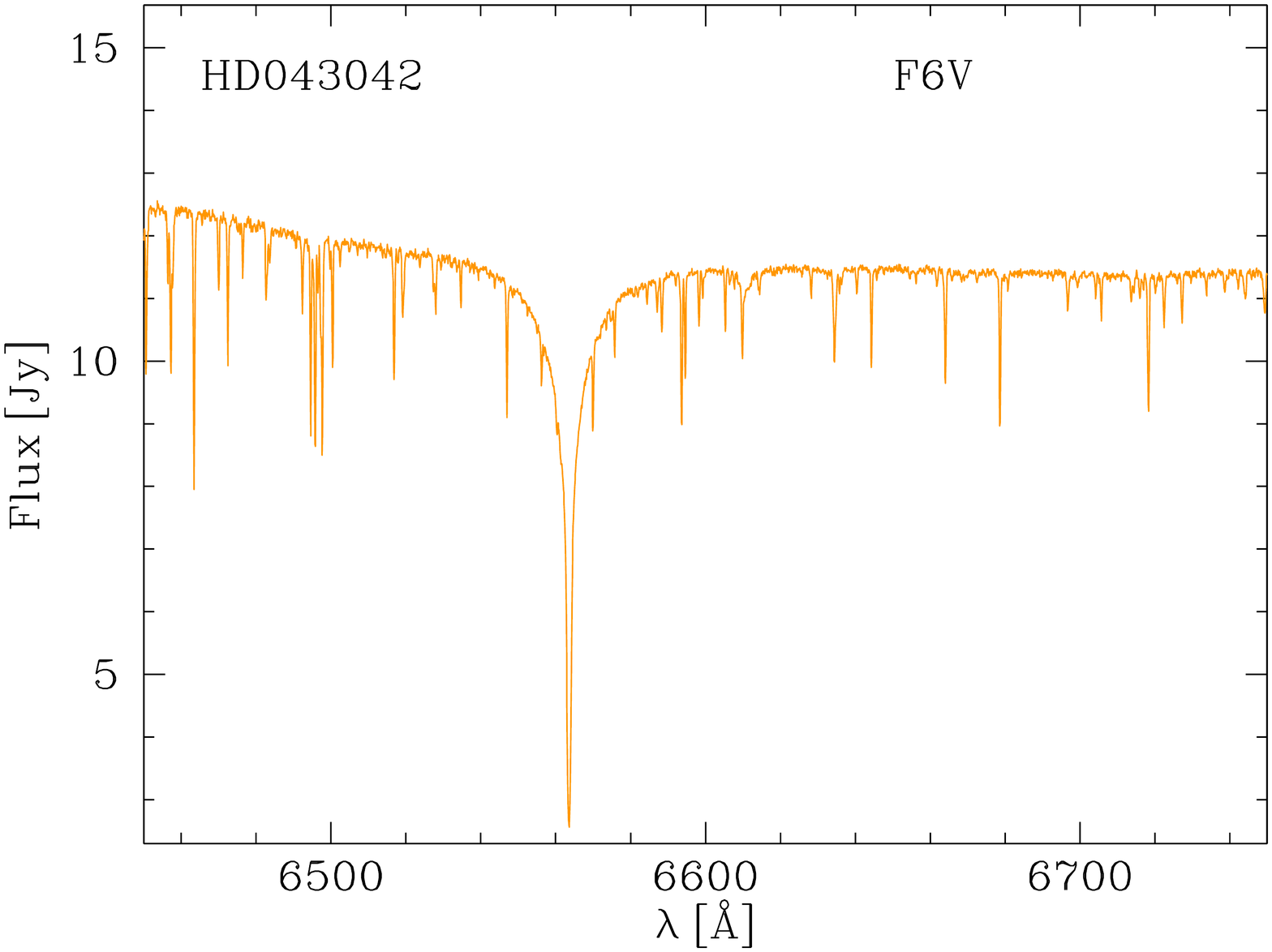}
\includegraphics[width=0.18\textwidth,angle=-90]{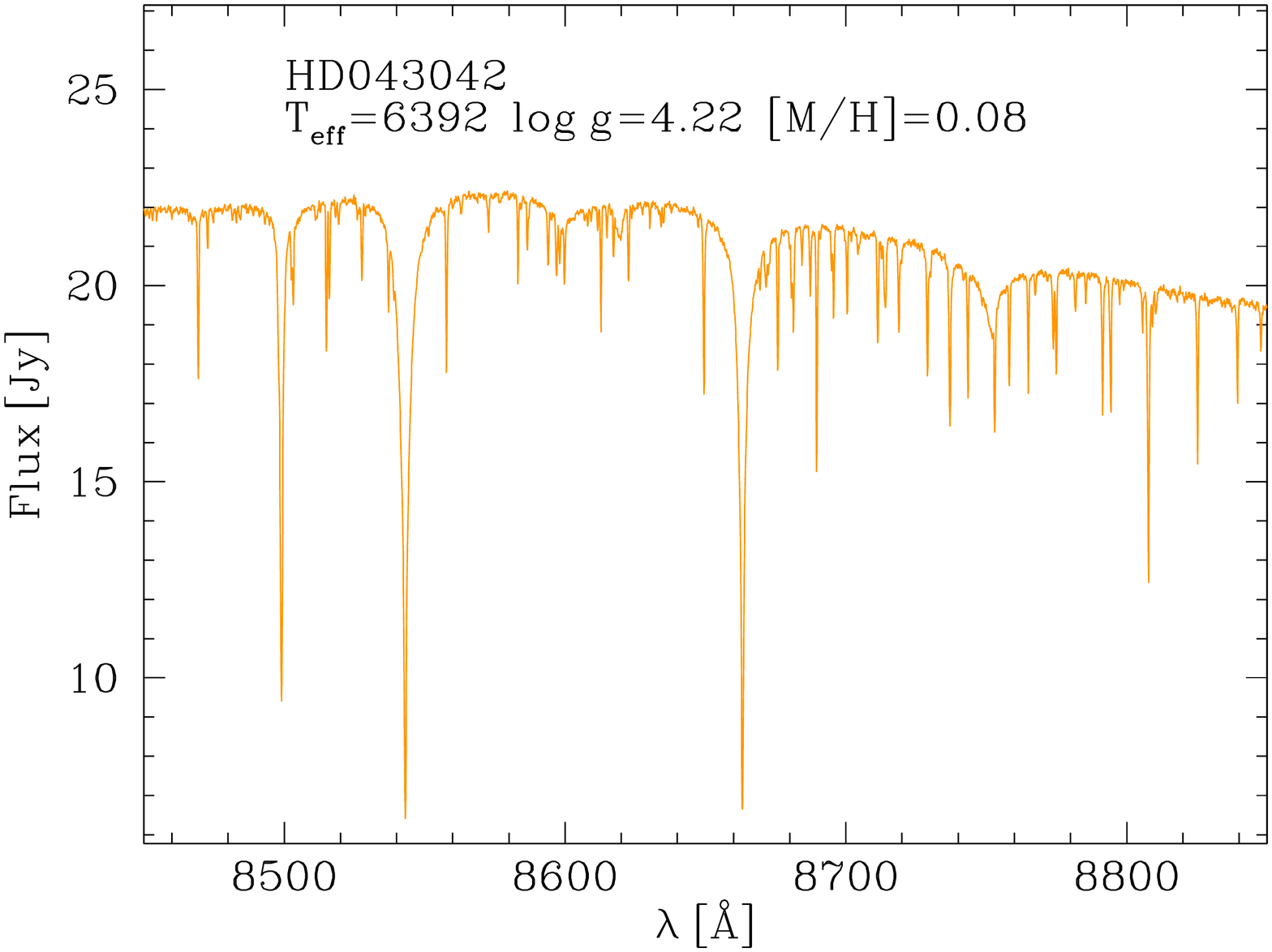}
\includegraphics[width=0.18\textwidth,angle=-90]{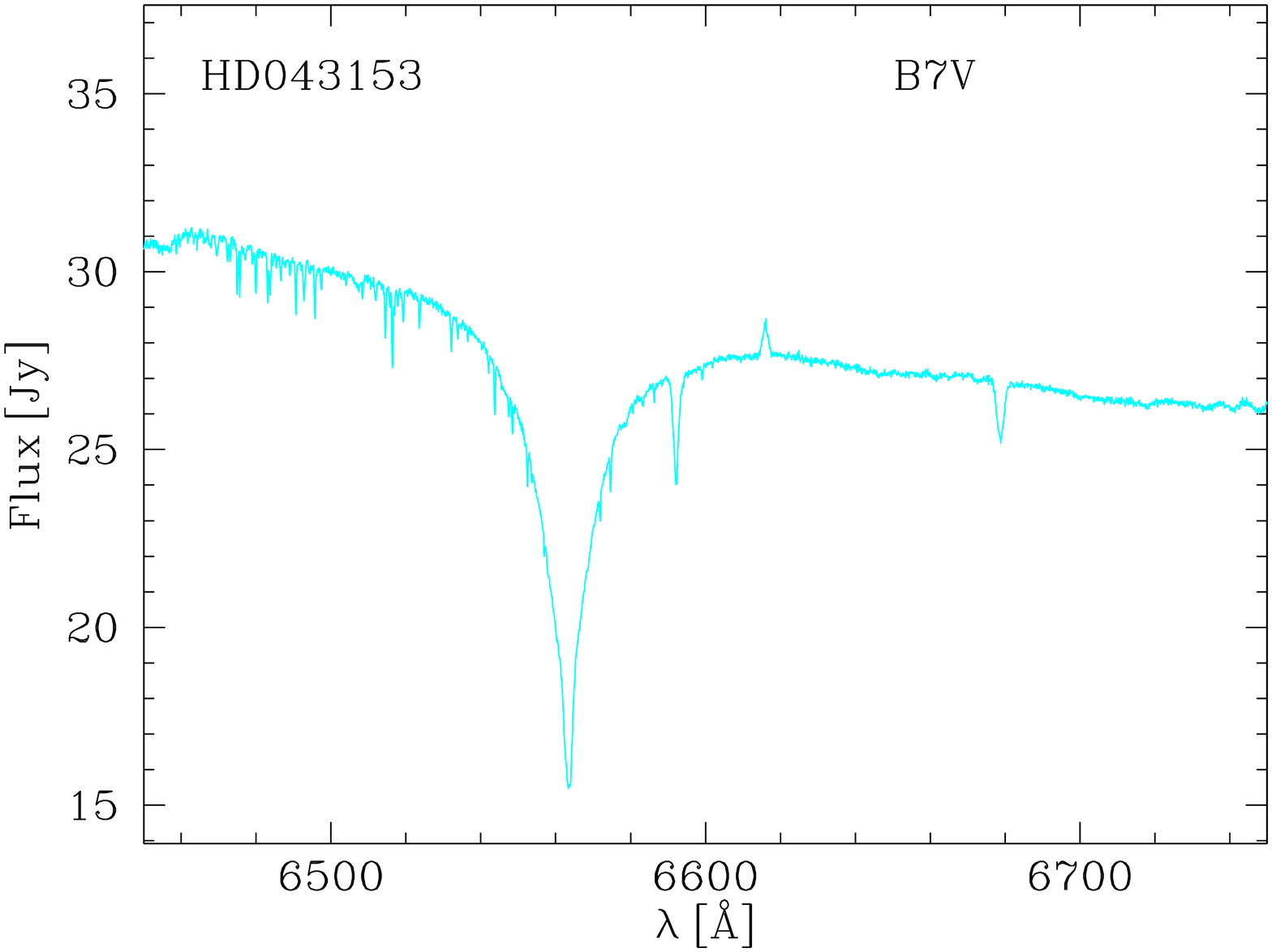}
\includegraphics[width=0.18\textwidth,angle=-90]{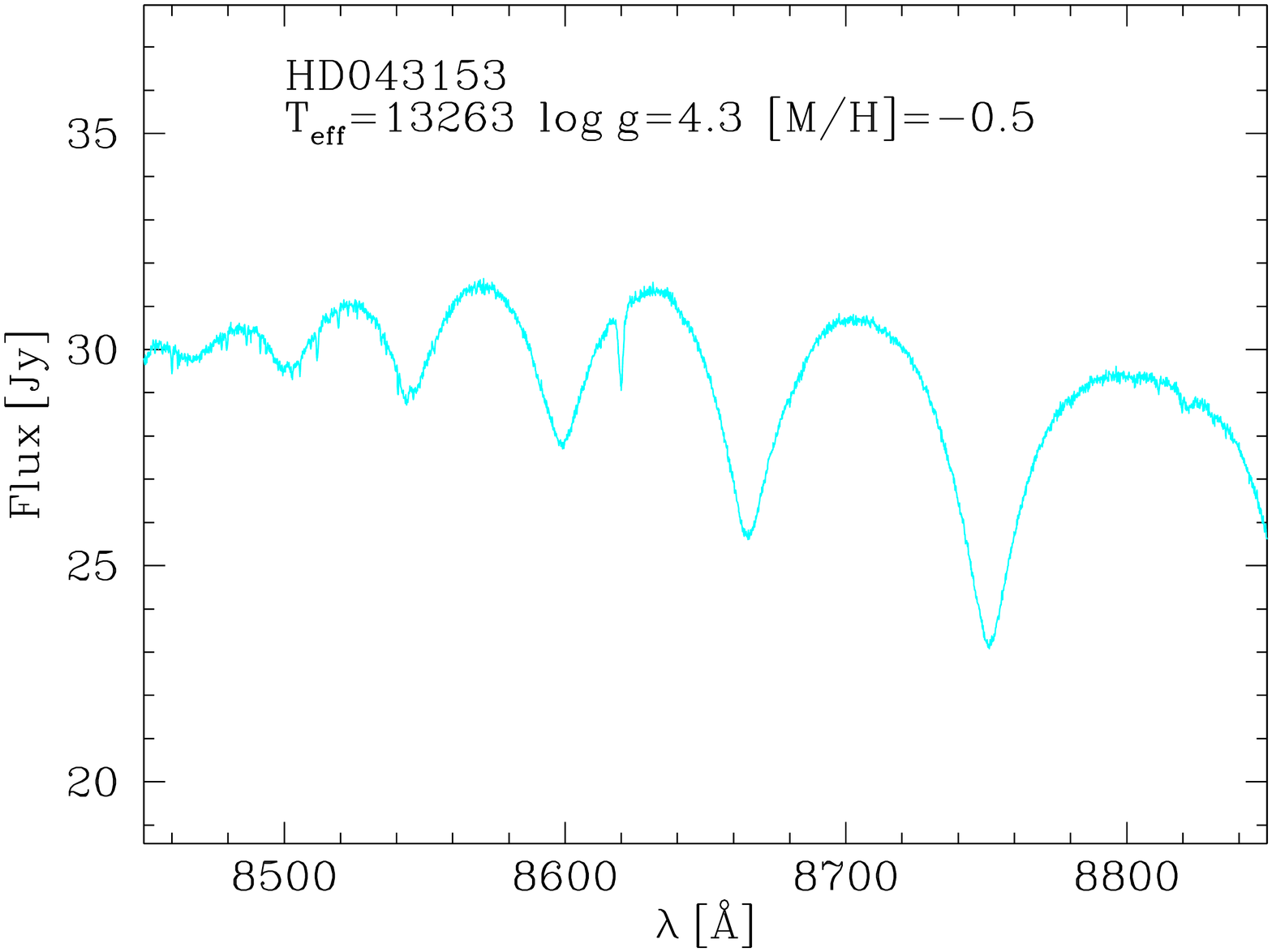}
\includegraphics[width=0.18\textwidth,angle=-90]{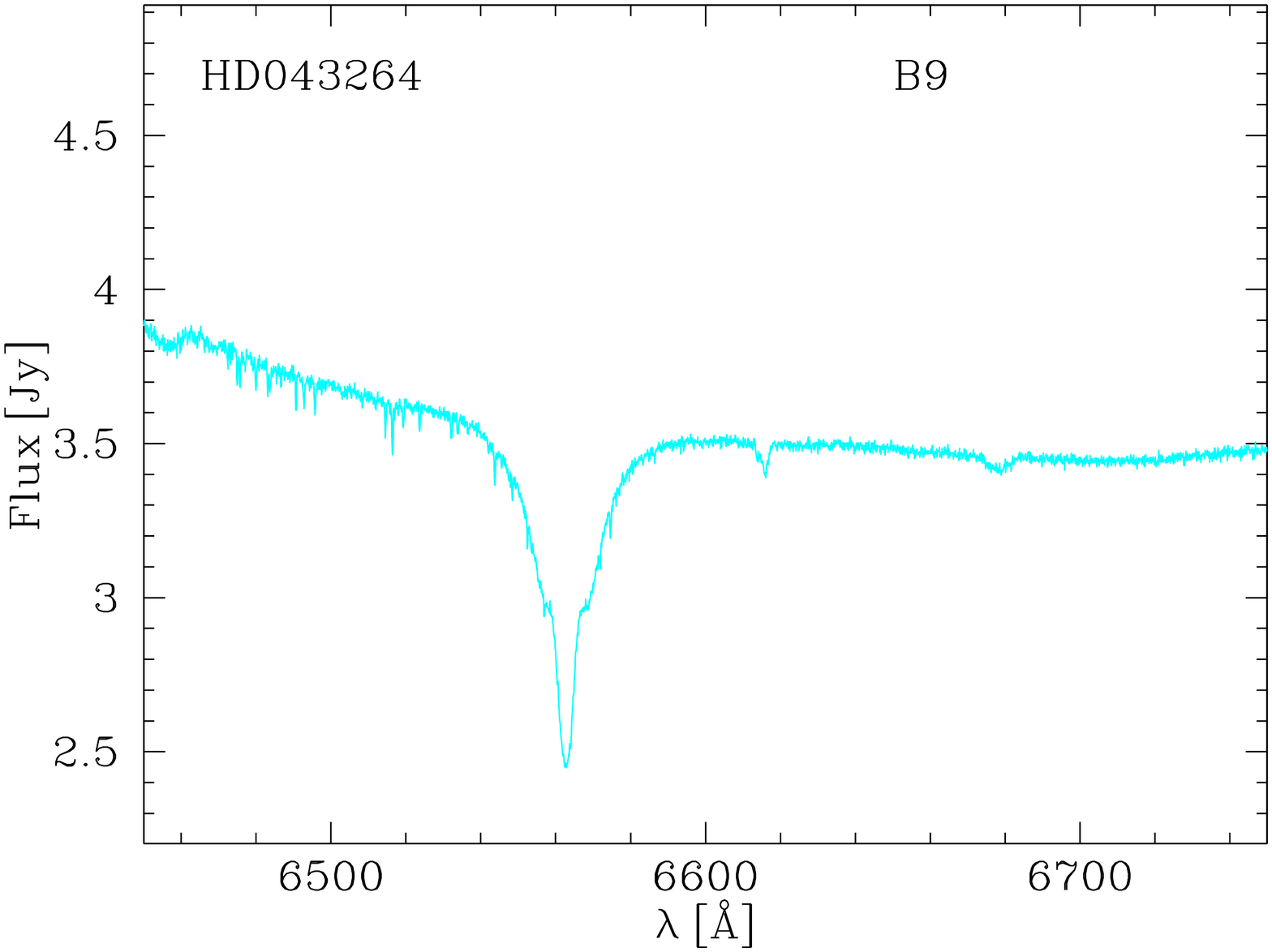}
\includegraphics[width=0.18\textwidth,angle=-90]{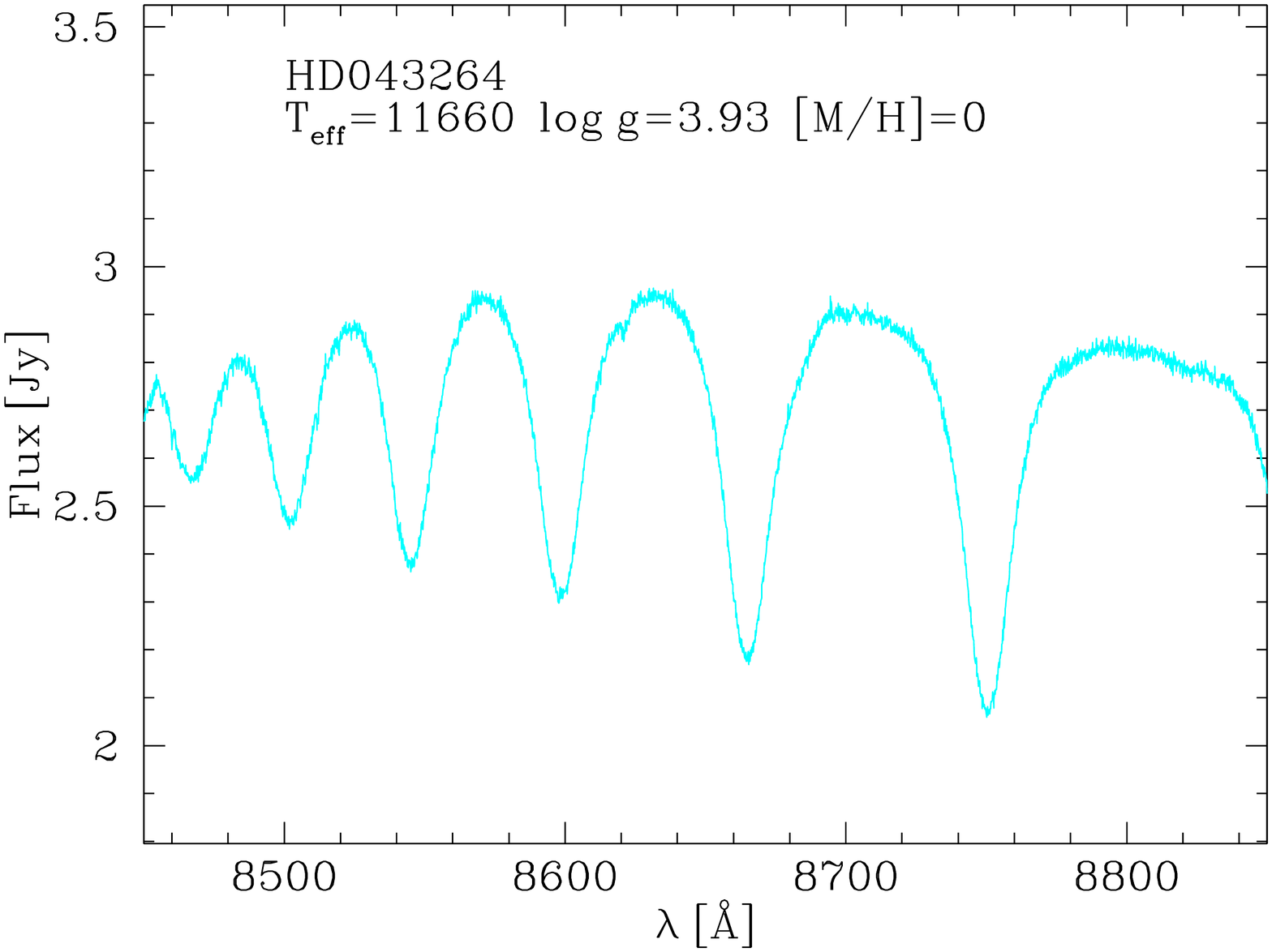}
\includegraphics[width=0.18\textwidth,angle=-90]{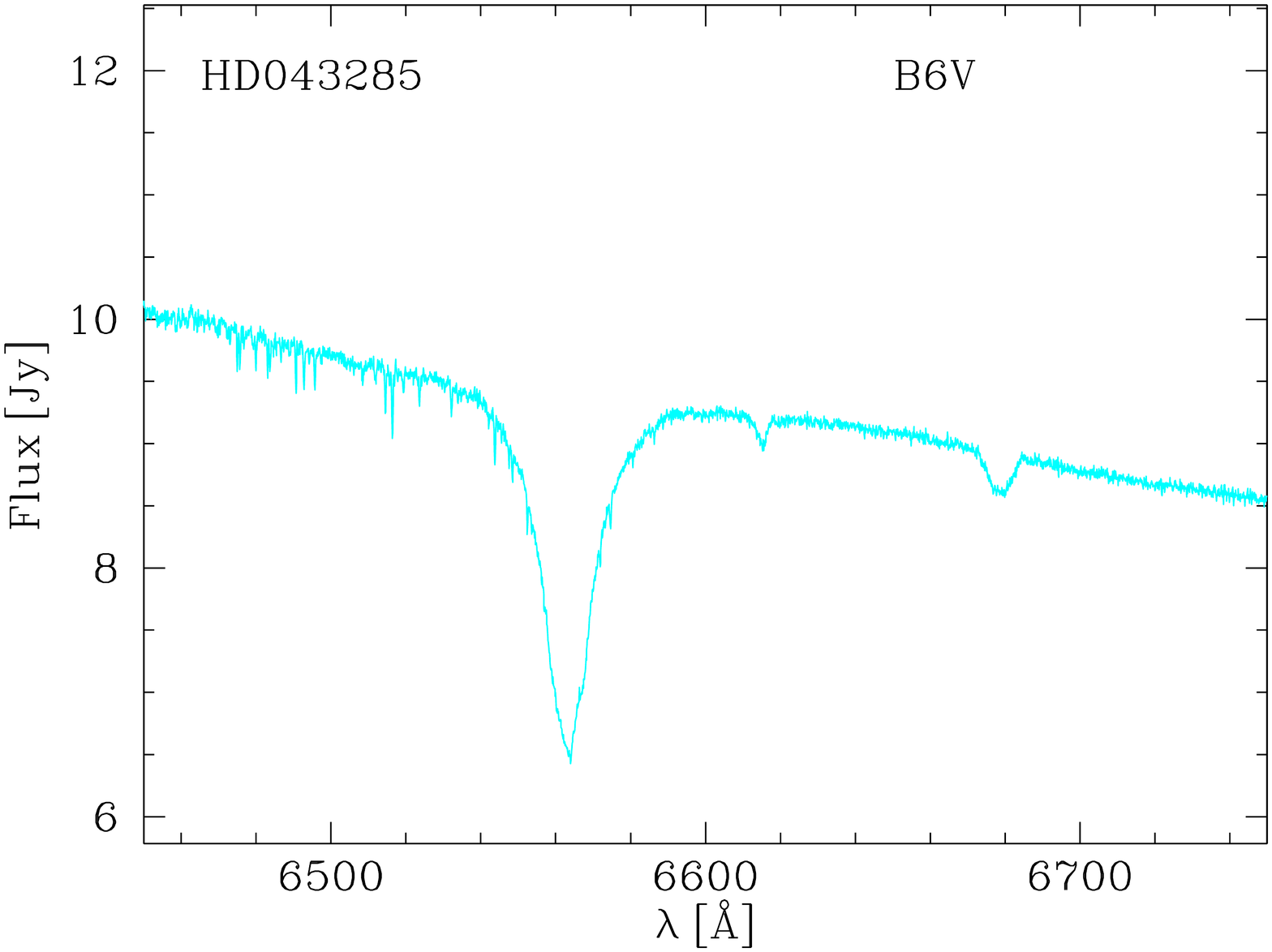}
\includegraphics[width=0.18\textwidth,angle=-90]{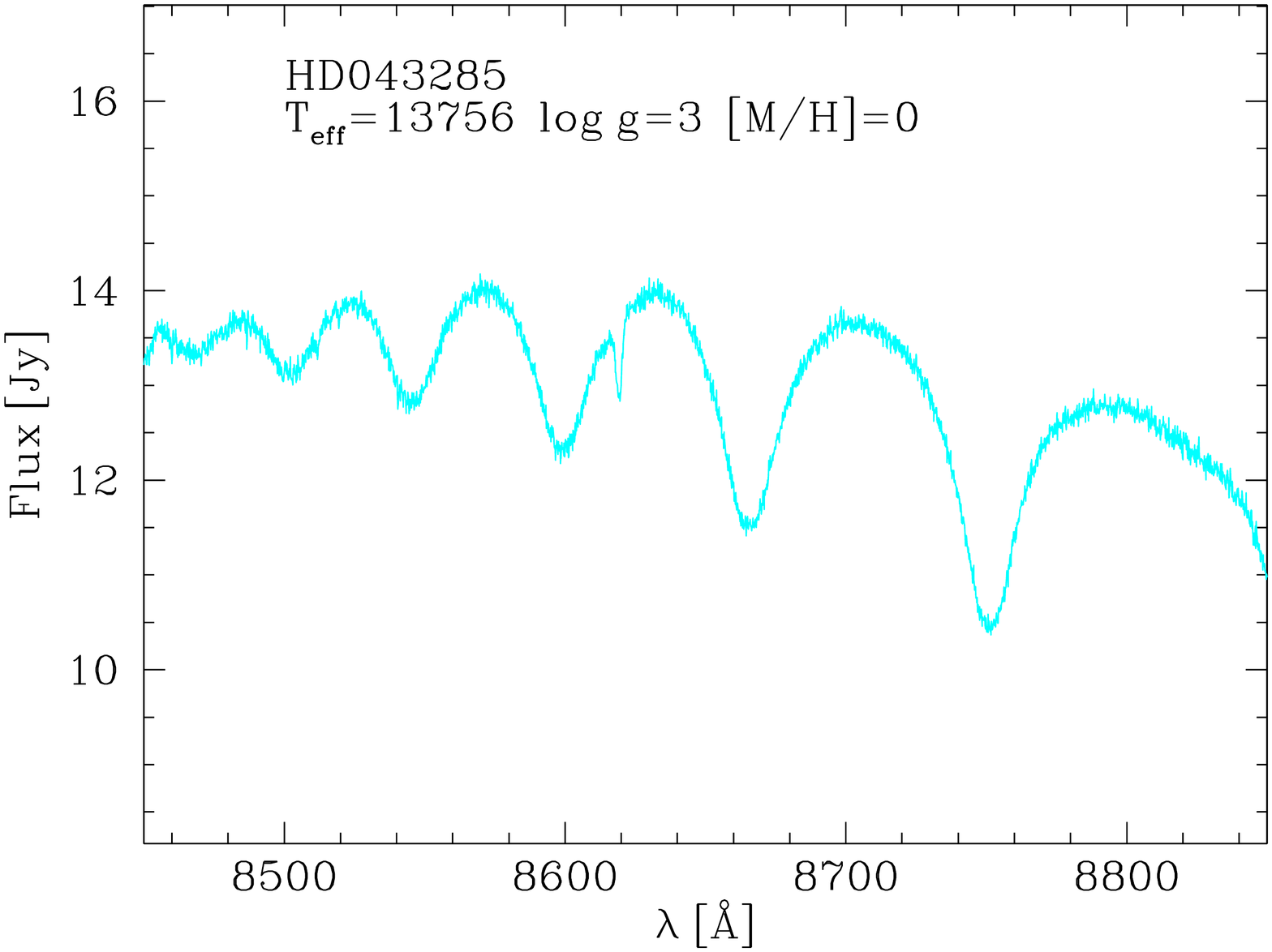}
\includegraphics[width=0.18\textwidth,angle=-90]{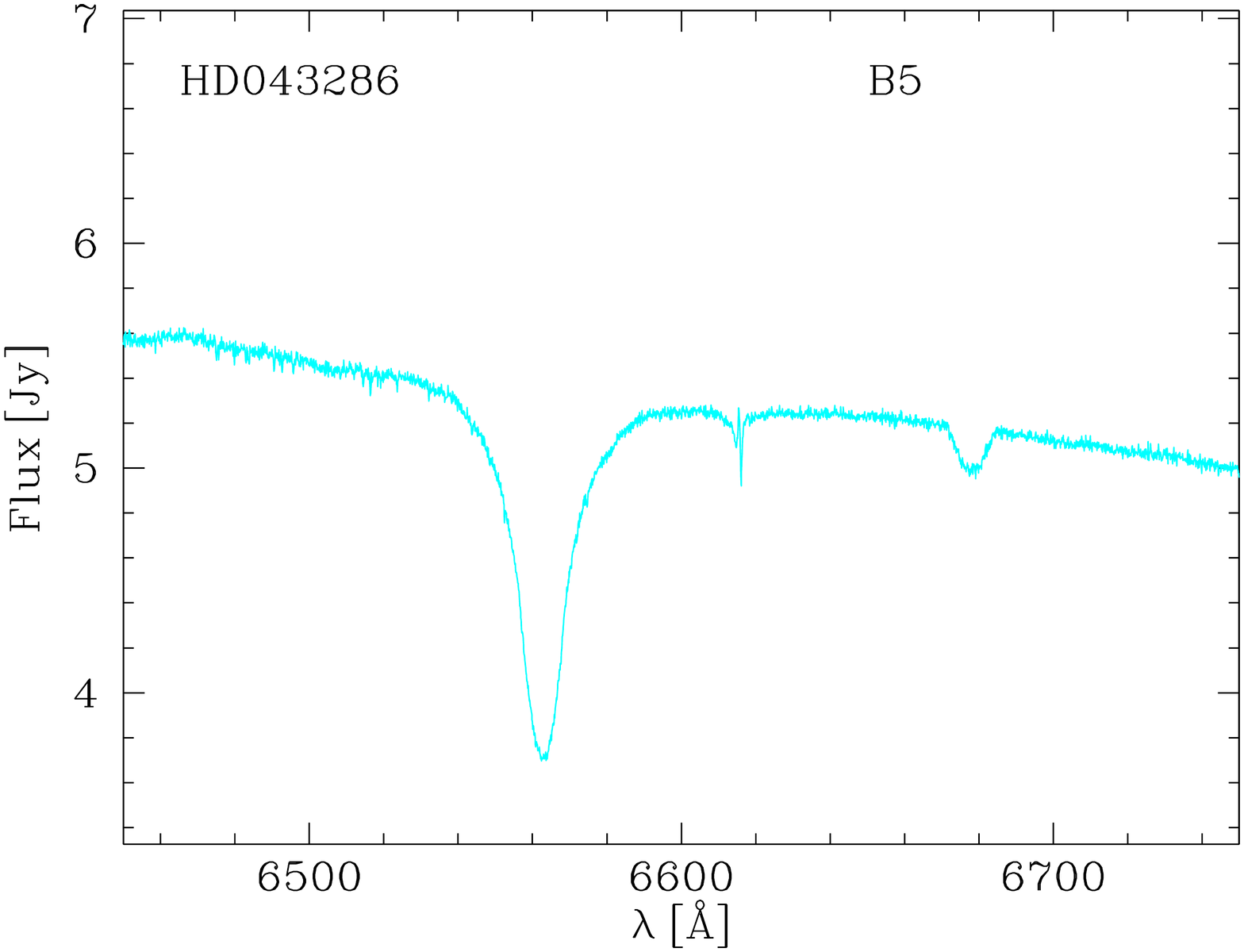}
\includegraphics[width=0.18\textwidth,angle=-90]{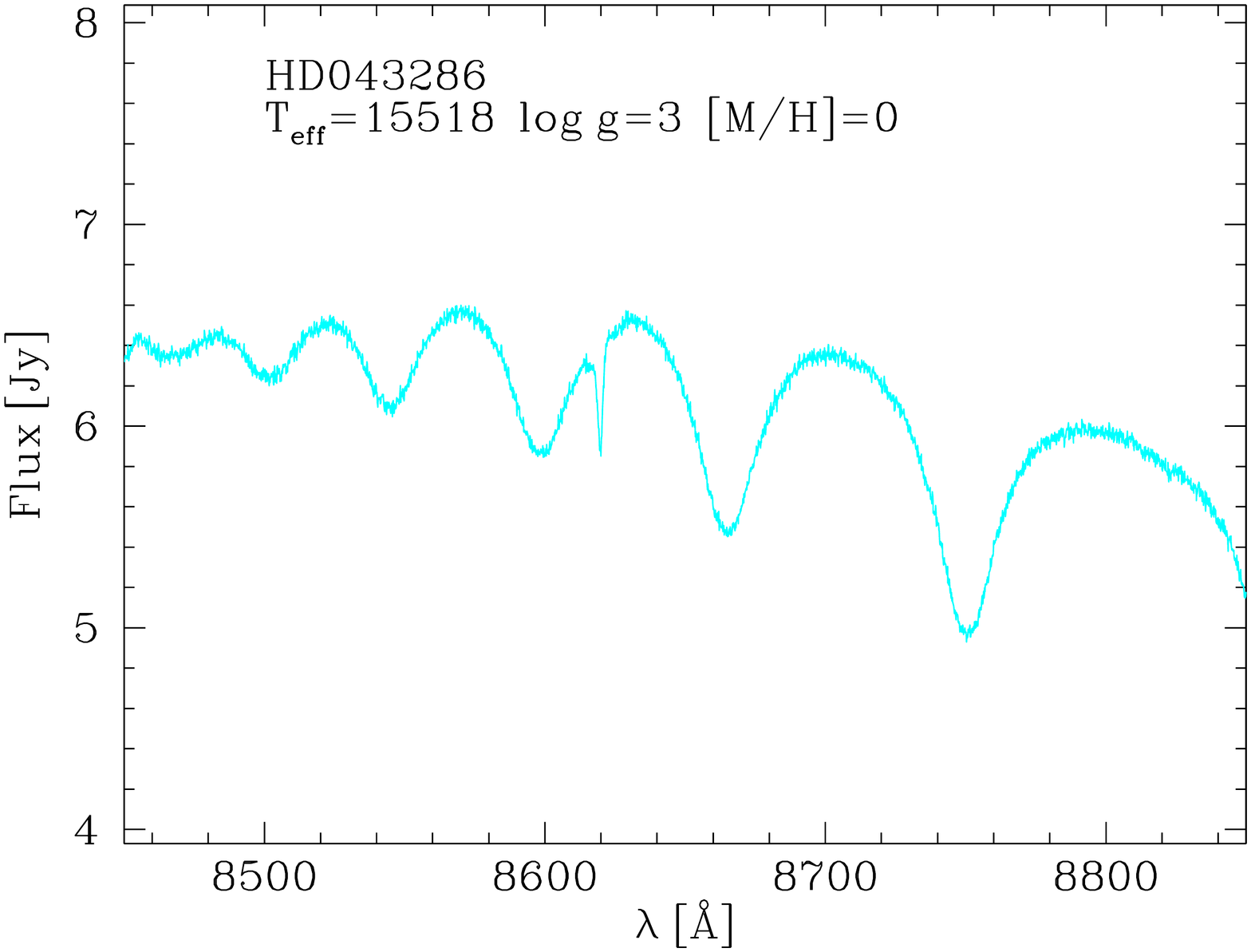}
\includegraphics[width=0.18\textwidth,angle=-90]{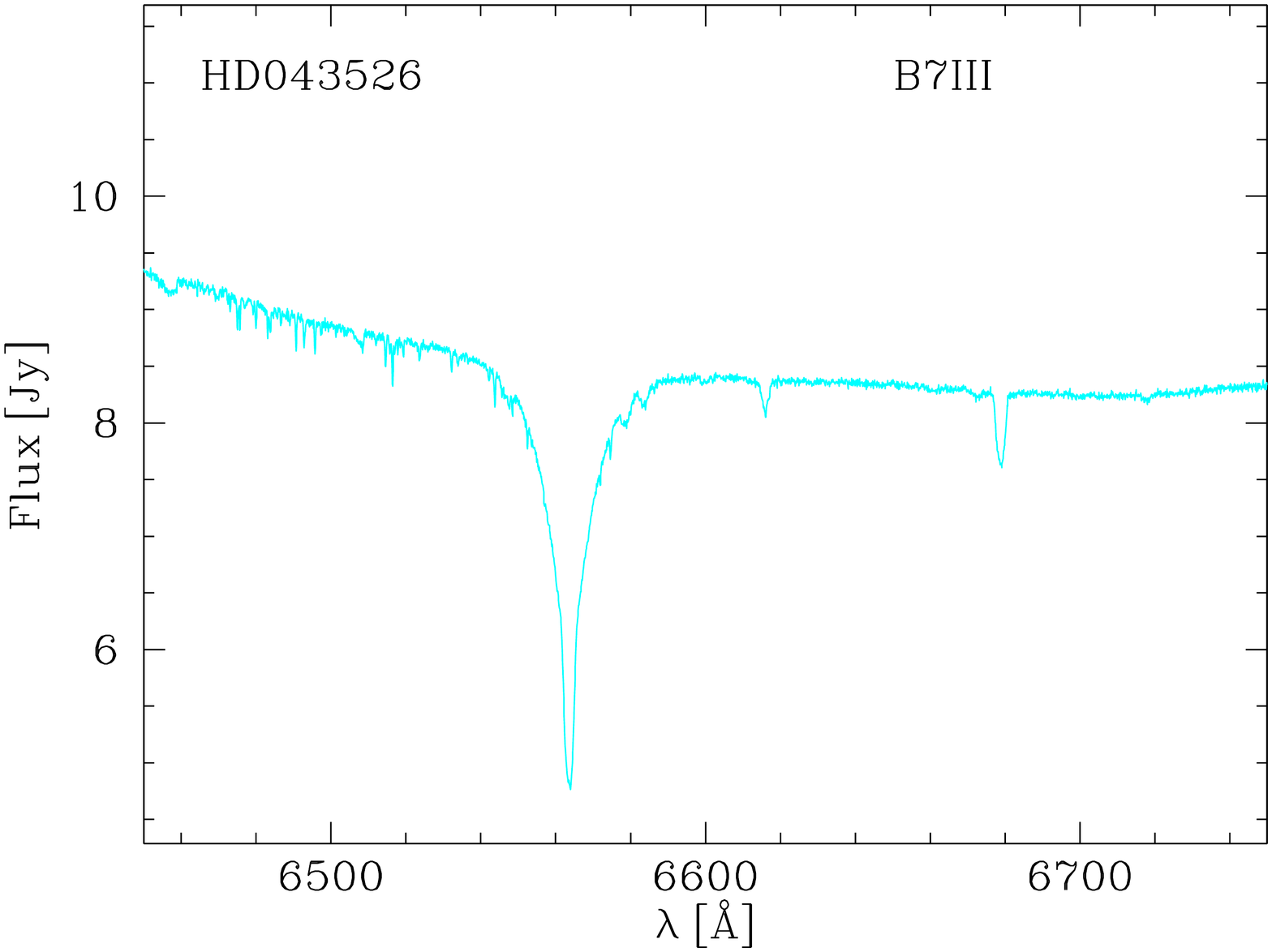}
\includegraphics[width=0.18\textwidth,angle=-90]{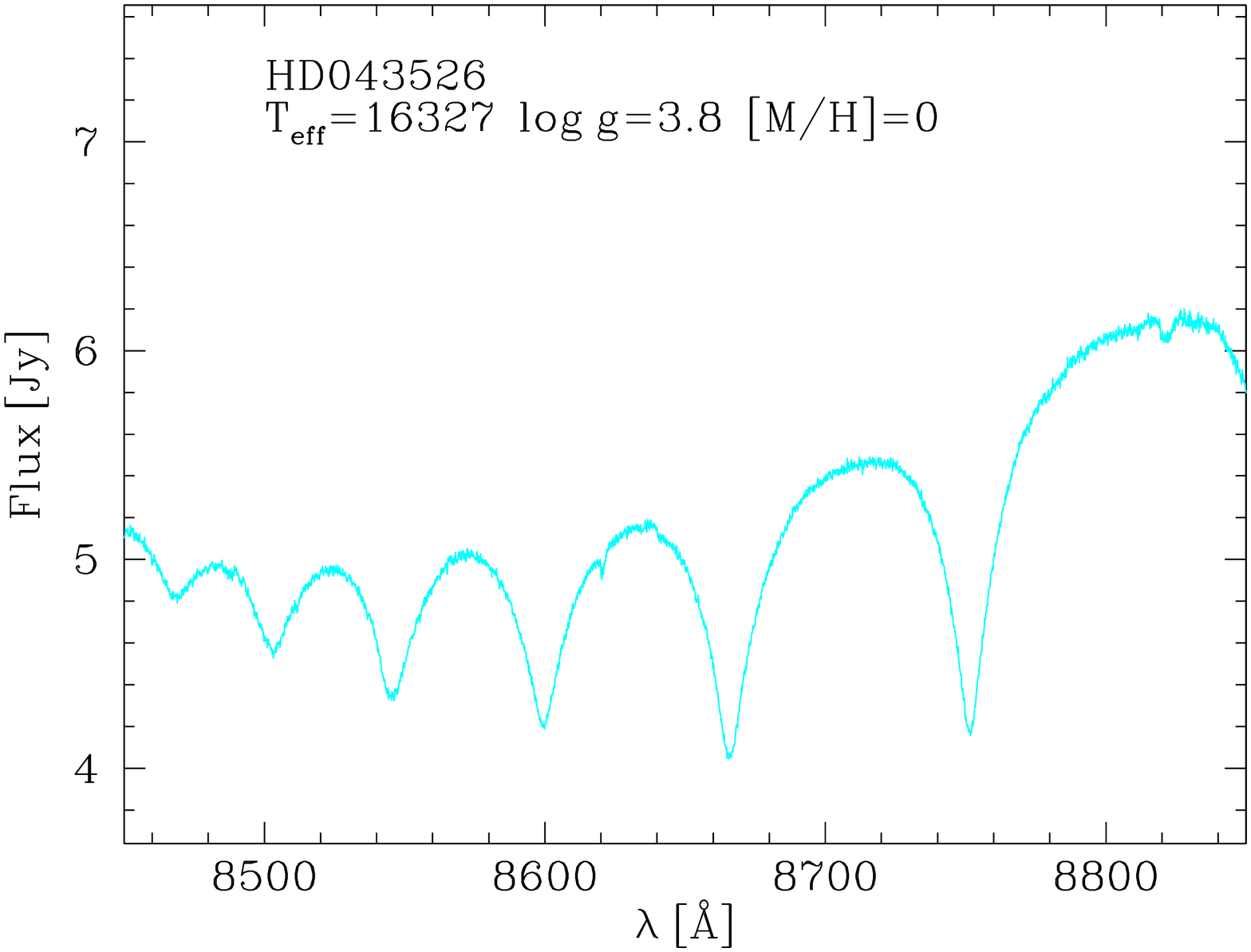}
\includegraphics[width=0.18\textwidth,angle=-90]{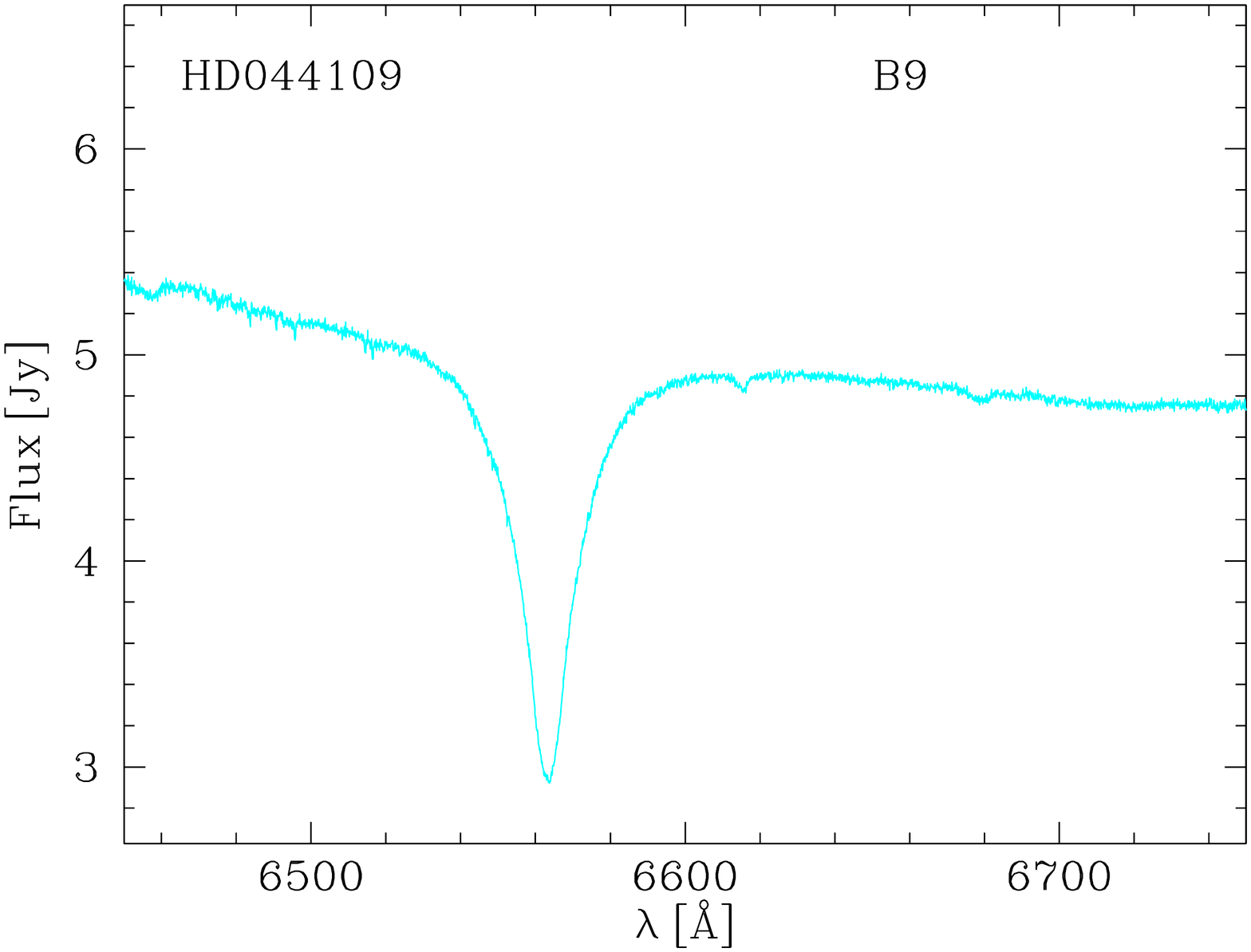}
\includegraphics[width=0.18\textwidth,angle=-90]{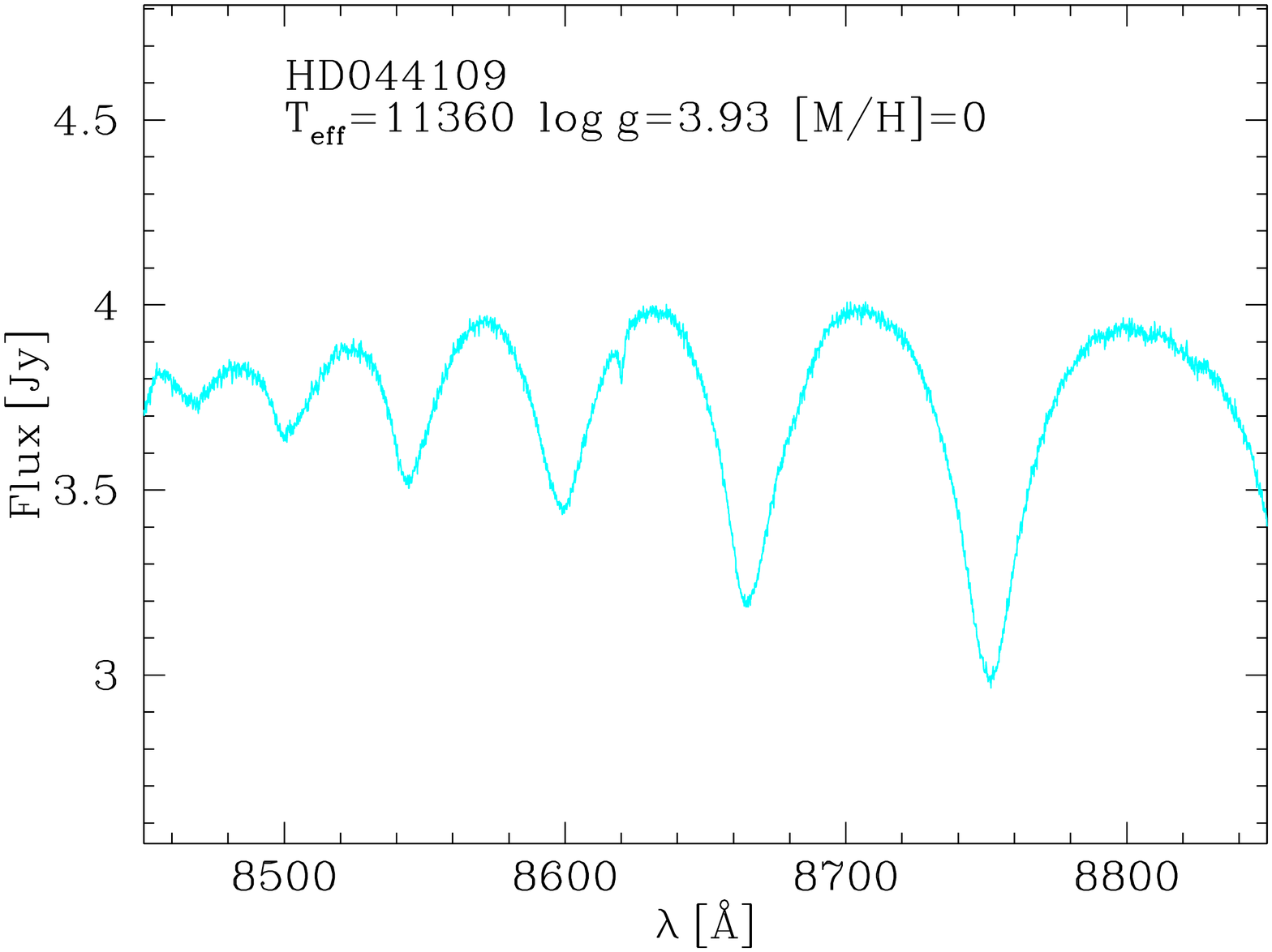}
\includegraphics[width=0.18\textwidth,angle=-90]{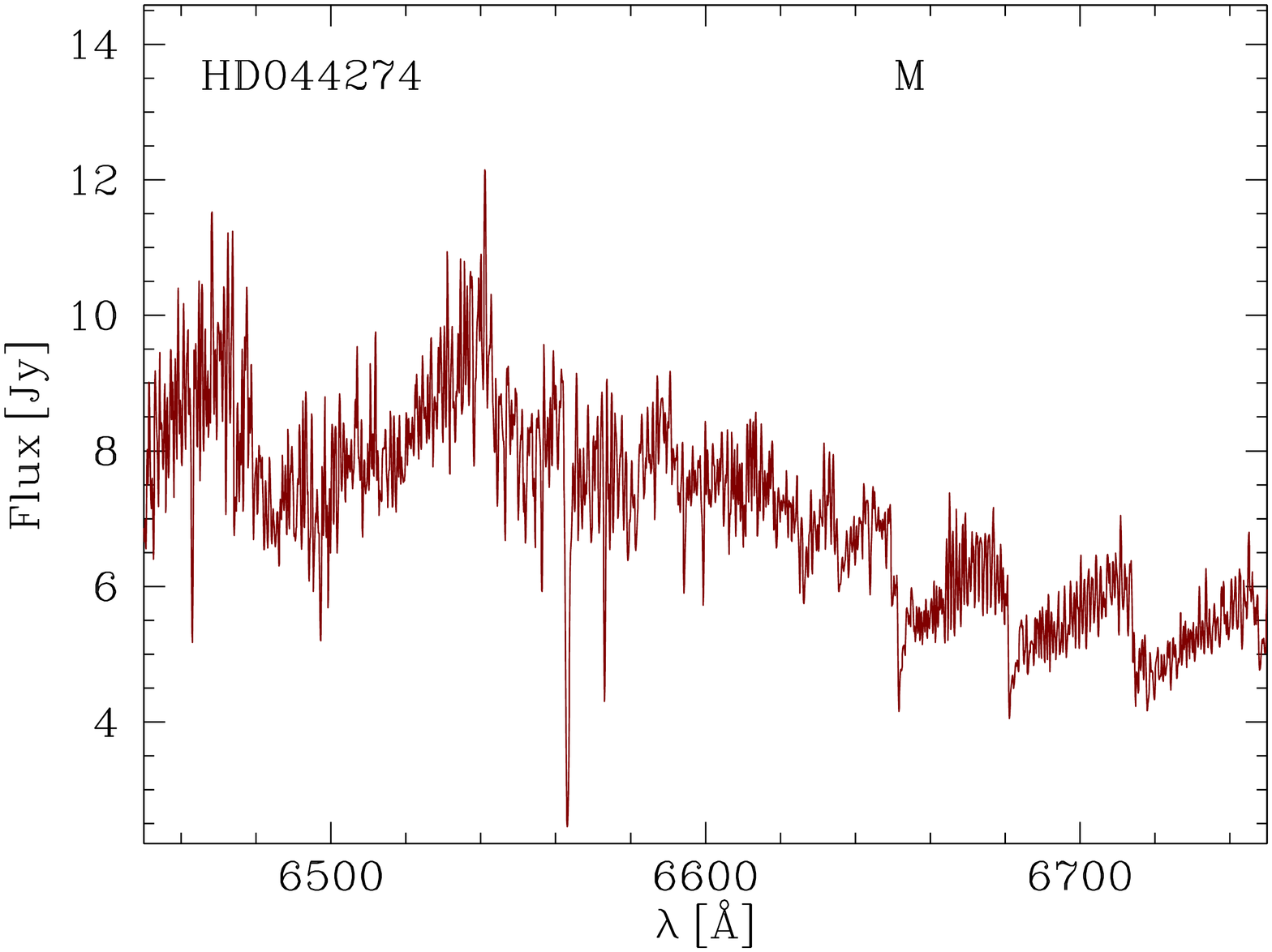}
\includegraphics[width=0.18\textwidth,angle=-90]{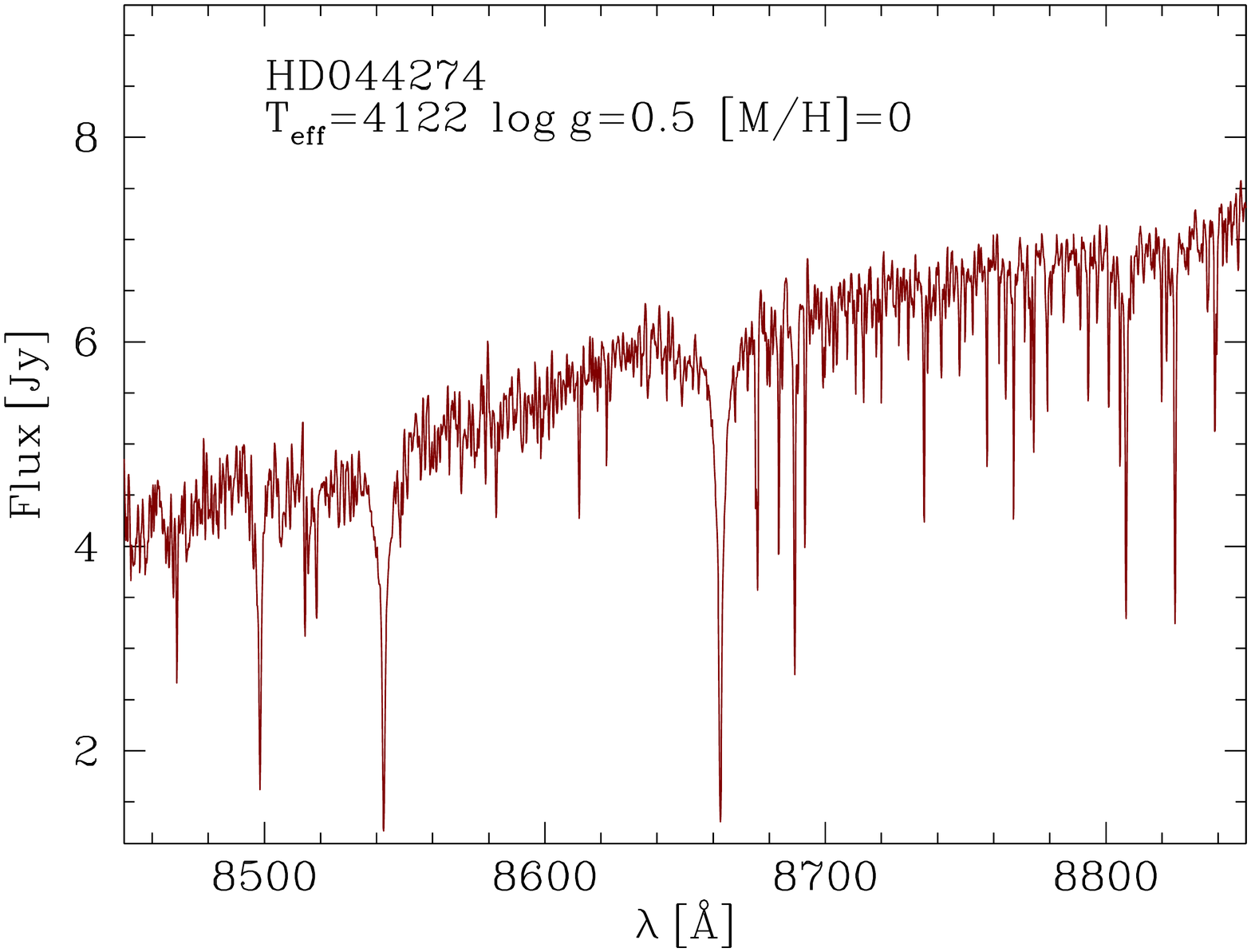}
\includegraphics[width=0.18\textwidth,angle=-90]{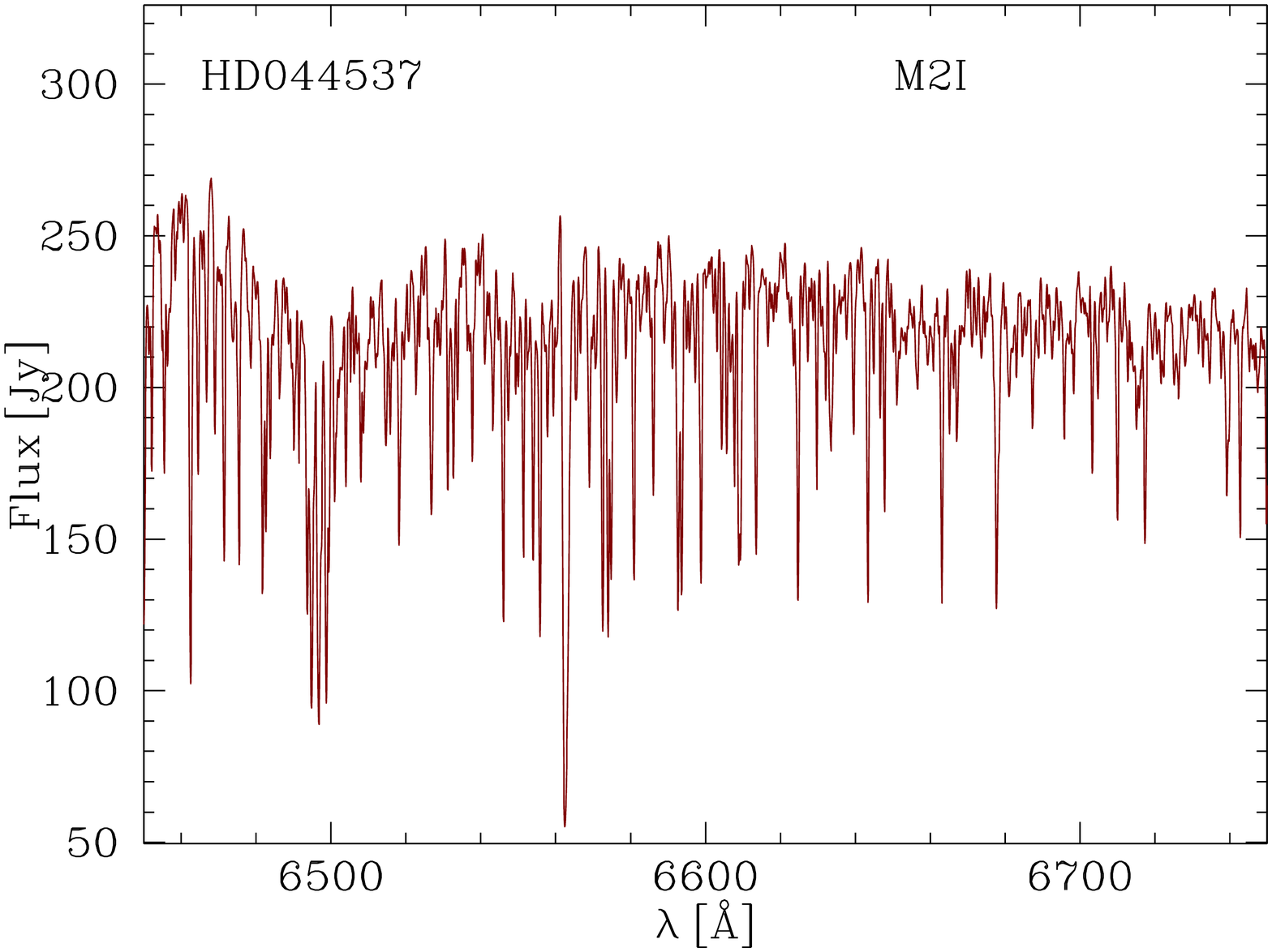}
\includegraphics[width=0.18\textwidth,angle=-90]{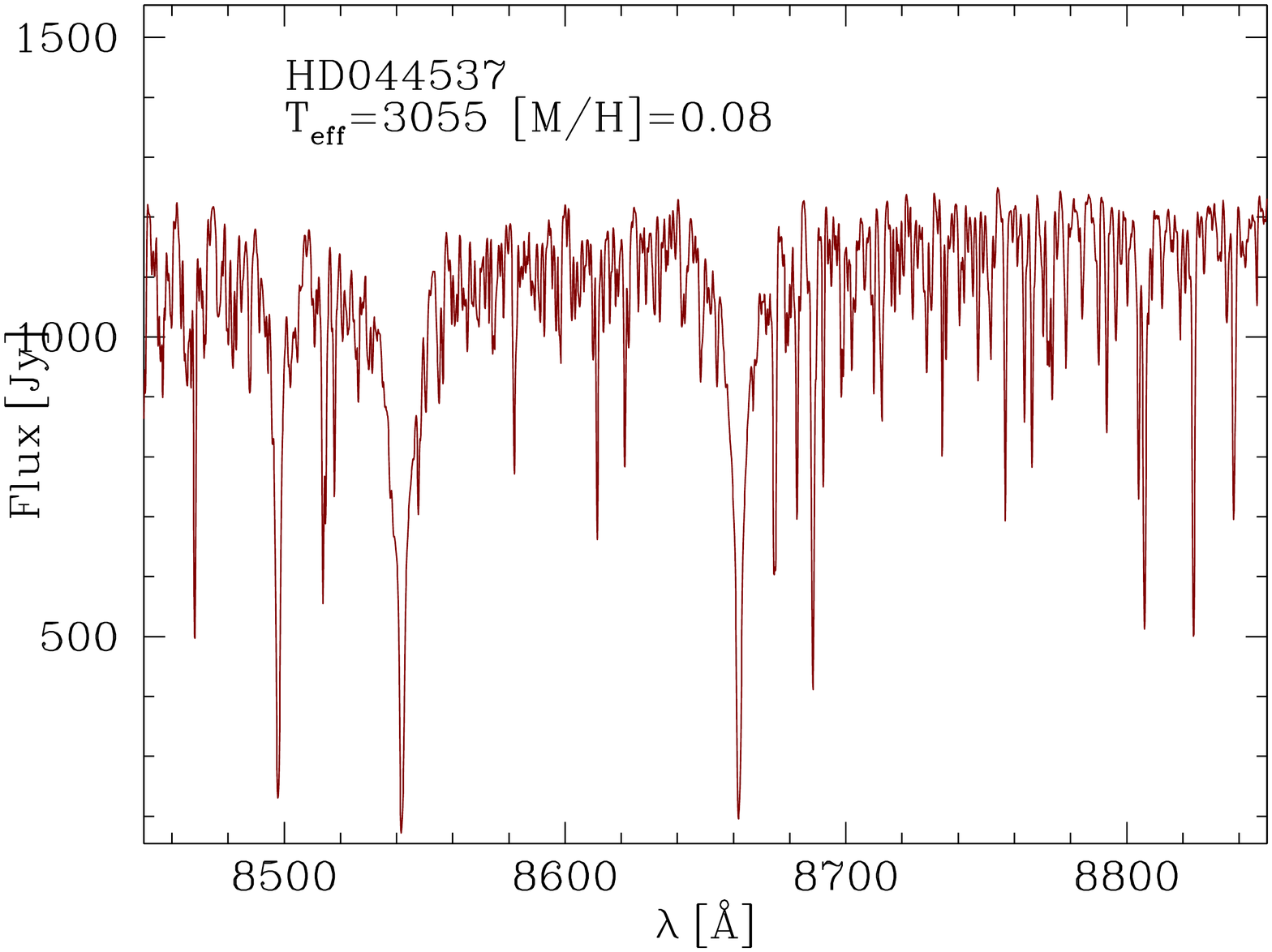}

\contcaption{10. Stars shown in this page are: HD042353, HD042543, HD042597, HD042807, HD042983, HD043042, HD043153, HD043264, HD043285, HD043286, HD043526, HD044109, HD044274 and HD044537.}
\end{figure*}

\begin{figure*}
\includegraphics[width=0.18\textwidth,angle=-90]{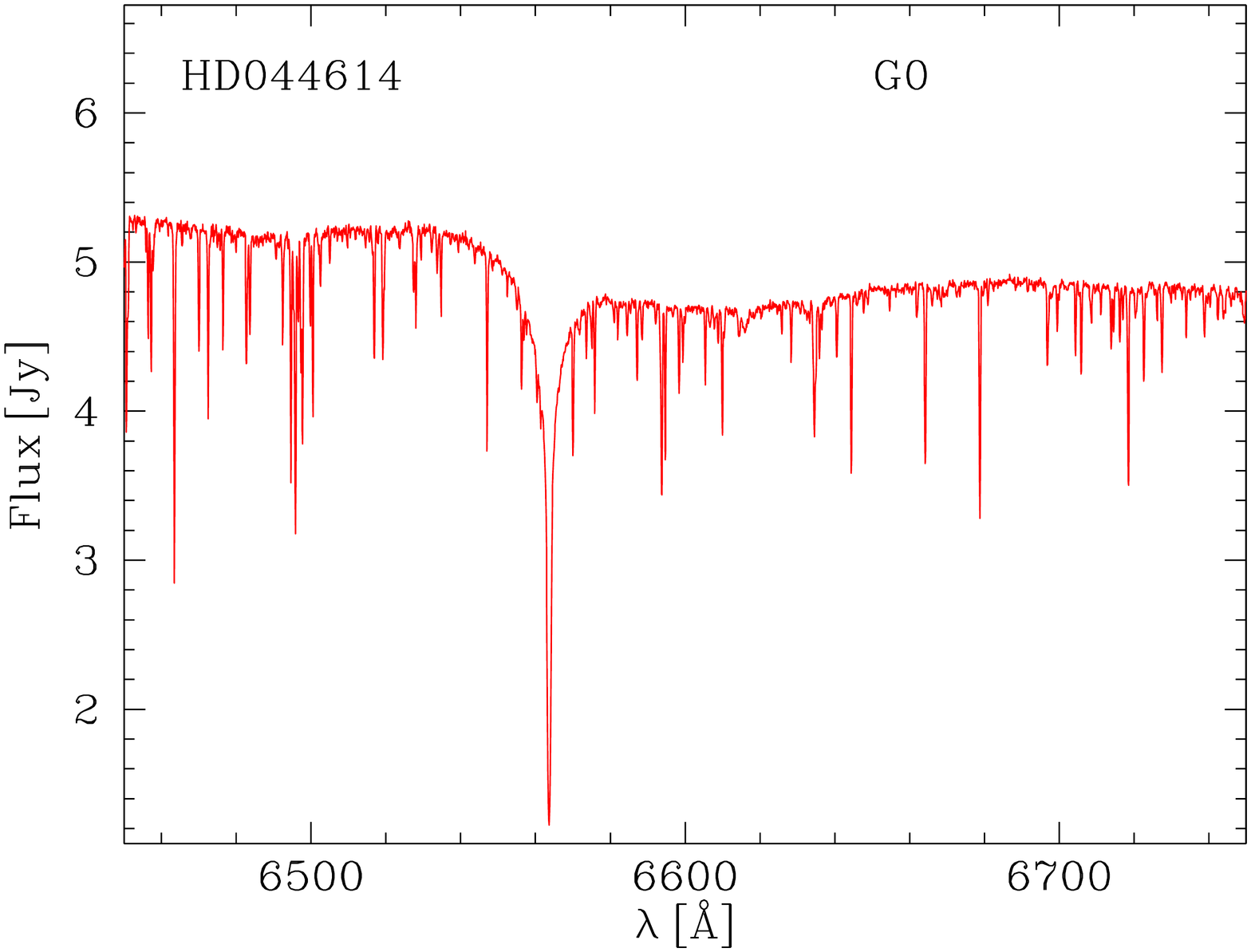}
\includegraphics[width=0.18\textwidth,angle=-90]{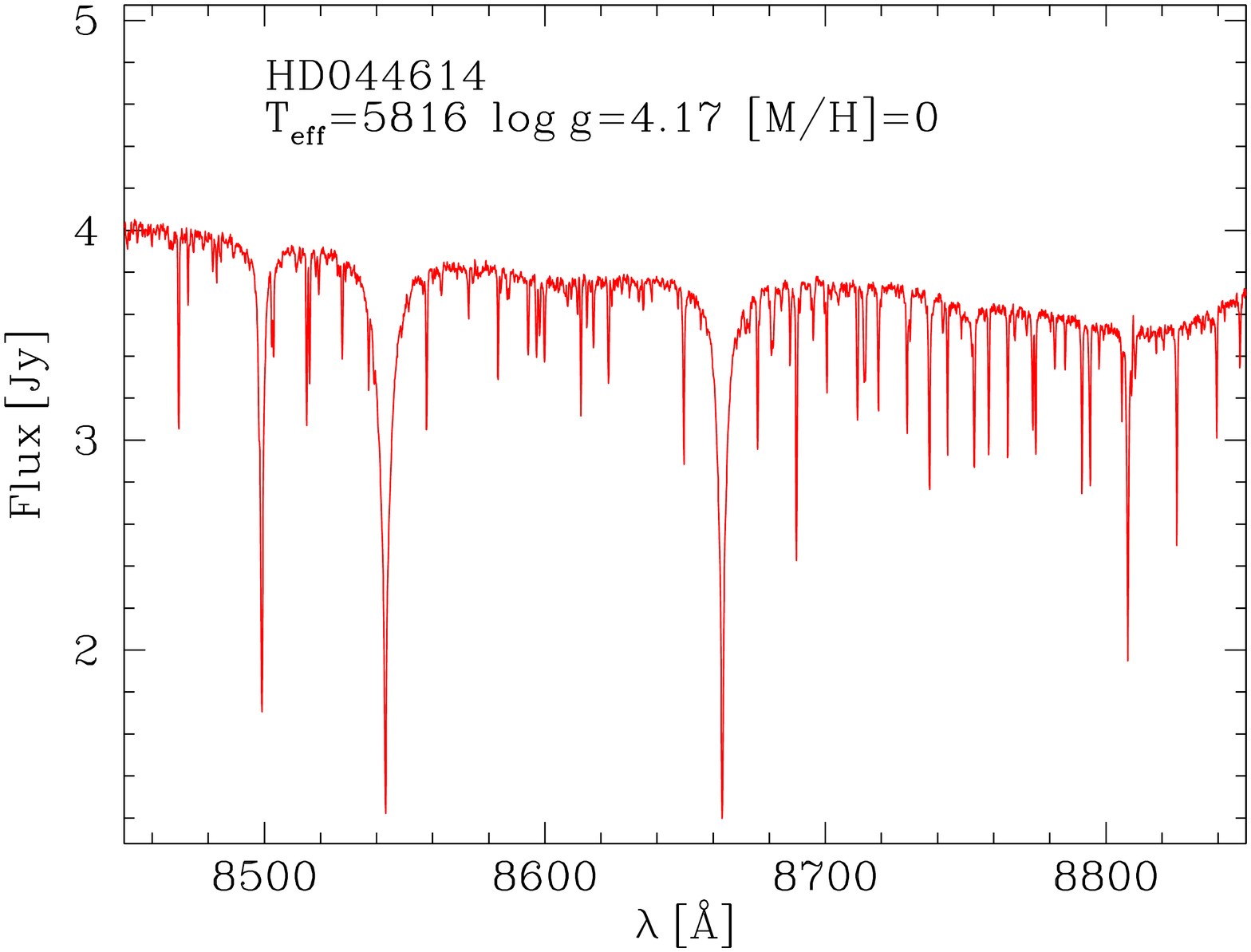}
\includegraphics[width=0.18\textwidth,angle=-90]{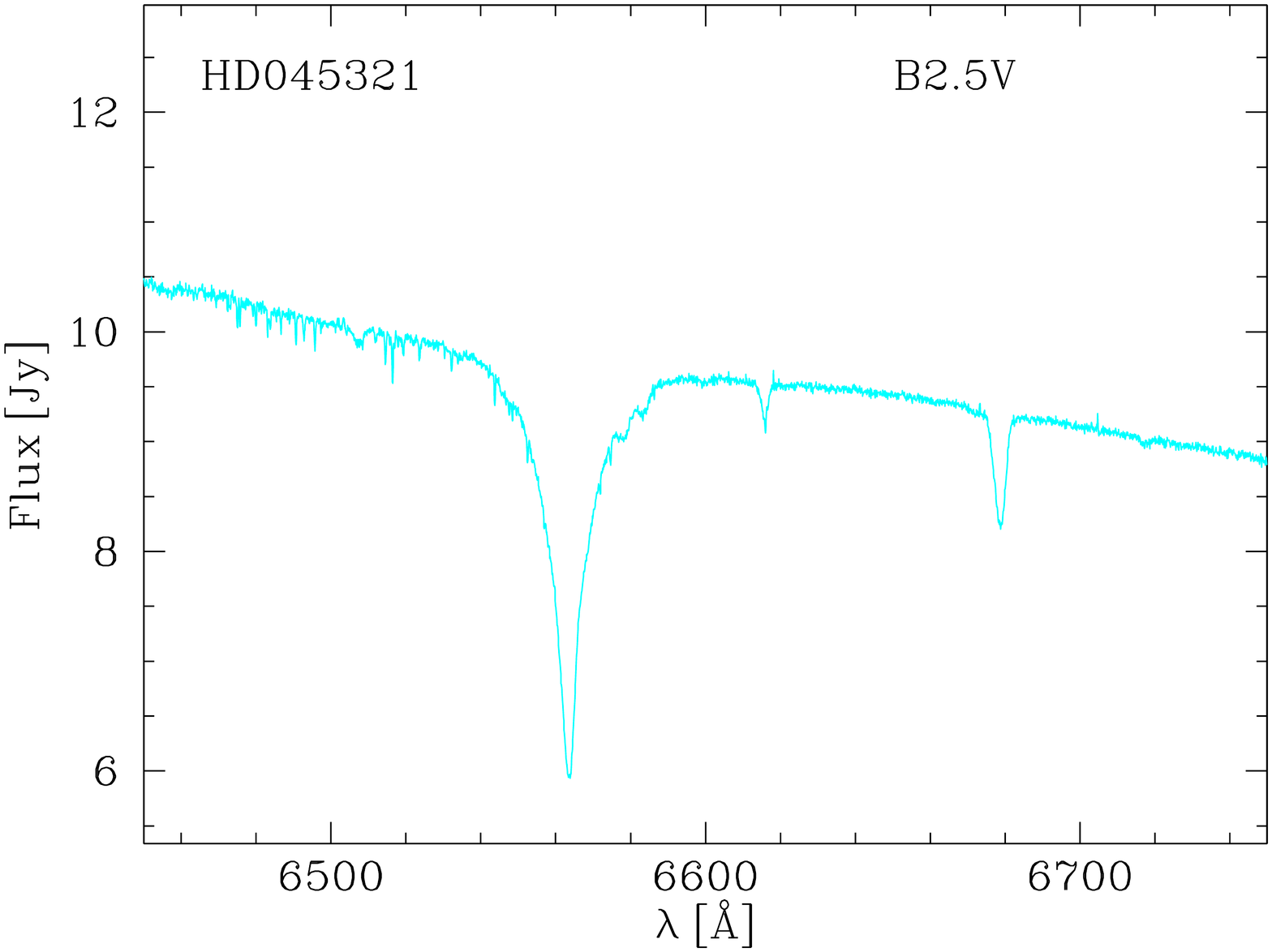}
\includegraphics[width=0.18\textwidth,angle=-90]{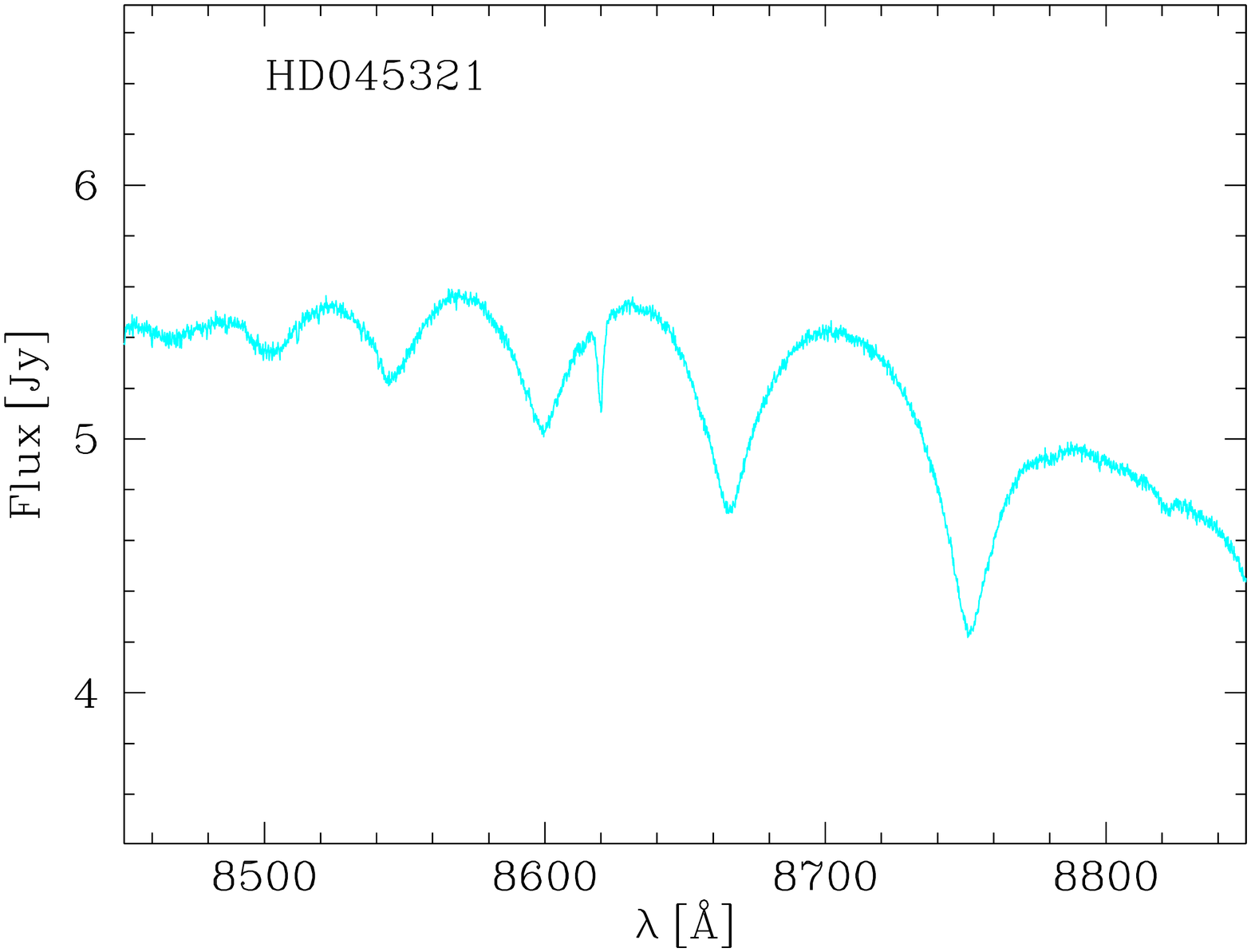}
\includegraphics[width=0.18\textwidth,angle=-90]{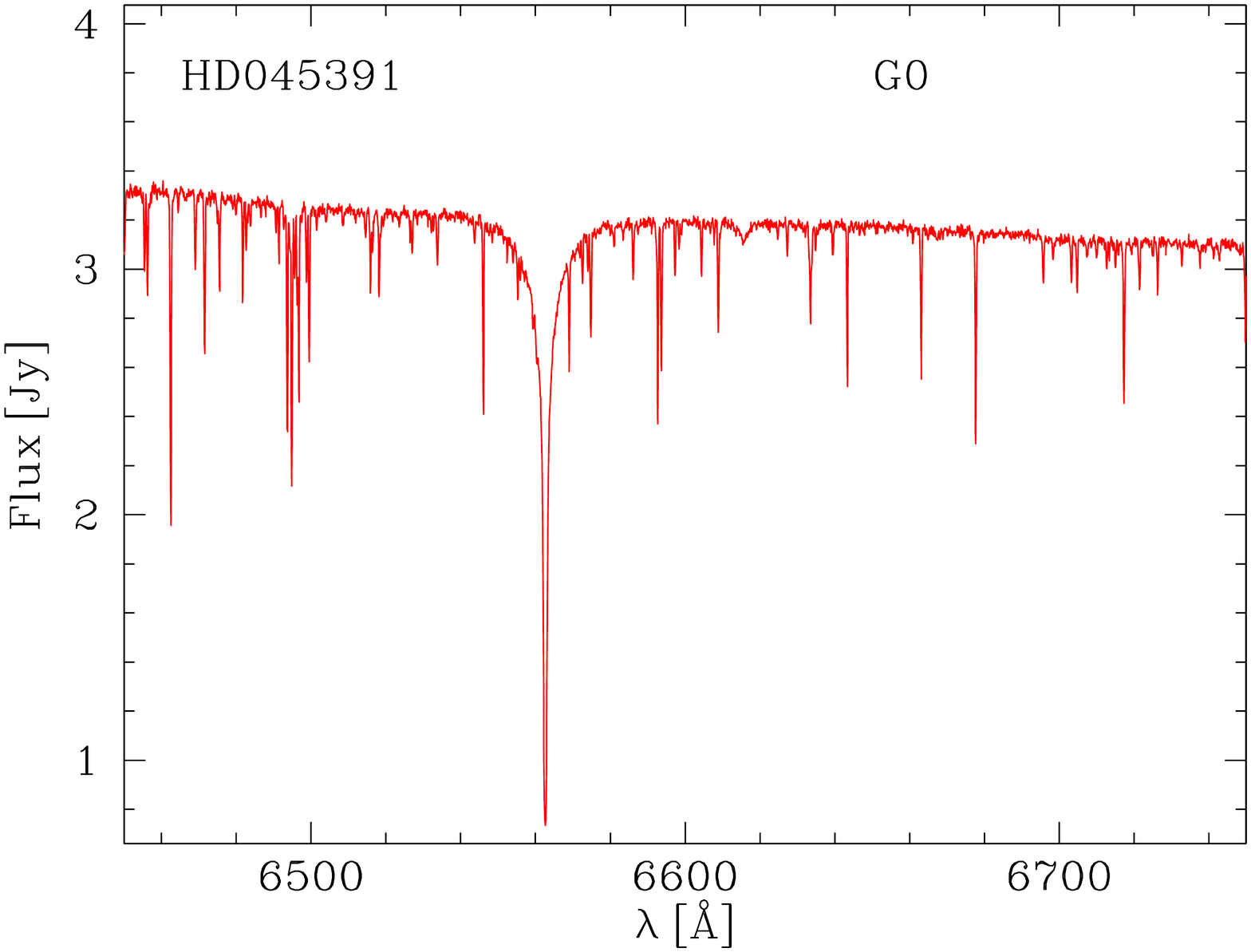}
\includegraphics[width=0.18\textwidth,angle=-90]{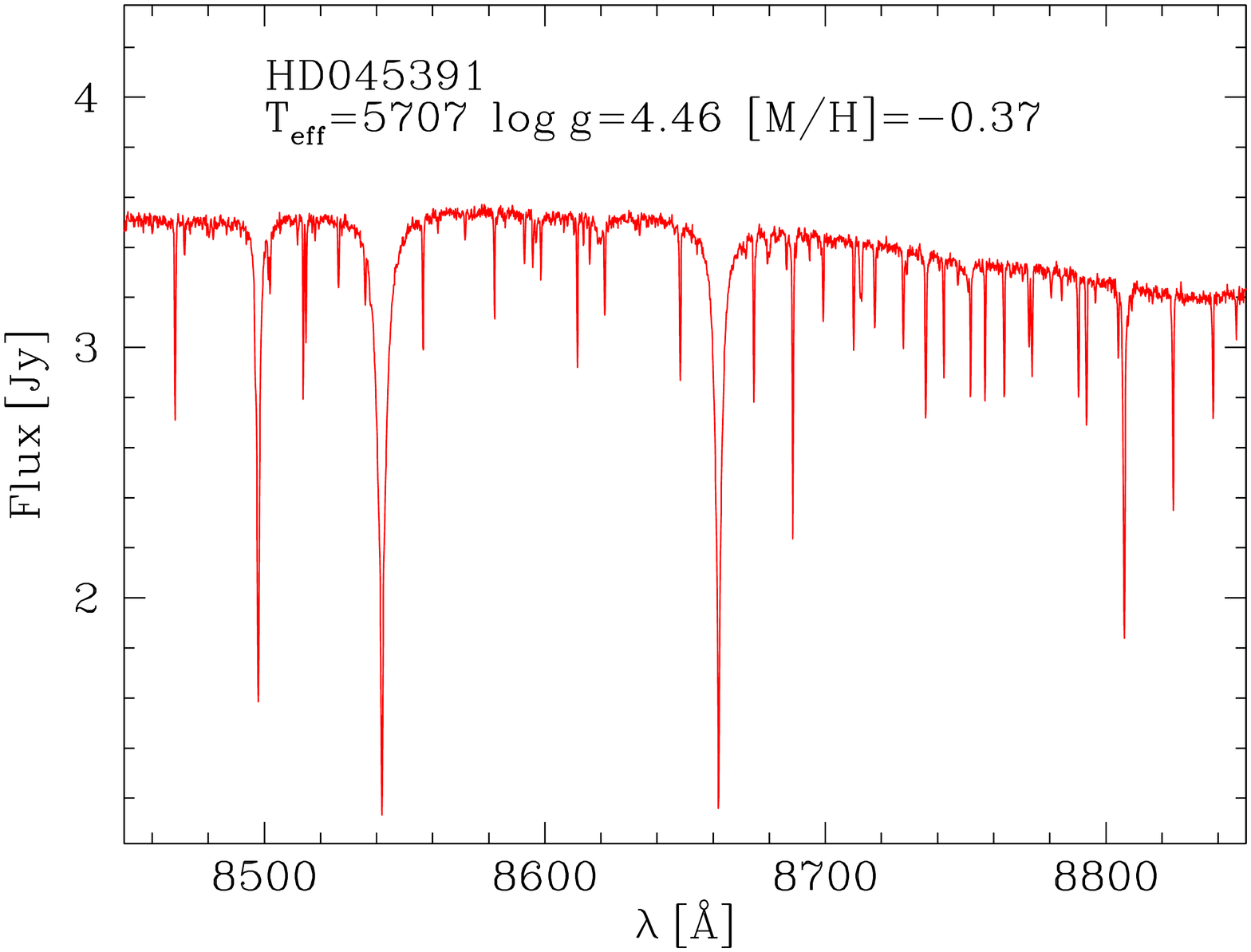}
\includegraphics[width=0.18\textwidth,angle=-90]{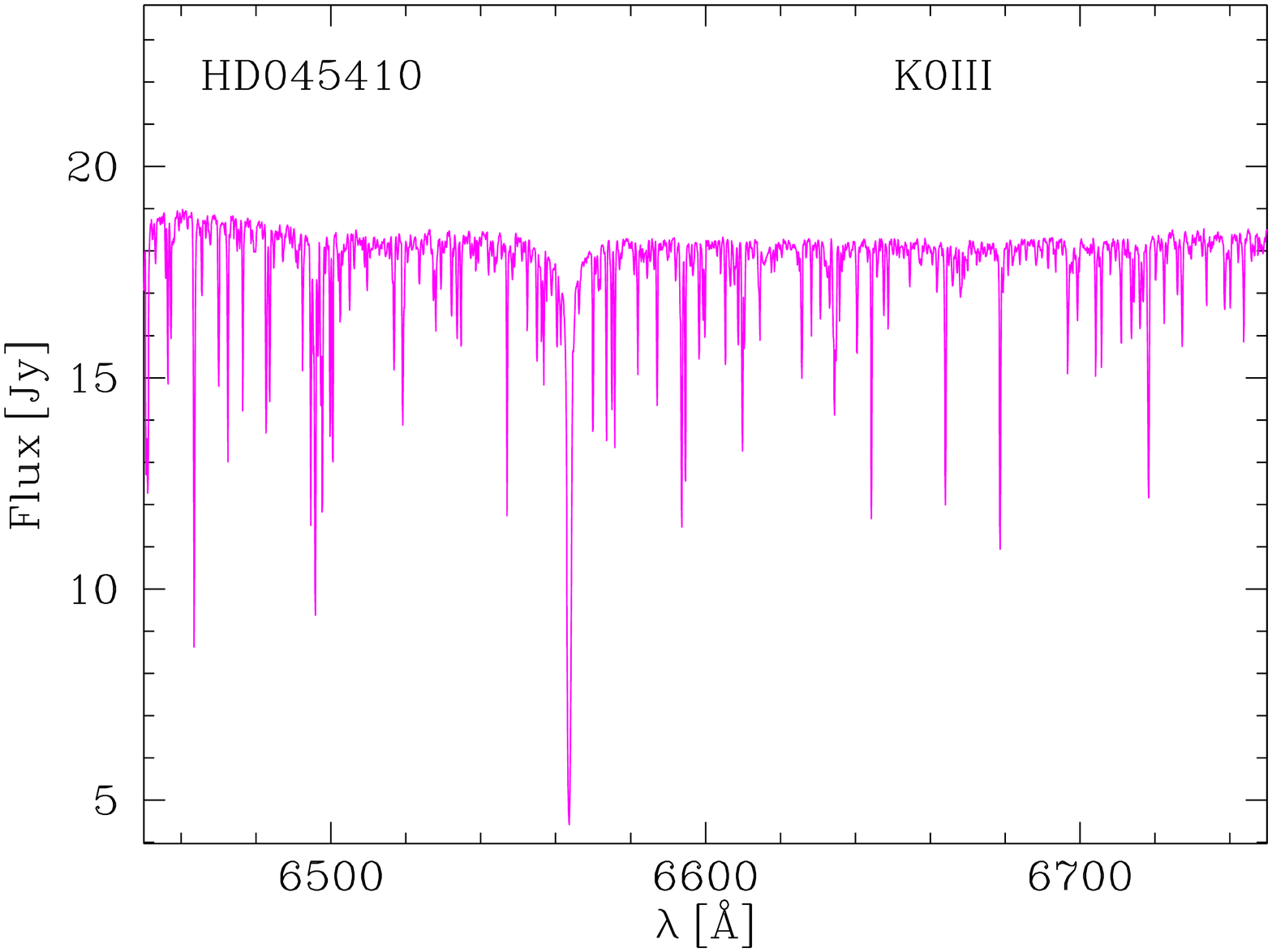}
\includegraphics[width=0.18\textwidth,angle=-90]{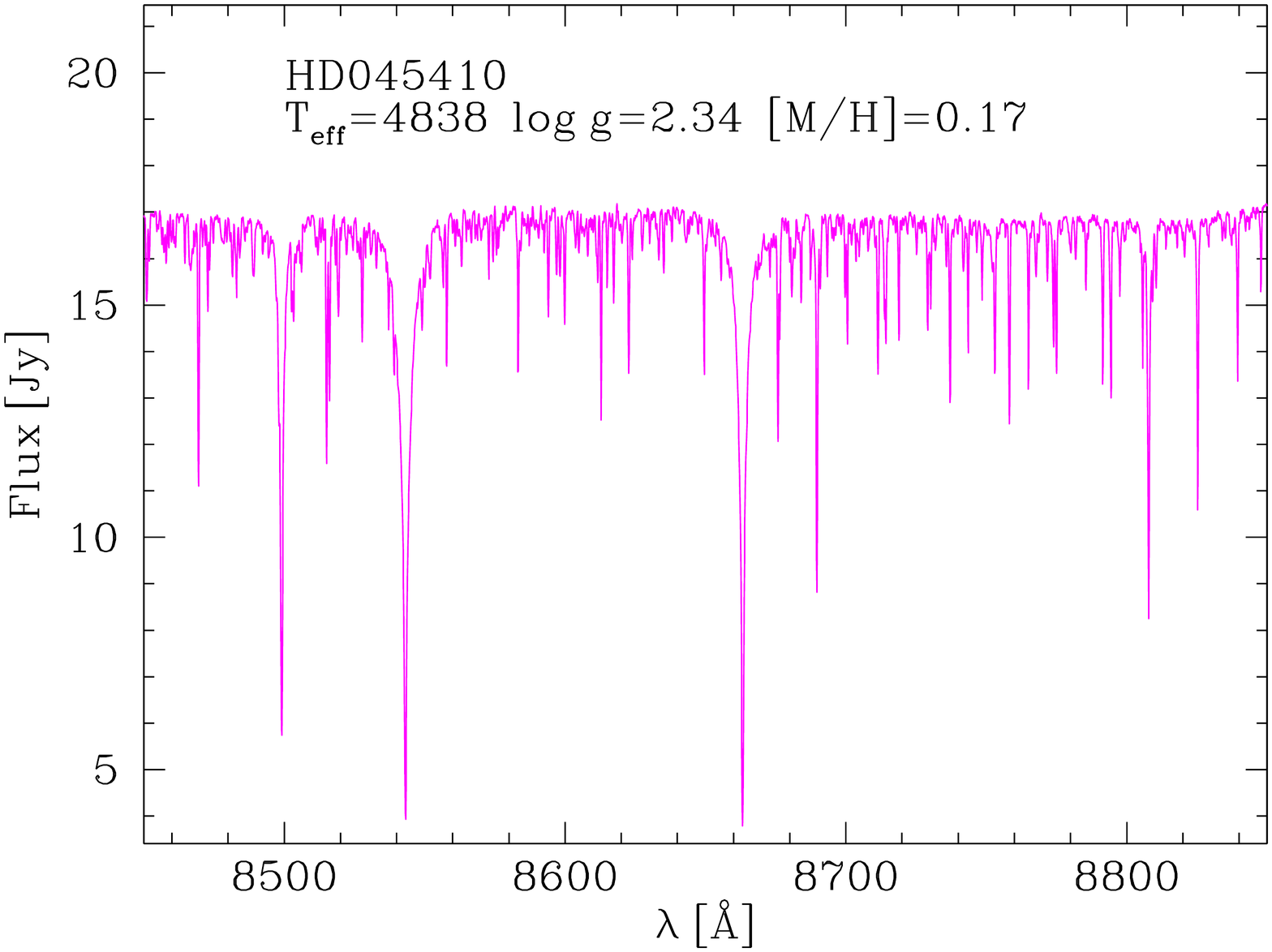}
\includegraphics[width=0.18\textwidth,angle=-90]{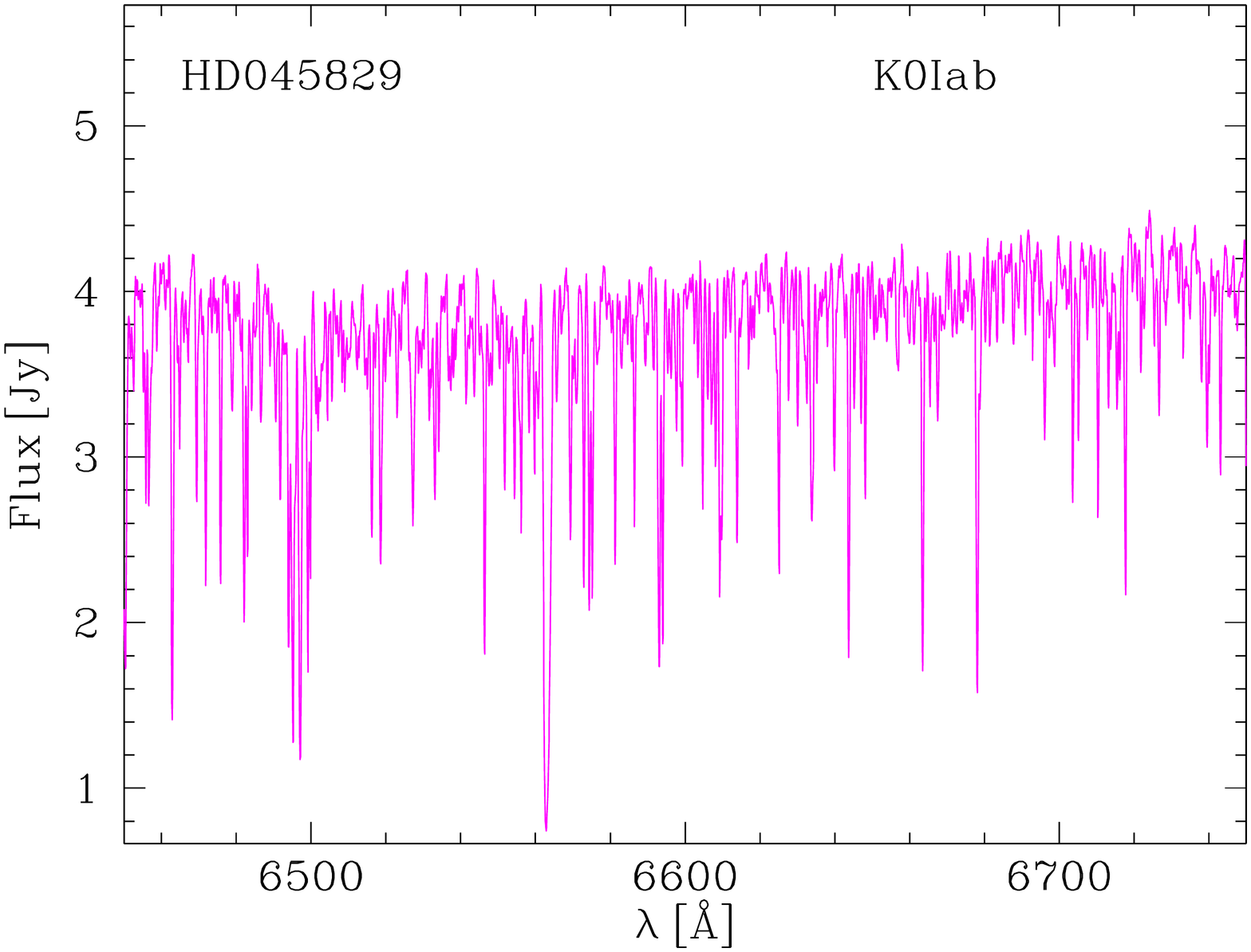}
\includegraphics[width=0.18\textwidth,angle=-90]{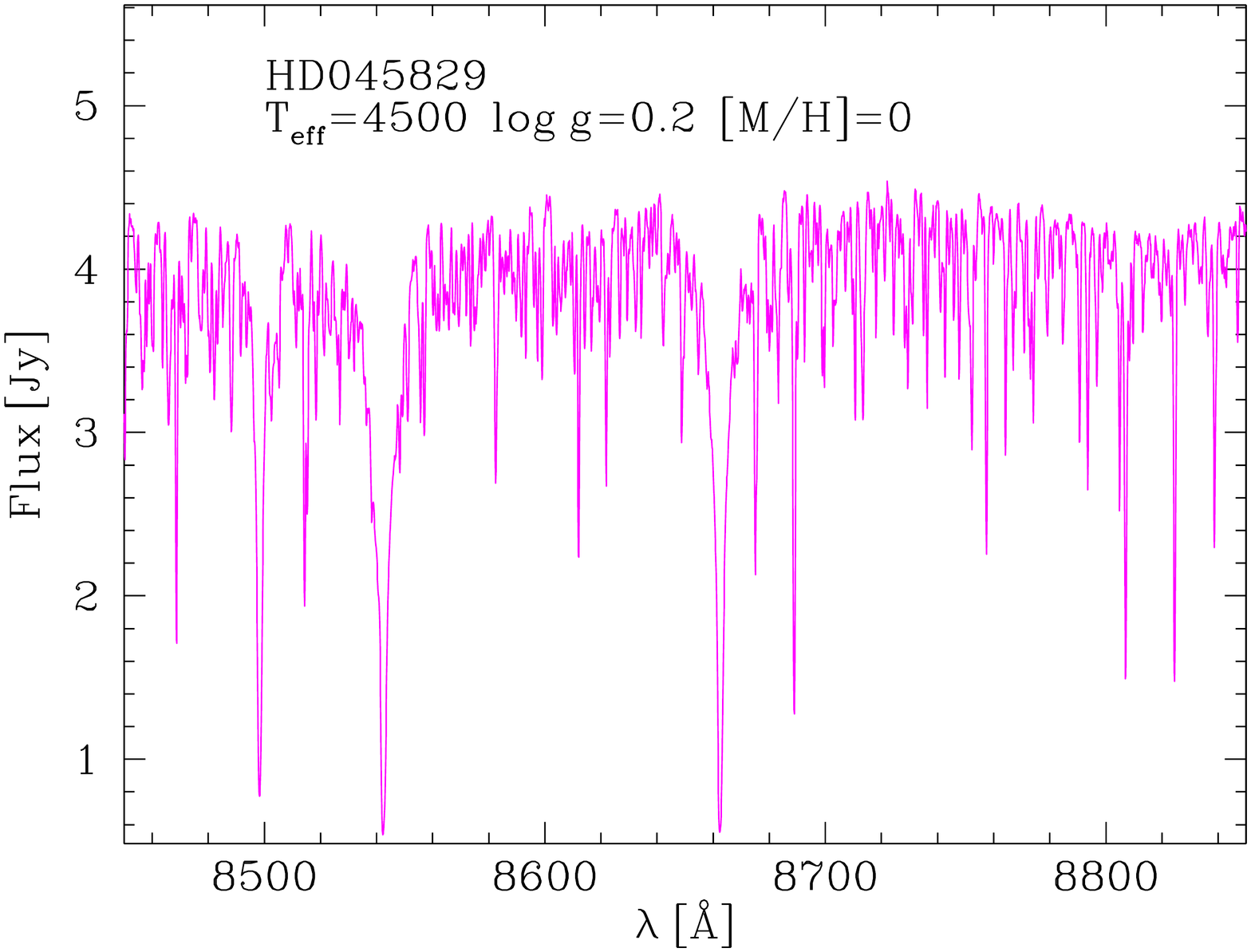}
\includegraphics[width=0.18\textwidth,angle=-90]{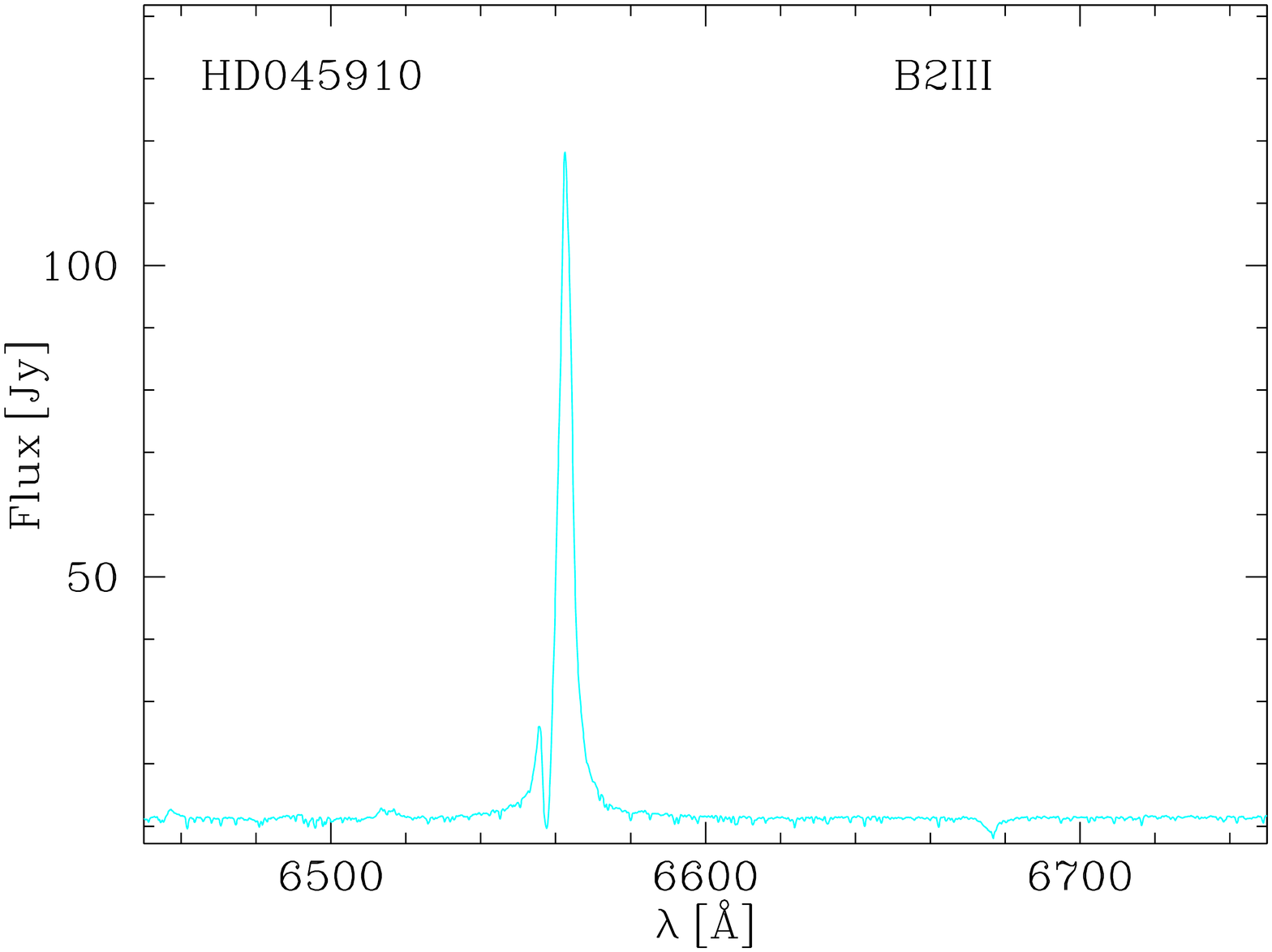}
\includegraphics[width=0.18\textwidth,angle=-90]{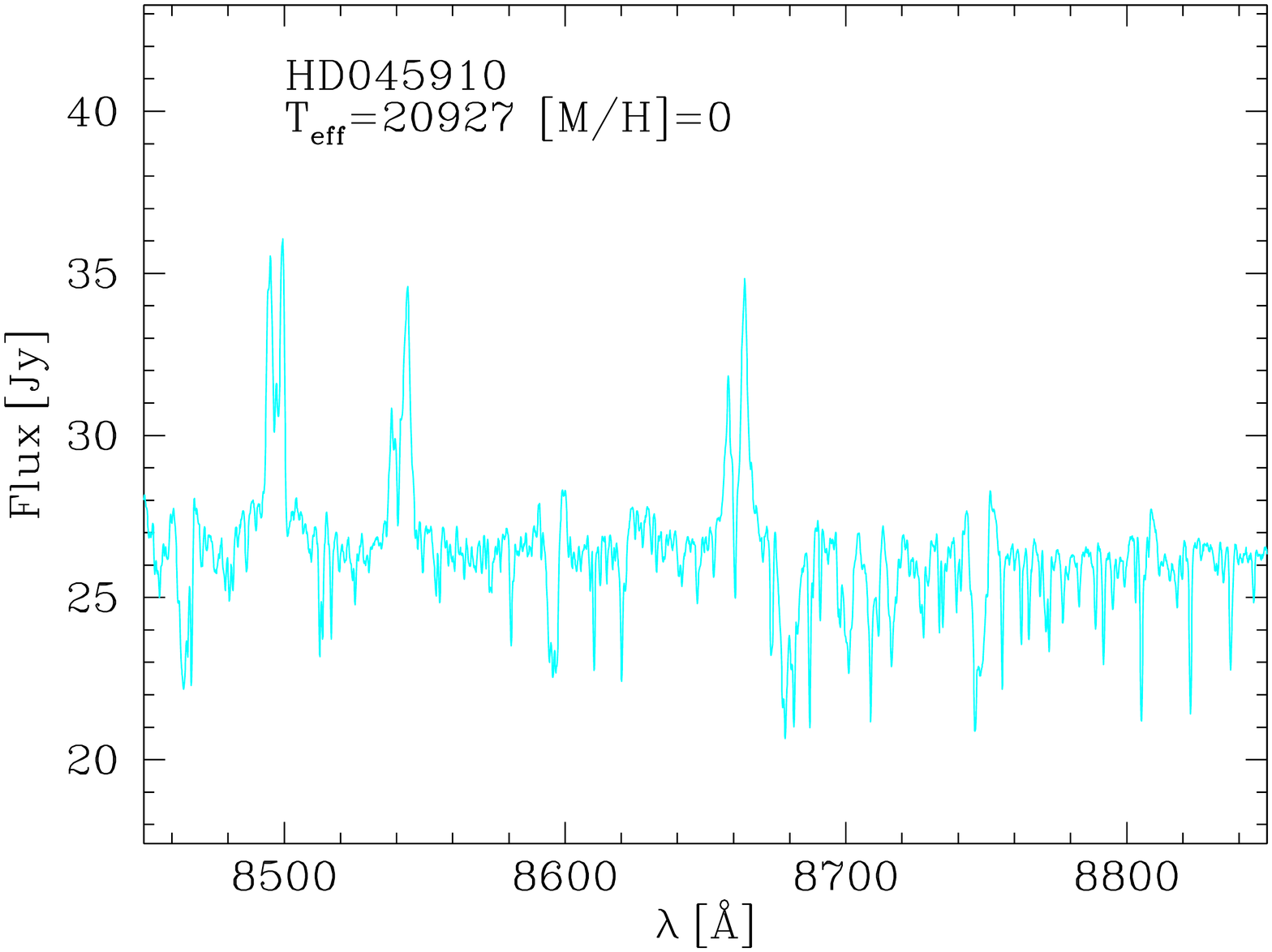}
\includegraphics[width=0.18\textwidth,angle=-90]{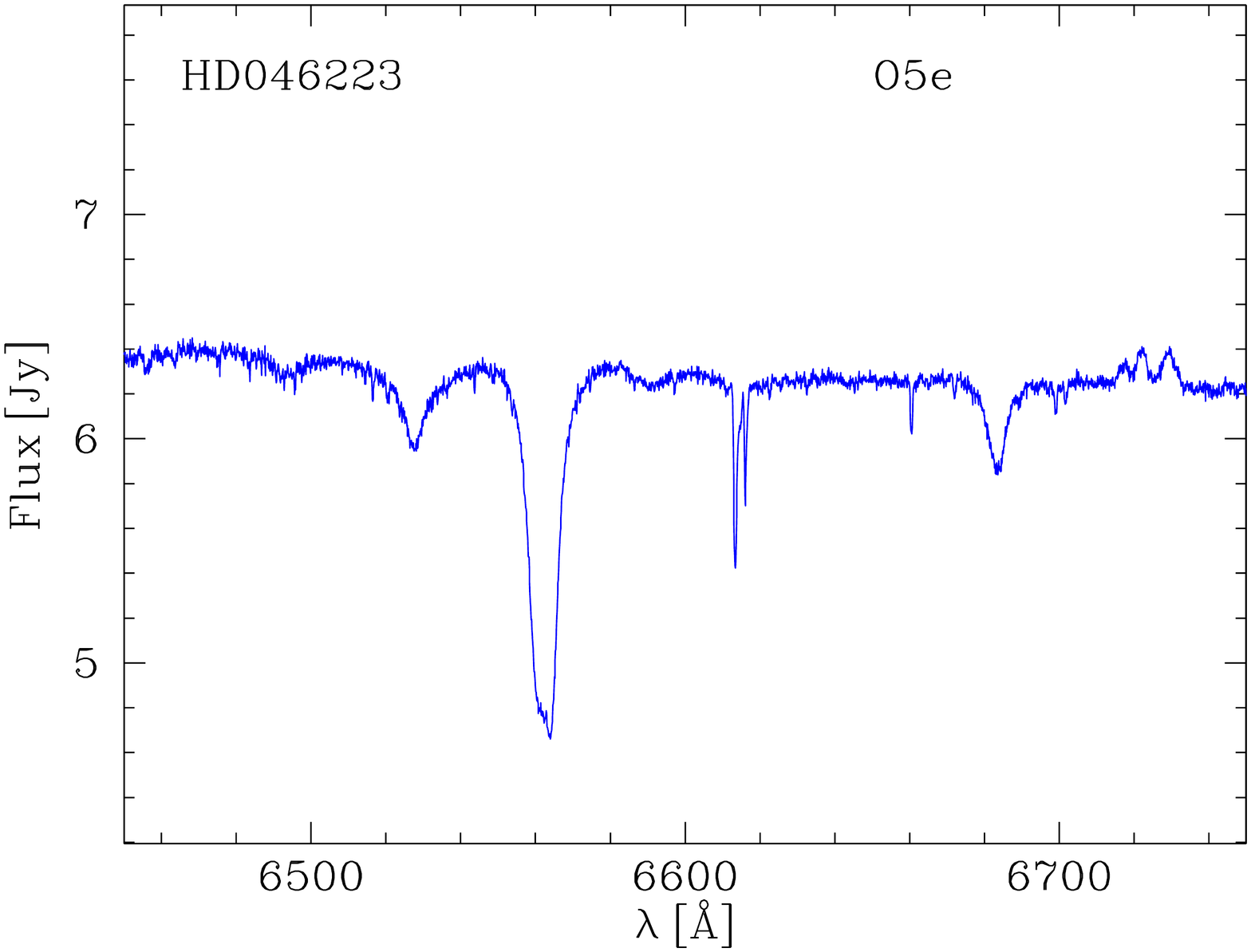}
\includegraphics[width=0.18\textwidth,angle=-90]{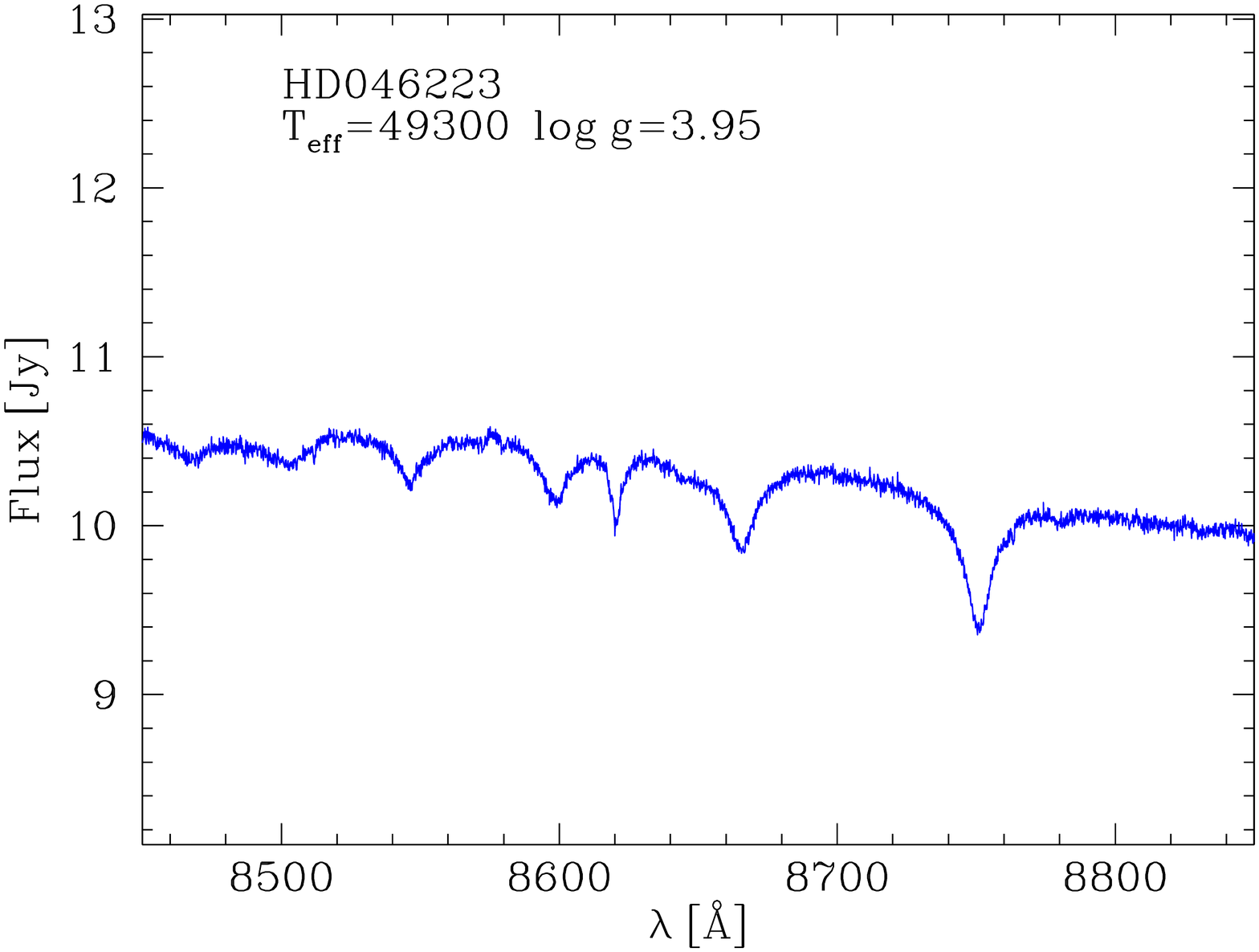}
\includegraphics[width=0.18\textwidth,angle=-90]{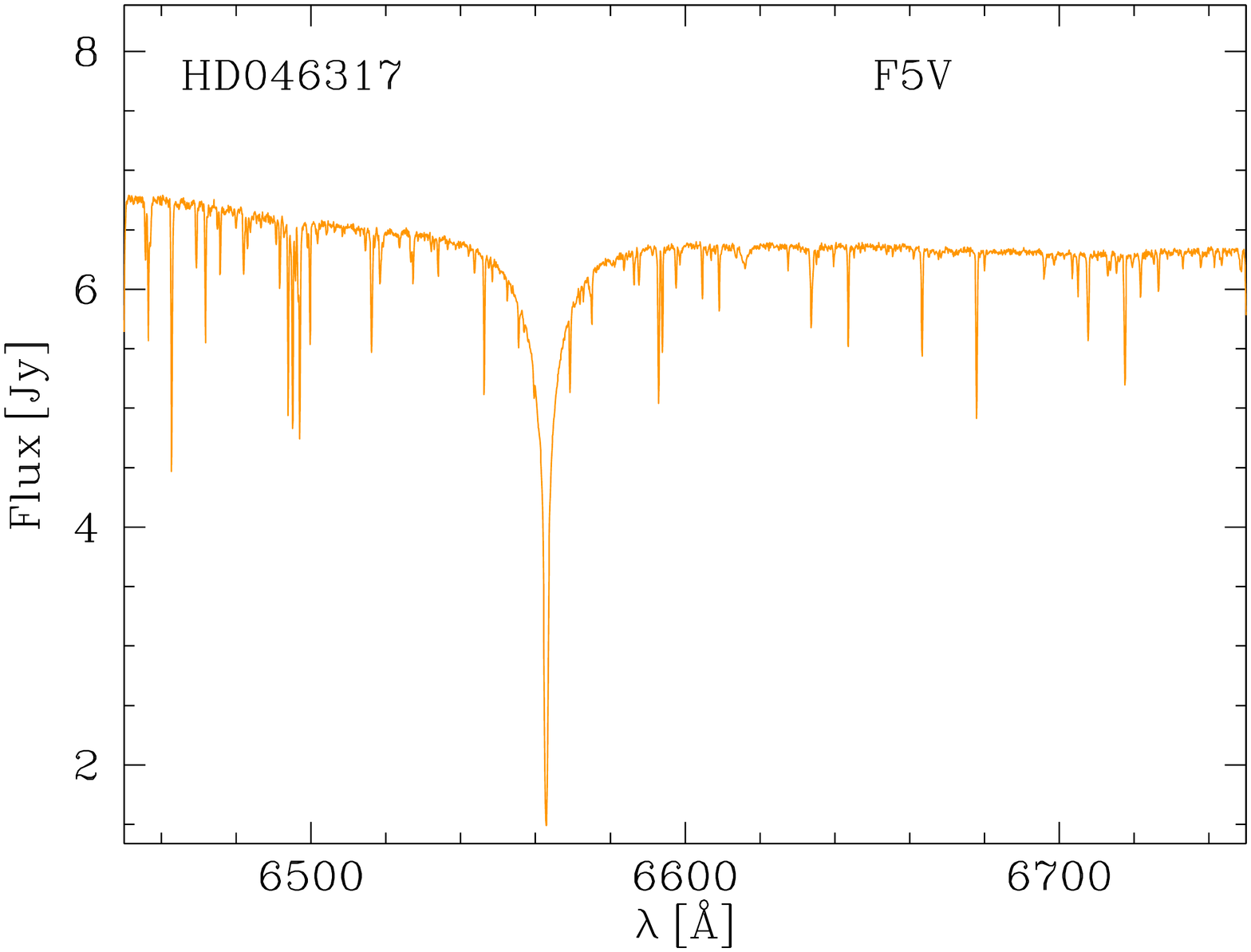}
\includegraphics[width=0.18\textwidth,angle=-90]{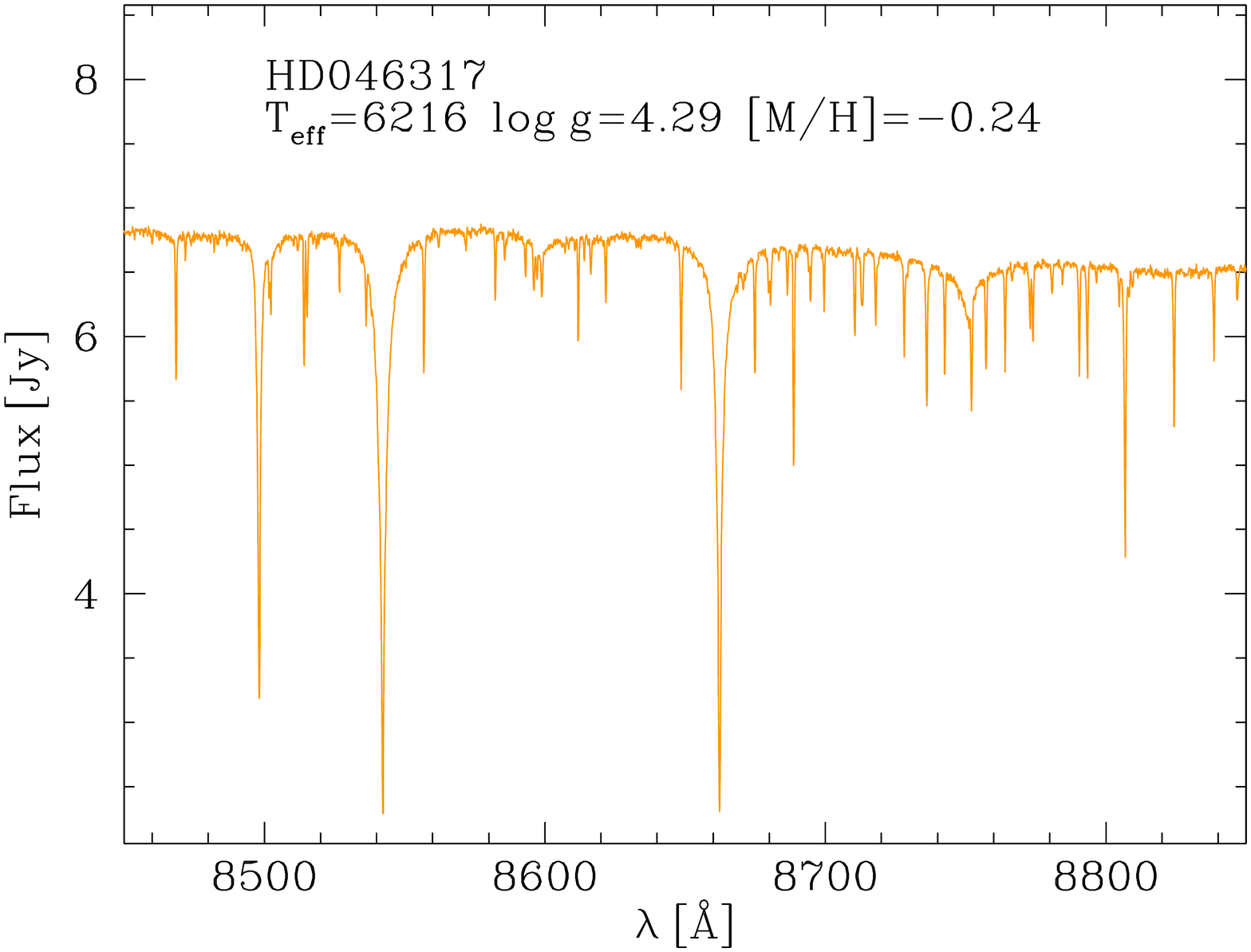}
\includegraphics[width=0.18\textwidth,angle=-90]{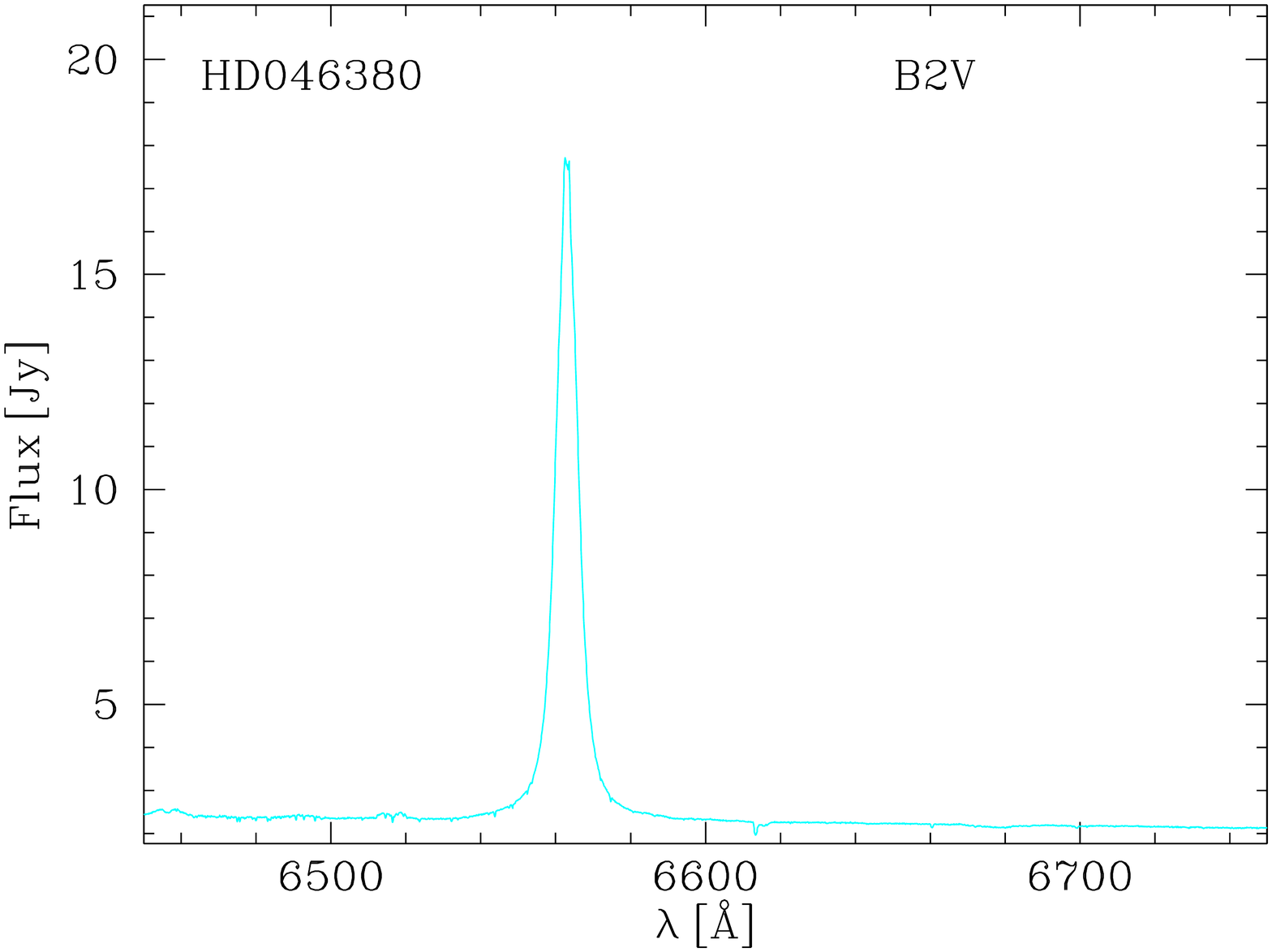}
\includegraphics[width=0.18\textwidth,angle=-90]{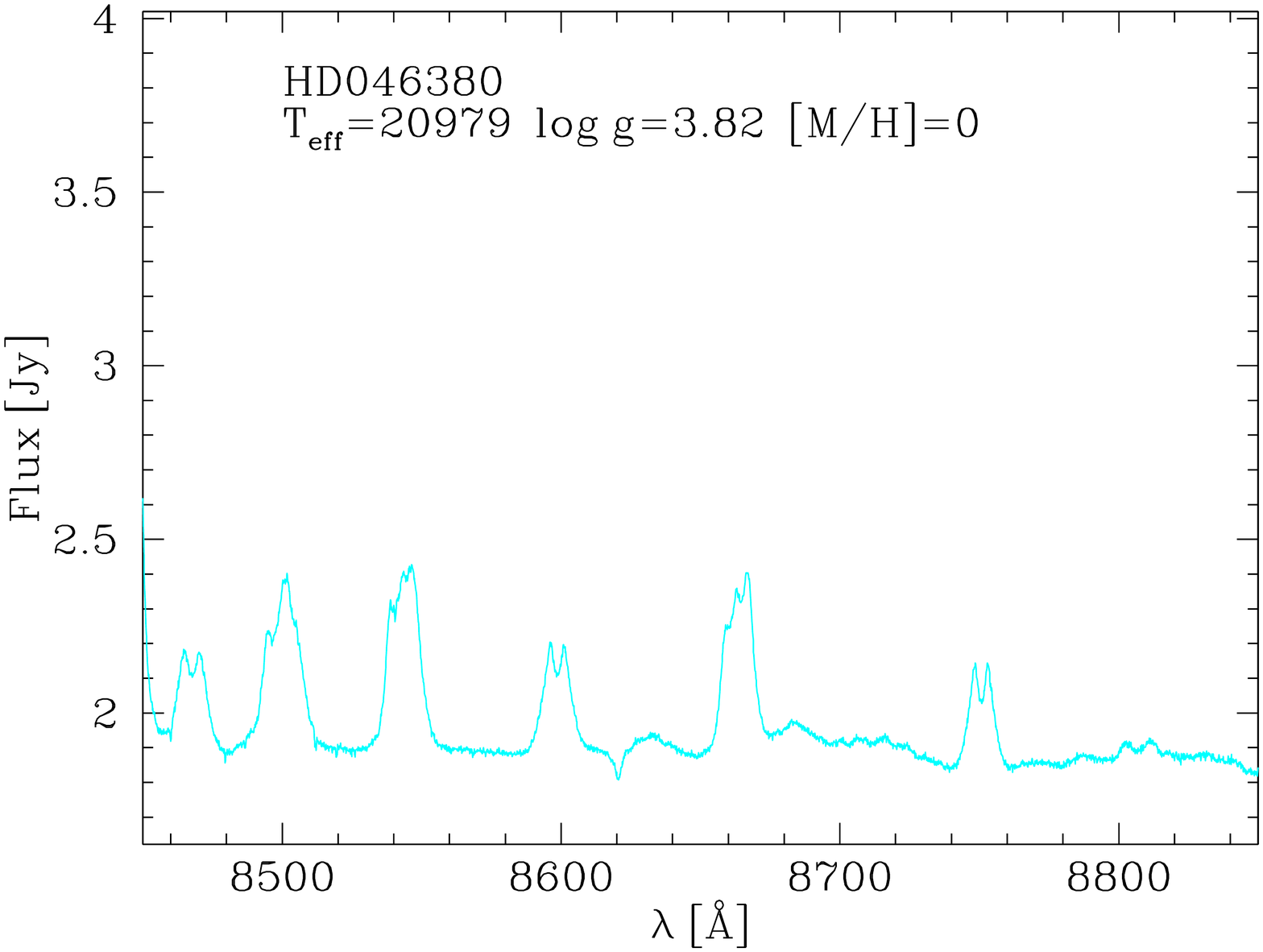}
\includegraphics[width=0.18\textwidth,angle=-90]{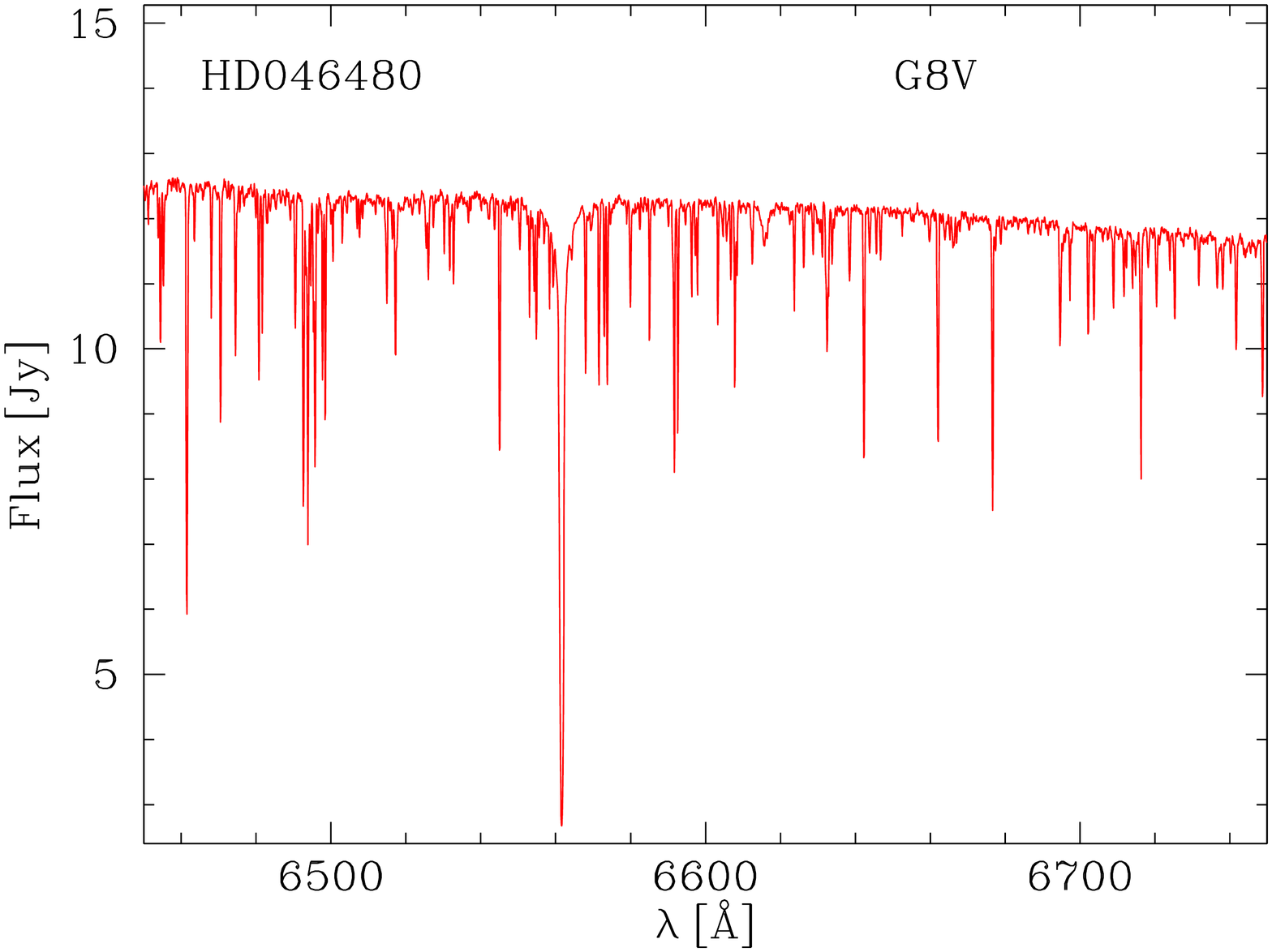}
\includegraphics[width=0.18\textwidth,angle=-90]{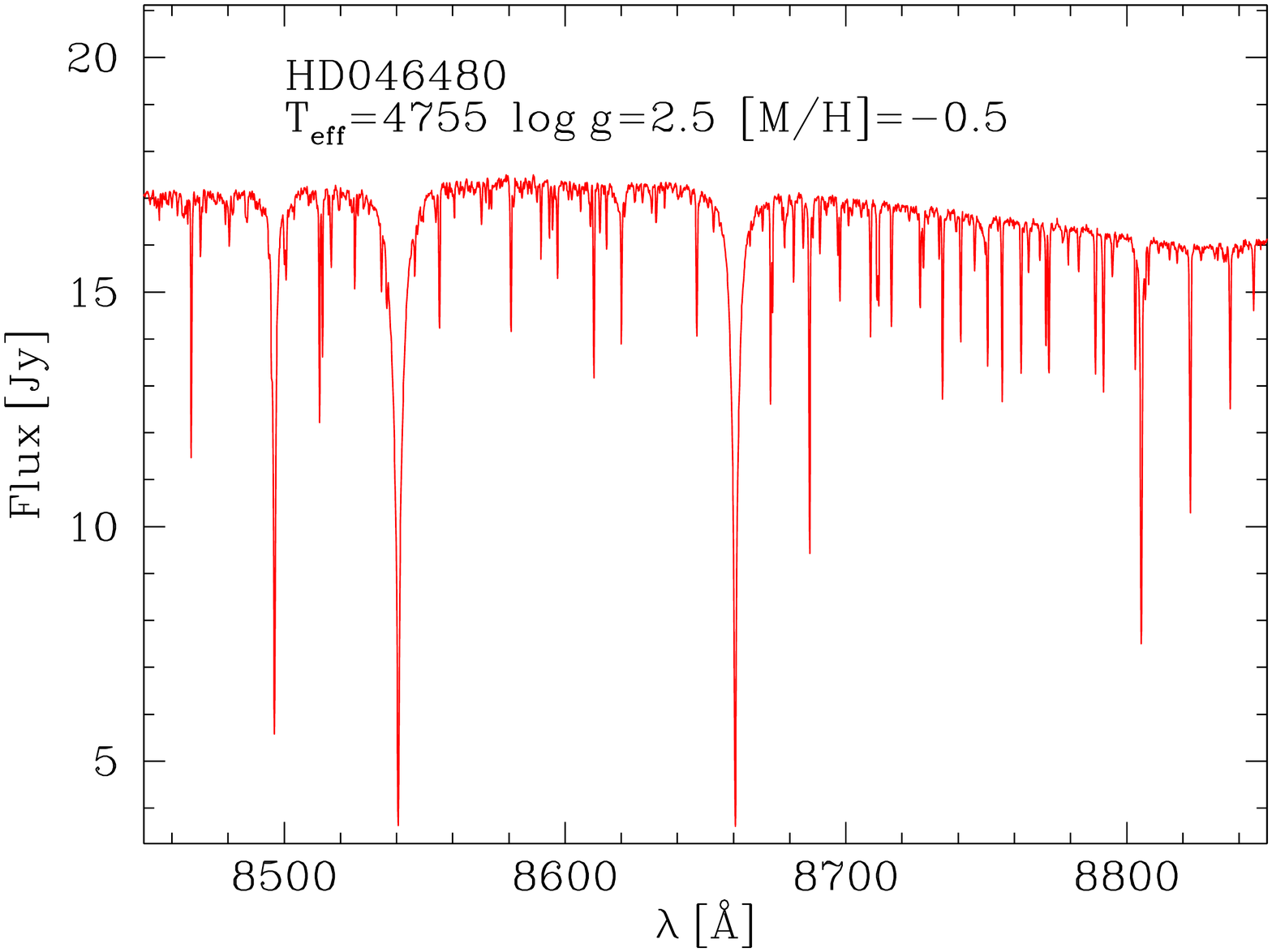}
\includegraphics[width=0.18\textwidth,angle=-90]{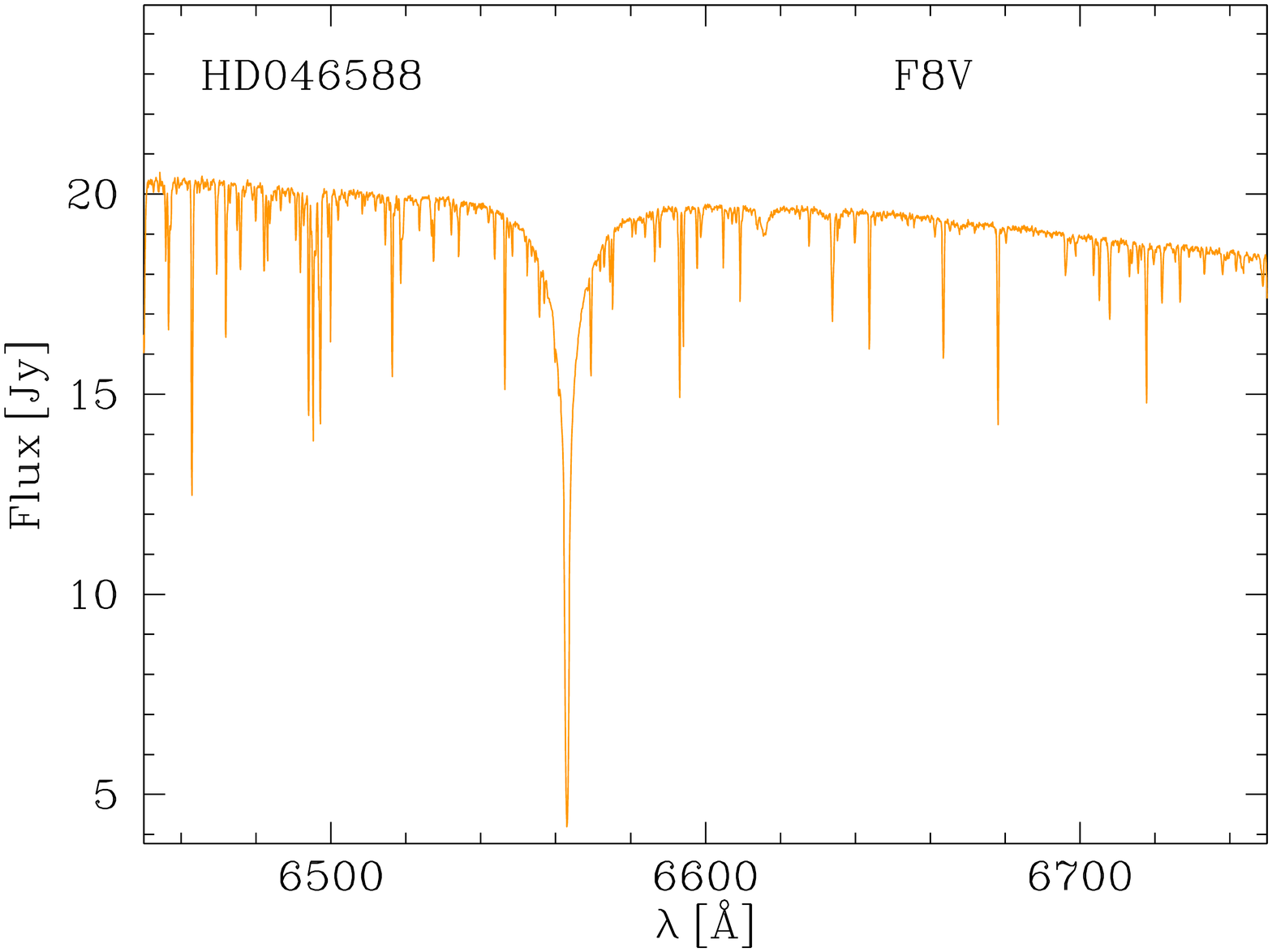}
\includegraphics[width=0.18\textwidth,angle=-90]{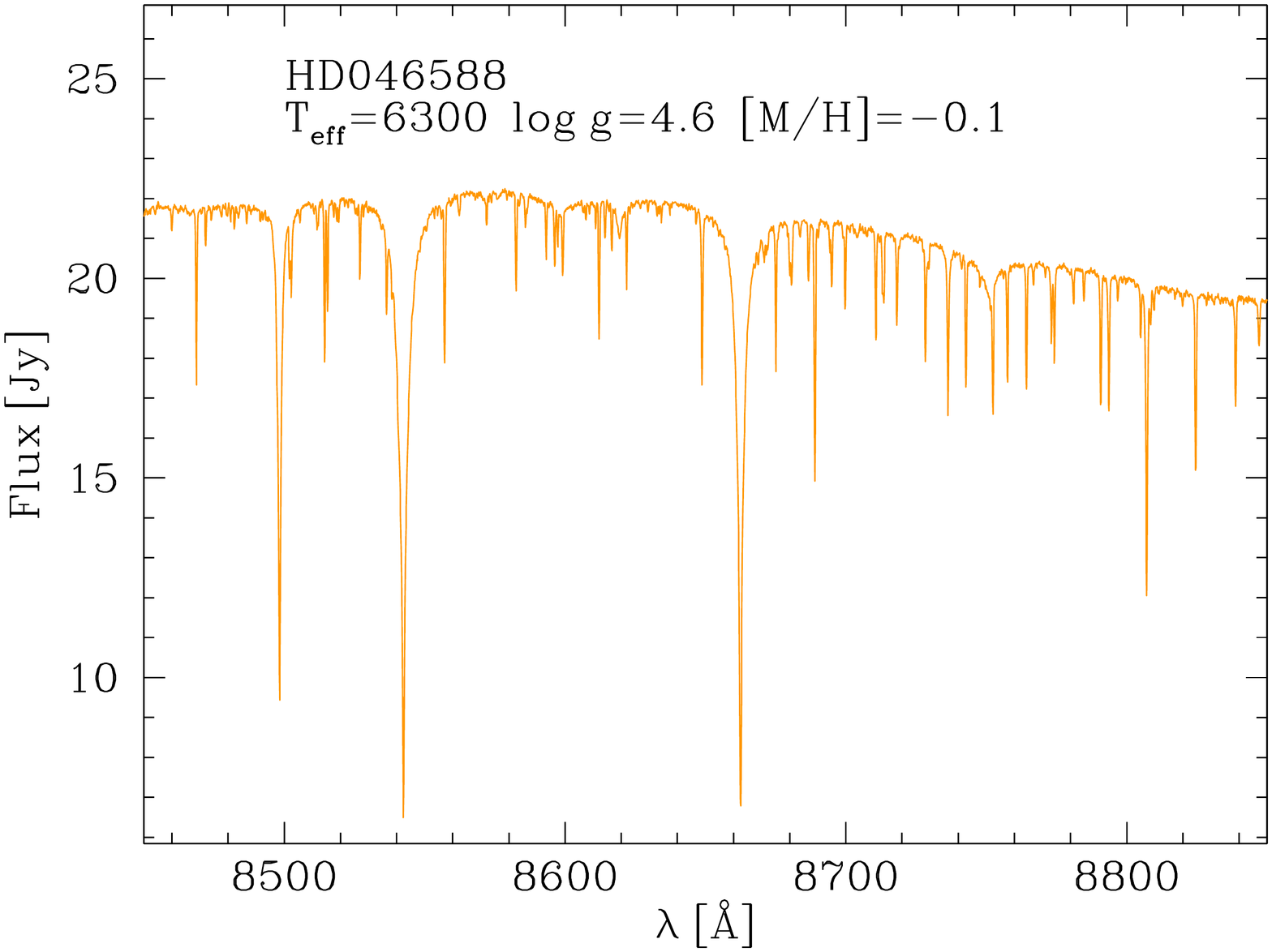}
\includegraphics[width=0.18\textwidth,angle=-90]{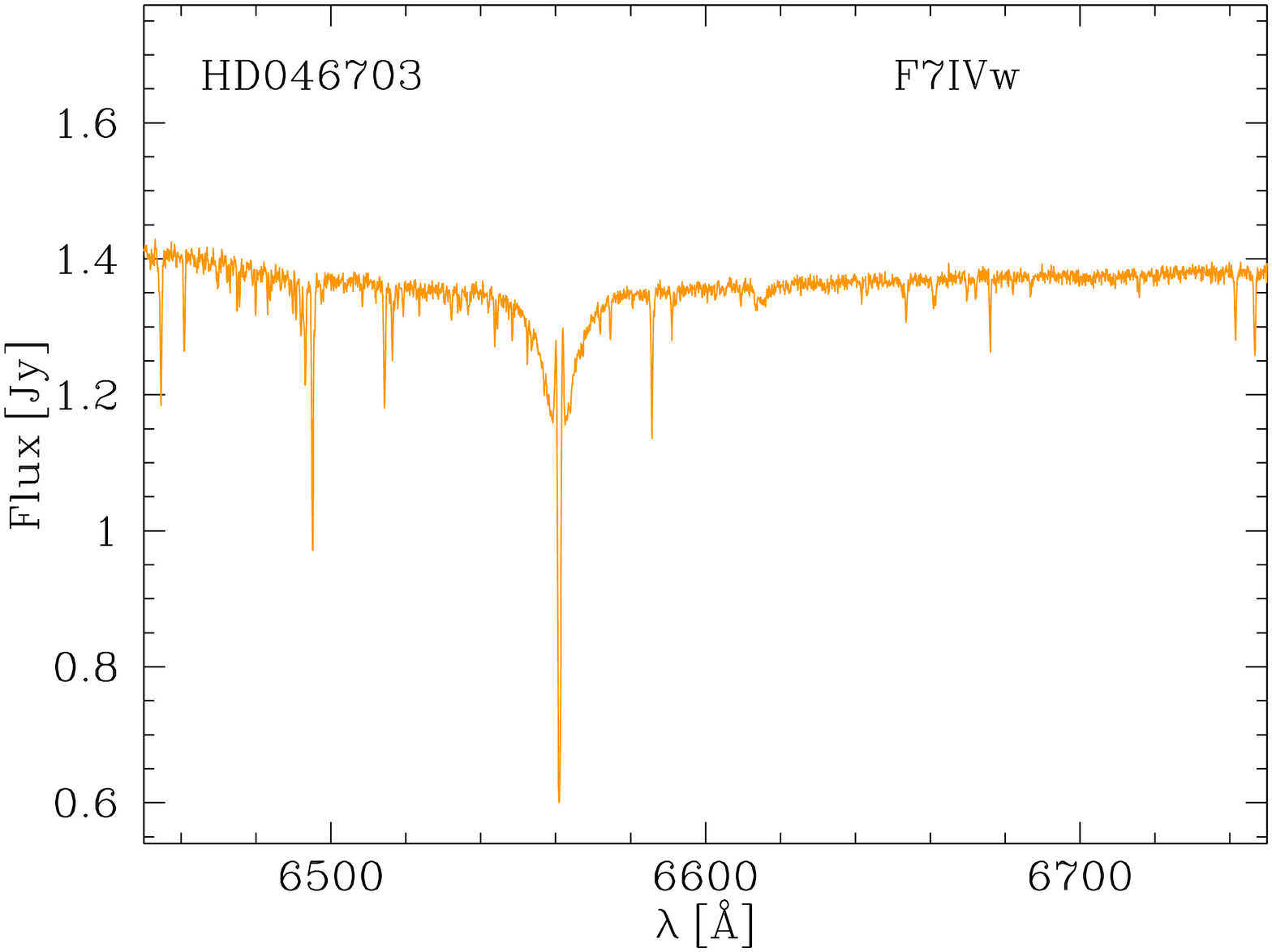}
\includegraphics[width=0.18\textwidth,angle=-90]{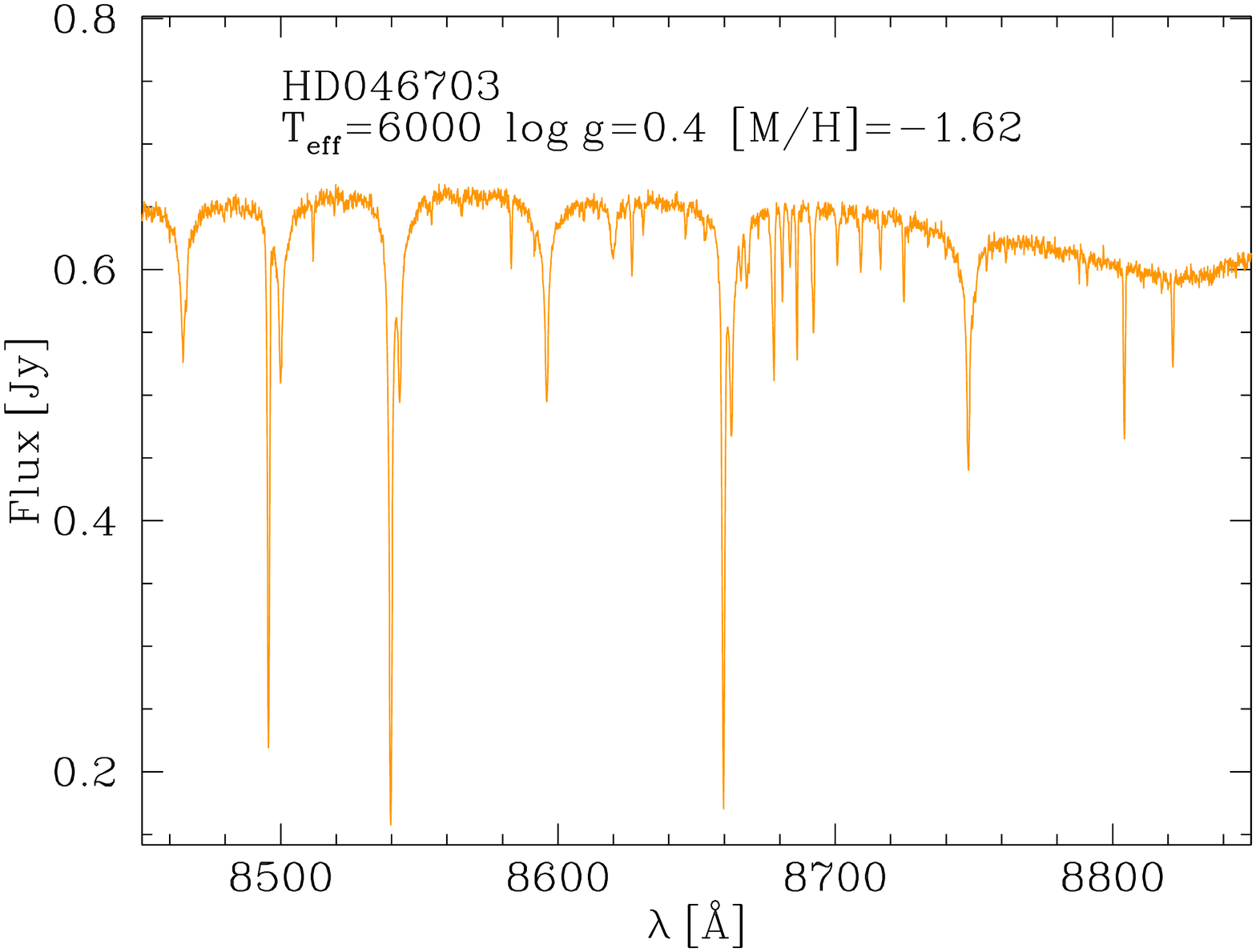}
\includegraphics[width=0.18\textwidth,angle=-90]{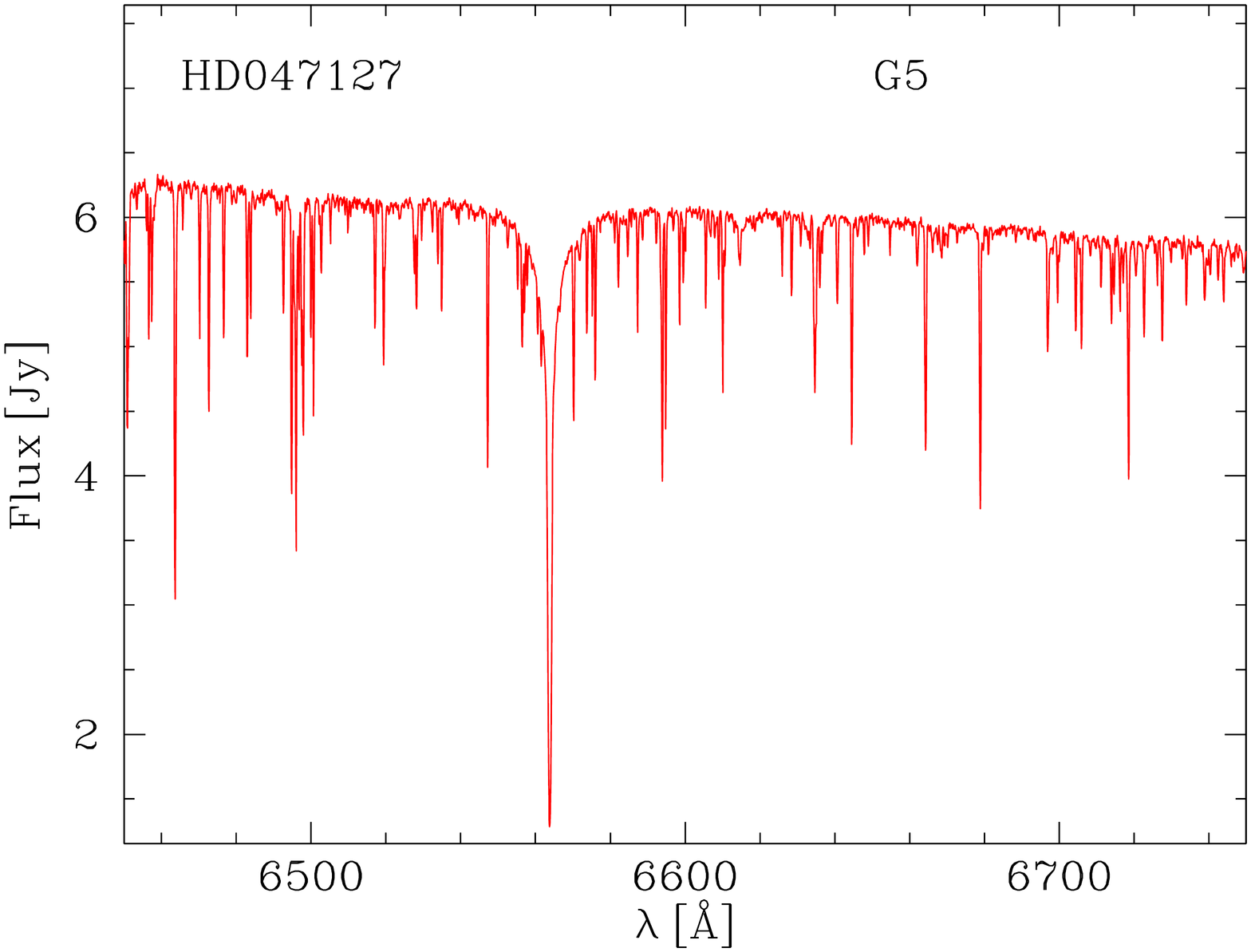}
\includegraphics[width=0.18\textwidth,angle=-90]{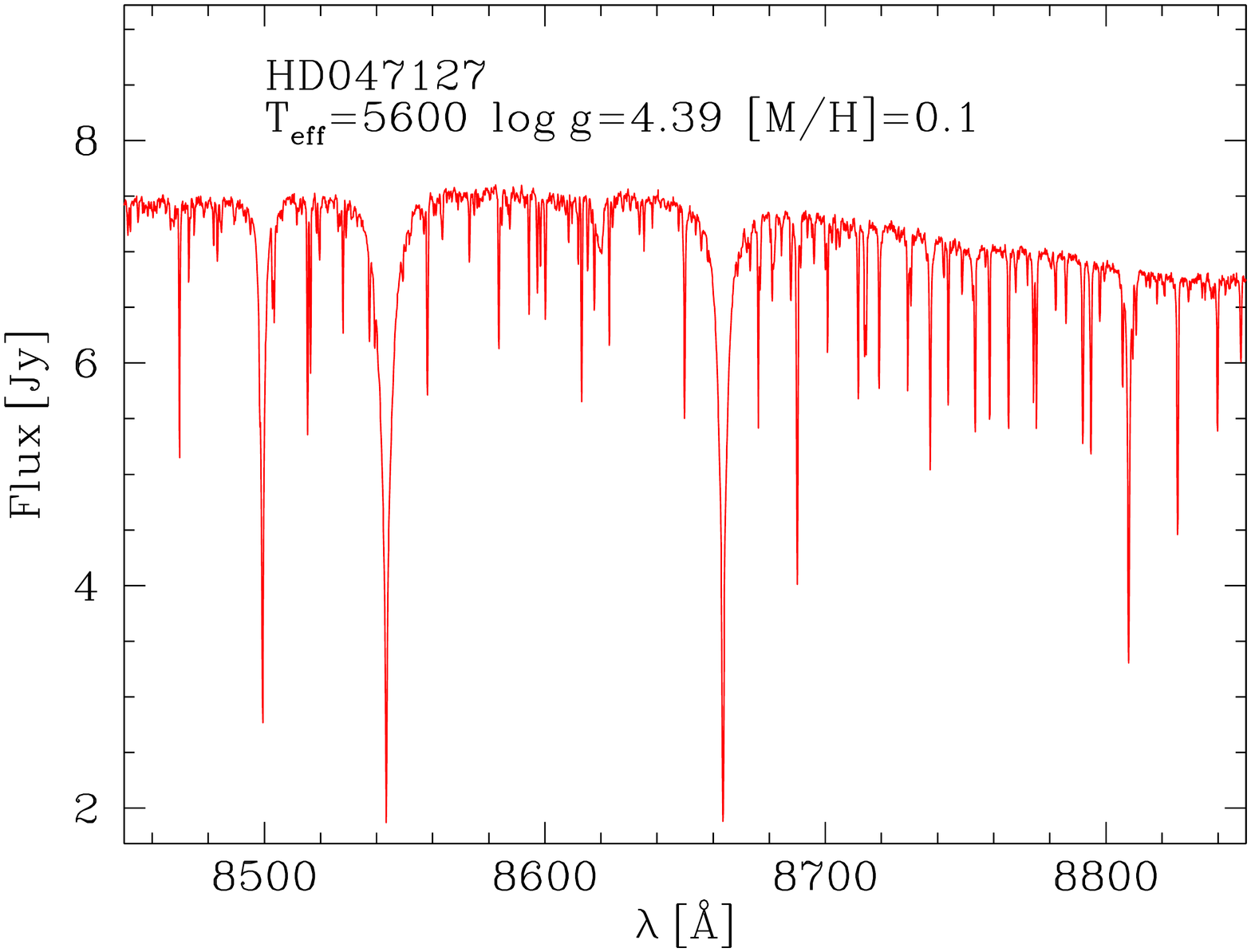}
\includegraphics[width=0.18\textwidth,angle=-90]{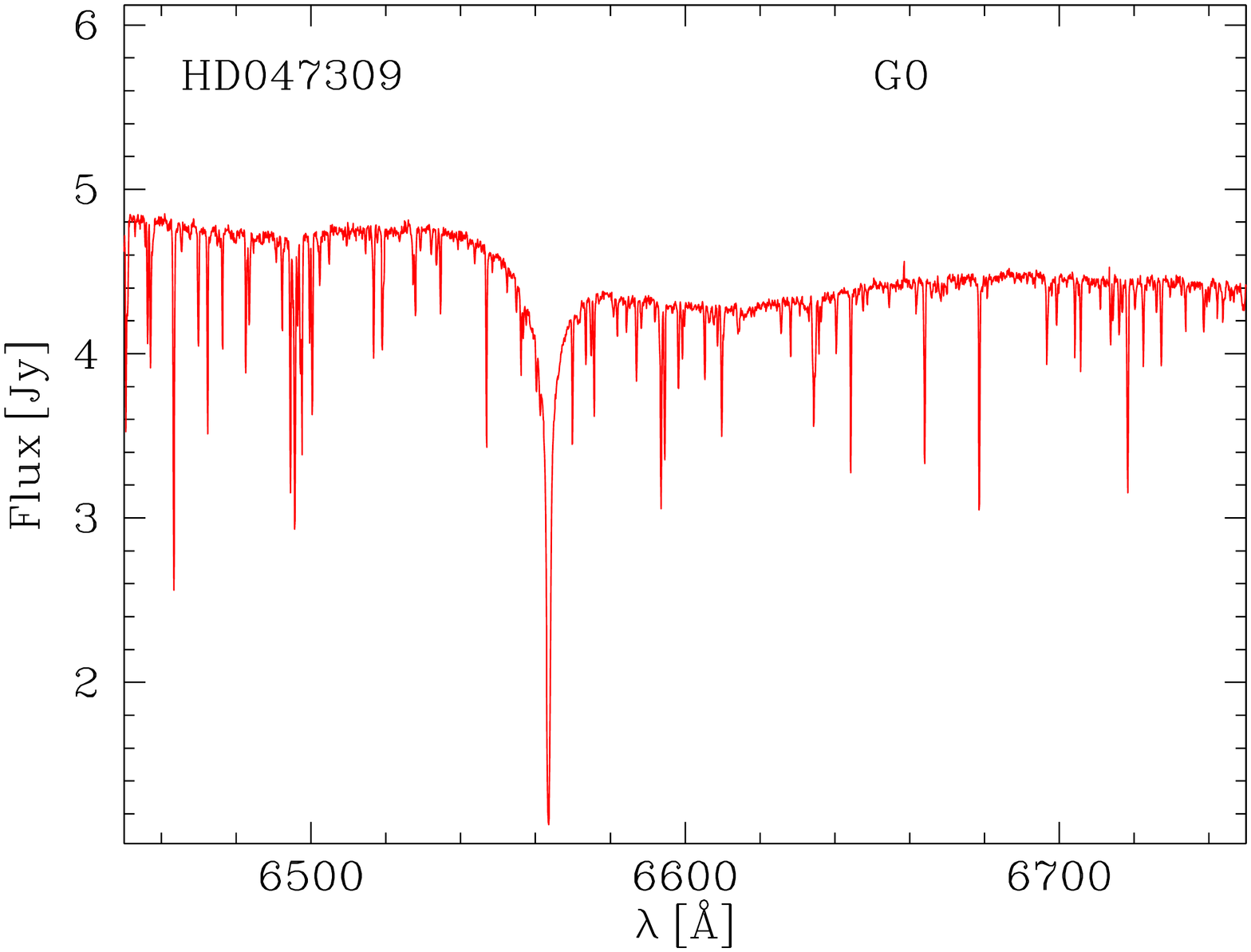}
\includegraphics[width=0.18\textwidth,angle=-90]{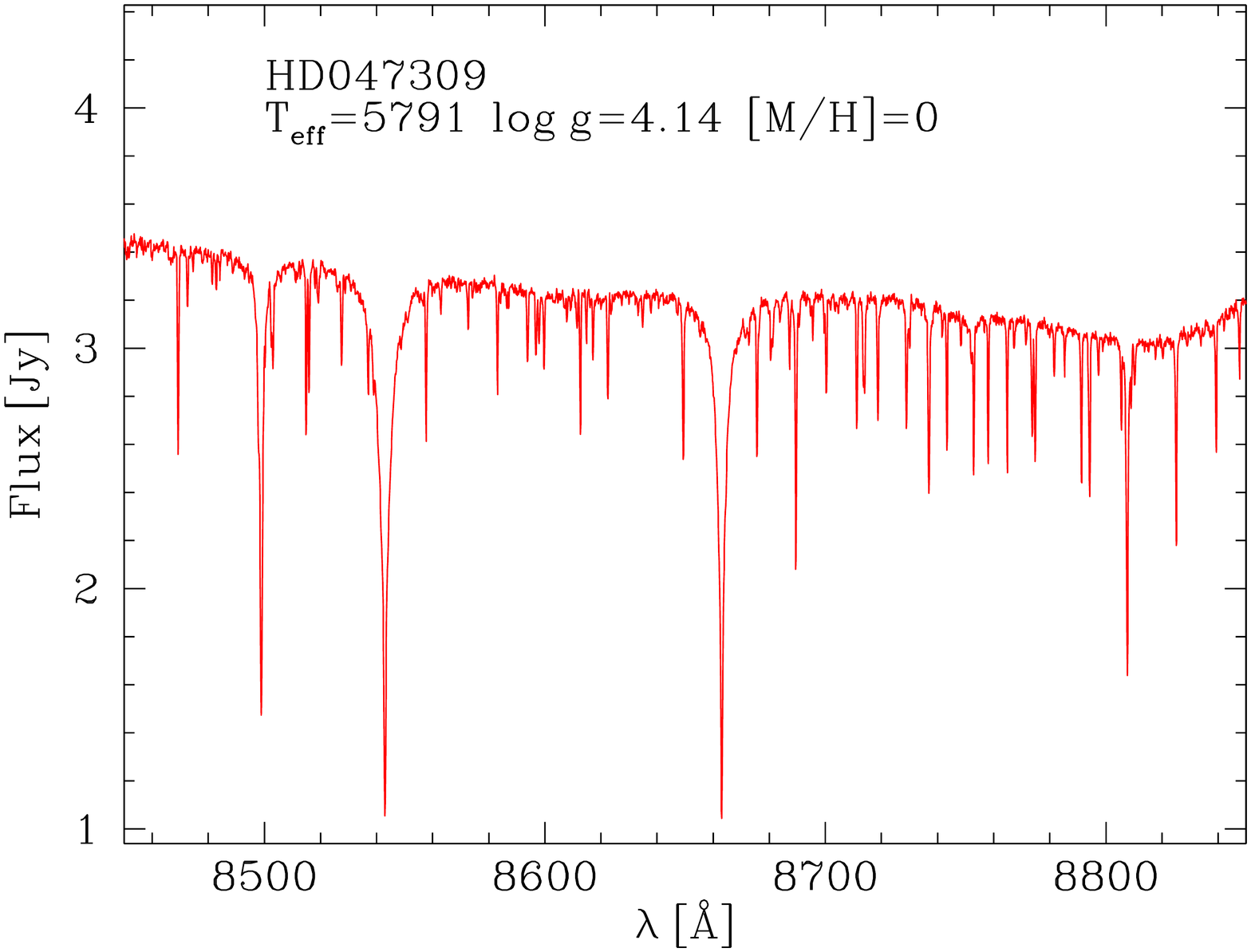}

\contcaption{11. Stars shown in this page are:  HD044614, HD045321, HD045391, HD045410, HD045829, HD045910, HD046223, HD046317, HD046380, HD046480, HD046588, HD046703, HD047127 and HD047309.}
\end{figure*}

\begin{figure*}
\includegraphics[width=0.18\textwidth,angle=-90]{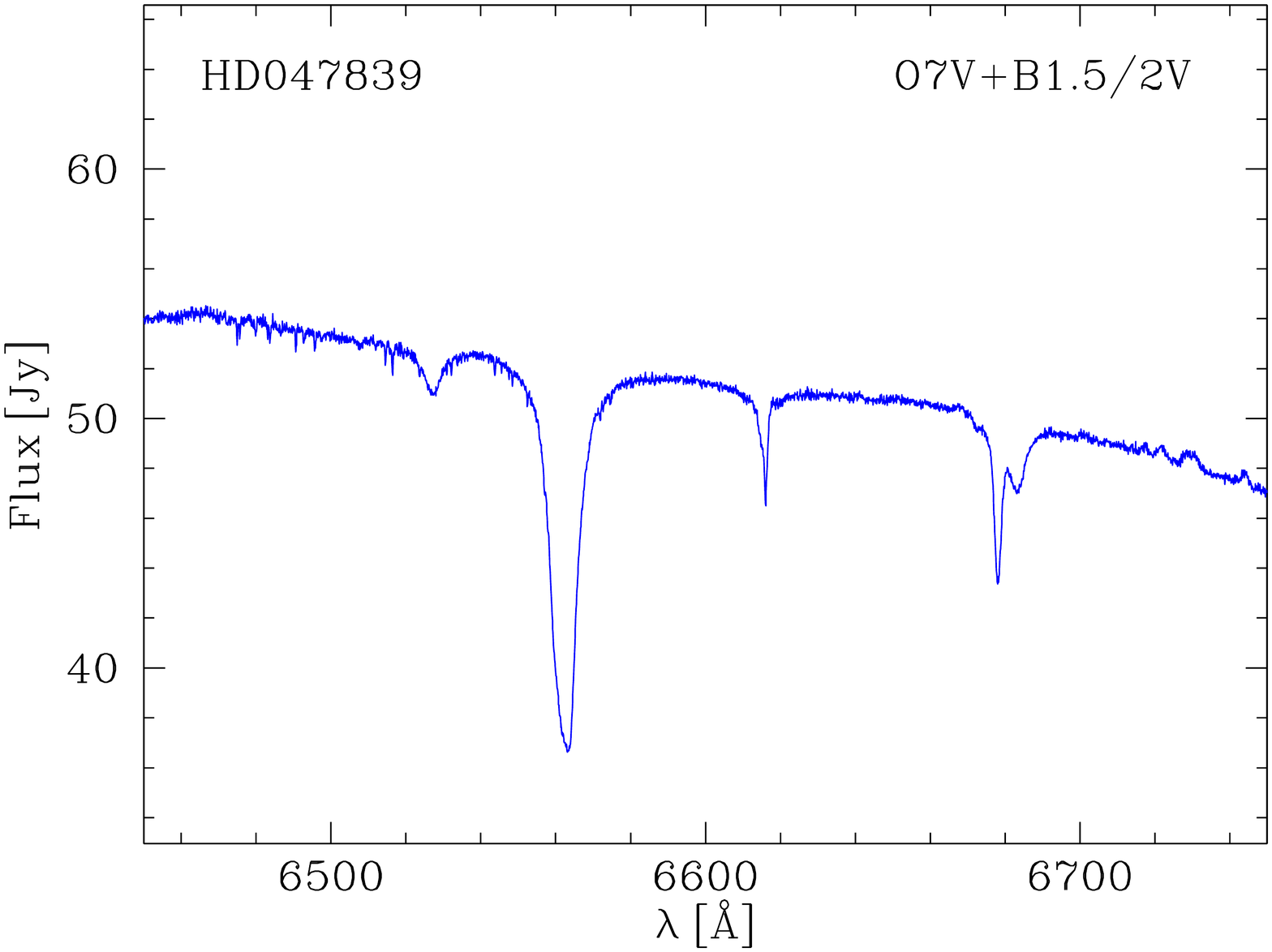}
\includegraphics[width=0.18\textwidth,angle=-90]{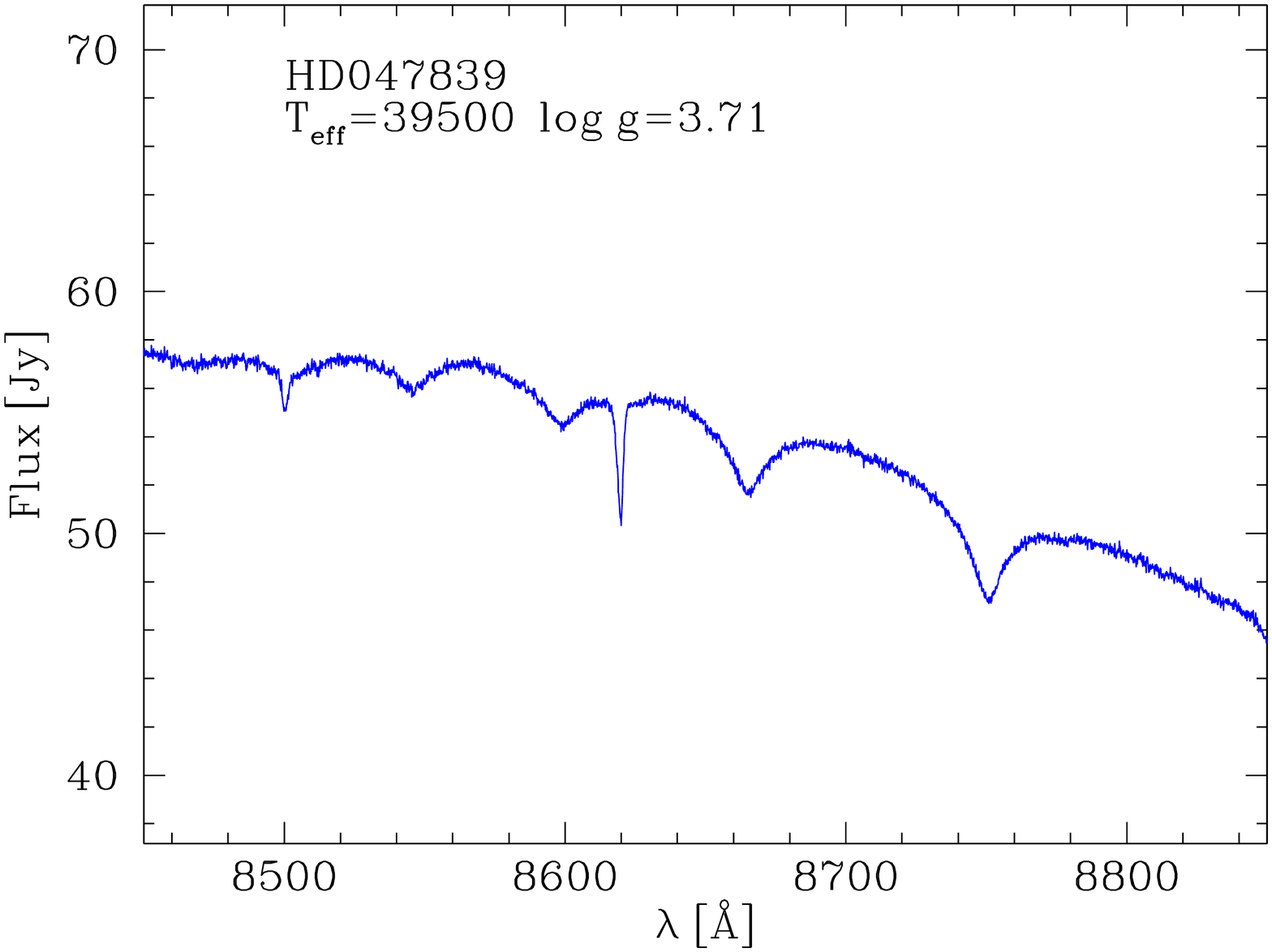}
\includegraphics[width=0.18\textwidth,angle=-90]{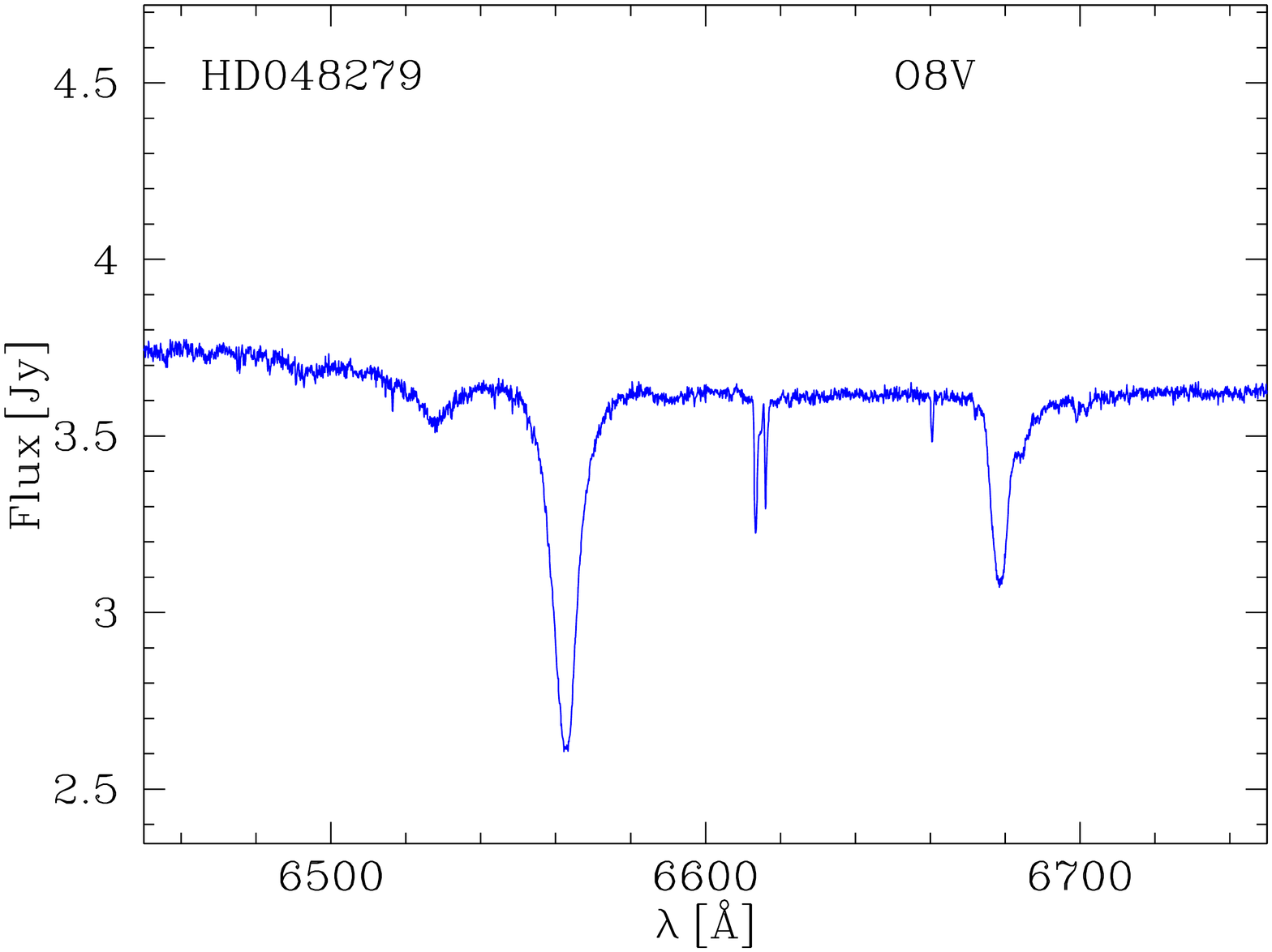}
\includegraphics[width=0.18\textwidth,angle=-90]{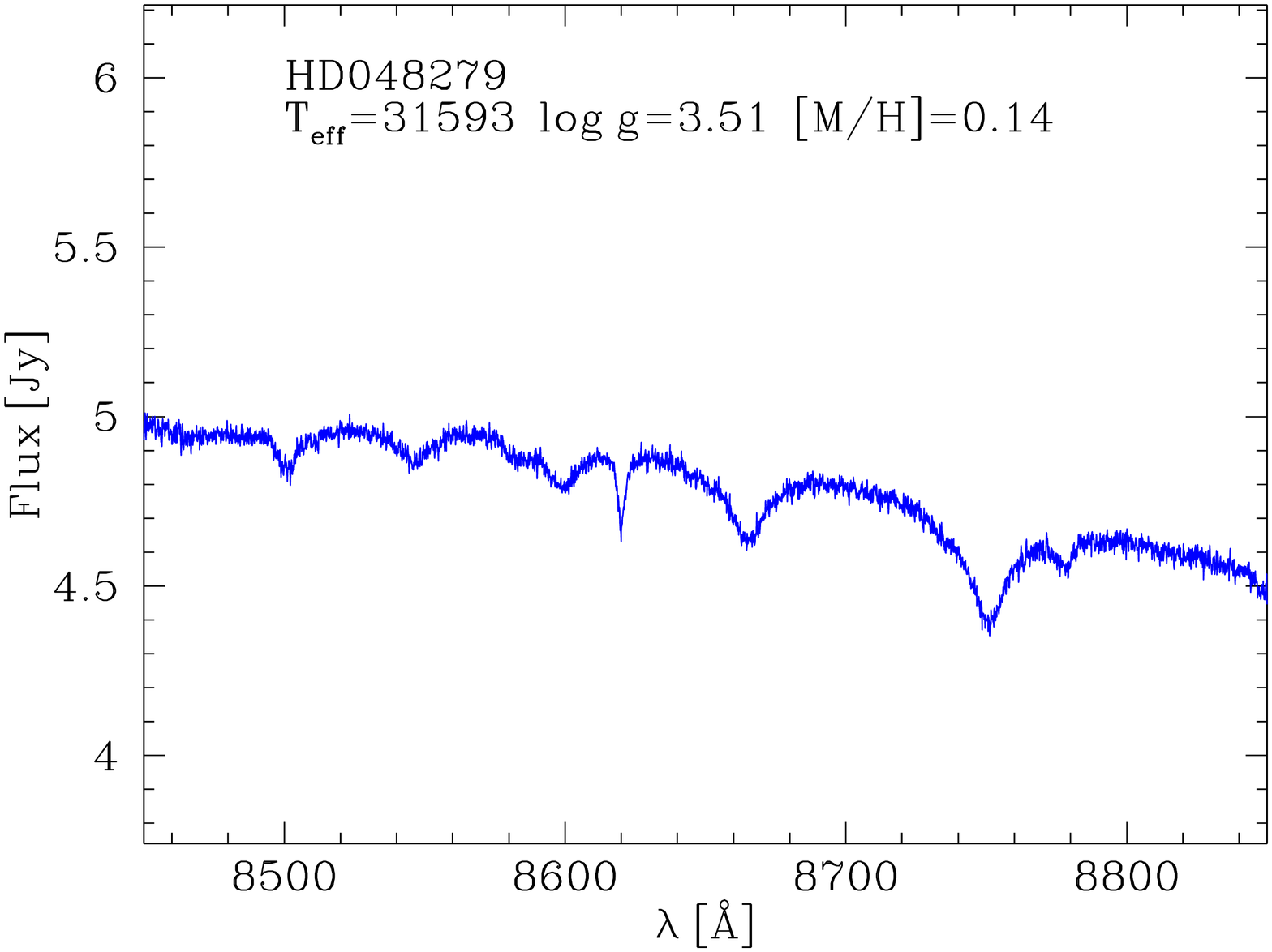}
\includegraphics[width=0.18\textwidth,angle=-90]{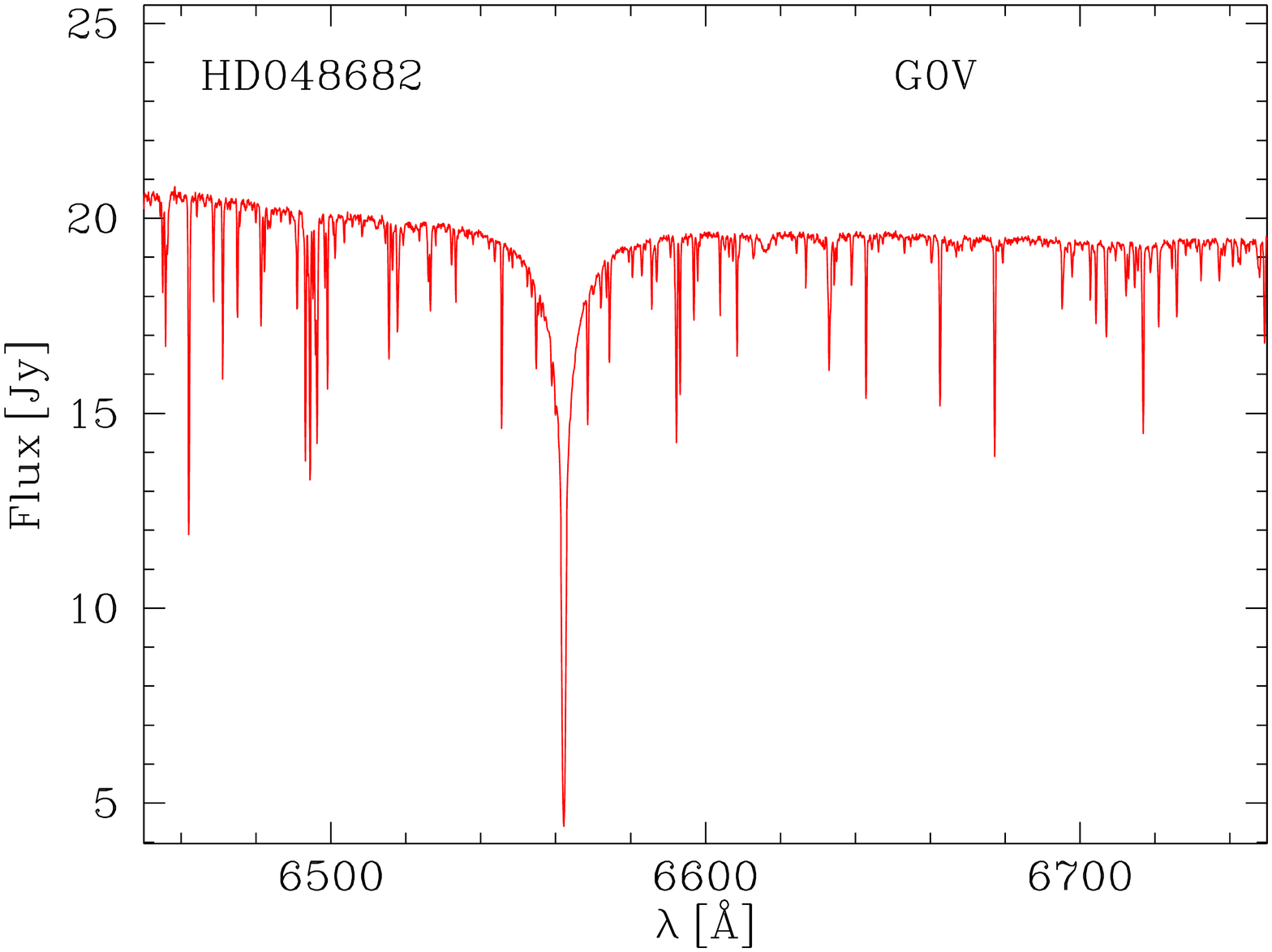}
\includegraphics[width=0.18\textwidth,angle=-90]{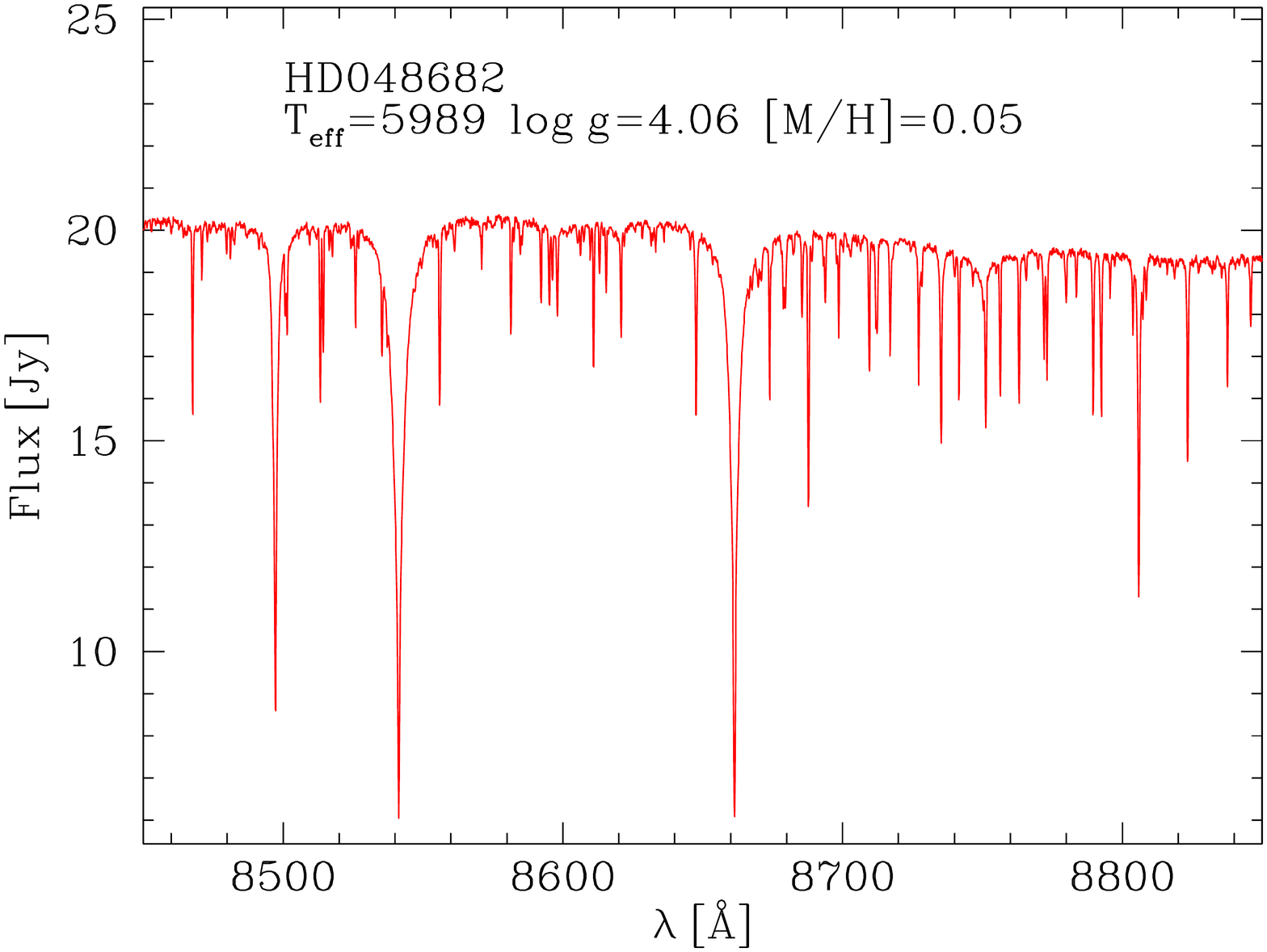}
\includegraphics[width=0.18\textwidth,angle=-90]{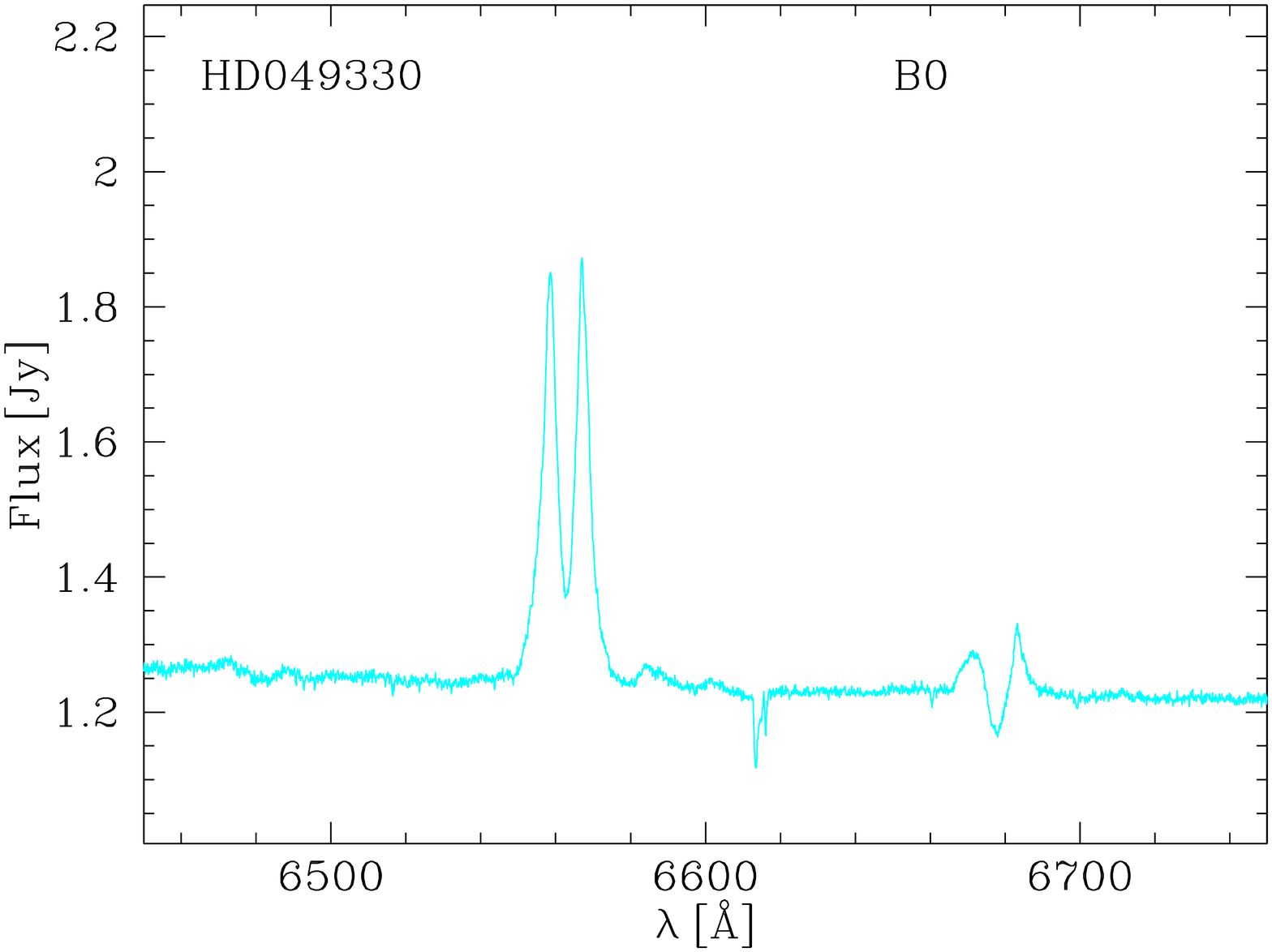}
\includegraphics[width=0.18\textwidth,angle=-90]{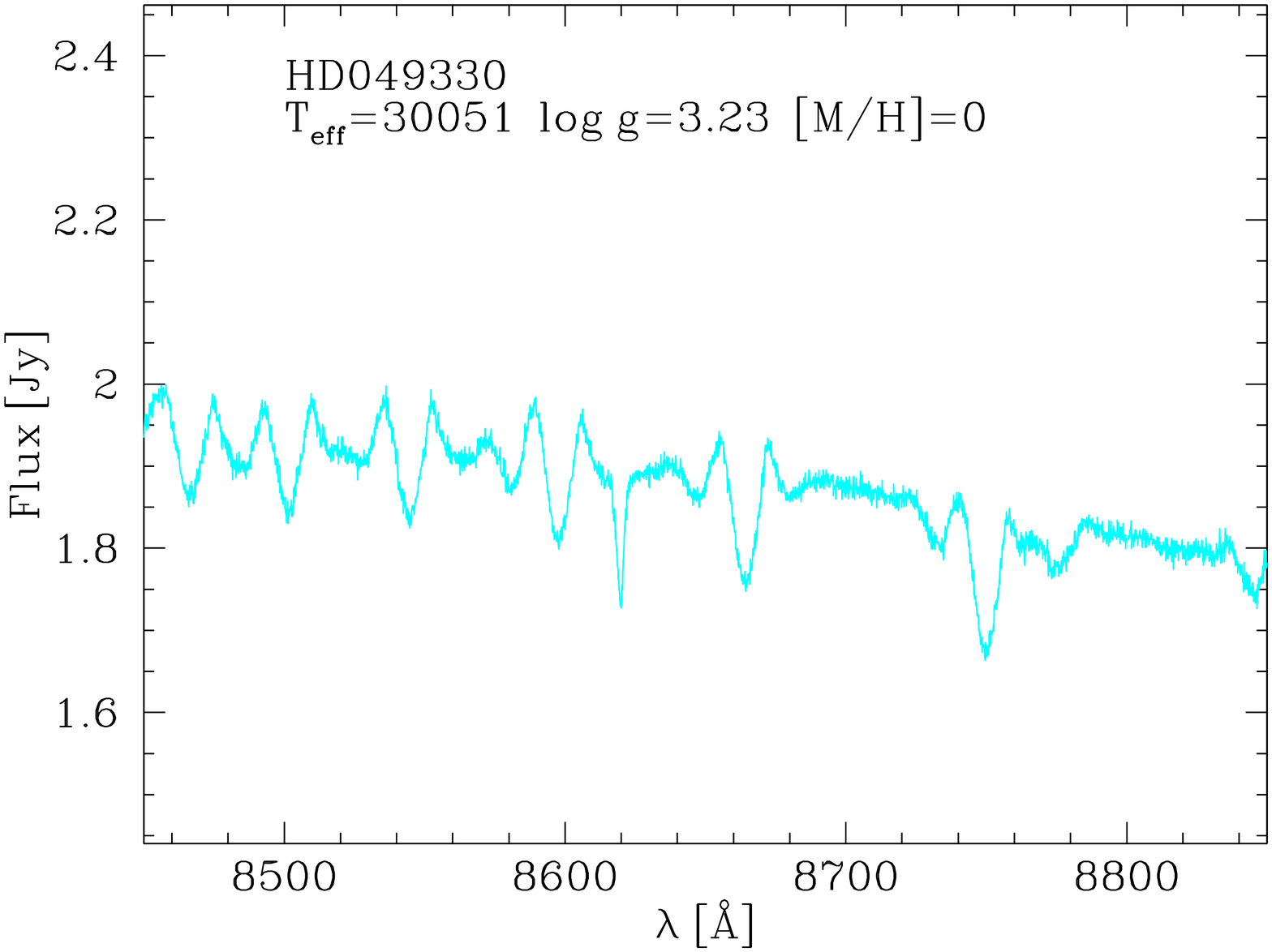}
\includegraphics[width=0.18\textwidth,angle=-90]{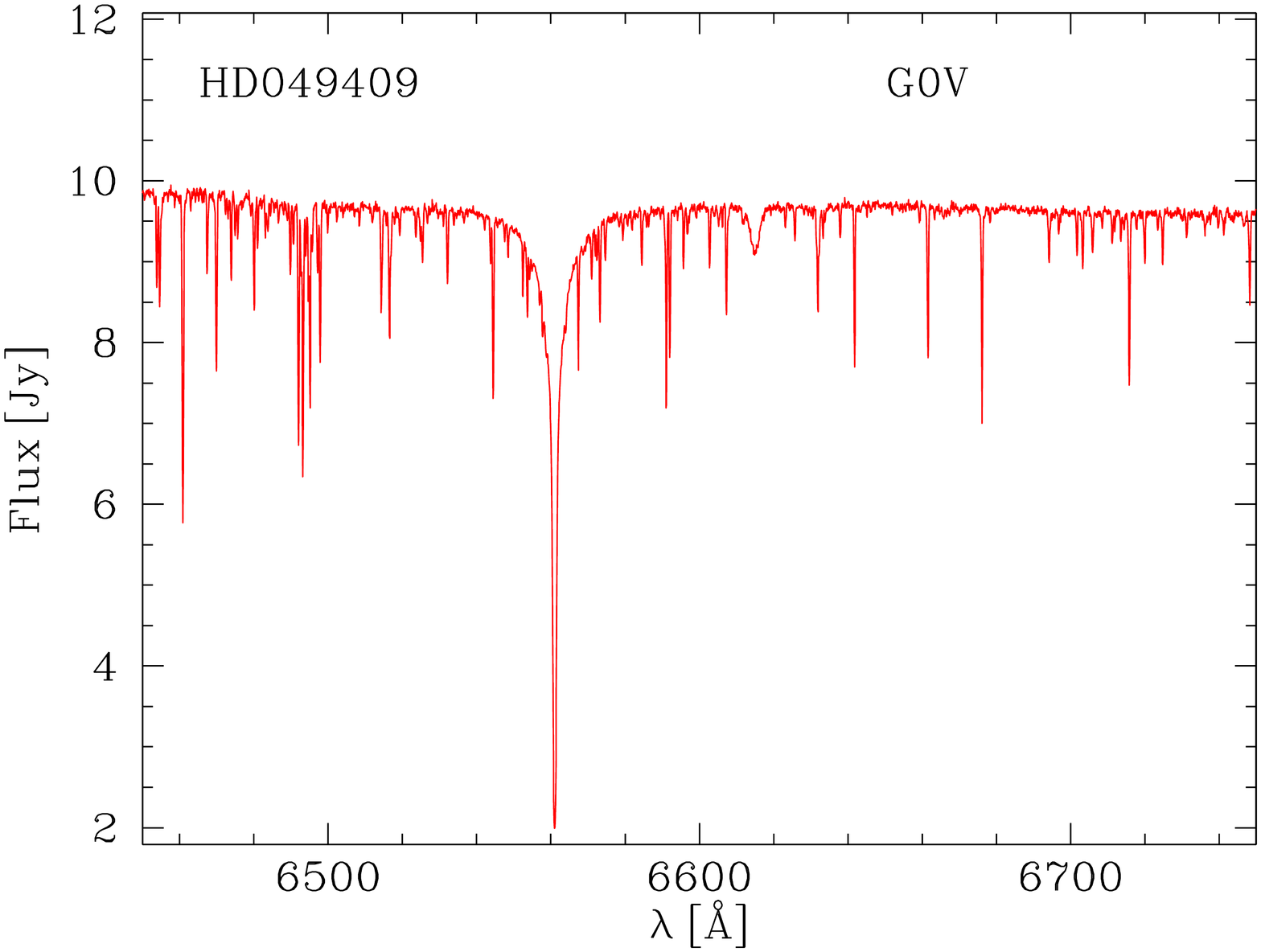}
\includegraphics[width=0.18\textwidth,angle=-90]{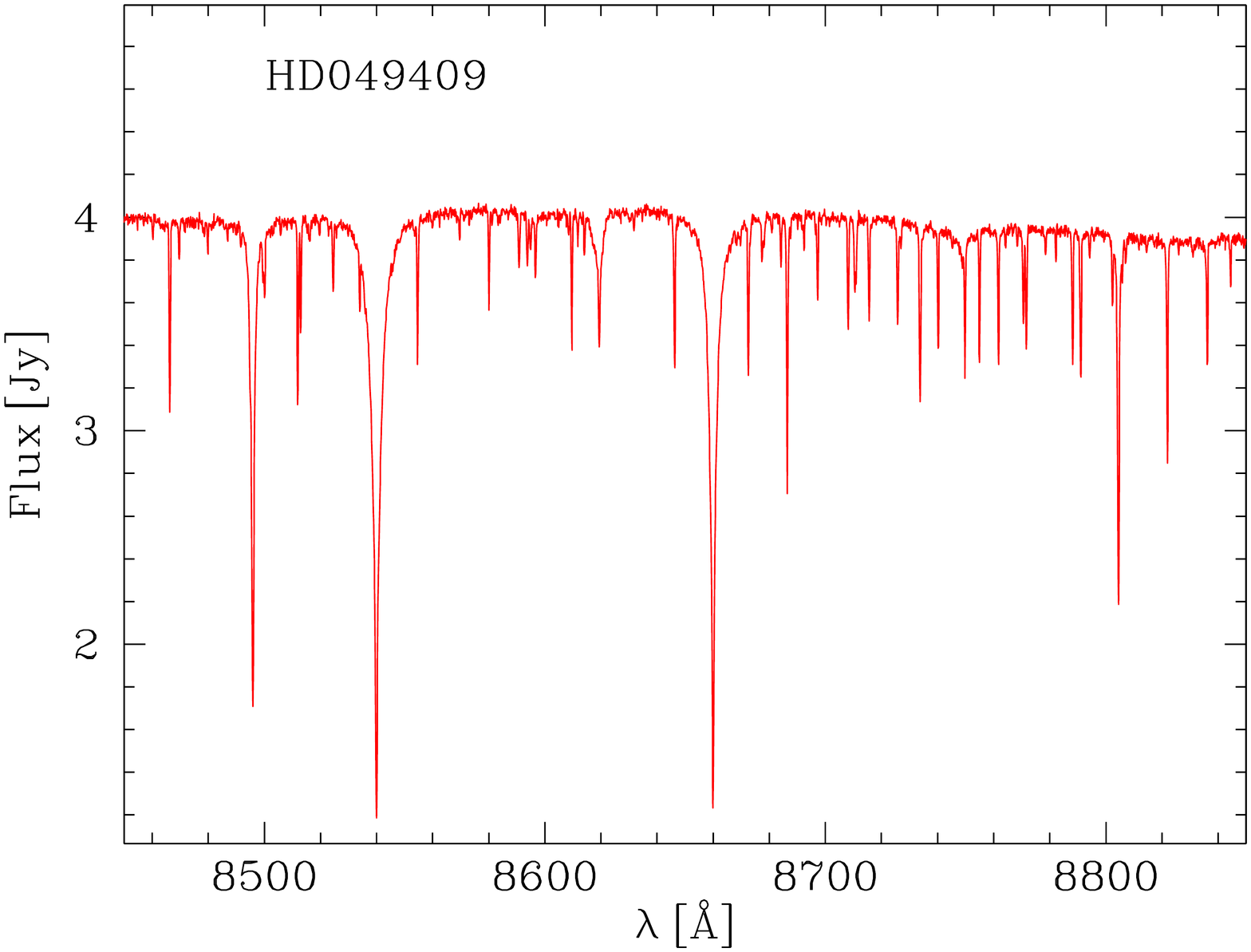}
\includegraphics[width=0.18\textwidth,angle=-90]{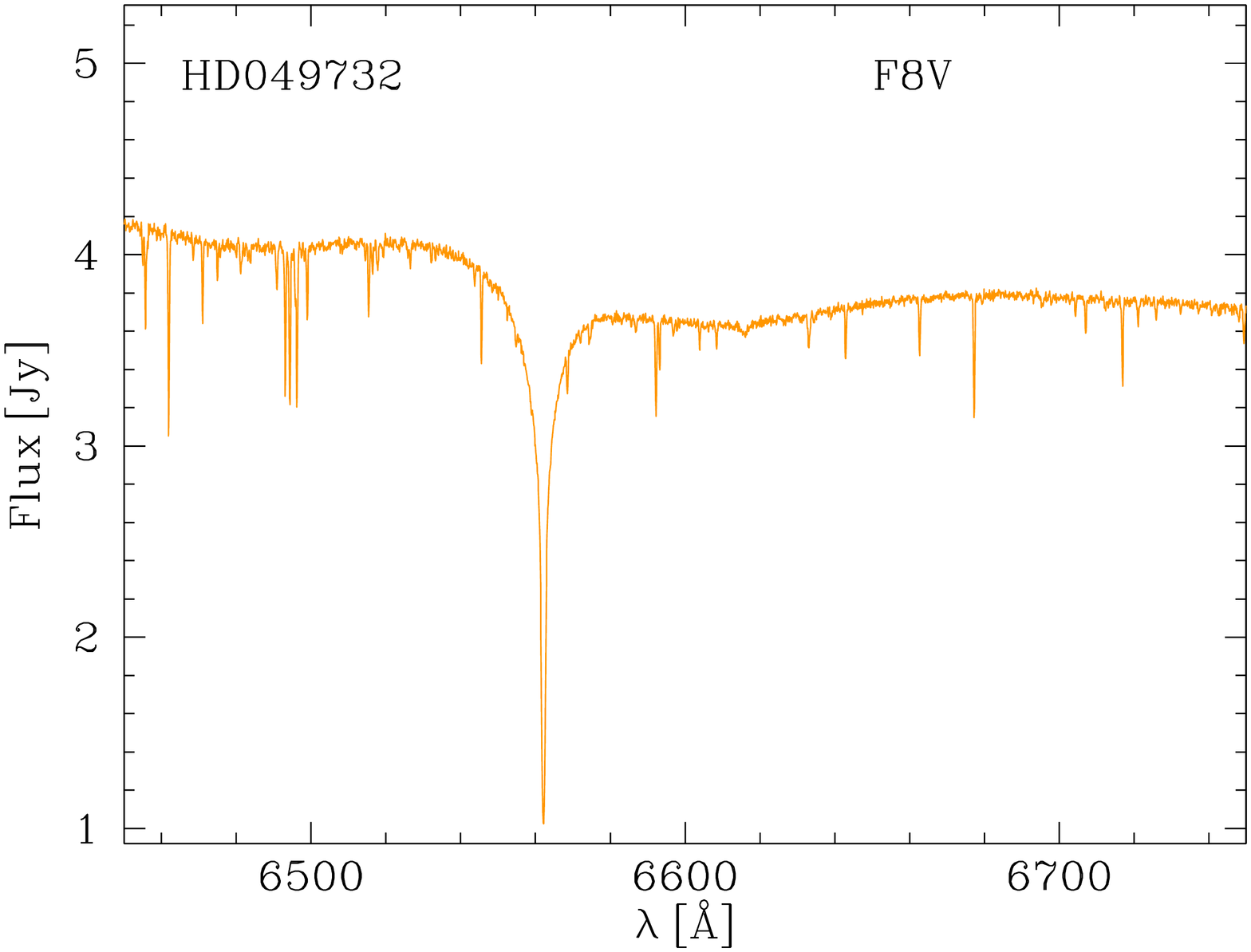}
\includegraphics[width=0.18\textwidth,angle=-90]{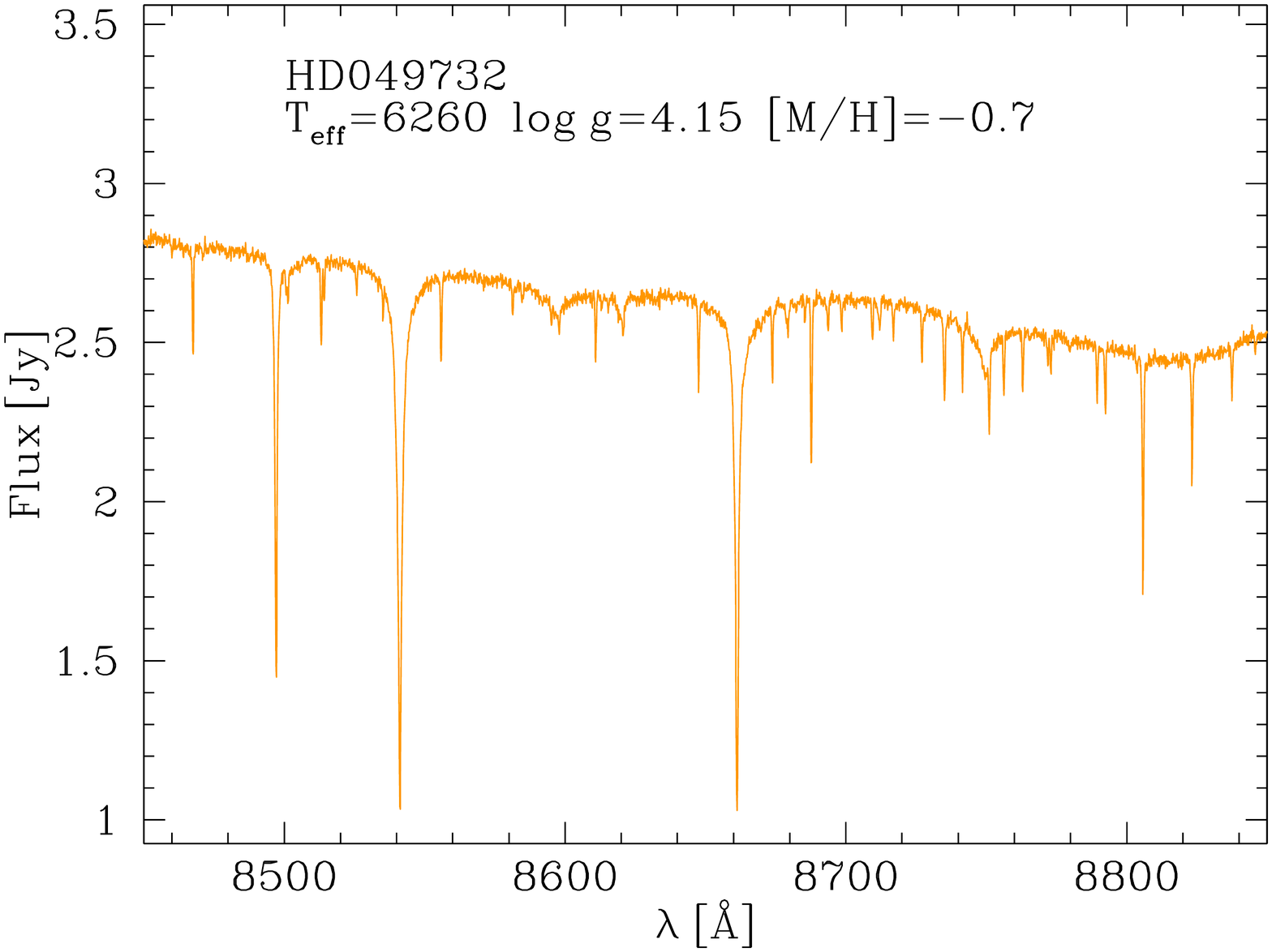}
\includegraphics[width=0.18\textwidth,angle=-90]{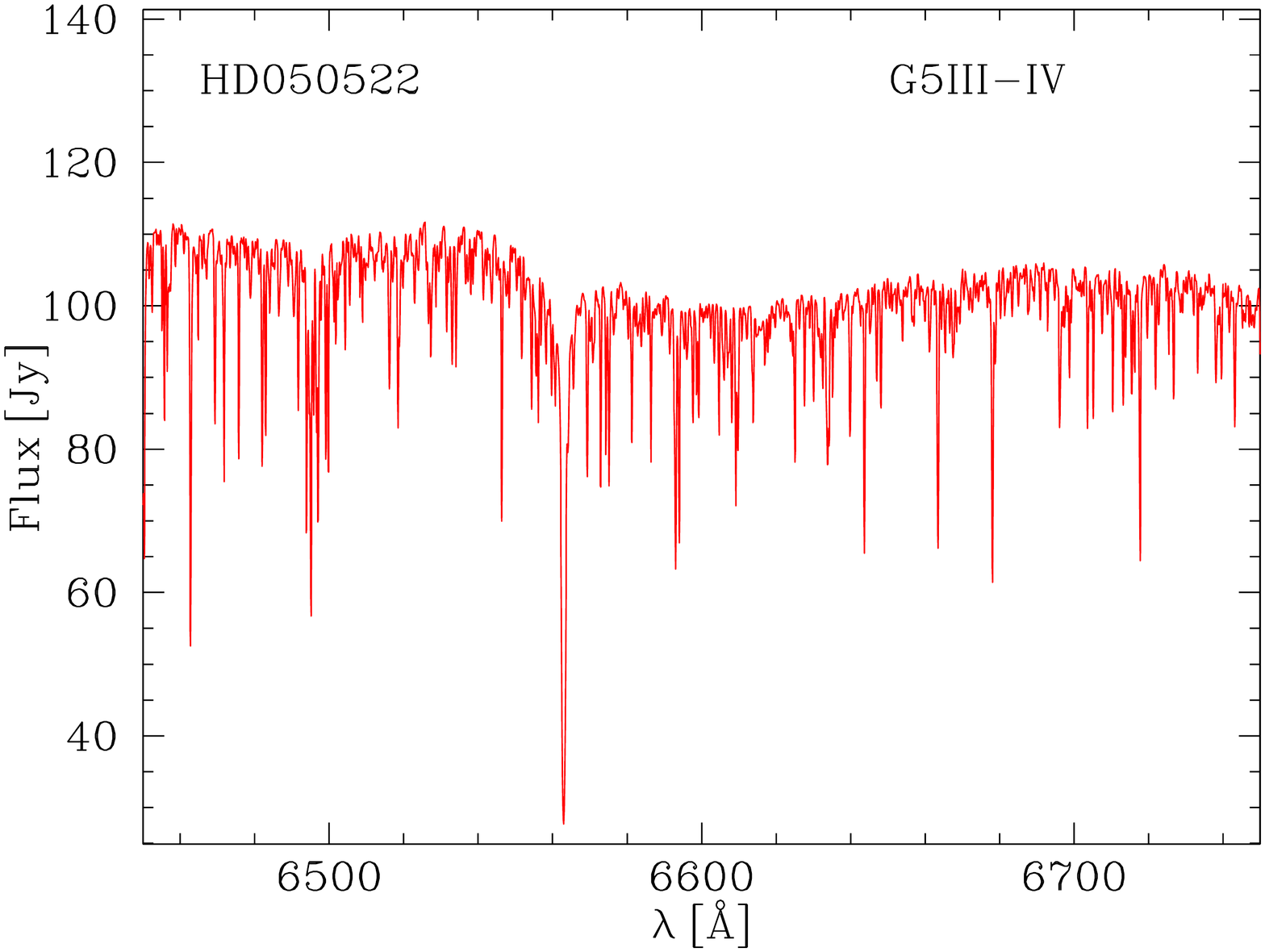}
\includegraphics[width=0.18\textwidth,angle=-90]{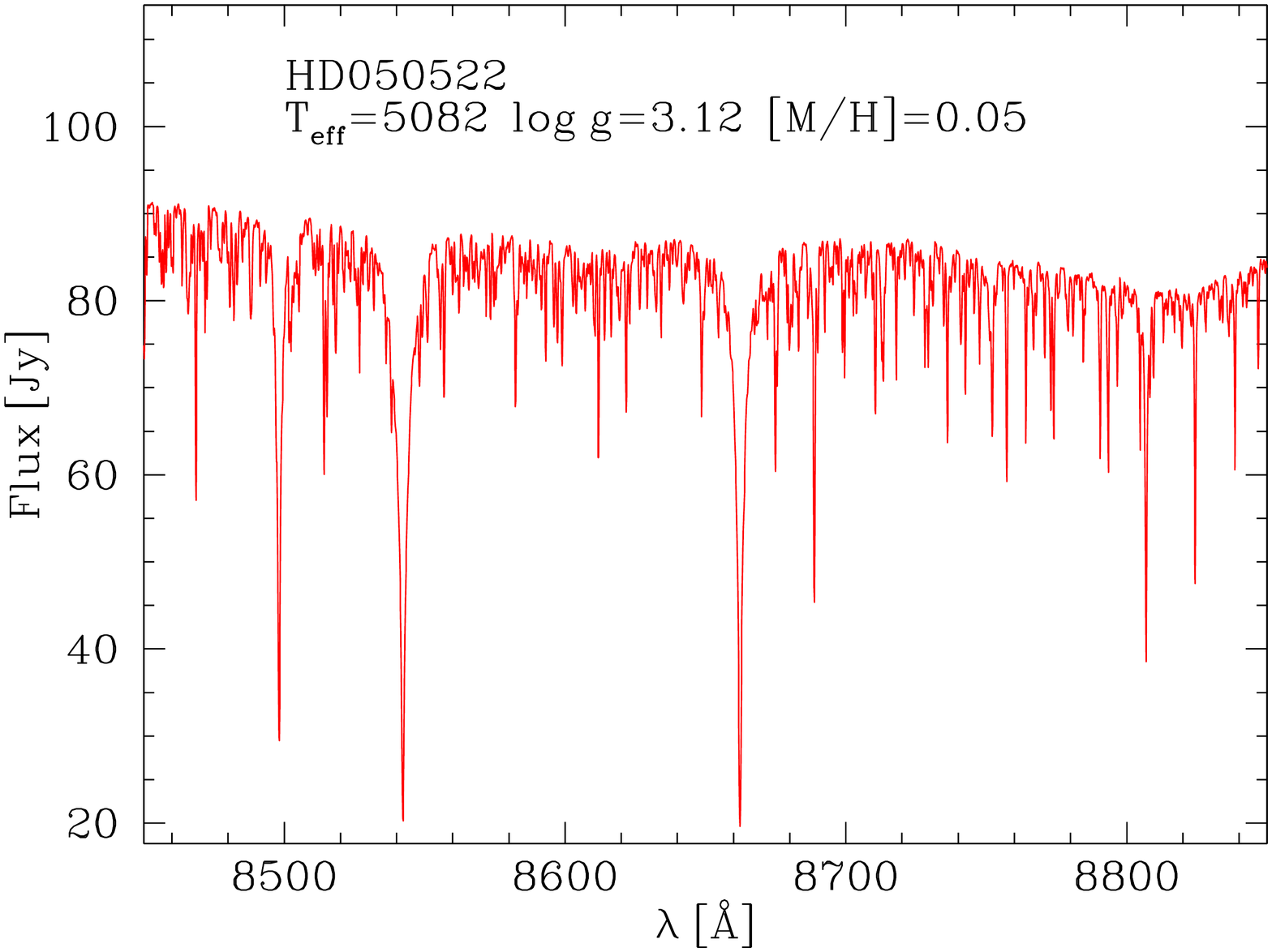}
\includegraphics[width=0.18\textwidth,angle=-90]{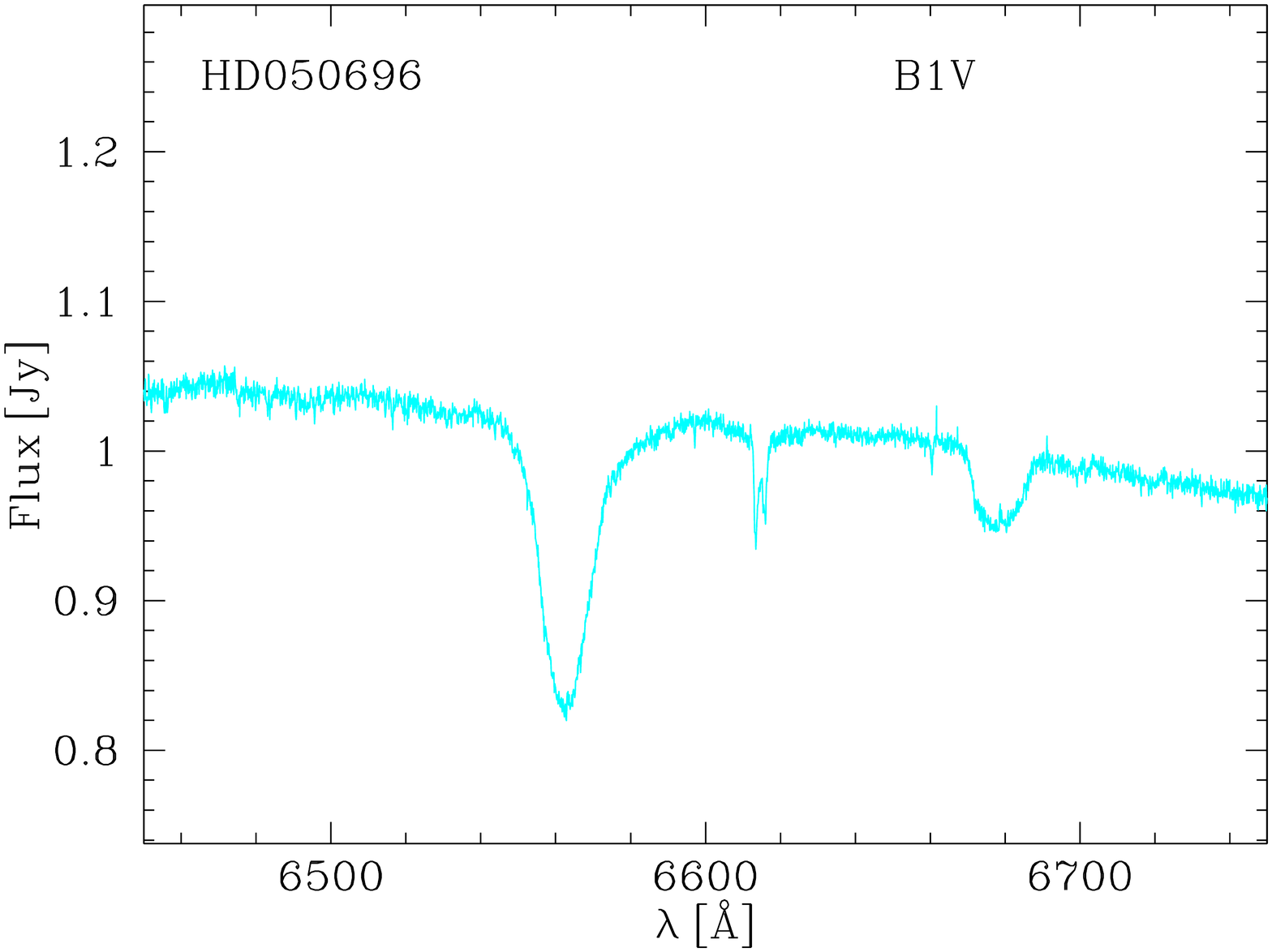}
\includegraphics[width=0.18\textwidth,angle=-90]{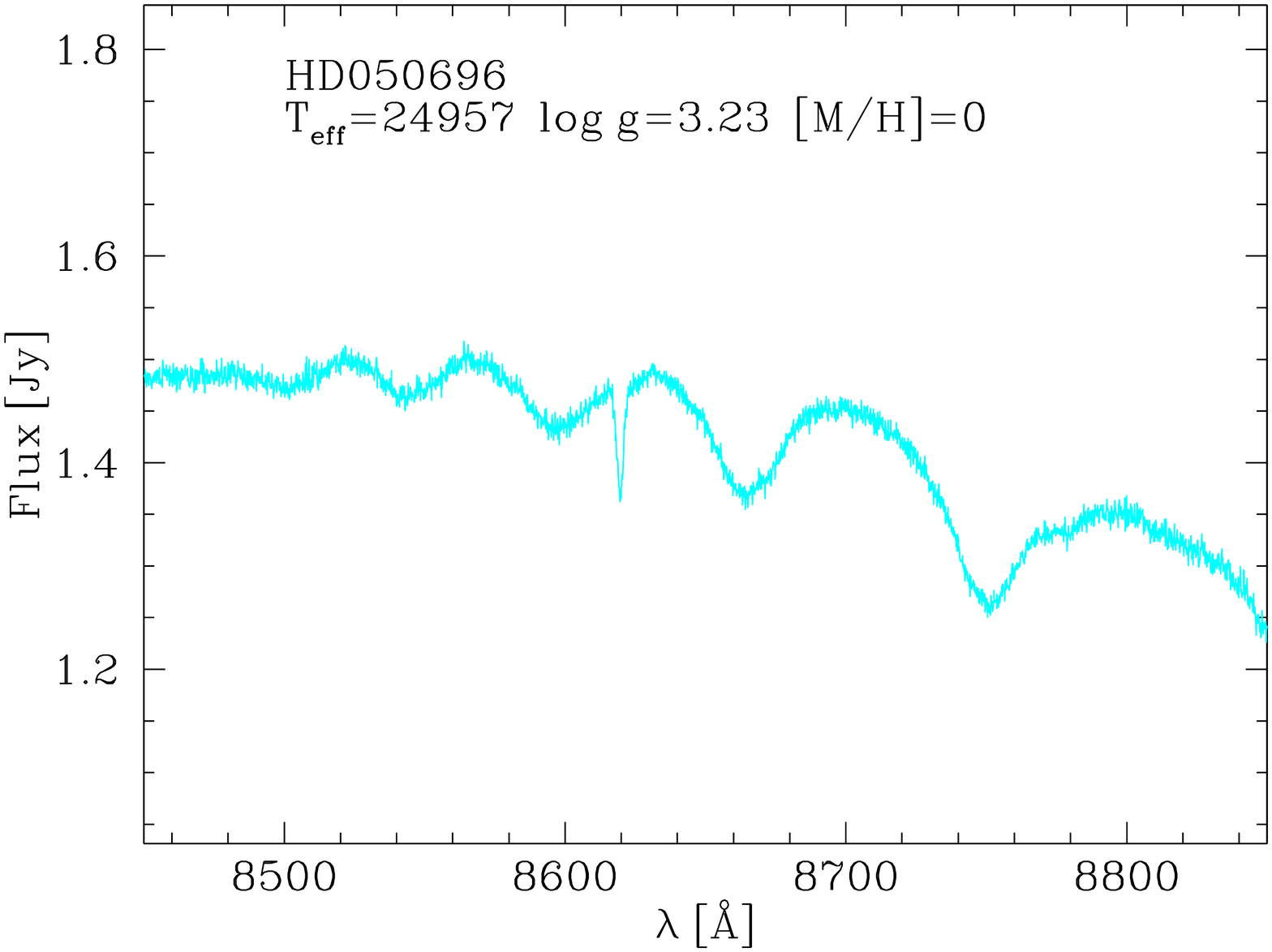}
\includegraphics[width=0.18\textwidth,angle=-90]{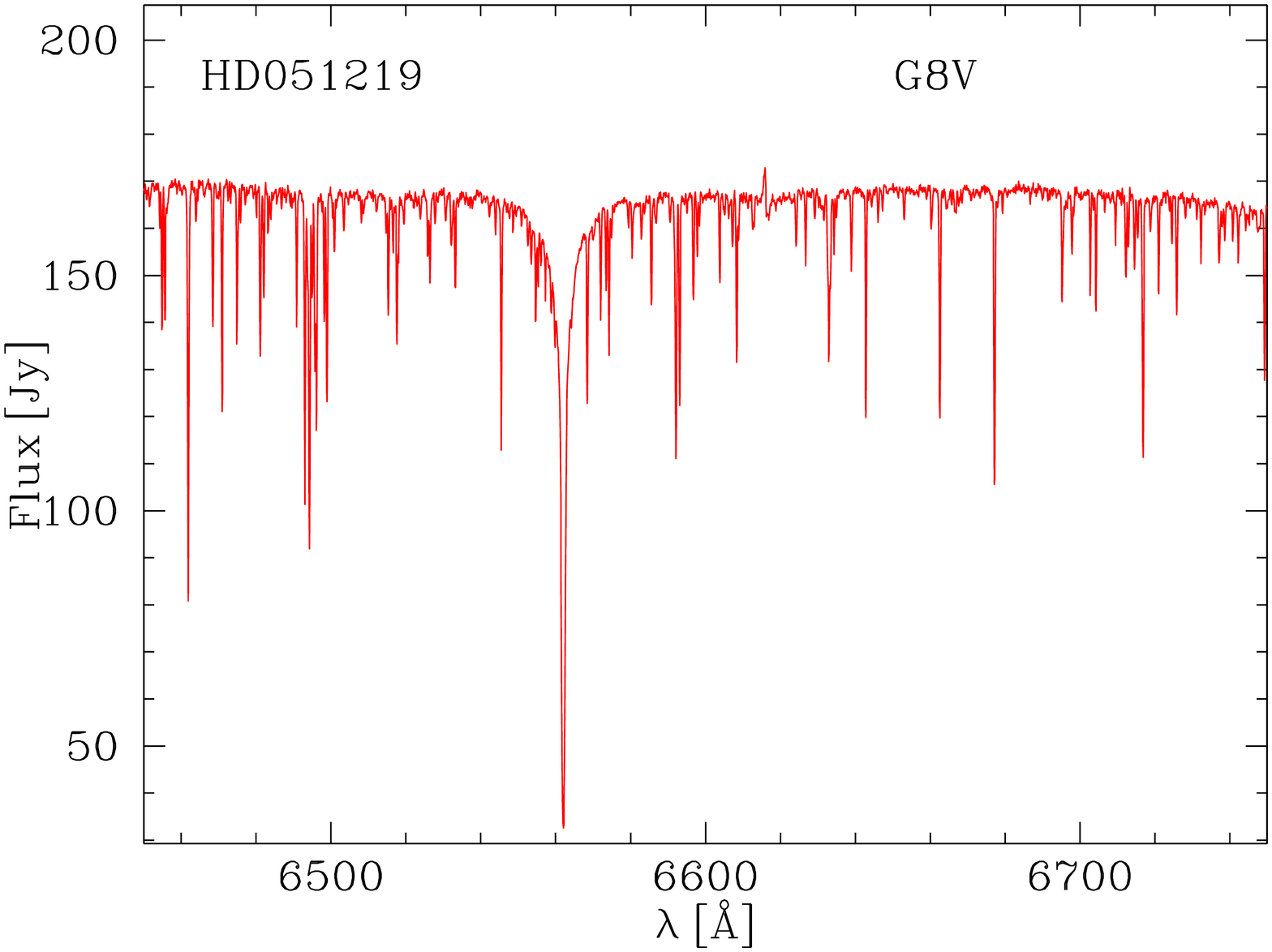}
\includegraphics[width=0.18\textwidth,angle=-90]{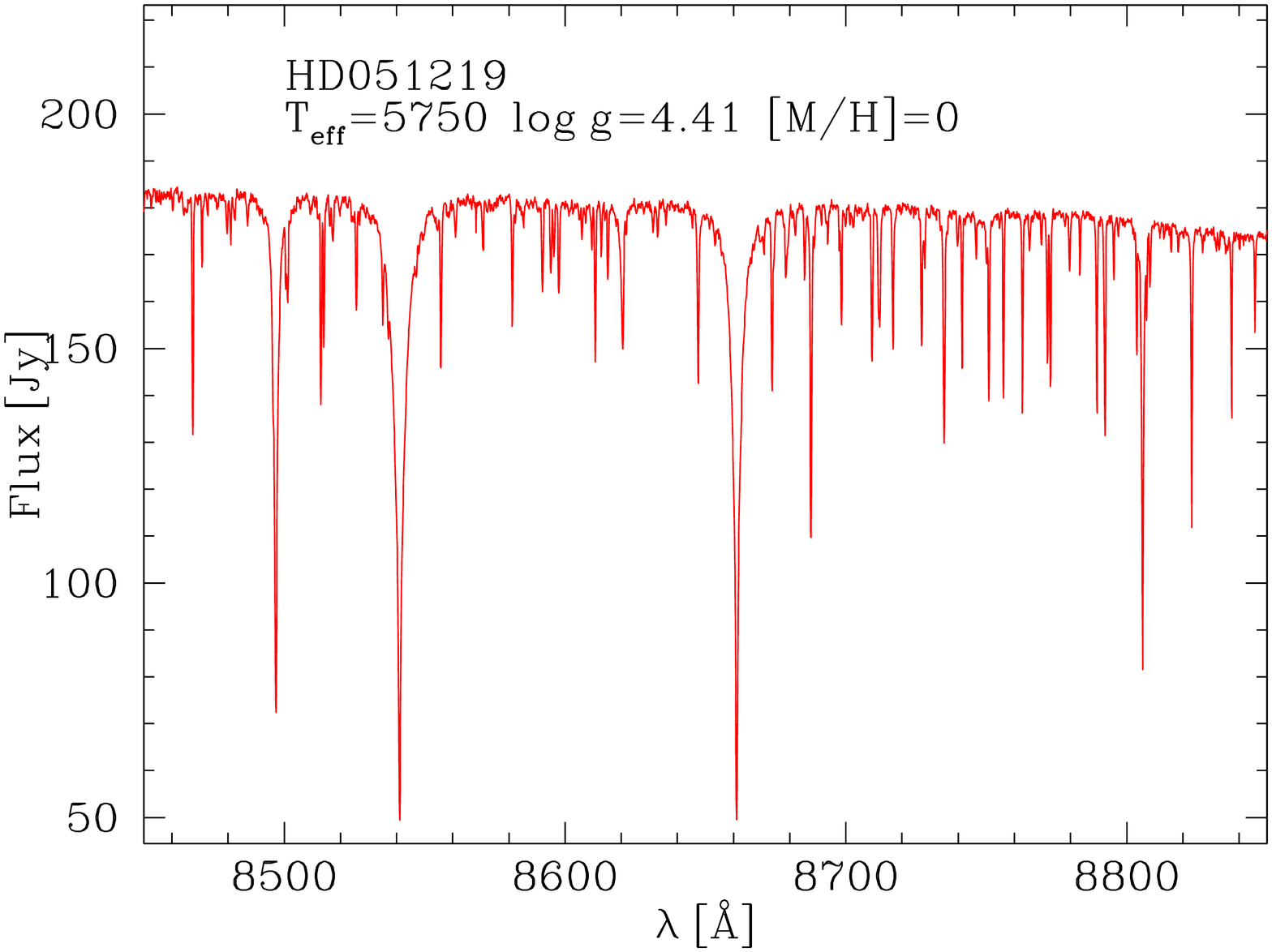}
\includegraphics[width=0.18\textwidth,angle=-90]{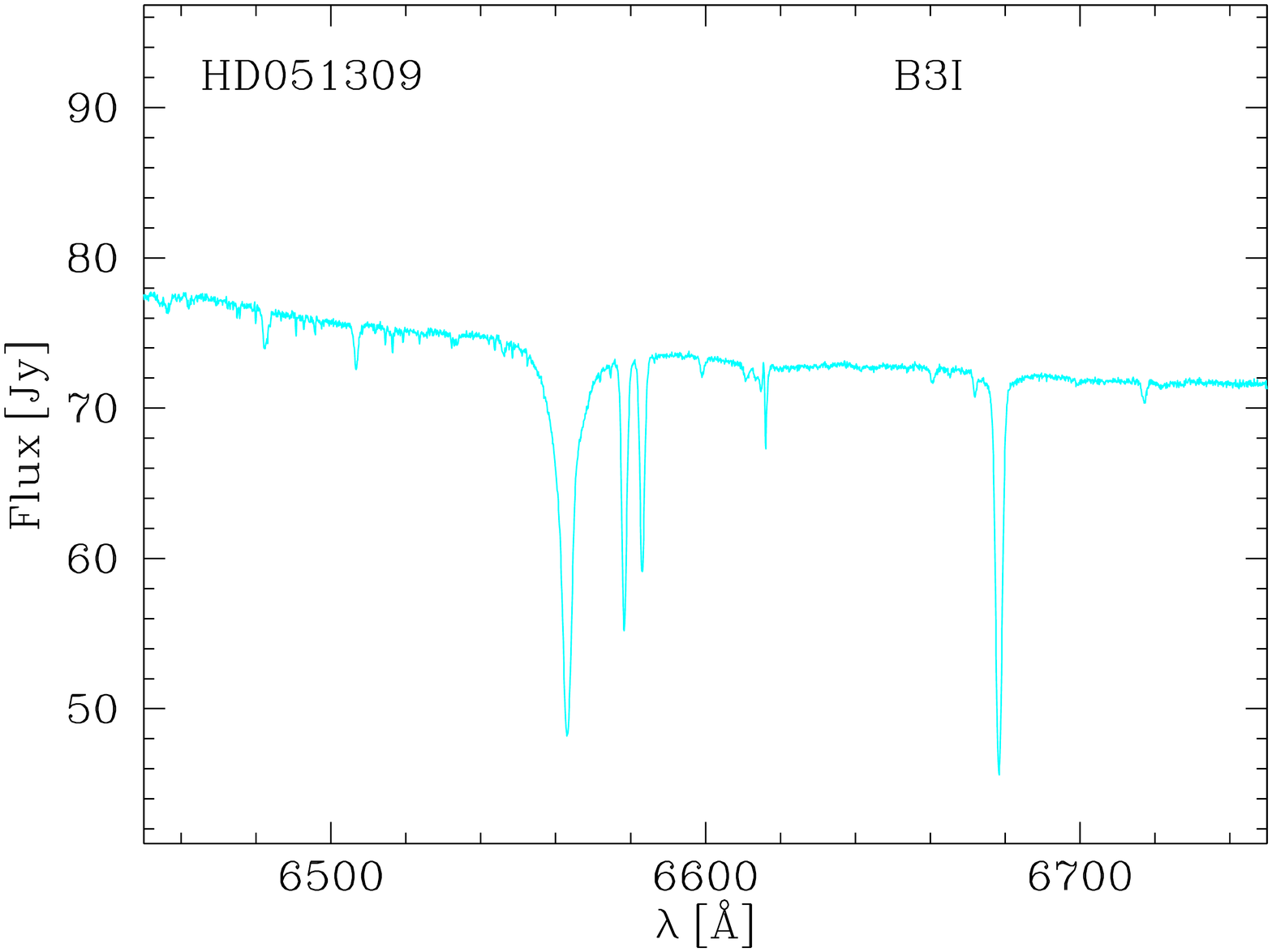}
\includegraphics[width=0.18\textwidth,angle=-90]{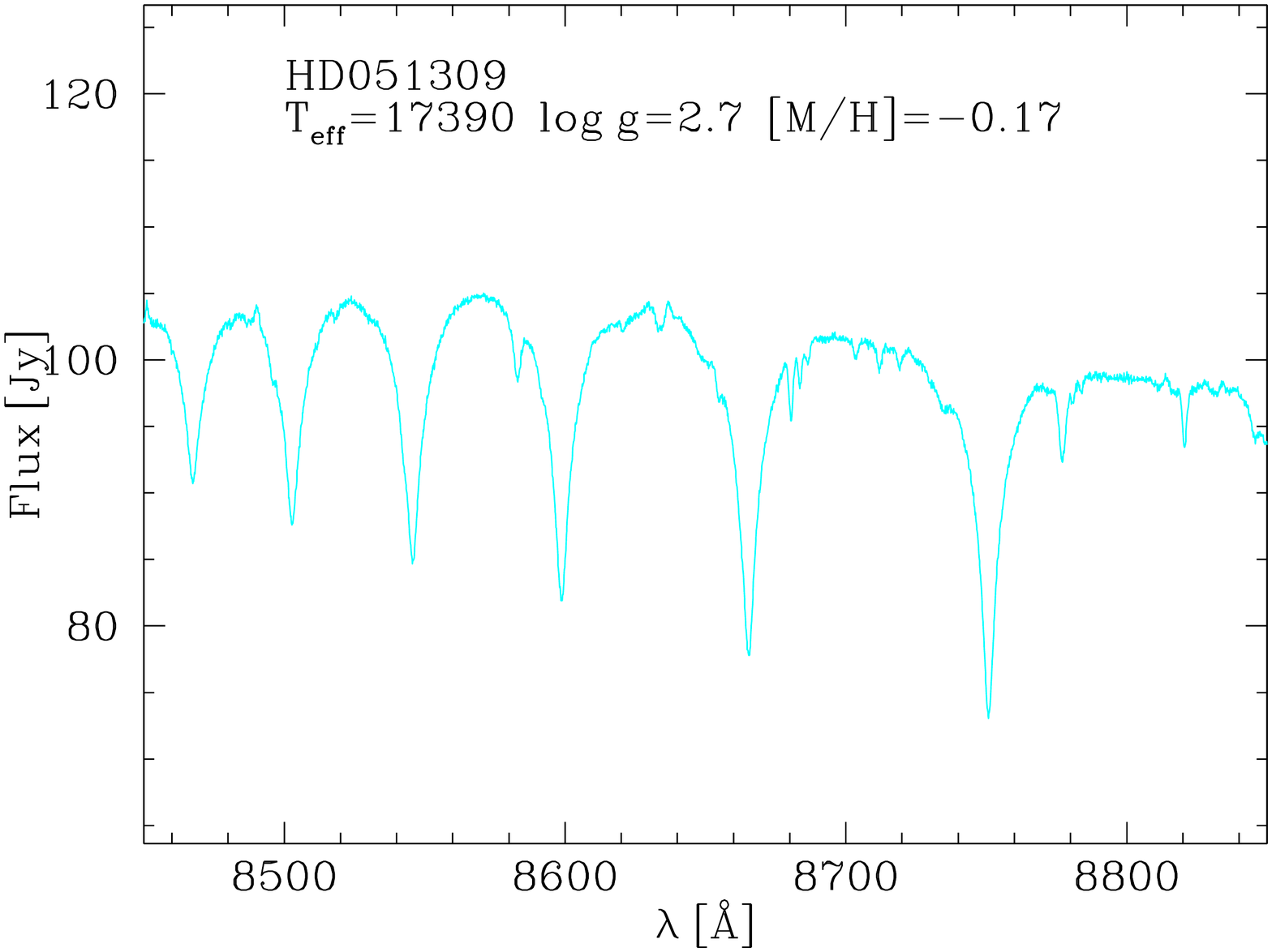}
\includegraphics[width=0.18\textwidth,angle=-90]{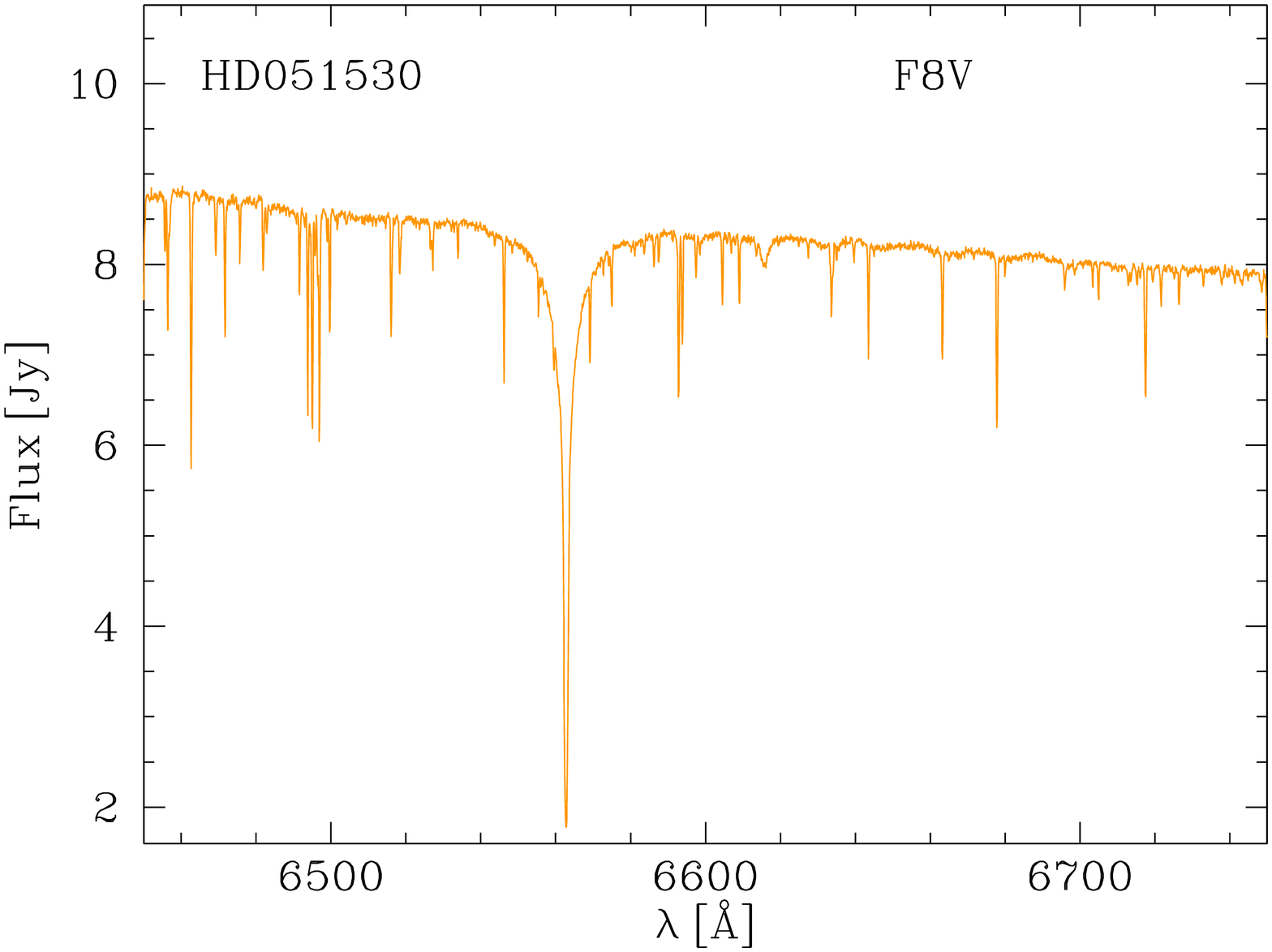}
\includegraphics[width=0.18\textwidth,angle=-90]{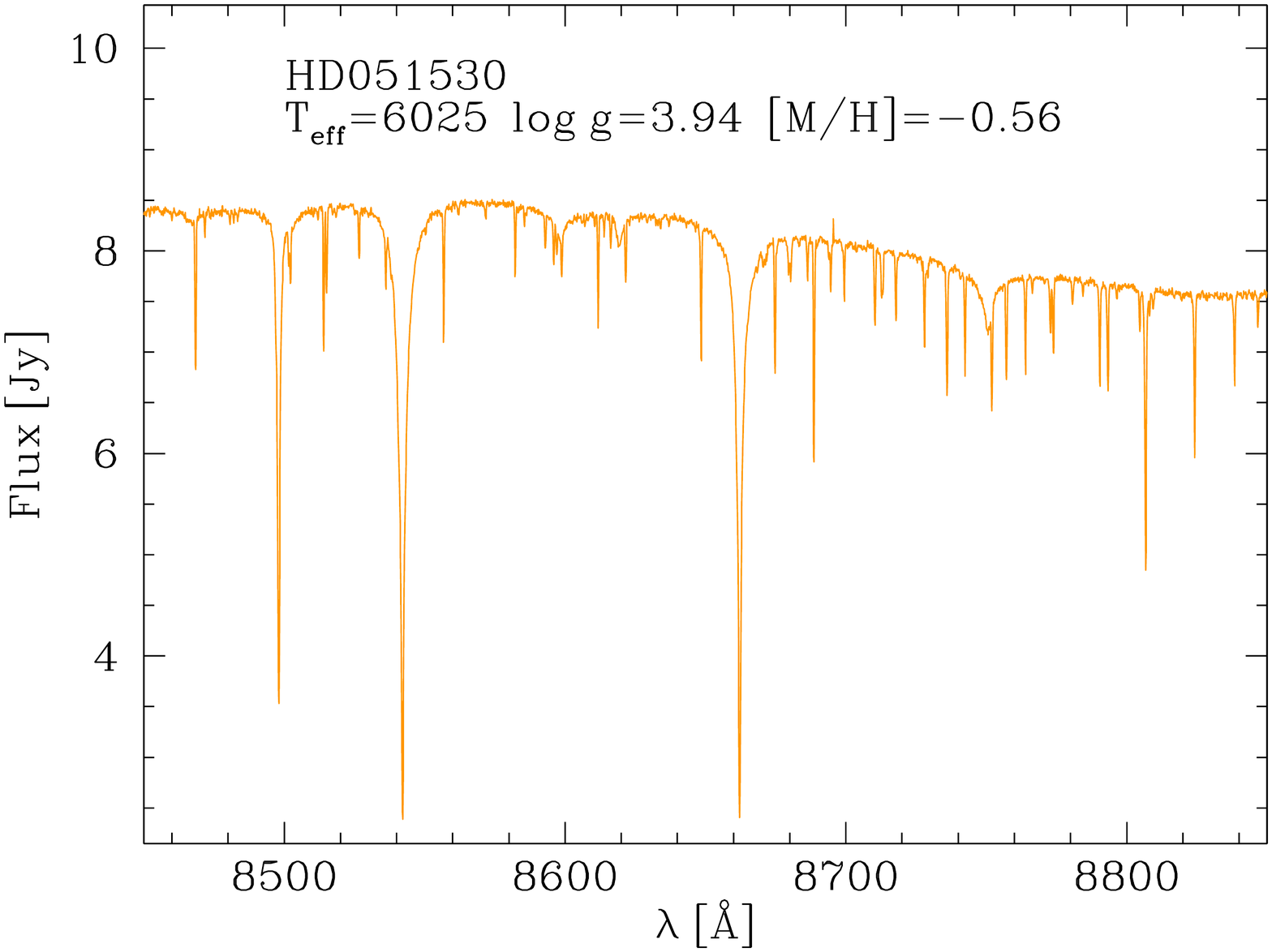}
\includegraphics[width=0.18\textwidth,angle=-90]{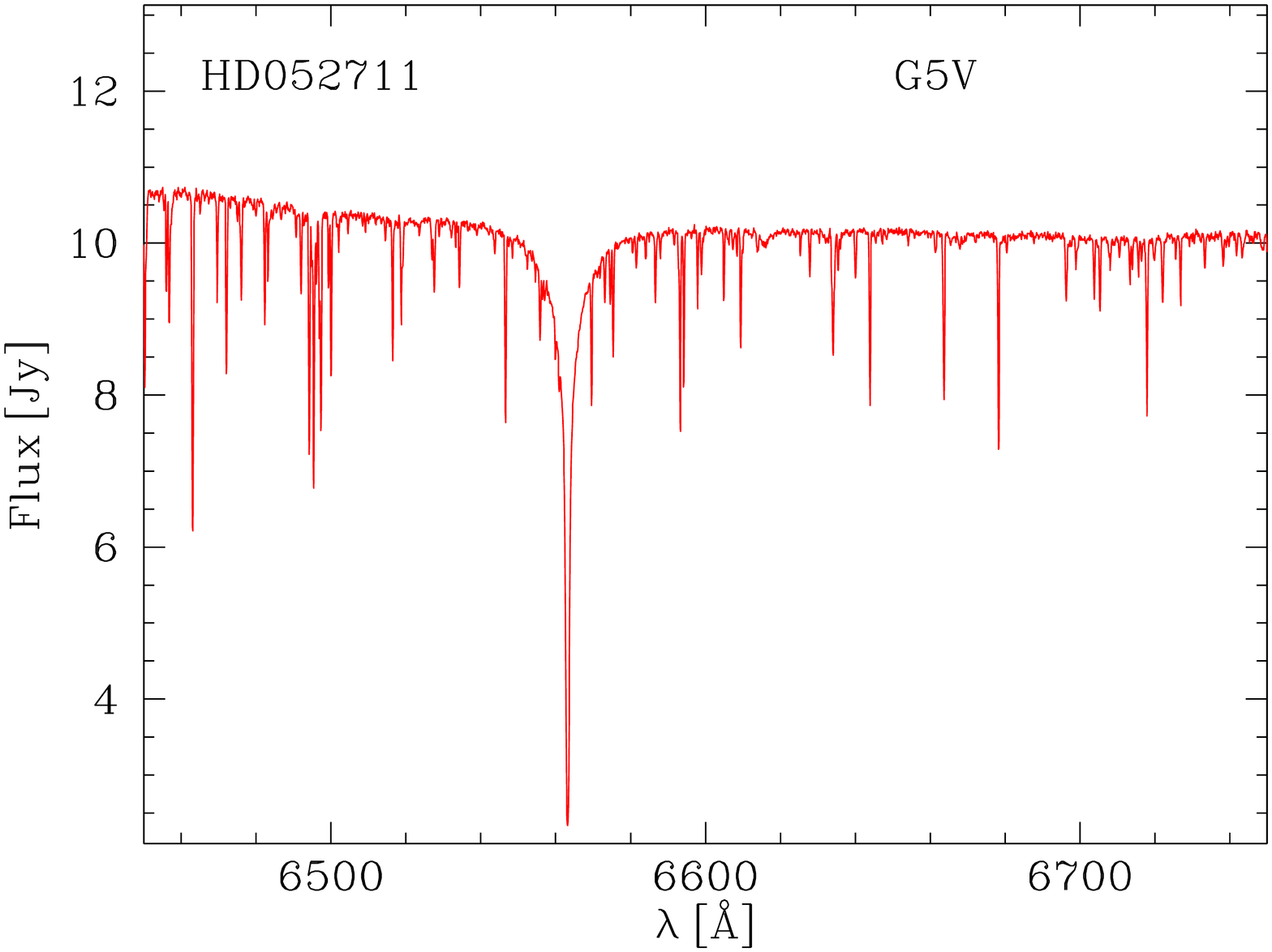}
\includegraphics[width=0.18\textwidth,angle=-90]{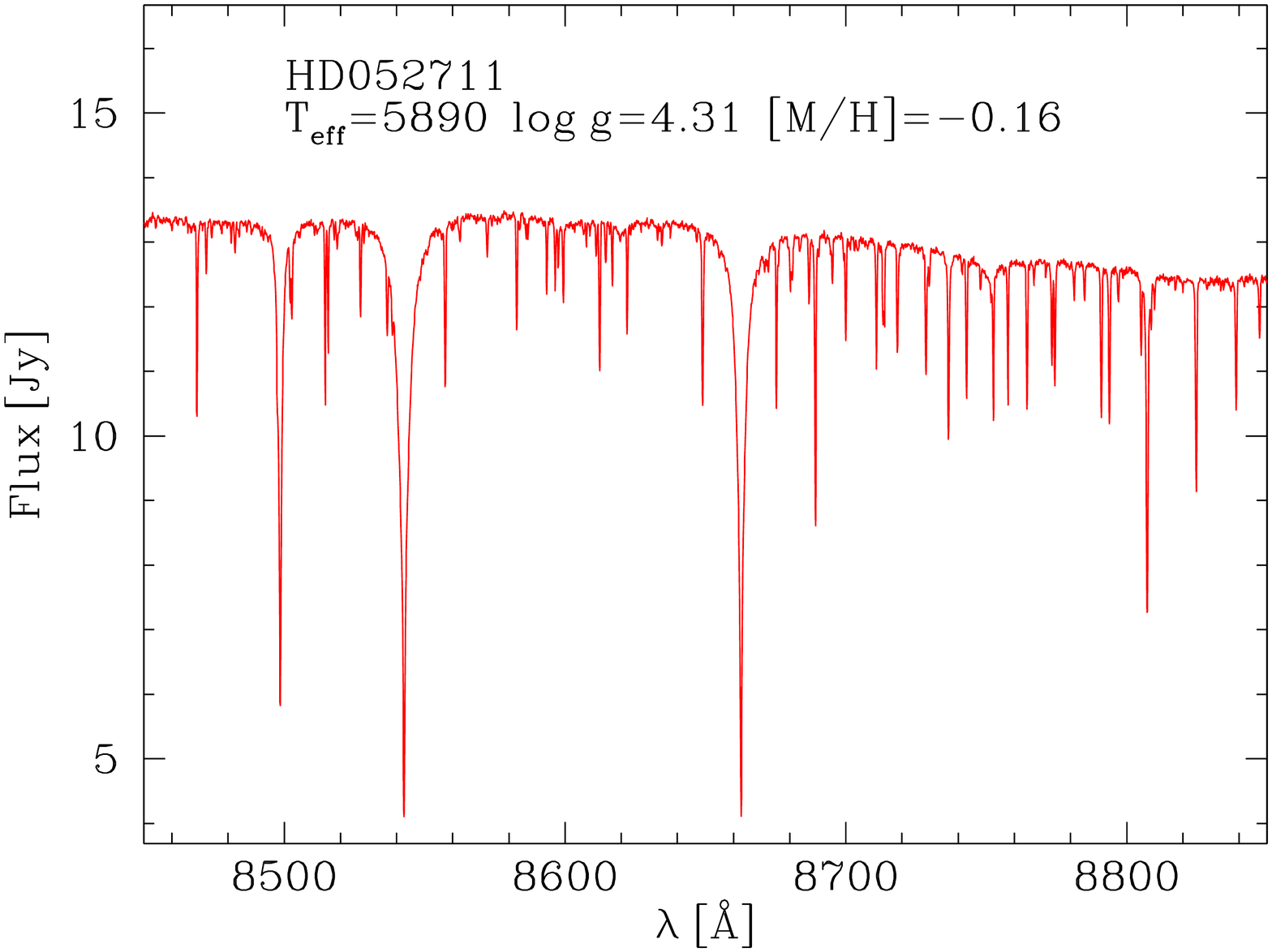}
\includegraphics[width=0.18\textwidth,angle=-90]{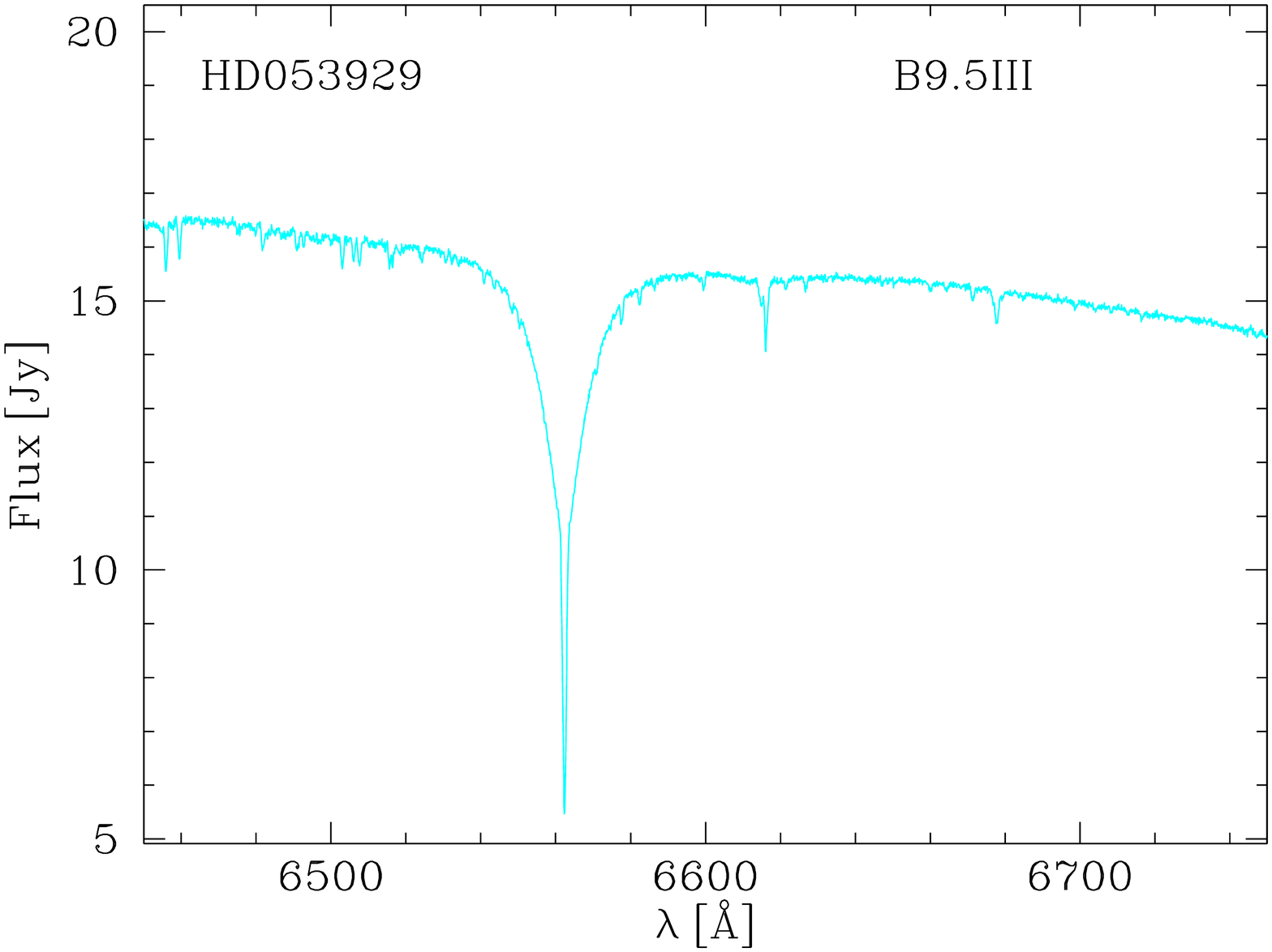}
\includegraphics[width=0.18\textwidth,angle=-90]{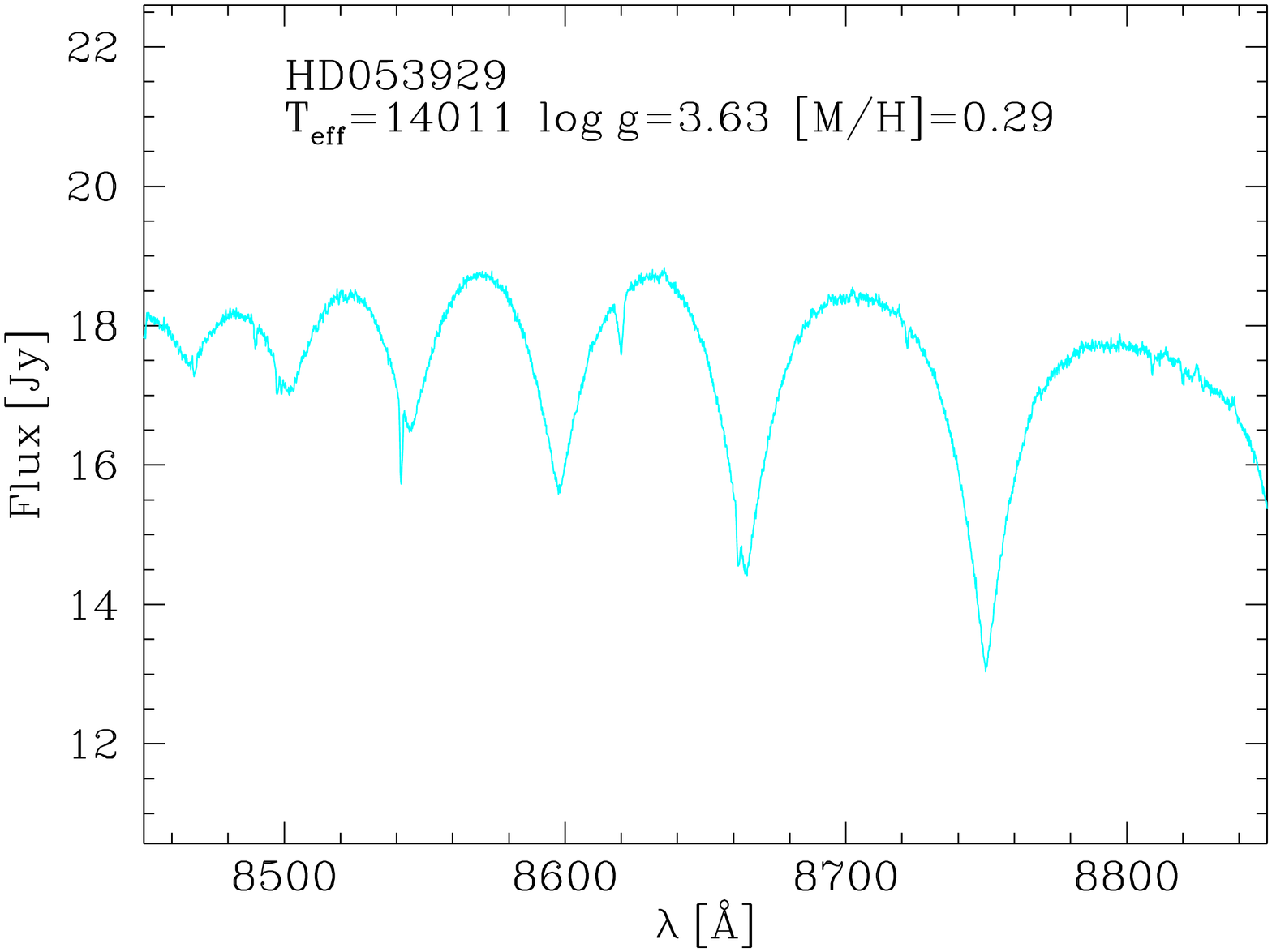}
\includegraphics[width=0.18\textwidth,angle=-90]{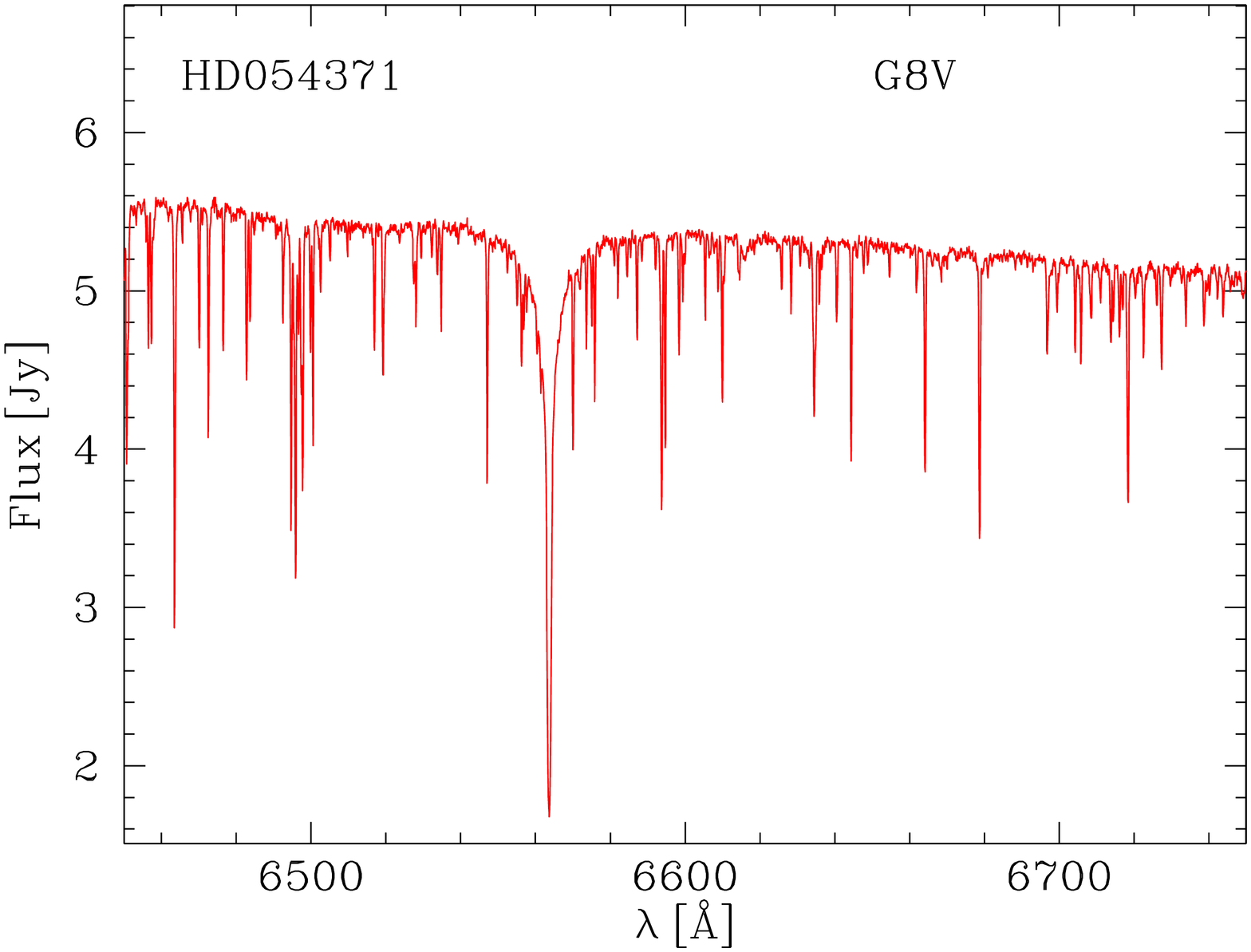}
\includegraphics[width=0.18\textwidth,angle=-90]{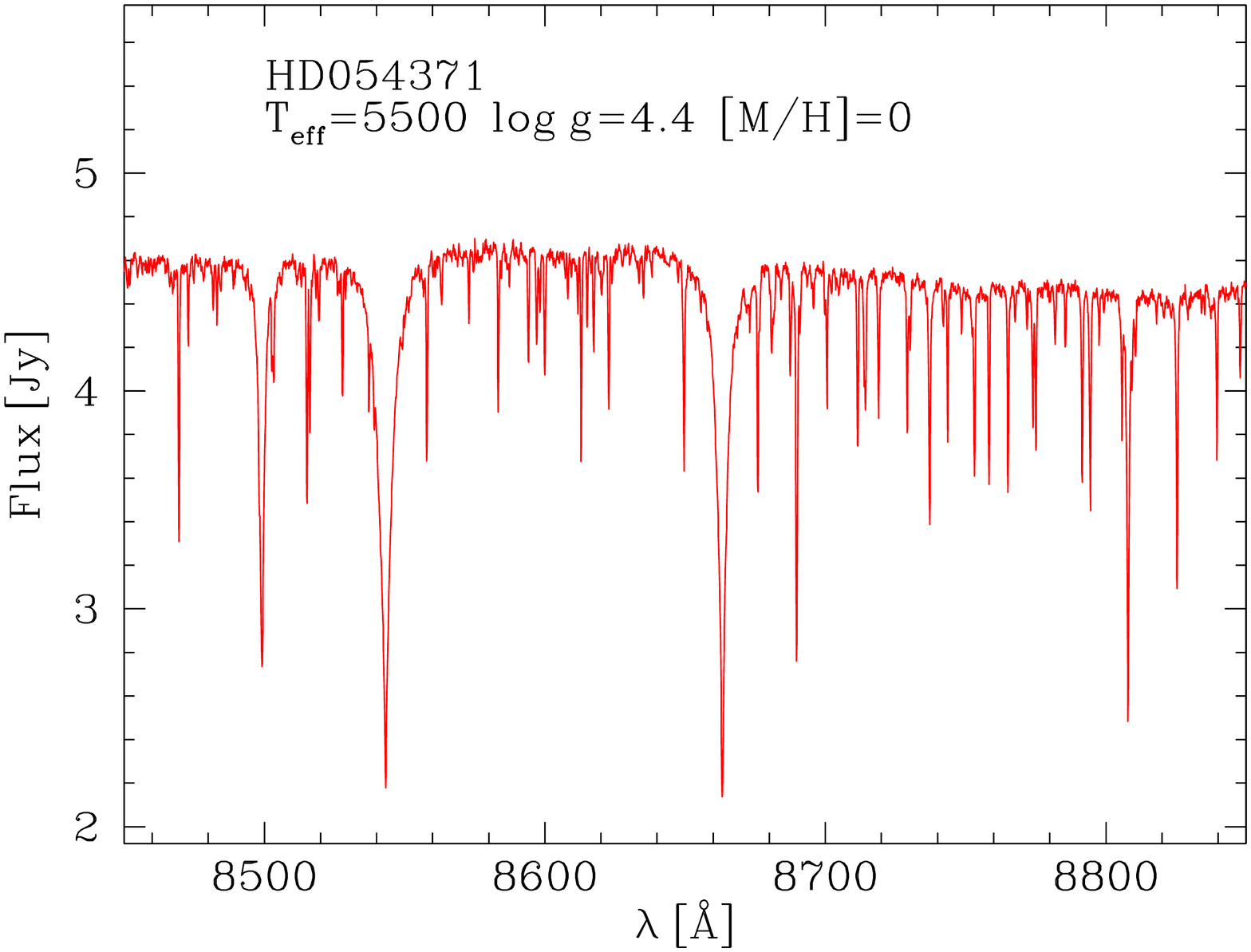}

\contcaption{12. Stars shown in this page are:  HD047839, HD048279, HD048682, HD049330, HD049409, HD049732, HD050522, HD050696, HD051219, HD051309, HD051530, HD052711, HD053929 and HD054371.}
\end{figure*}

\begin{figure*}
\includegraphics[width=0.18\textwidth,angle=-90]{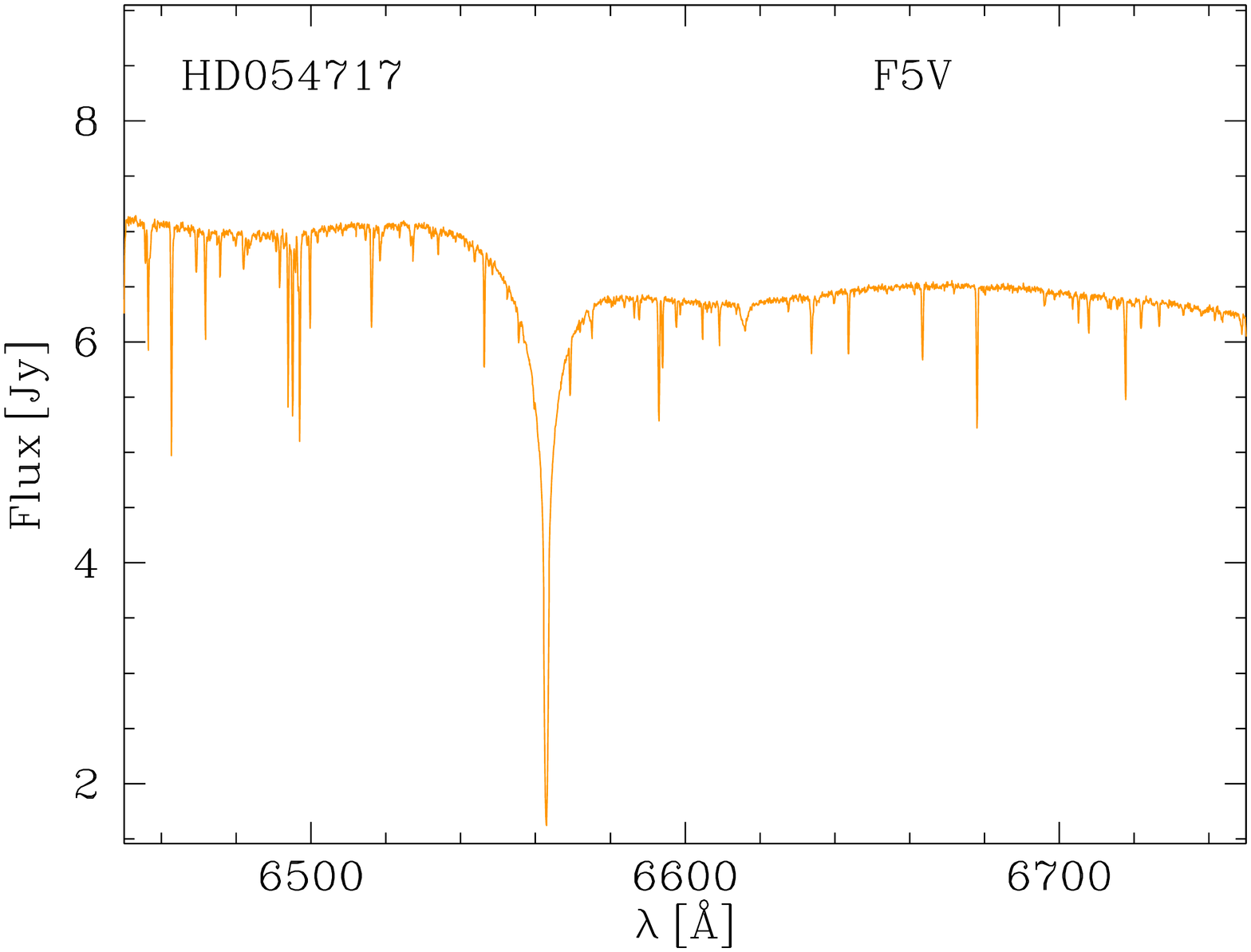}
\includegraphics[width=0.18\textwidth,angle=-90]{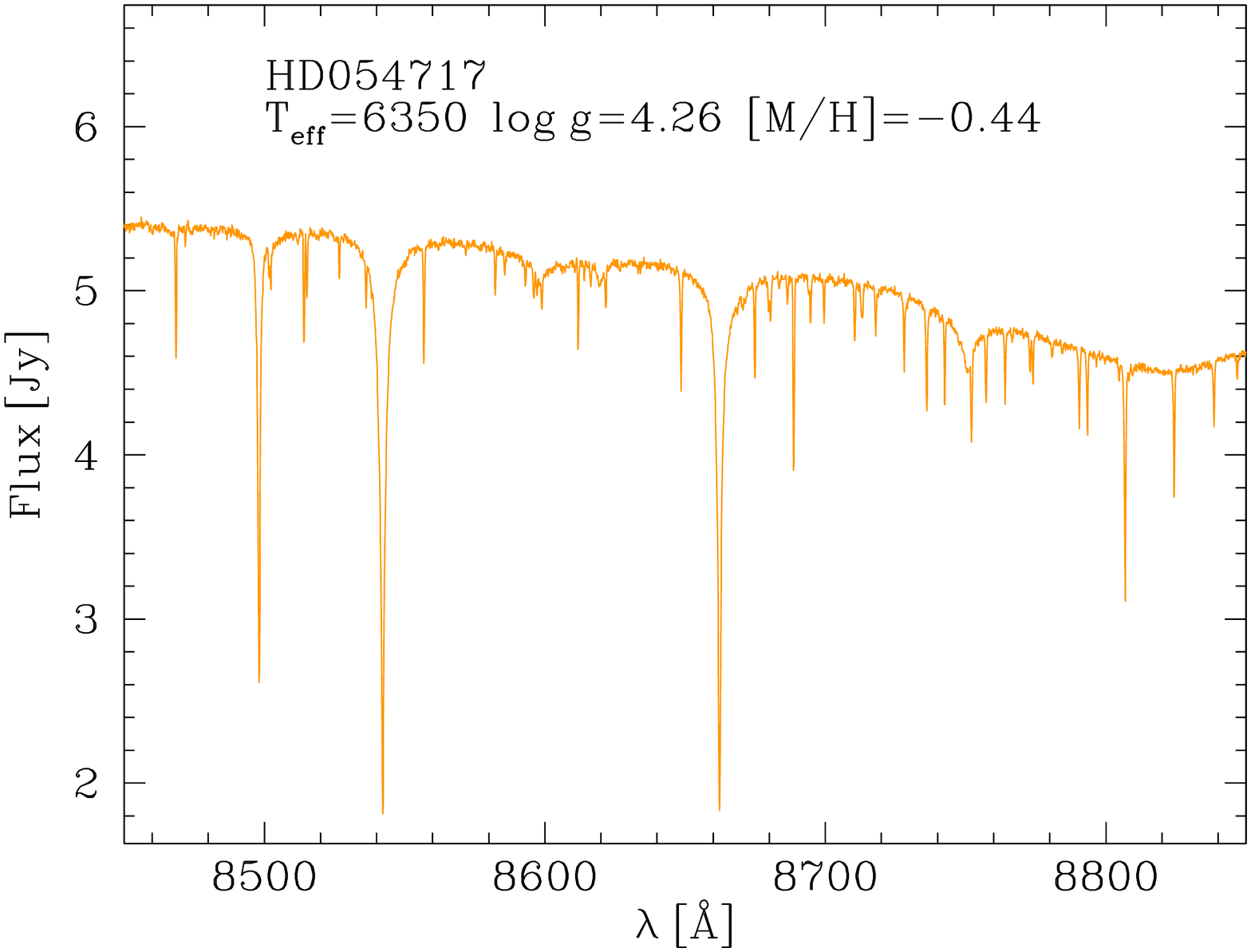}
\includegraphics[width=0.18\textwidth,angle=-90]{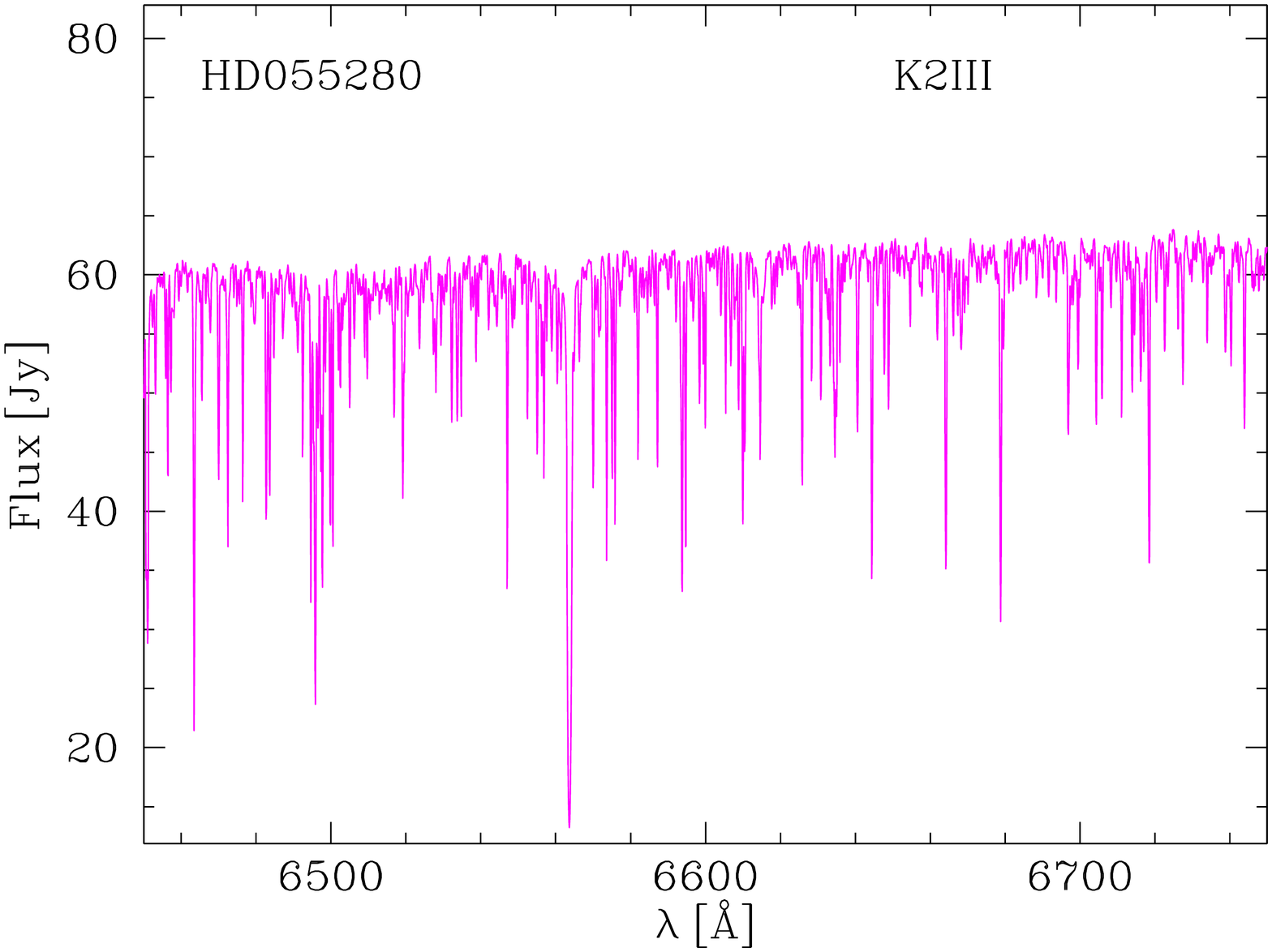}
\includegraphics[width=0.18\textwidth,angle=-90]{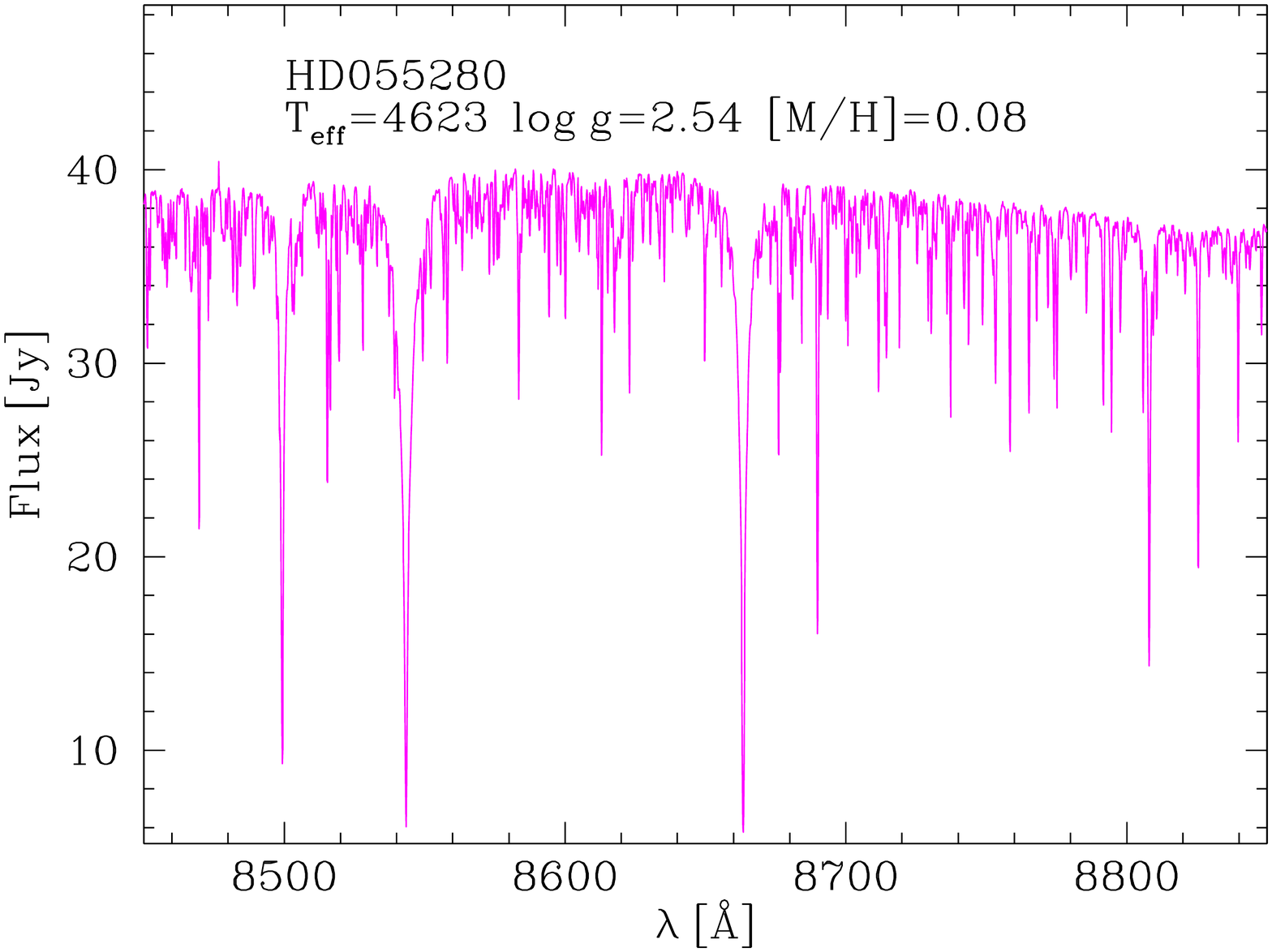}
\includegraphics[width=0.18\textwidth,angle=-90]{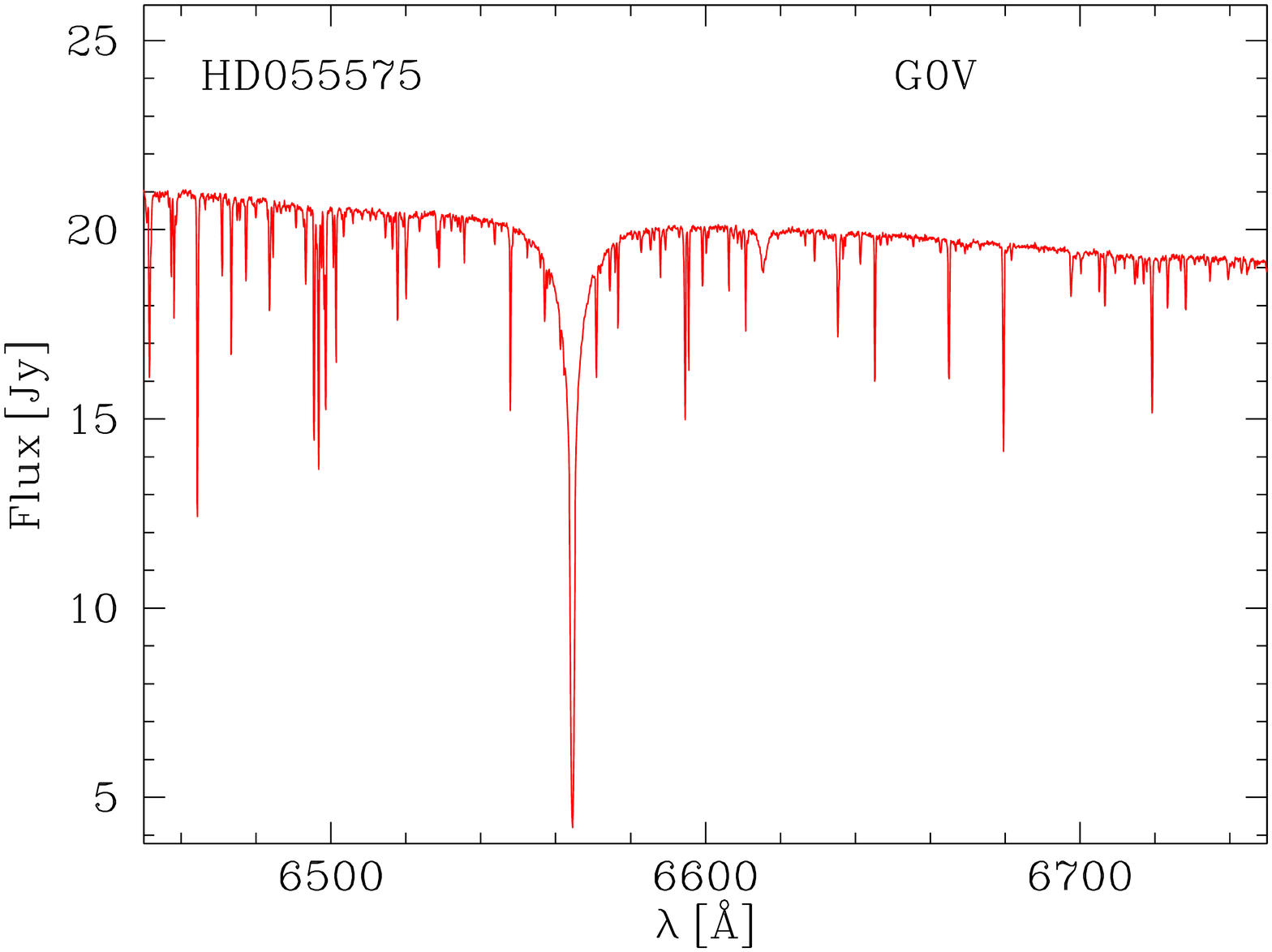}
\includegraphics[width=0.18\textwidth,angle=-90]{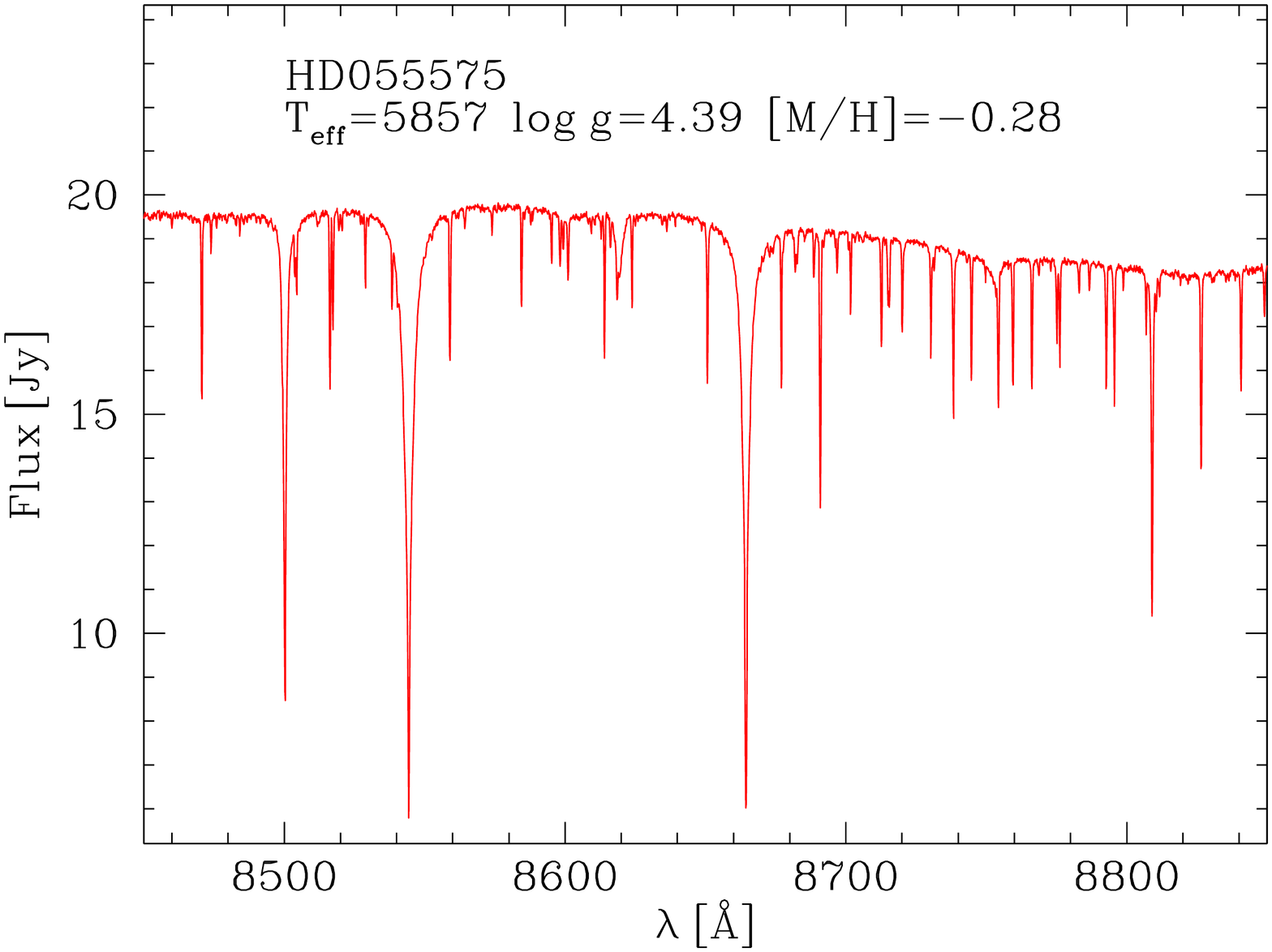}
\includegraphics[width=0.18\textwidth,angle=-90]{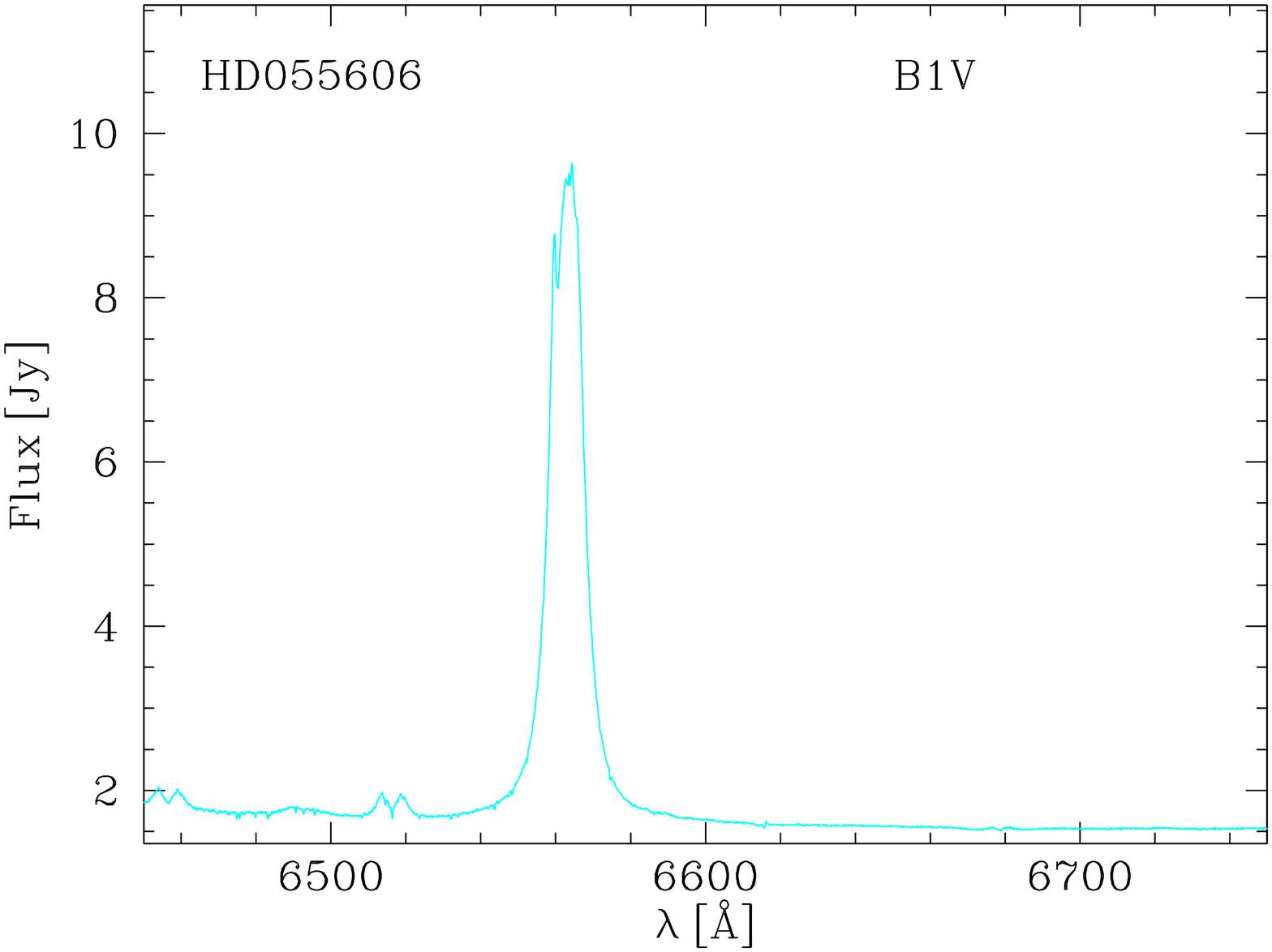}
\includegraphics[width=0.18\textwidth,angle=-90]{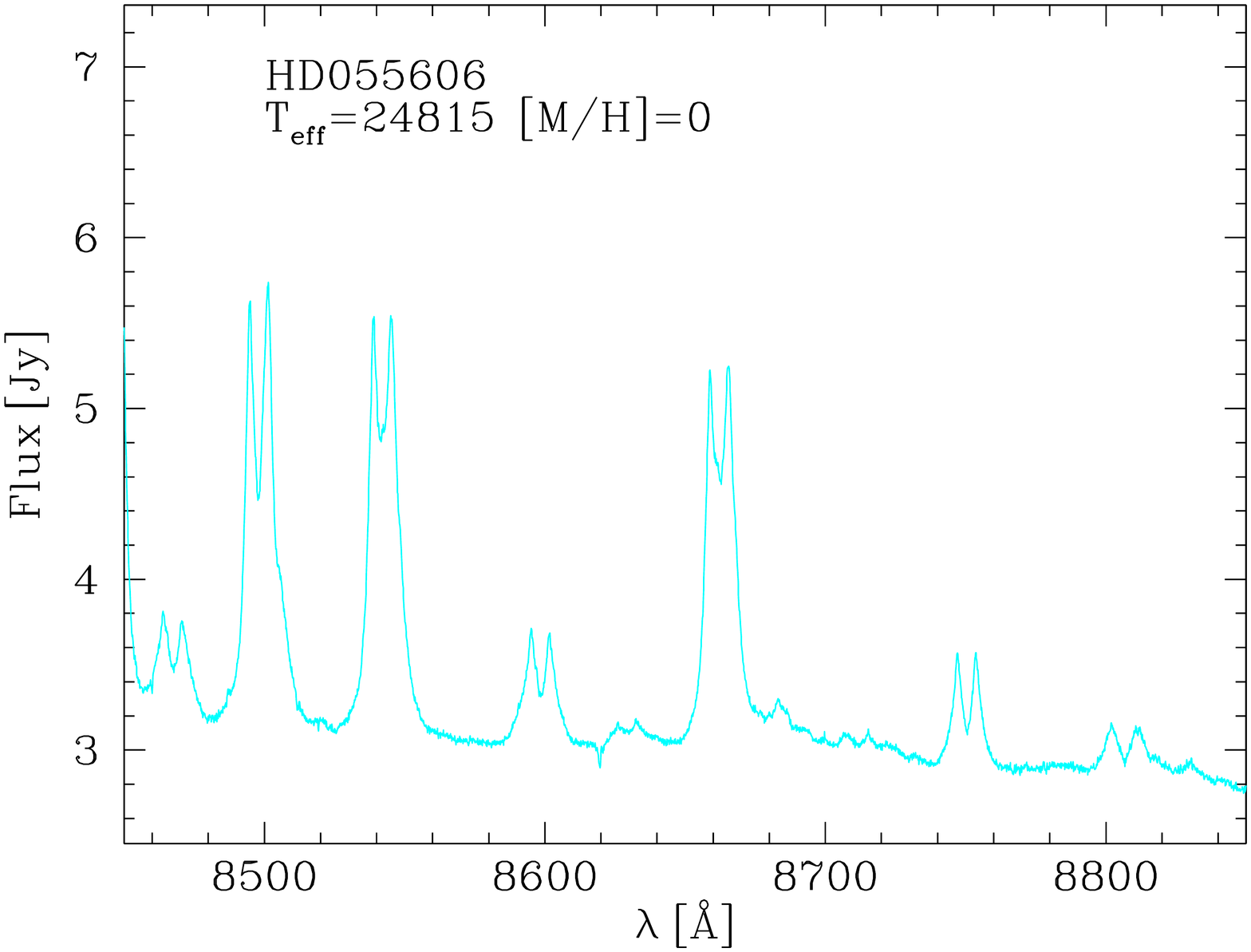}
\includegraphics[width=0.18\textwidth,angle=-90]{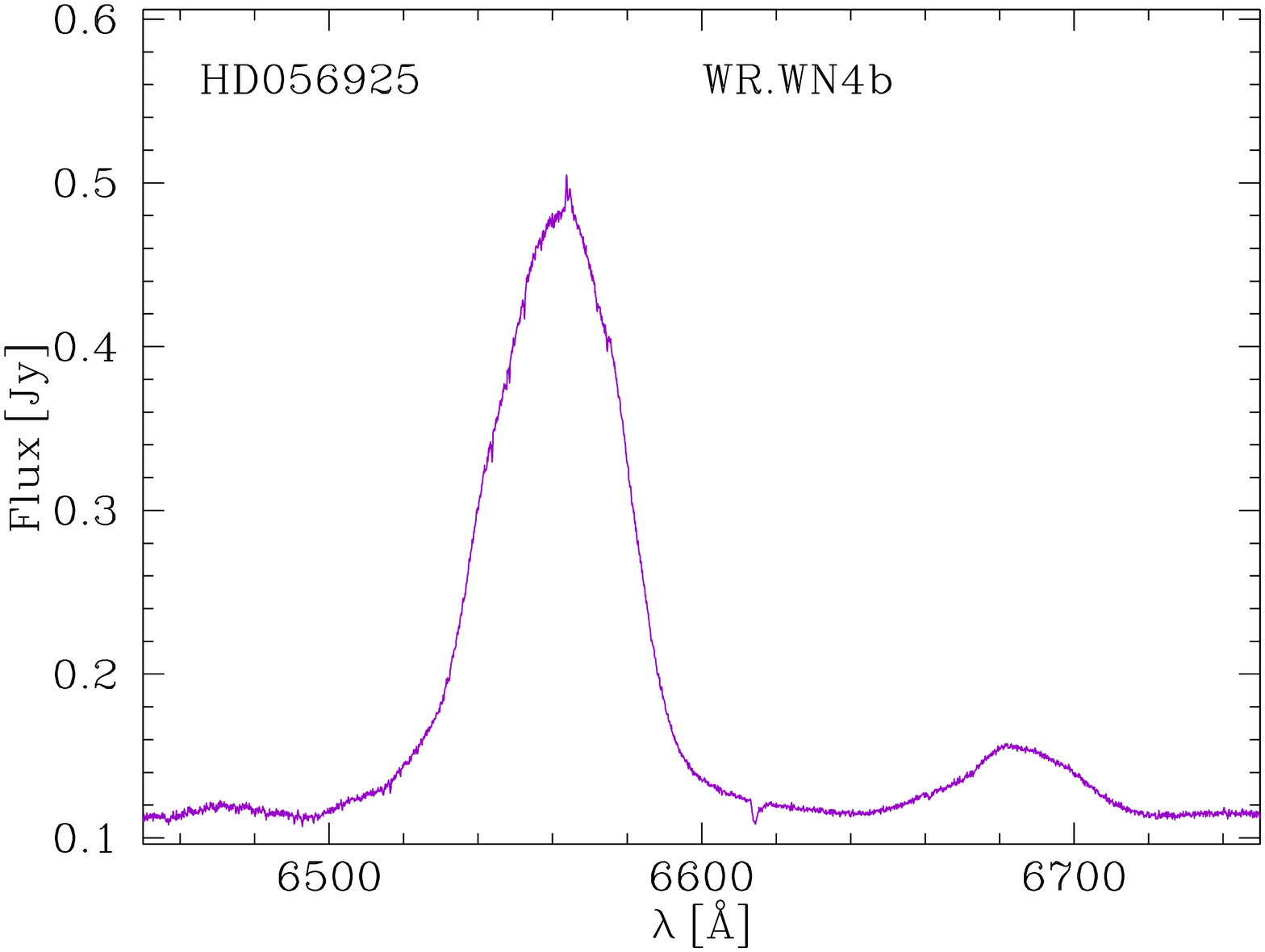}
\includegraphics[width=0.18\textwidth,angle=-90]{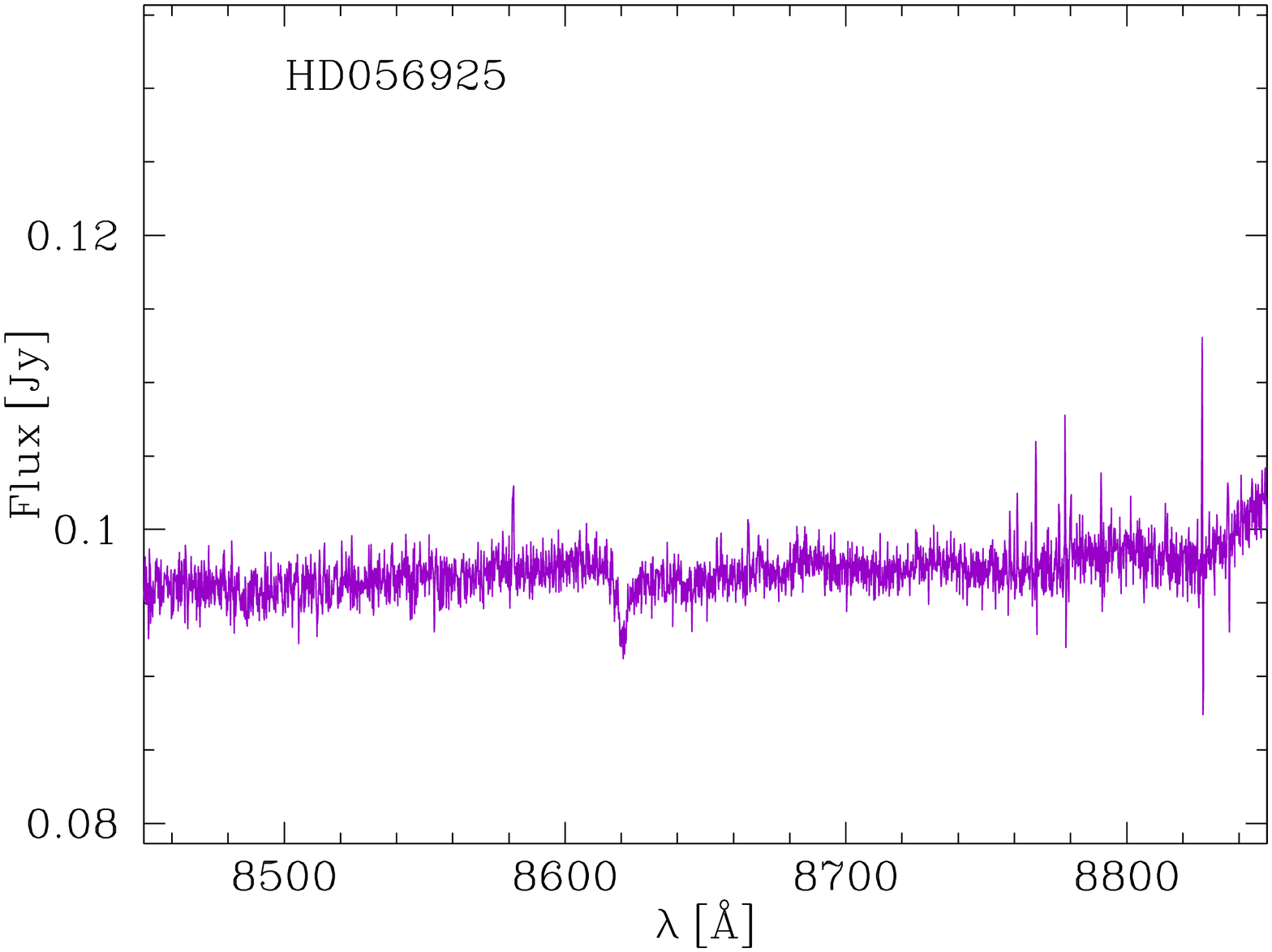}
\includegraphics[width=0.18\textwidth,angle=-90]{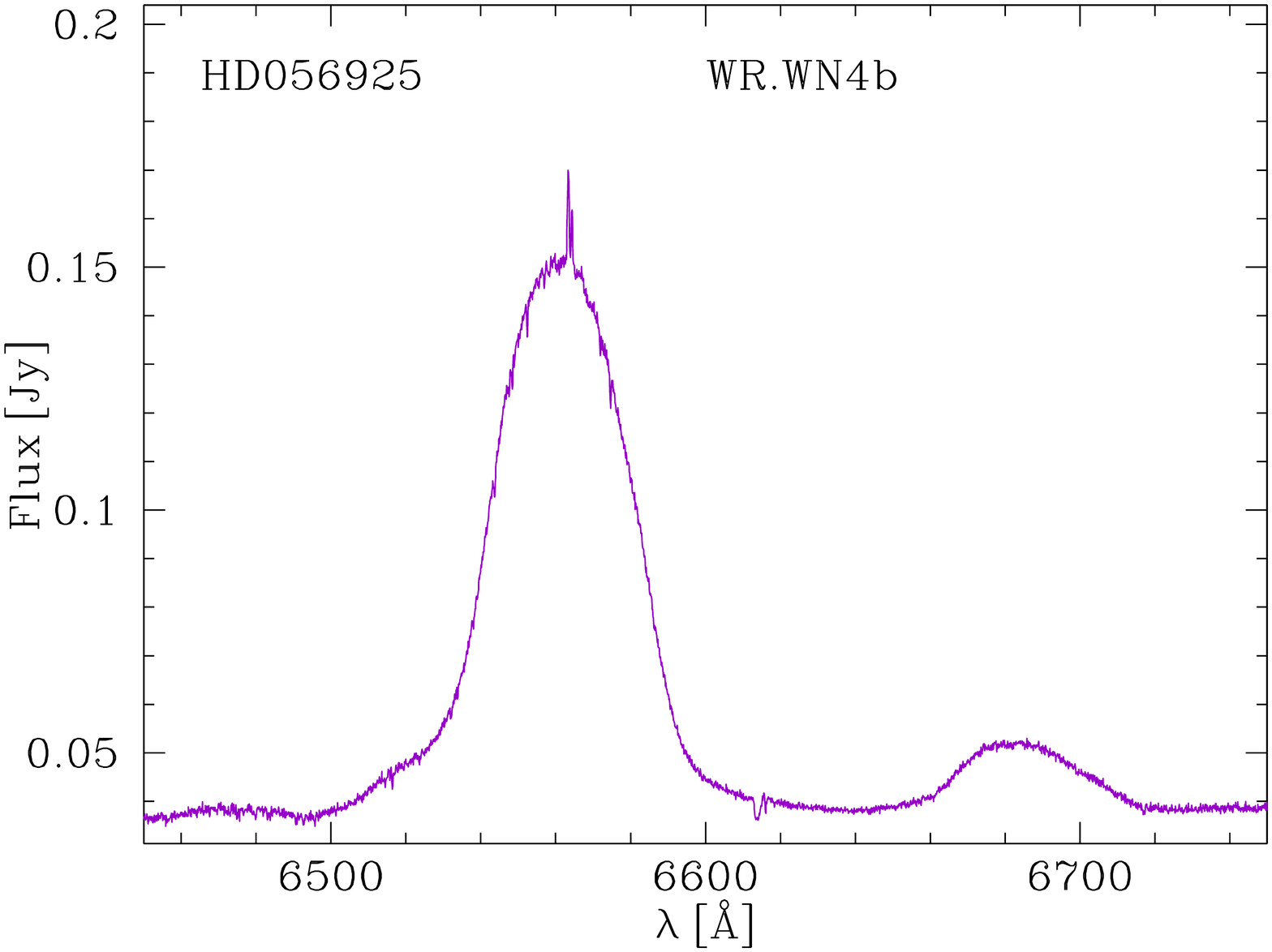}
\includegraphics[width=0.18\textwidth,angle=-90]{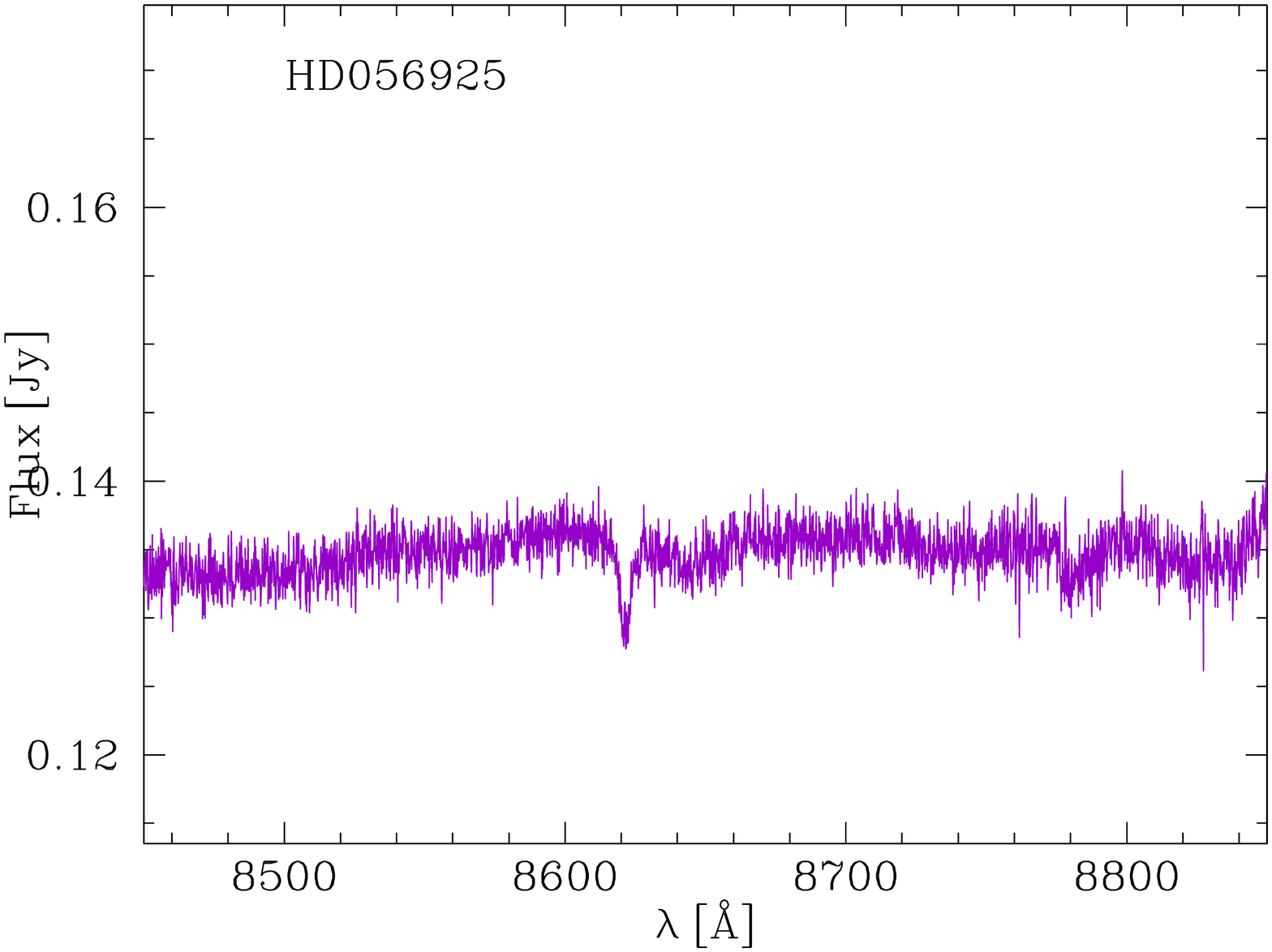}
\includegraphics[width=0.18\textwidth,angle=-90]{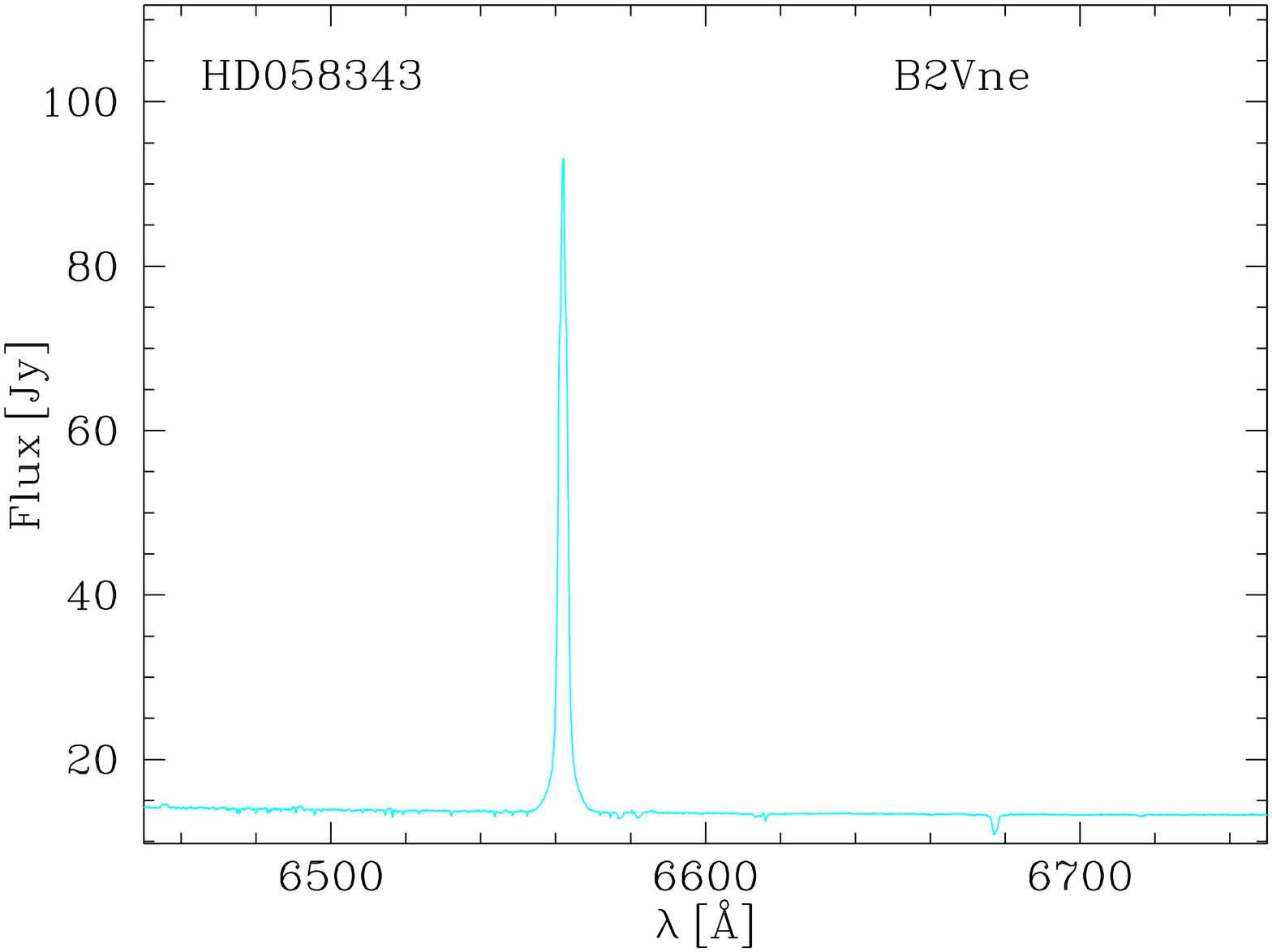}
\includegraphics[width=0.18\textwidth,angle=-90]{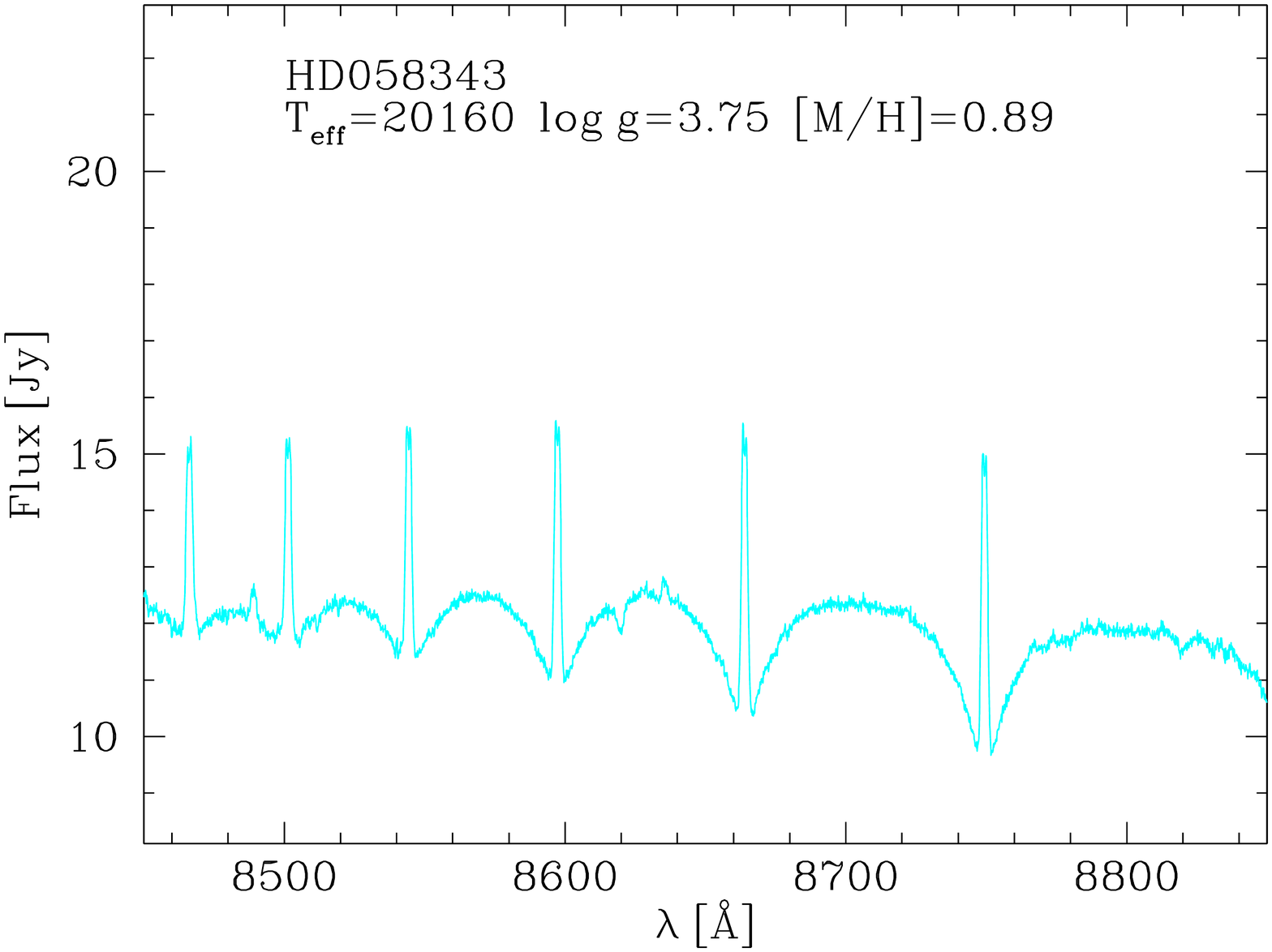}
\includegraphics[width=0.18\textwidth,angle=-90]{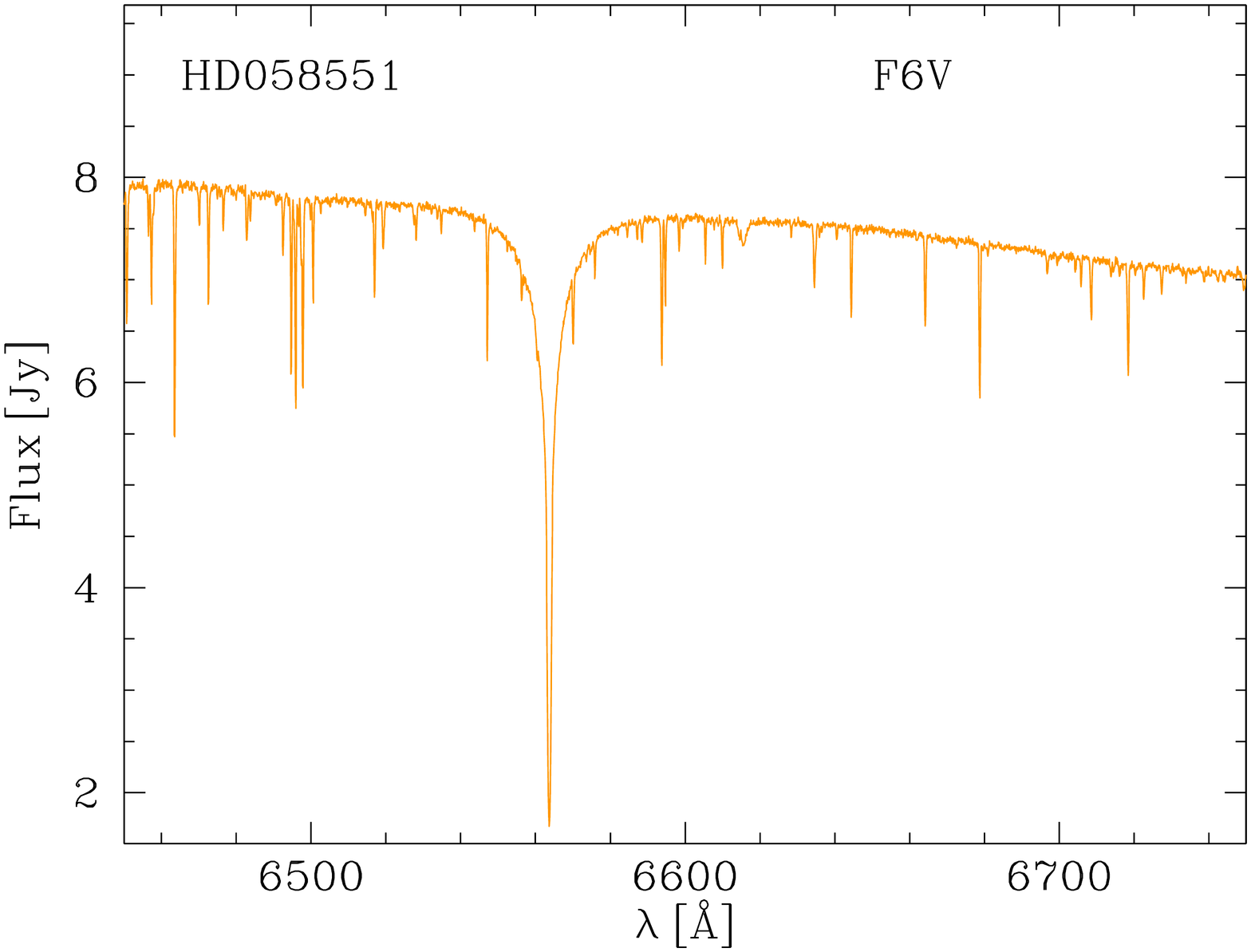}
\includegraphics[width=0.18\textwidth,angle=-90]{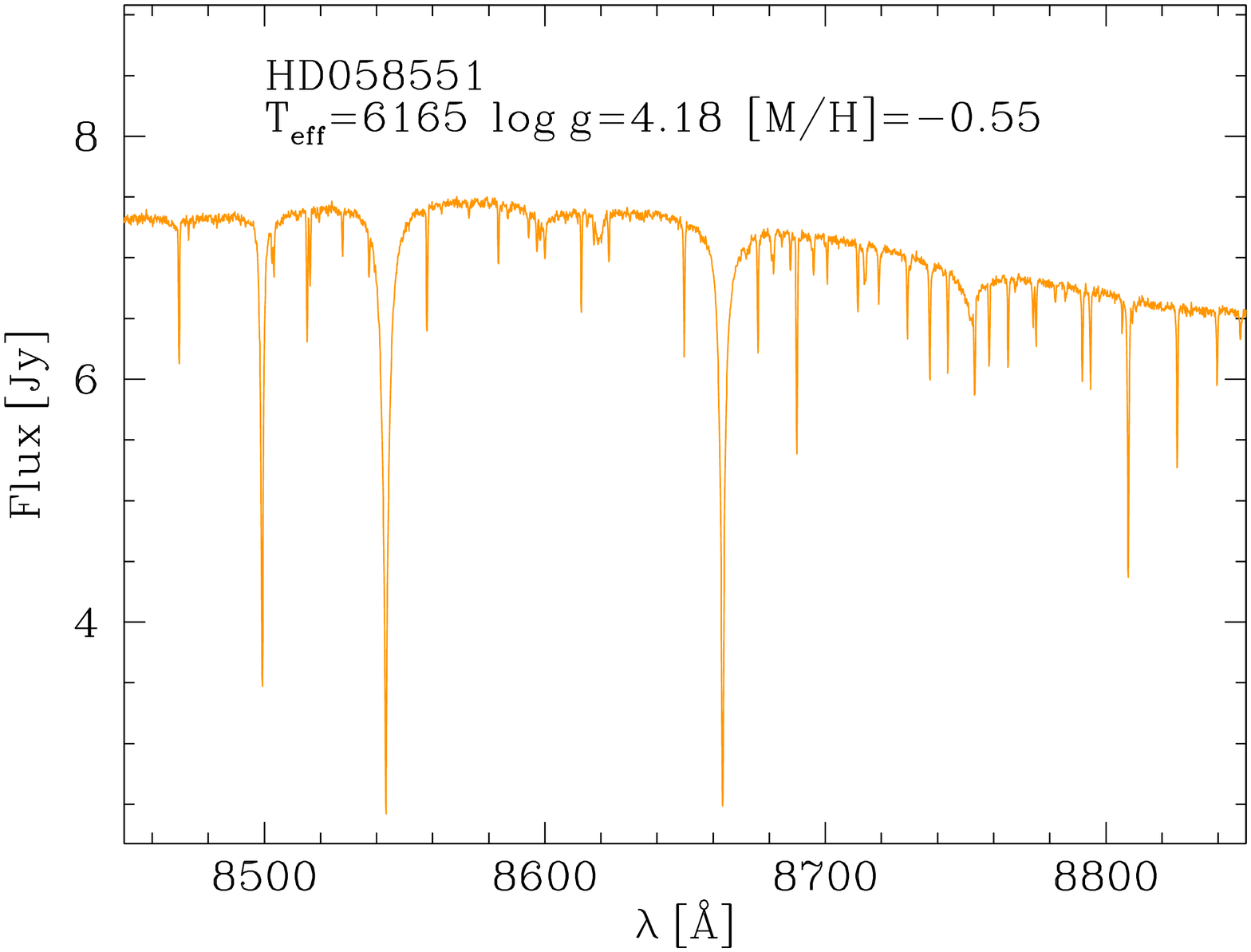}
\includegraphics[width=0.18\textwidth,angle=-90]{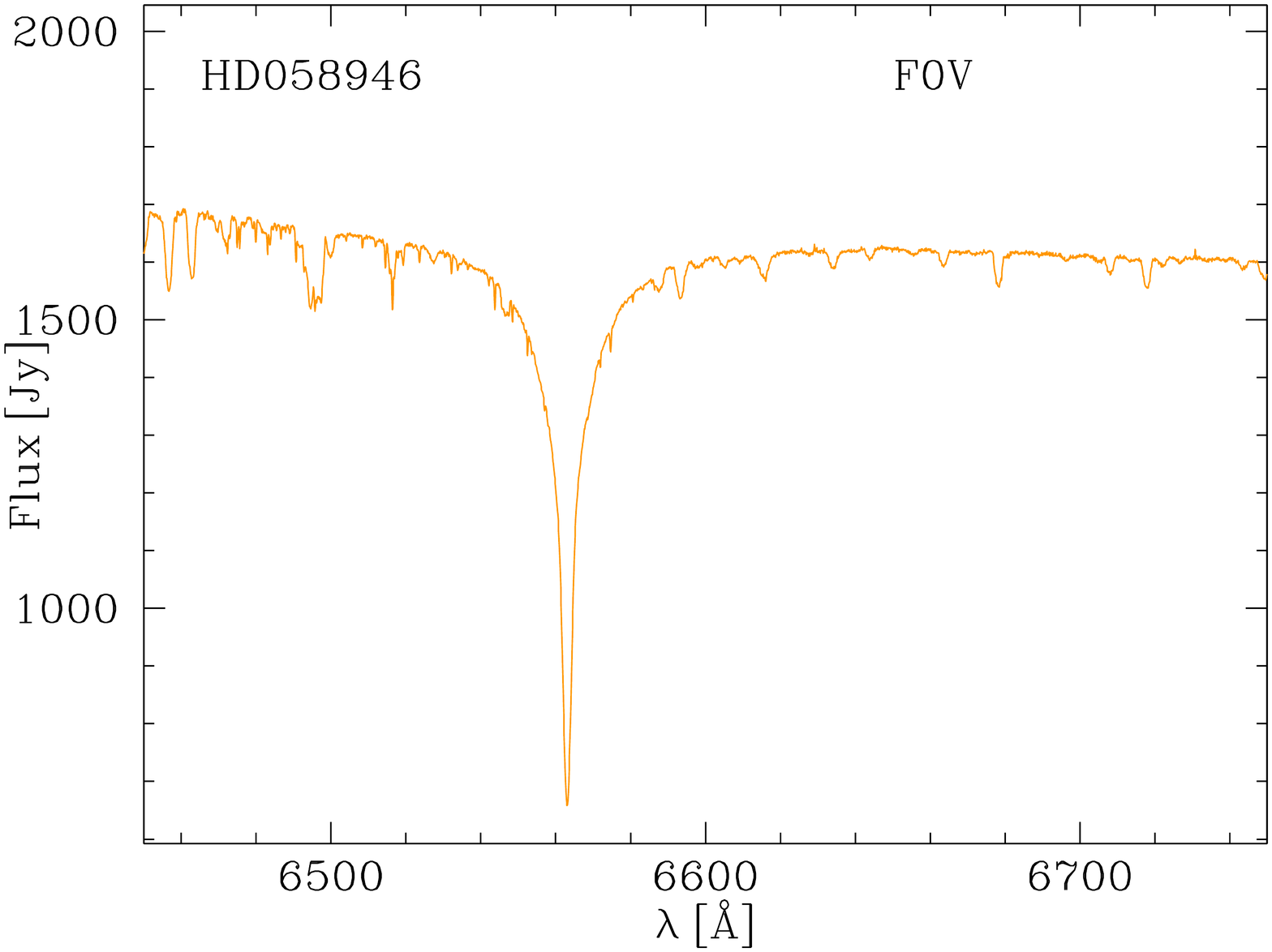}
\includegraphics[width=0.18\textwidth,angle=-90]{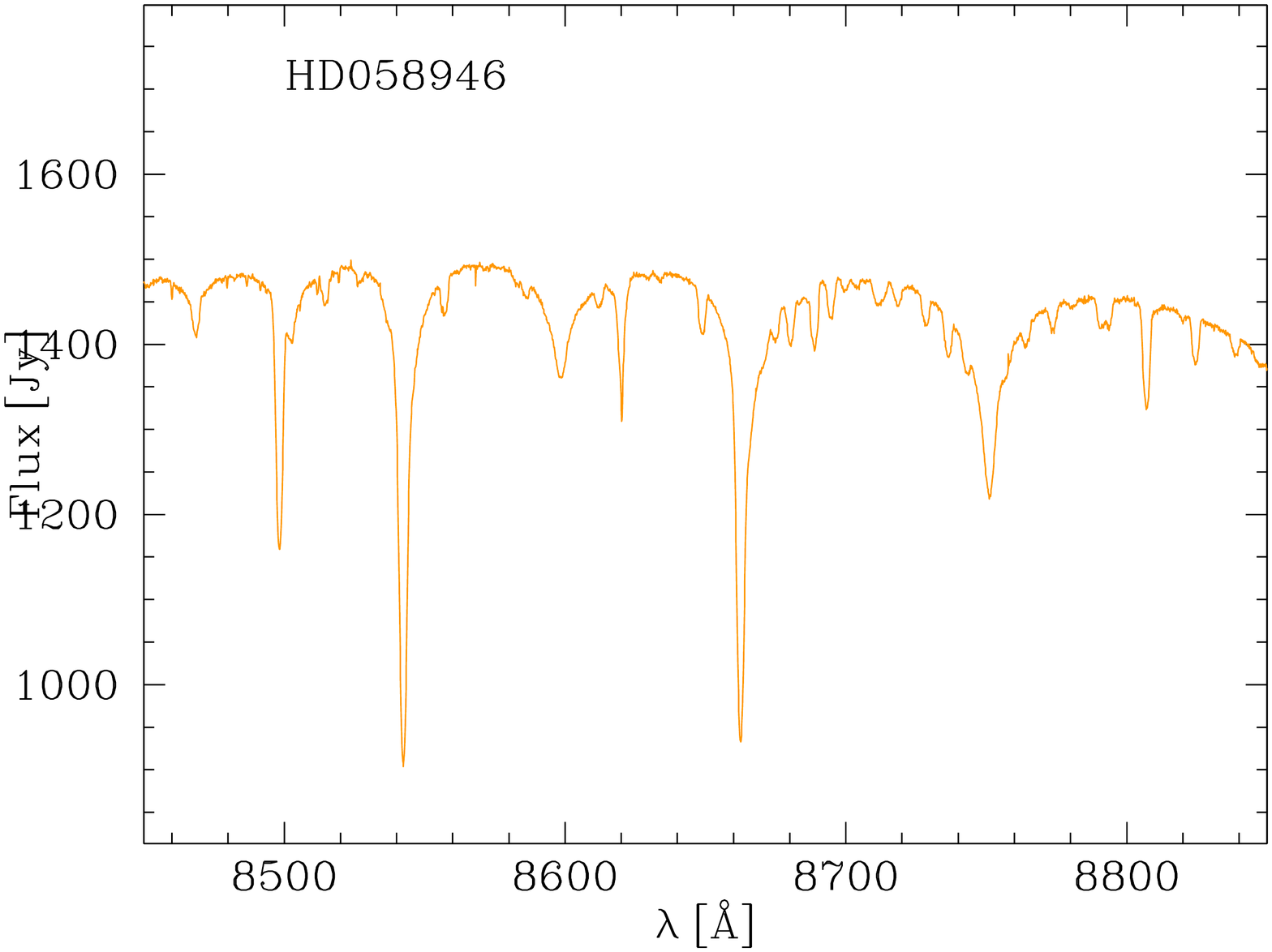}
\includegraphics[width=0.18\textwidth,angle=-90]{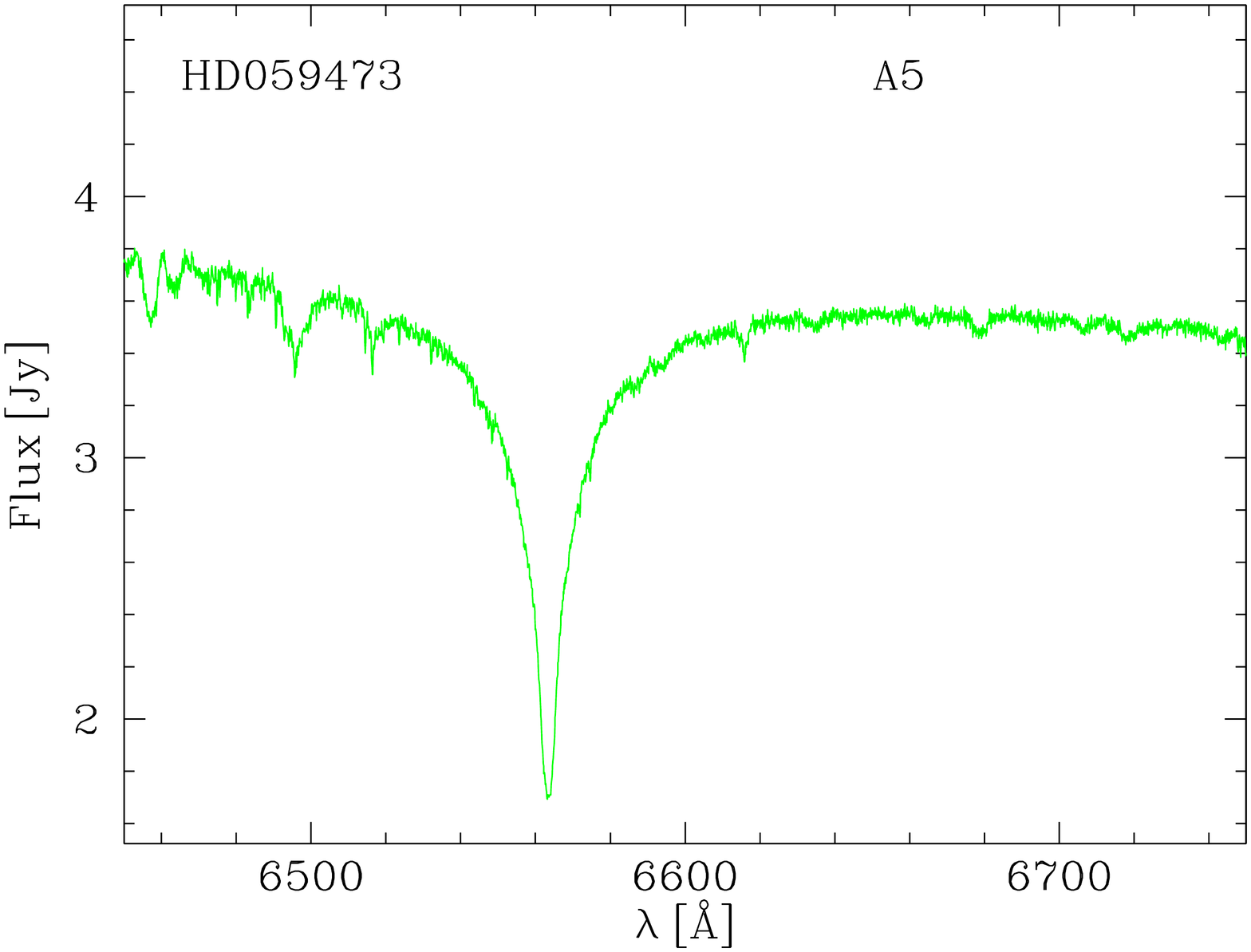}
\includegraphics[width=0.18\textwidth,angle=-90]{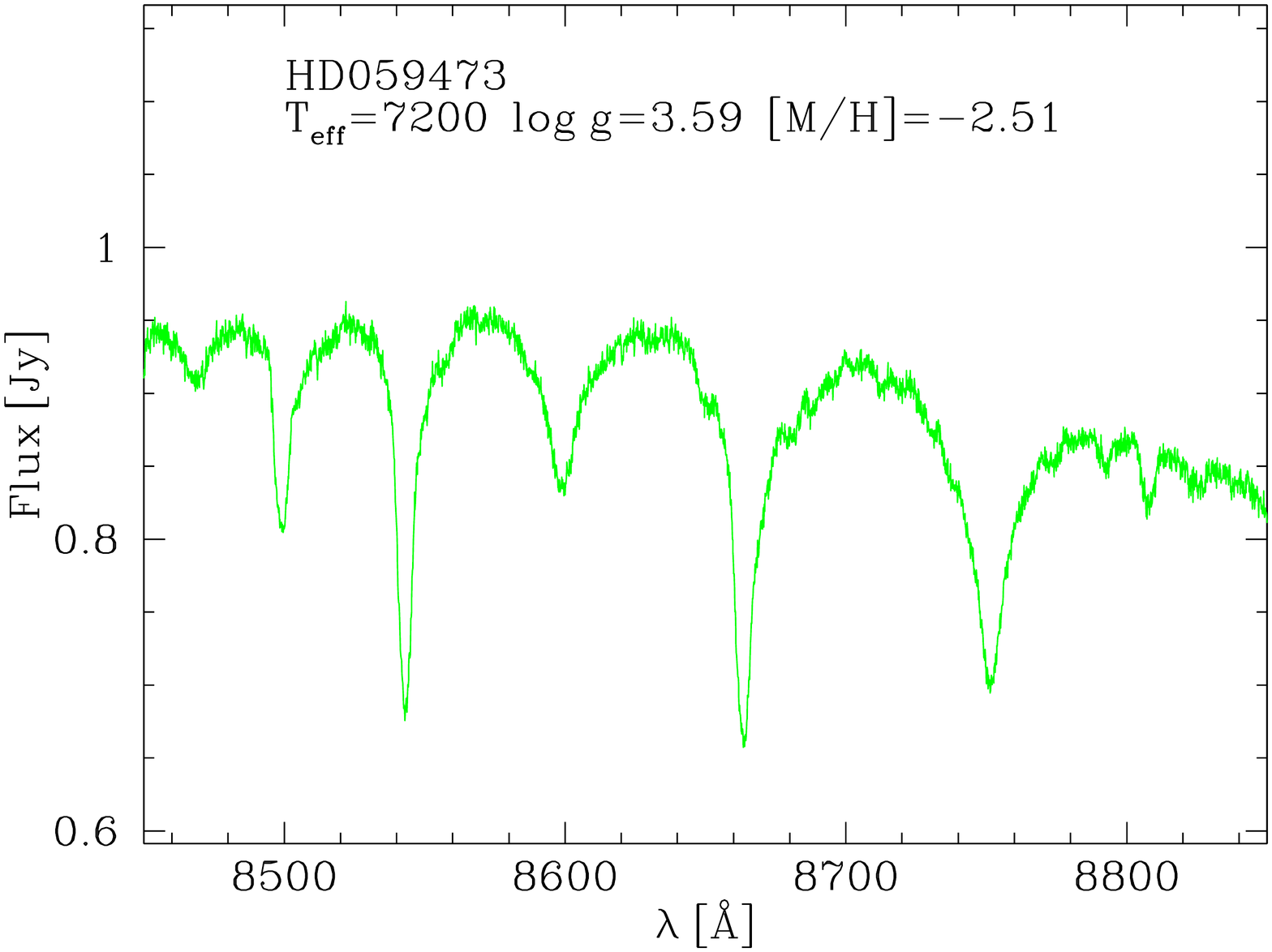}
\includegraphics[width=0.18\textwidth,angle=-90]{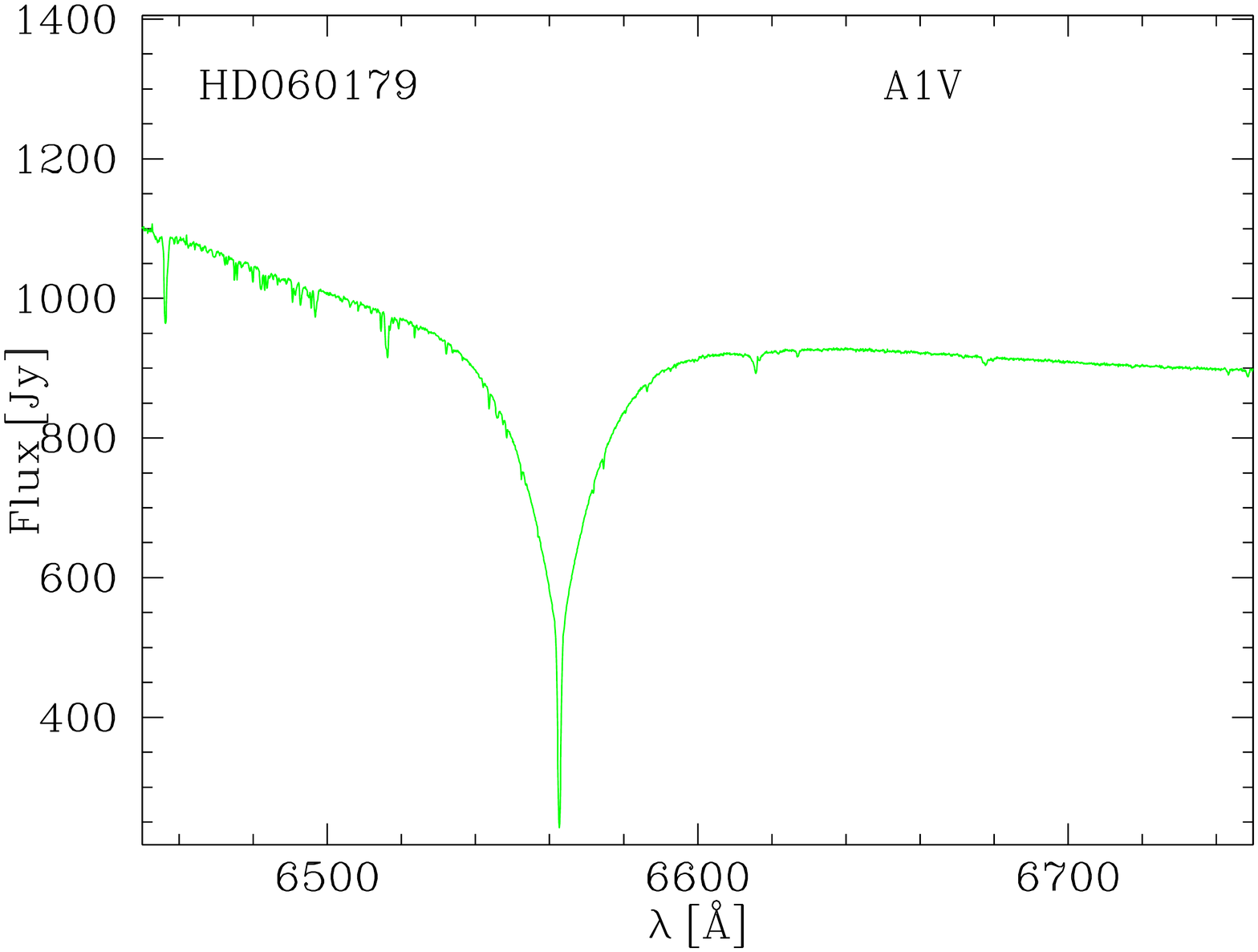}
\includegraphics[width=0.18\textwidth,angle=-90]{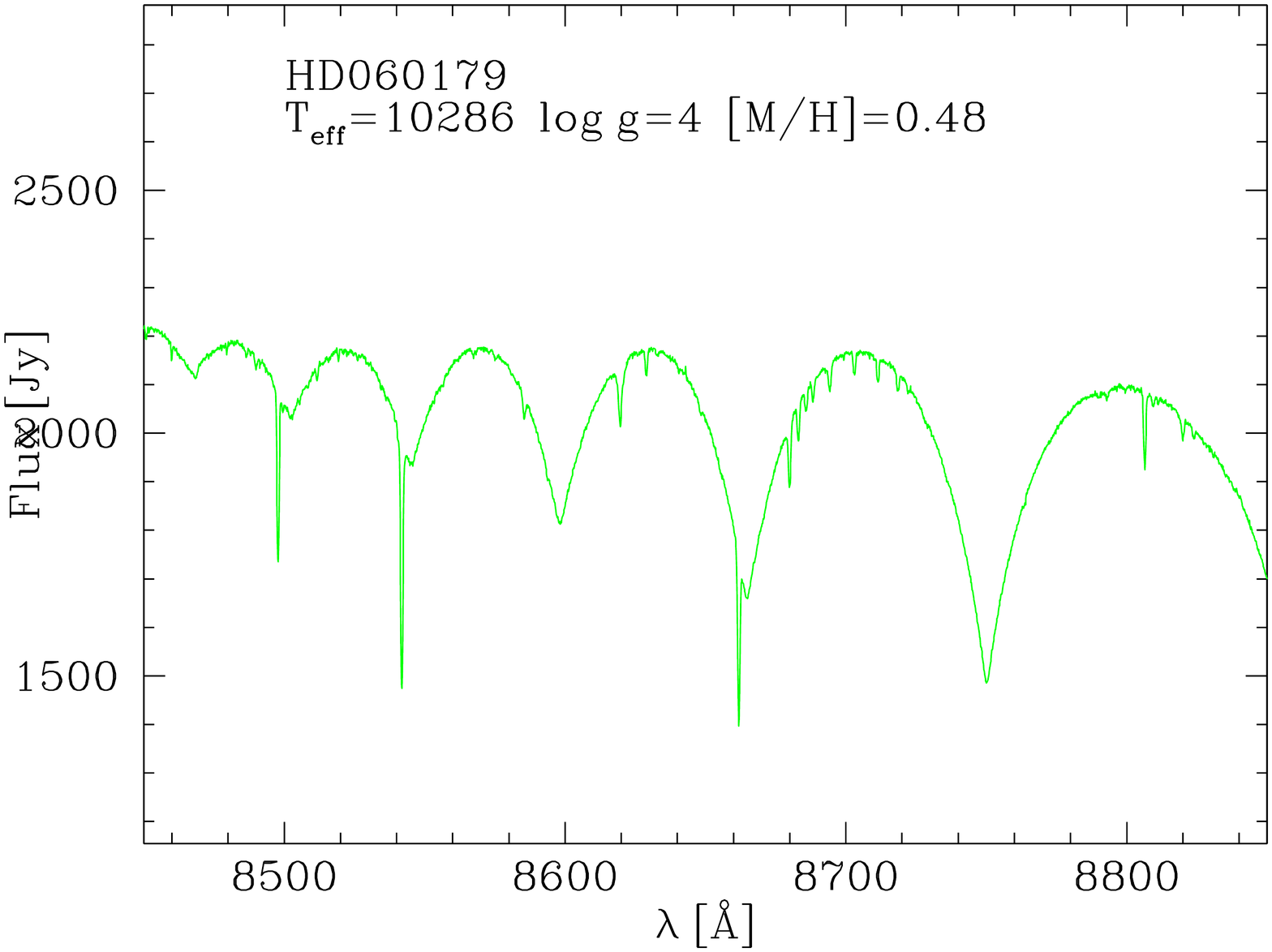}
\includegraphics[width=0.18\textwidth,angle=-90]{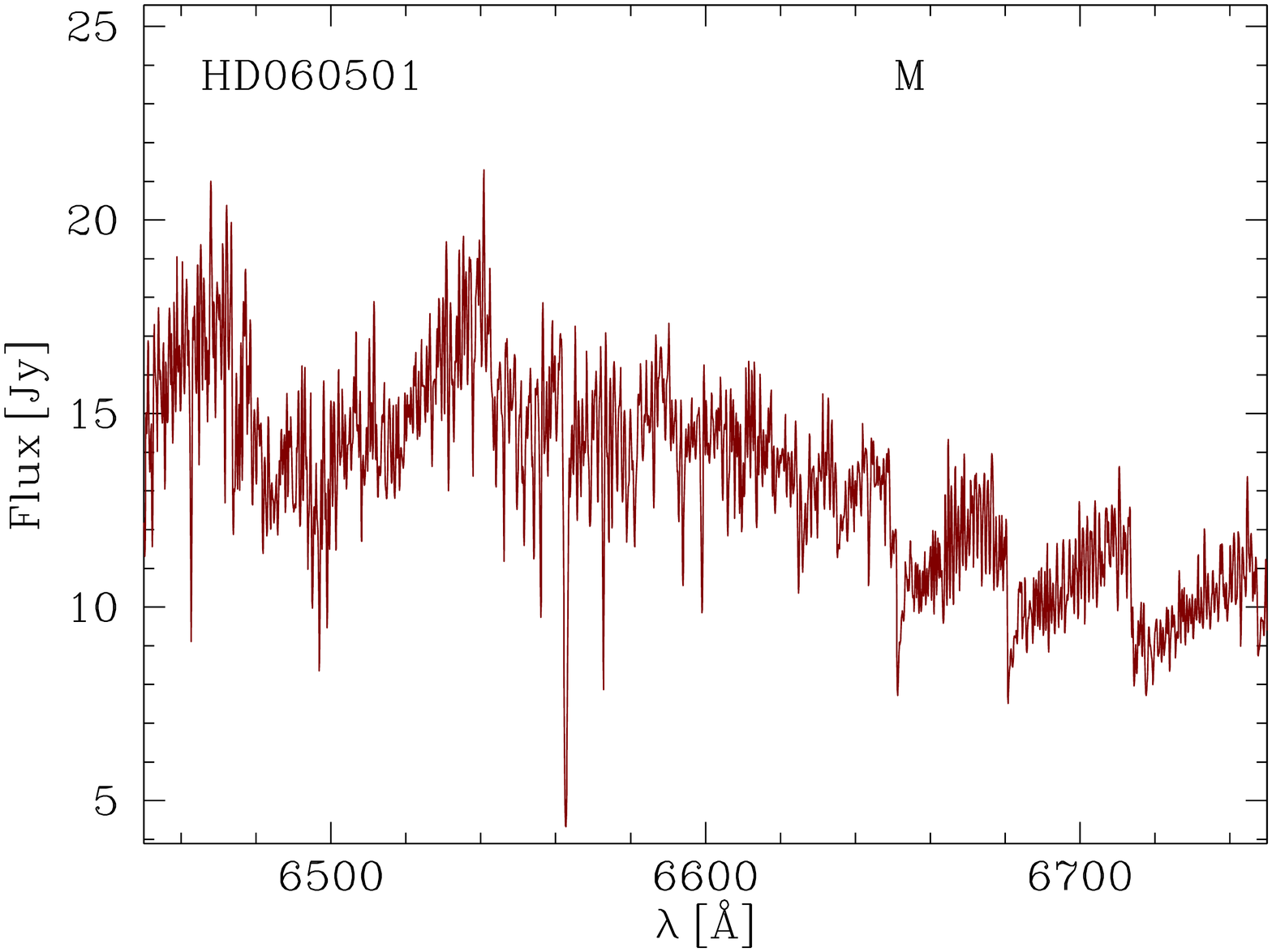}
\includegraphics[width=0.18\textwidth,angle=-90]{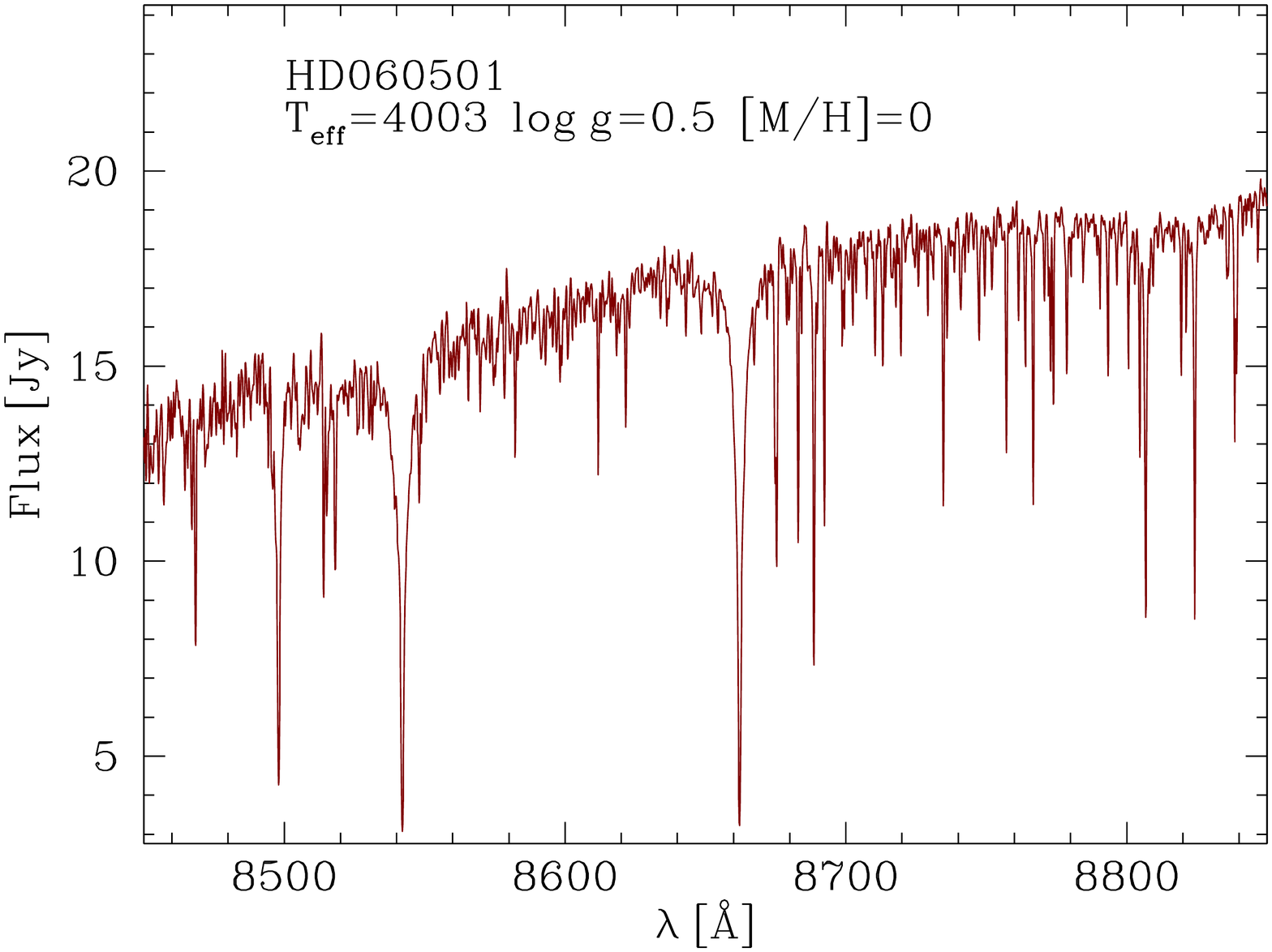}
\includegraphics[width=0.18\textwidth,angle=-90]{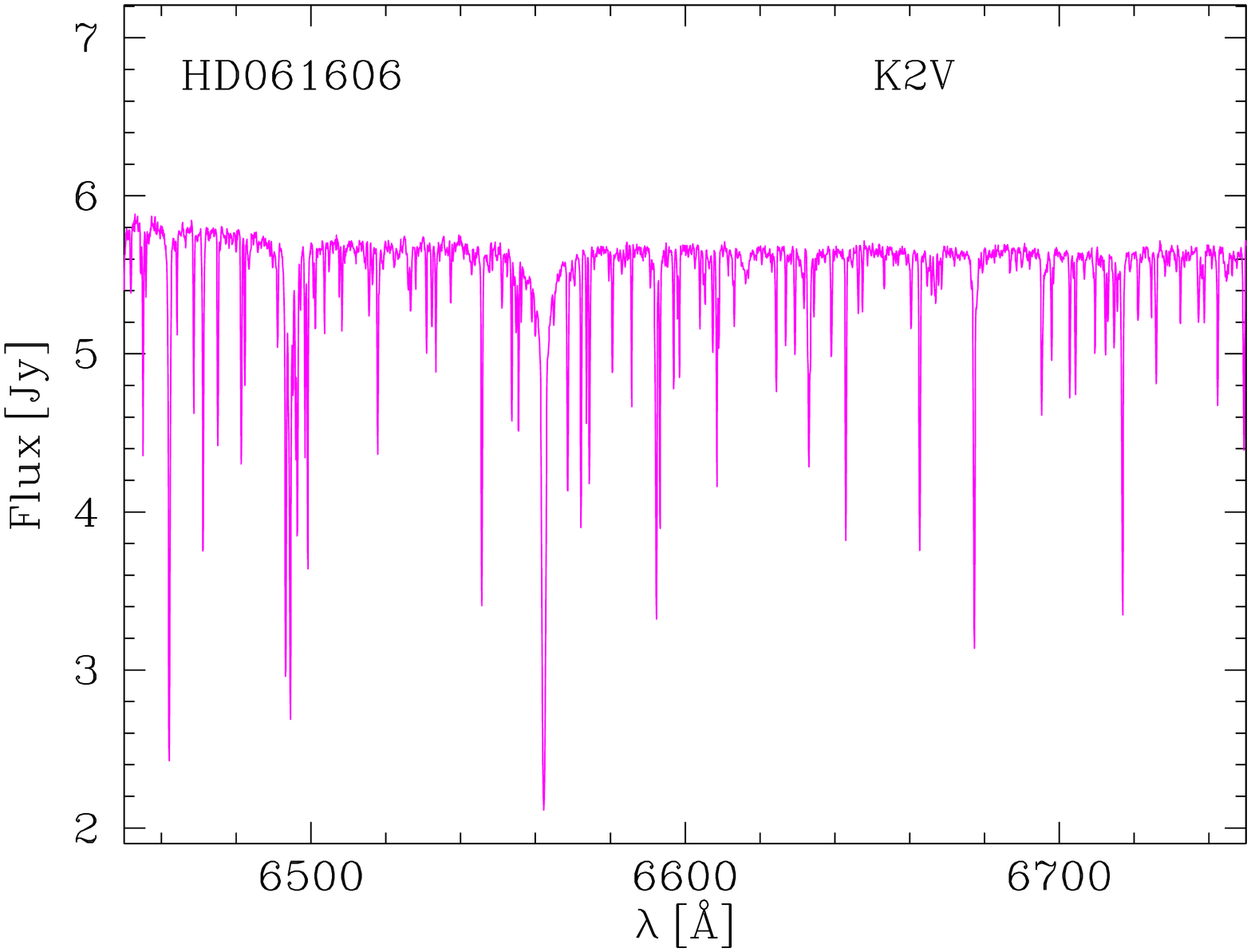}
\includegraphics[width=0.18\textwidth,angle=-90]{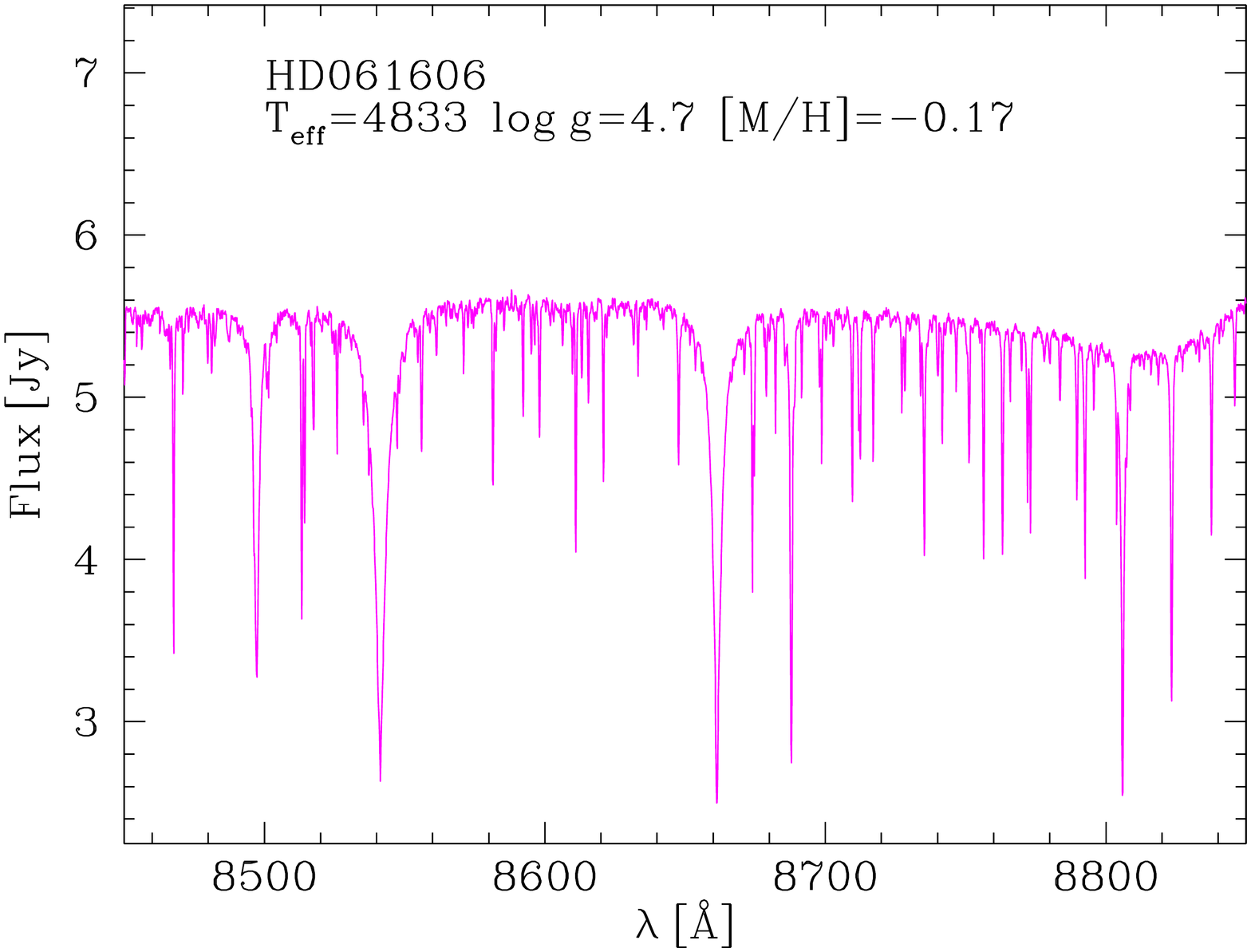}
\includegraphics[width=0.18\textwidth,angle=-90]{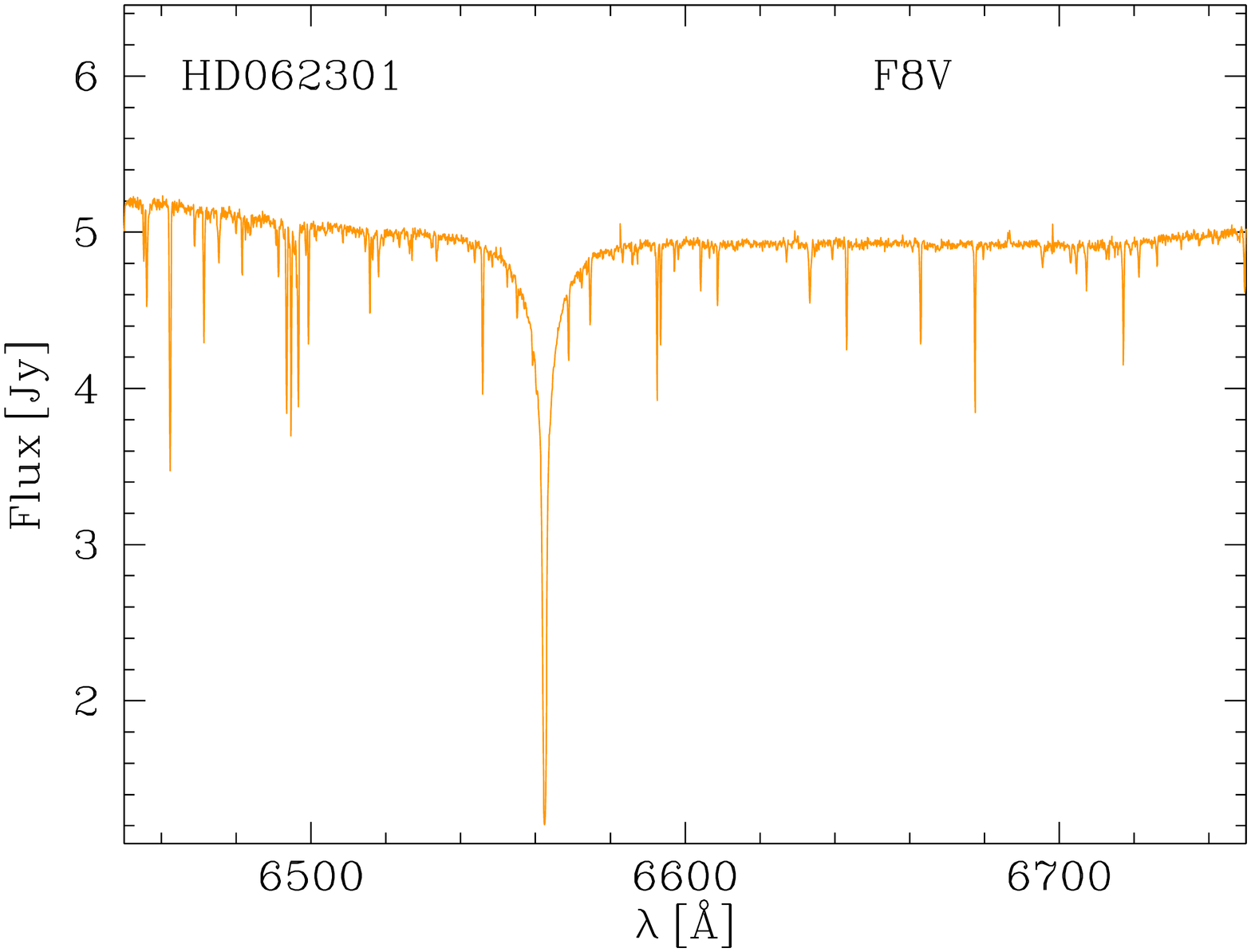}
\includegraphics[width=0.18\textwidth,angle=-90]{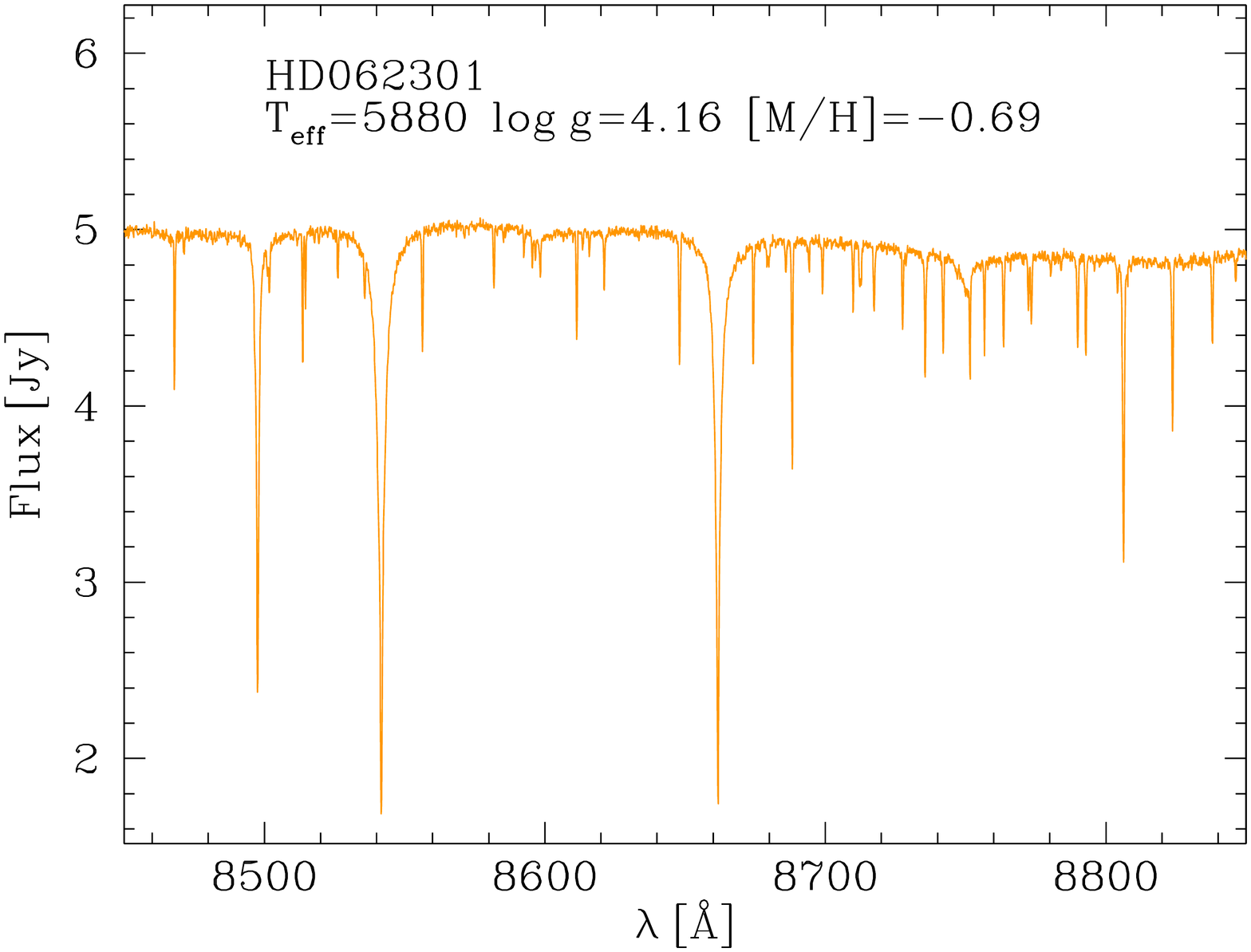}

\contcaption{13. Stars shown in this page are:  HD054717, HD055280, HD055575, HD055606, HD056925, HD056925, HD058343, HD058551, HD058946, HD059473, HD060179, HD060501, HD061606 and HD062301.}
\end{figure*}

\begin{figure*}
\includegraphics[width=0.18\textwidth,angle=-90]{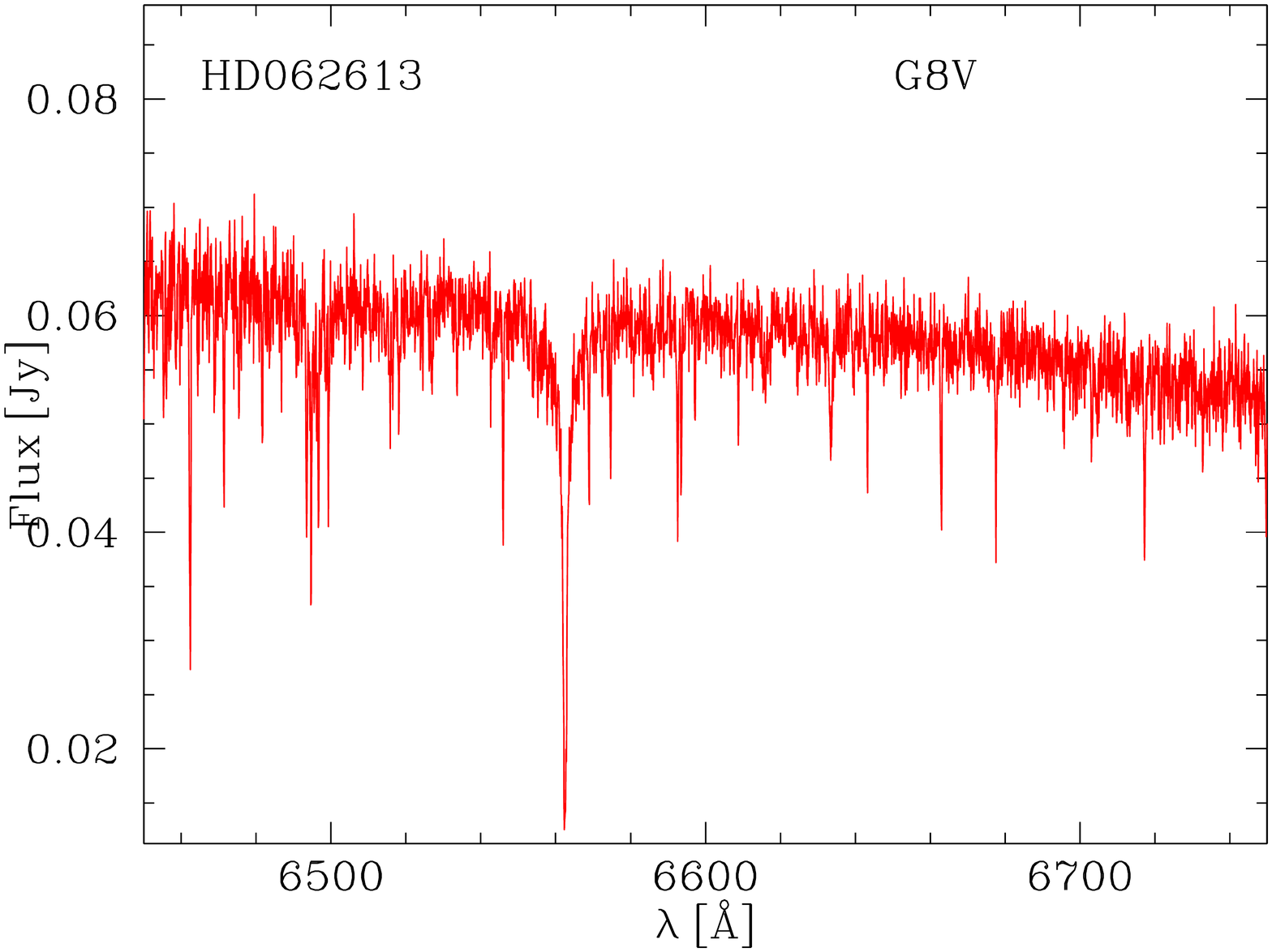}
\includegraphics[width=0.18\textwidth,angle=-90]{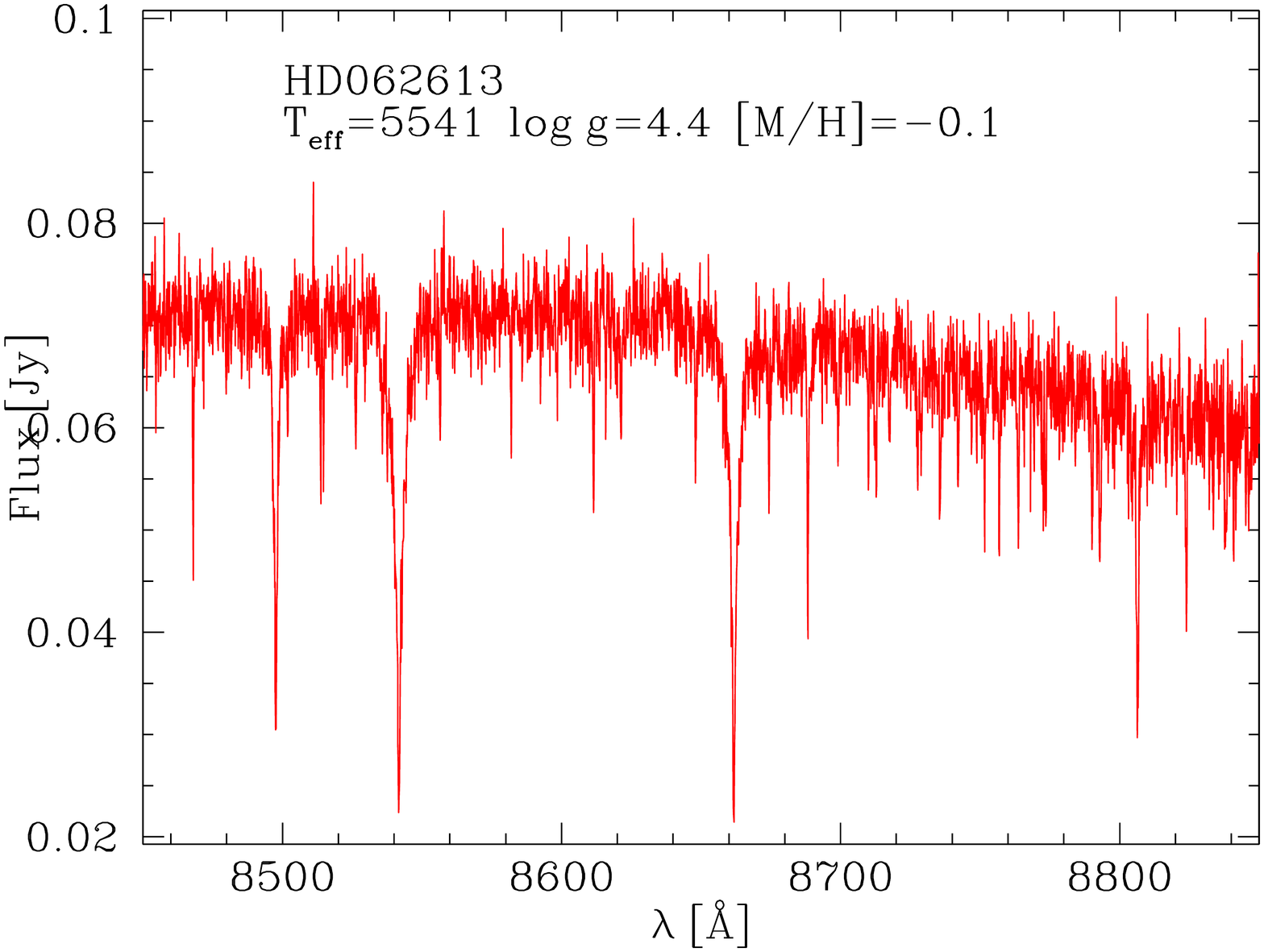}
\includegraphics[width=0.18\textwidth,angle=-90]{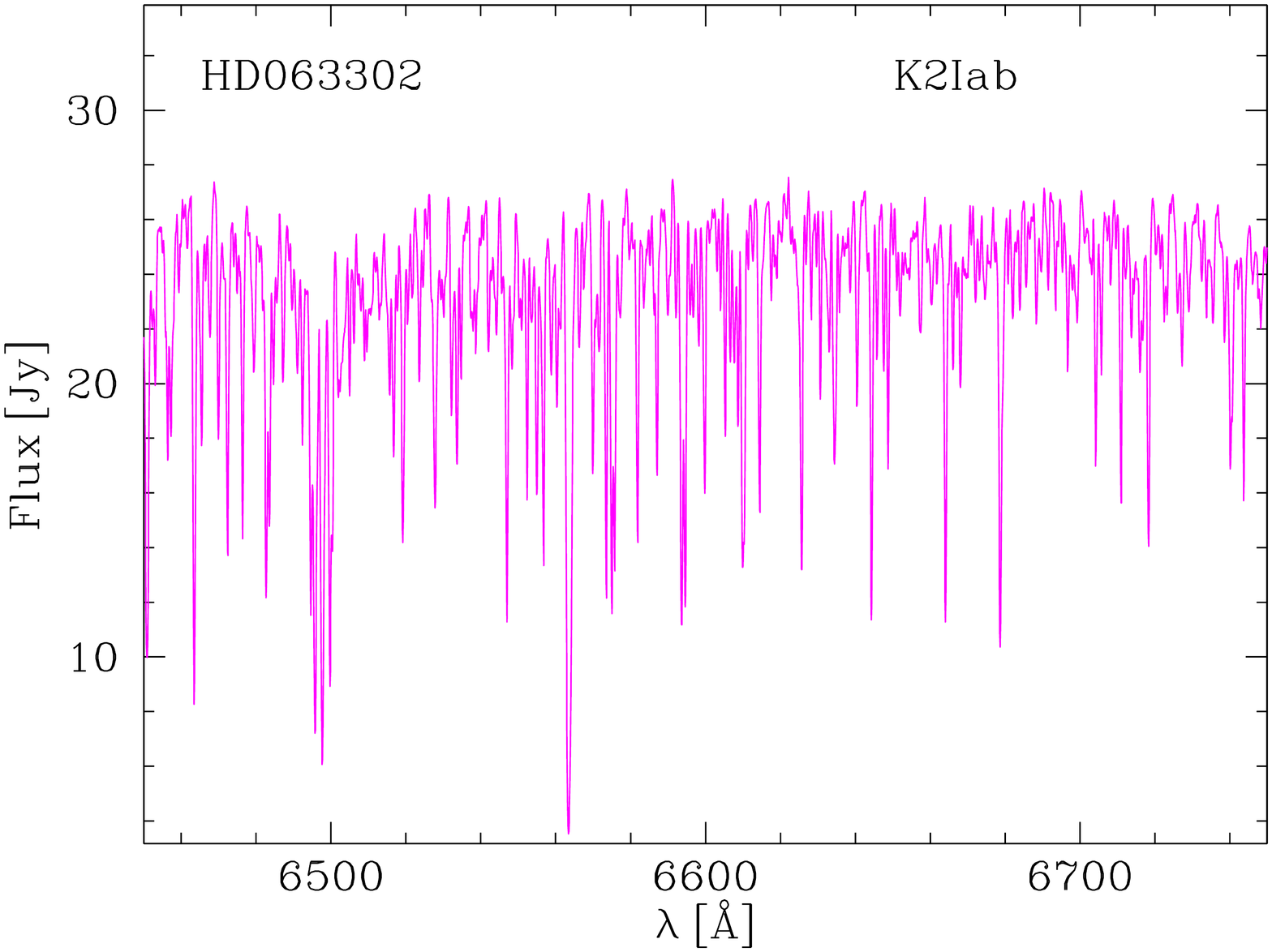}
\includegraphics[width=0.18\textwidth,angle=-90]{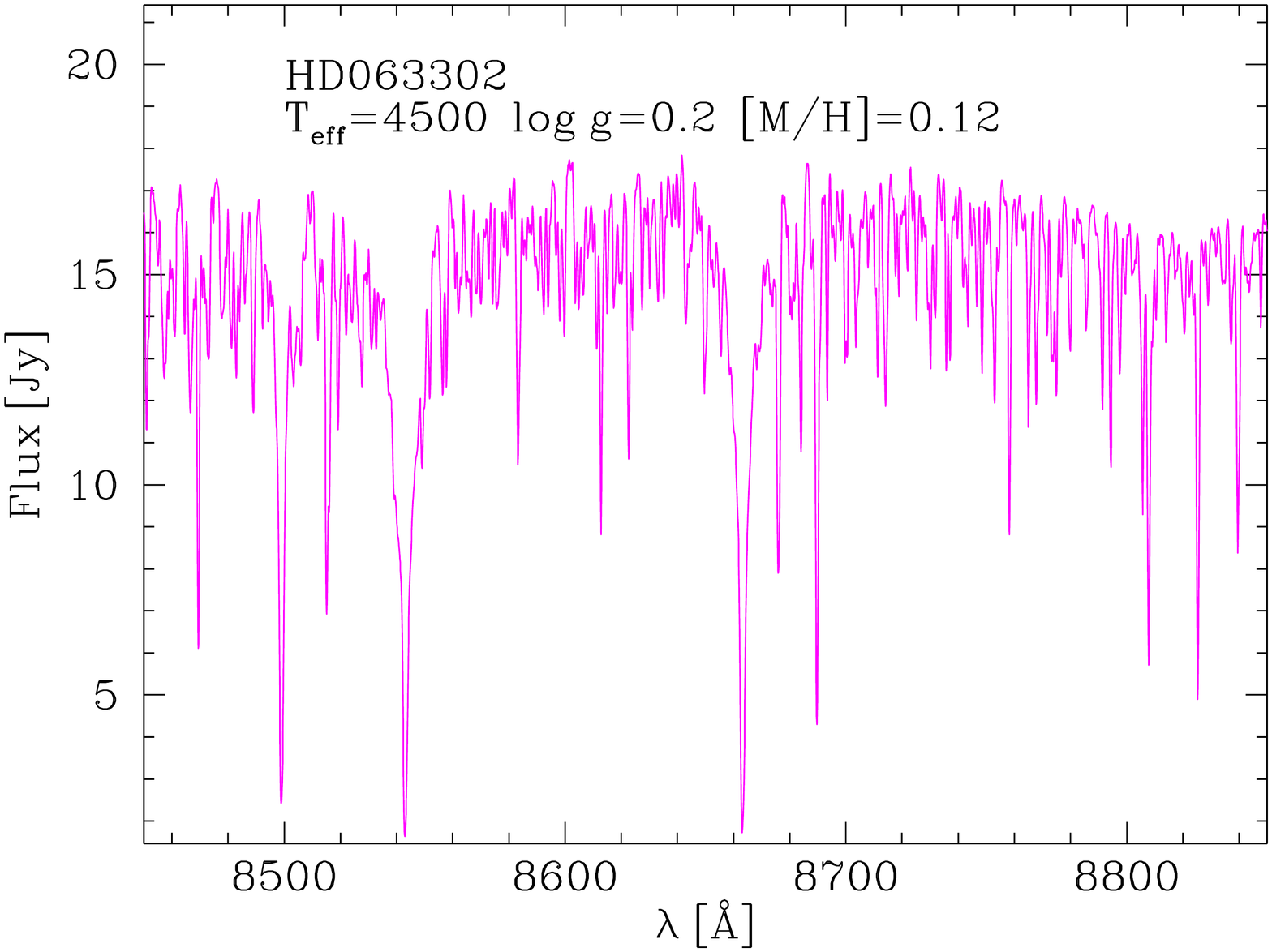}
\includegraphics[width=0.18\textwidth,angle=-90]{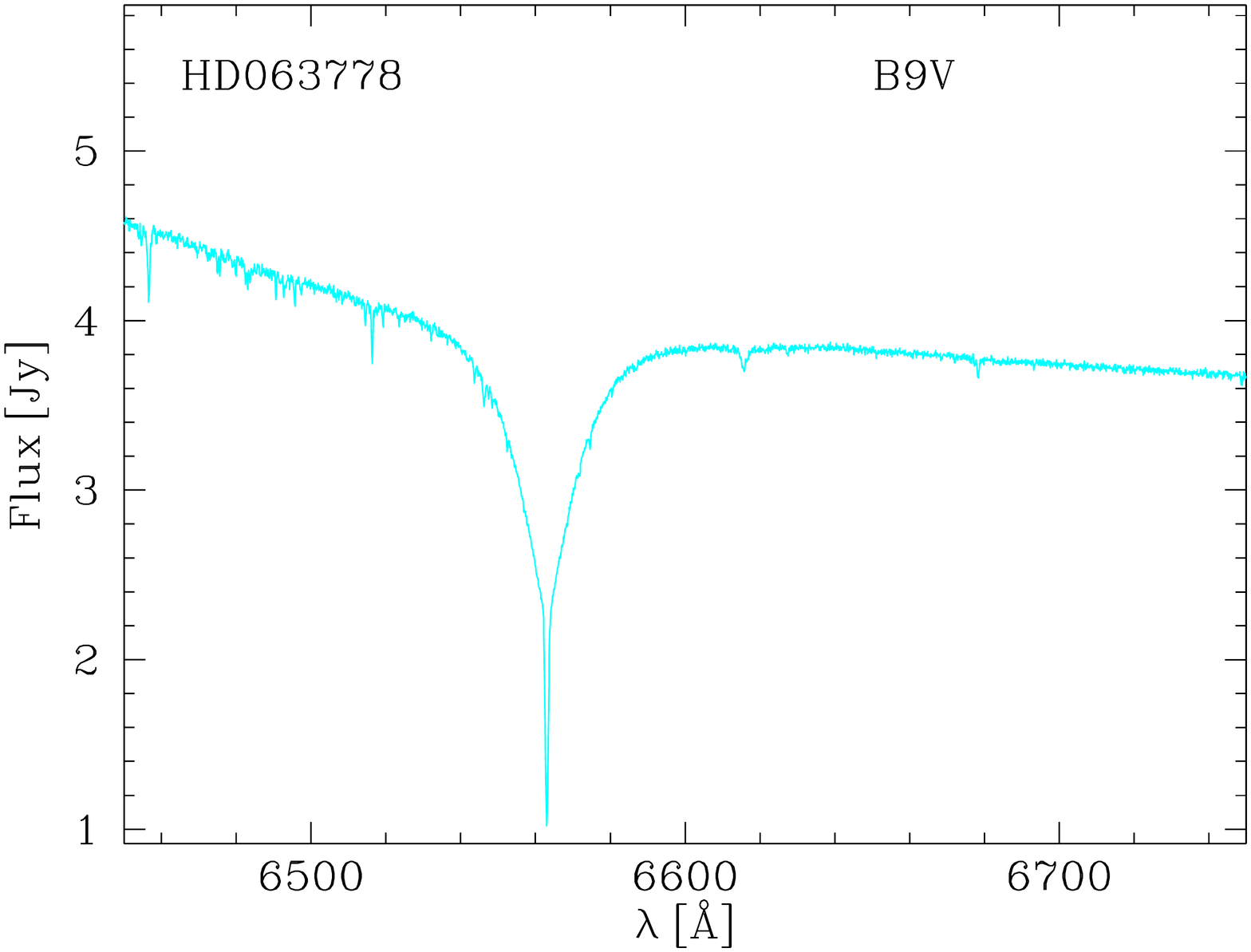}
\includegraphics[width=0.18\textwidth,angle=-90]{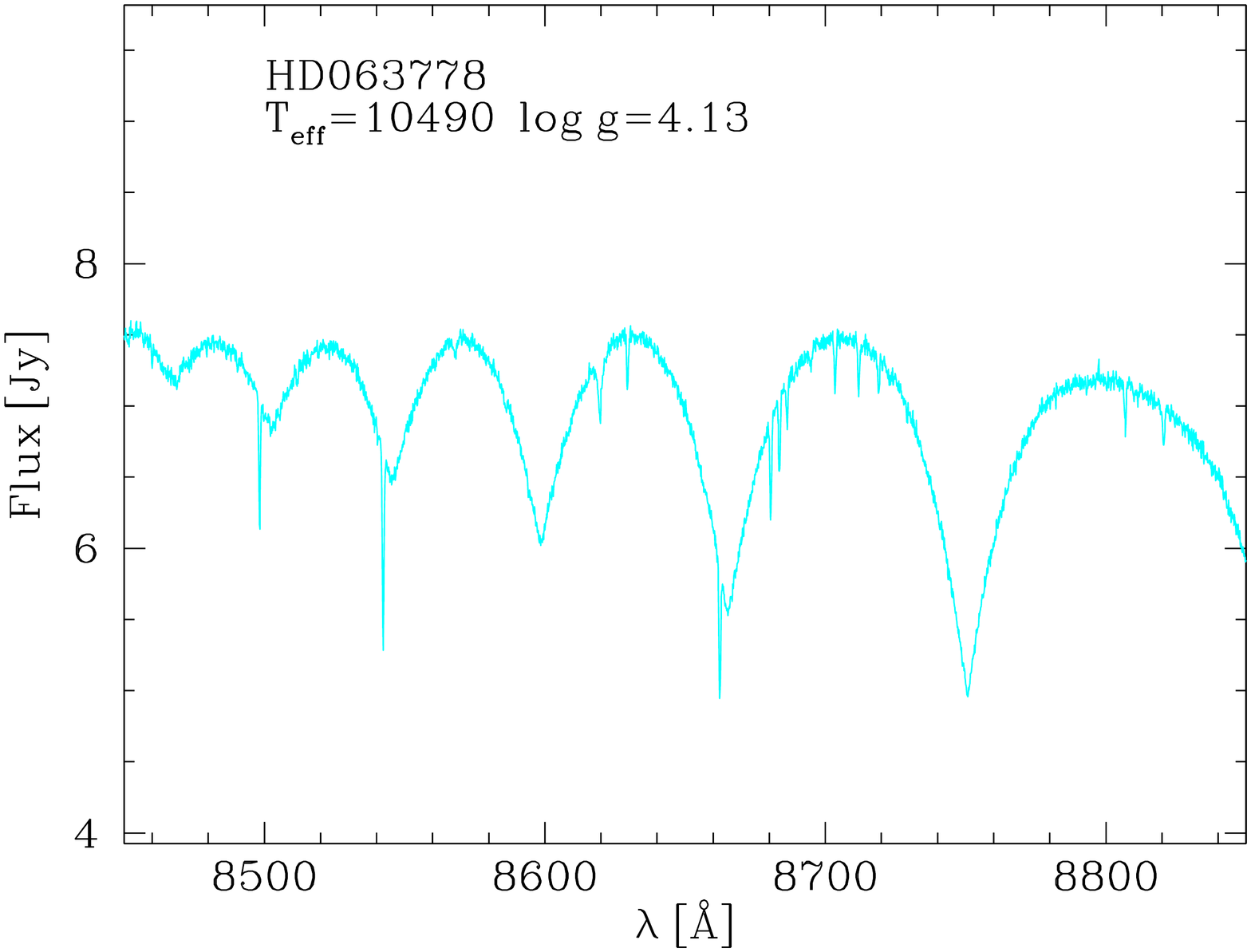}
\includegraphics[width=0.18\textwidth,angle=-90]{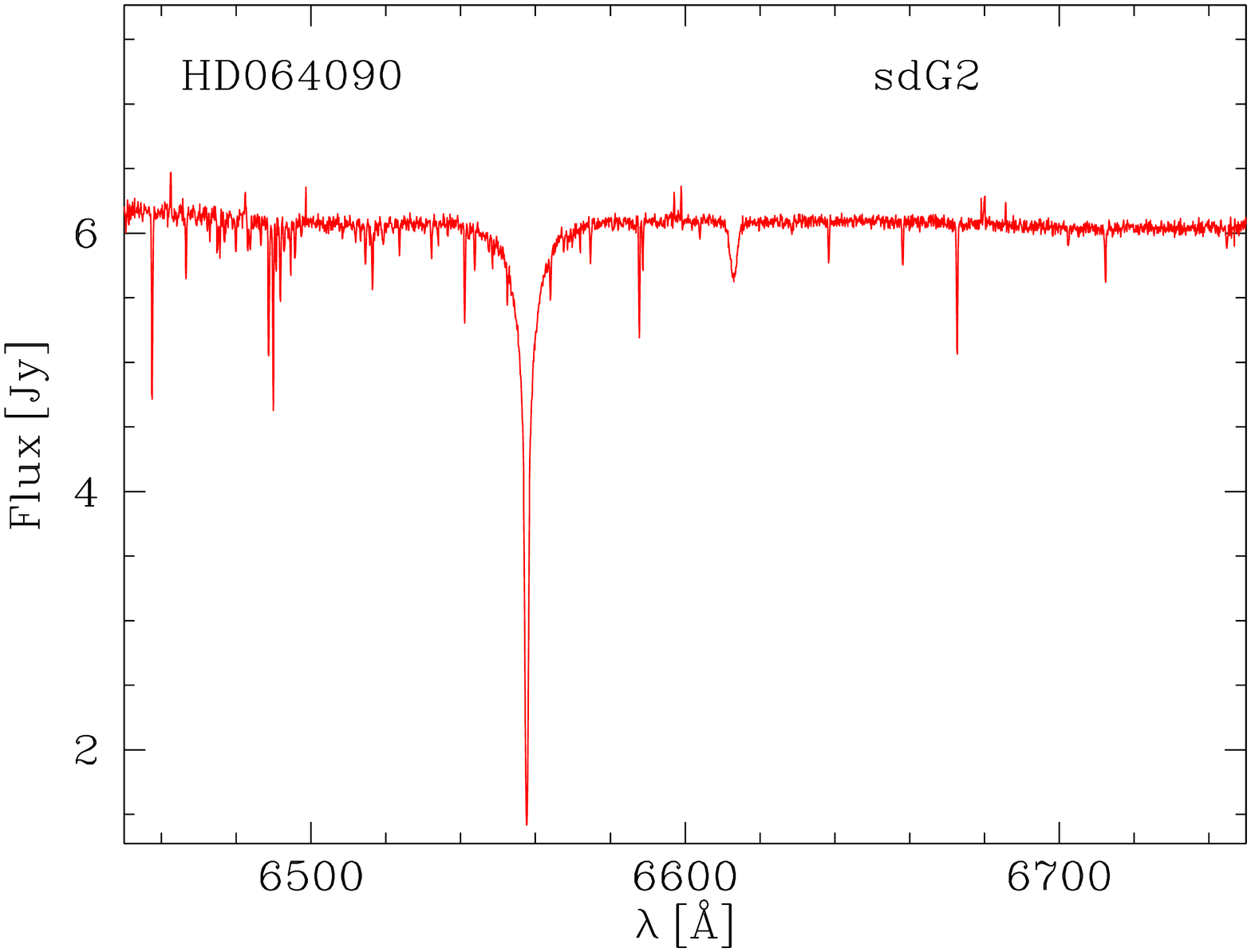}
\includegraphics[width=0.18\textwidth,angle=-90]{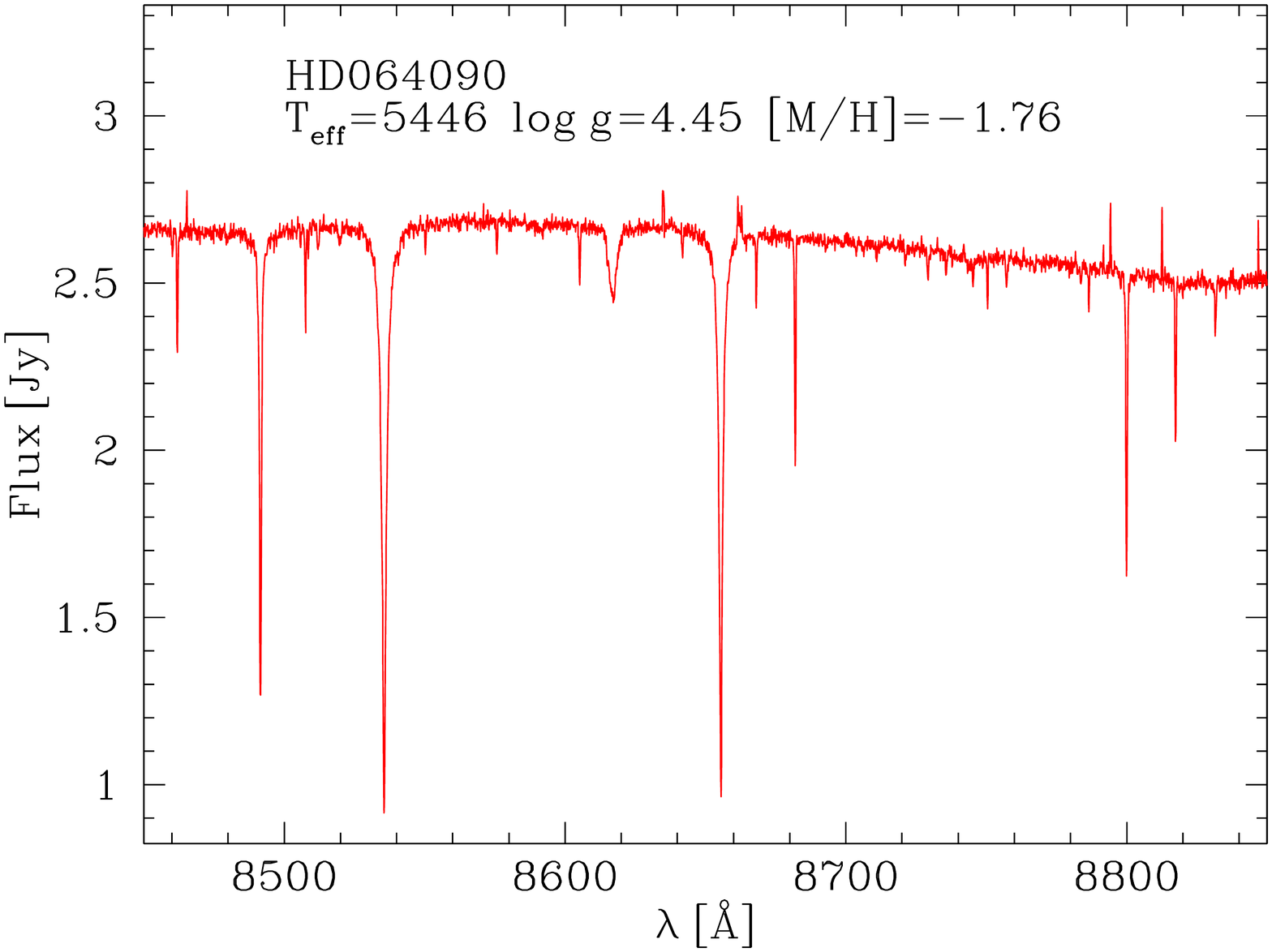}
\includegraphics[width=0.18\textwidth,angle=-90]{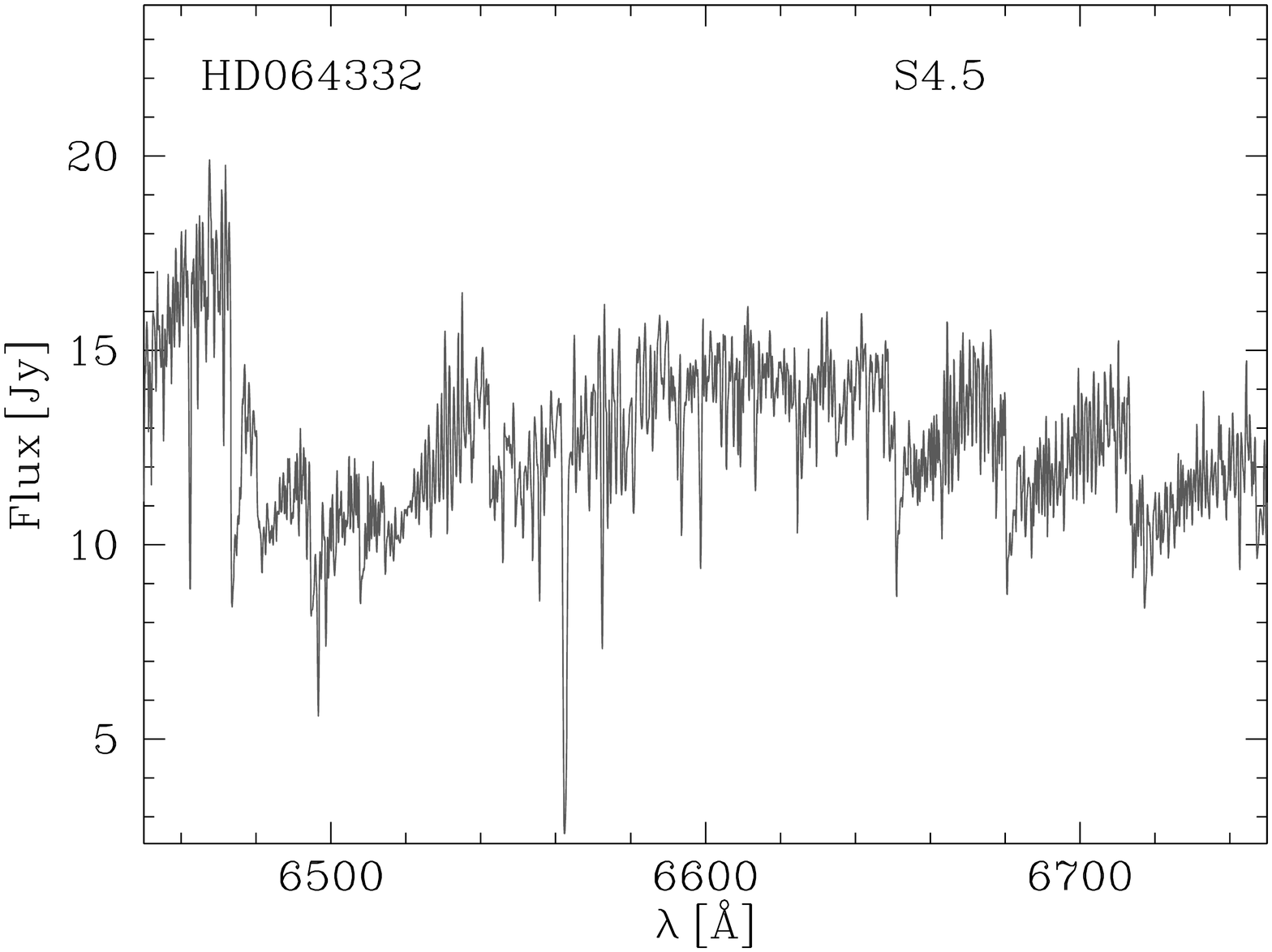}
\includegraphics[width=0.18\textwidth,angle=-90]{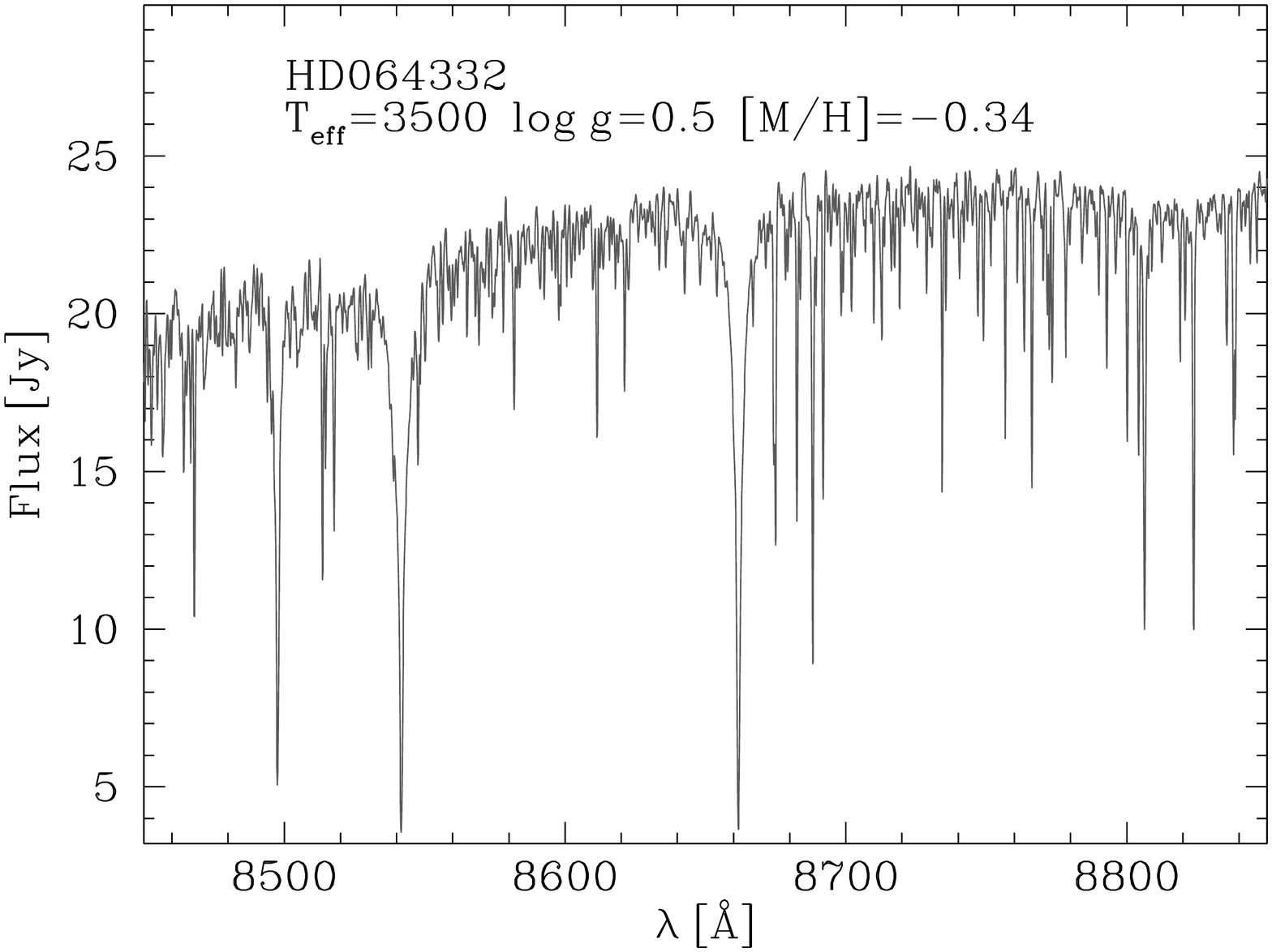}
\includegraphics[width=0.18\textwidth,angle=-90]{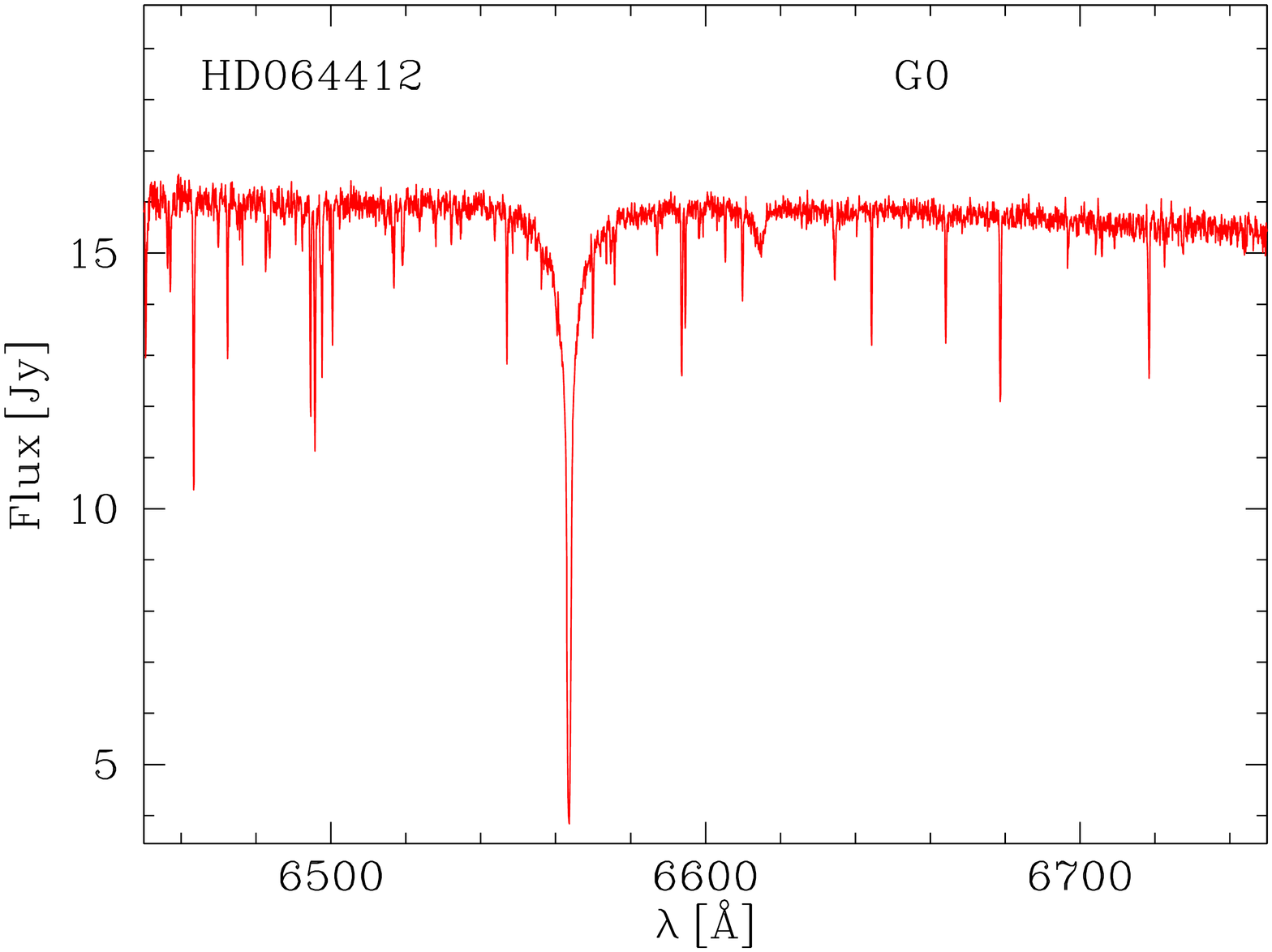} 
\includegraphics[width=0.18\textwidth,angle=-90]{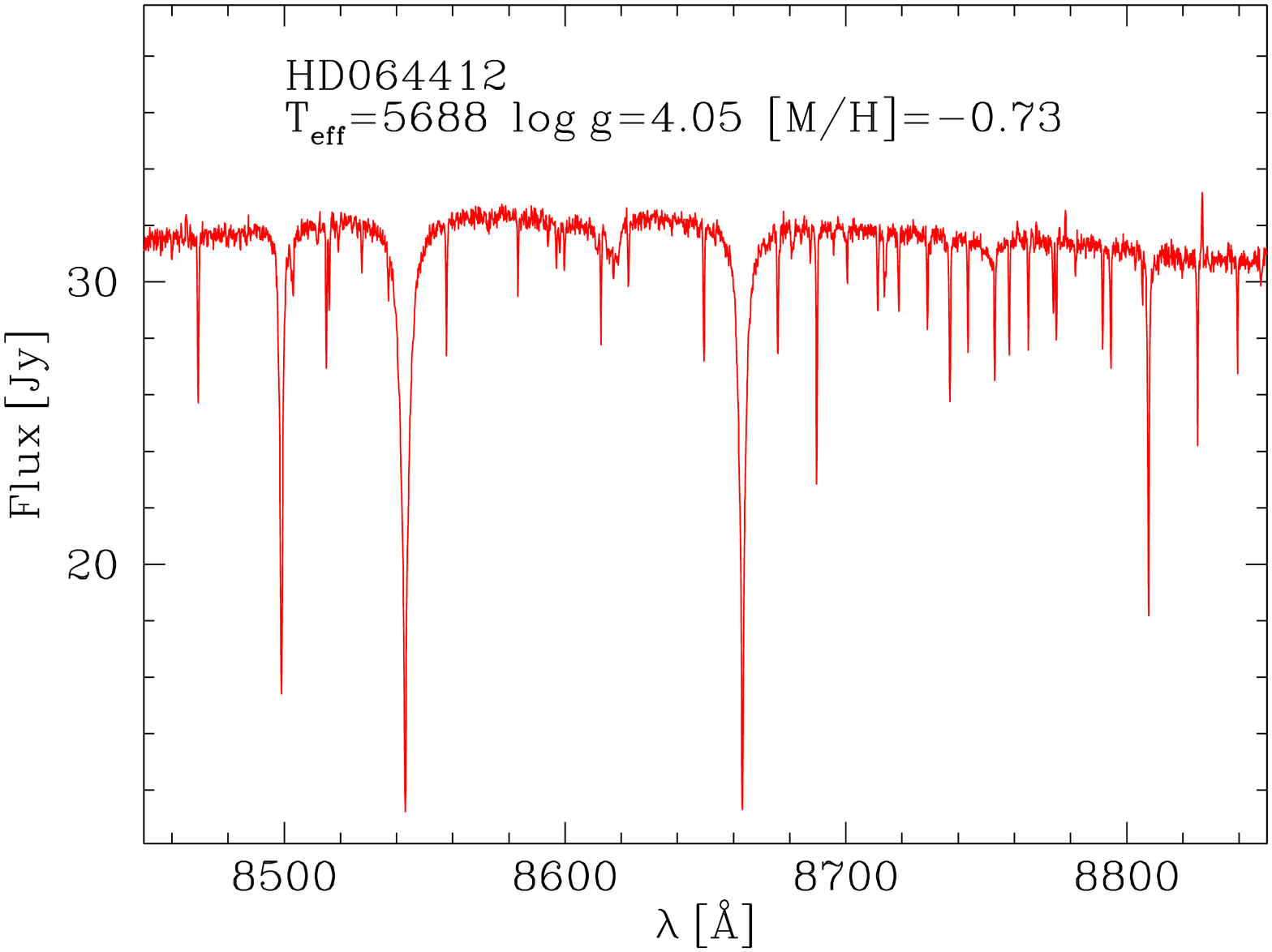}
\includegraphics[width=0.18\textwidth,angle=-90]{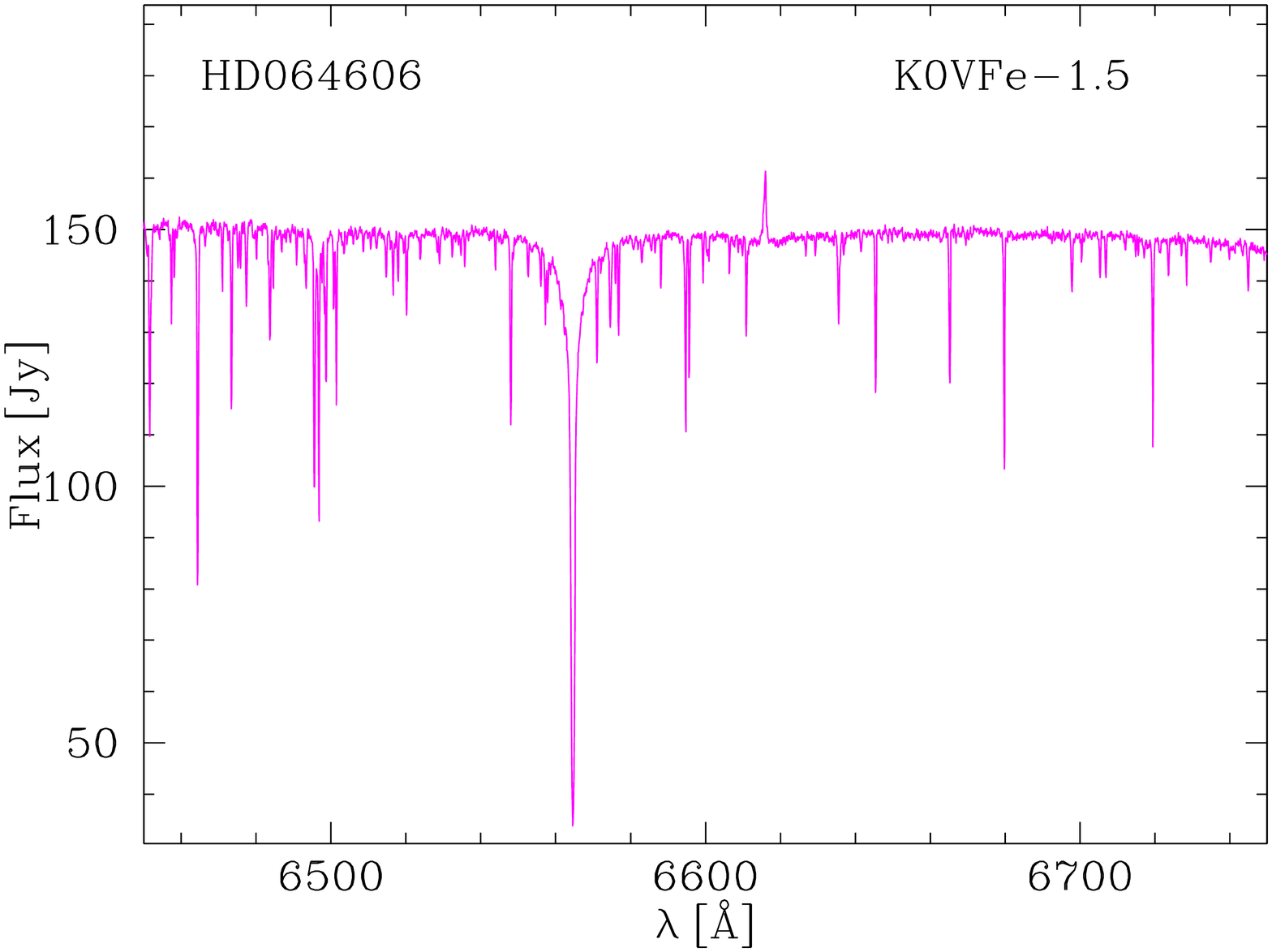}
\includegraphics[width=0.18\textwidth,angle=-90]{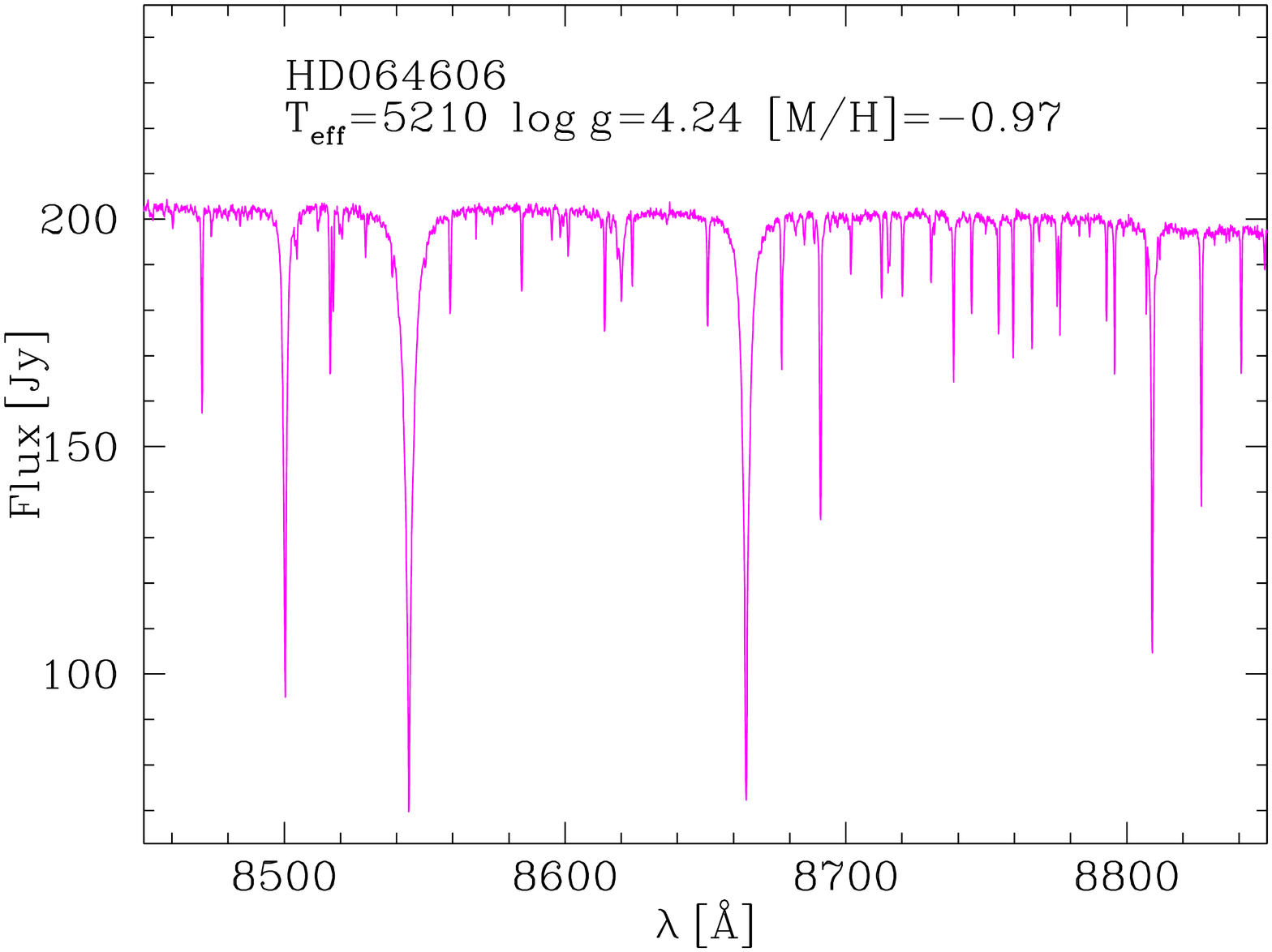}
\includegraphics[width=0.18\textwidth,angle=-90]{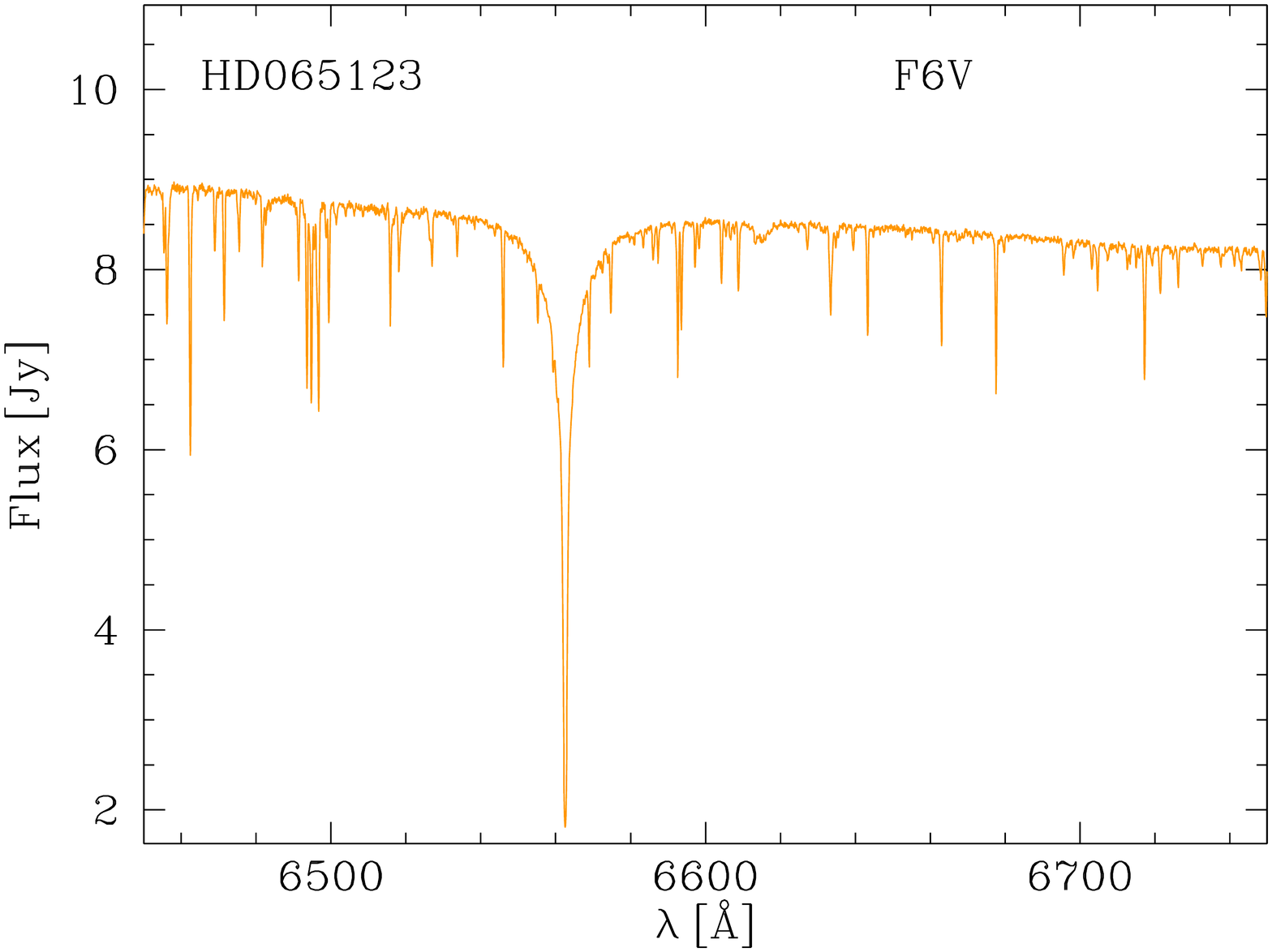}
\includegraphics[width=0.18\textwidth,angle=-90]{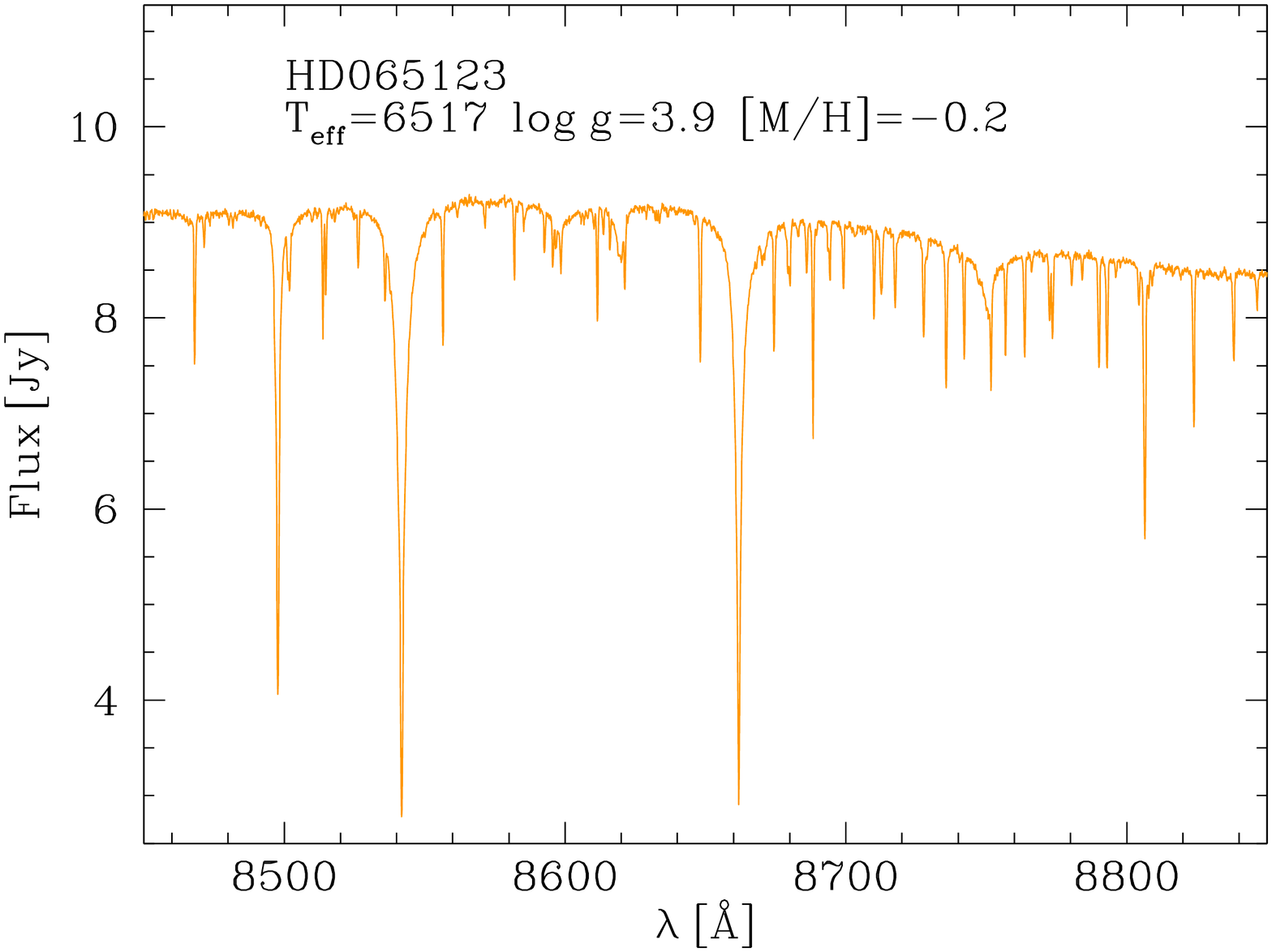}
\includegraphics[width=0.18\textwidth,angle=-90]{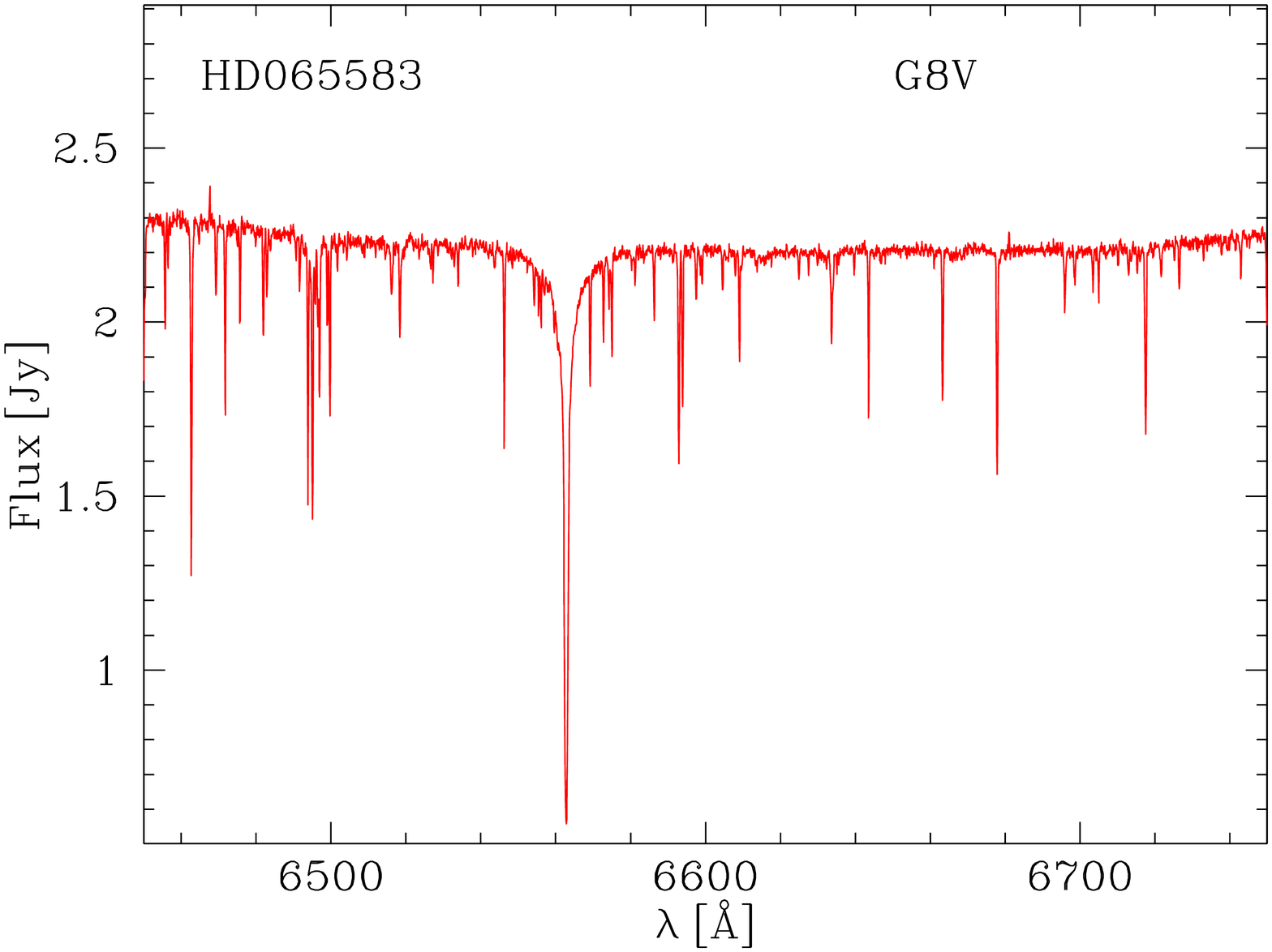}
\includegraphics[width=0.18\textwidth,angle=-90]{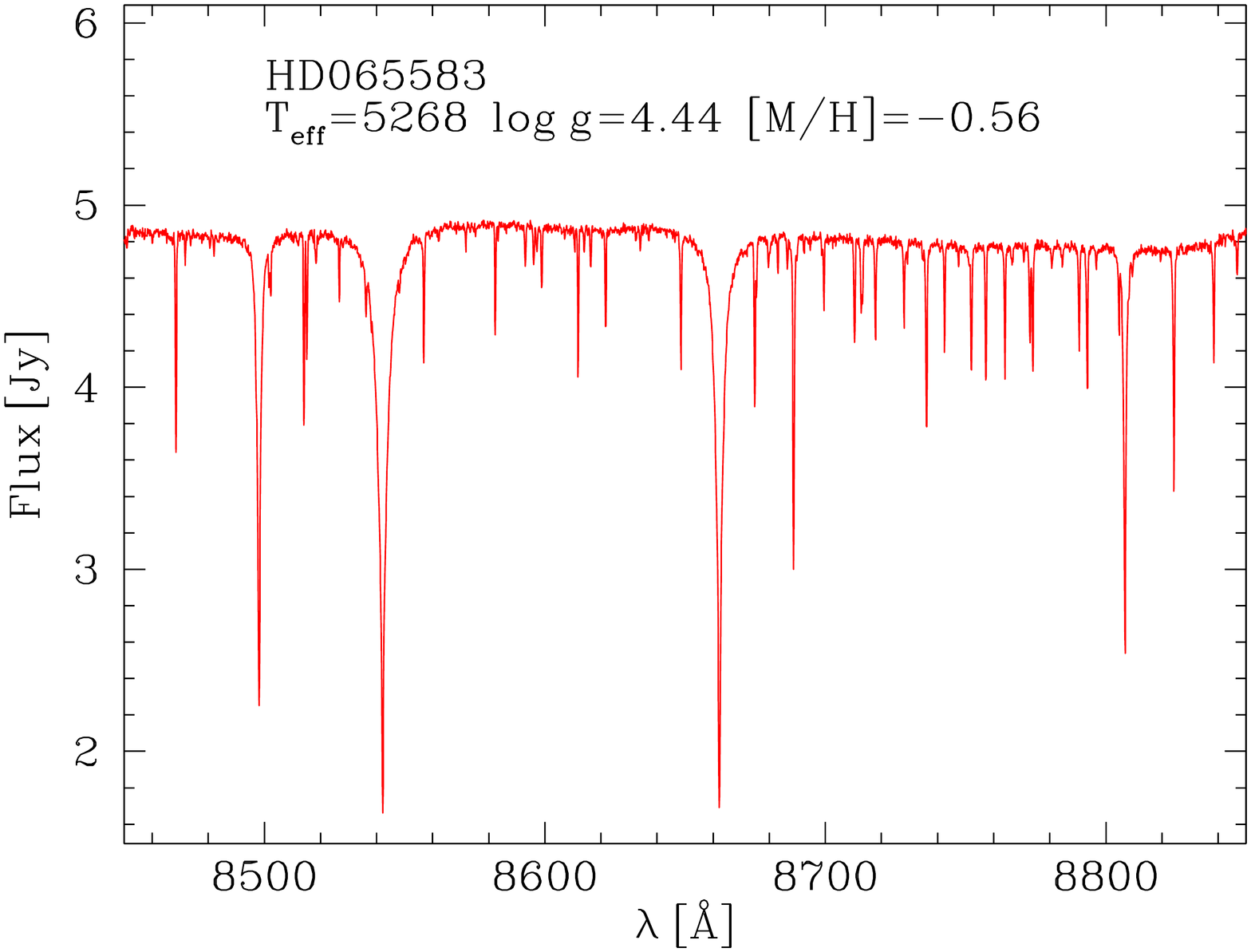}
\includegraphics[width=0.18\textwidth,angle=-90]{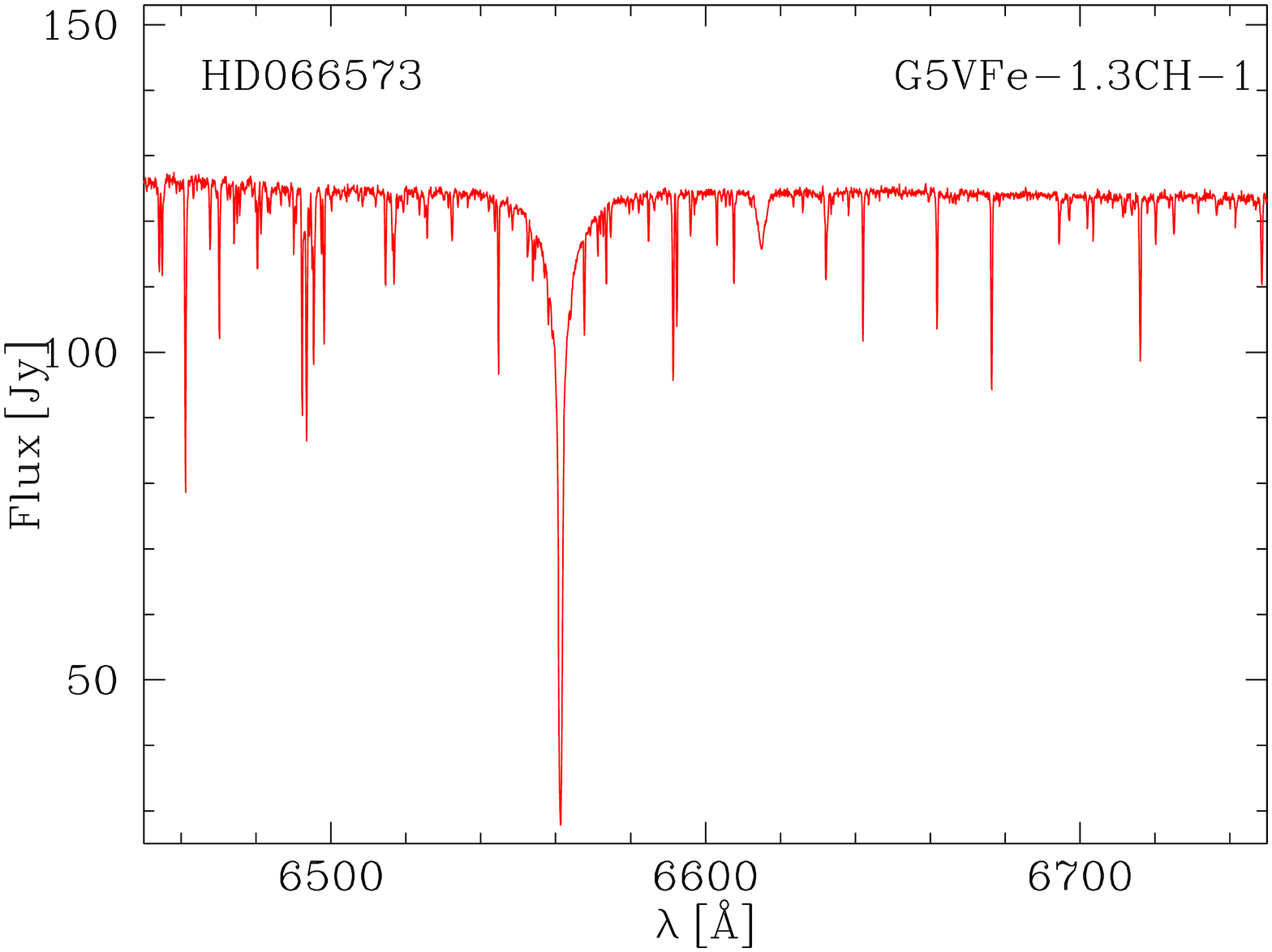}
\includegraphics[width=0.18\textwidth,angle=-90]{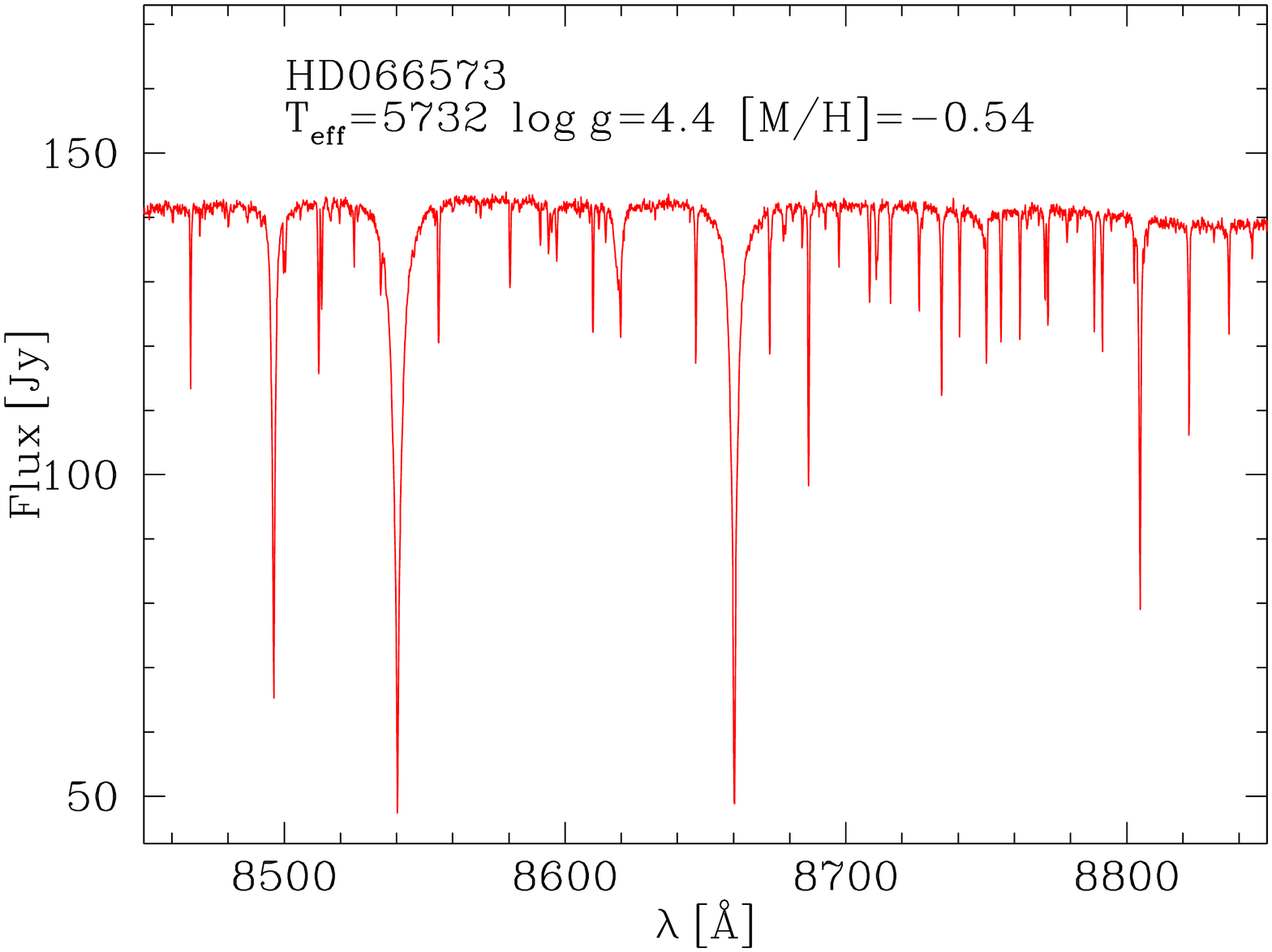}
\includegraphics[width=0.18\textwidth,angle=-90]{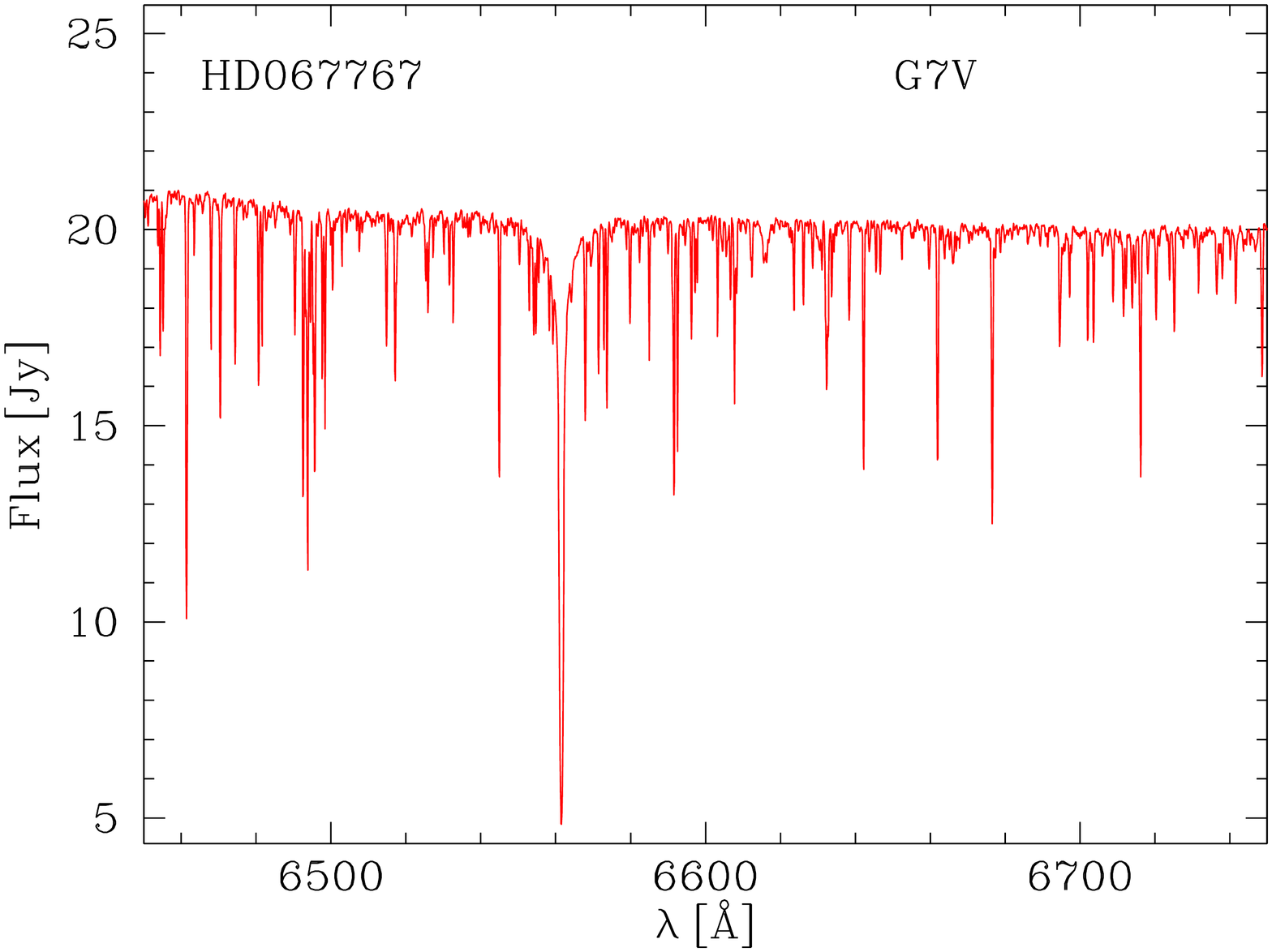}
\includegraphics[width=0.18\textwidth,angle=-90]{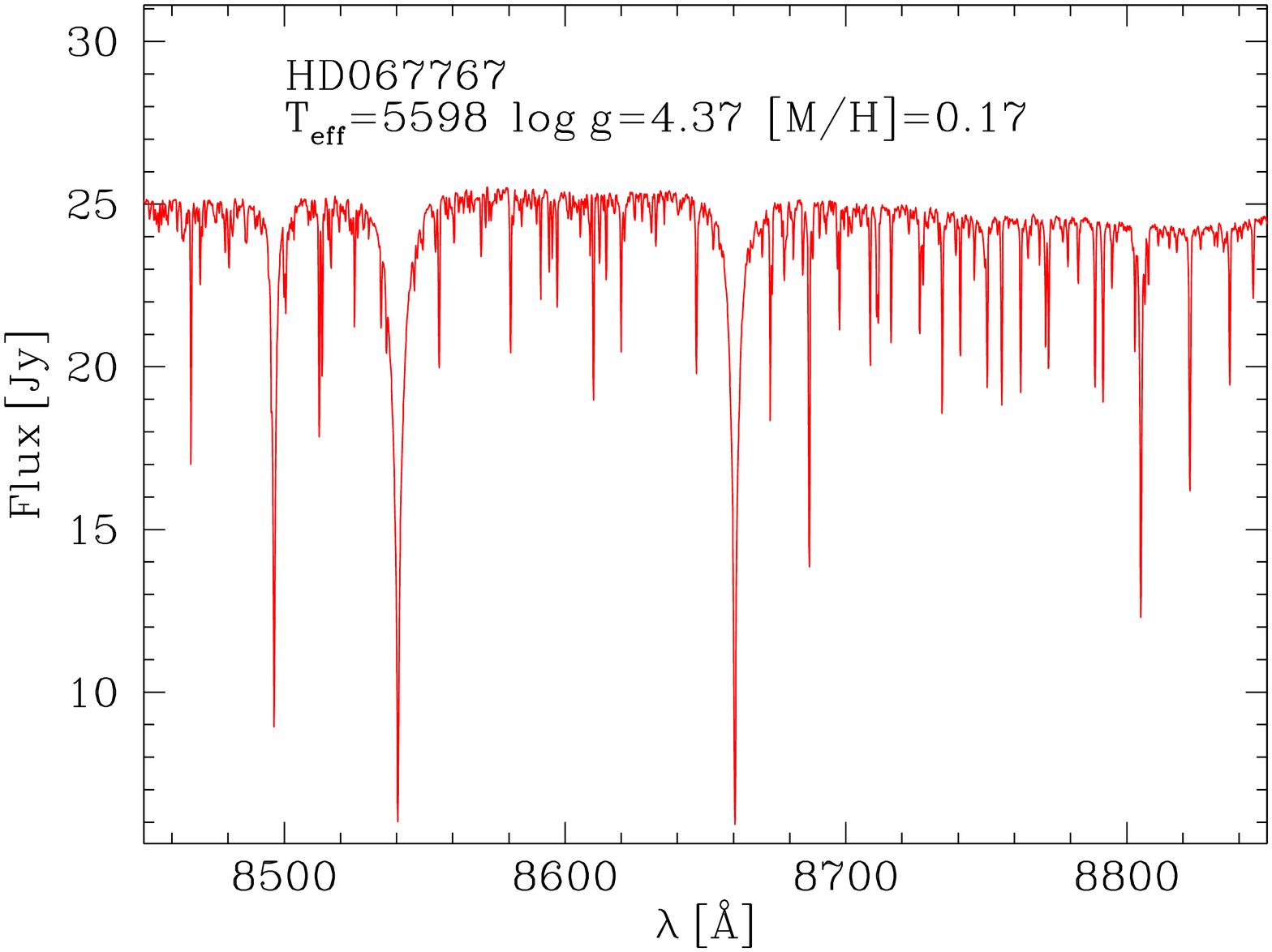}
\includegraphics[width=0.18\textwidth,angle=-90]{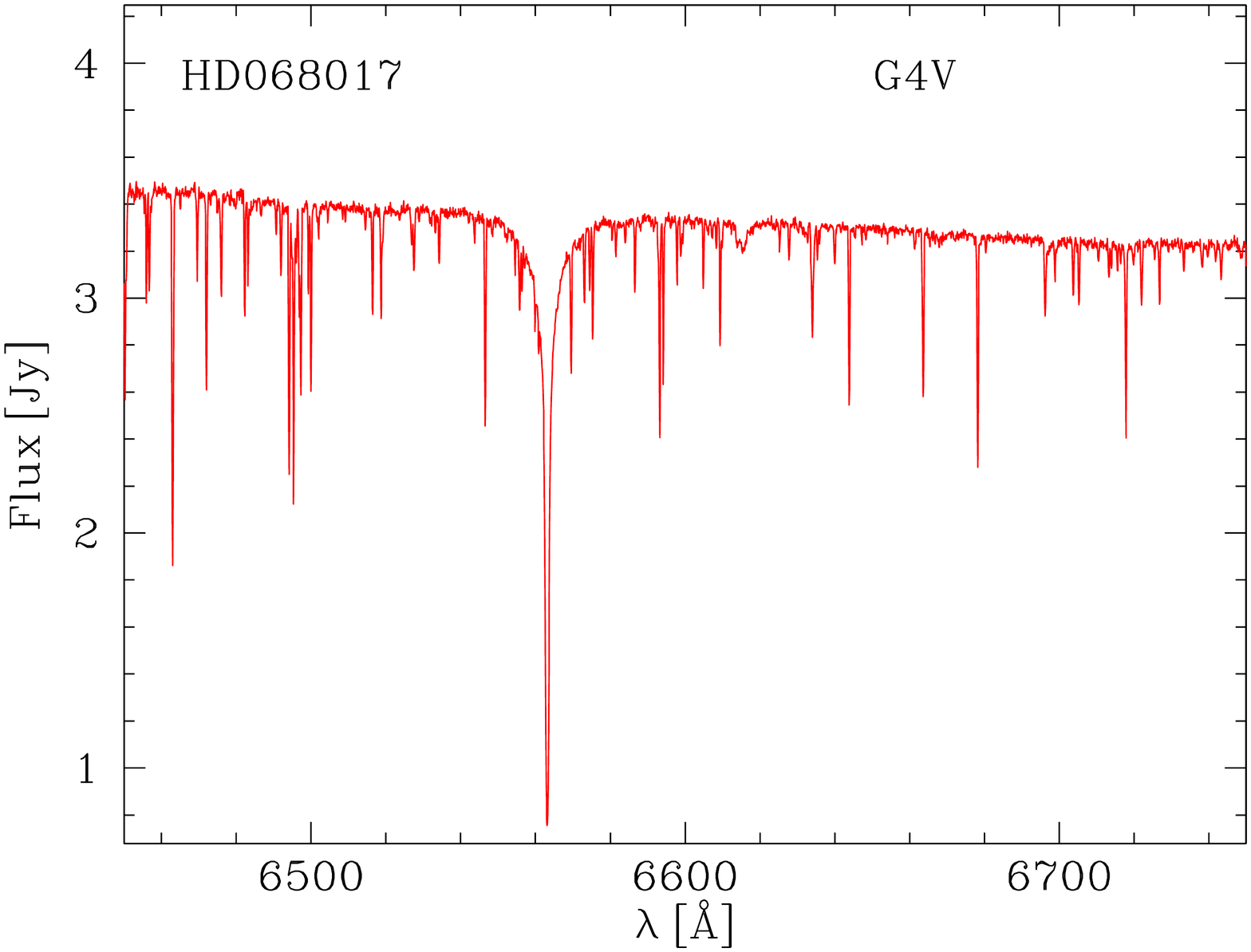}
\includegraphics[width=0.18\textwidth,angle=-90]{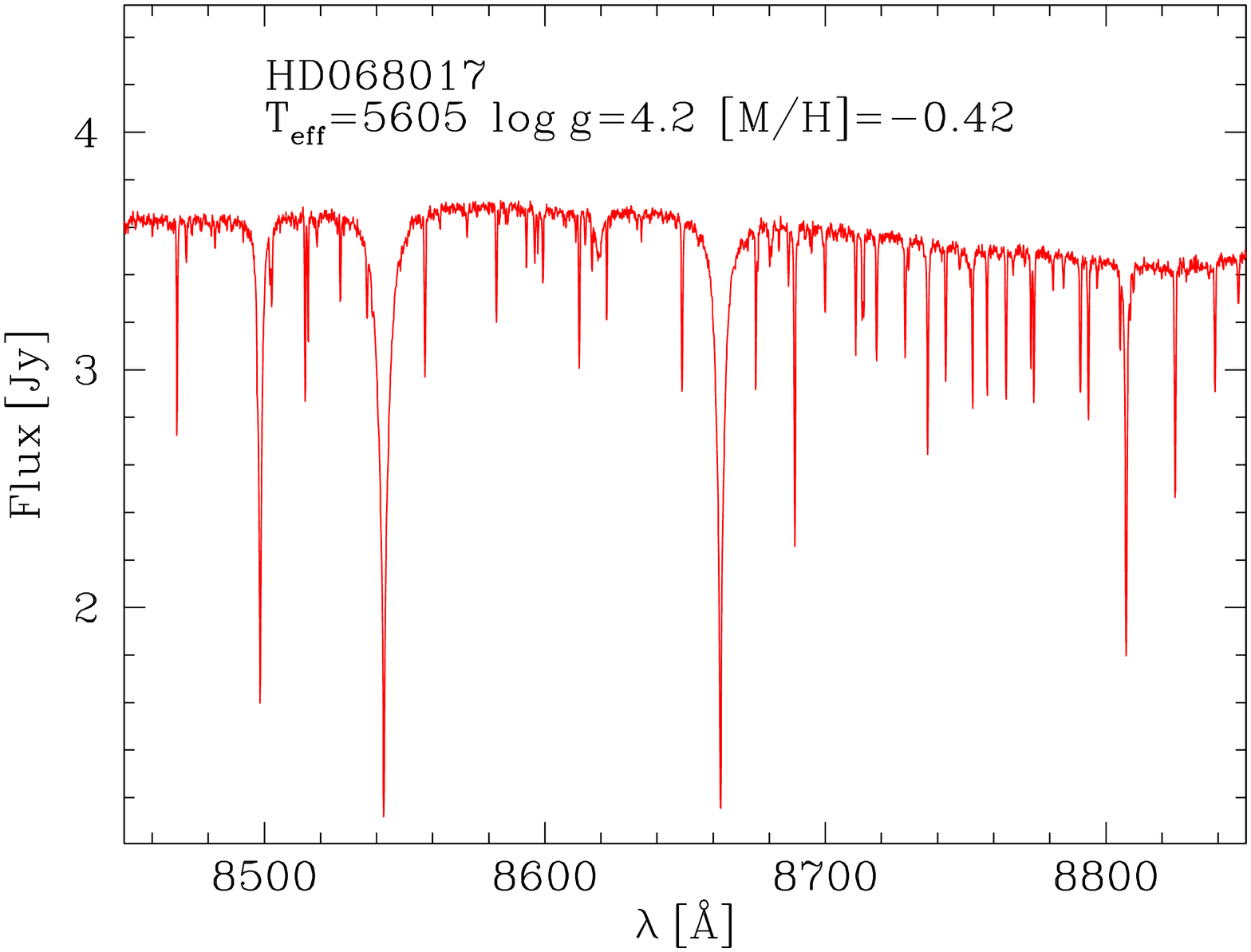}
\includegraphics[width=0.18\textwidth,angle=-90]{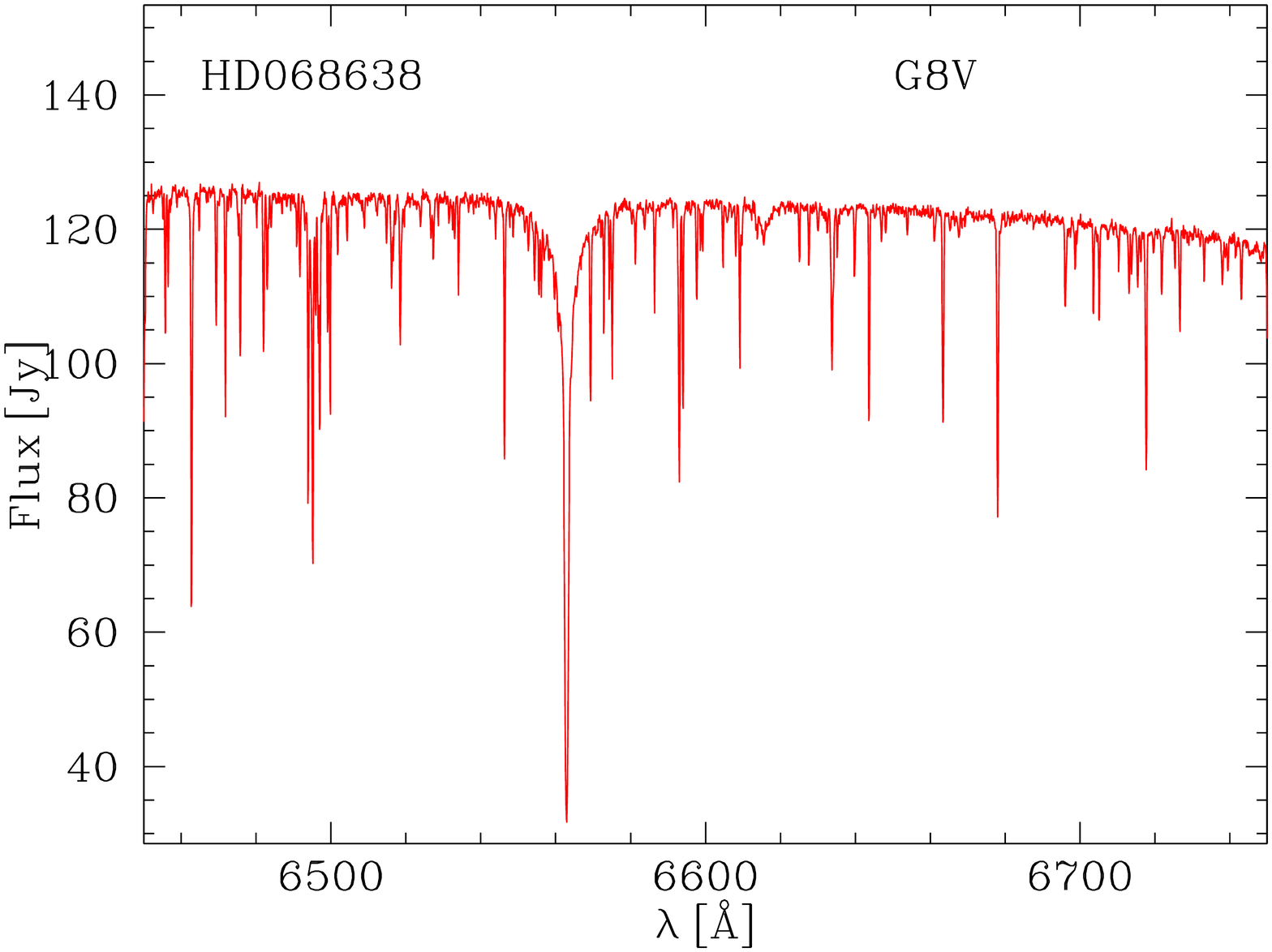}
\includegraphics[width=0.18\textwidth,angle=-90]{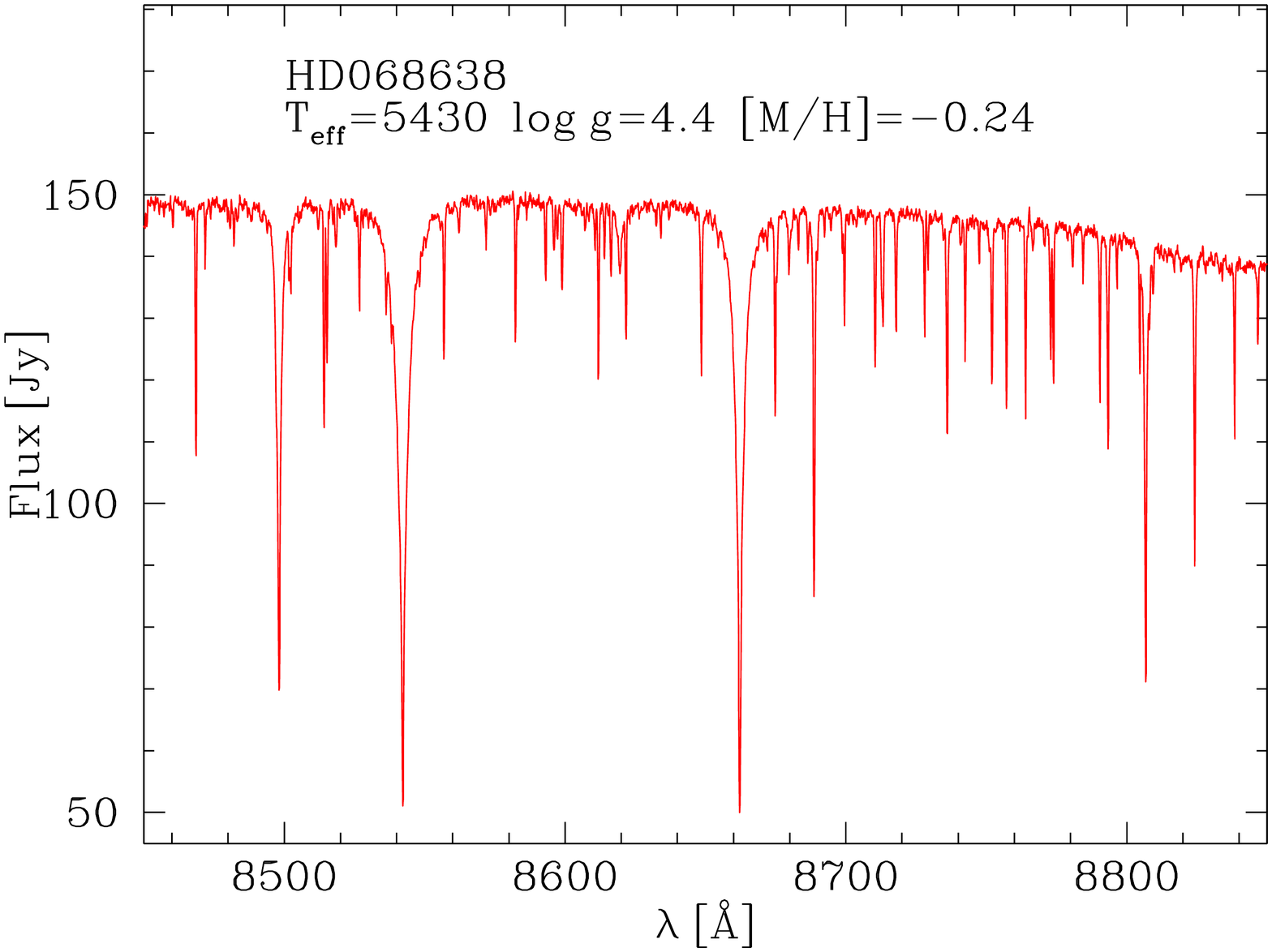}
\includegraphics[width=0.18\textwidth,angle=-90]{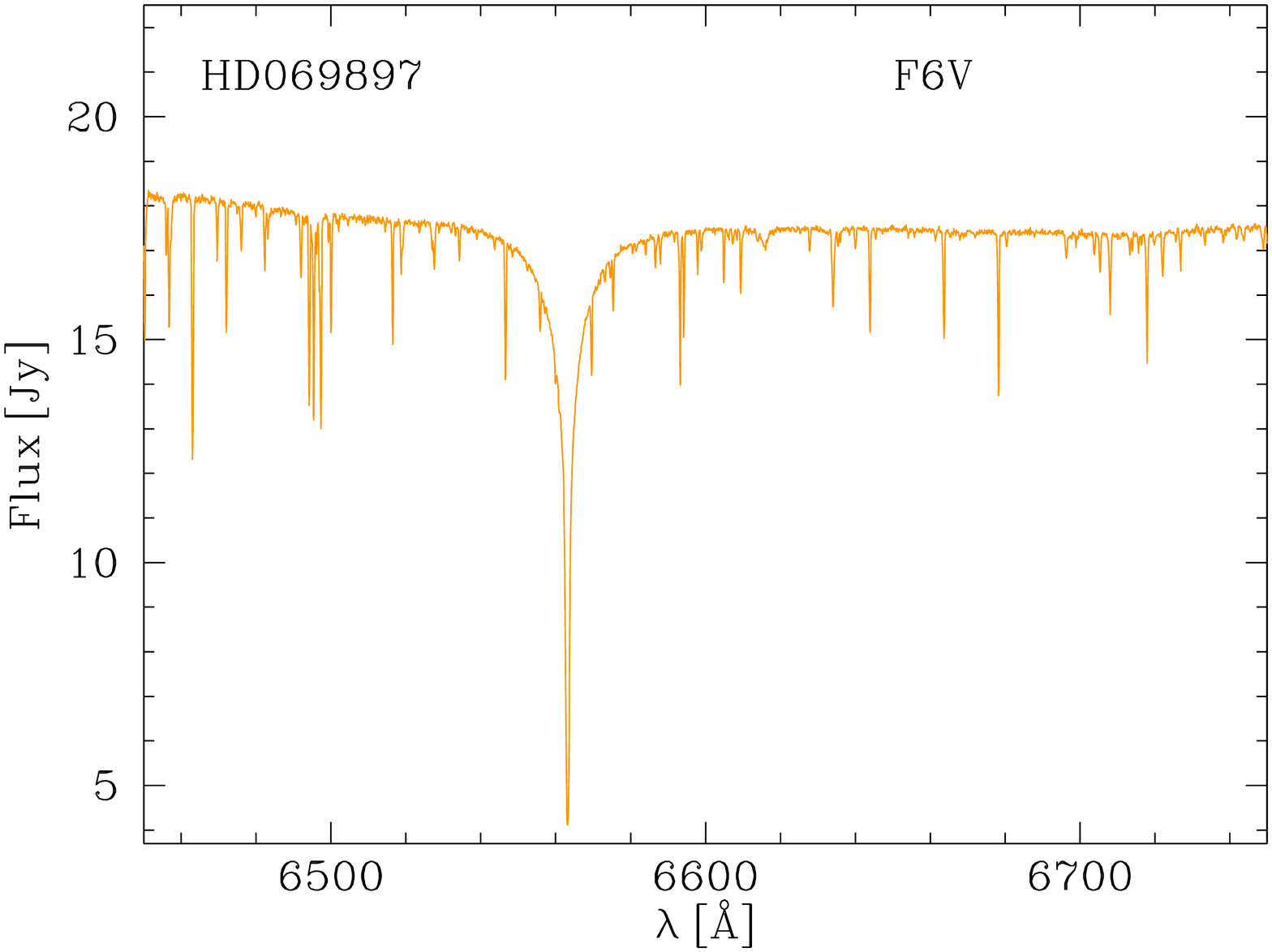}
\includegraphics[width=0.18\textwidth,angle=-90]{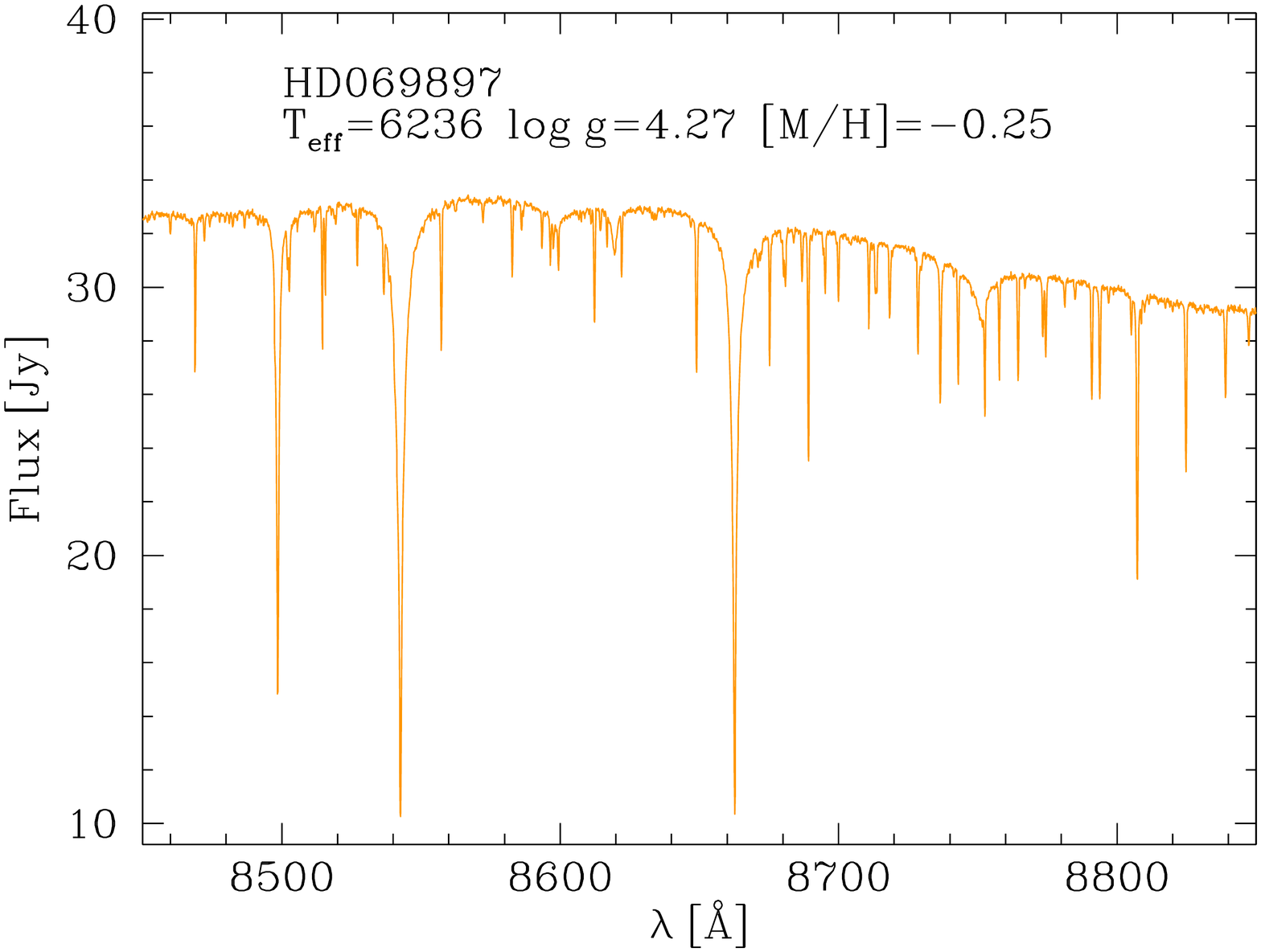}

\contcaption{14. Stars shown in this page are: HD062613, HD063302, HD063778, HD064090, HD064332, HD064412, HD064606, HD065123, HD065583, HD066573, HD067767, HD068017, HD068638 and HD069897.}
\end{figure*}

\begin{figure*}
\includegraphics[width=0.18\textwidth,angle=-90]{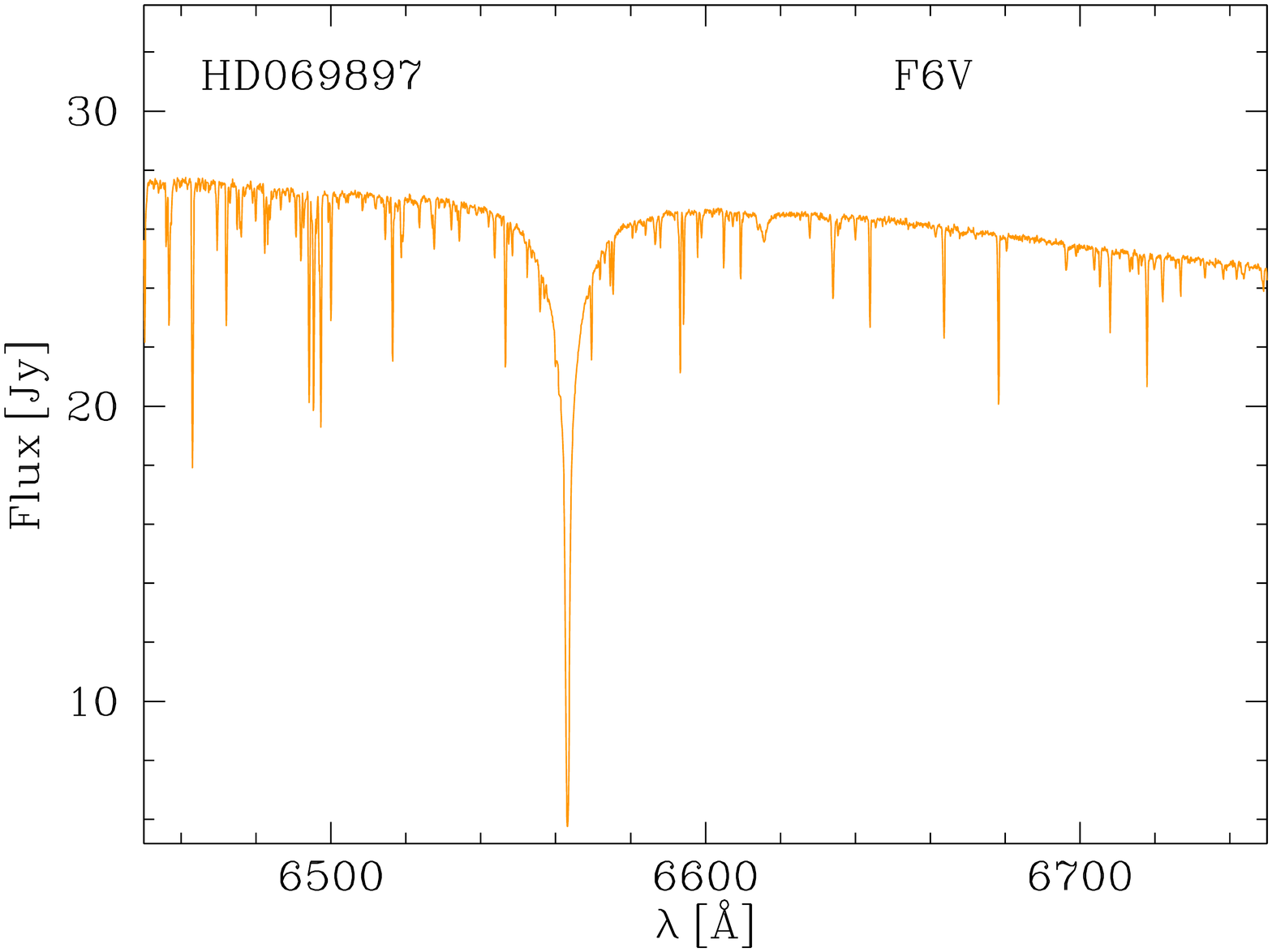}
\includegraphics[width=0.18\textwidth,angle=-90]{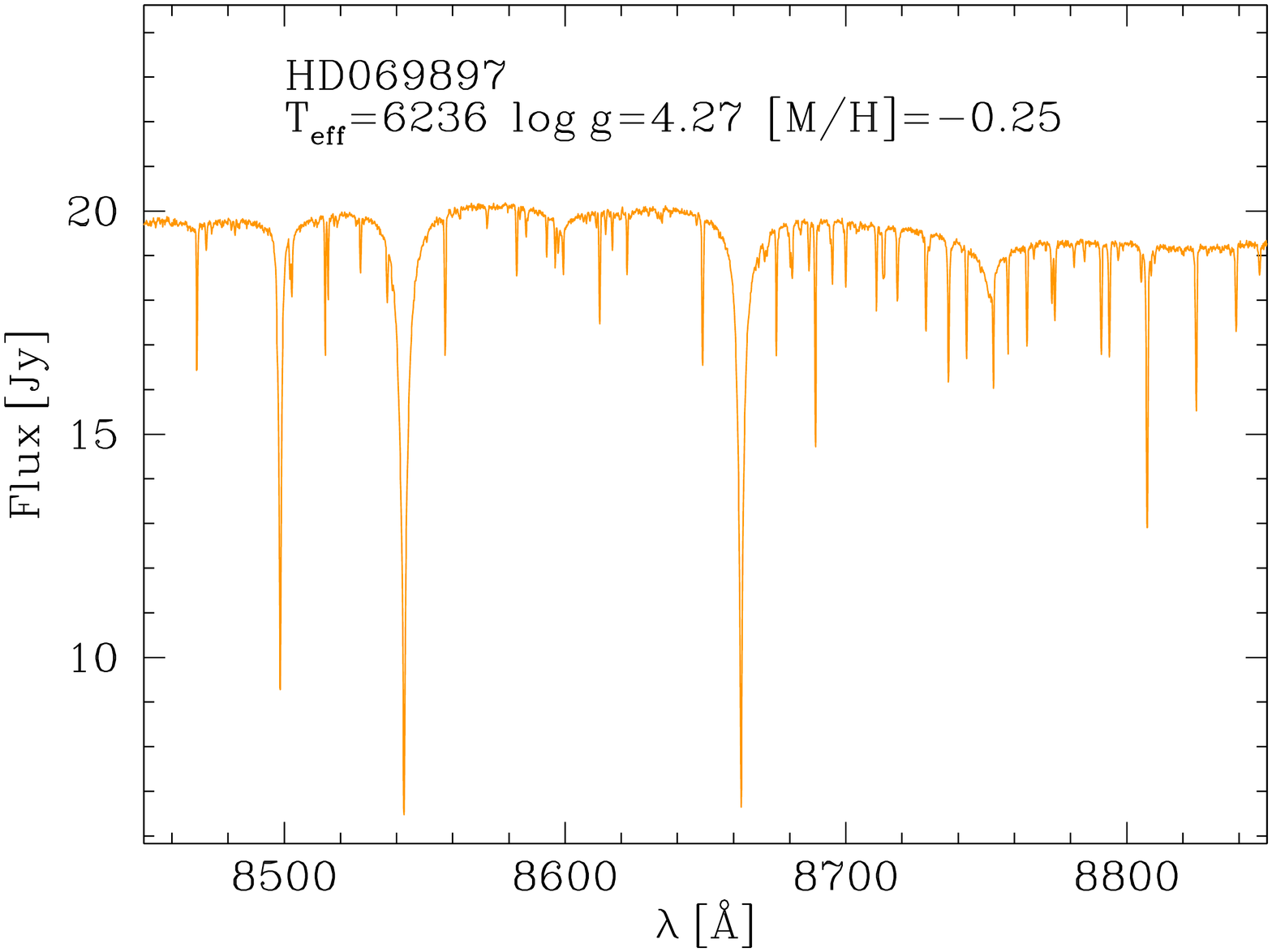}
\includegraphics[width=0.18\textwidth,angle=-90]{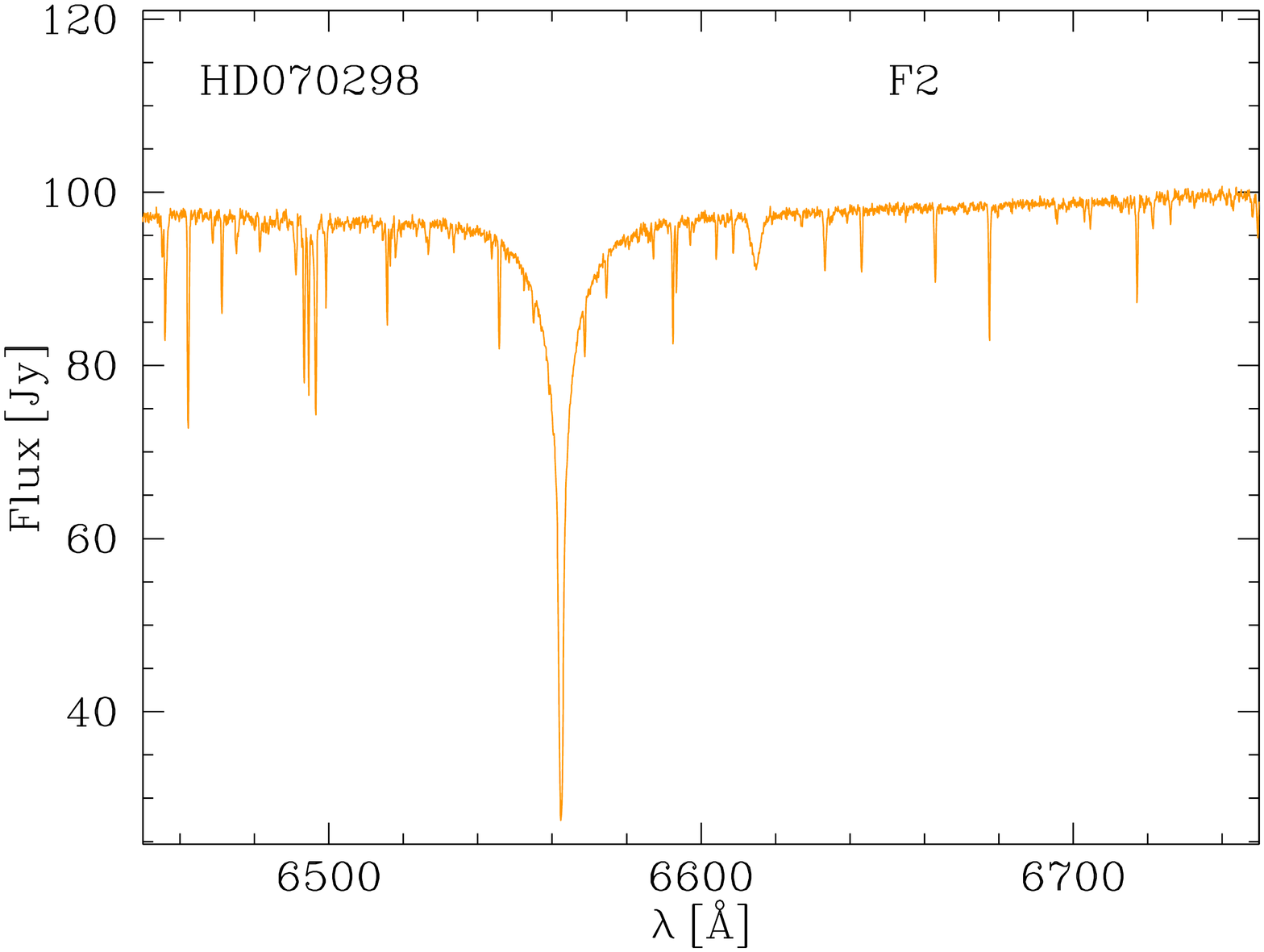}
\includegraphics[width=0.18\textwidth,angle=-90]{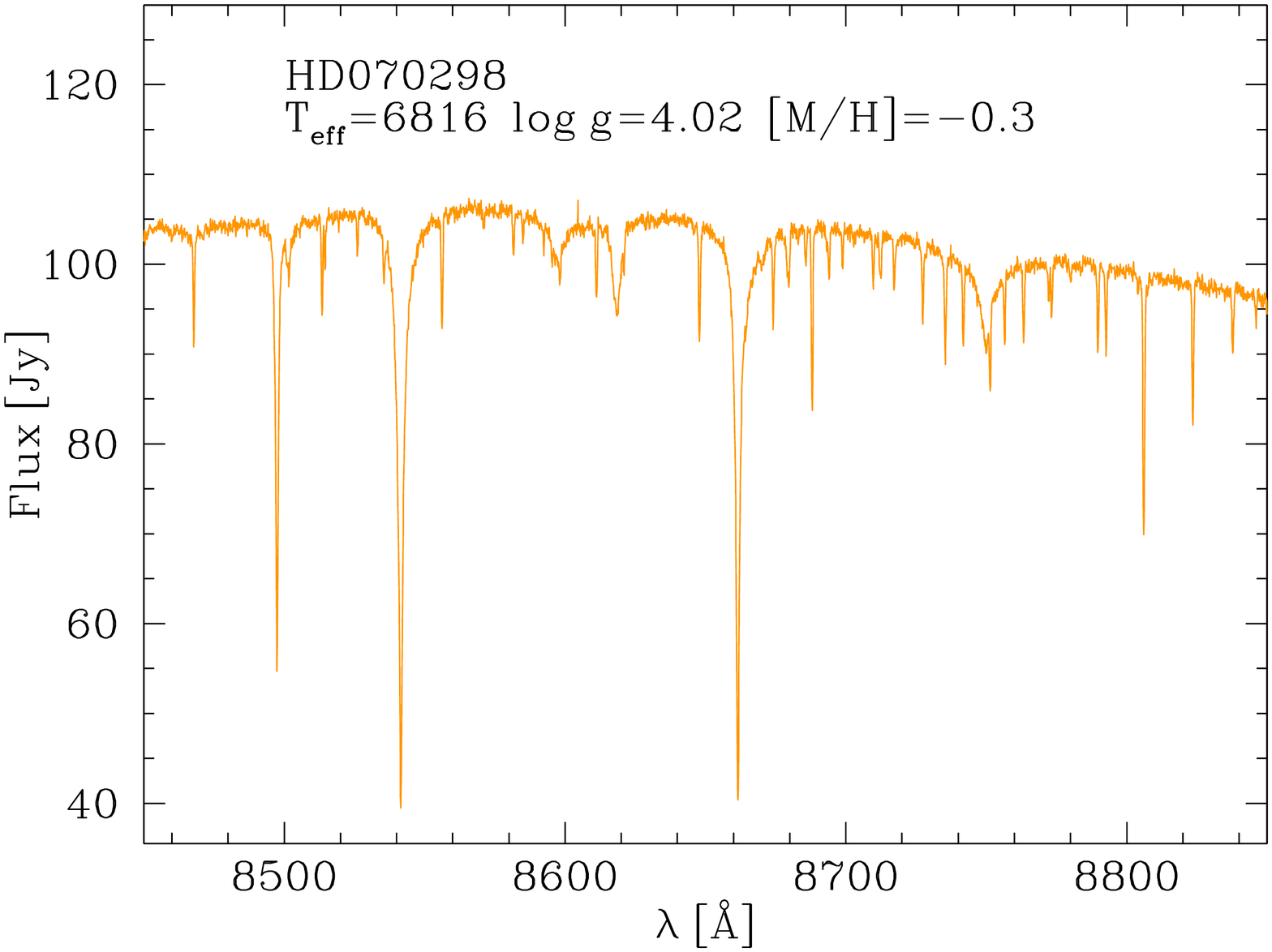}
\includegraphics[width=0.18\textwidth,angle=-90]{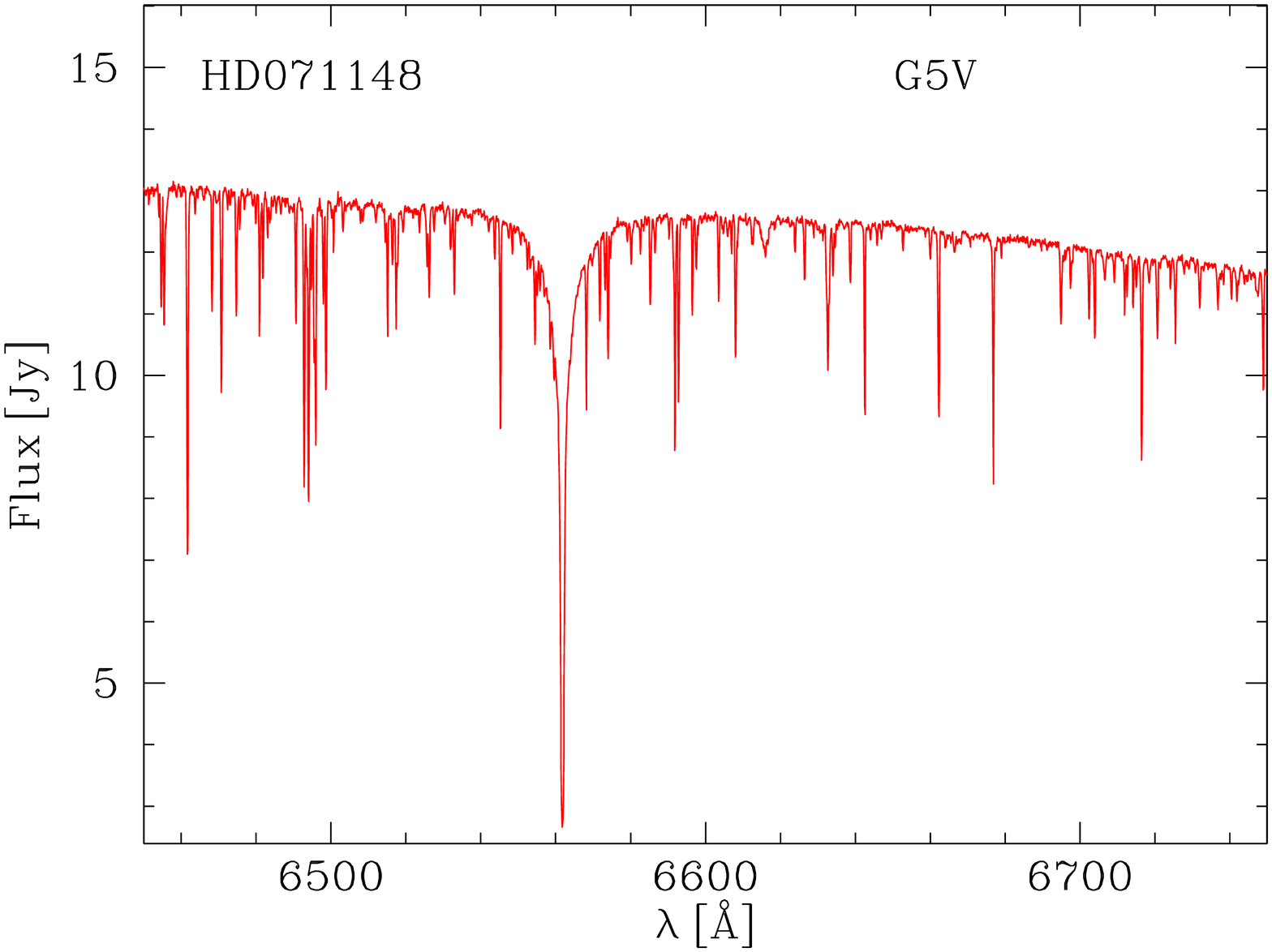}
\includegraphics[width=0.18\textwidth,angle=-90]{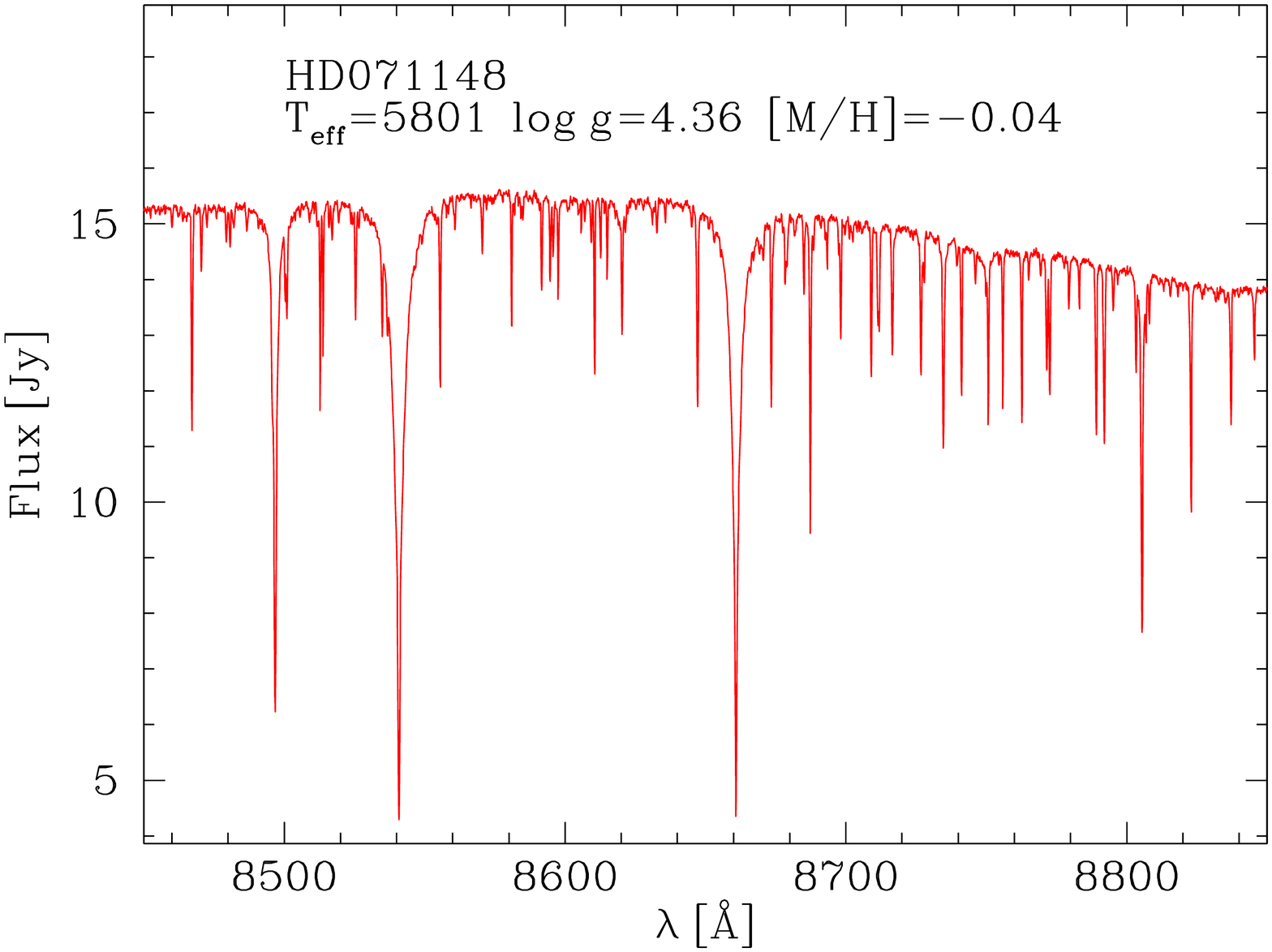}
\includegraphics[width=0.18\textwidth,angle=-90]{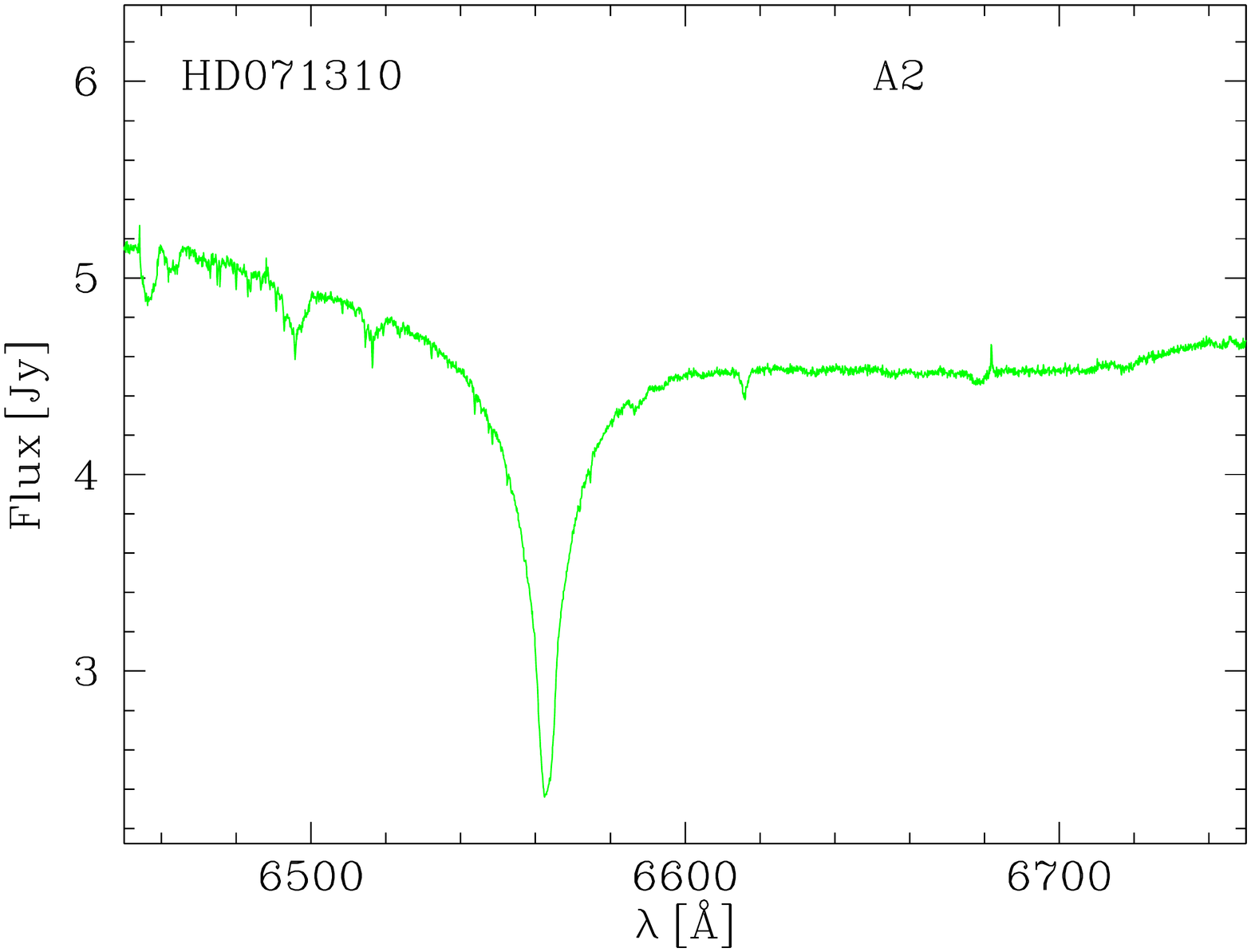}
\includegraphics[width=0.18\textwidth,angle=-90]{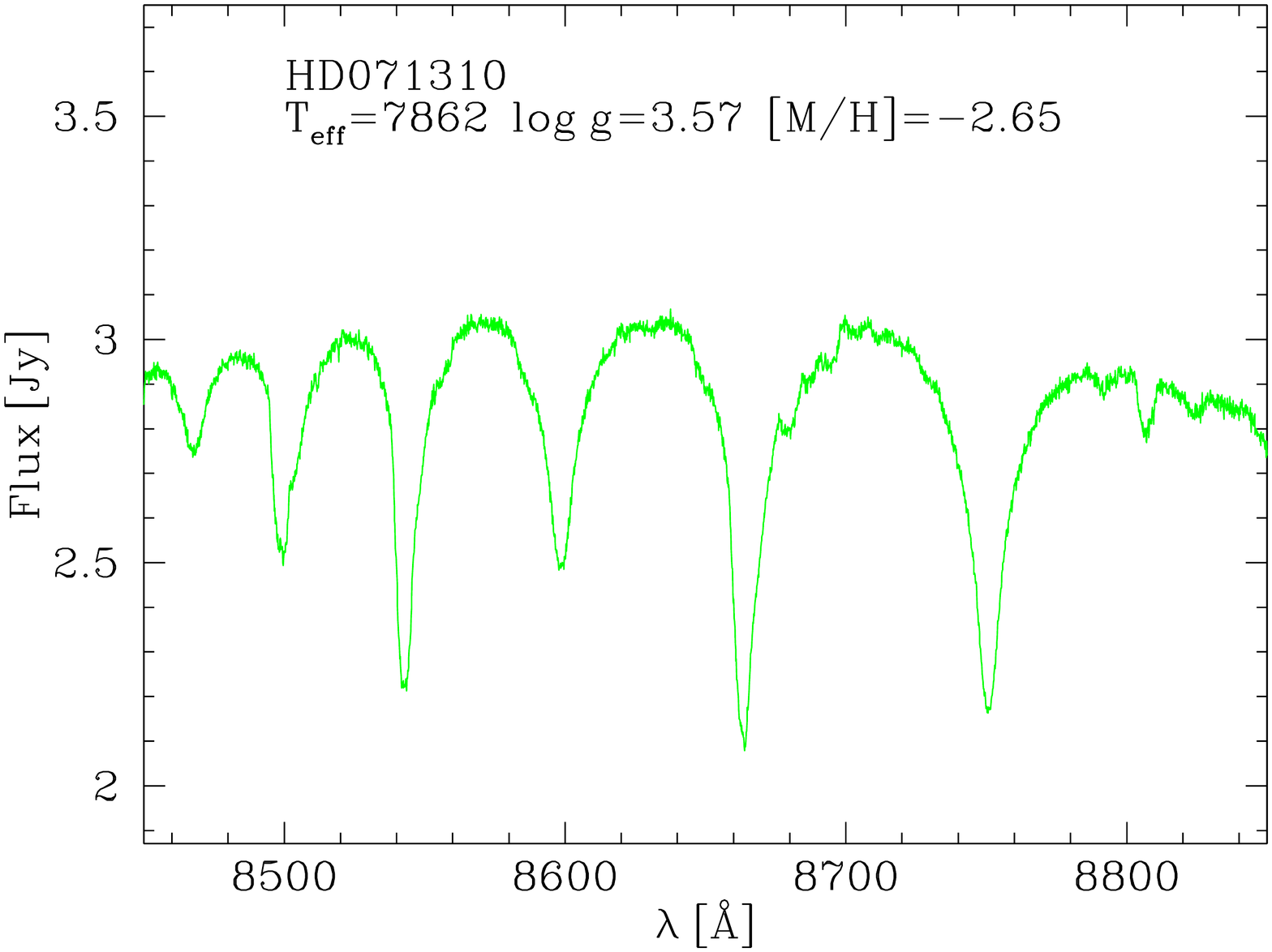}
\includegraphics[width=0.18\textwidth,angle=-90]{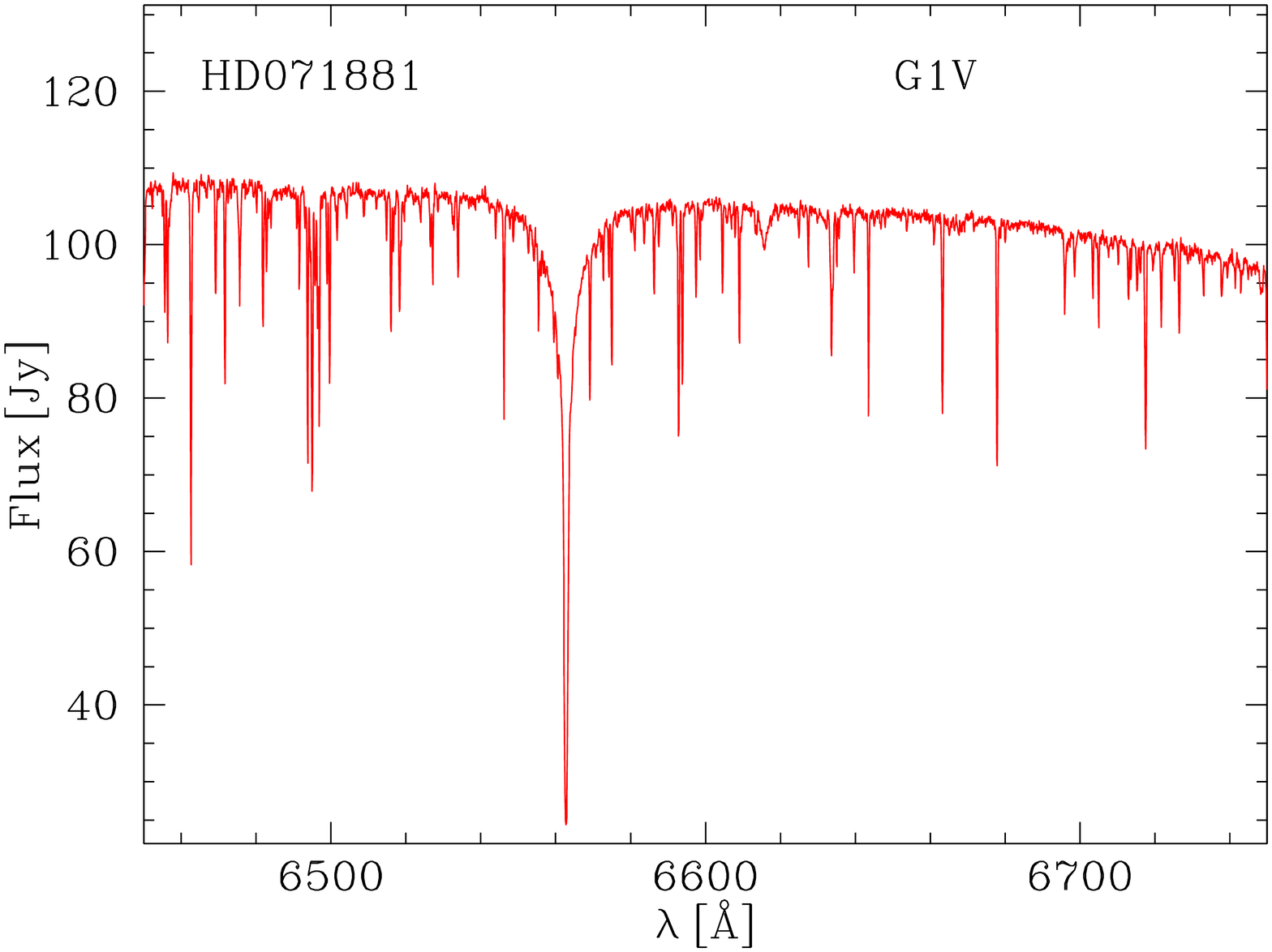}
\includegraphics[width=0.18\textwidth,angle=-90]{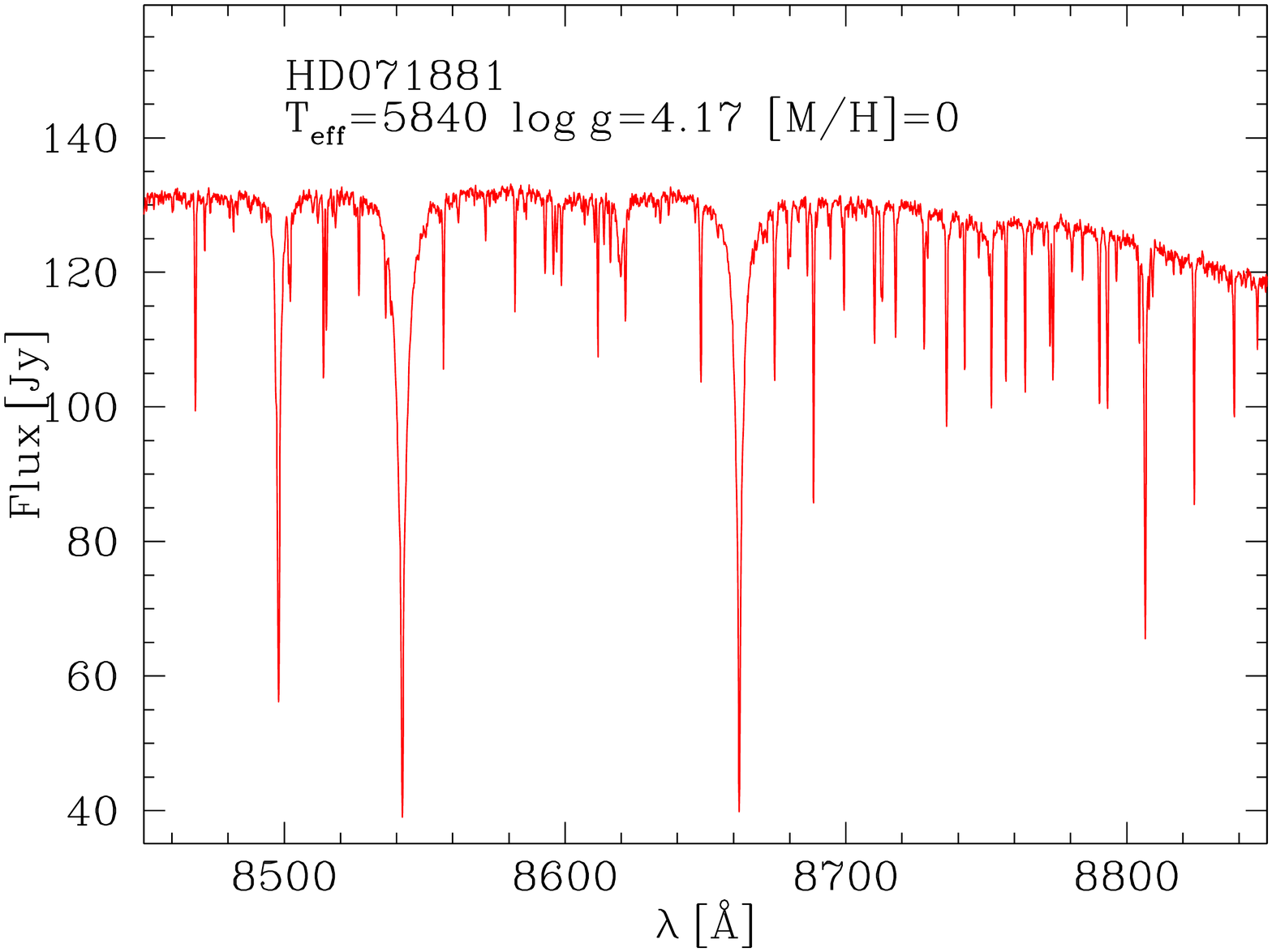}
\includegraphics[width=0.18\textwidth,angle=-90]{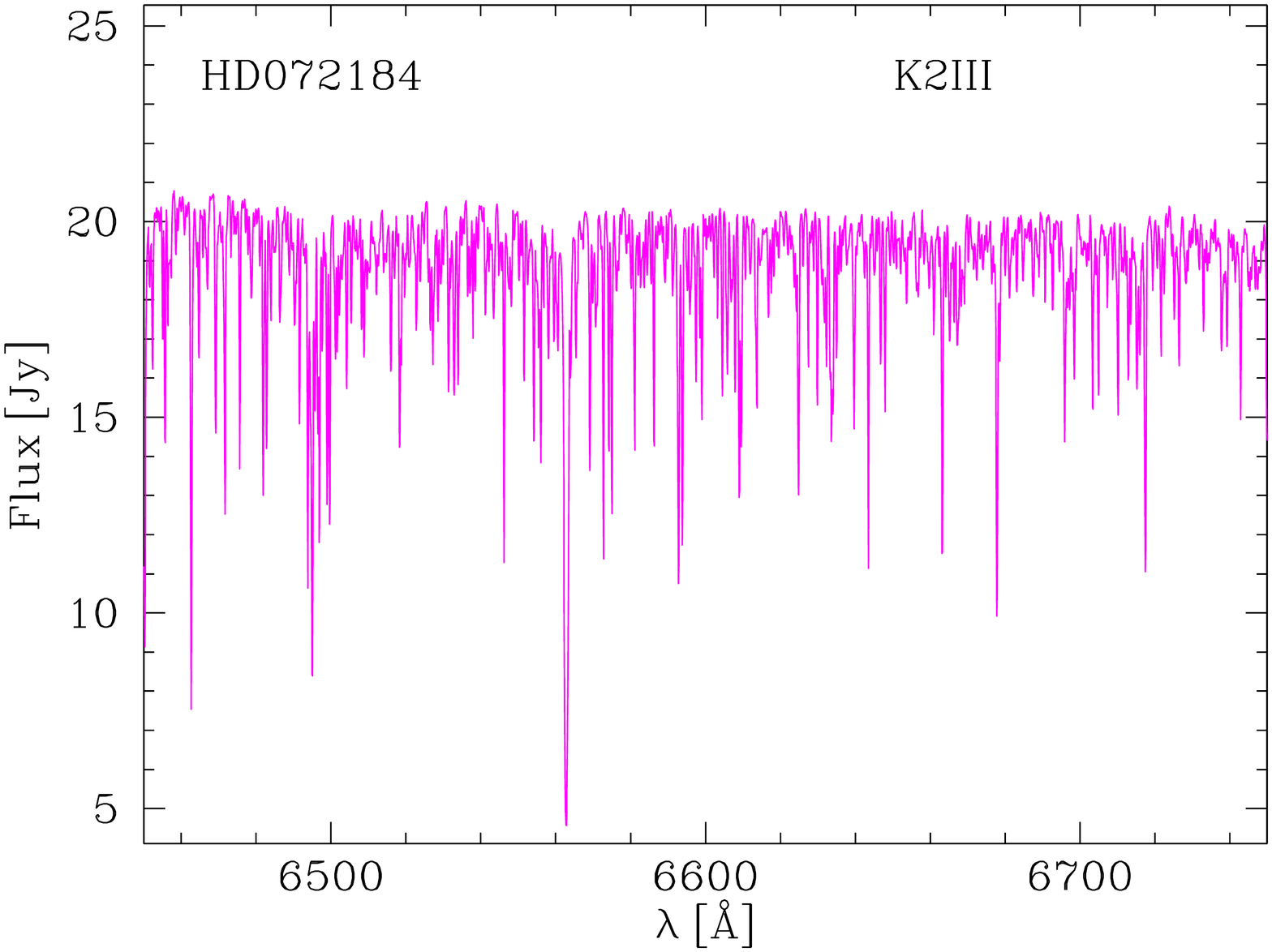}
\includegraphics[width=0.18\textwidth,angle=-90]{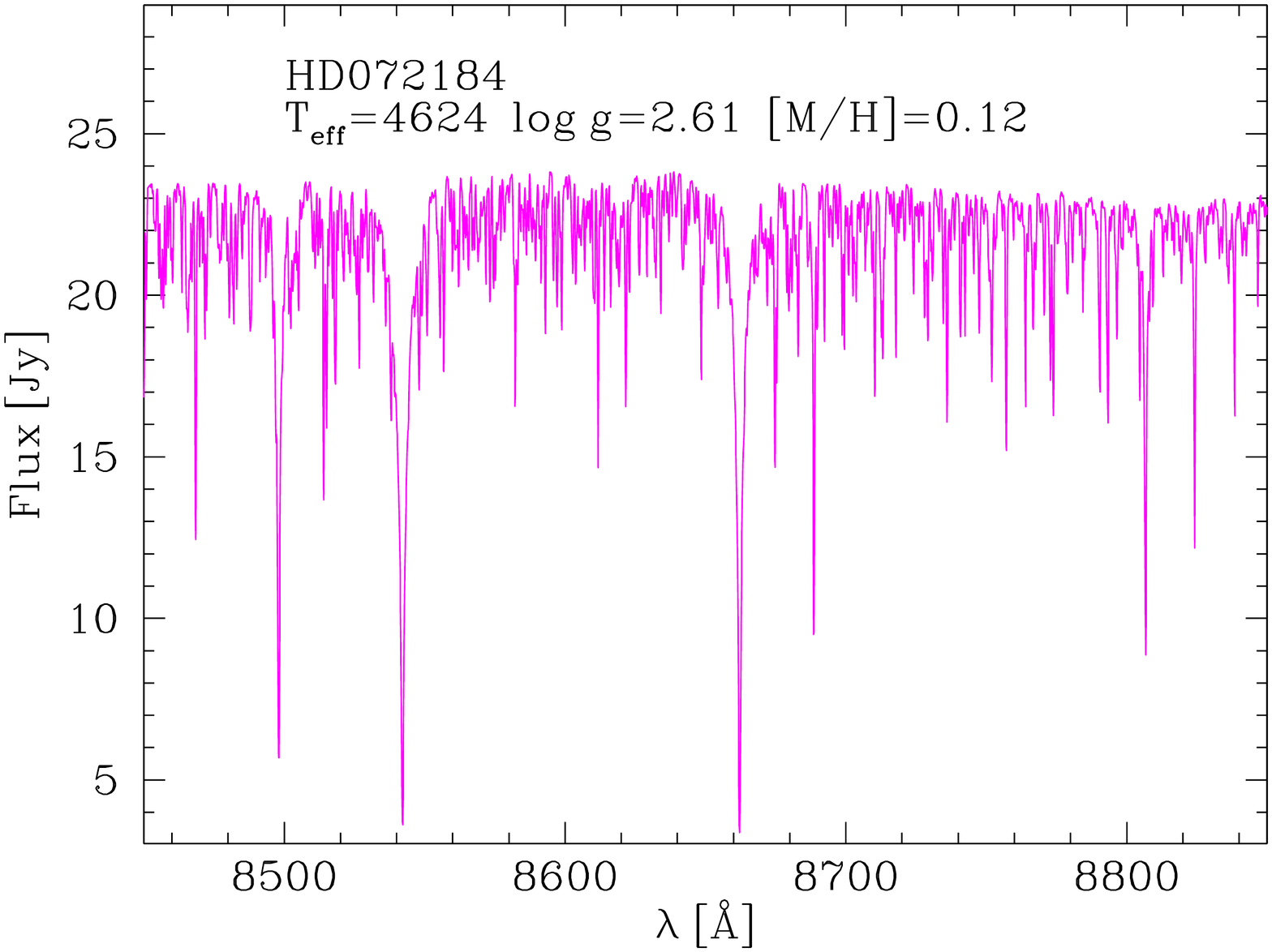}
\includegraphics[width=0.18\textwidth,angle=-90]{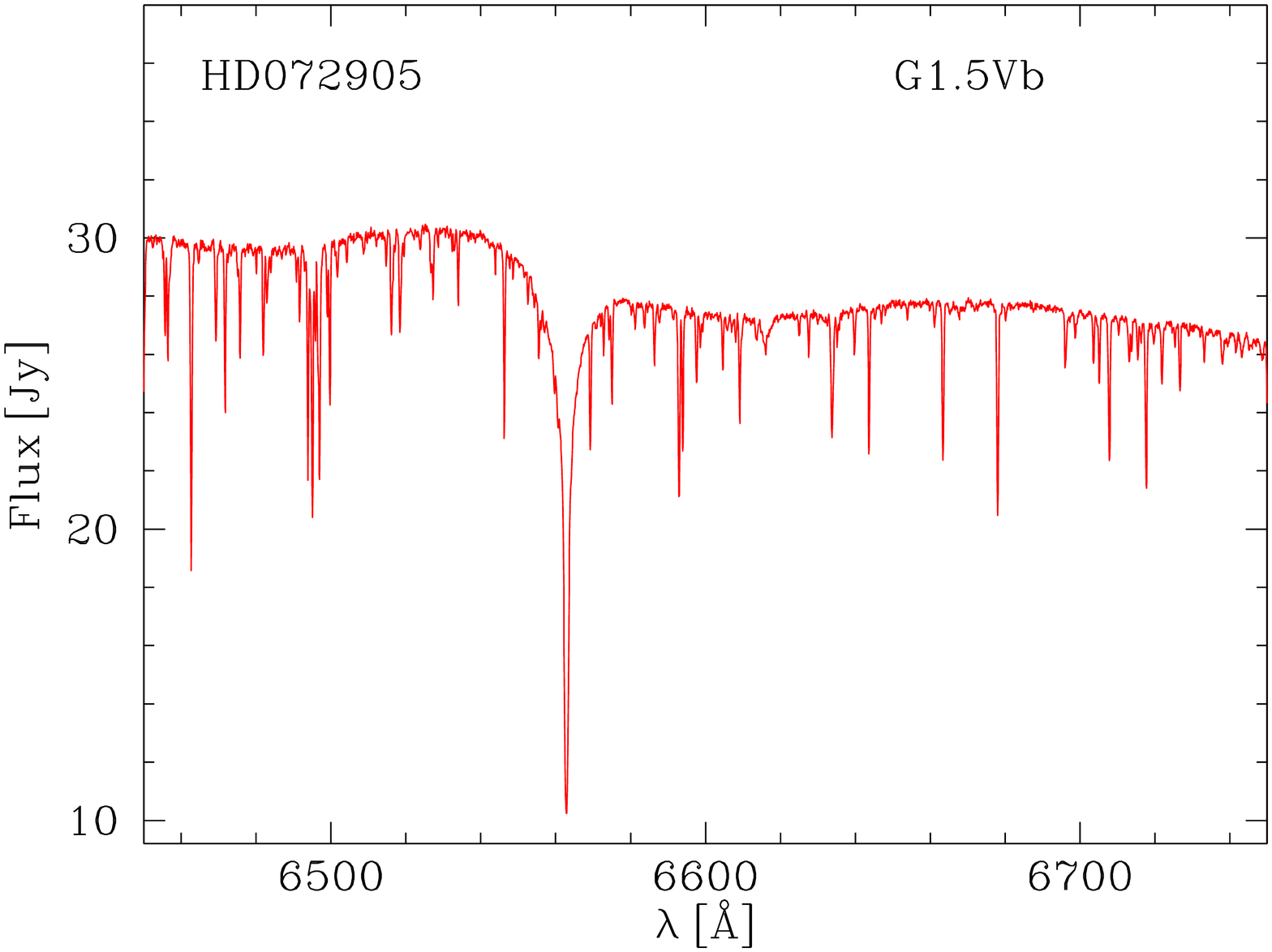}
\includegraphics[width=0.18\textwidth,angle=-90]{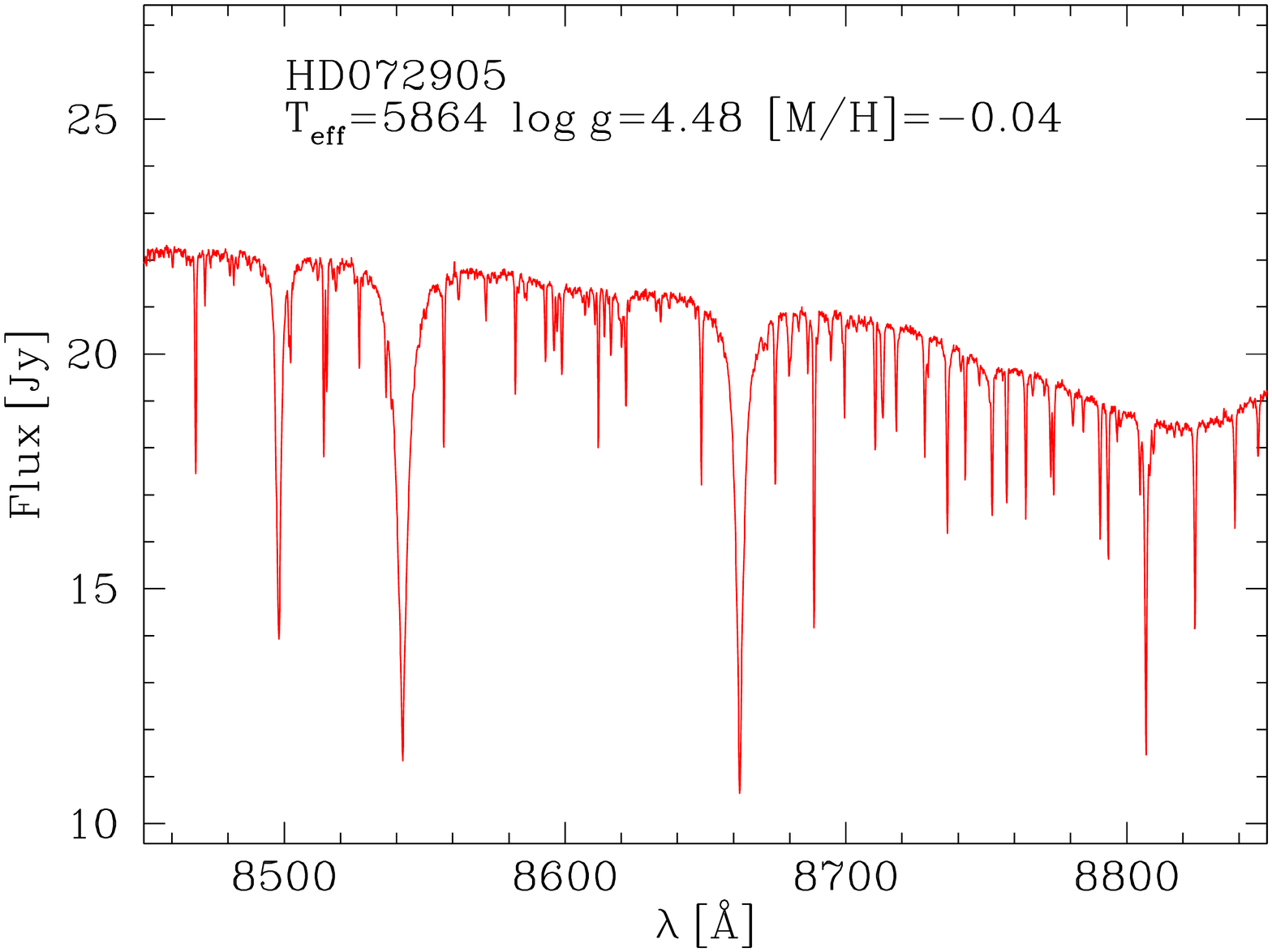}
\includegraphics[width=0.18\textwidth,angle=-90]{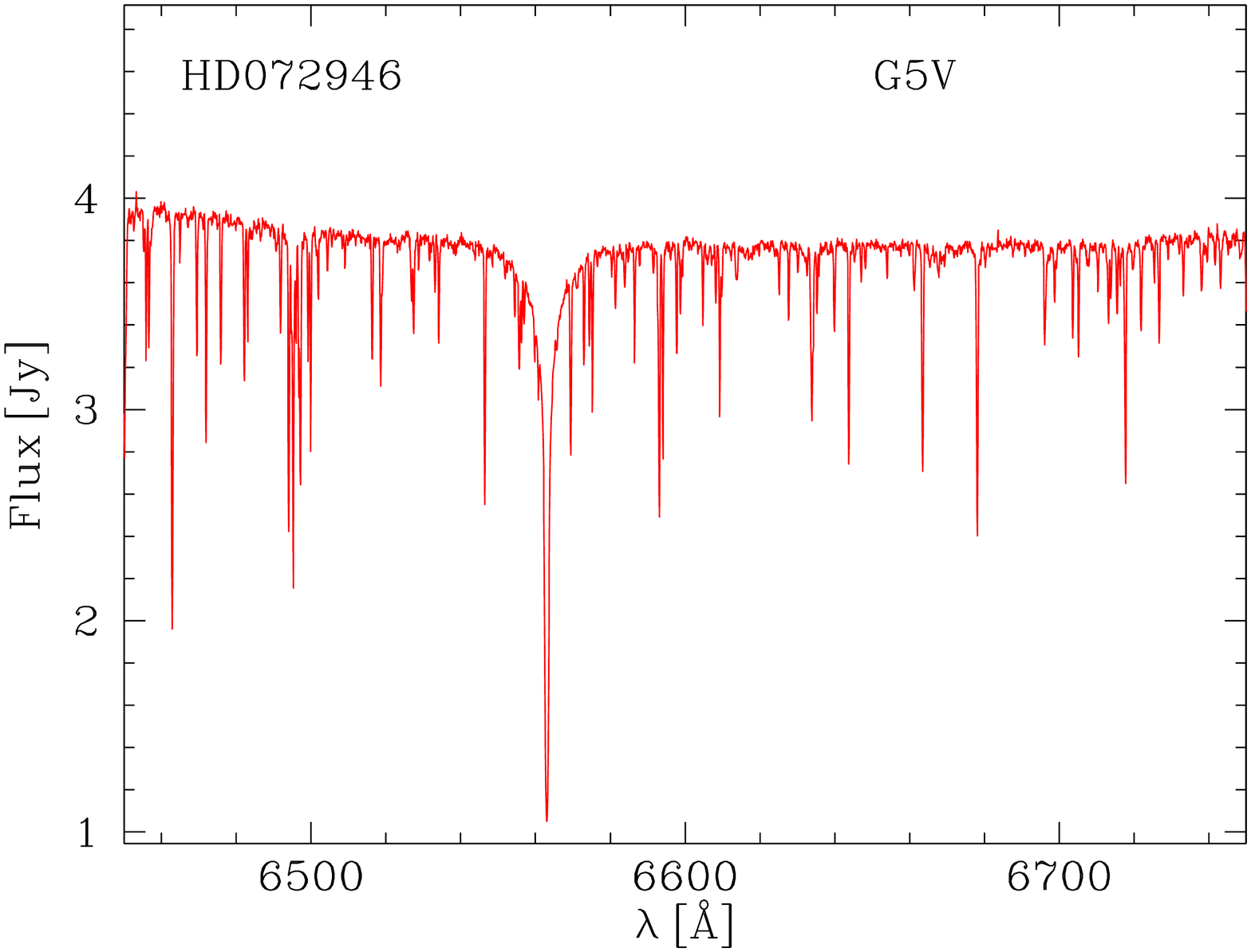}
\includegraphics[width=0.18\textwidth,angle=-90]{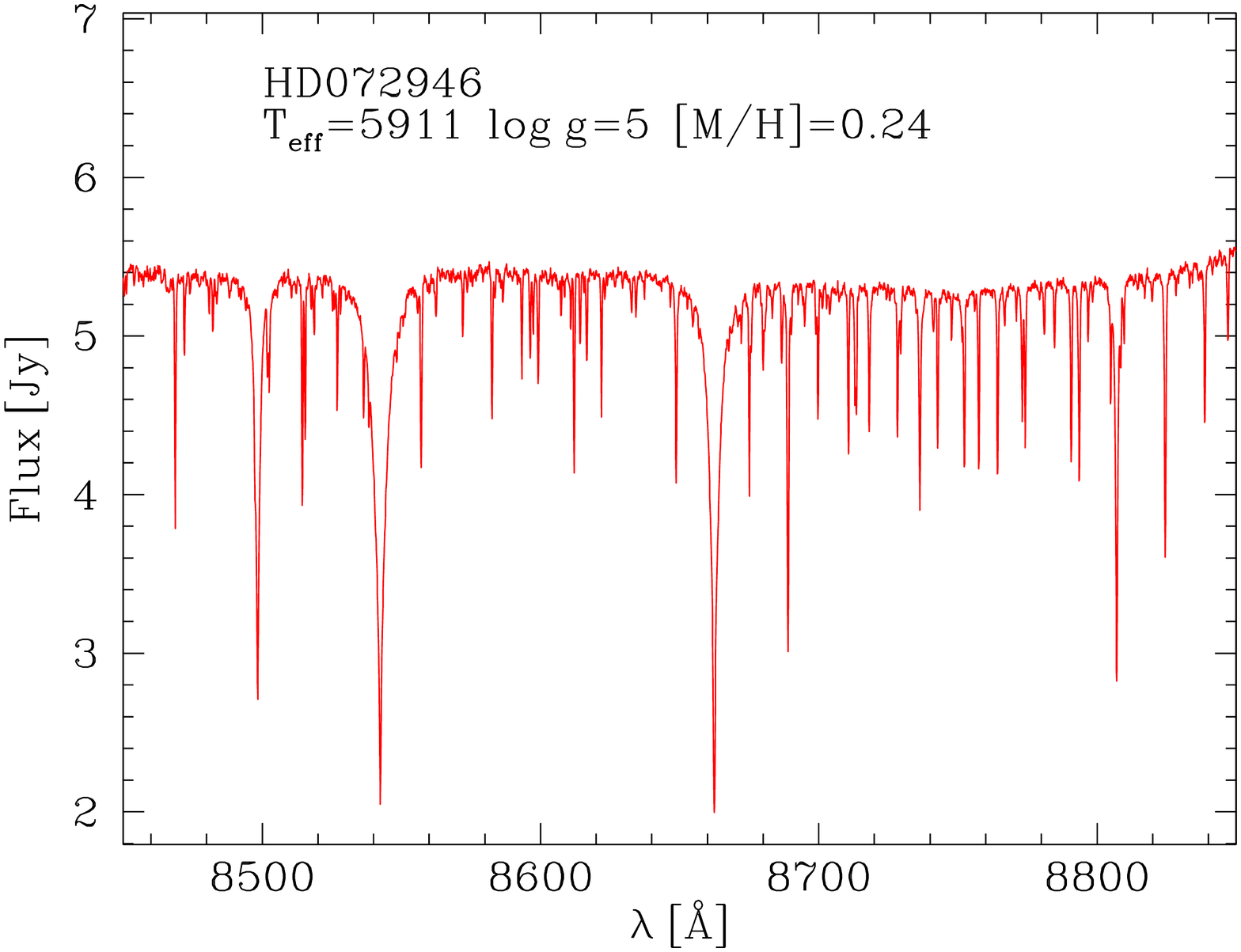}
\includegraphics[width=0.18\textwidth,angle=-90]{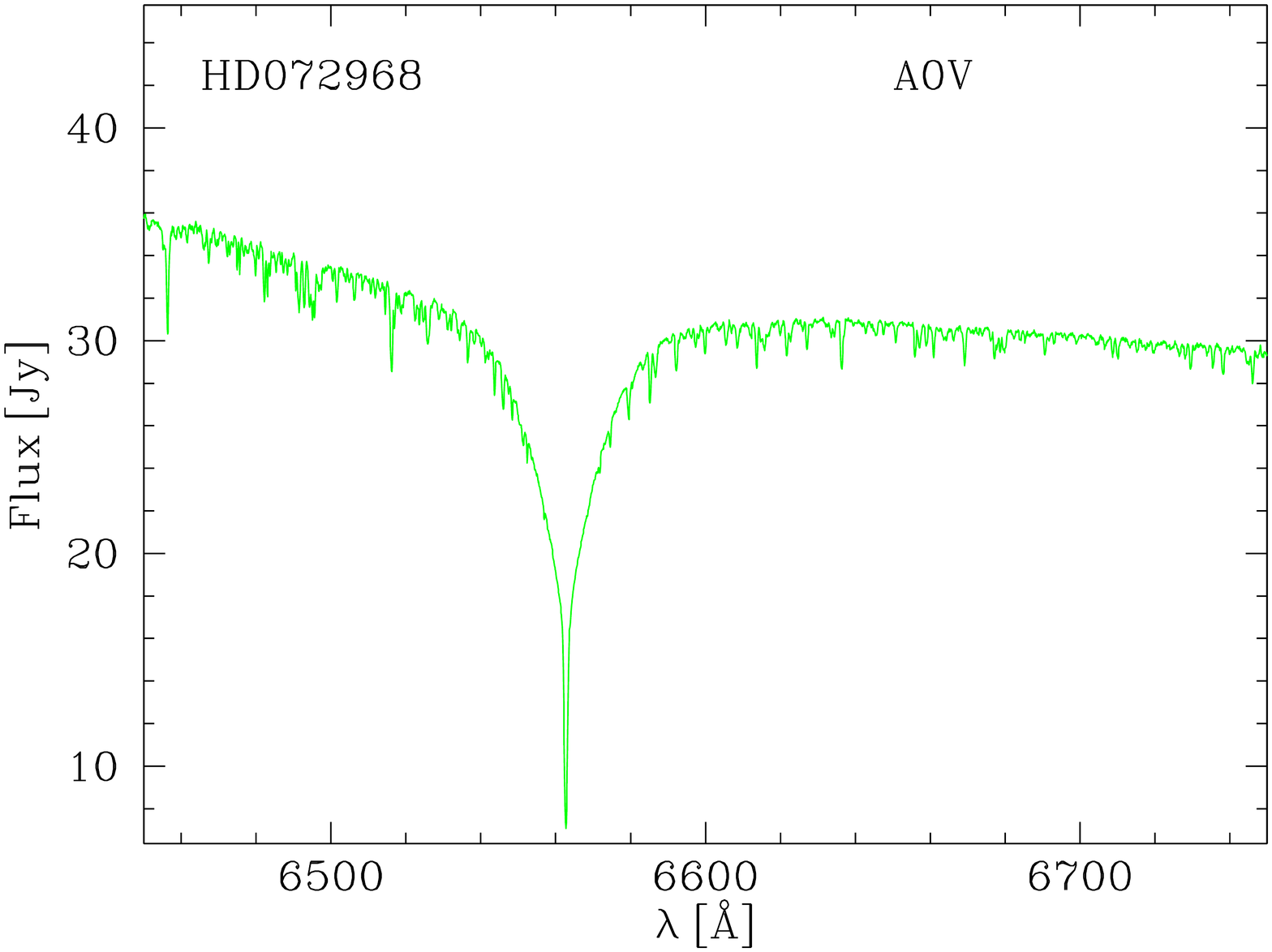}
\includegraphics[width=0.18\textwidth,angle=-90]{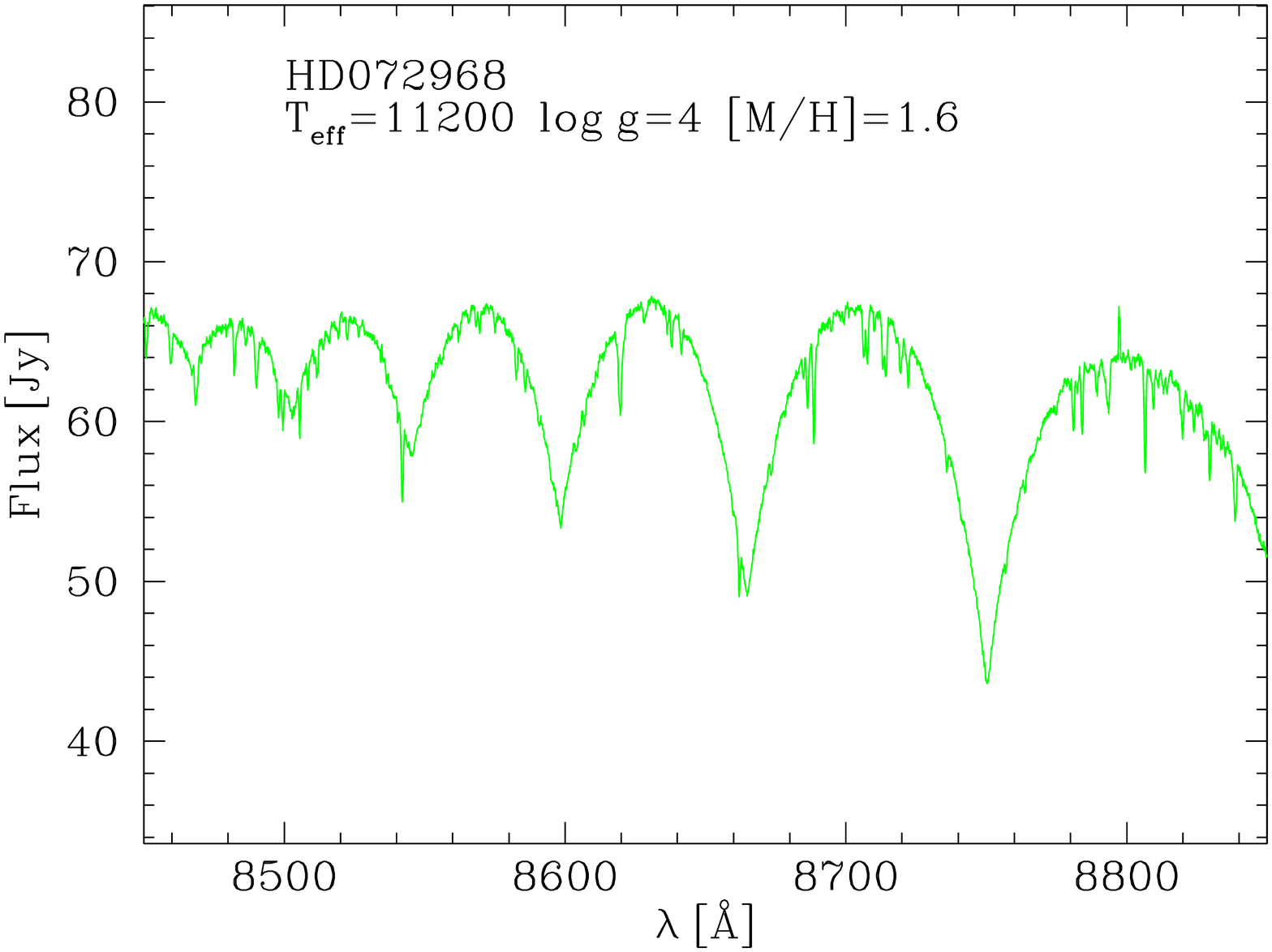}
\includegraphics[width=0.18\textwidth,angle=-90]{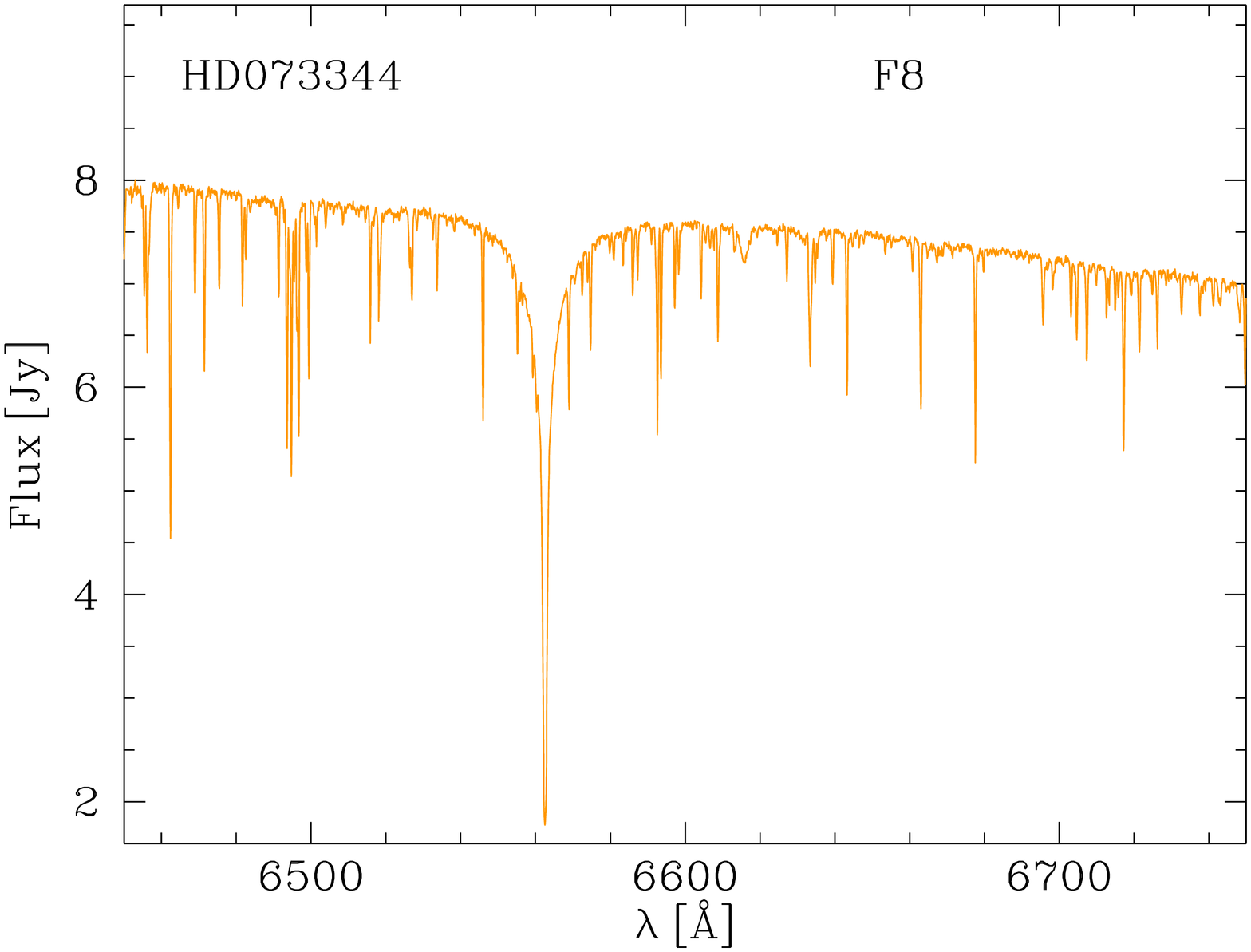}
\includegraphics[width=0.18\textwidth,angle=-90]{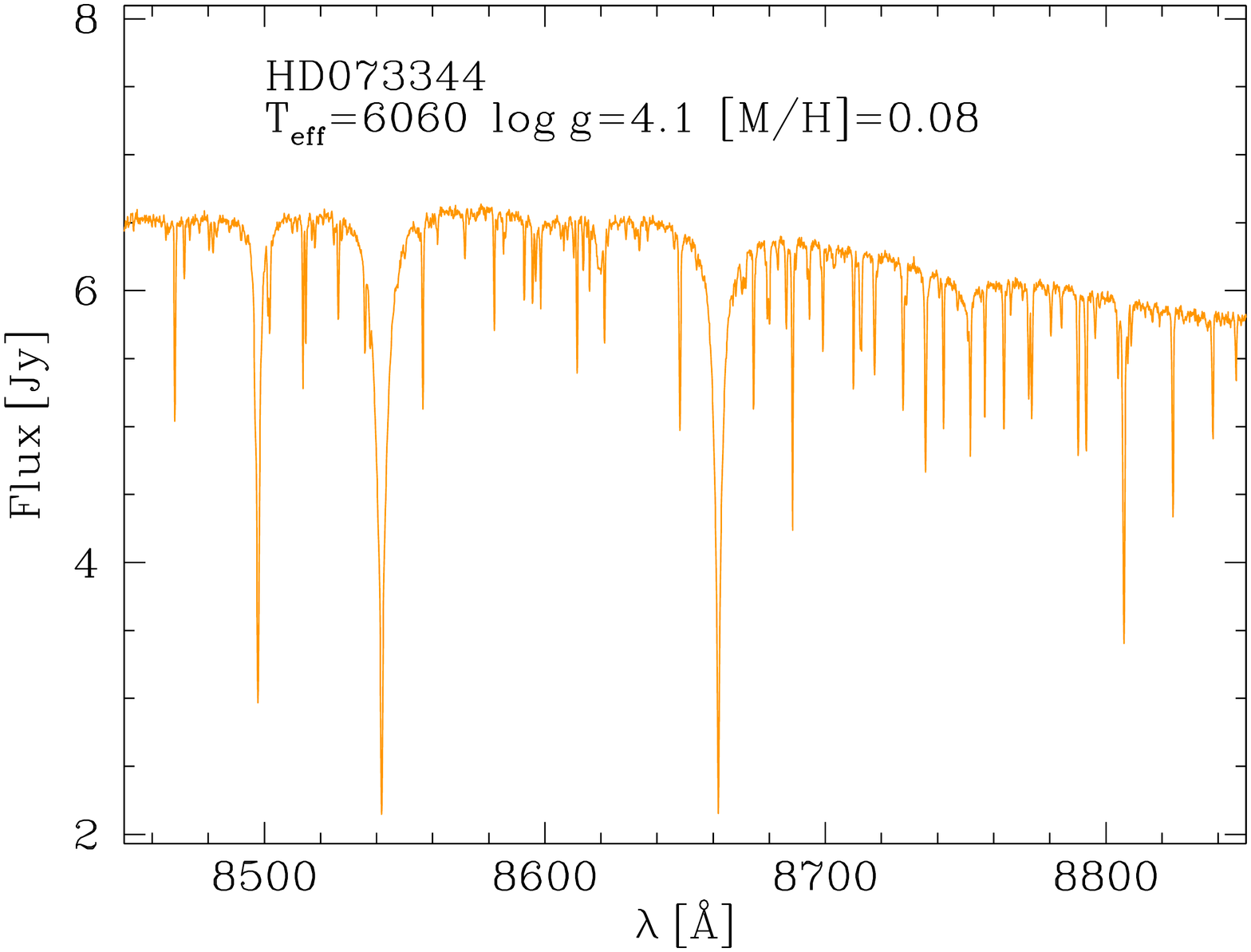}
\includegraphics[width=0.18\textwidth,angle=-90]{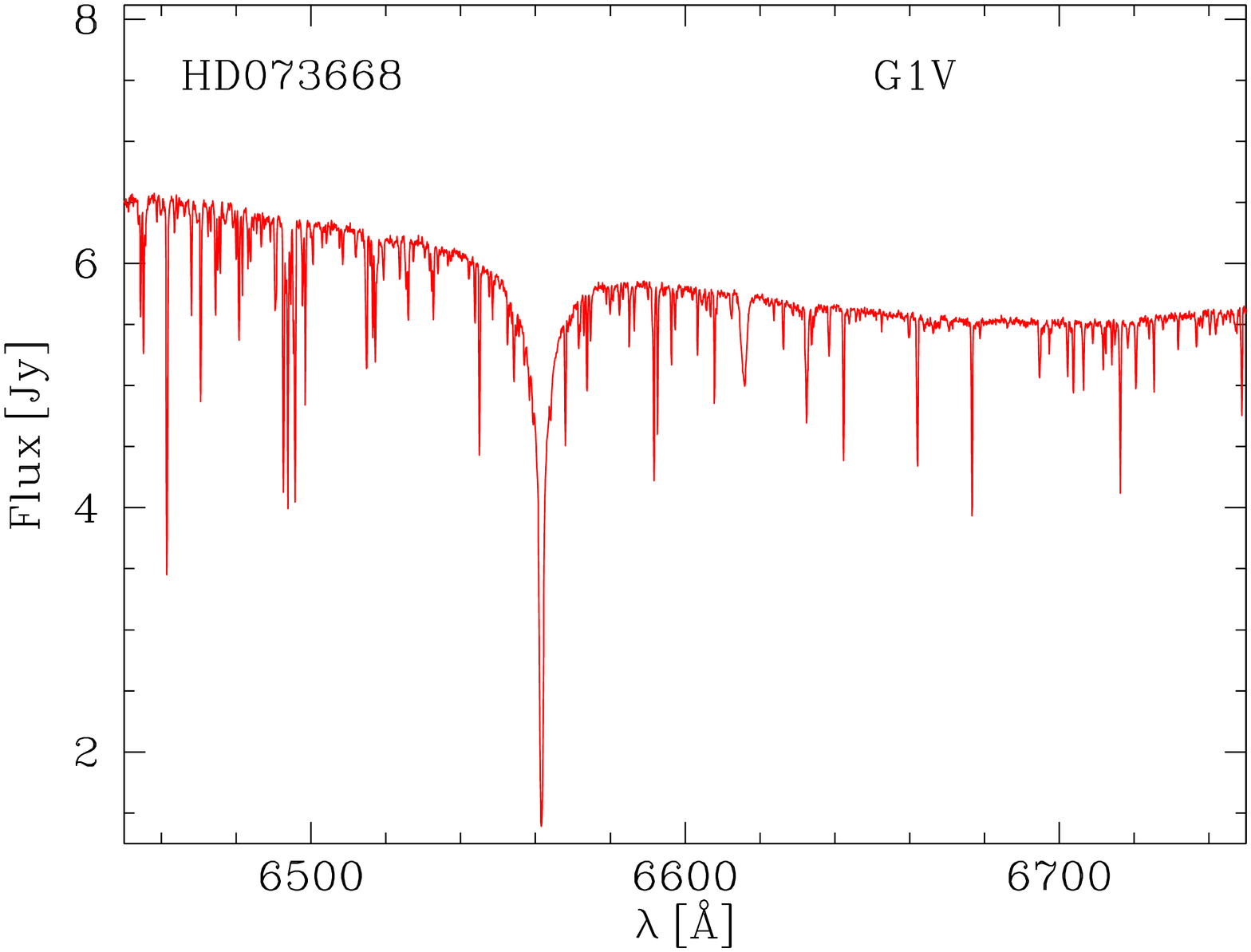}
\includegraphics[width=0.18\textwidth,angle=-90]{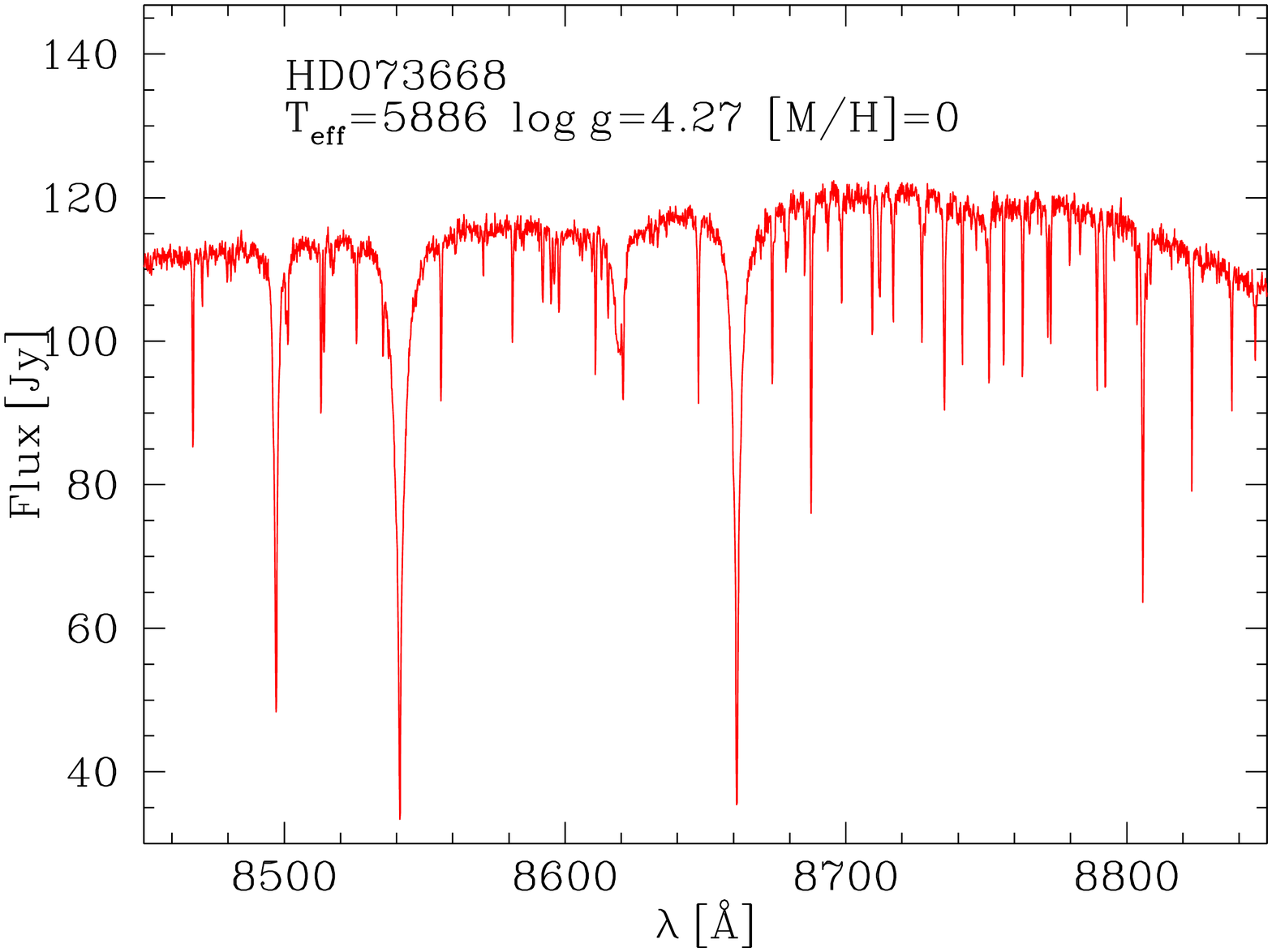}
\includegraphics[width=0.18\textwidth,angle=-90]{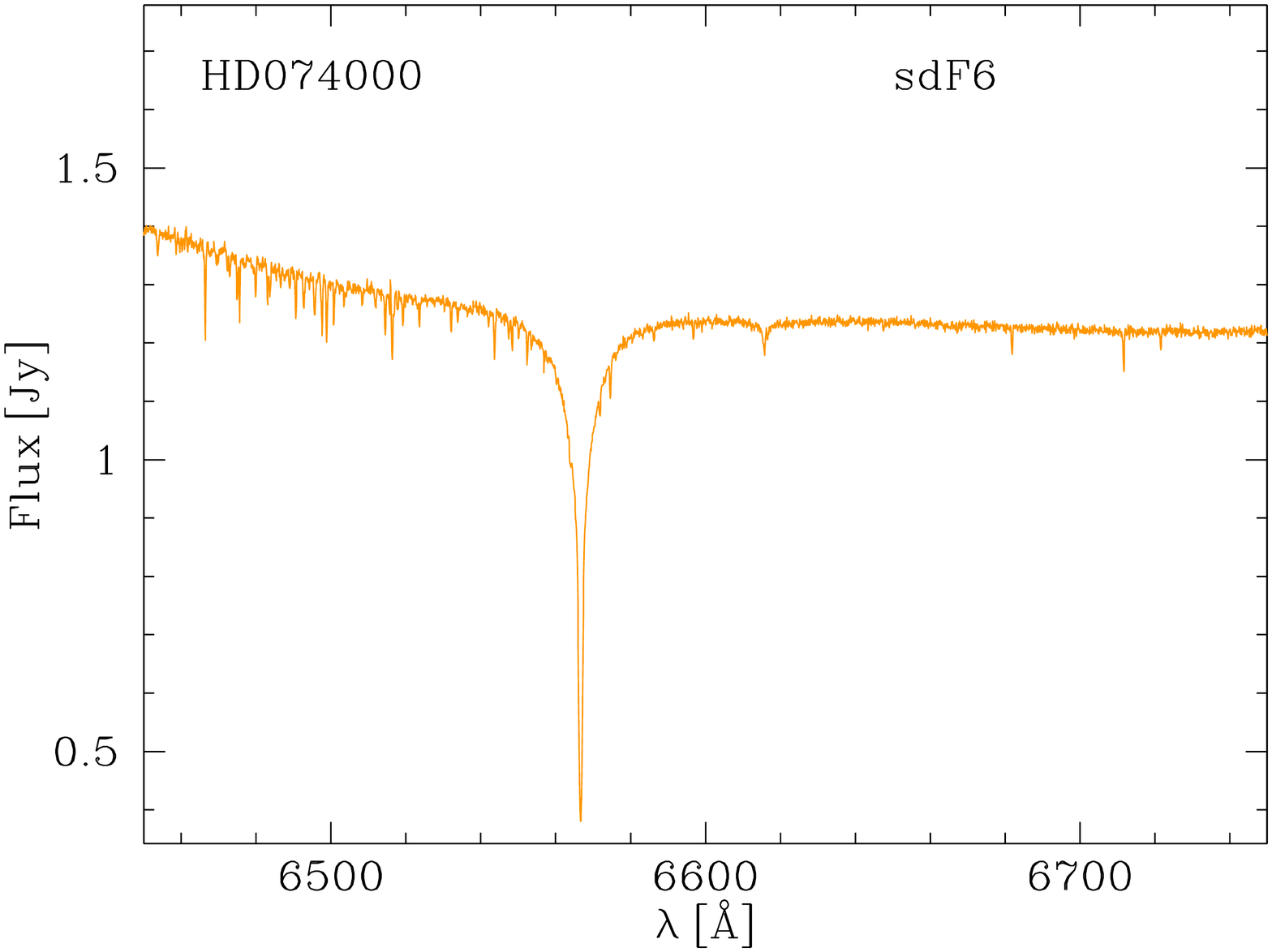}
\includegraphics[width=0.18\textwidth,angle=-90]{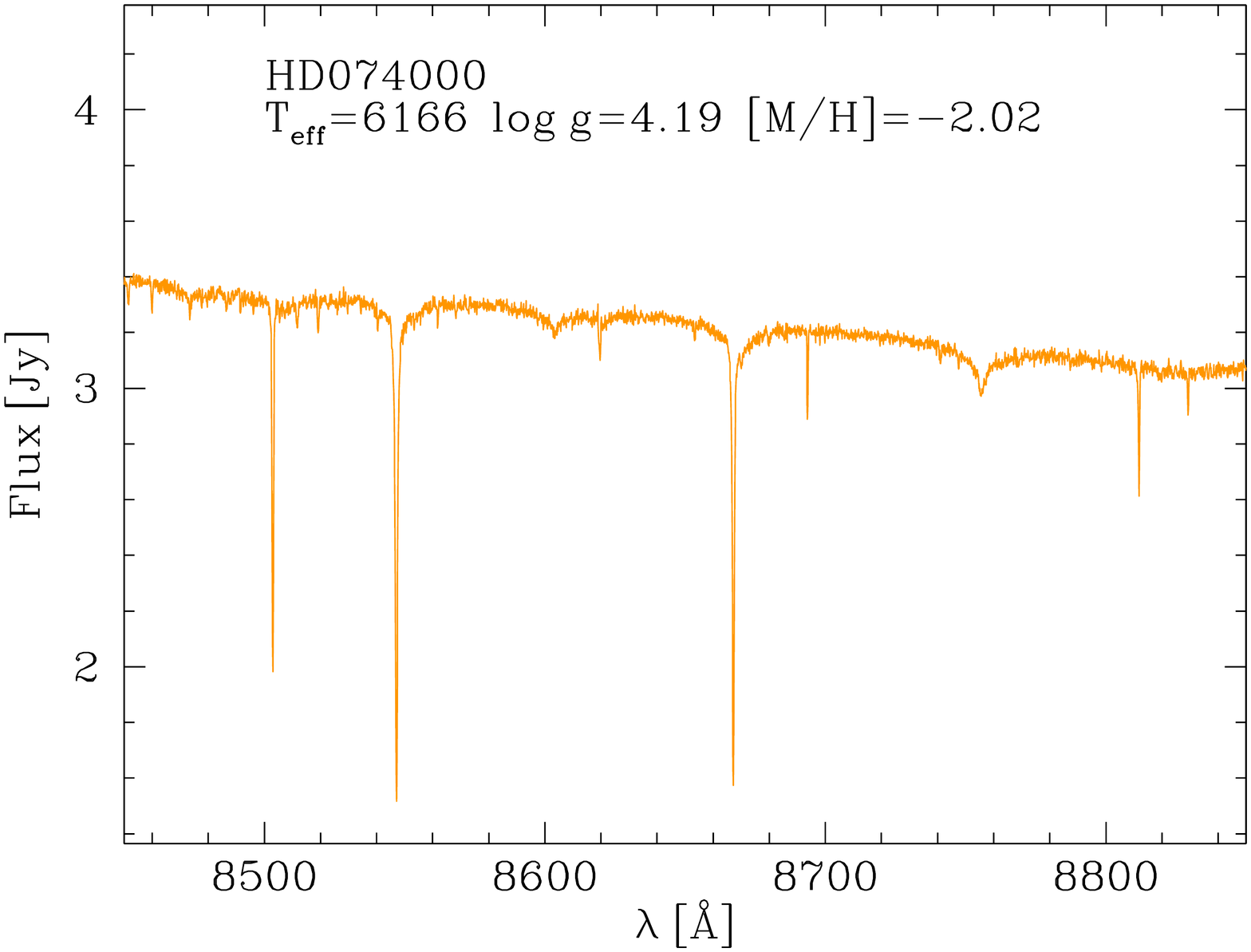}
\includegraphics[width=0.18\textwidth,angle=-90]{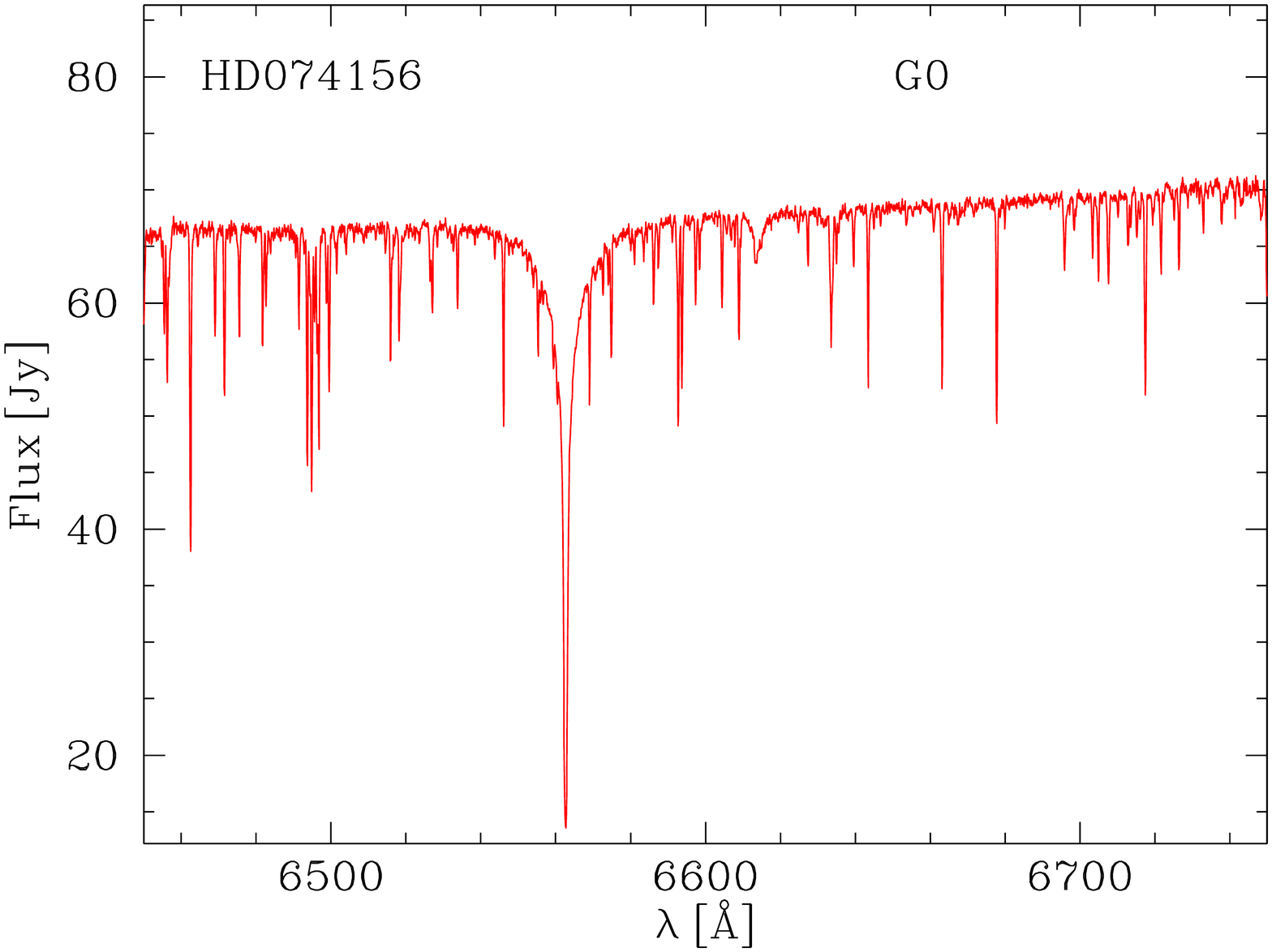}
\includegraphics[width=0.18\textwidth,angle=-90]{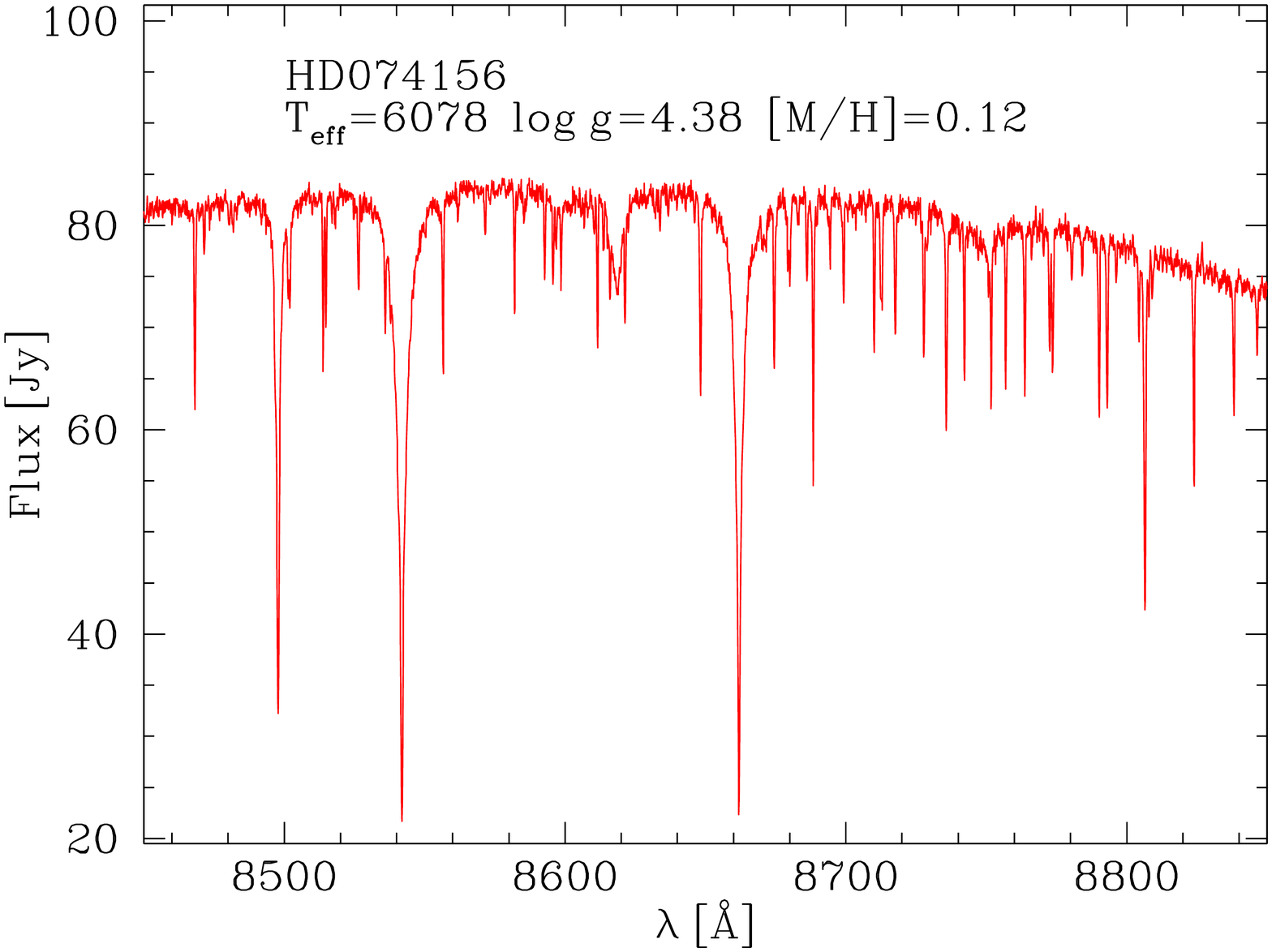}
\includegraphics[width=0.18\textwidth,angle=-90]{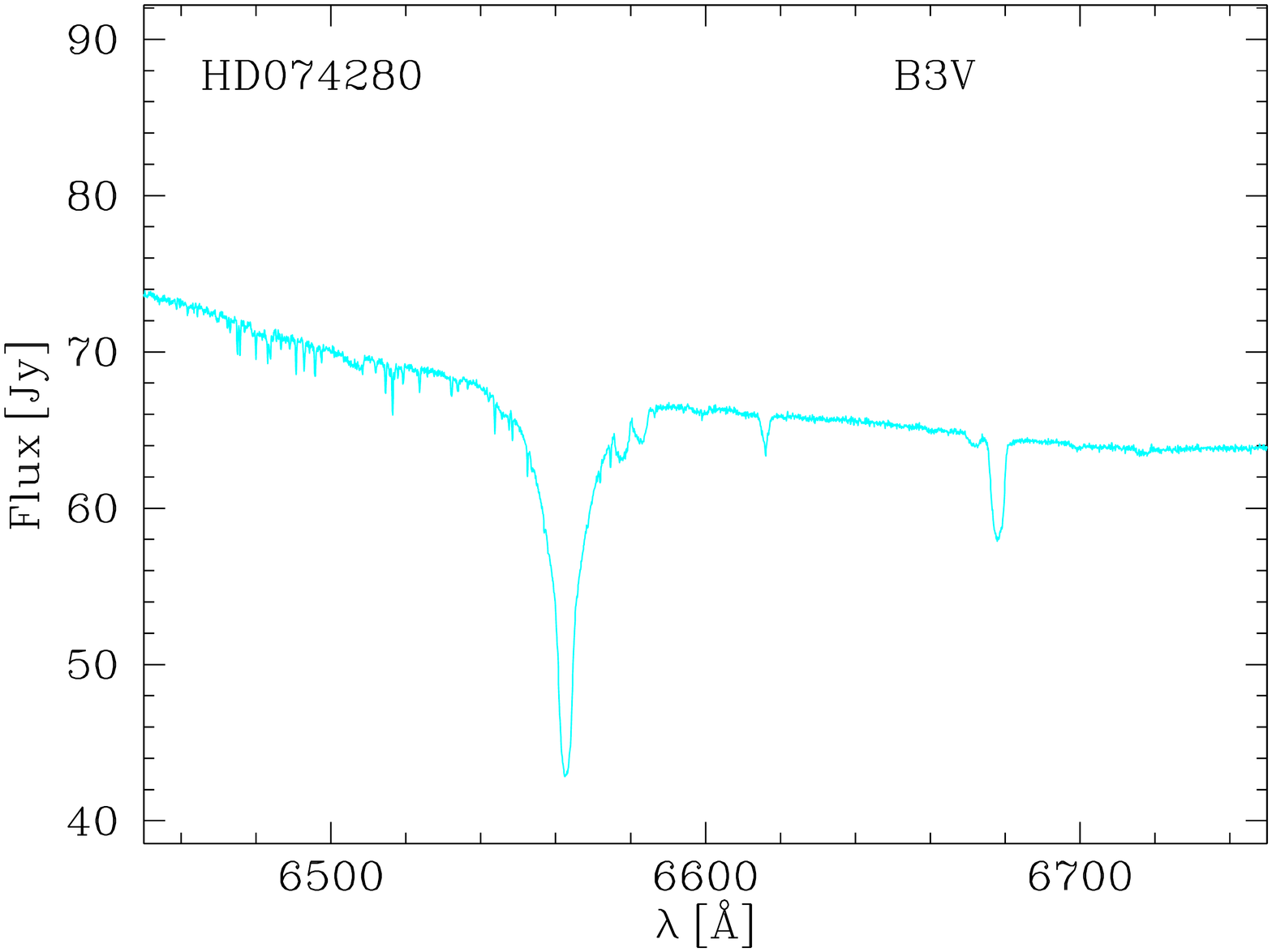}
\includegraphics[width=0.18\textwidth,angle=-90]{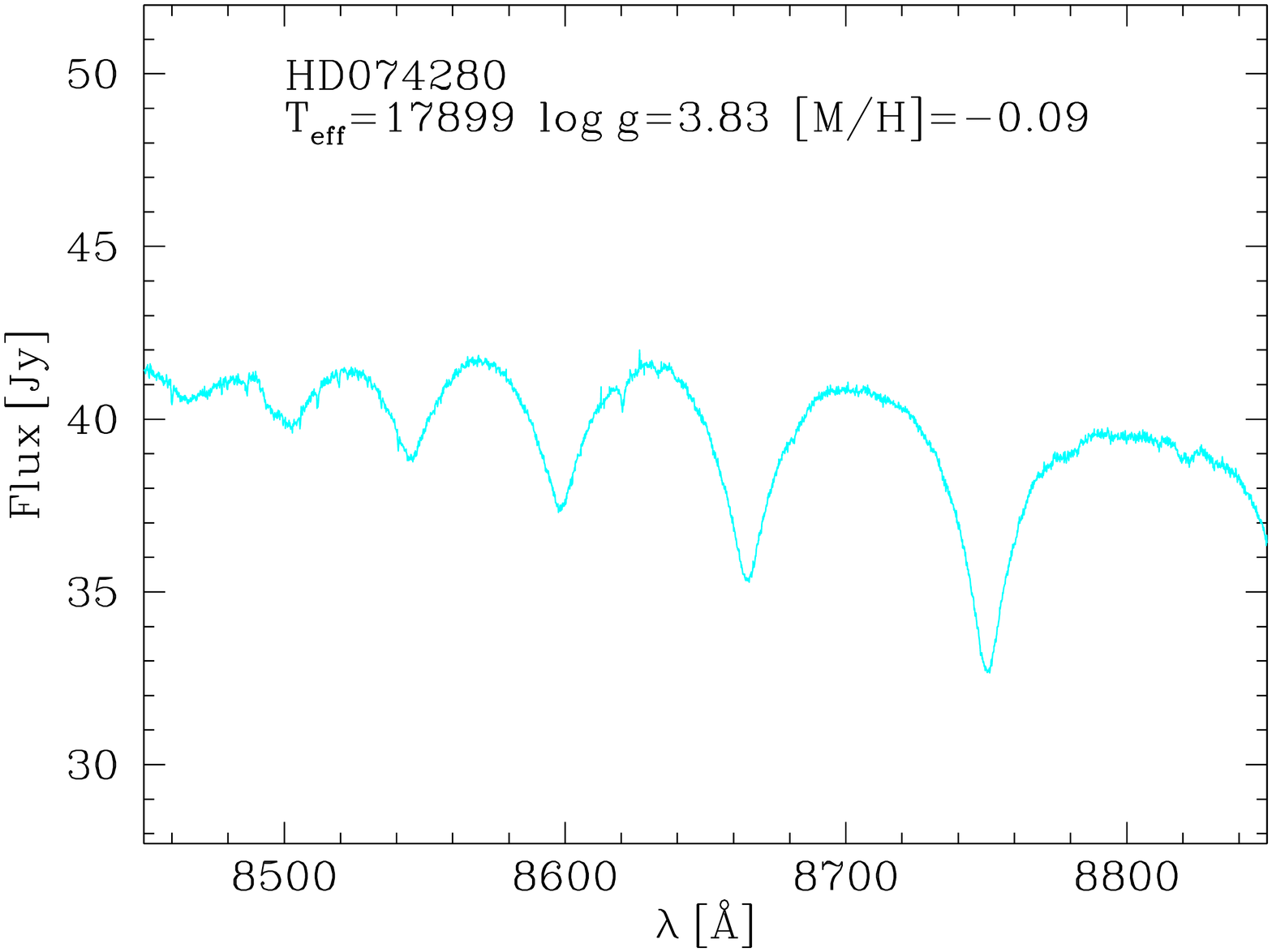}

\contcaption{15. Stars shown in this page are: HD069897, HD070298, HD071148, HD071310, HD071881, HD072184, HD072905, HD072946, HD072968, HD073344, HD073668, HD074000, HD074156 and HD074280.}
\end{figure*}

\begin{figure*}
\includegraphics[width=0.18\textwidth,angle=-90]{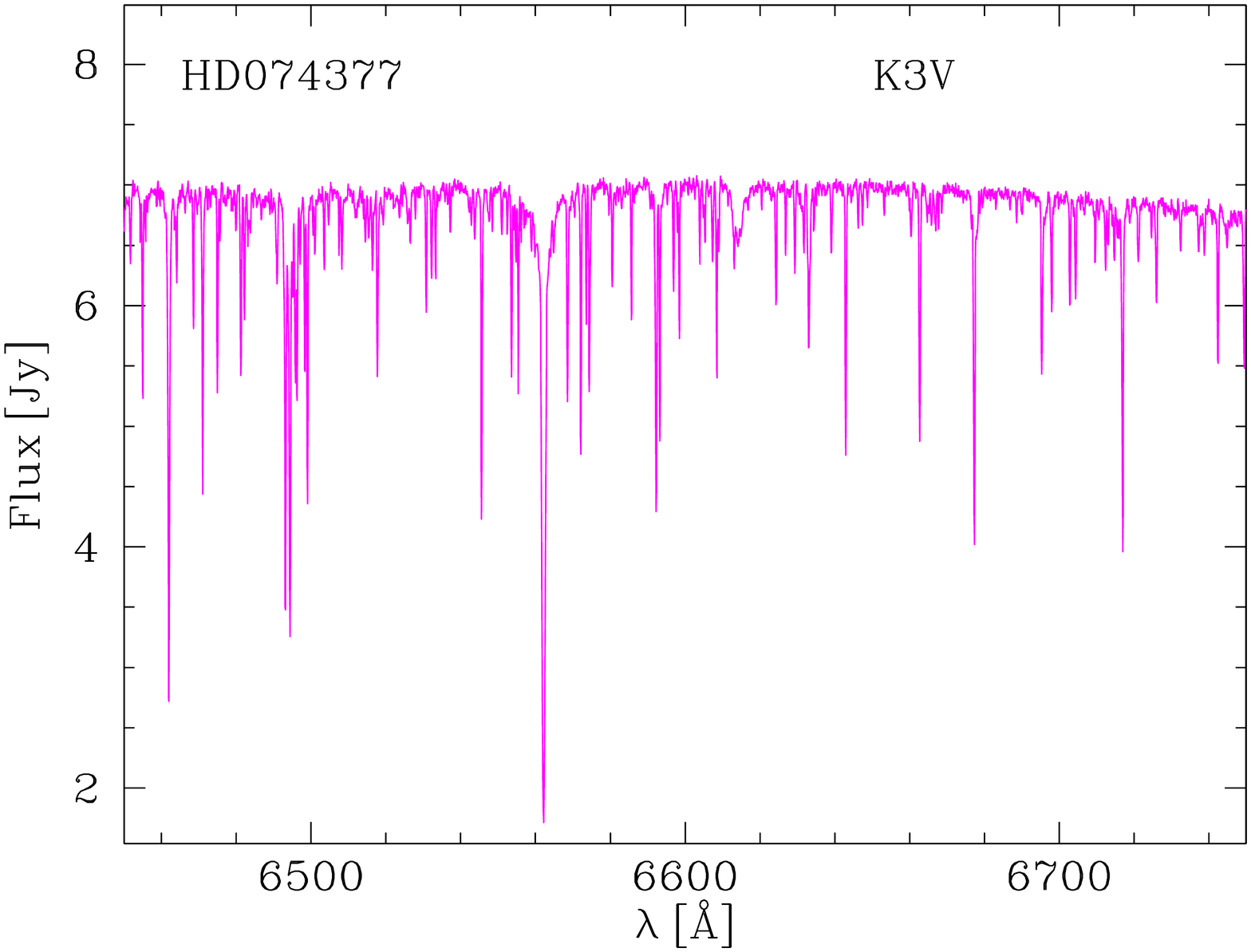}
\includegraphics[width=0.18\textwidth,angle=-90]{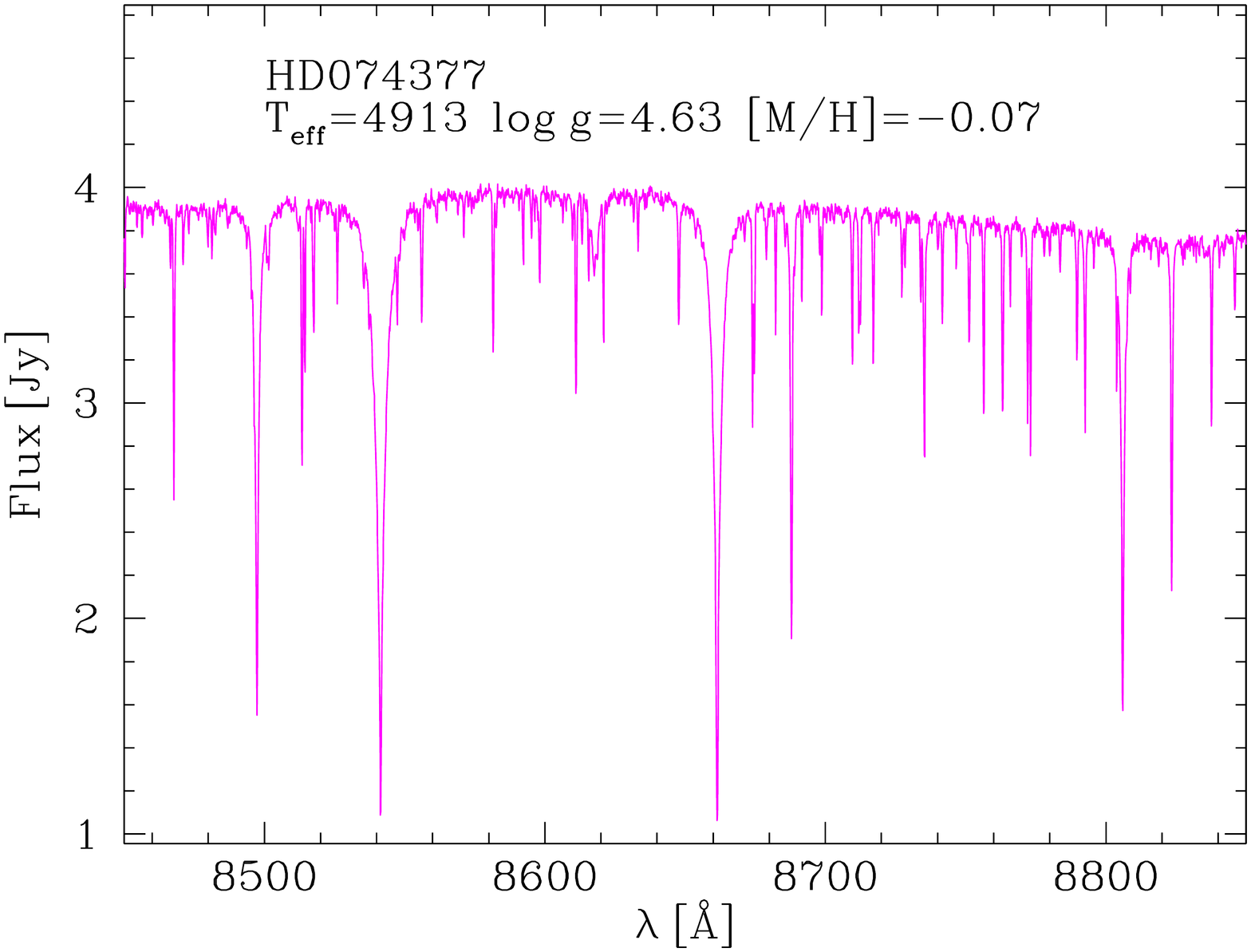}
\includegraphics[width=0.18\textwidth,angle=-90]{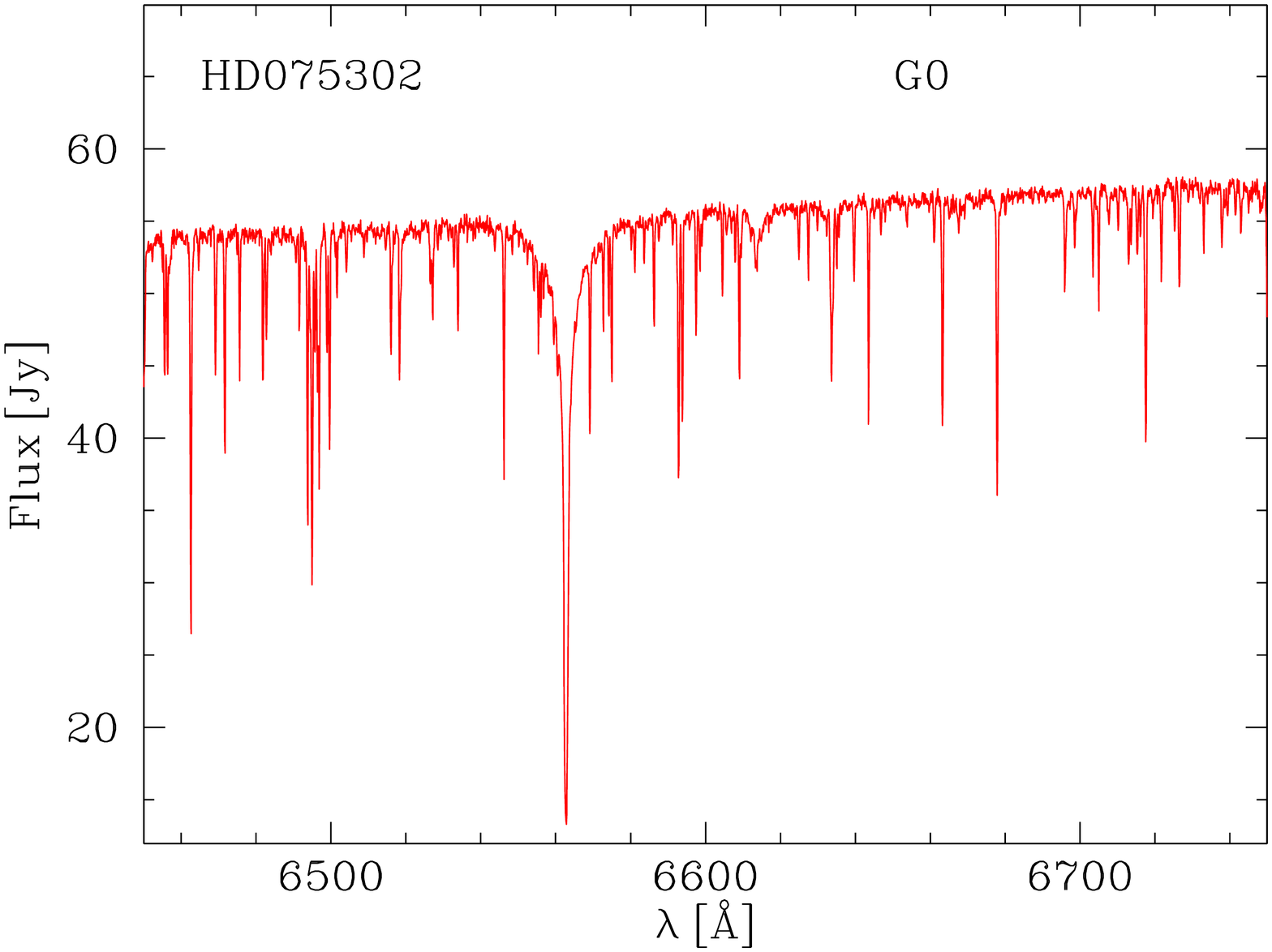}
\includegraphics[width=0.18\textwidth,angle=-90]{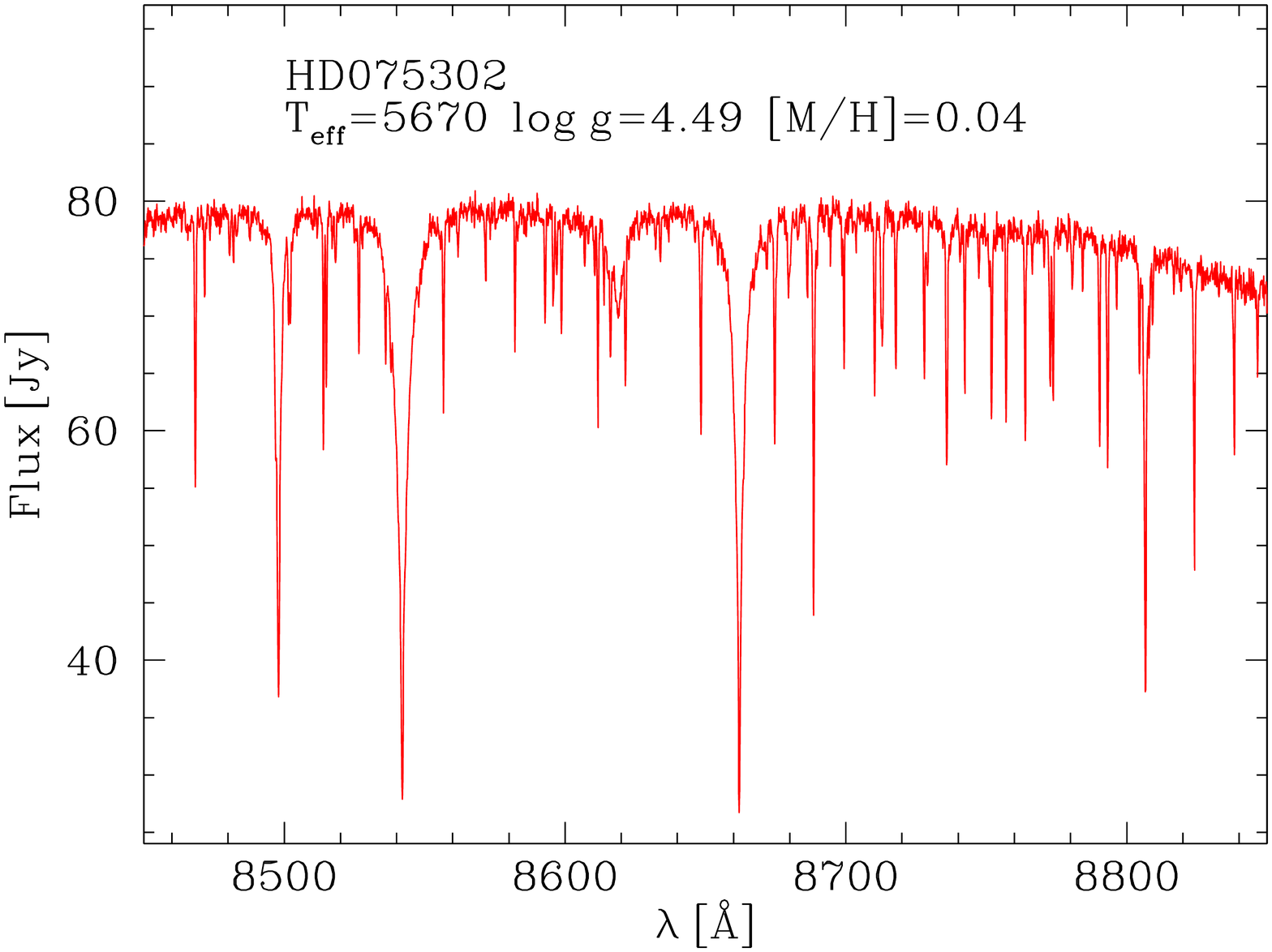}
\includegraphics[width=0.18\textwidth,angle=-90]{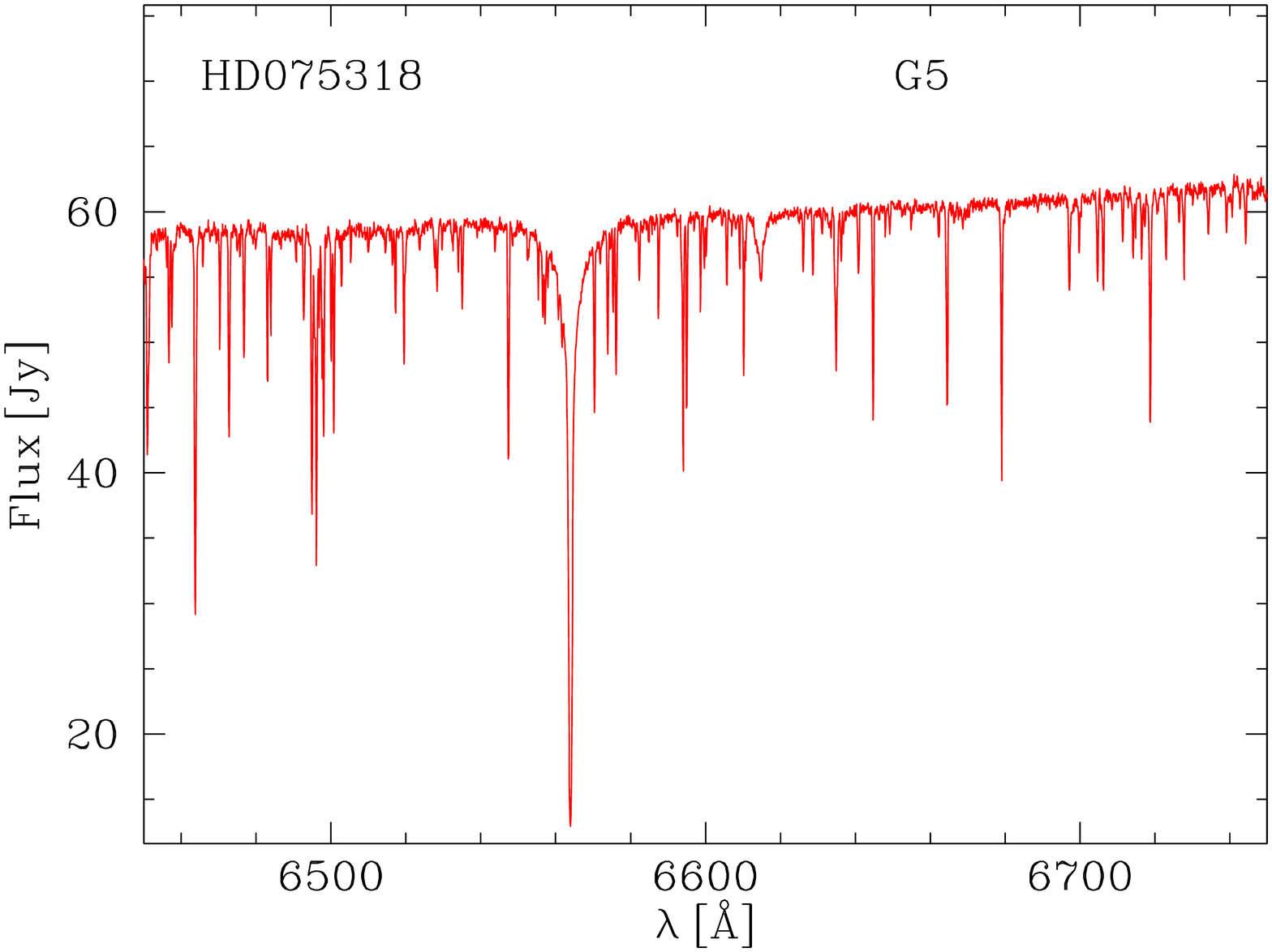}
\includegraphics[width=0.18\textwidth,angle=-90]{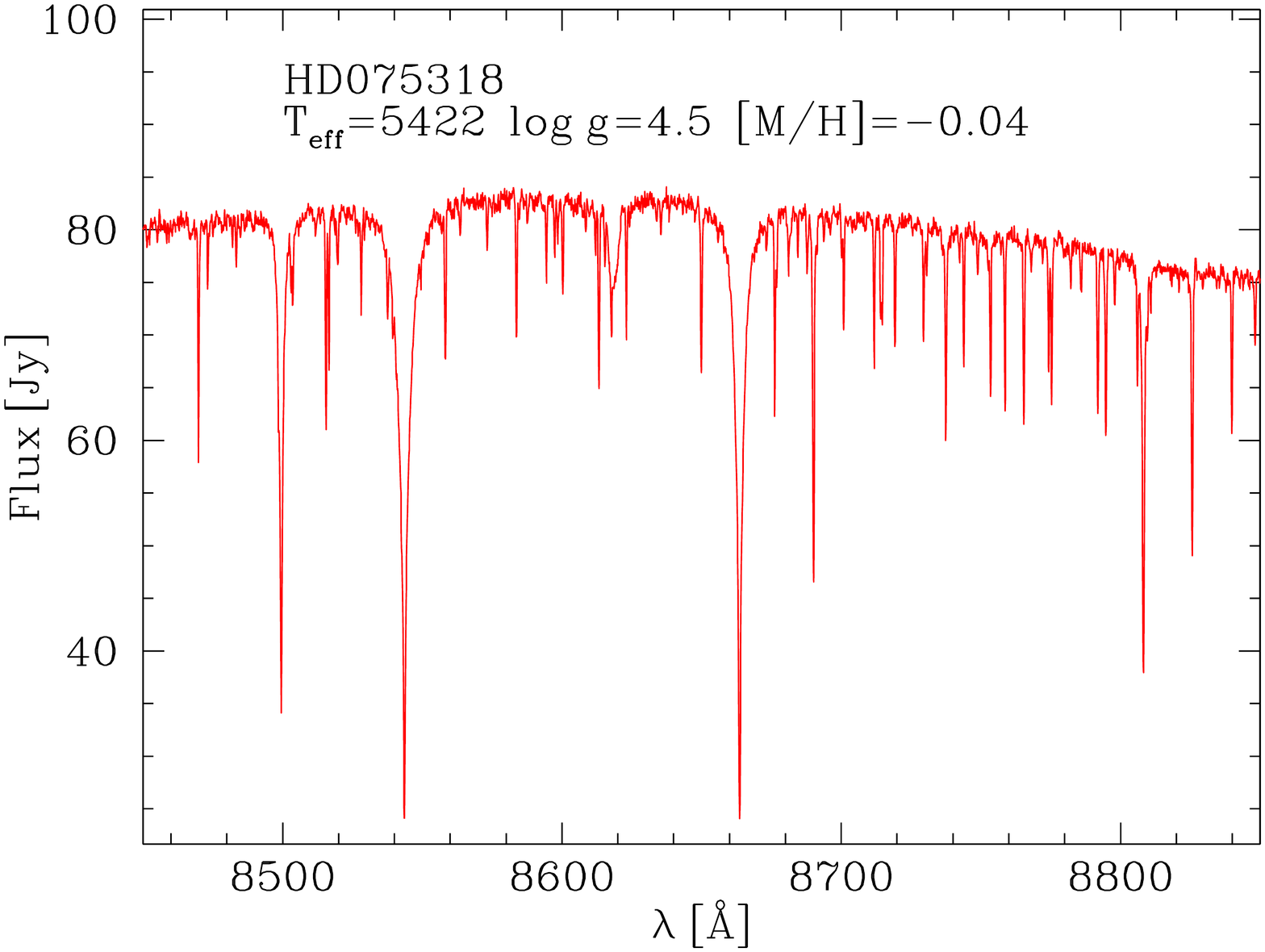}
\includegraphics[width=0.18\textwidth,angle=-90]{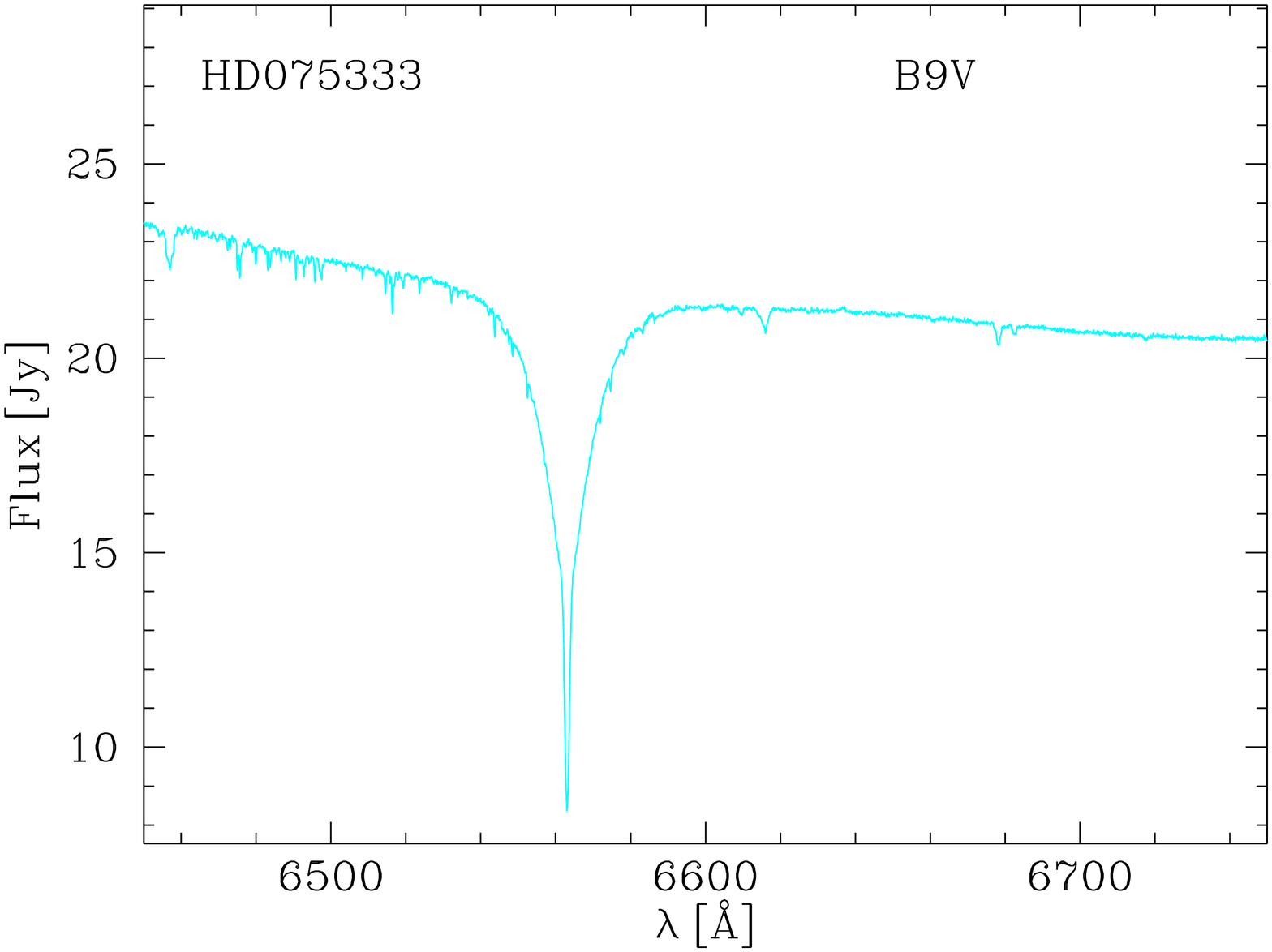}
\includegraphics[width=0.18\textwidth,angle=-90]{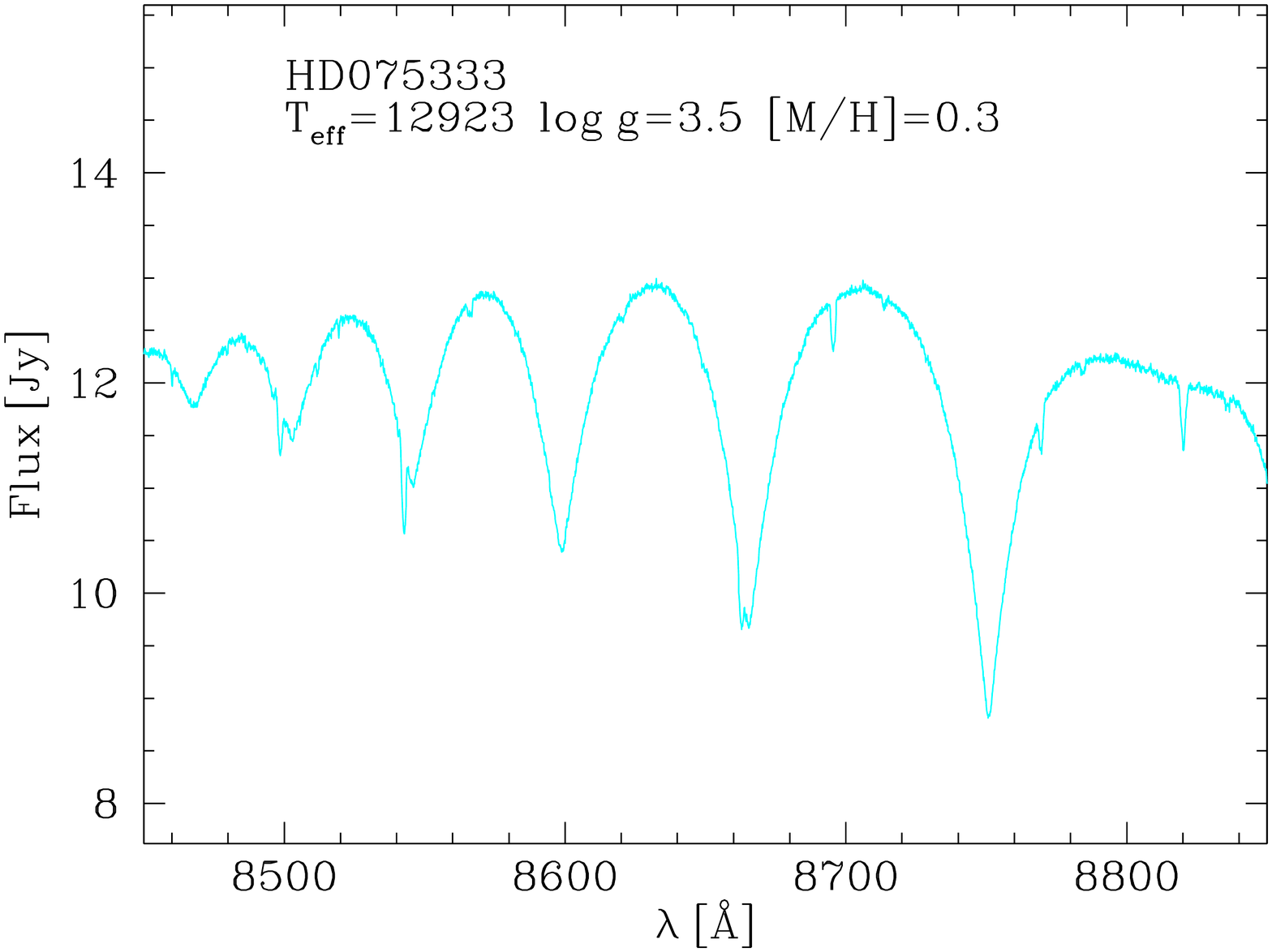}
\includegraphics[width=0.18\textwidth,angle=-90]{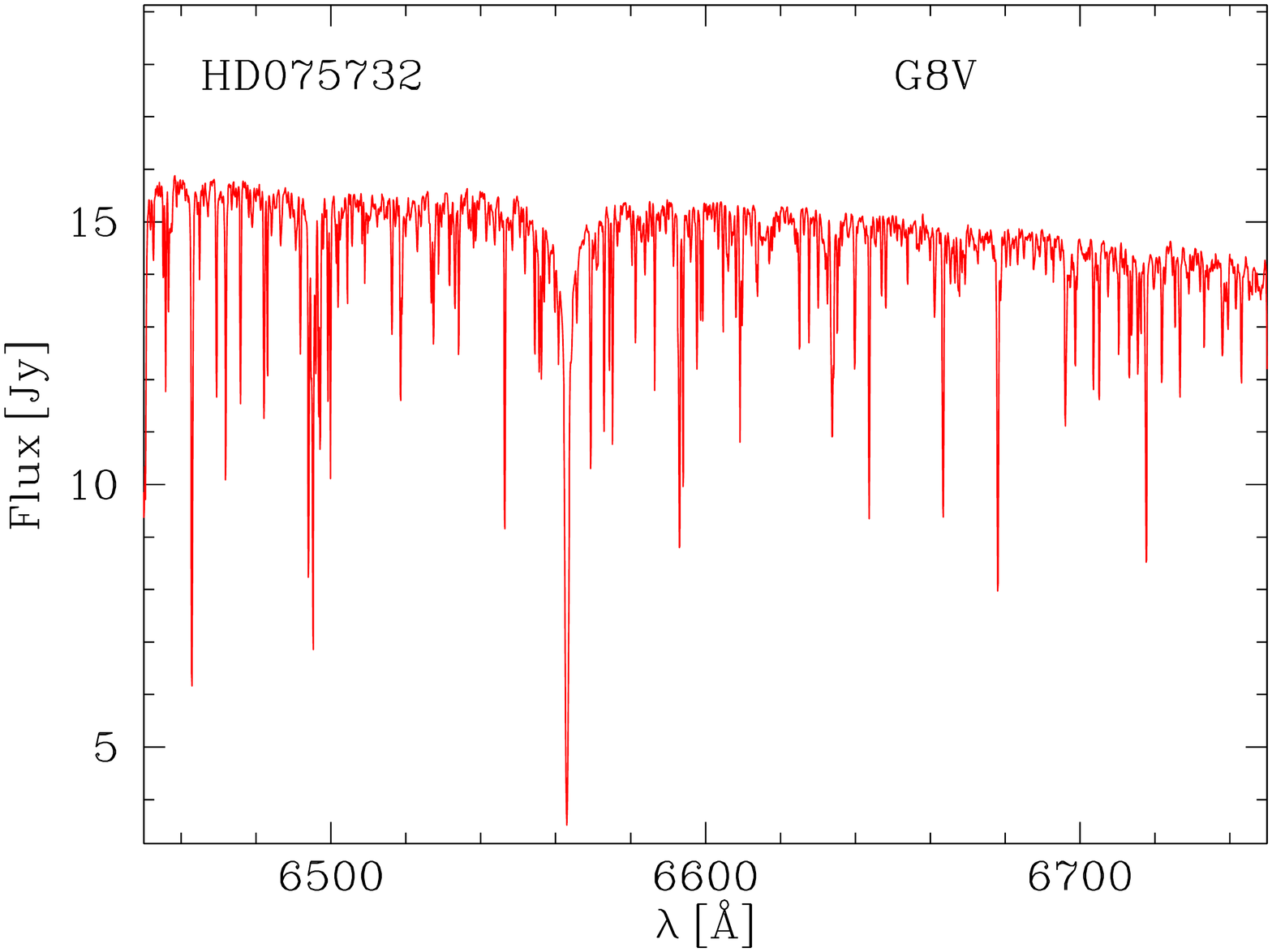}
\includegraphics[width=0.18\textwidth,angle=-90]{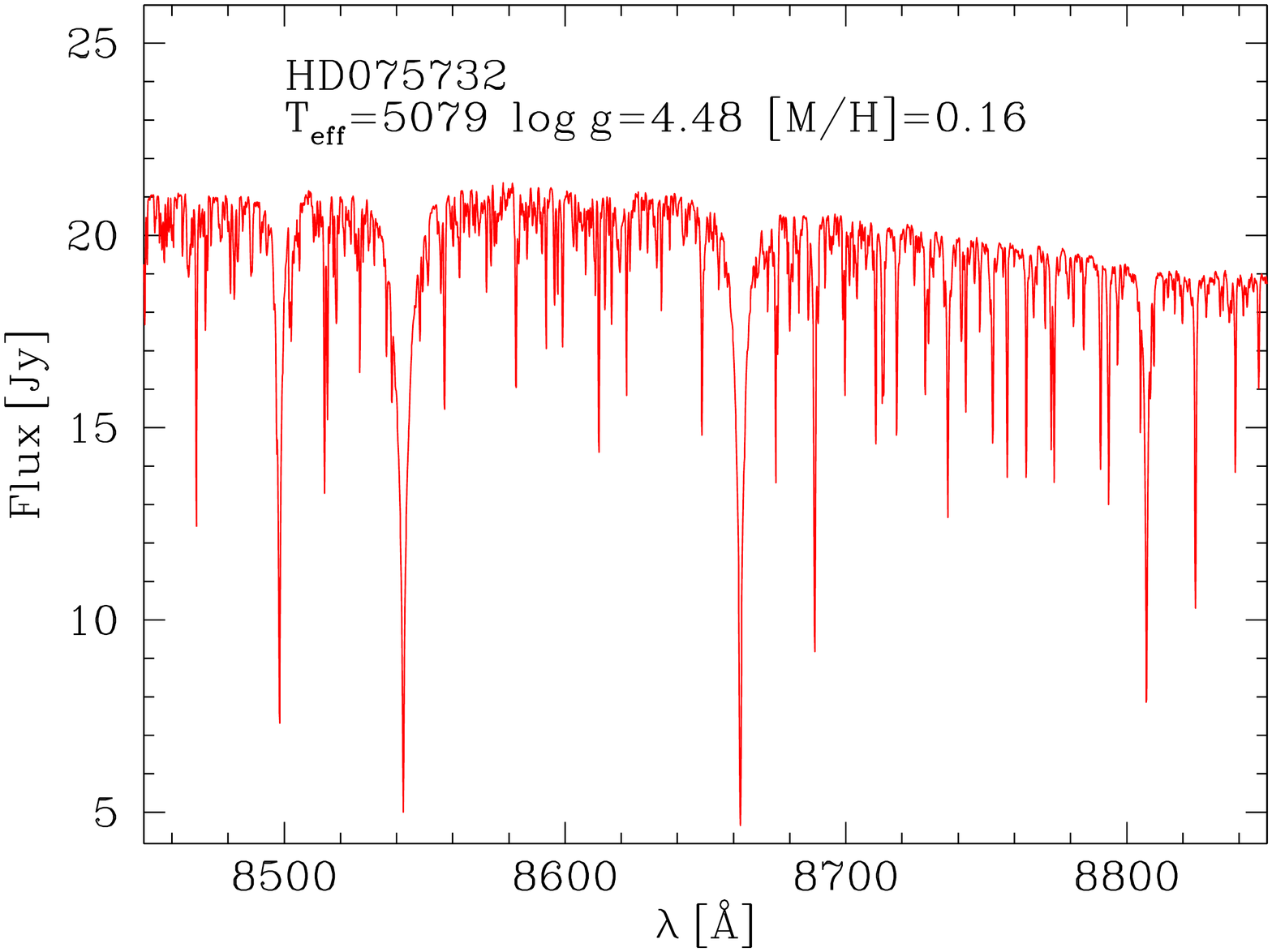}
\includegraphics[width=0.18\textwidth,angle=-90]{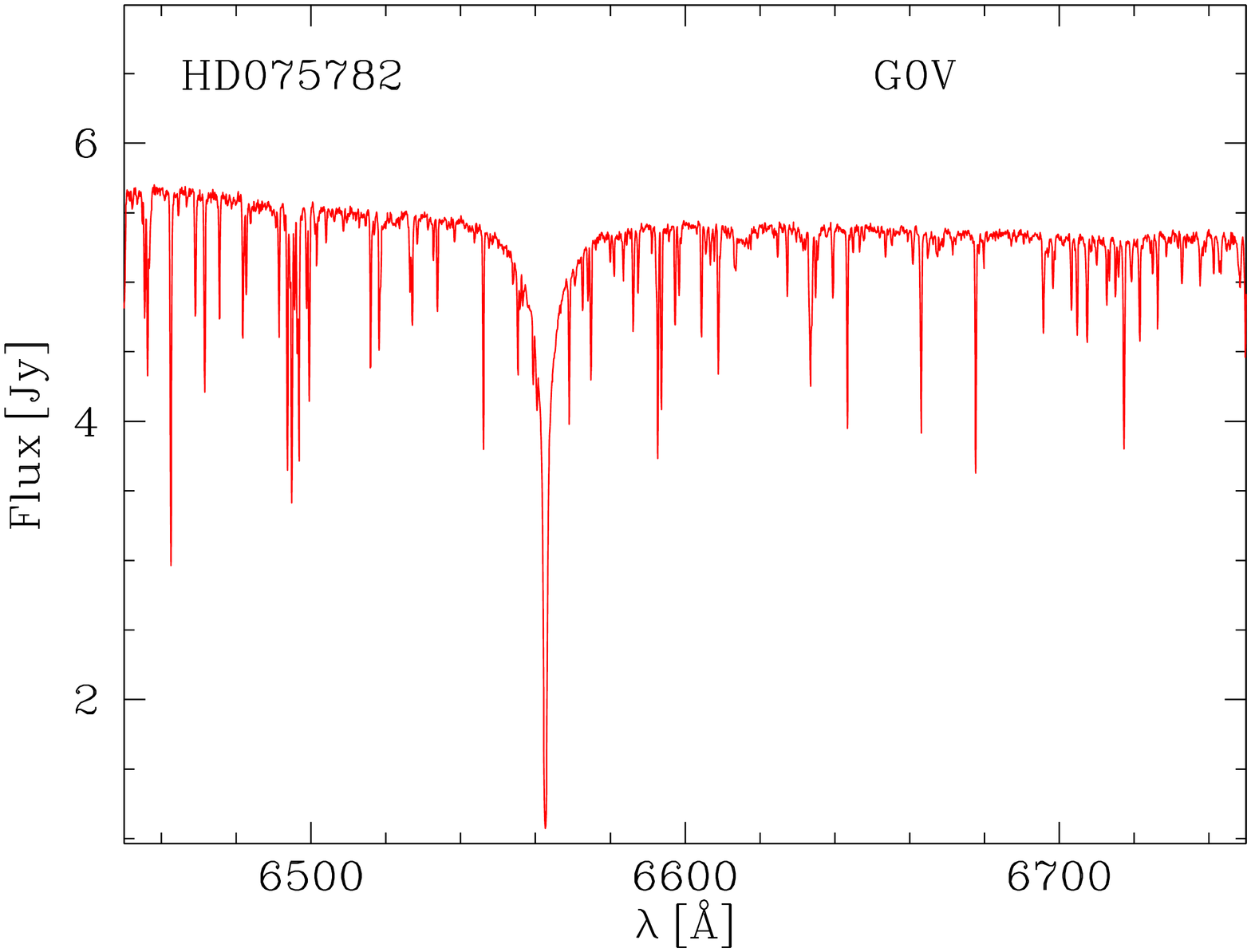}
\includegraphics[width=0.18\textwidth,angle=-90]{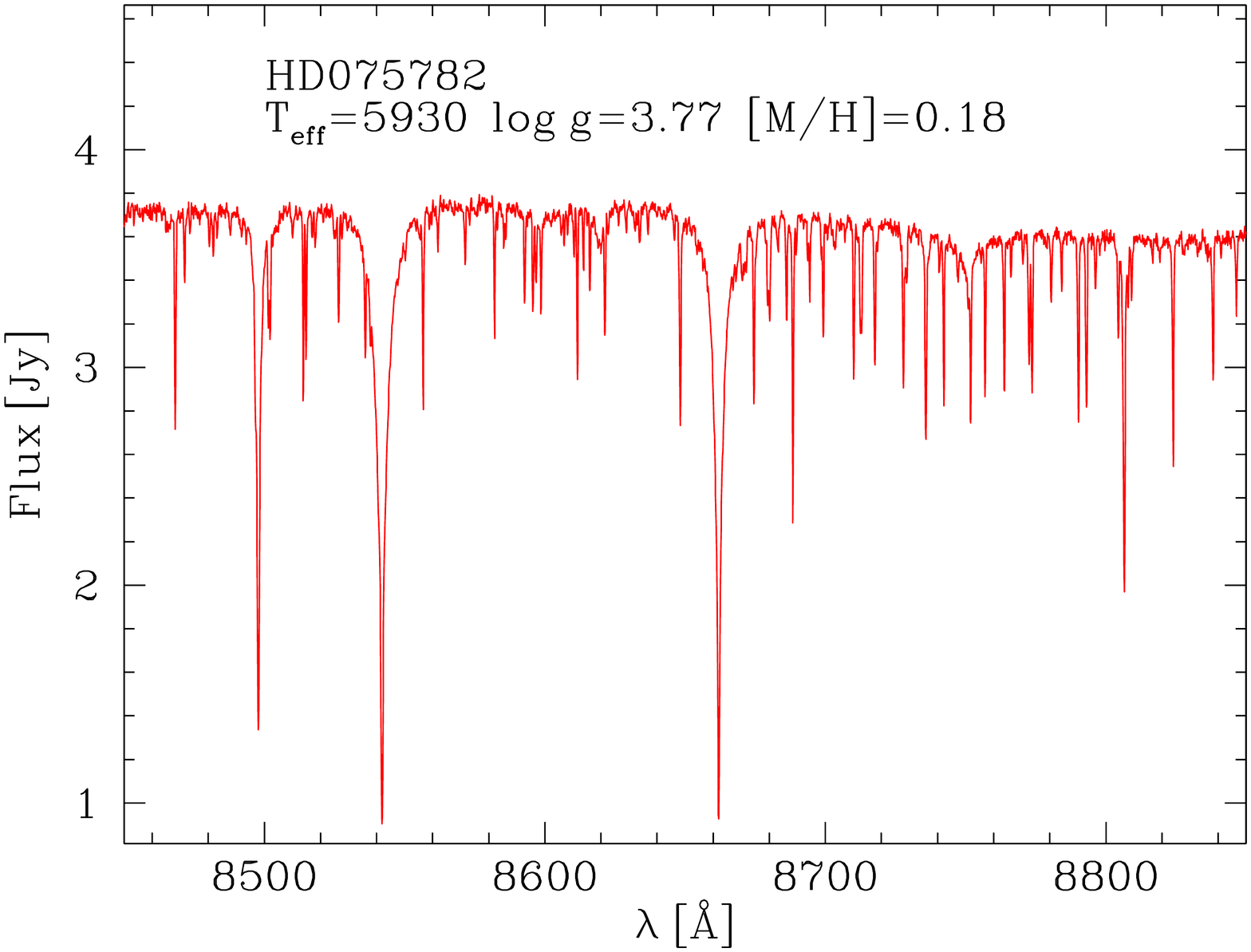}
\includegraphics[width=0.18\textwidth,angle=-90]{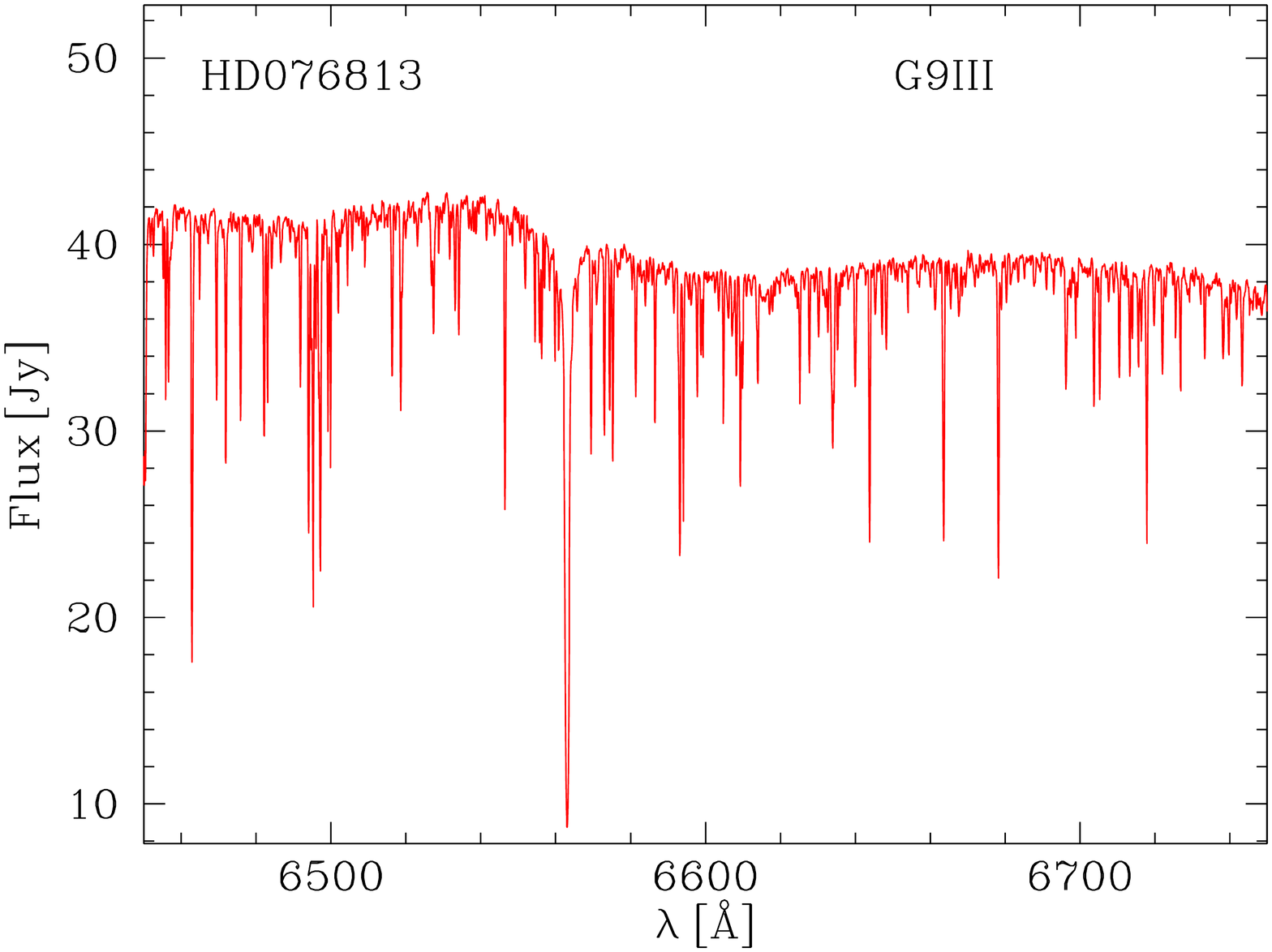}
\includegraphics[width=0.18\textwidth,angle=-90]{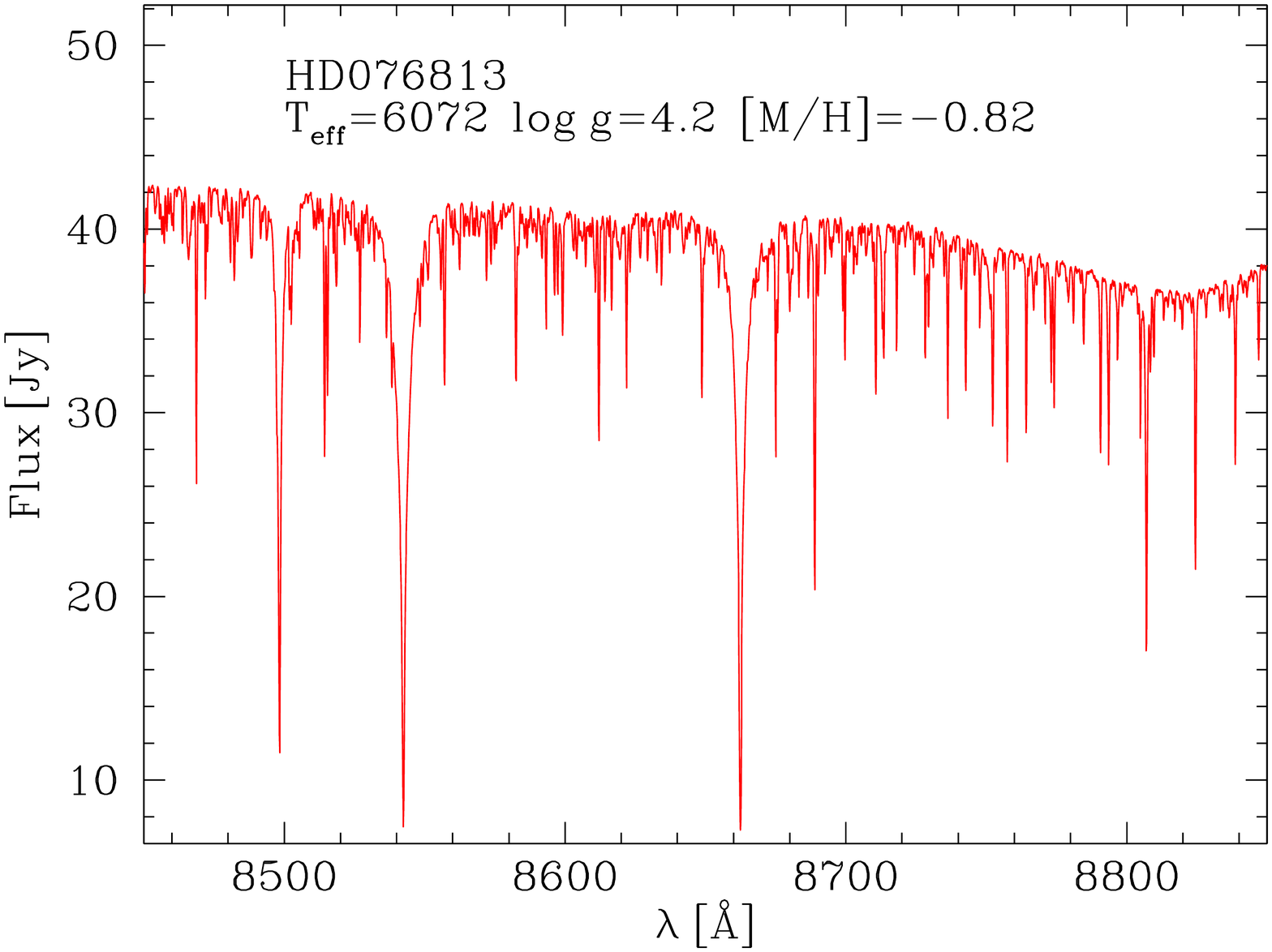}
\includegraphics[width=0.18\textwidth,angle=-90]{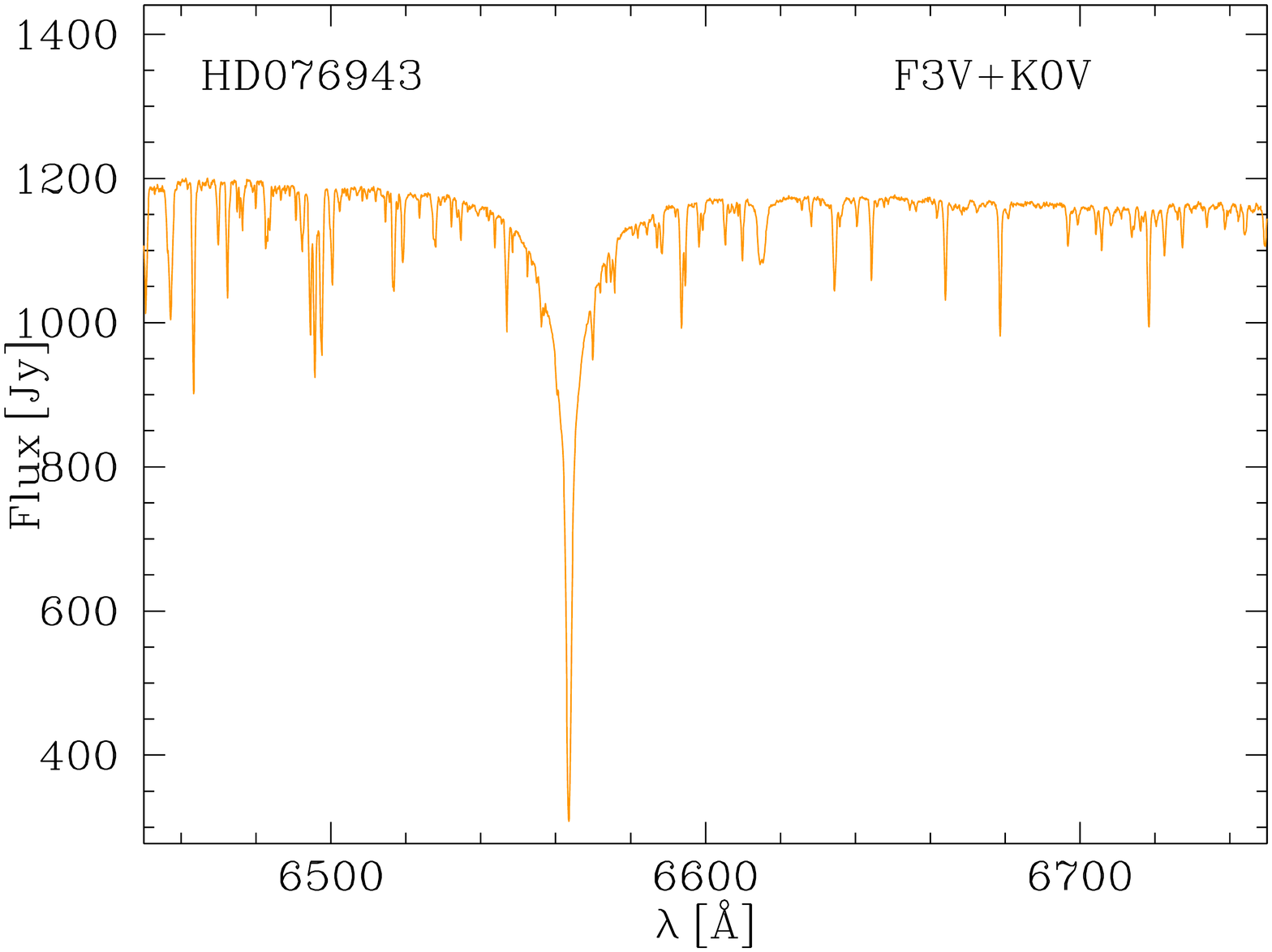}
\includegraphics[width=0.18\textwidth,angle=-90]{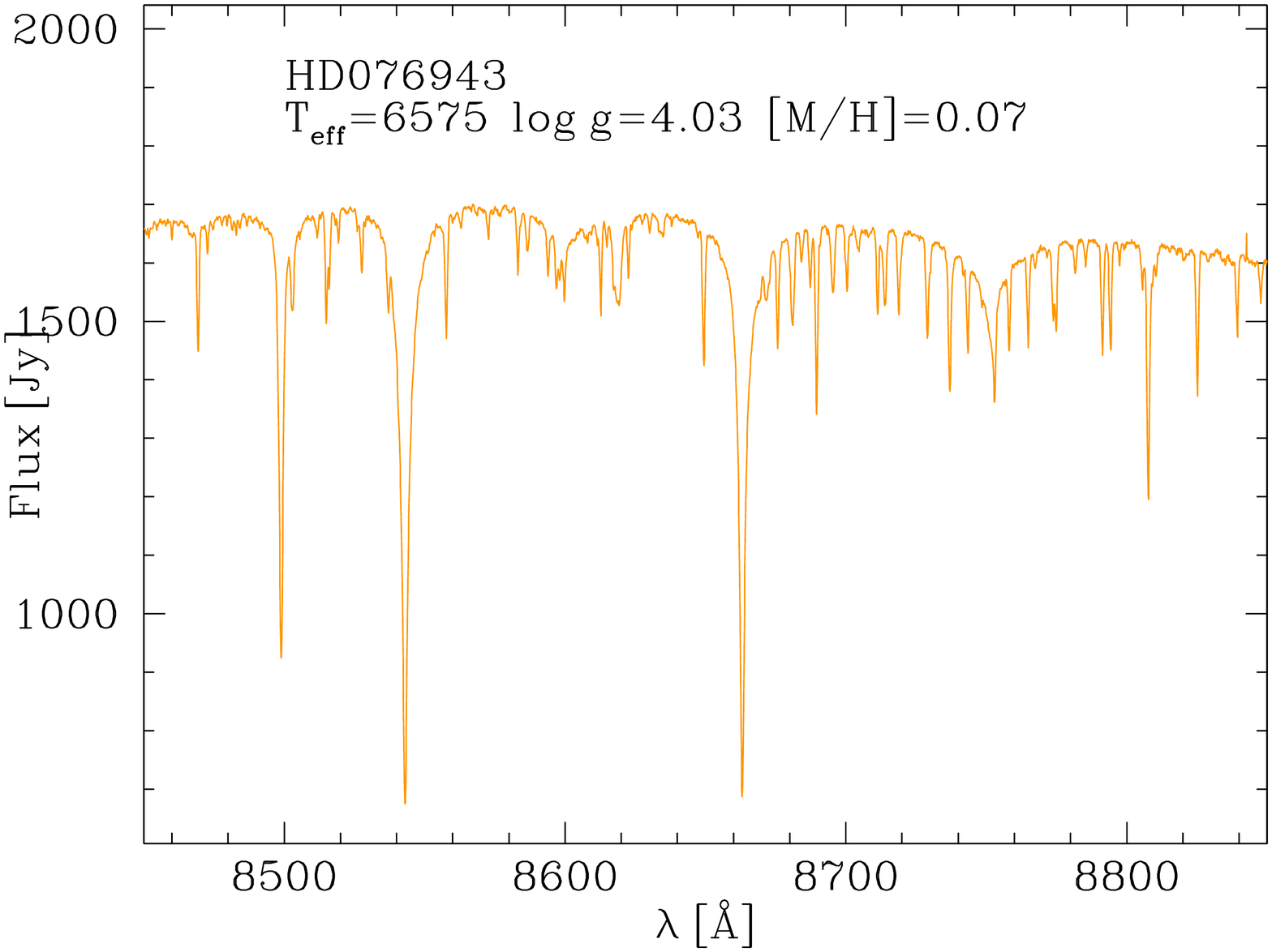}
\includegraphics[width=0.18\textwidth,angle=-90]{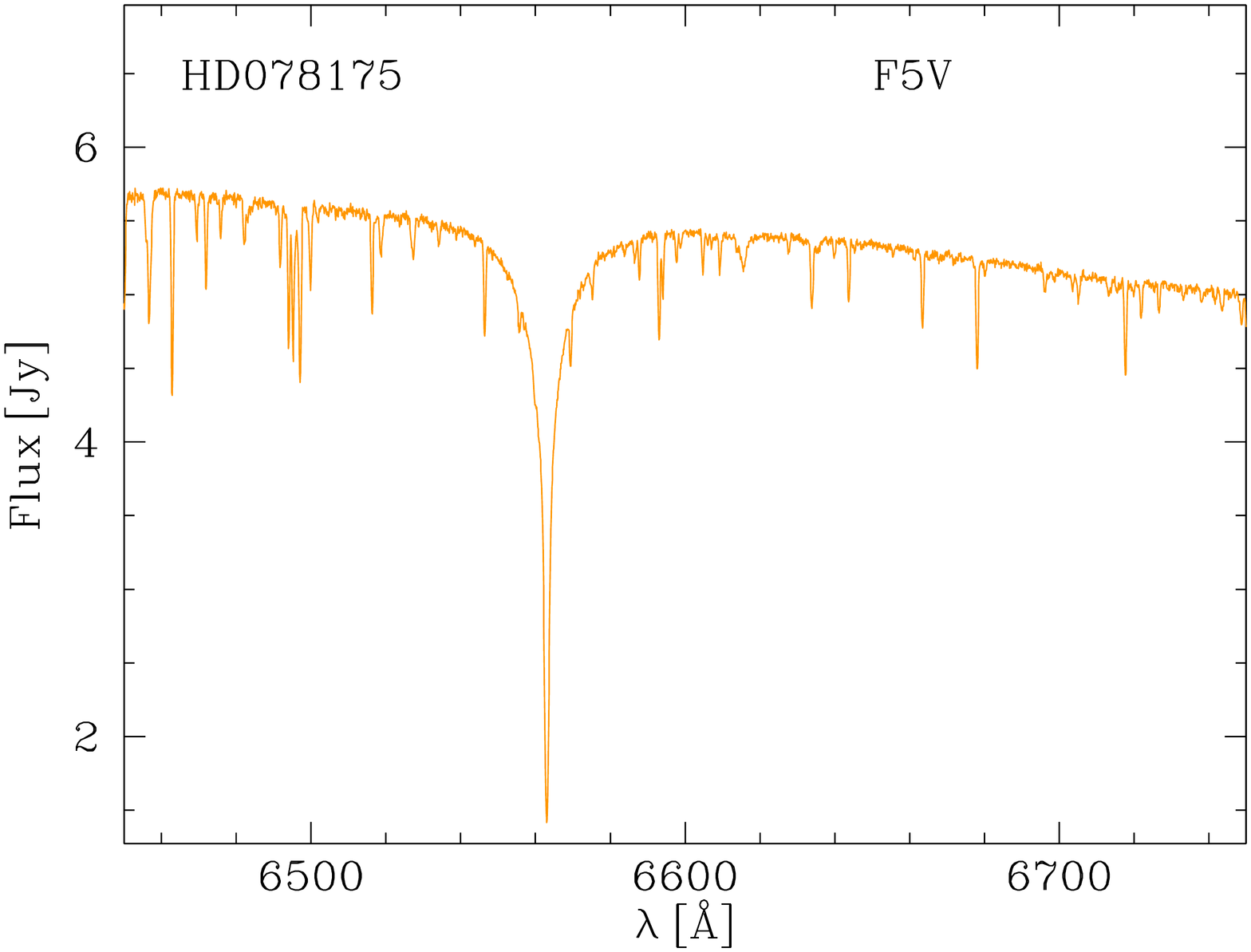}
\includegraphics[width=0.18\textwidth,angle=-90]{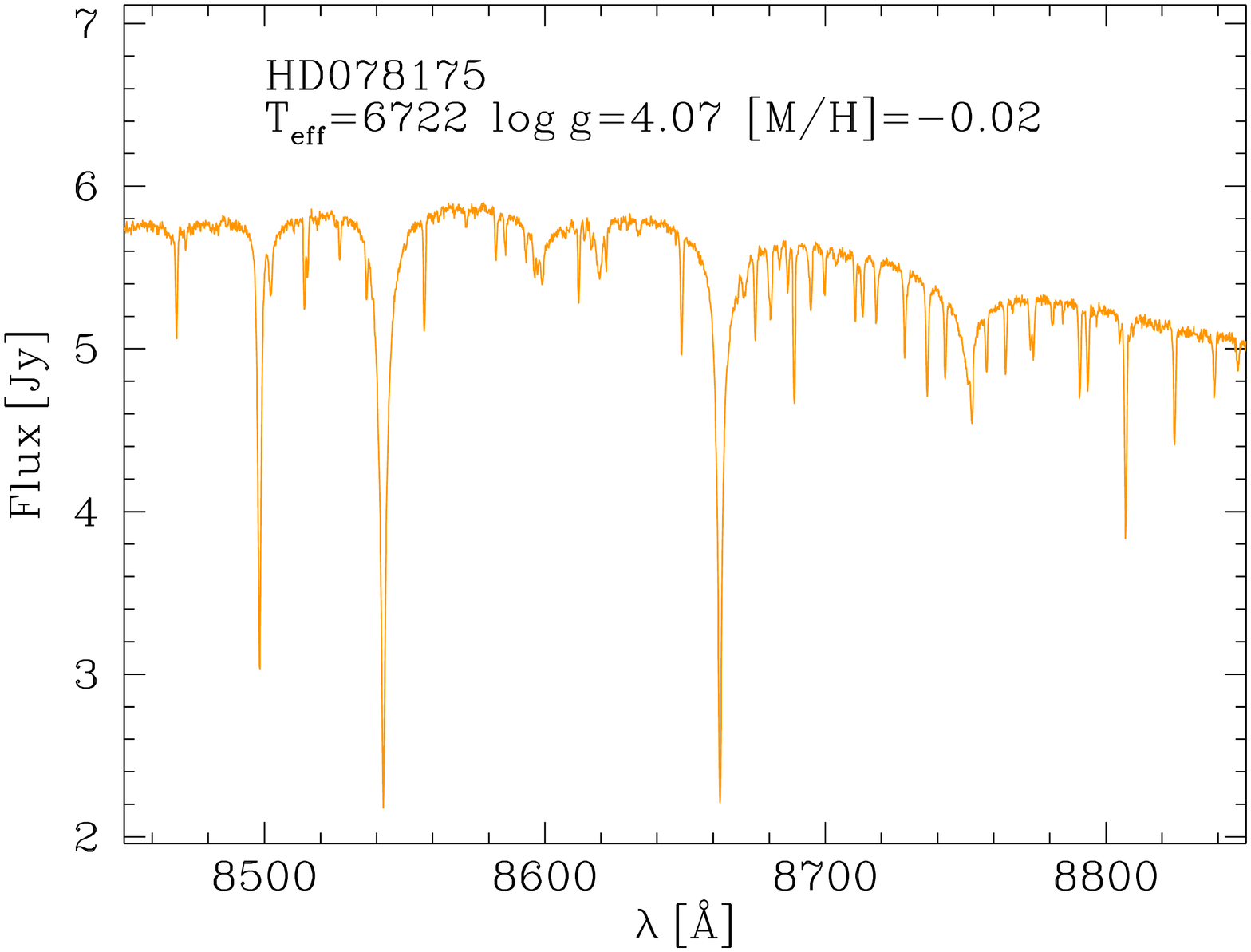}
\includegraphics[width=0.18\textwidth,angle=-90]{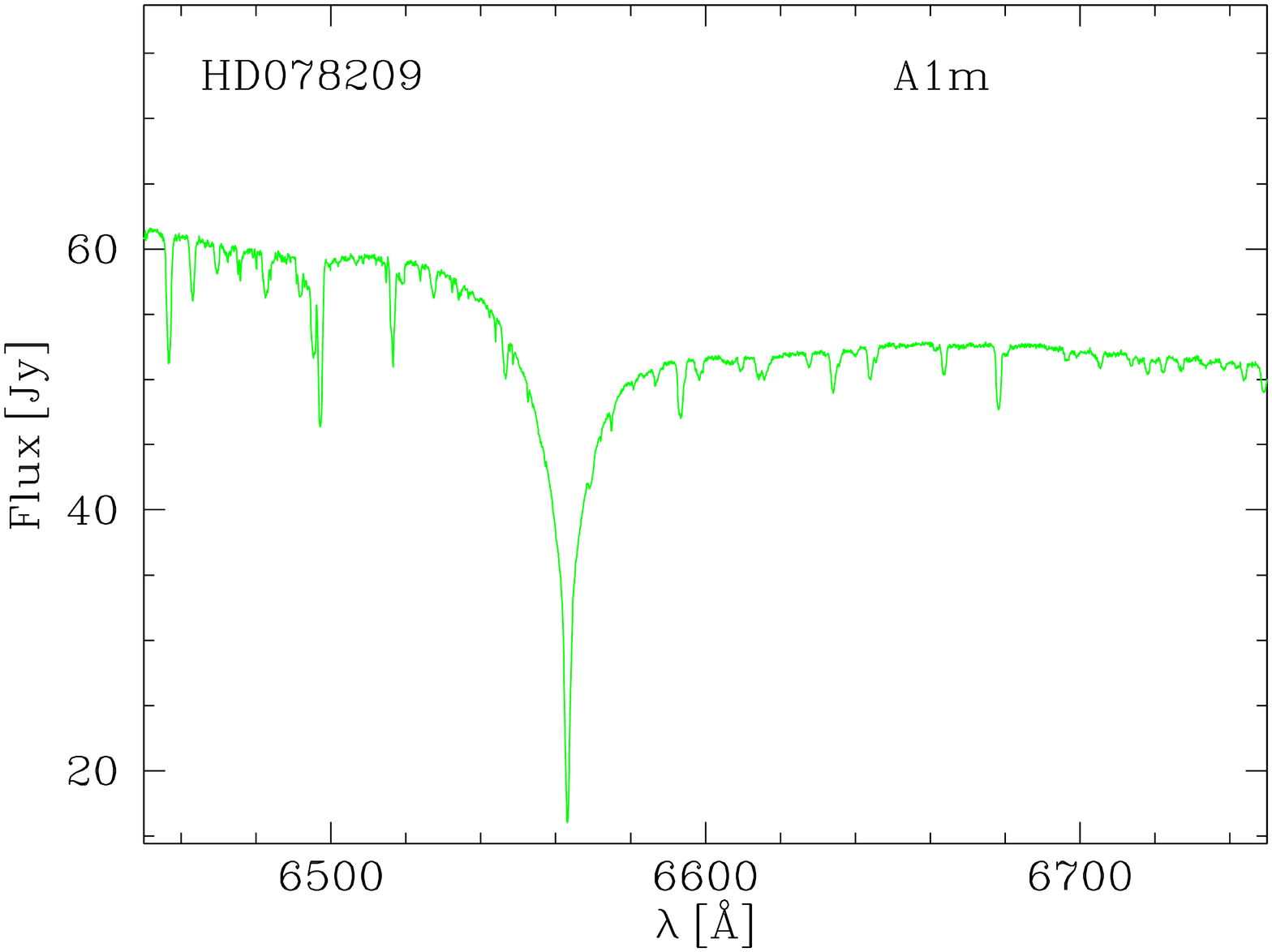}
\includegraphics[width=0.18\textwidth,angle=-90]{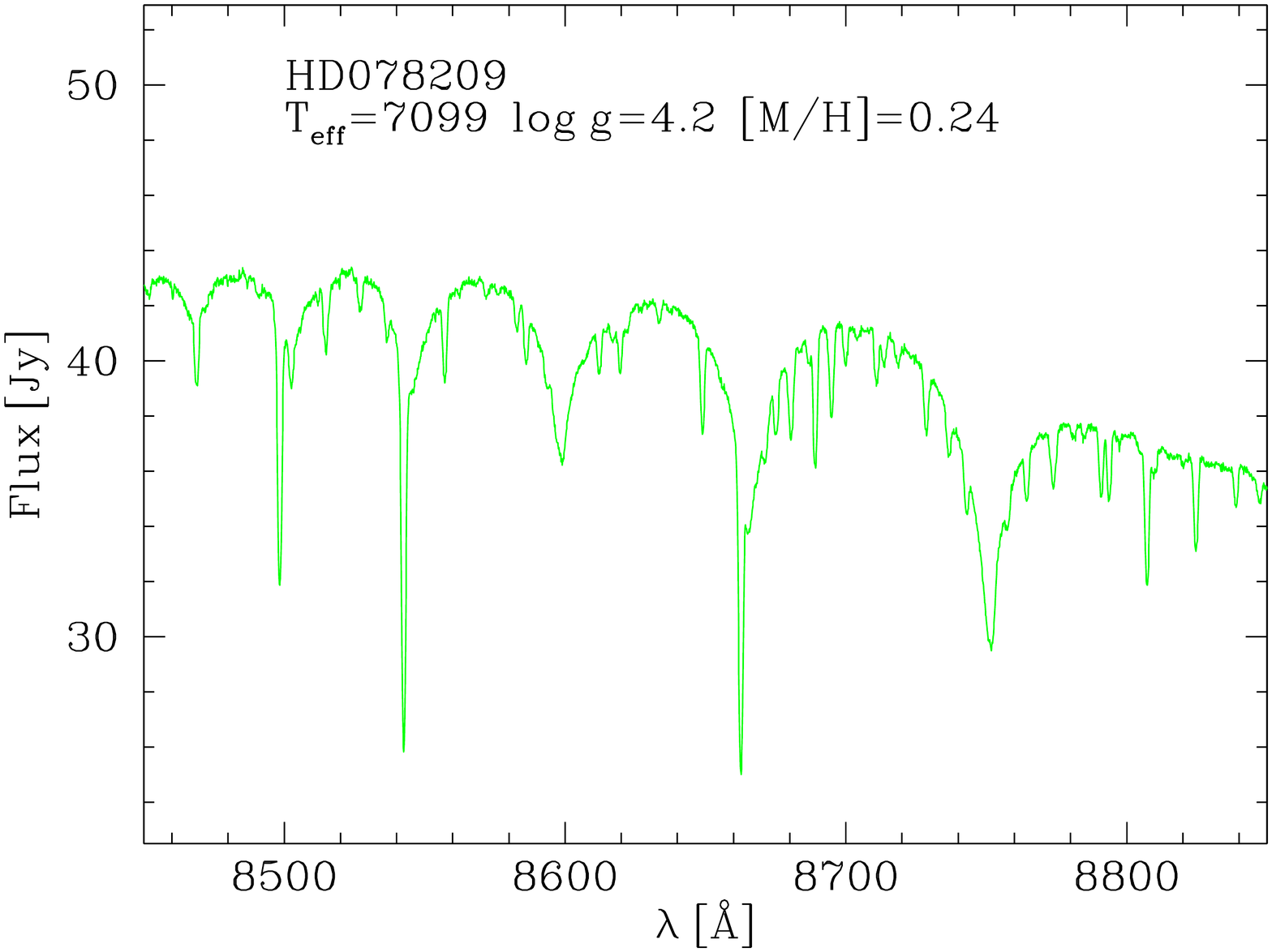}
\includegraphics[width=0.18\textwidth,angle=-90]{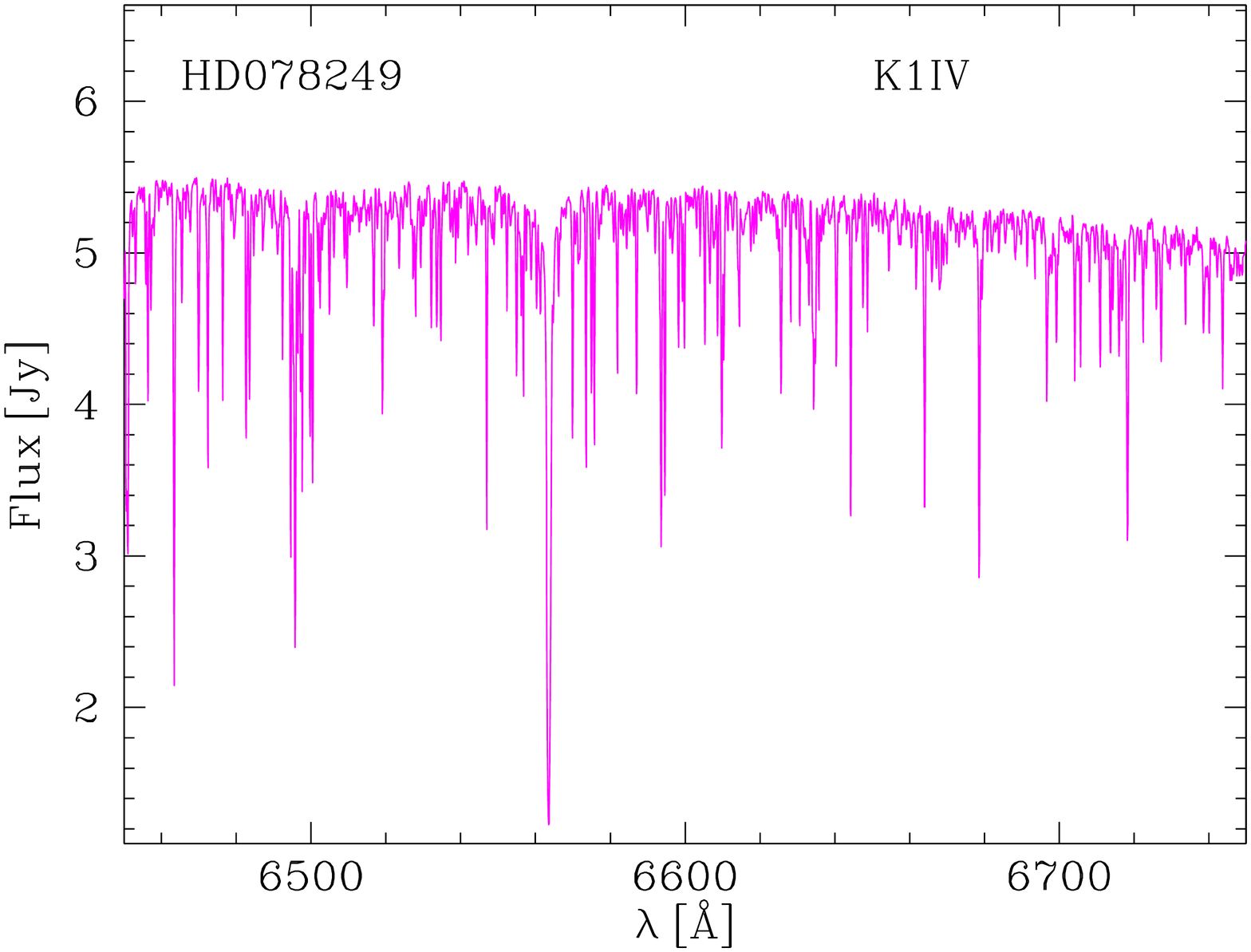}
\includegraphics[width=0.18\textwidth,angle=-90]{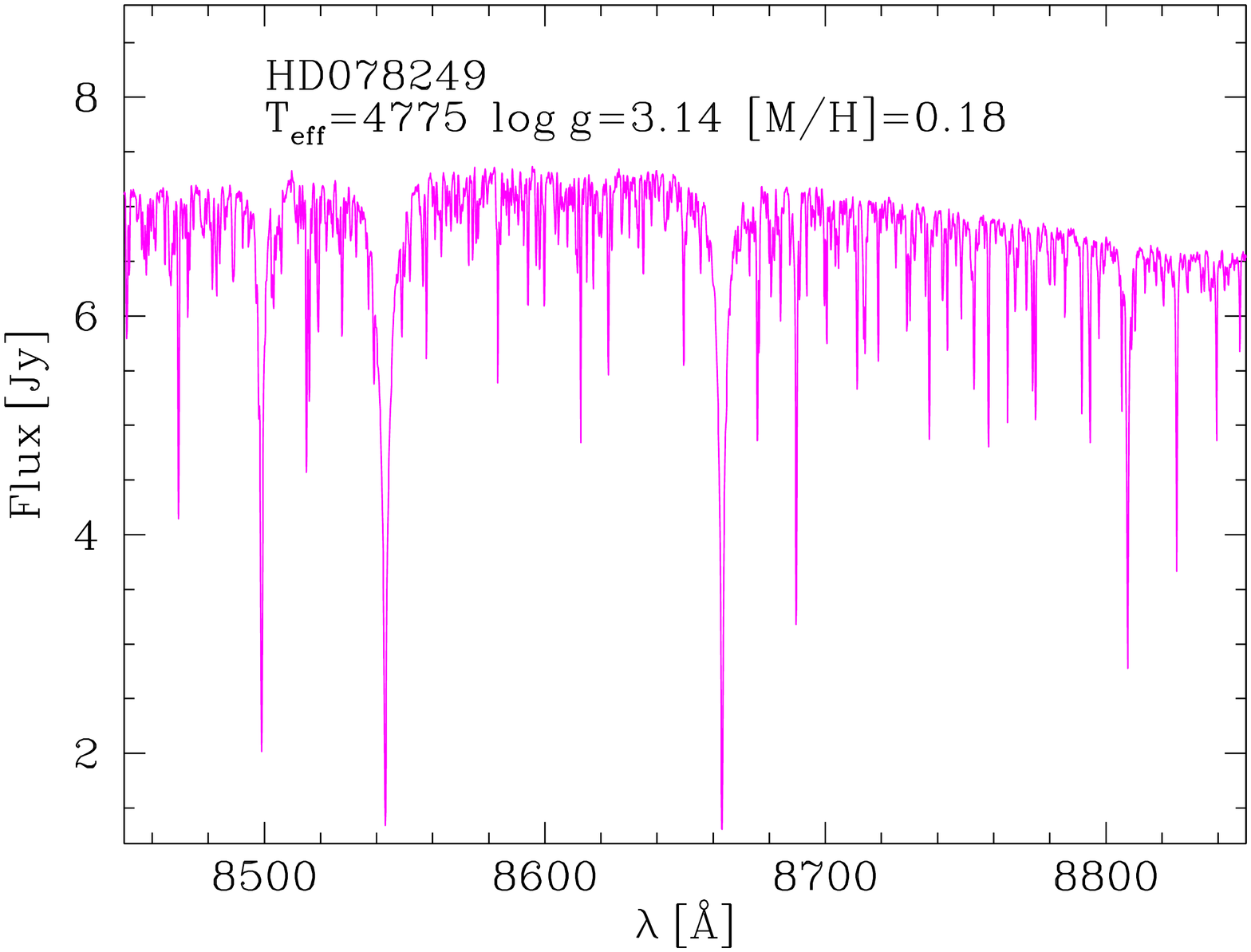}
\includegraphics[width=0.18\textwidth,angle=-90]{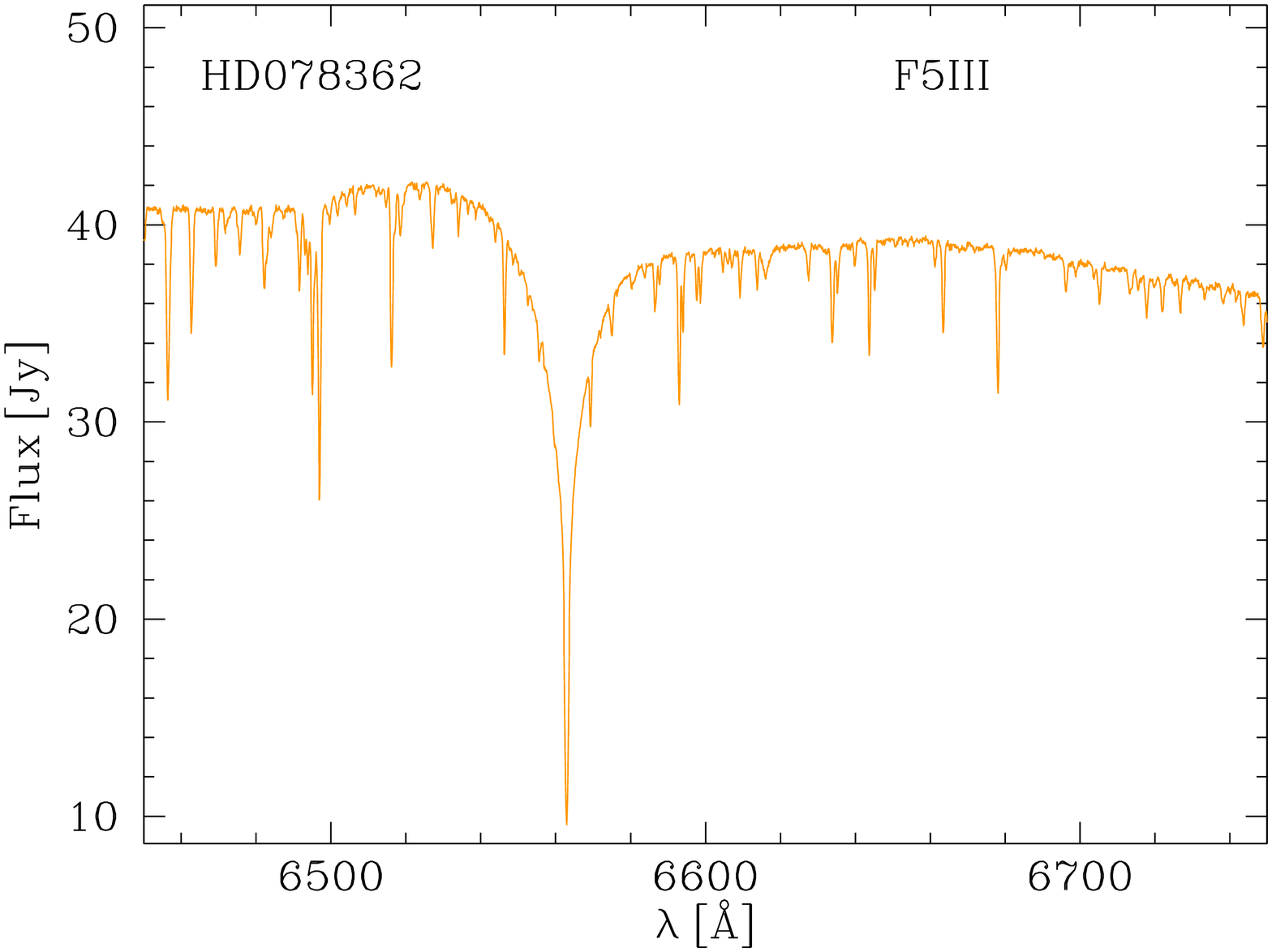}
\includegraphics[width=0.18\textwidth,angle=-90]{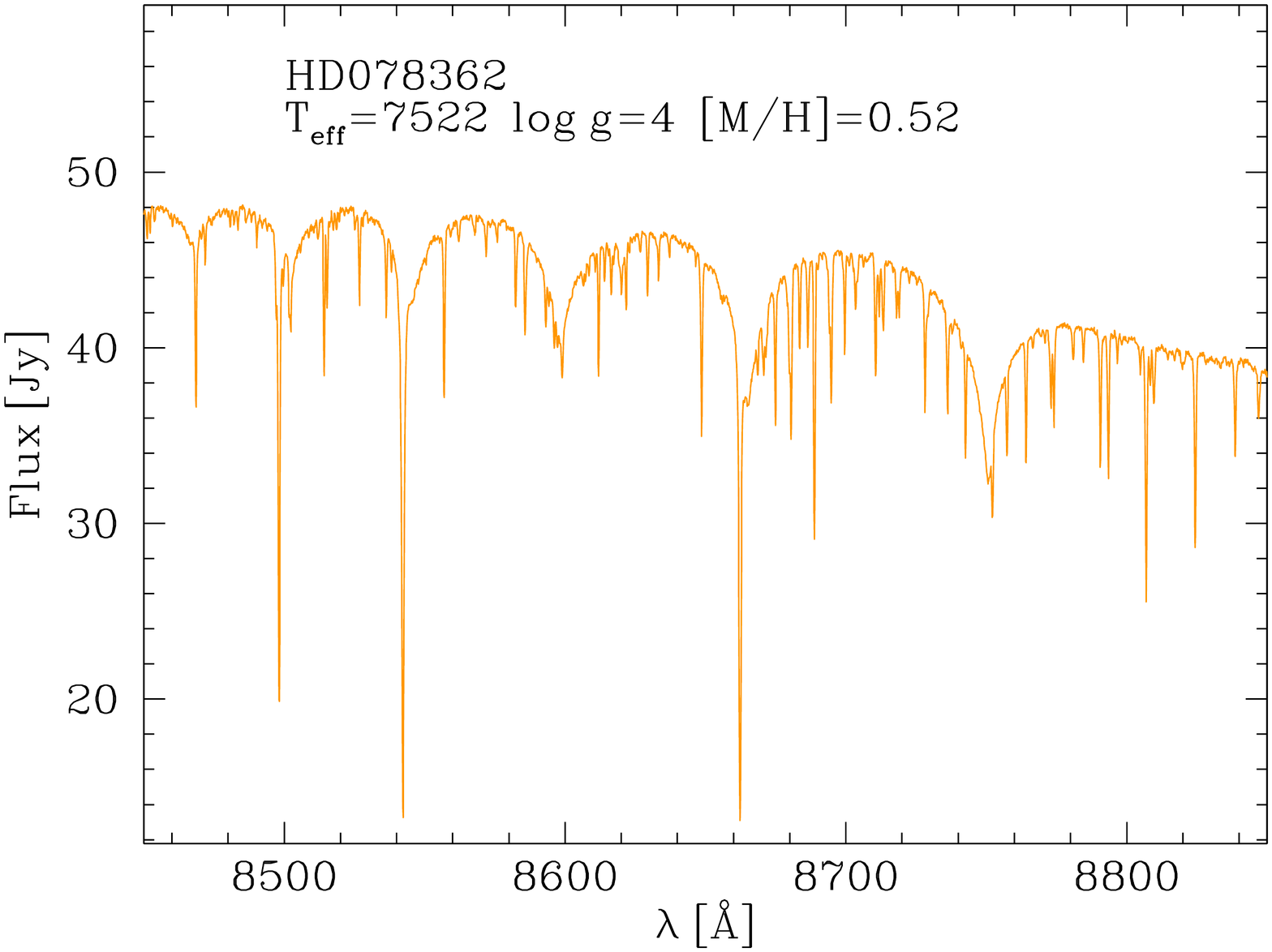}
\includegraphics[width=0.18\textwidth,angle=-90]{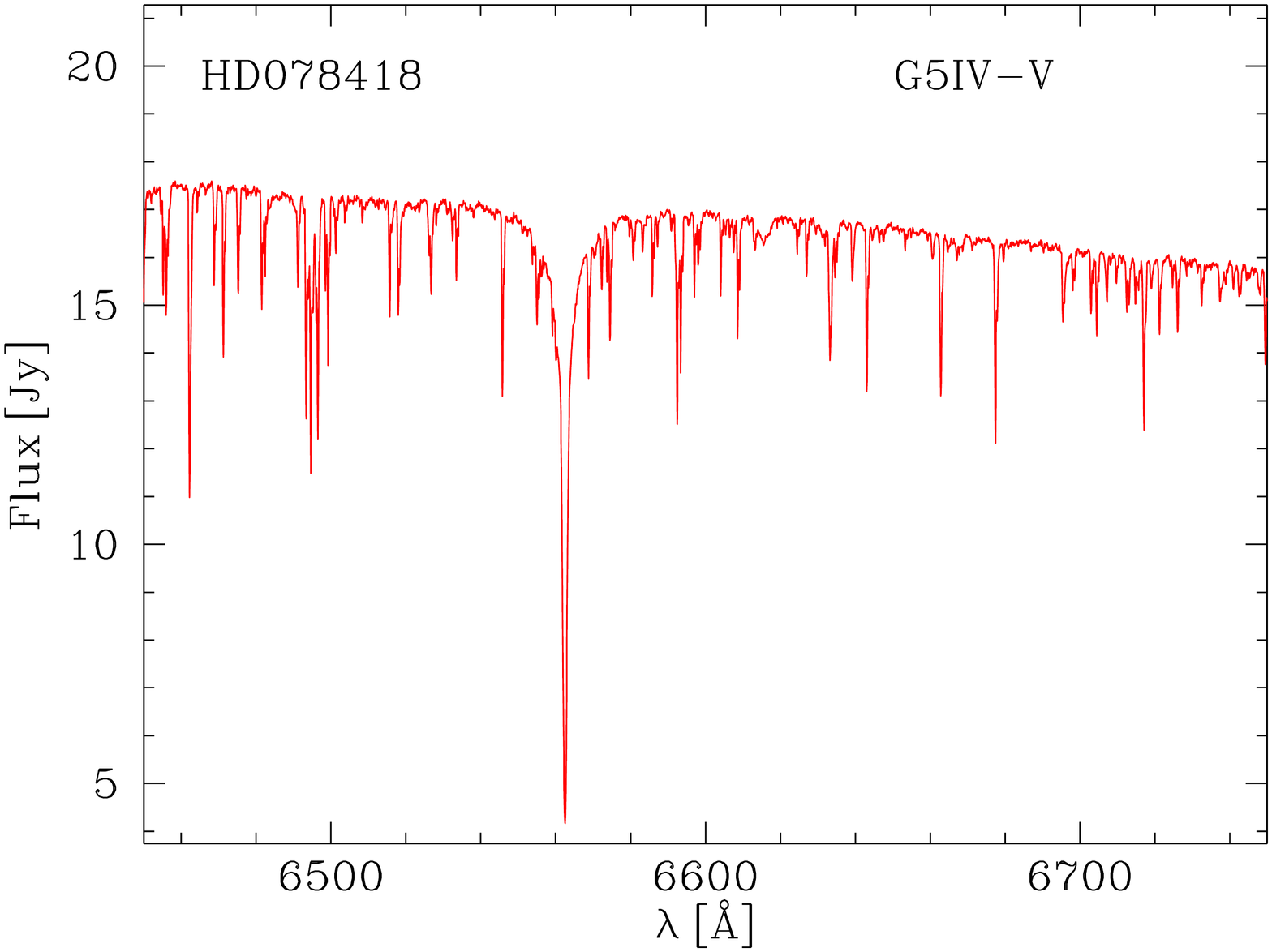}
\includegraphics[width=0.18\textwidth,angle=-90]{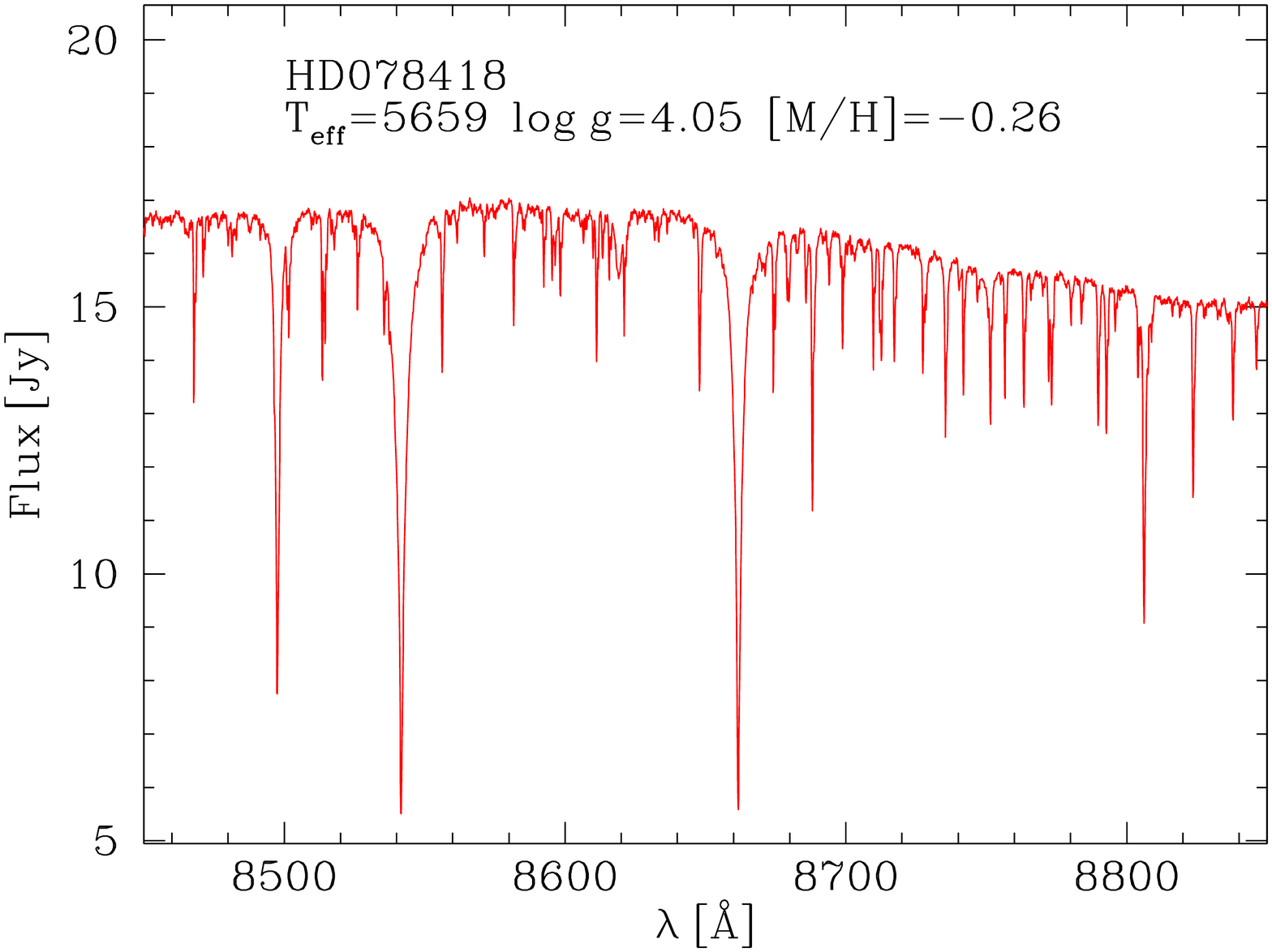}
\includegraphics[width=0.18\textwidth,angle=-90]{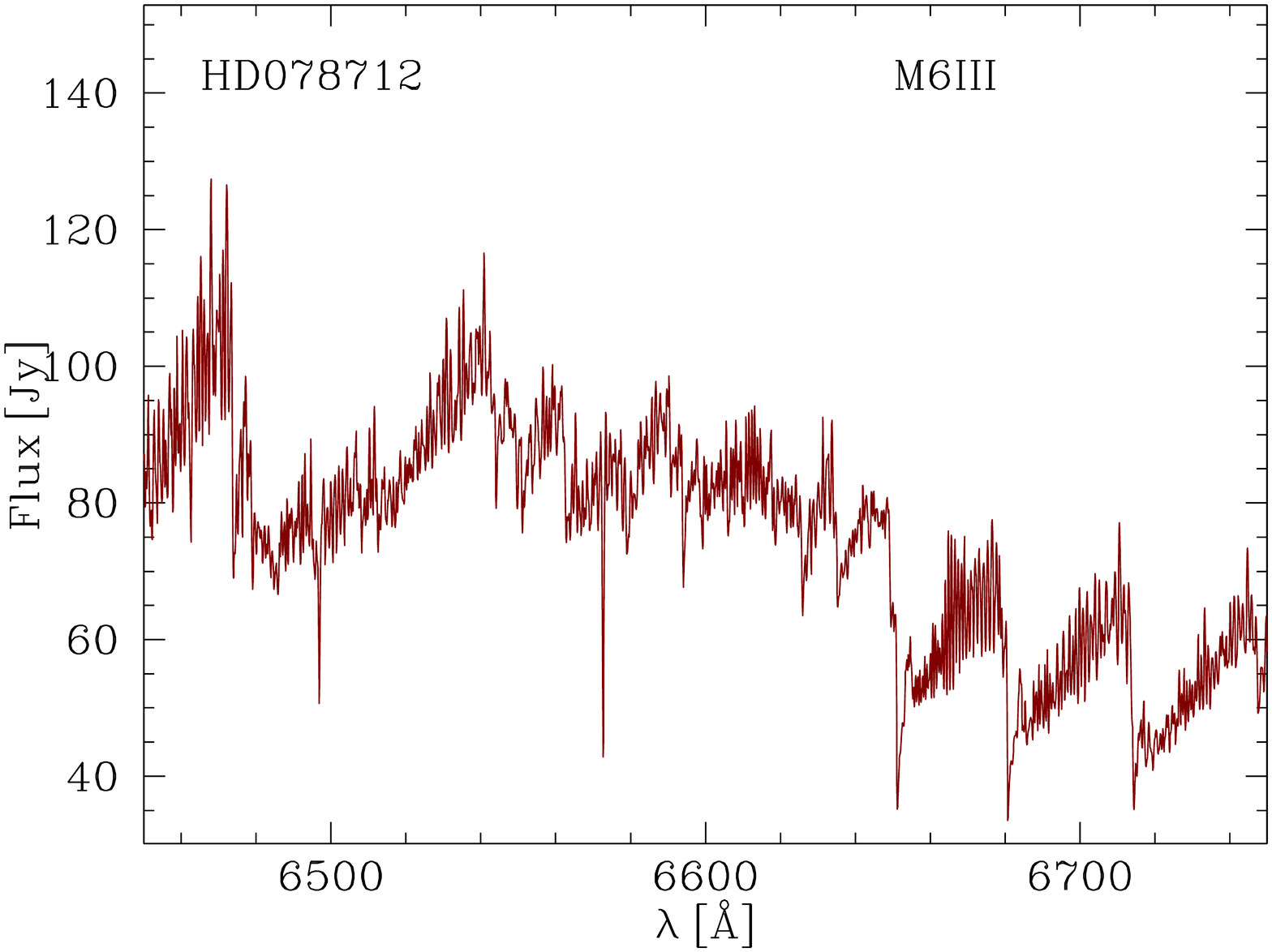}
\includegraphics[width=0.18\textwidth,angle=-90]{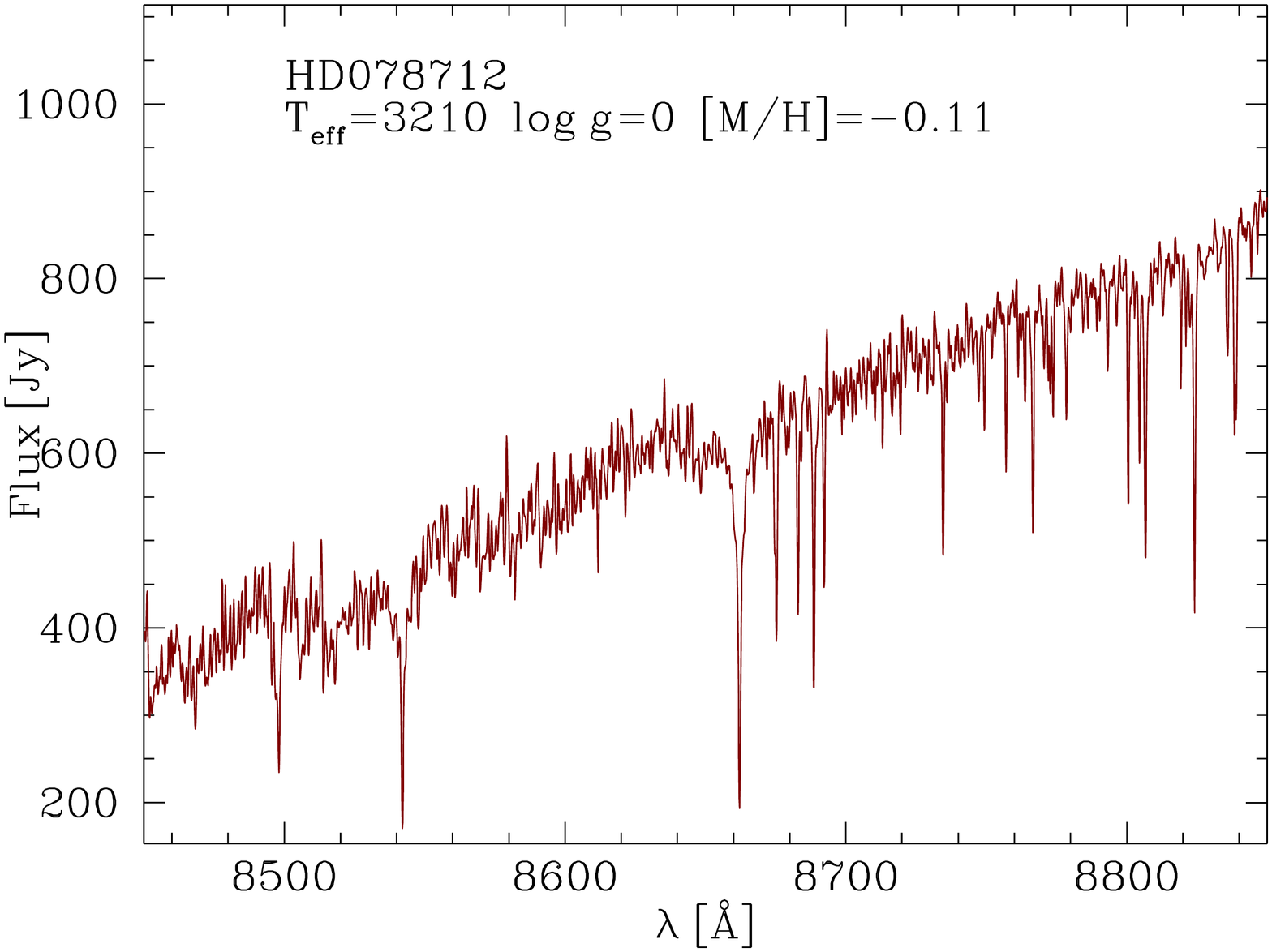}

\contcaption{16. Stars shown in this page are: HD074377, HD075302, HD075318, HD075333, HD075732, HD075782, HD076813, HD076943, HD078175, HD078209, HD078249, HD078362, HD078418 and HD078712.}
\end{figure*}

\begin{figure*}
\includegraphics[width=0.18\textwidth,angle=-90]{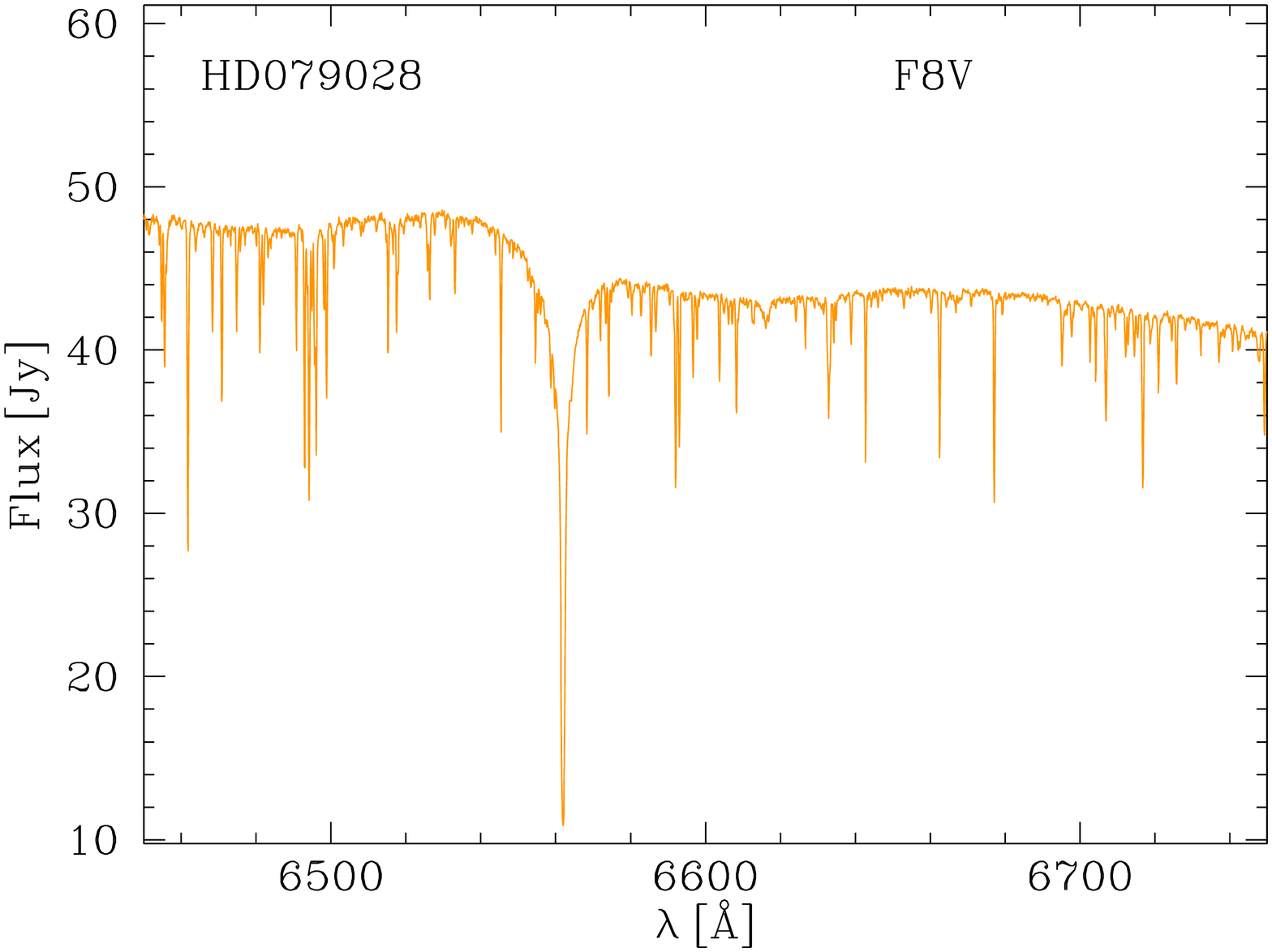}
\includegraphics[width=0.18\textwidth,angle=-90]{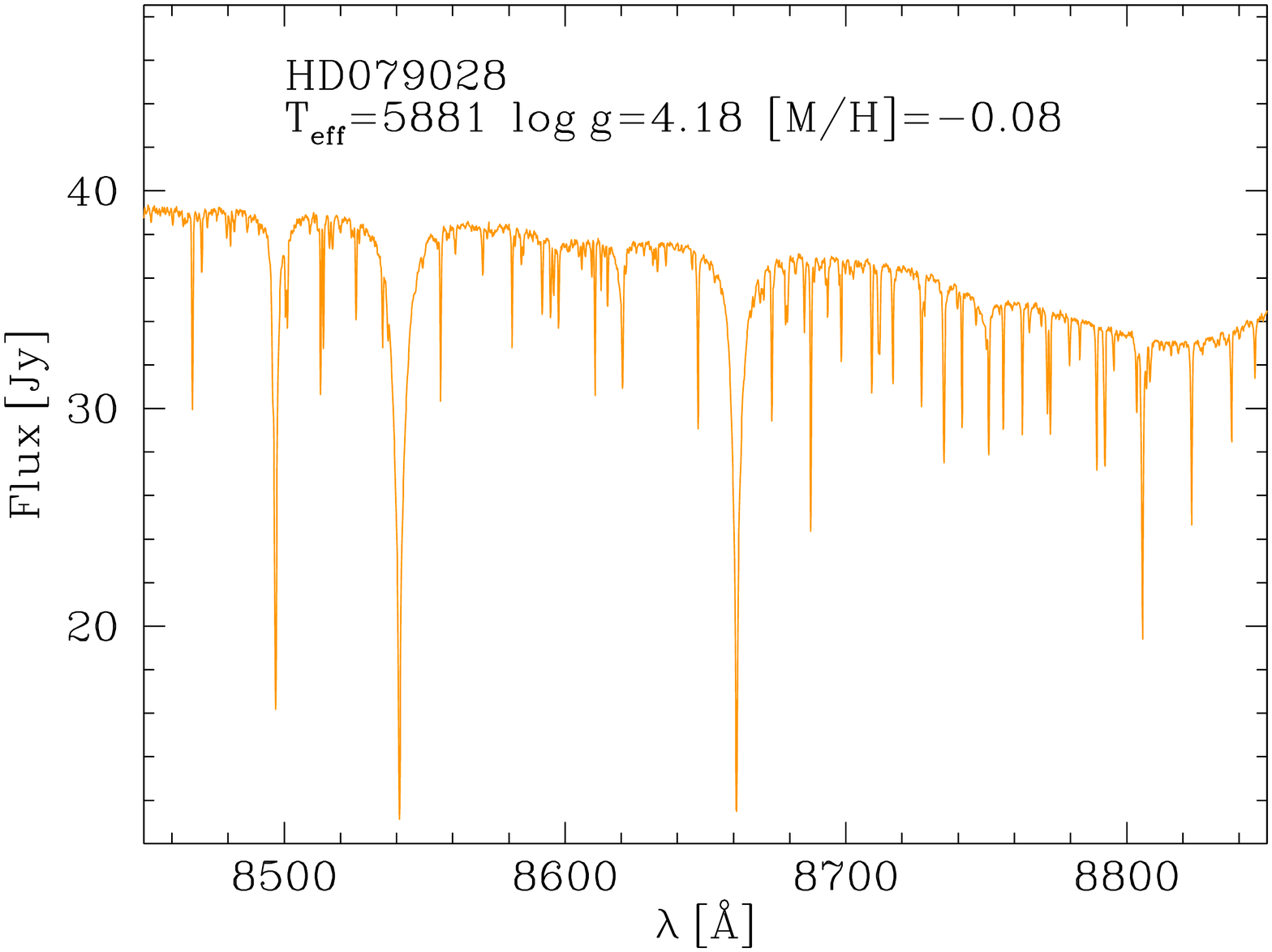}
\includegraphics[width=0.18\textwidth,angle=-90]{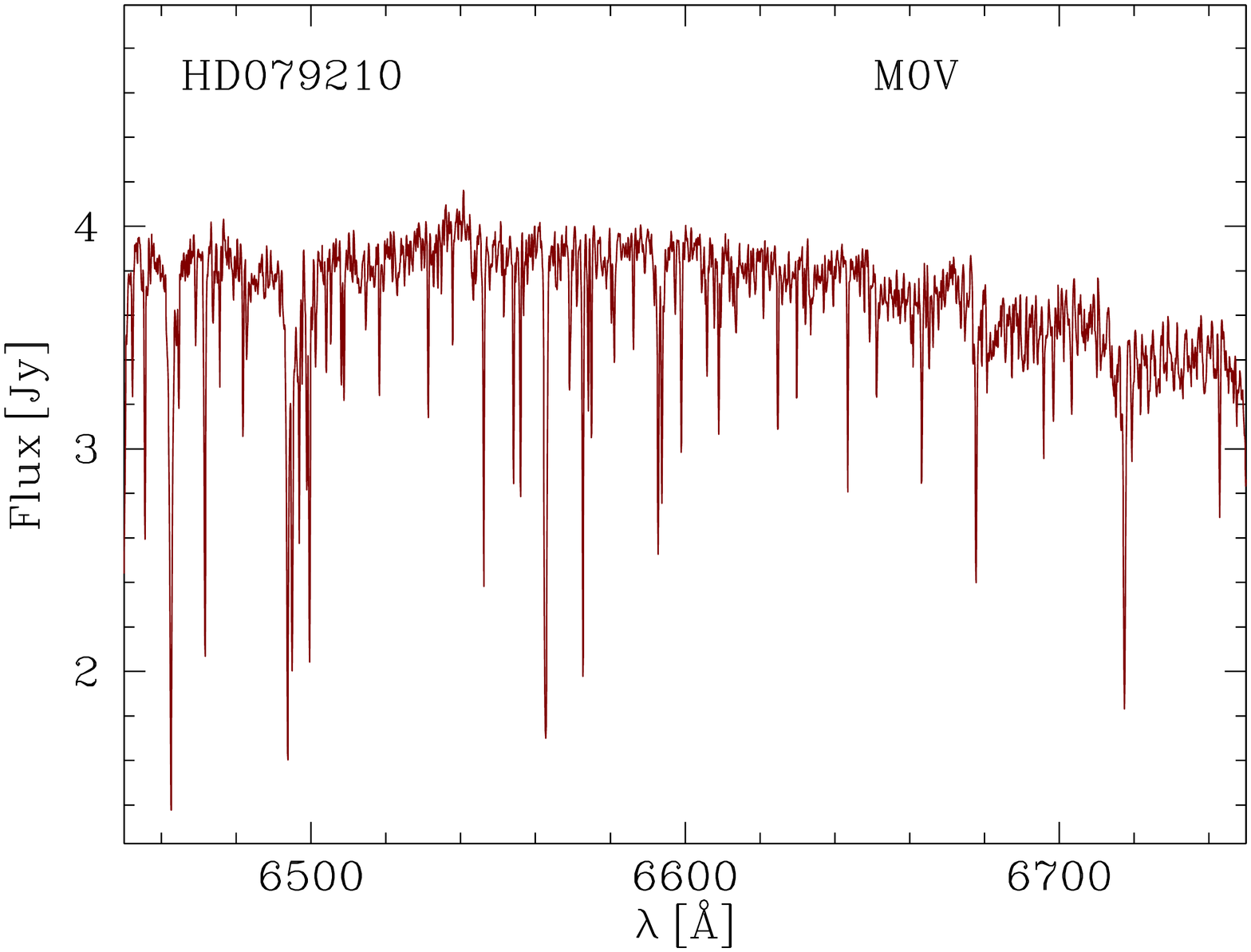}
\includegraphics[width=0.18\textwidth,angle=-90]{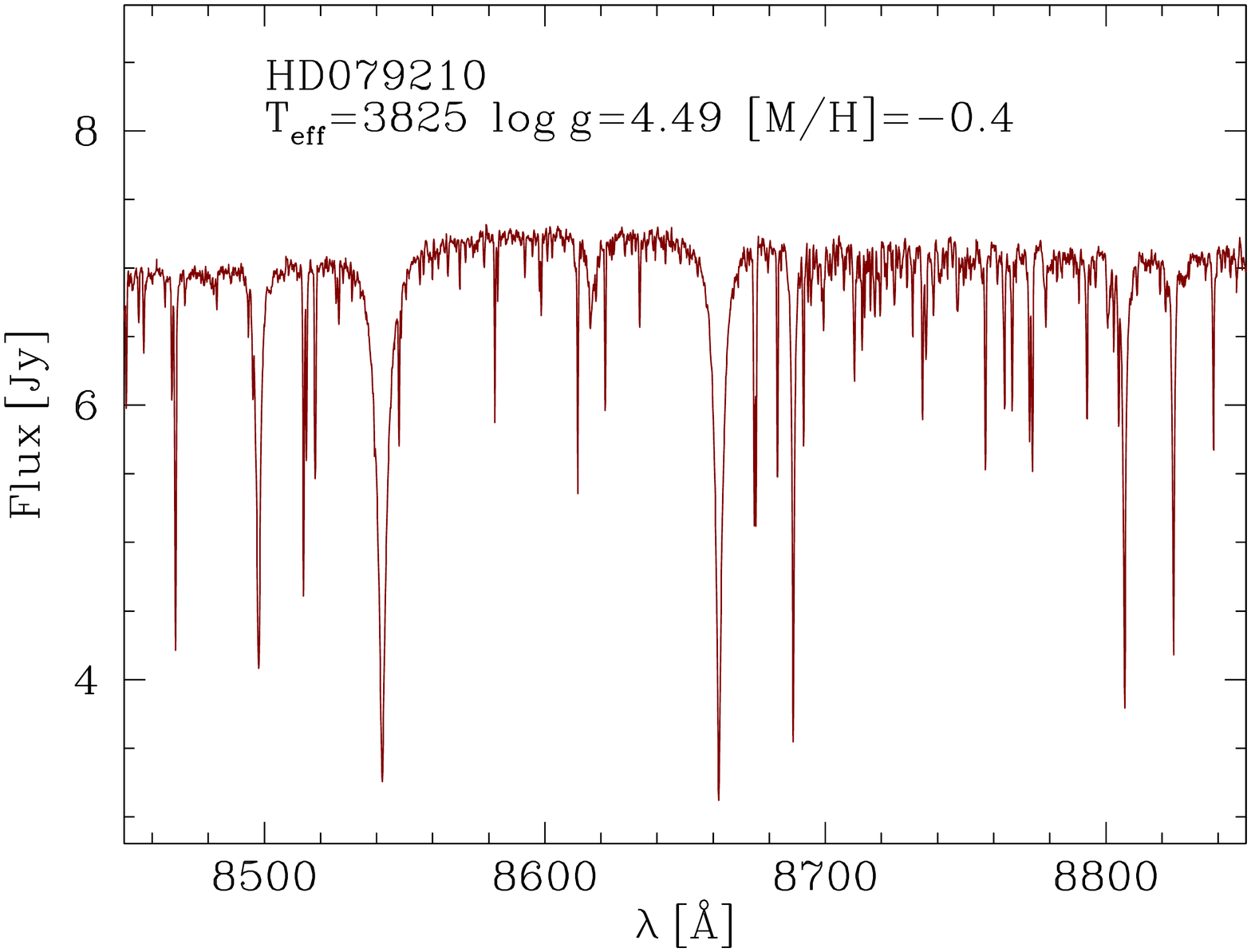}
\includegraphics[width=0.18\textwidth,angle=-90]{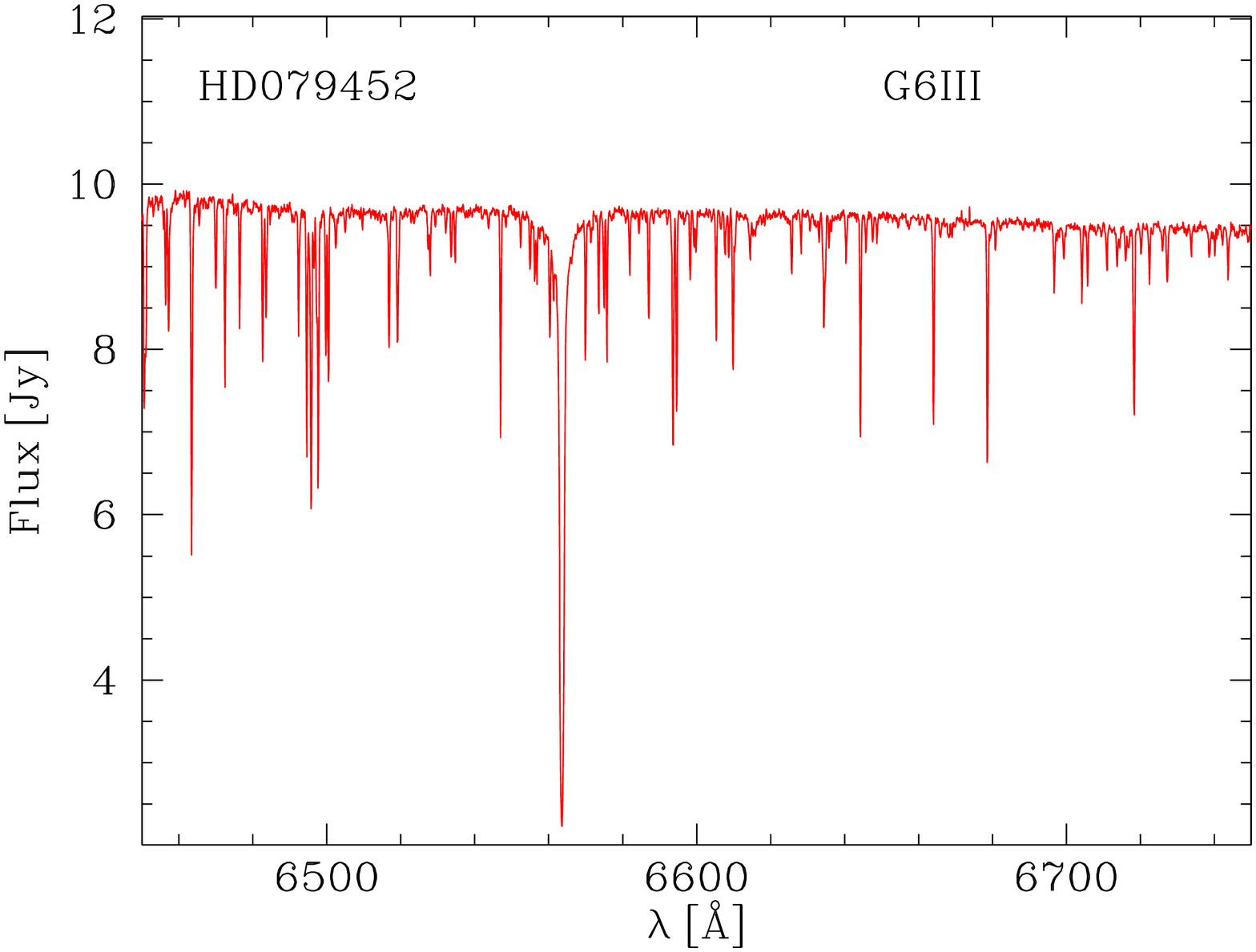}
\includegraphics[width=0.18\textwidth,angle=-90]{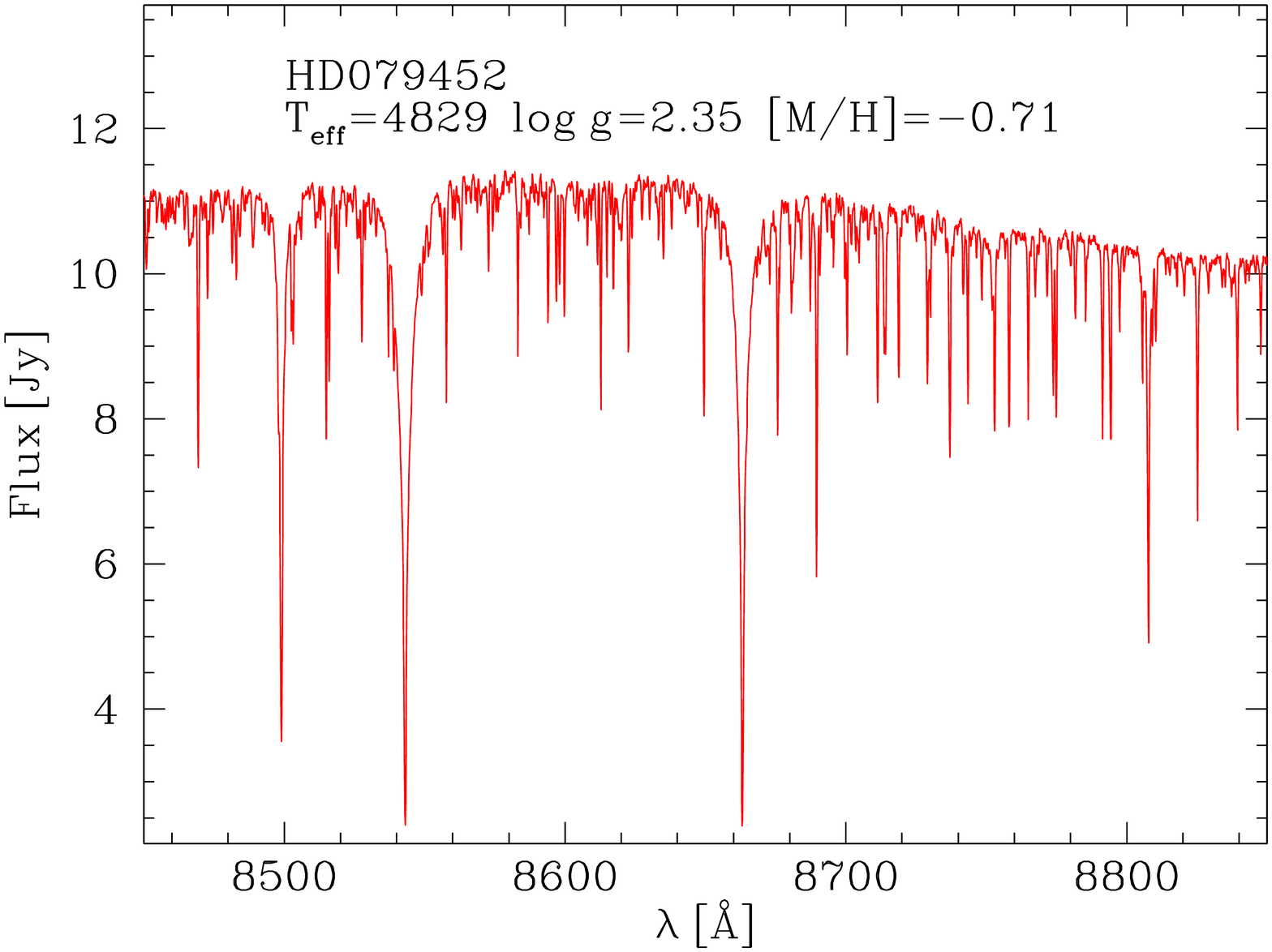}
\includegraphics[width=0.18\textwidth,angle=-90]{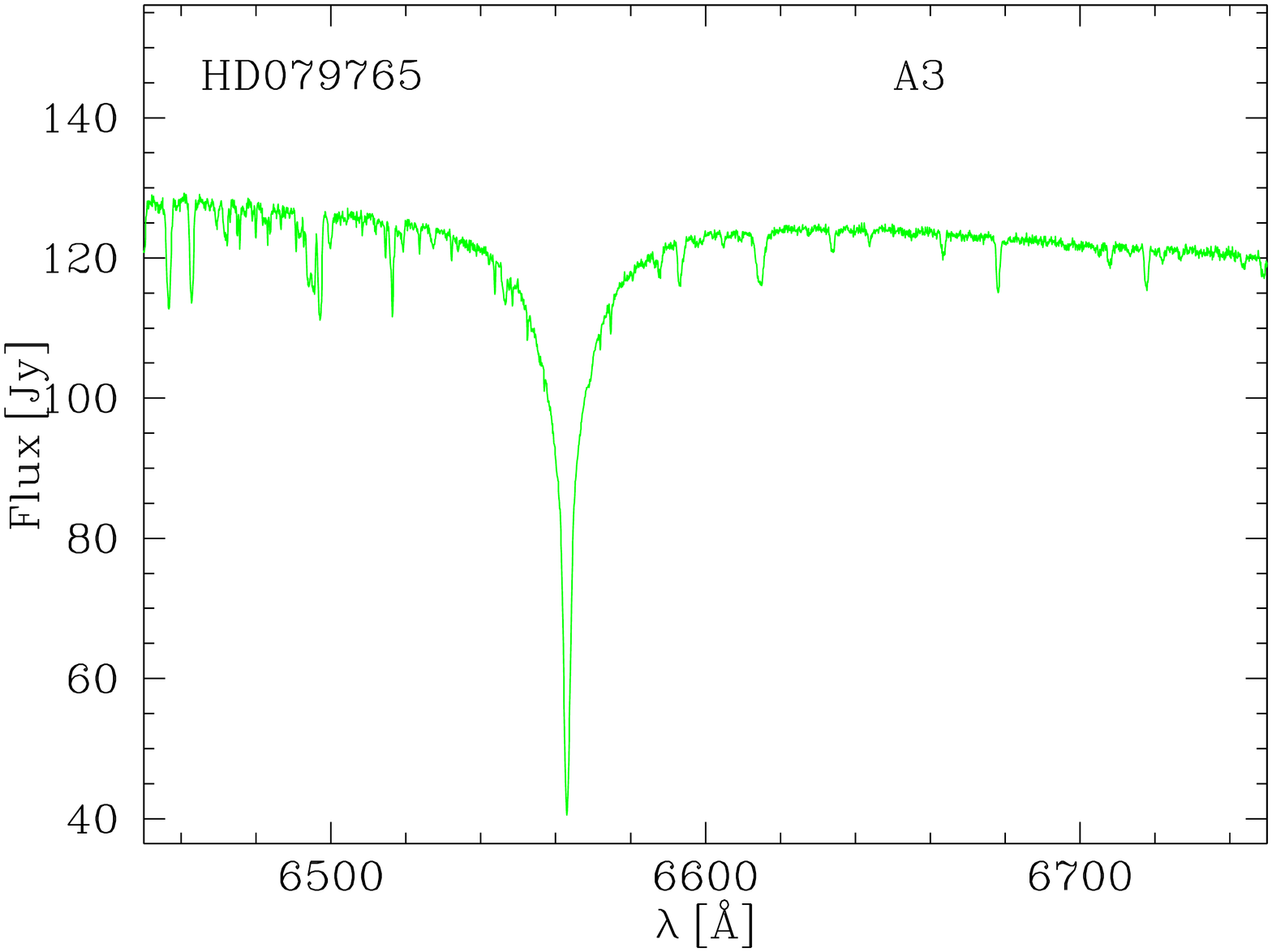}
\includegraphics[width=0.18\textwidth,angle=-90]{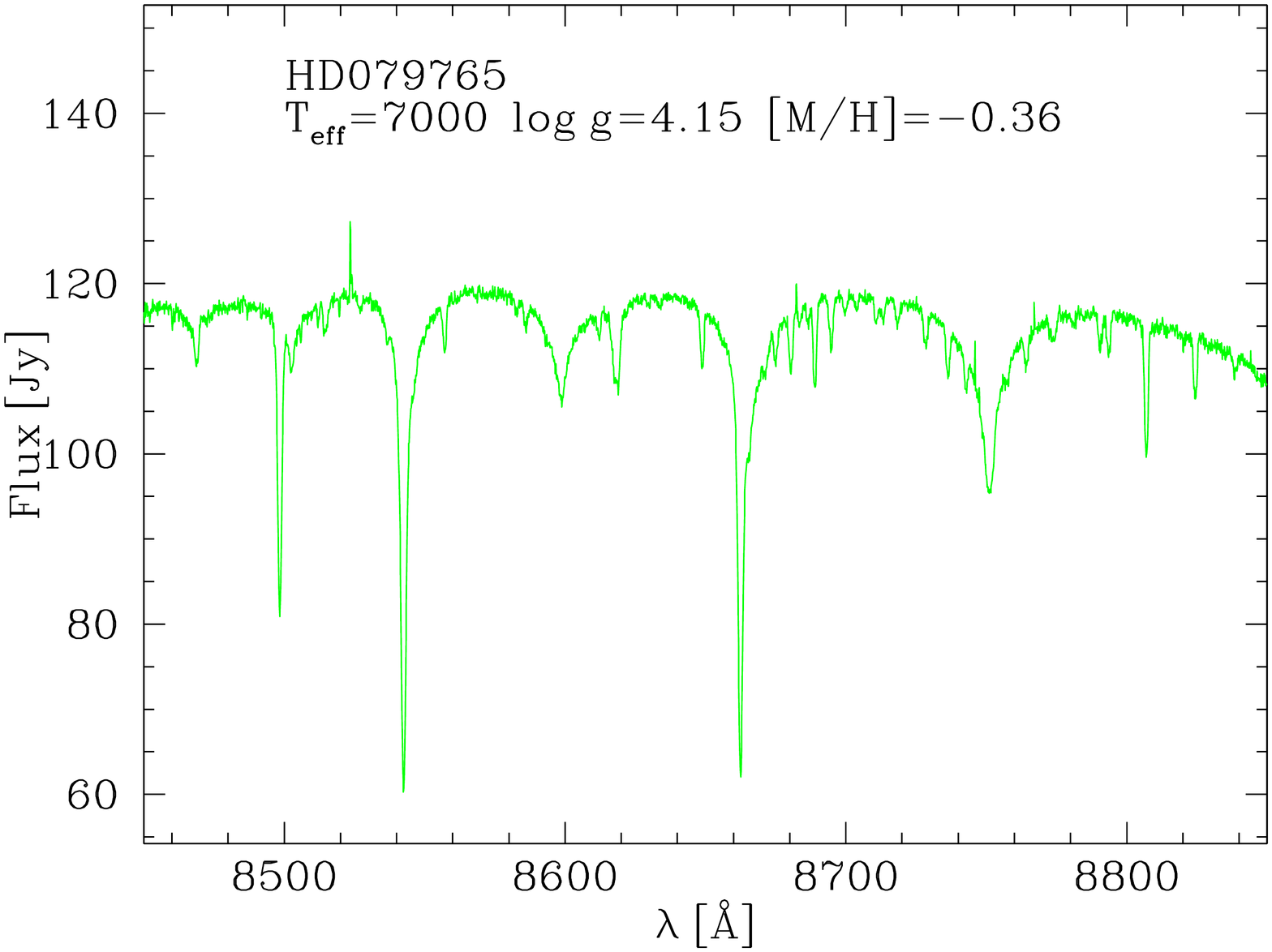}
\includegraphics[width=0.18\textwidth,angle=-90]{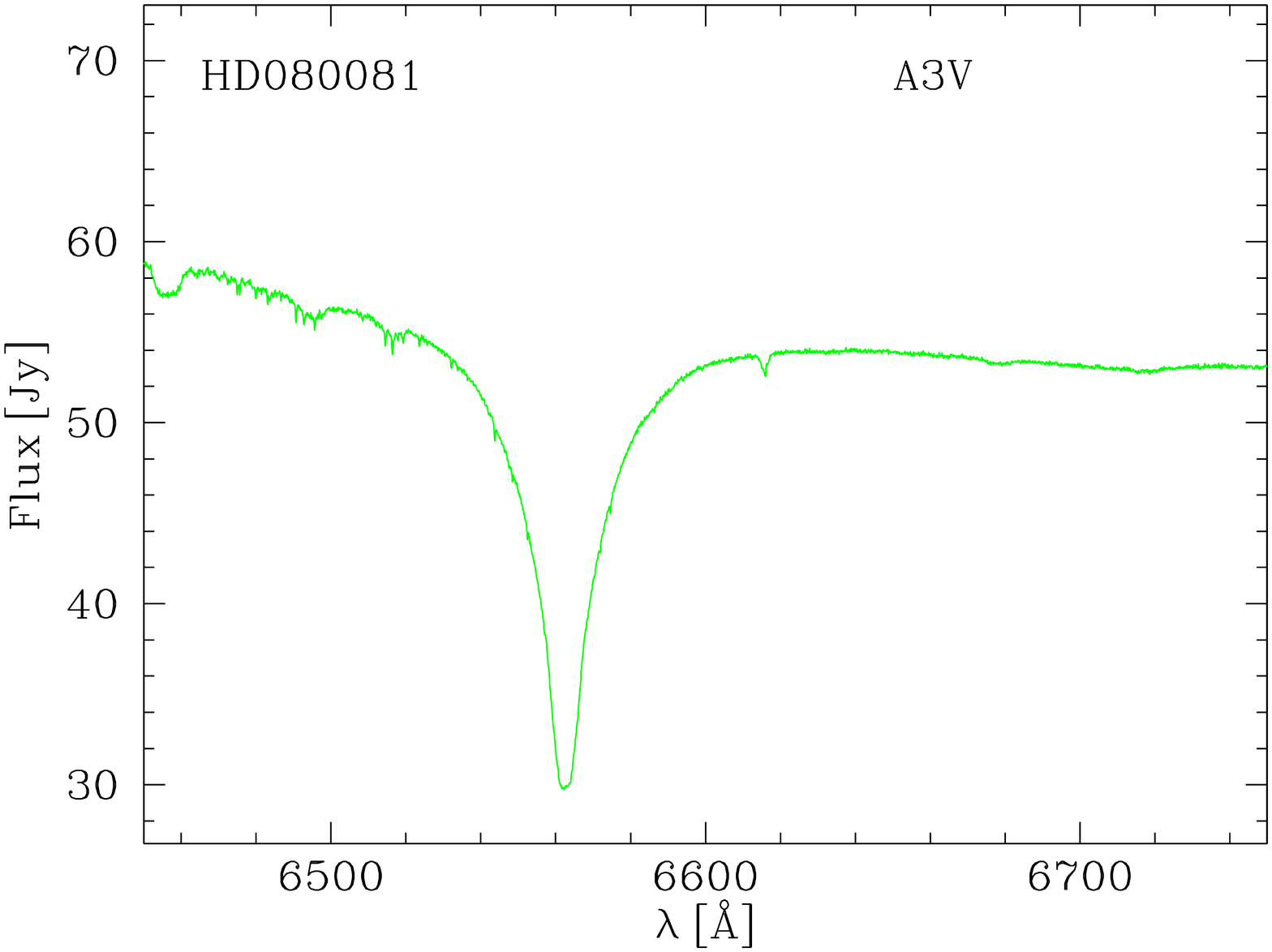}
\includegraphics[width=0.18\textwidth,angle=-90]{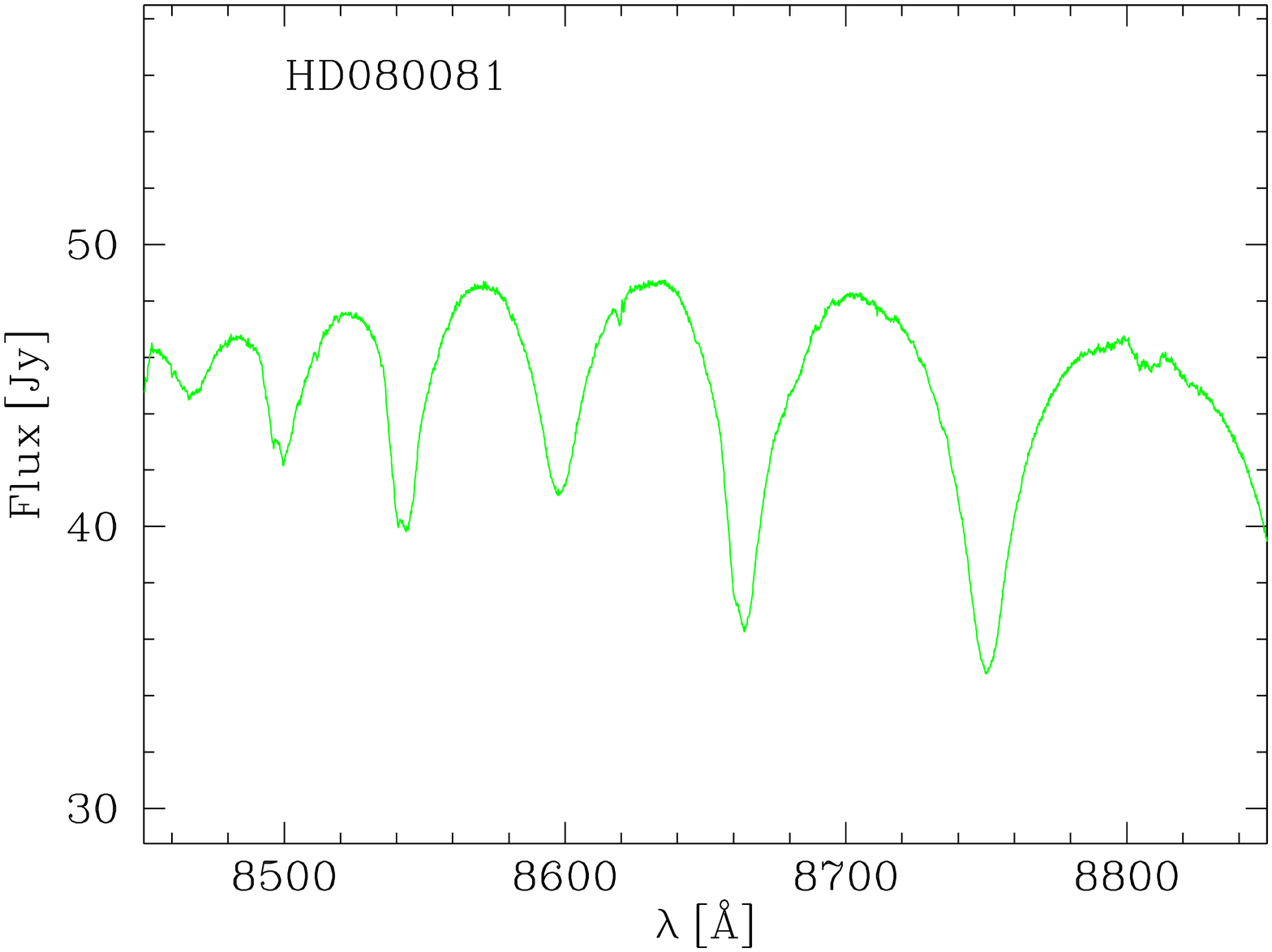}
\includegraphics[width=0.18\textwidth,angle=-90]{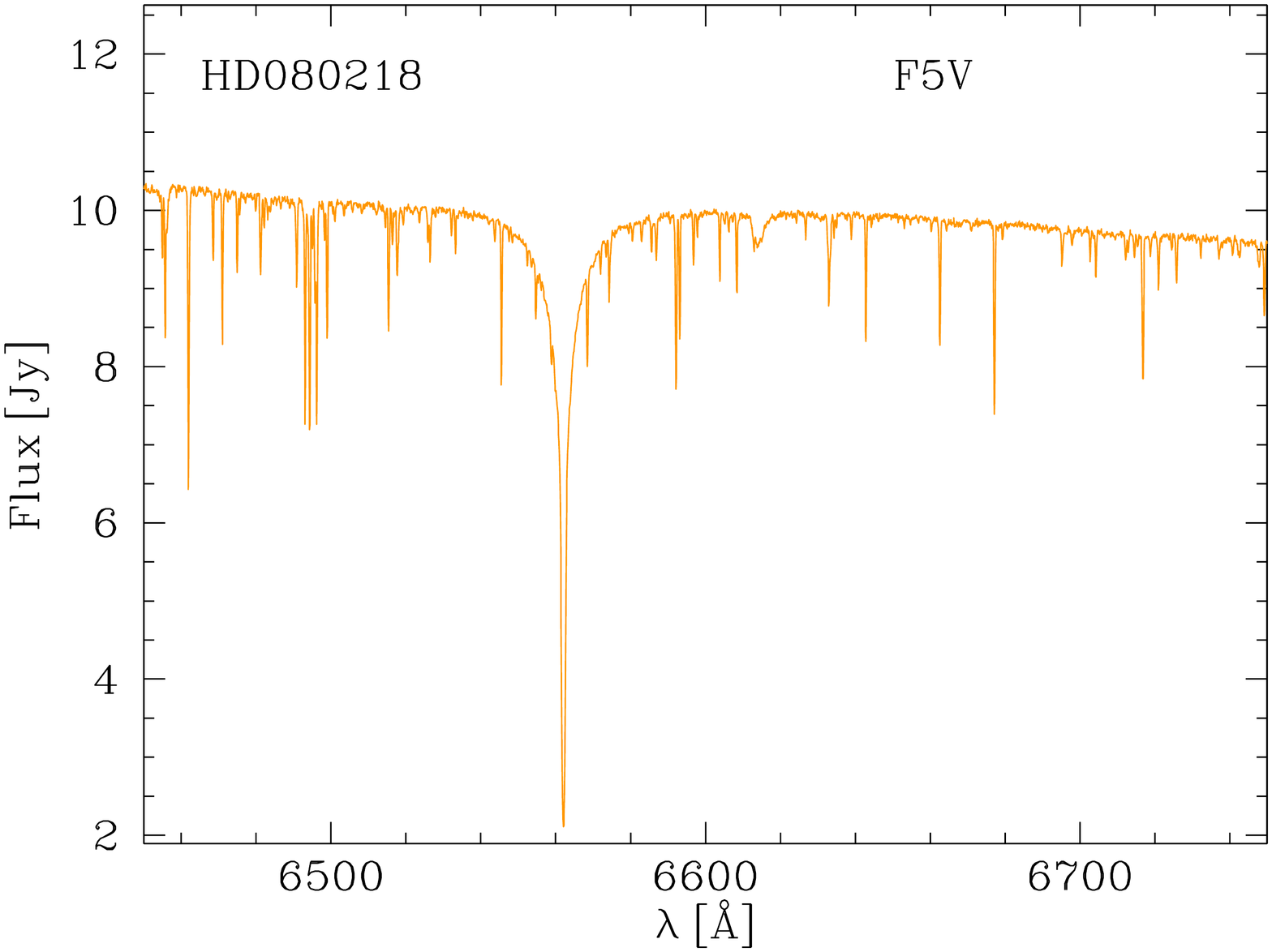}
\includegraphics[width=0.18\textwidth,angle=-90]{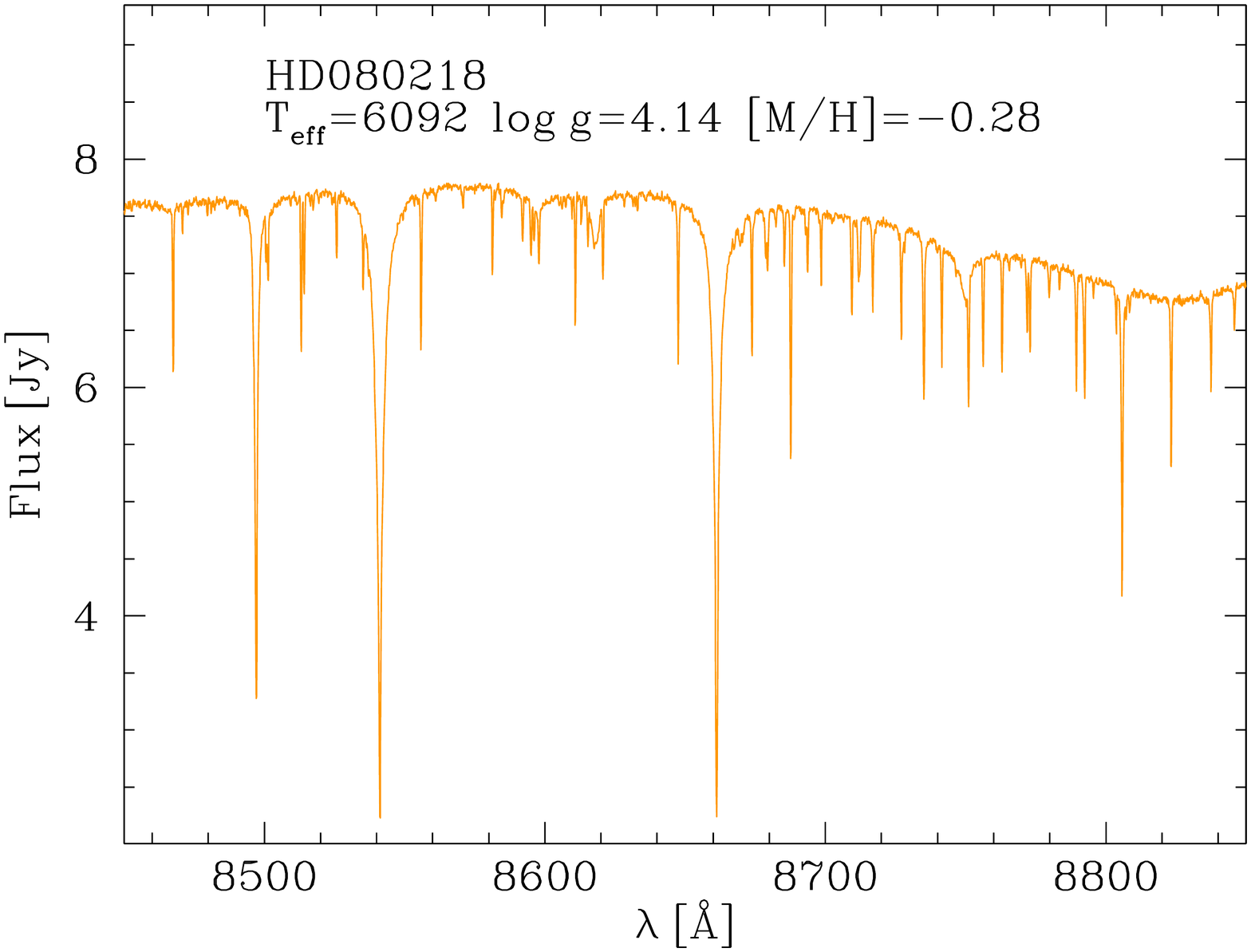}
\includegraphics[width=0.18\textwidth,angle=-90]{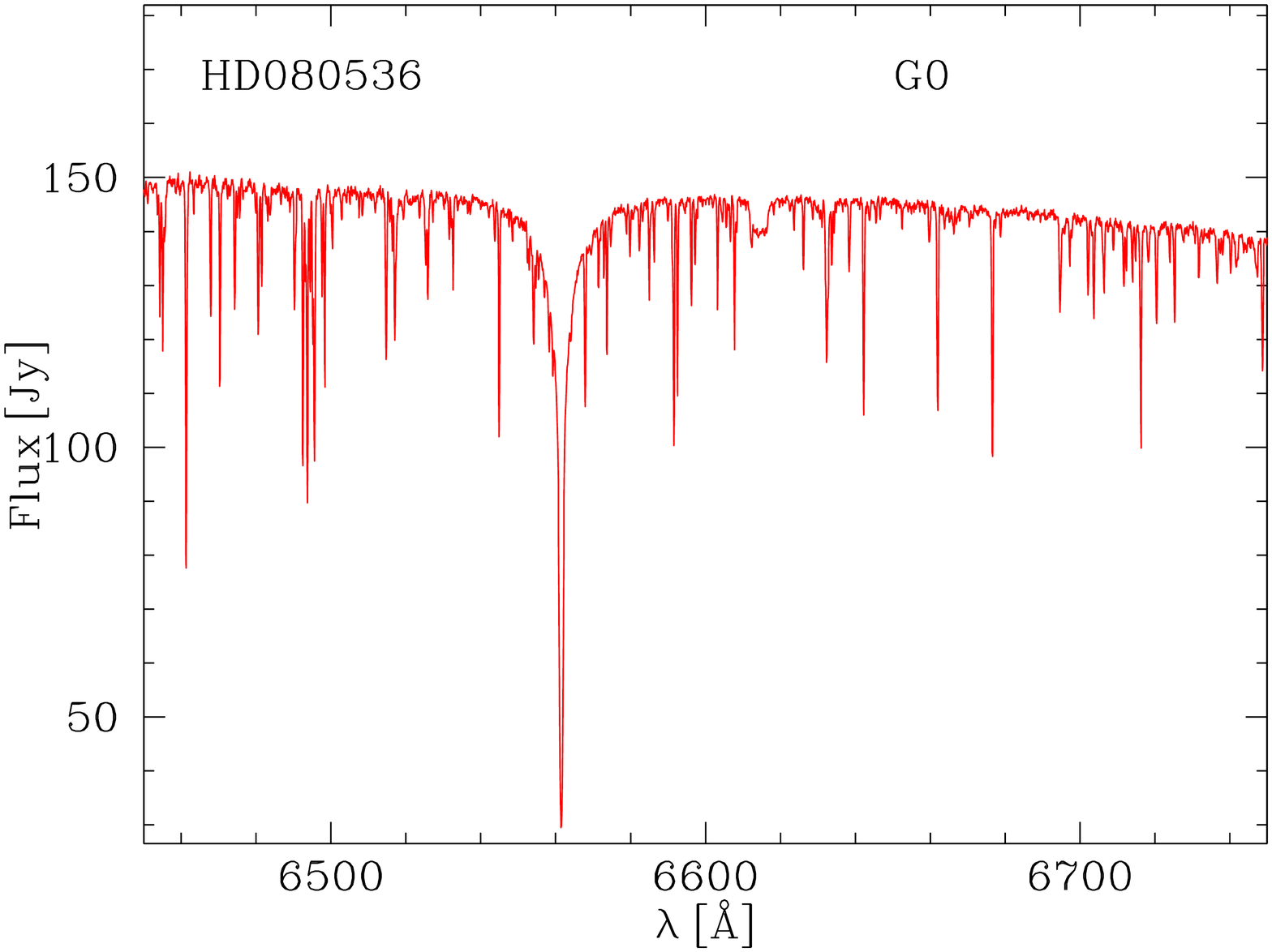}
\includegraphics[width=0.18\textwidth,angle=-90]{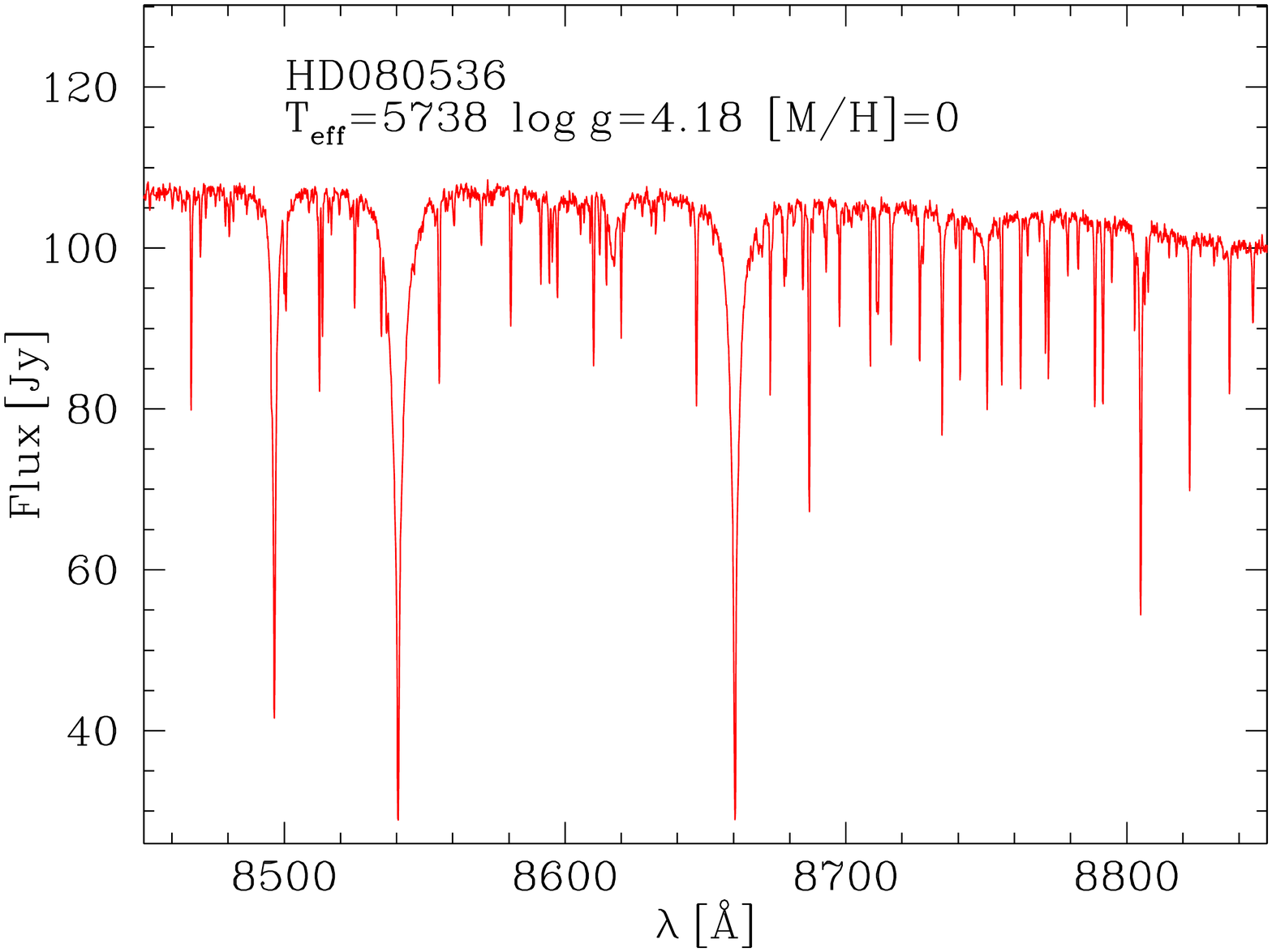}
\includegraphics[width=0.18\textwidth,angle=-90]{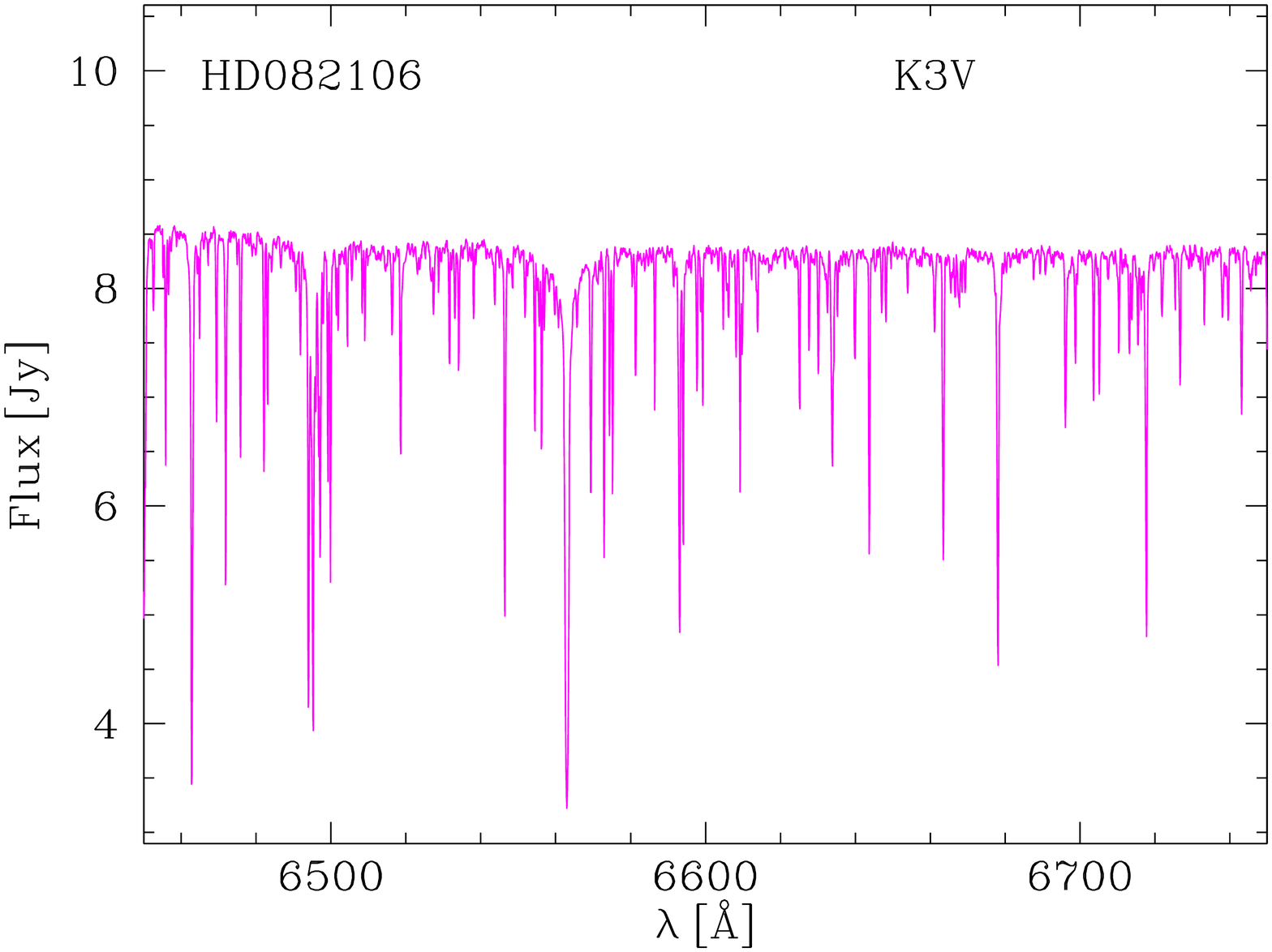}
\includegraphics[width=0.18\textwidth,angle=-90]{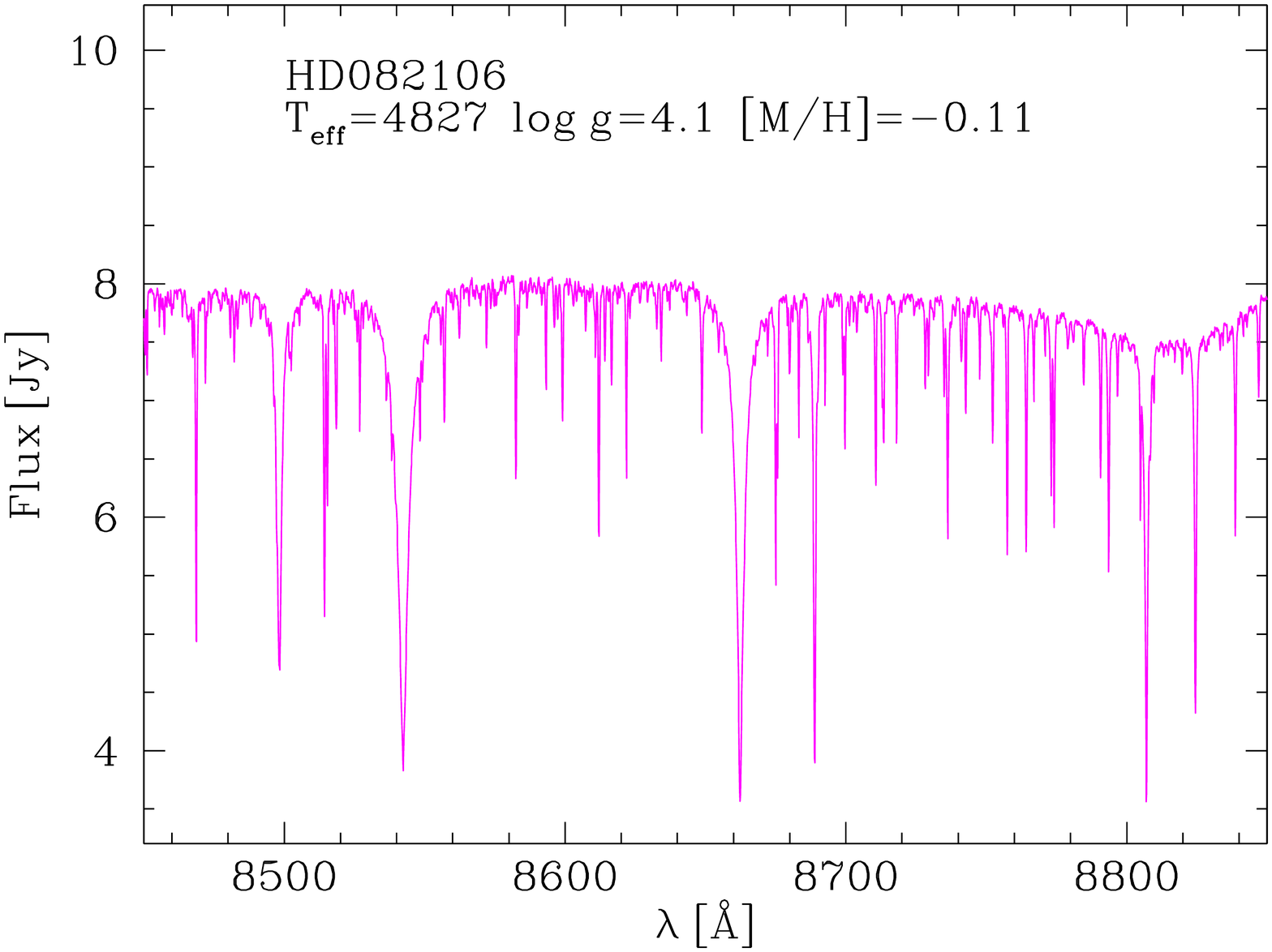}
\includegraphics[width=0.18\textwidth,angle=-90]{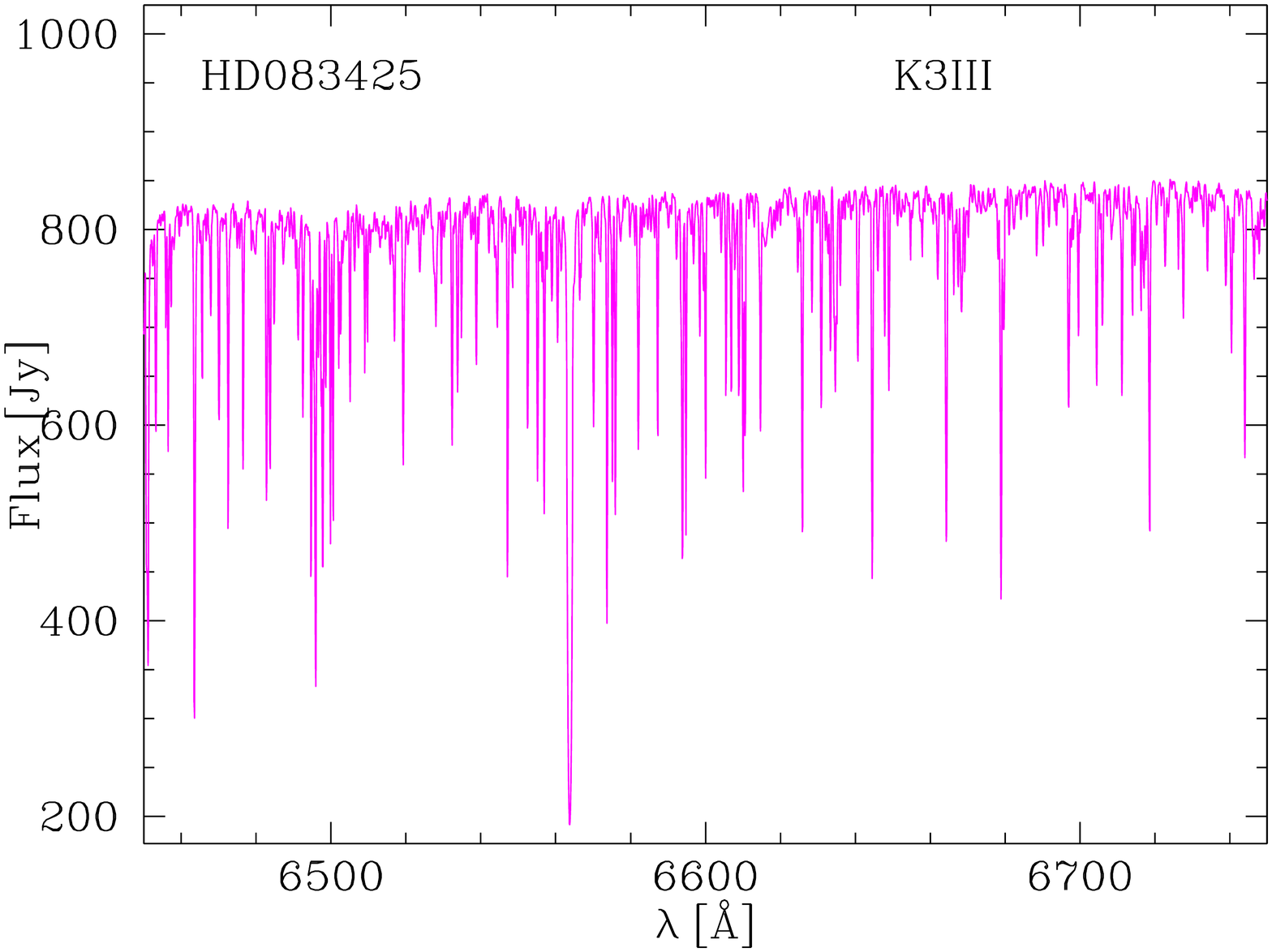}
\includegraphics[width=0.18\textwidth,angle=-90]{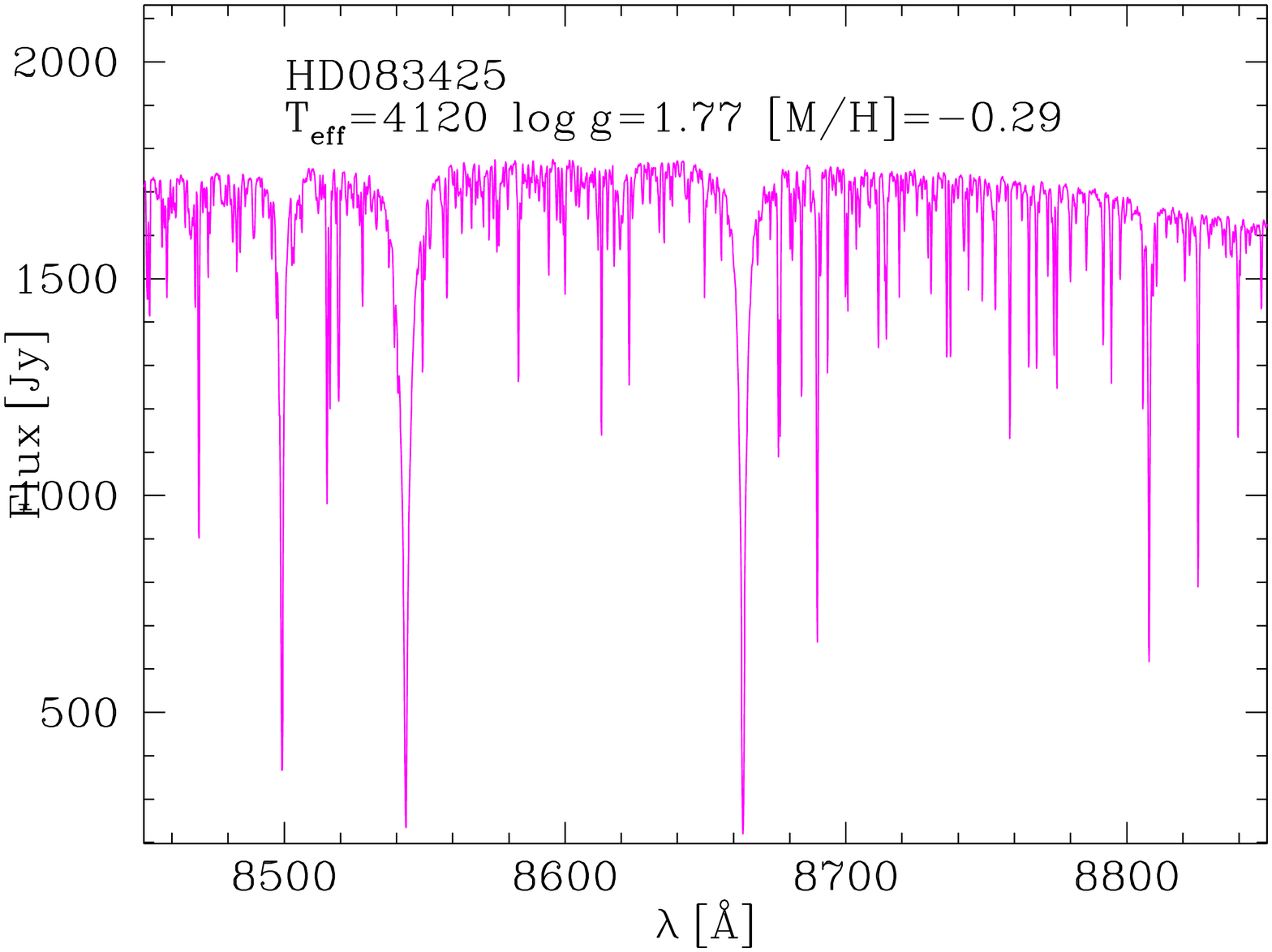}
\includegraphics[width=0.18\textwidth,angle=-90]{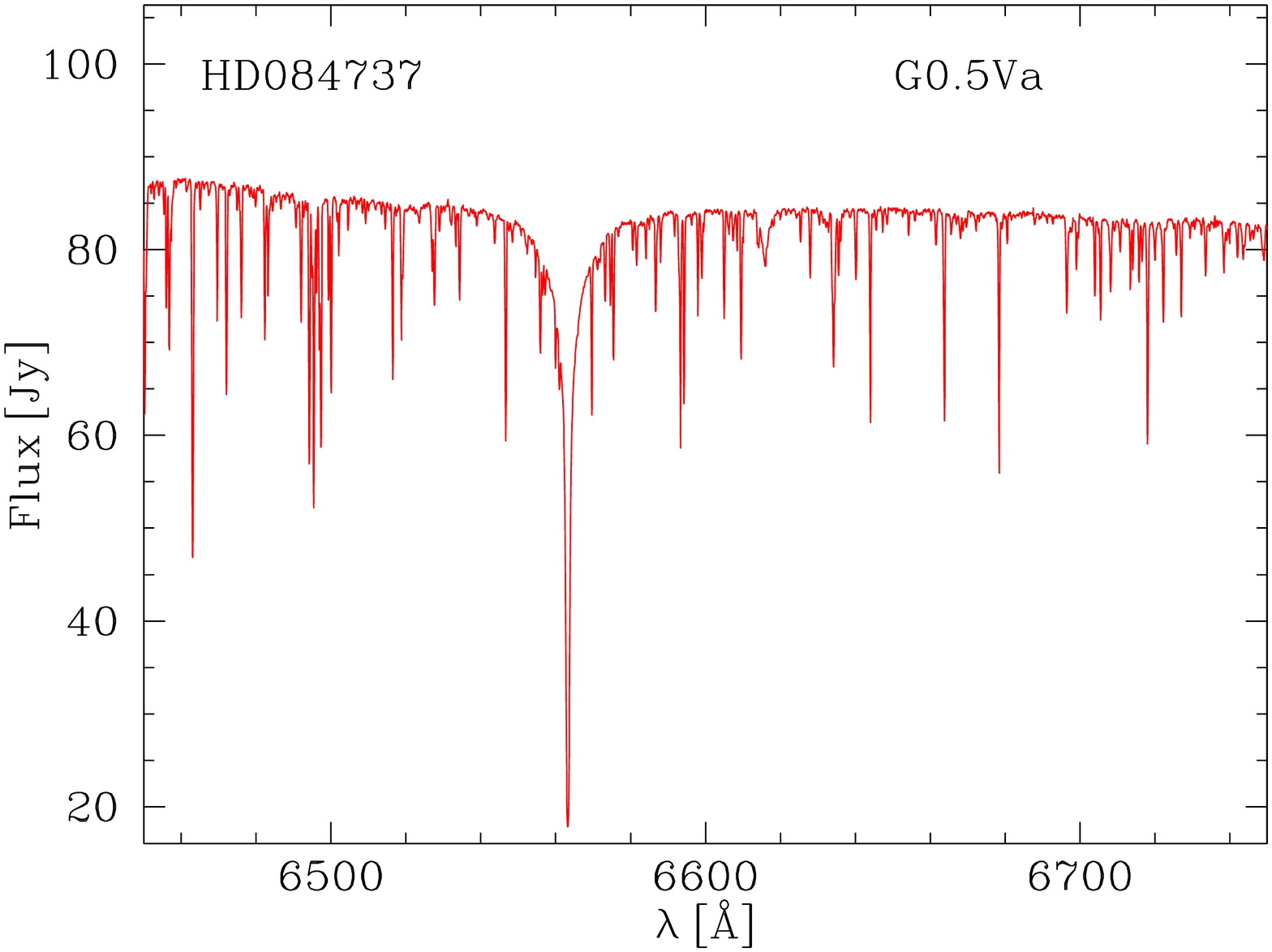}
\includegraphics[width=0.18\textwidth,angle=-90]{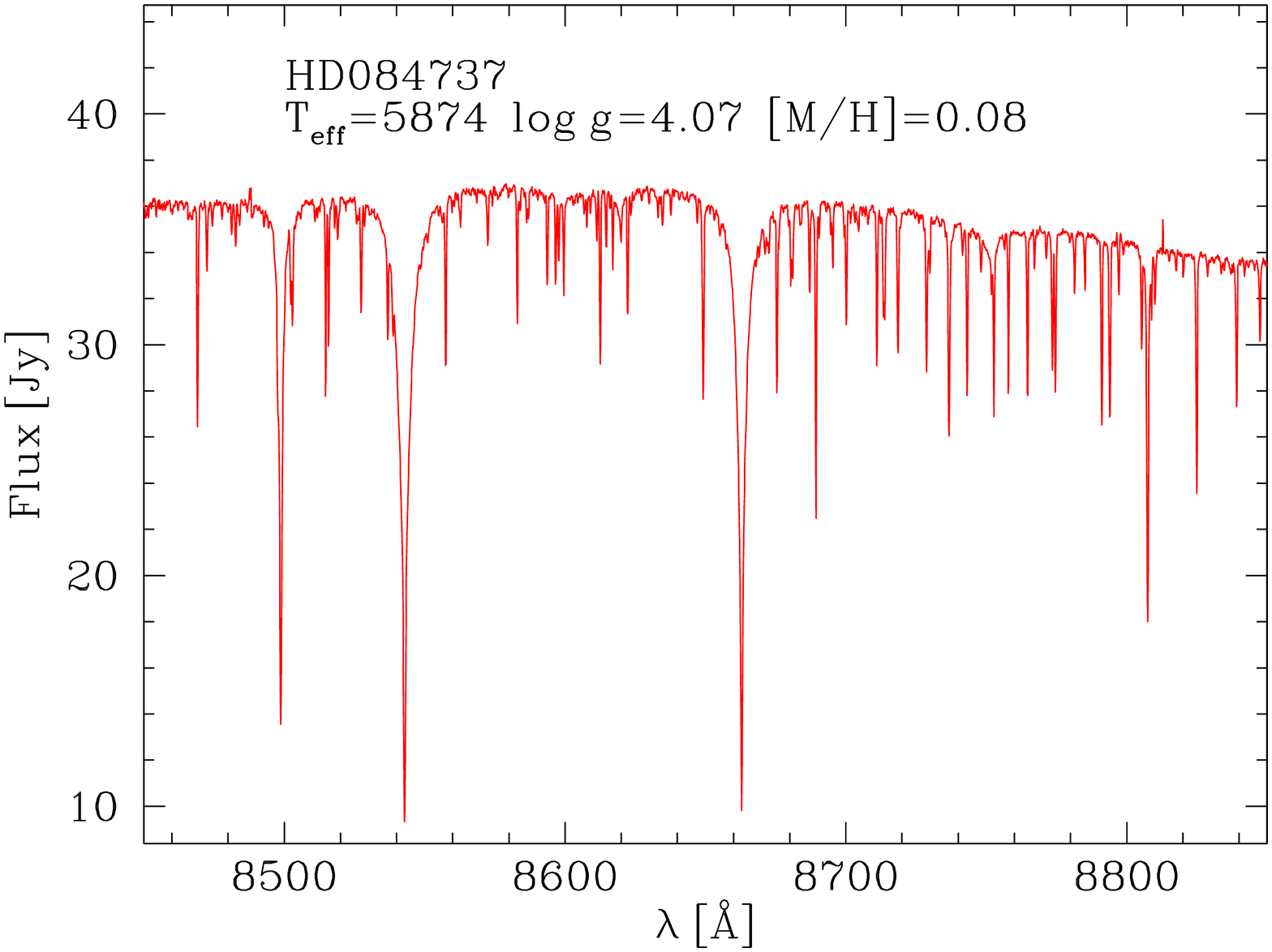}
\includegraphics[width=0.18\textwidth,angle=-90]{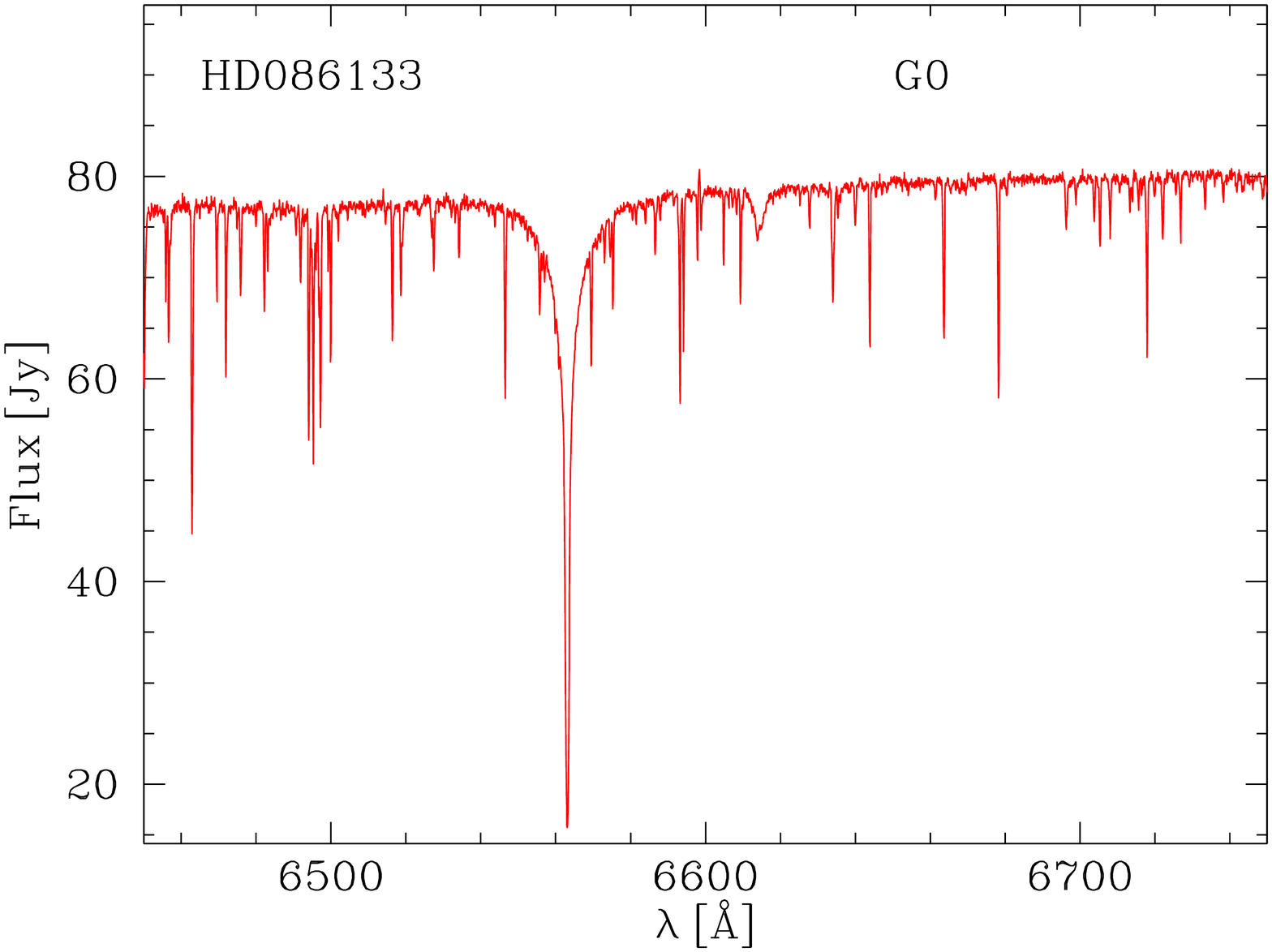}
\includegraphics[width=0.18\textwidth,angle=-90]{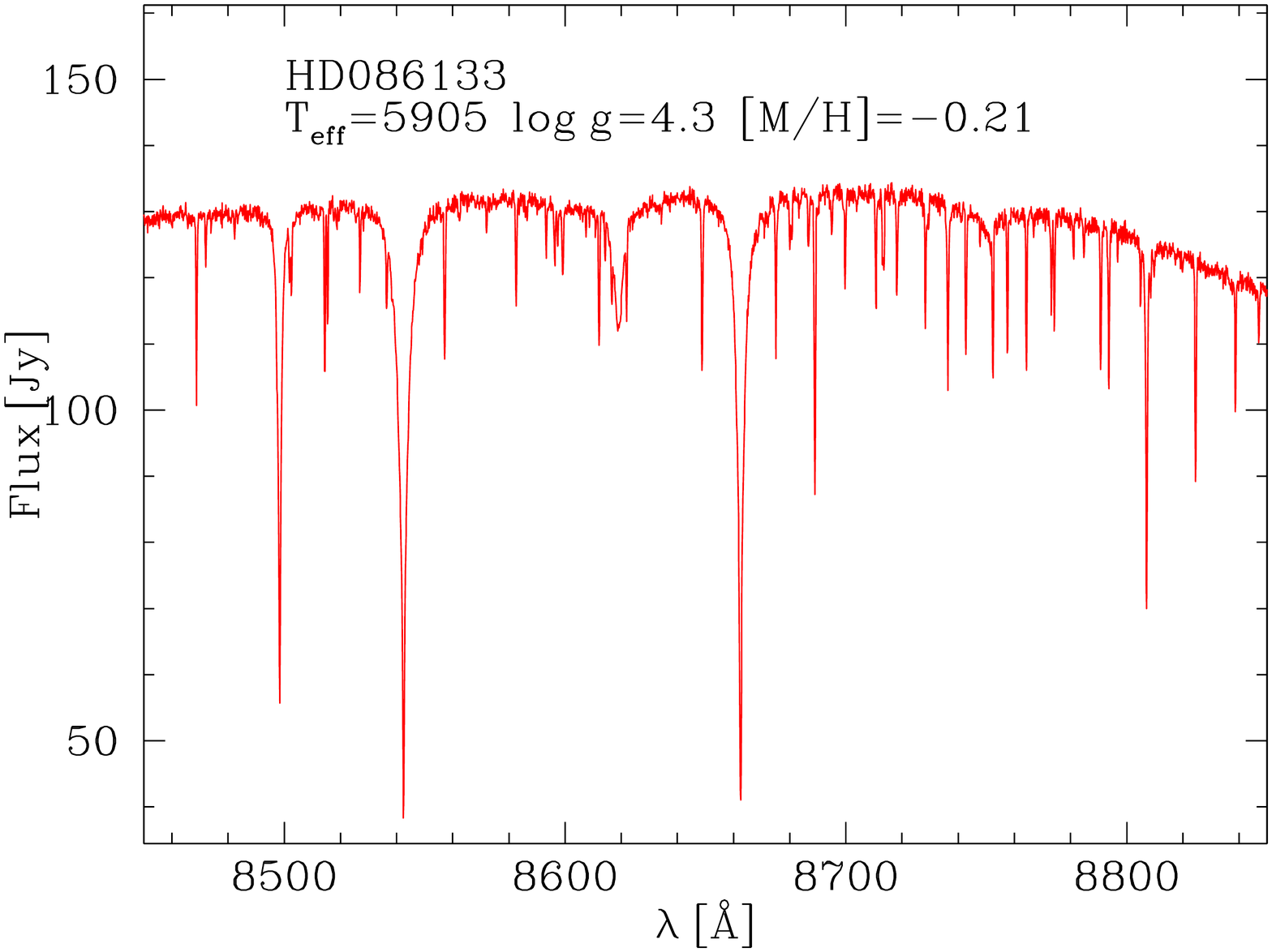}
\includegraphics[width=0.18\textwidth,angle=-90]{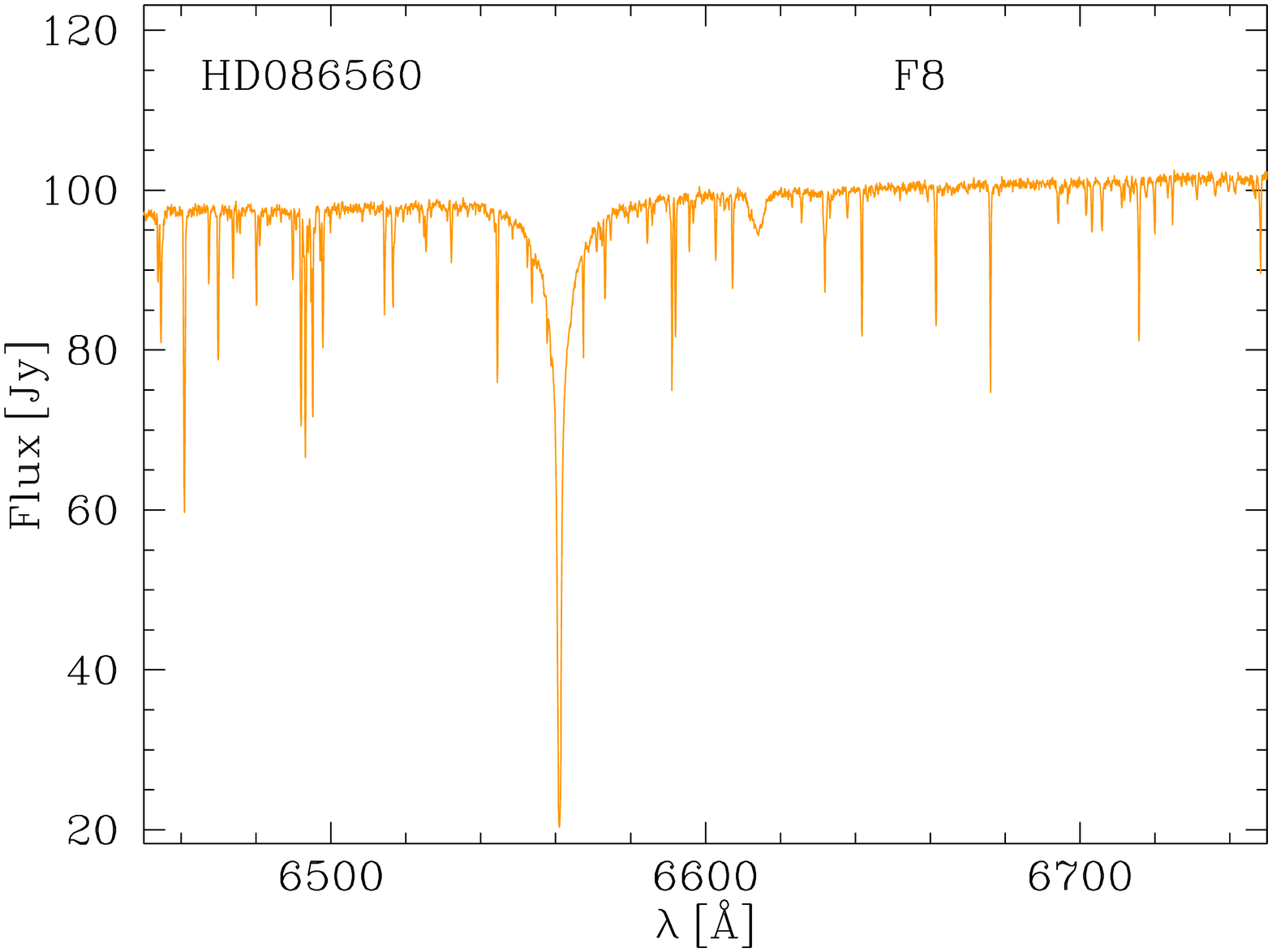}
\includegraphics[width=0.18\textwidth,angle=-90]{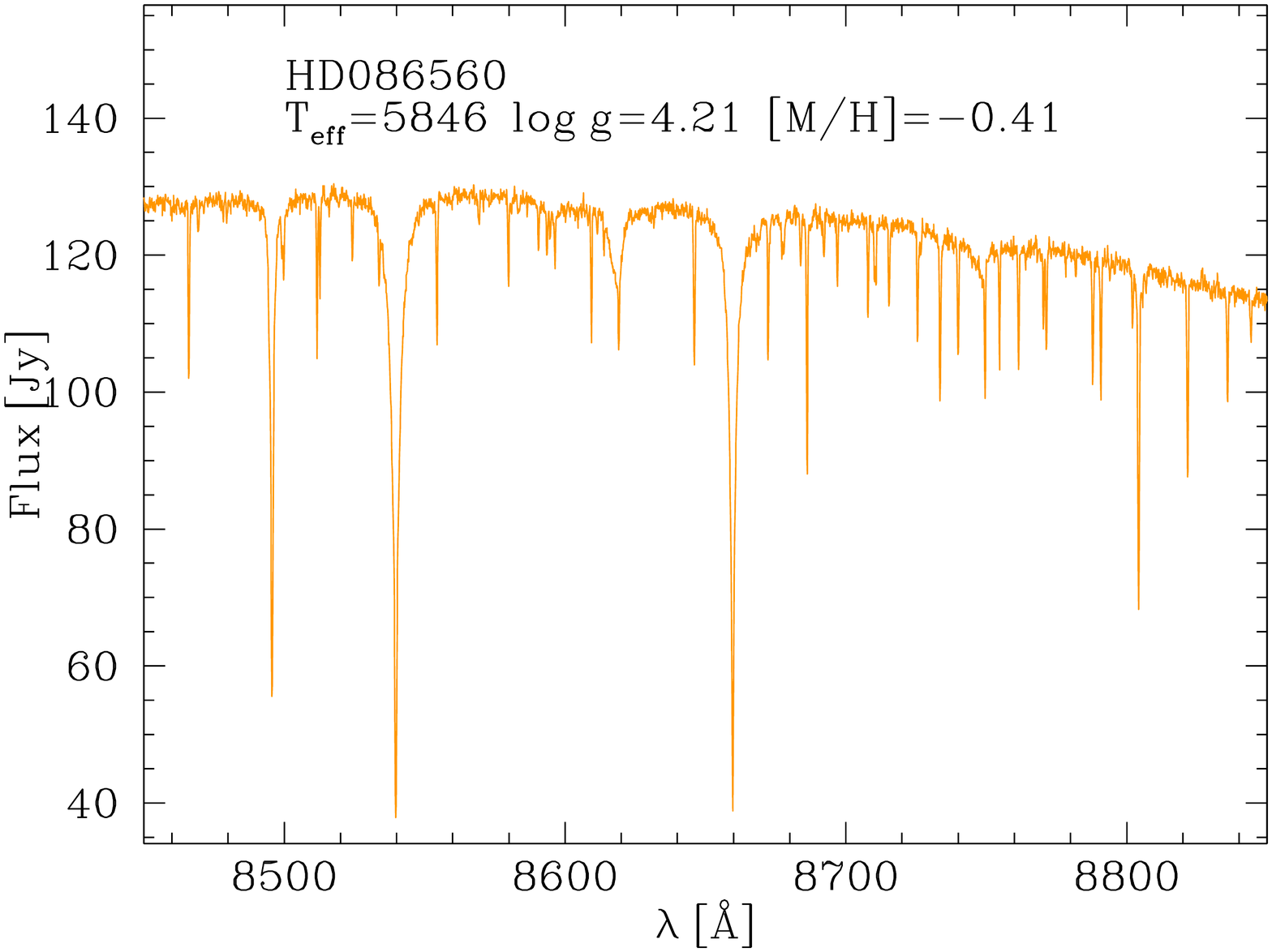}
\includegraphics[width=0.18\textwidth,angle=-90]{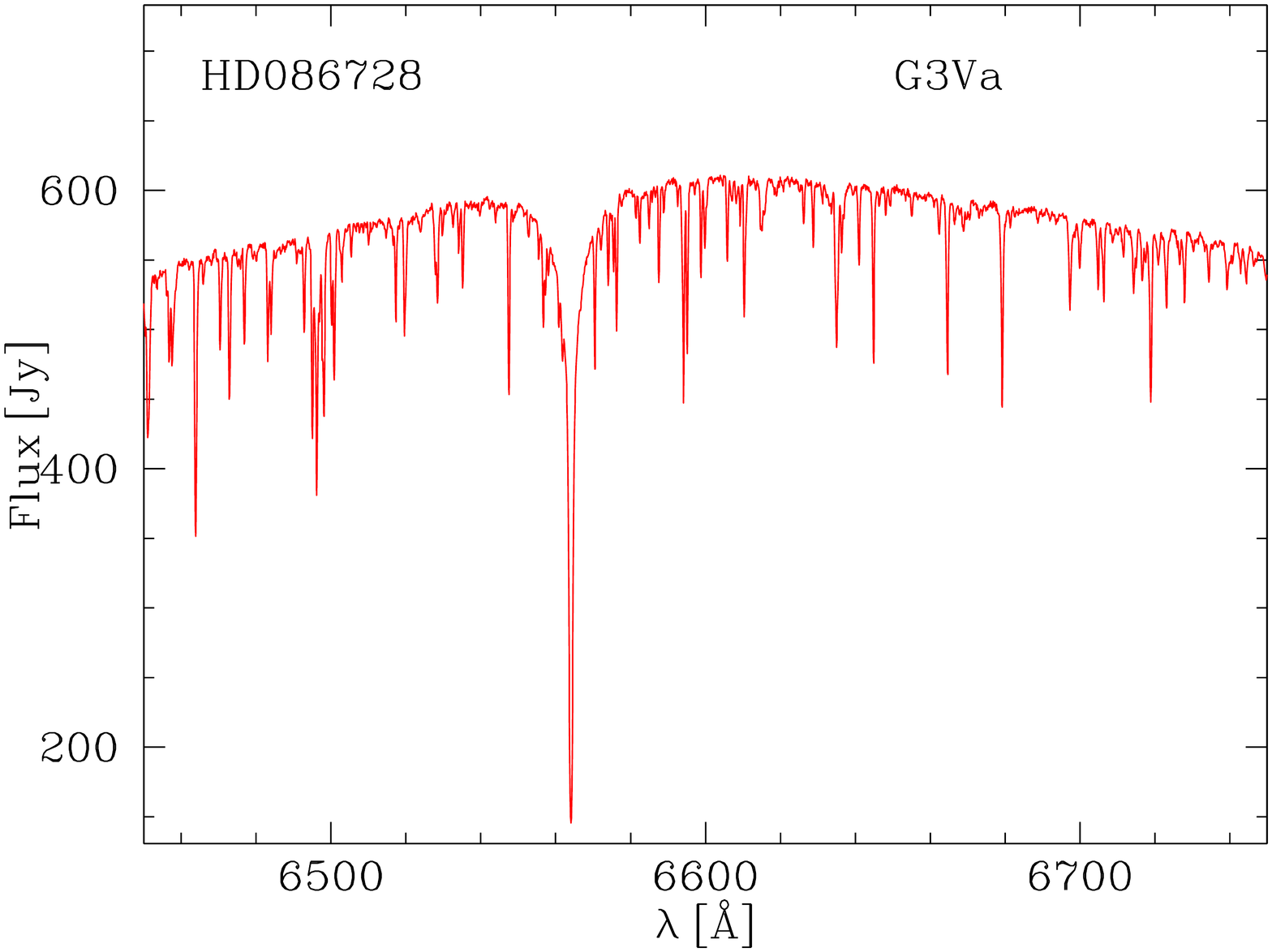}
\includegraphics[width=0.18\textwidth,angle=-90]{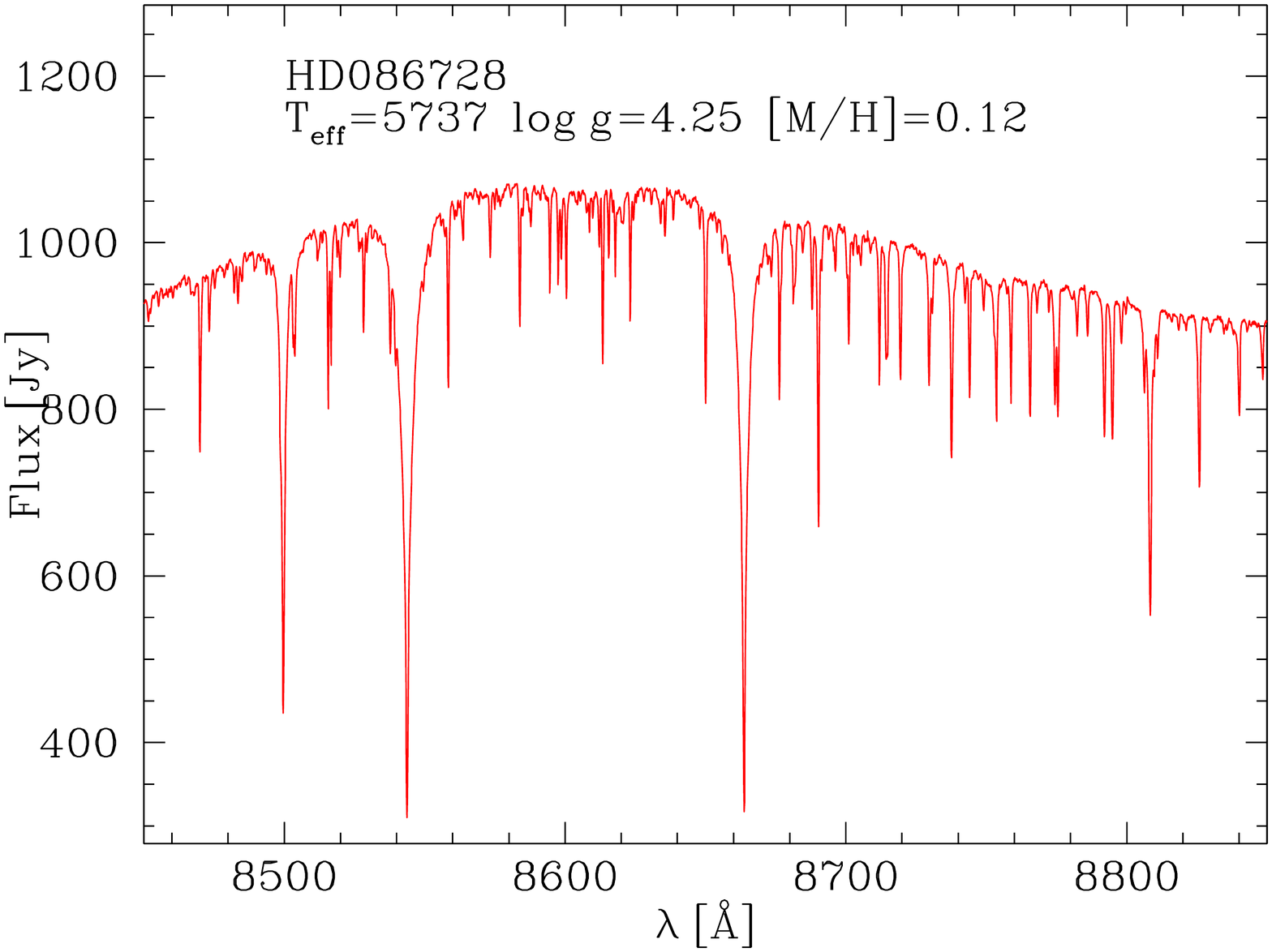}
\includegraphics[width=0.18\textwidth,angle=-90]{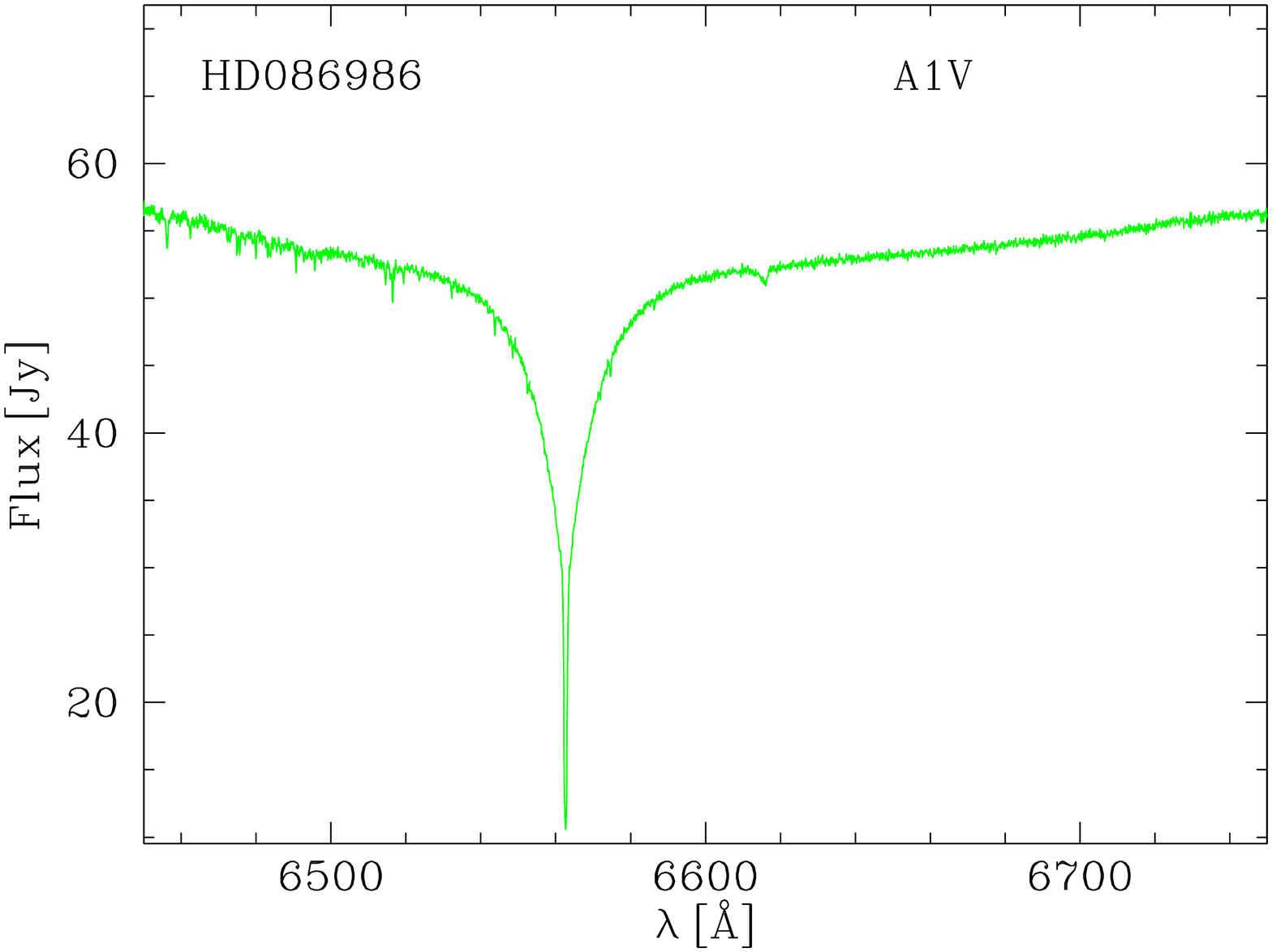}
\includegraphics[width=0.18\textwidth,angle=-90]{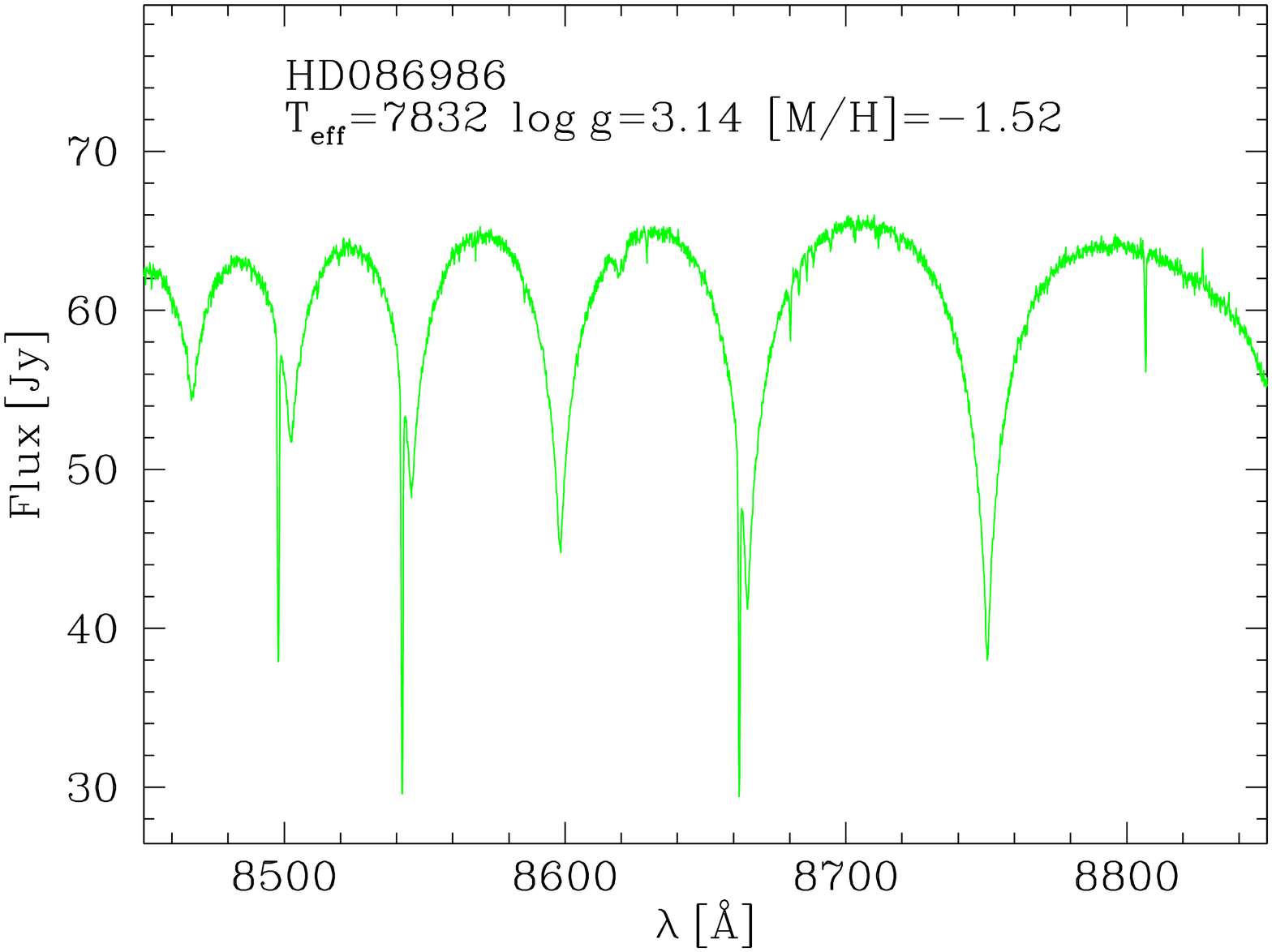}

\contcaption{17. Stars shown in this page are: HD079028, HD079210, HD079452, HD079765, HD080081, HD080218, HD080536, HD082106, HD083425, HD084737, HD086133, HD086560, HD086728 and HD086986.}
\end{figure*}

\begin{figure*}
\includegraphics[width=0.18\textwidth,angle=-90]{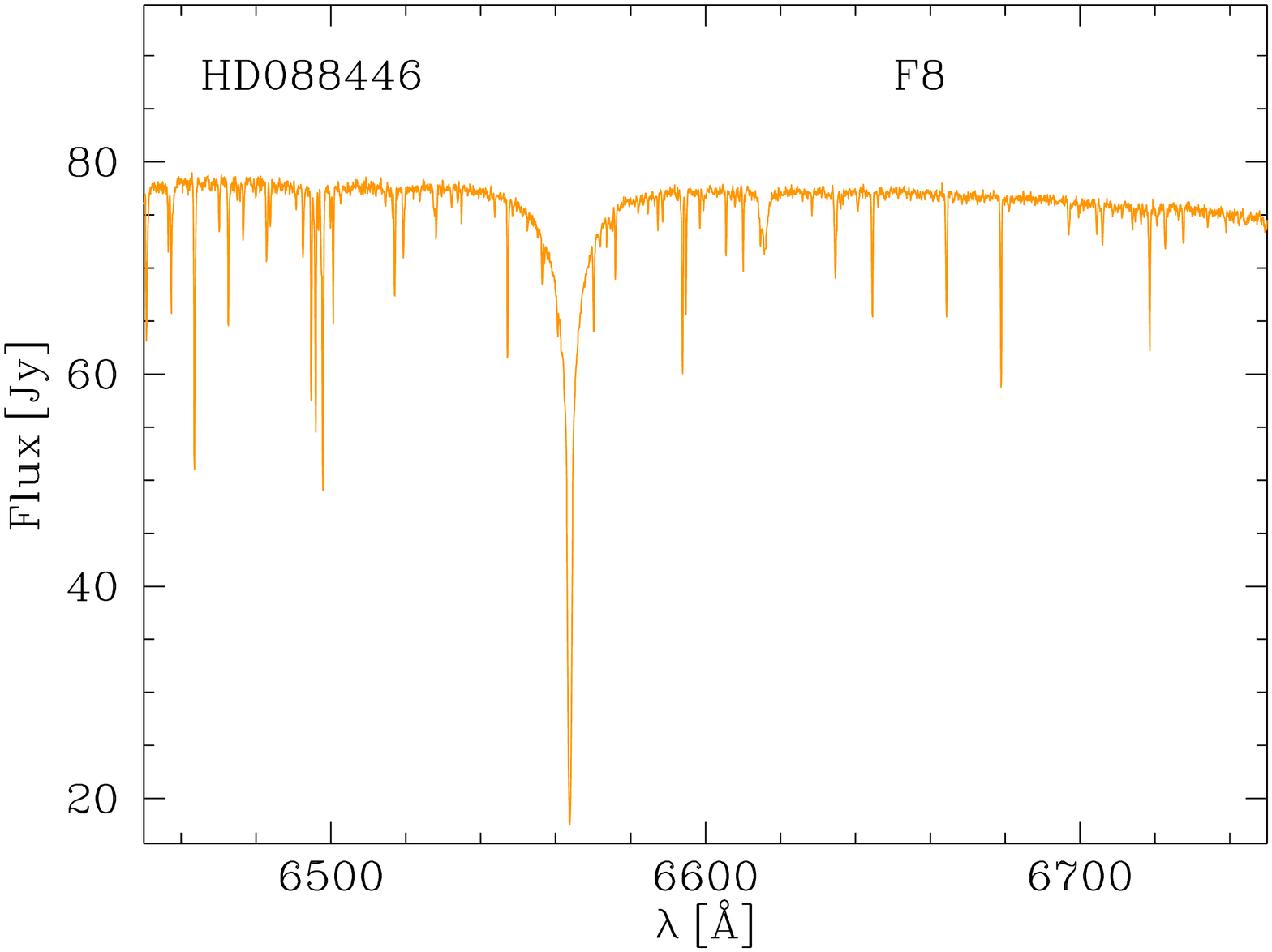}
\includegraphics[width=0.18\textwidth,angle=-90]{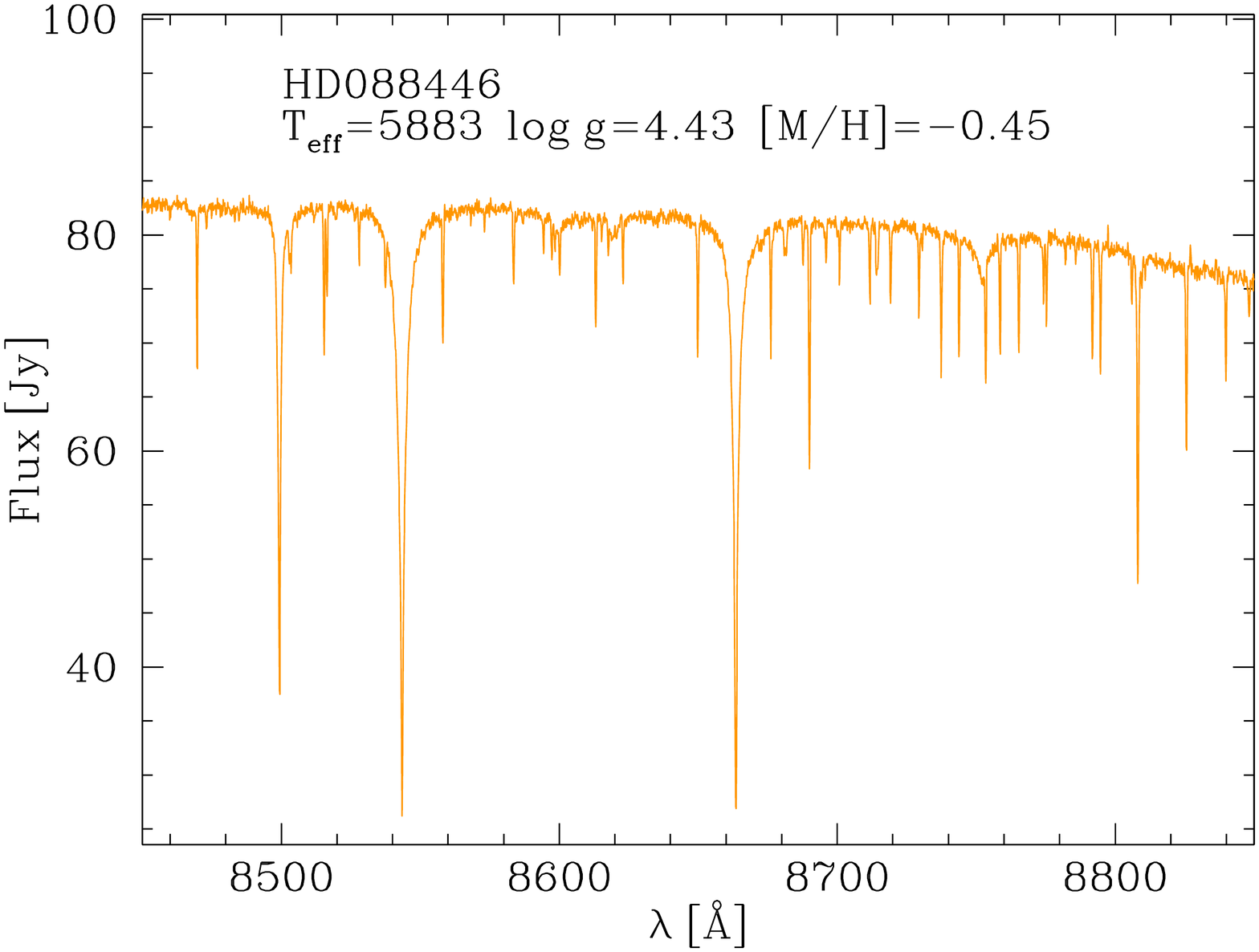}
\includegraphics[width=0.18\textwidth,angle=-90]{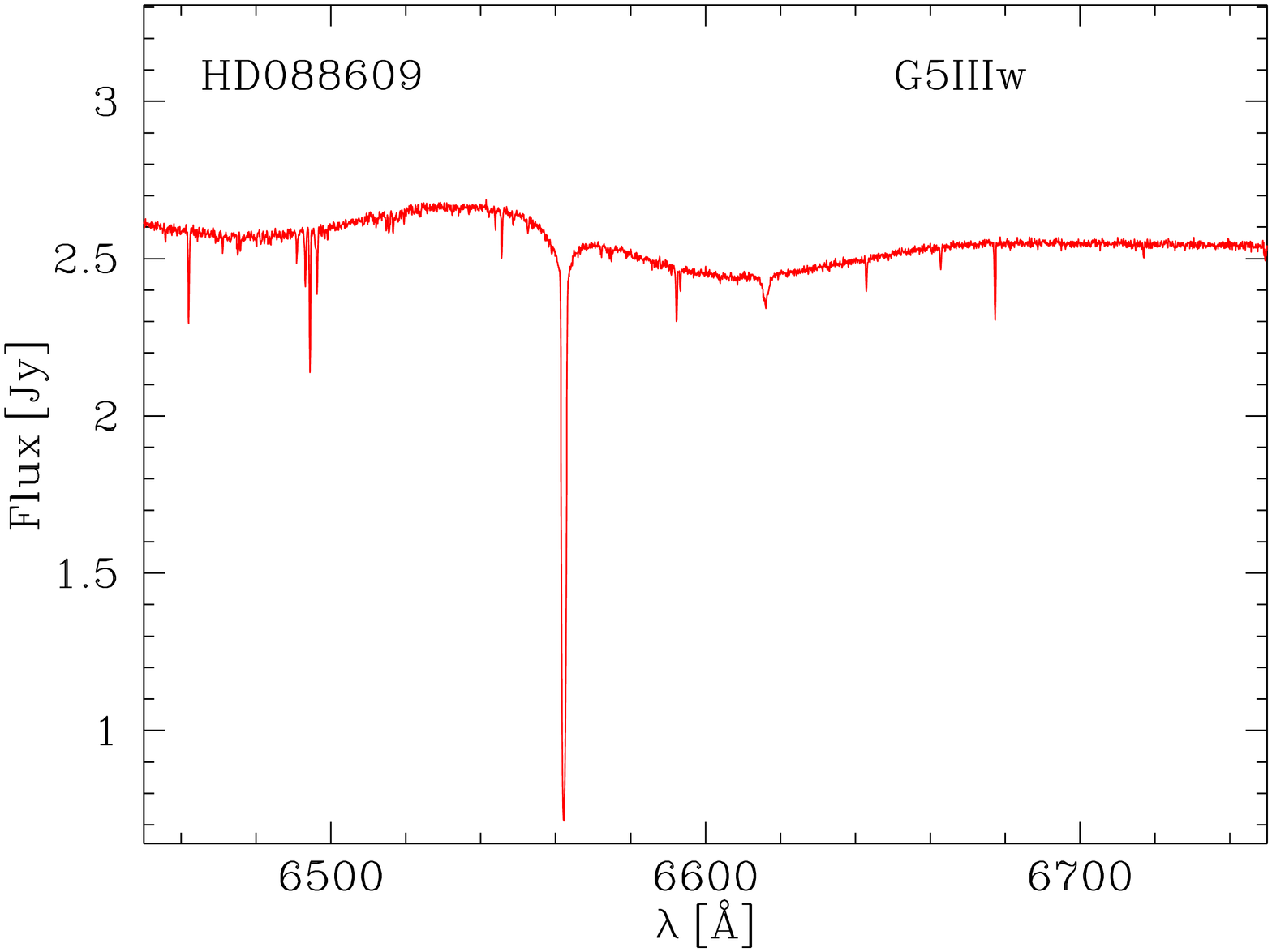}
\includegraphics[width=0.18\textwidth,angle=-90]{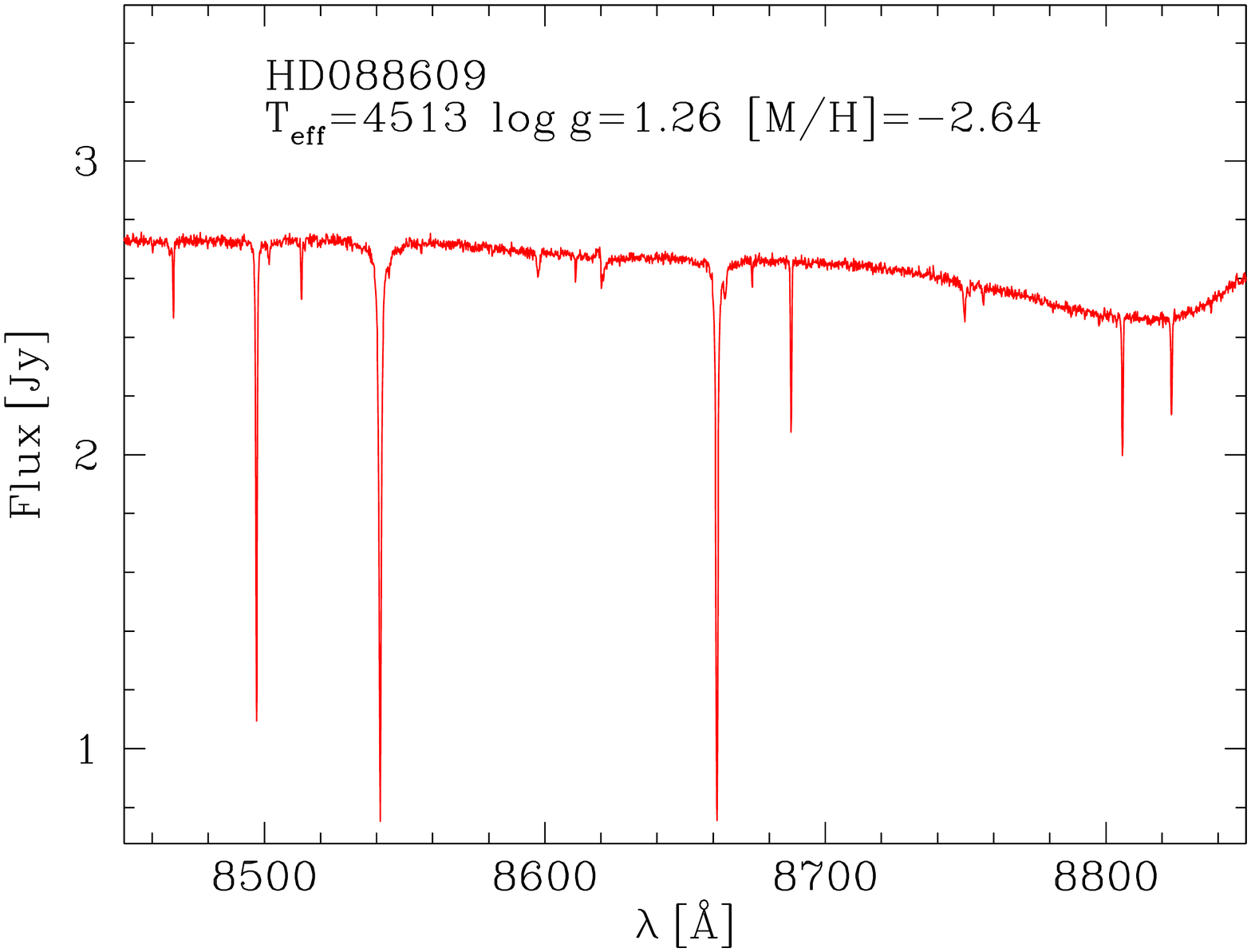}
\includegraphics[width=0.18\textwidth,angle=-90]{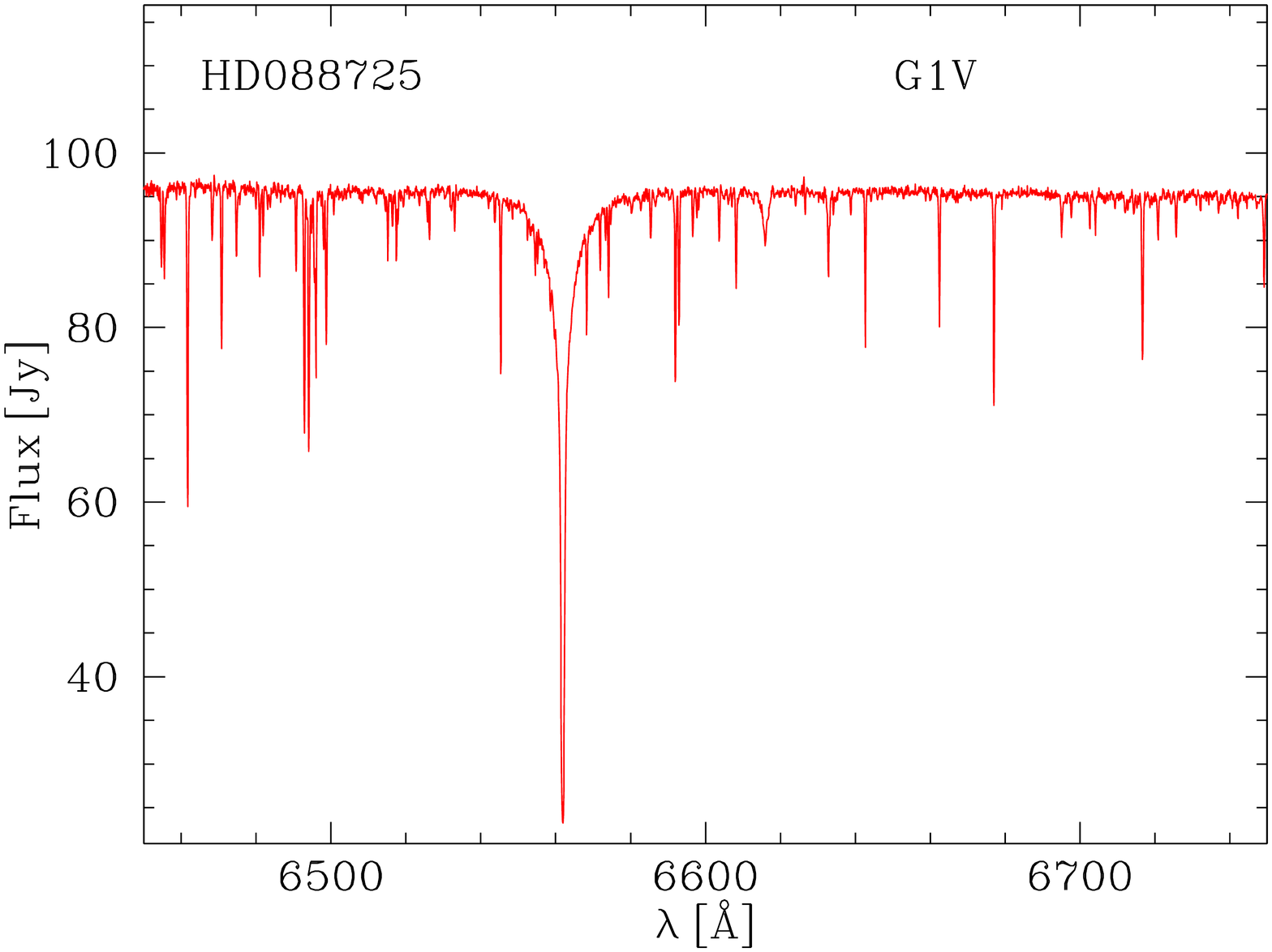}
\includegraphics[width=0.18\textwidth,angle=-90]{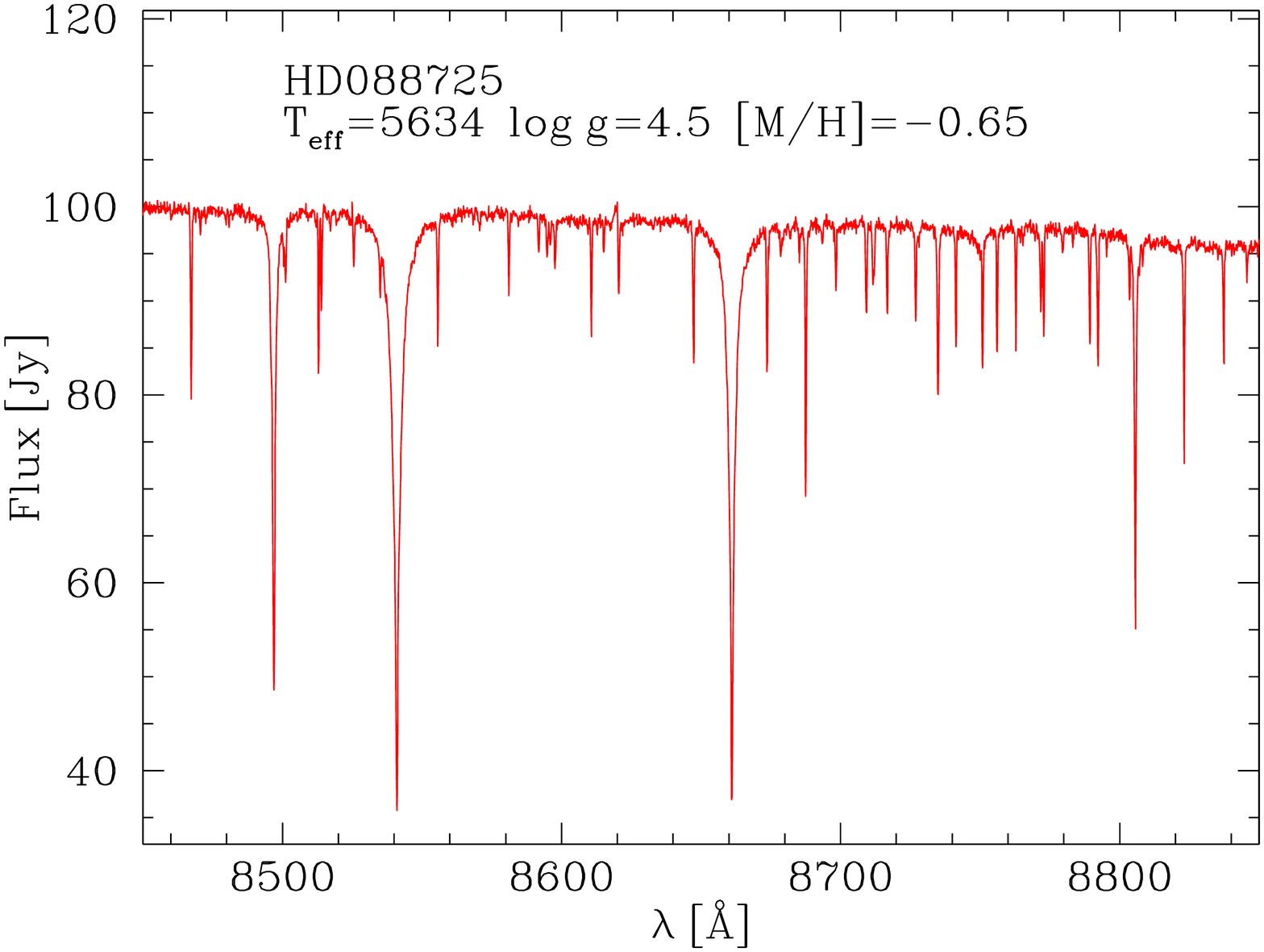}
\includegraphics[width=0.18\textwidth,angle=-90]{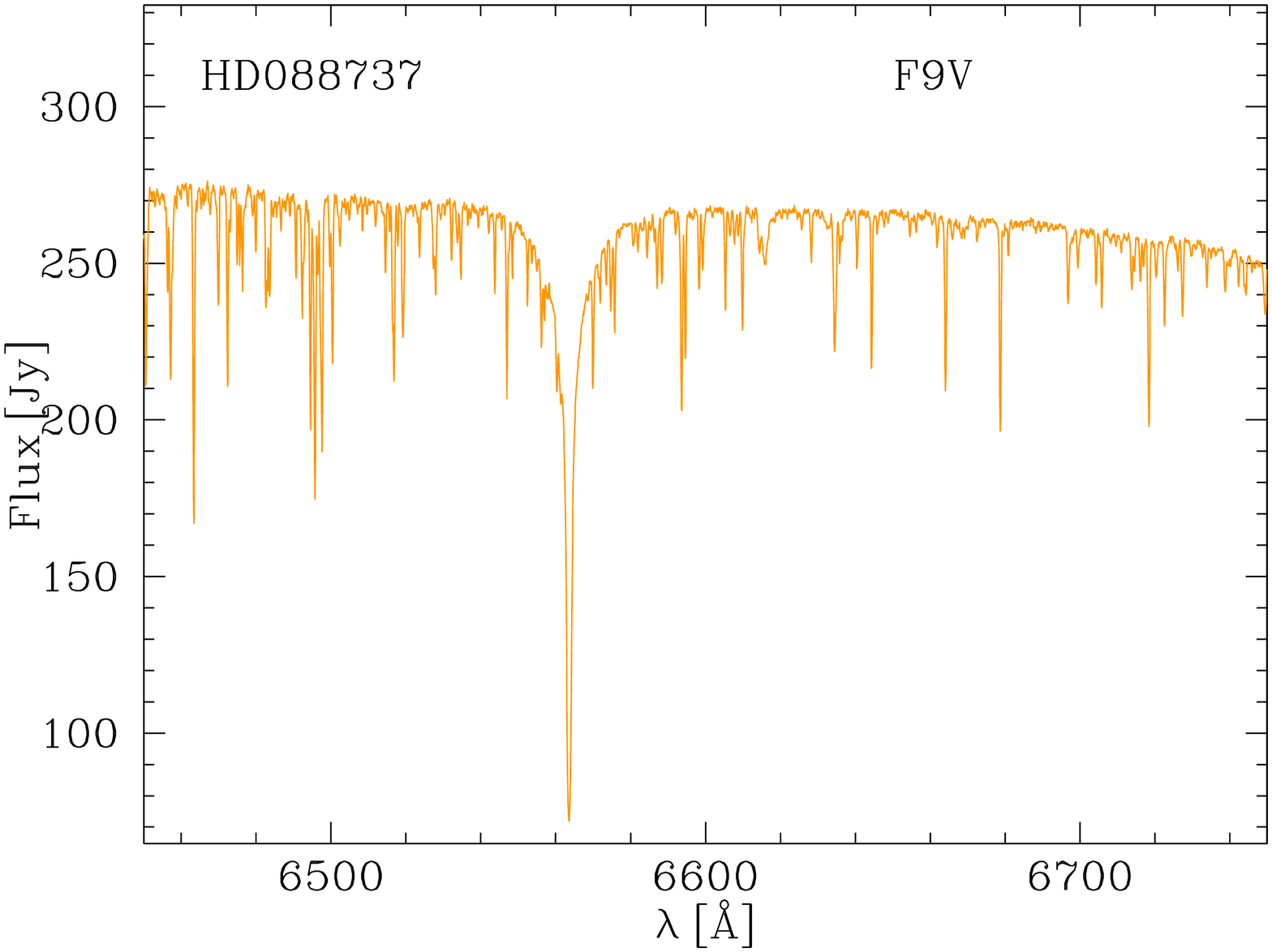}
\includegraphics[width=0.18\textwidth,angle=-90]{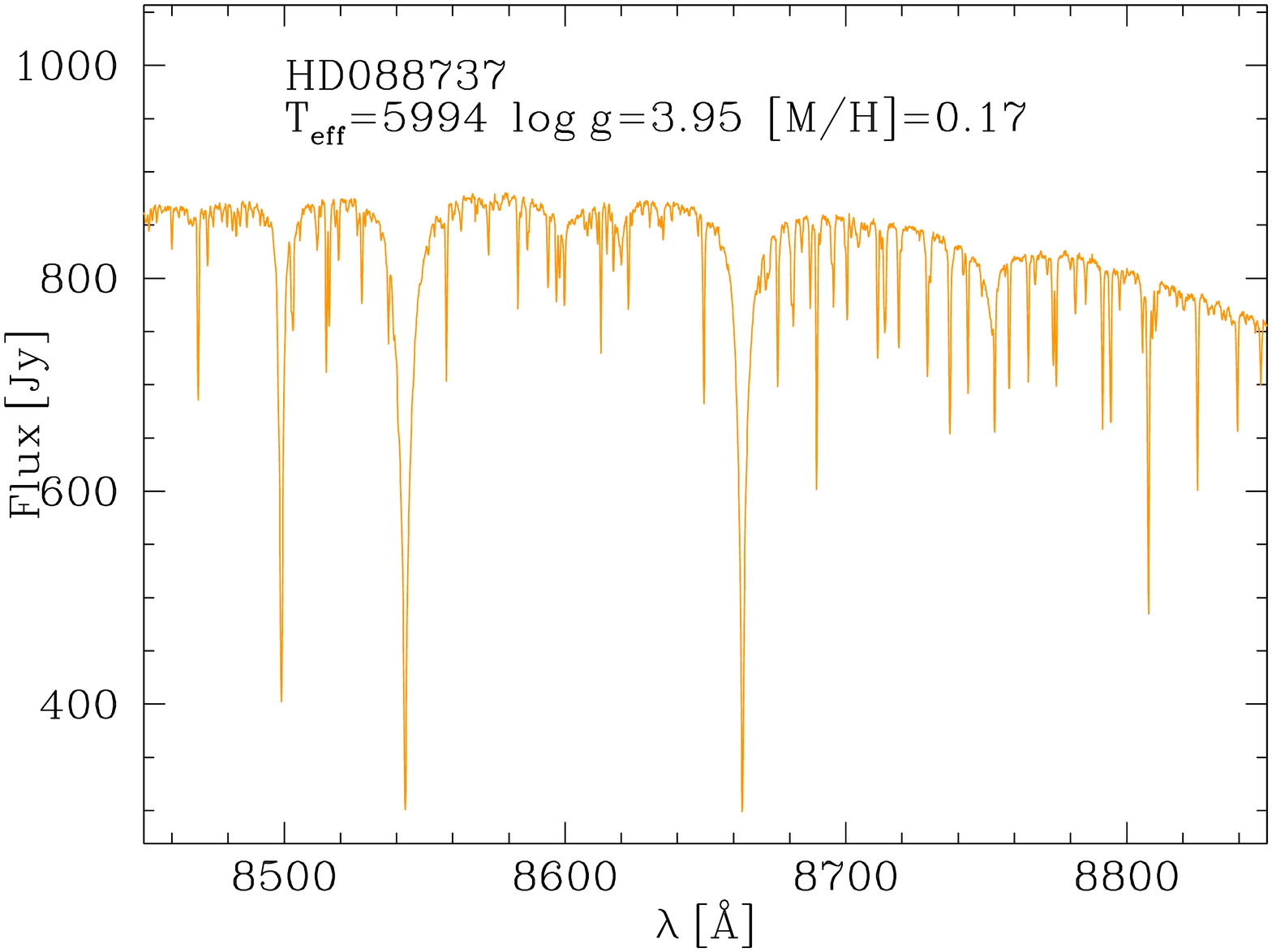}
\includegraphics[width=0.18\textwidth,angle=-90]{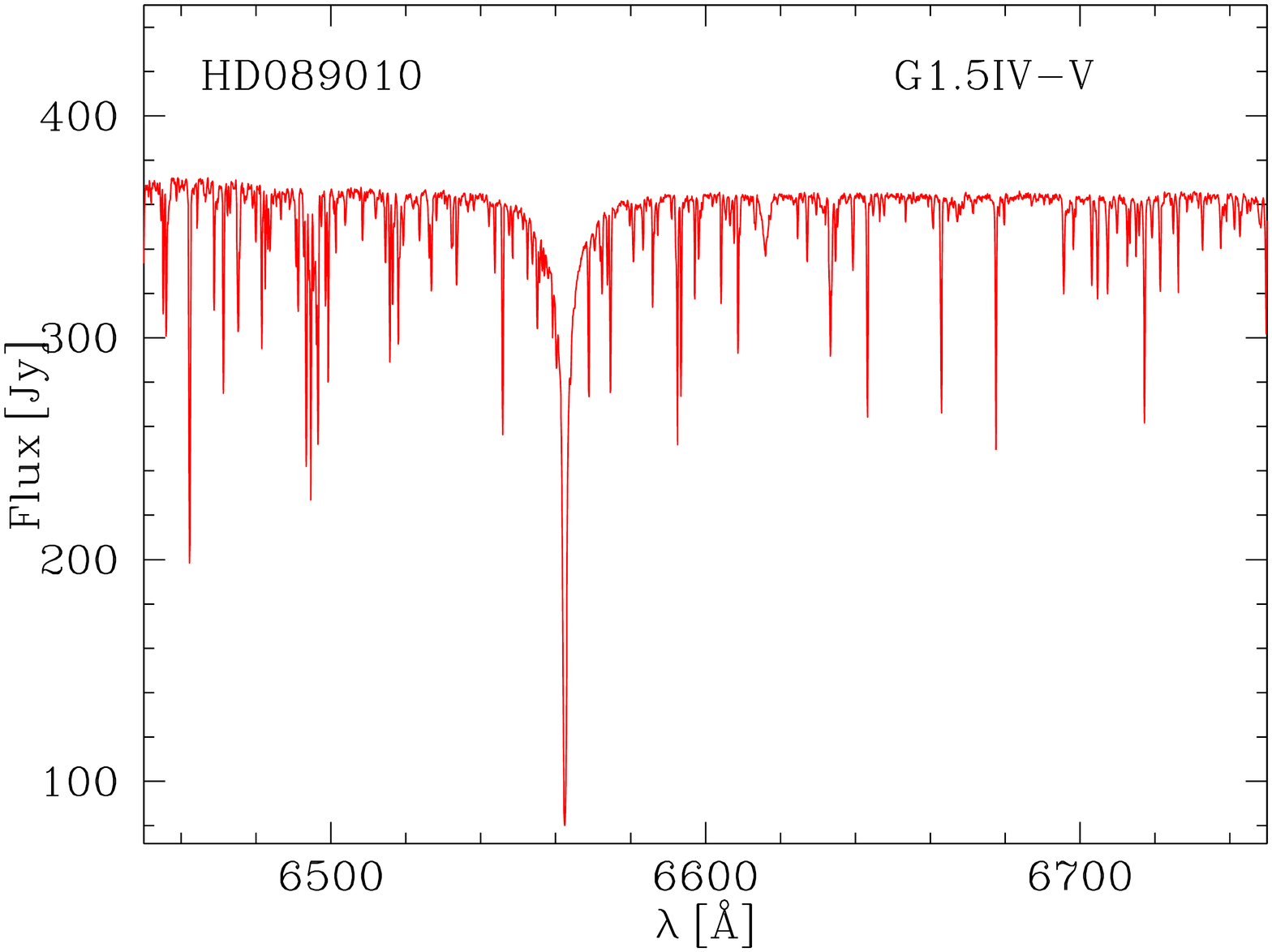}
\includegraphics[width=0.18\textwidth,angle=-90]{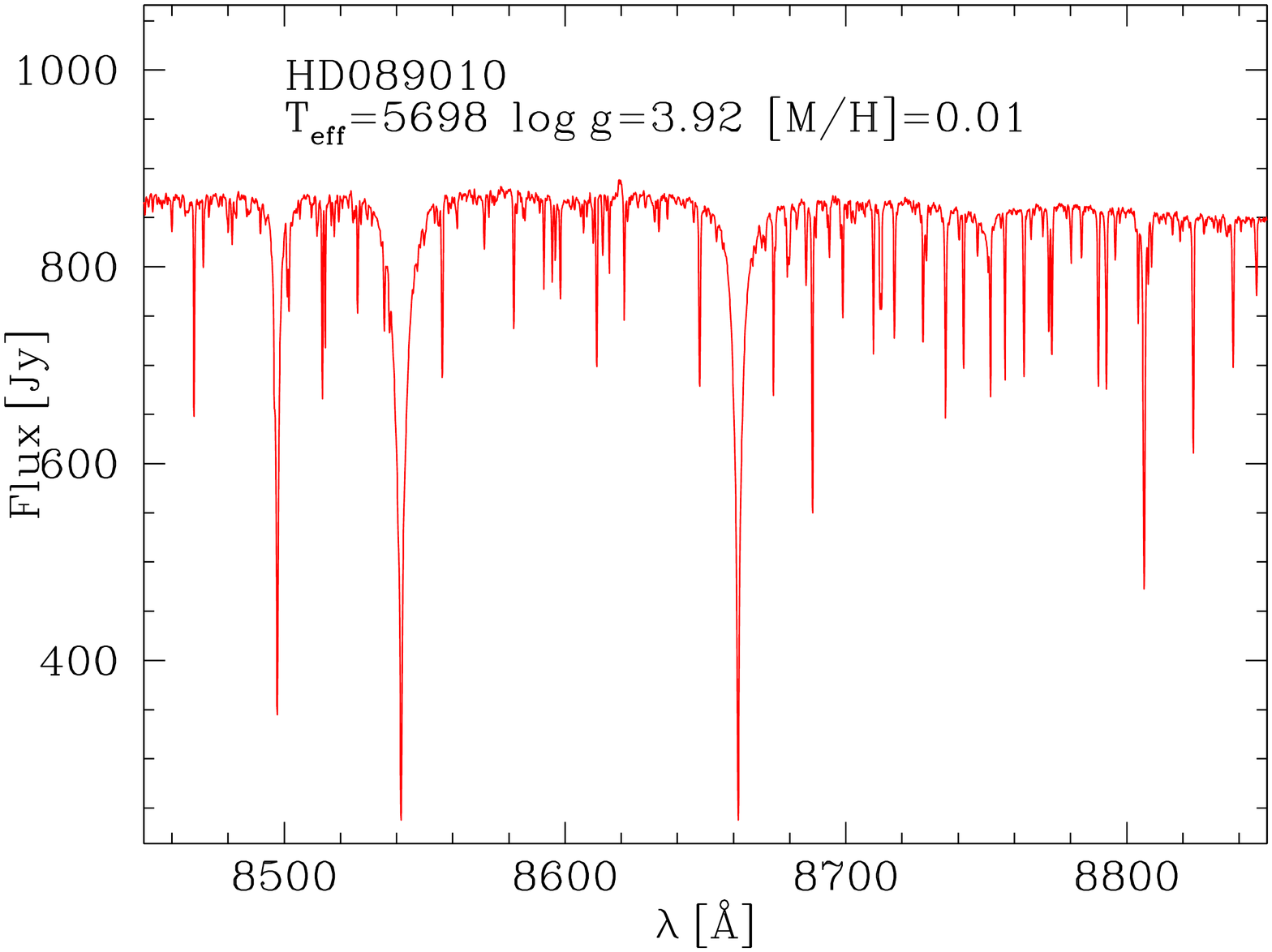}
\includegraphics[width=0.18\textwidth,angle=-90]{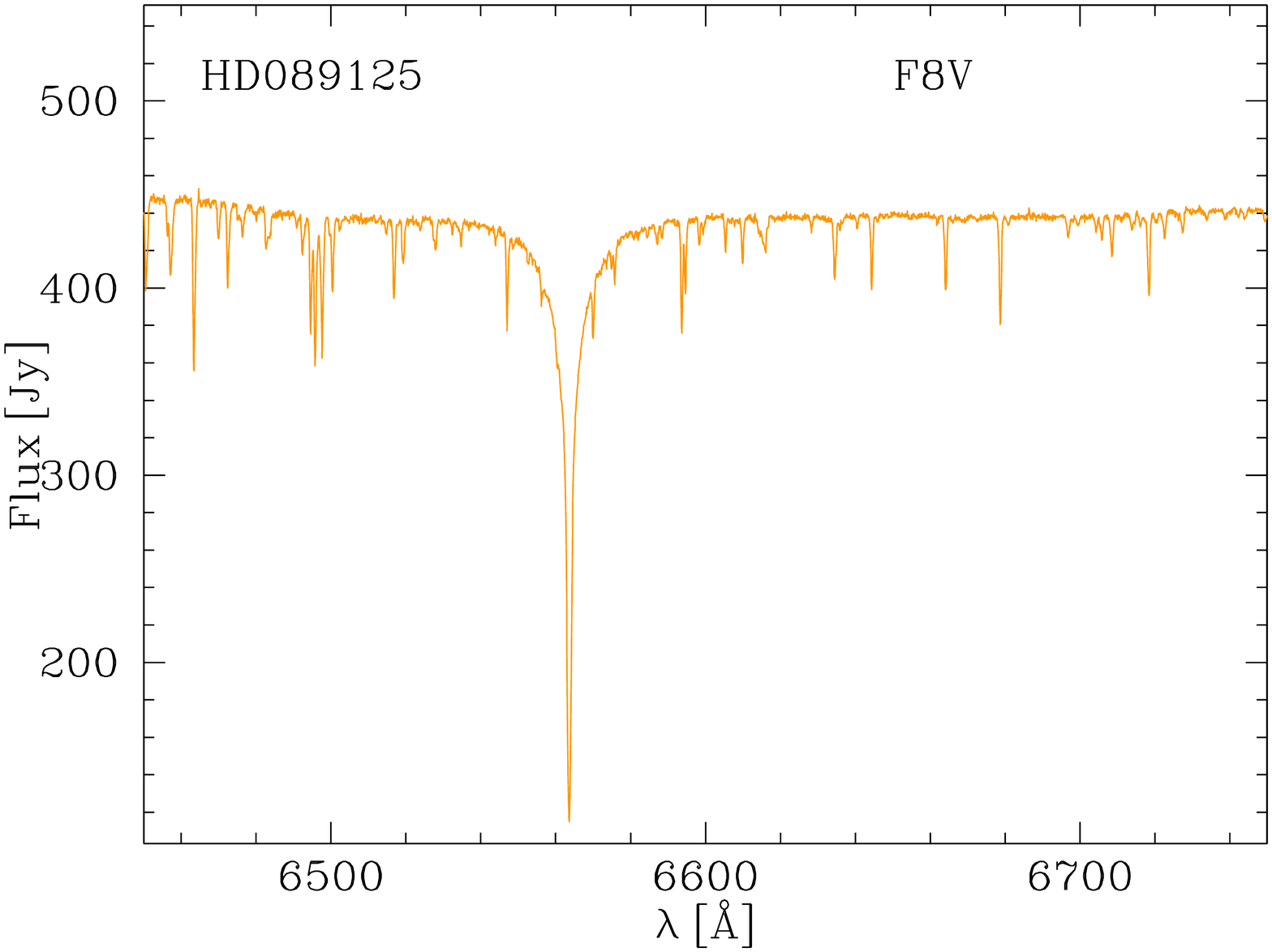}
\includegraphics[width=0.18\textwidth,angle=-90]{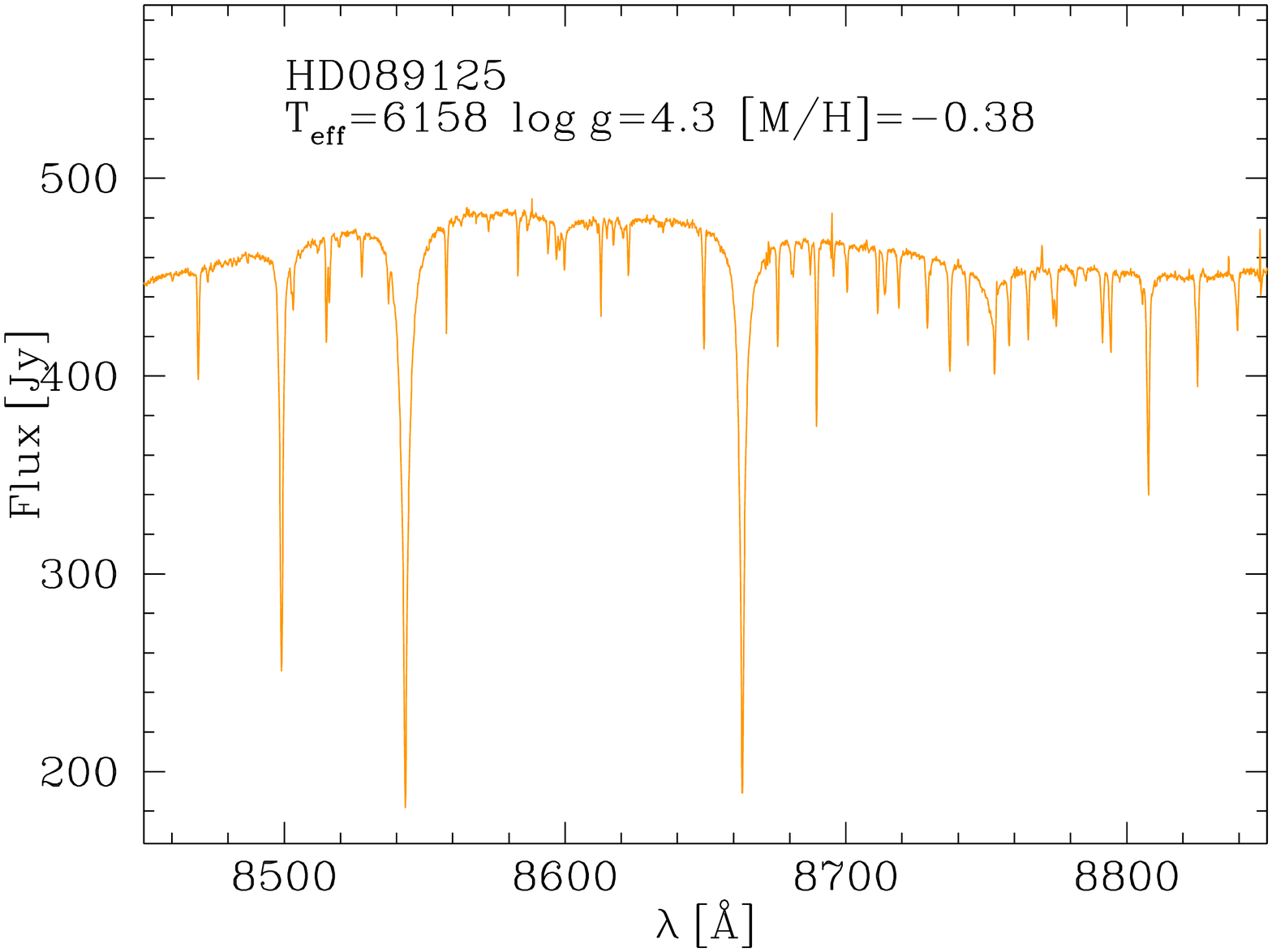}
\includegraphics[width=0.18\textwidth,angle=-90]{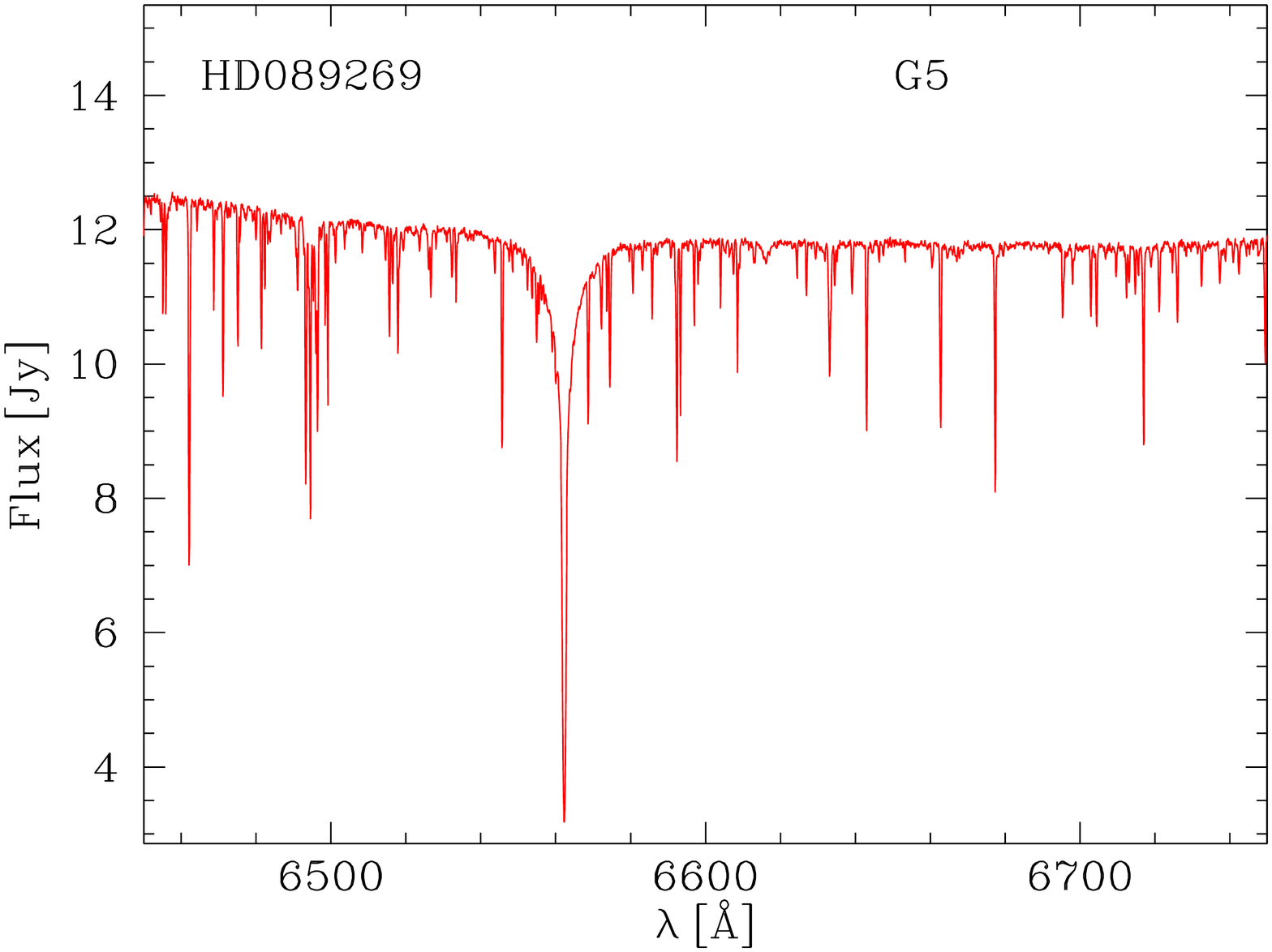}
\includegraphics[width=0.18\textwidth,angle=-90]{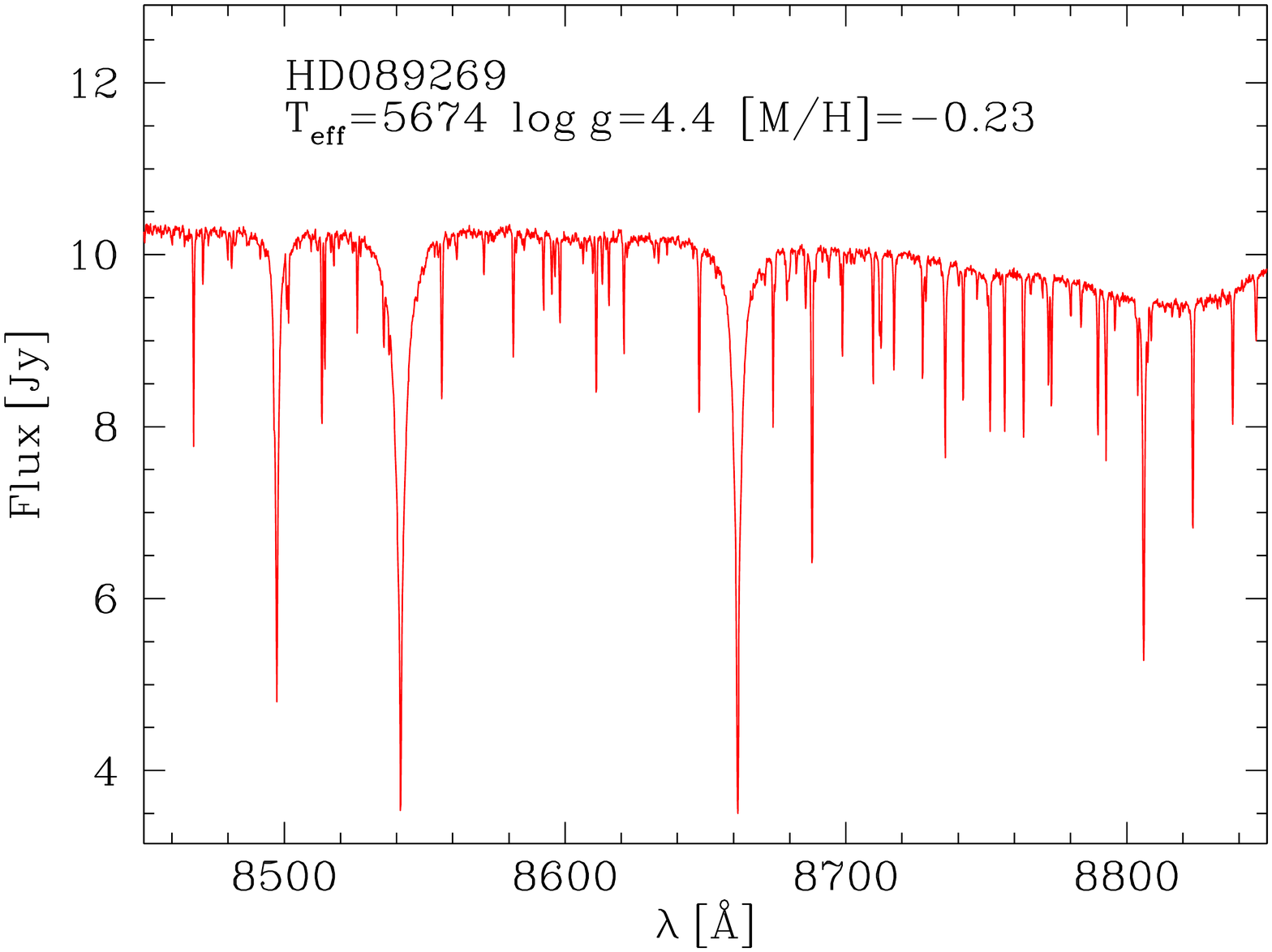}
\includegraphics[width=0.18\textwidth,angle=-90]{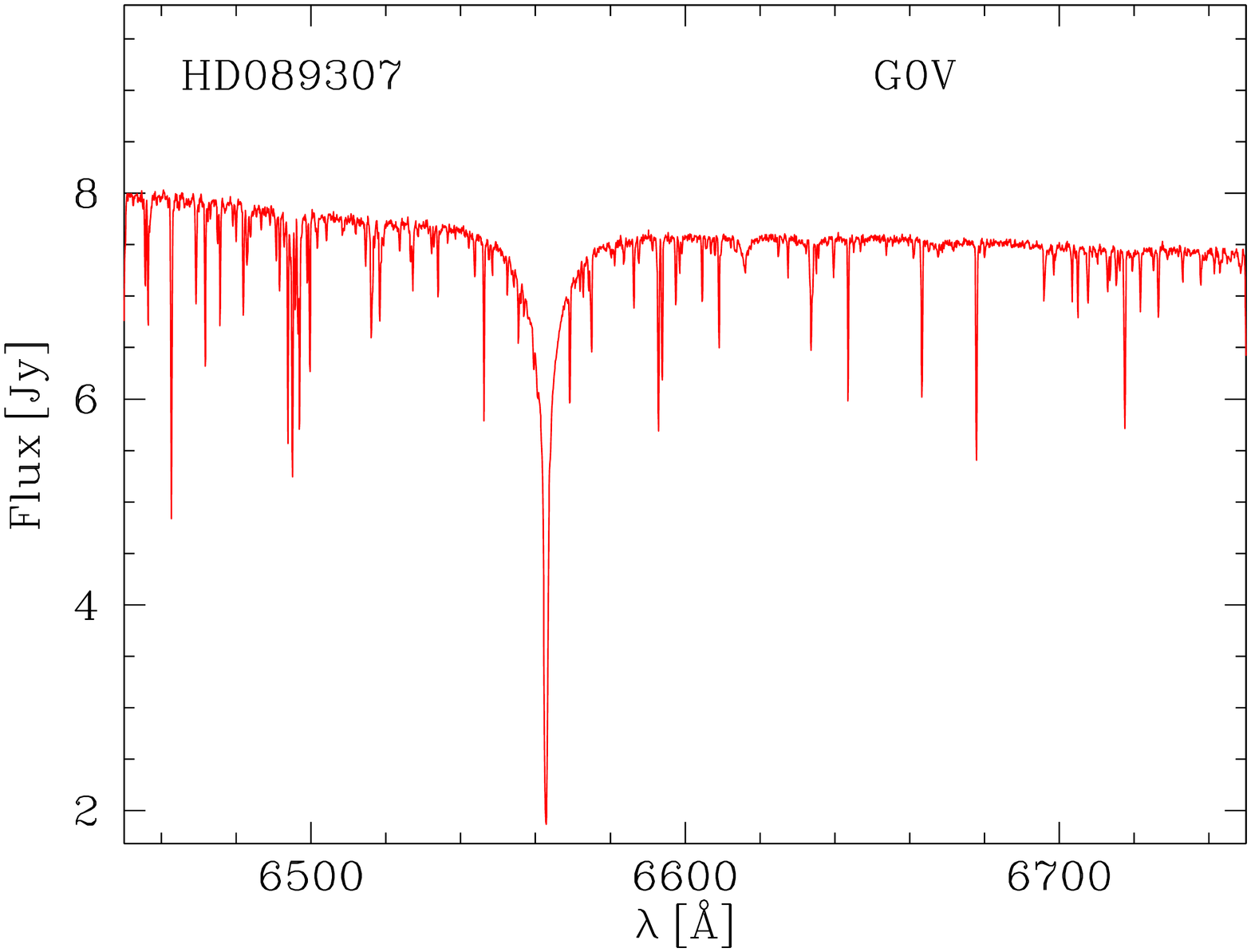}
\includegraphics[width=0.18\textwidth,angle=-90]{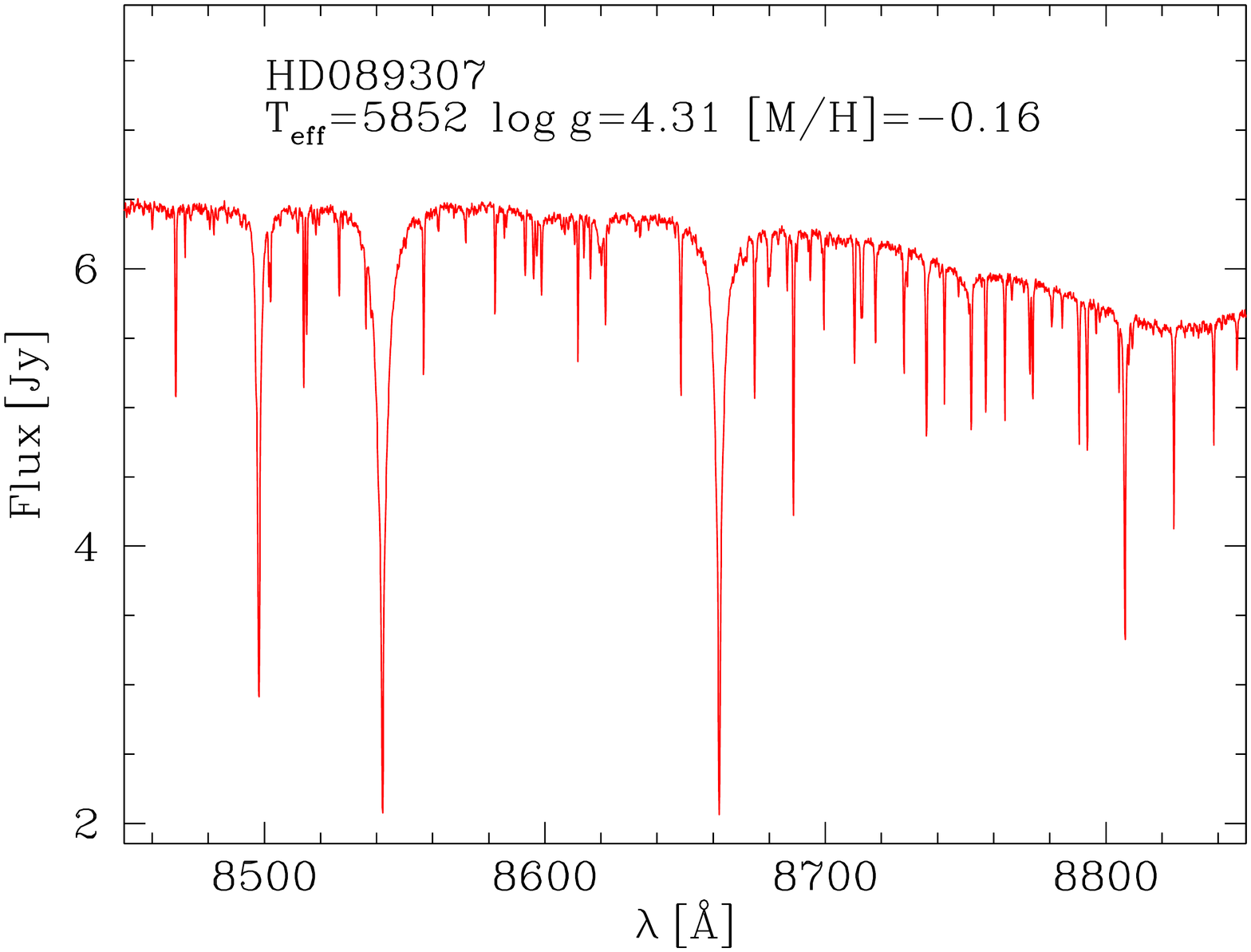}
\includegraphics[width=0.18\textwidth,angle=-90]{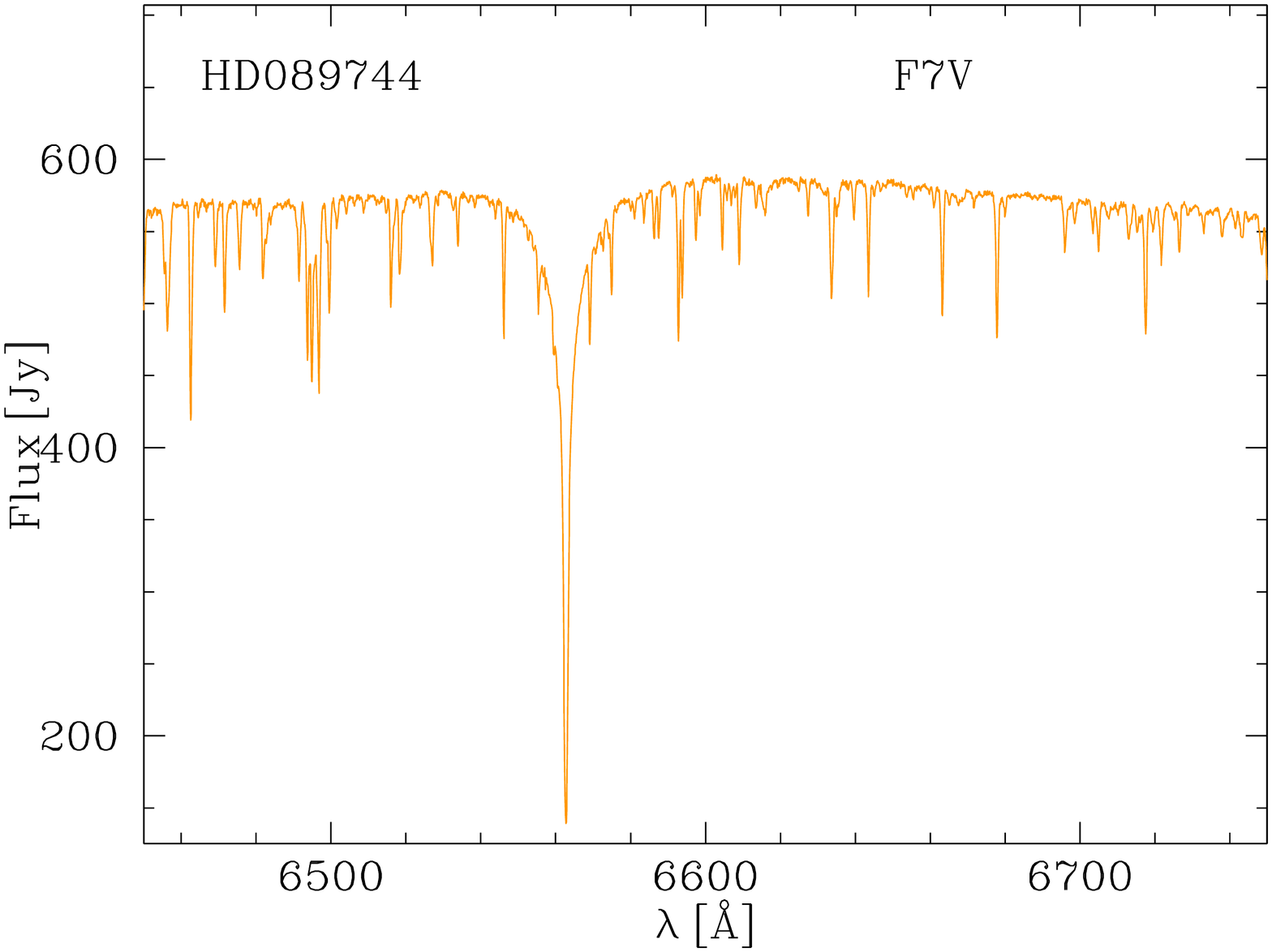}
\includegraphics[width=0.18\textwidth,angle=-90]{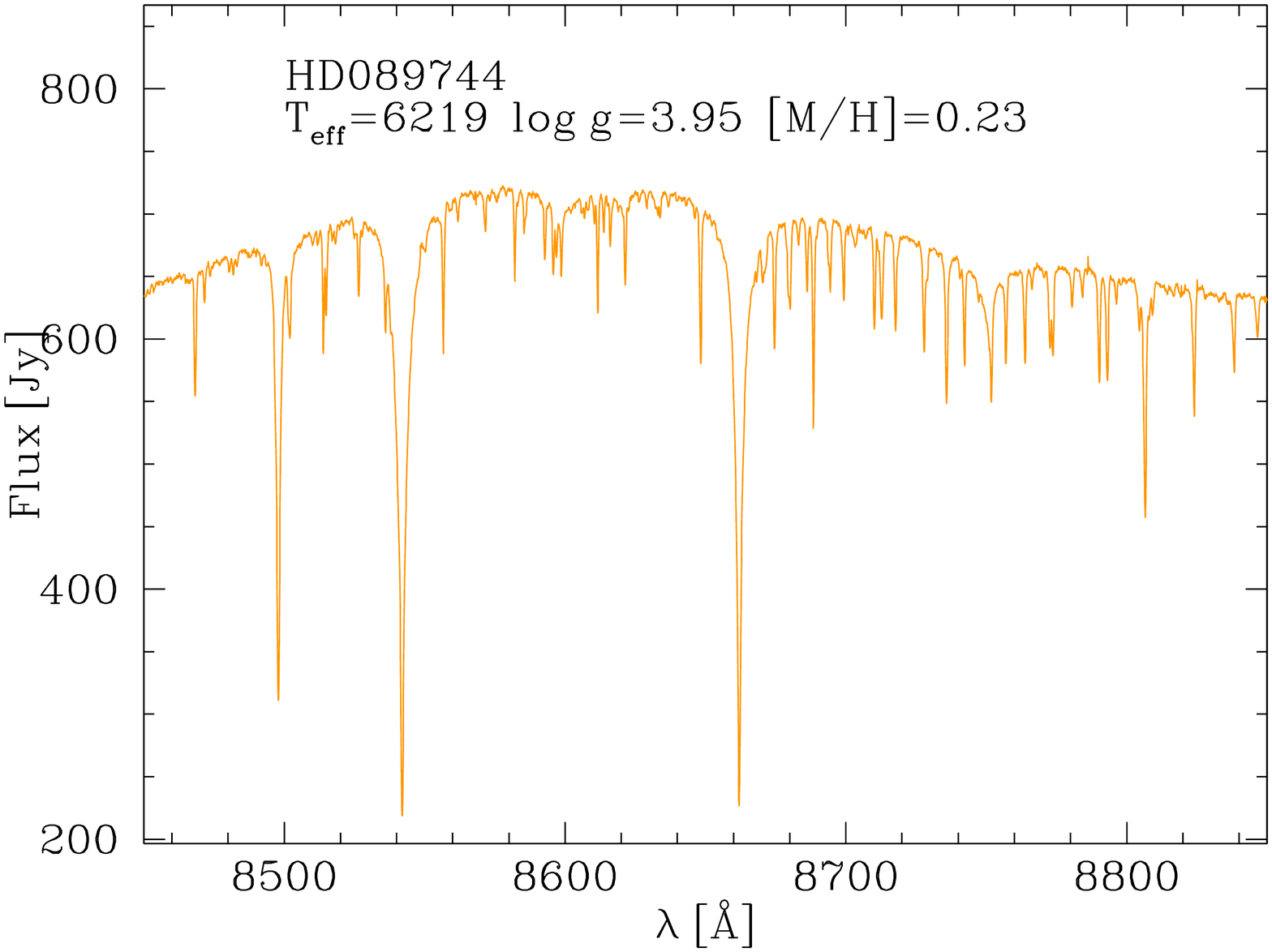}
\includegraphics[width=0.18\textwidth,angle=-90]{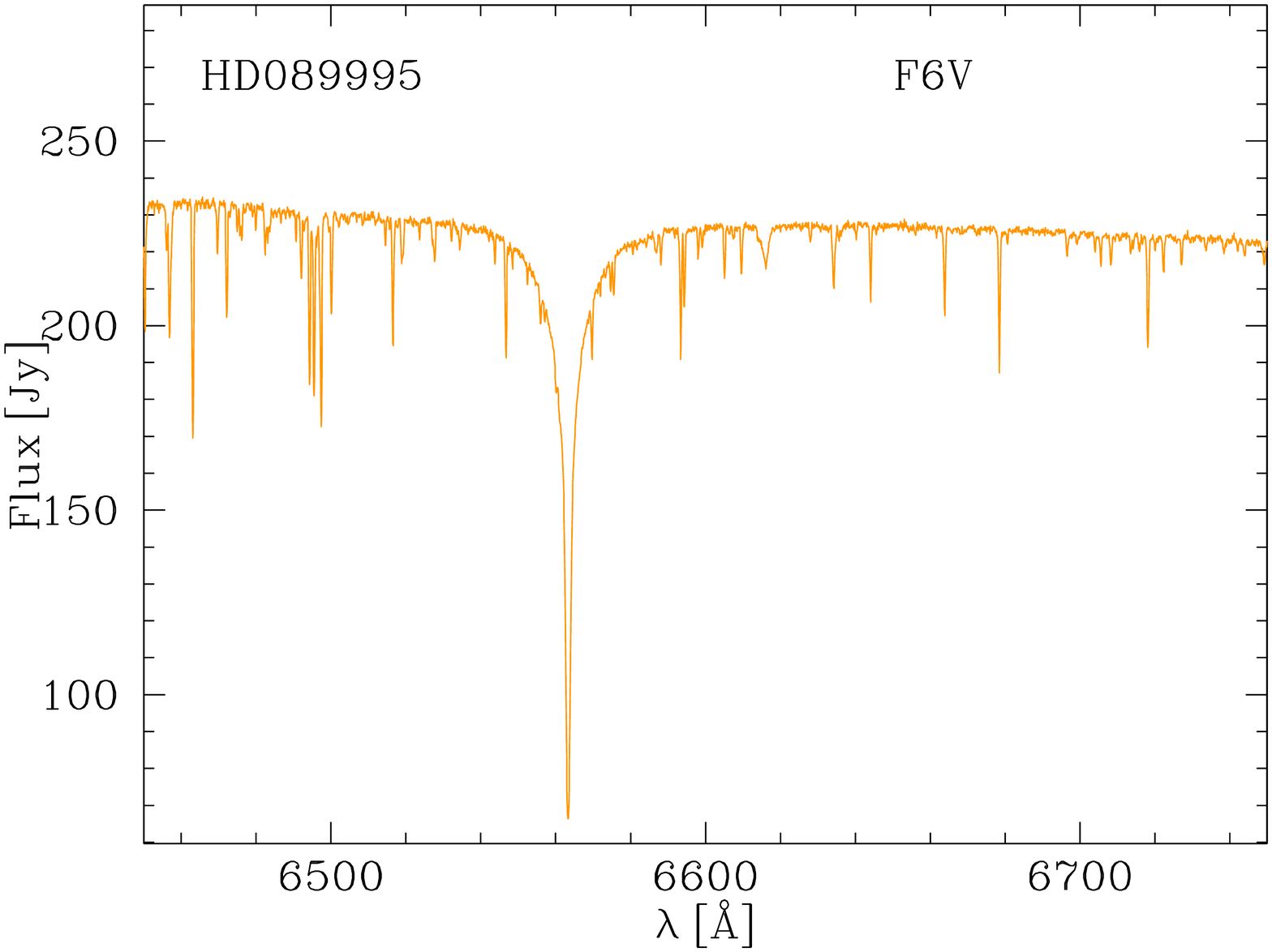}
\includegraphics[width=0.18\textwidth,angle=-90]{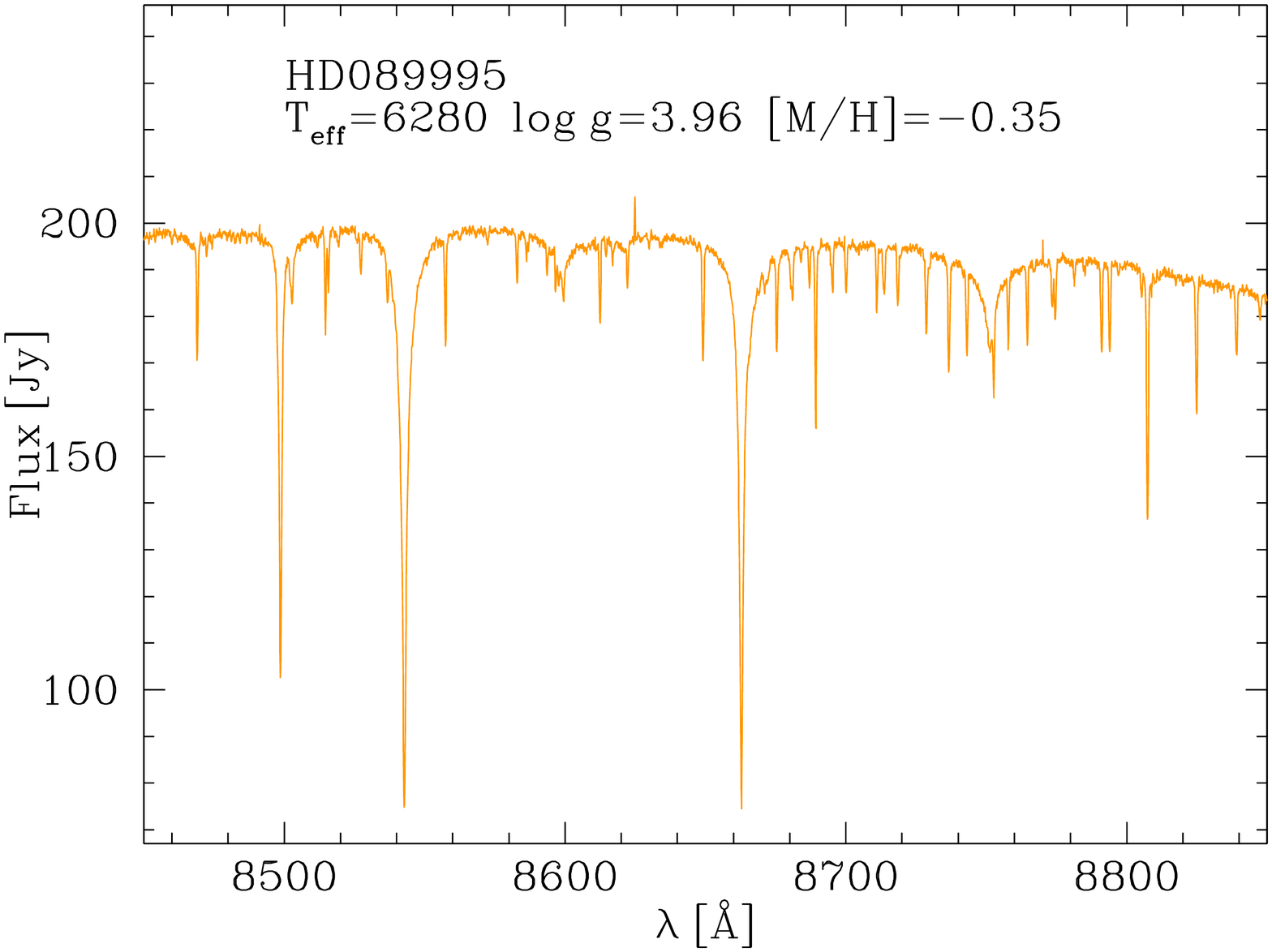}
\includegraphics[width=0.18\textwidth,angle=-90]{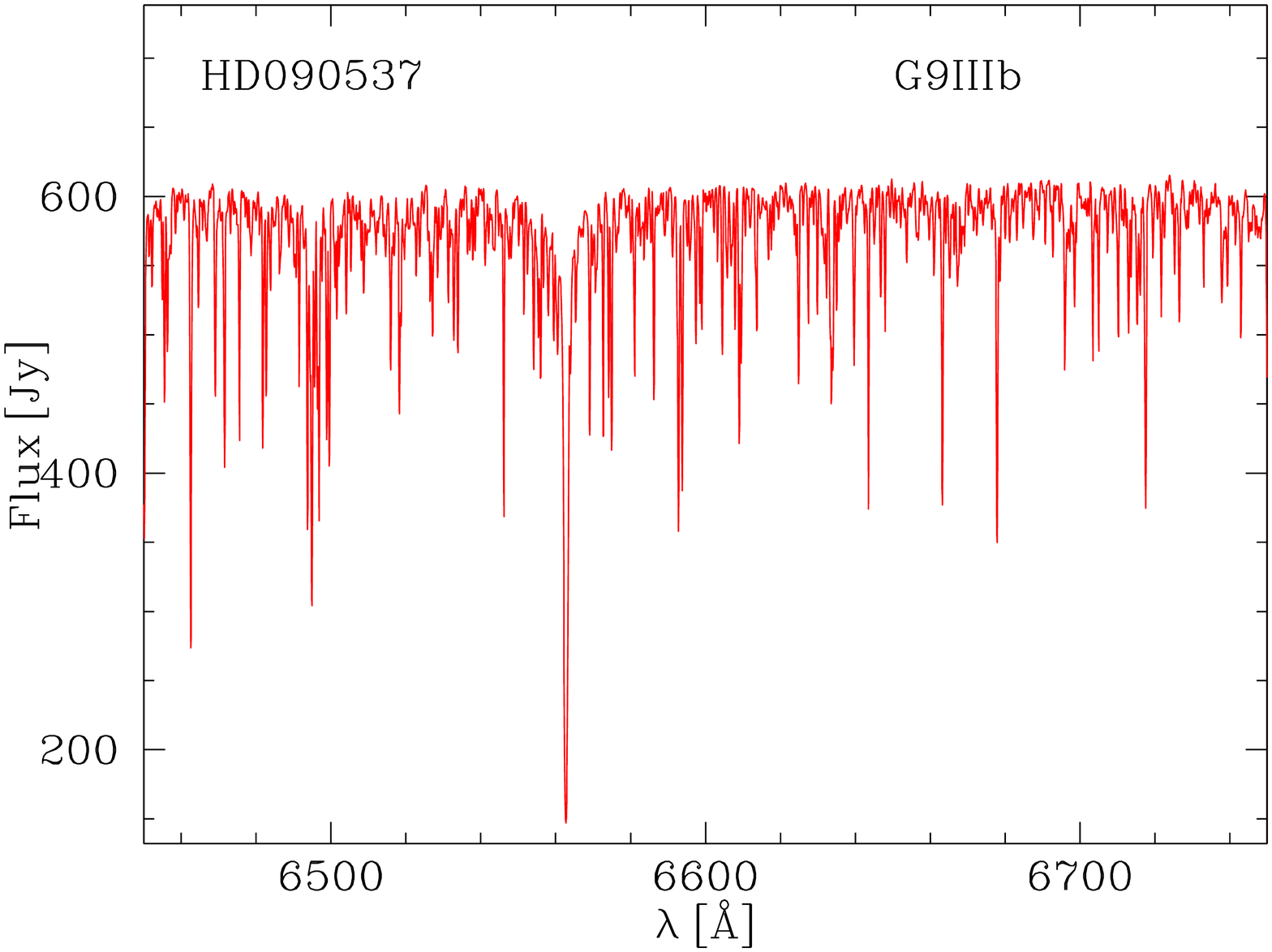}
\includegraphics[width=0.18\textwidth,angle=-90]{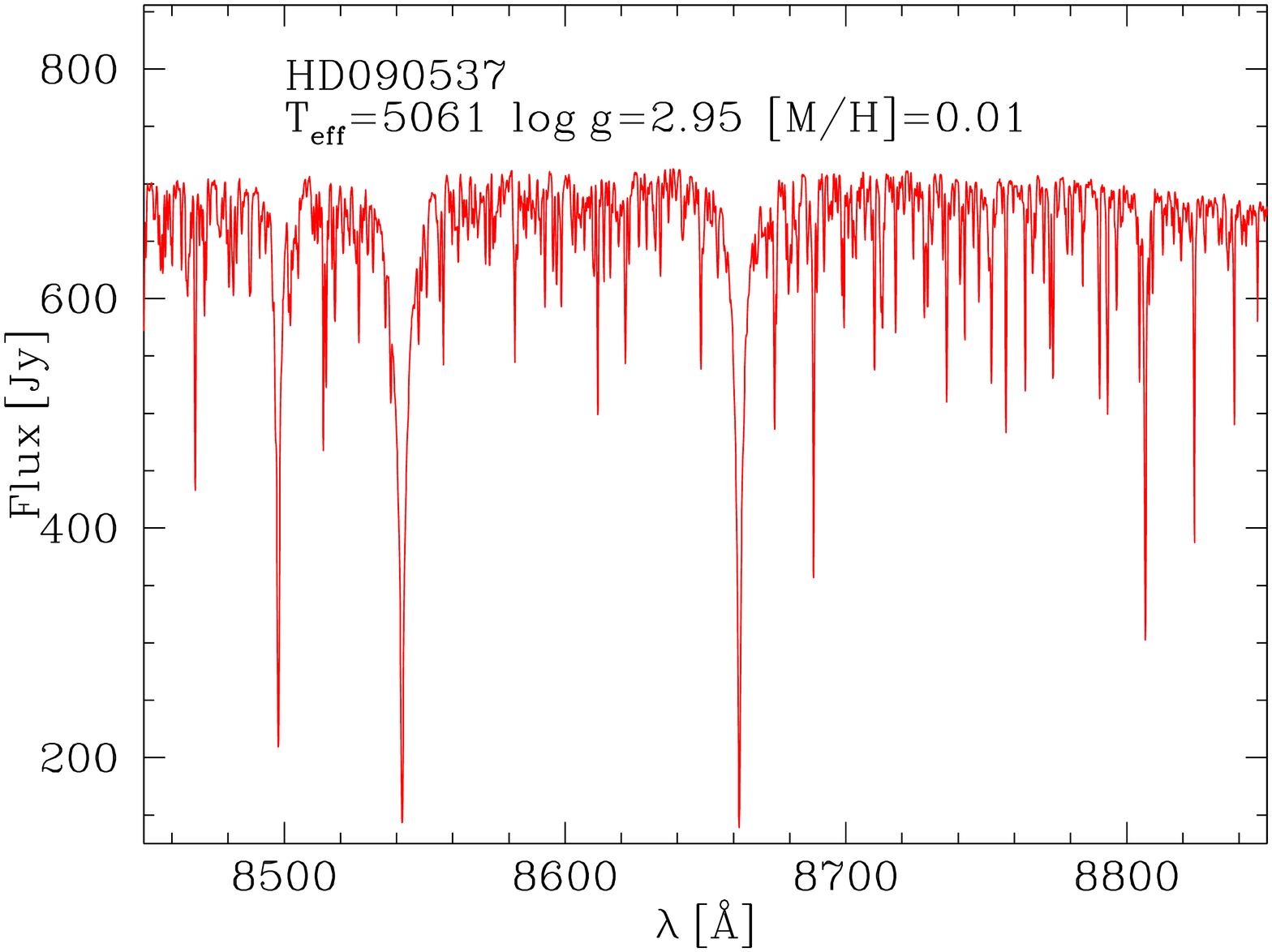}
\includegraphics[width=0.18\textwidth,angle=-90]{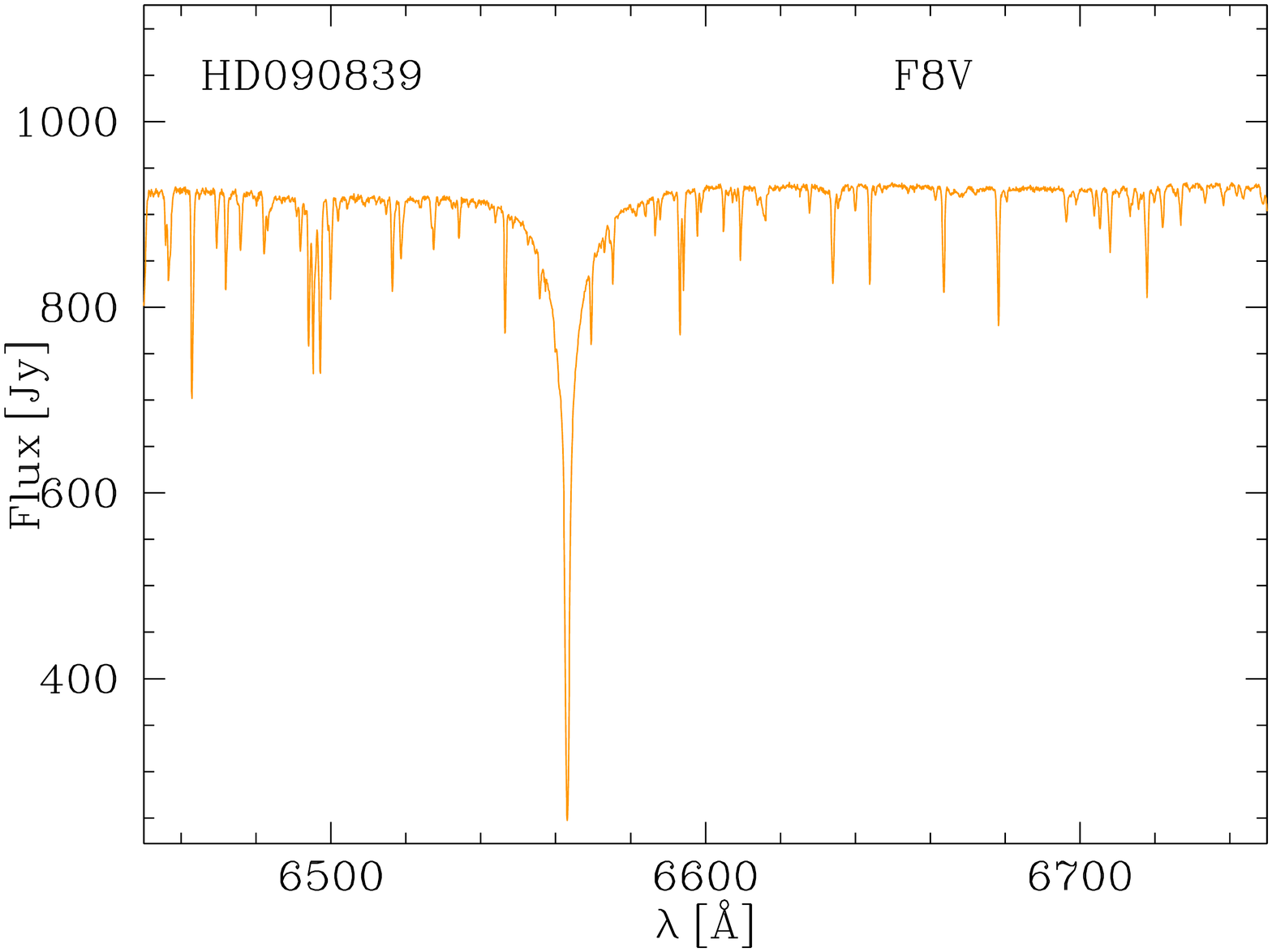}
\includegraphics[width=0.18\textwidth,angle=-90]{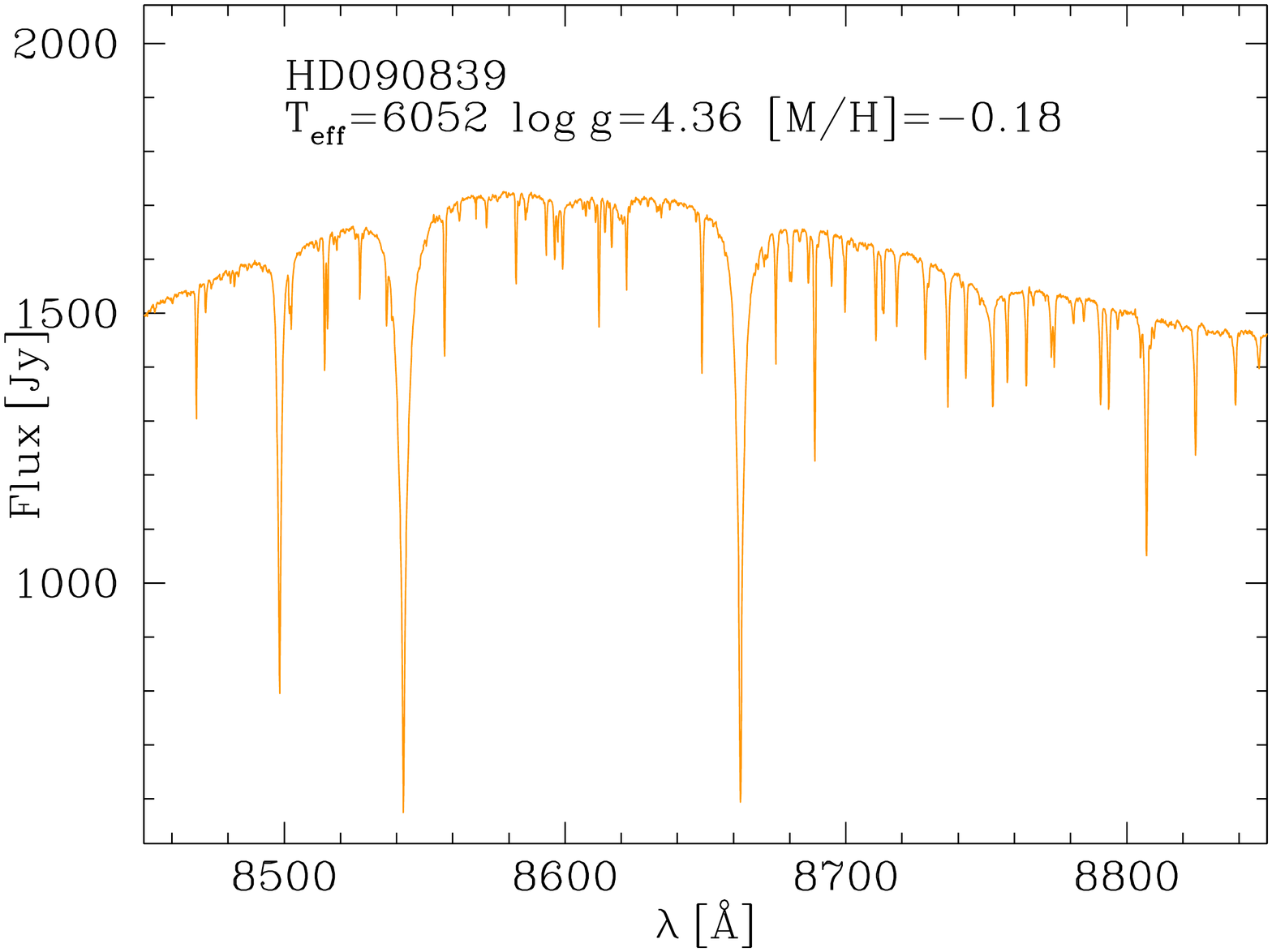}
\includegraphics[width=0.18\textwidth,angle=-90]{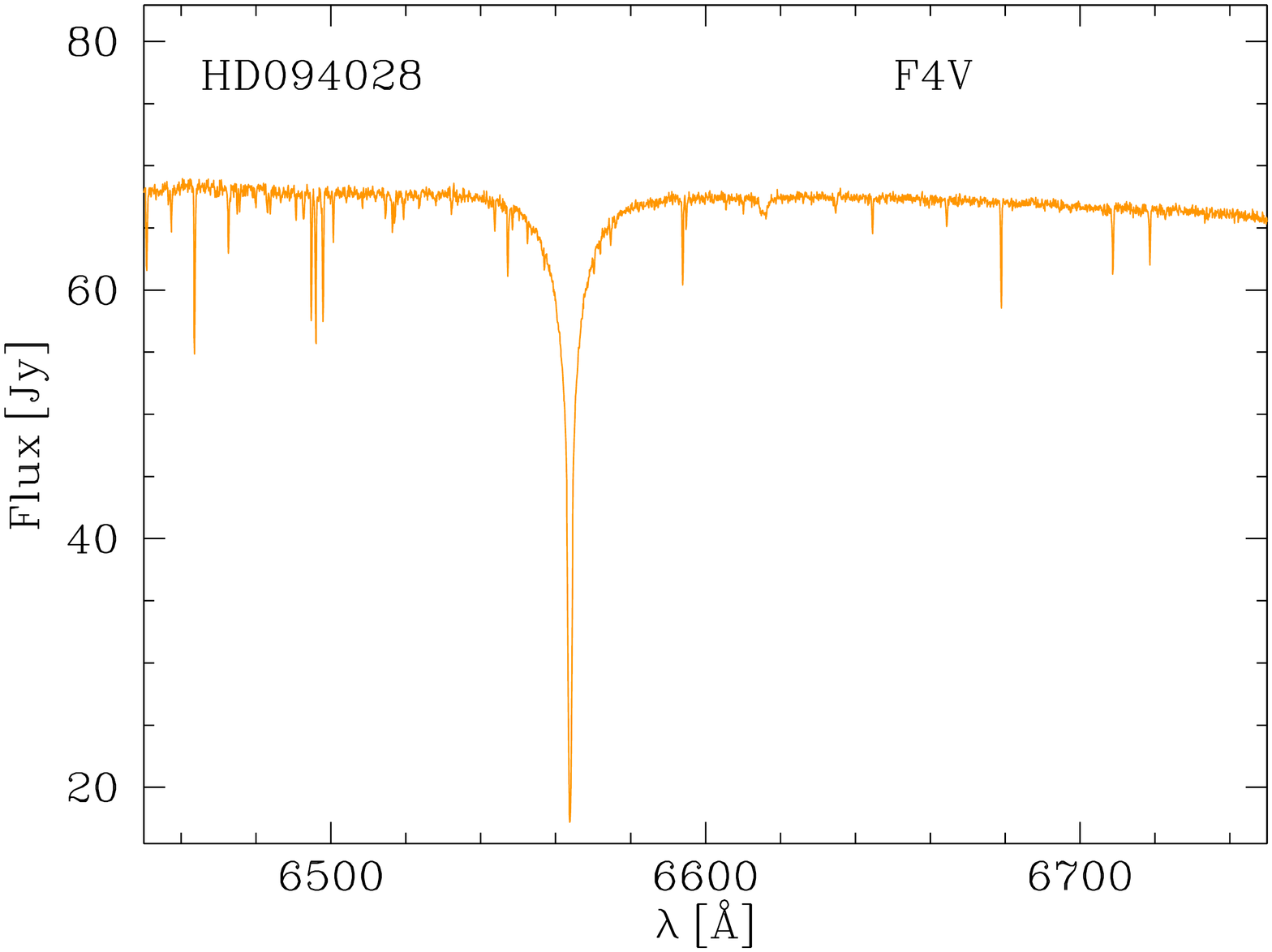}
\includegraphics[width=0.18\textwidth,angle=-90]{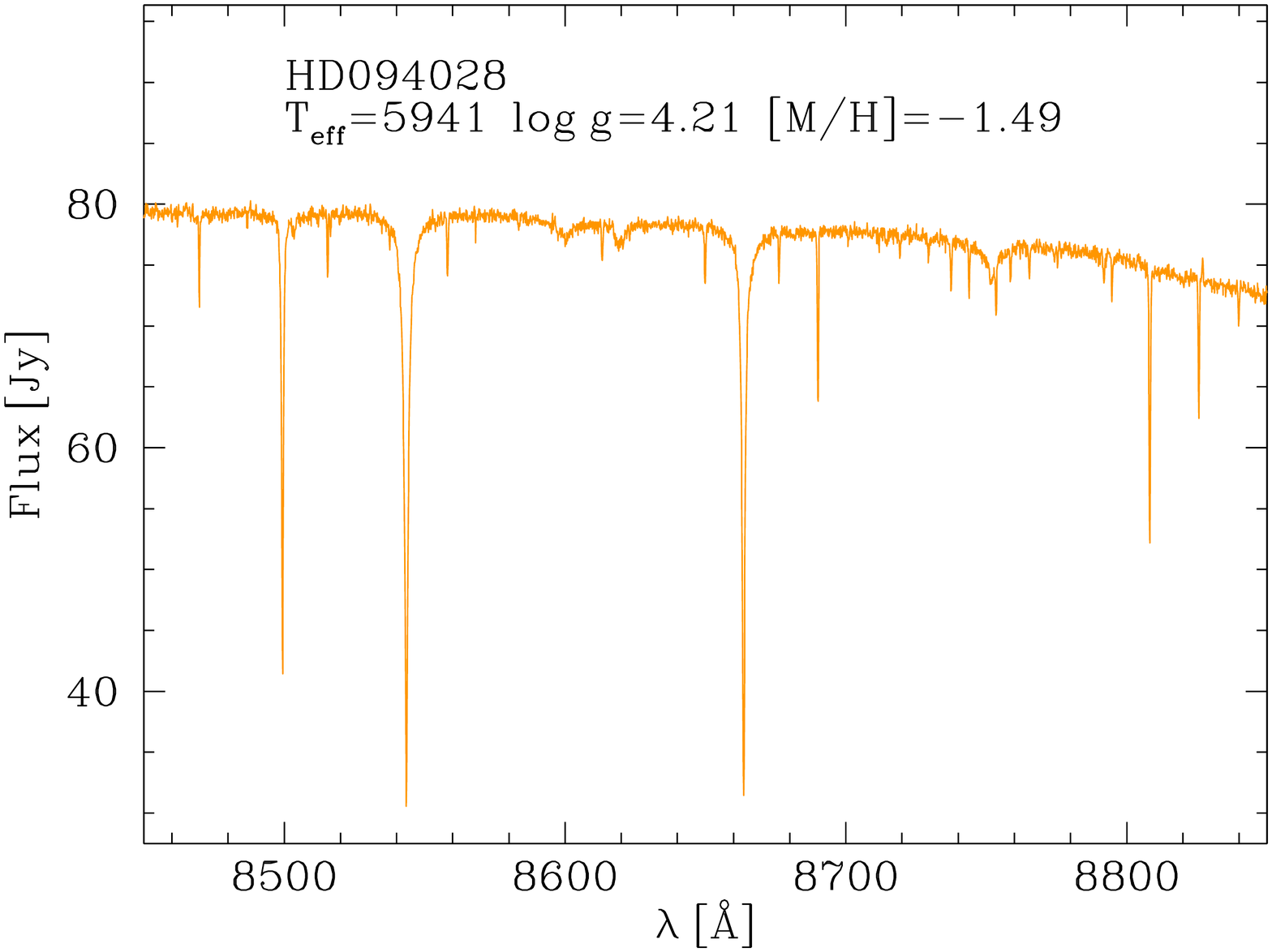}
\includegraphics[width=0.18\textwidth,angle=-90]{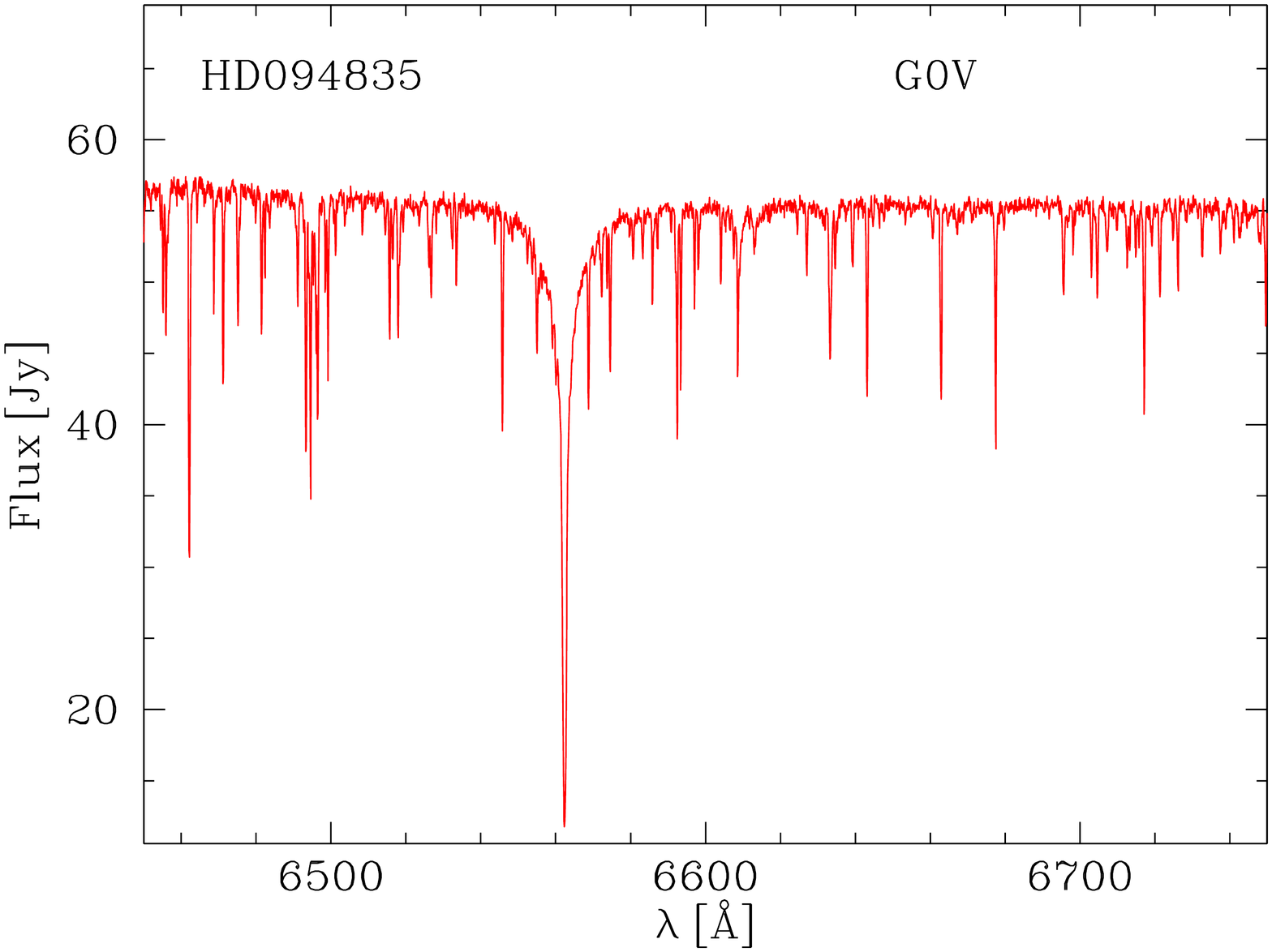}
\includegraphics[width=0.18\textwidth,angle=-90]{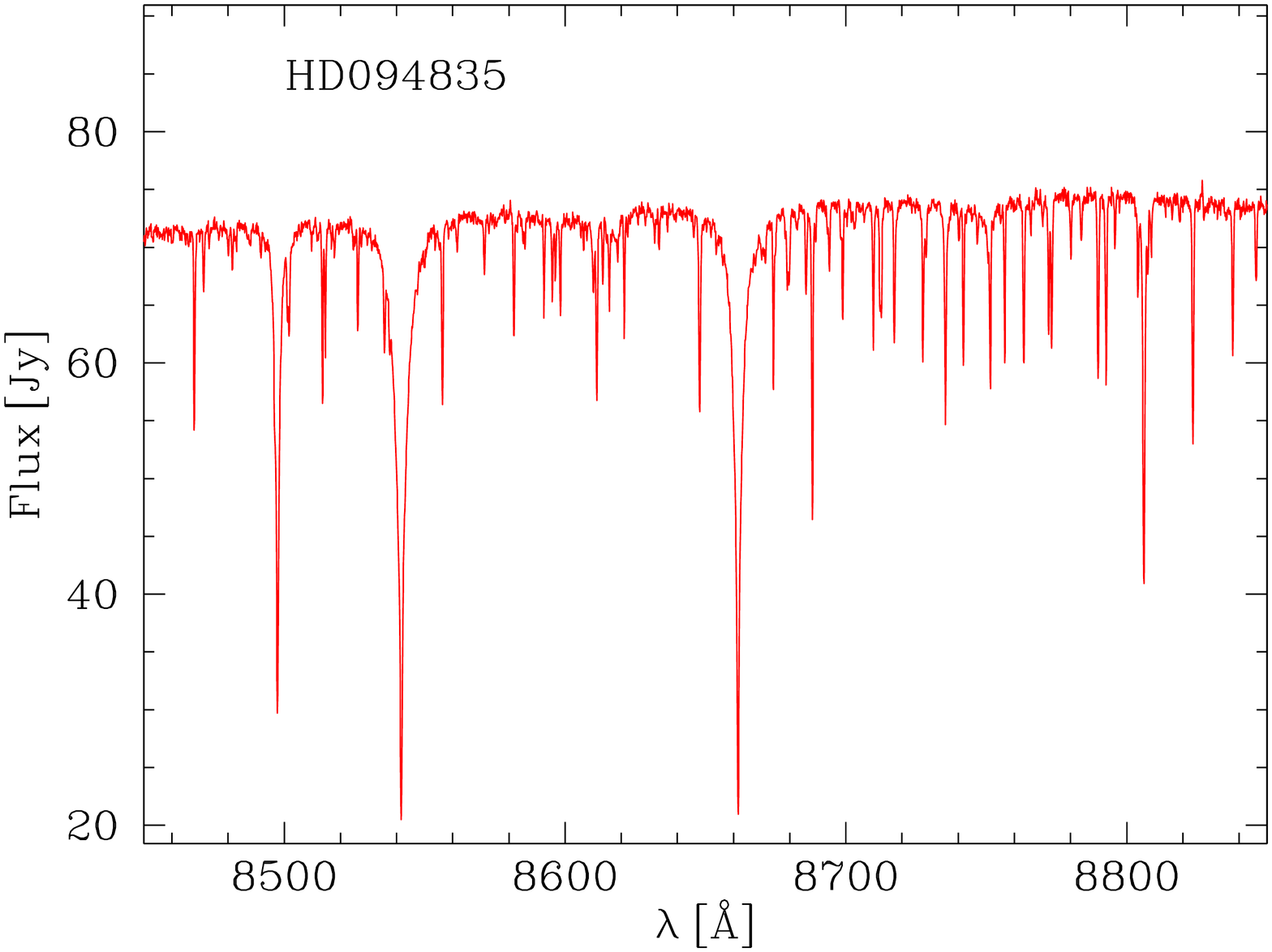}

\contcaption{18. Stars shown in this page are: HD088446, HD088609, HD088725, HD088737, HD089010, HD089125, HD089269, HD089307, HD089744, HD089995, HD090537, HD090839, HD094028 and HD094835.}
\end{figure*}

\begin{figure*}
\includegraphics[width=0.18\textwidth,angle=-90]{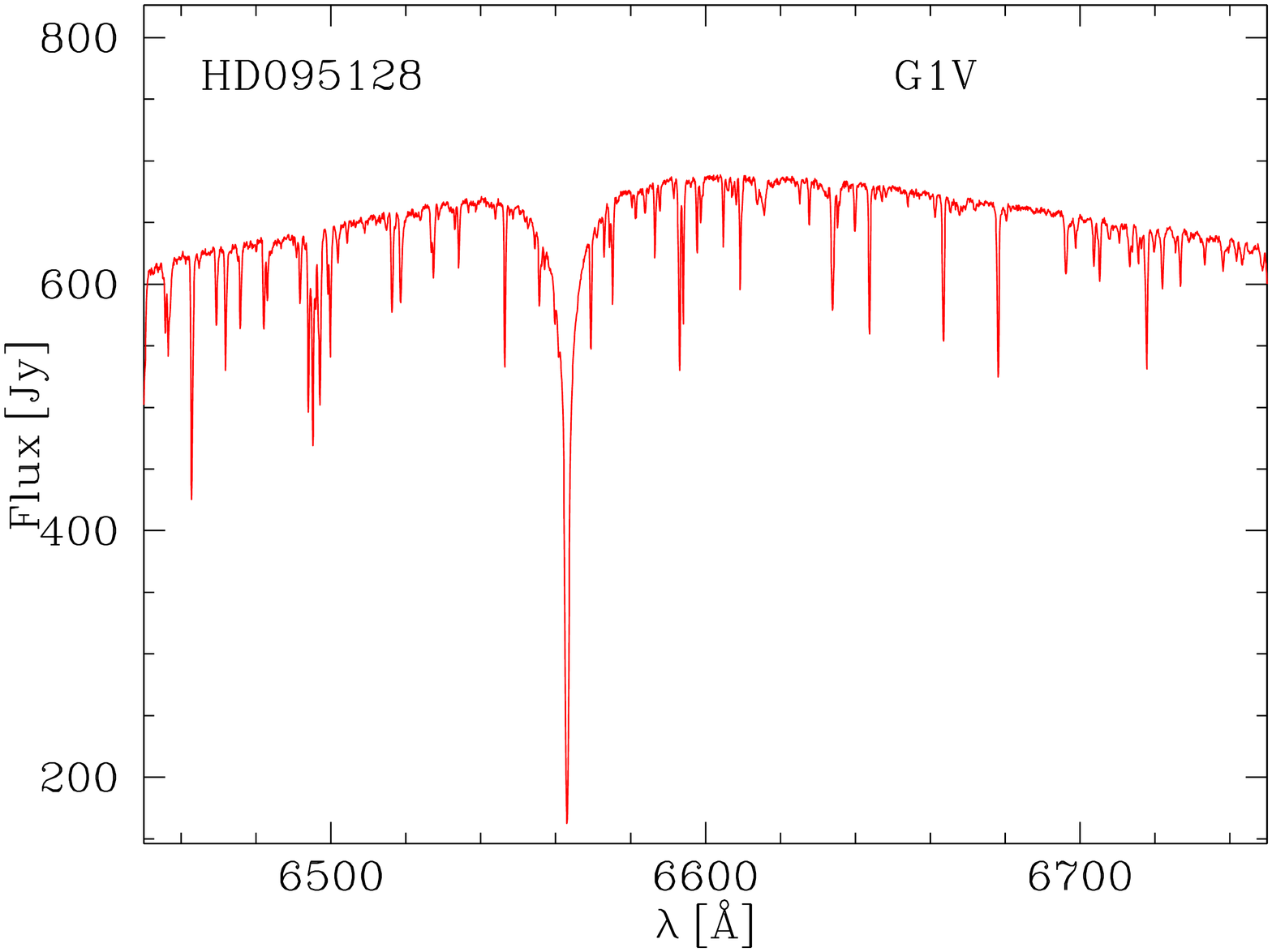}
\includegraphics[width=0.18\textwidth,angle=-90]{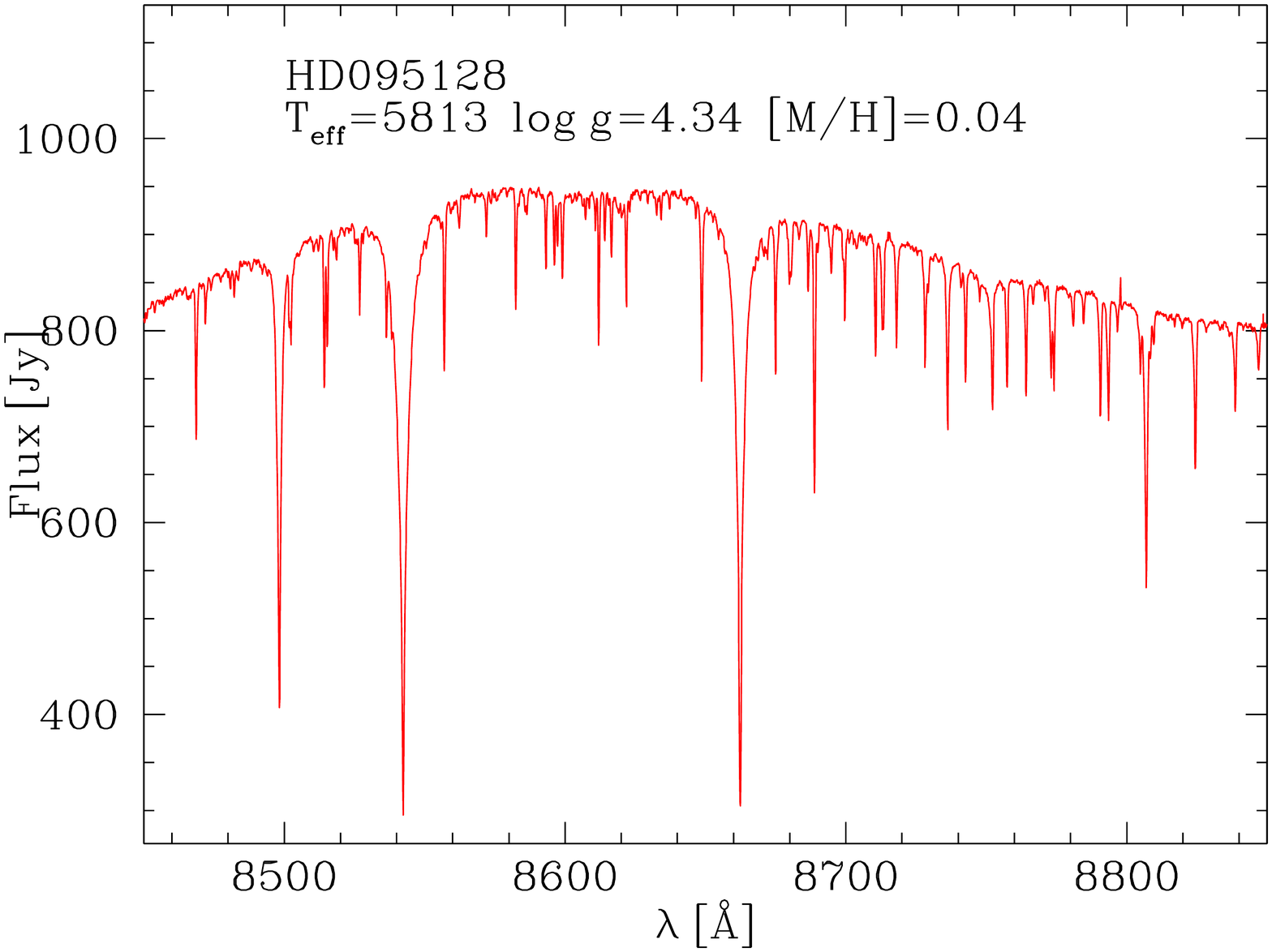}
\includegraphics[width=0.18\textwidth,angle=-90]{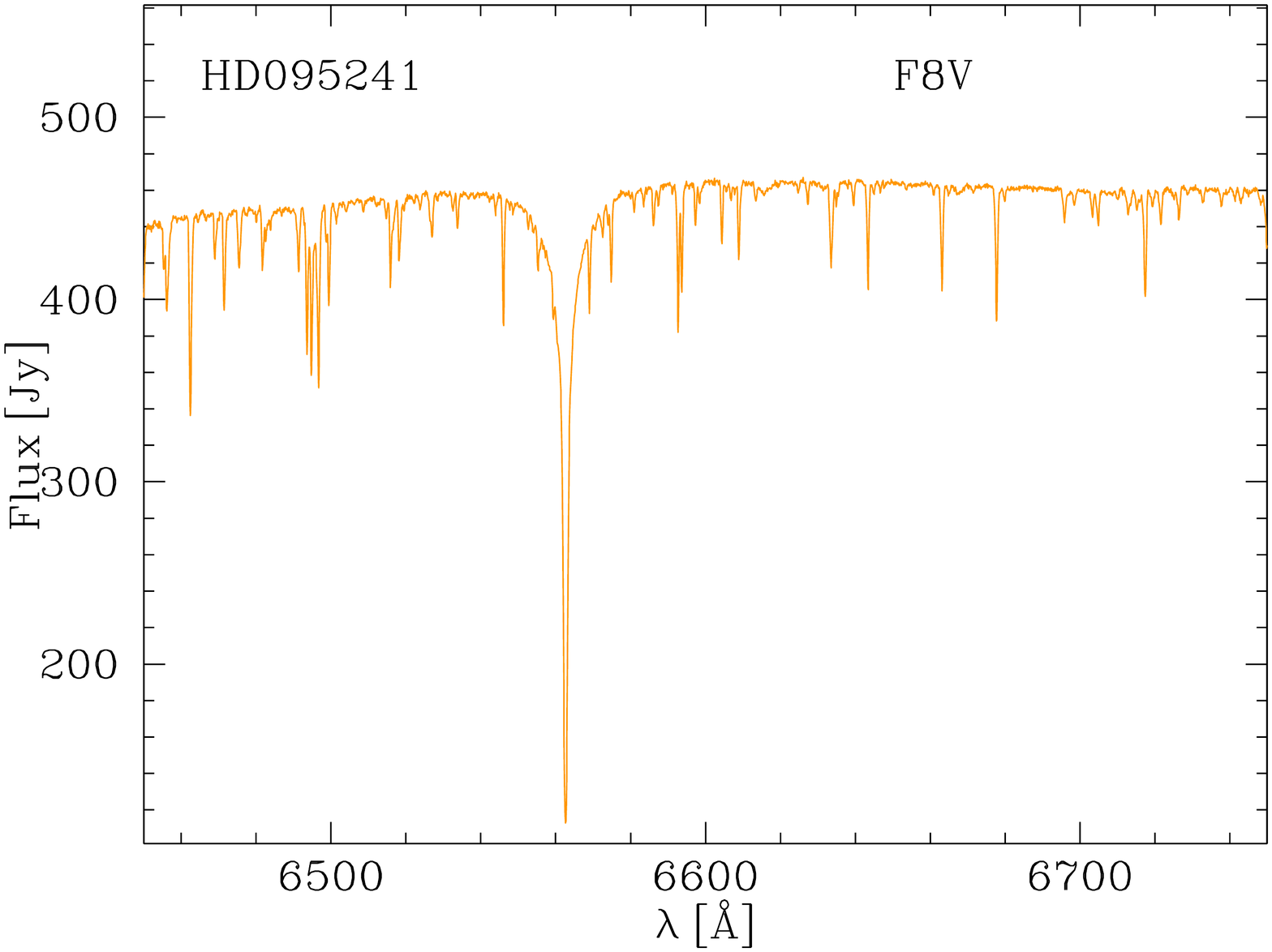}
\includegraphics[width=0.18\textwidth,angle=-90]{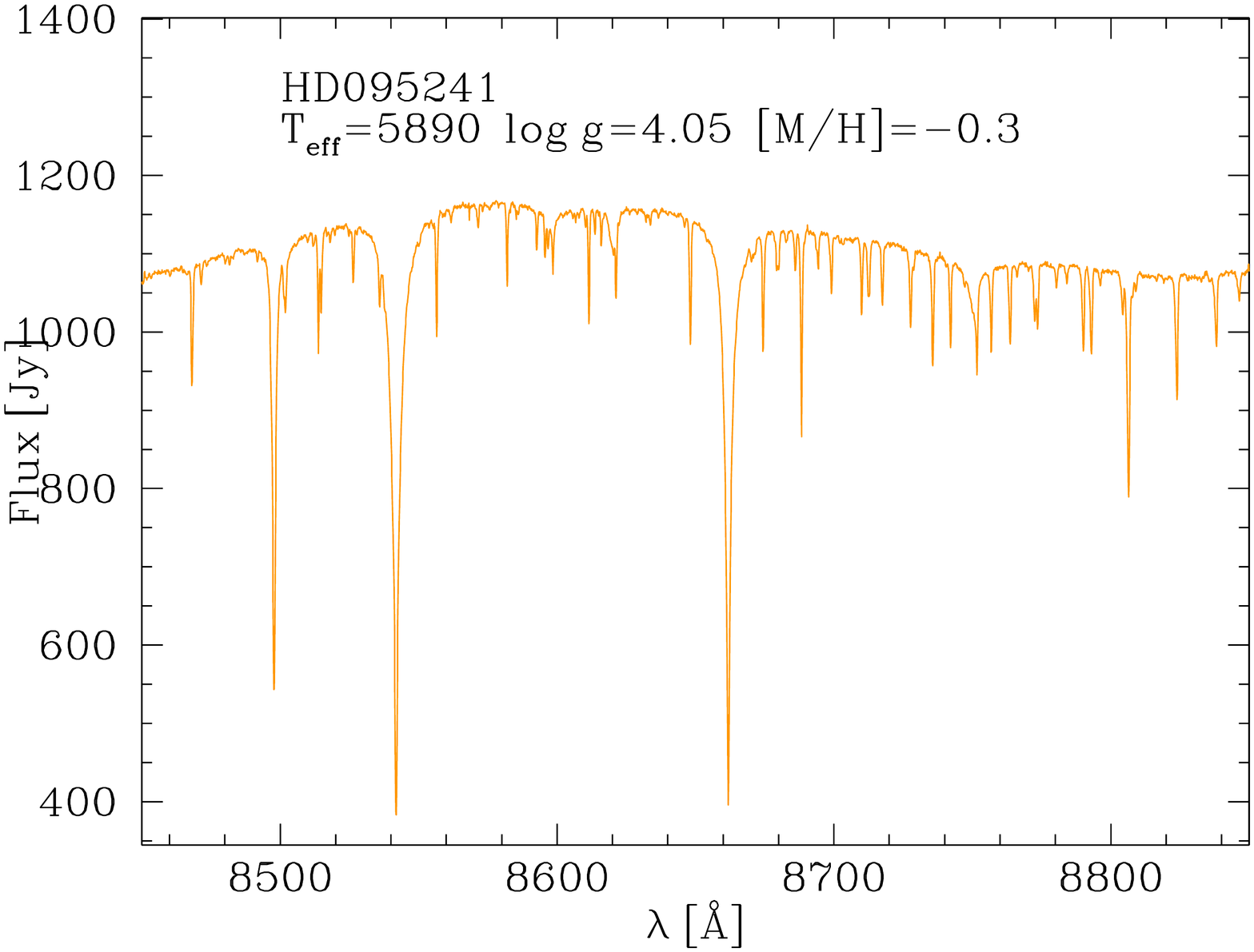}
\includegraphics[width=0.18\textwidth,angle=-90]{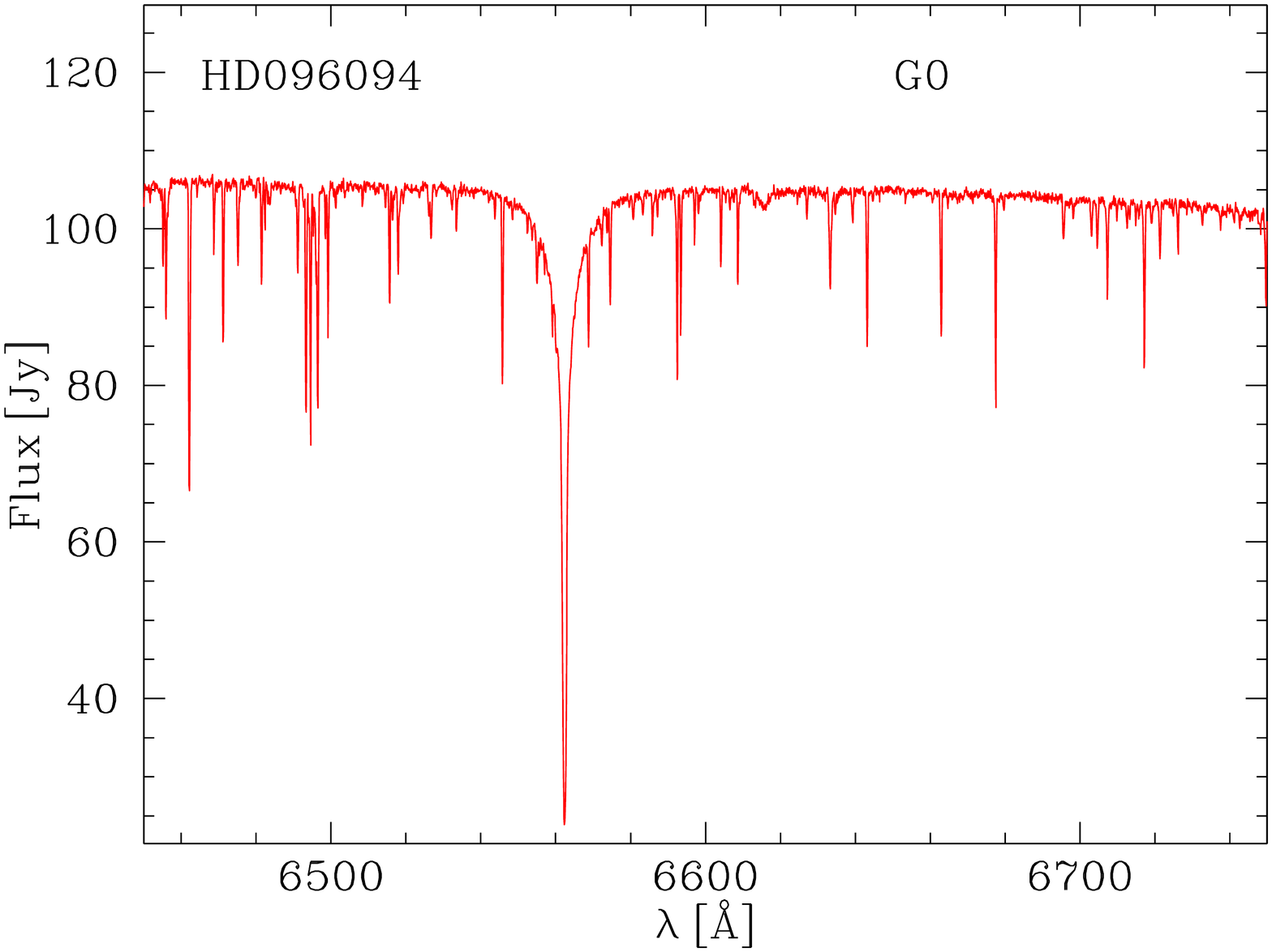}
\includegraphics[width=0.18\textwidth,angle=-90]{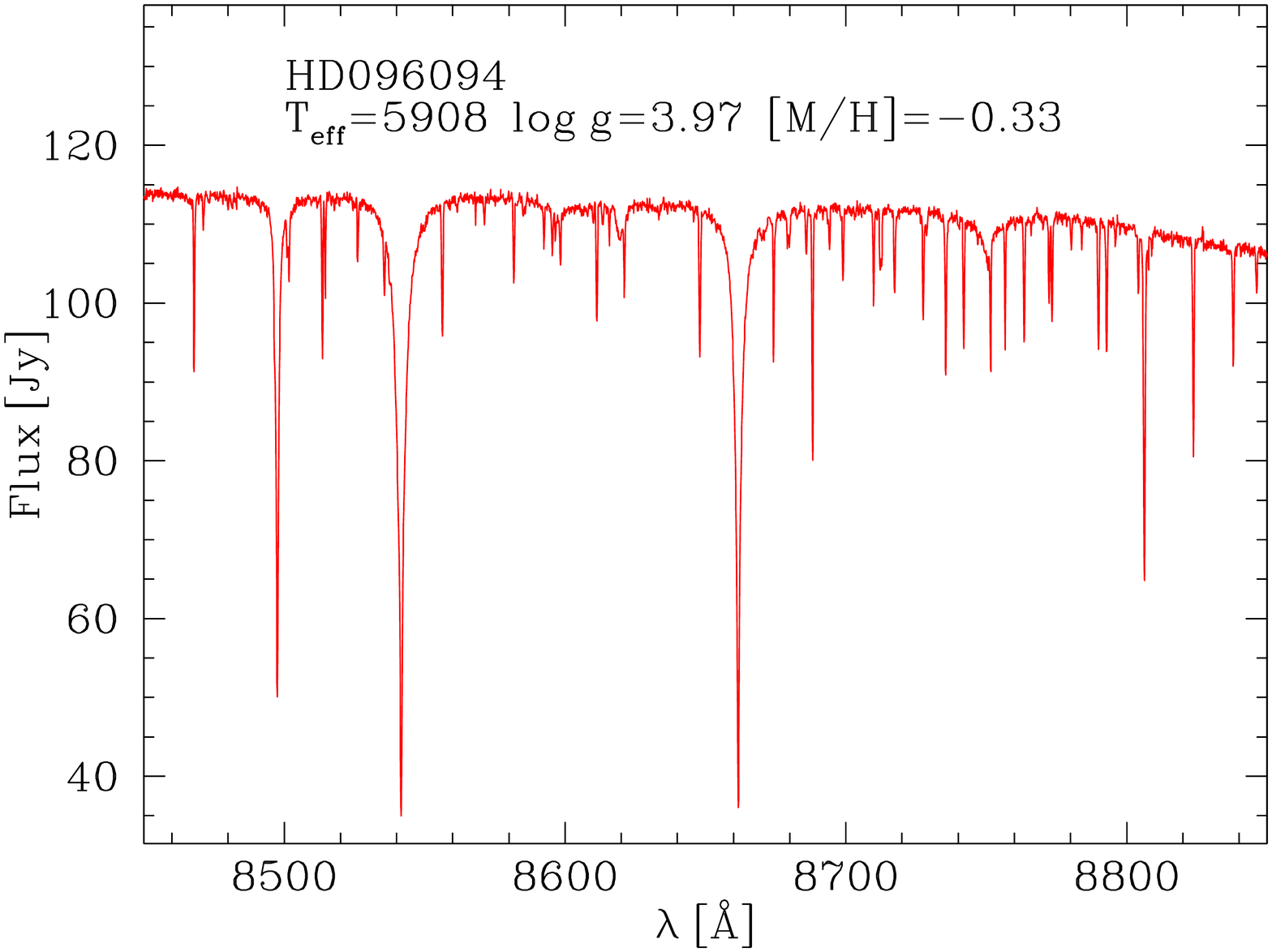}
\includegraphics[width=0.18\textwidth,angle=-90]{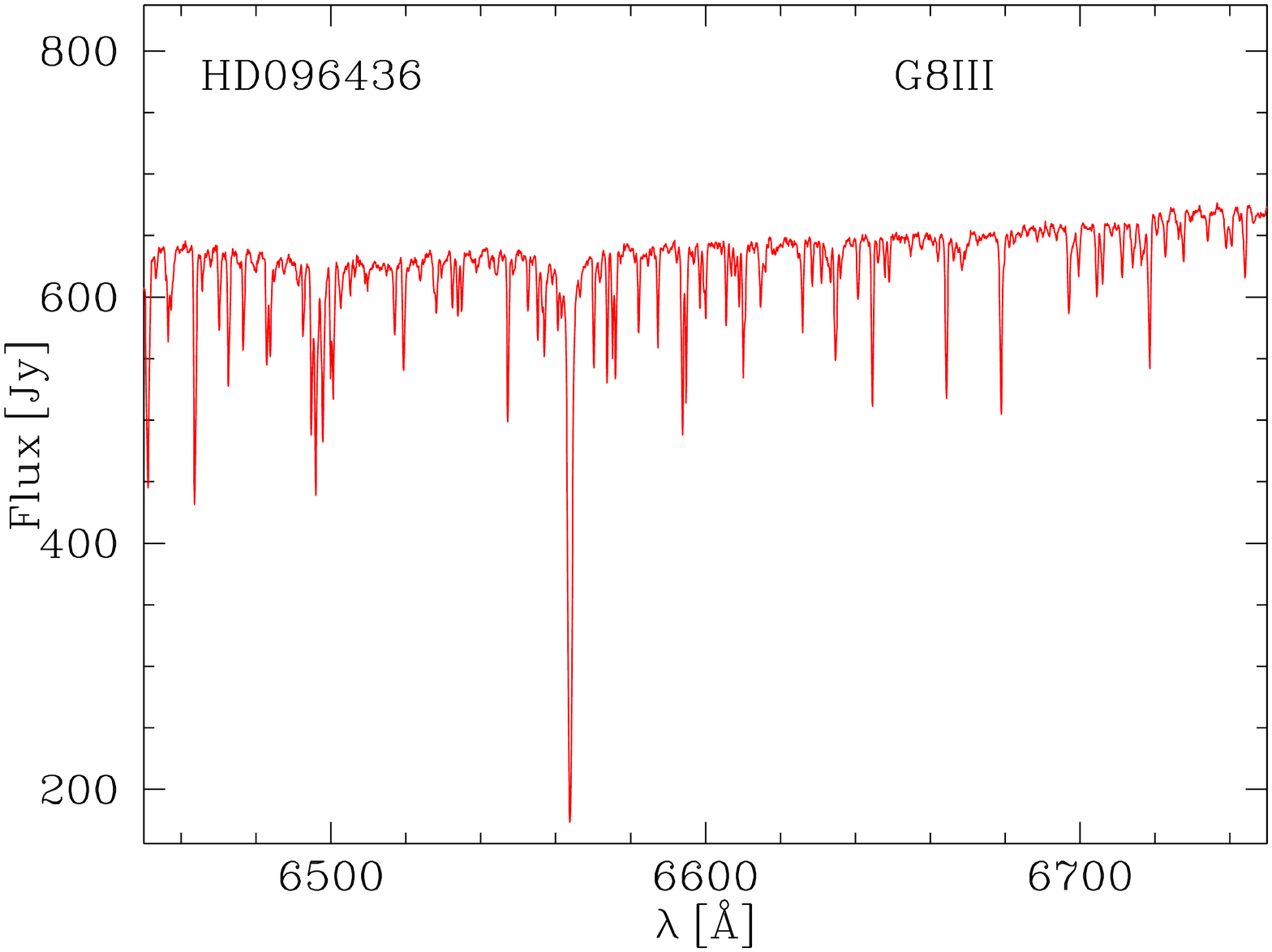}
\includegraphics[width=0.18\textwidth,angle=-90]{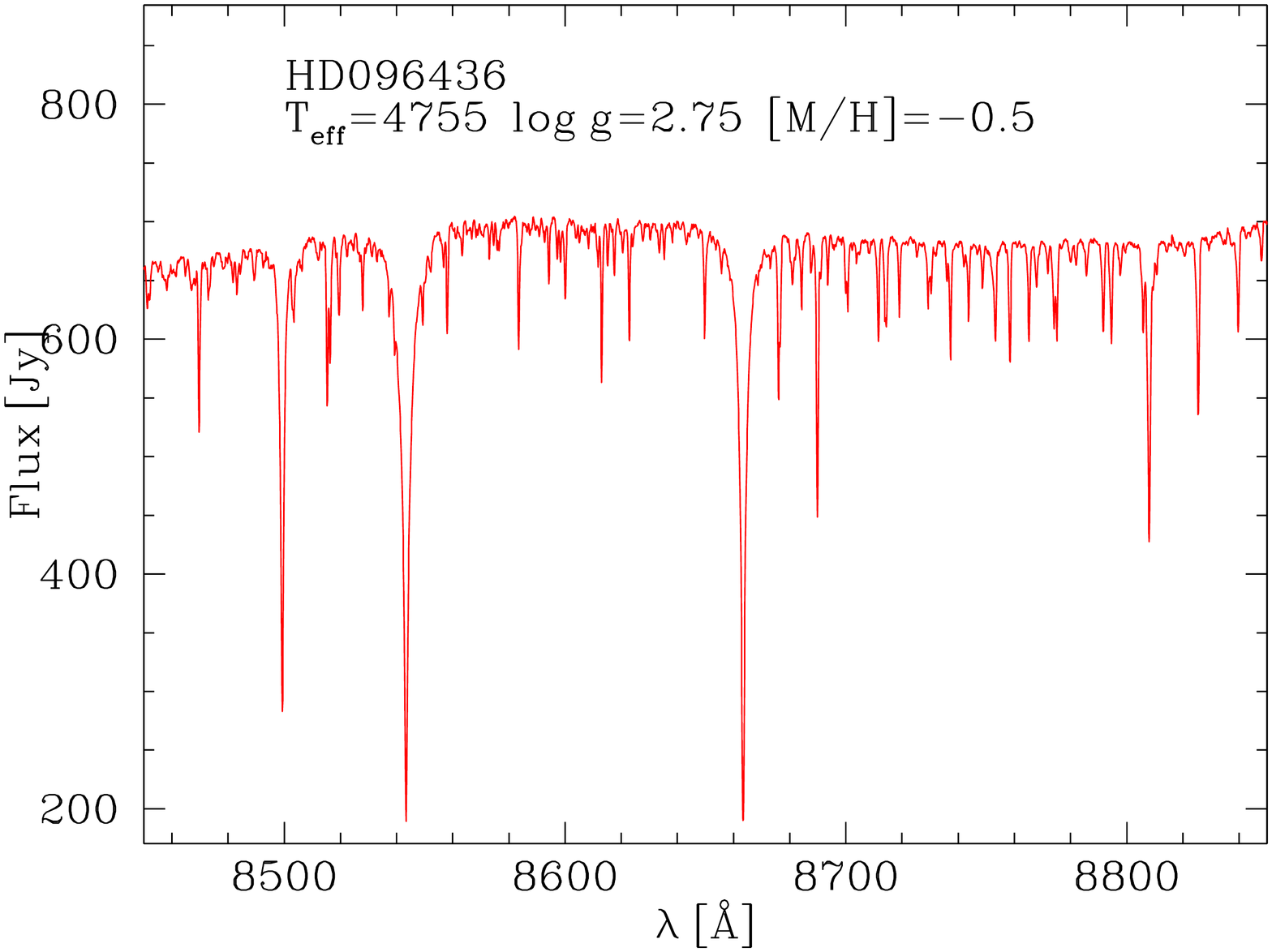}
\includegraphics[width=0.18\textwidth,angle=-90]{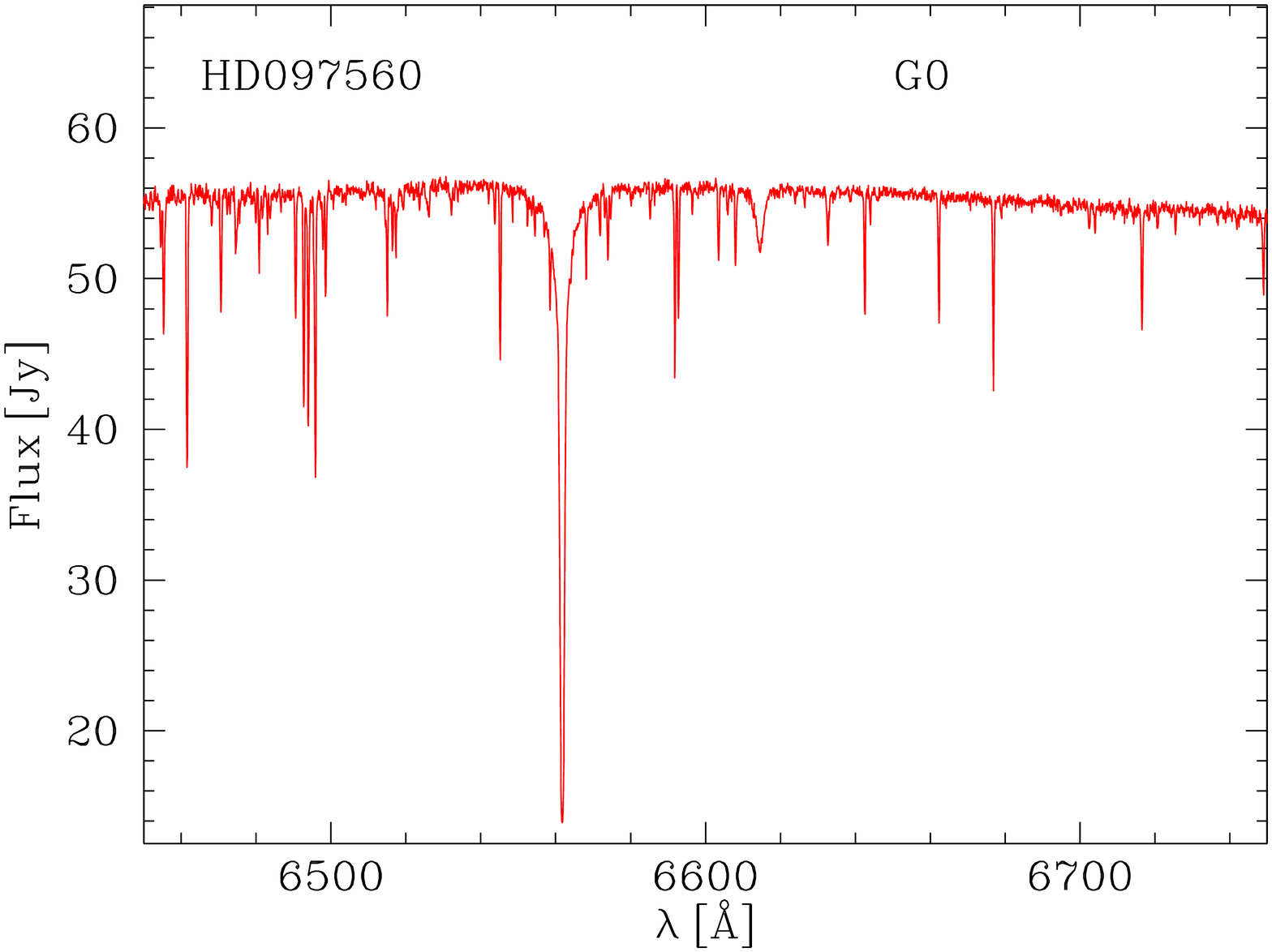}
\includegraphics[width=0.18\textwidth,angle=-90]{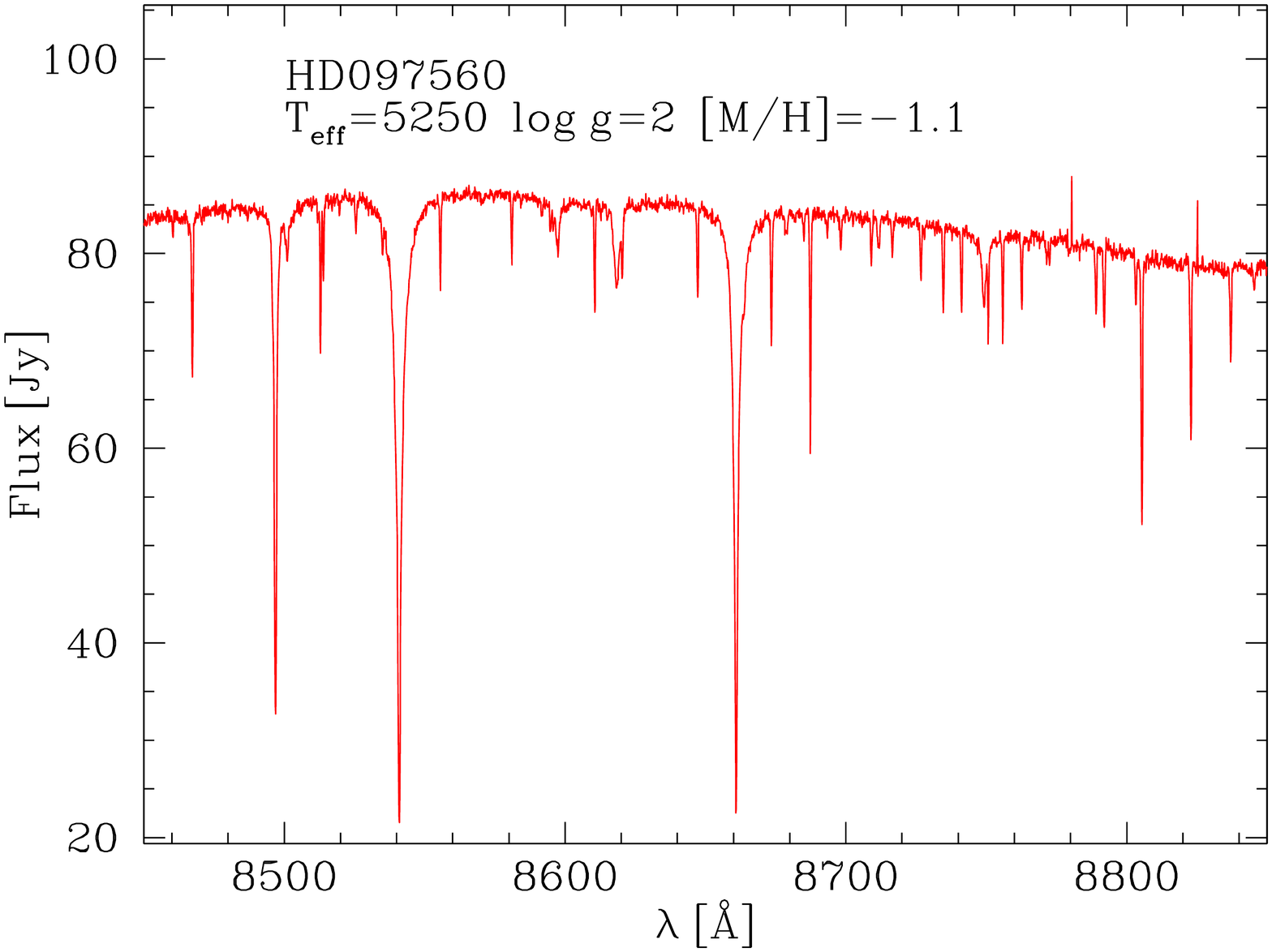}
\includegraphics[width=0.18\textwidth,angle=-90]{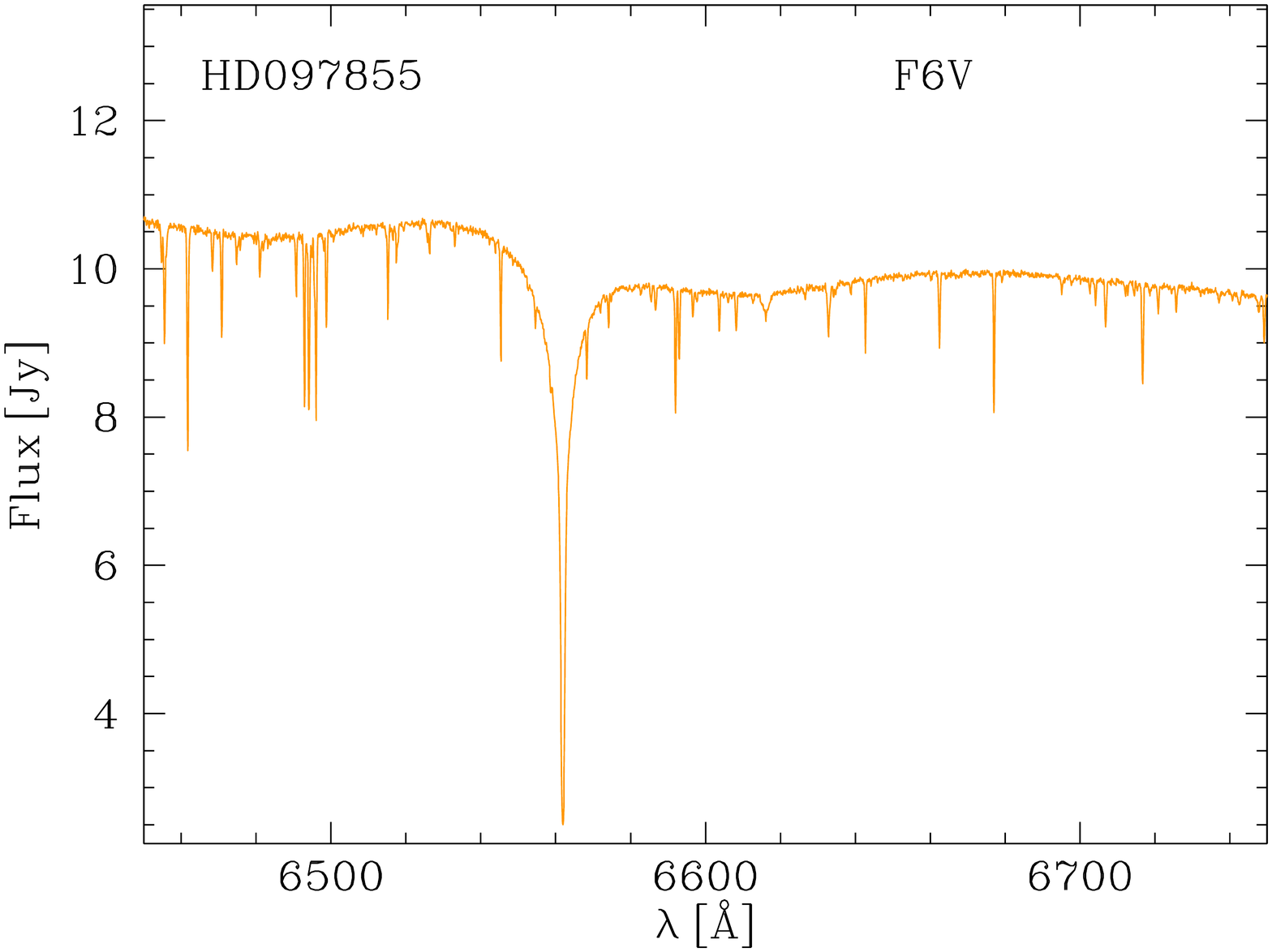}
\includegraphics[width=0.18\textwidth,angle=-90]{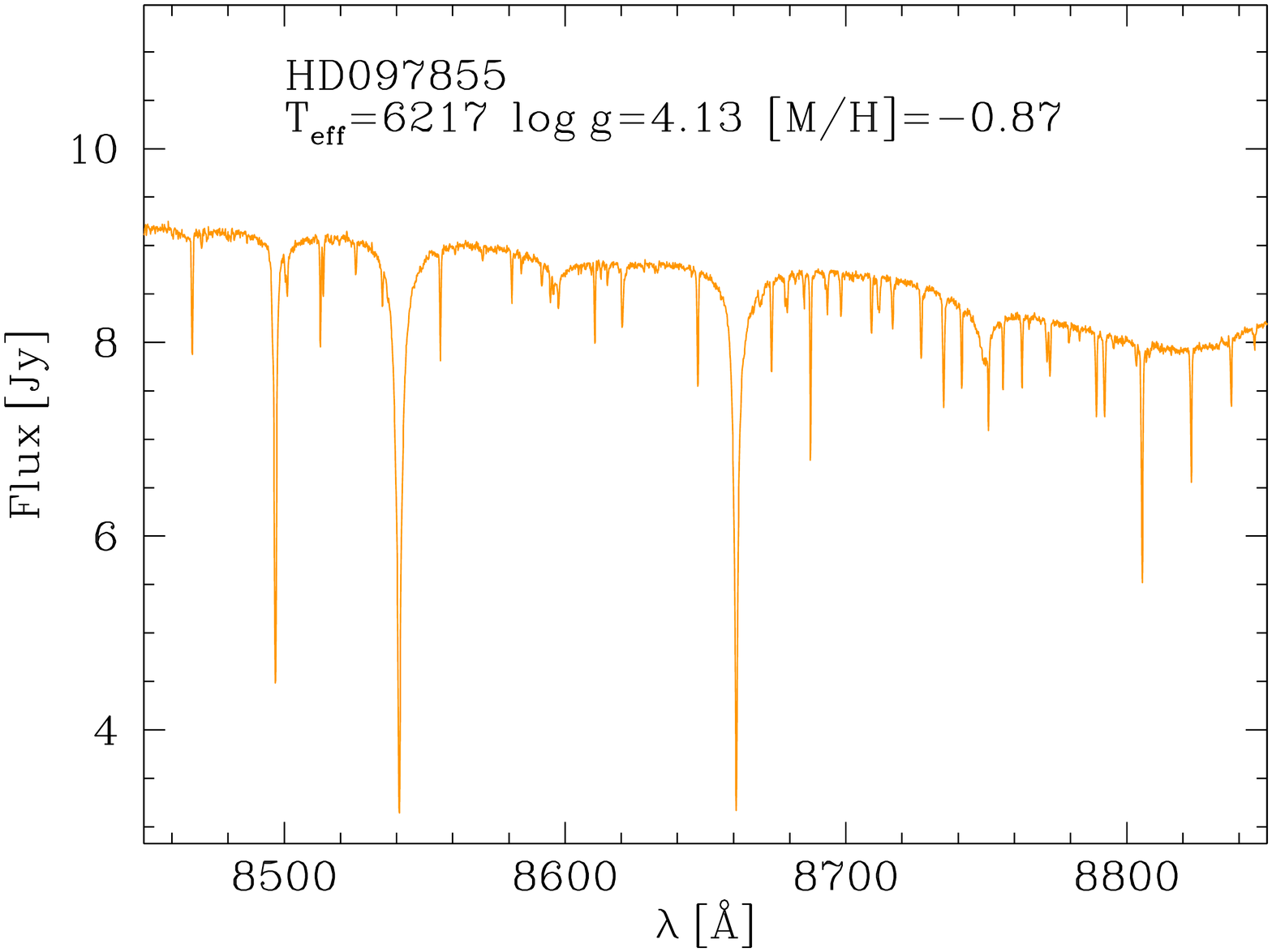}
\includegraphics[width=0.18\textwidth,angle=-90]{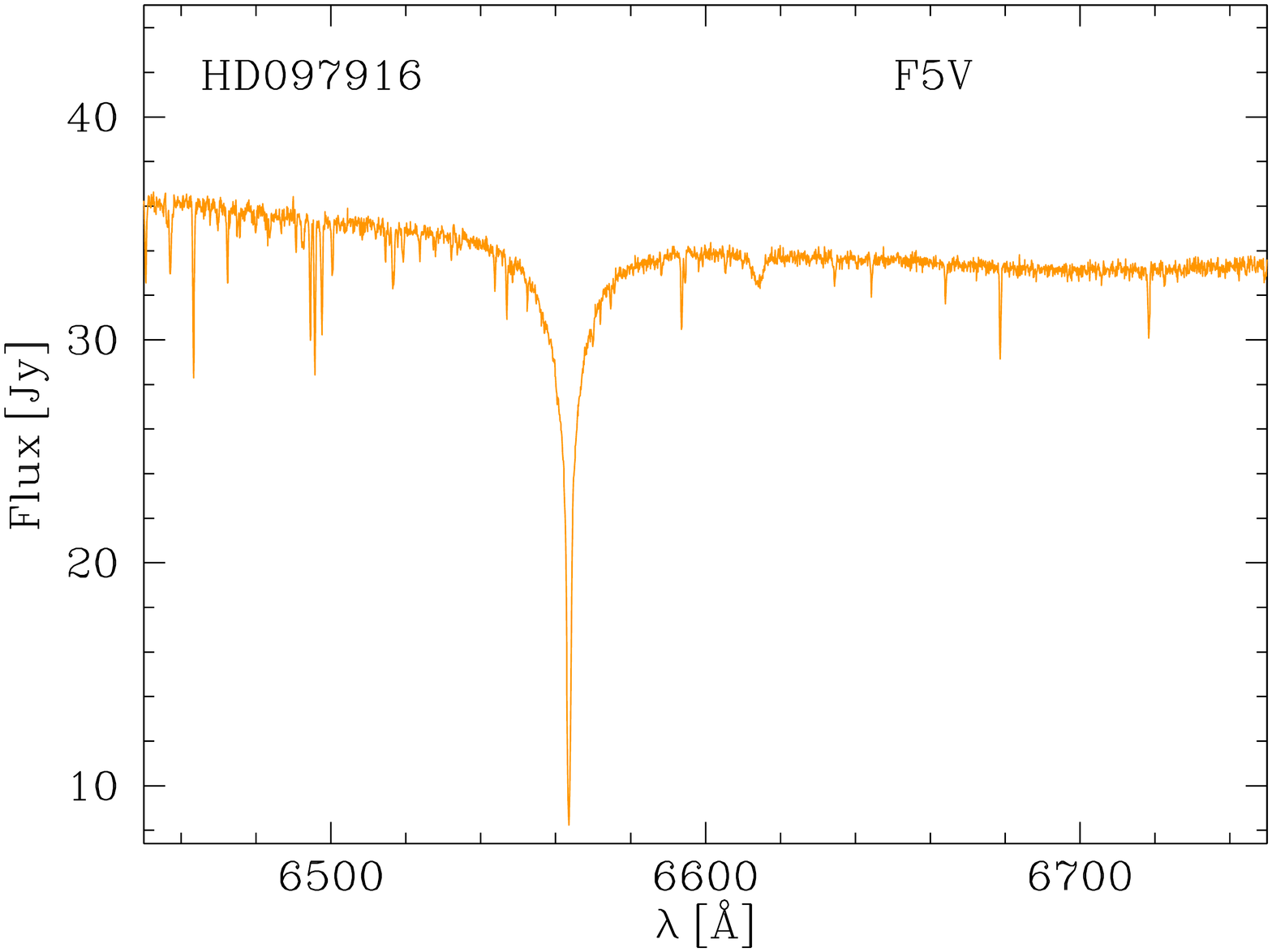}
\includegraphics[width=0.18\textwidth,angle=-90]{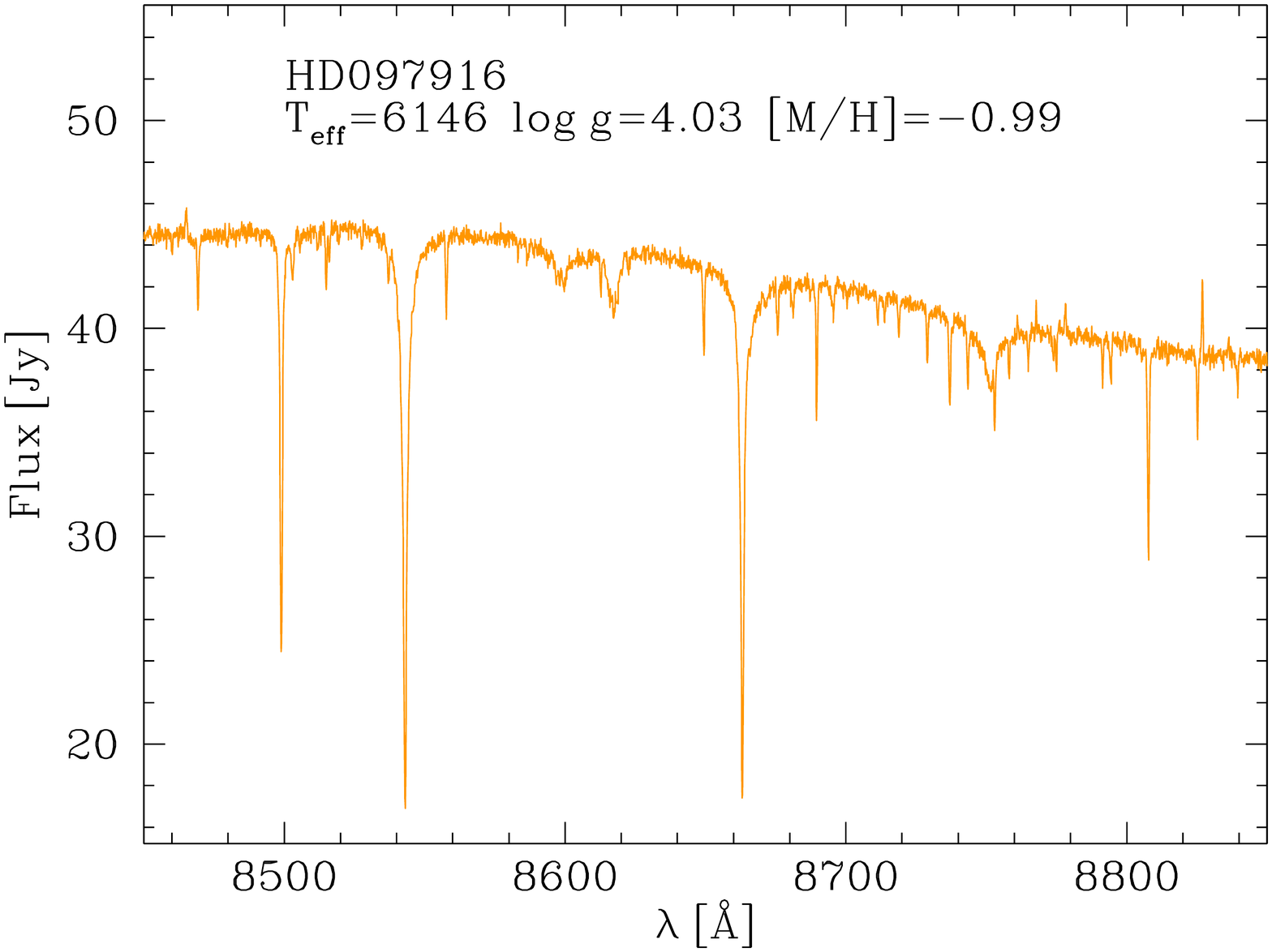}
\includegraphics[width=0.18\textwidth,angle=-90]{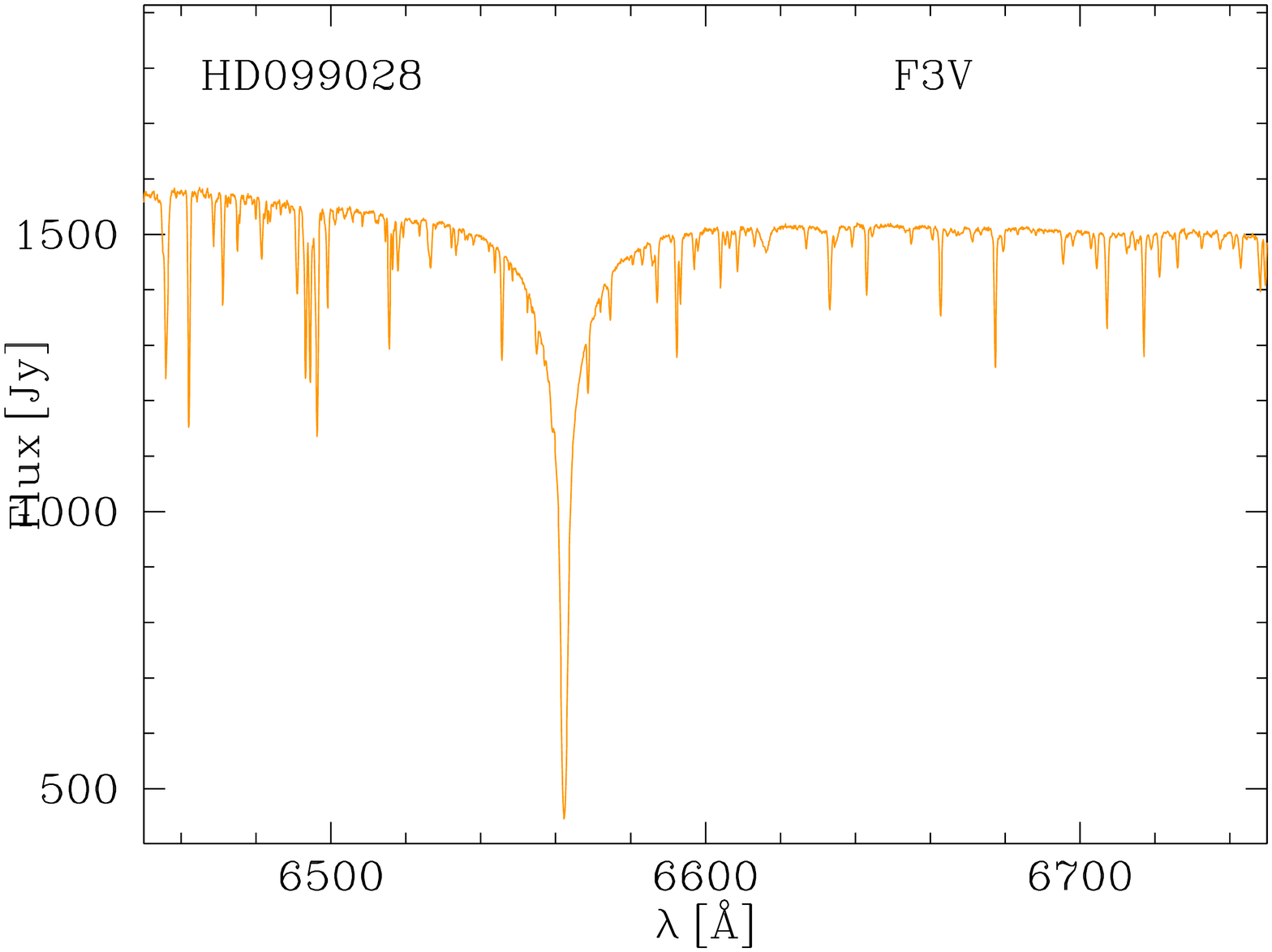}
\includegraphics[width=0.18\textwidth,angle=-90]{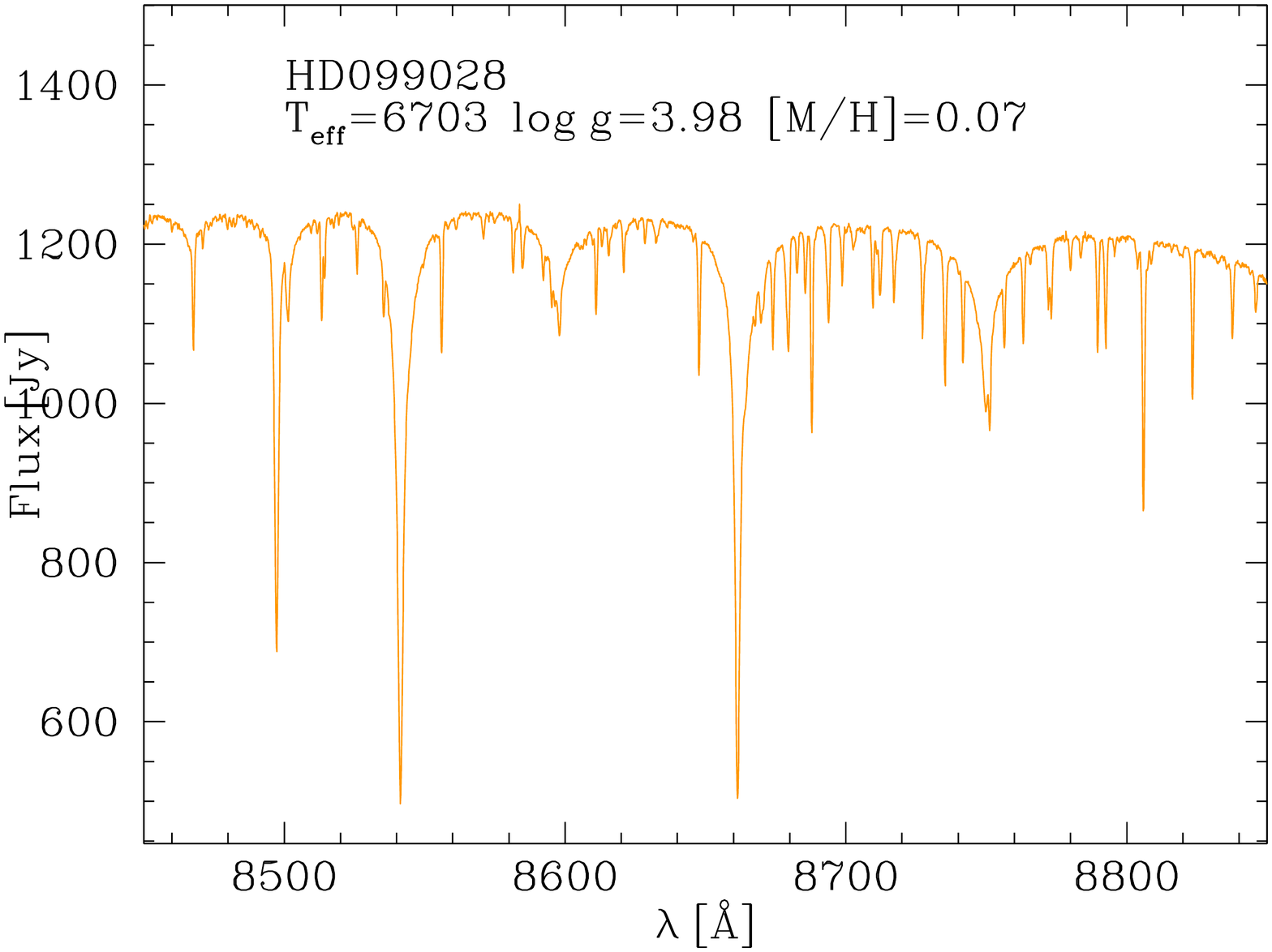}
\includegraphics[width=0.18\textwidth,angle=-90]{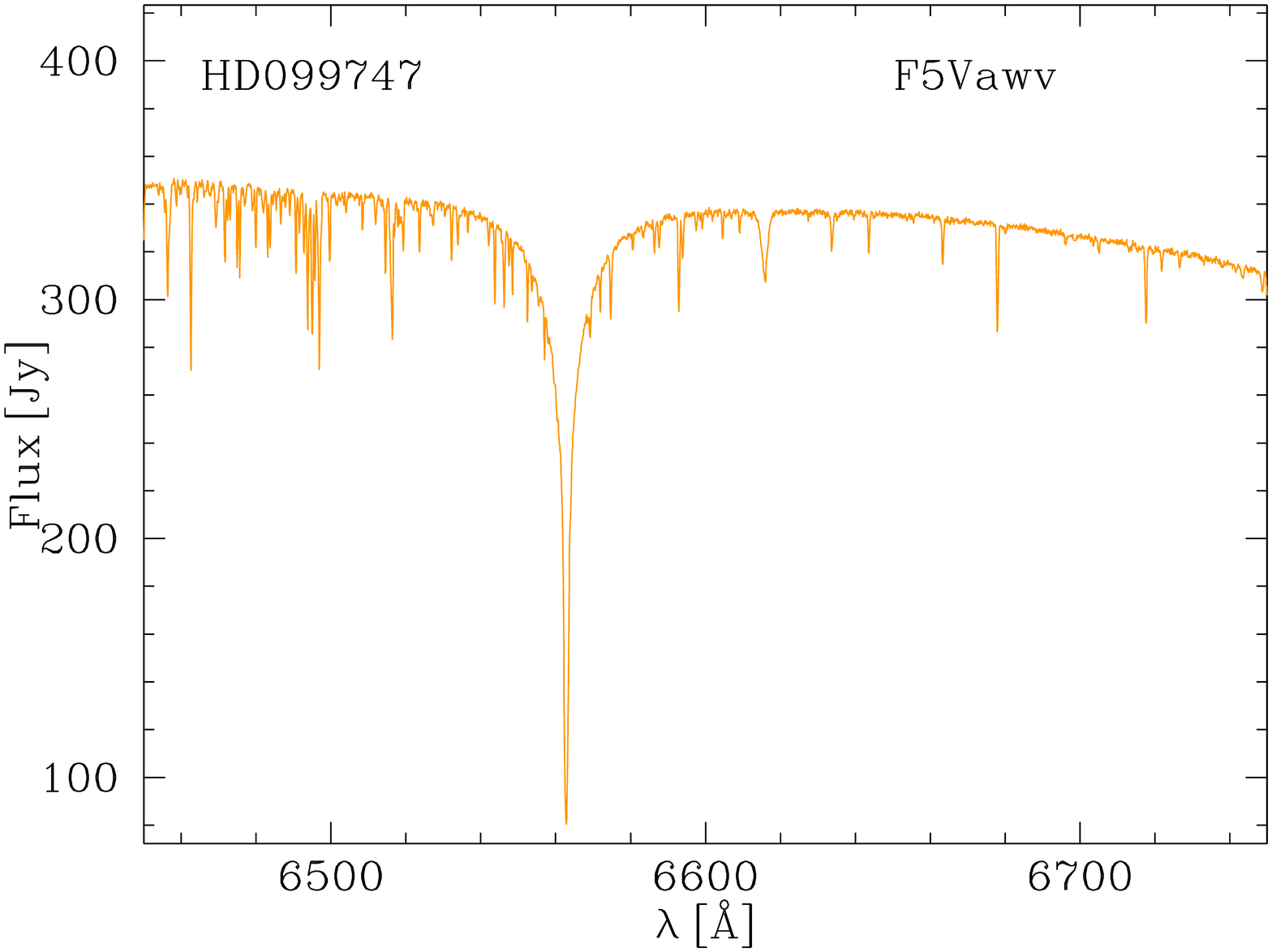}
\includegraphics[width=0.18\textwidth,angle=-90]{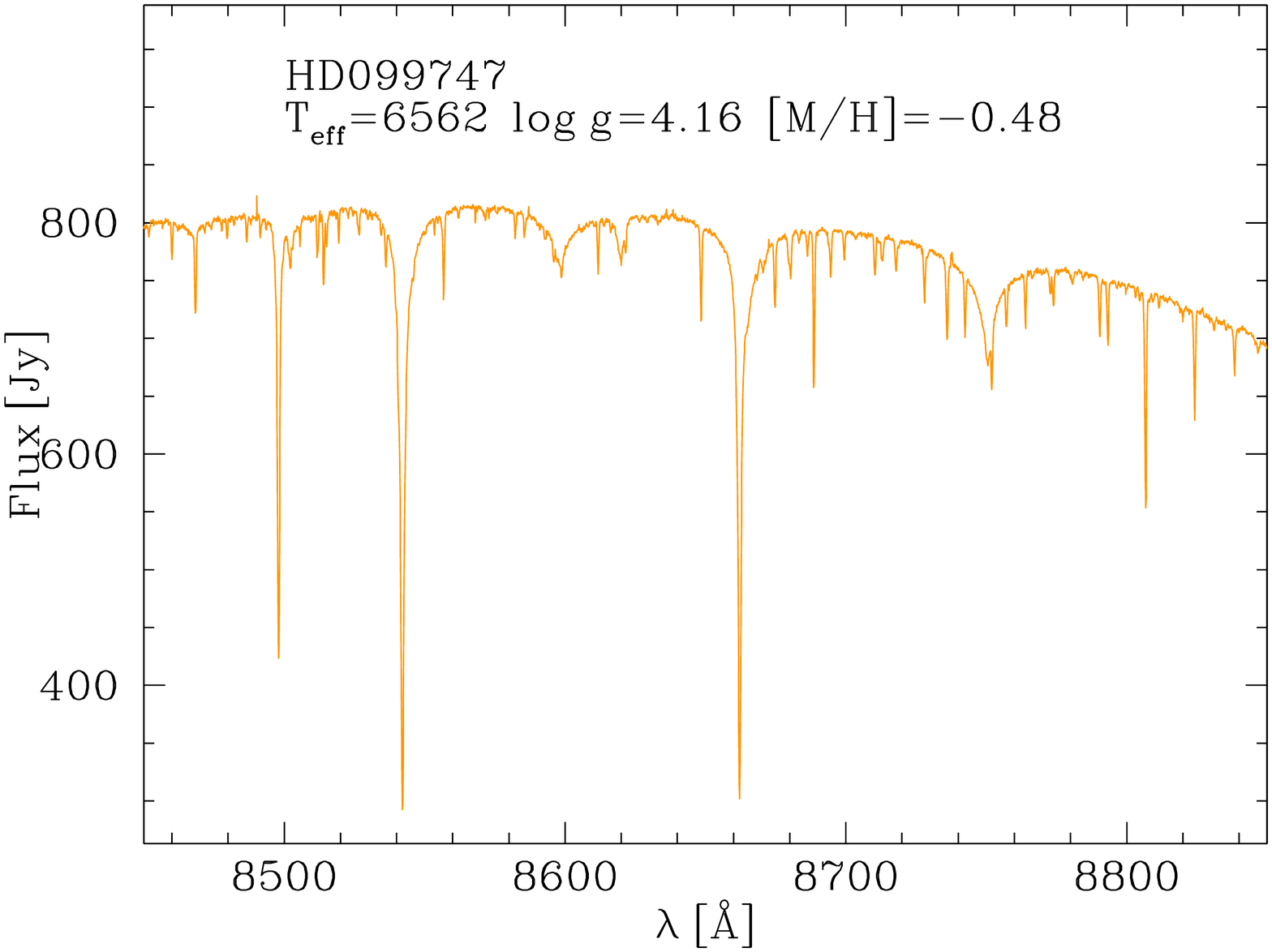}
\includegraphics[width=0.18\textwidth,angle=-90]{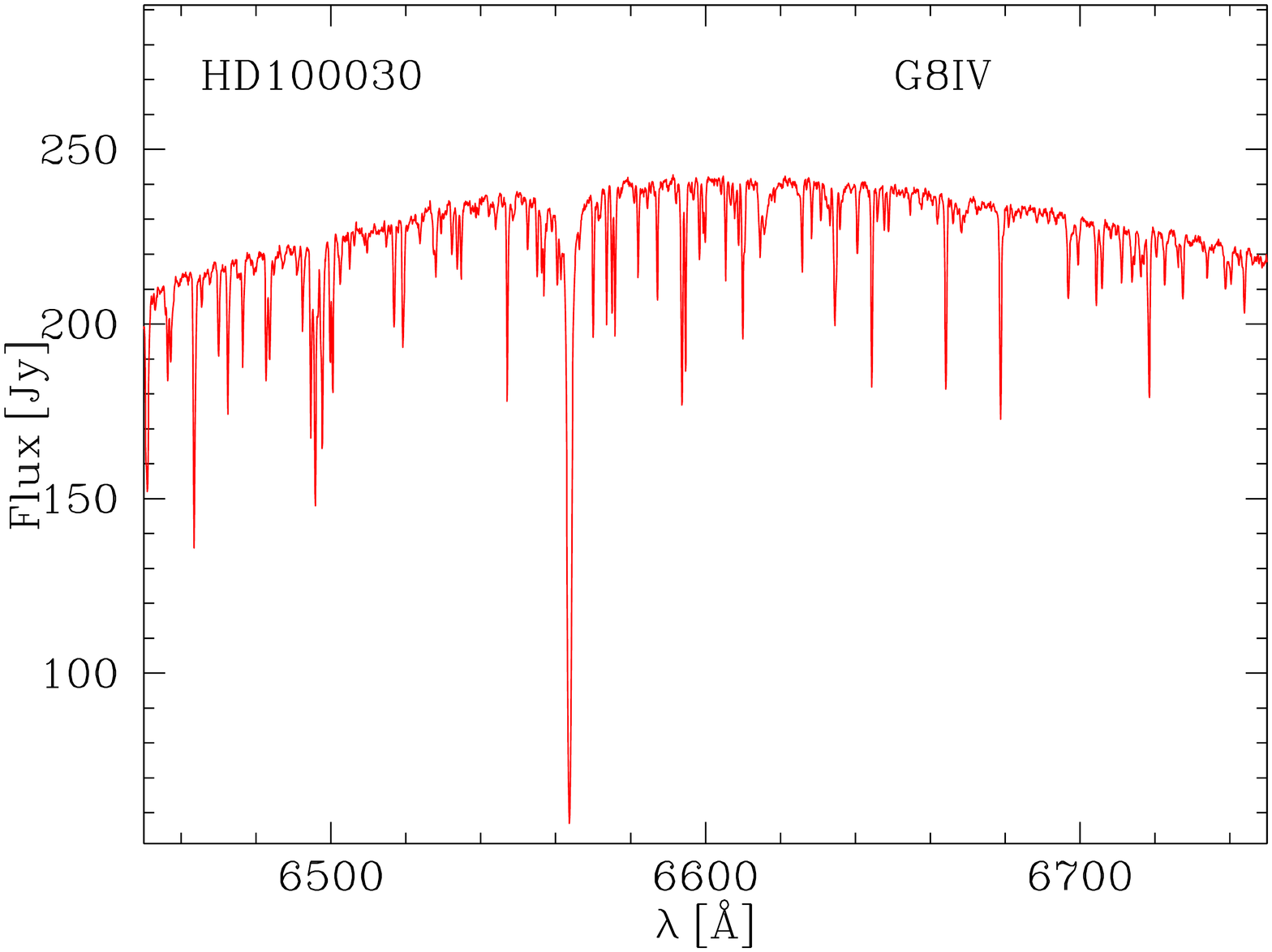}
\includegraphics[width=0.18\textwidth,angle=-90]{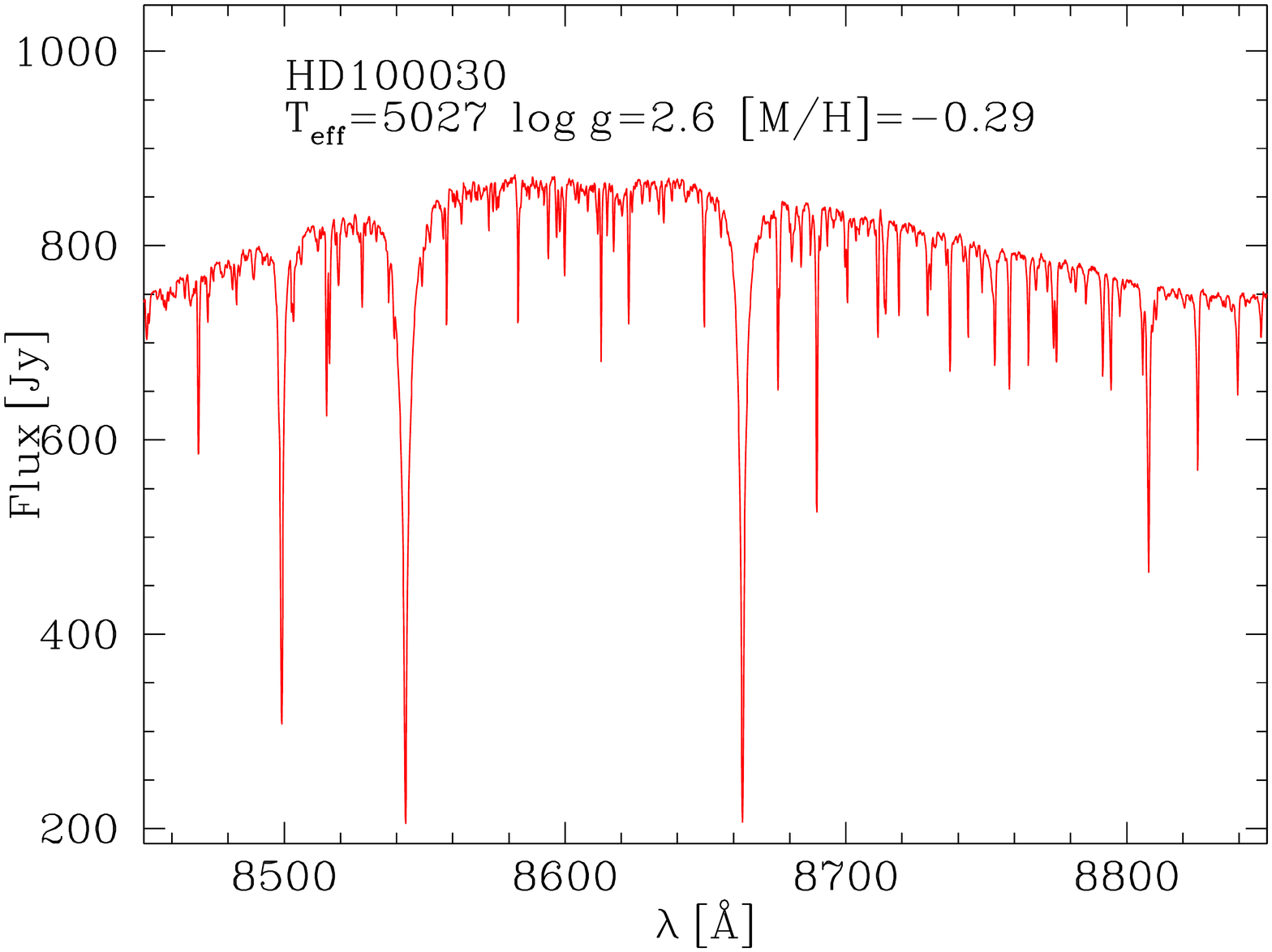}
\includegraphics[width=0.18\textwidth,angle=-90]{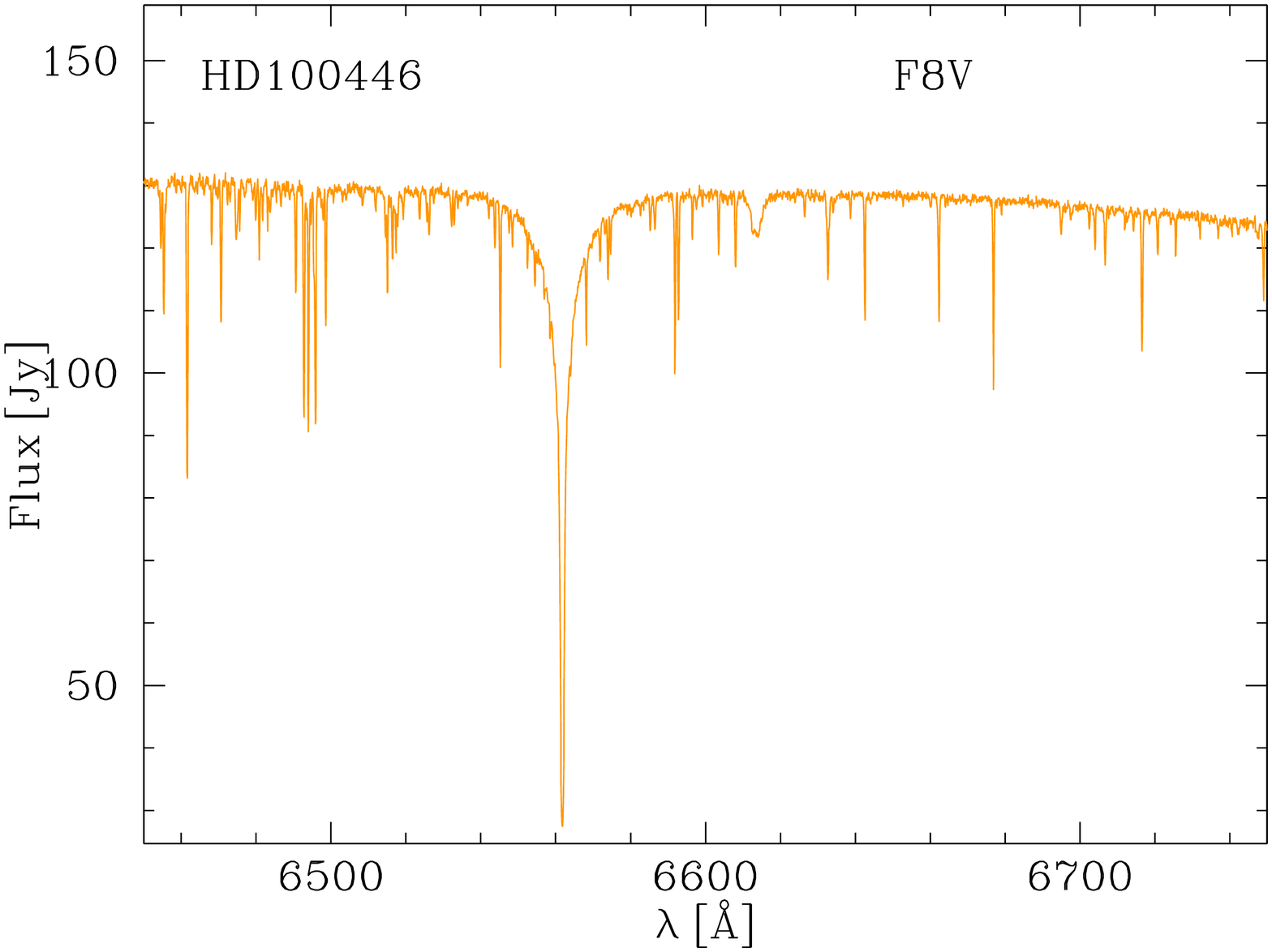}
\includegraphics[width=0.18\textwidth,angle=-90]{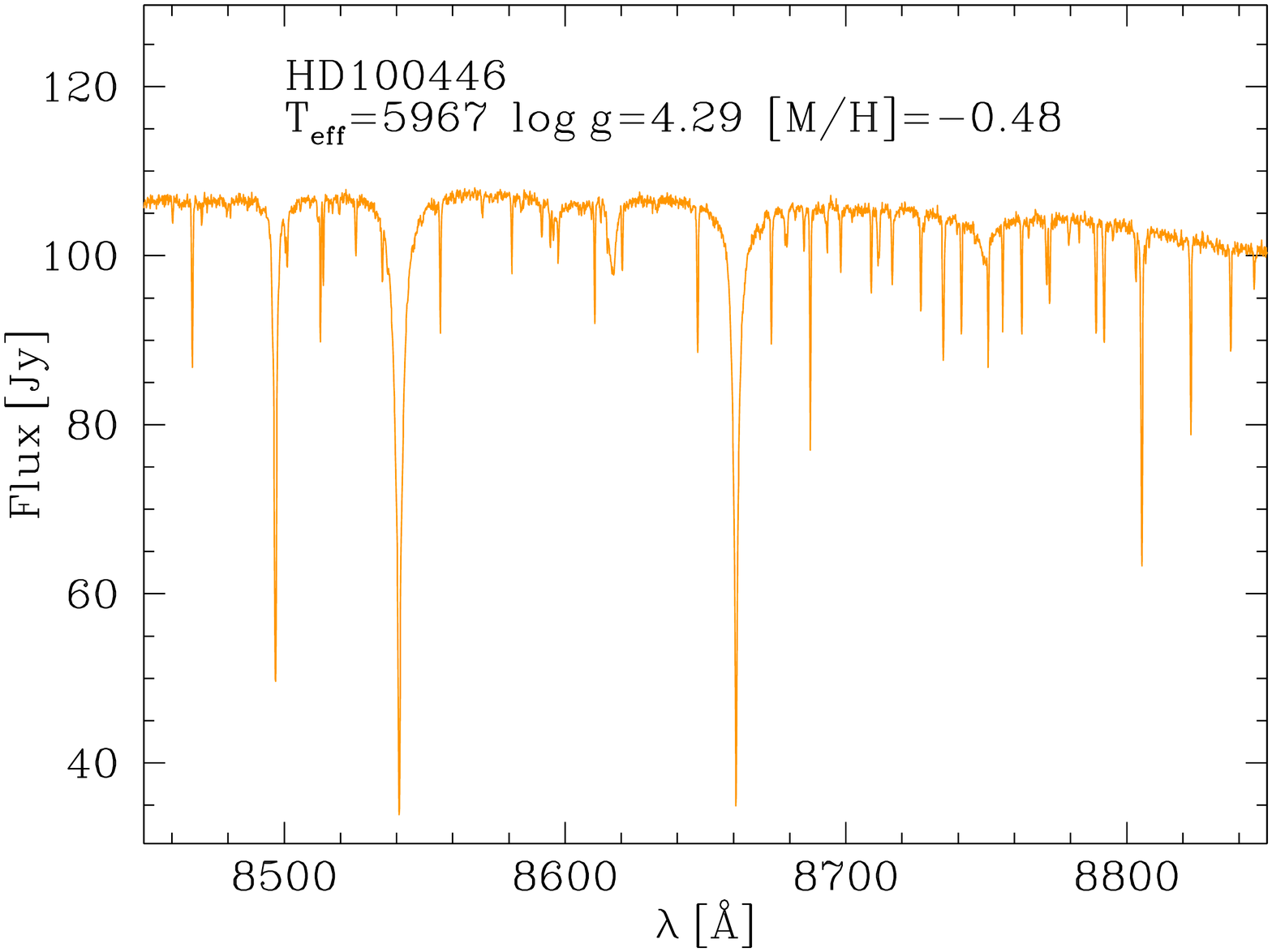}
\includegraphics[width=0.18\textwidth,angle=-90]{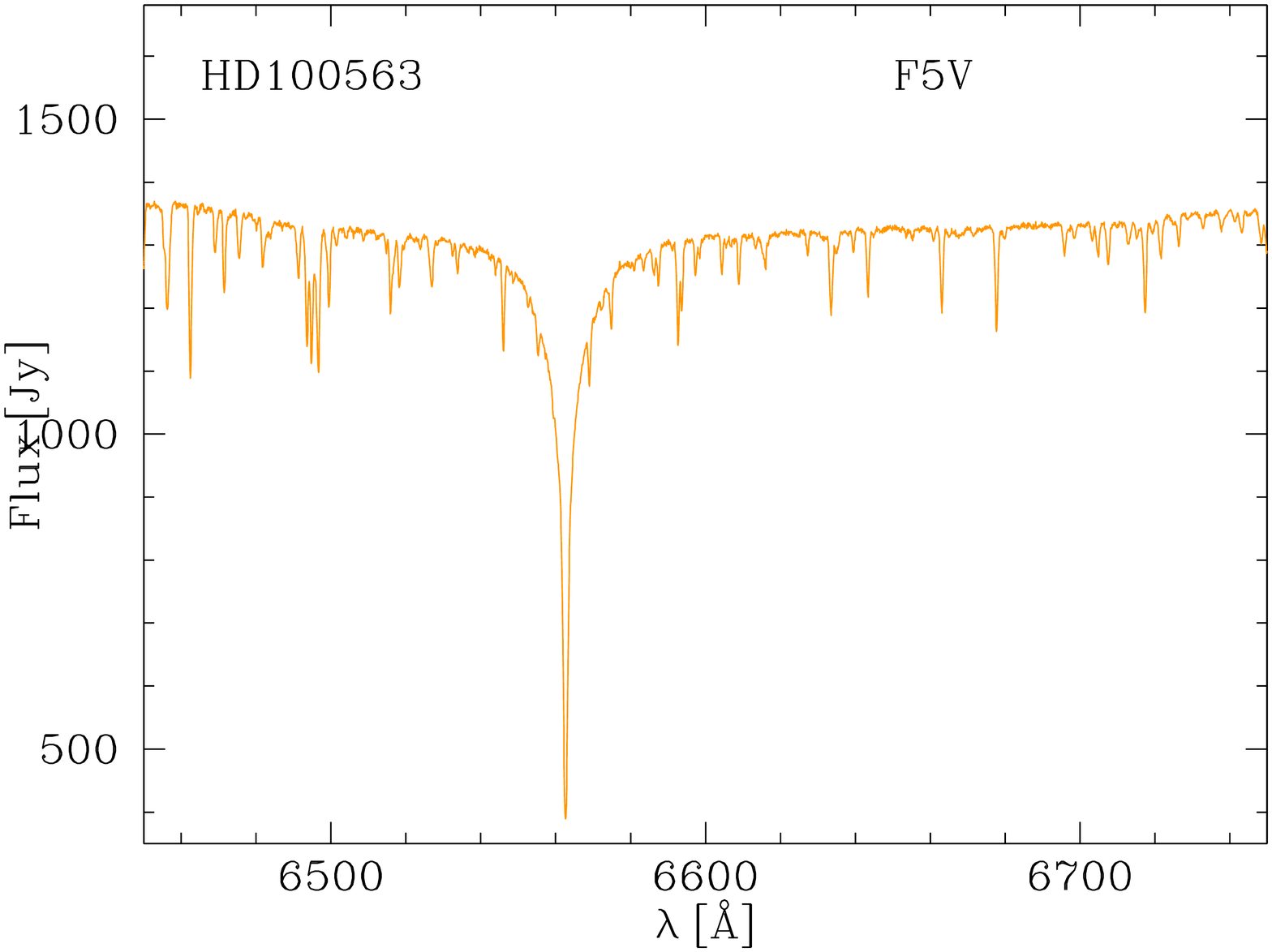}
\includegraphics[width=0.18\textwidth,angle=-90]{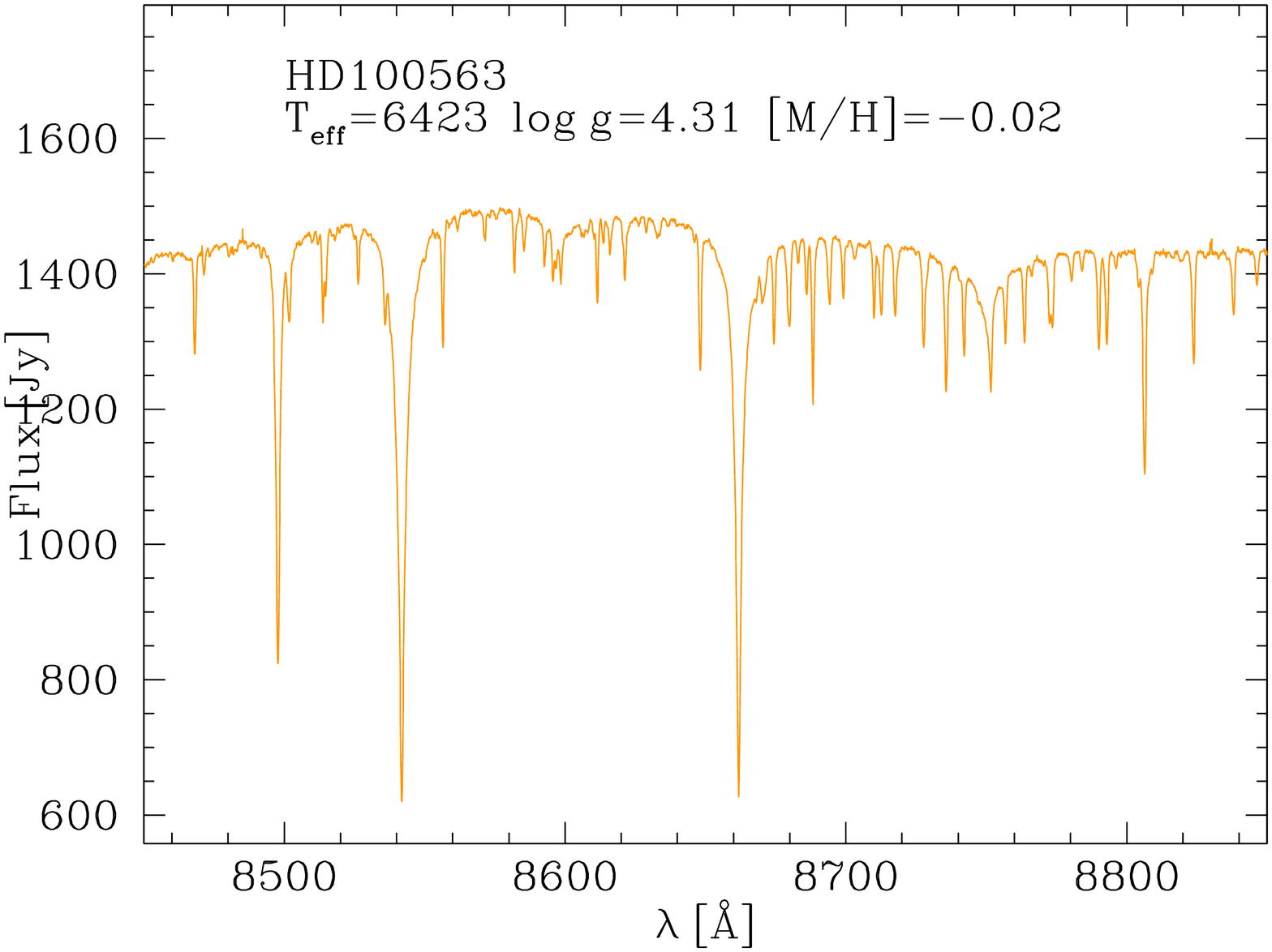}
\includegraphics[width=0.18\textwidth,angle=-90]{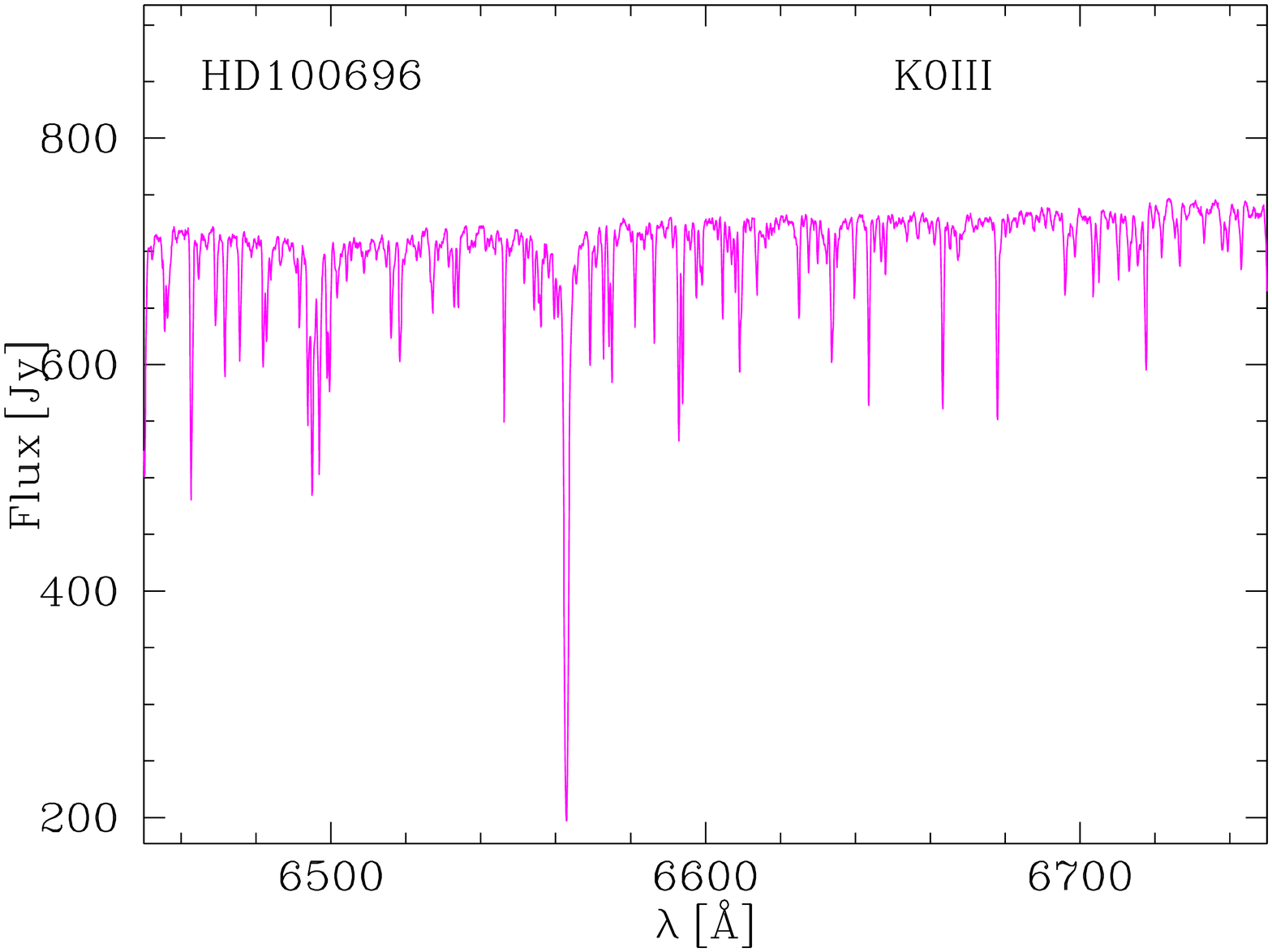}
\includegraphics[width=0.18\textwidth,angle=-90]{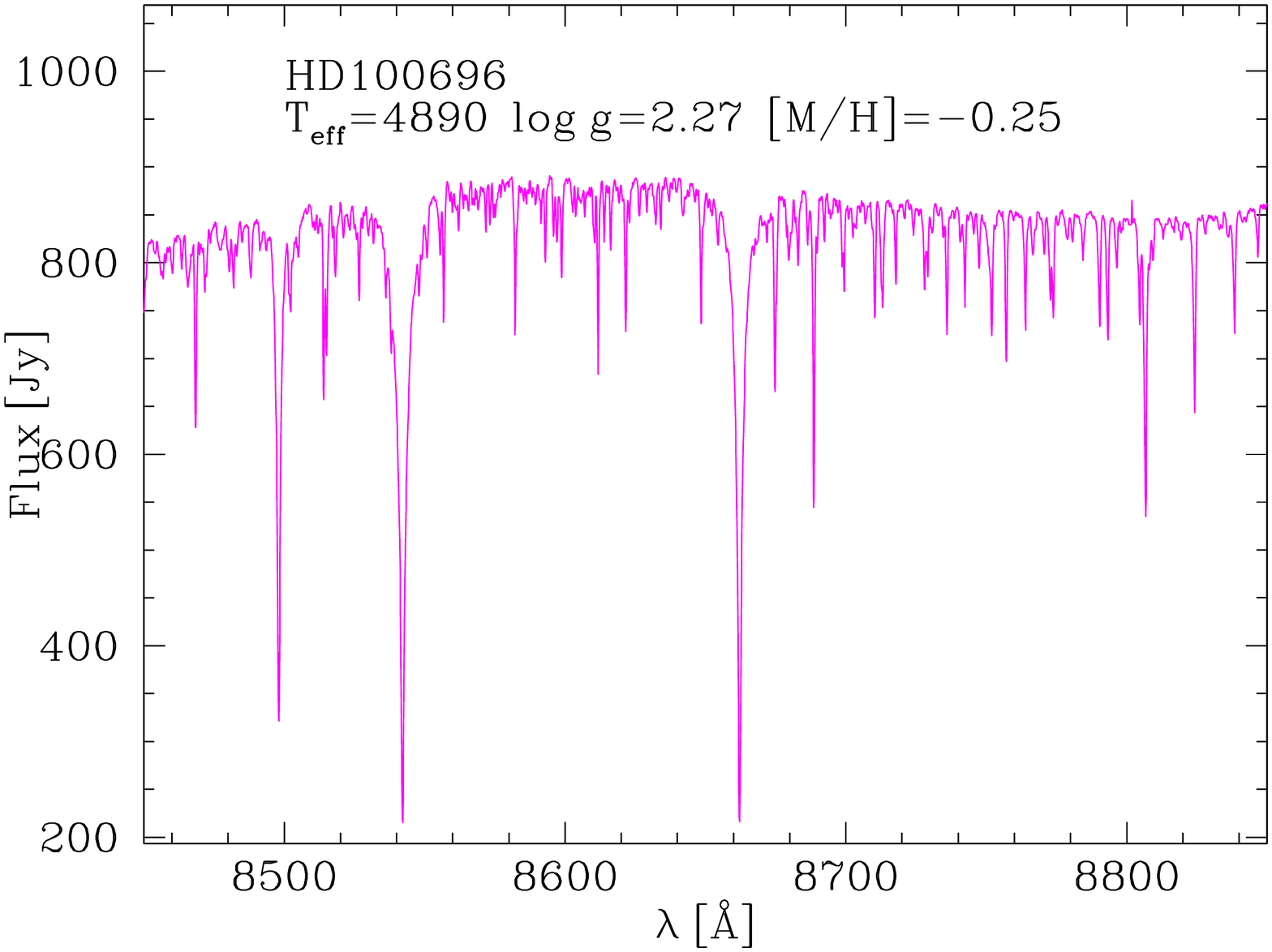}
\includegraphics[width=0.18\textwidth,angle=-90]{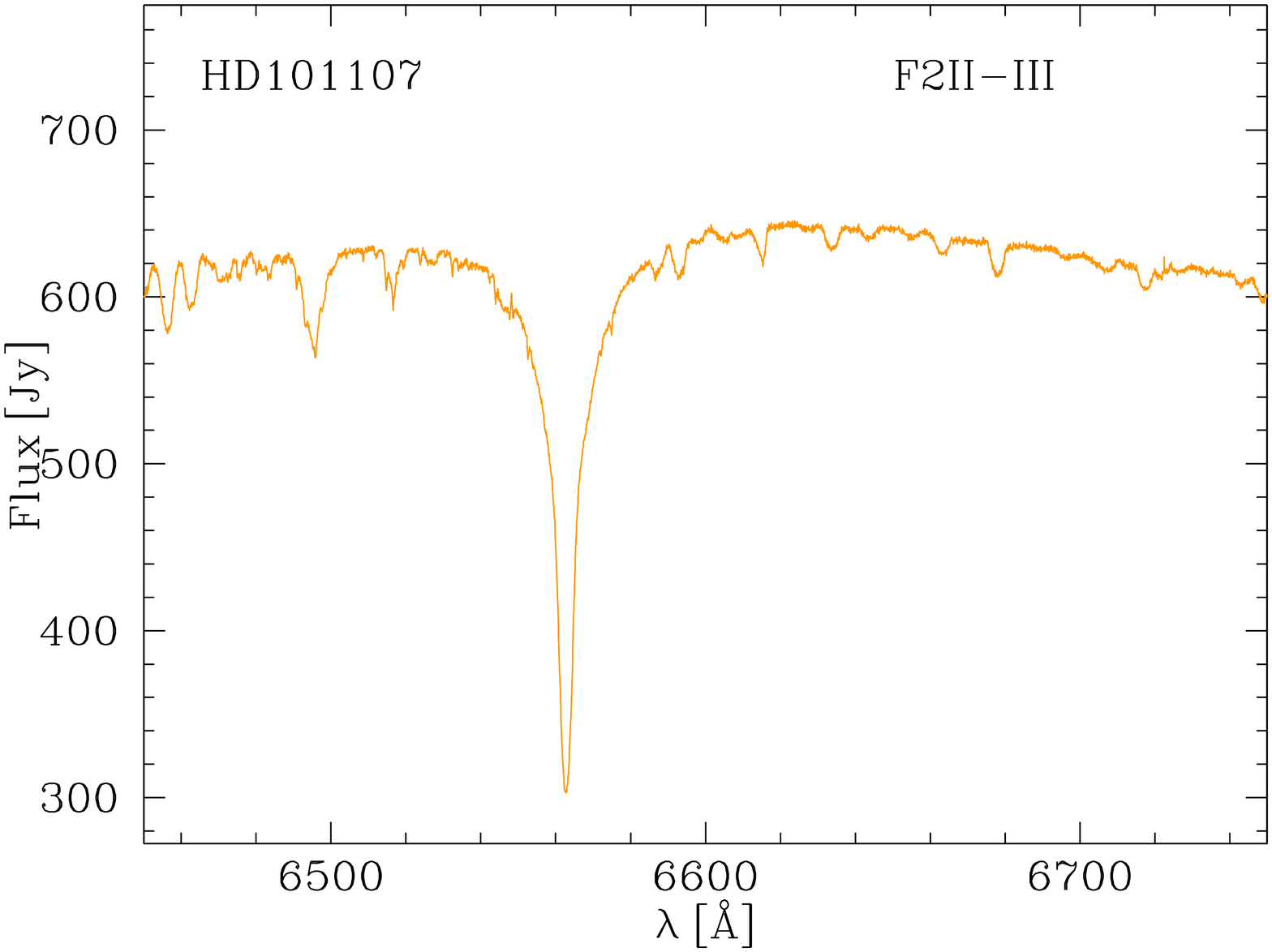}
\includegraphics[width=0.18\textwidth,angle=-90]{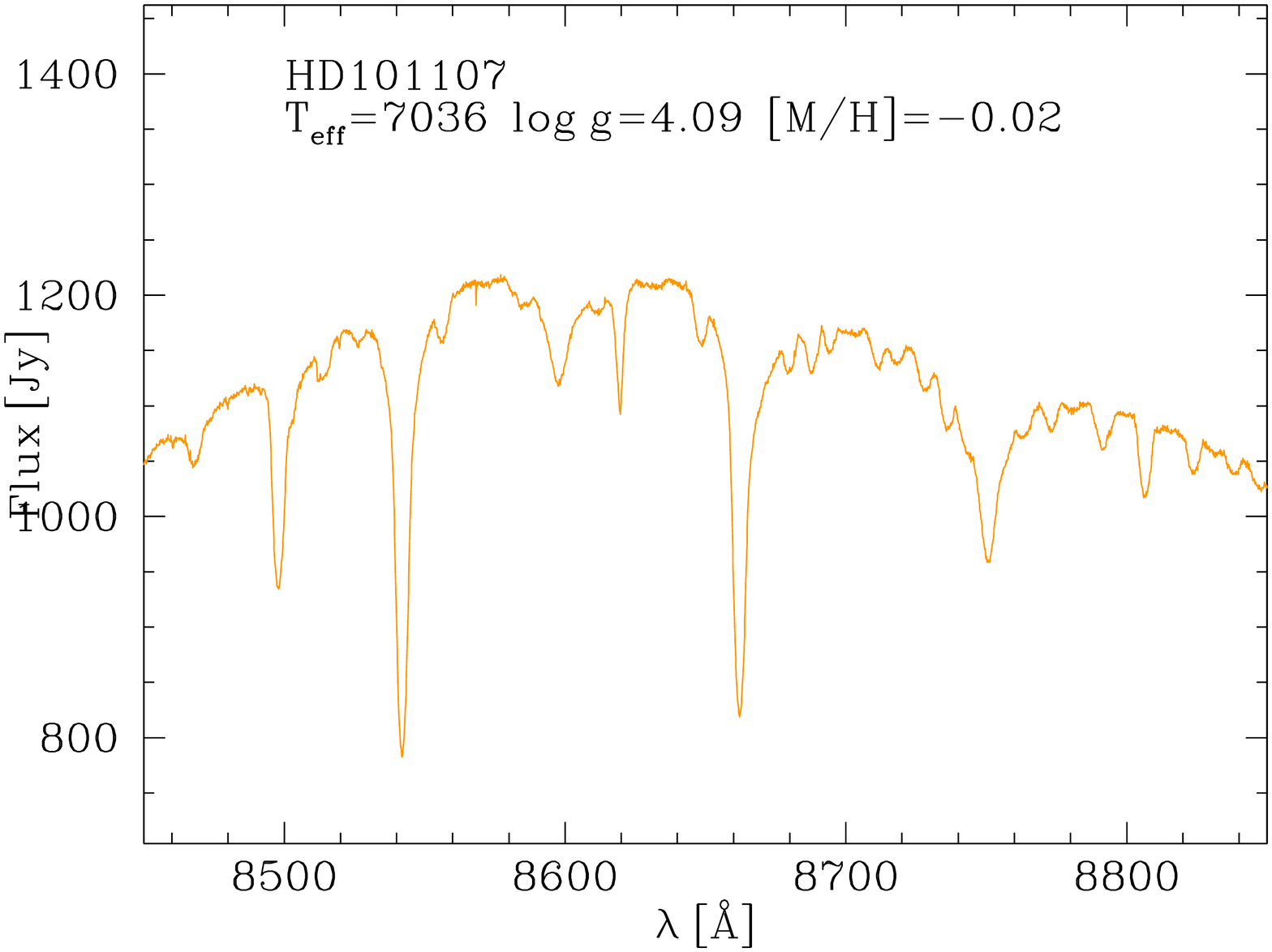}

\contcaption{19. Stars shown in this page are: HD095128, HD095241, HD096094, HD096436, HD097560, HD097855, HD097916, HD099028, HD099747, HD100030, HD100446, HD100563, HD100696 and HD101107.}
\end{figure*}

\begin{figure*}
\includegraphics[width=0.18\textwidth,angle=-90]{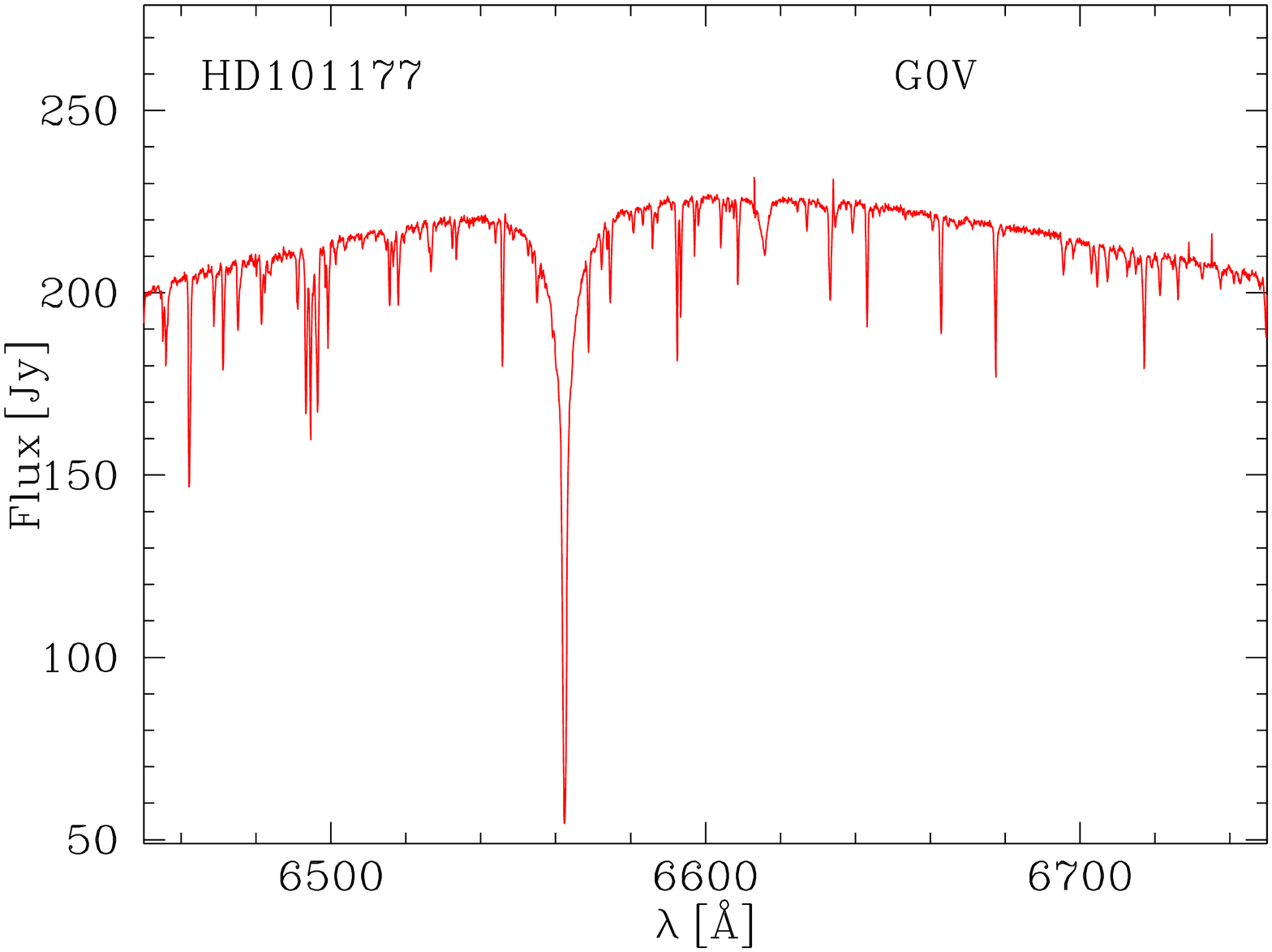}
\includegraphics[width=0.18\textwidth,angle=-90]{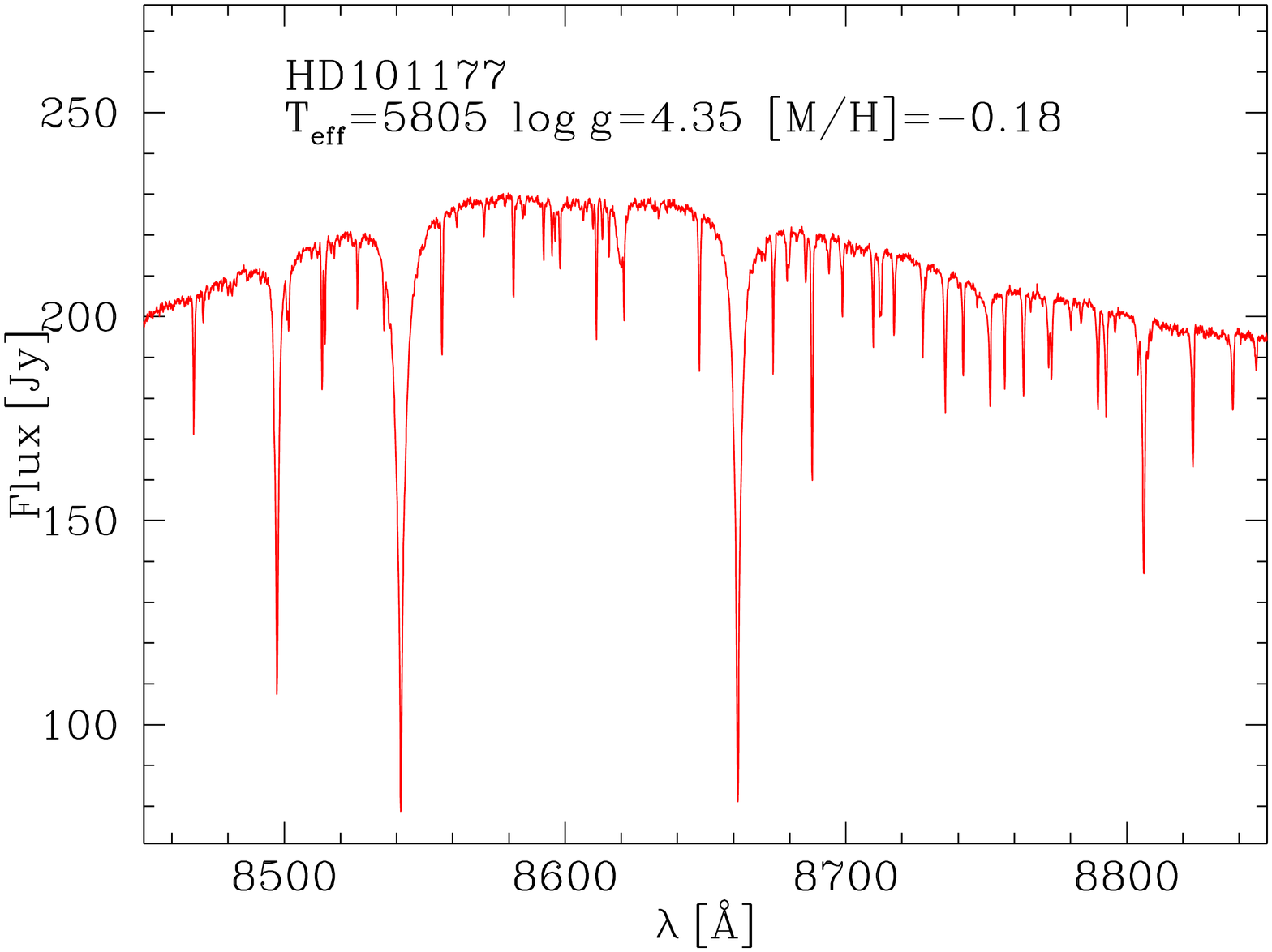}
\includegraphics[width=0.18\textwidth,angle=-90]{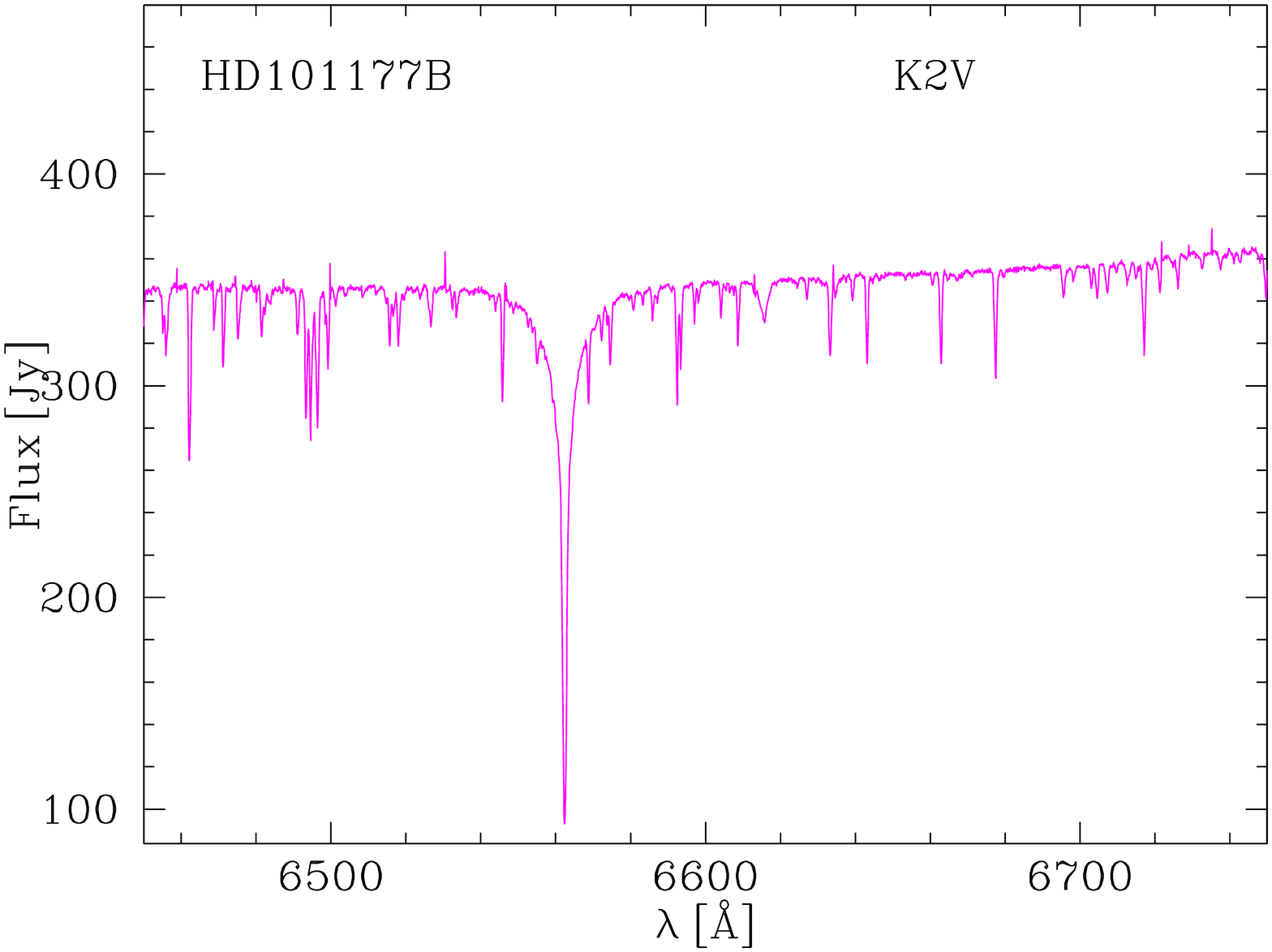}
\includegraphics[width=0.18\textwidth,angle=-90]{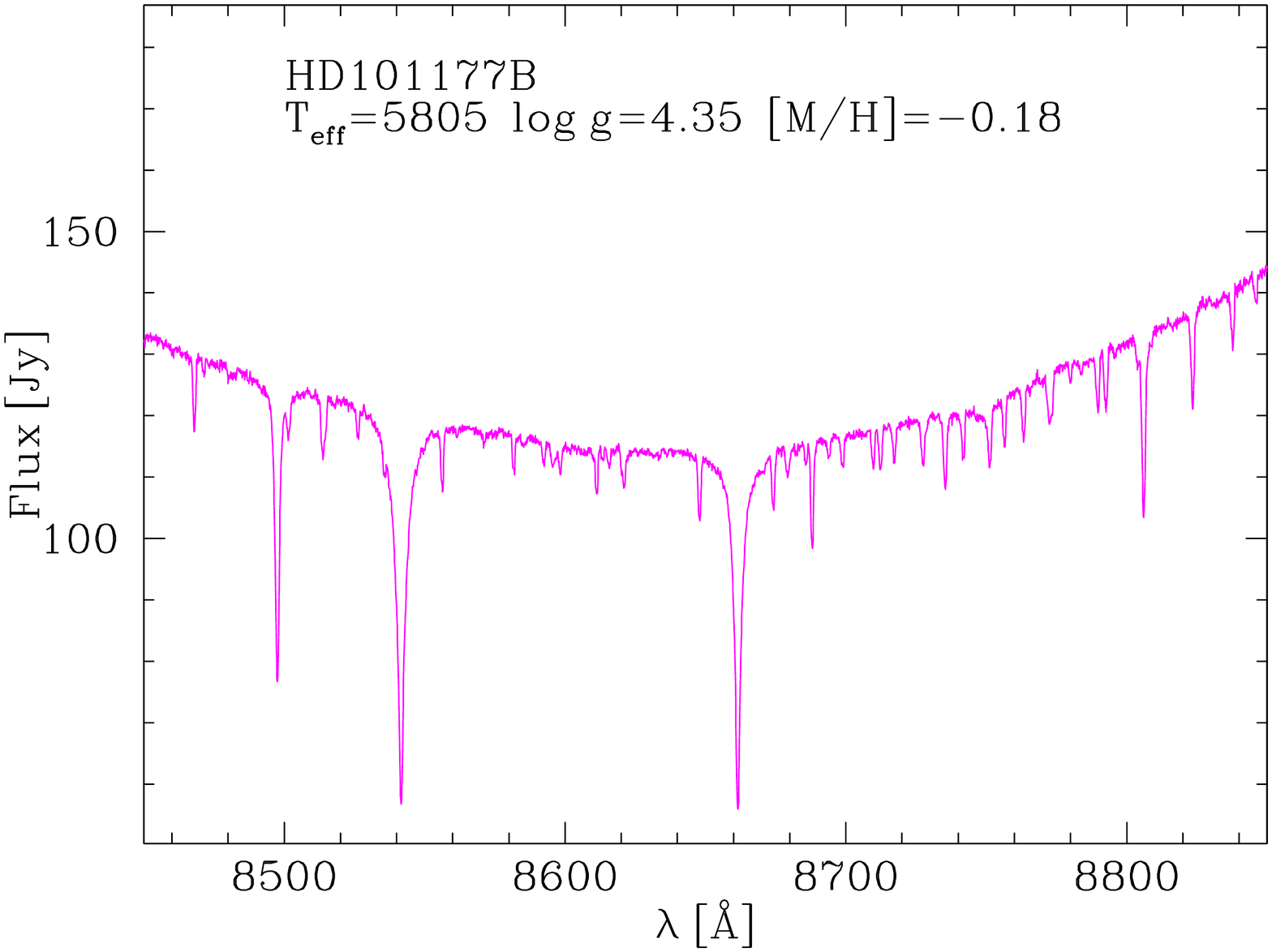}
\includegraphics[width=0.18\textwidth,angle=-90]{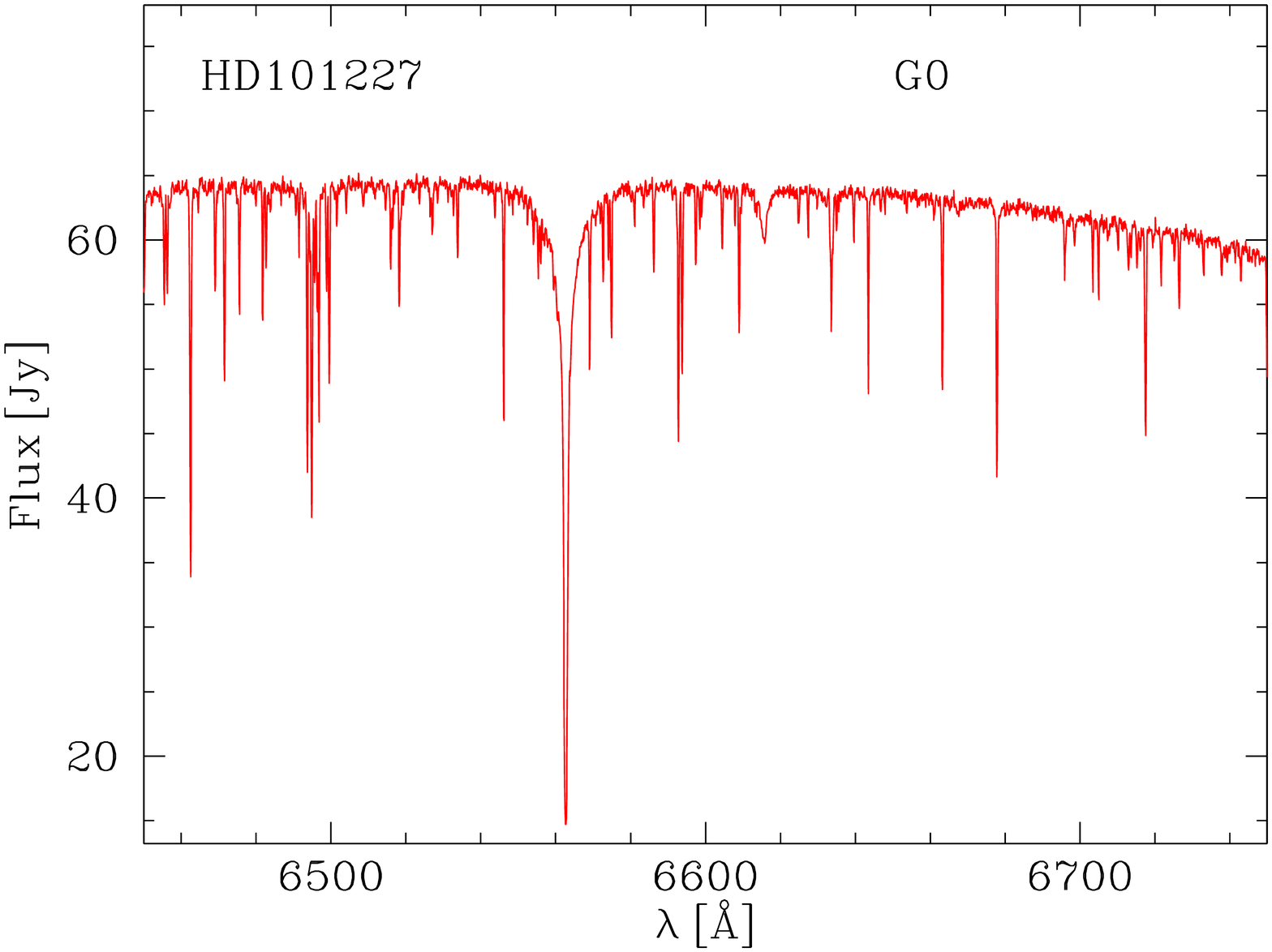}
\includegraphics[width=0.18\textwidth,angle=-90]{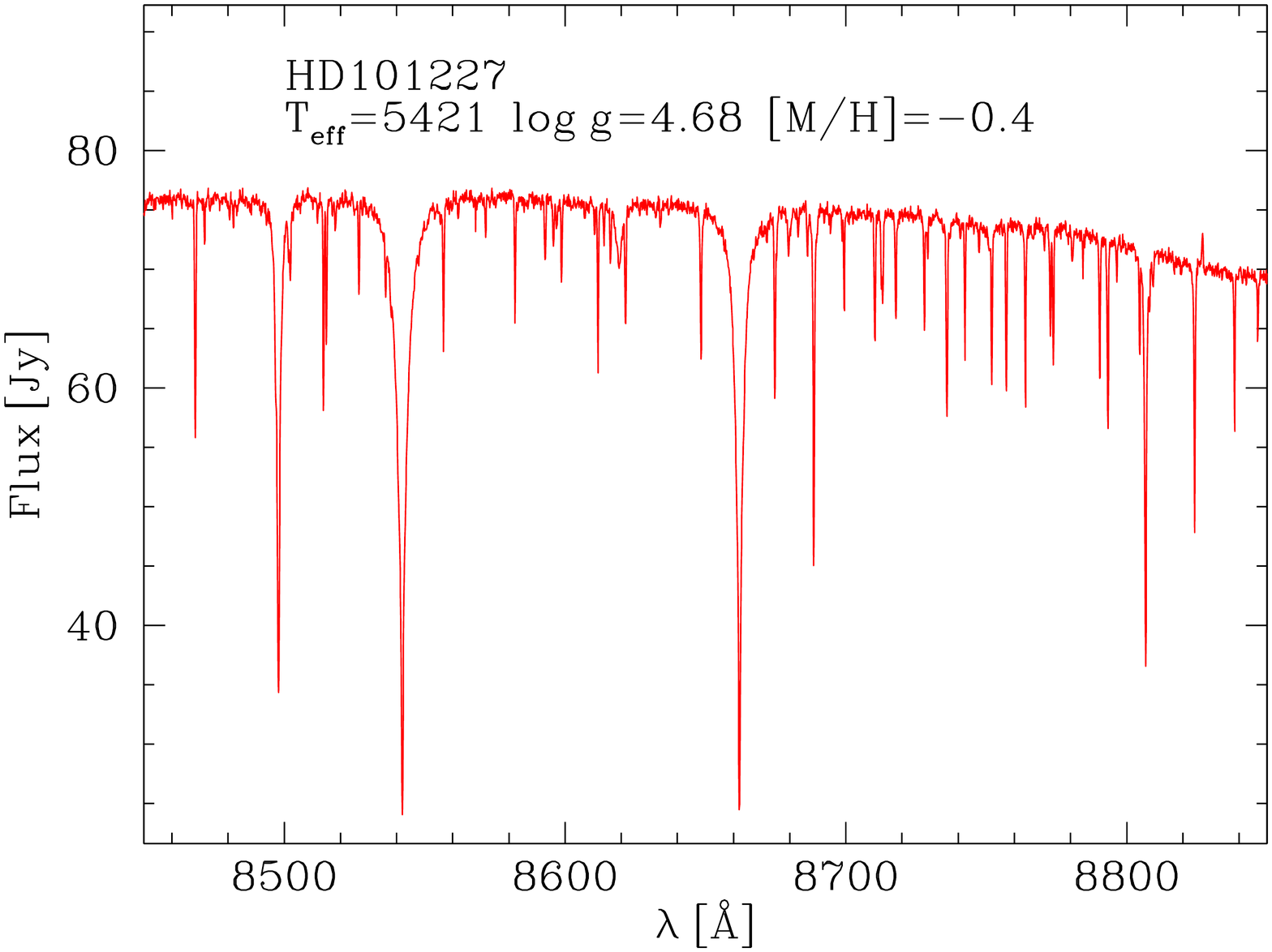}
\includegraphics[width=0.18\textwidth,angle=-90]{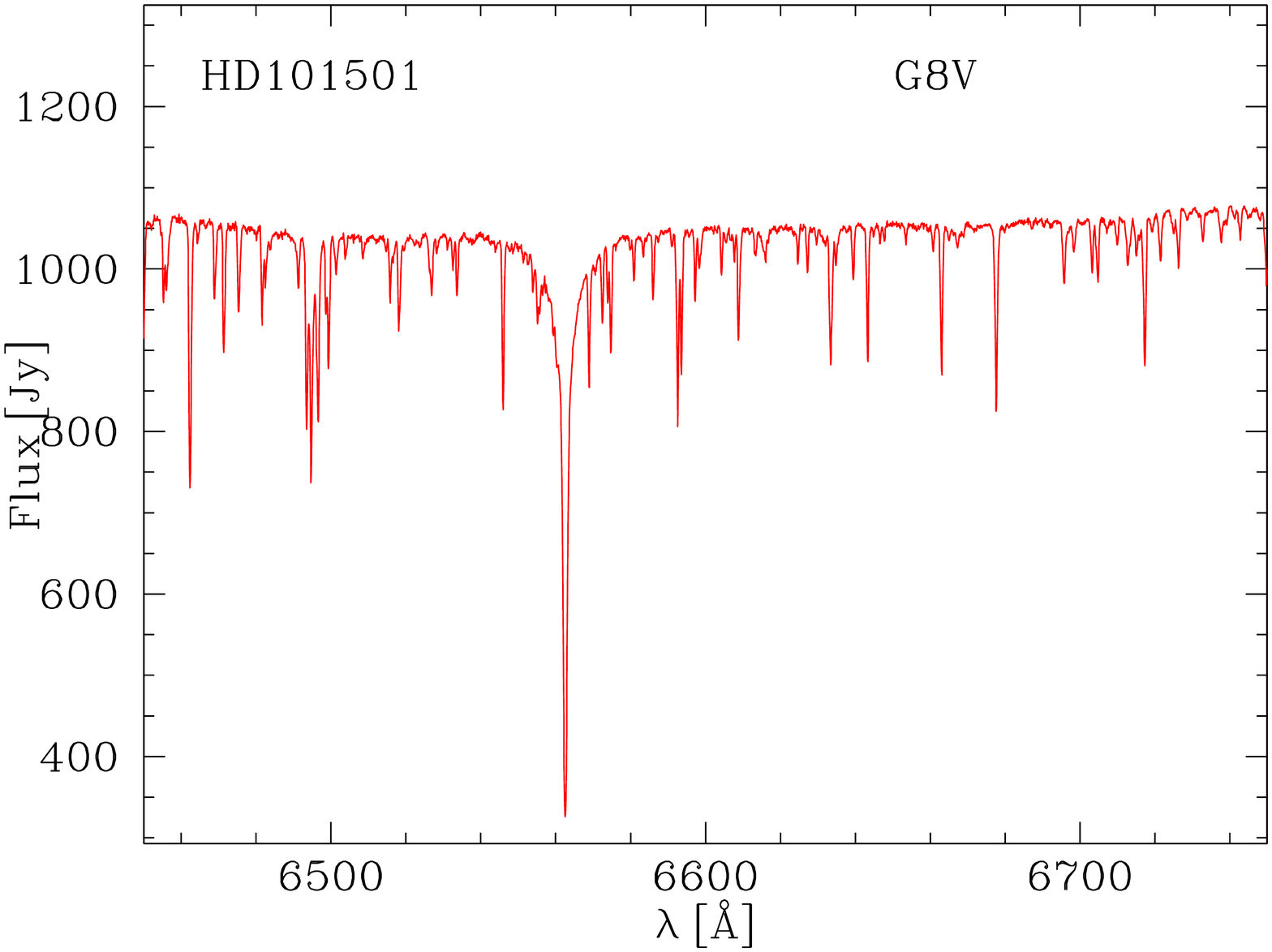}
\includegraphics[width=0.18\textwidth,angle=-90]{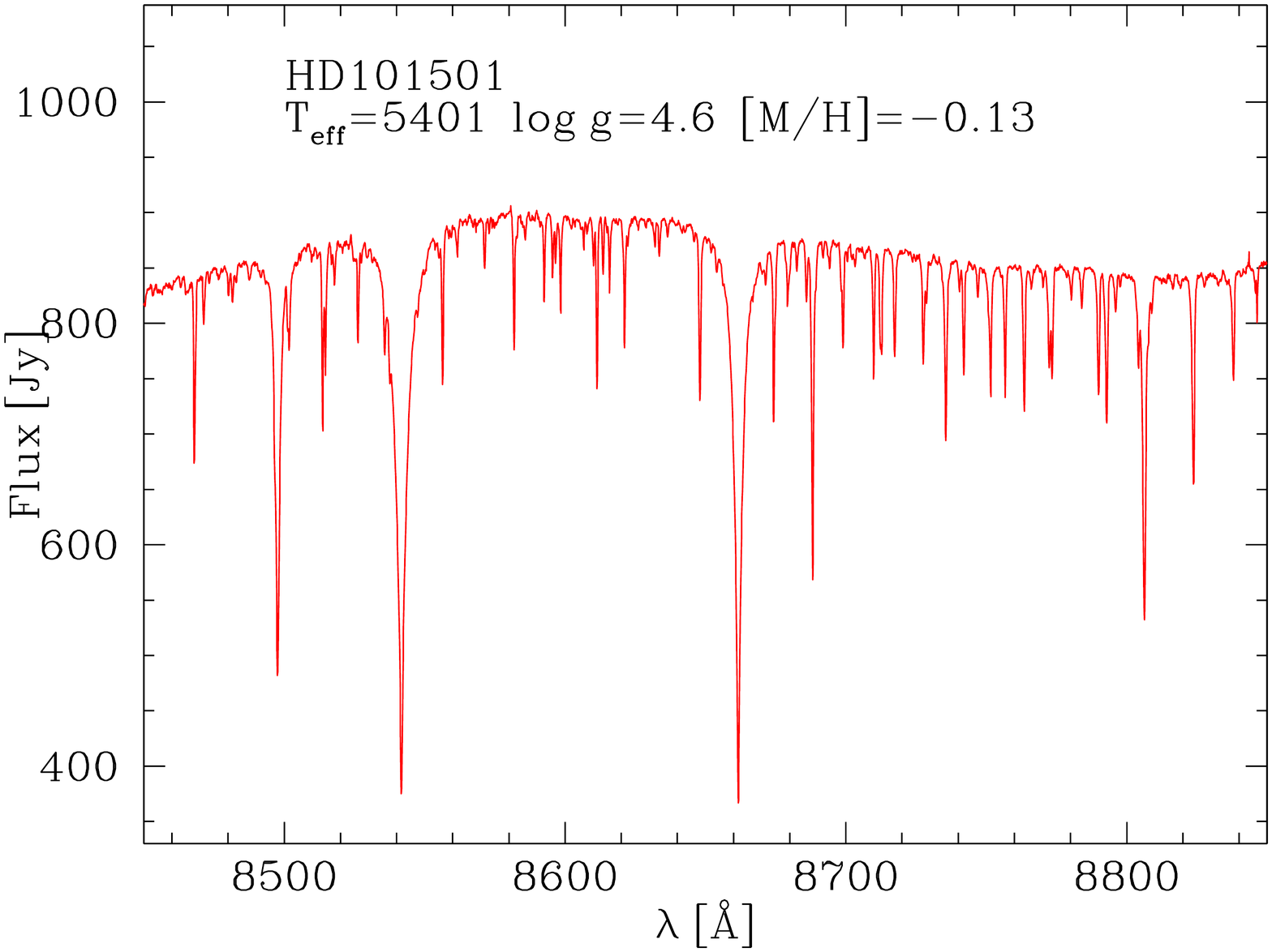}
\includegraphics[width=0.18\textwidth,angle=-90]{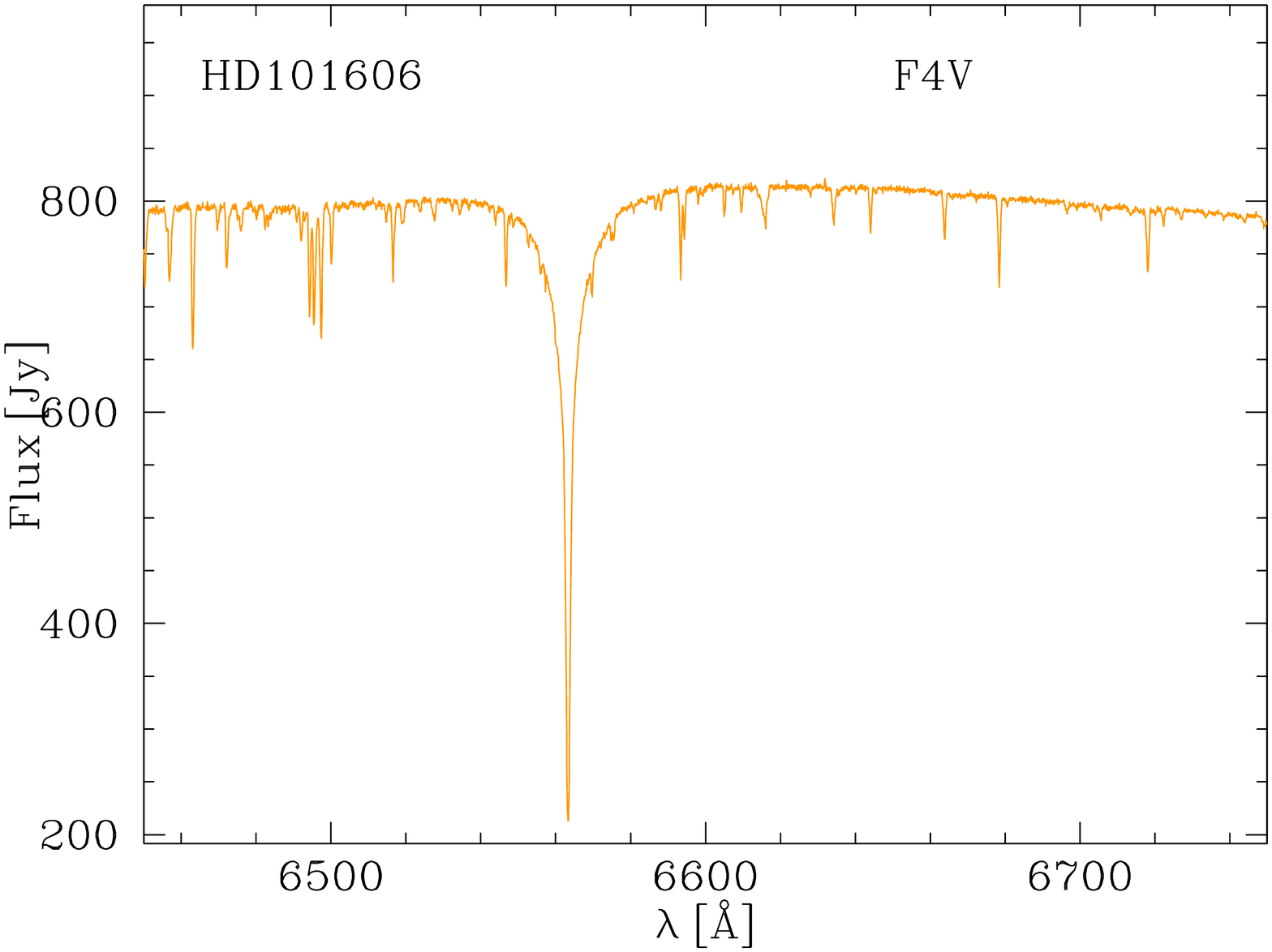}
\includegraphics[width=0.18\textwidth,angle=-90]{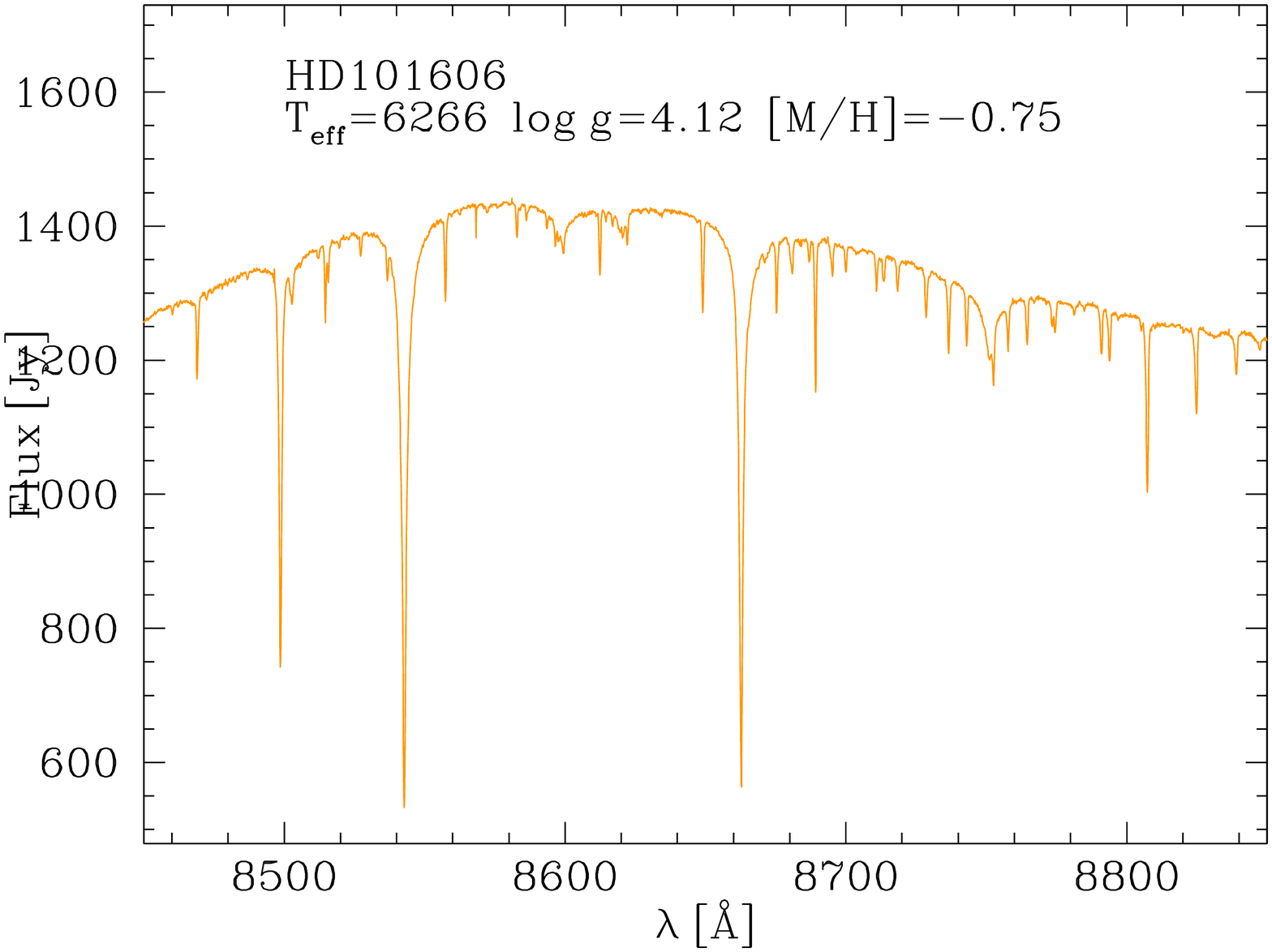}
\includegraphics[width=0.18\textwidth,angle=-90]{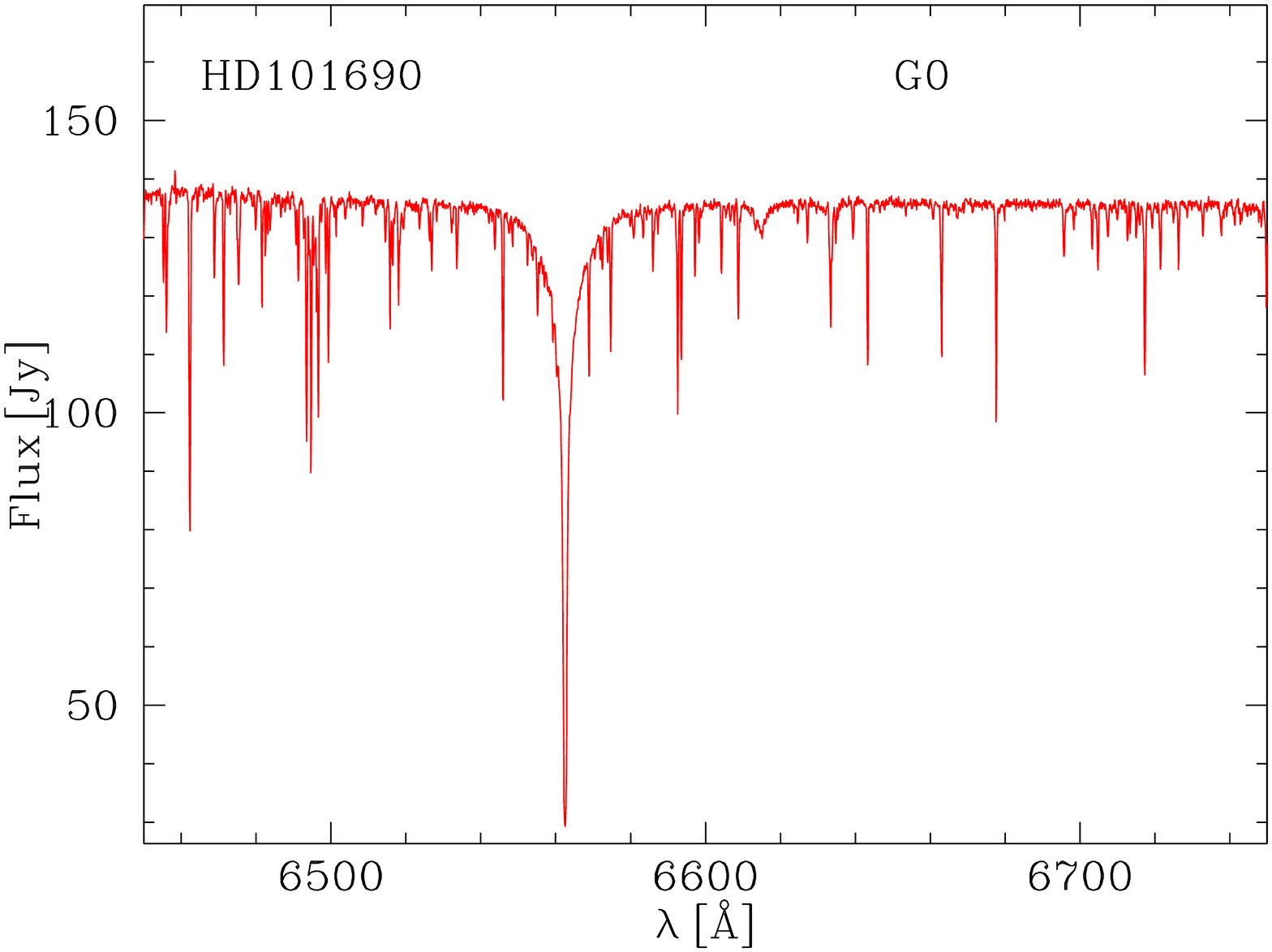}
\includegraphics[width=0.18\textwidth,angle=-90]{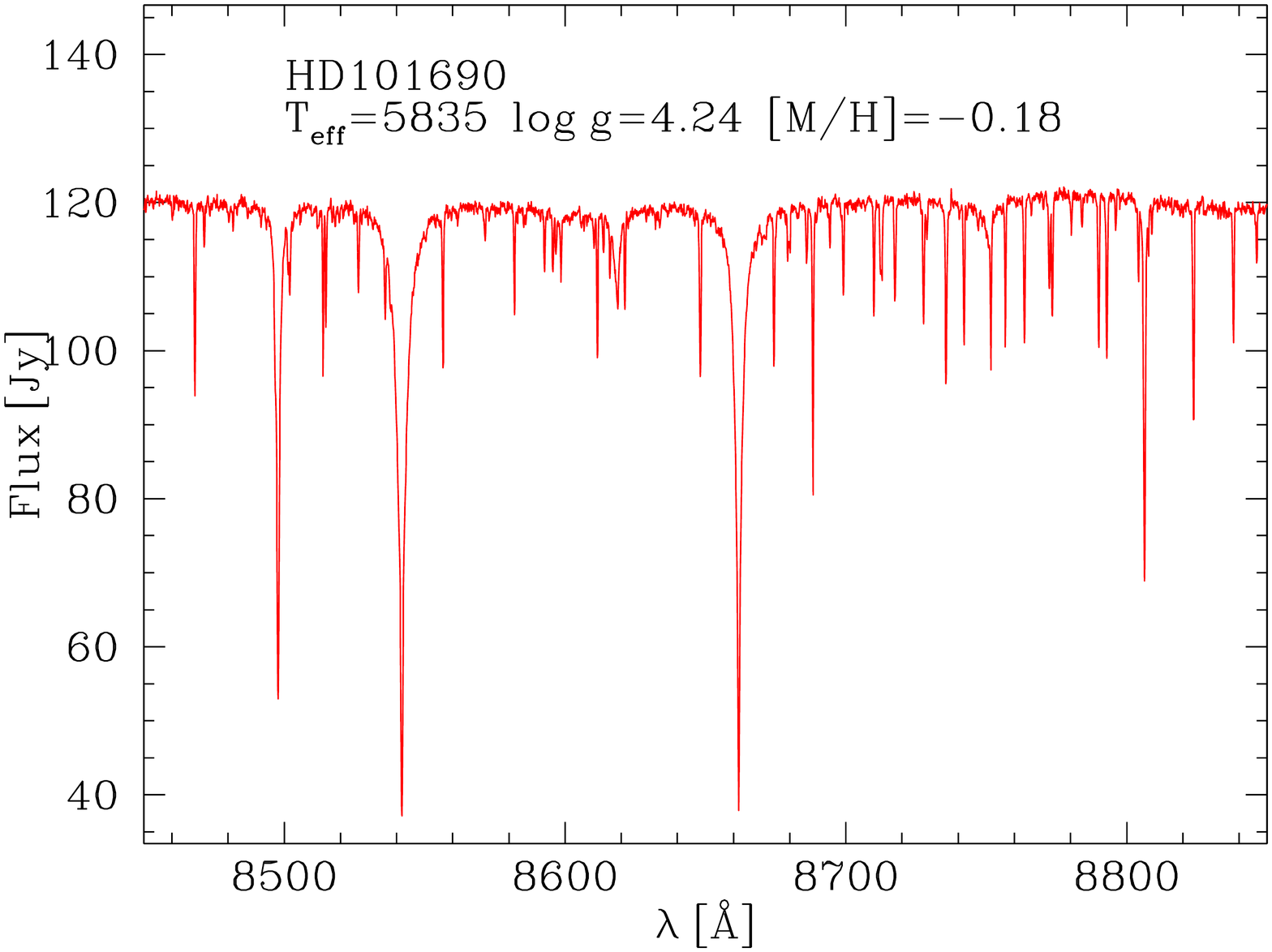}
\includegraphics[width=0.18\textwidth,angle=-90]{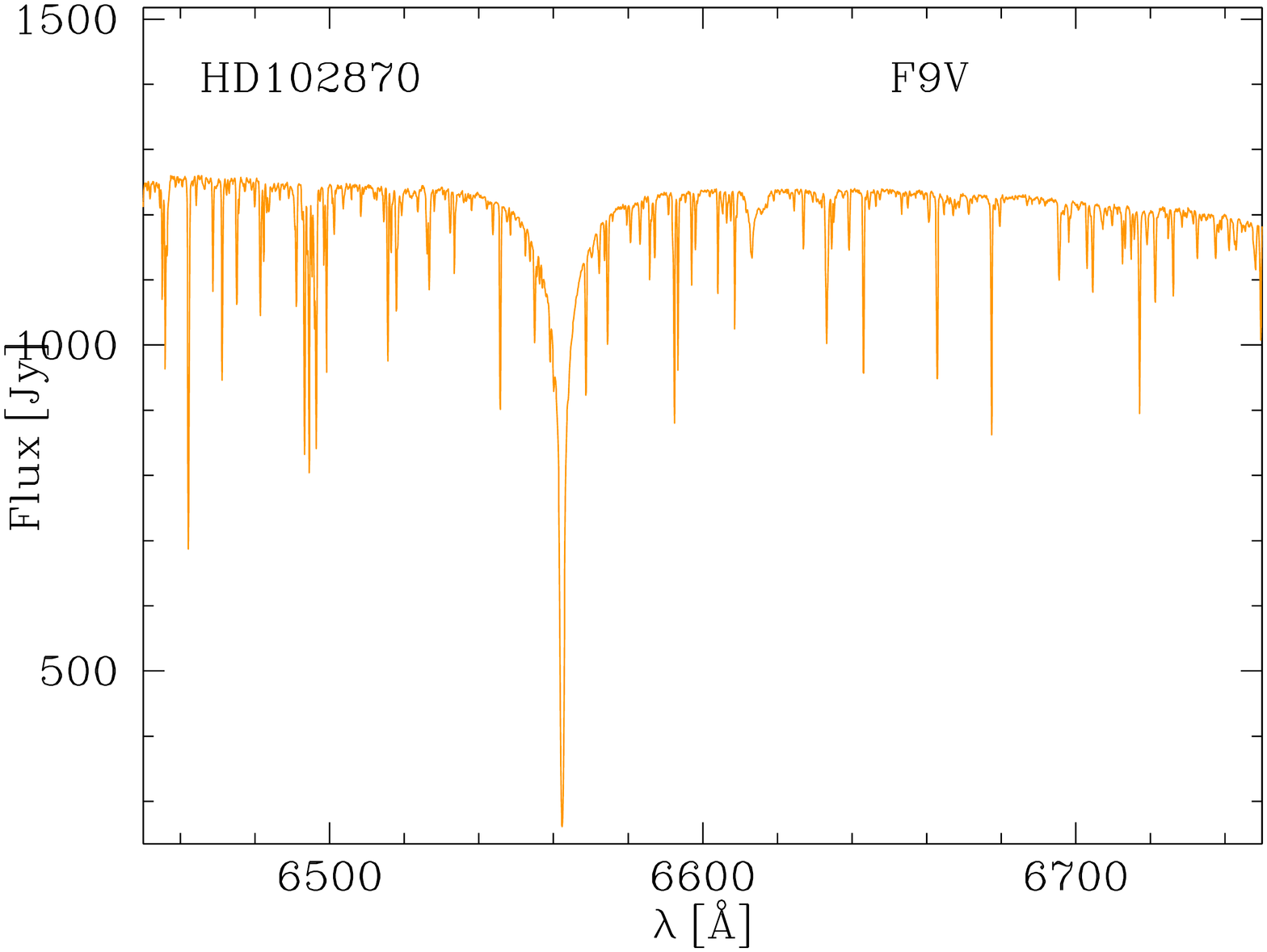}
\includegraphics[width=0.18\textwidth,angle=-90]{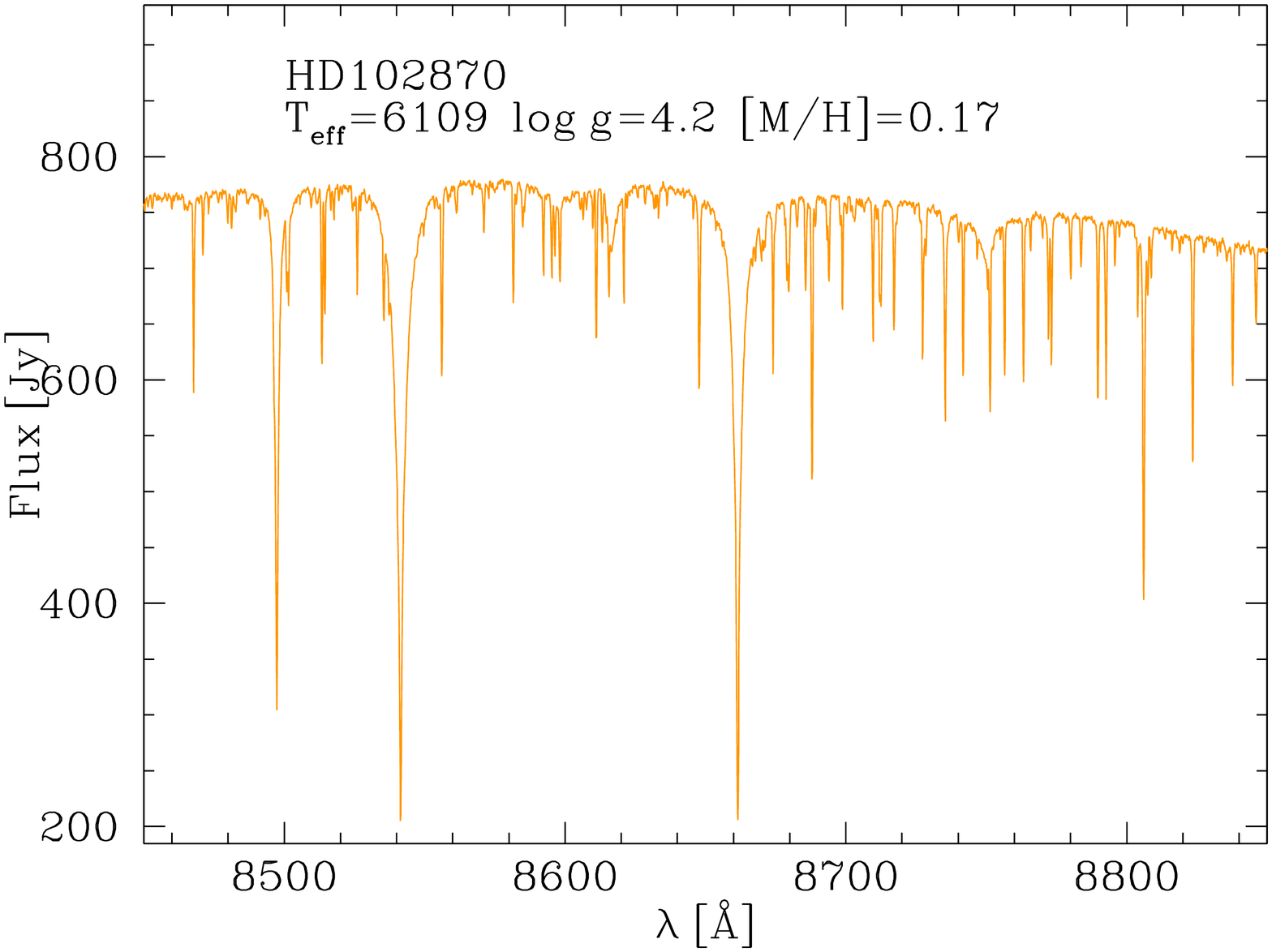}
\includegraphics[width=0.18\textwidth,angle=-90]{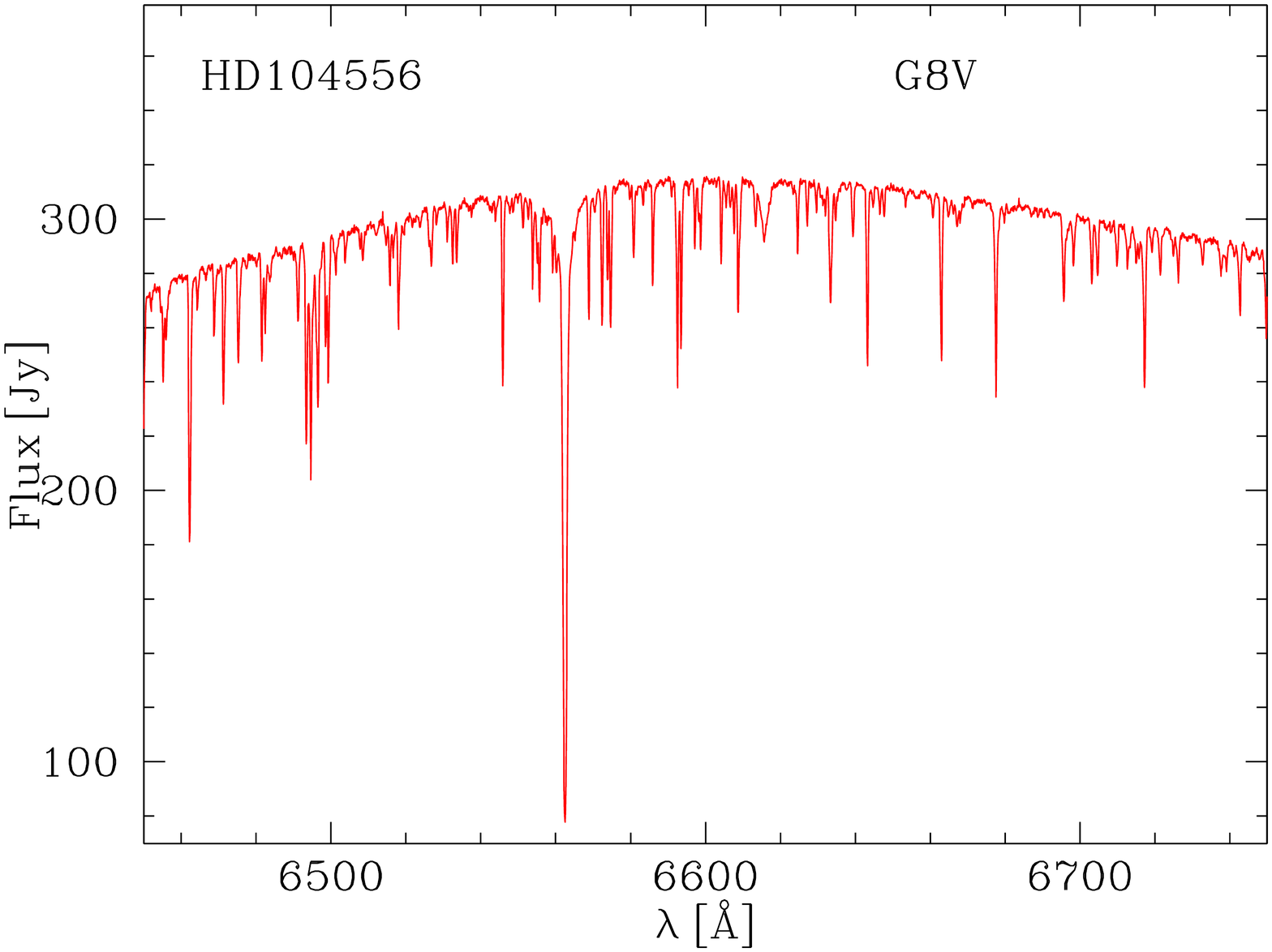}
\includegraphics[width=0.18\textwidth,angle=-90]{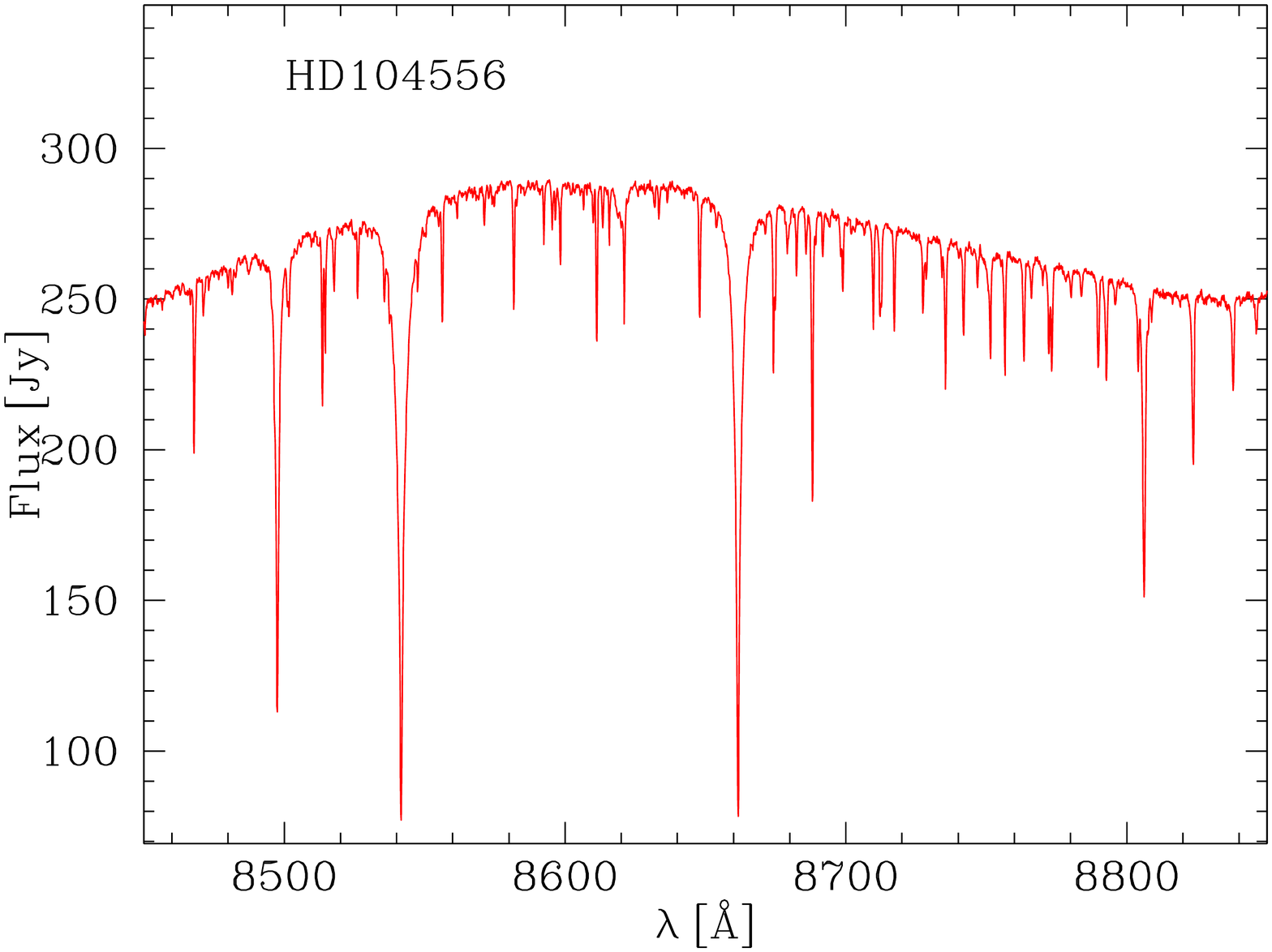}
\includegraphics[width=0.18\textwidth,angle=-90]{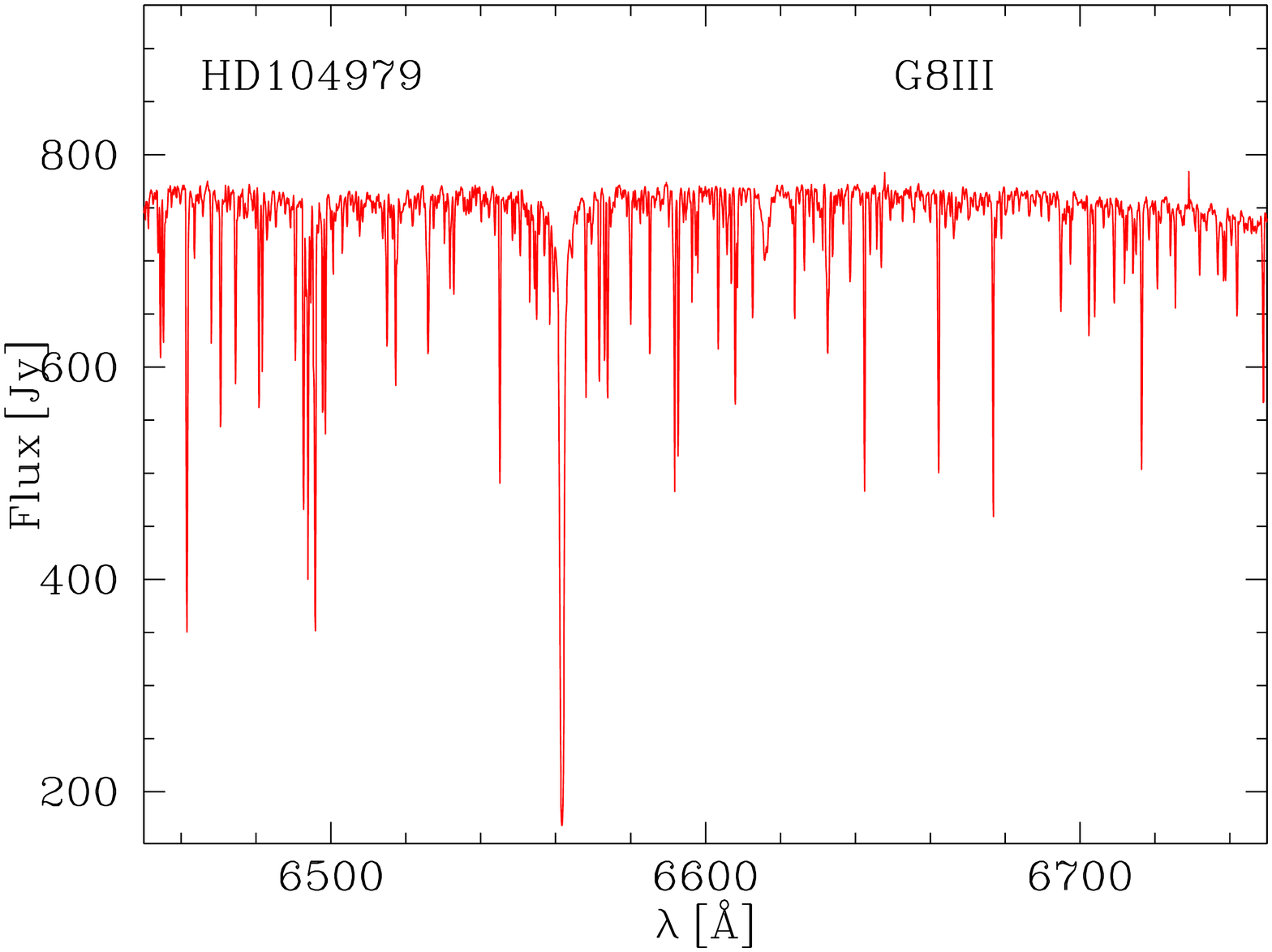}
\includegraphics[width=0.18\textwidth,angle=-90]{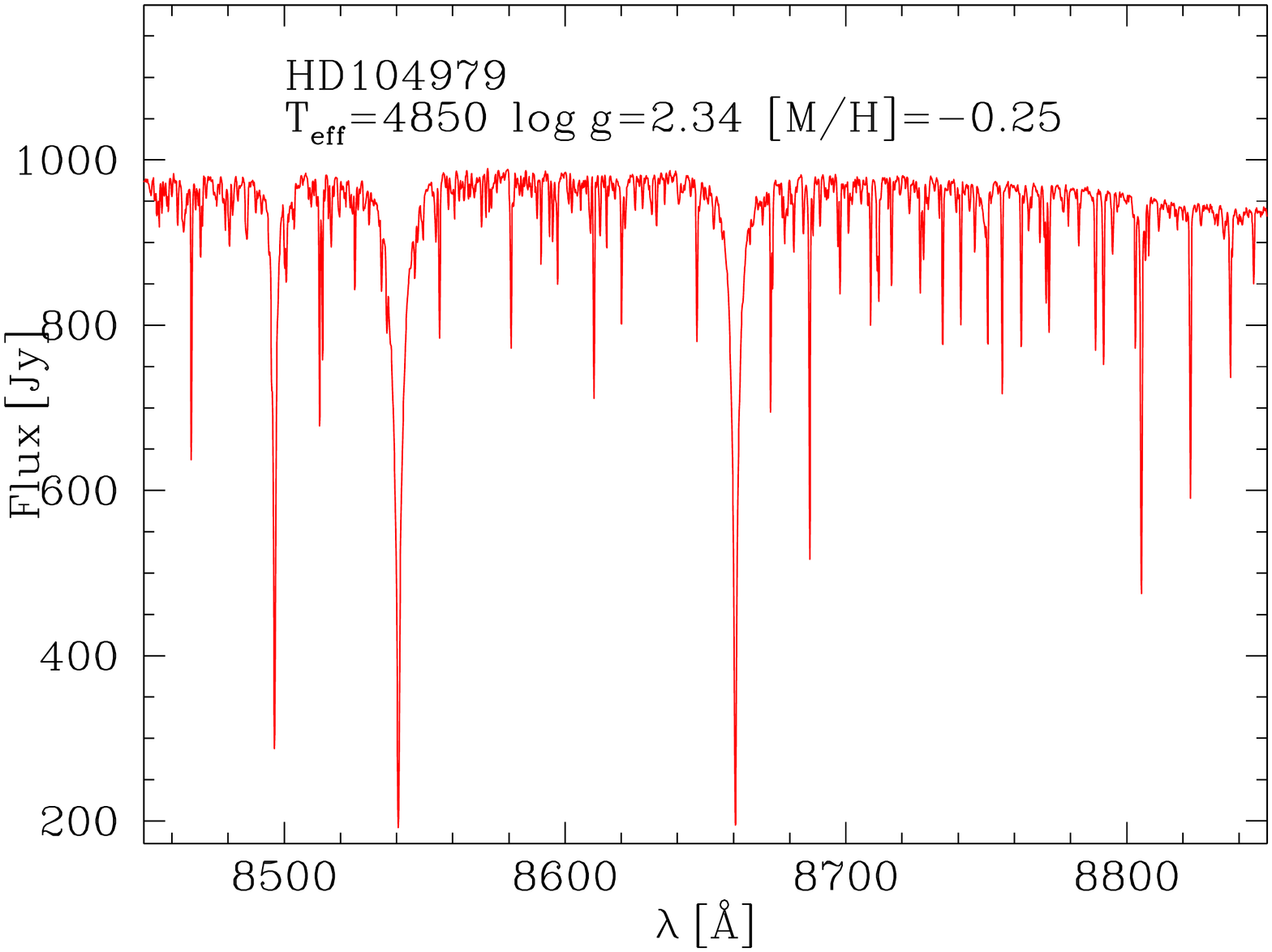}
\includegraphics[width=0.18\textwidth,angle=-90]{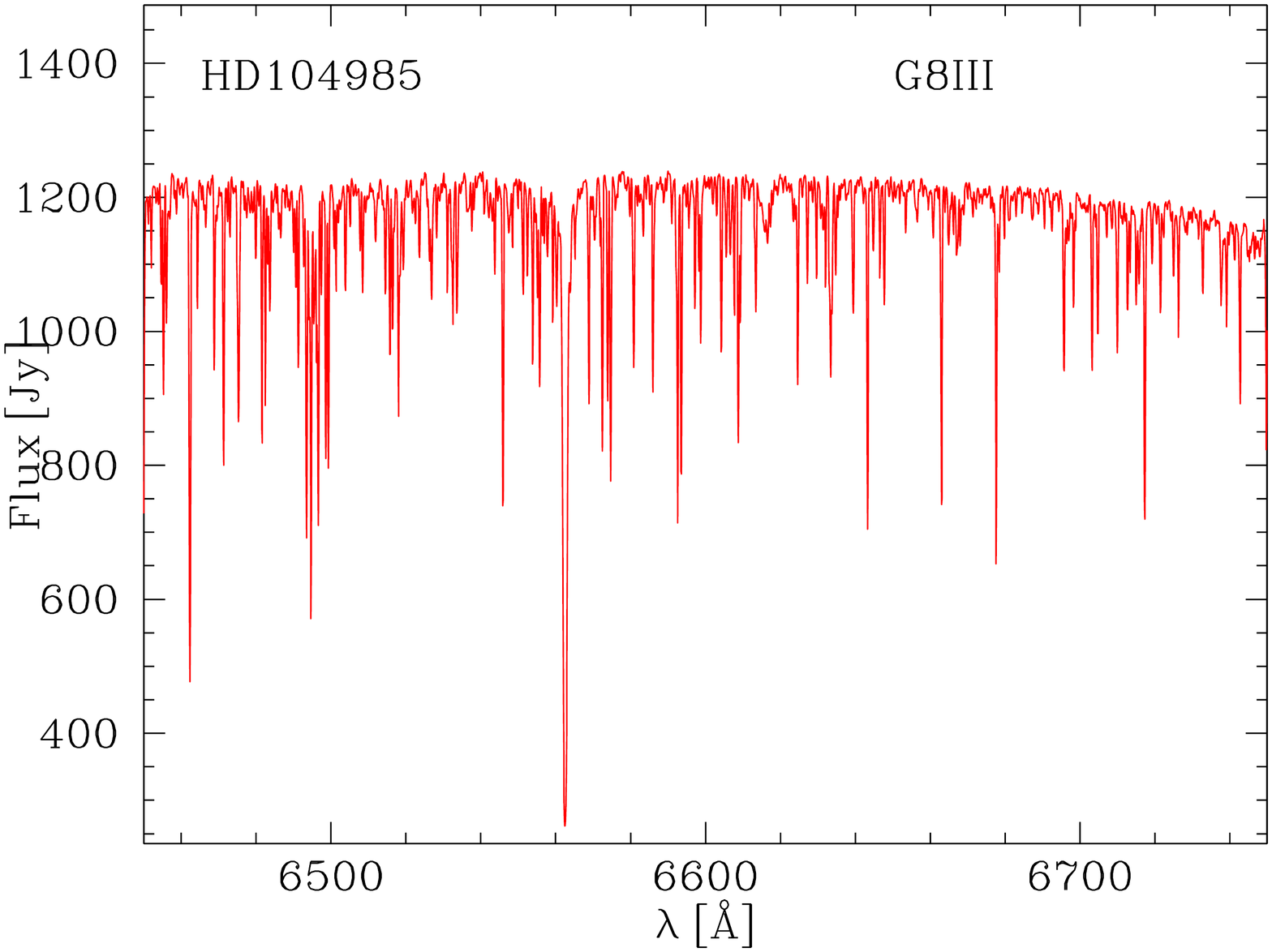}
\includegraphics[width=0.18\textwidth,angle=-90]{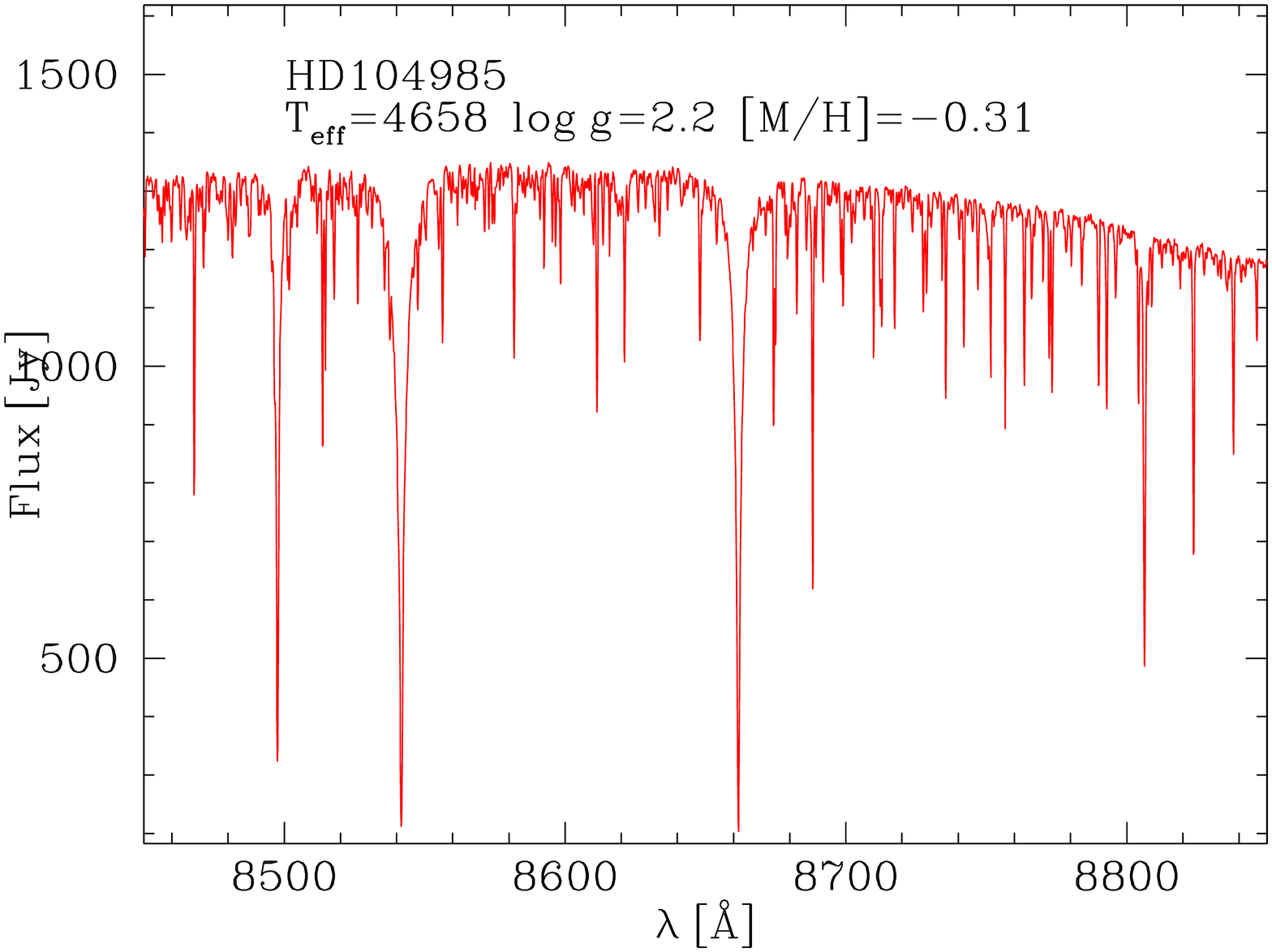}
\includegraphics[width=0.18\textwidth,angle=-90]{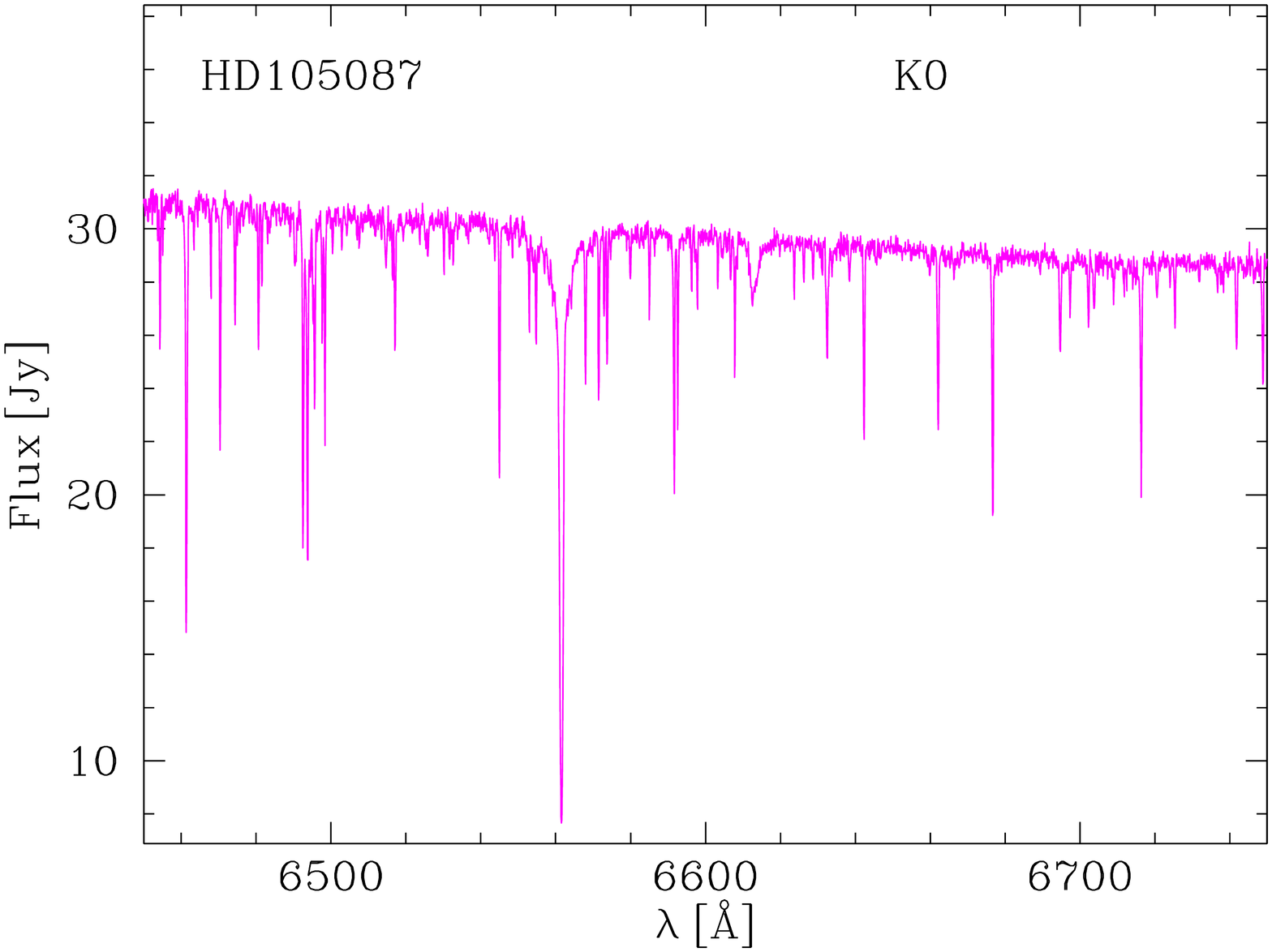}
\includegraphics[width=0.18\textwidth,angle=-90]{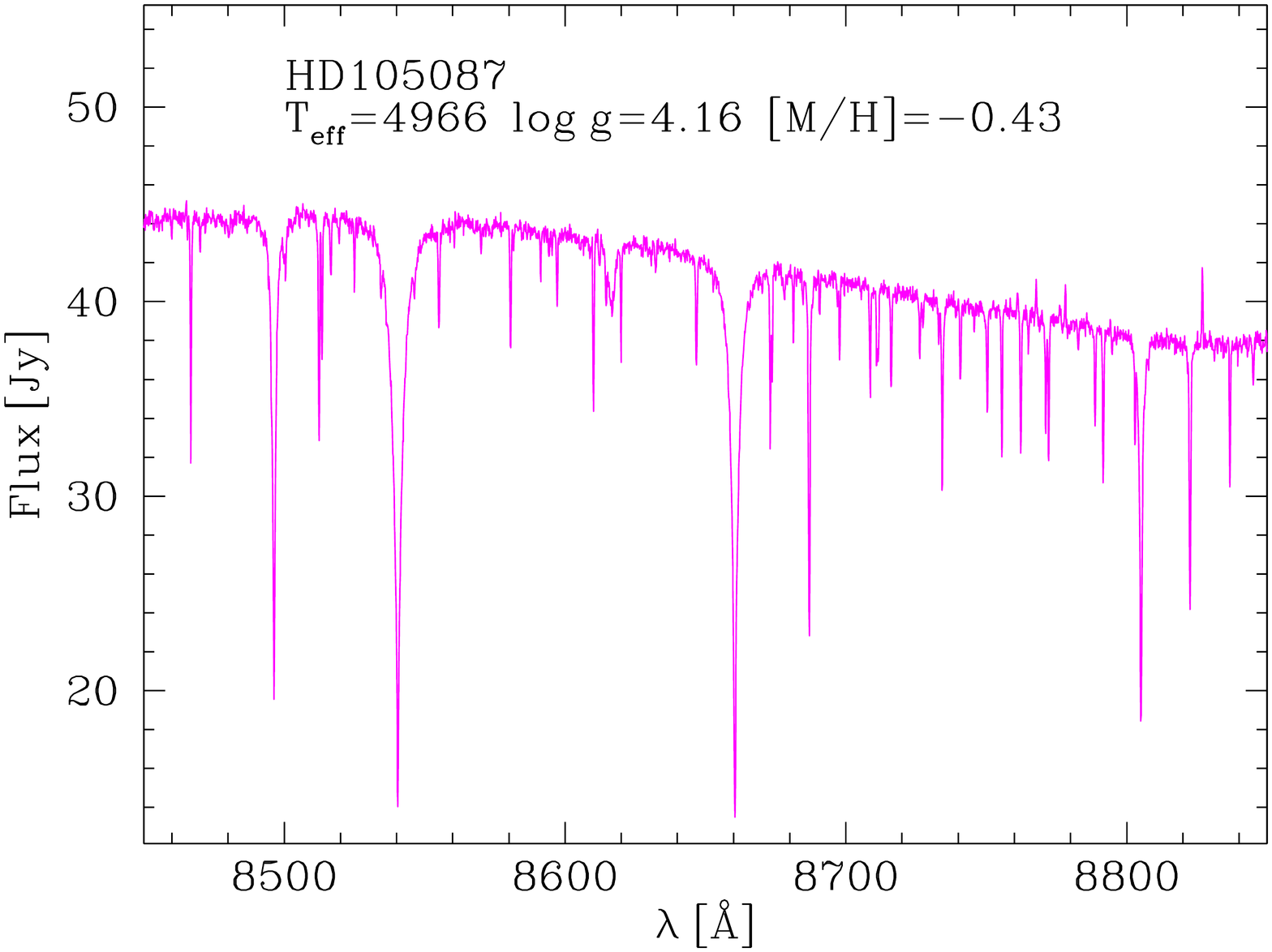}
\includegraphics[width=0.18\textwidth,angle=-90]{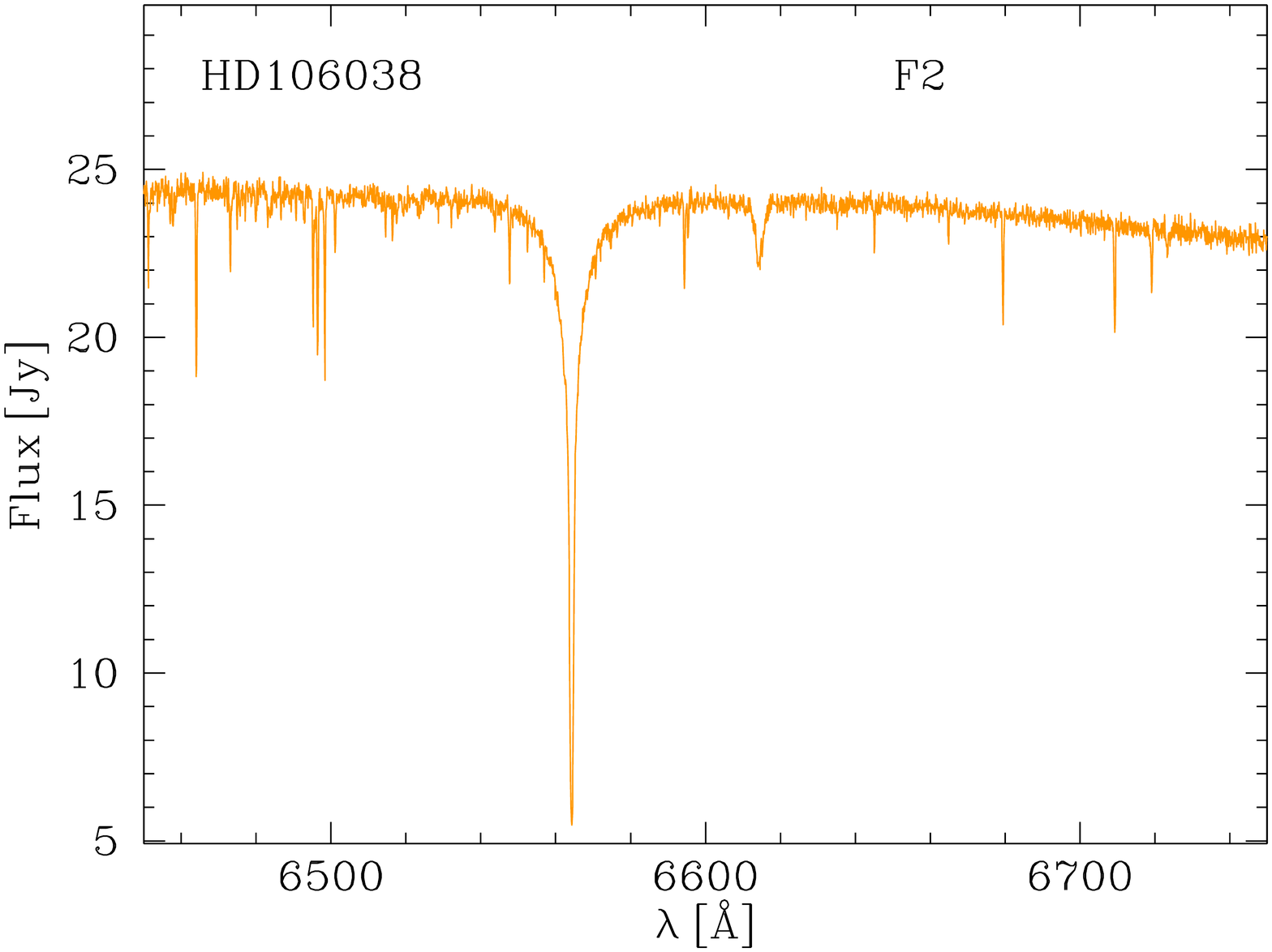}
\includegraphics[width=0.18\textwidth,angle=-90]{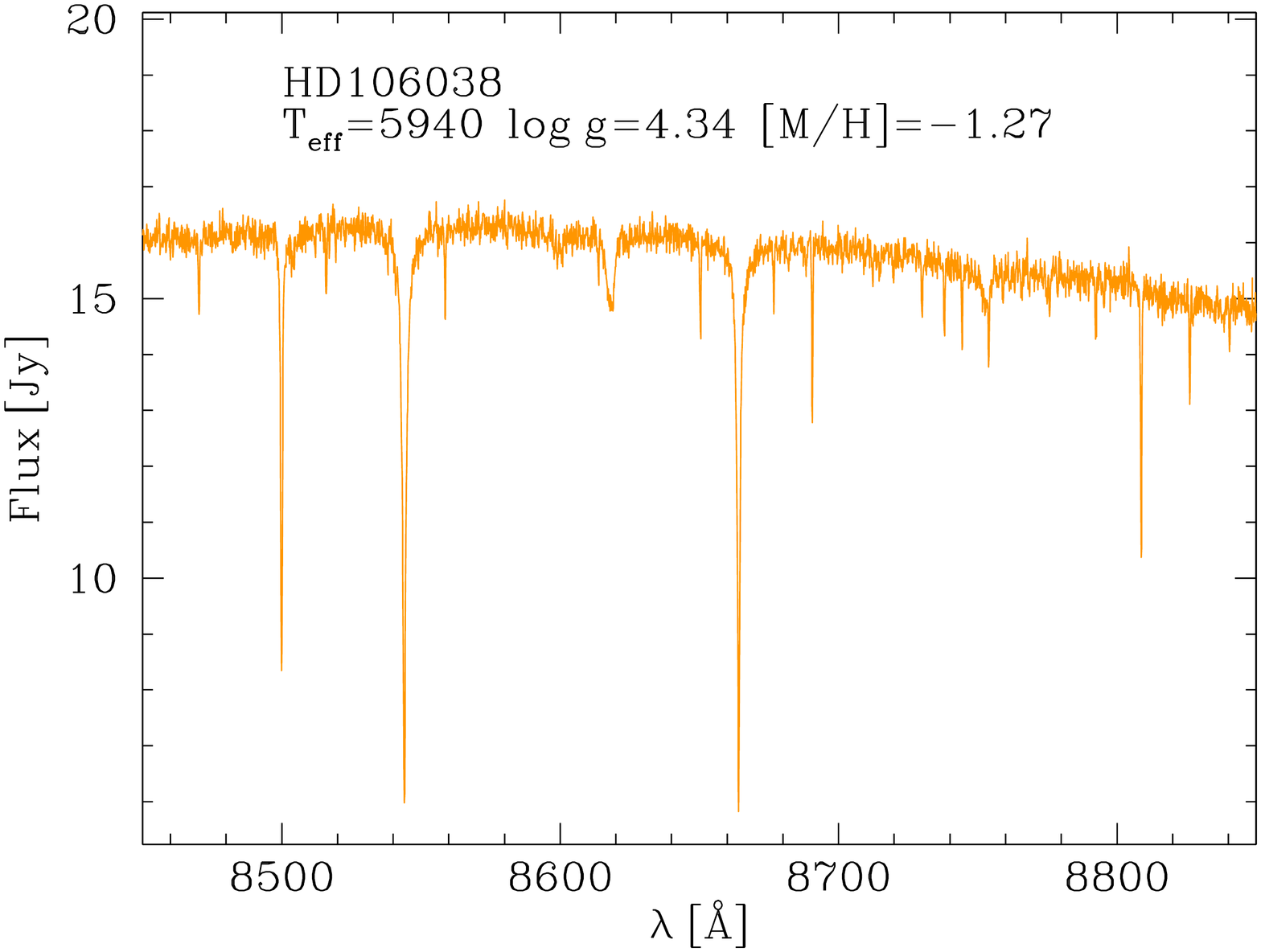}
\includegraphics[width=0.18\textwidth,angle=-90]{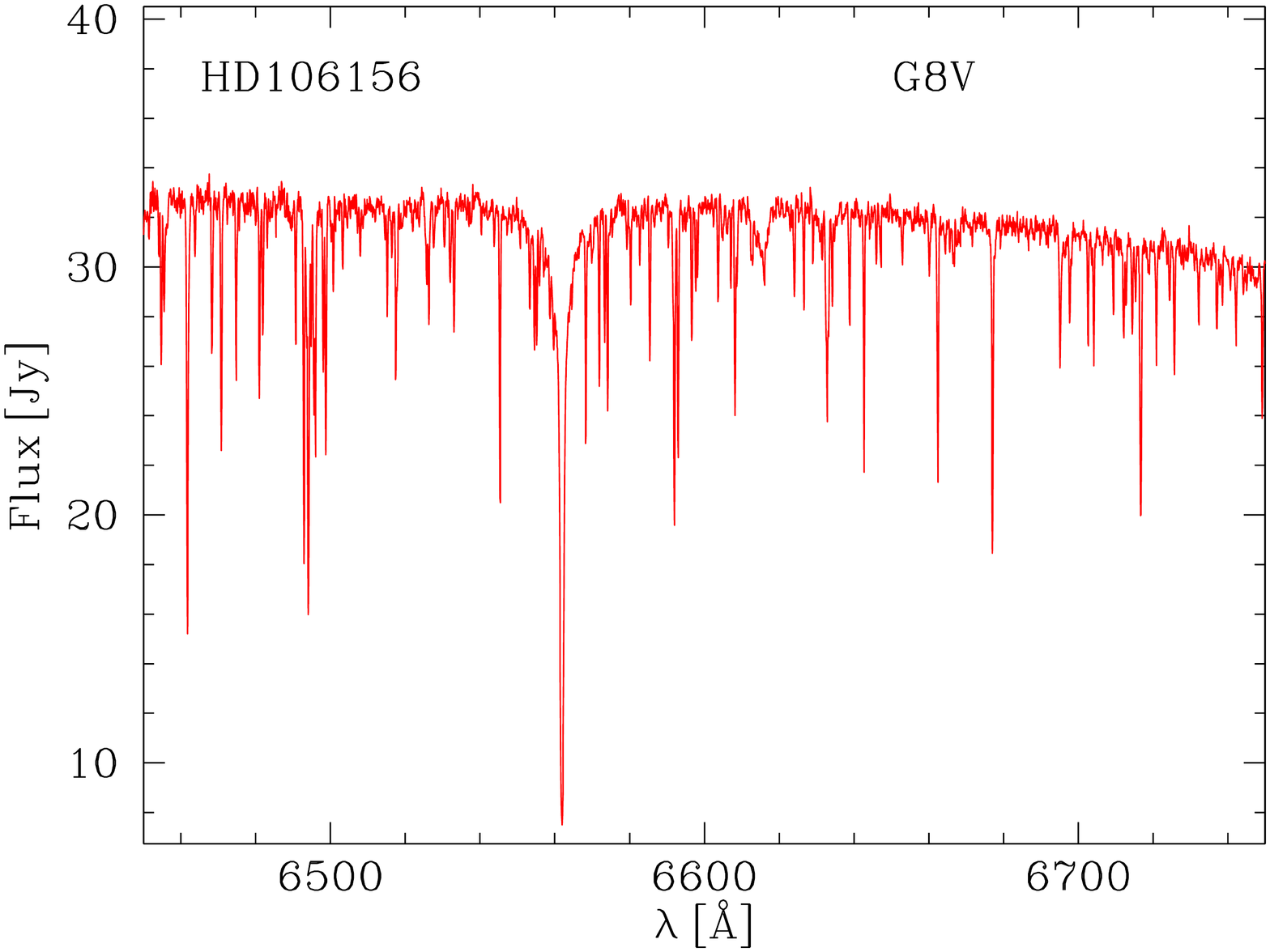} 
\includegraphics[width=0.18\textwidth,angle=-90]{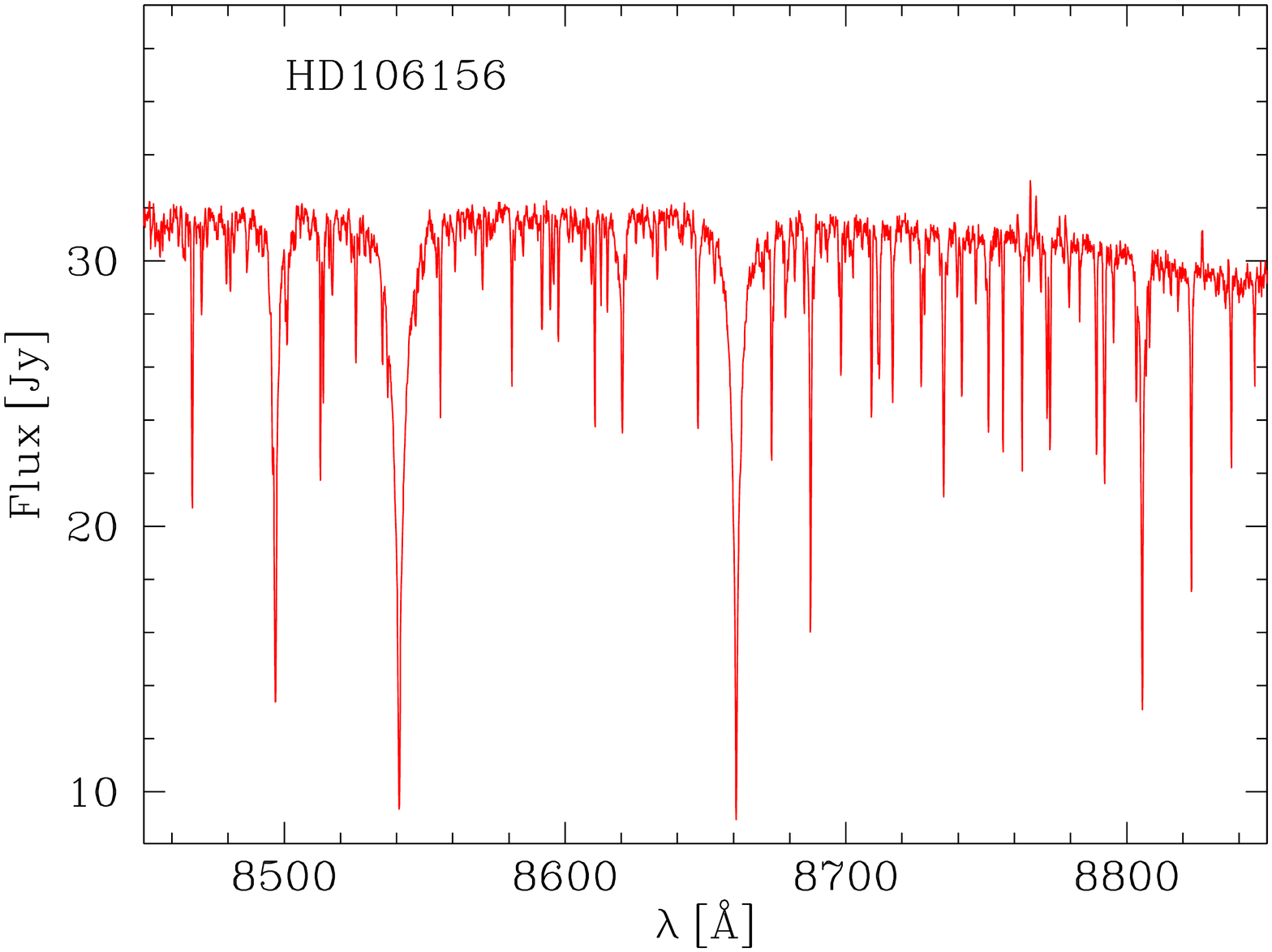}
\includegraphics[width=0.18\textwidth,angle=-90]{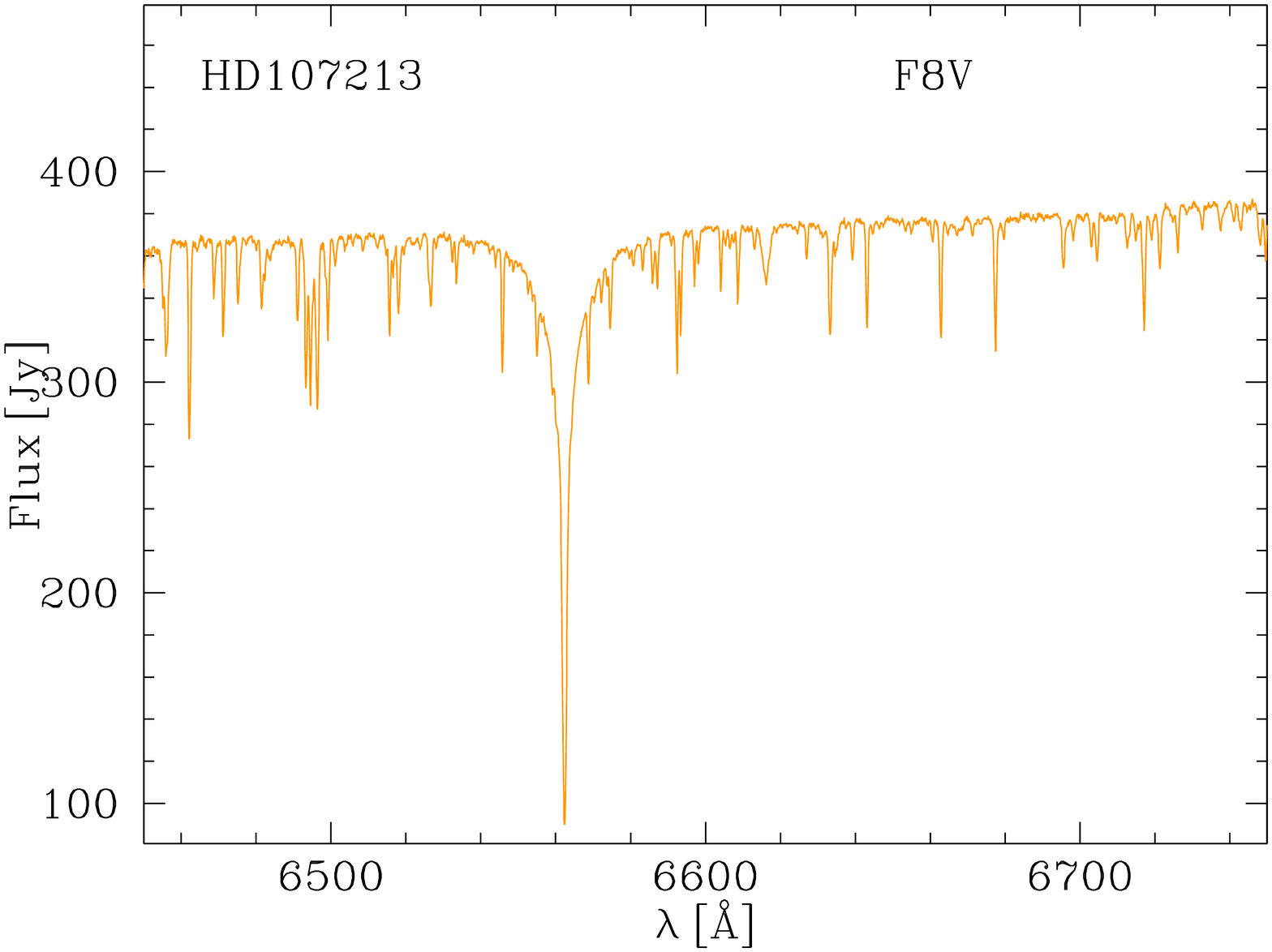}
\includegraphics[width=0.18\textwidth,angle=-90]{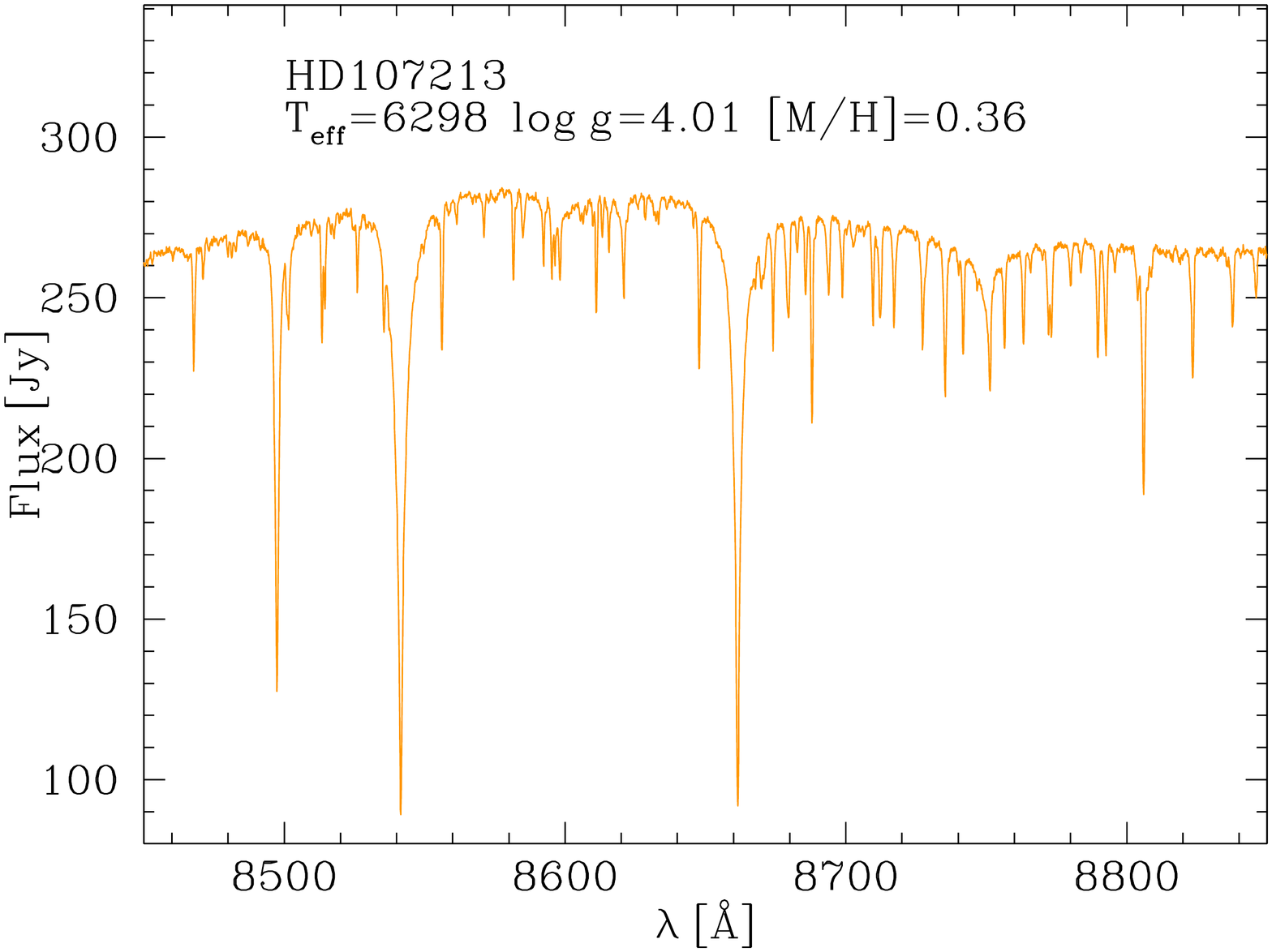}

\contcaption{20. Stars shown in this page are: HD101177, HD101177B, HD101227, HD101501, HD101606, HD101690, HD102870, HD104556, HD104979, HD104985, HD105087, HD106038, HD106156 and HD107213}
\end{figure*}

\begin{figure*}
\includegraphics[width=0.18\textwidth,angle=-90]{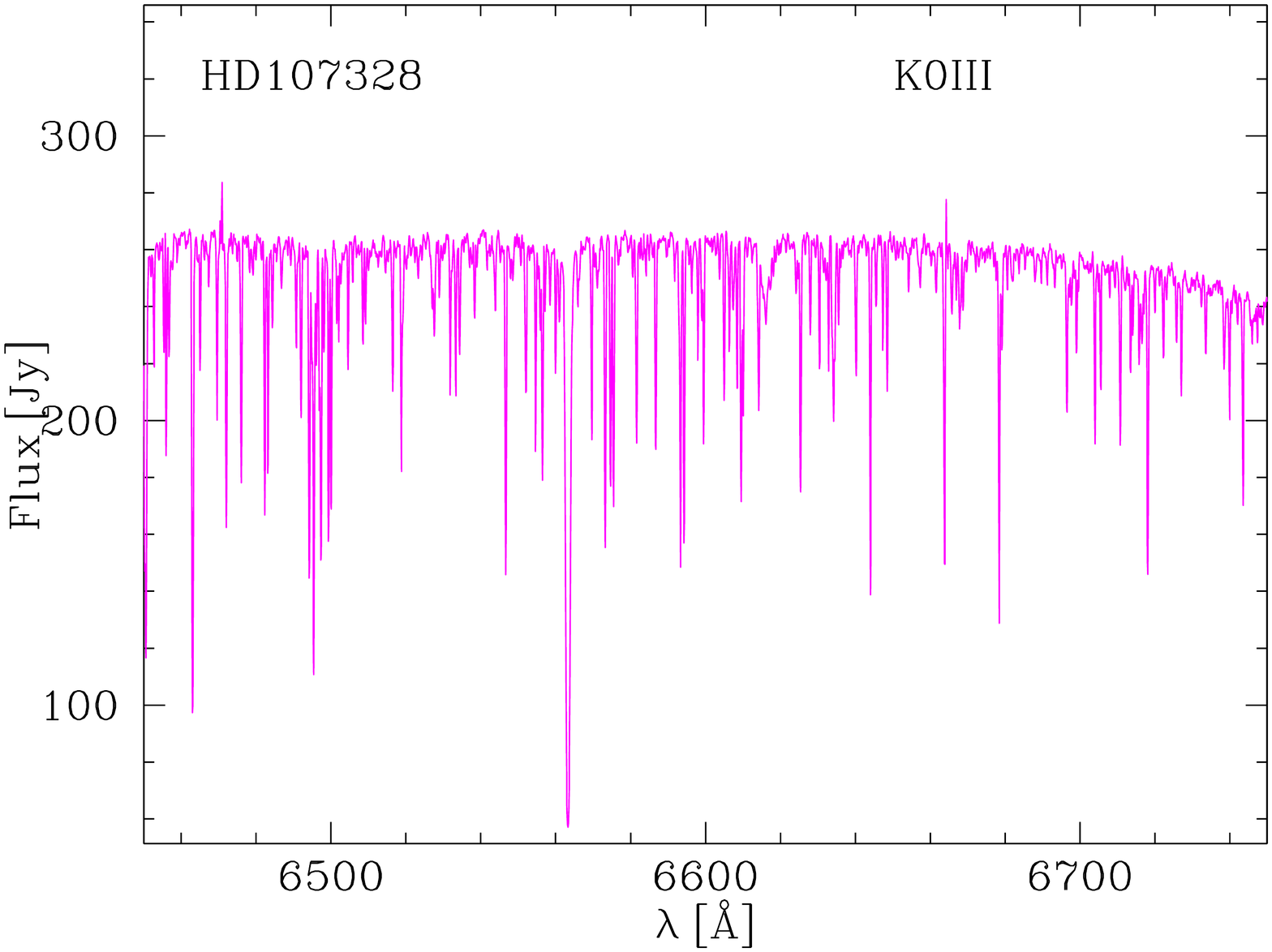}
\includegraphics[width=0.18\textwidth,angle=-90]{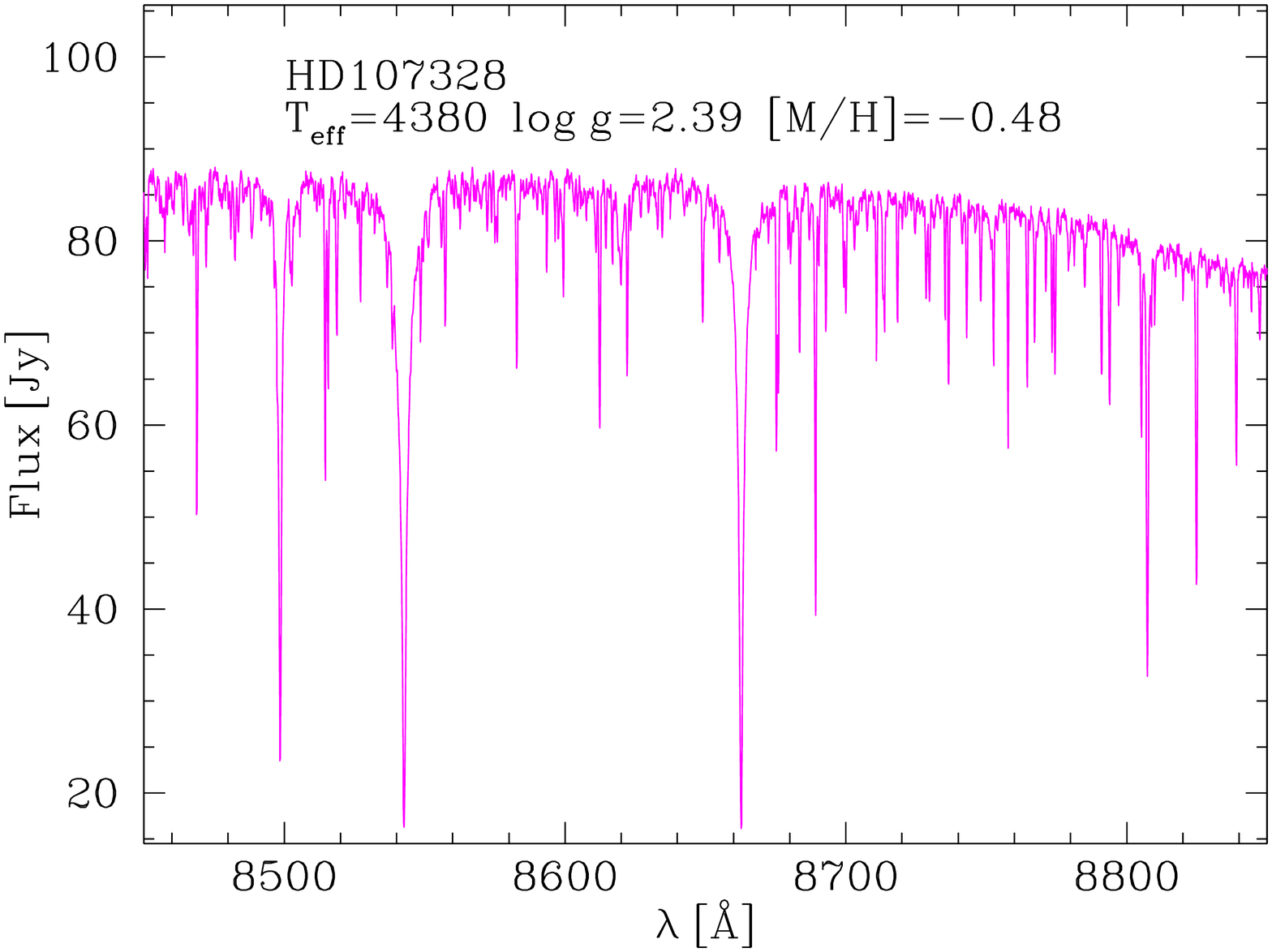}
\includegraphics[width=0.18\textwidth,angle=-90]{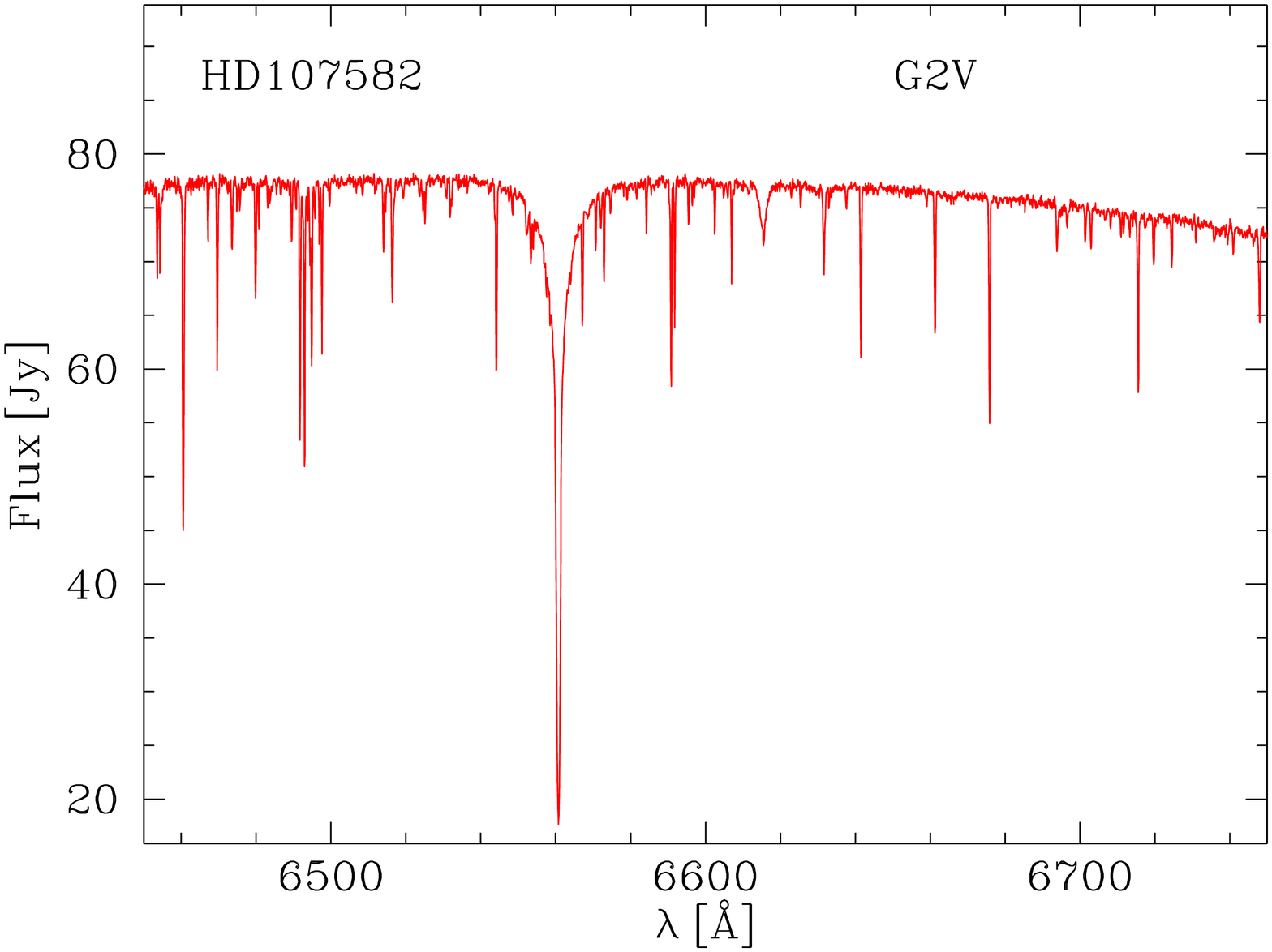}
\includegraphics[width=0.18\textwidth,angle=-90]{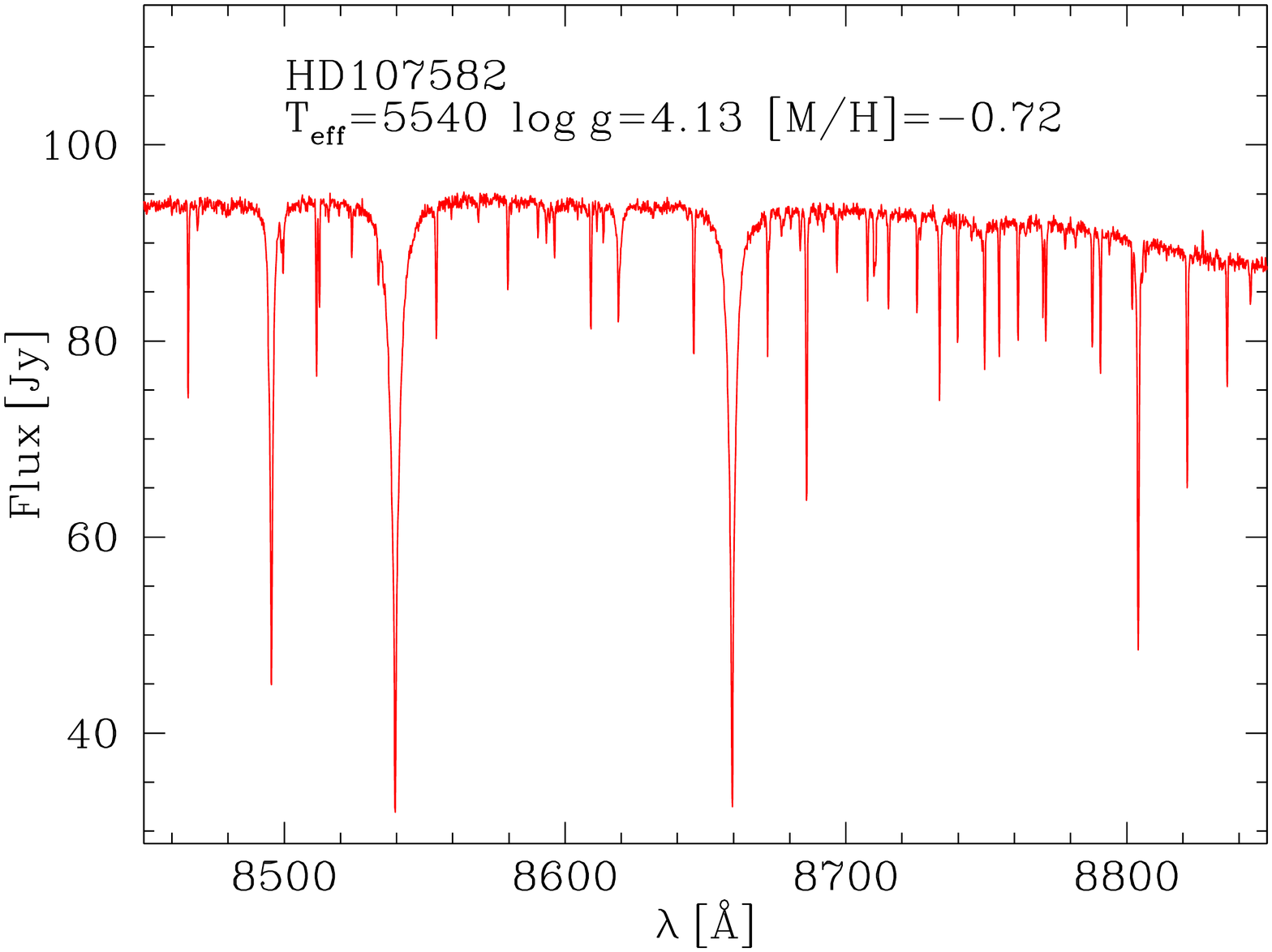}
\includegraphics[width=0.18\textwidth,angle=-90]{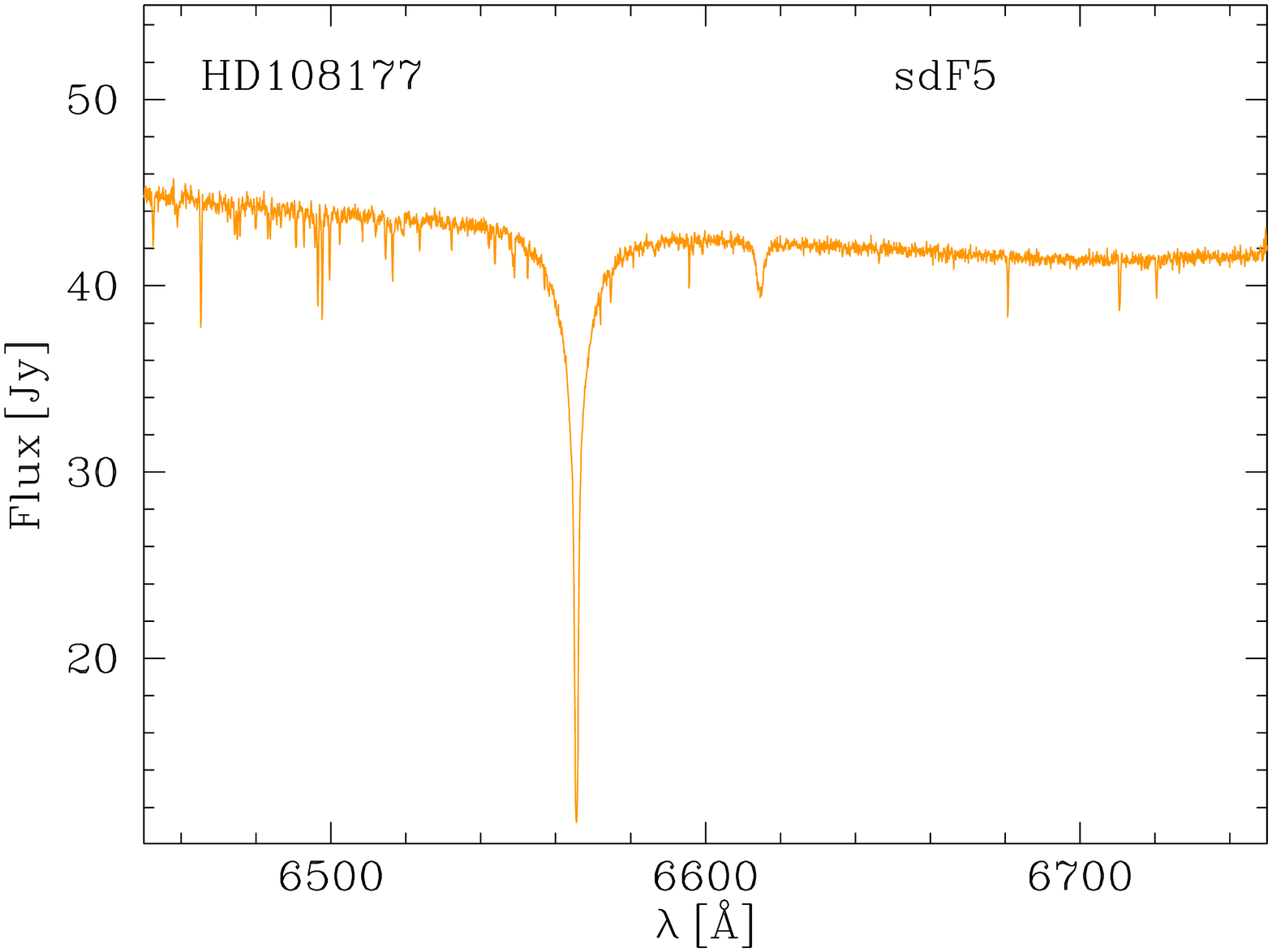}
\includegraphics[width=0.18\textwidth,angle=-90]{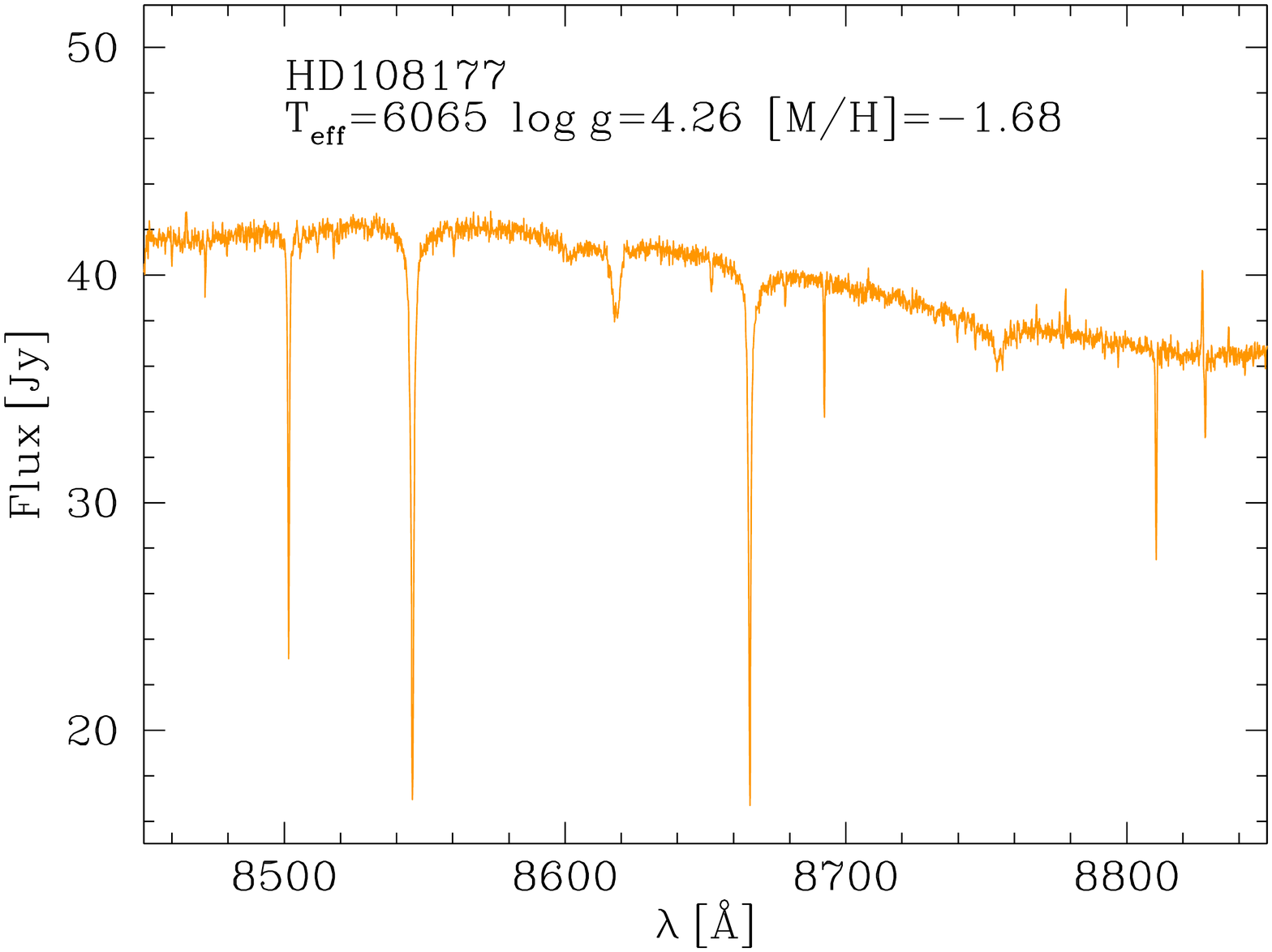}
\includegraphics[width=0.18\textwidth,angle=-90]{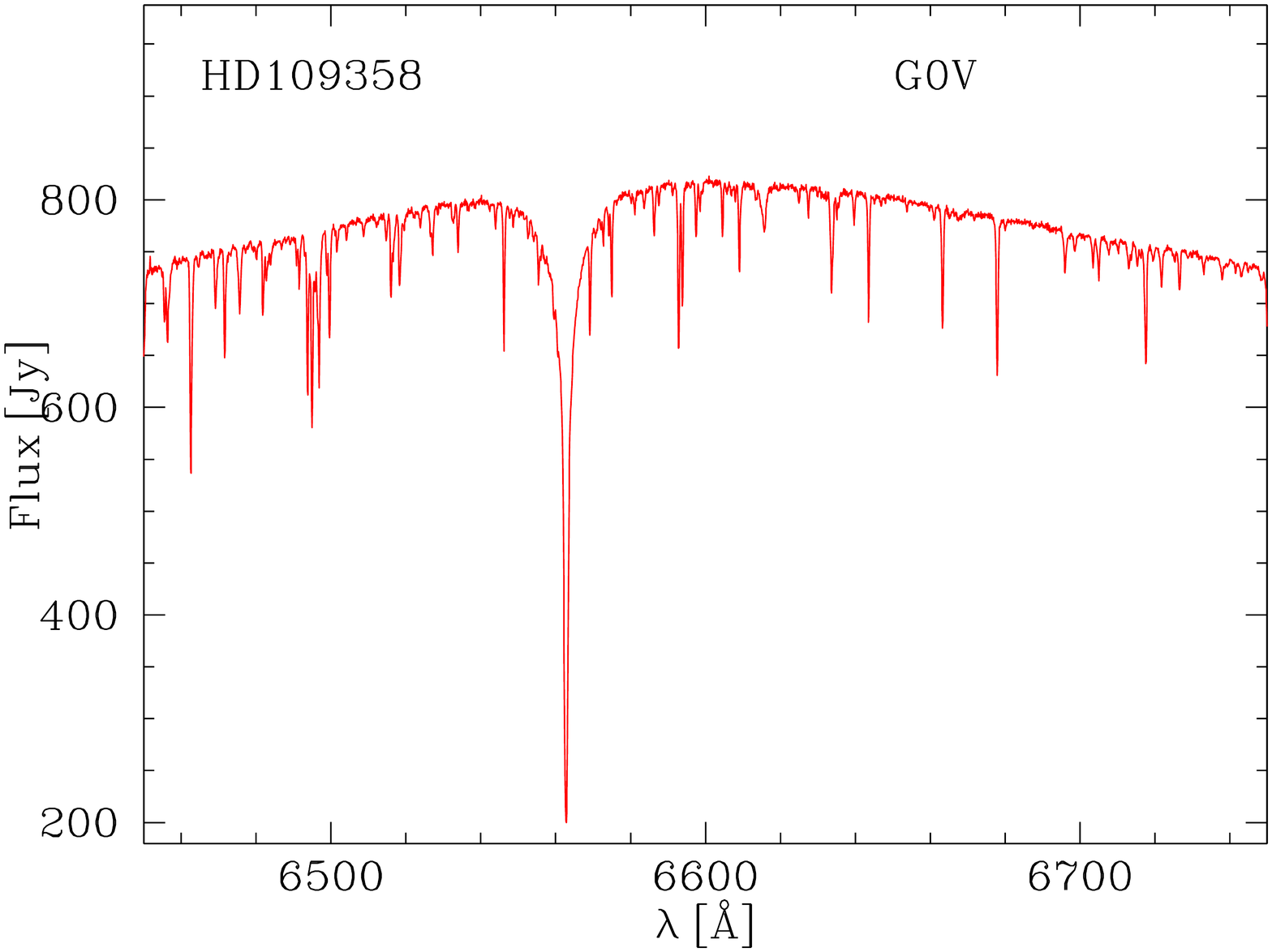}
\includegraphics[width=0.18\textwidth,angle=-90]{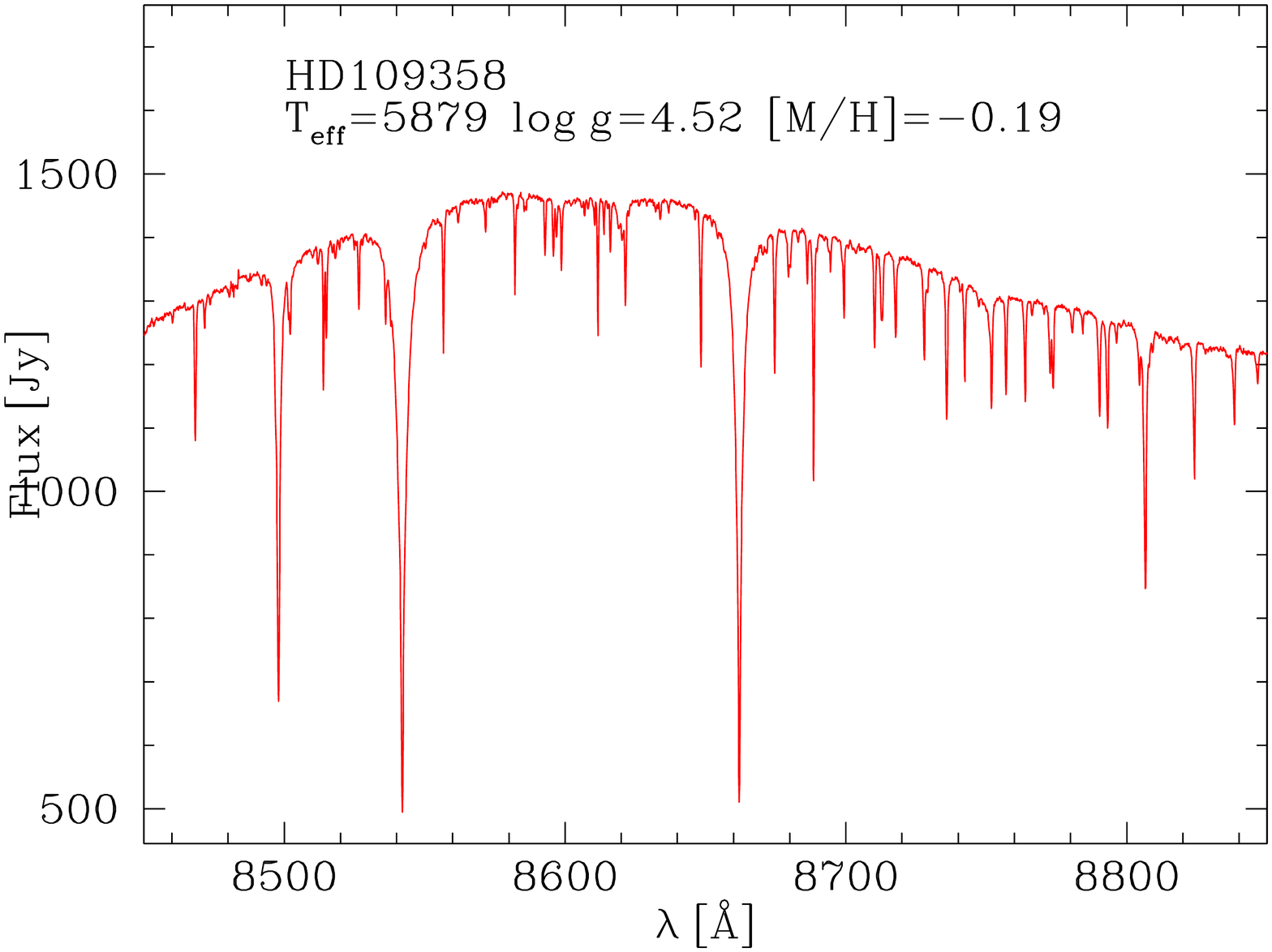}
\includegraphics[width=0.18\textwidth,angle=-90]{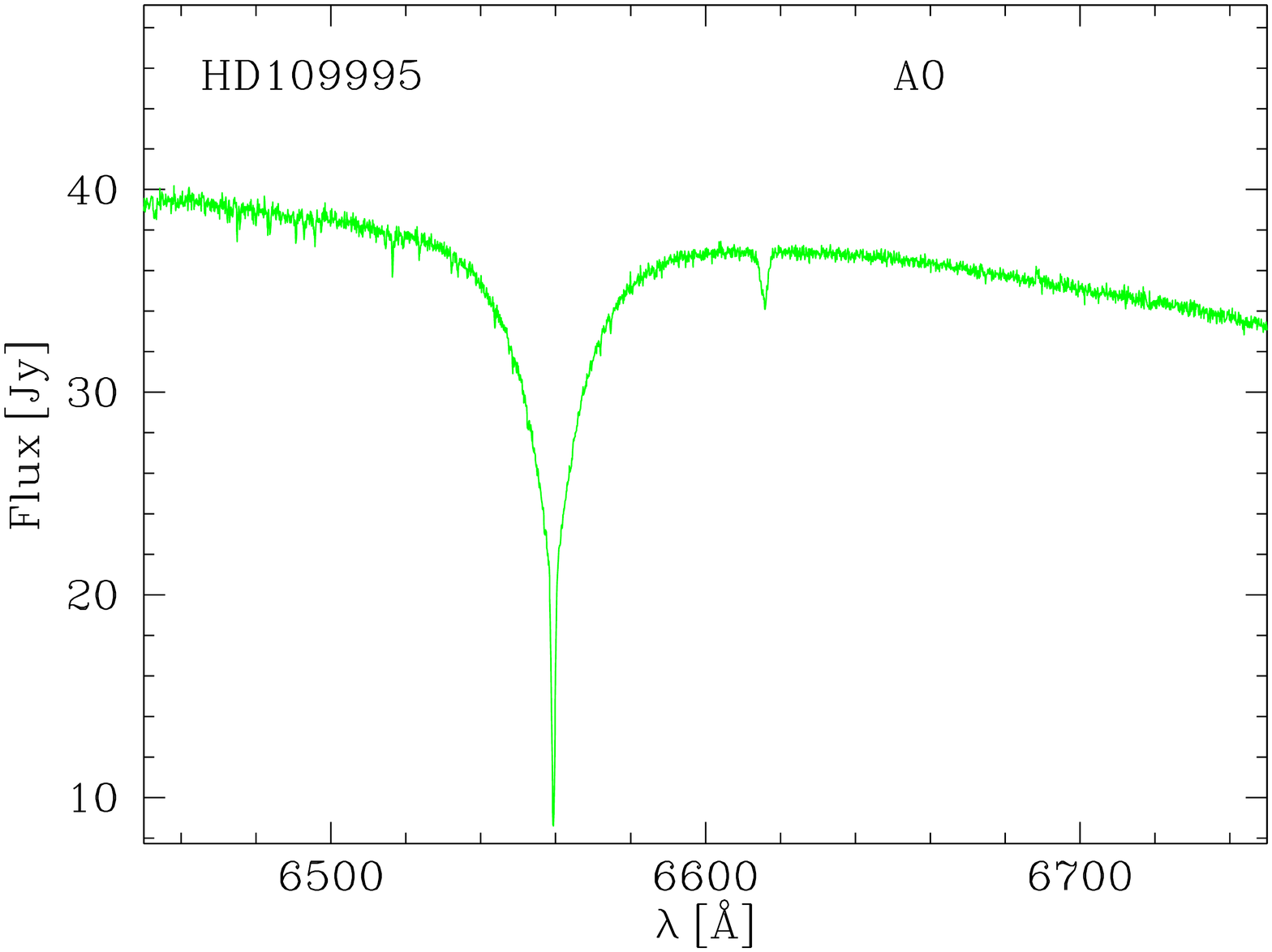}
\includegraphics[width=0.18\textwidth,angle=-90]{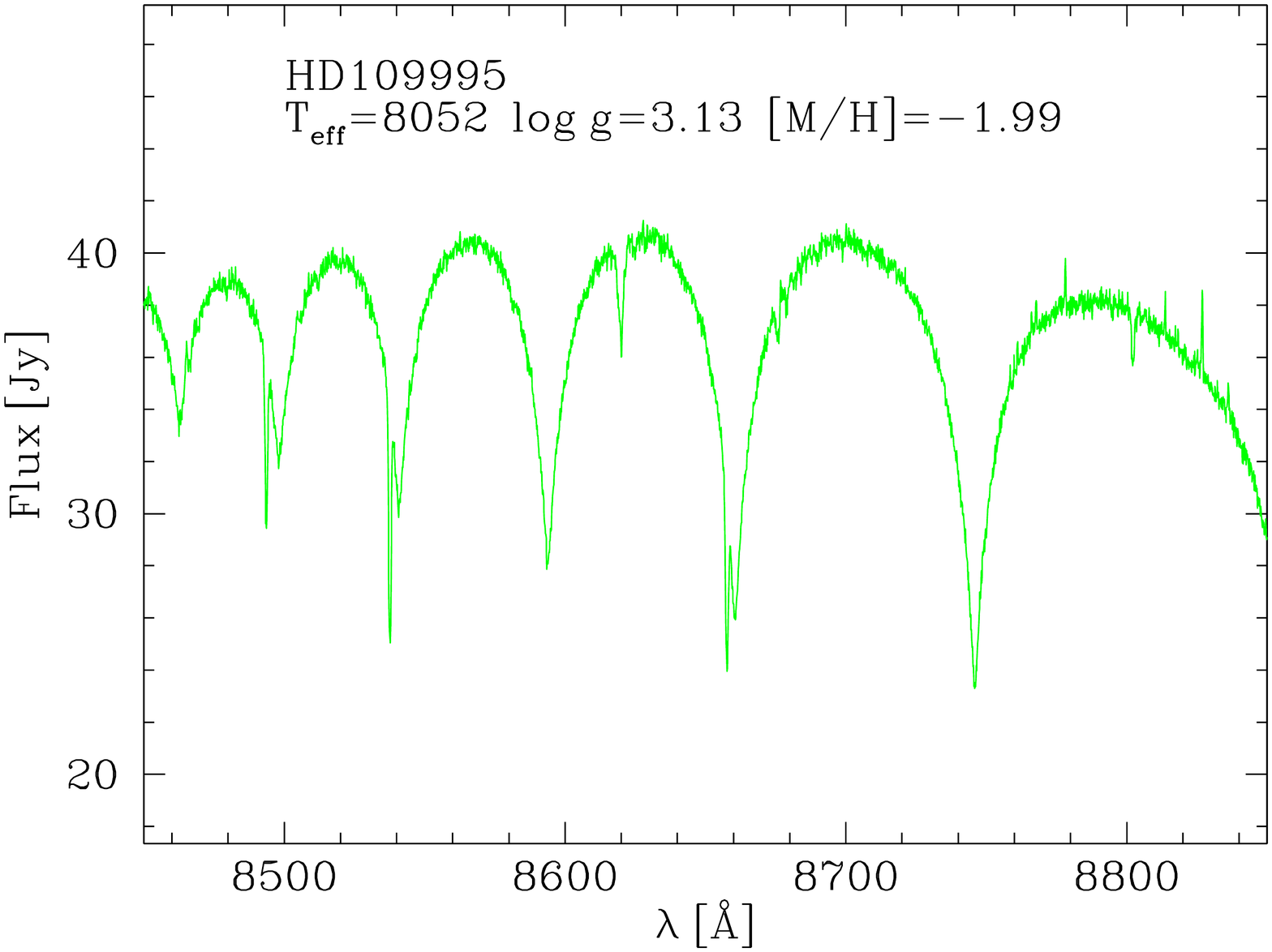}
\includegraphics[width=0.18\textwidth,angle=-90]{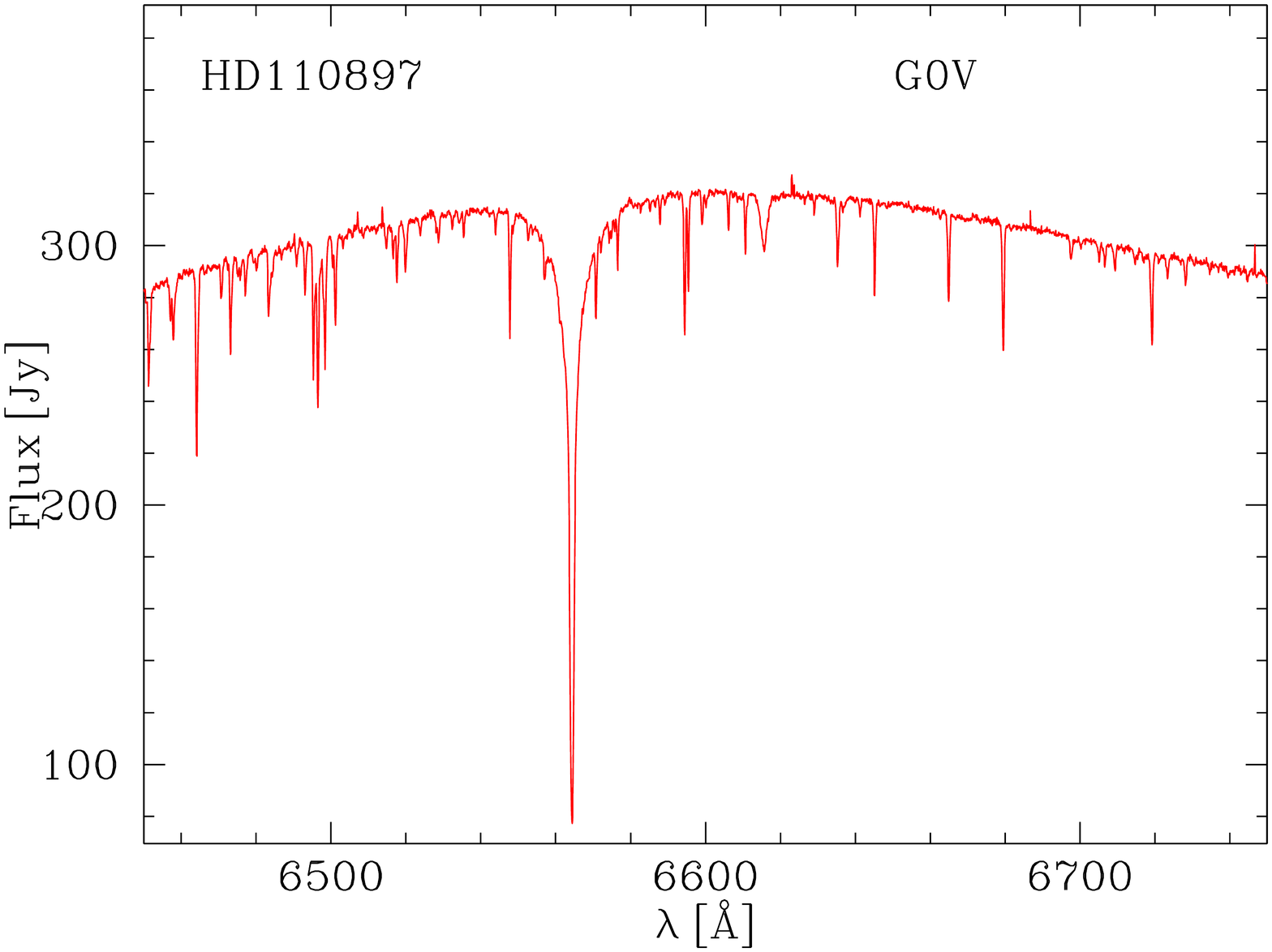}
\includegraphics[width=0.18\textwidth,angle=-90]{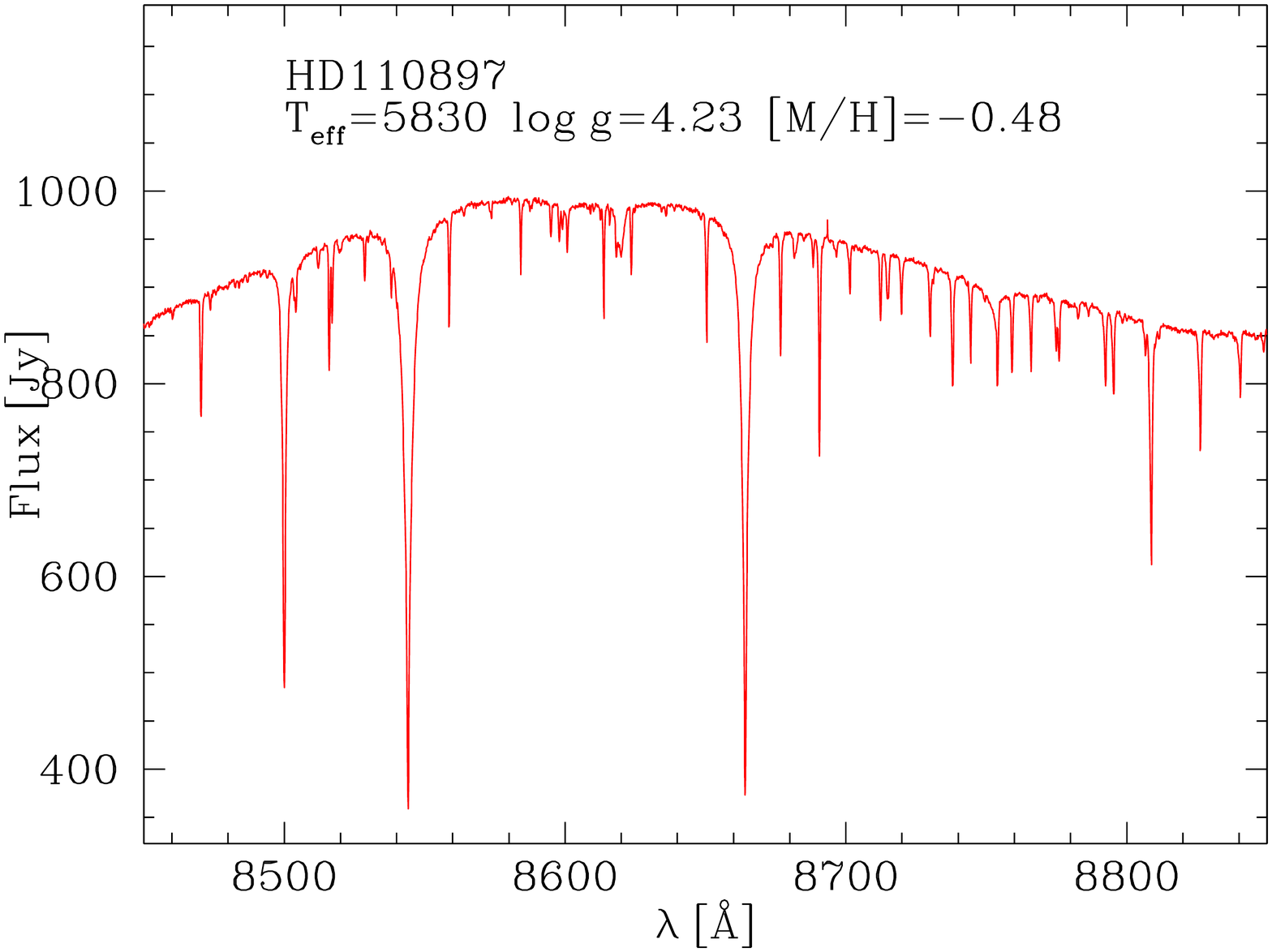}
\includegraphics[width=0.18\textwidth,angle=-90]{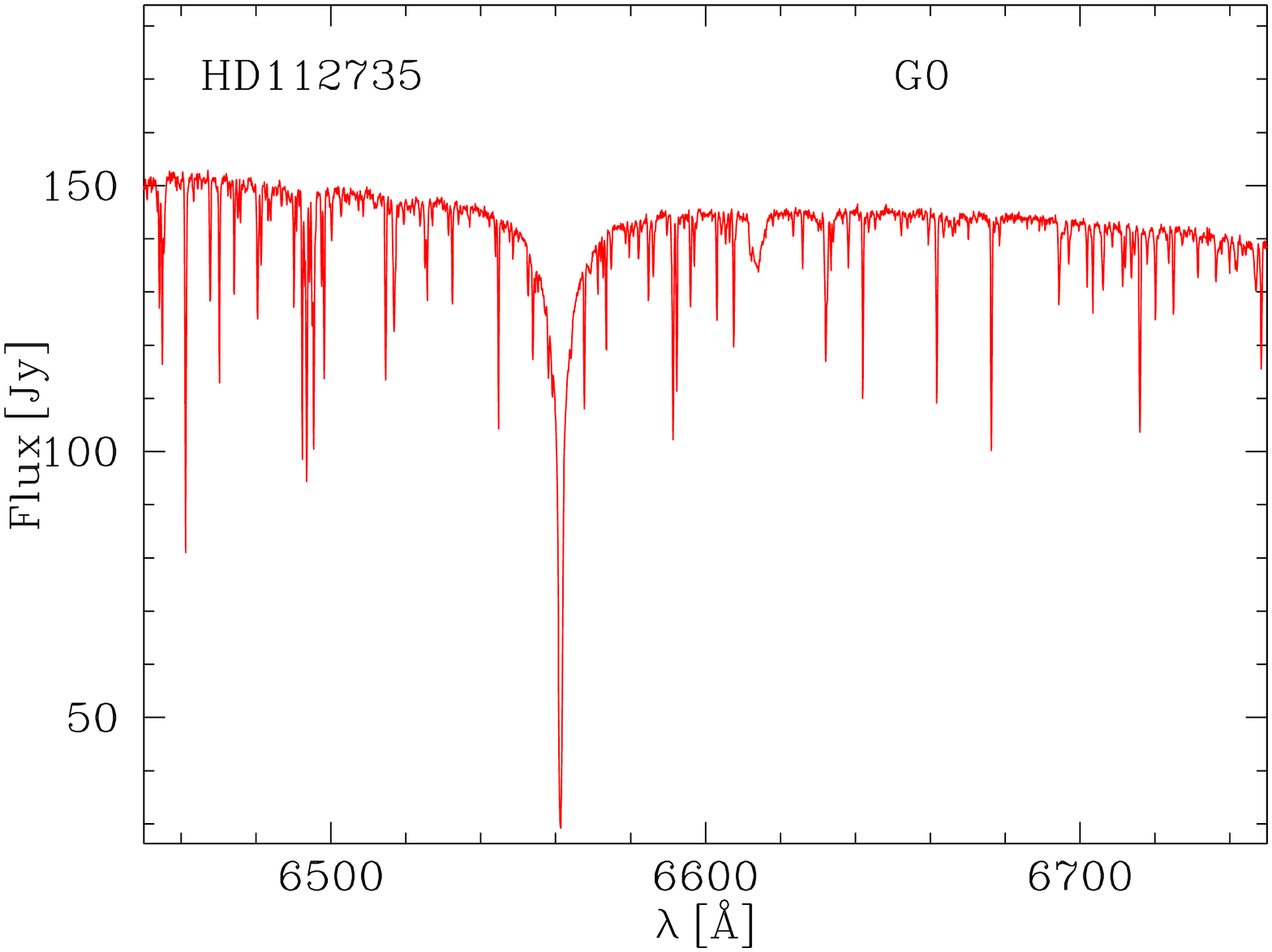}
\includegraphics[width=0.18\textwidth,angle=-90]{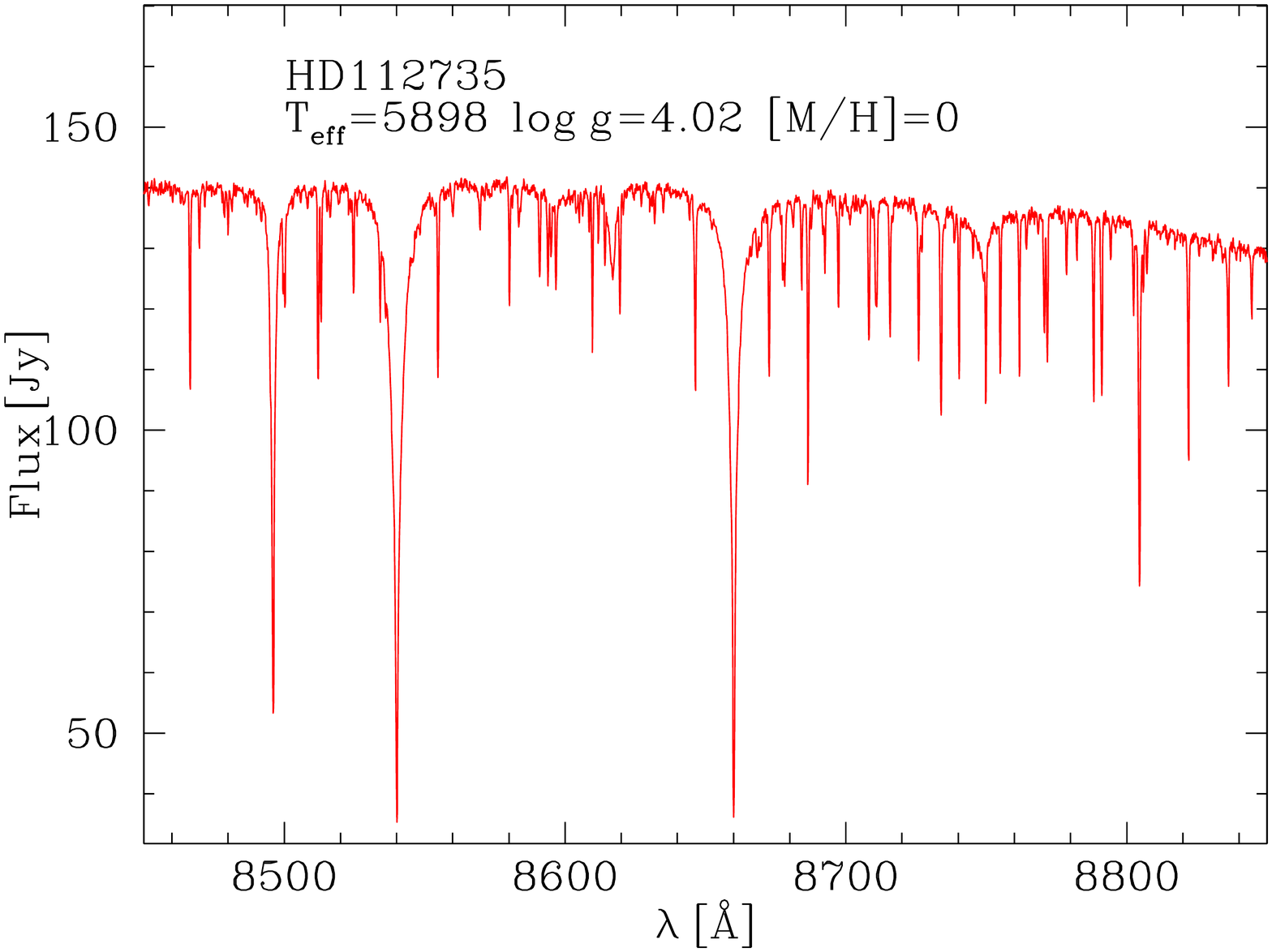}
\includegraphics[width=0.18\textwidth,angle=-90]{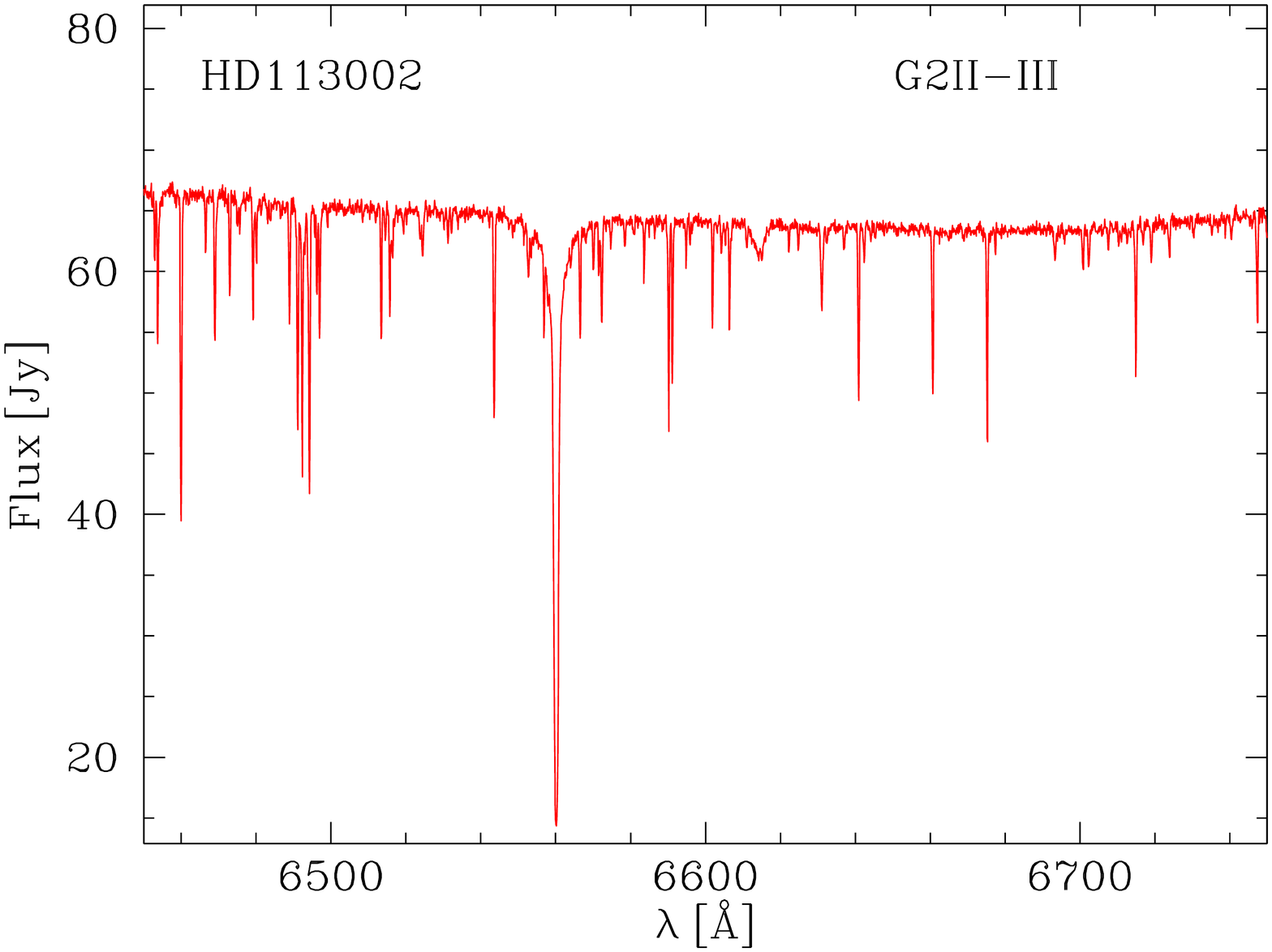}
\includegraphics[width=0.18\textwidth,angle=-90]{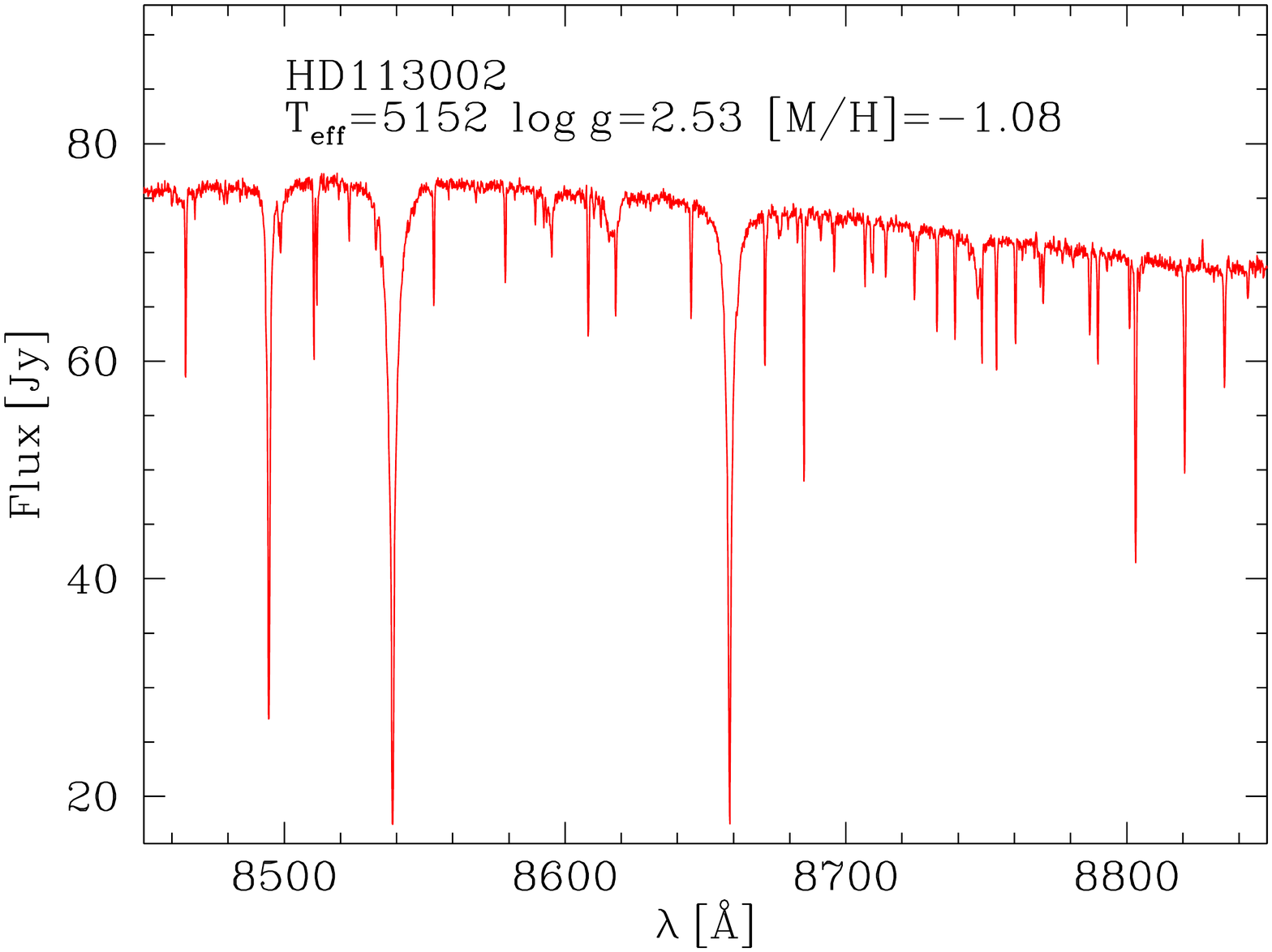}
\includegraphics[width=0.18\textwidth,angle=-90]{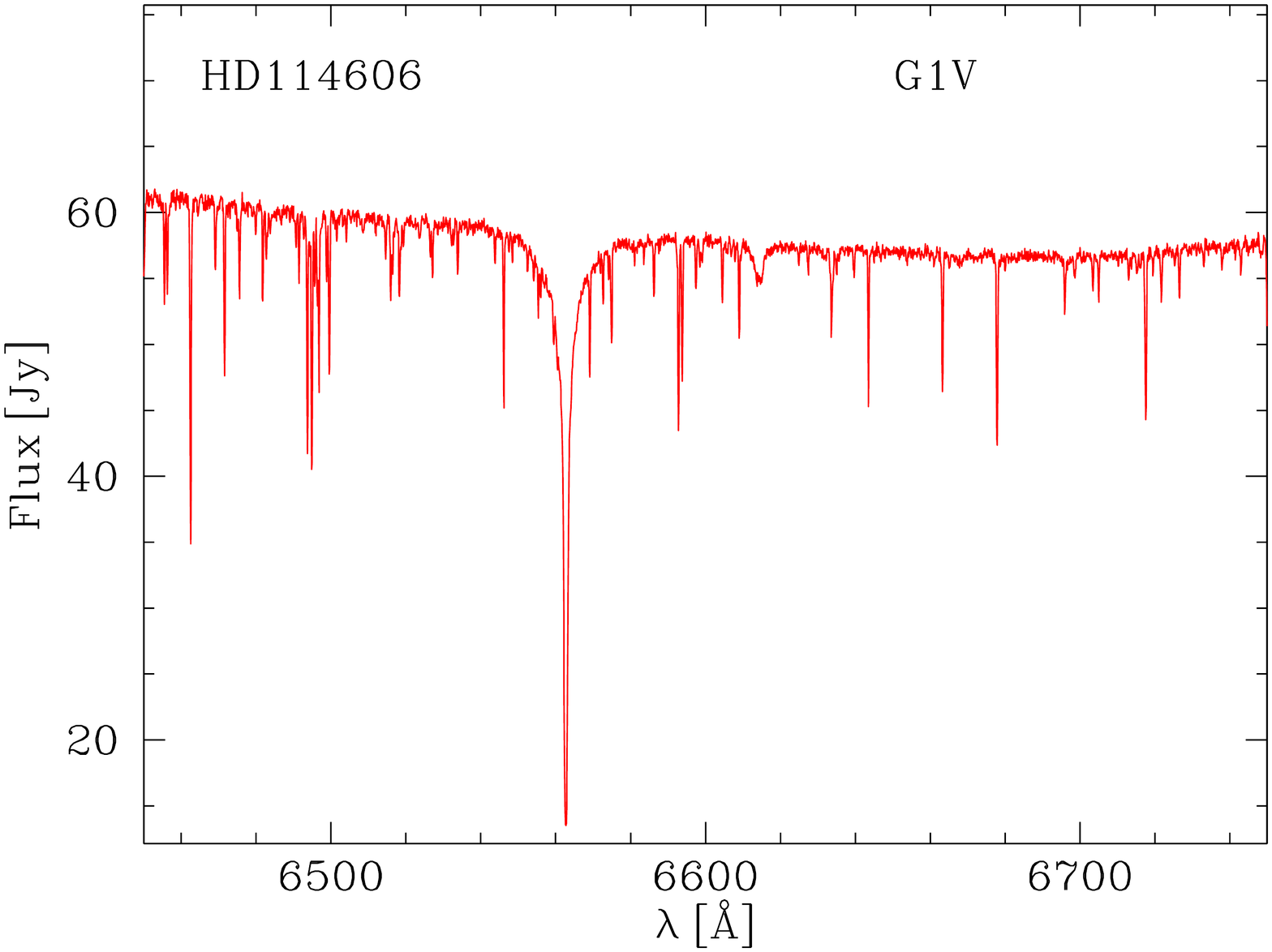}
\includegraphics[width=0.18\textwidth,angle=-90]{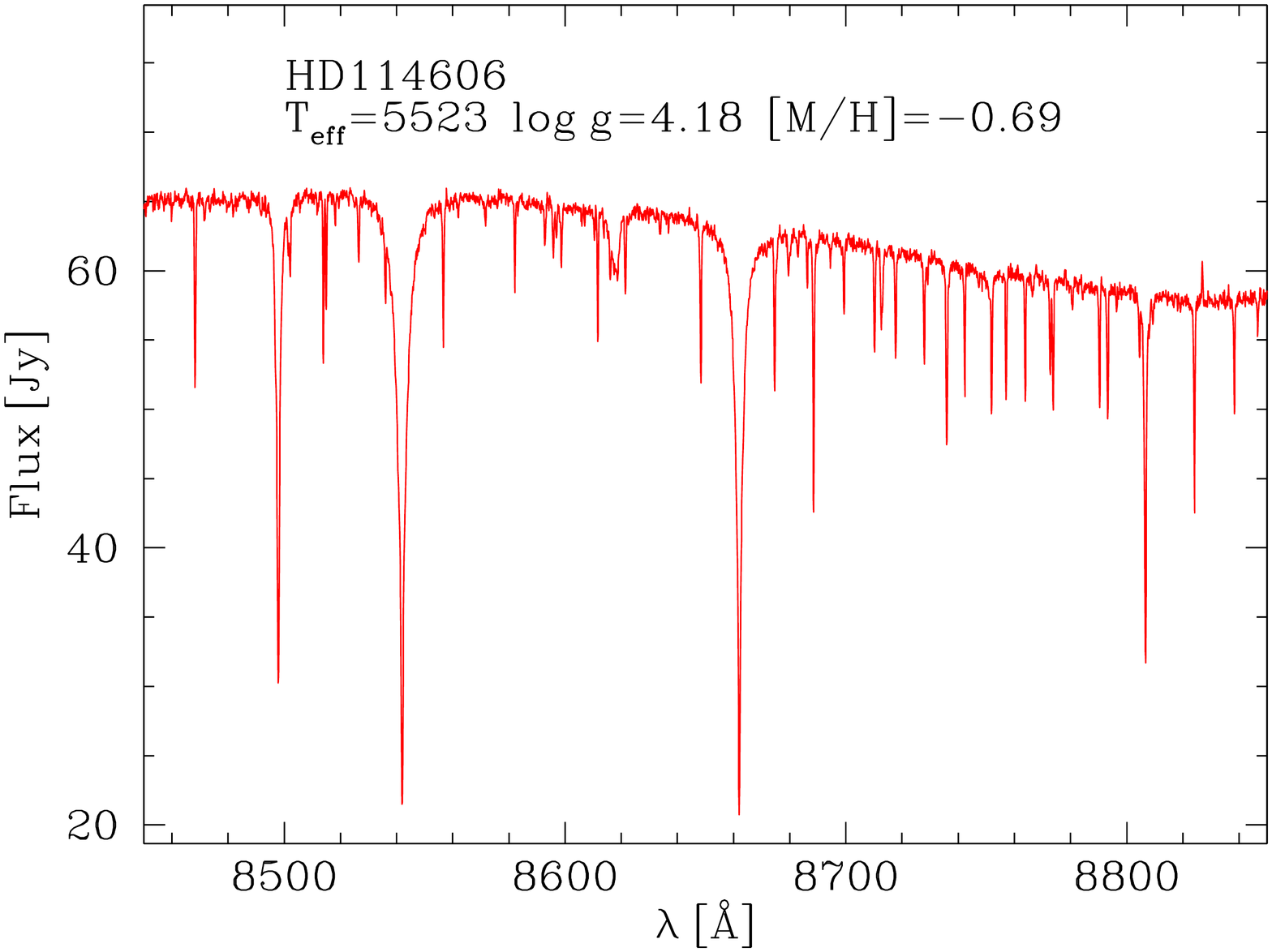}
\includegraphics[width=0.18\textwidth,angle=-90]{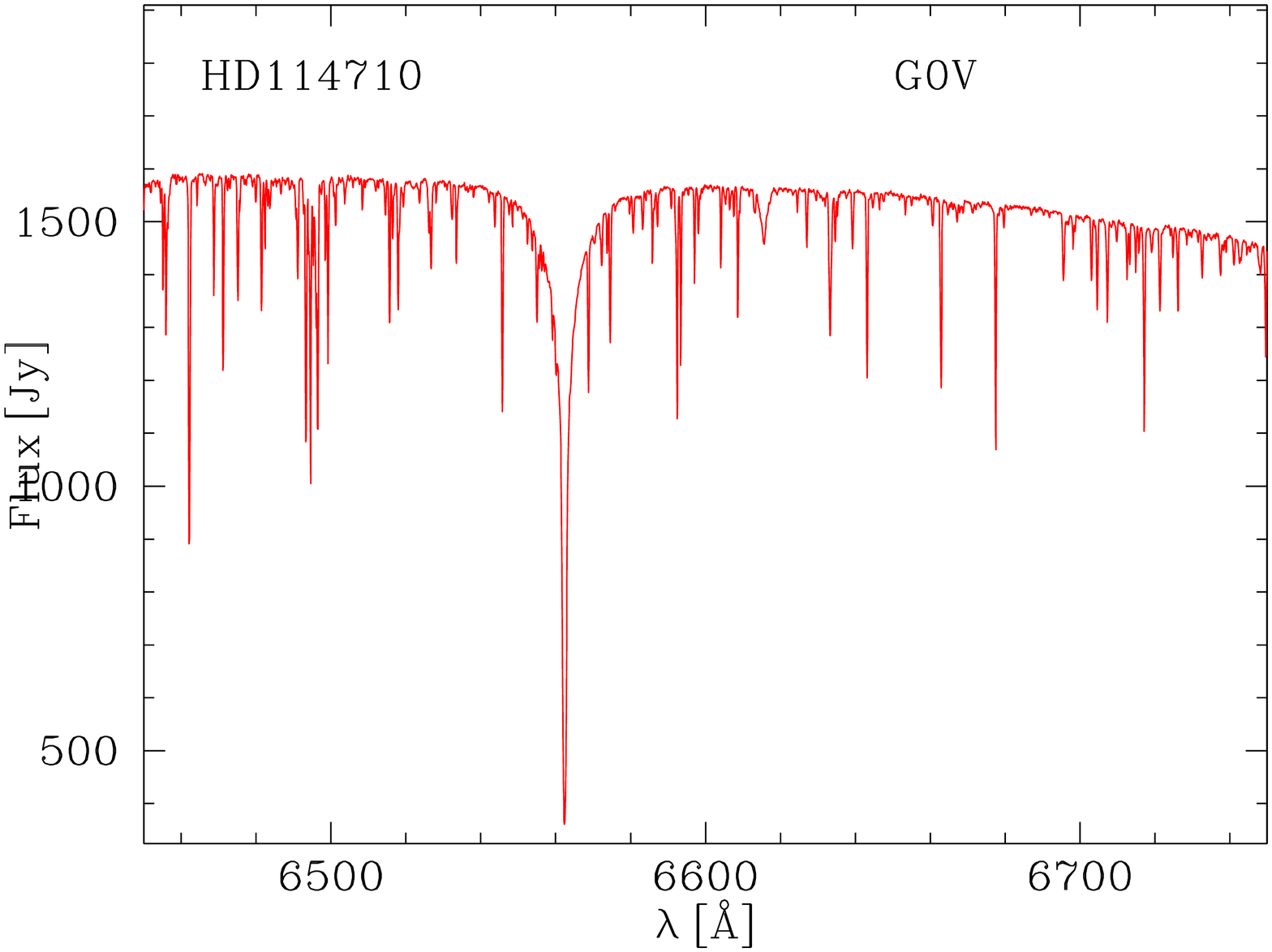}
\includegraphics[width=0.18\textwidth,angle=-90]{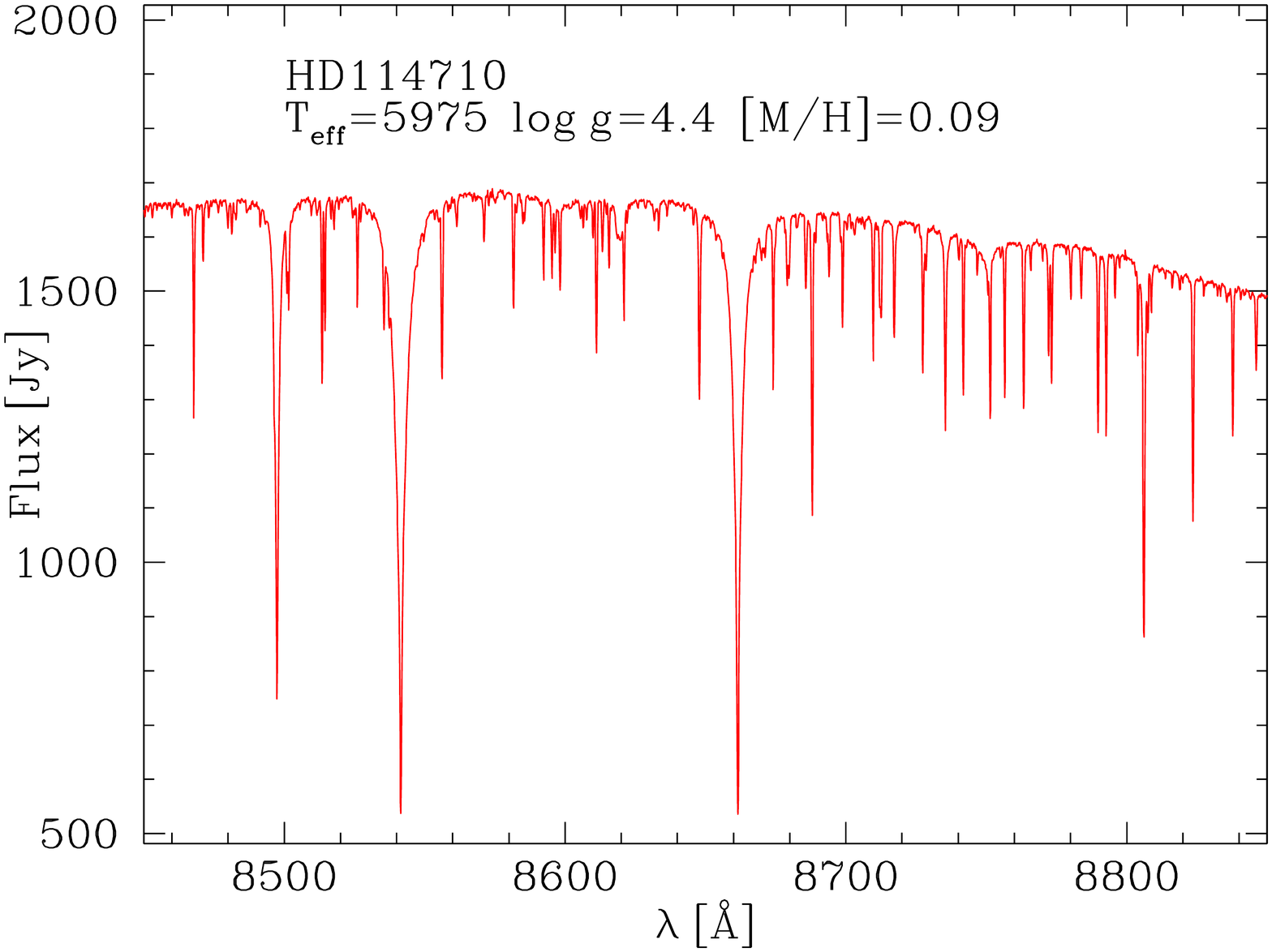}
\includegraphics[width=0.18\textwidth,angle=-90]{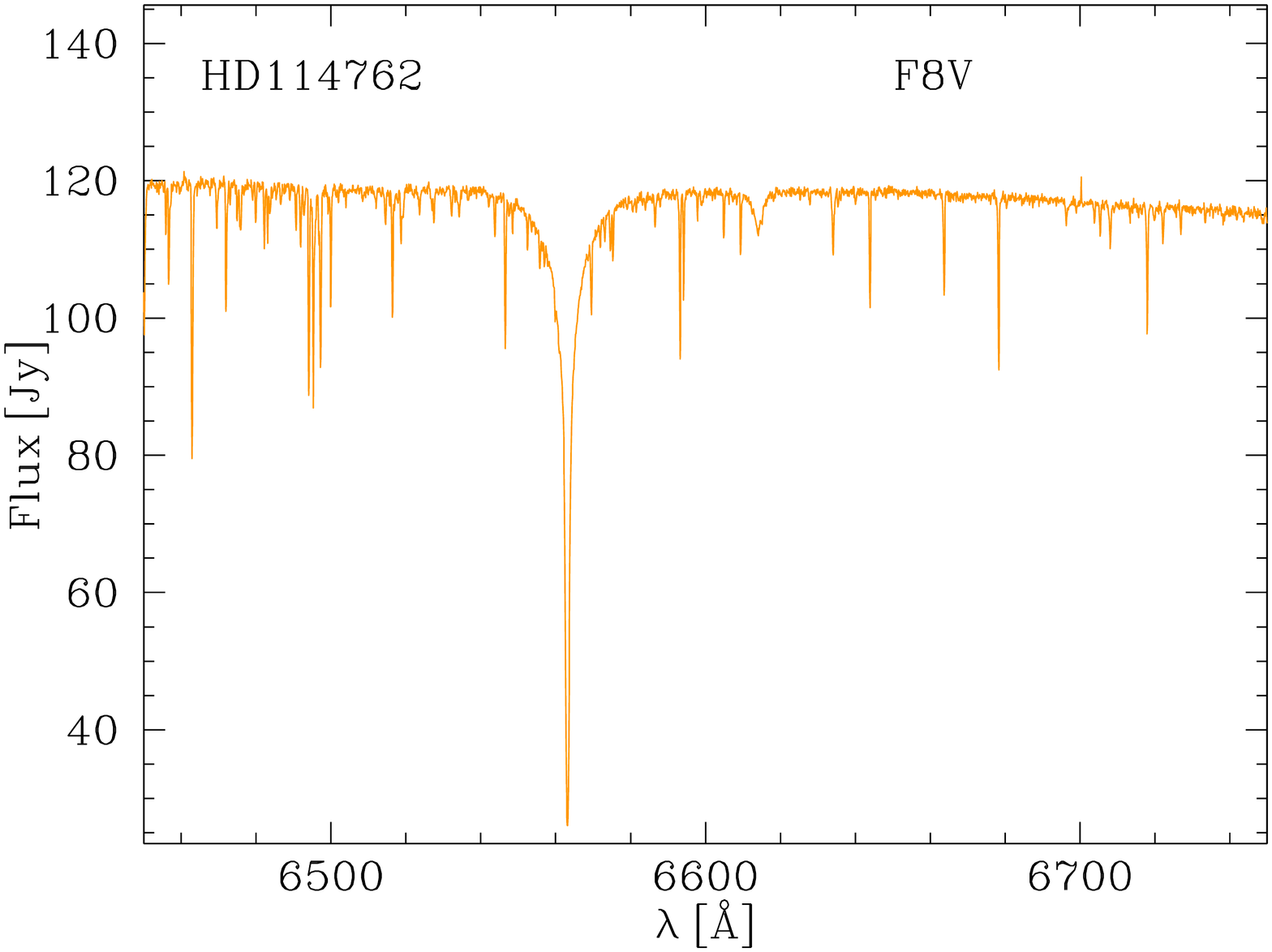}
\includegraphics[width=0.18\textwidth,angle=-90]{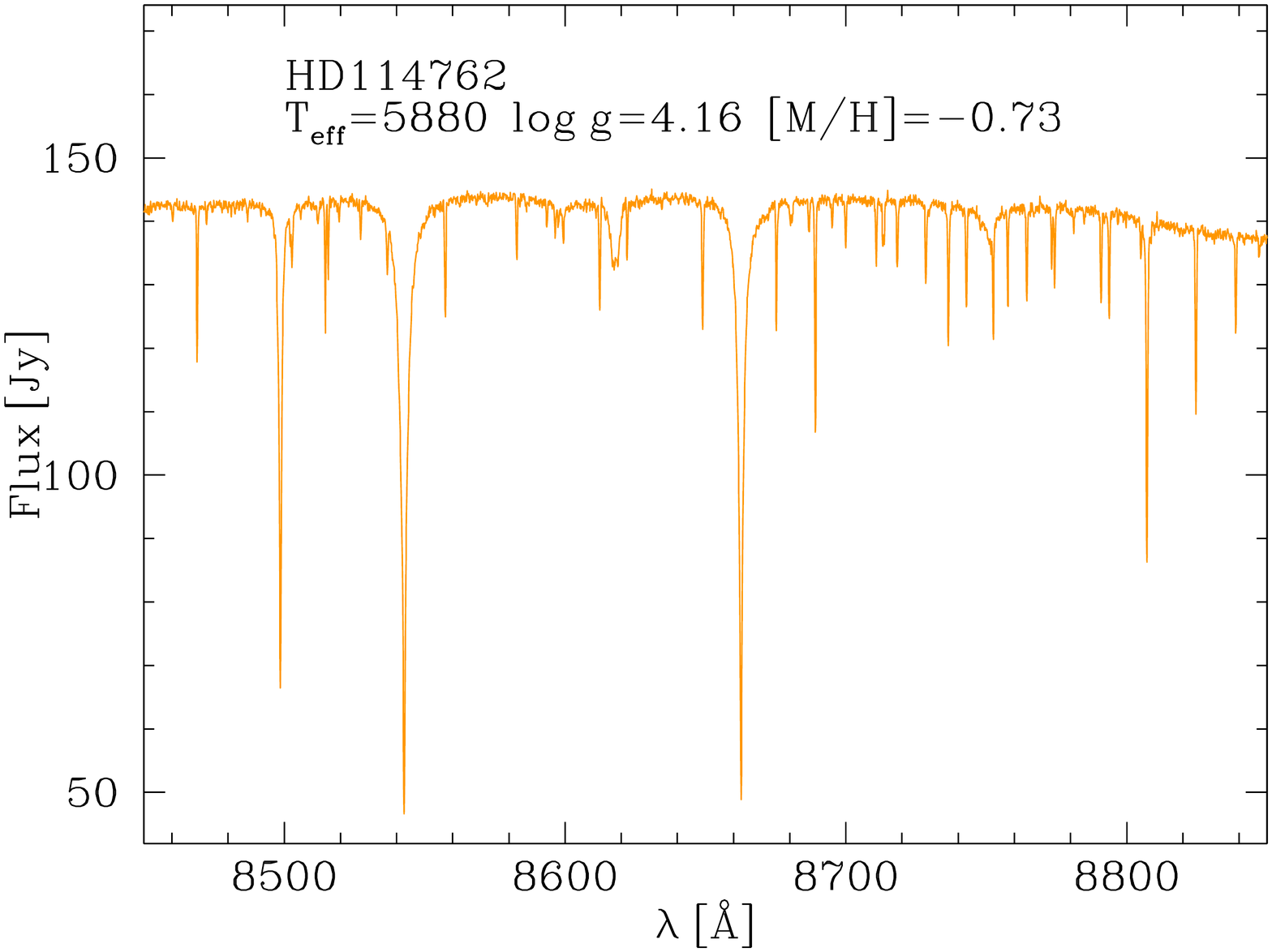}
\includegraphics[width=0.18\textwidth,angle=-90]{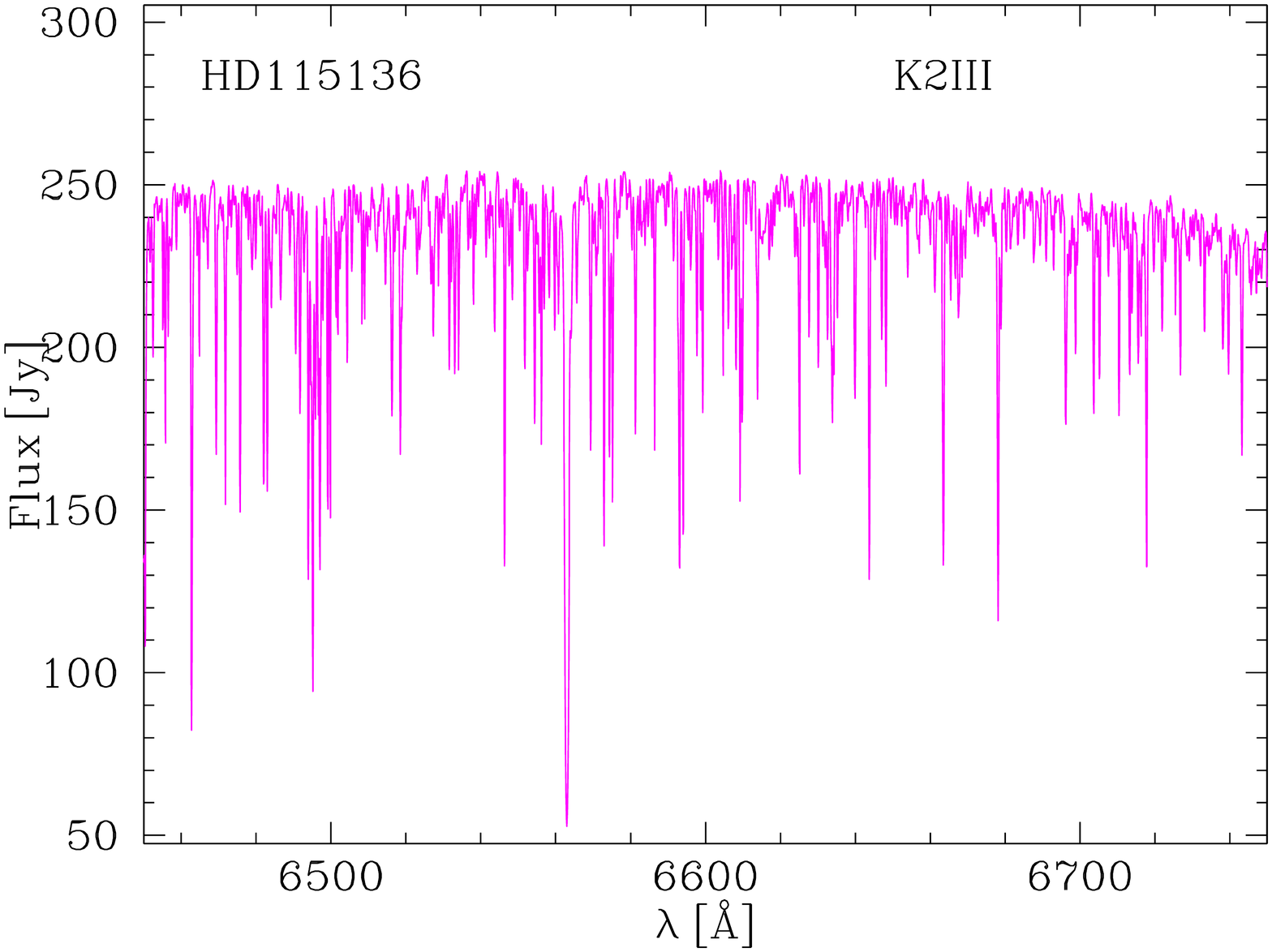}
\includegraphics[width=0.18\textwidth,angle=-90]{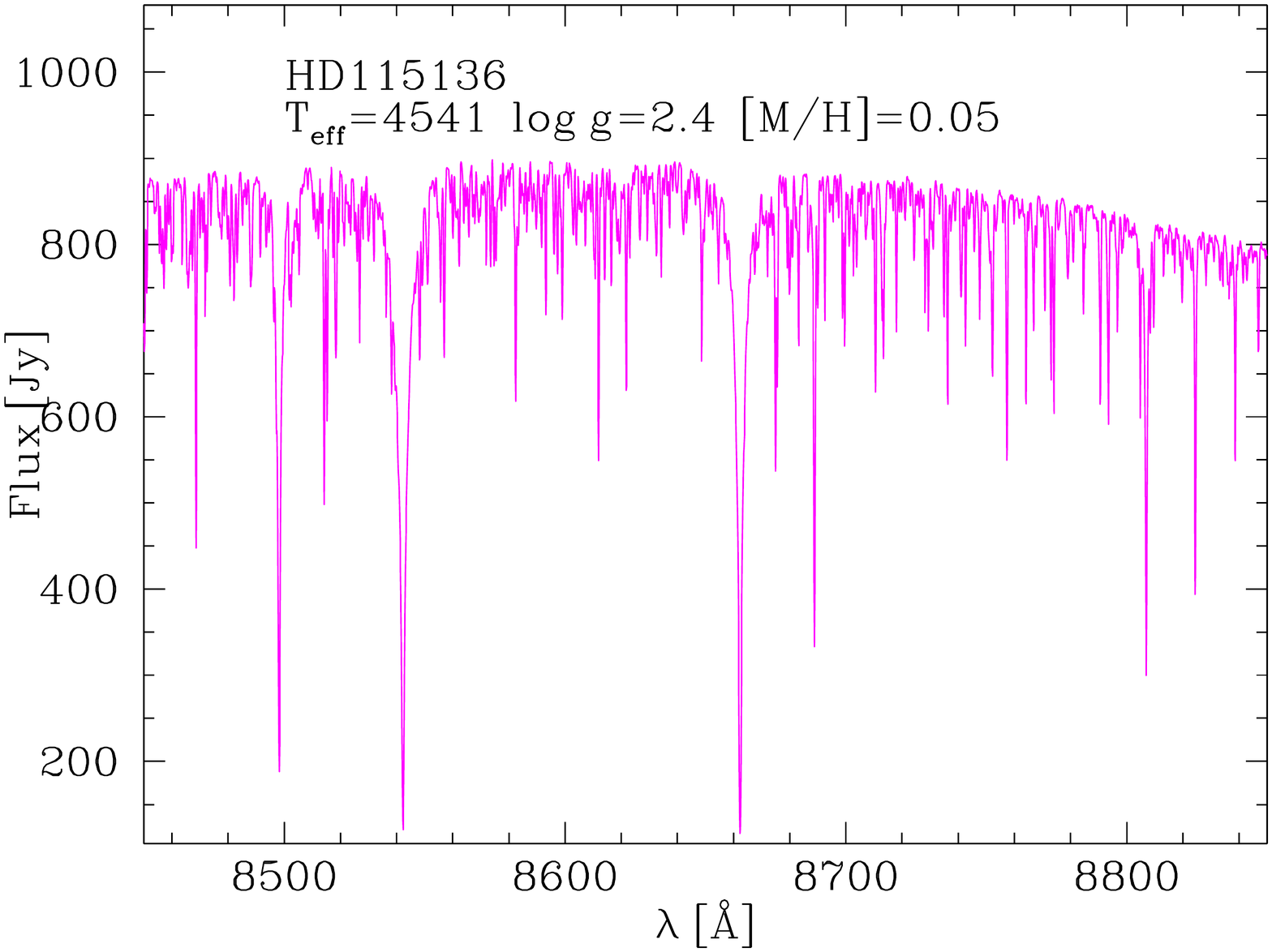}
\includegraphics[width=0.18\textwidth,angle=-90]{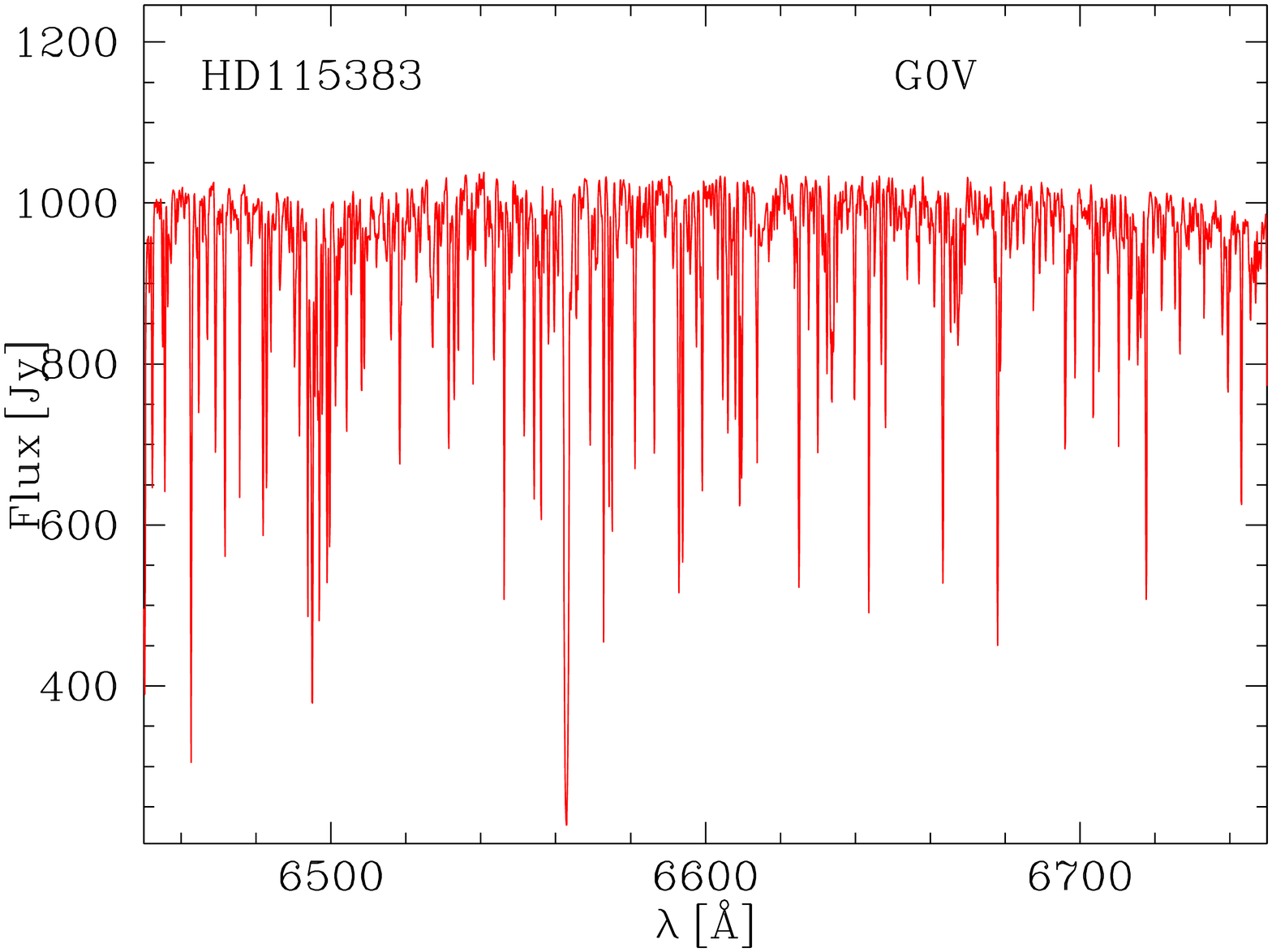}
\includegraphics[width=0.18\textwidth,angle=-90]{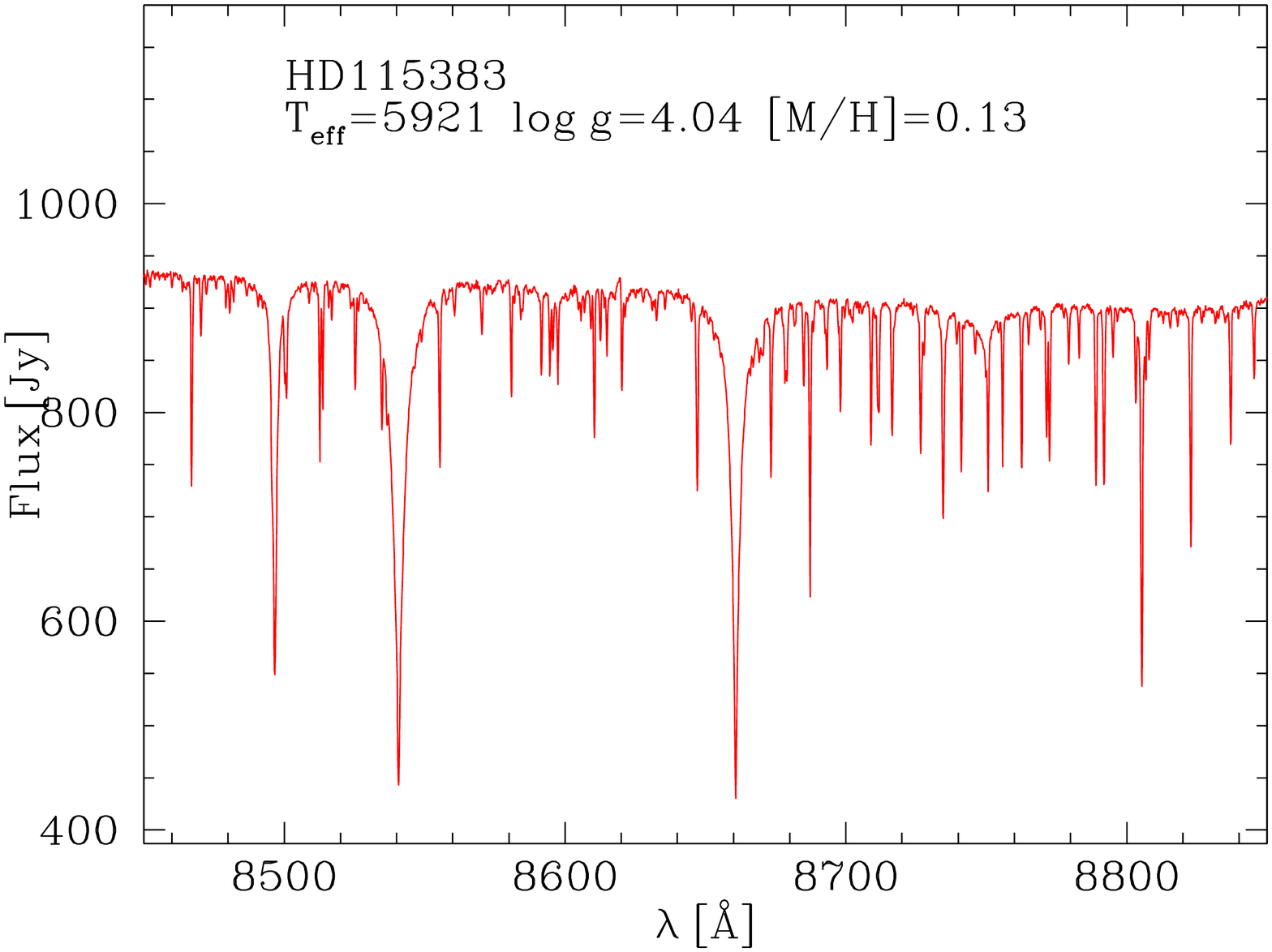}
\includegraphics[width=0.18\textwidth,angle=-90]{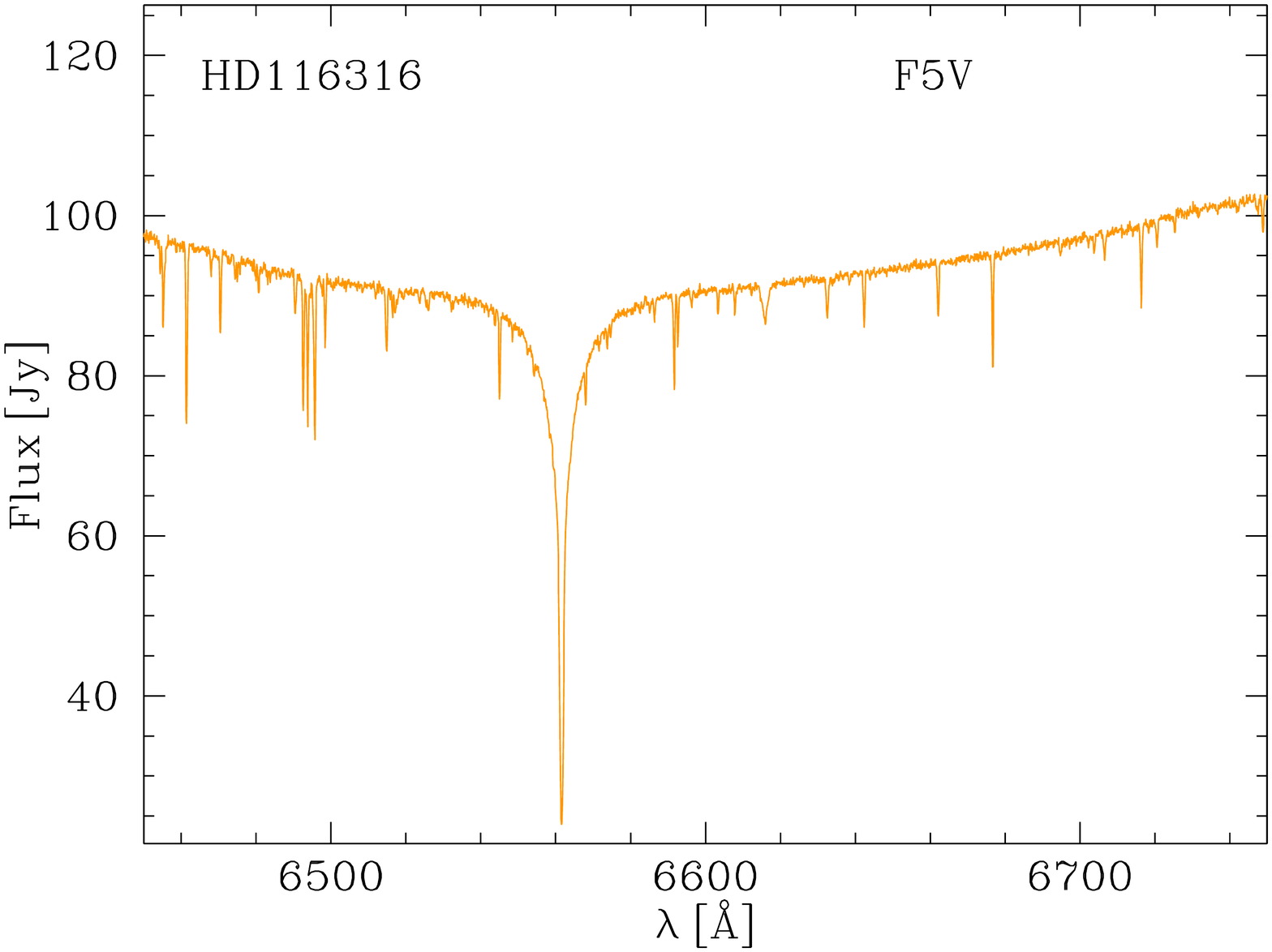}
\includegraphics[width=0.18\textwidth,angle=-90]{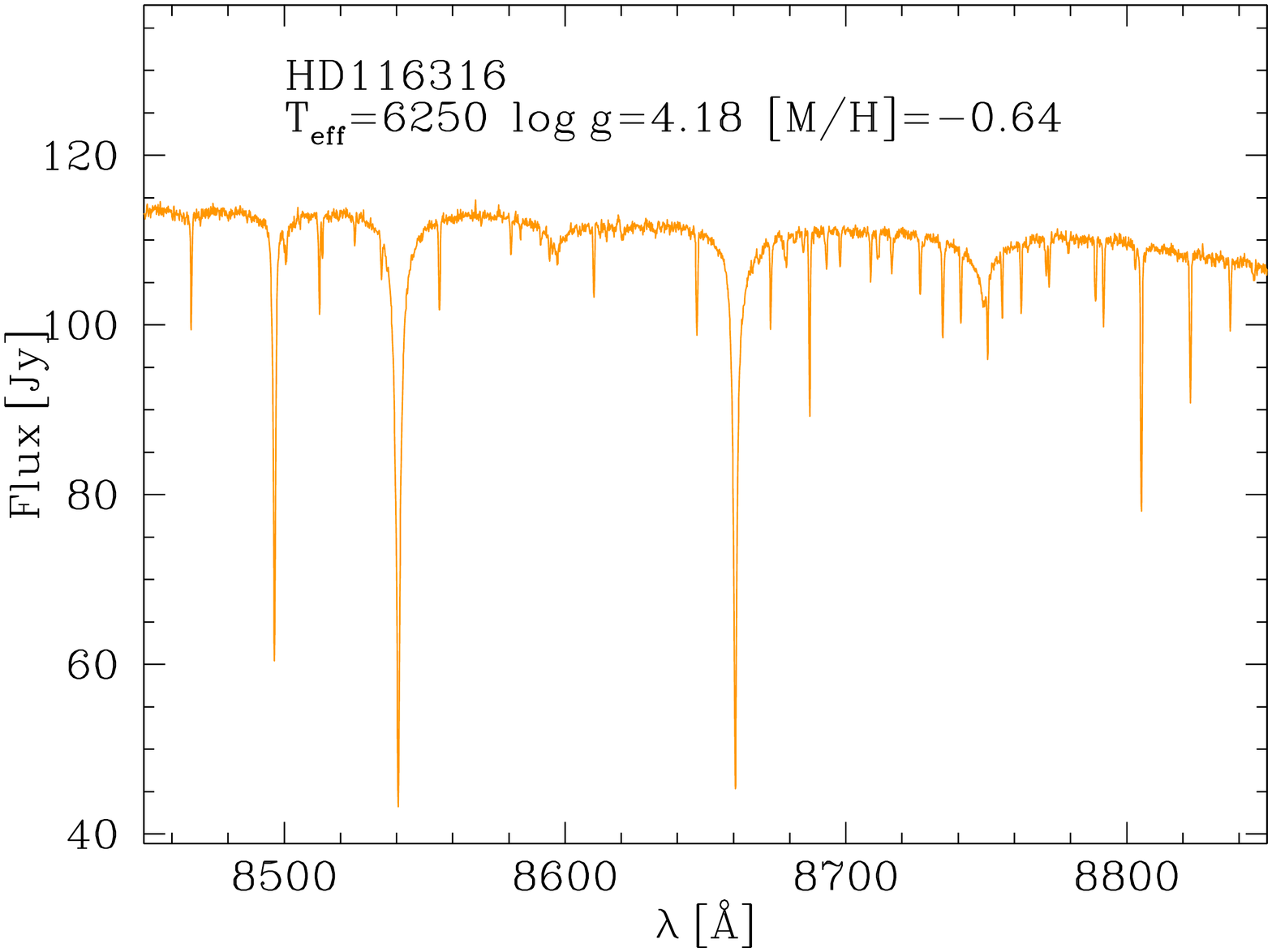}

\contcaption{21. Stars shown in this page are: HD107328, HD107582, HD108177, HD109358, HD109995, HD110897, HD112735, HD113002, HD114606, HD114710, HD114762, HD115136, HD115383 and HD116316.}
\end{figure*}

\begin{figure*}
\includegraphics[width=0.18\textwidth,angle=-90]{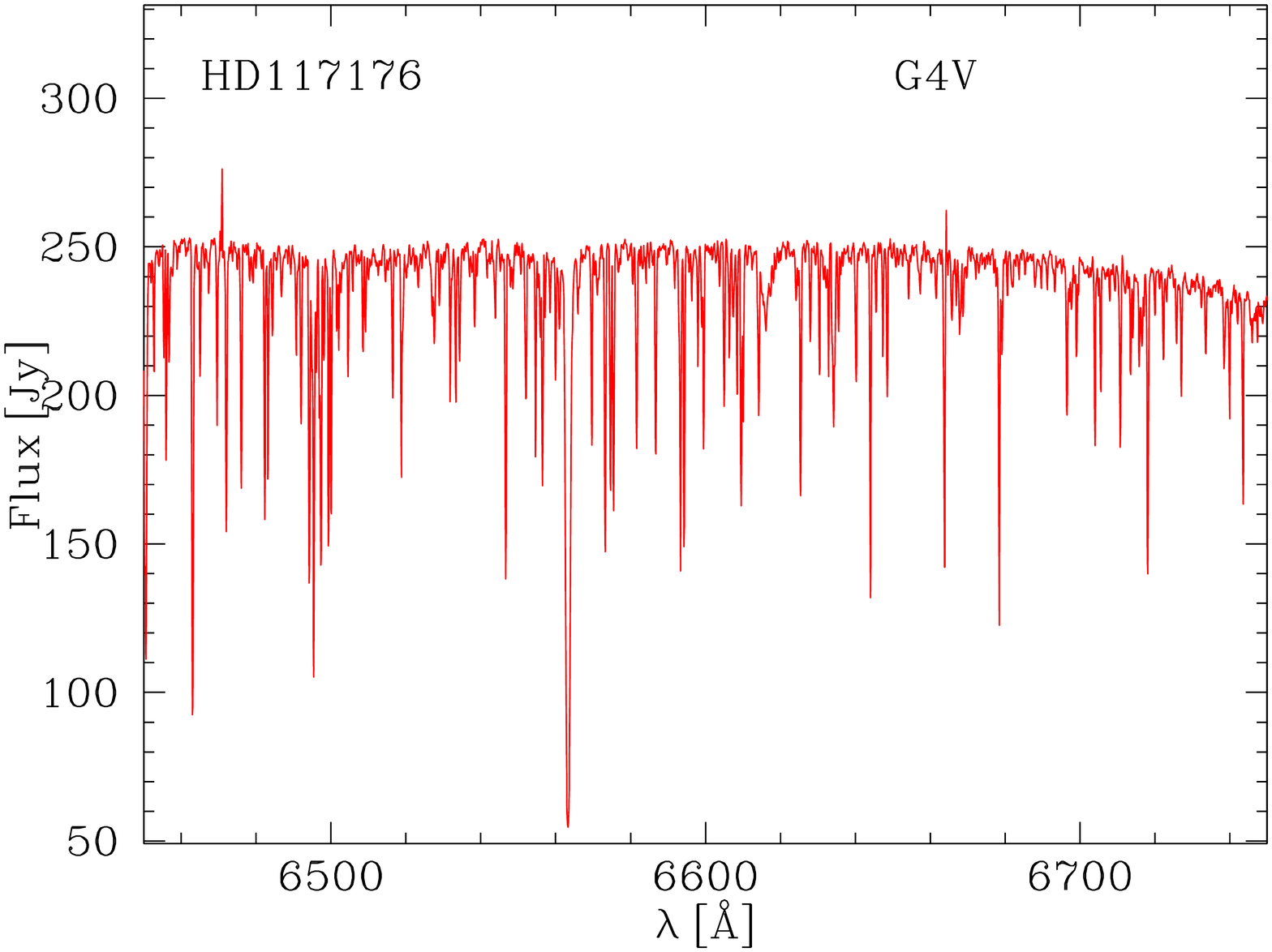}
\includegraphics[width=0.18\textwidth,angle=-90]{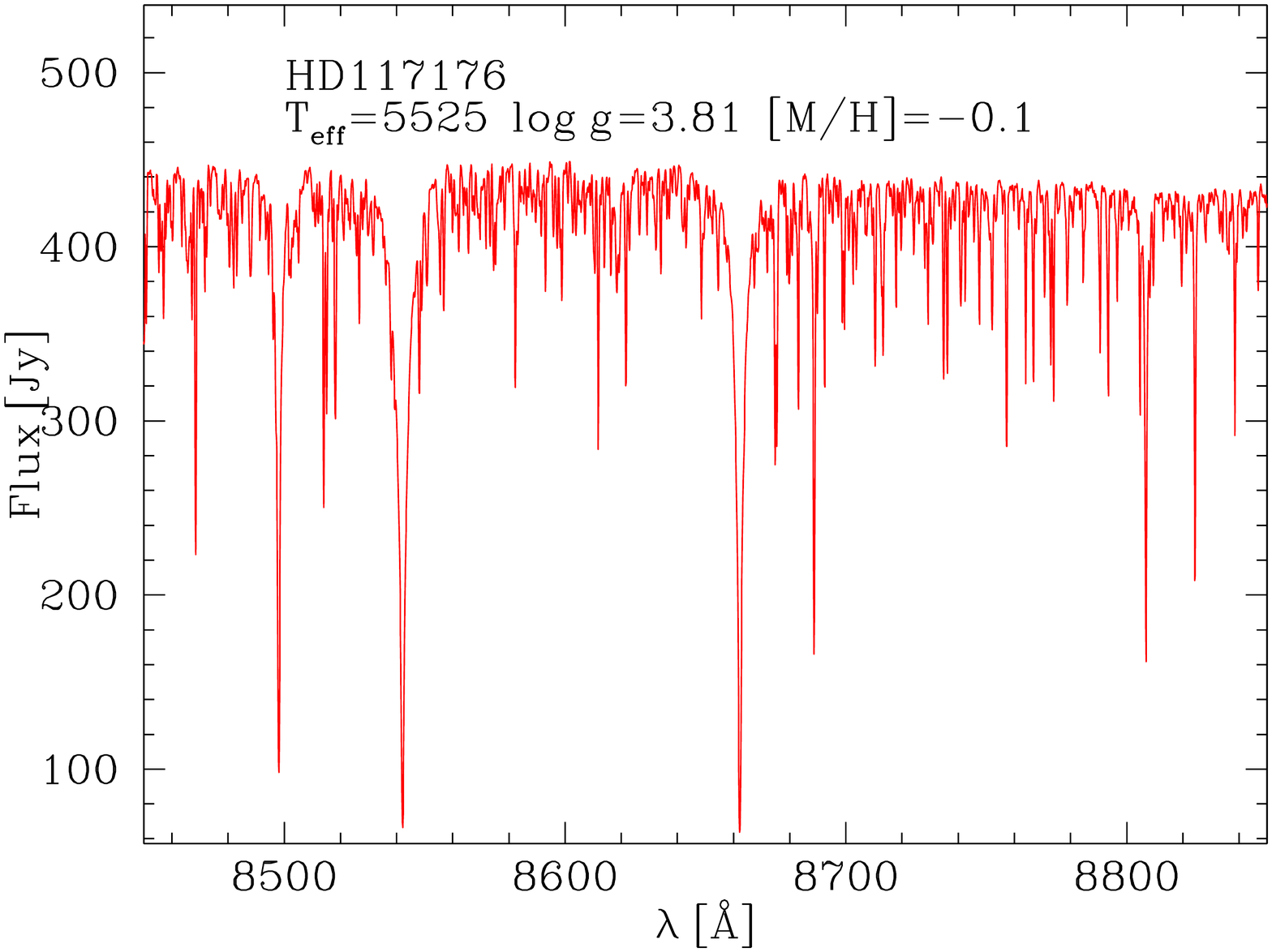}
\includegraphics[width=0.18\textwidth,angle=-90]{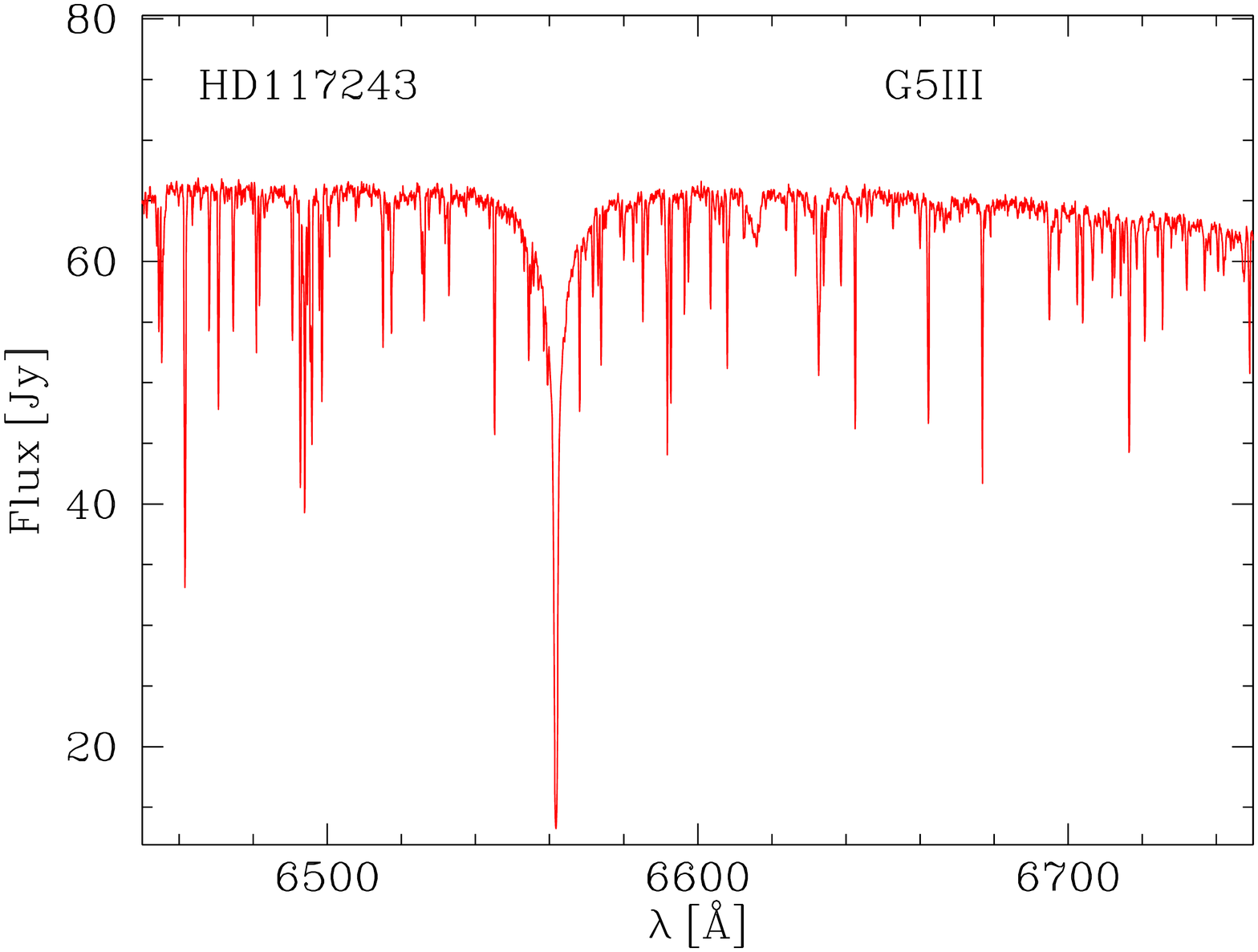}
\includegraphics[width=0.18\textwidth,angle=-90]{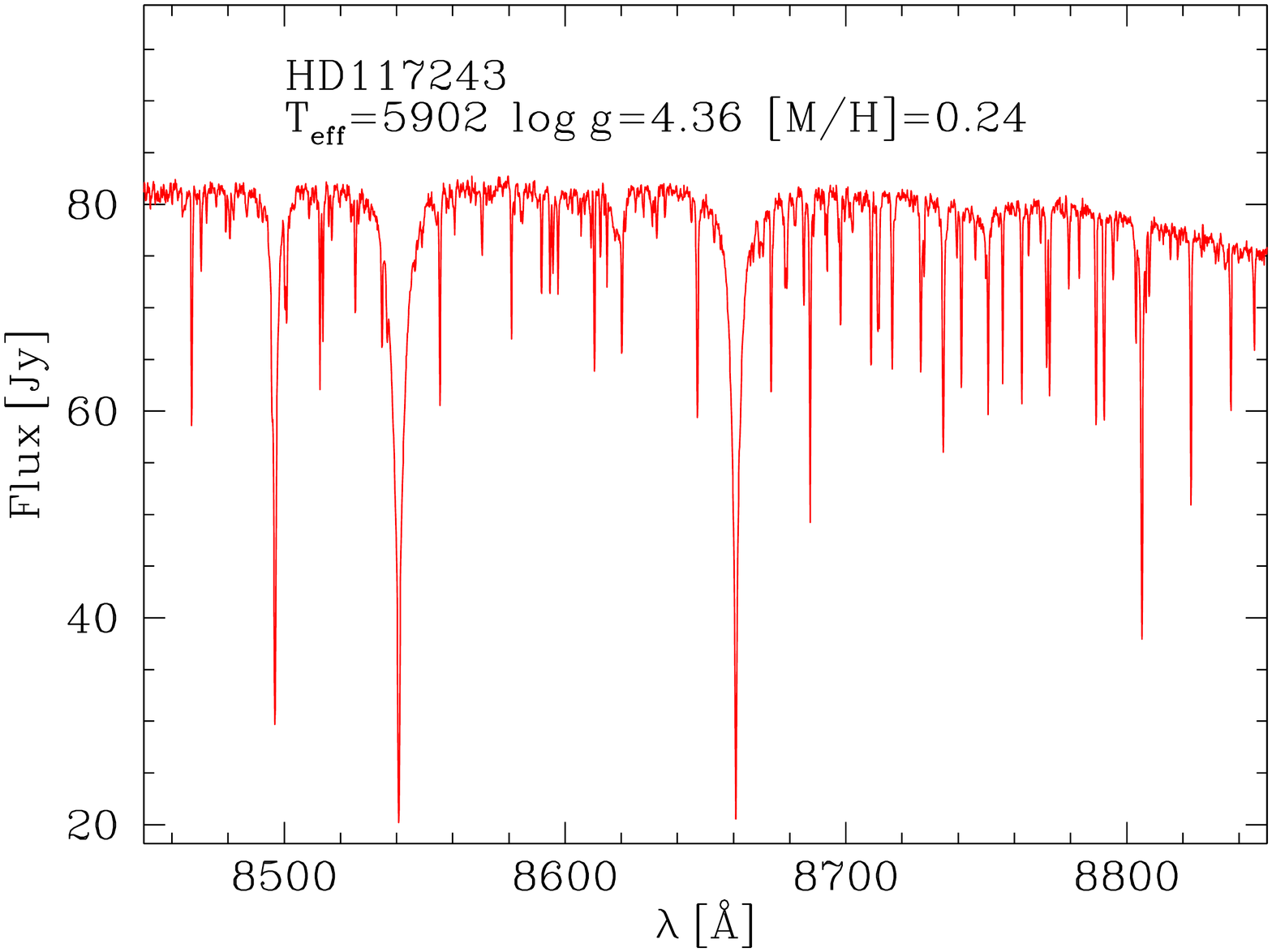}
\includegraphics[width=0.18\textwidth,angle=-90]{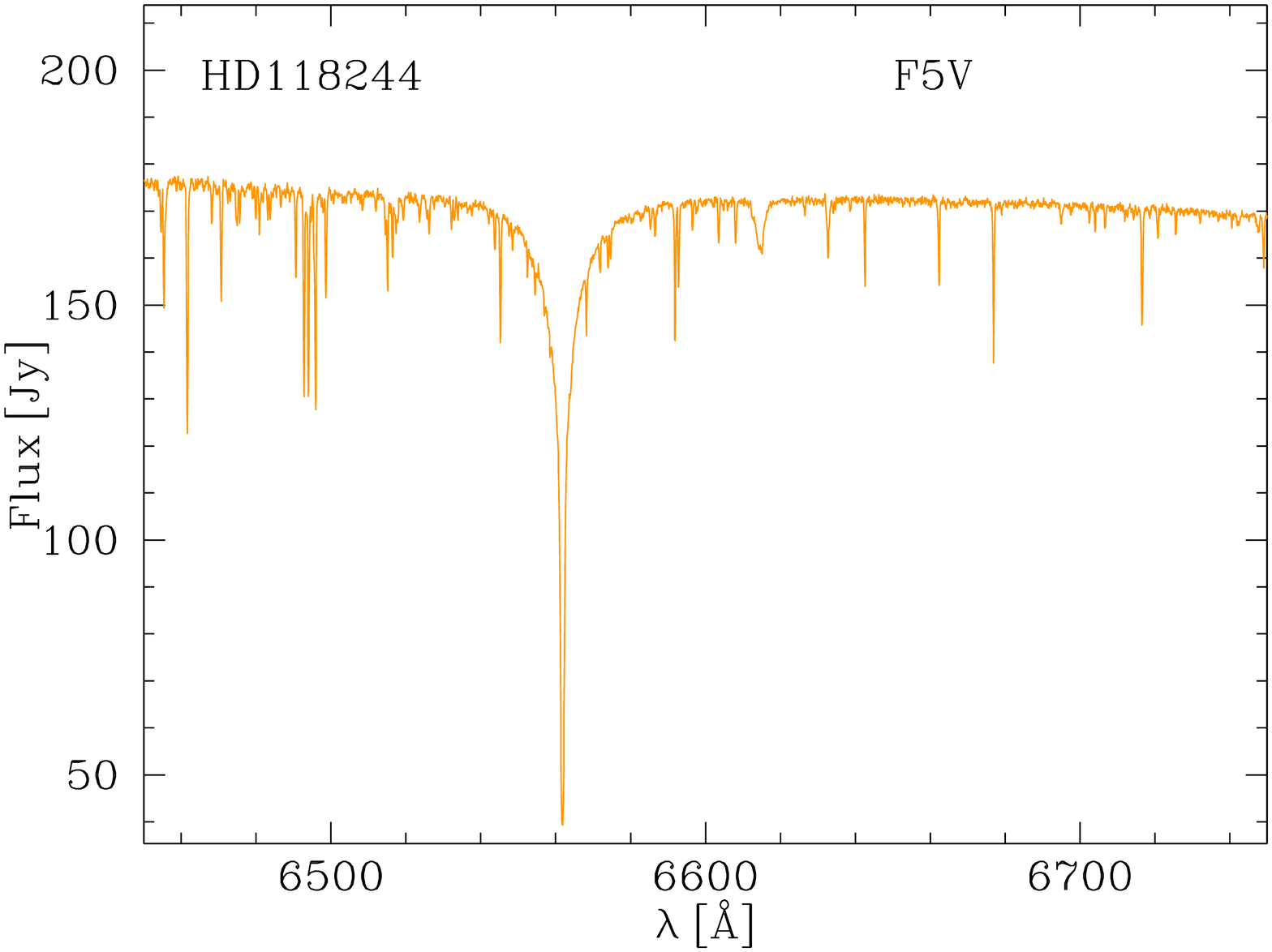}
\includegraphics[width=0.18\textwidth,angle=-90]{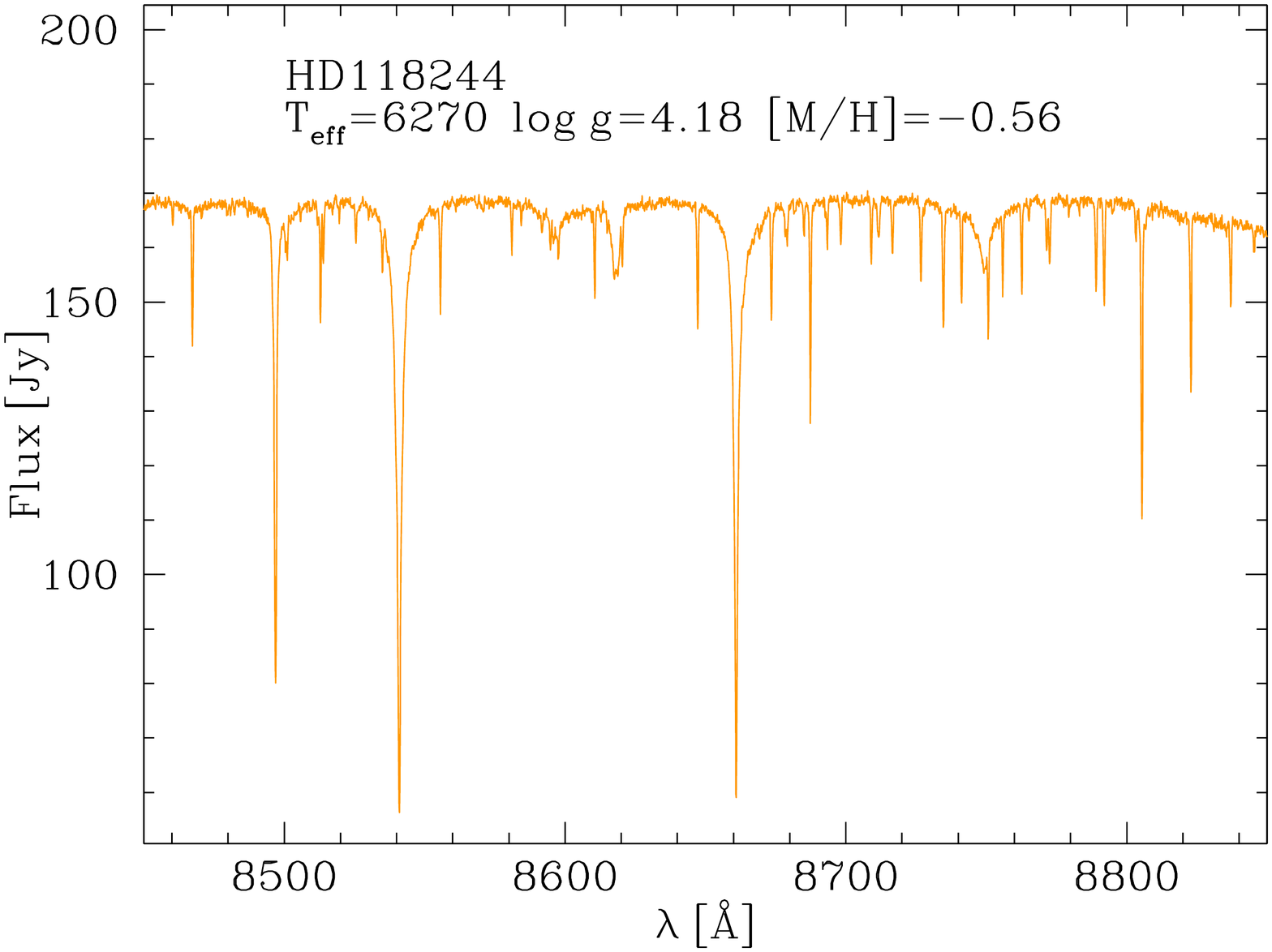}
\includegraphics[width=0.18\textwidth,angle=-90]{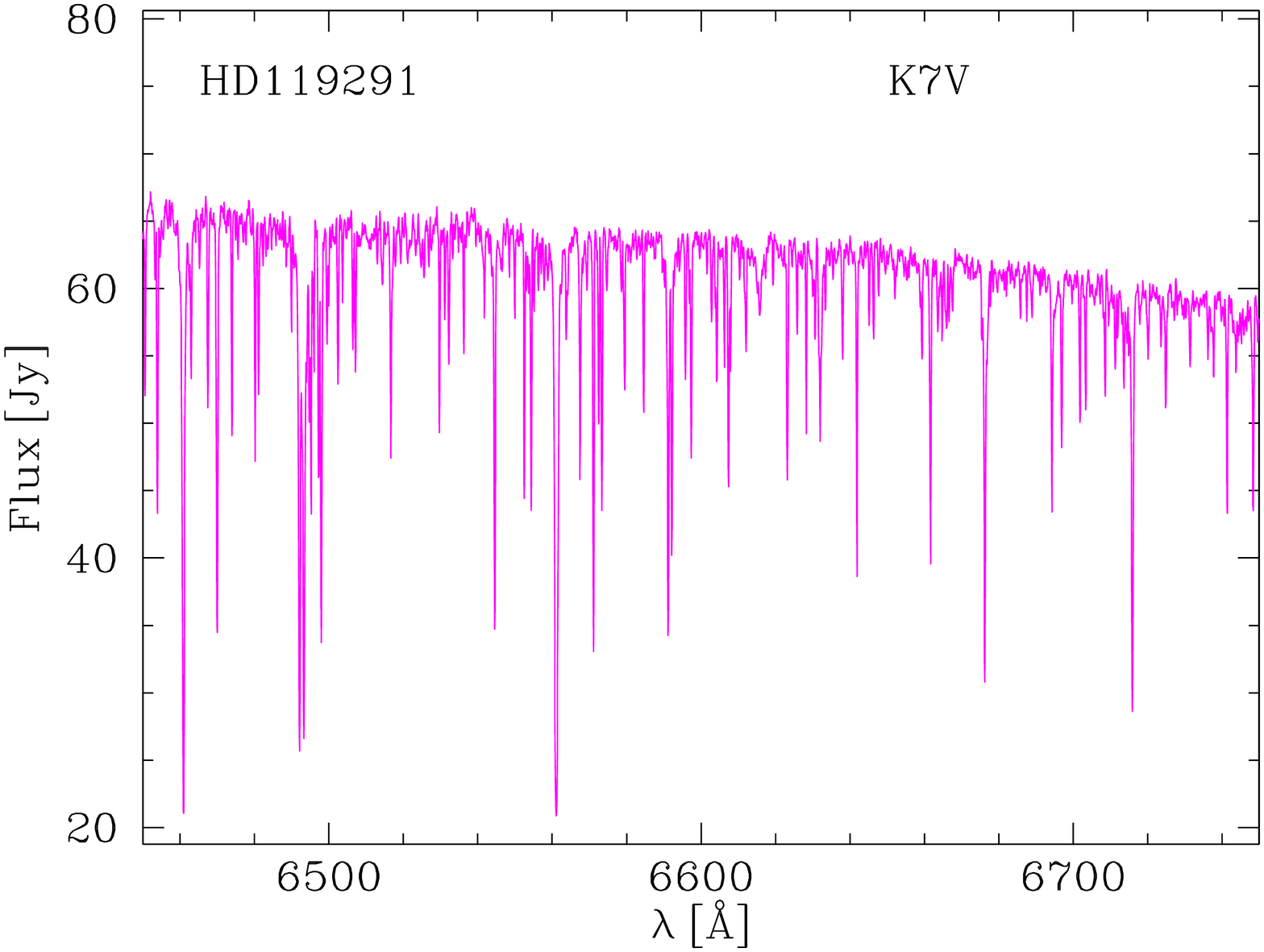}
\includegraphics[width=0.18\textwidth,angle=-90]{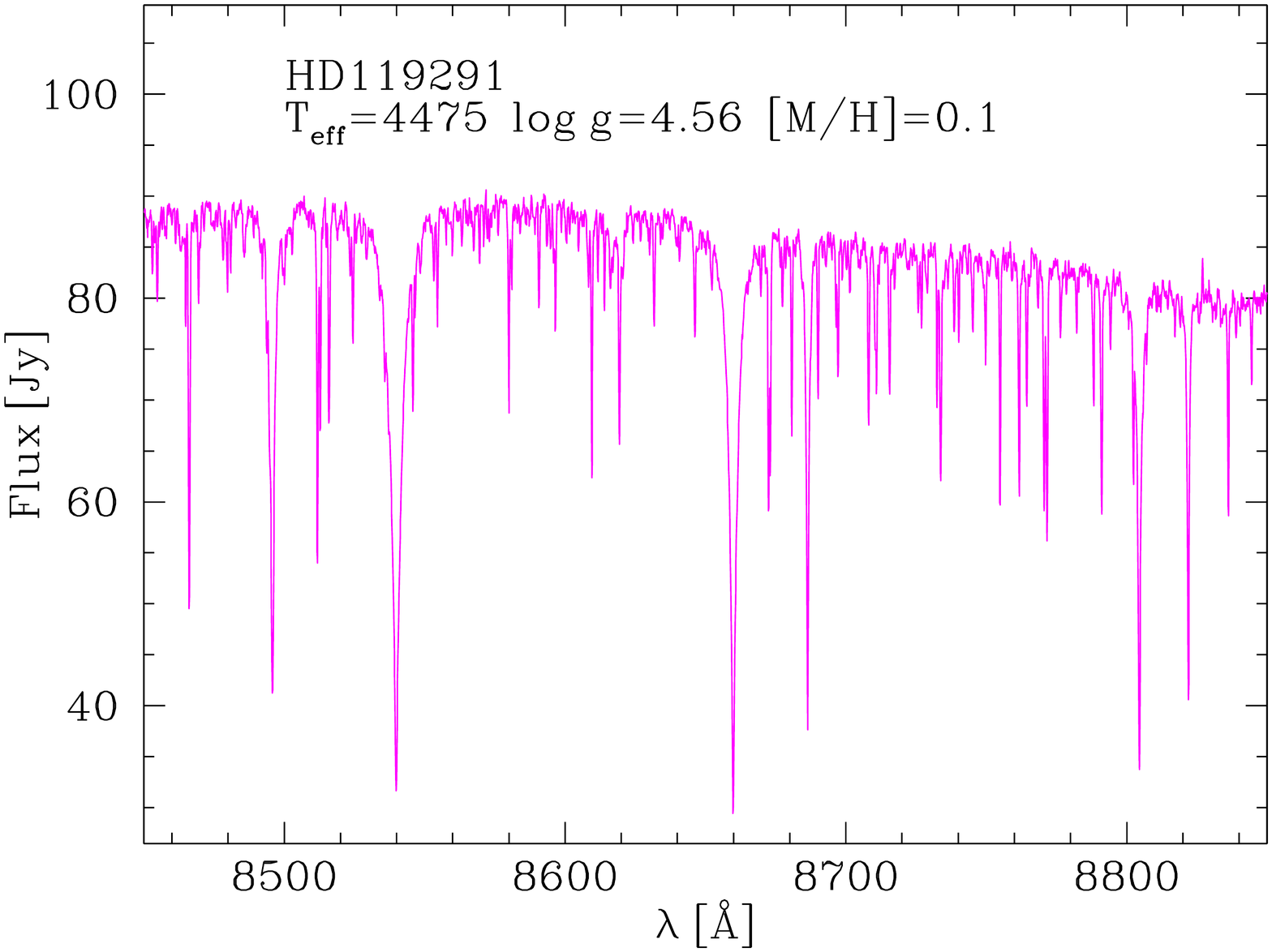}
\includegraphics[width=0.18\textwidth,angle=-90]{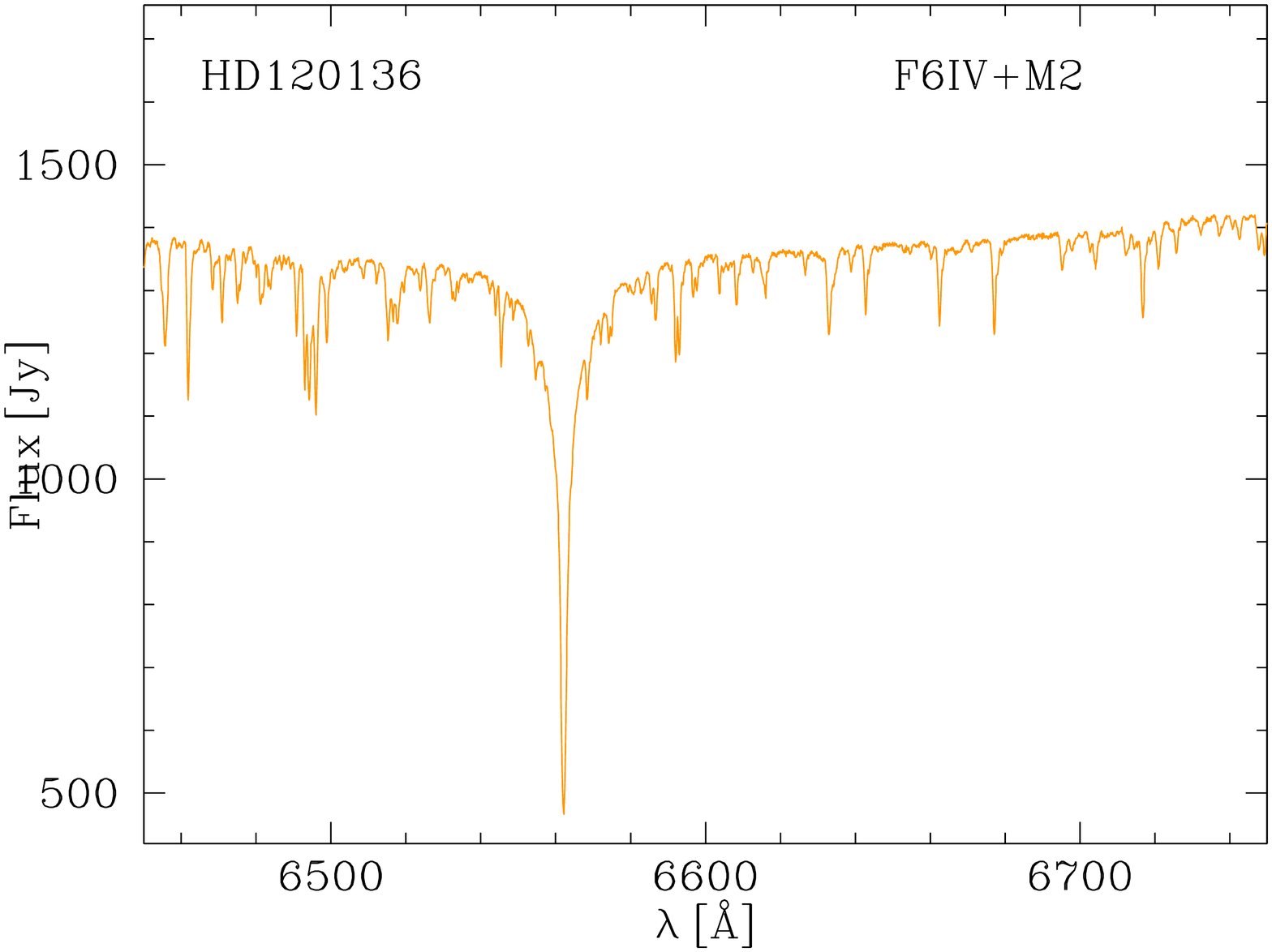}
\includegraphics[width=0.18\textwidth,angle=-90]{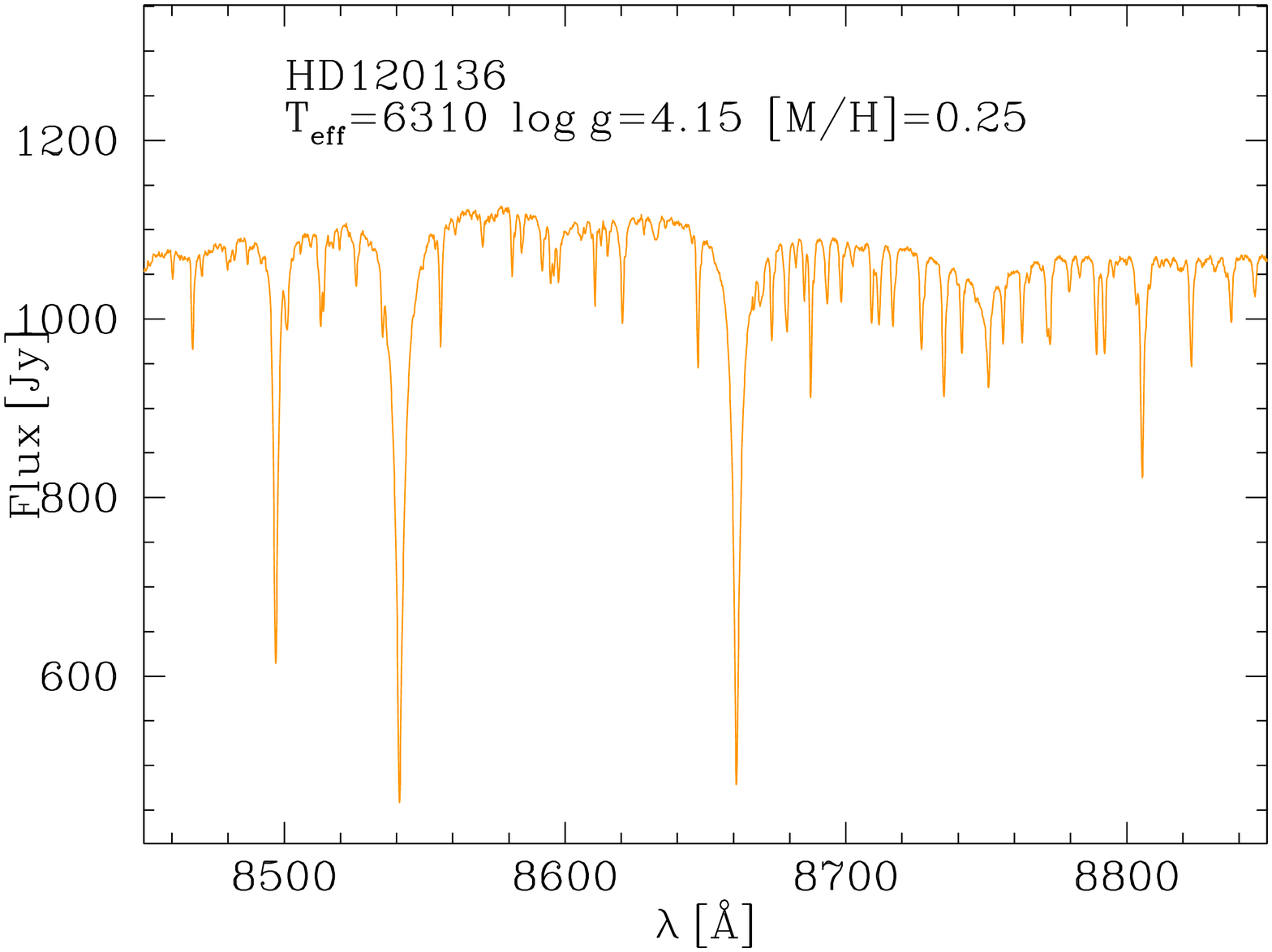}
\includegraphics[width=0.18\textwidth,angle=-90]{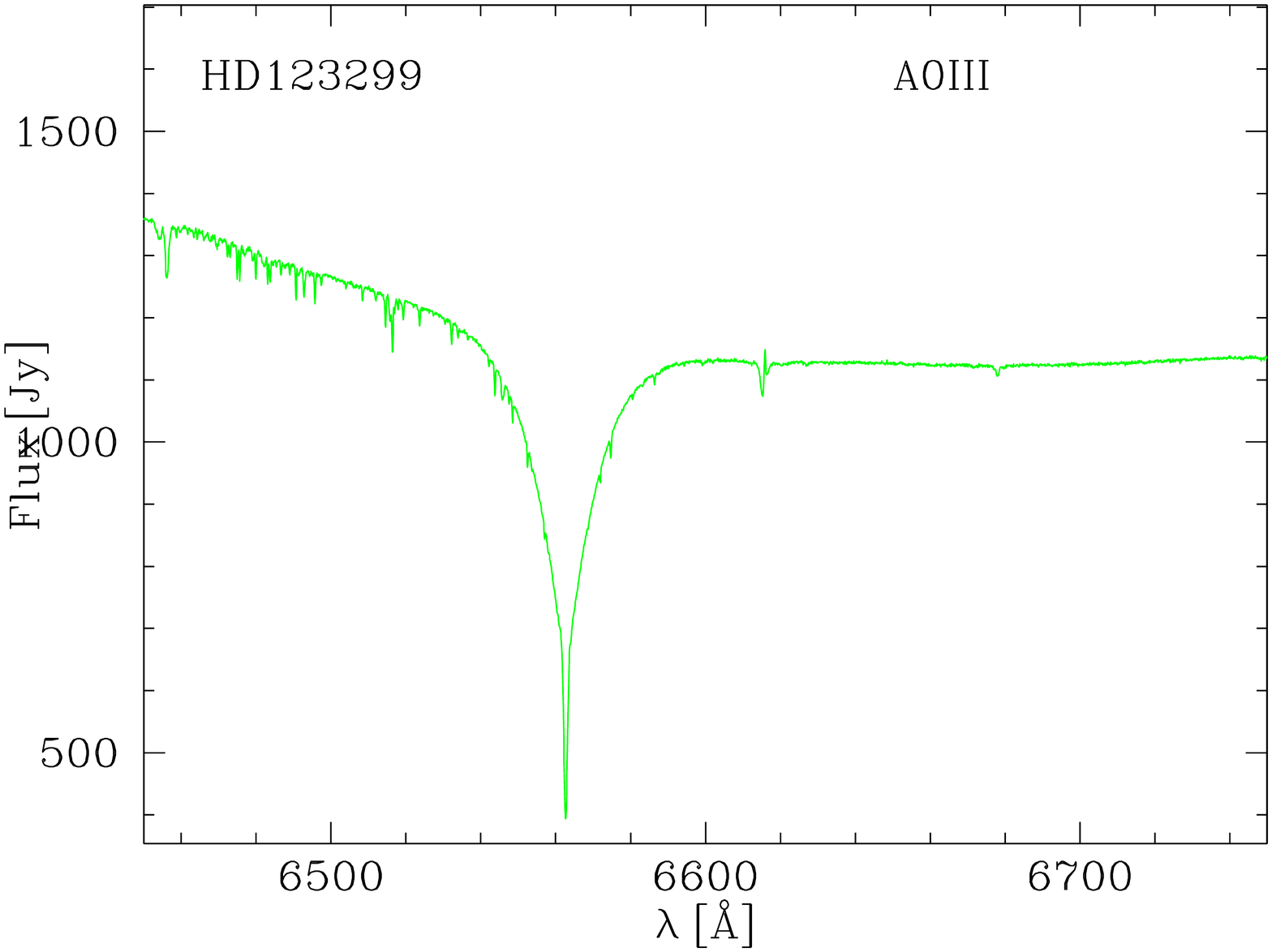}
\includegraphics[width=0.18\textwidth,angle=-90]{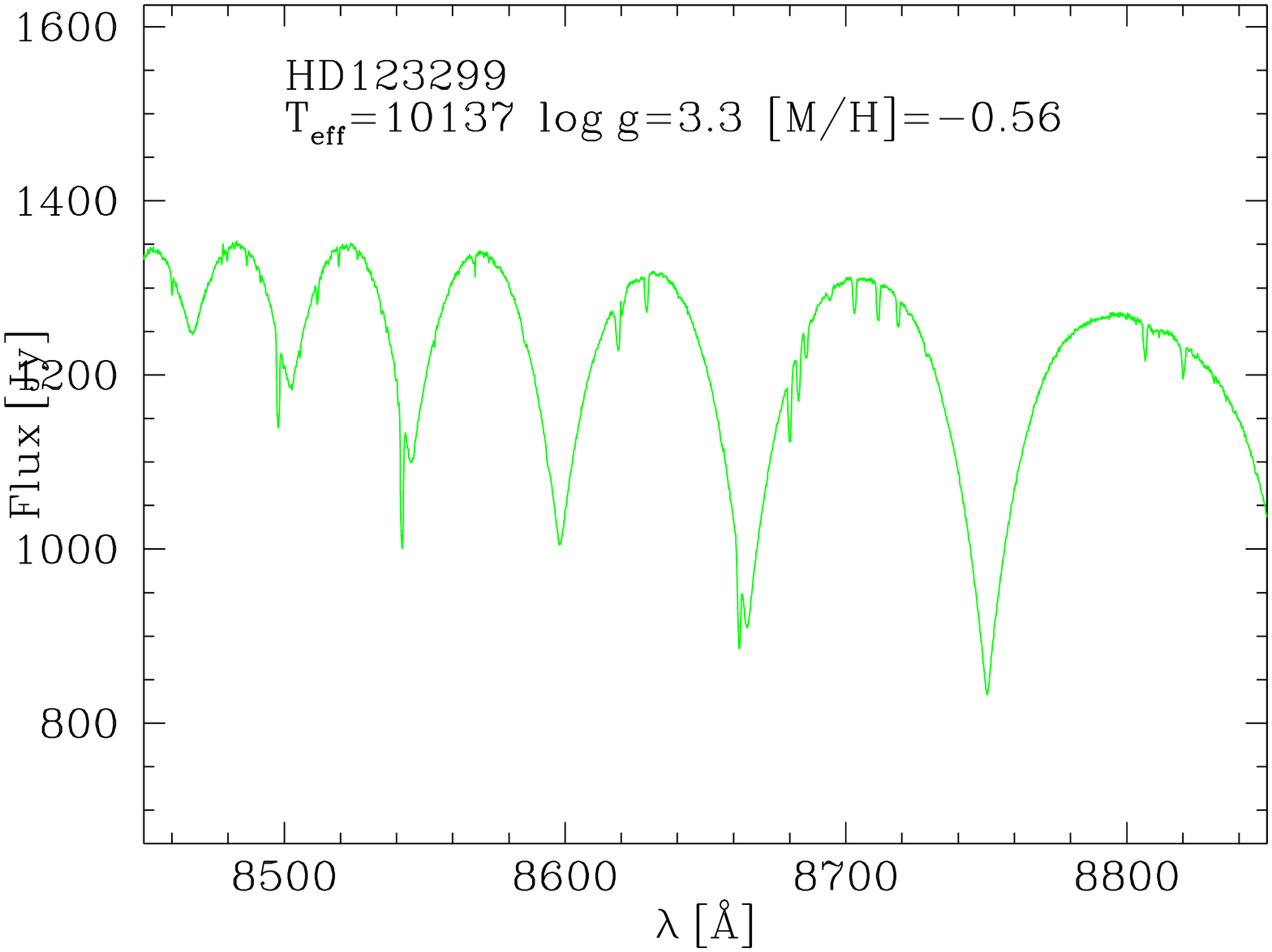}
\includegraphics[width=0.18\textwidth,angle=-90]{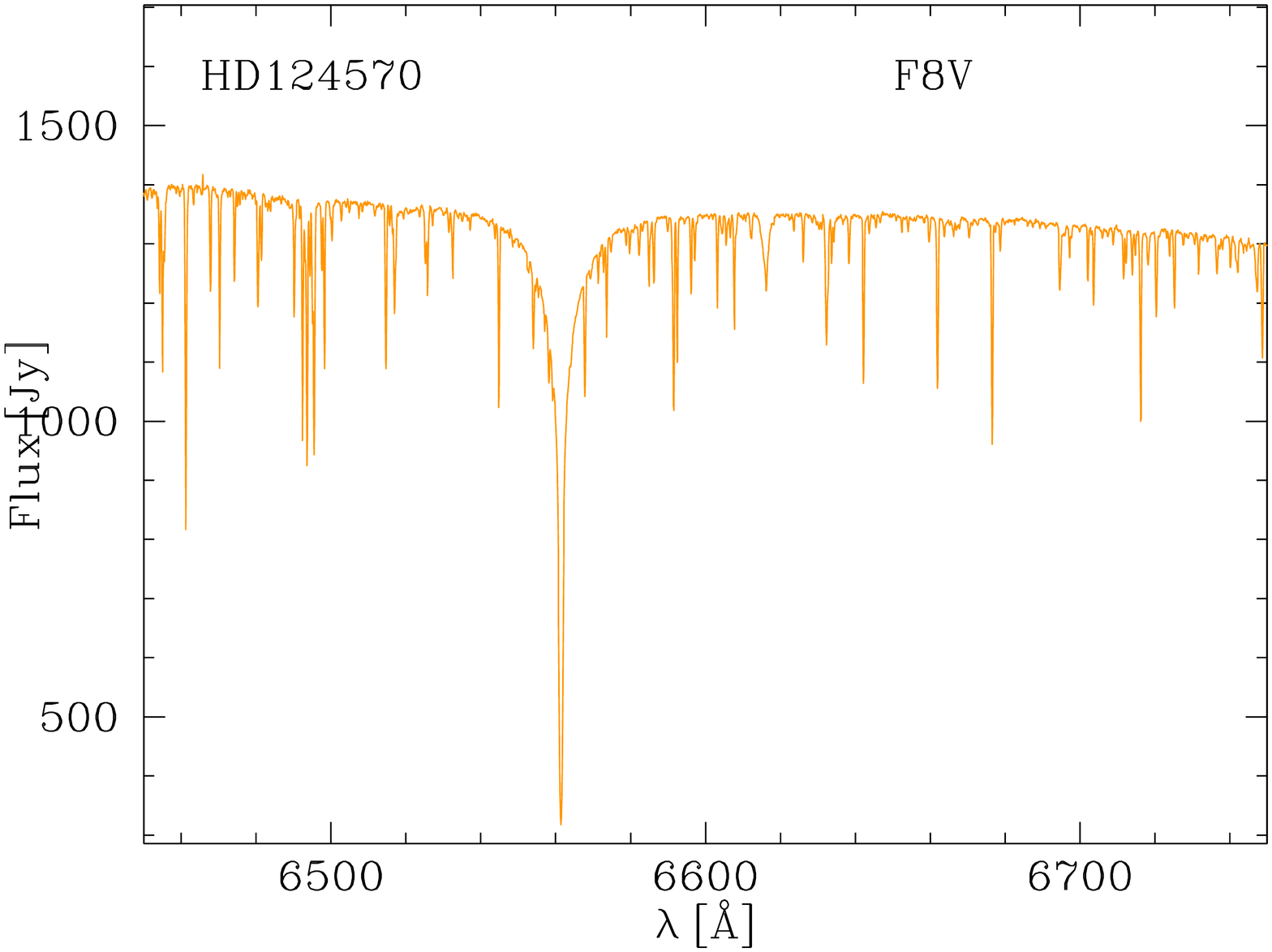}
\includegraphics[width=0.18\textwidth,angle=-90]{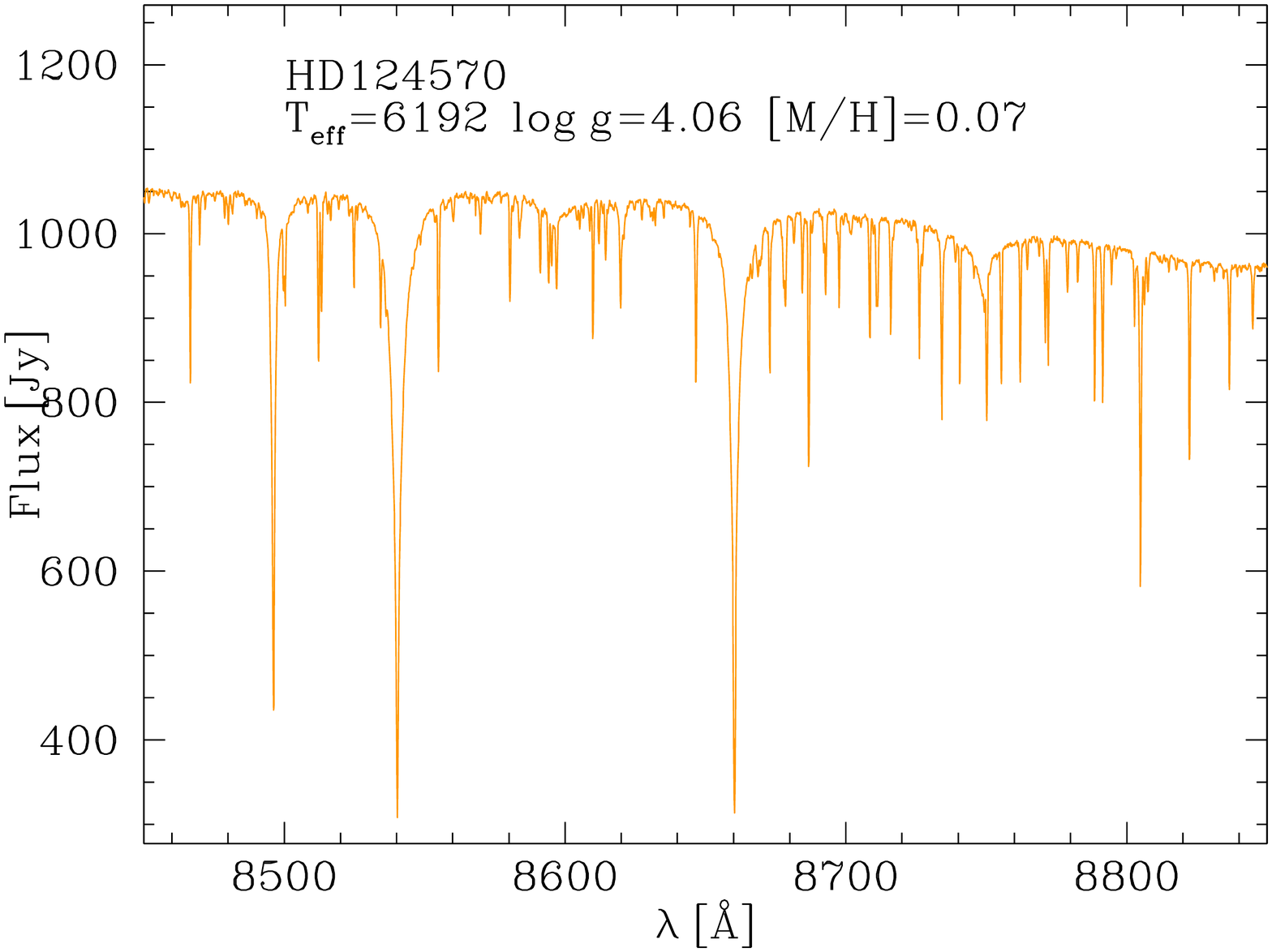}
\includegraphics[width=0.18\textwidth,angle=-90]{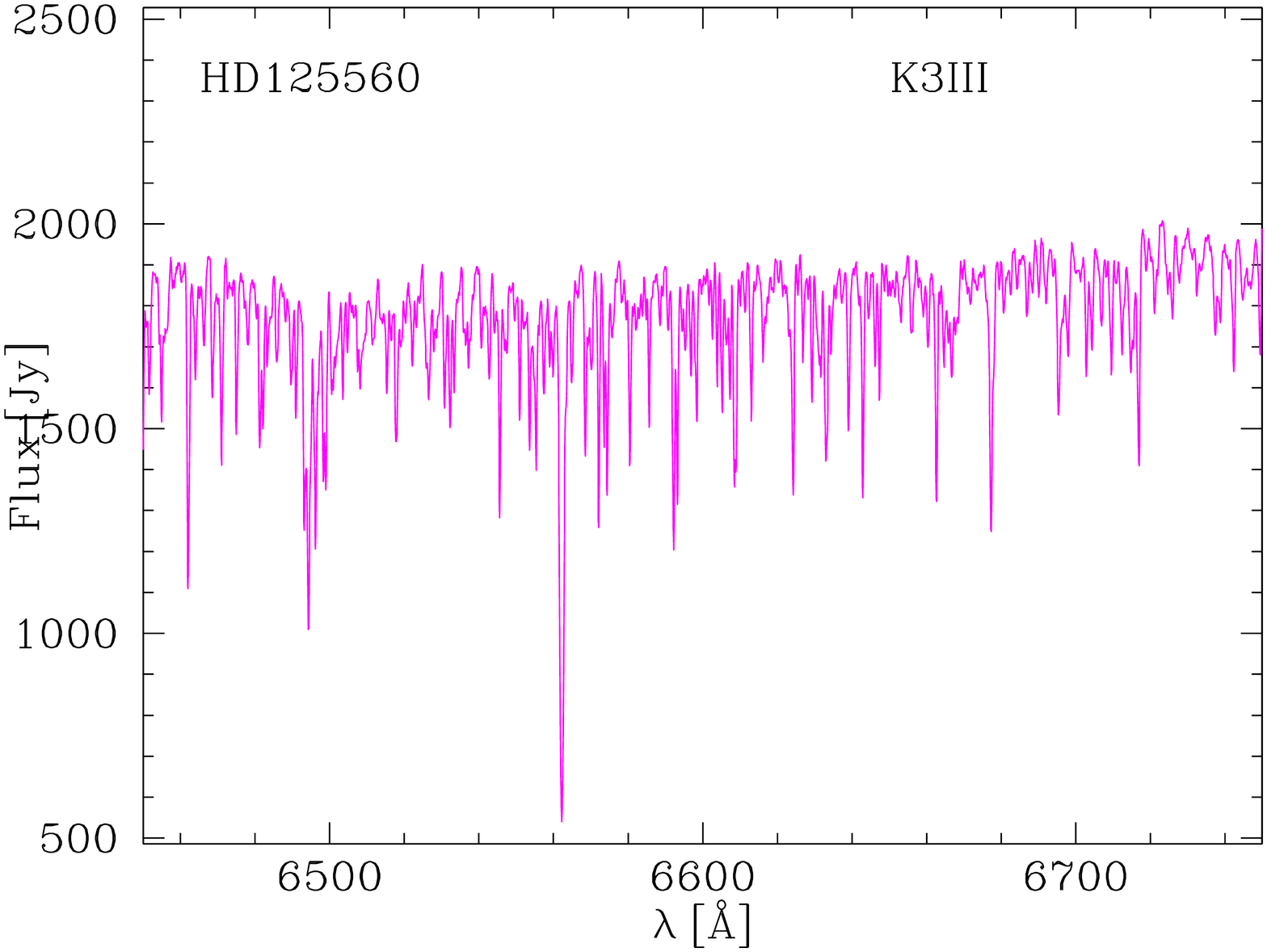}
\includegraphics[width=0.18\textwidth,angle=-90]{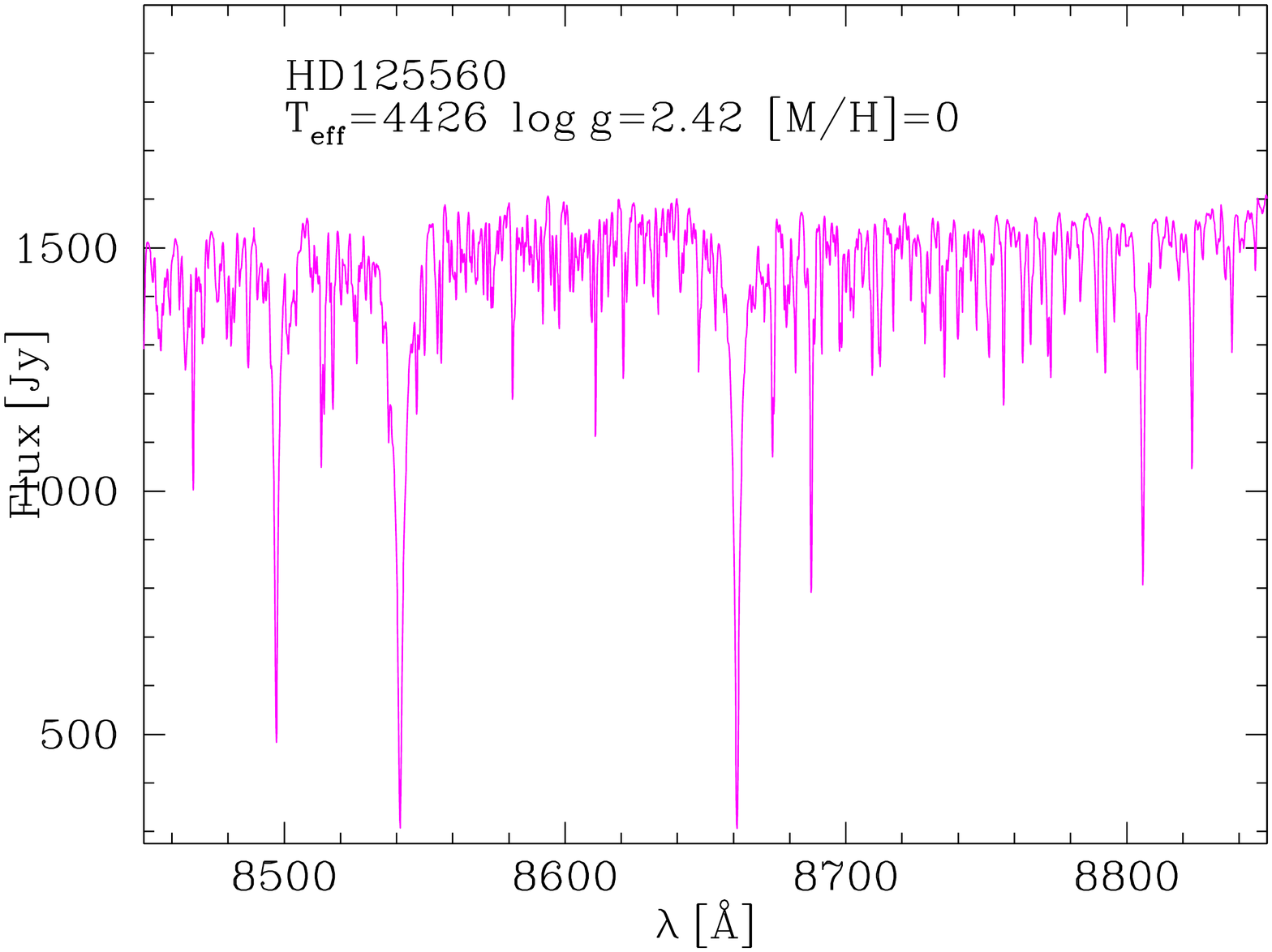}
\includegraphics[width=0.18\textwidth,angle=-90]{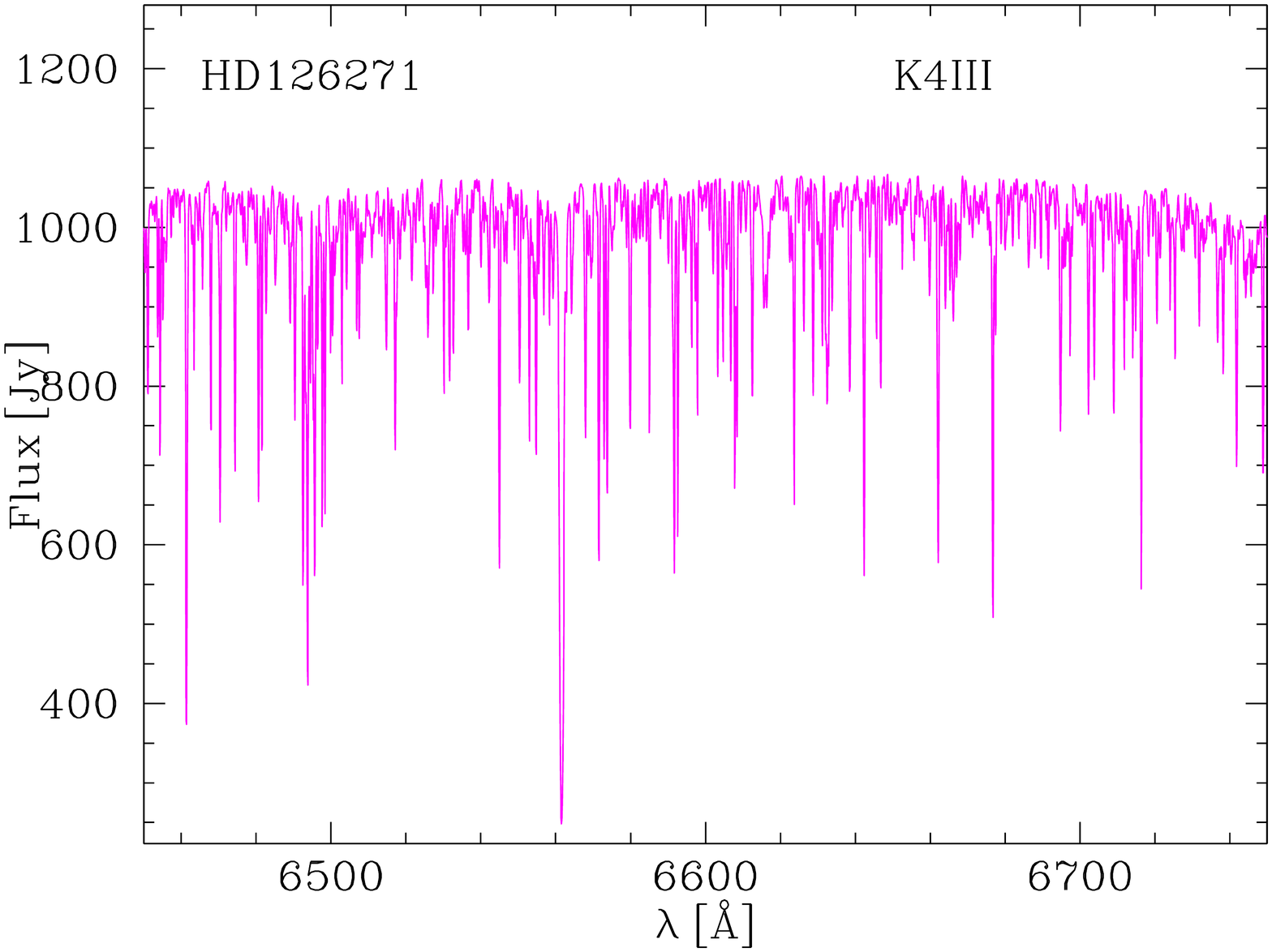}
\includegraphics[width=0.18\textwidth,angle=-90]{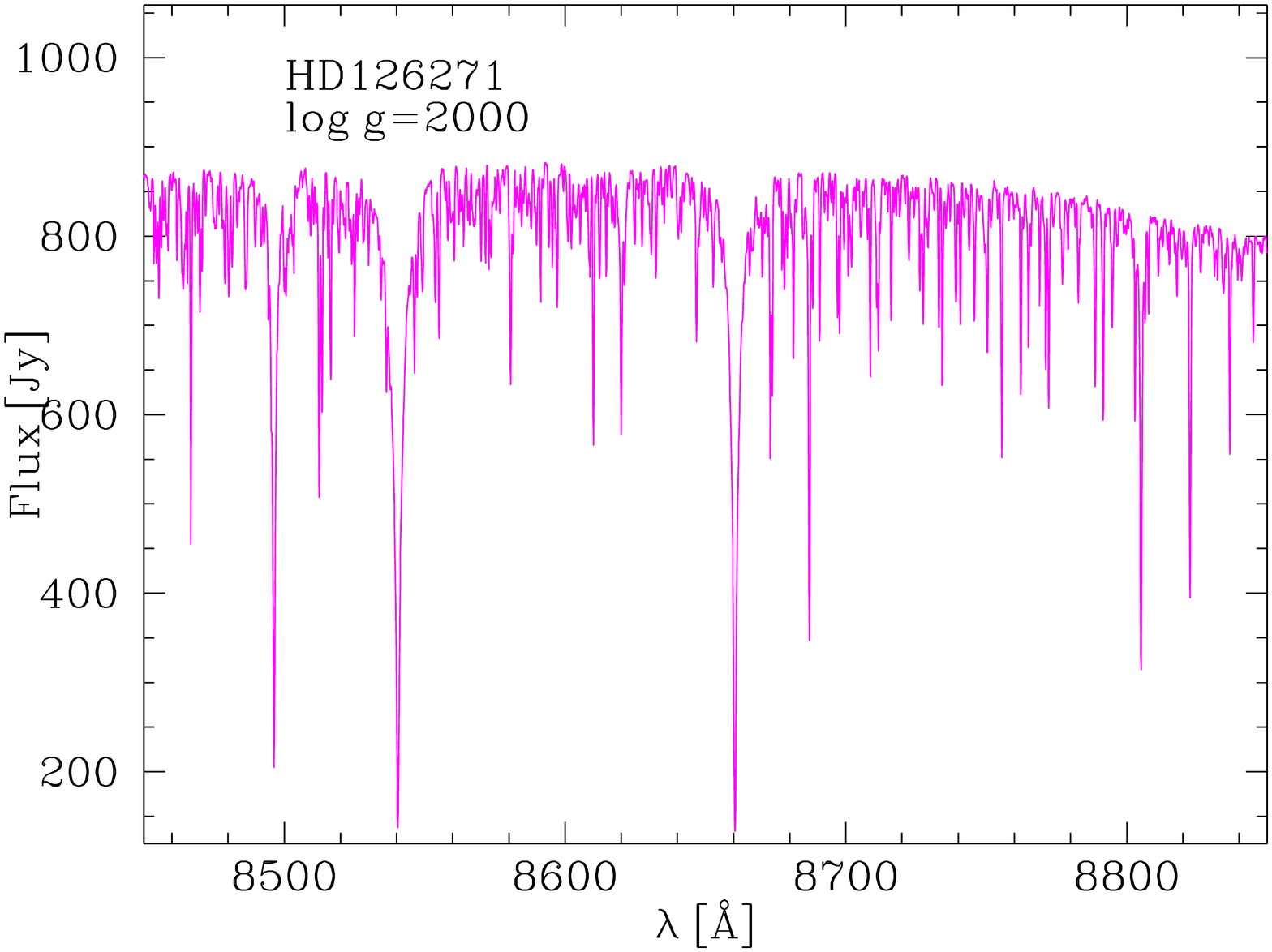}
\includegraphics[width=0.18\textwidth,angle=-90]{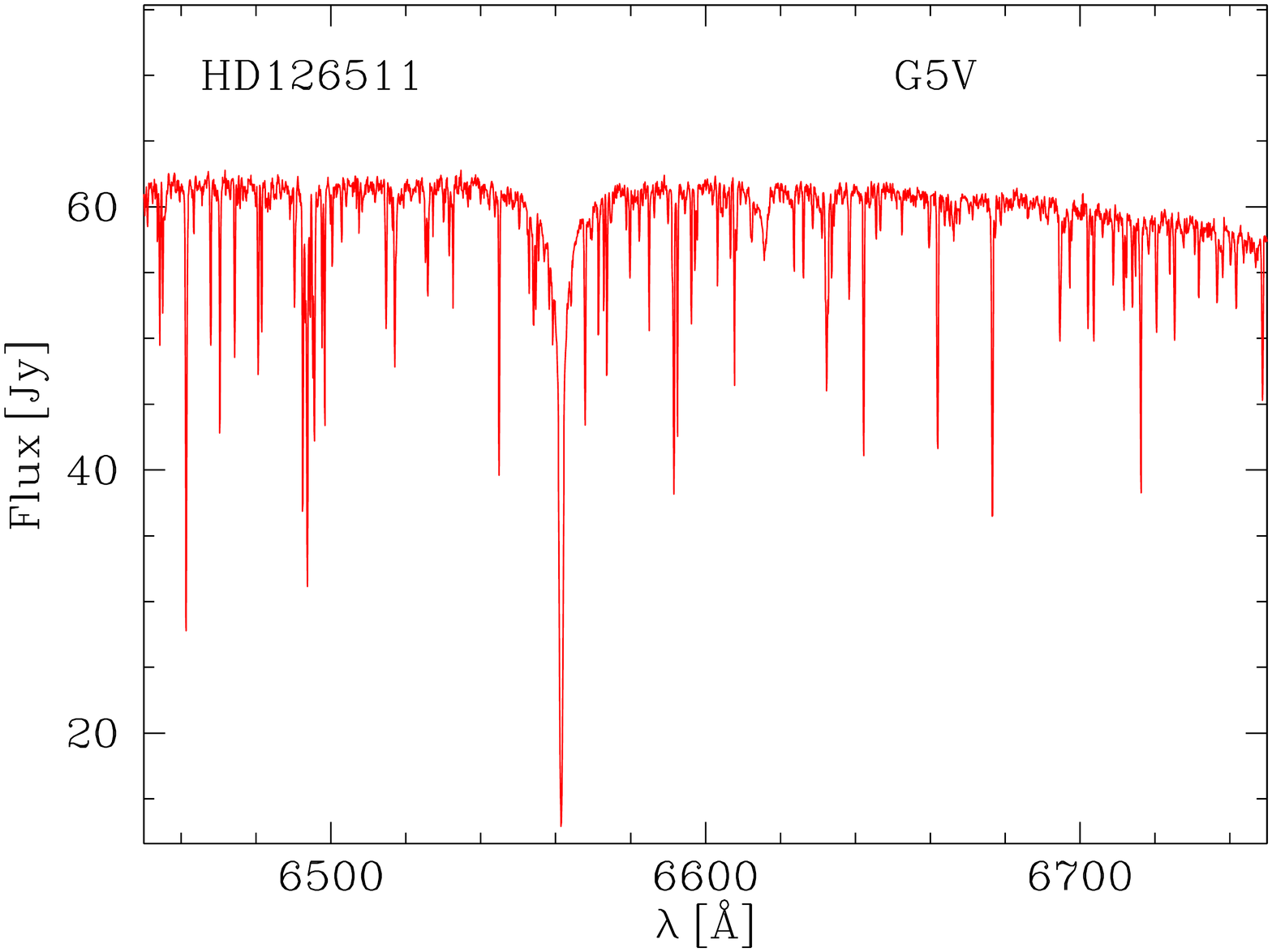}
\includegraphics[width=0.18\textwidth,angle=-90]{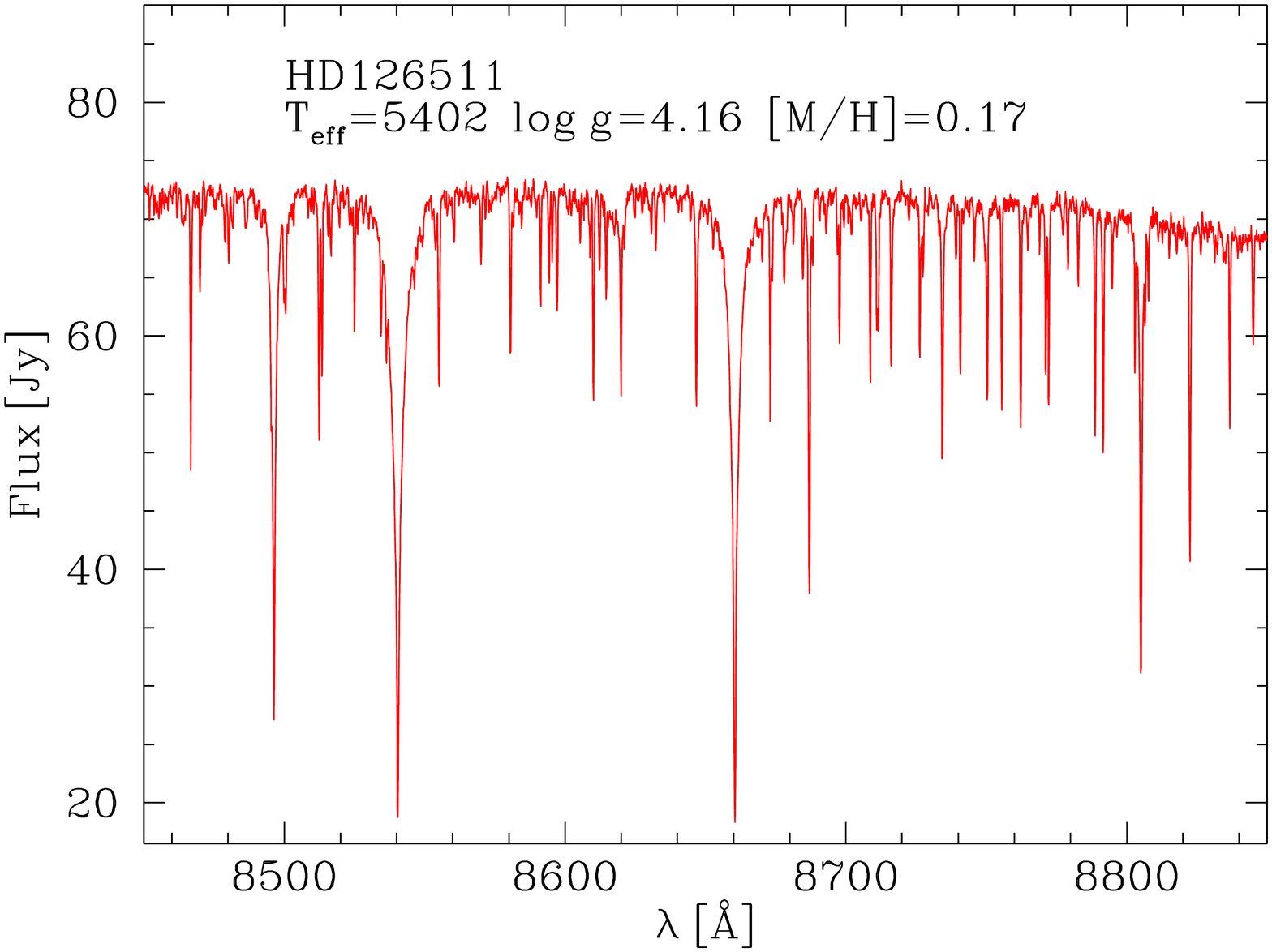}
\includegraphics[width=0.18\textwidth,angle=-90]{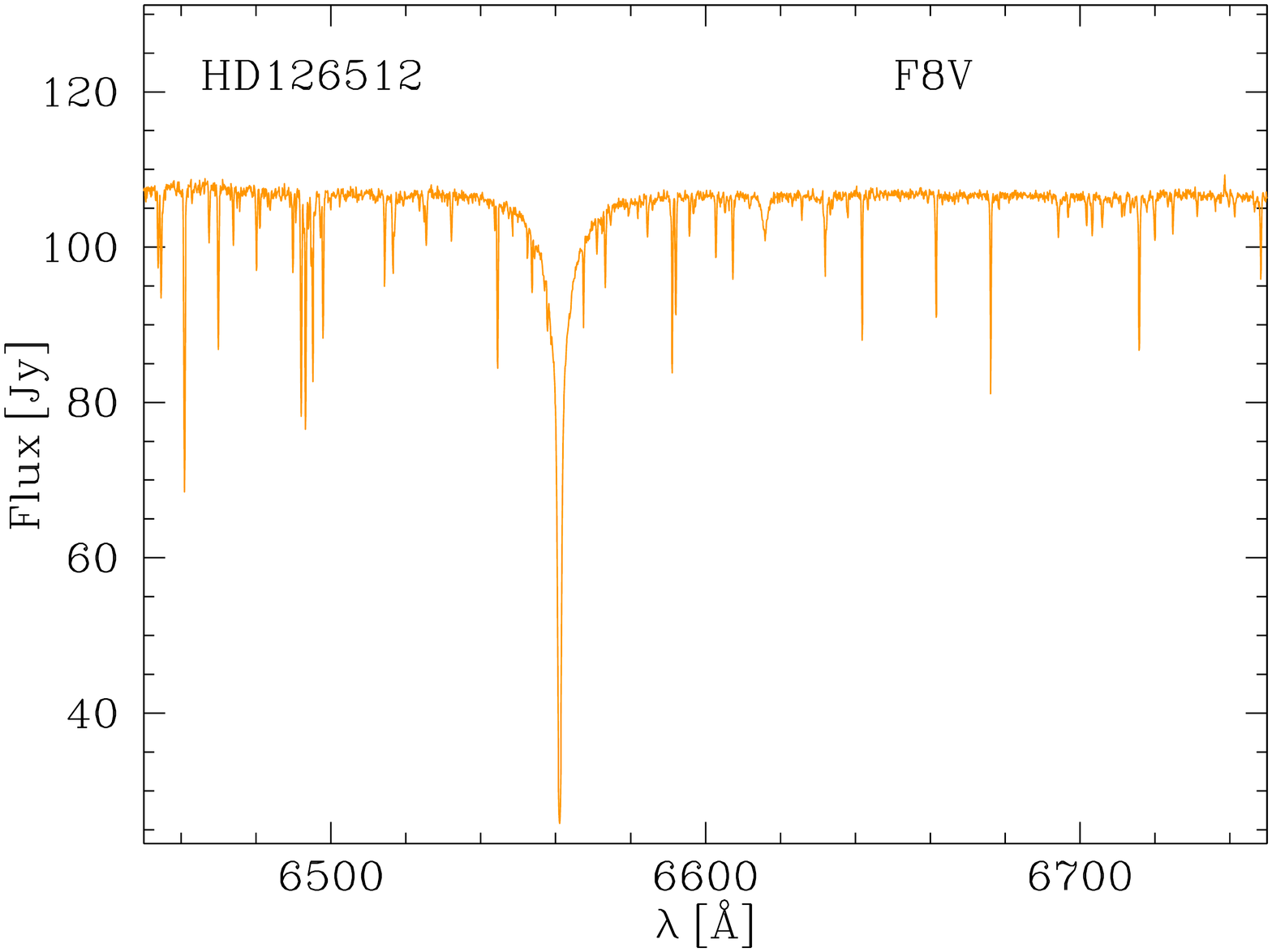}
\includegraphics[width=0.18\textwidth,angle=-90]{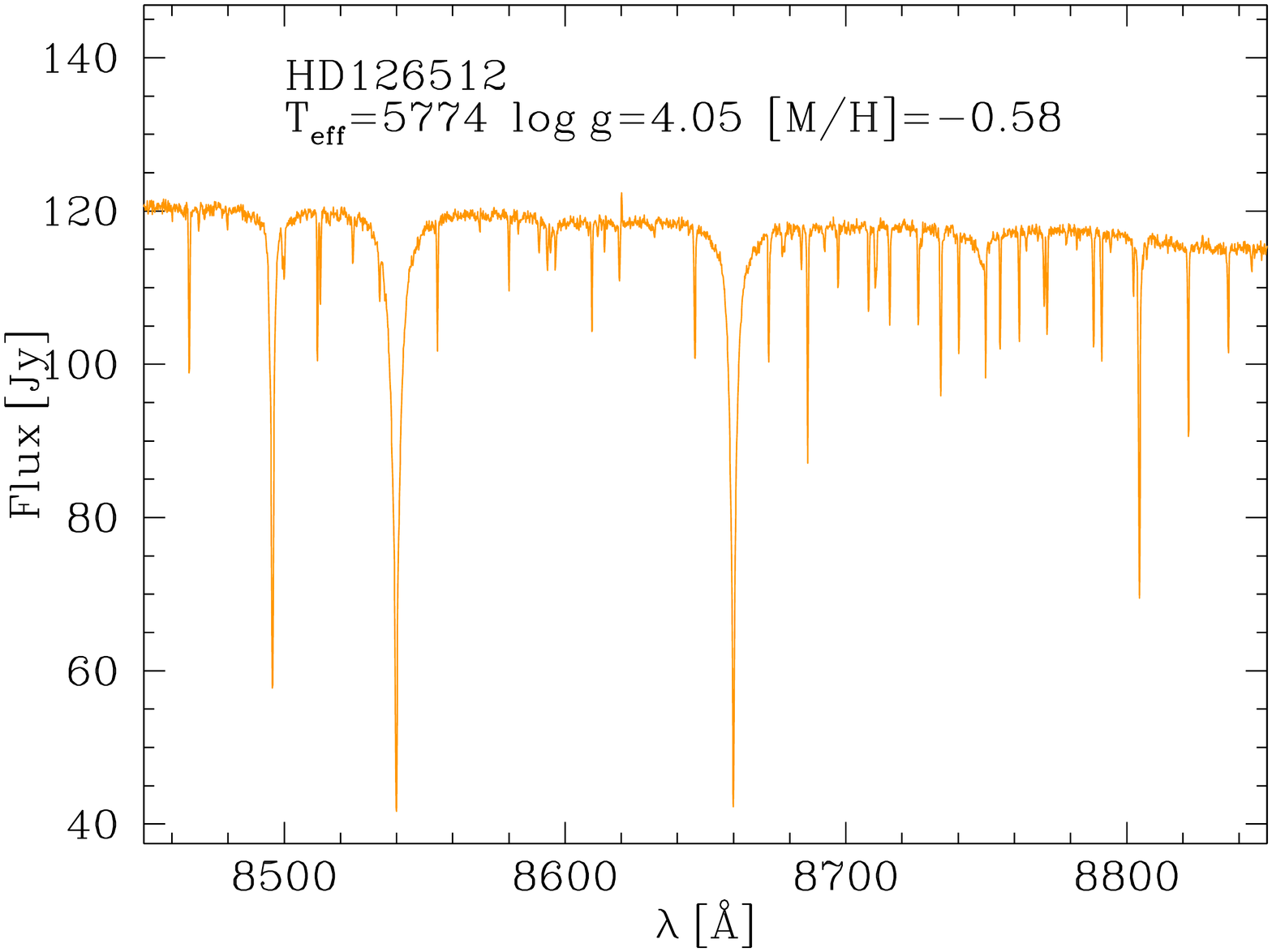}
\includegraphics[width=0.18\textwidth,angle=-90]{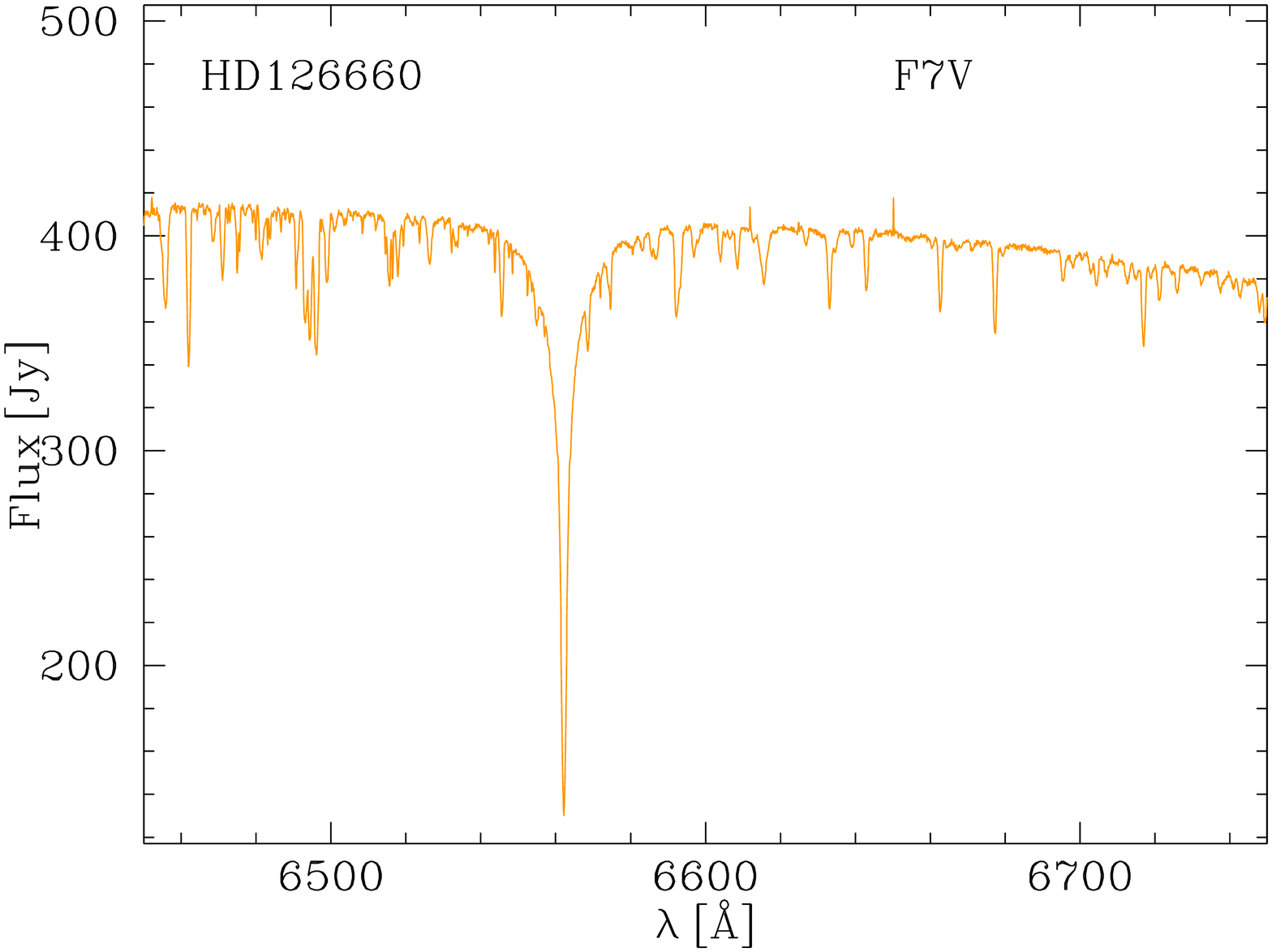}
\includegraphics[width=0.18\textwidth,angle=-90]{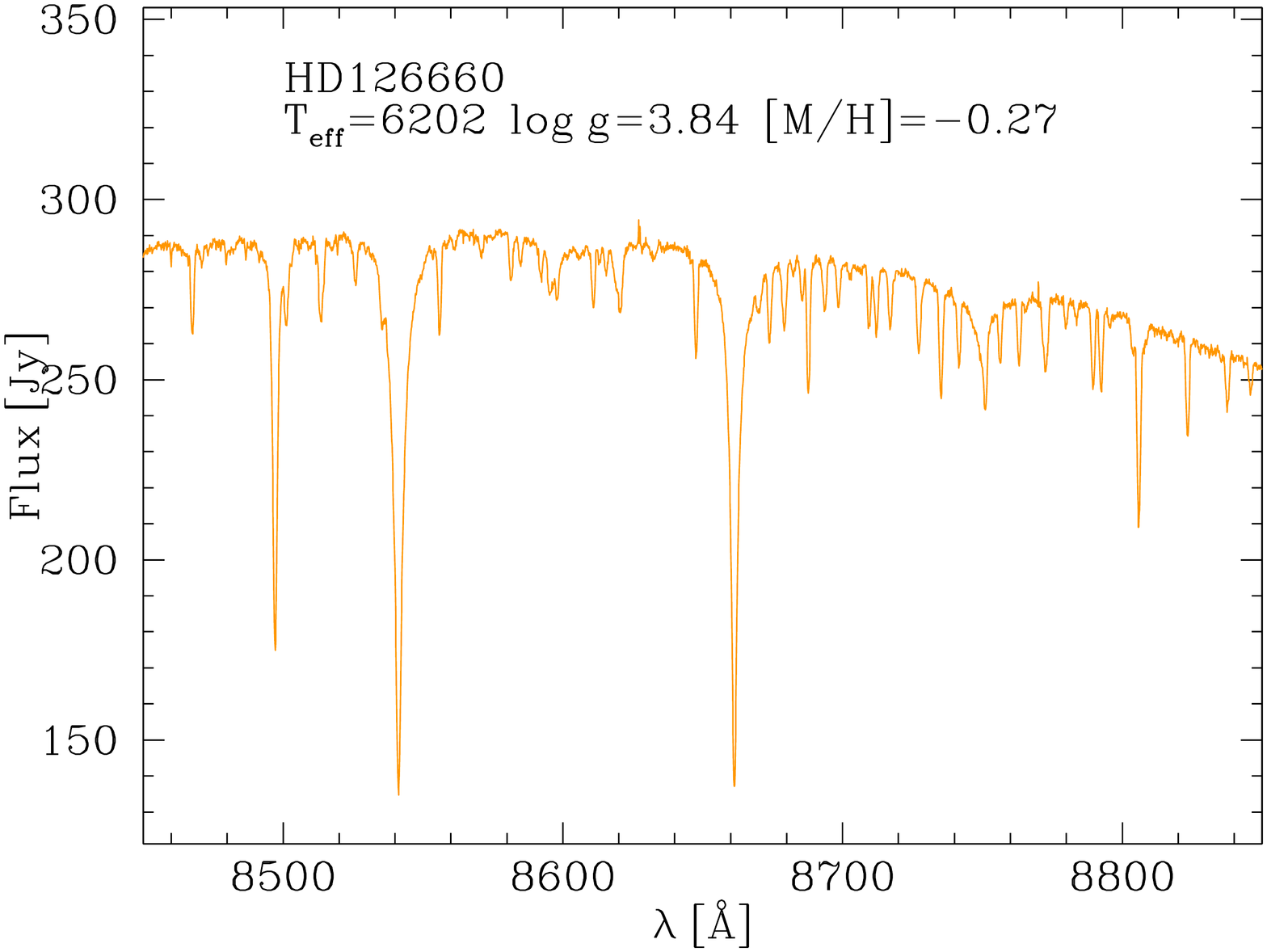}
\includegraphics[width=0.18\textwidth,angle=-90]{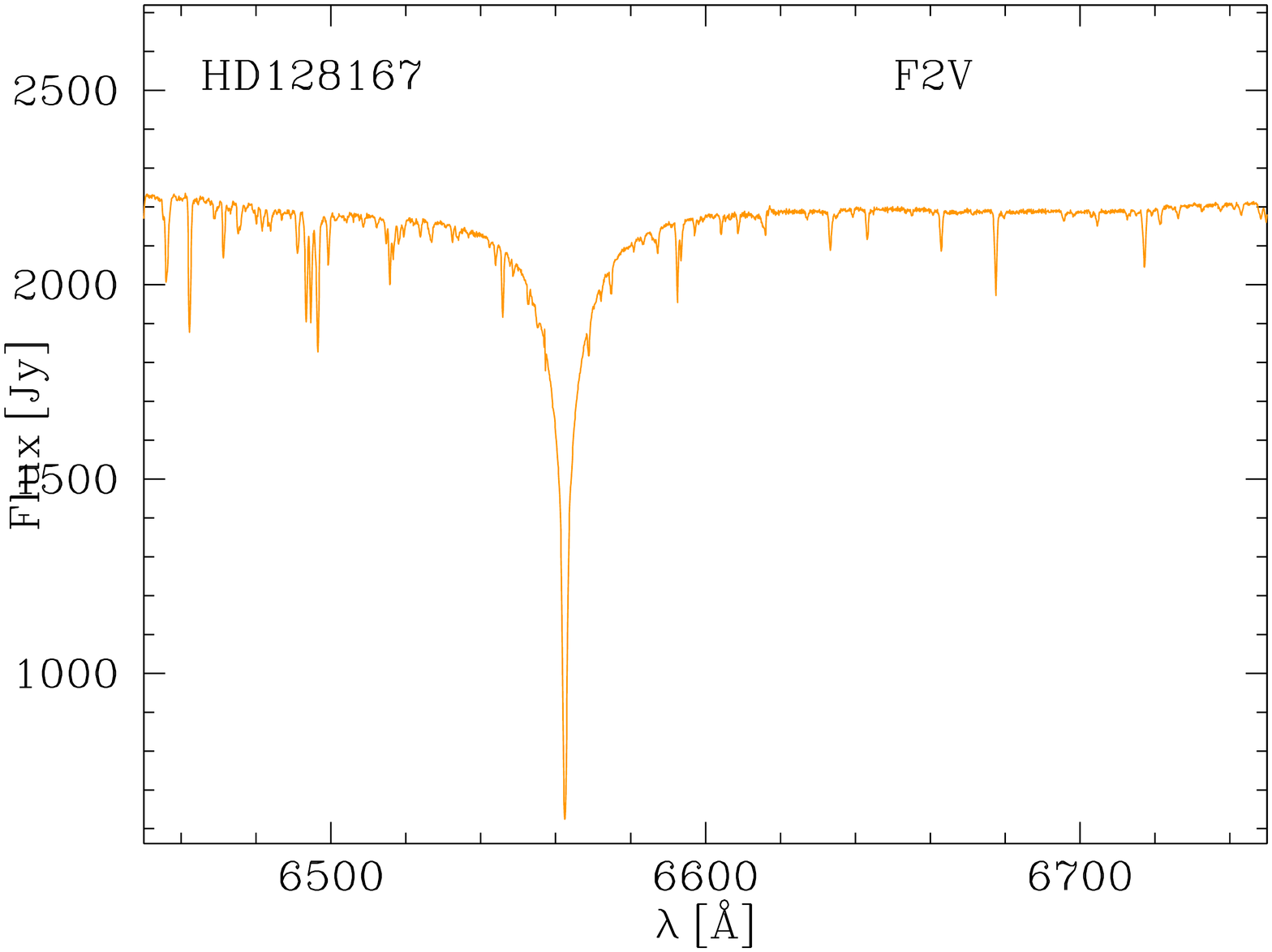}
\includegraphics[width=0.18\textwidth,angle=-90]{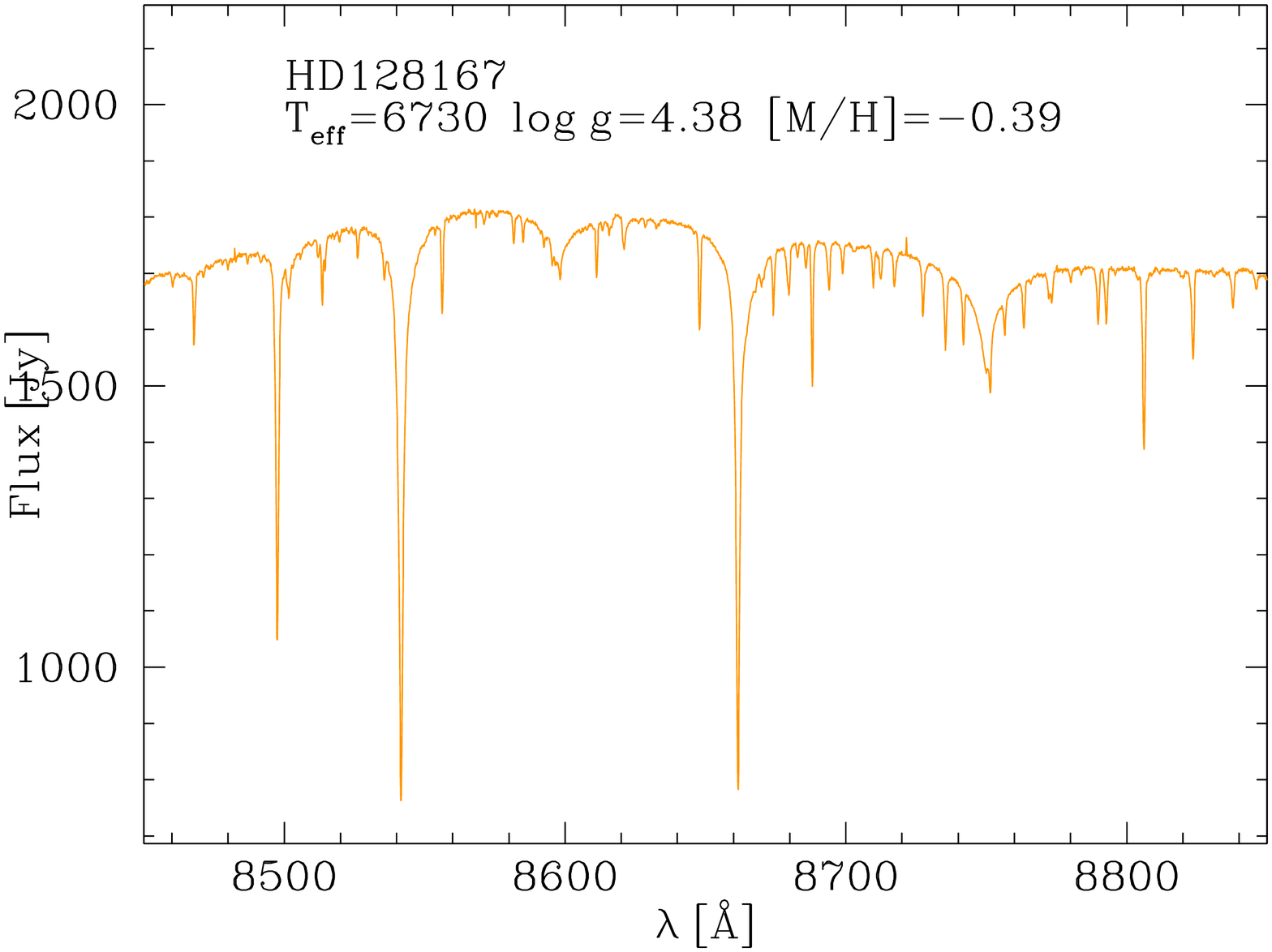}
\includegraphics[width=0.18\textwidth,angle=-90]{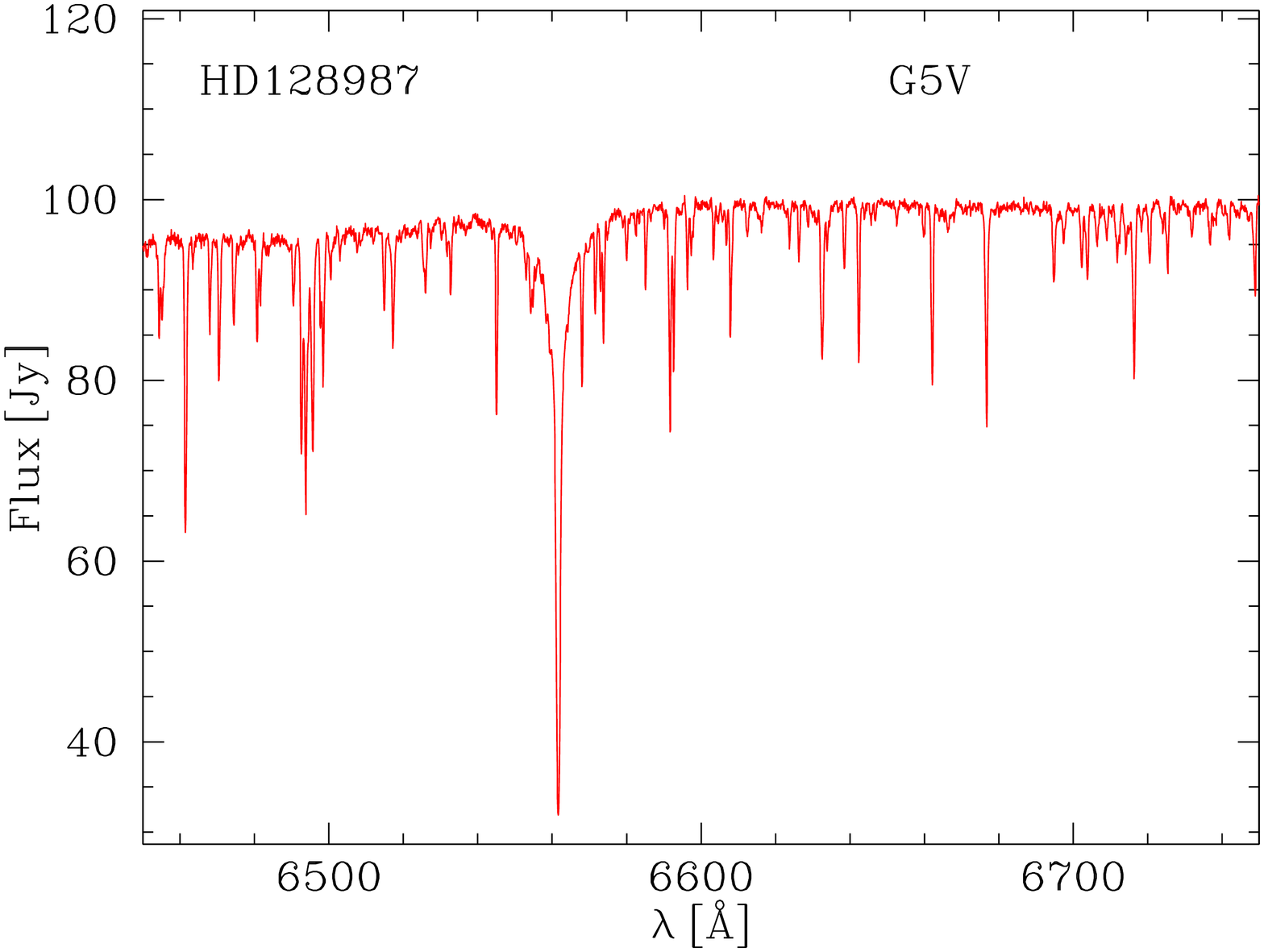}
\includegraphics[width=0.18\textwidth,angle=-90]{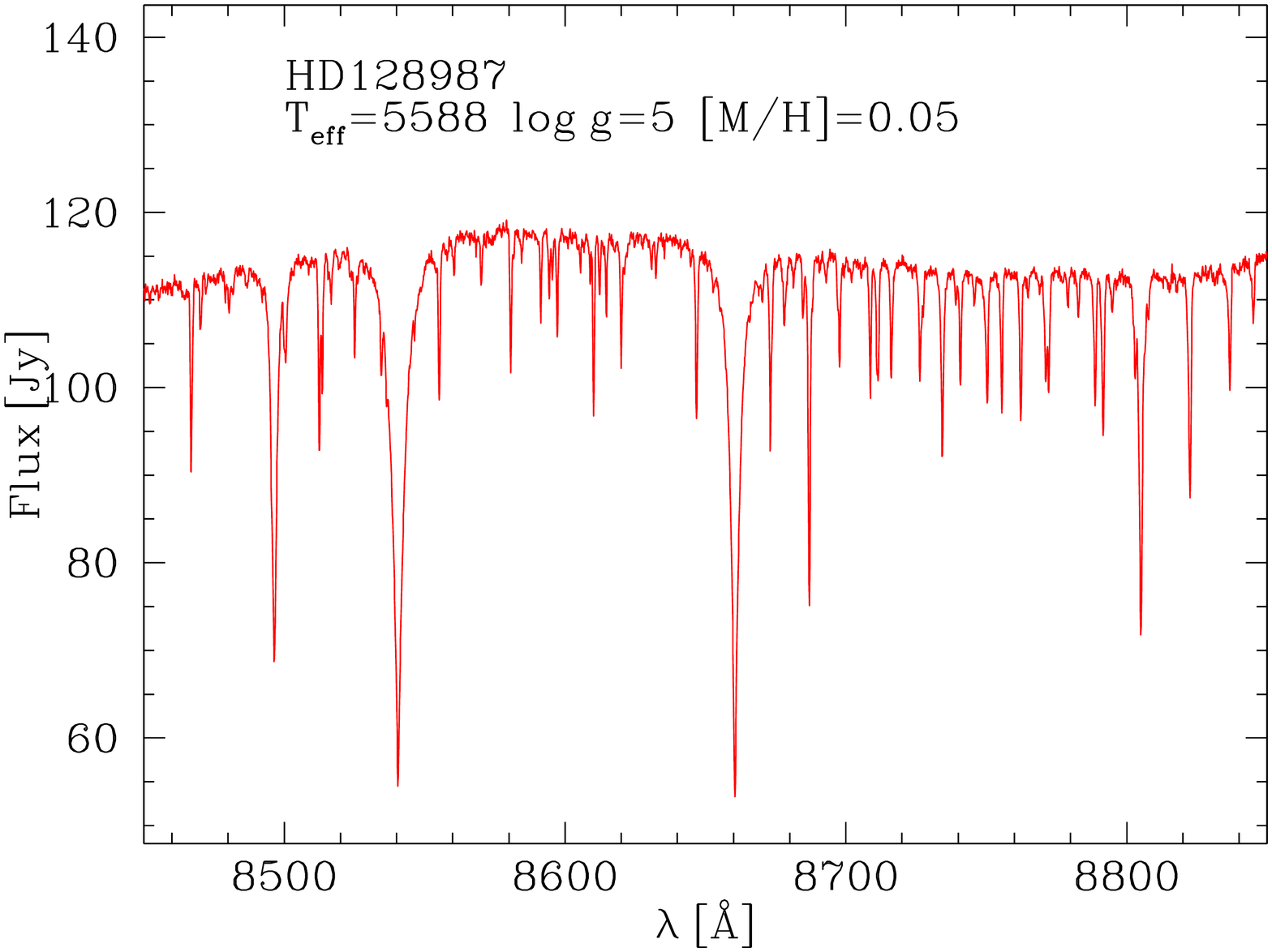}

\contcaption{22. Stars shown in this page are: HD117176, HD117243, HD118244, HD119291, HD120136, HD123299, HD124570, HD125560, HD126271, HD126511, HD126512, HD126660, HD128167 and HD128987.}
\end{figure*}

\begin{figure*}
\includegraphics[width=0.18\textwidth,angle=-90]{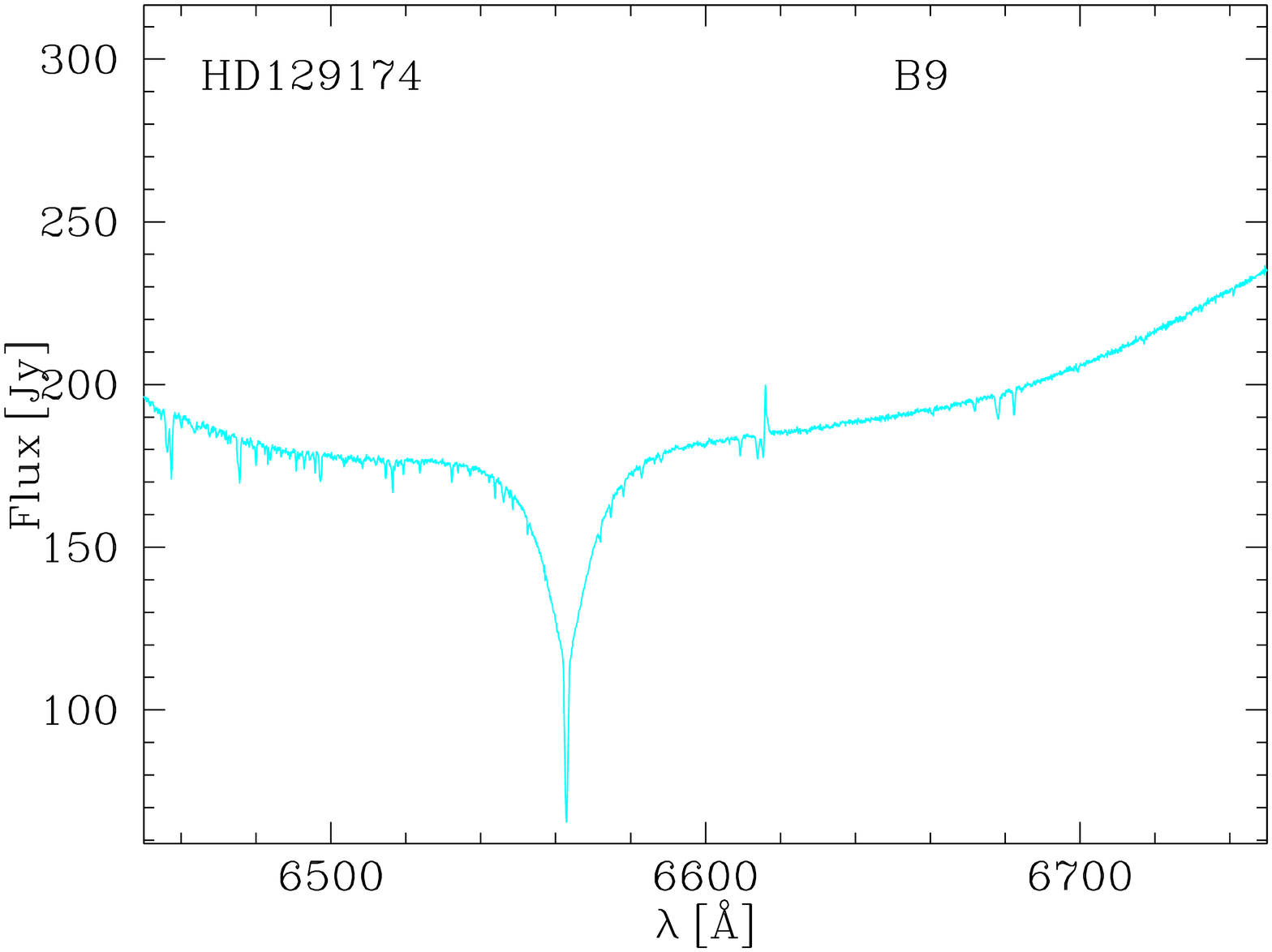}
\includegraphics[width=0.18\textwidth,angle=-90]{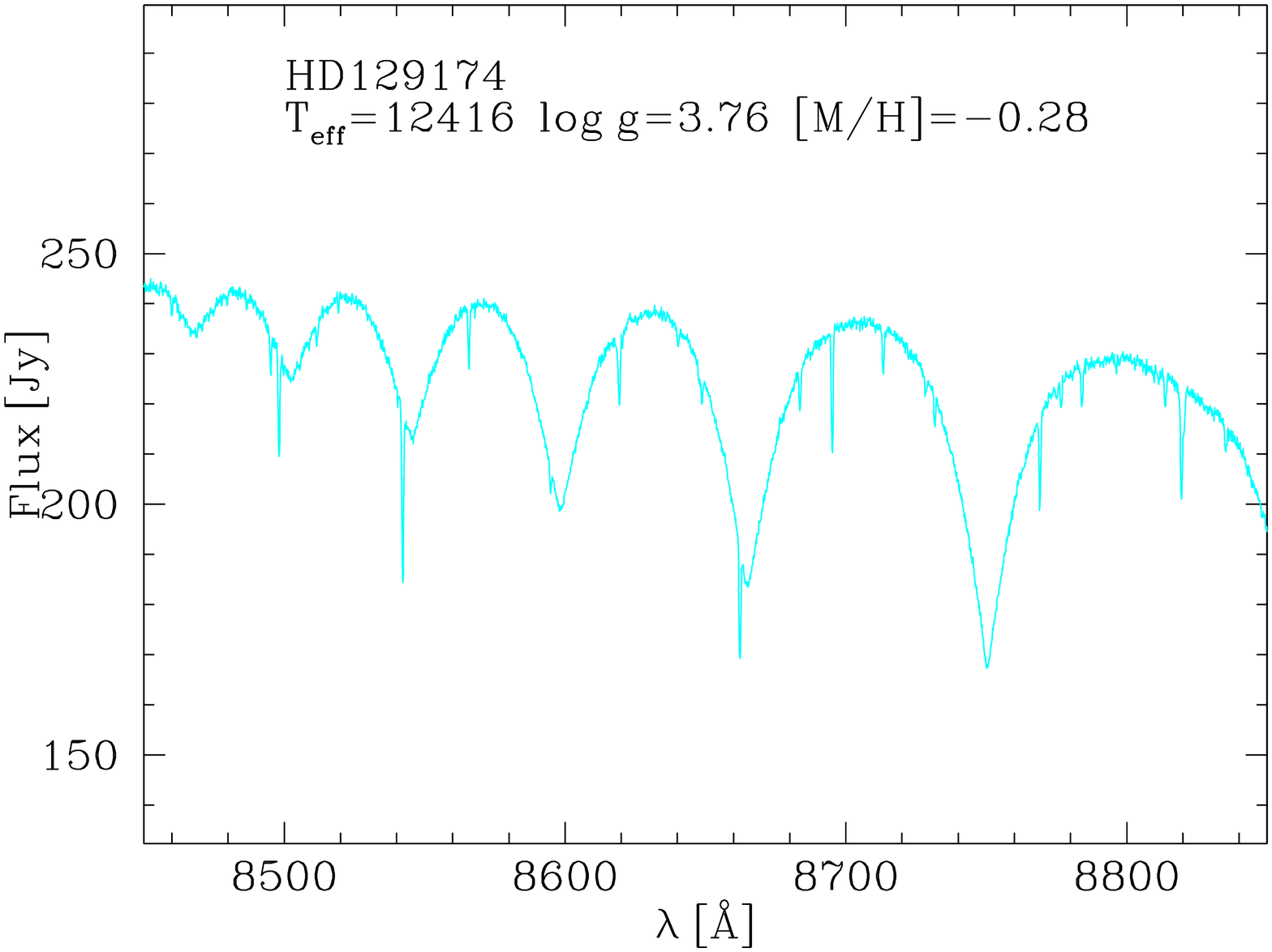}
\includegraphics[width=0.18\textwidth,angle=-90]{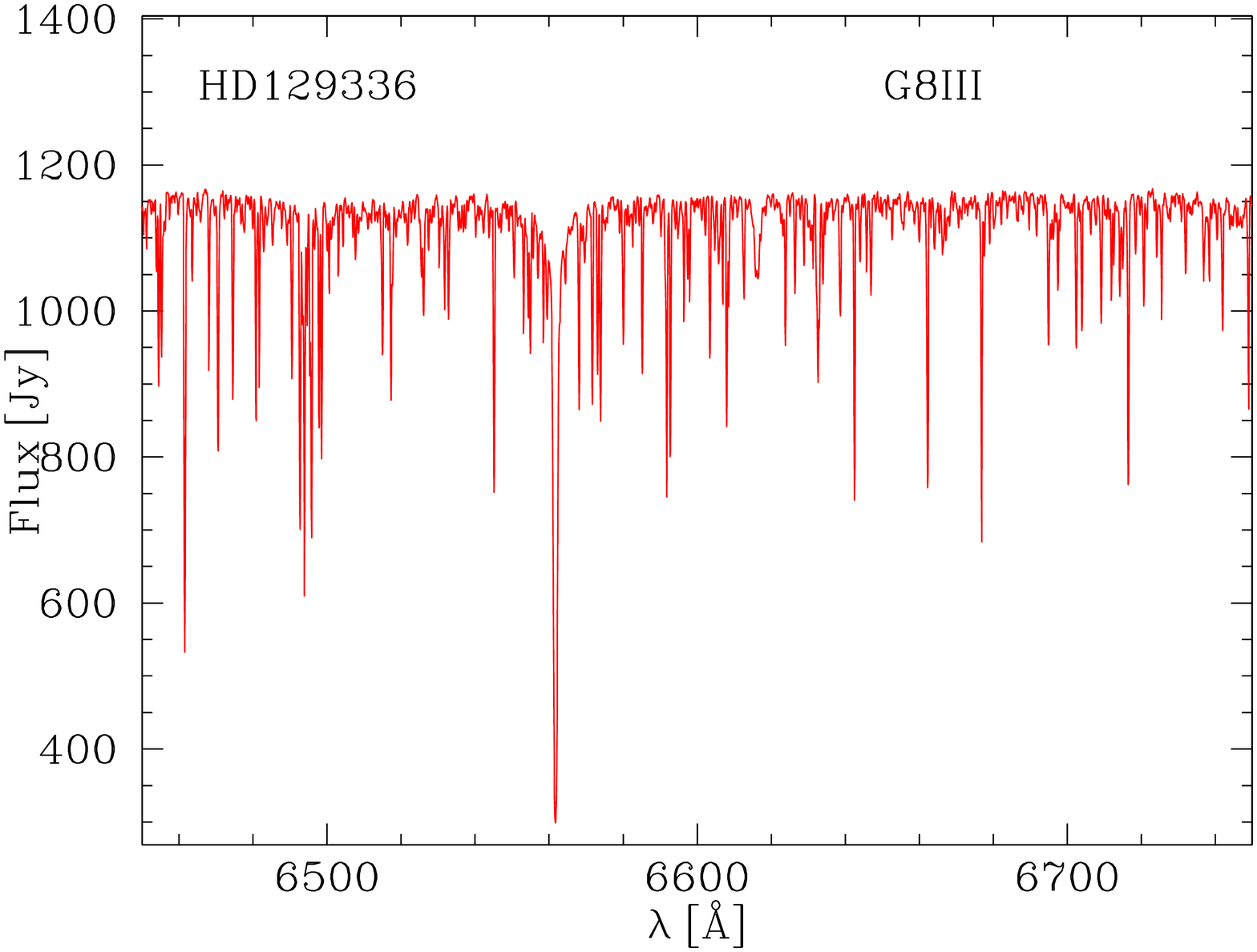}
\includegraphics[width=0.18\textwidth,angle=-90]{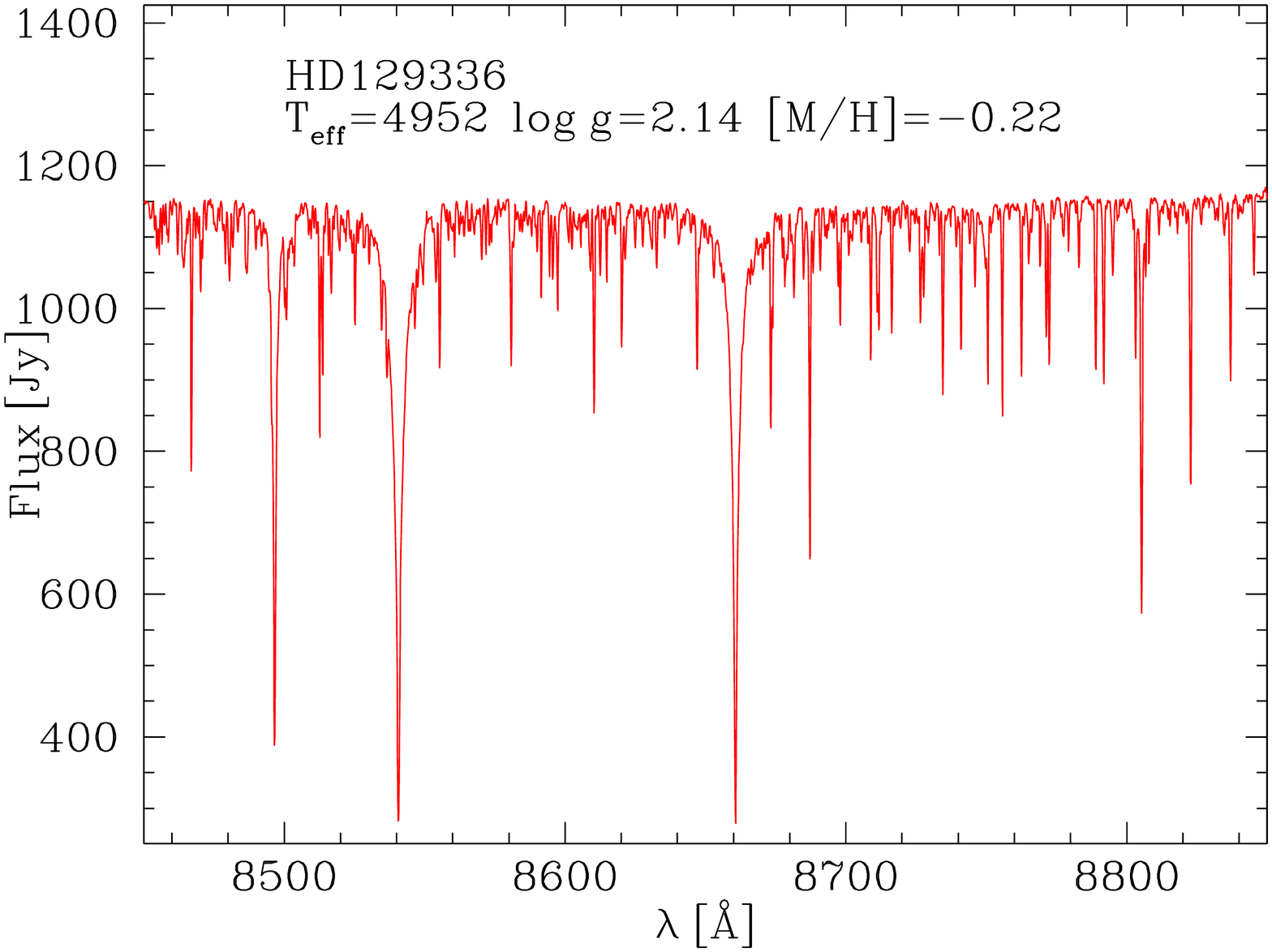}
\includegraphics[width=0.18\textwidth,angle=-90]{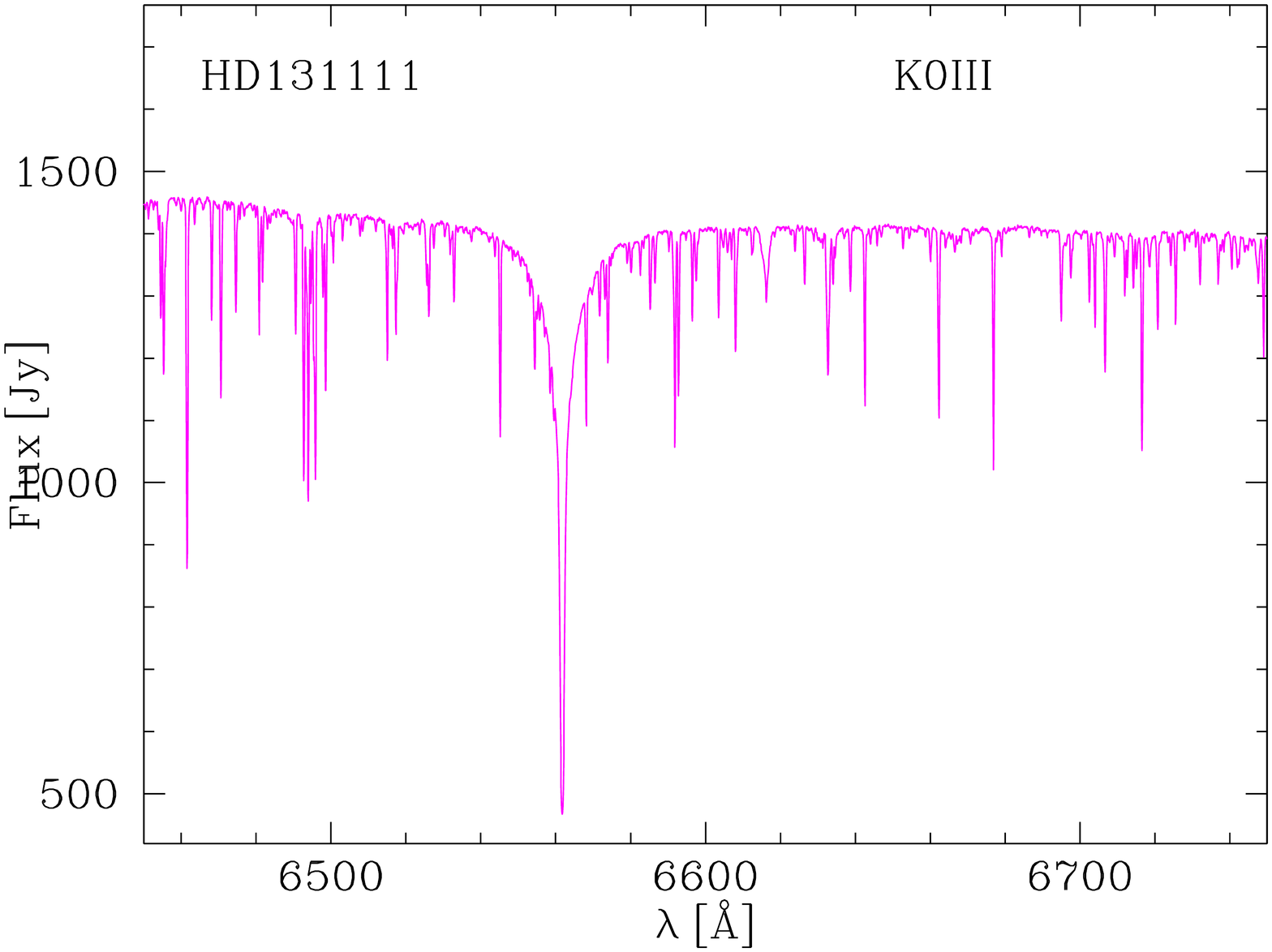}
\includegraphics[width=0.18\textwidth,angle=-90]{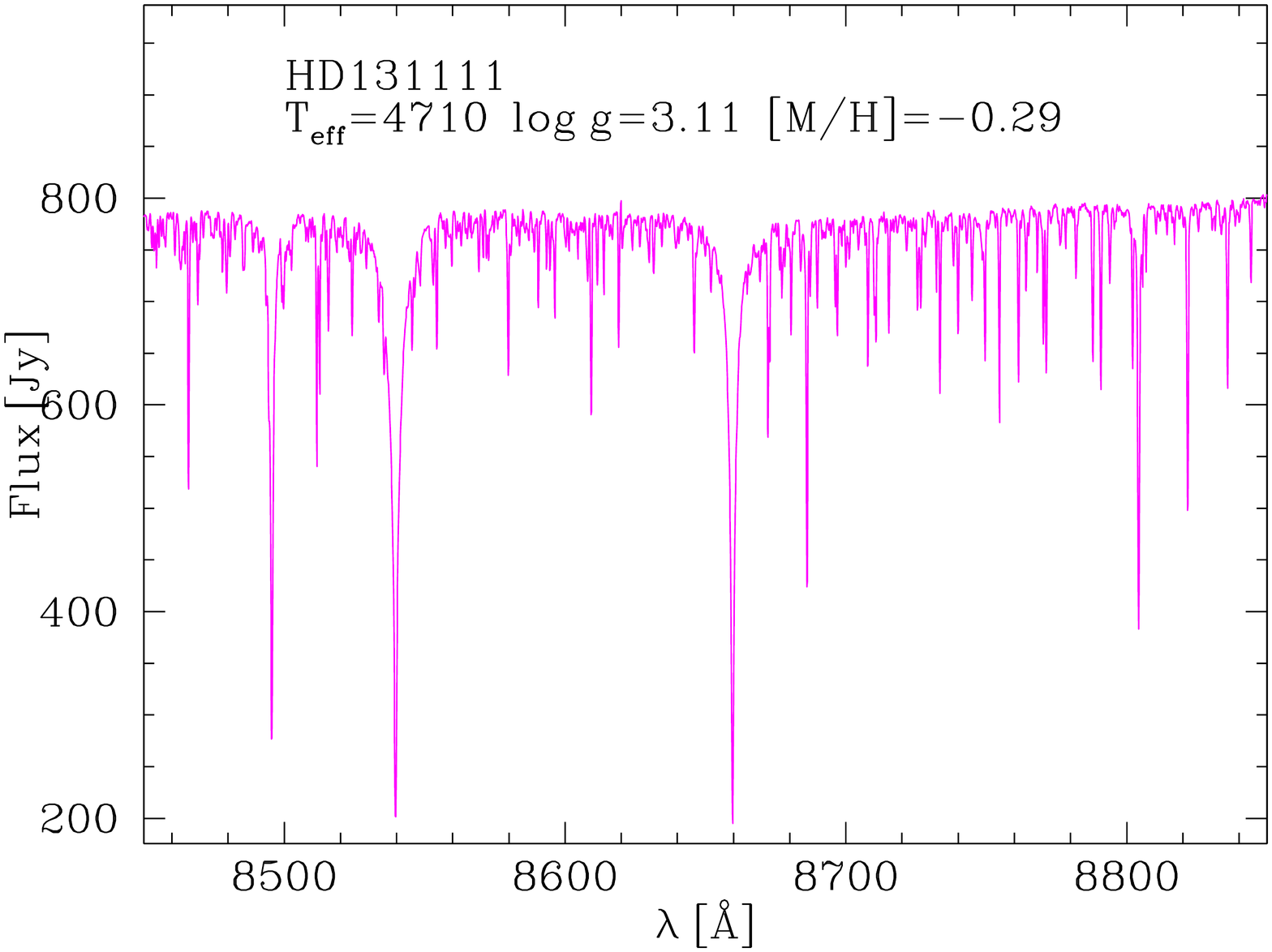}
\includegraphics[width=0.18\textwidth,angle=-90]{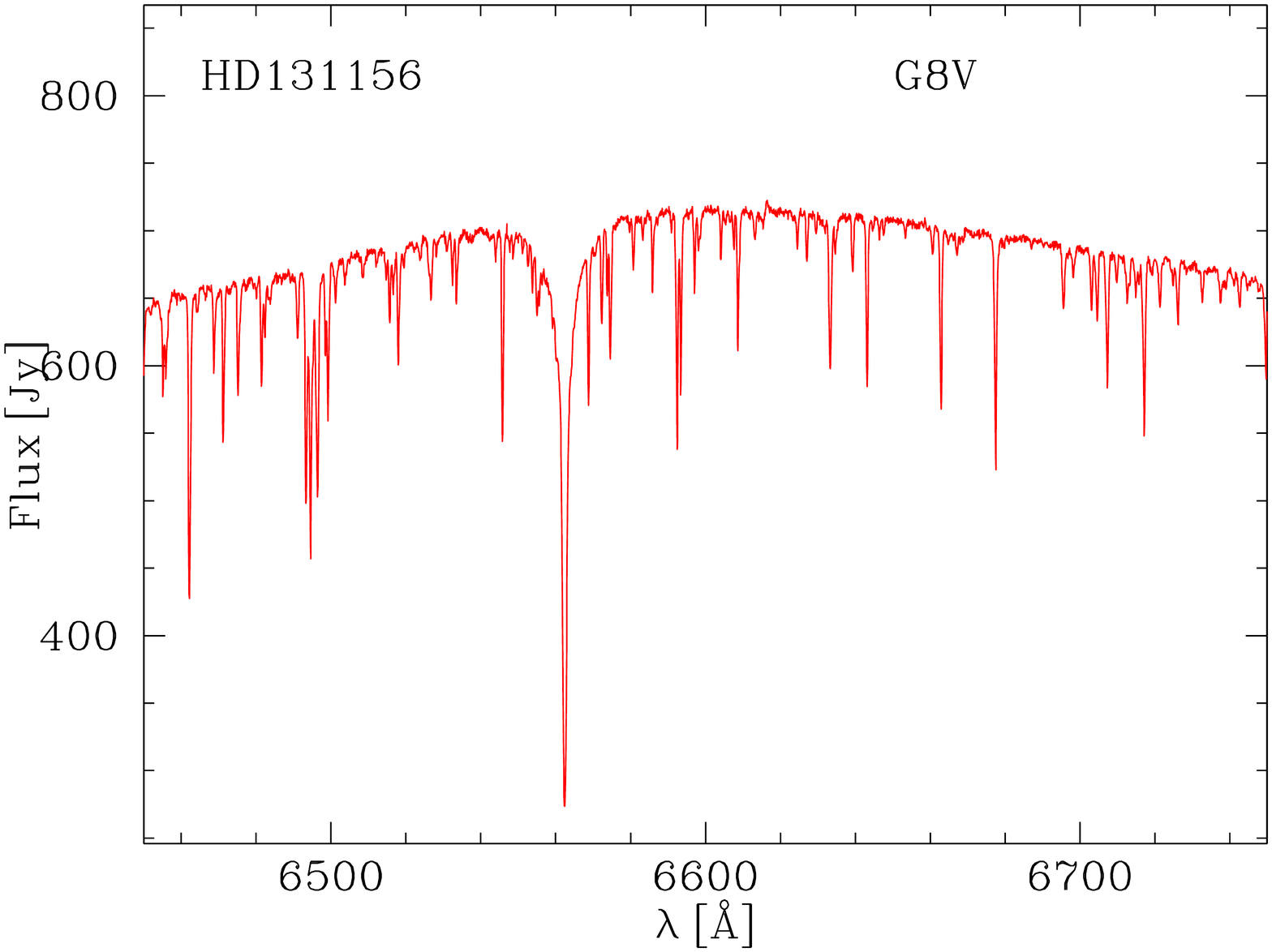}
\includegraphics[width=0.18\textwidth,angle=-90]{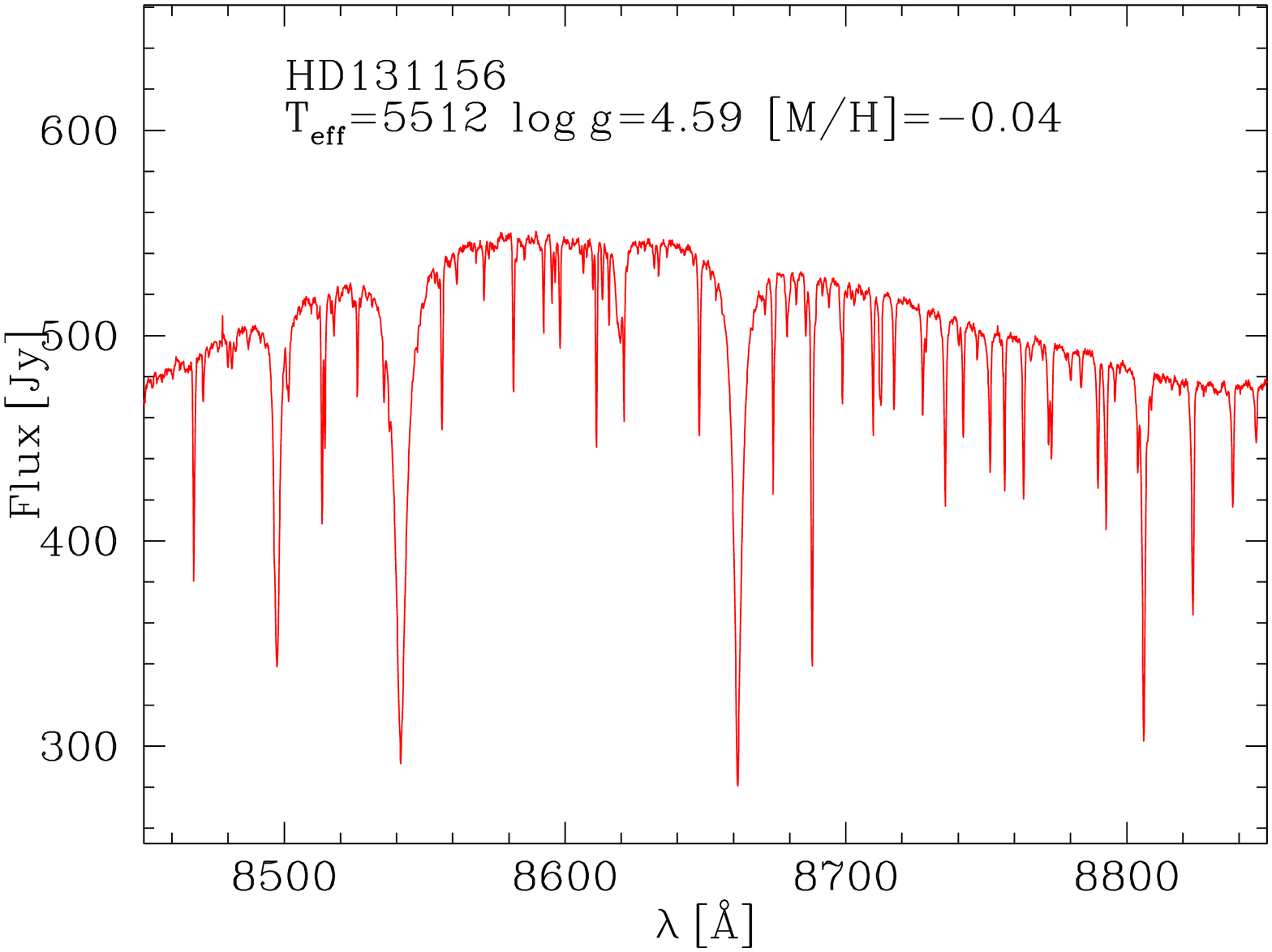}
\includegraphics[width=0.18\textwidth,angle=-90]{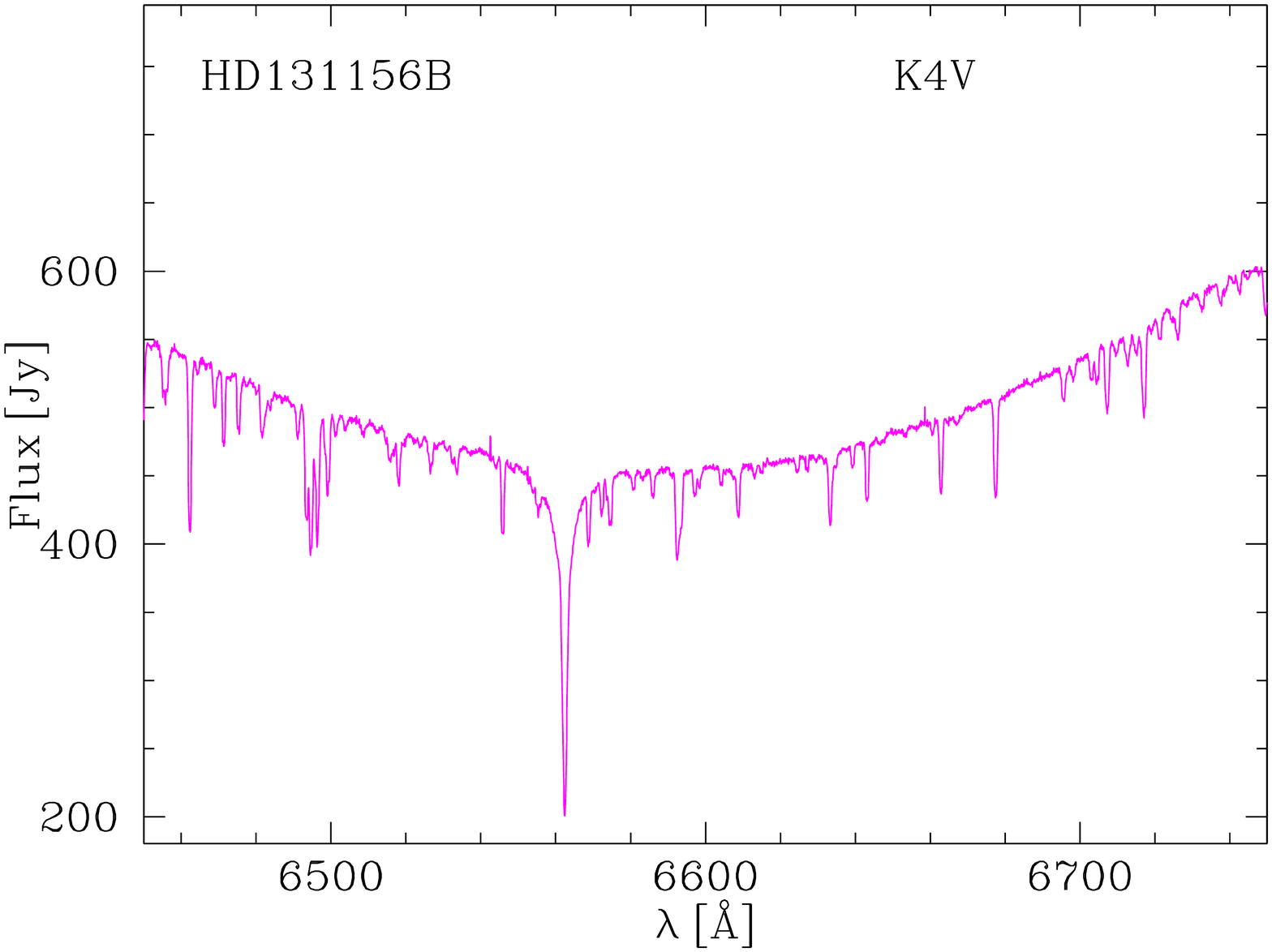}
\includegraphics[width=0.18\textwidth,angle=-90]{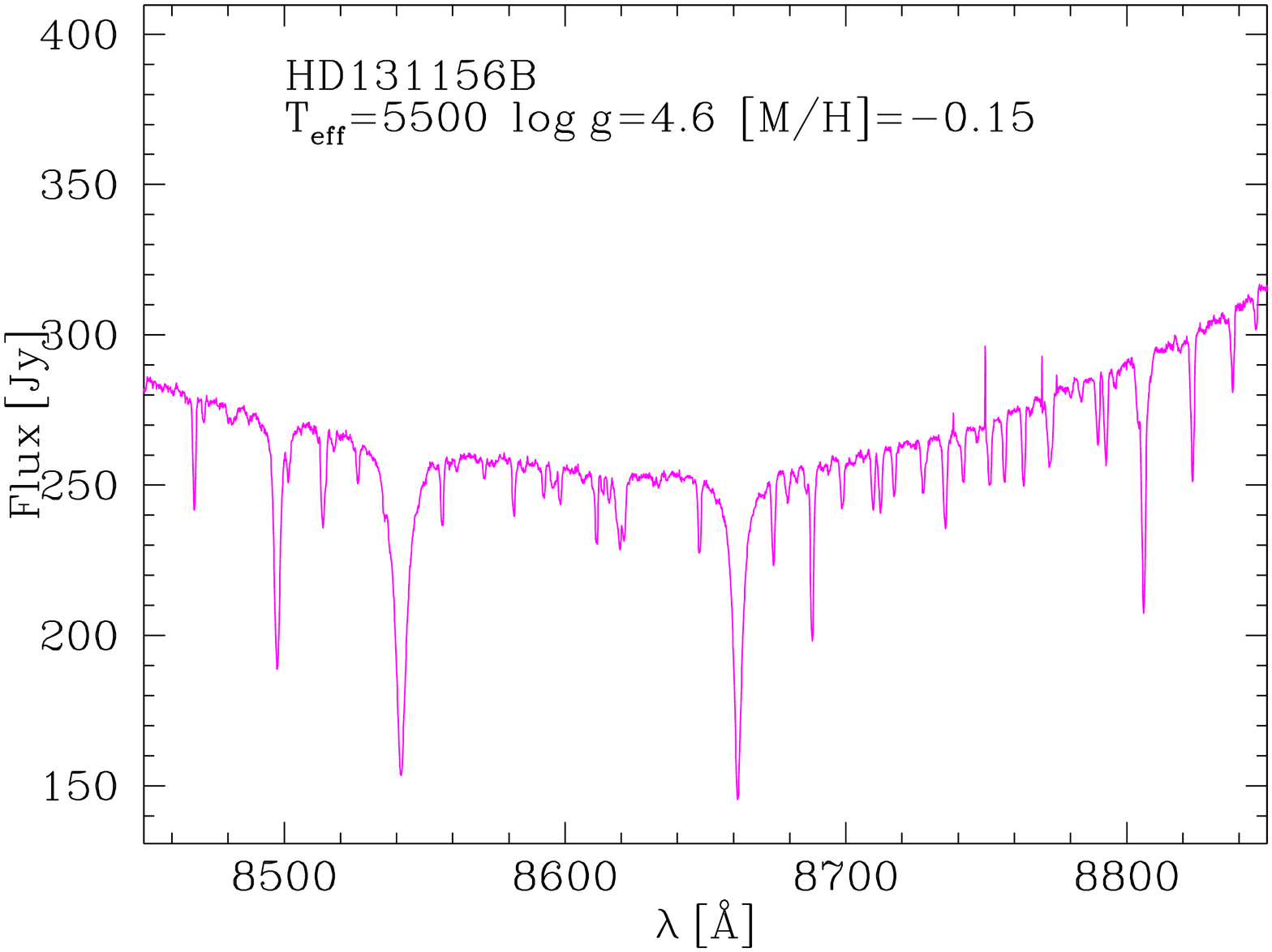}
\includegraphics[width=0.18\textwidth,angle=-90]{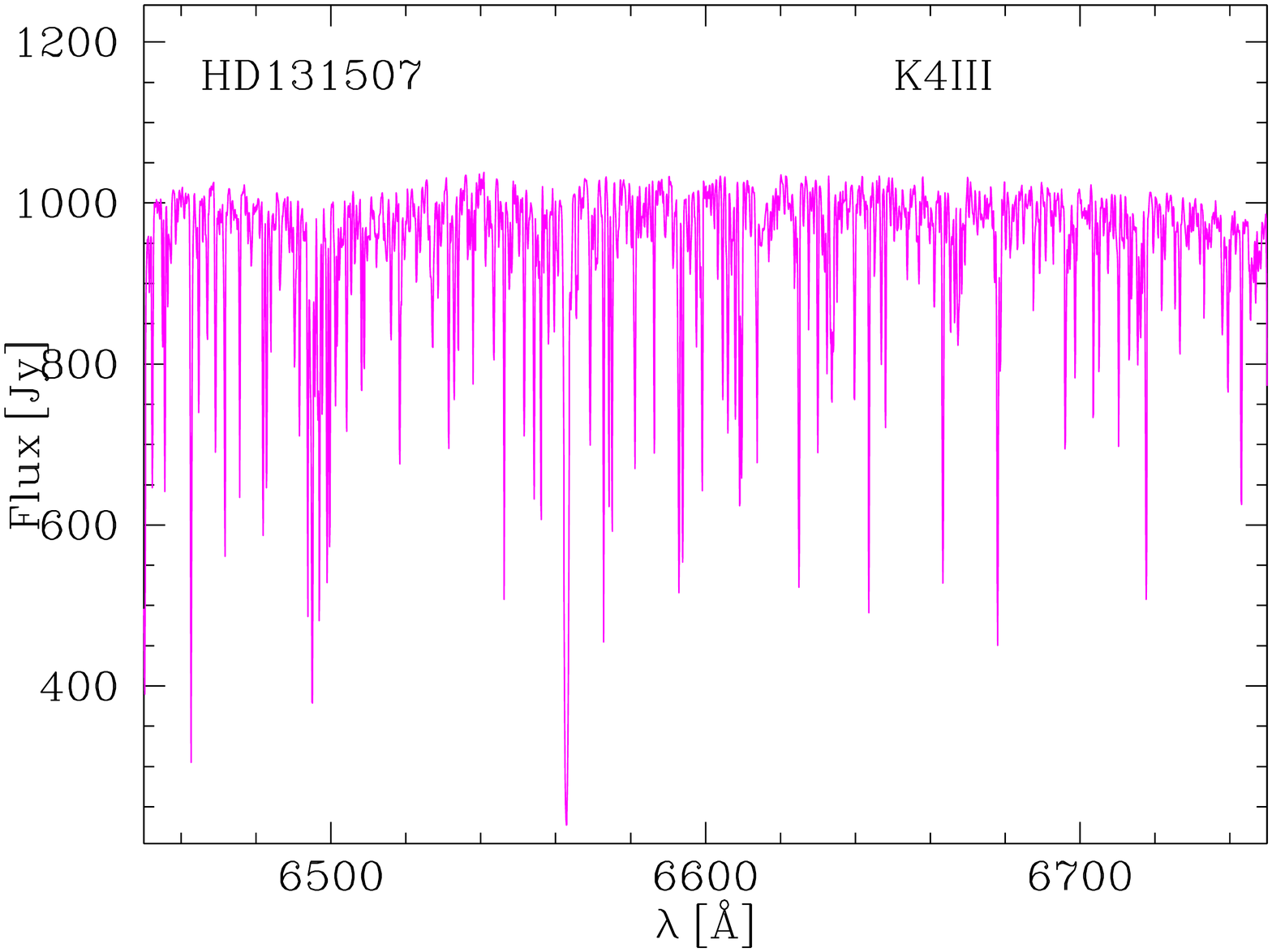}
\includegraphics[width=0.18\textwidth,angle=-90]{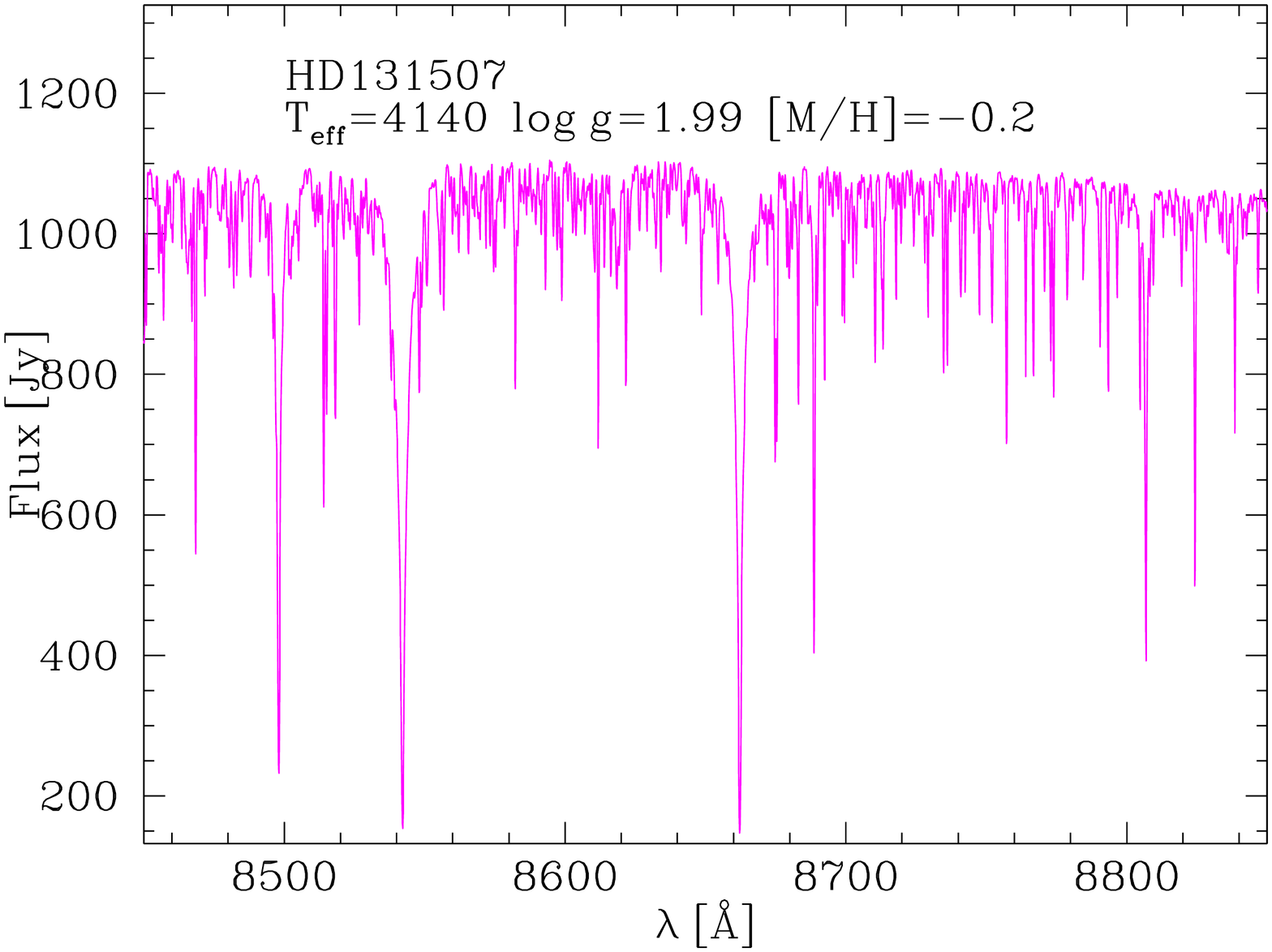}
\includegraphics[width=0.18\textwidth,angle=-90]{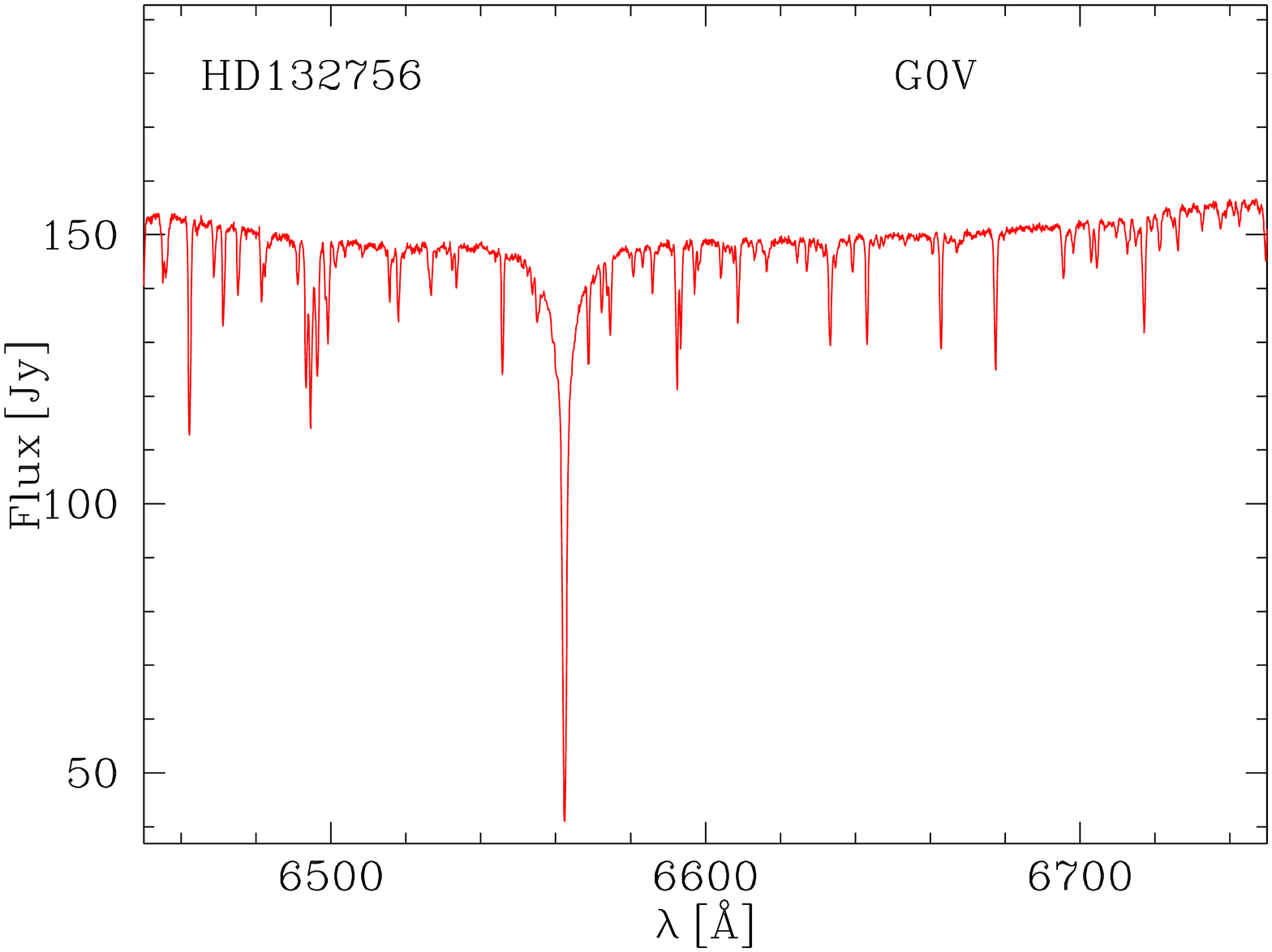}
\includegraphics[width=0.18\textwidth,angle=-90]{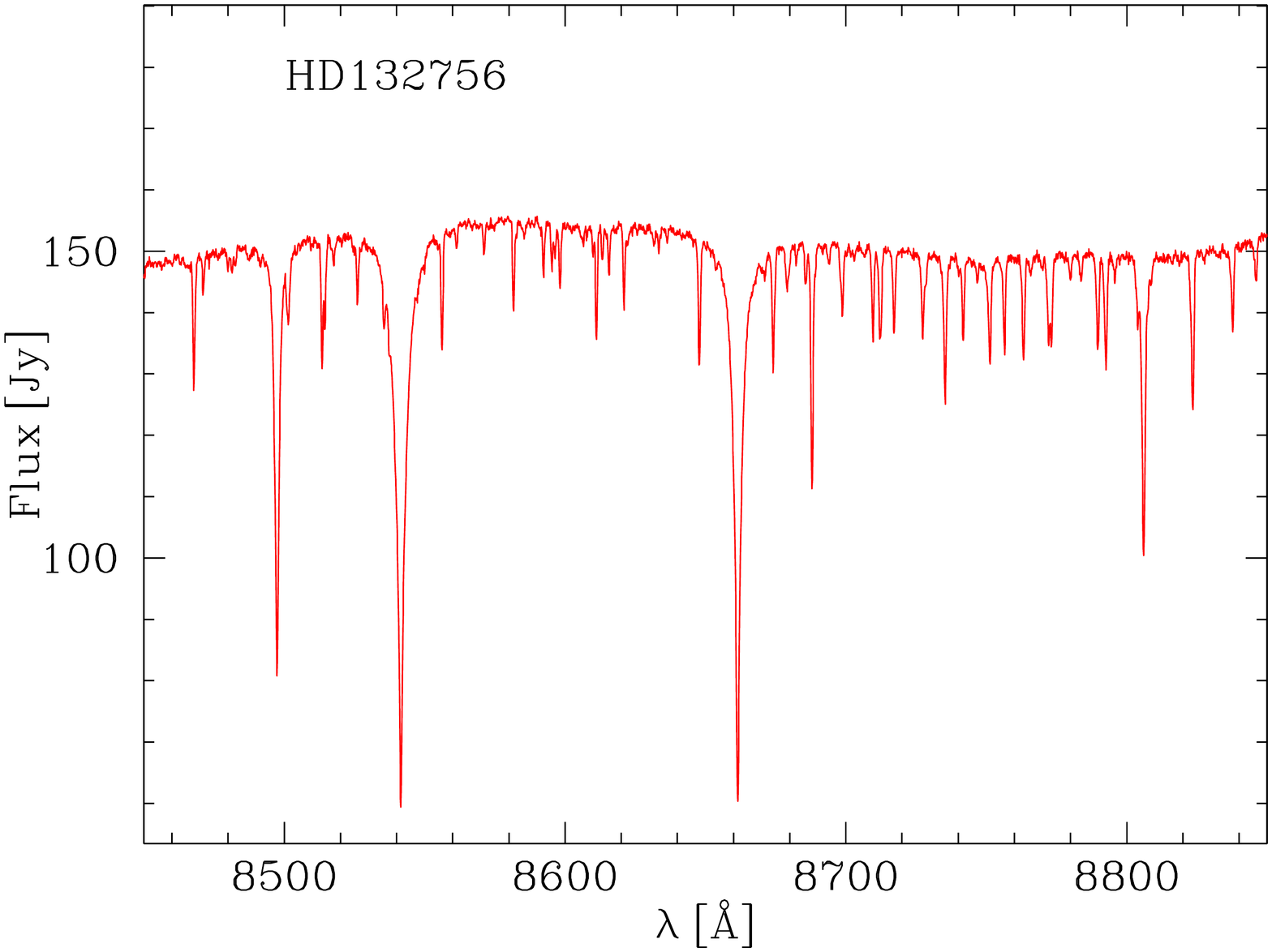}
\includegraphics[width=0.18\textwidth,angle=-90]{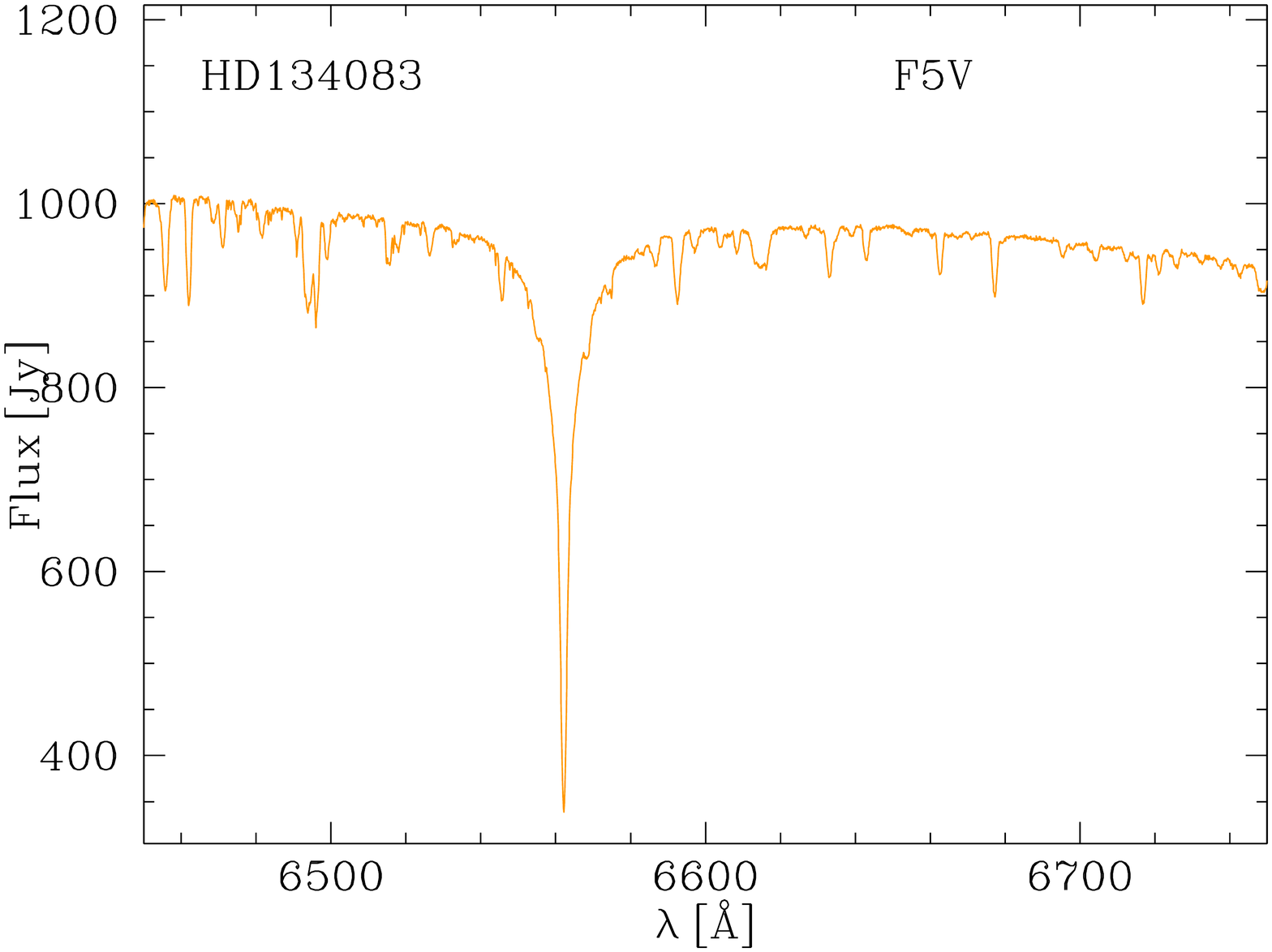}
\includegraphics[width=0.18\textwidth,angle=-90]{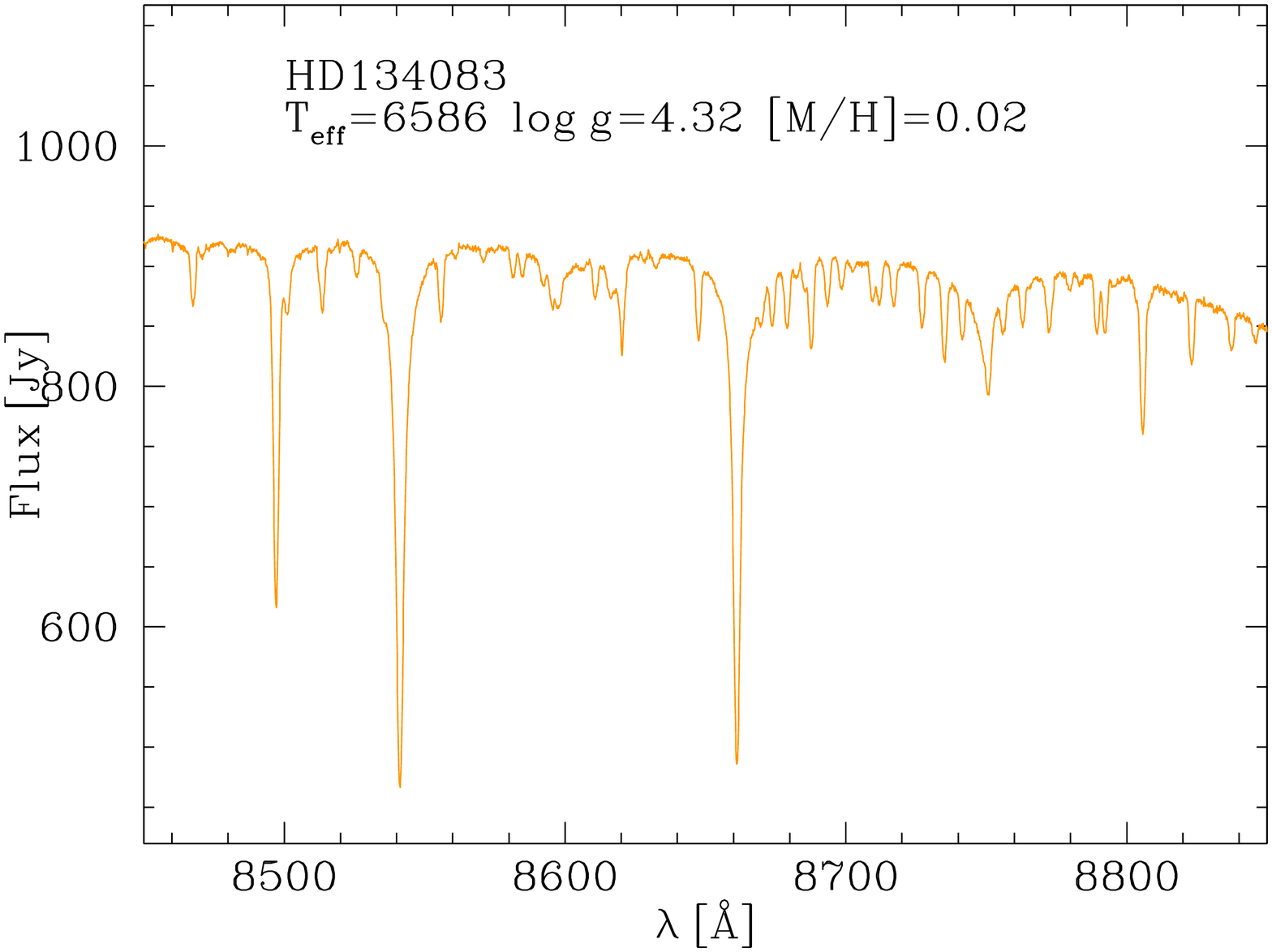}
\includegraphics[width=0.18\textwidth,angle=-90]{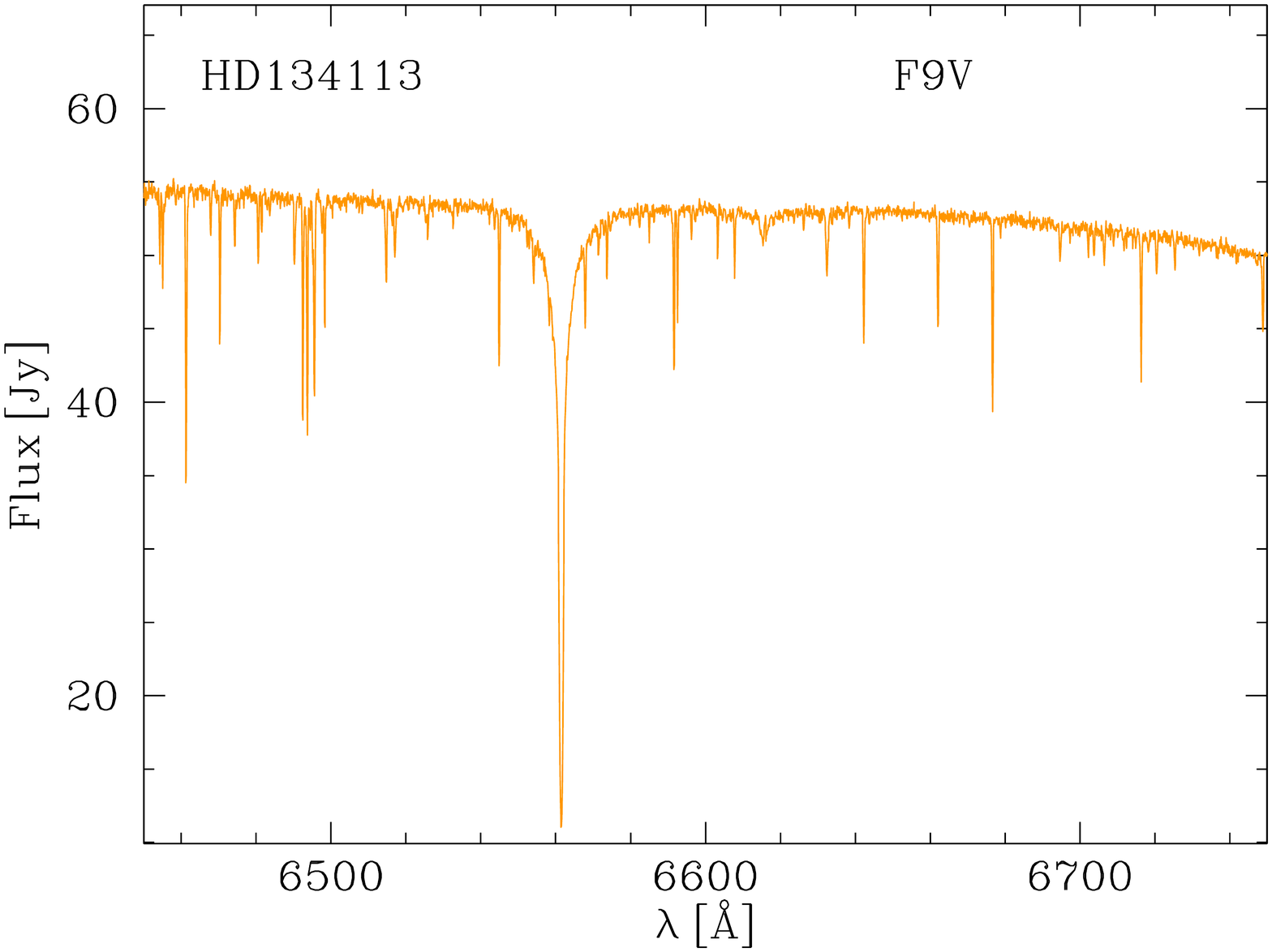}
\includegraphics[width=0.18\textwidth,angle=-90]{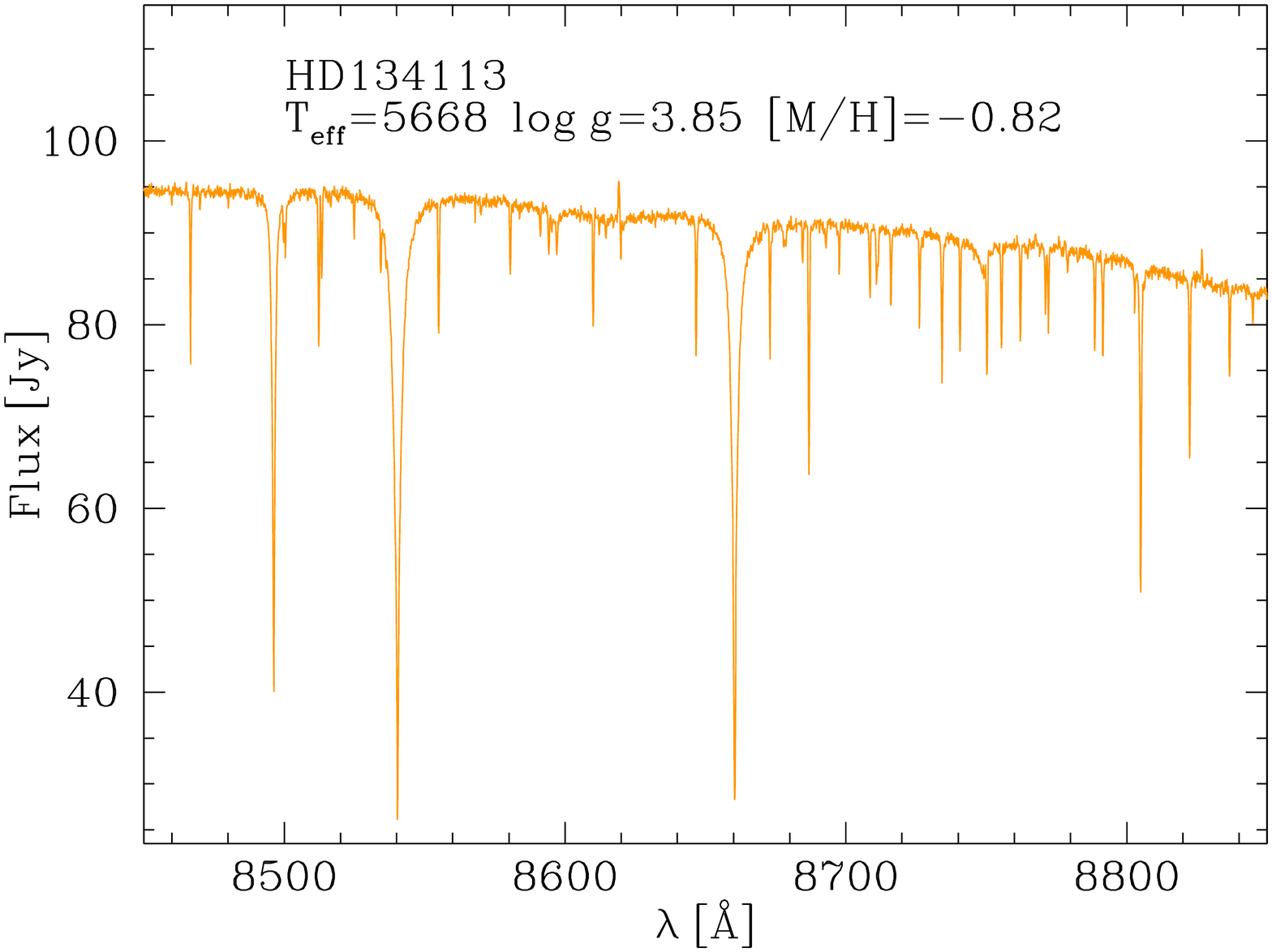}
\includegraphics[width=0.18\textwidth,angle=-90]{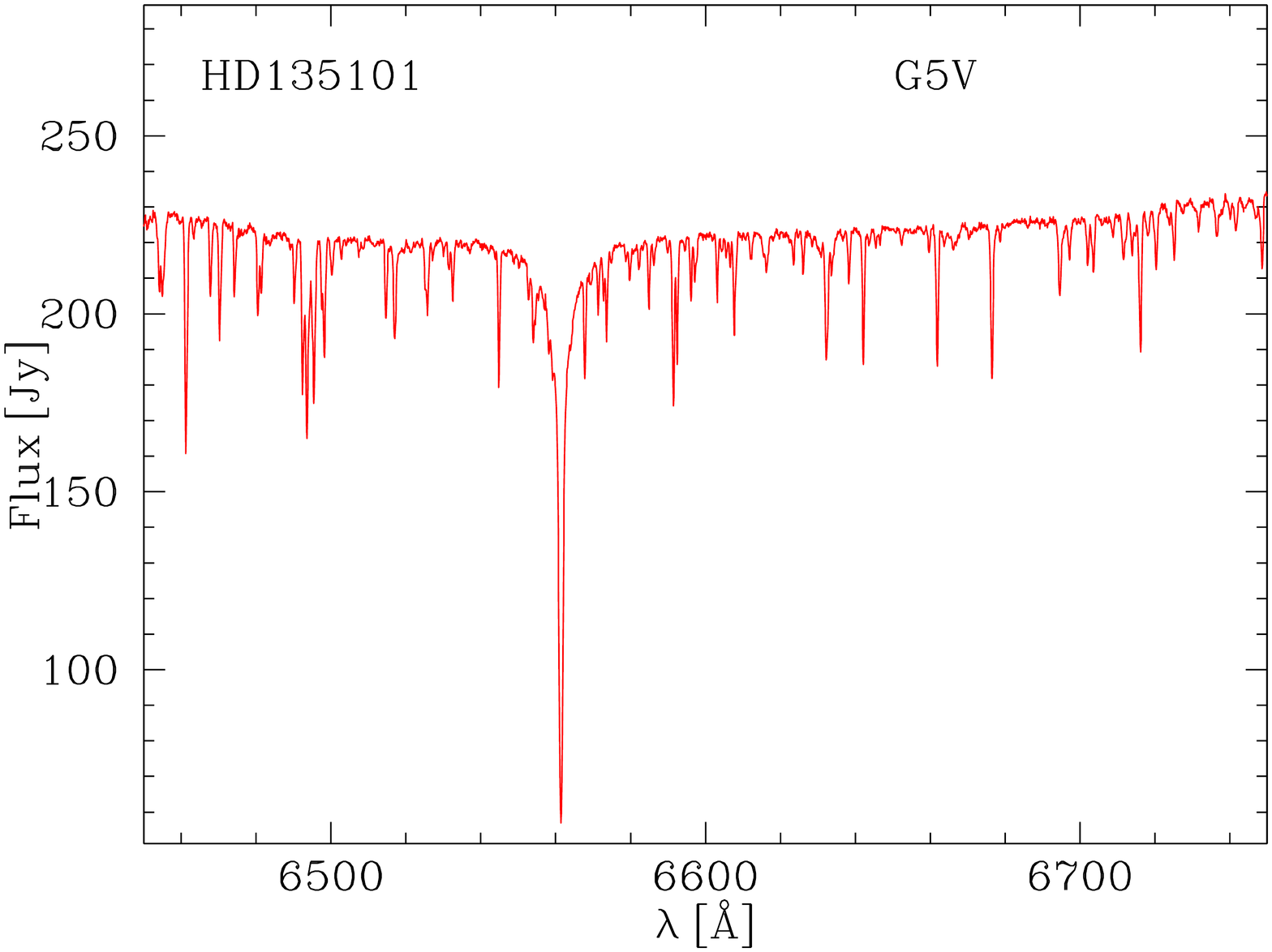}
\includegraphics[width=0.18\textwidth,angle=-90]{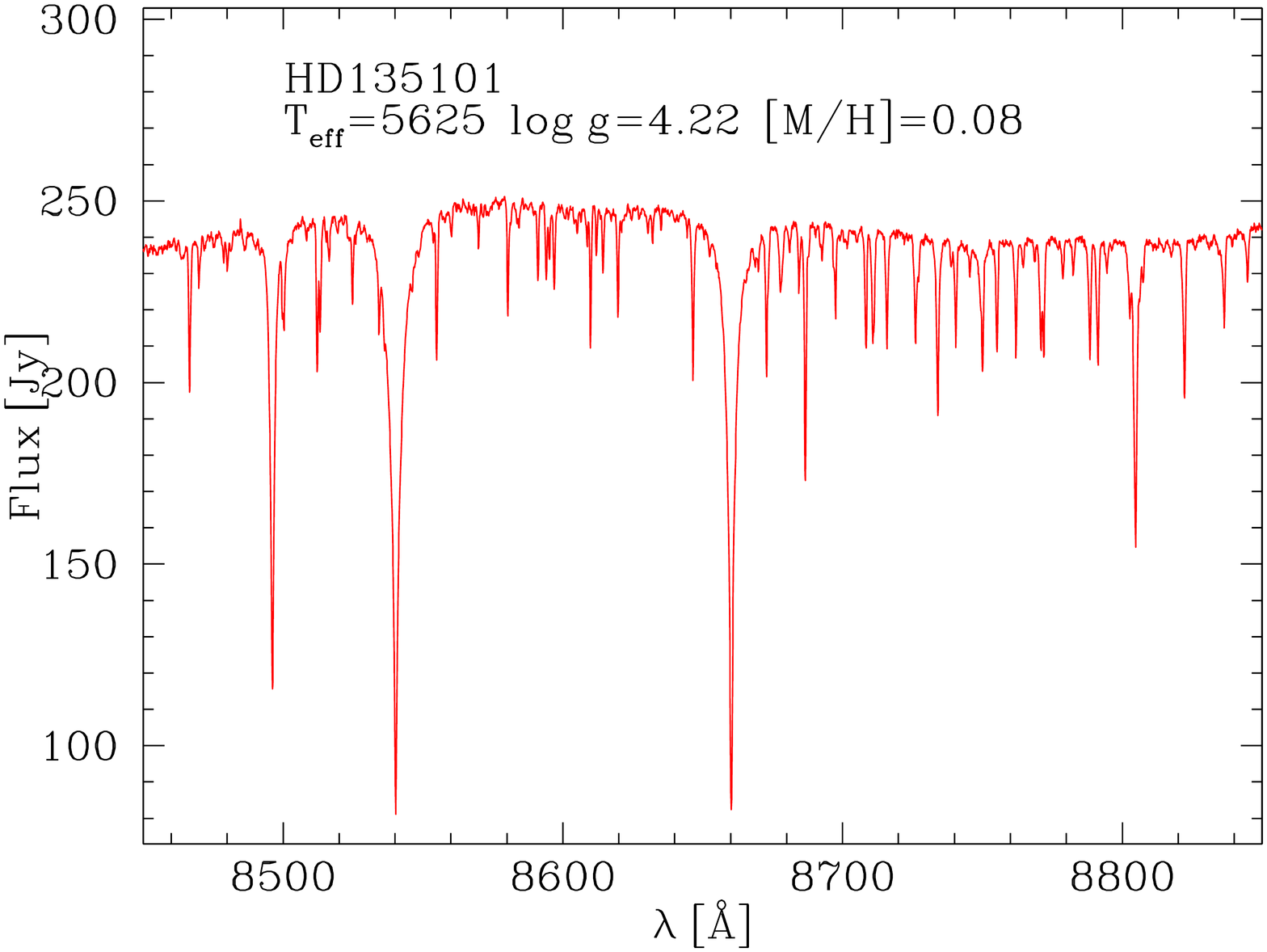}
\includegraphics[width=0.18\textwidth,angle=-90]{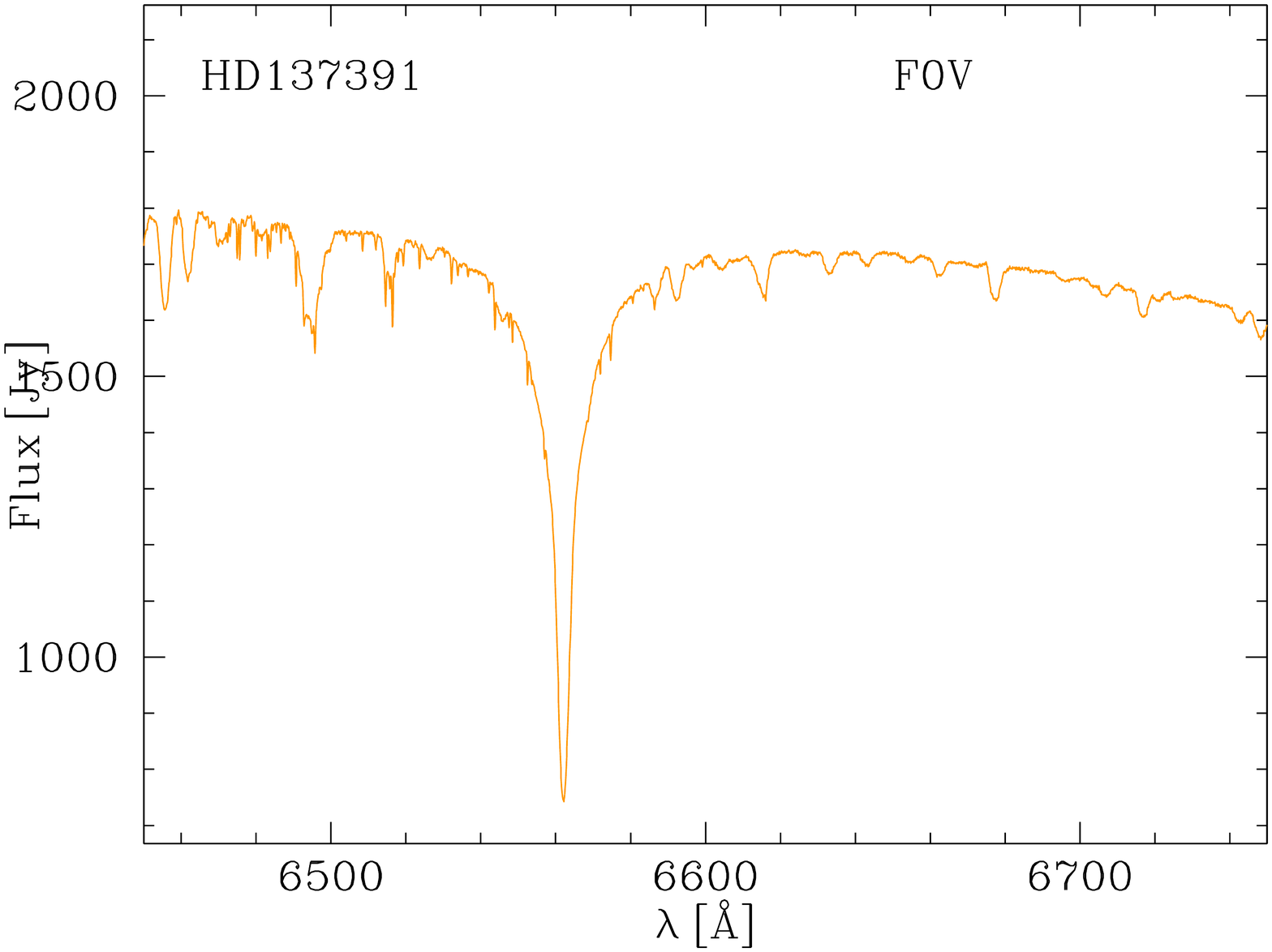}
\includegraphics[width=0.18\textwidth,angle=-90]{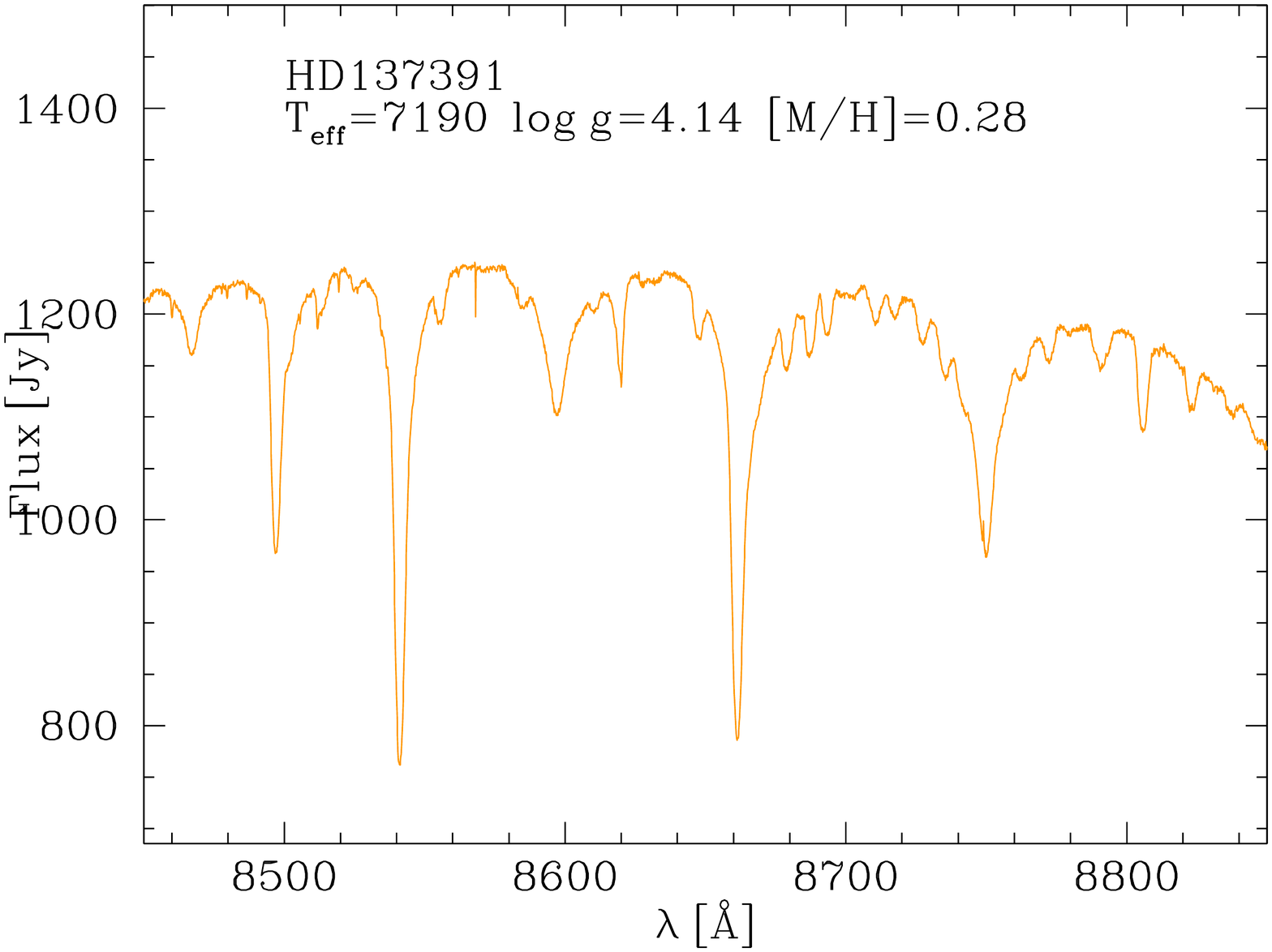}
\includegraphics[width=0.18\textwidth,angle=-90]{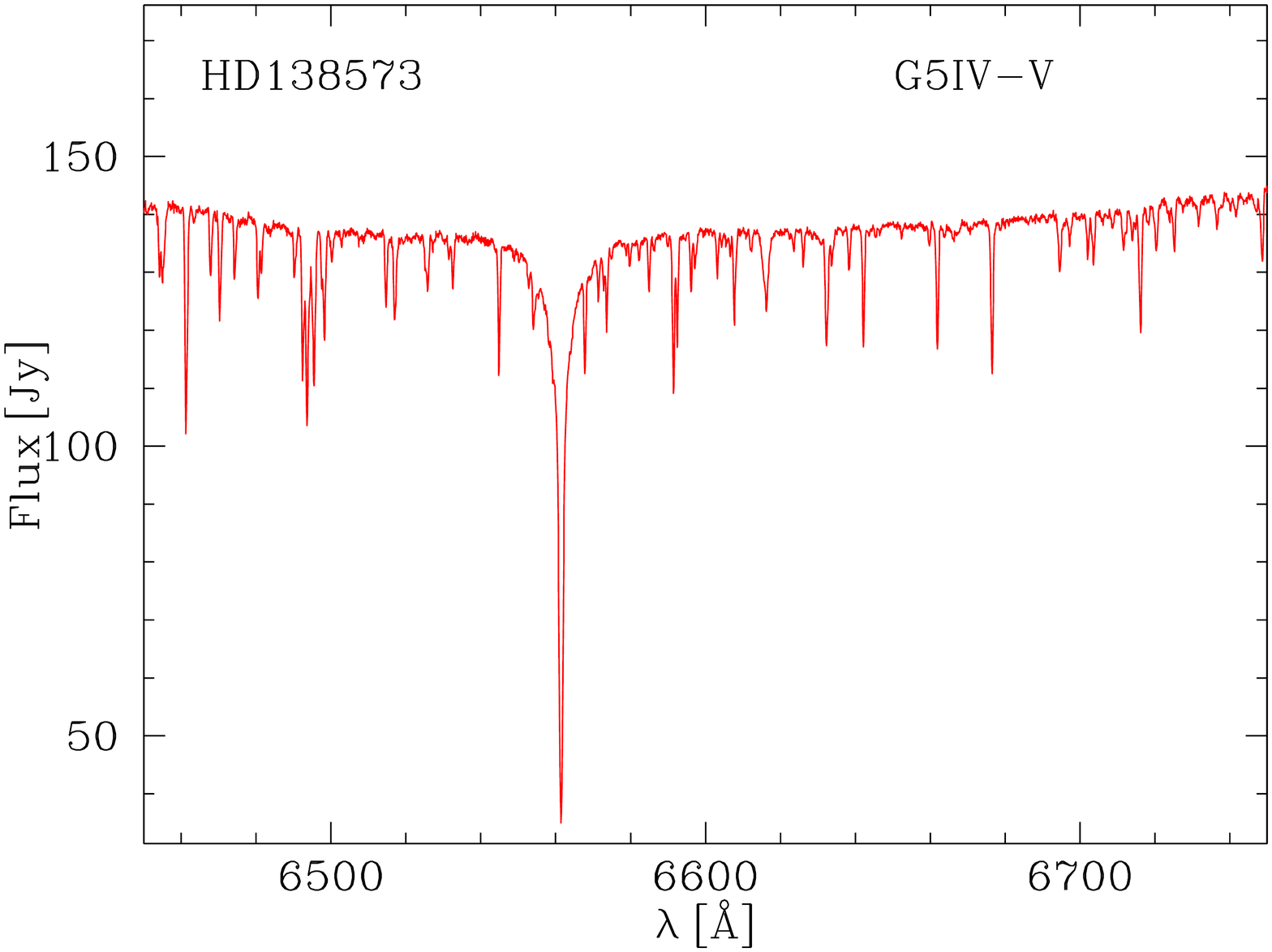}
\includegraphics[width=0.18\textwidth,angle=-90]{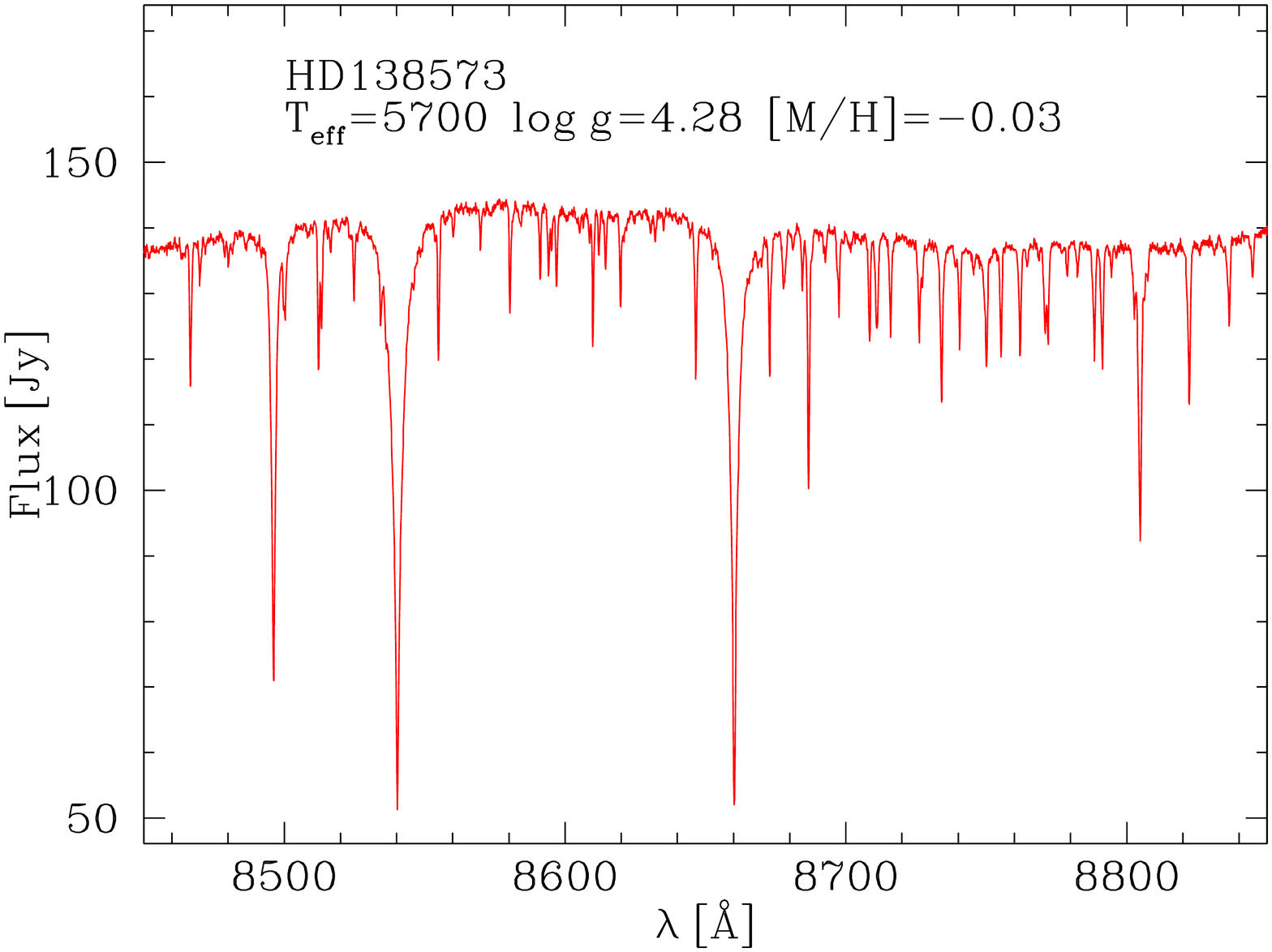}
\includegraphics[width=0.18\textwidth,angle=-90]{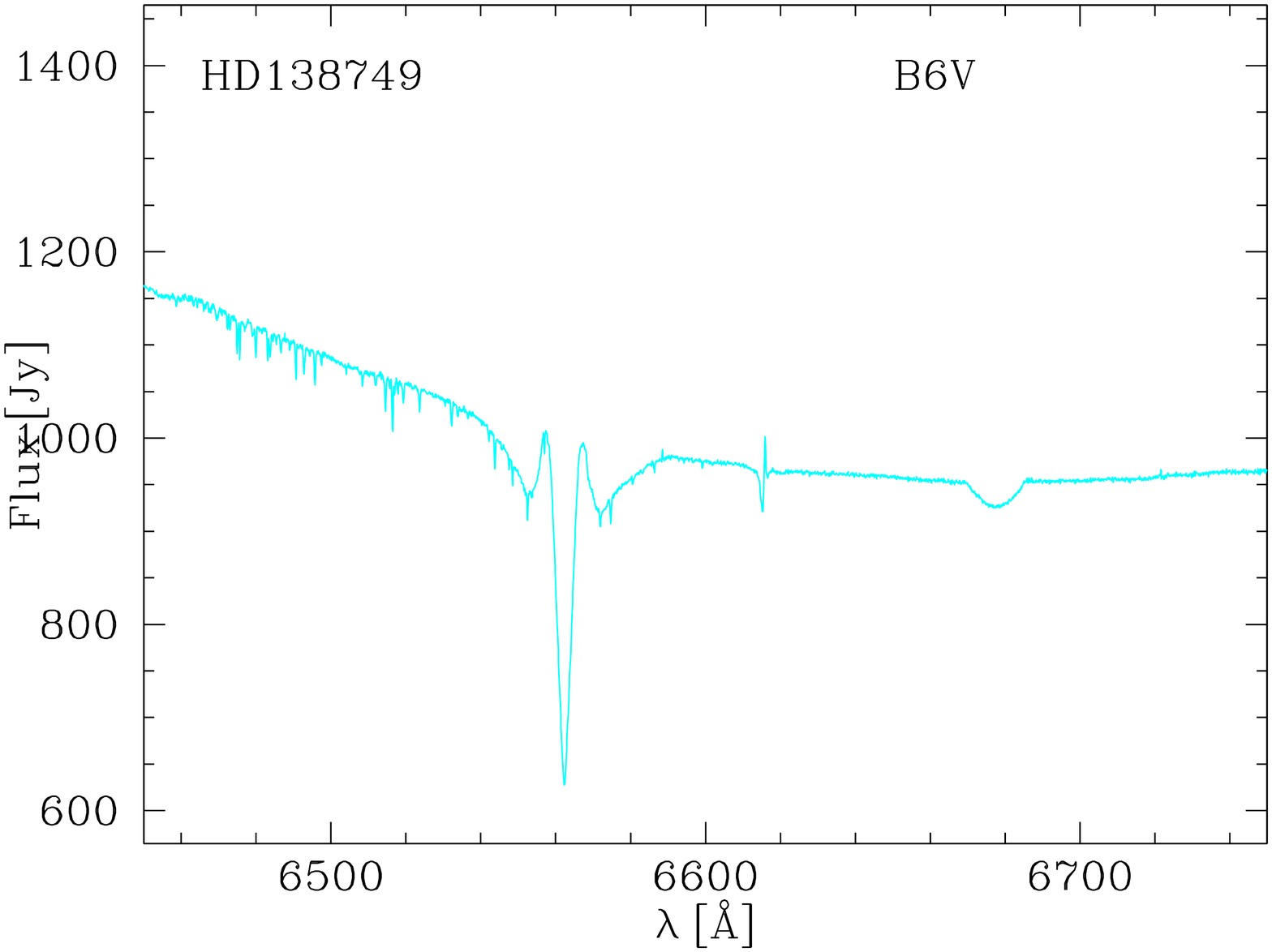}
\includegraphics[width=0.18\textwidth,angle=-90]{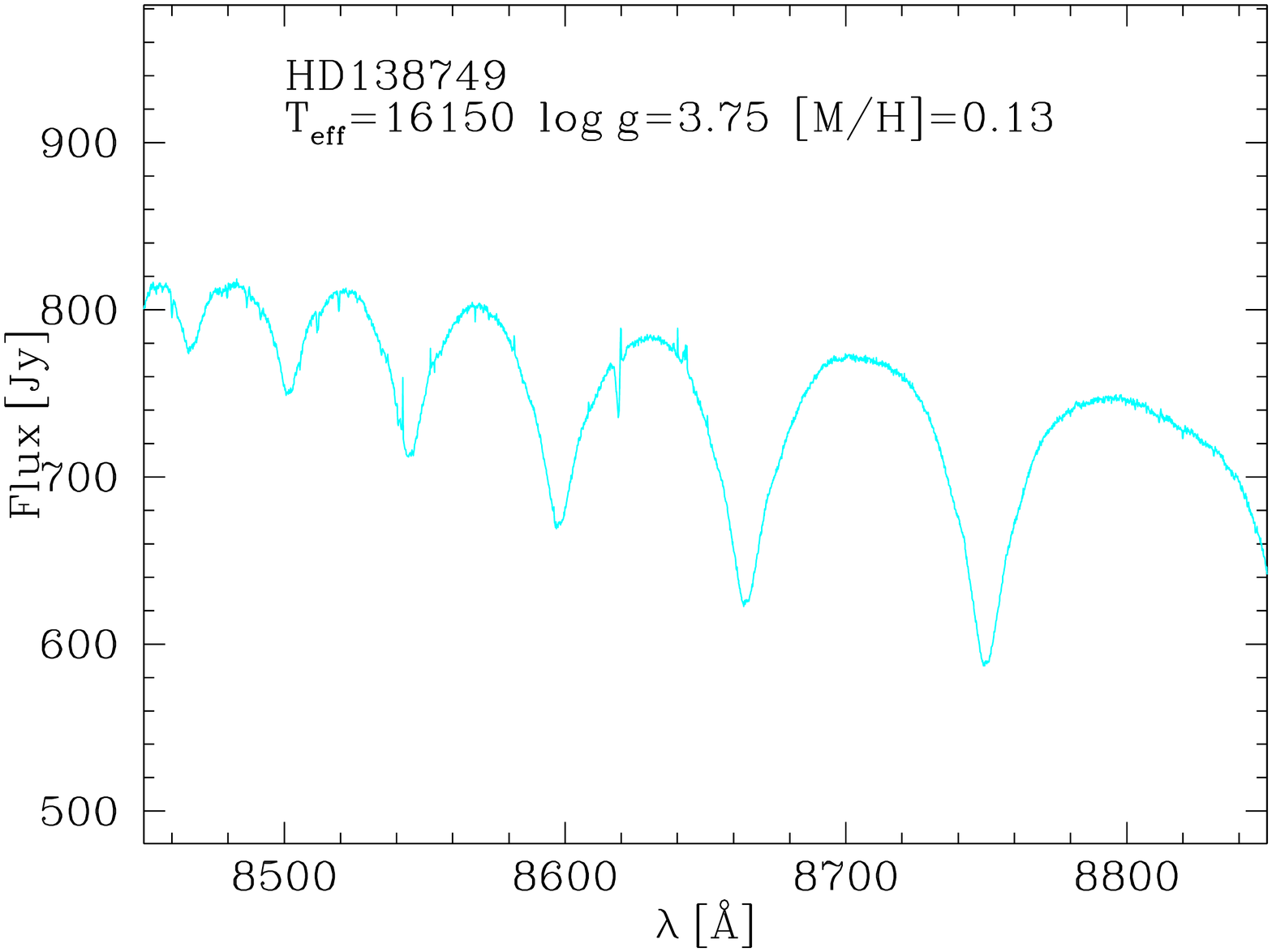}
\includegraphics[width=0.18\textwidth,angle=-90]{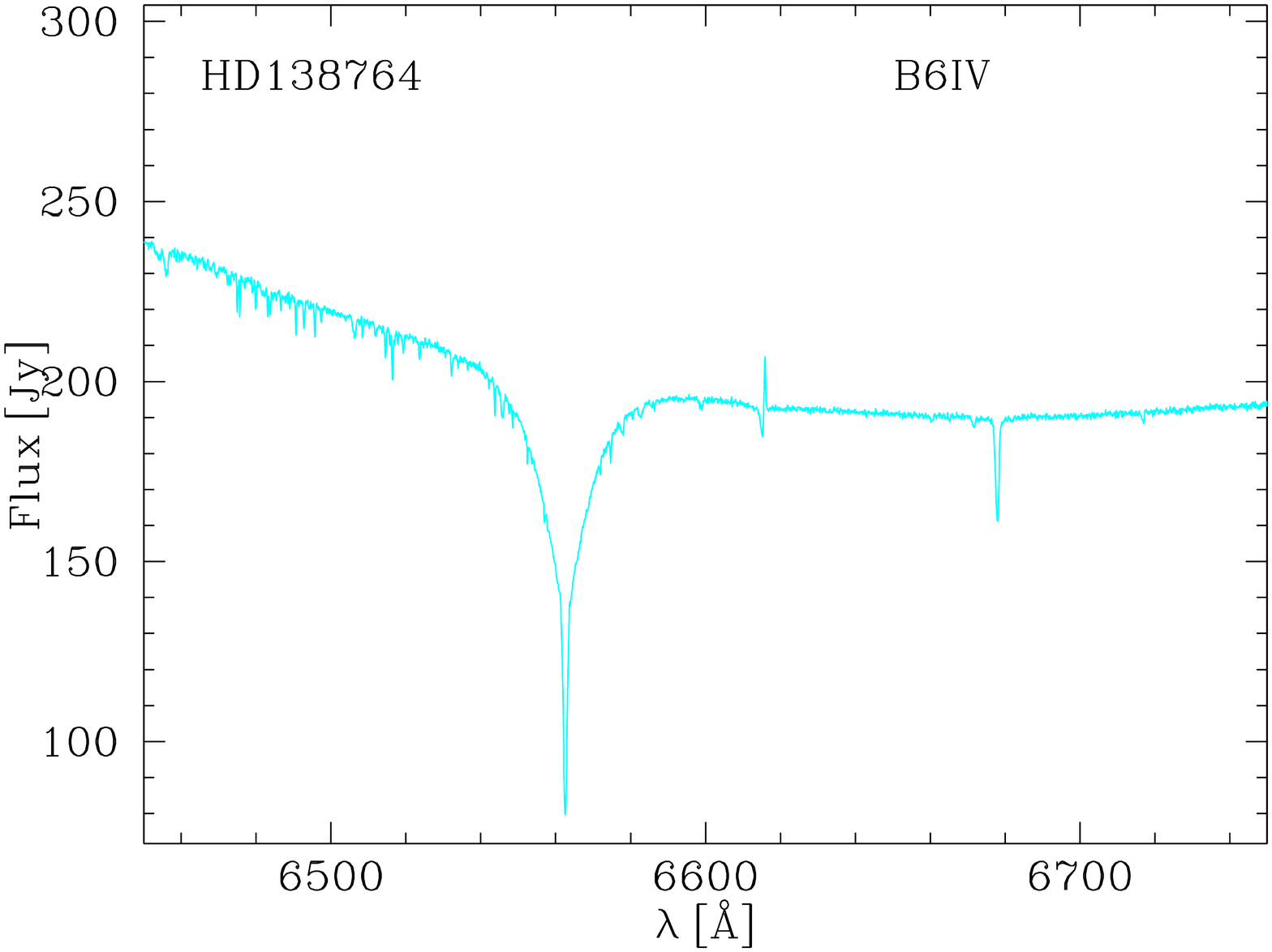}
\includegraphics[width=0.18\textwidth,angle=-90]{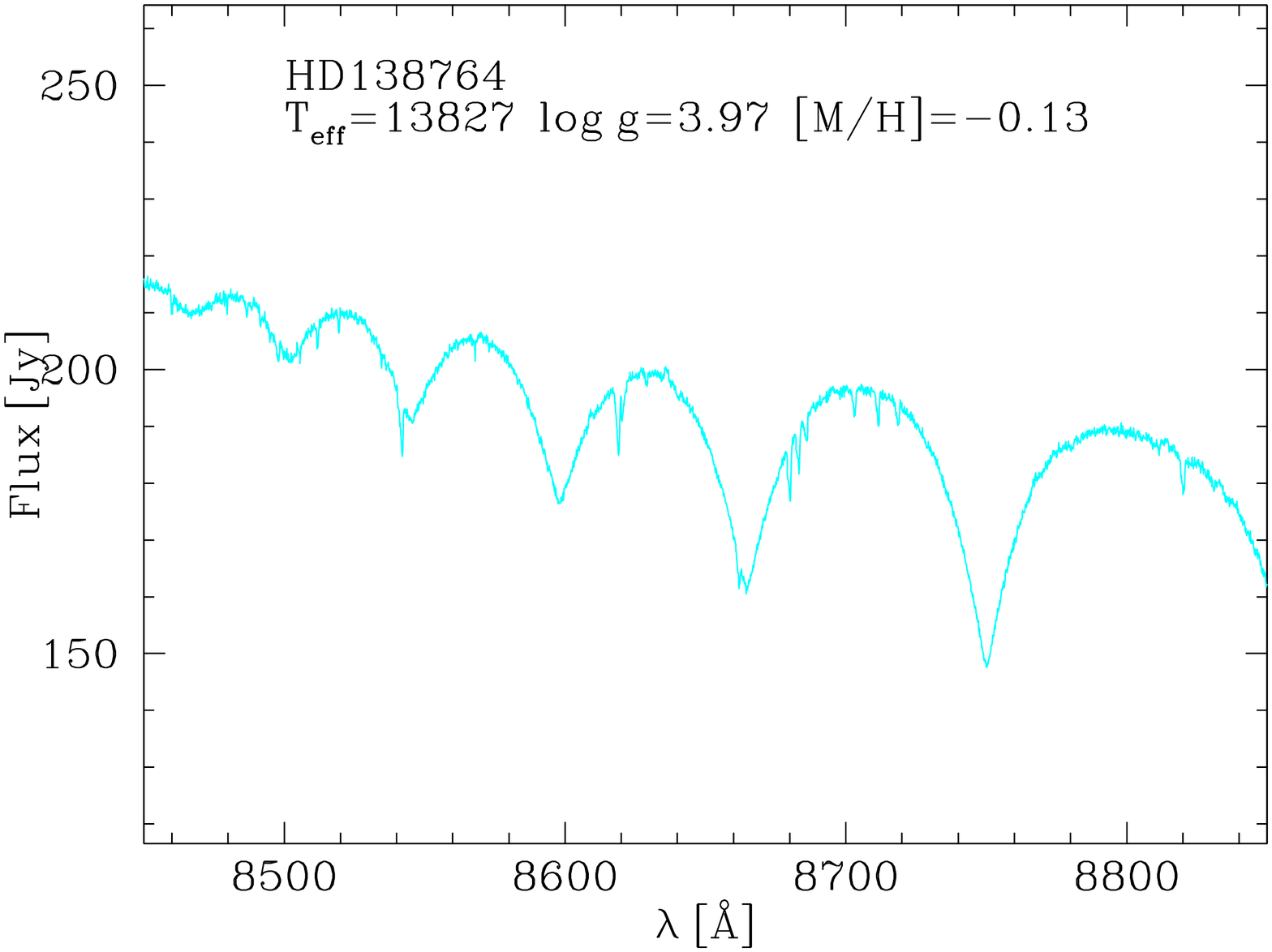}

\contcaption{23. Stars shown in this page are: HD129174, HD129336, HD131111, HD131156, HD131156B, HD131507, HD132756, HD134083, HD134113, HD135101, HD137391, HD138573, HD138749 and HD138764.}
\end{figure*}

\begin{figure*}
\includegraphics[width=0.18\textwidth,angle=-90]{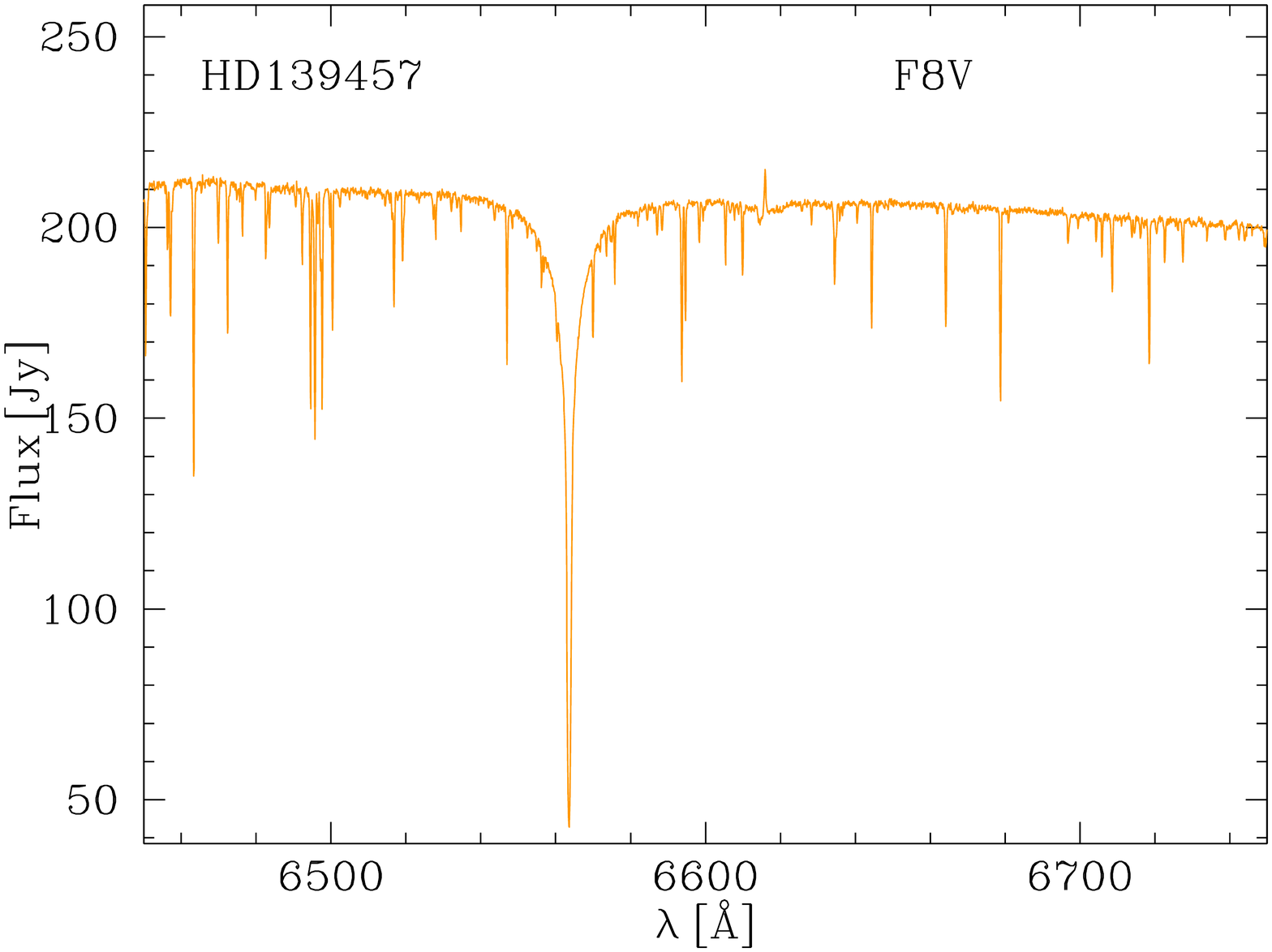}
\includegraphics[width=0.18\textwidth,angle=-90]{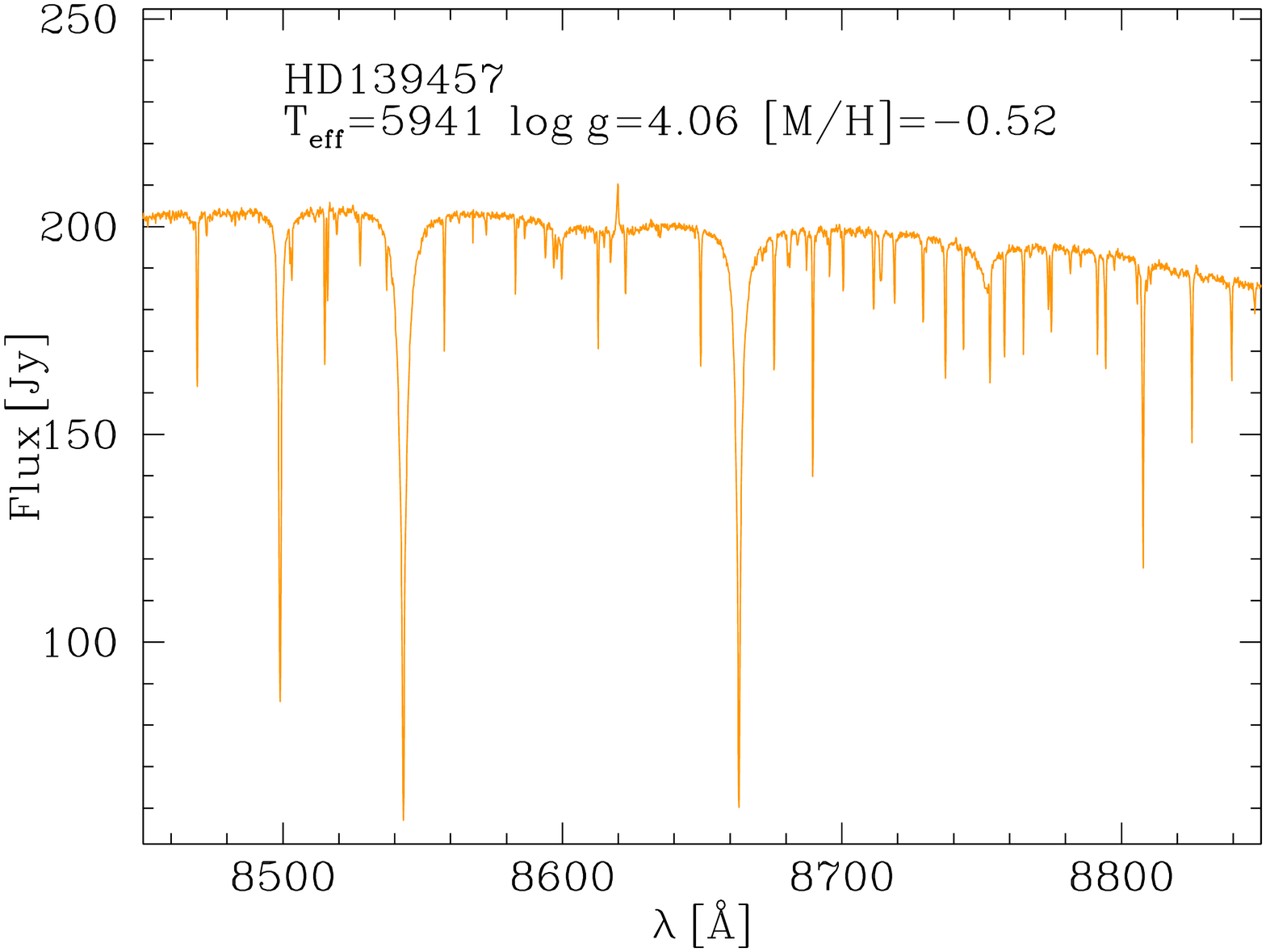}
\includegraphics[width=0.18\textwidth,angle=-90]{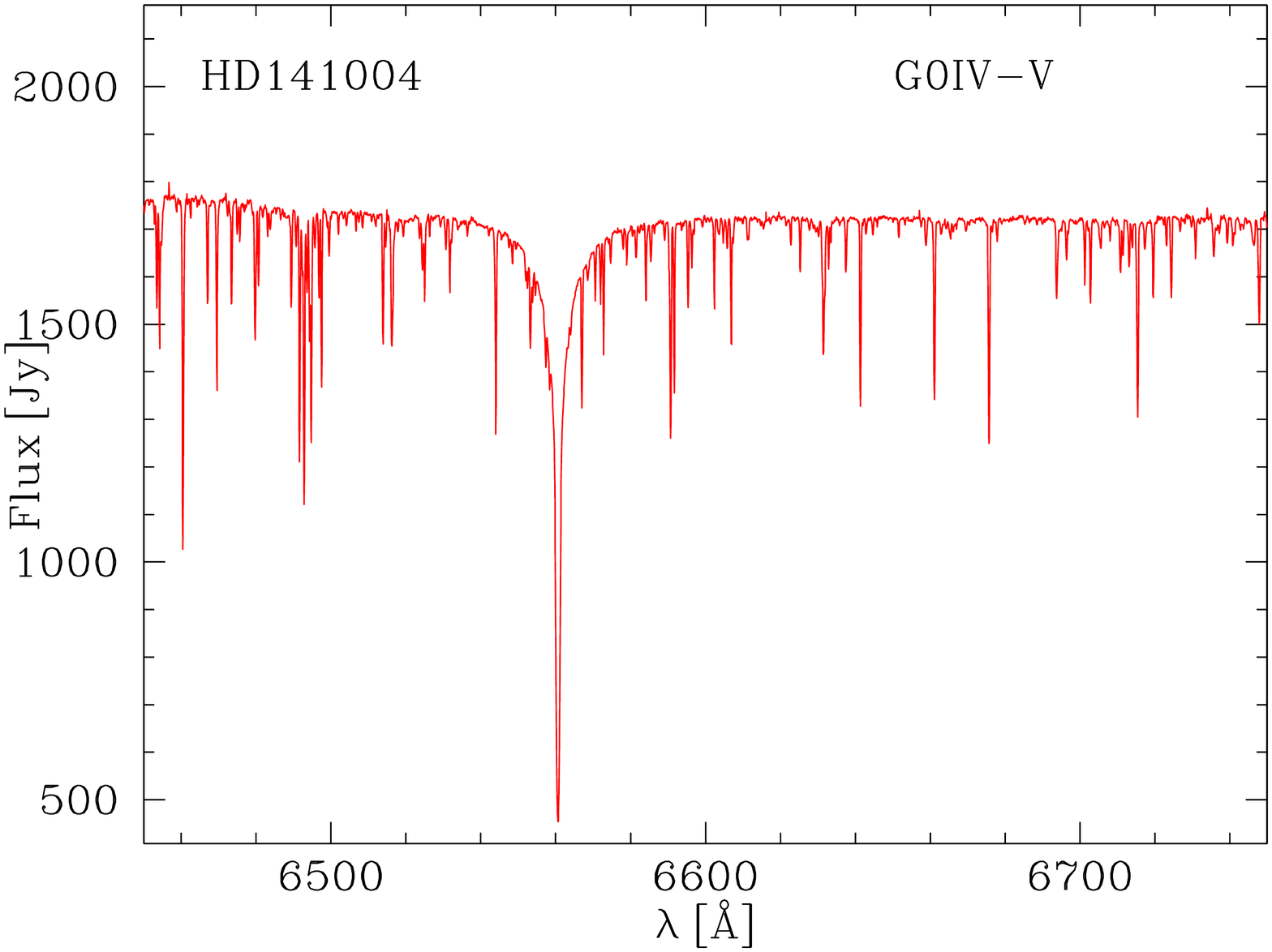}
\includegraphics[width=0.18\textwidth,angle=-90]{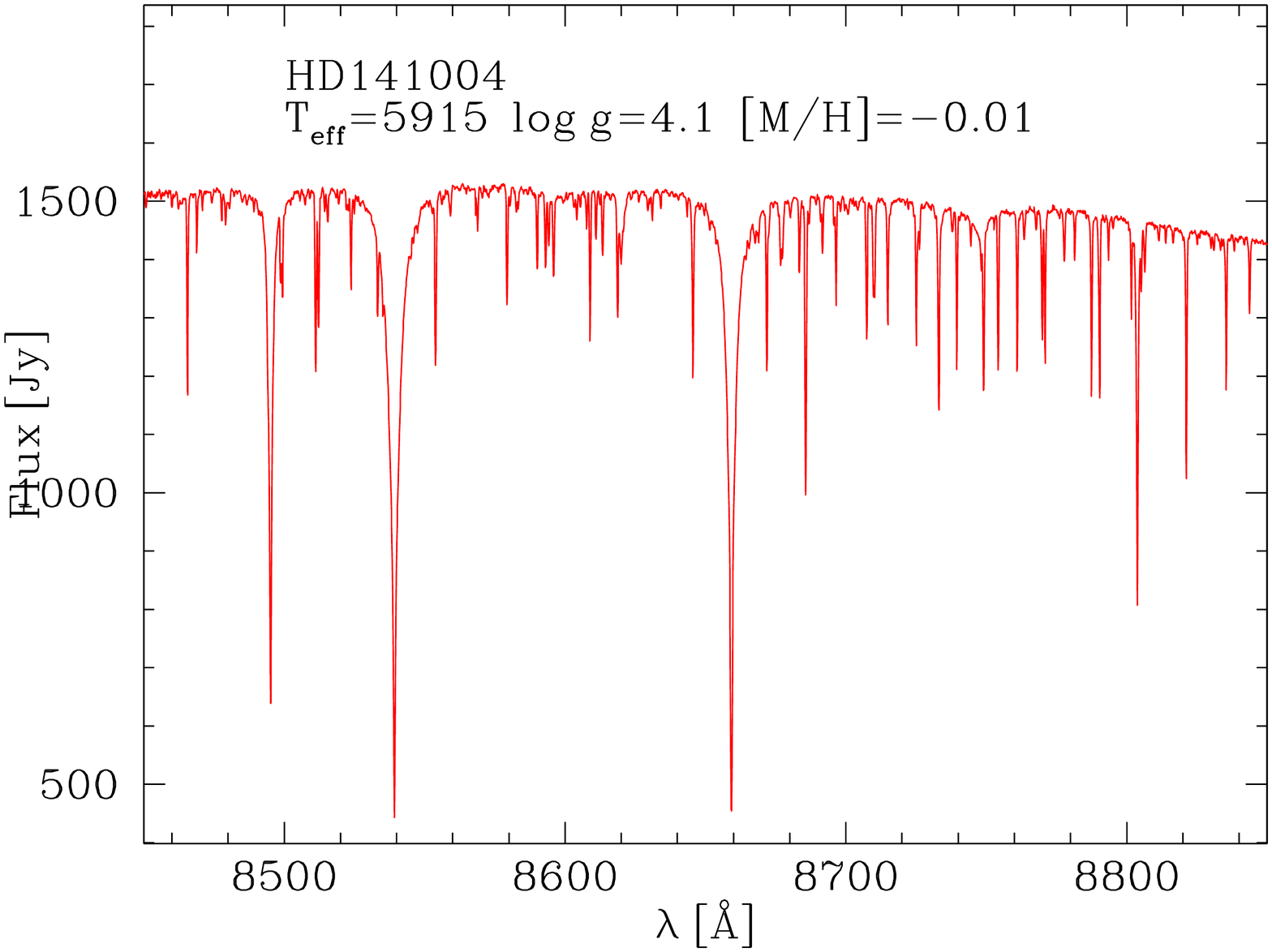}
\includegraphics[width=0.18\textwidth,angle=-90]{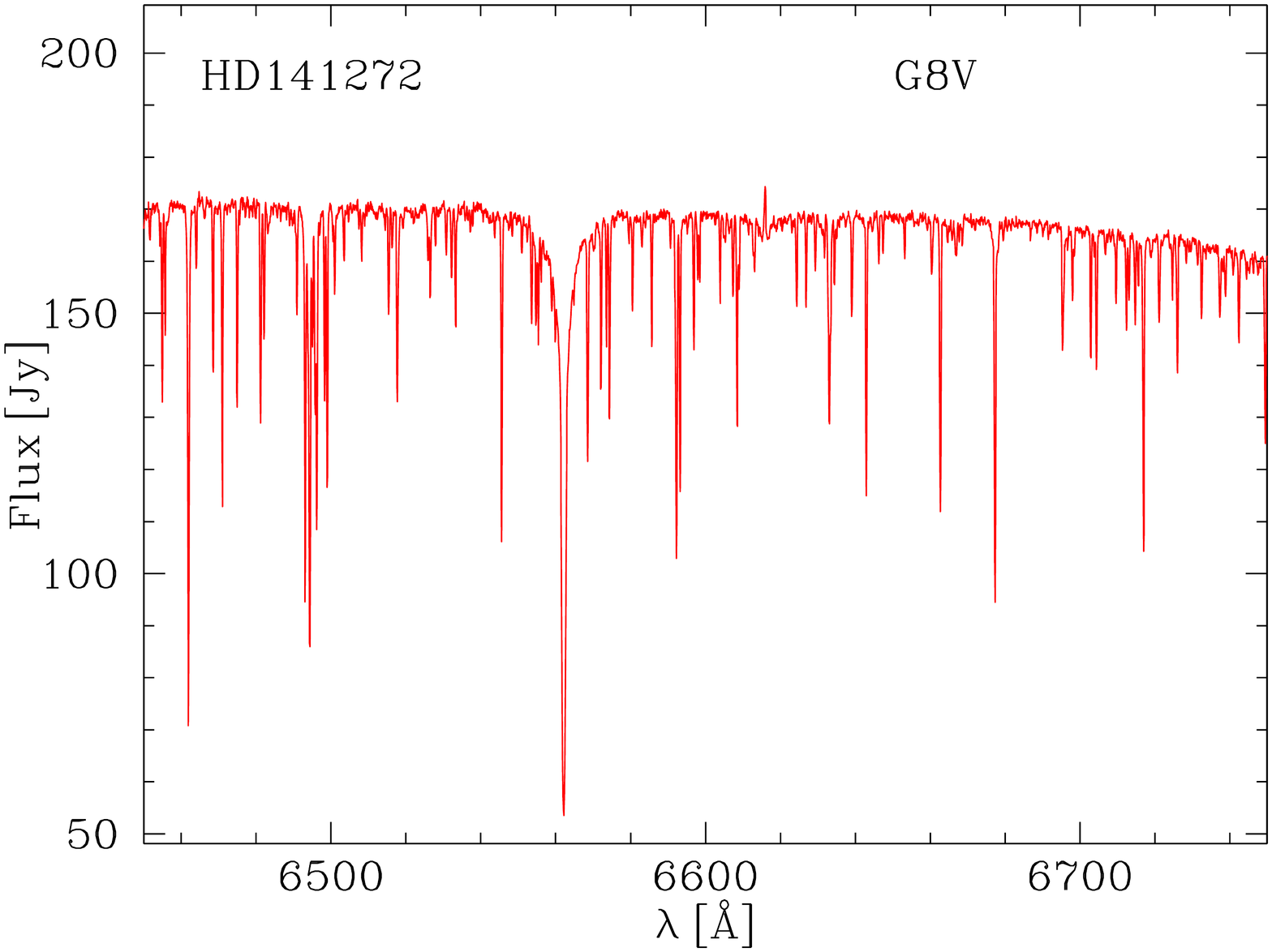}
\includegraphics[width=0.18\textwidth,angle=-90]{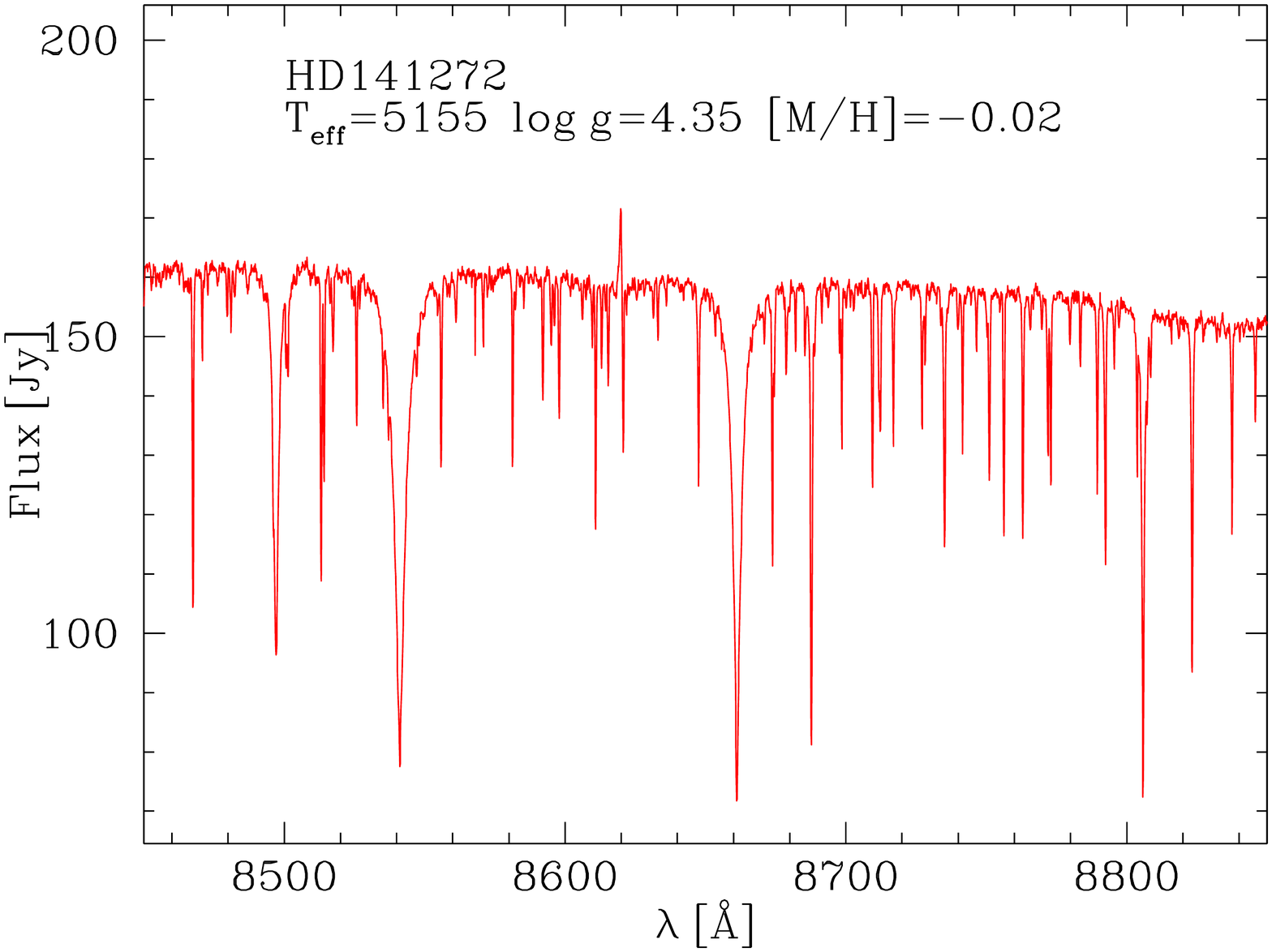}
\includegraphics[width=0.18\textwidth,angle=-90]{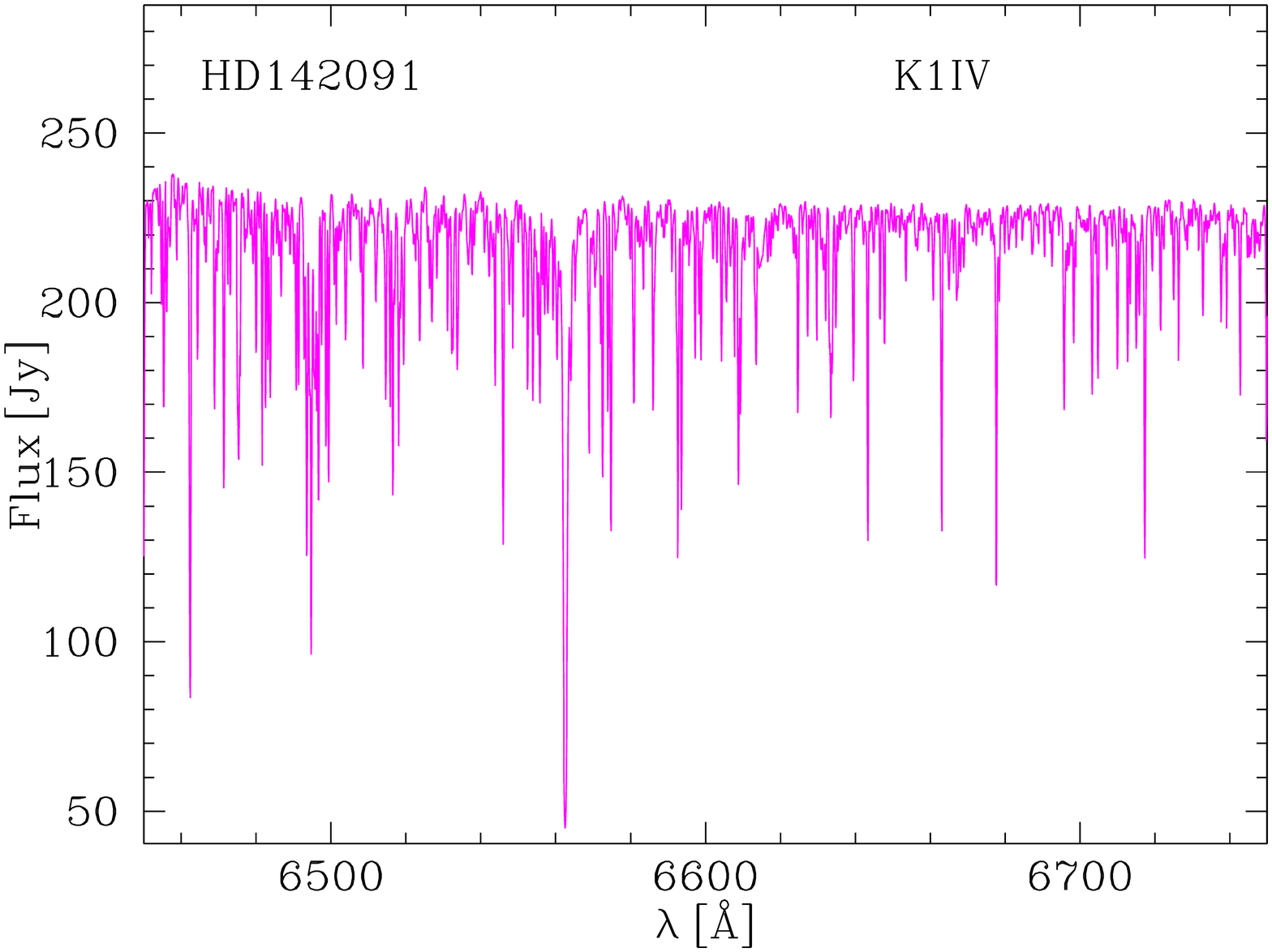}
\includegraphics[width=0.18\textwidth,angle=-90]{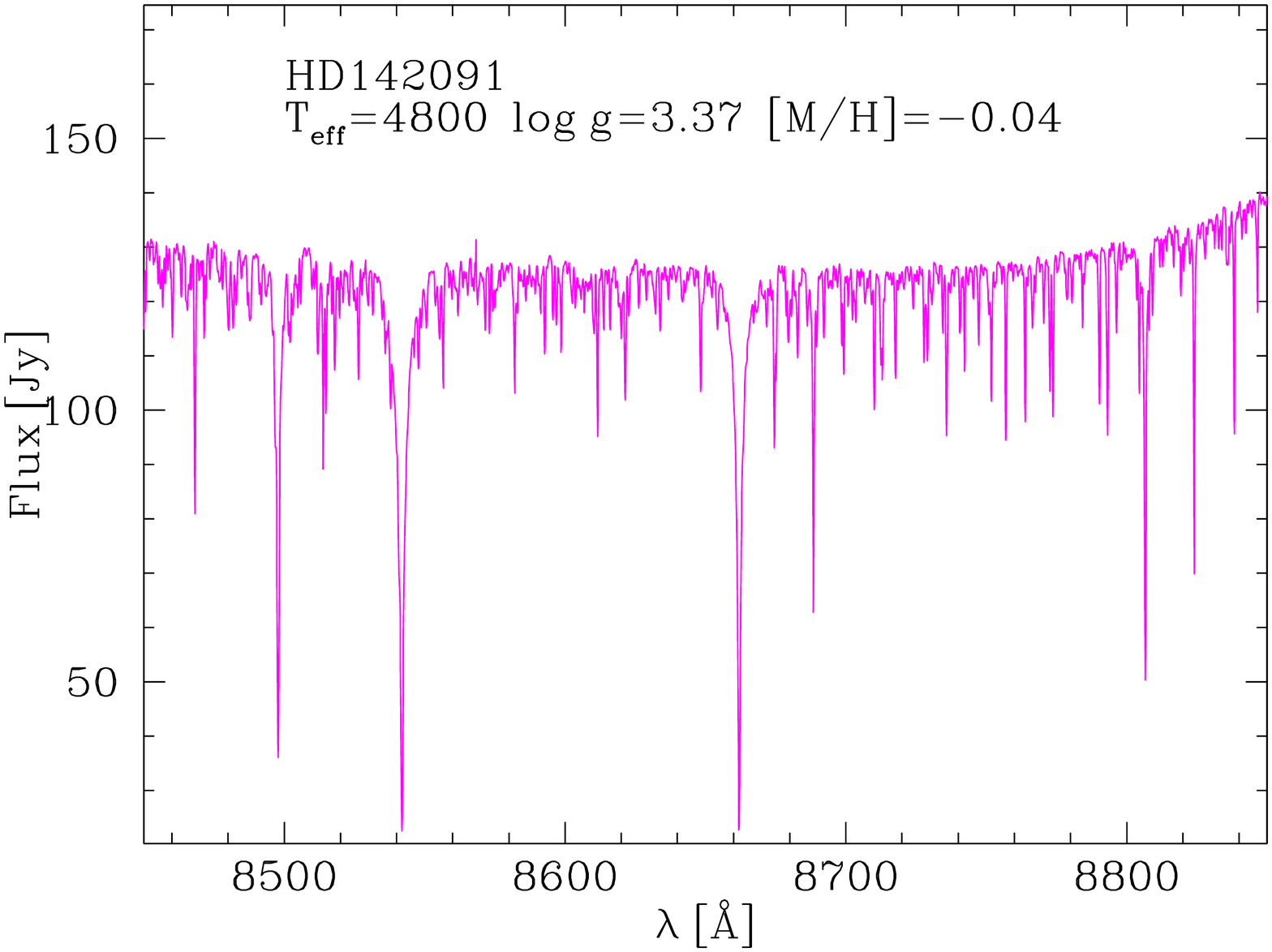}
\includegraphics[width=0.18\textwidth,angle=-90]{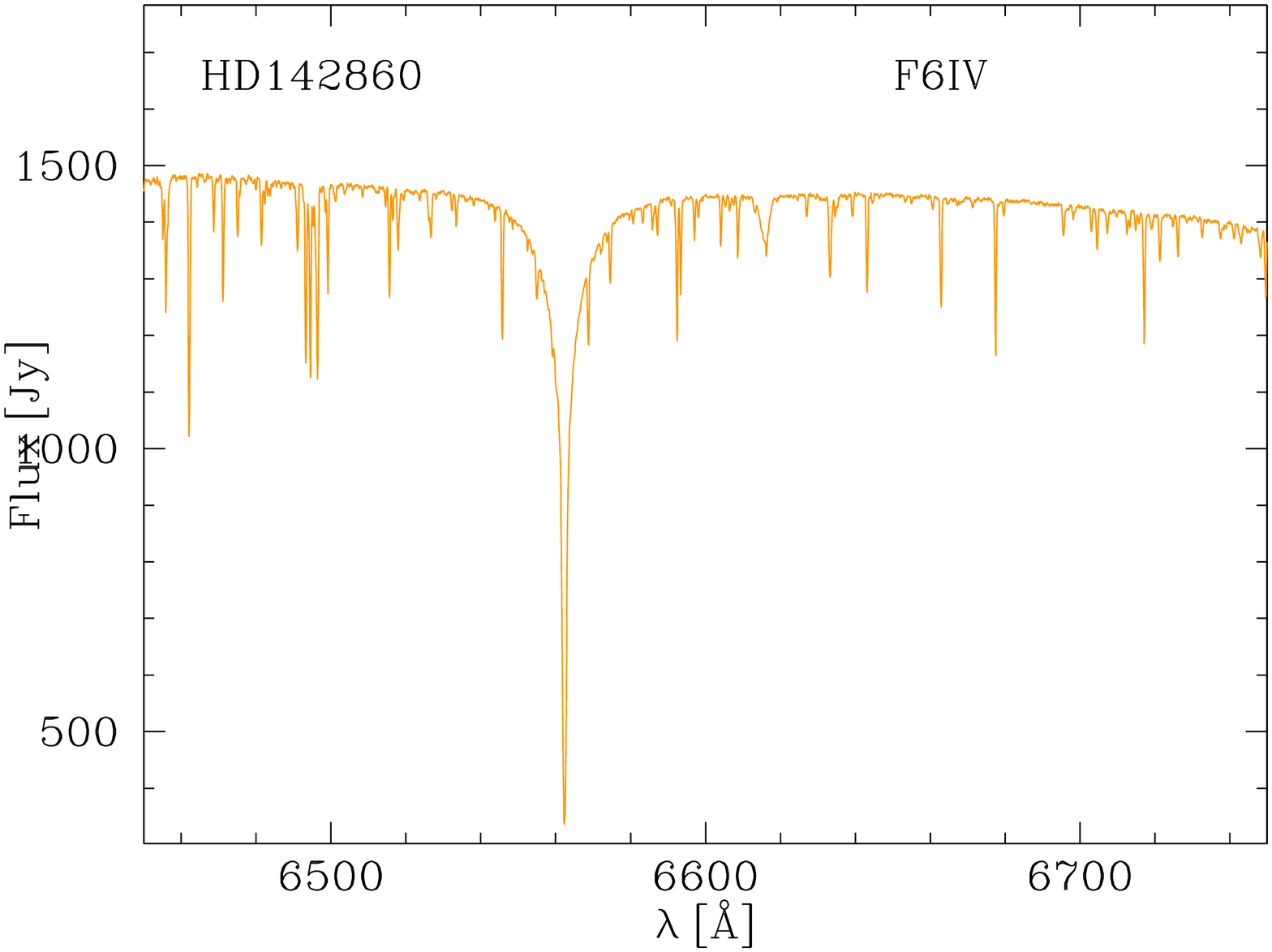}
\includegraphics[width=0.18\textwidth,angle=-90]{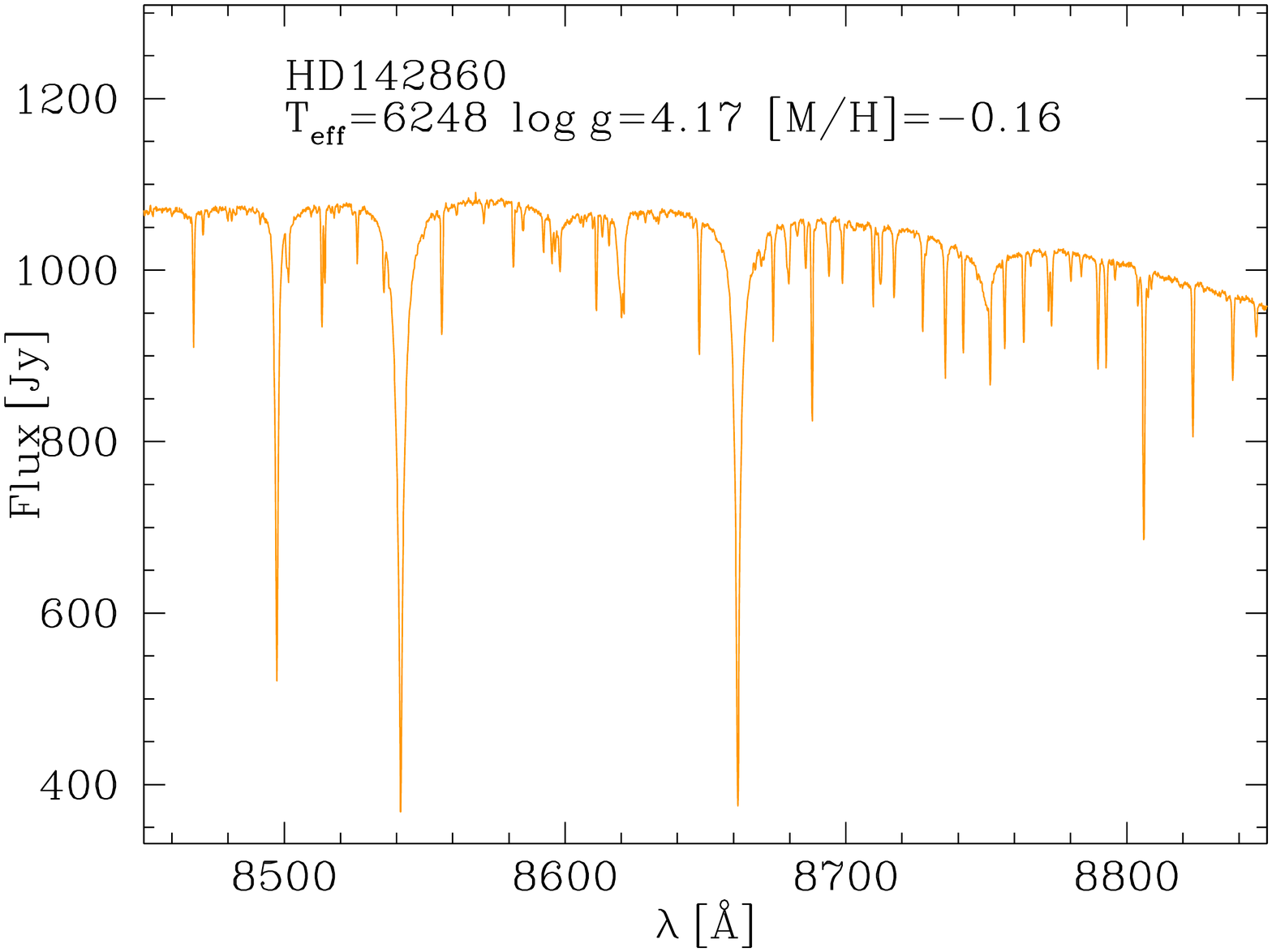}
\includegraphics[width=0.18\textwidth,angle=-90]{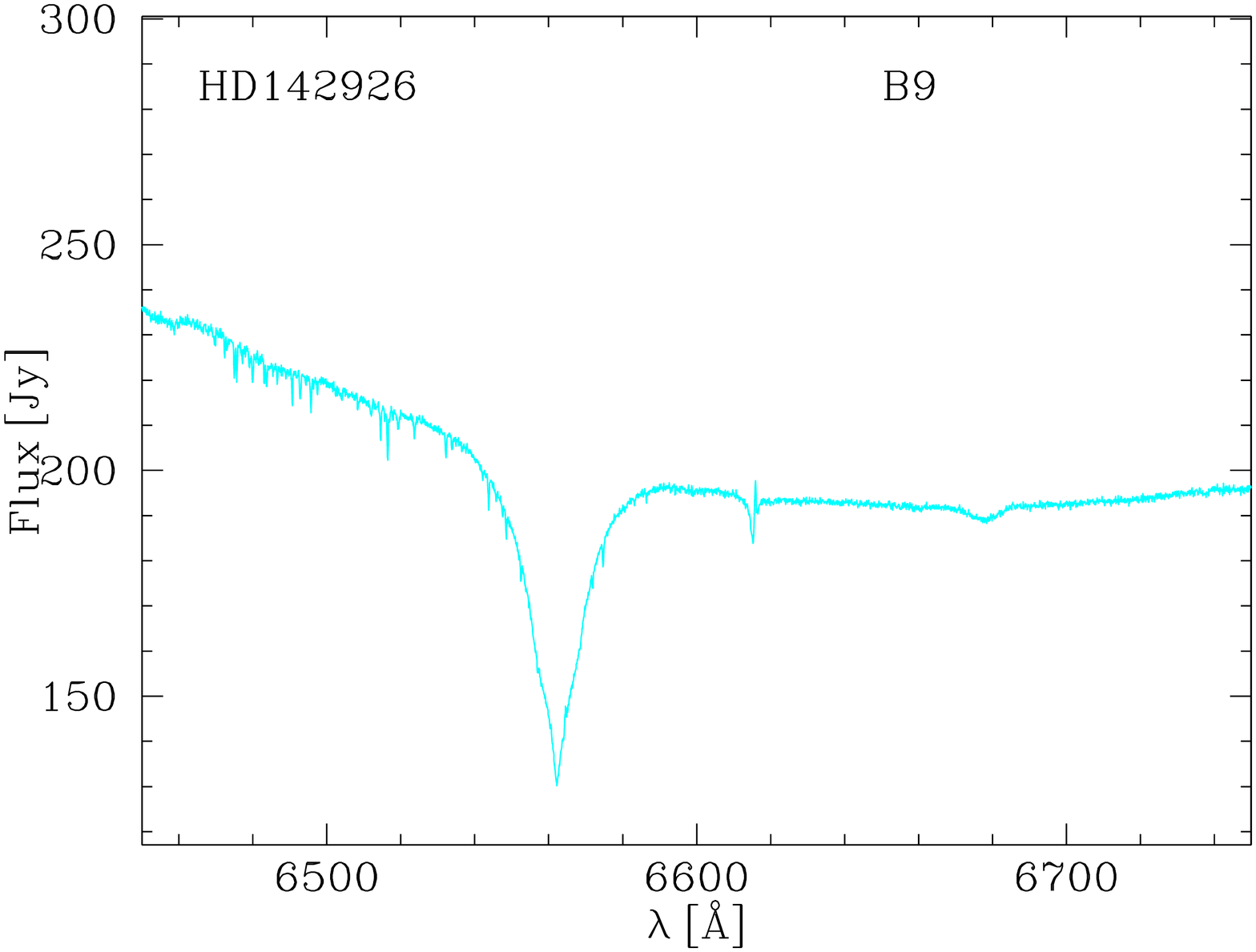}
\includegraphics[width=0.18\textwidth,angle=-90]{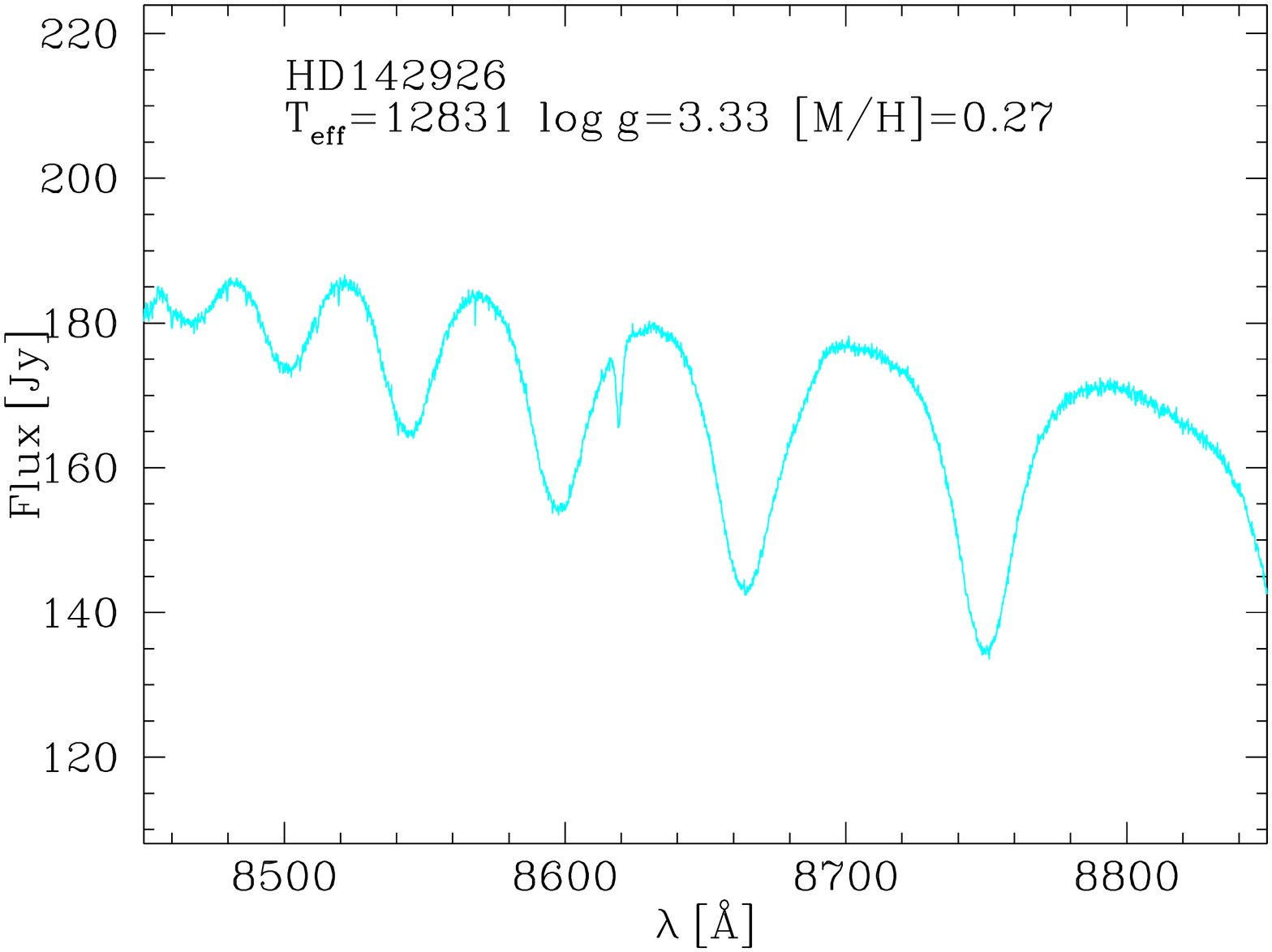}
\includegraphics[width=0.18\textwidth,angle=-90]{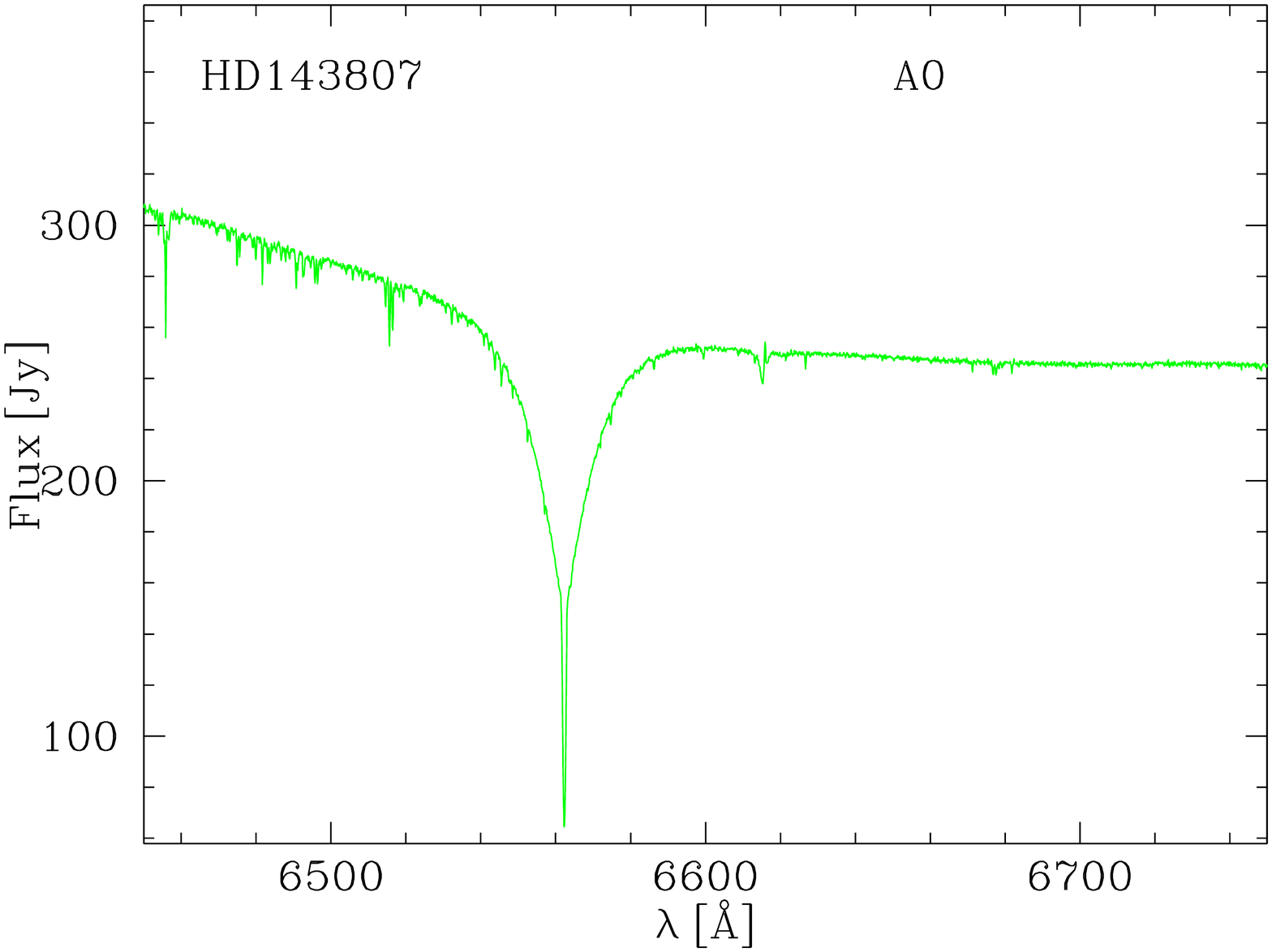}
\includegraphics[width=0.18\textwidth,angle=-90]{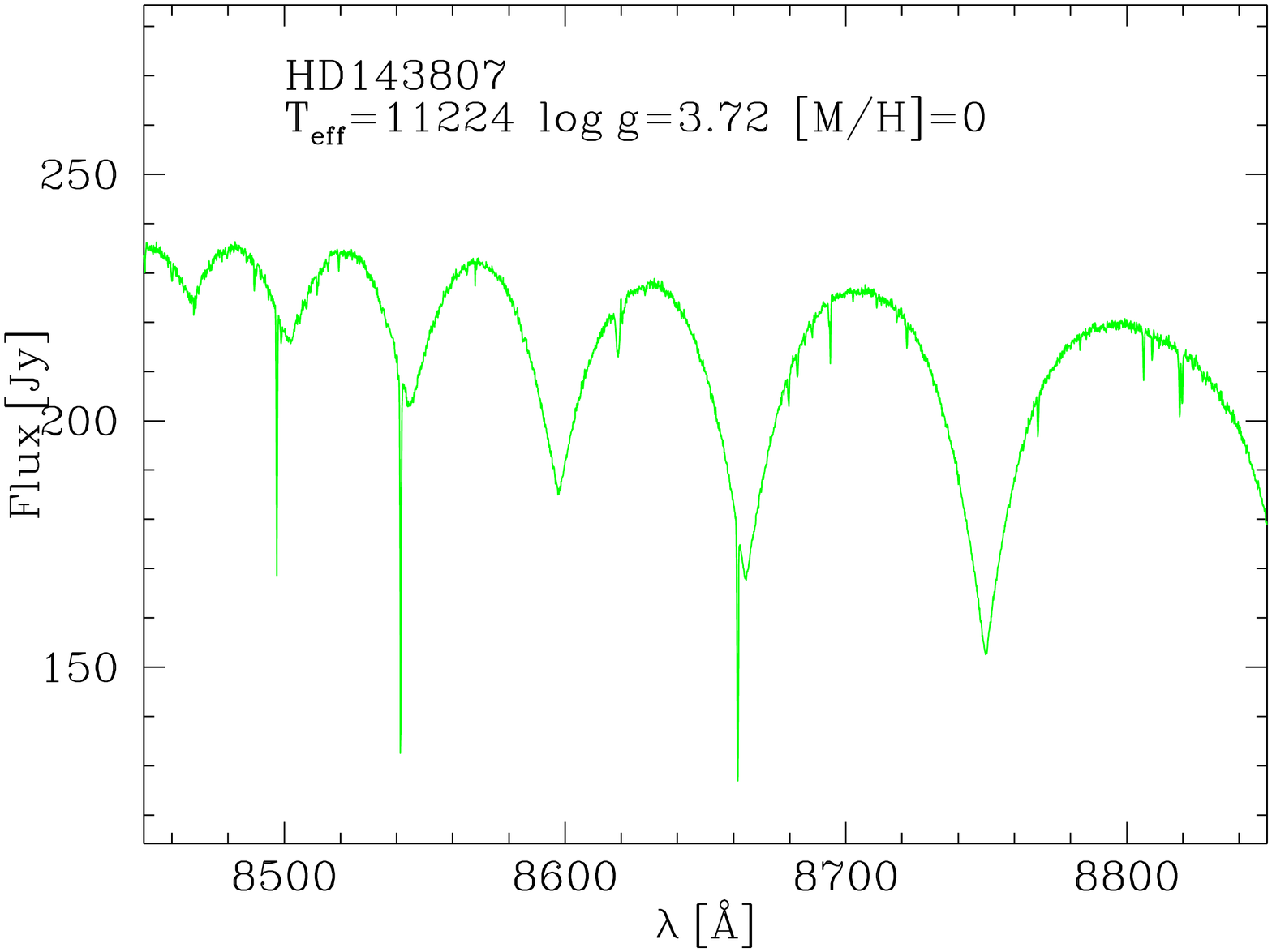}
\includegraphics[width=0.18\textwidth,angle=-90]{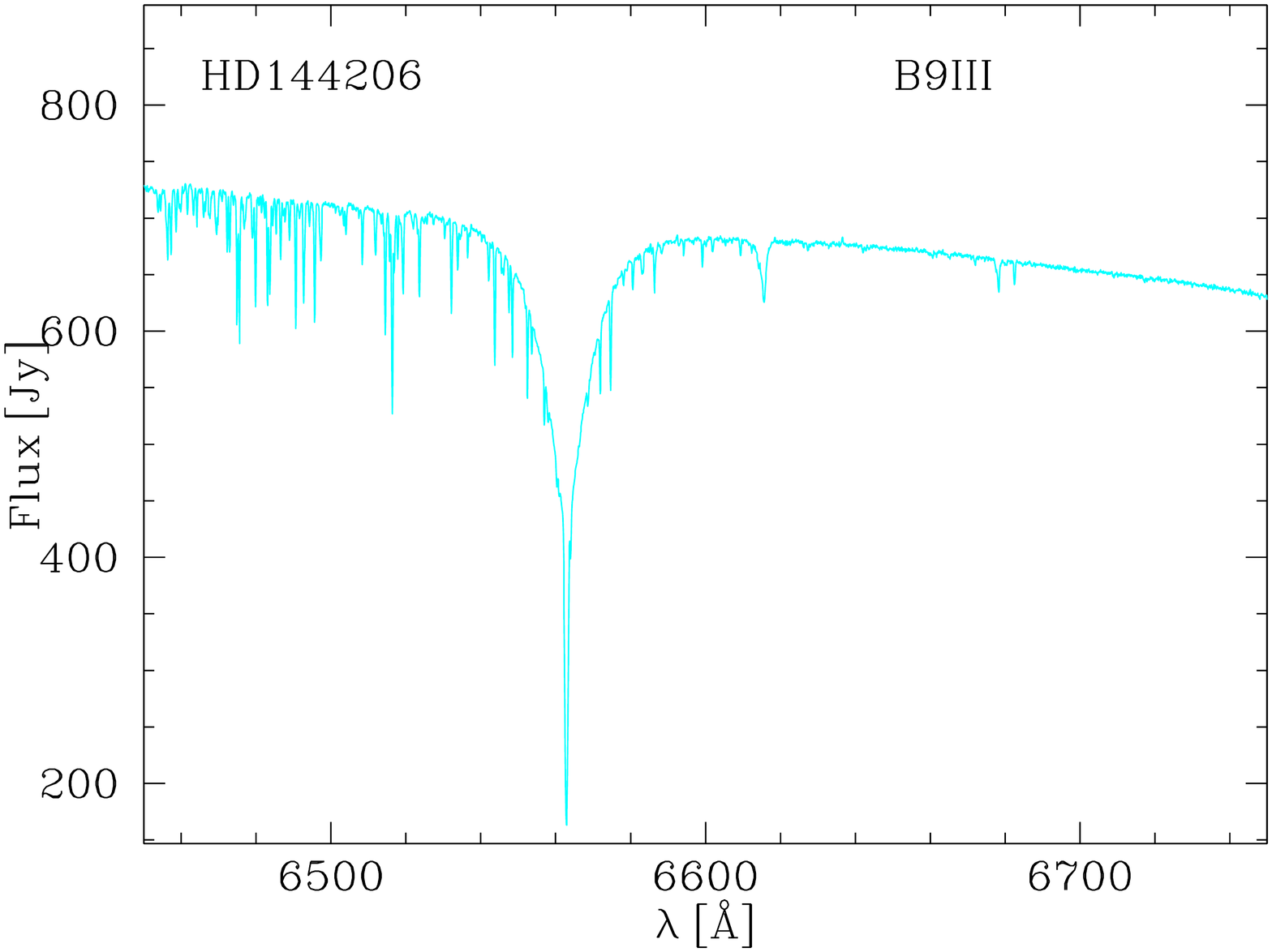}
\includegraphics[width=0.18\textwidth,angle=-90]{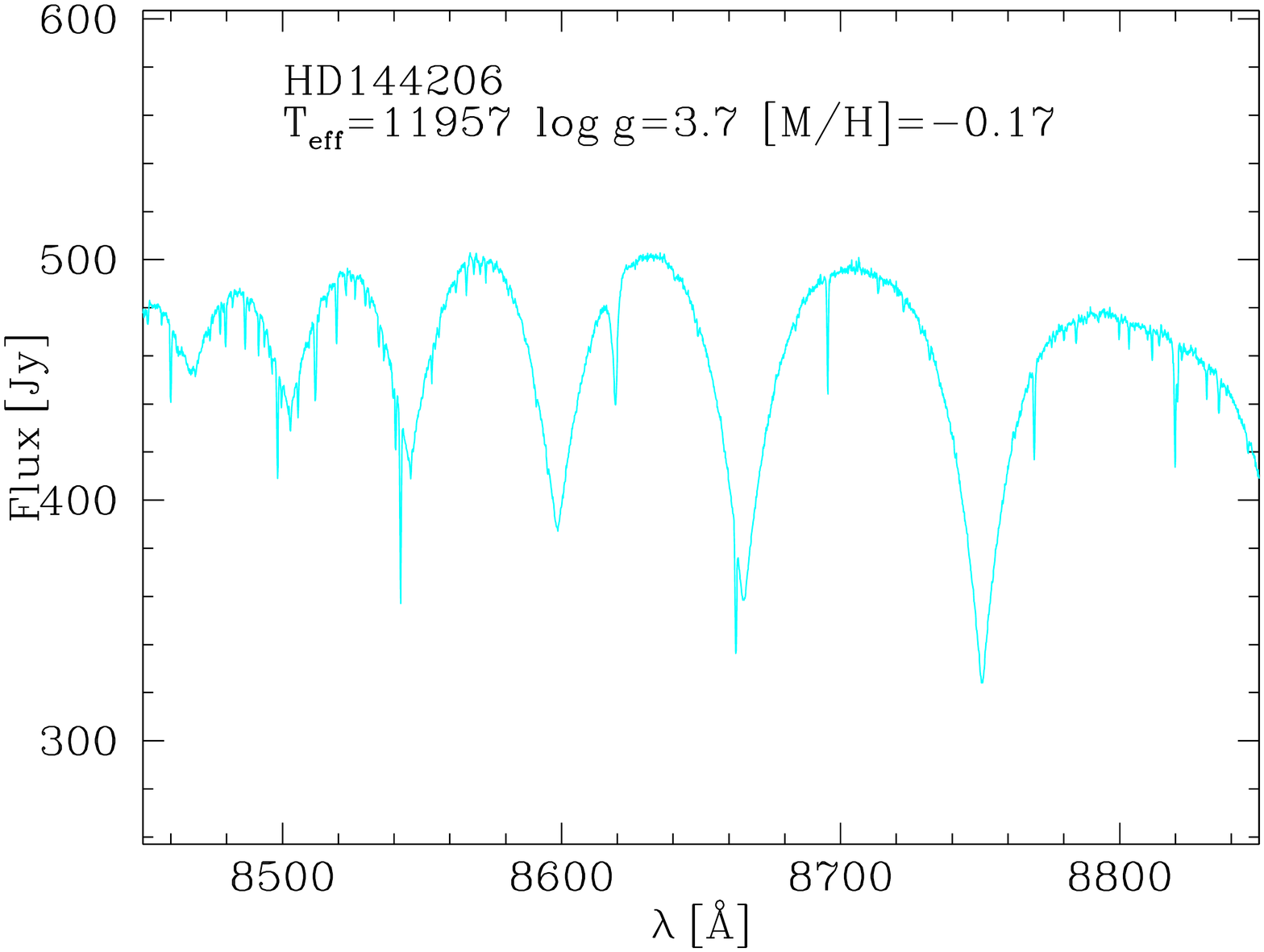}
\includegraphics[width=0.18\textwidth,angle=-90]{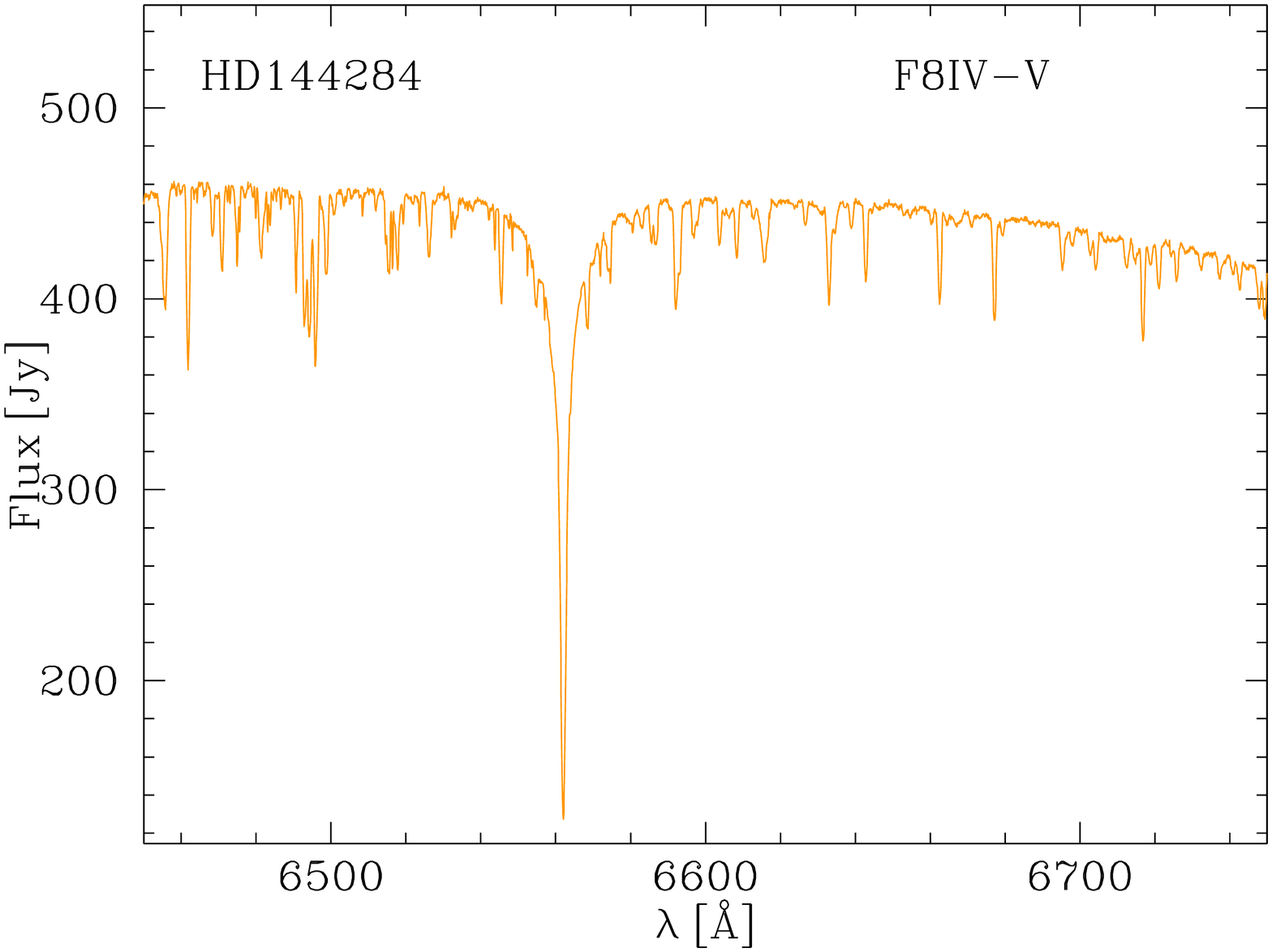}
\includegraphics[width=0.18\textwidth,angle=-90]{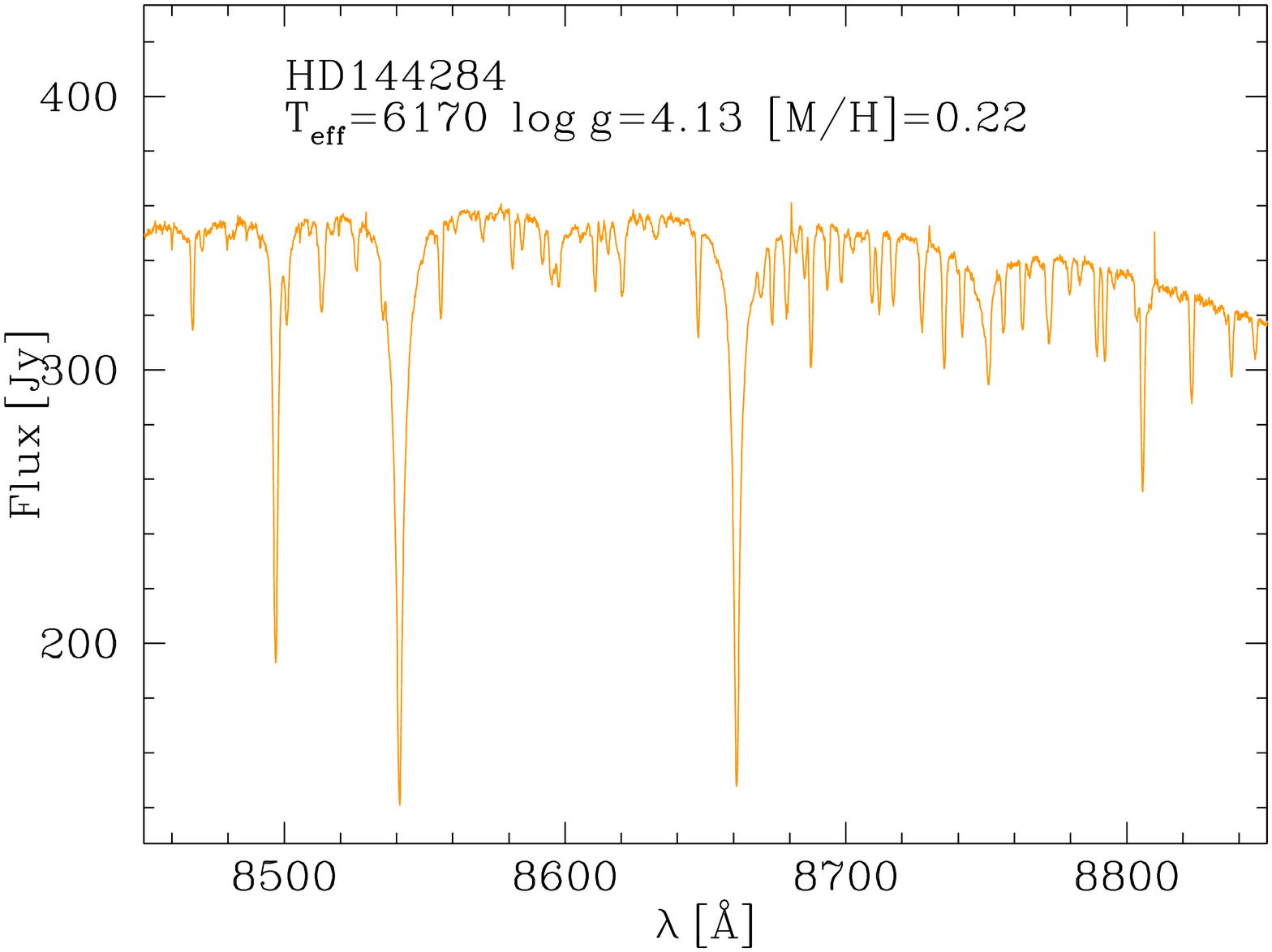}
\includegraphics[width=0.18\textwidth,angle=-90]{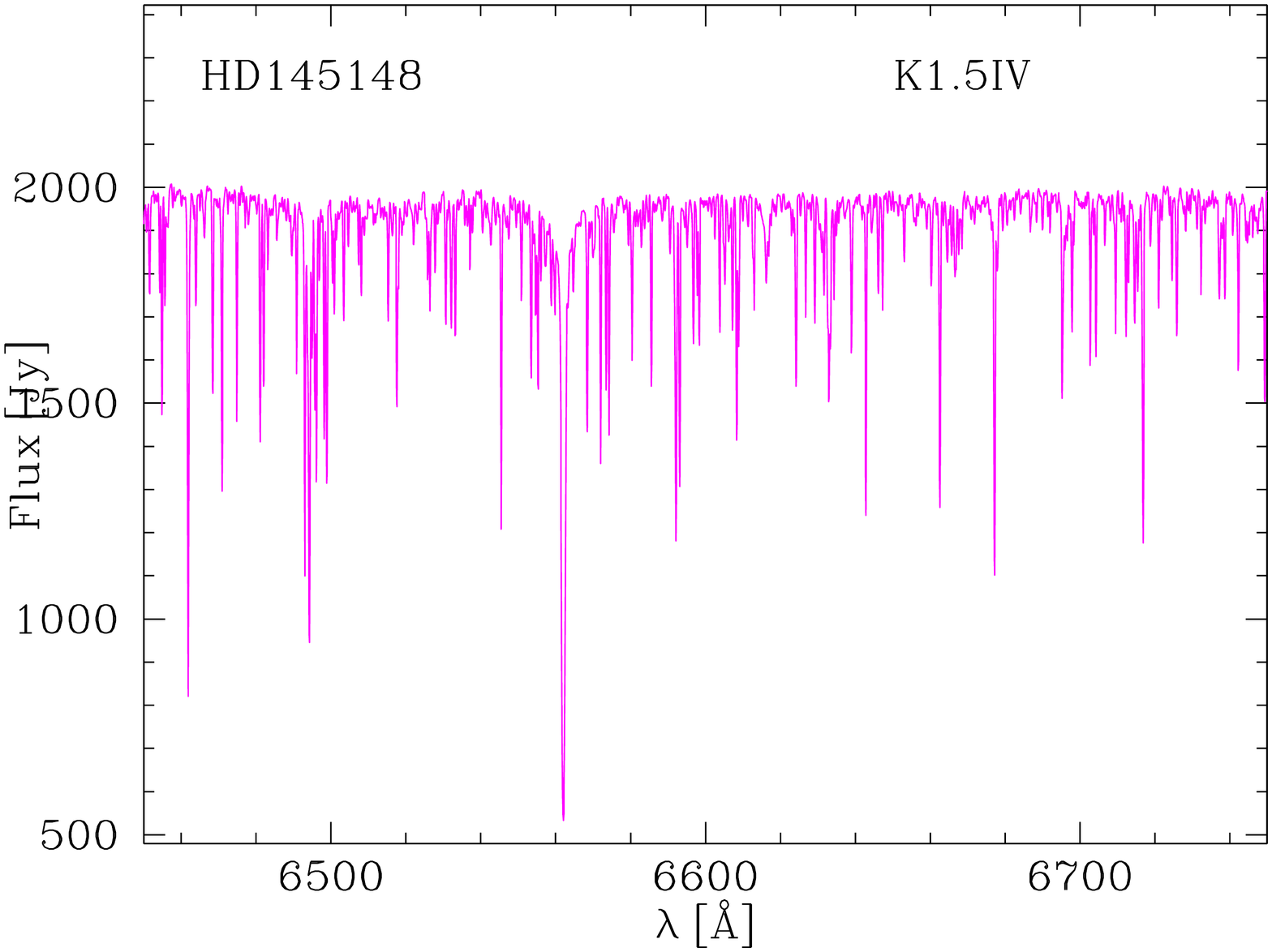}
\includegraphics[width=0.18\textwidth,angle=-90]{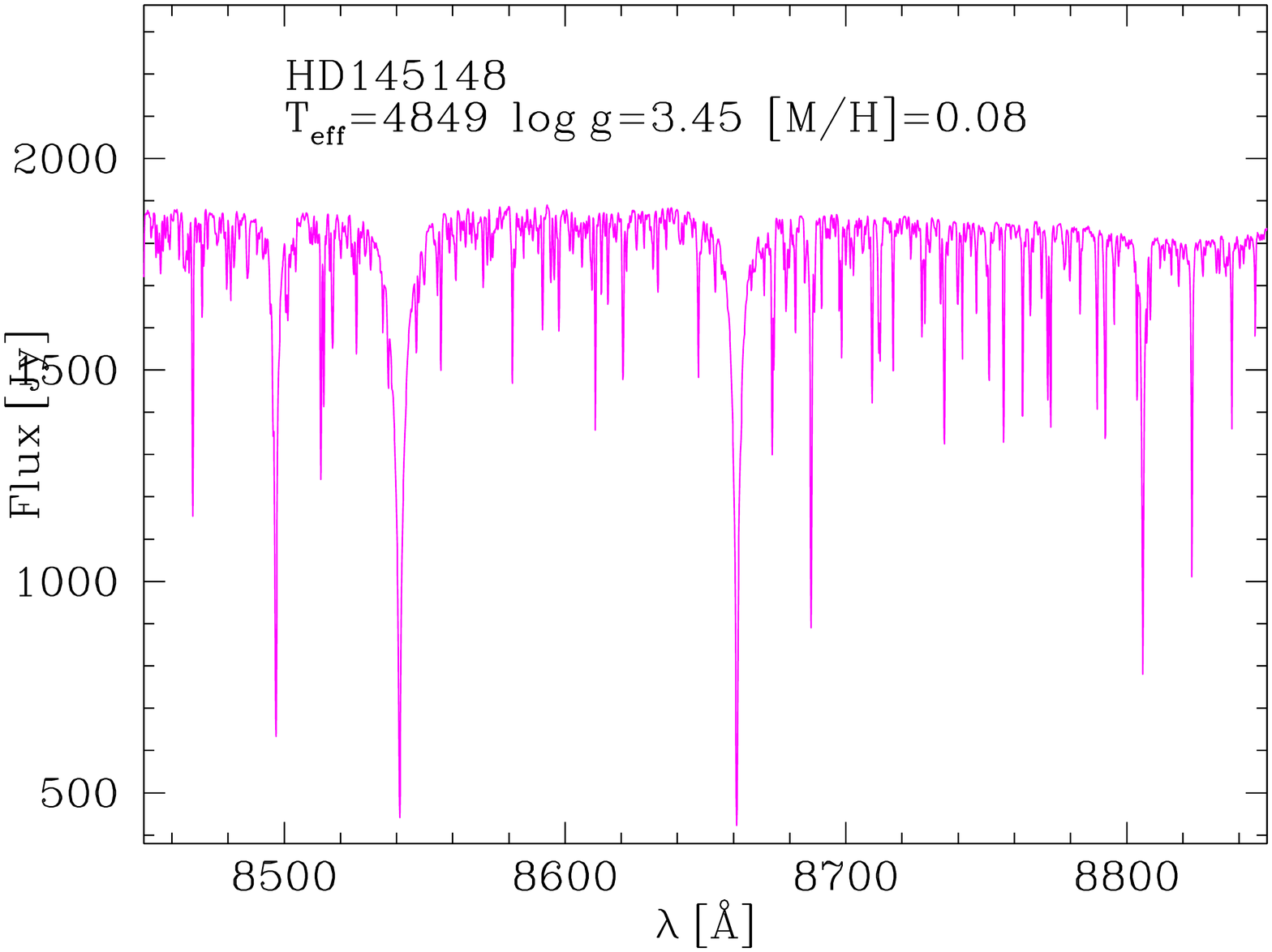}
\includegraphics[width=0.18\textwidth,angle=-90]{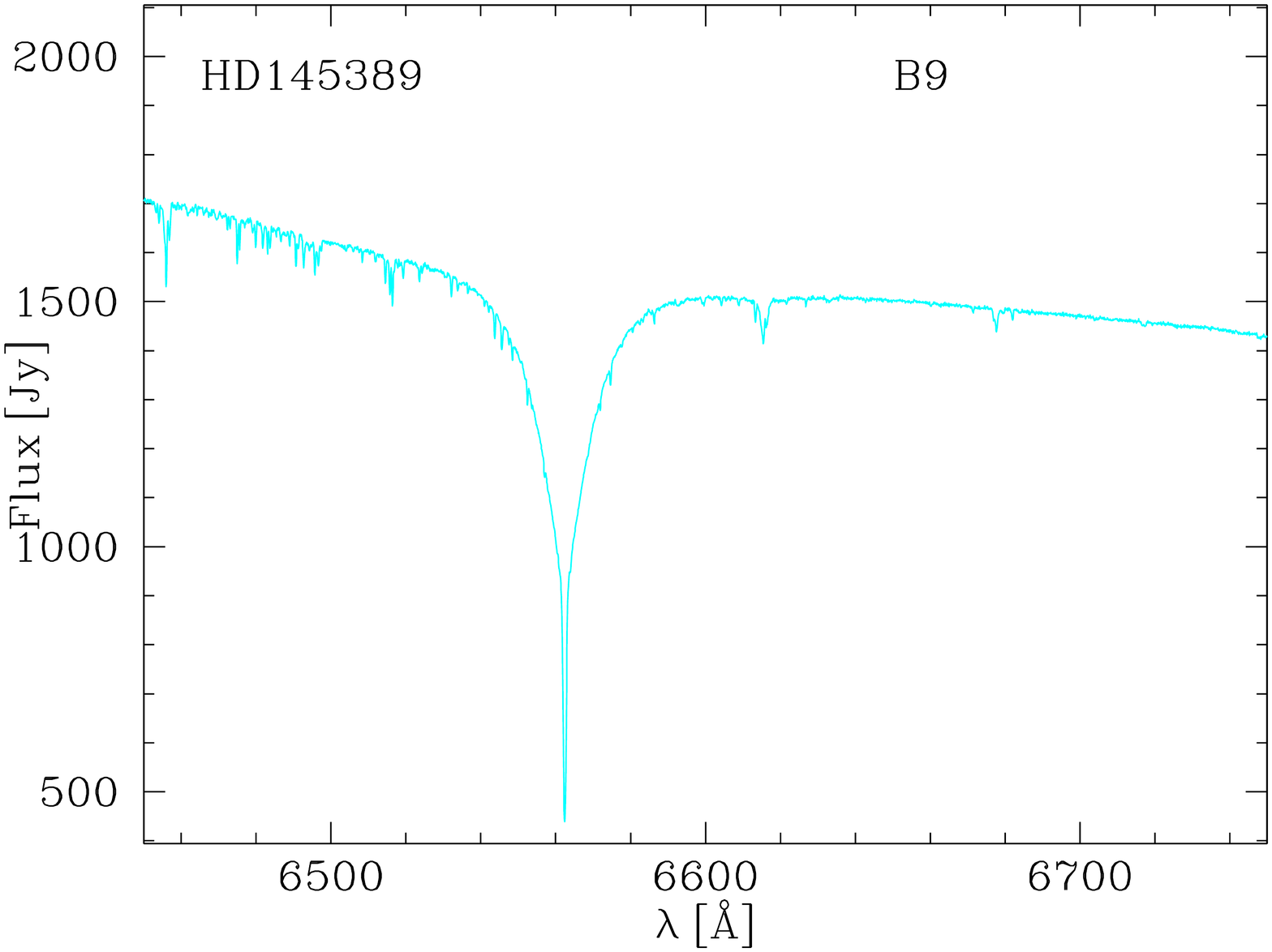}
\includegraphics[width=0.18\textwidth,angle=-90]{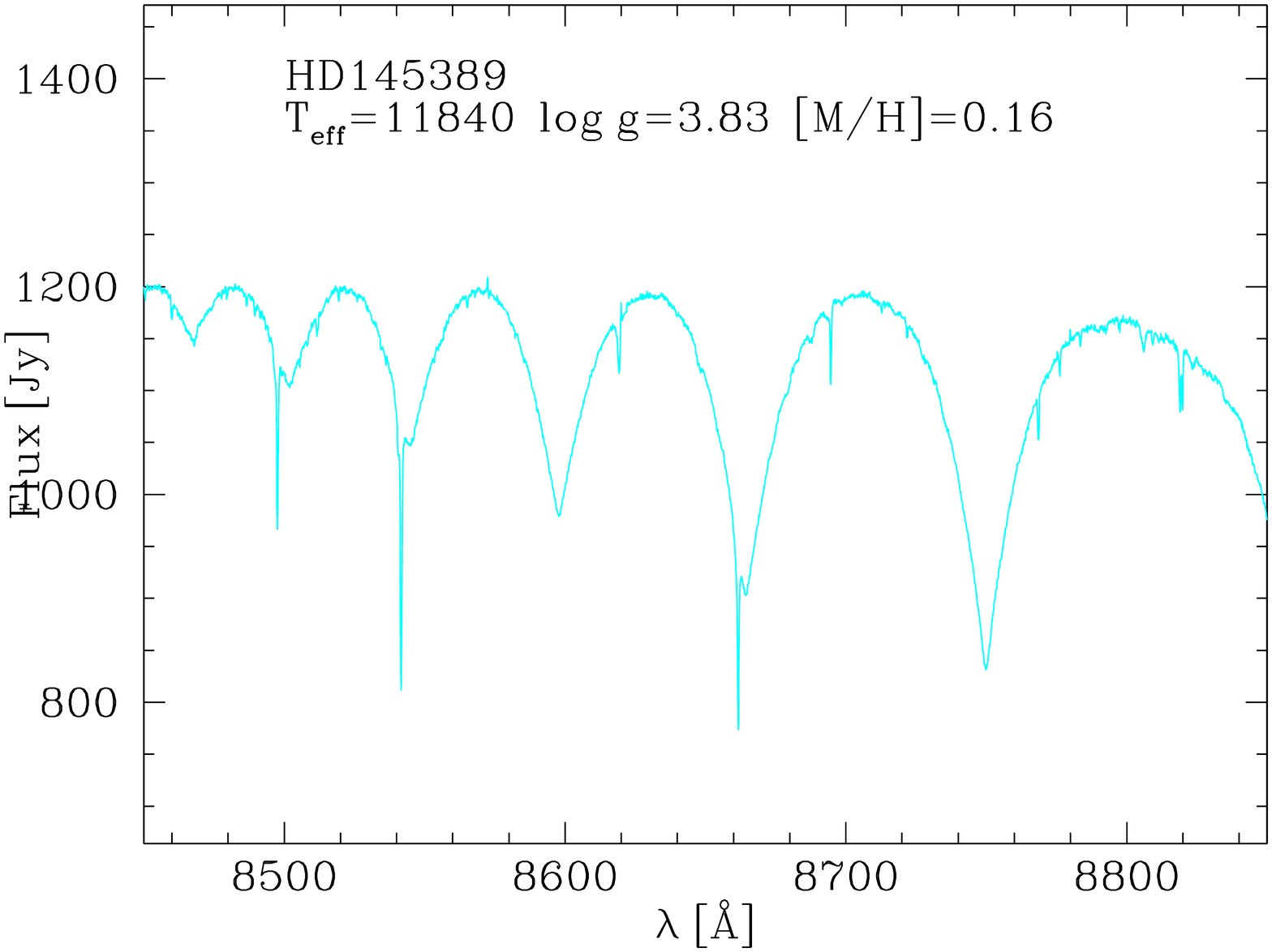}
\includegraphics[width=0.18\textwidth,angle=-90]{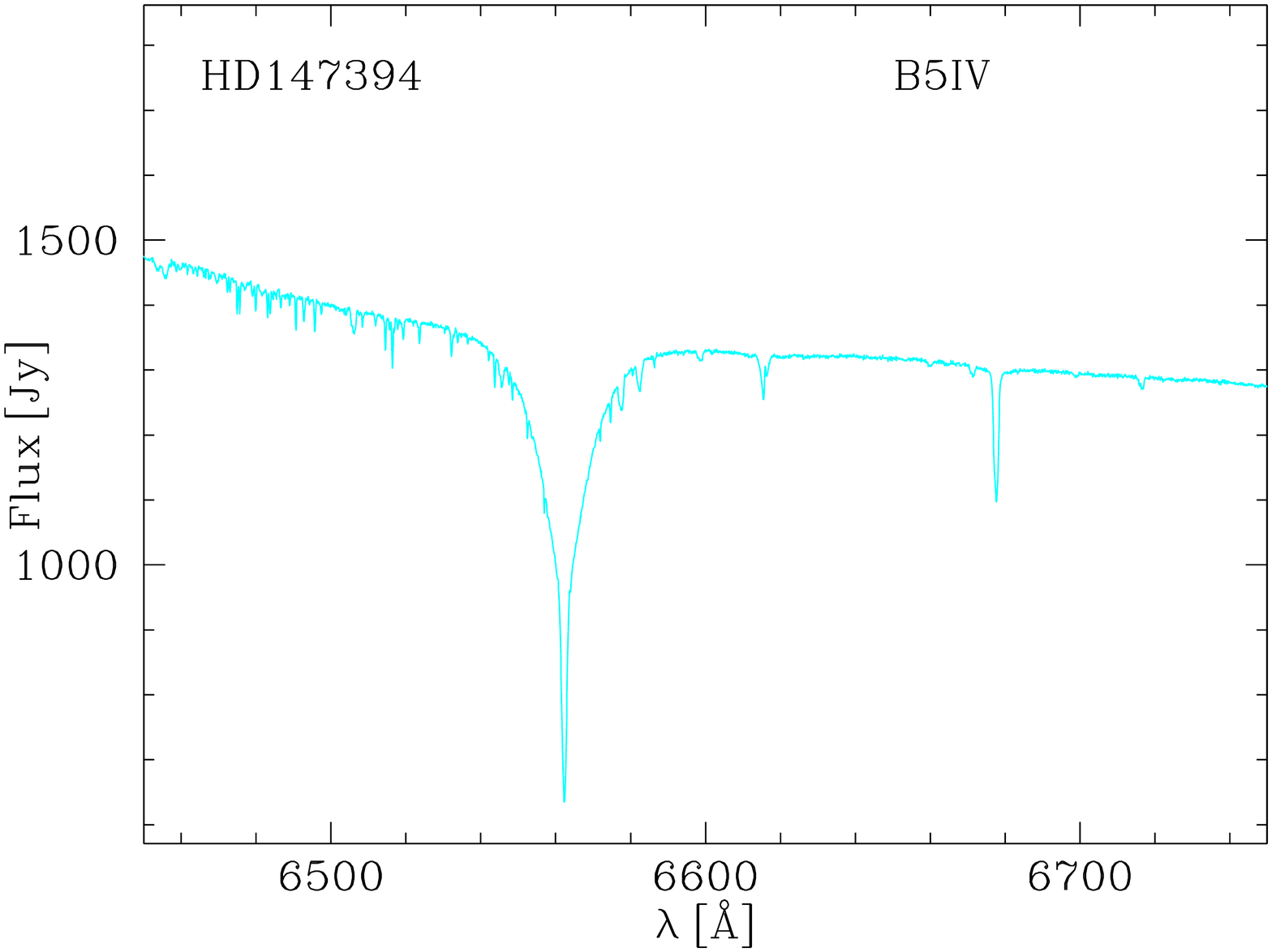}
\includegraphics[width=0.18\textwidth,angle=-90]{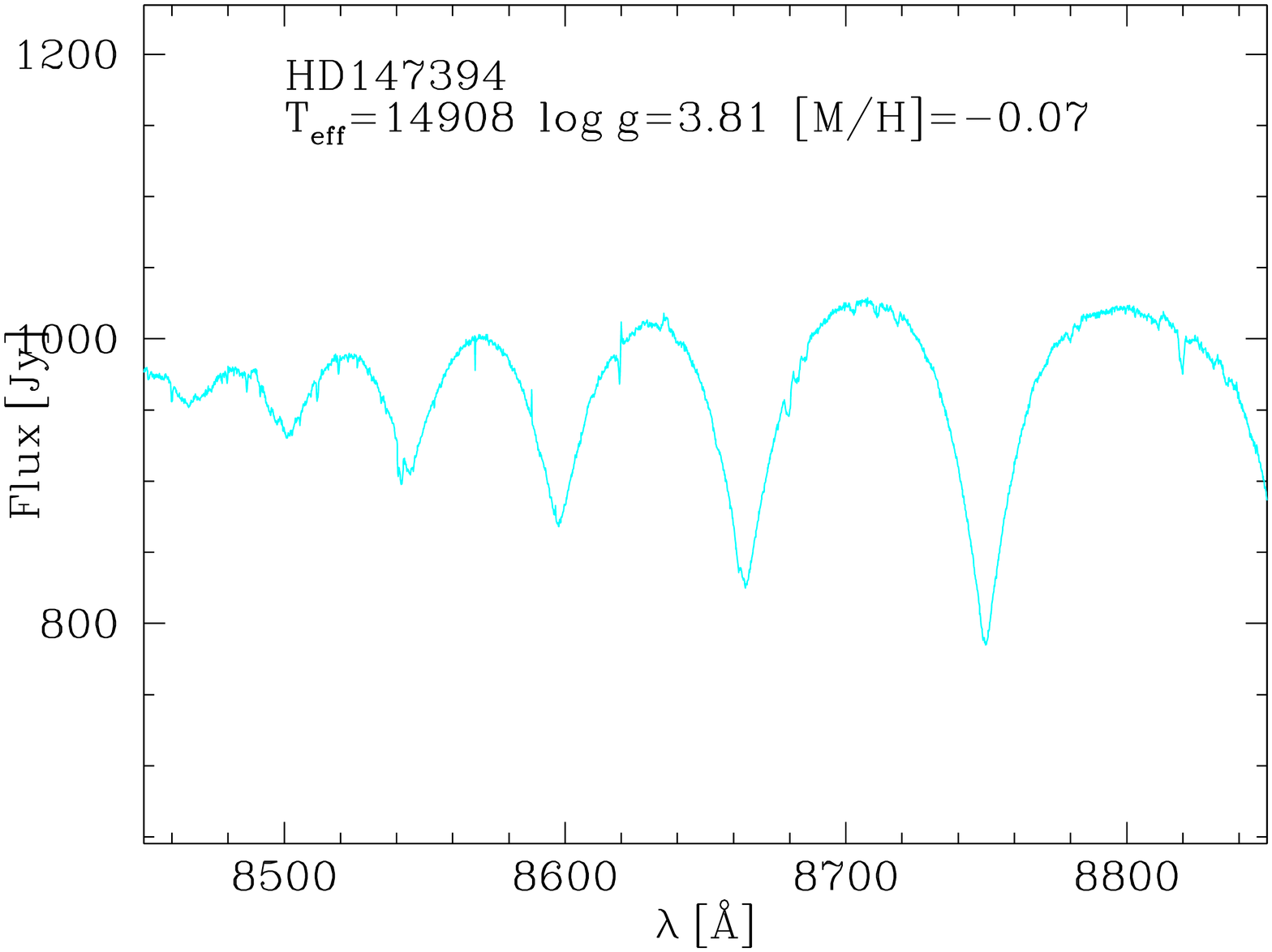}
\includegraphics[width=0.18\textwidth,angle=-90]{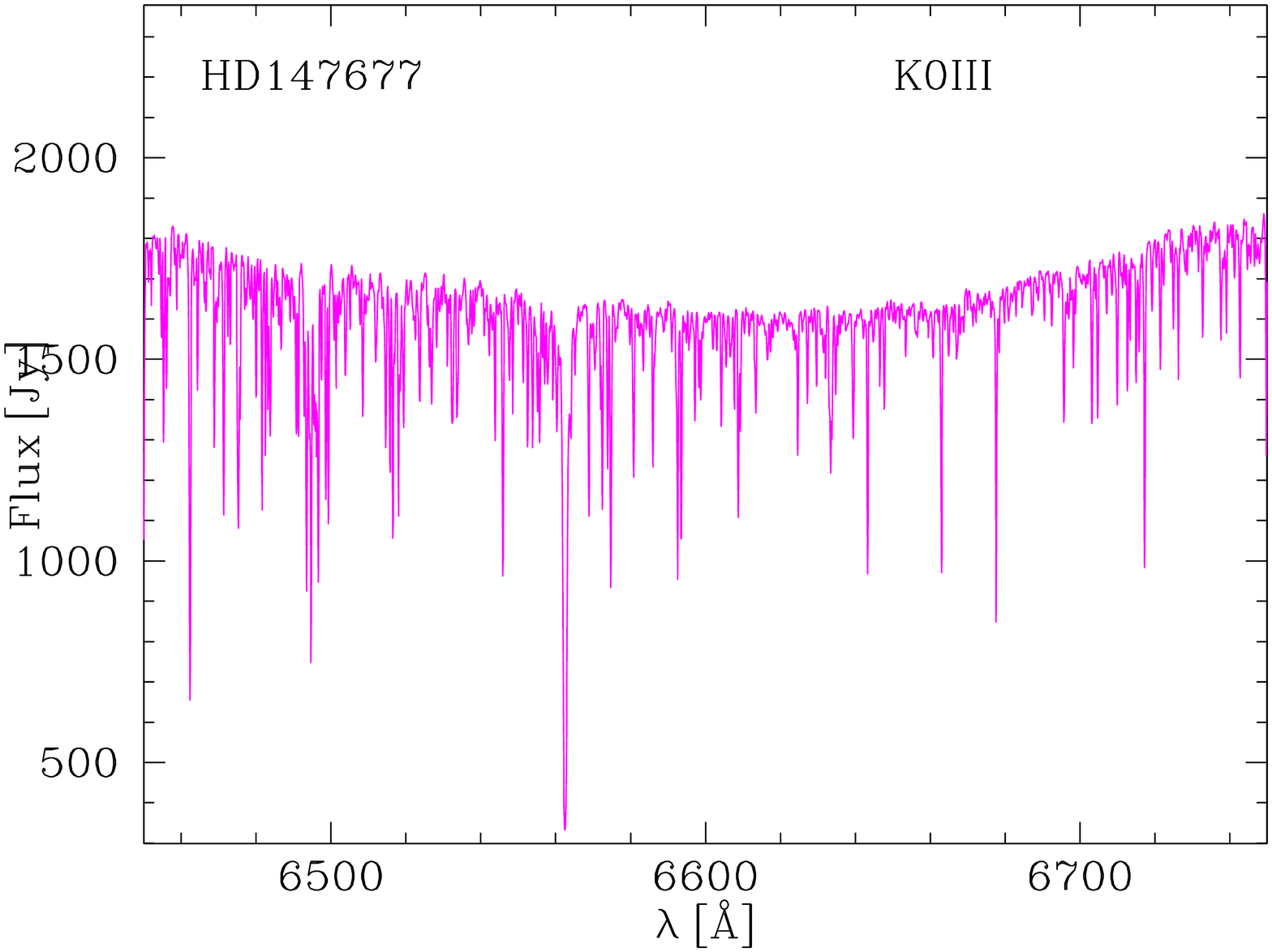}
\includegraphics[width=0.18\textwidth,angle=-90]{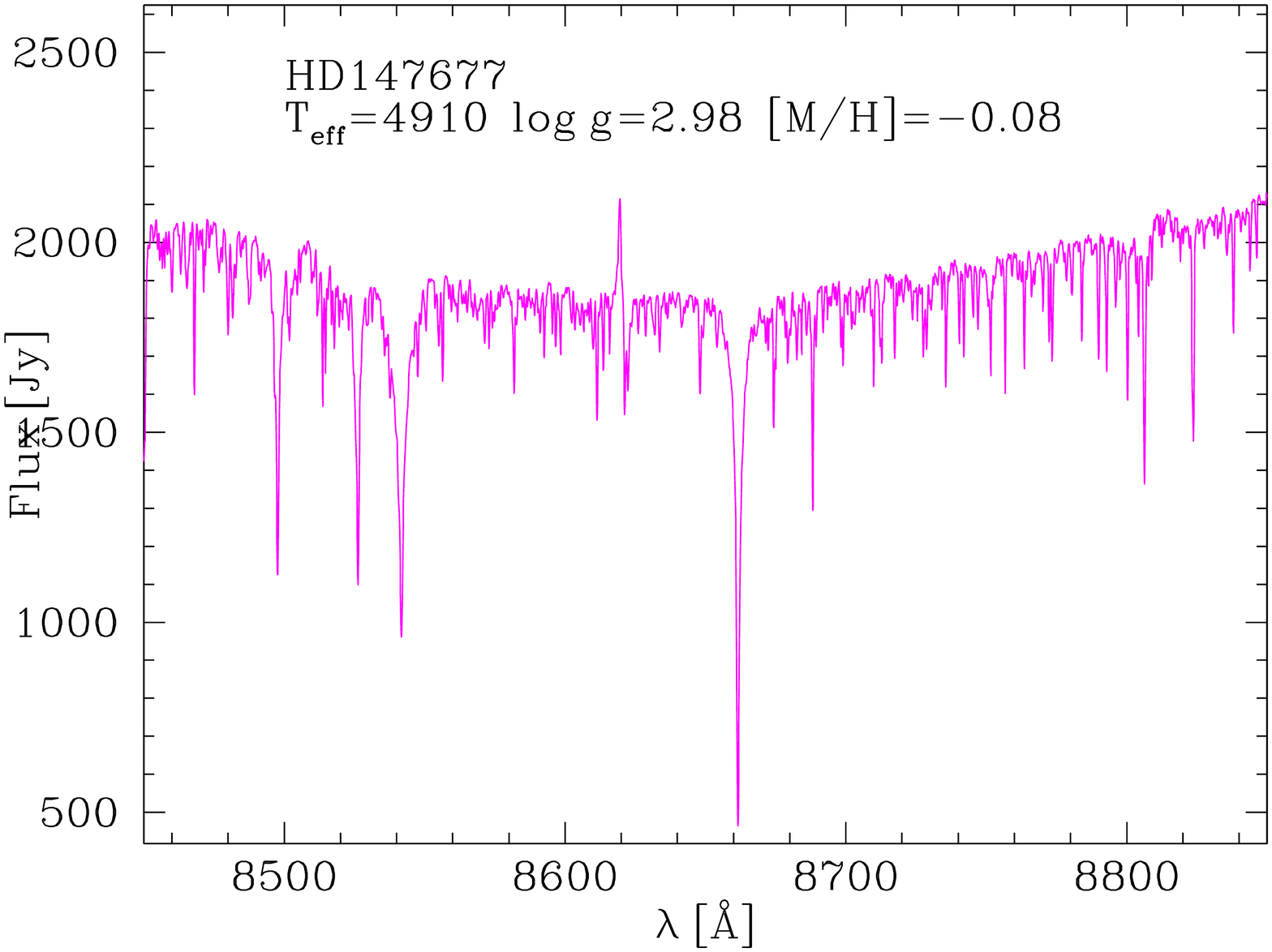}
\includegraphics[width=0.18\textwidth,angle=-90]{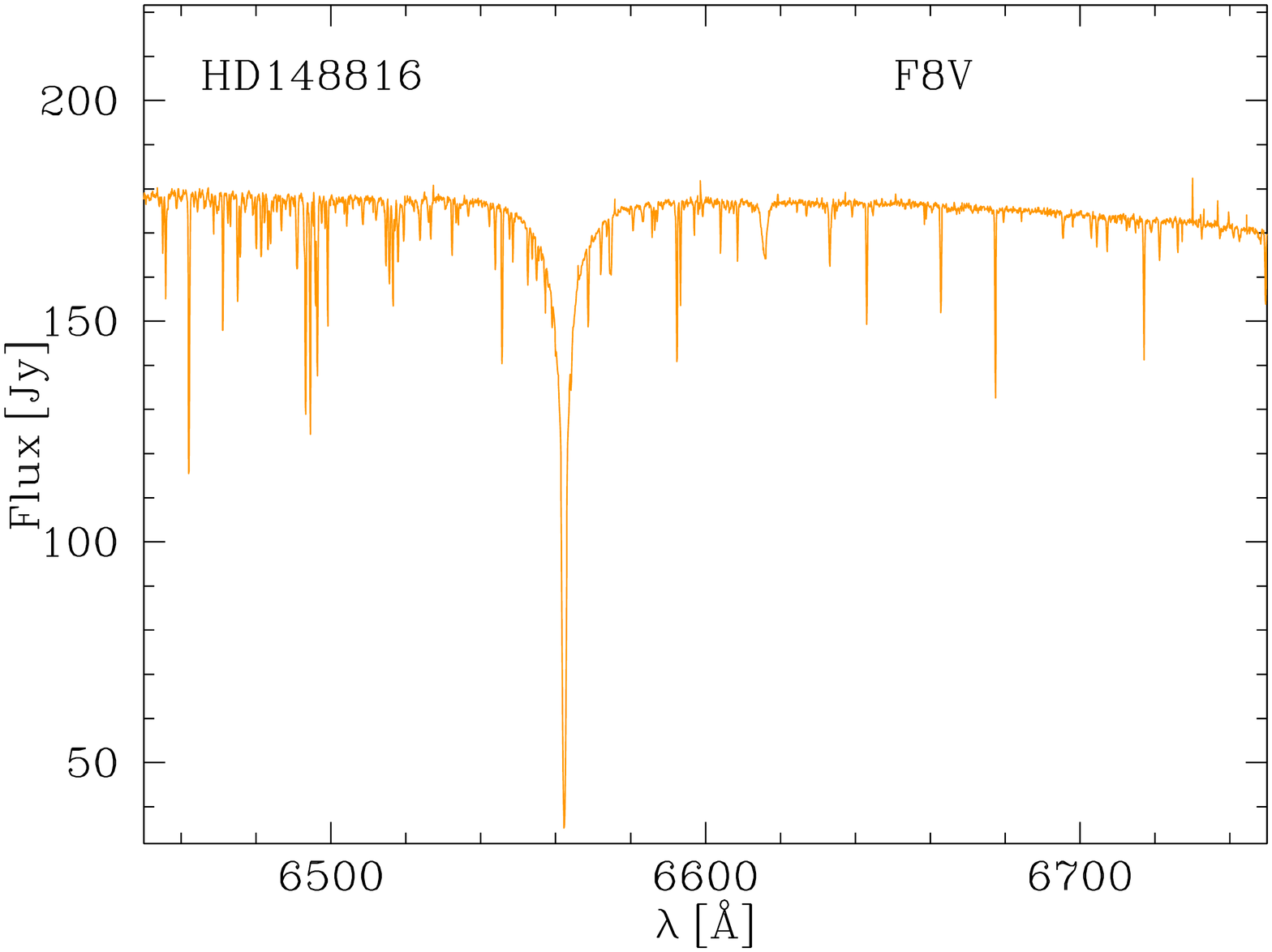}
\includegraphics[width=0.18\textwidth,angle=-90]{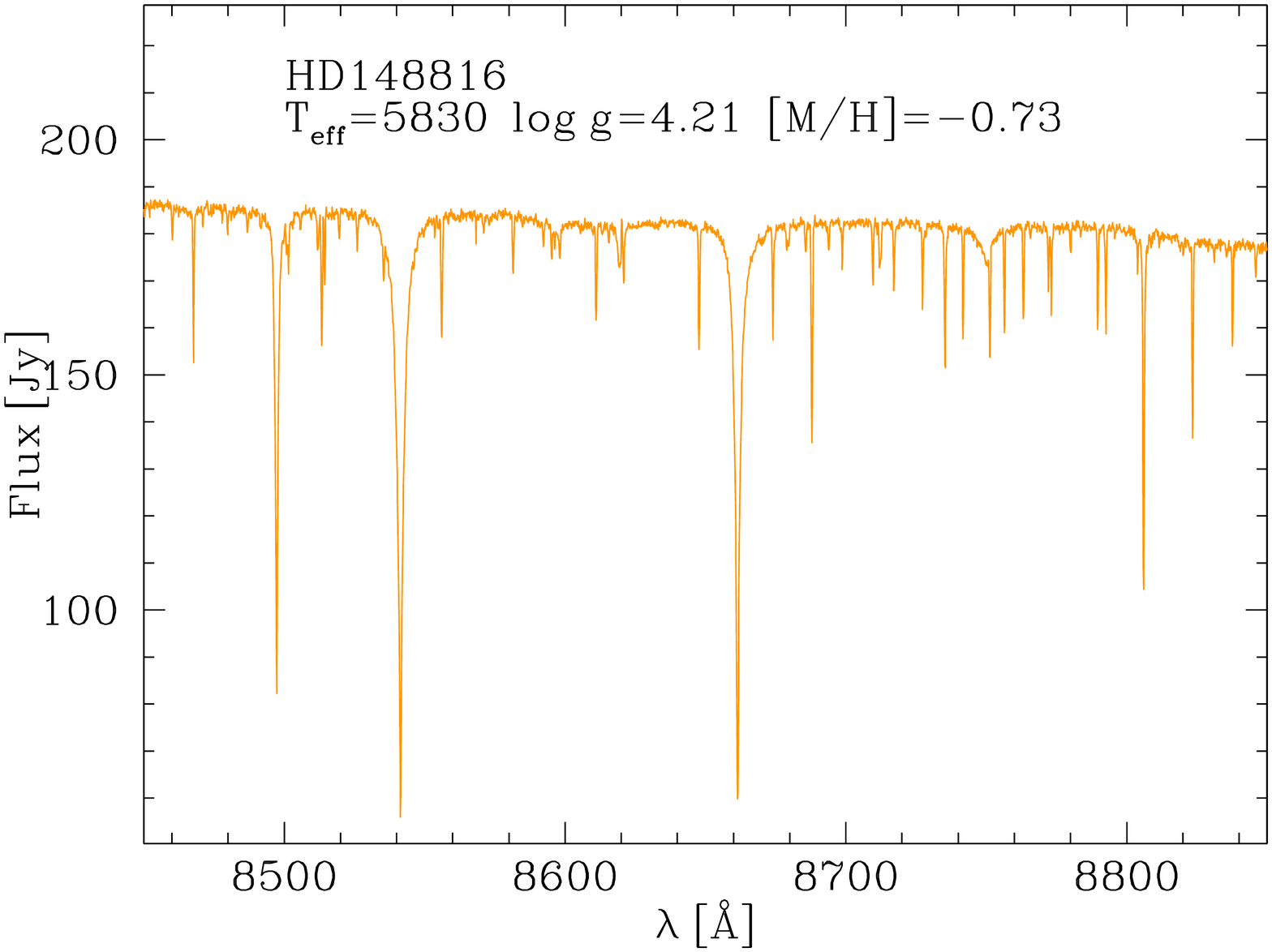}

\contcaption{24. Stars shown in this page are: HD139457, HD141004, HD141272, HD142091, HD142860, HD142926, HD143807, HD144206, HD144284, HD145148, HD145389, HD147394, HD147677 and HD148816.}
\end{figure*}

\begin{figure*}
\includegraphics[width=0.18\textwidth,angle=-90]{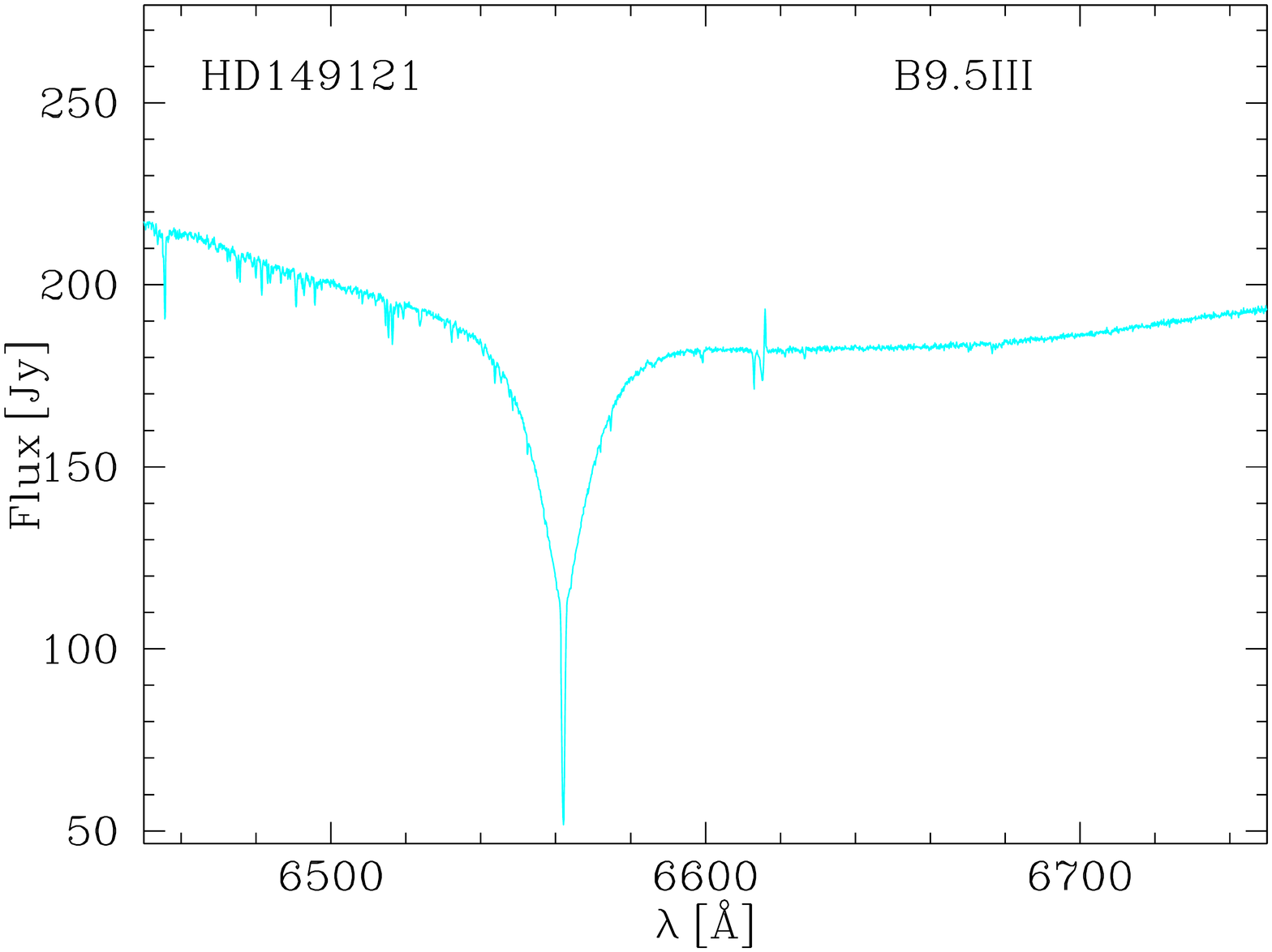}
\includegraphics[width=0.18\textwidth,angle=-90]{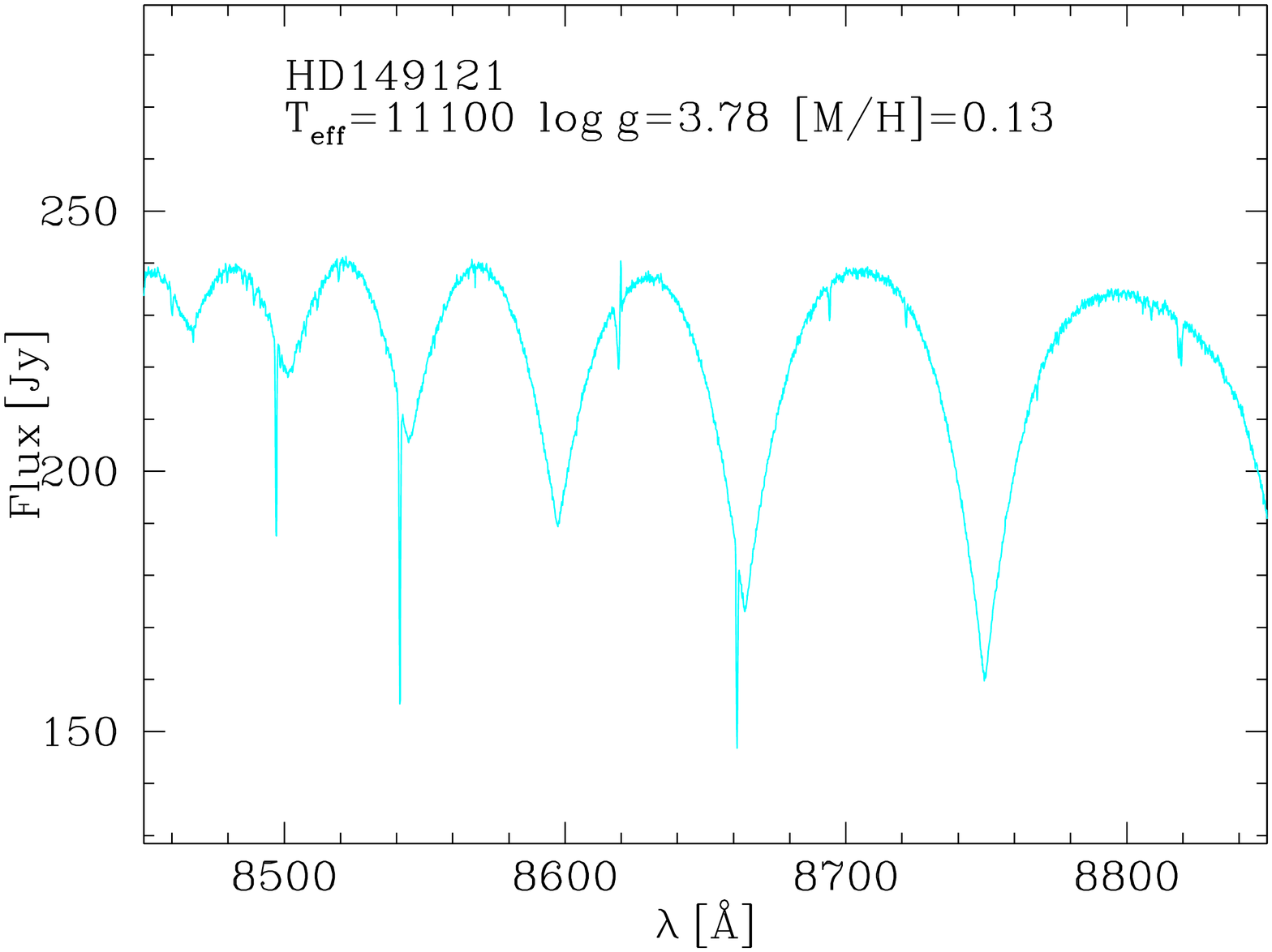}
\includegraphics[width=0.18\textwidth,angle=-90]{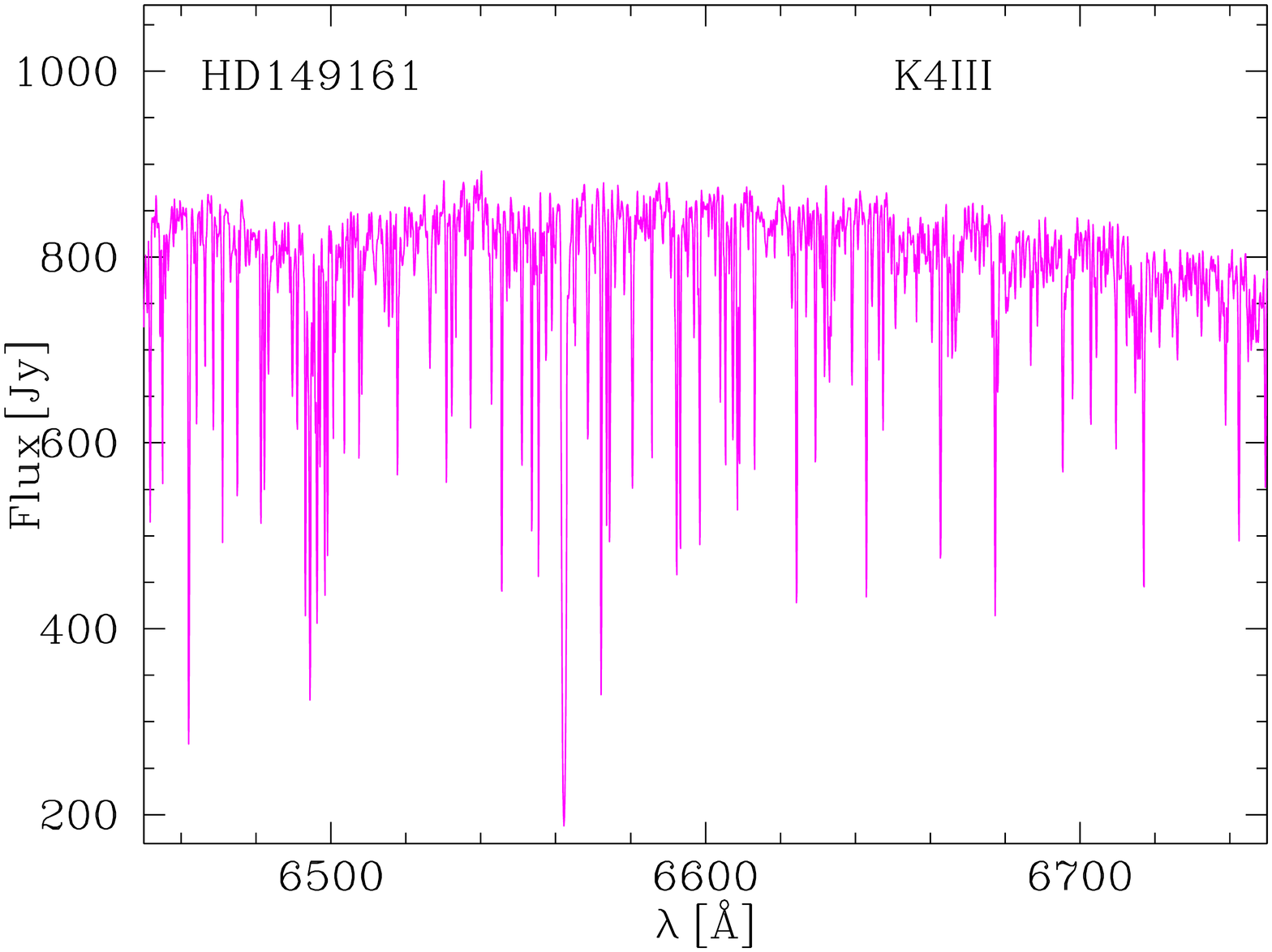}
\includegraphics[width=0.18\textwidth,angle=-90]{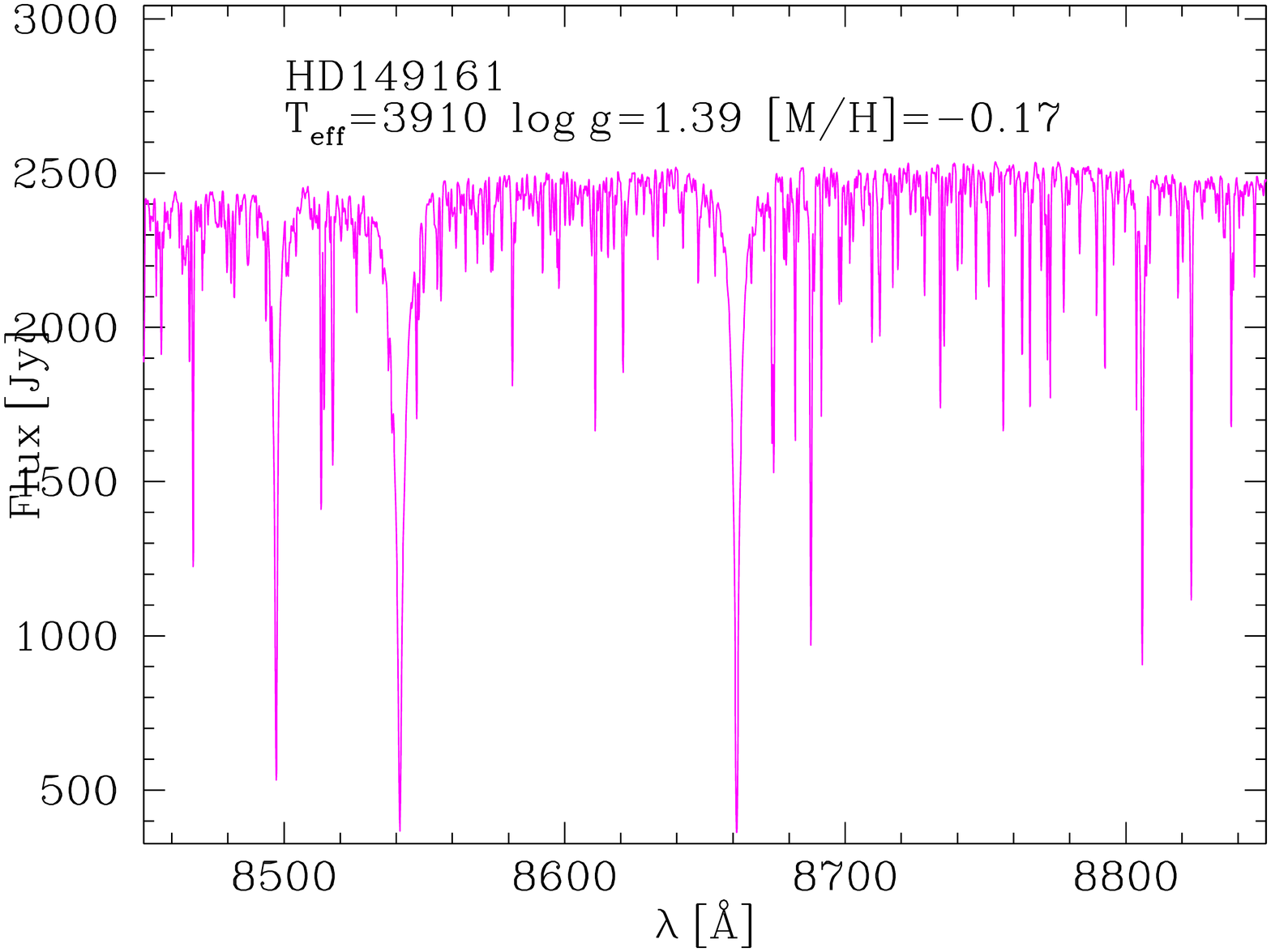}
\includegraphics[width=0.18\textwidth,angle=-90]{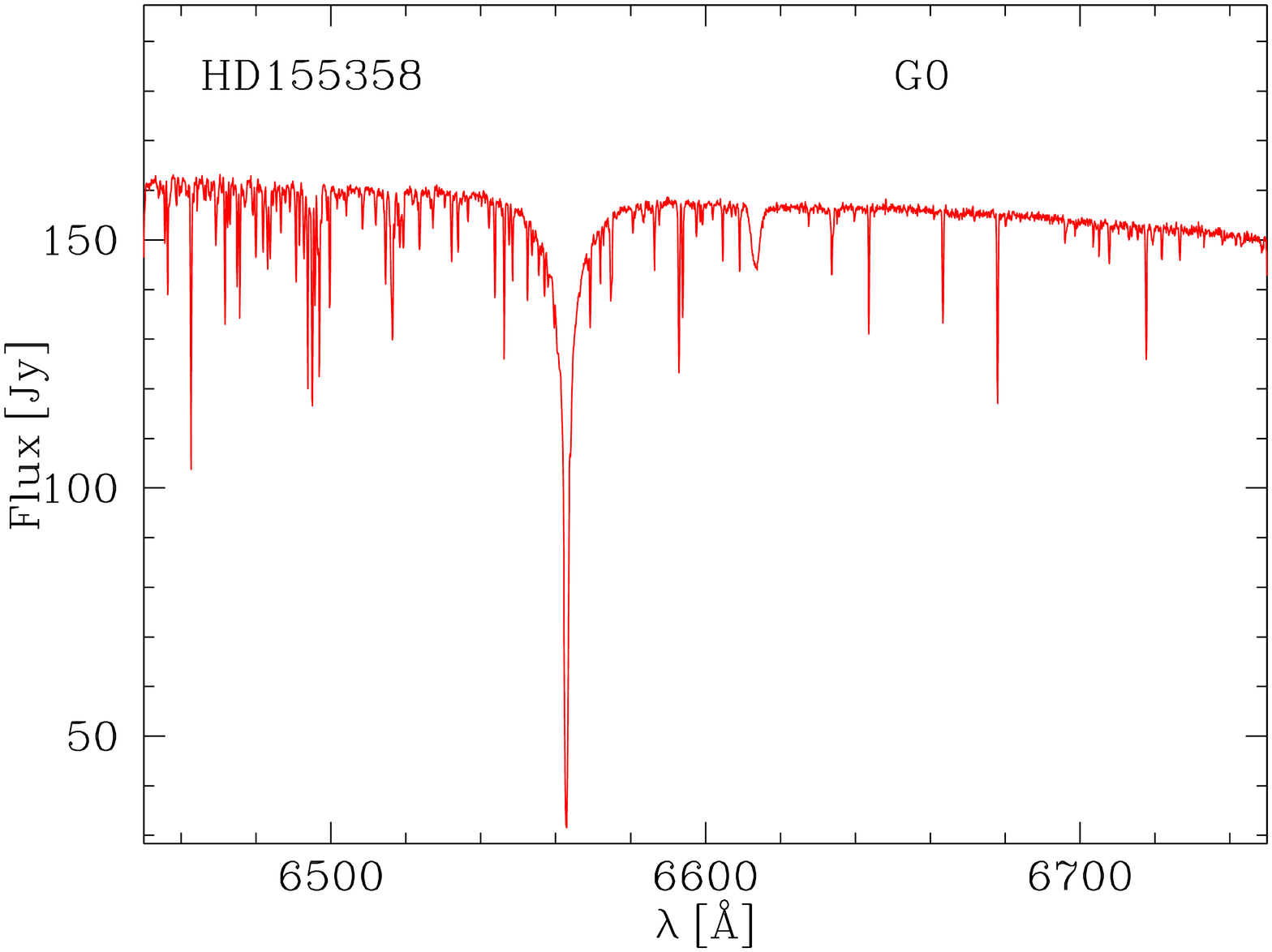}
\includegraphics[width=0.18\textwidth,angle=-90]{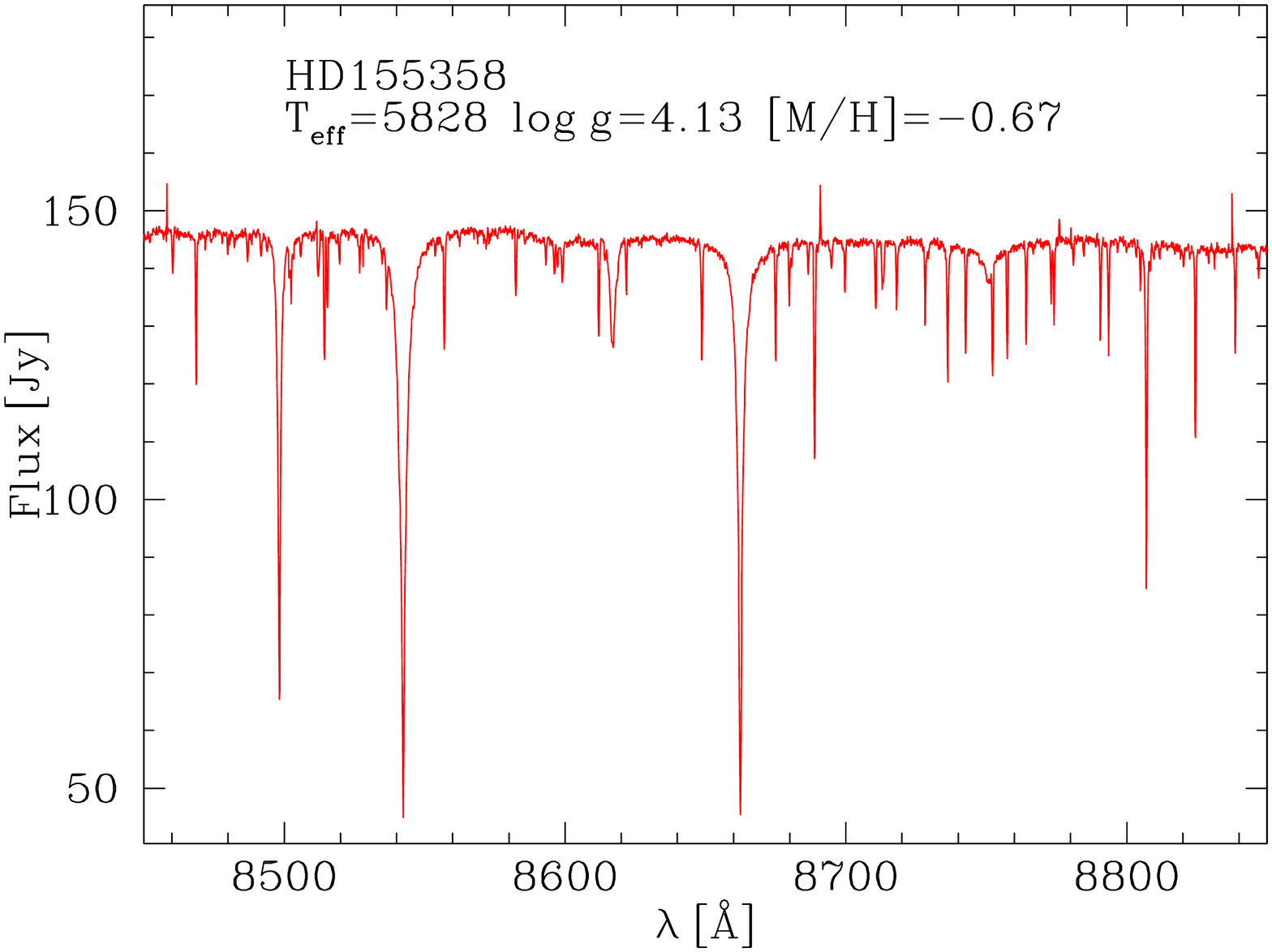}
\includegraphics[width=0.18\textwidth,angle=-90]{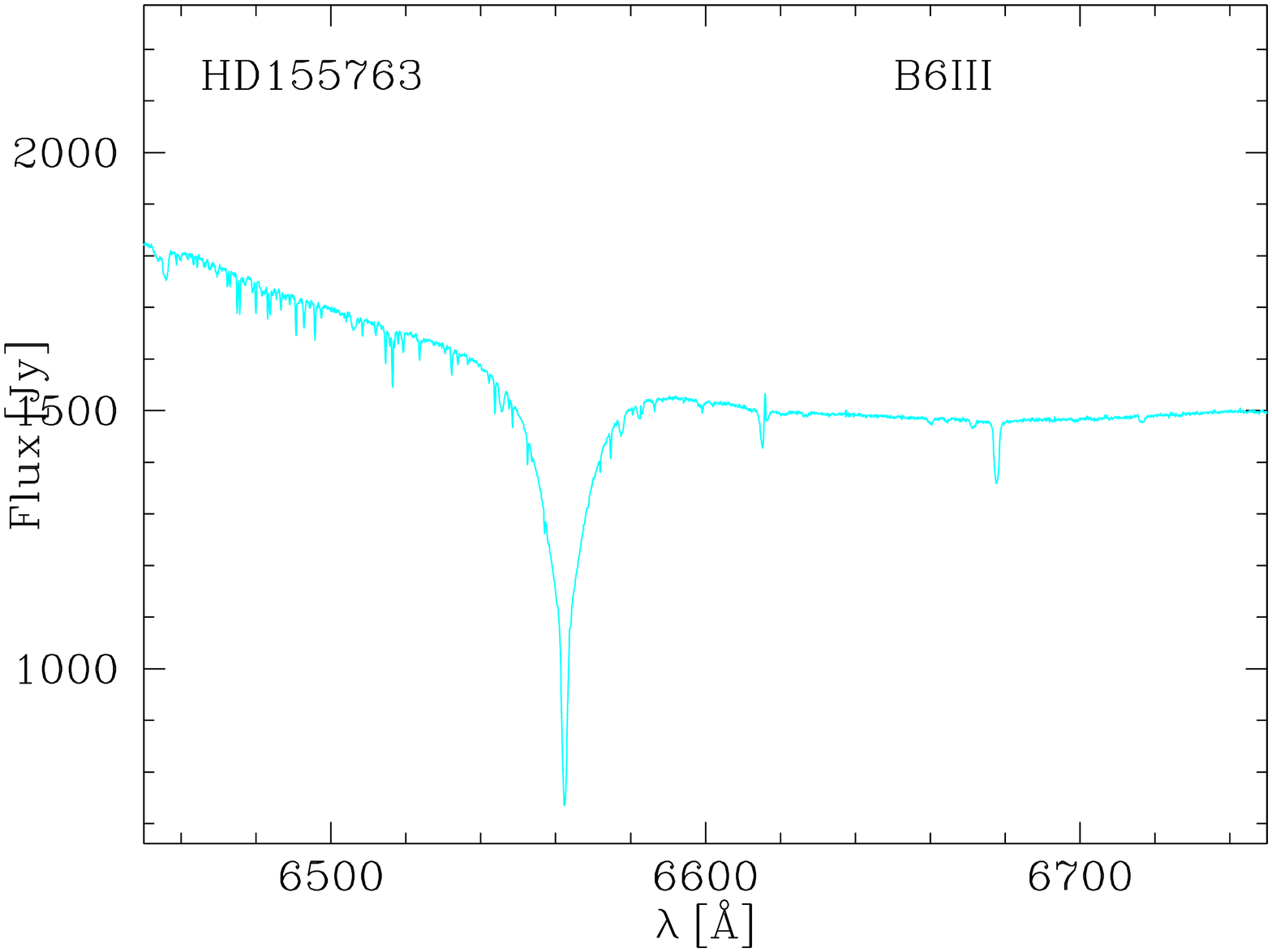}
\includegraphics[width=0.18\textwidth,angle=-90]{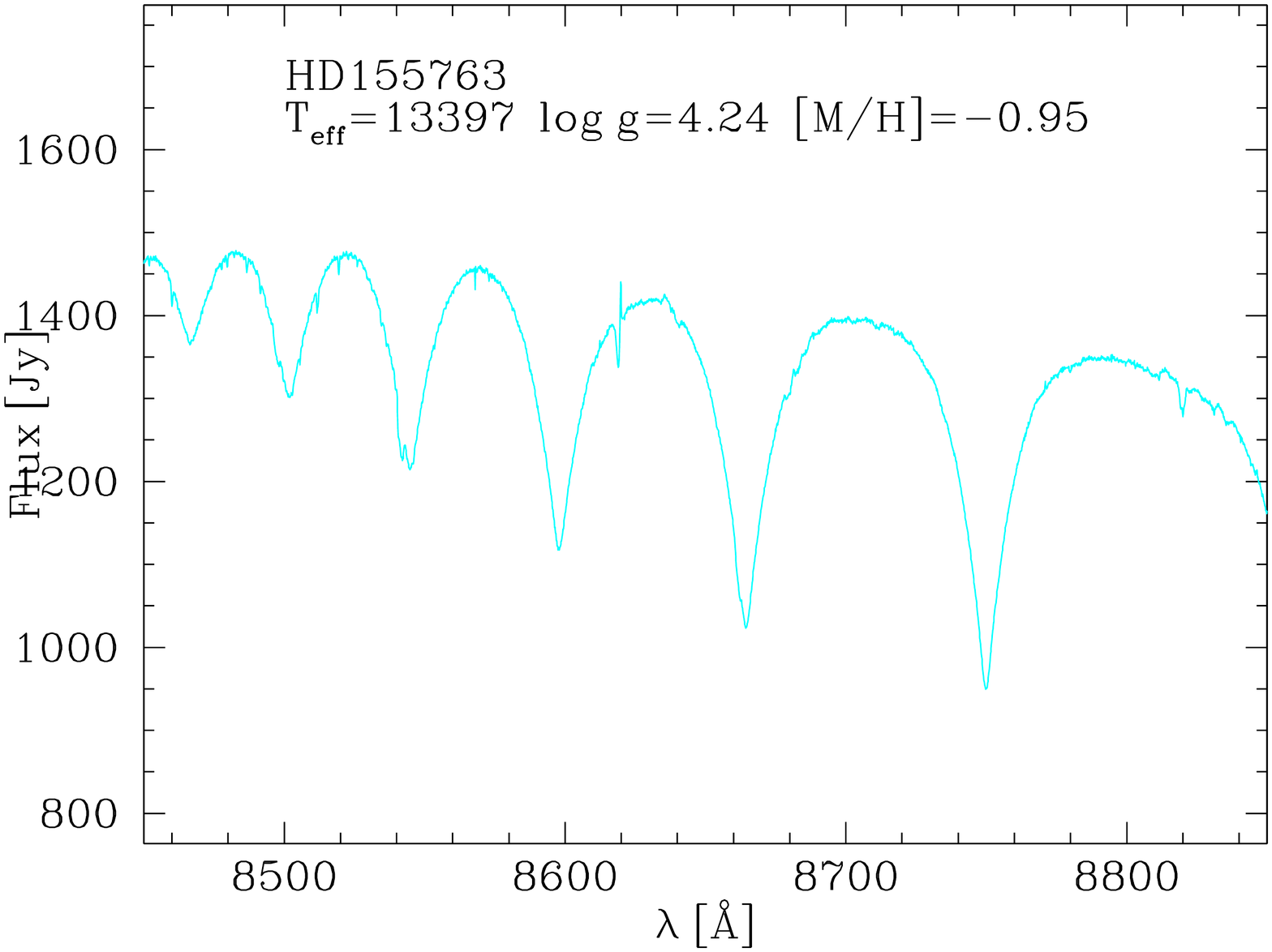}
\includegraphics[width=0.18\textwidth,angle=-90]{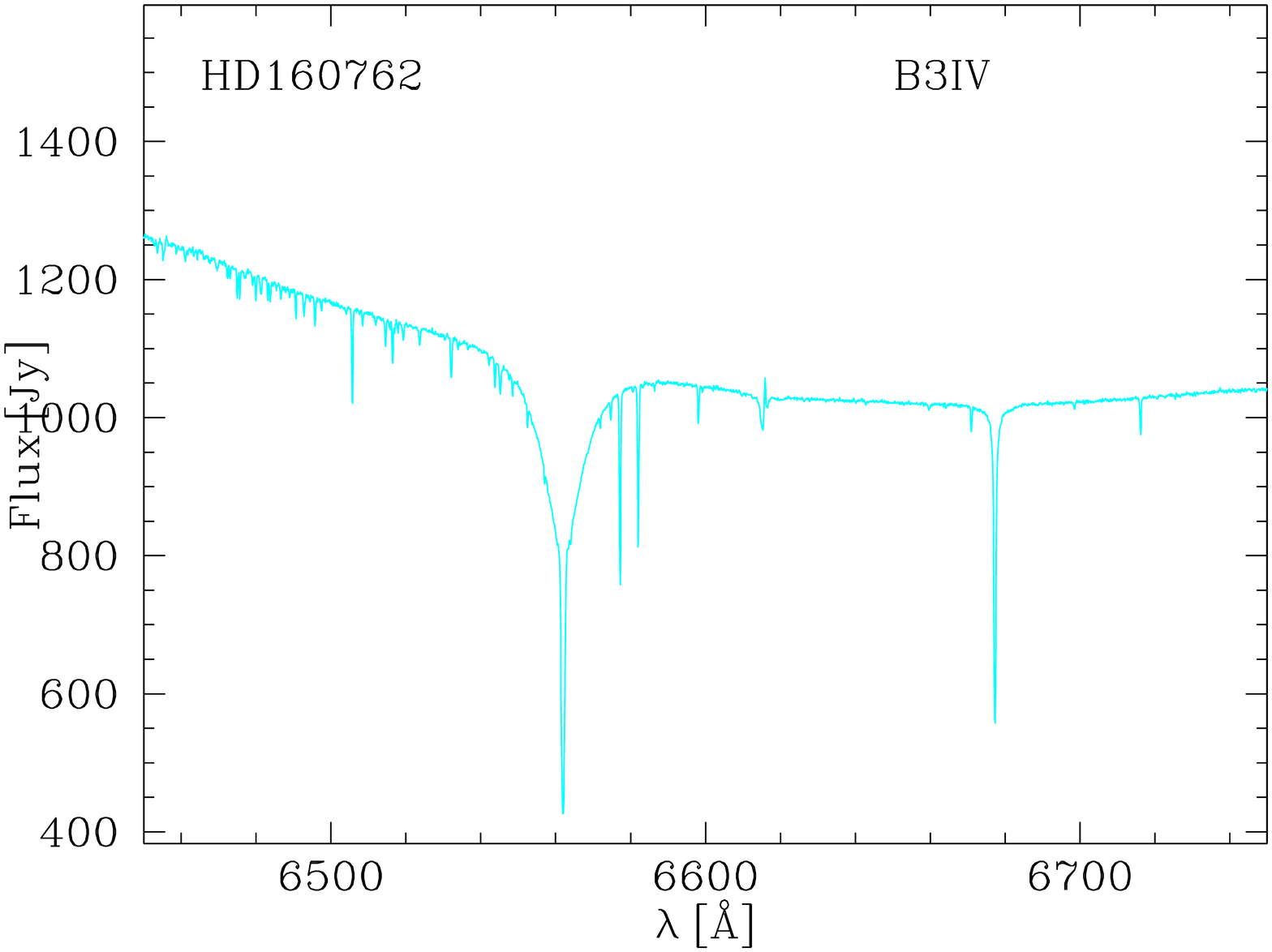}
\includegraphics[width=0.18\textwidth,angle=-90]{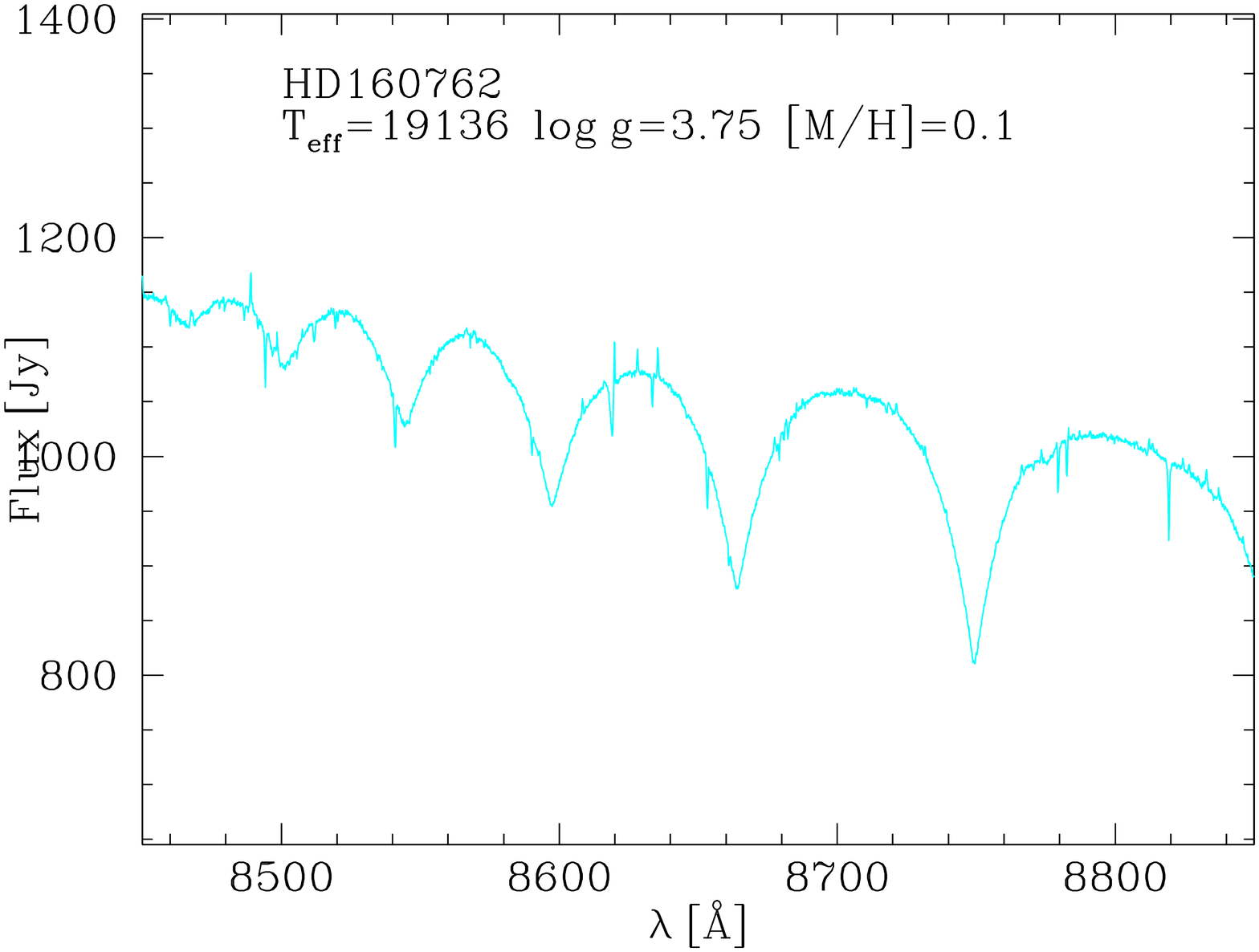}
\includegraphics[width=0.18\textwidth,angle=-90]{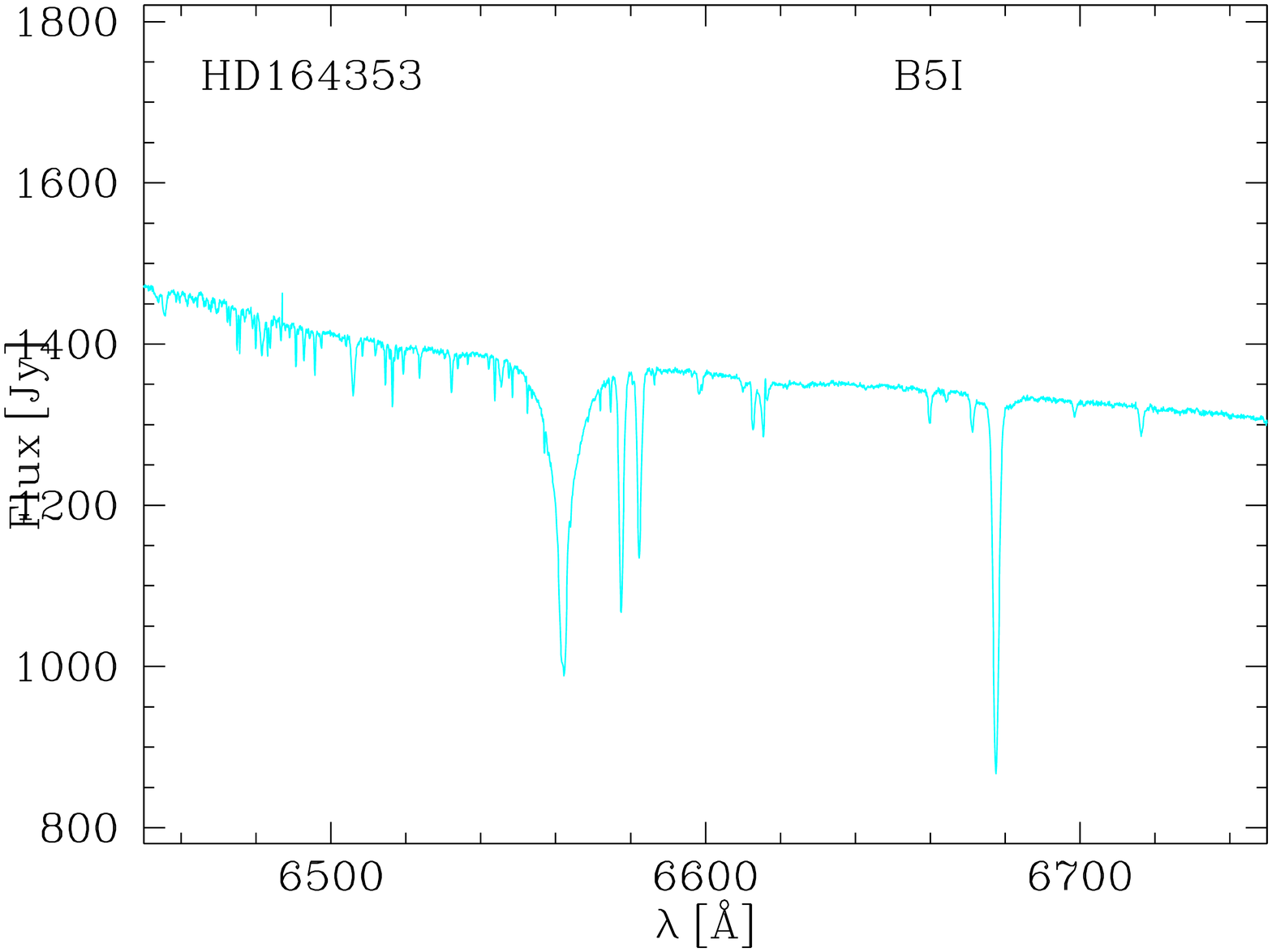}
\includegraphics[width=0.18\textwidth,angle=-90]{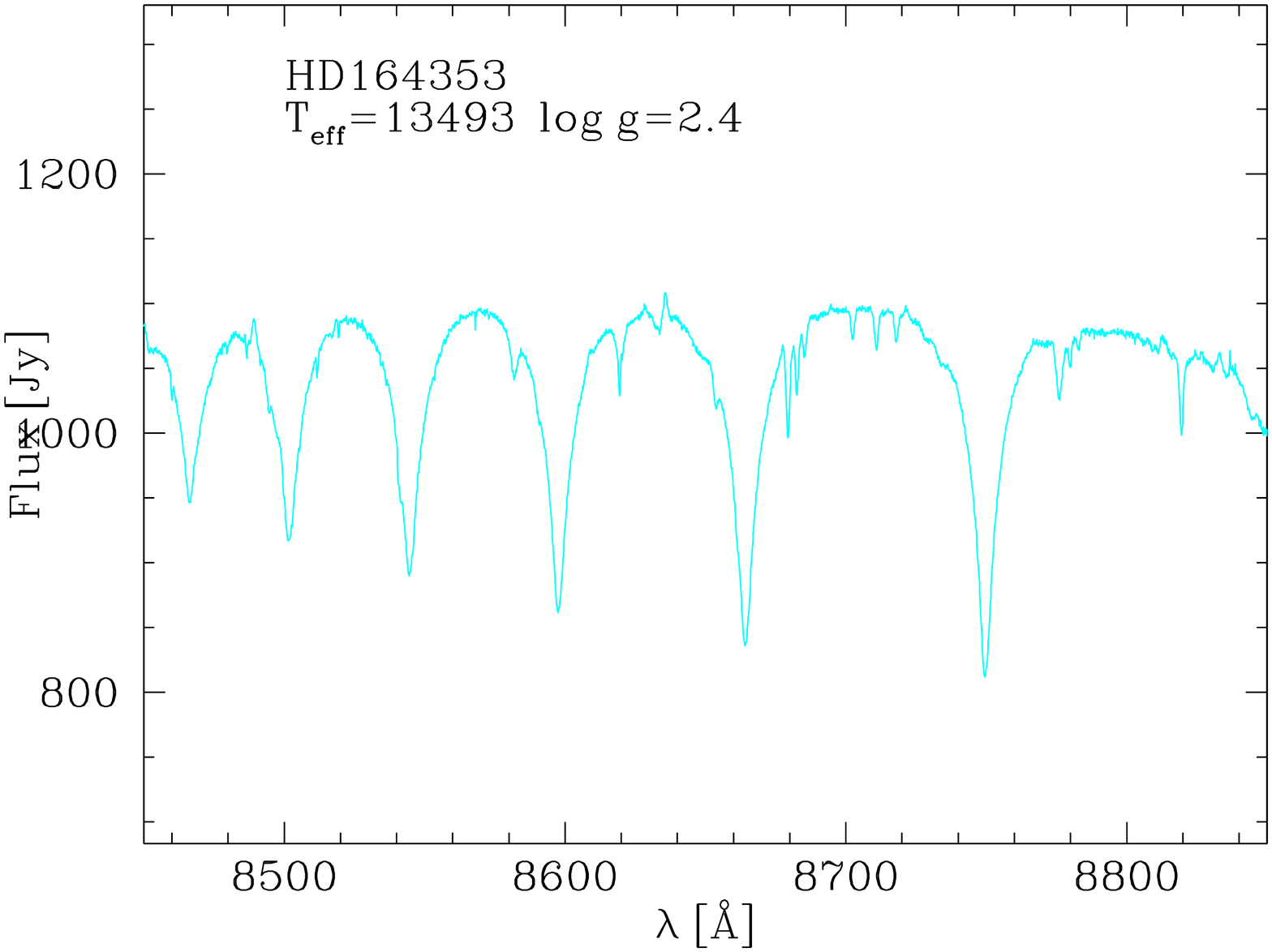}
\includegraphics[width=0.18\textwidth,angle=-90]{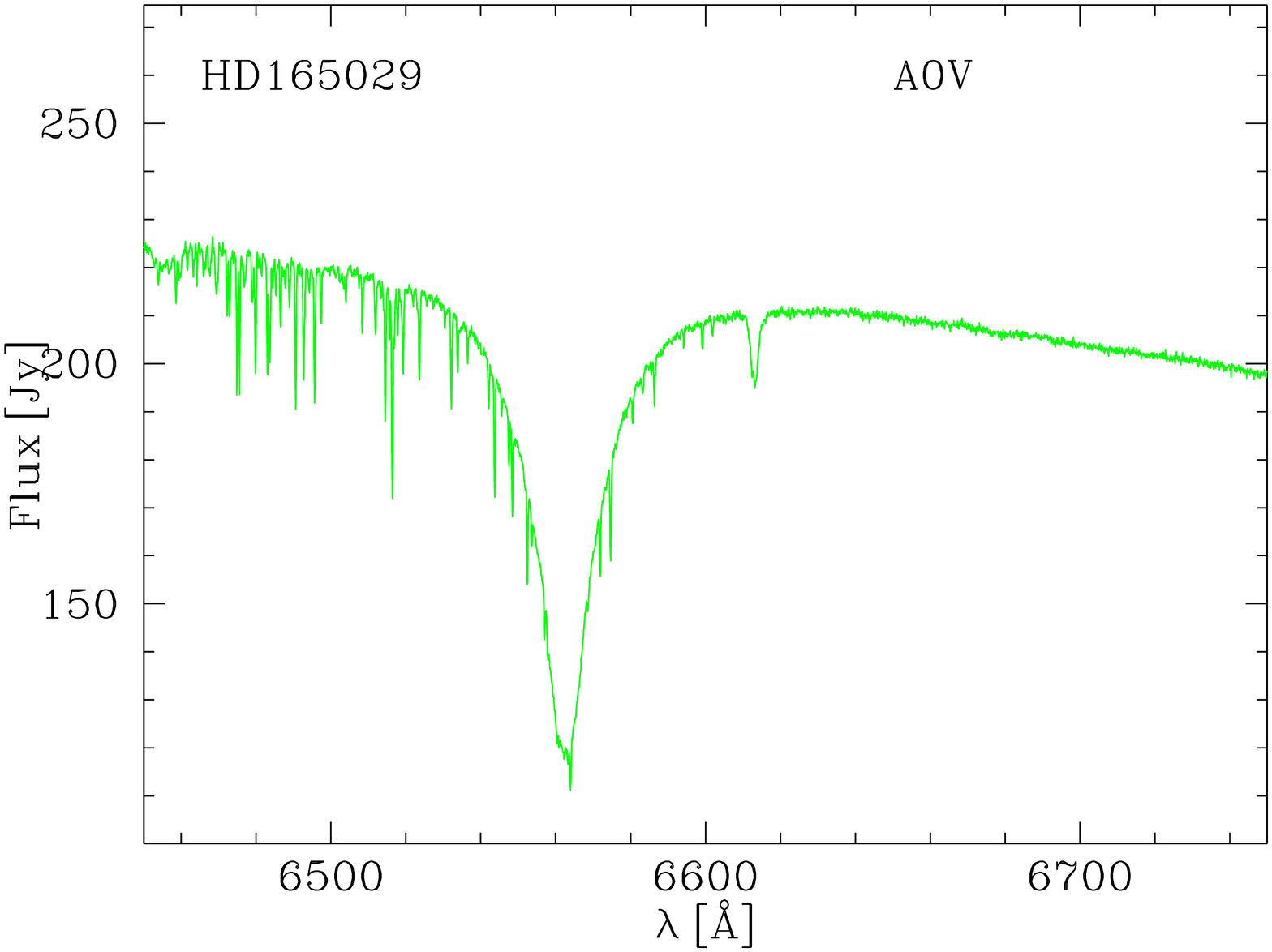}
\includegraphics[width=0.18\textwidth,angle=-90]{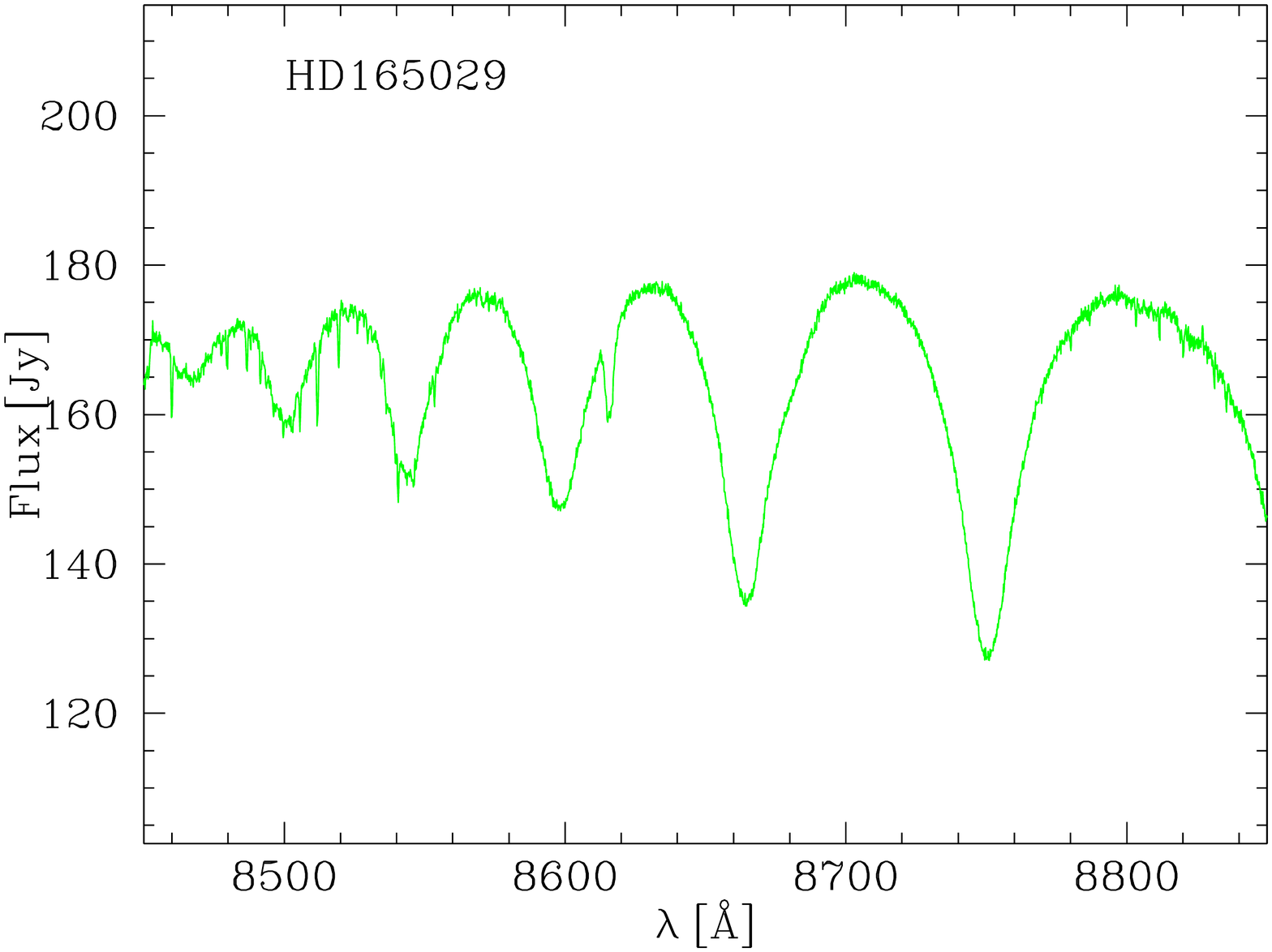}
\includegraphics[width=0.18\textwidth,angle=-90]{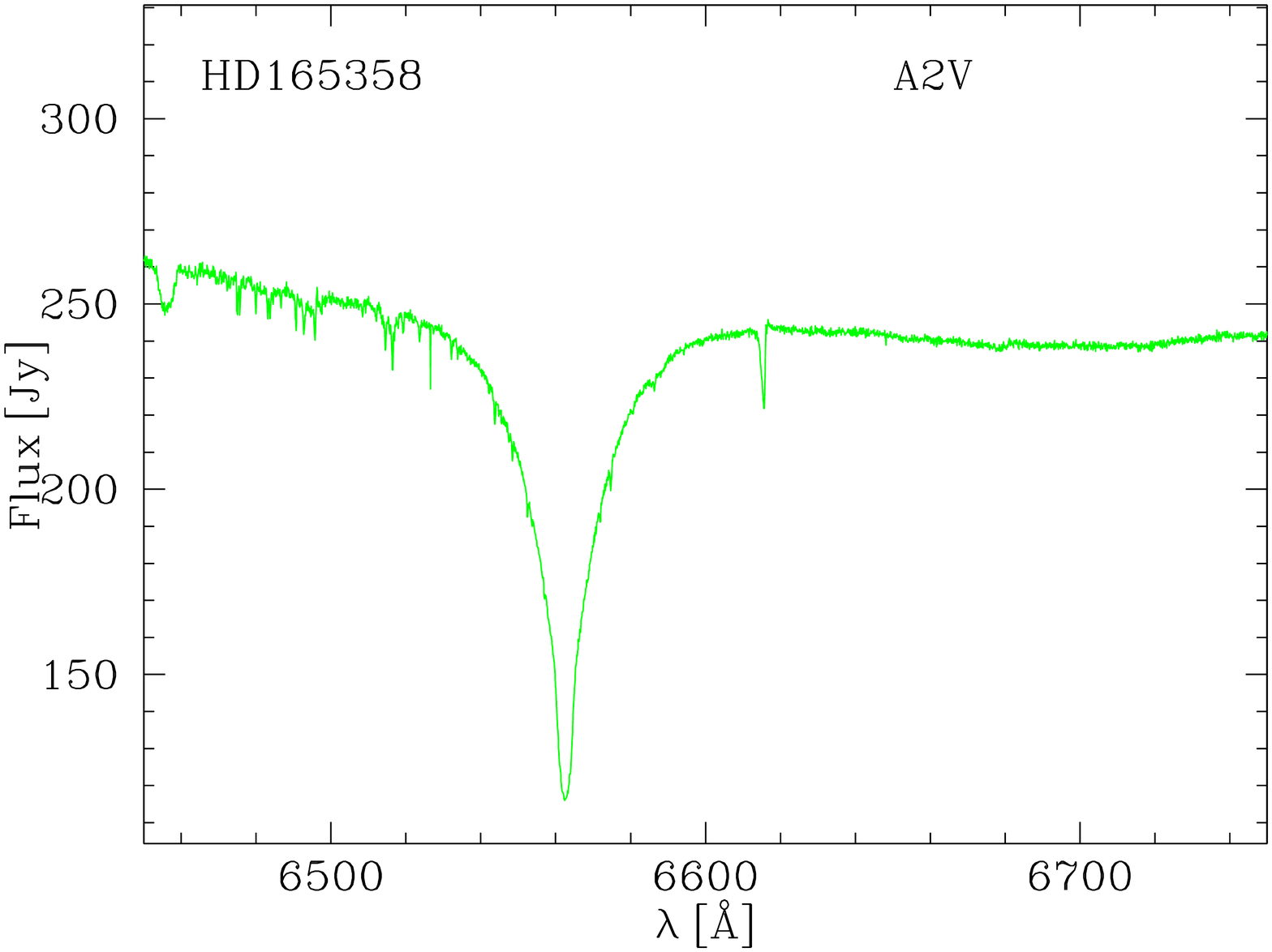}
\includegraphics[width=0.18\textwidth,angle=-90]{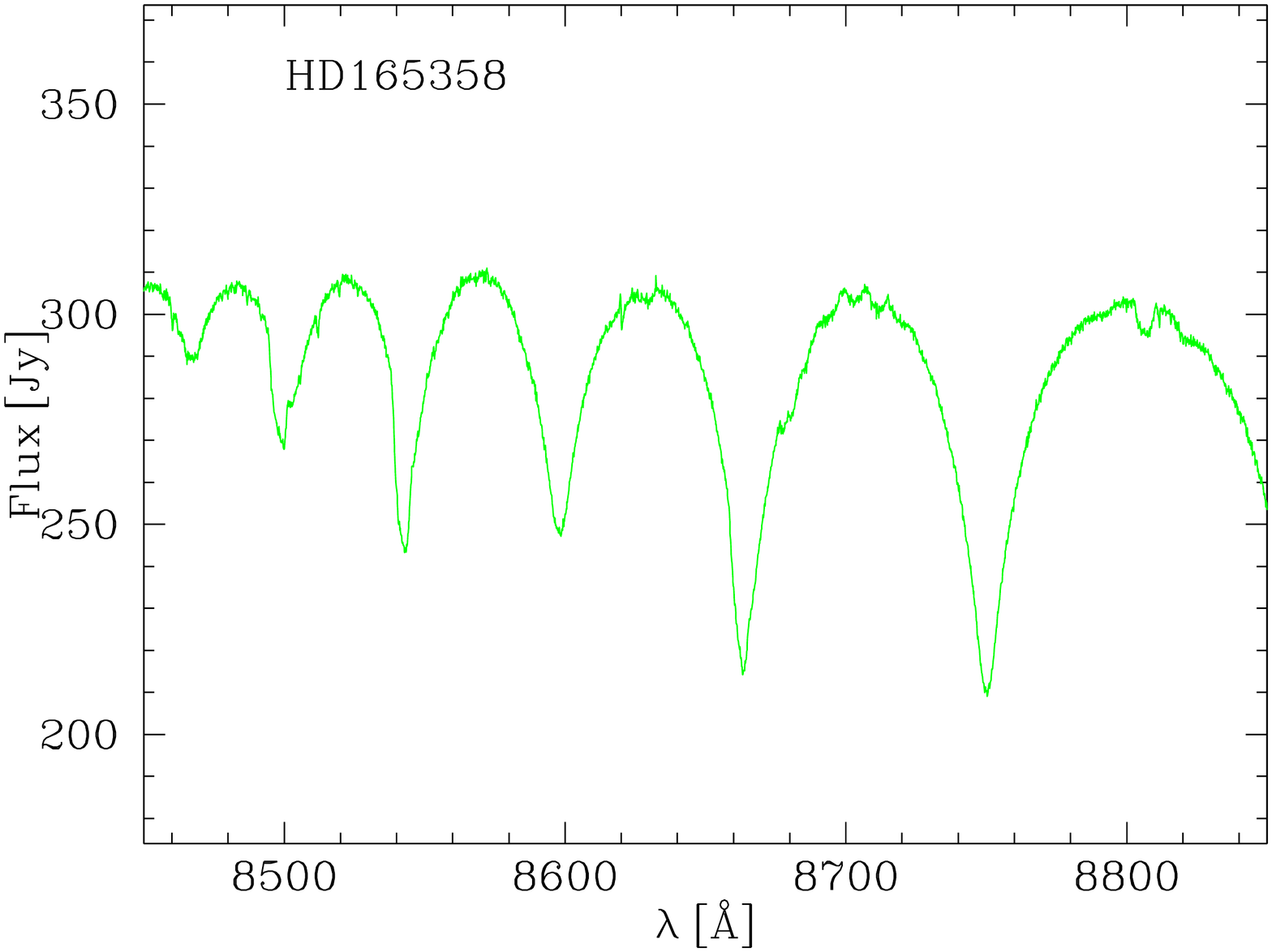}
\includegraphics[width=0.18\textwidth,angle=-90]{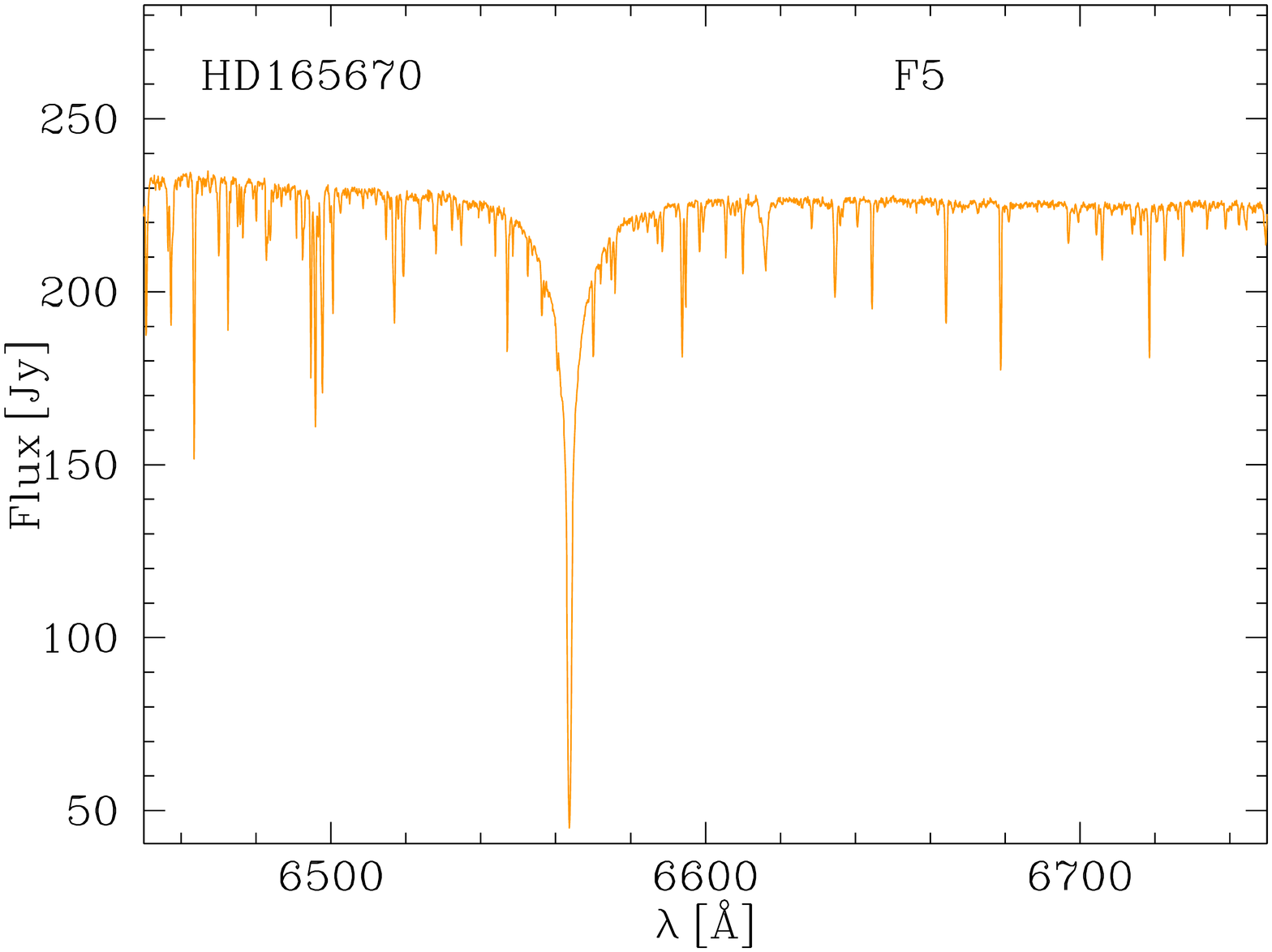}
\includegraphics[width=0.18\textwidth,angle=-90]{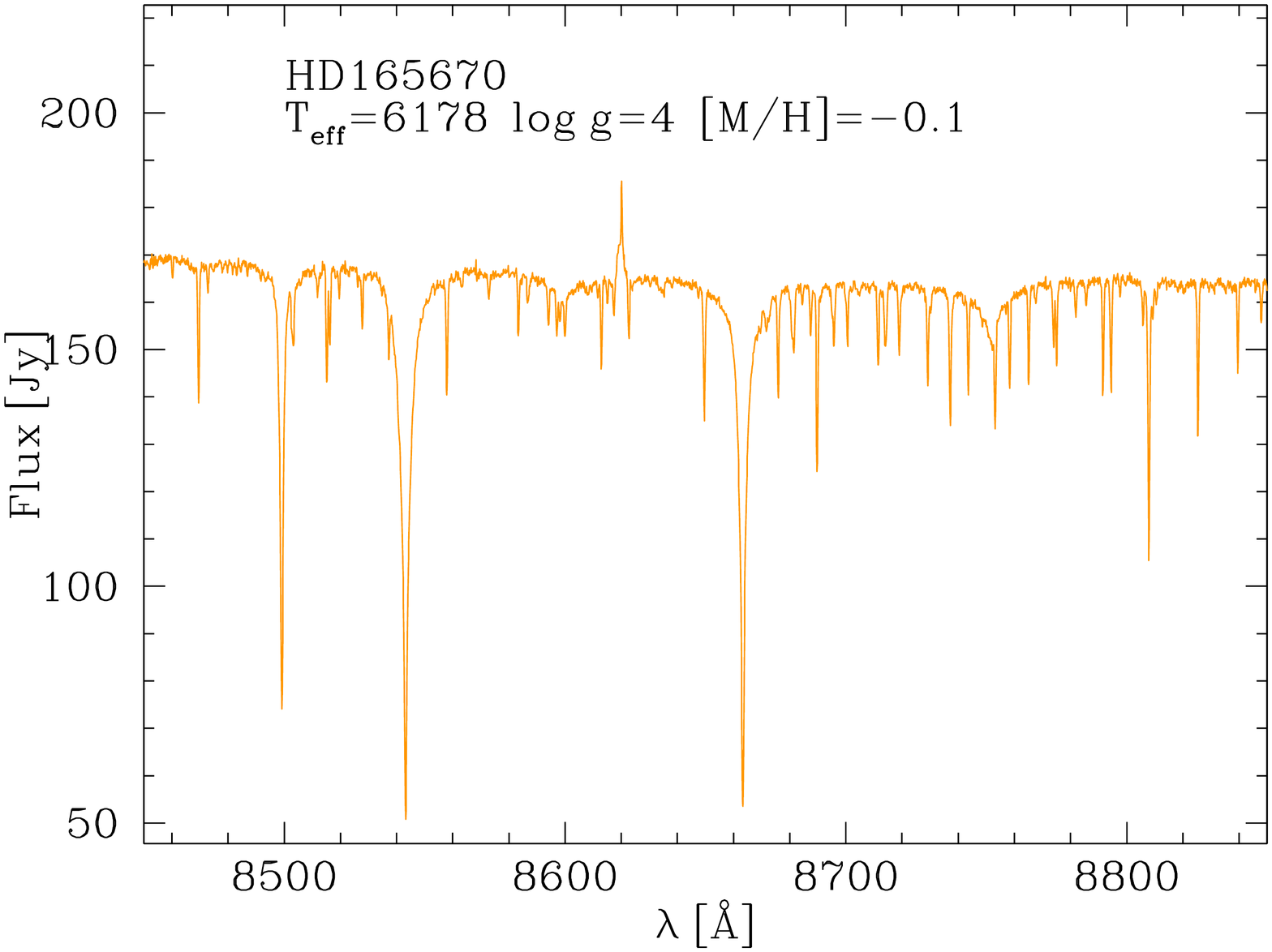}
\includegraphics[width=0.18\textwidth,angle=-90]{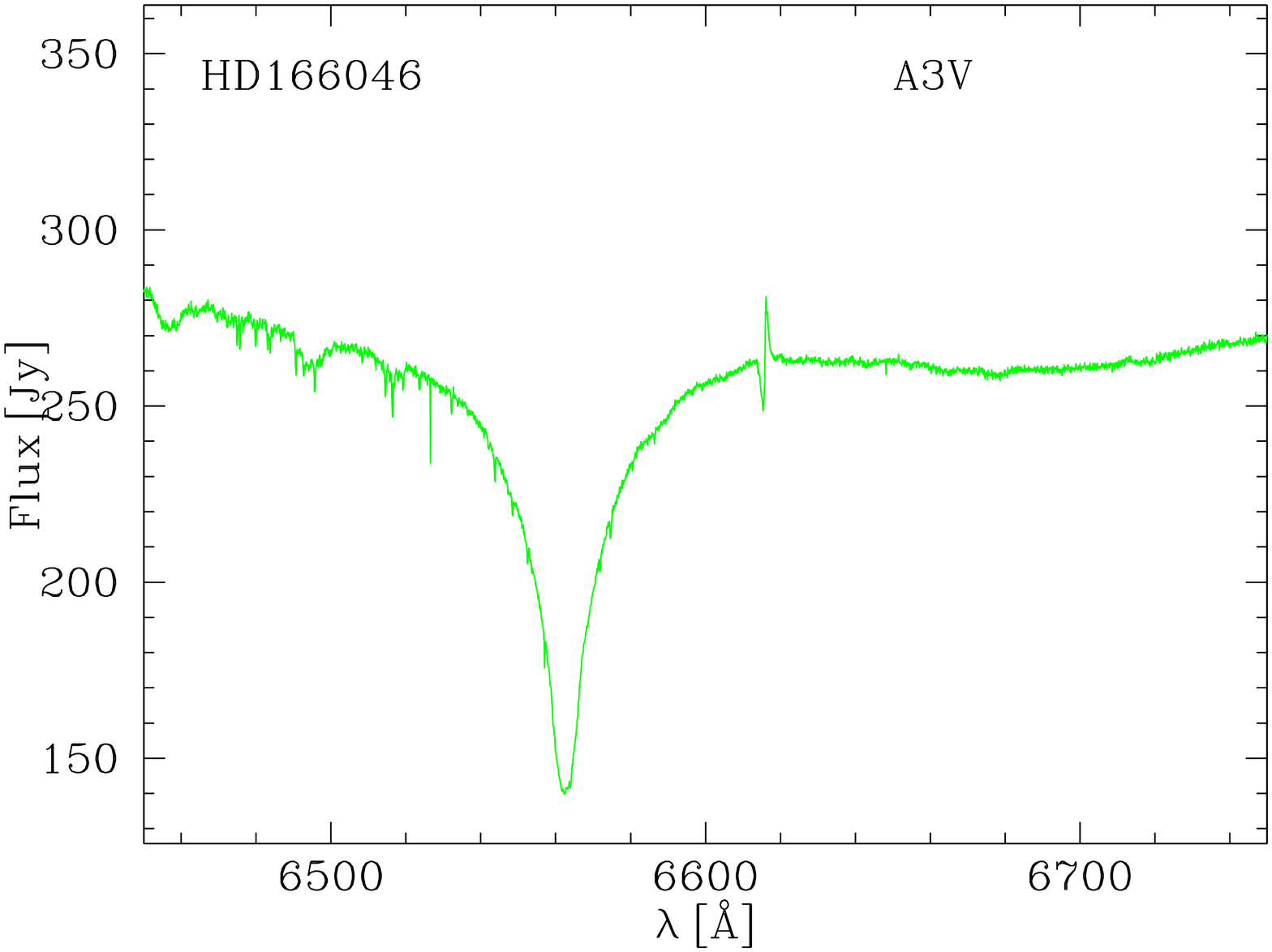}
\includegraphics[width=0.18\textwidth,angle=-90]{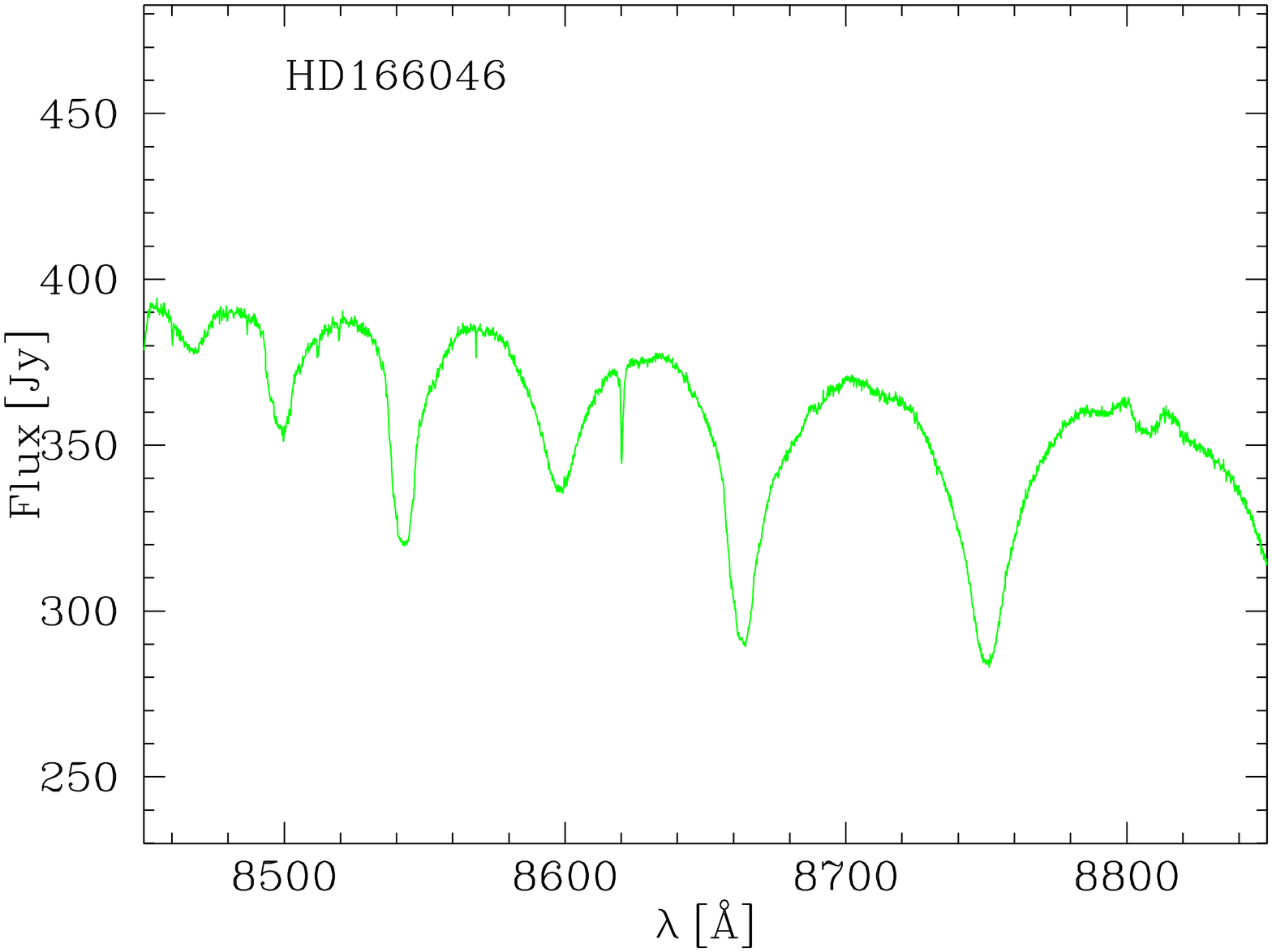}
\includegraphics[width=0.18\textwidth,angle=-90]{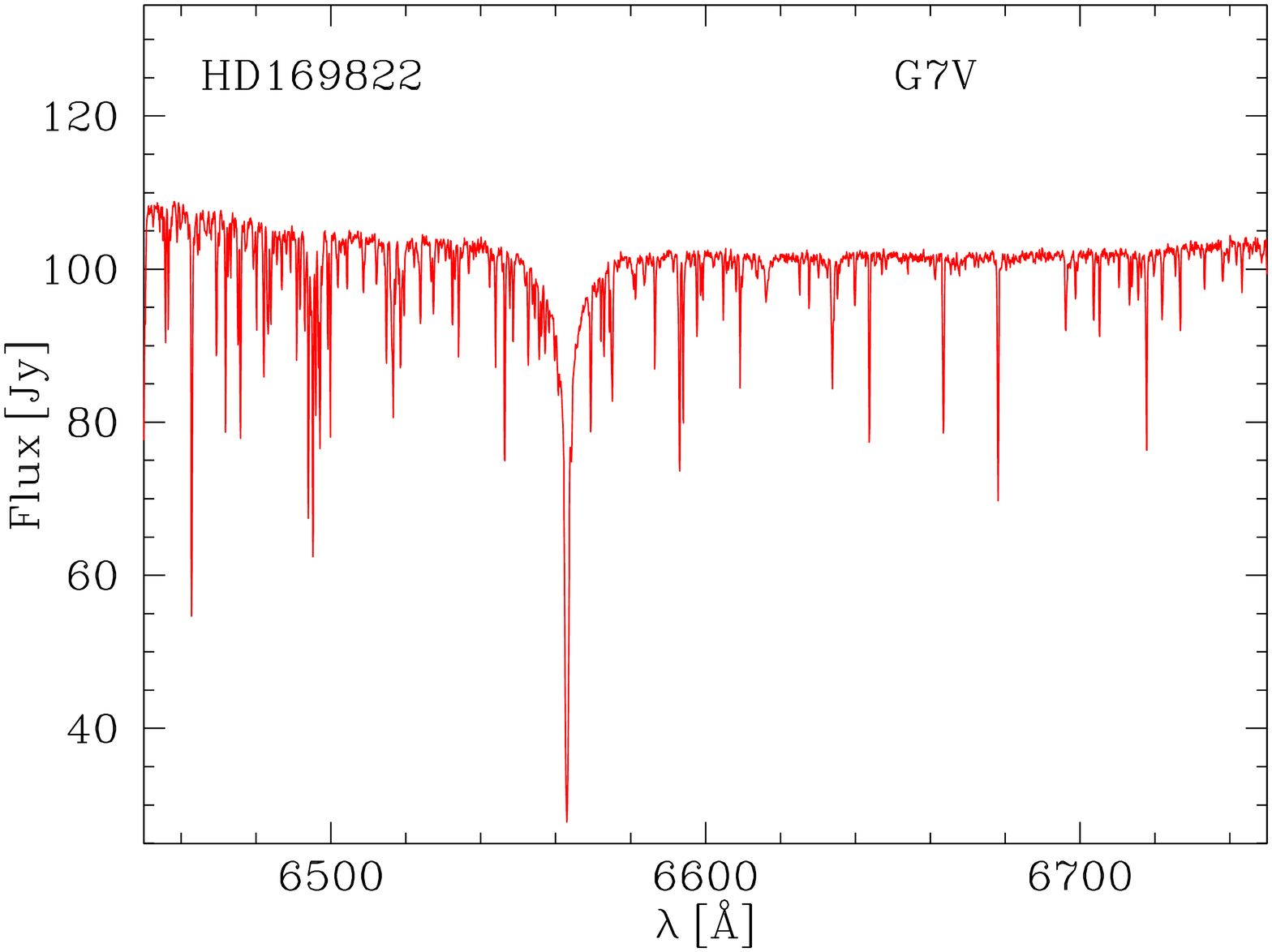}
\includegraphics[width=0.18\textwidth,angle=-90]{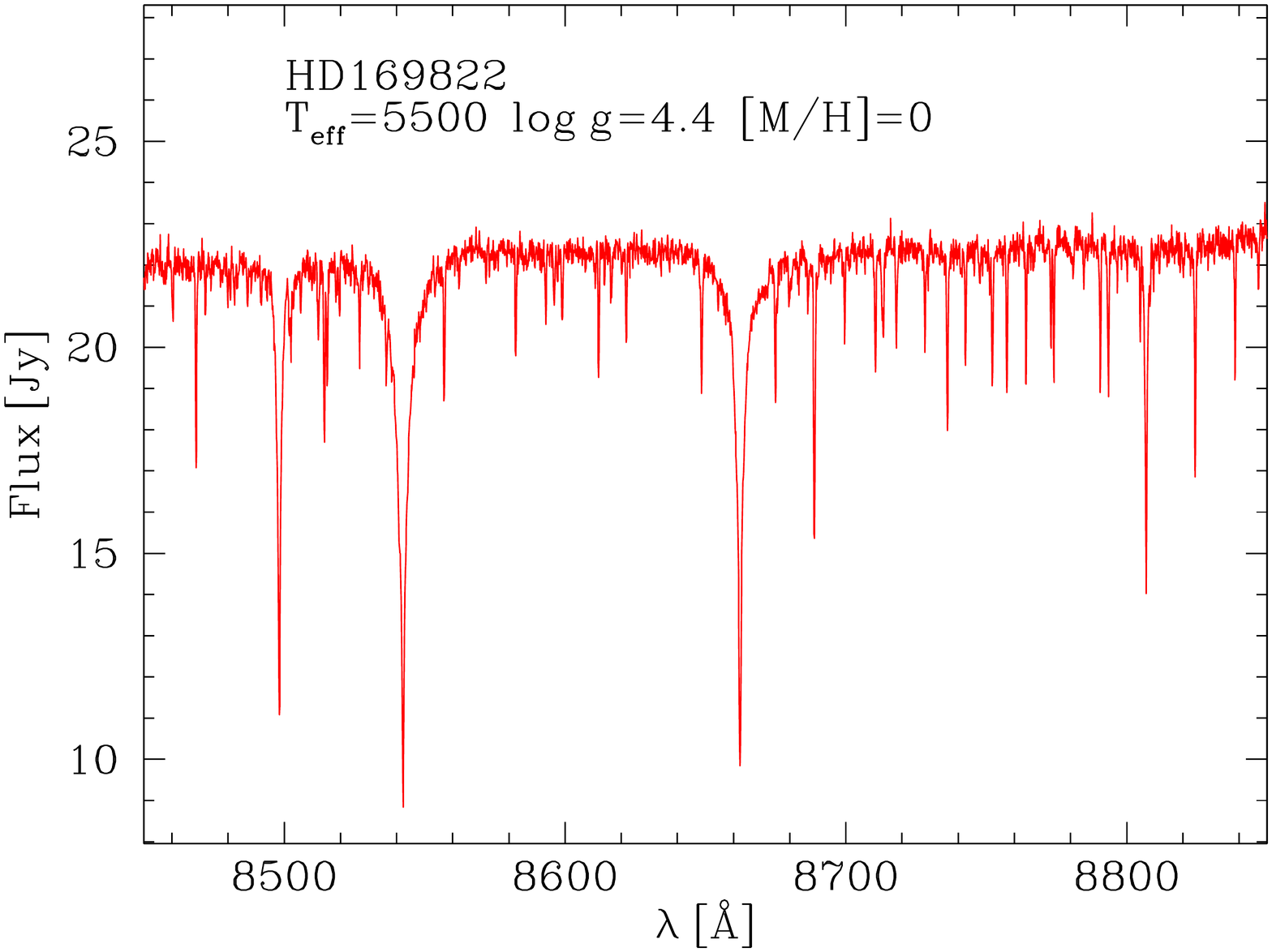}
\includegraphics[width=0.18\textwidth,angle=-90]{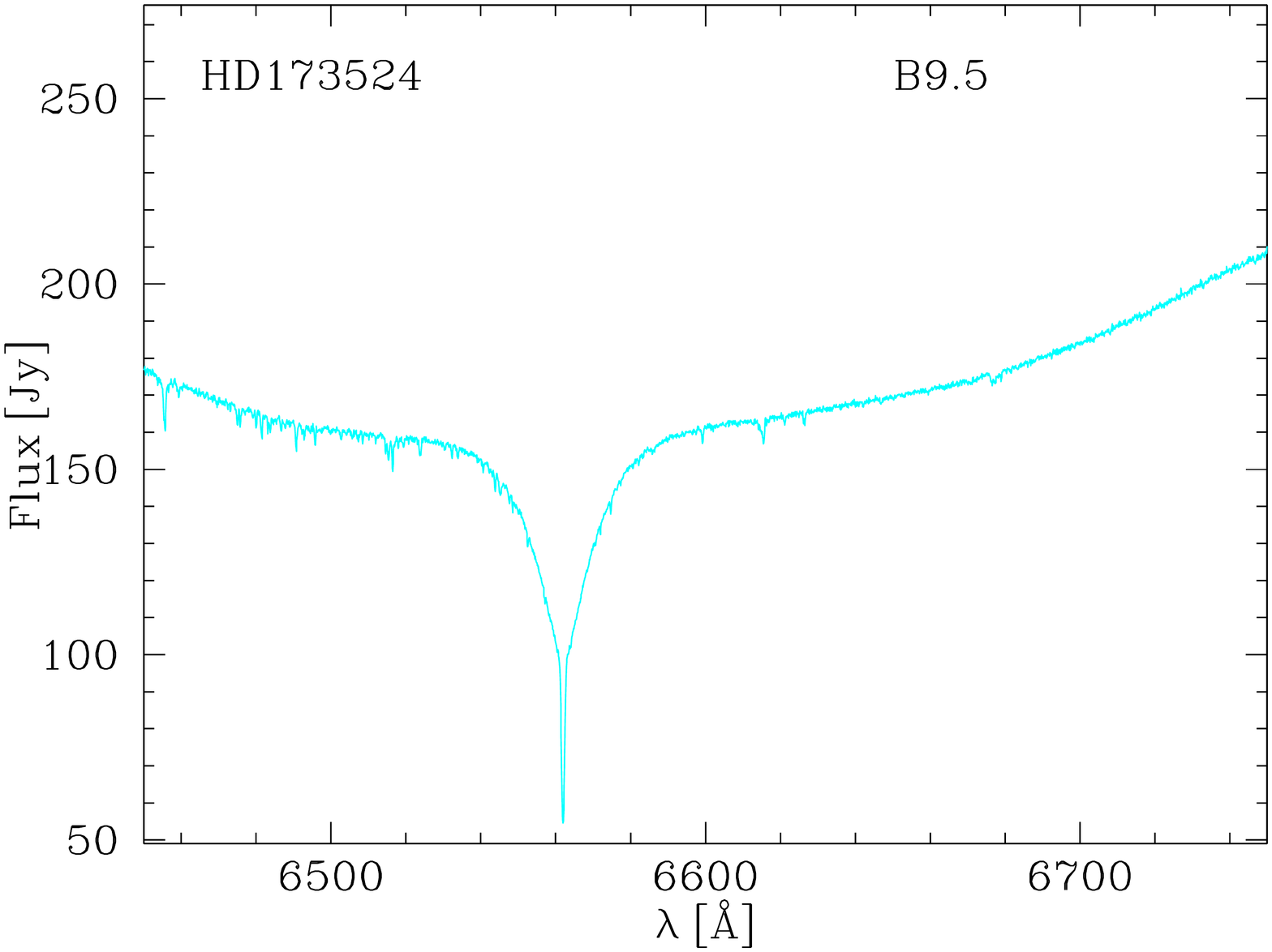}
\includegraphics[width=0.18\textwidth,angle=-90]{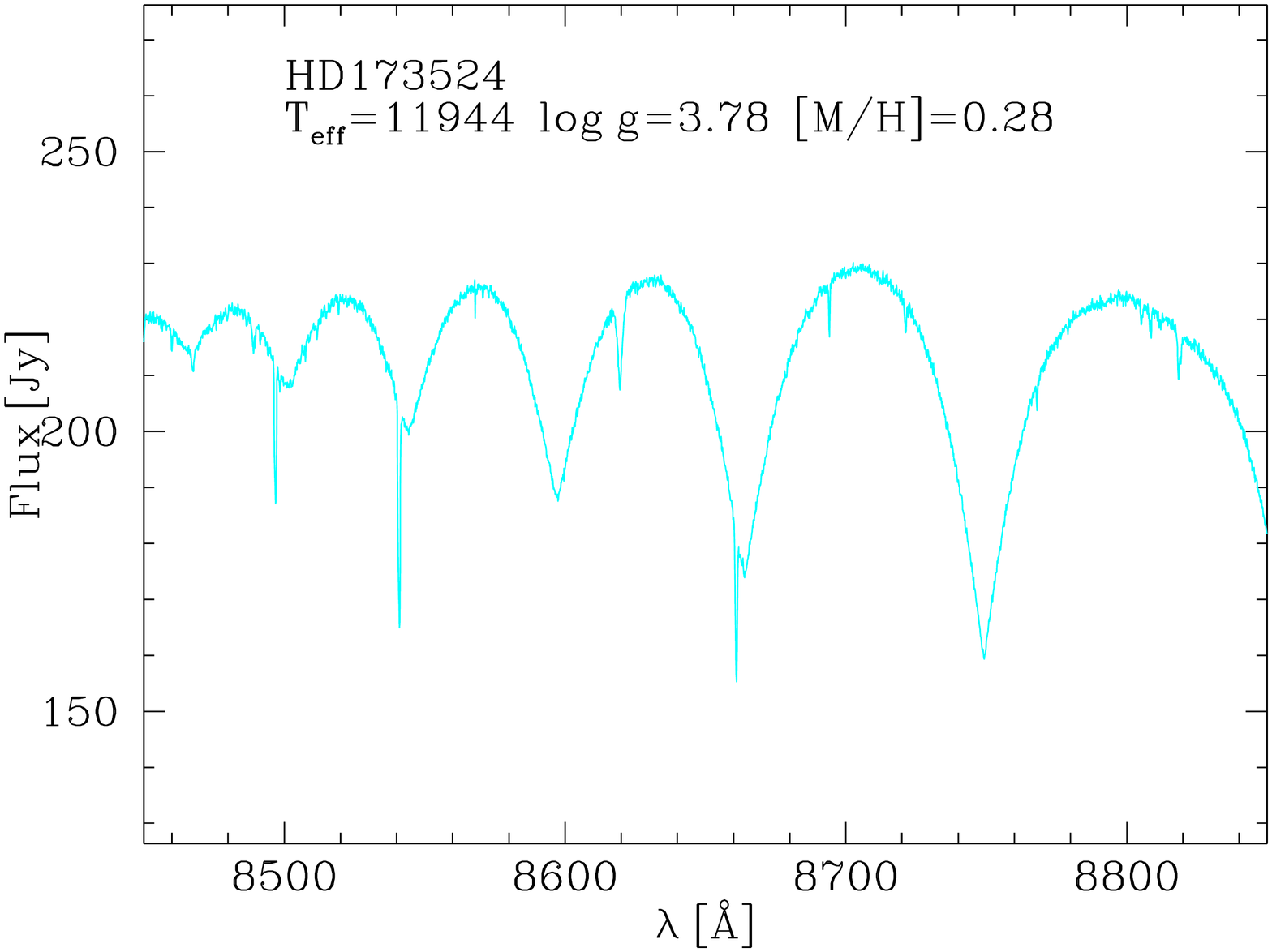}
\includegraphics[width=0.18\textwidth,angle=-90]{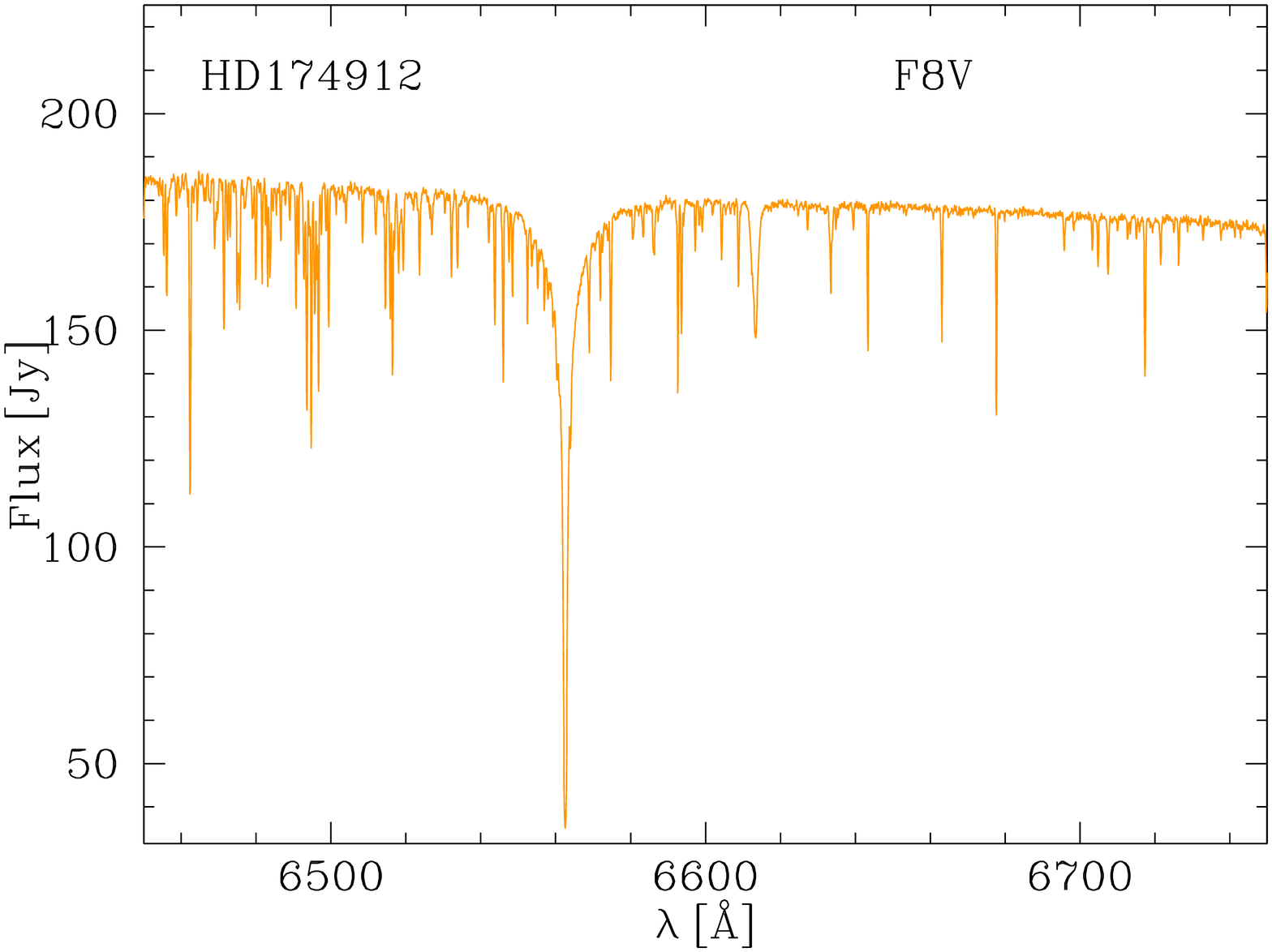}
\includegraphics[width=0.18\textwidth,angle=-90]{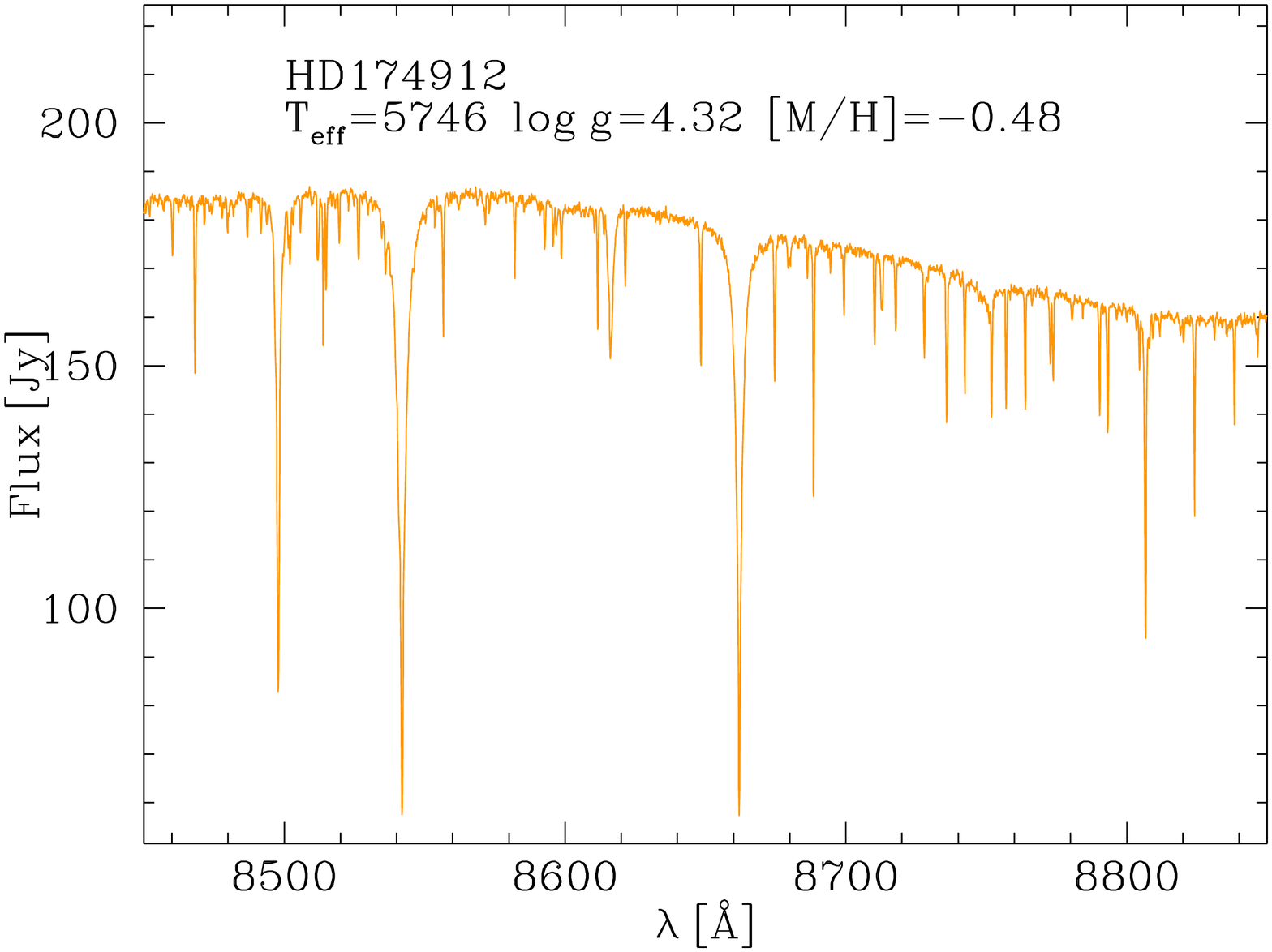}
\includegraphics[width=0.18\textwidth,angle=-90]{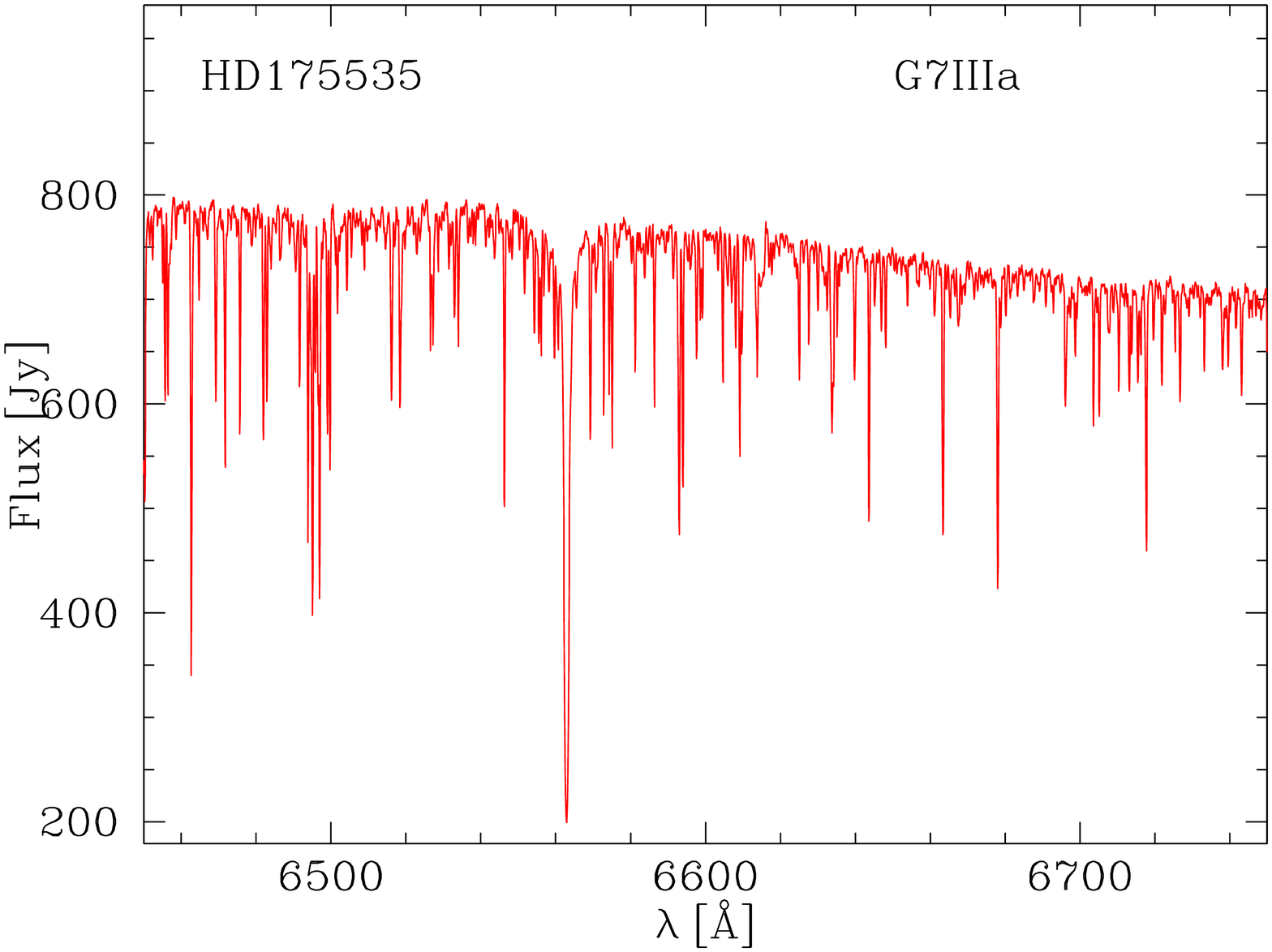}
\includegraphics[width=0.18\textwidth,angle=-90]{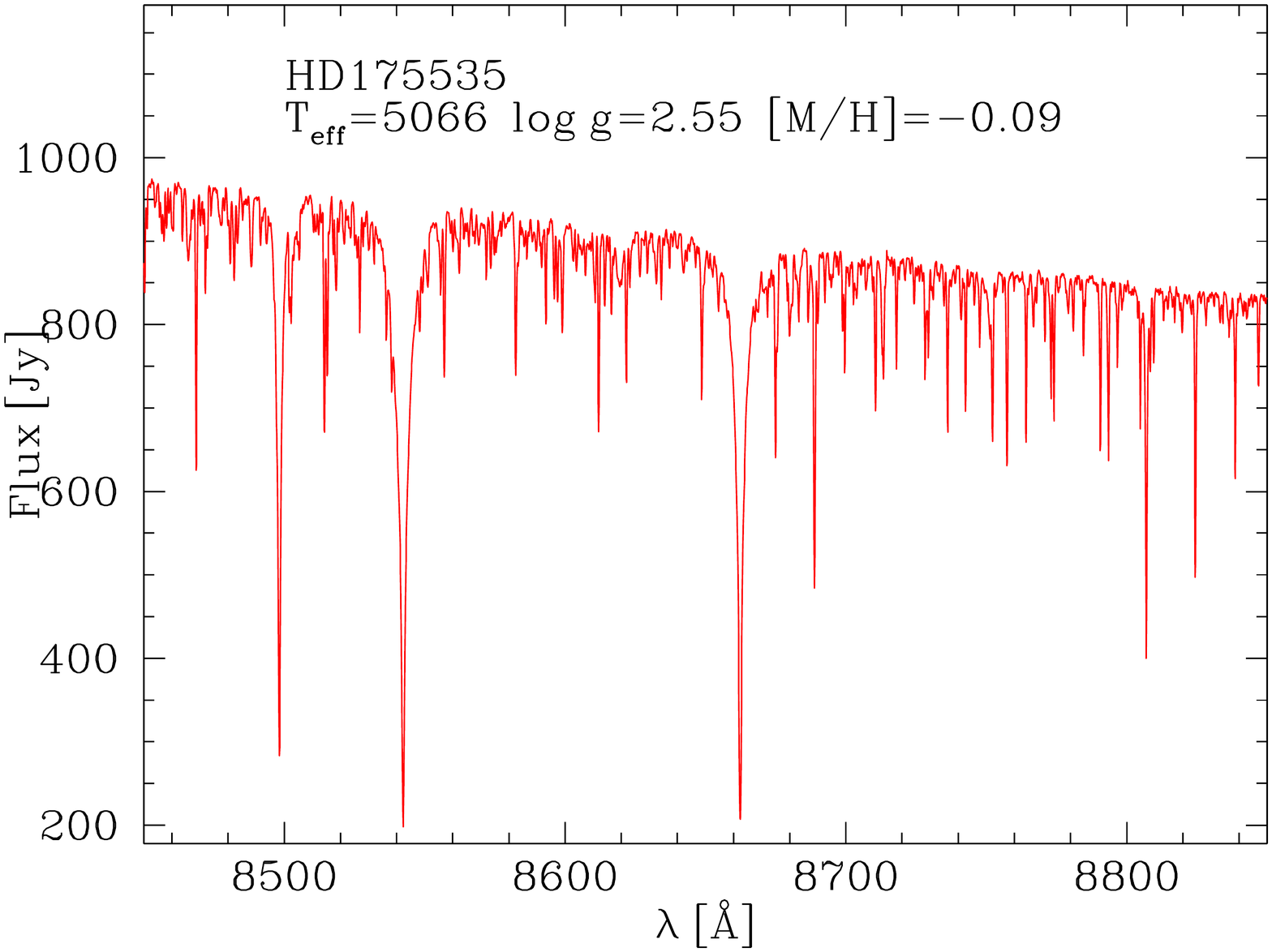}

\contcaption{25. Stars shown in this page are: HD149121, HD149161, HD155358, HD155763, HD160762, HD164353, HD165029, HD165358, HD165670, HD166046, HD169822, HD173524, HD174912 and HD175535.}
\end{figure*}

\begin{figure*}
\includegraphics[width=0.18\textwidth,angle=-90]{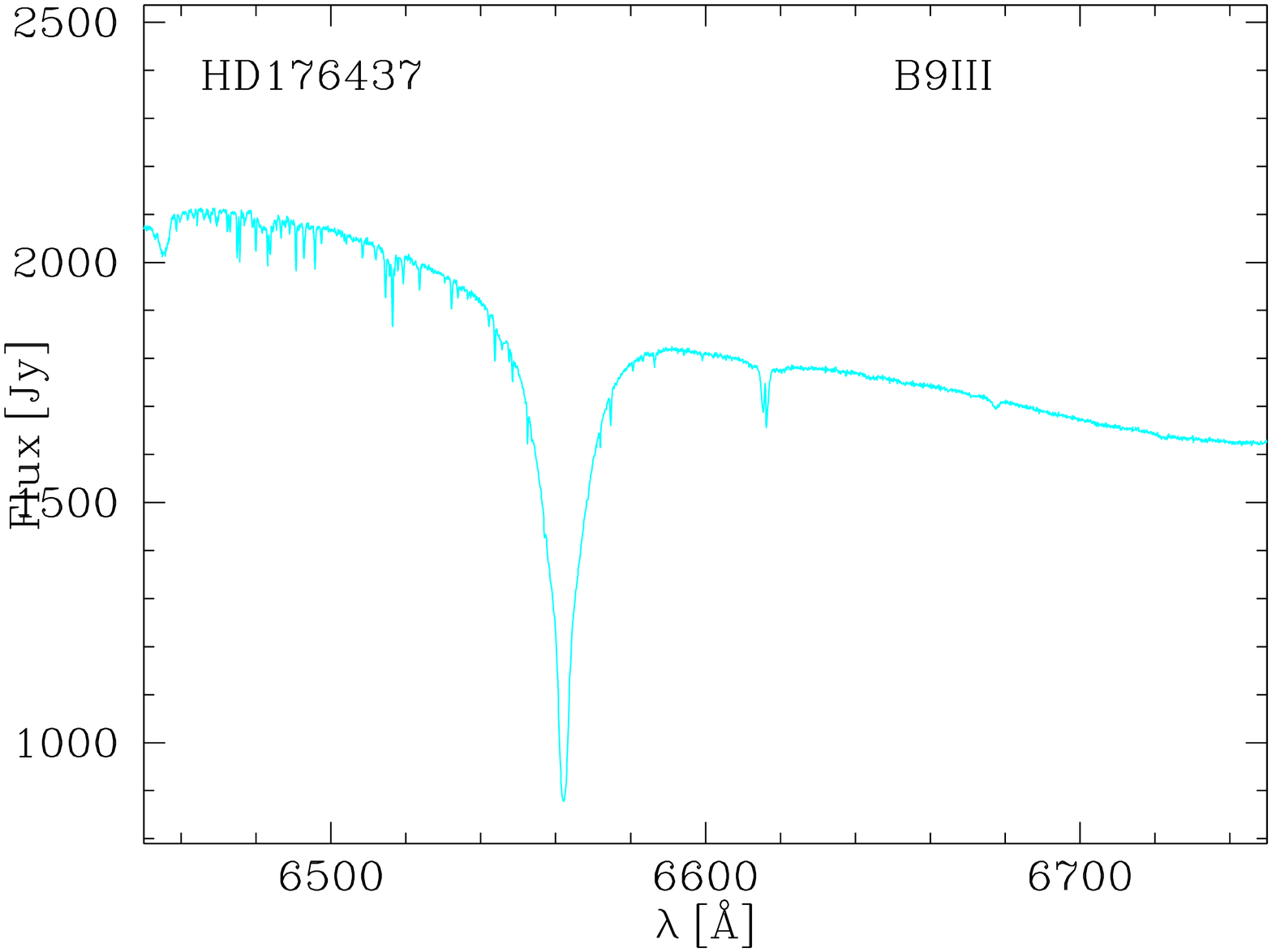}
\includegraphics[width=0.18\textwidth,angle=-90]{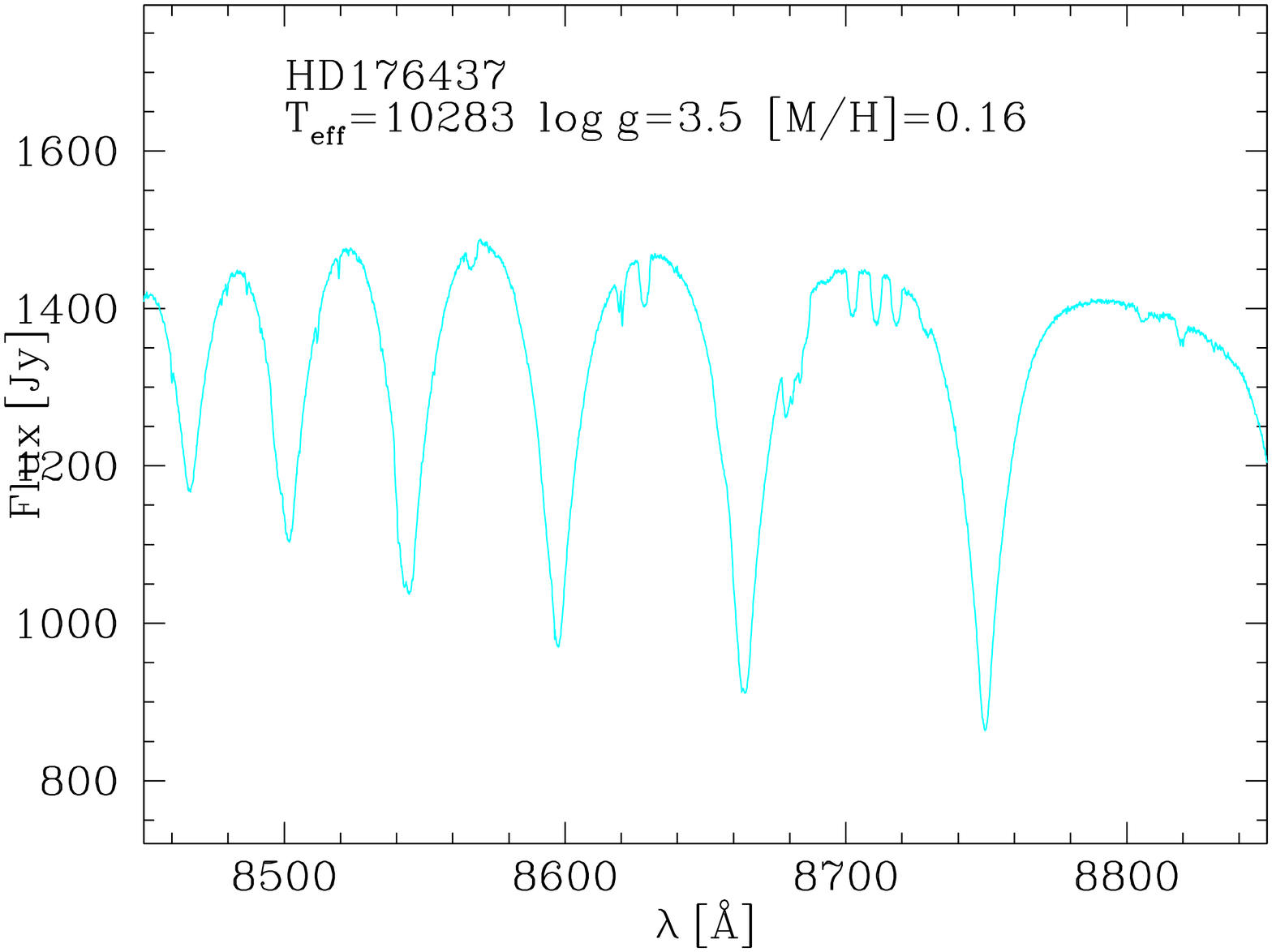}
\includegraphics[width=0.18\textwidth,angle=-90]{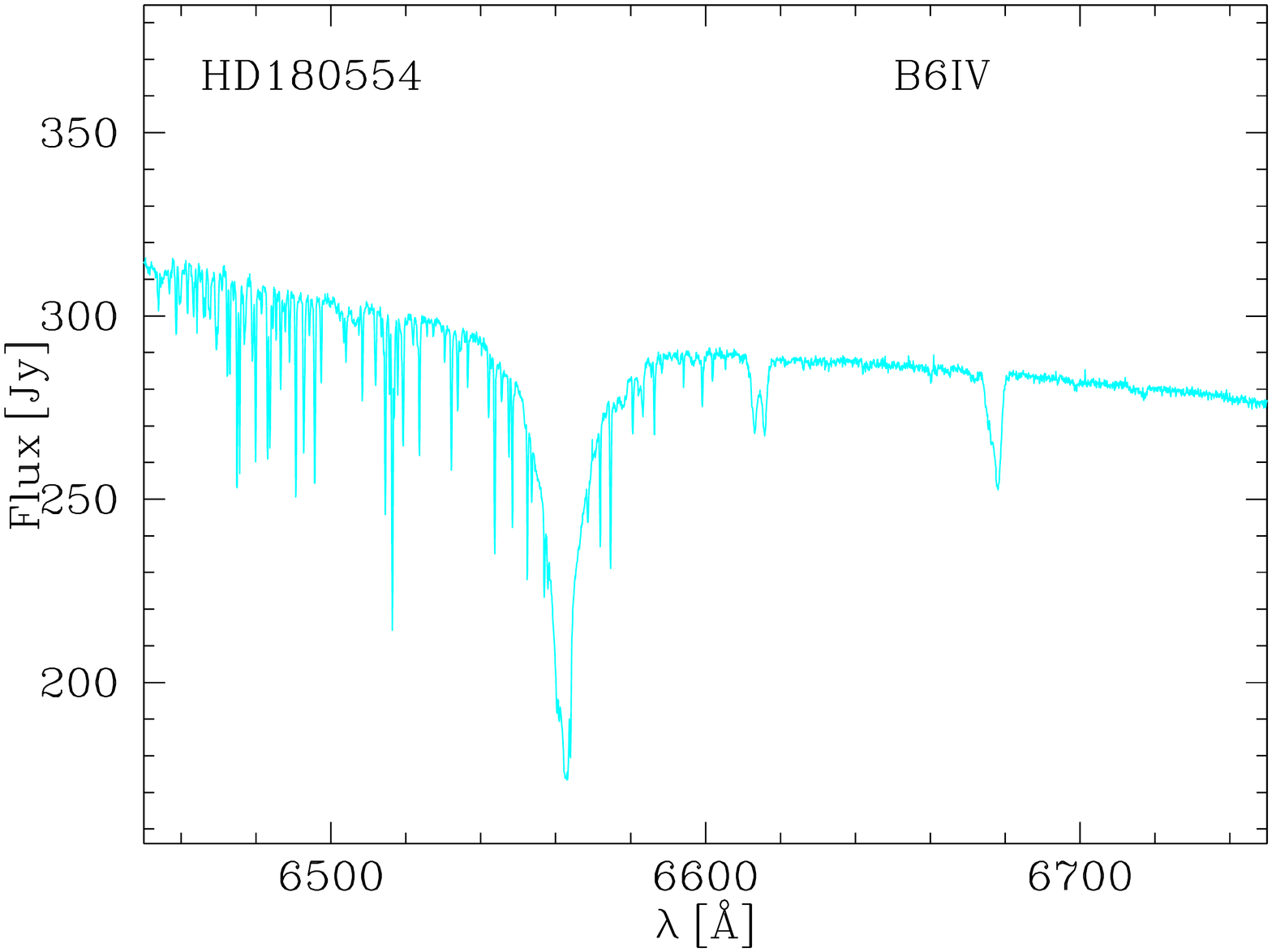}
\includegraphics[width=0.18\textwidth,angle=-90]{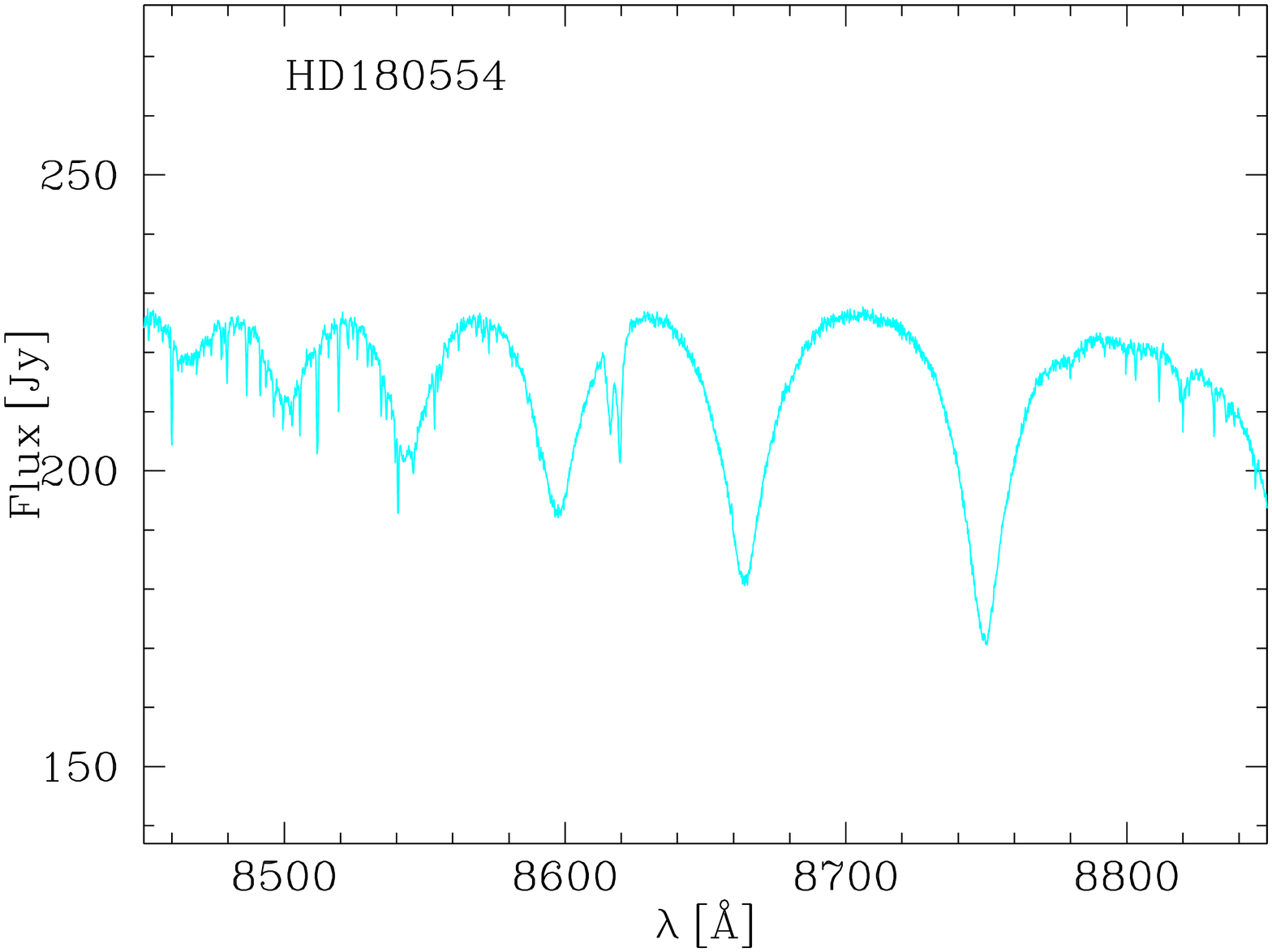}
\includegraphics[width=0.18\textwidth,angle=-90]{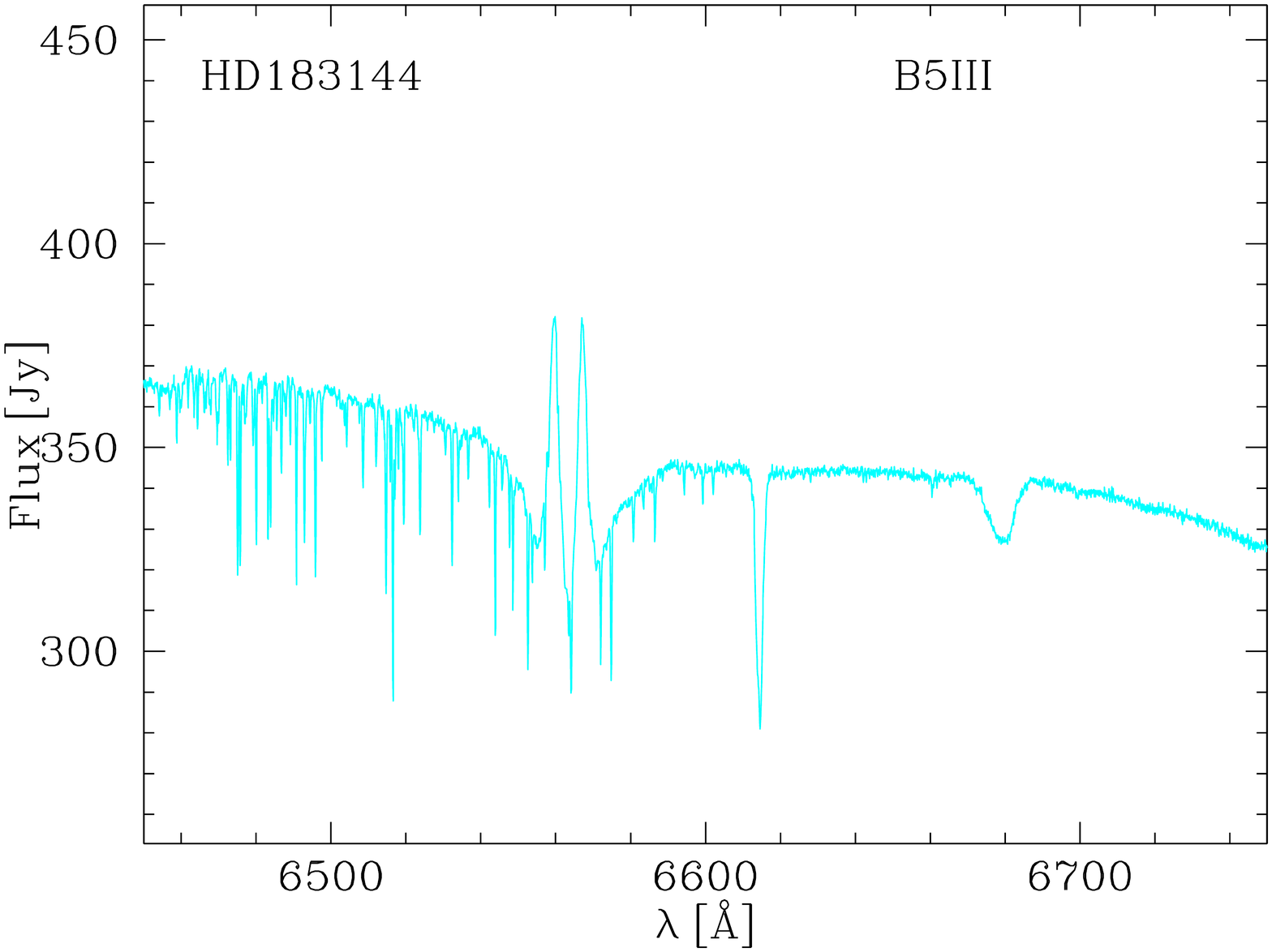}
\includegraphics[width=0.18\textwidth,angle=-90]{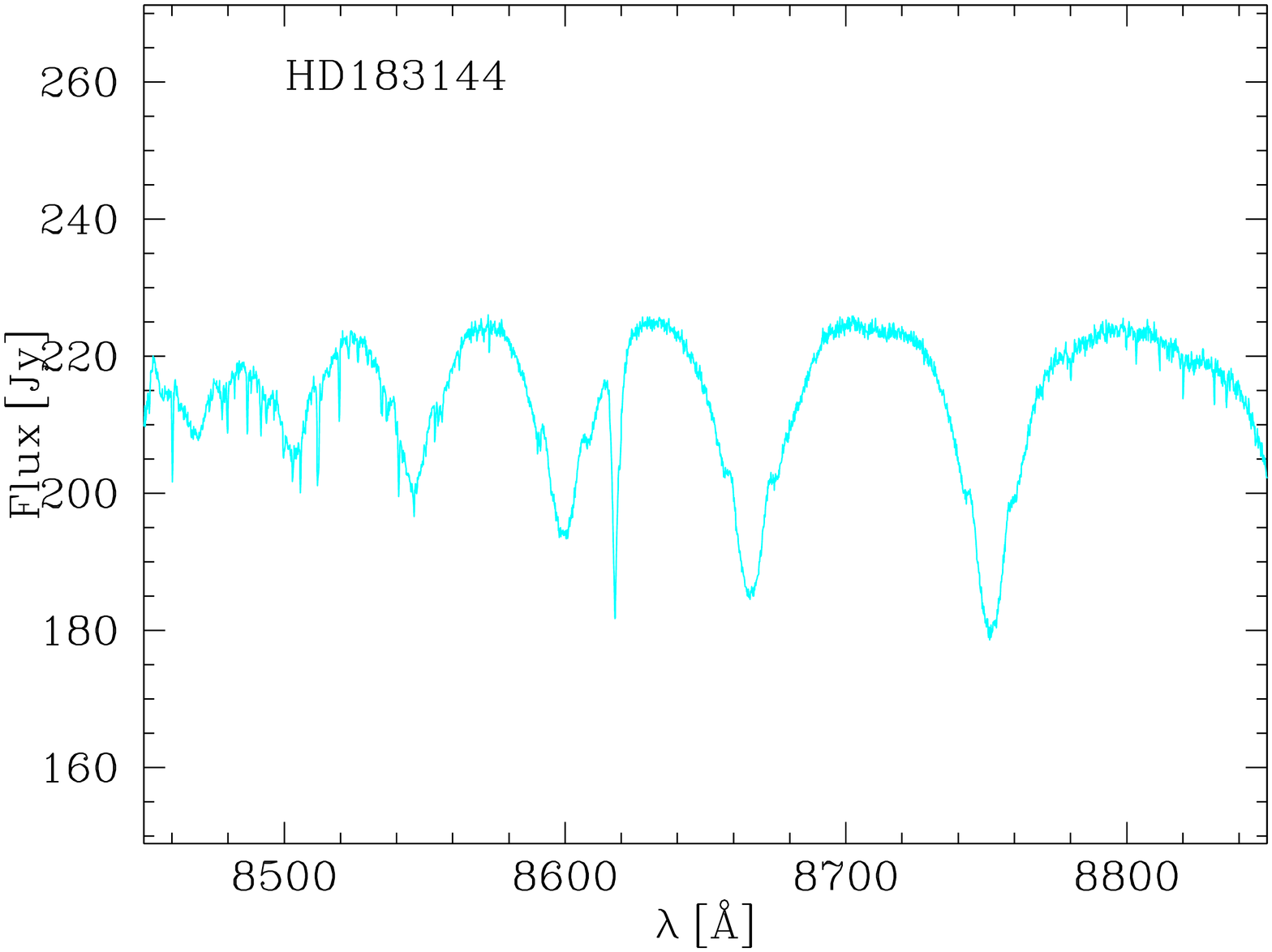}
\includegraphics[width=0.18\textwidth,angle=-90]{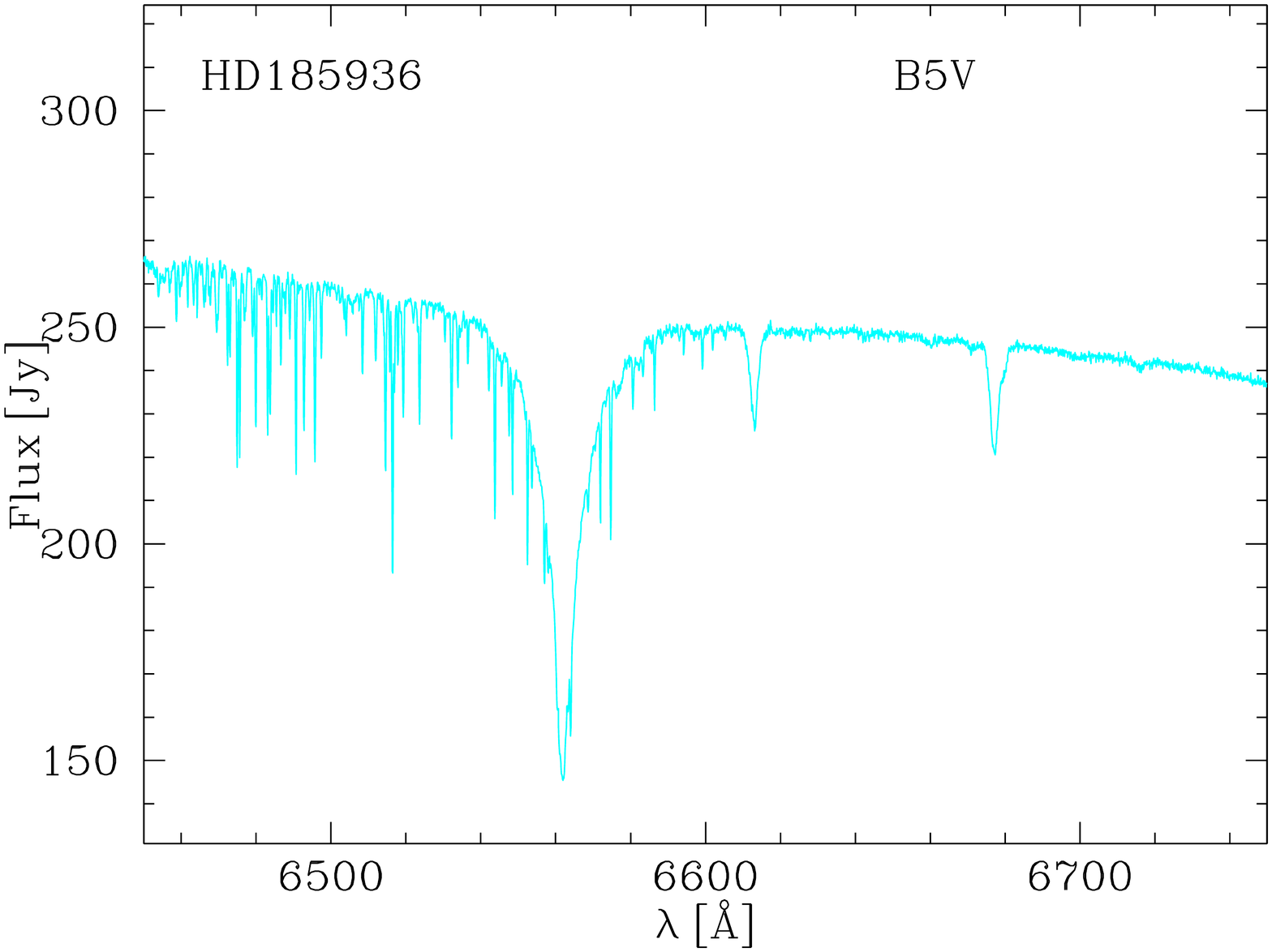}
\includegraphics[width=0.18\textwidth,angle=-90]{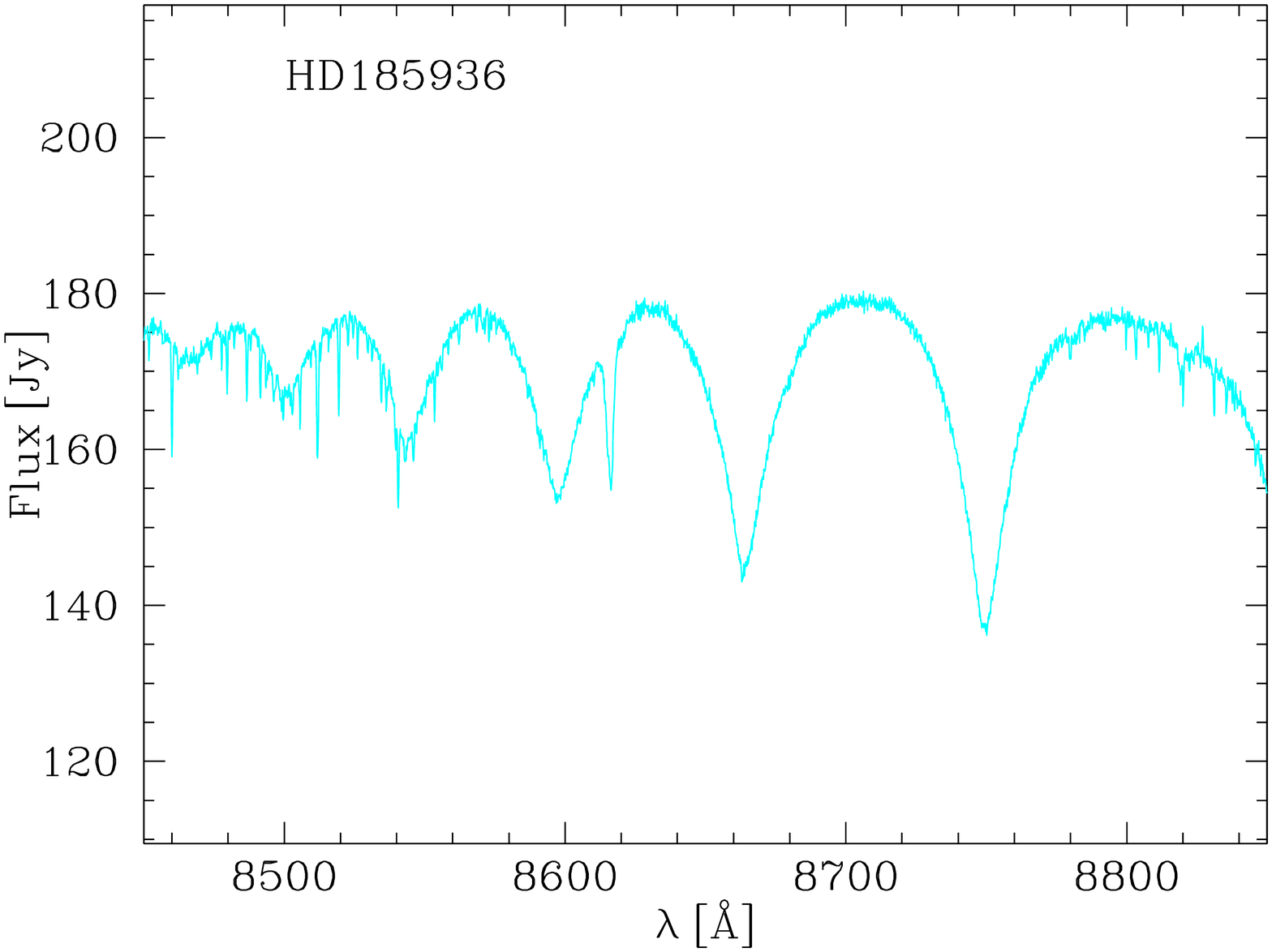}
\includegraphics[width=0.18\textwidth,angle=-90]{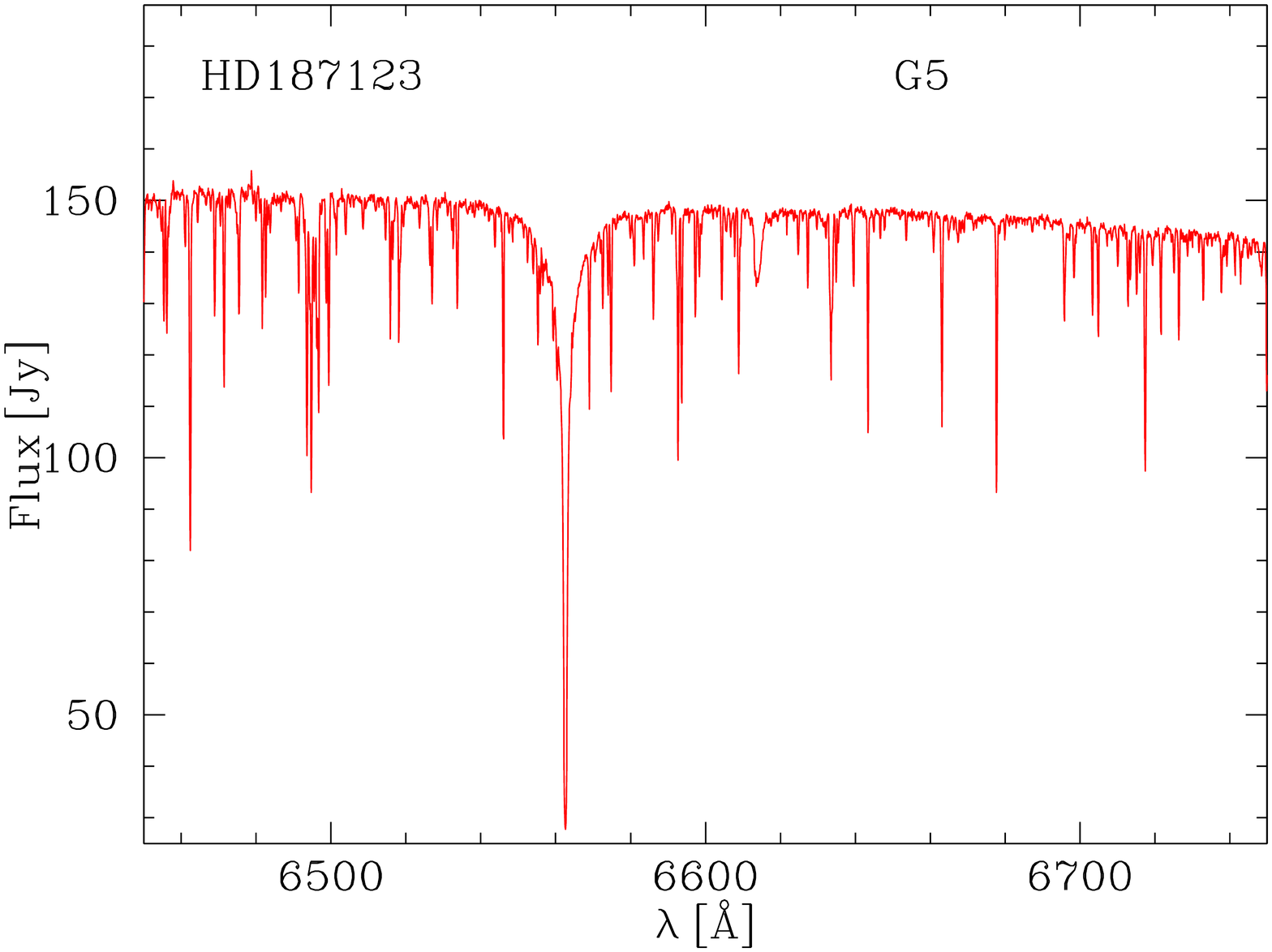}
\includegraphics[width=0.18\textwidth,angle=-90]{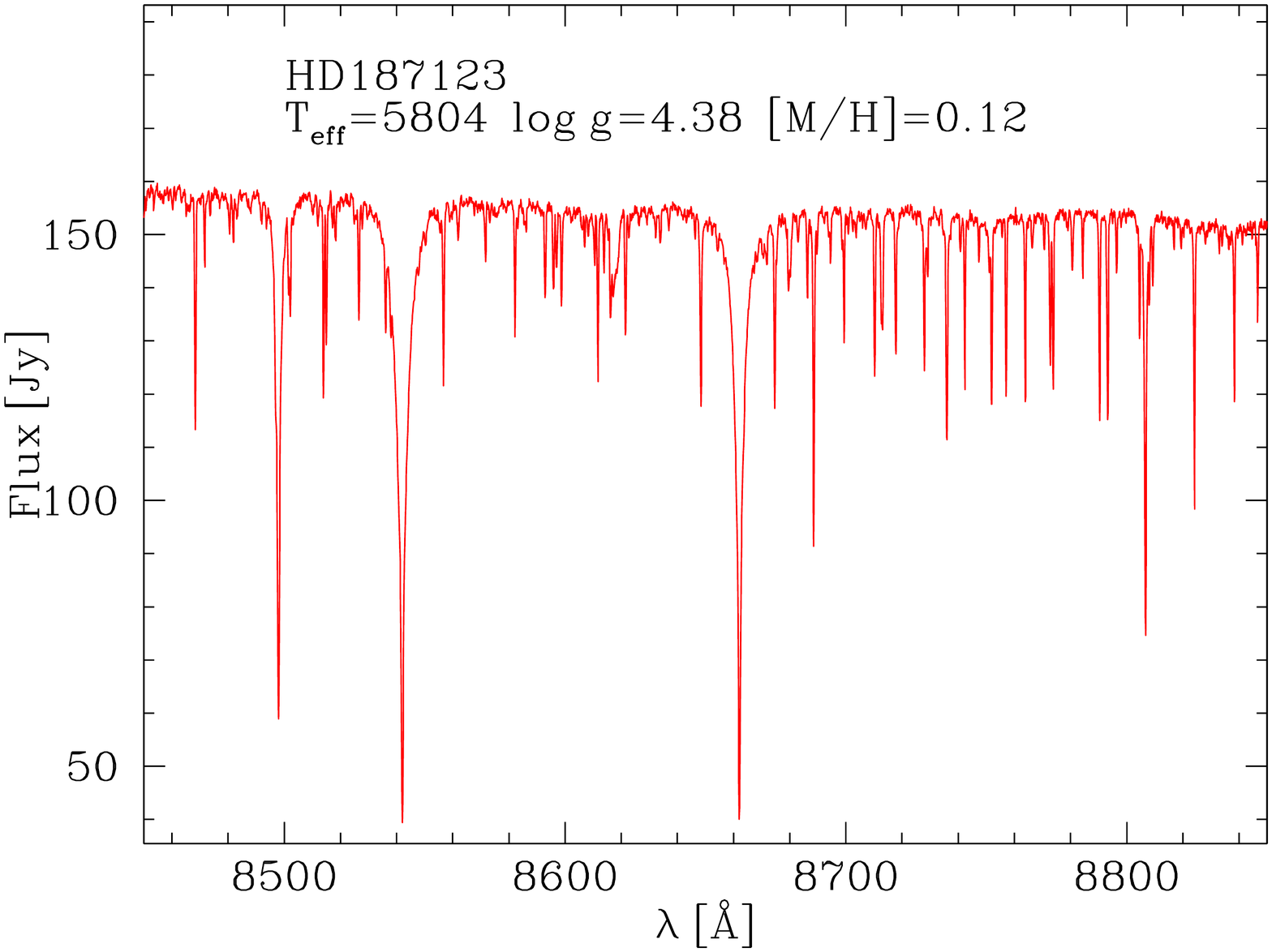}
\includegraphics[width=0.18\textwidth,angle=-90]{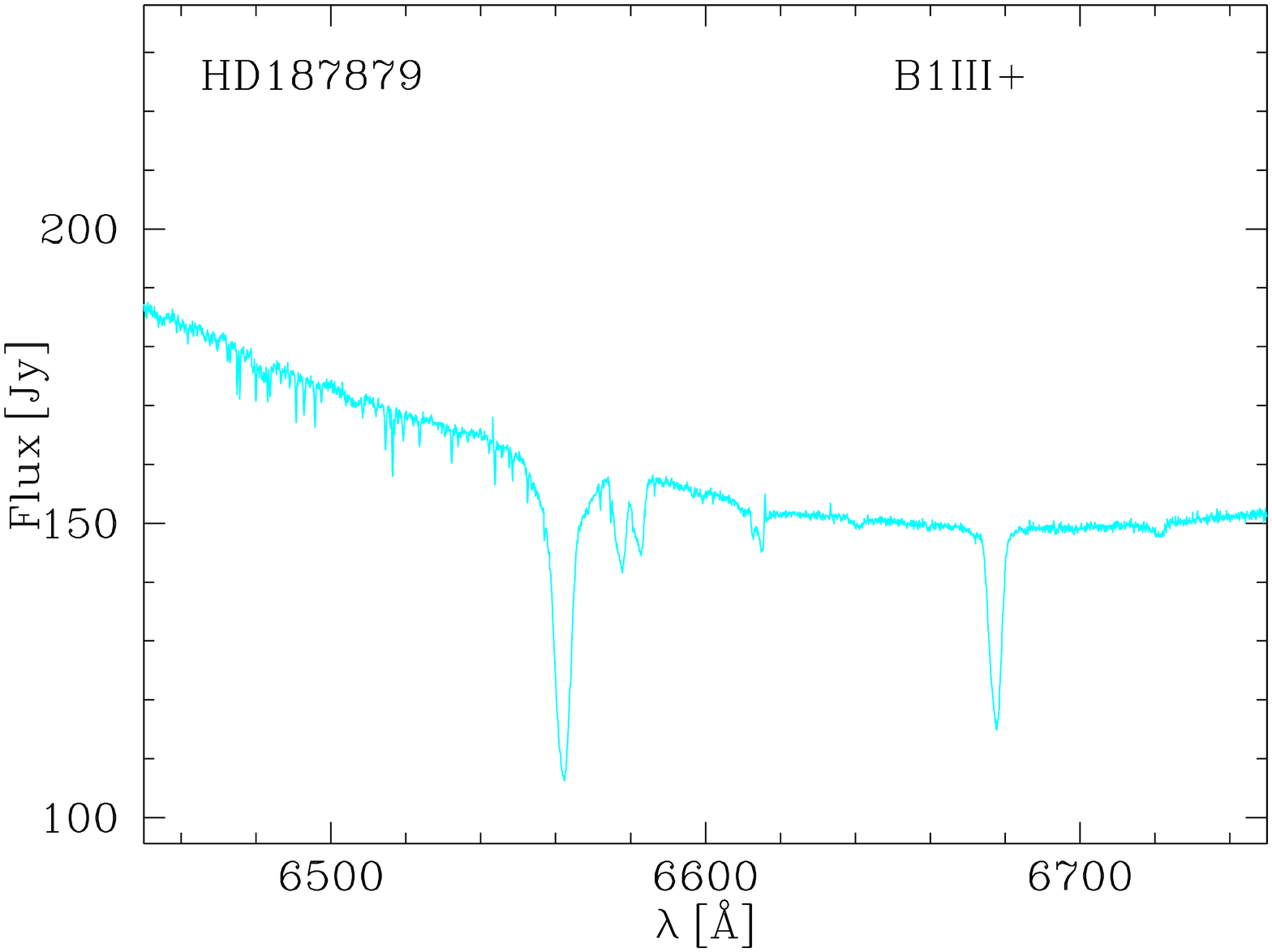} 
\includegraphics[width=0.18\textwidth,angle=-90]{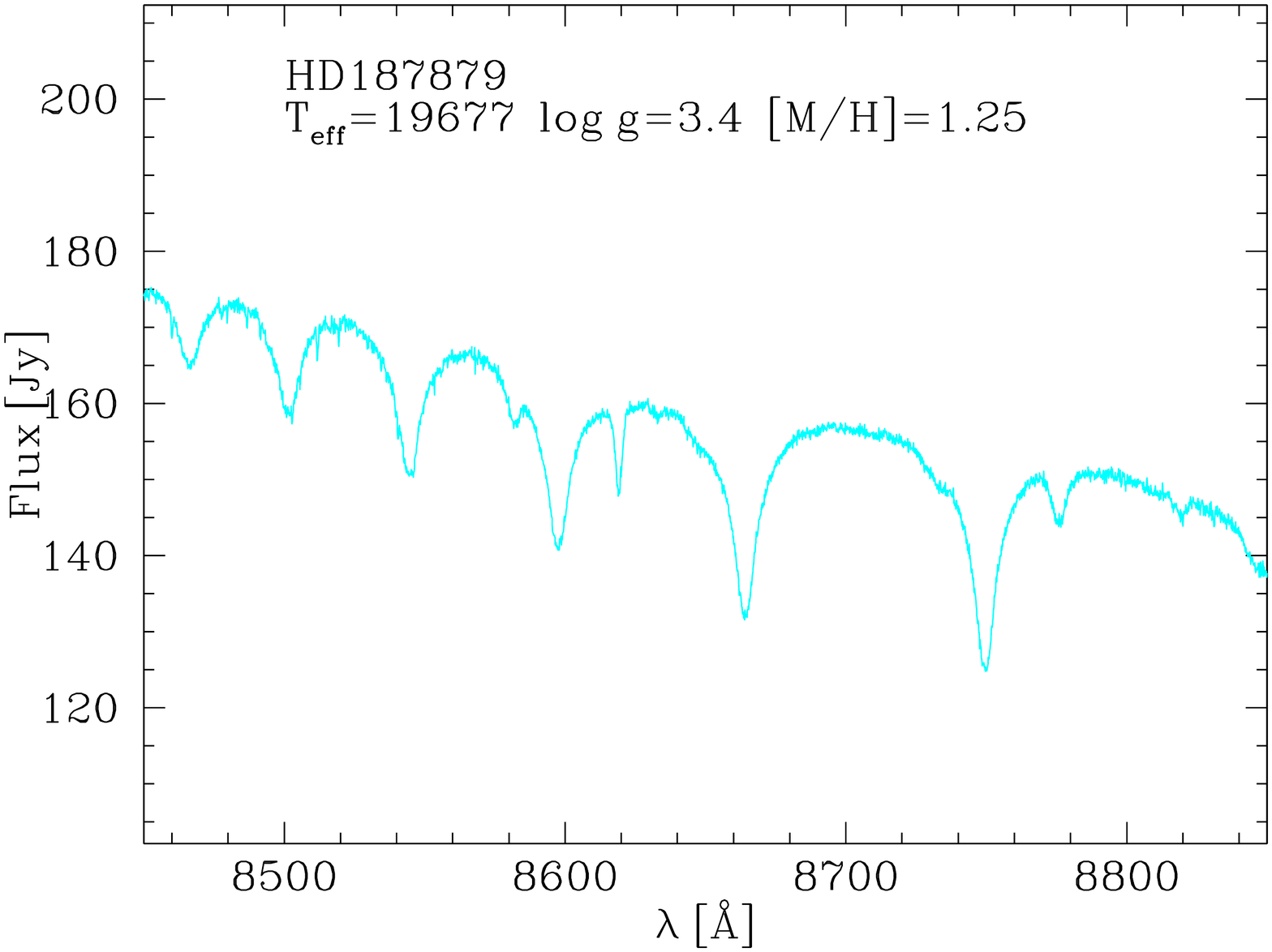}
\includegraphics[width=0.18\textwidth,angle=-90]{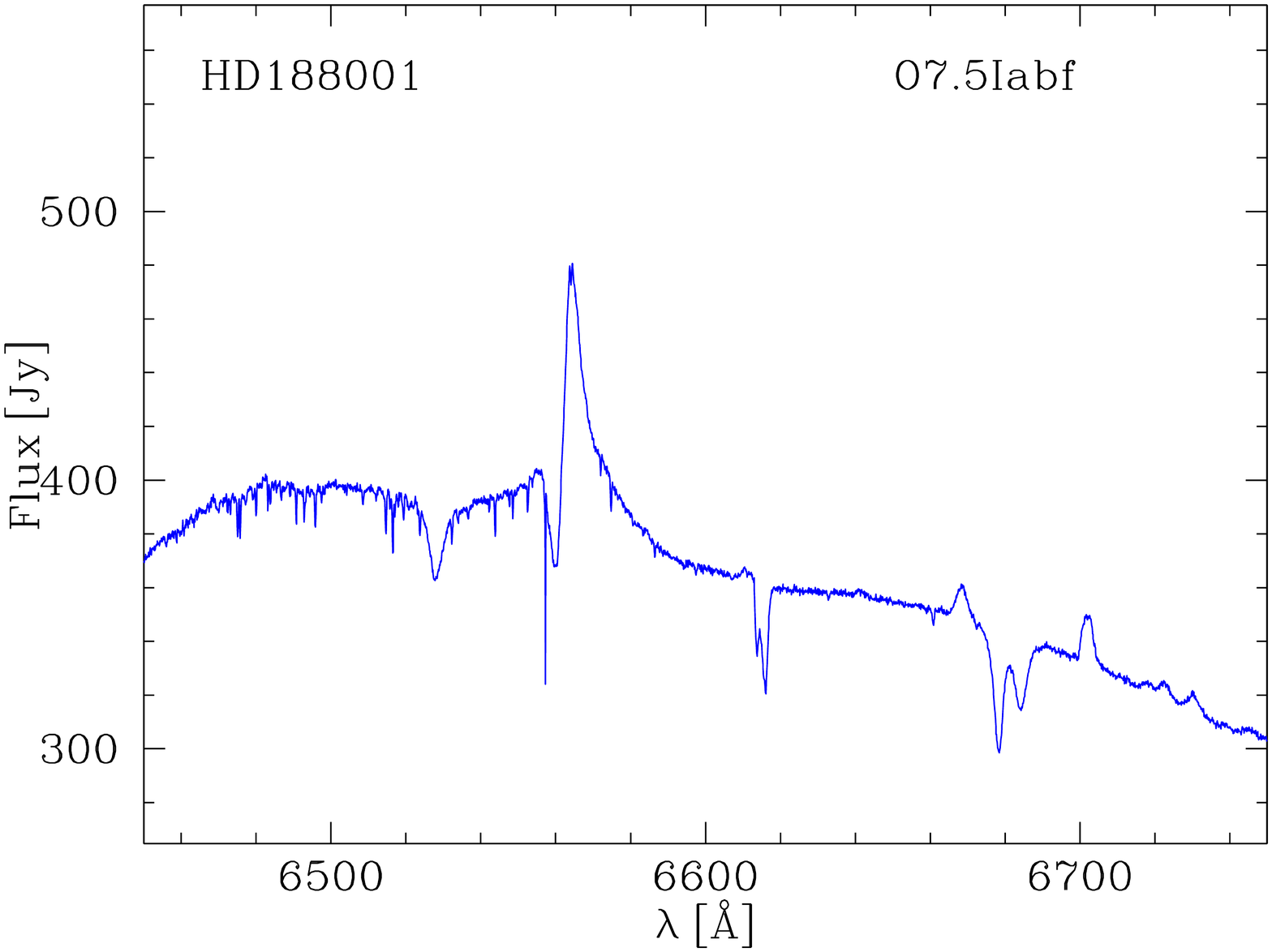}
\includegraphics[width=0.18\textwidth,angle=-90]{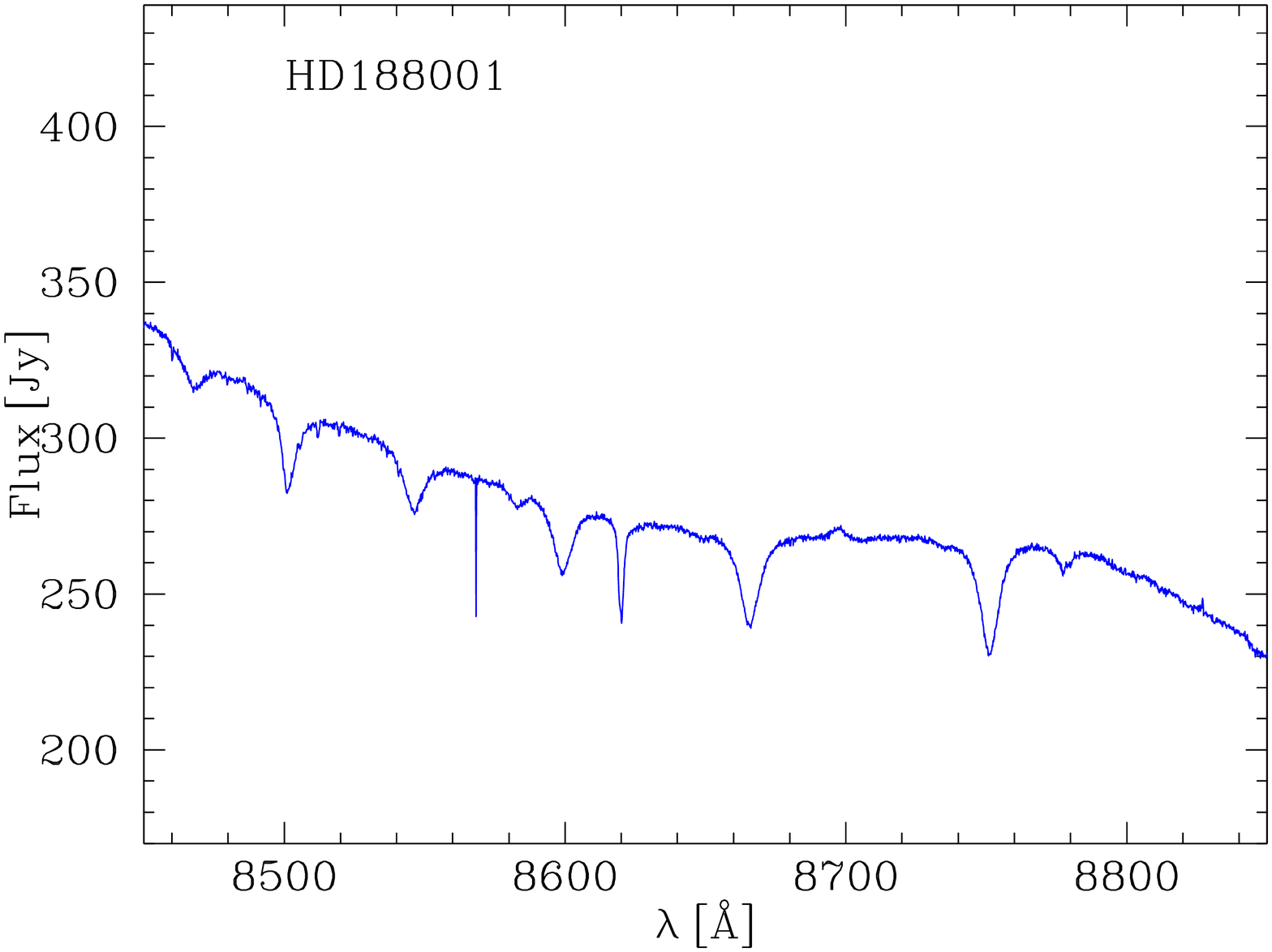}
\includegraphics[width=0.18\textwidth,angle=-90]{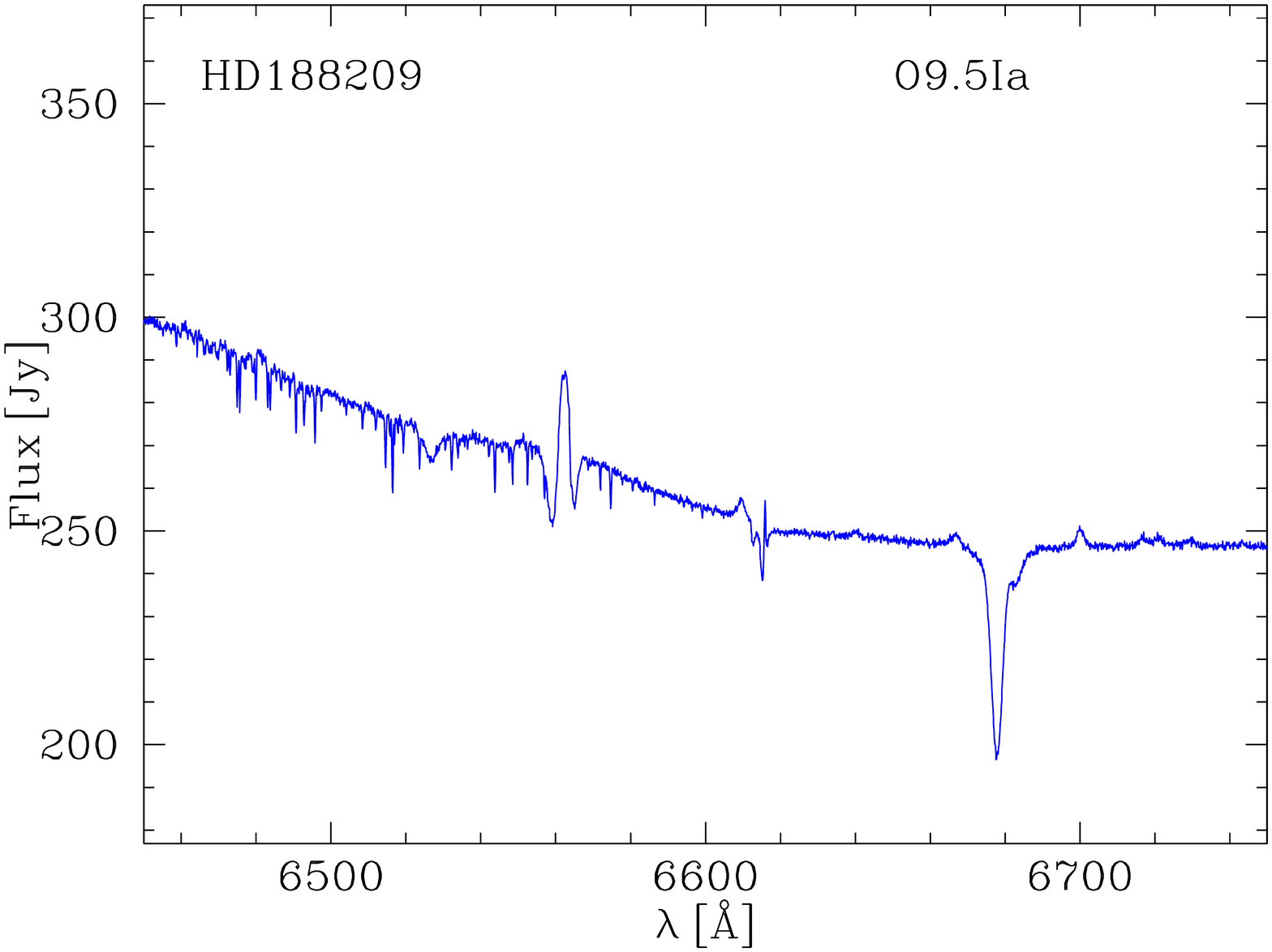}
\includegraphics[width=0.18\textwidth,angle=-90]{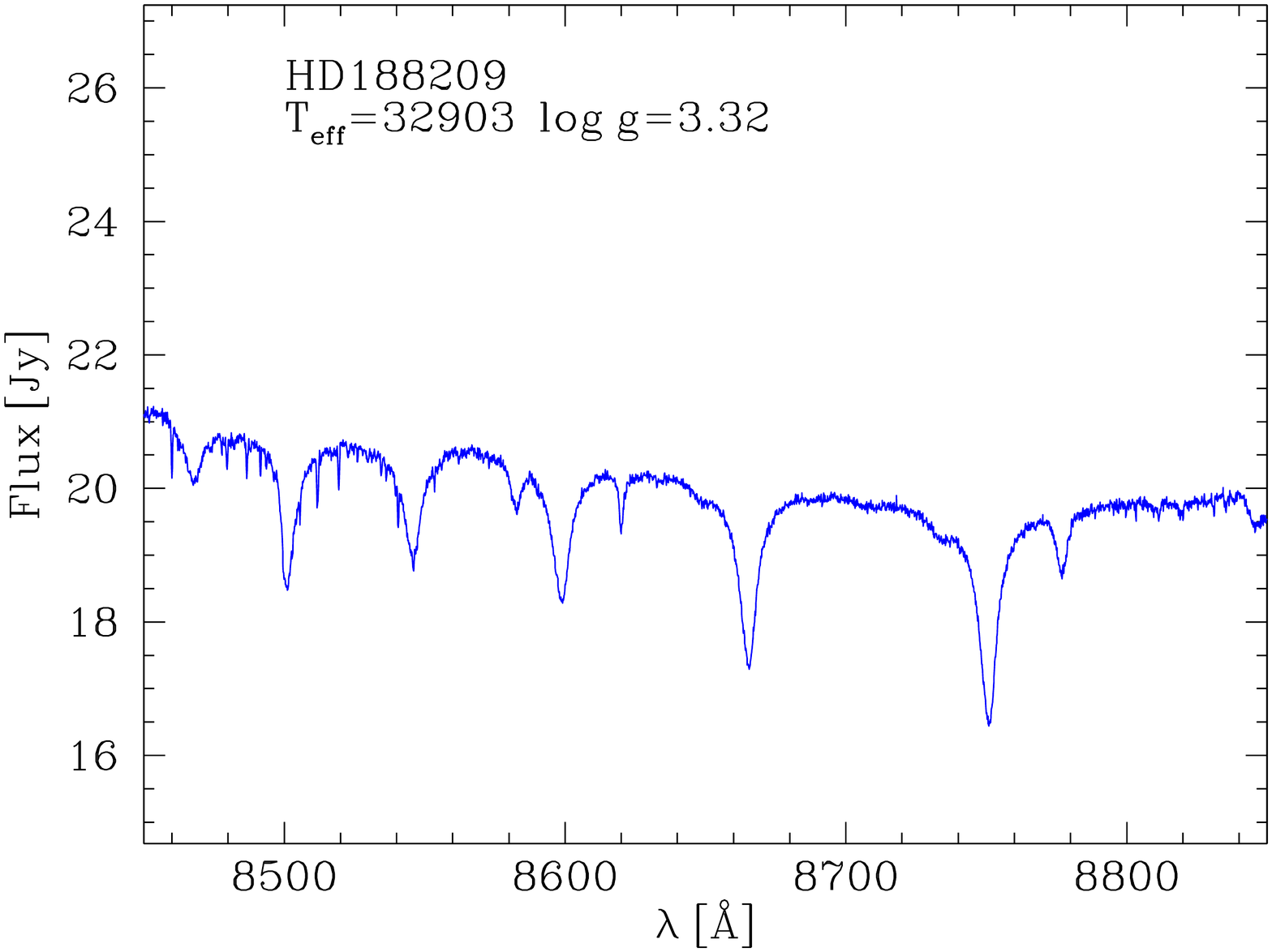}
\includegraphics[width=0.18\textwidth,angle=-90]{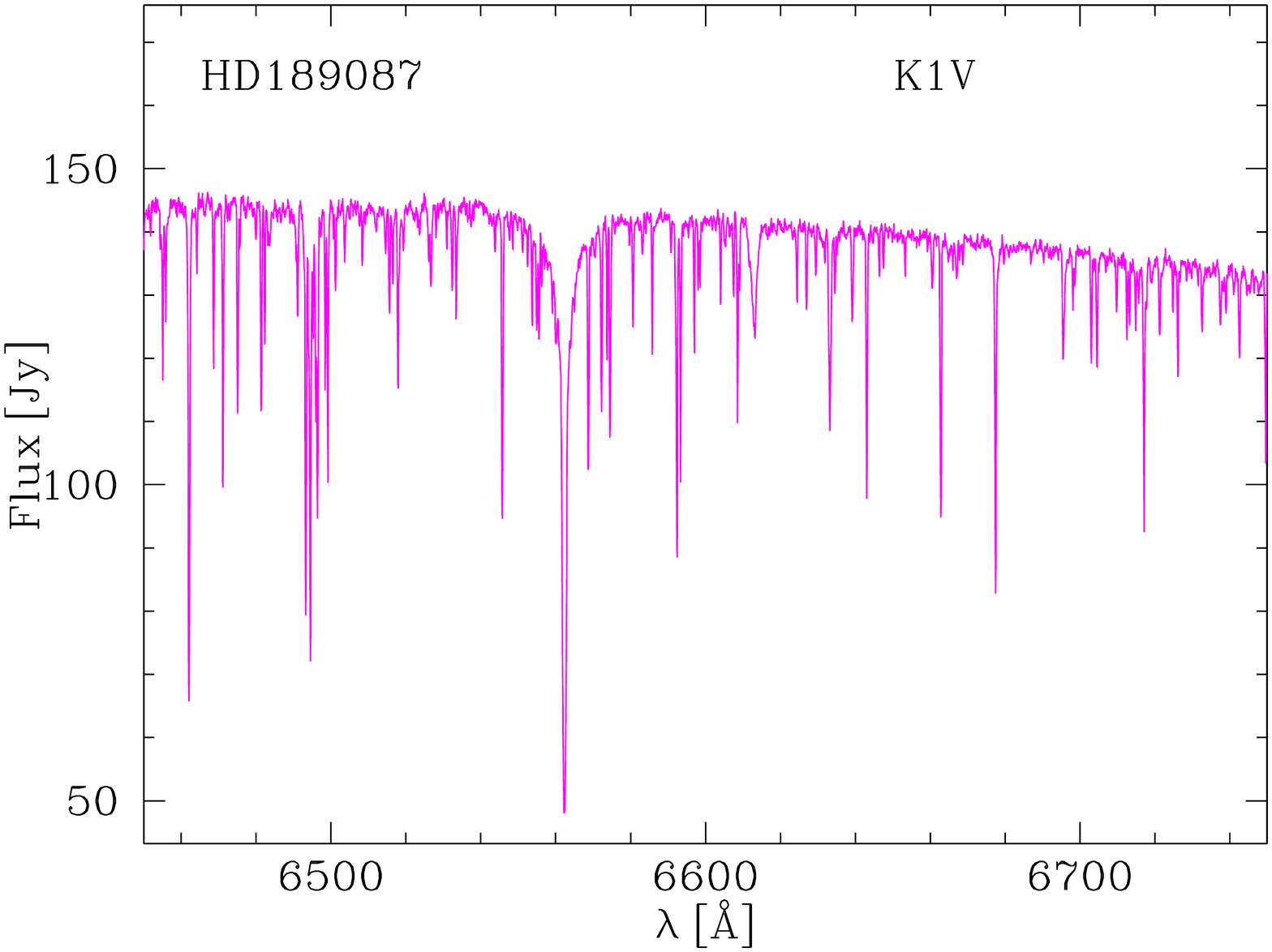}
\includegraphics[width=0.18\textwidth,angle=-90]{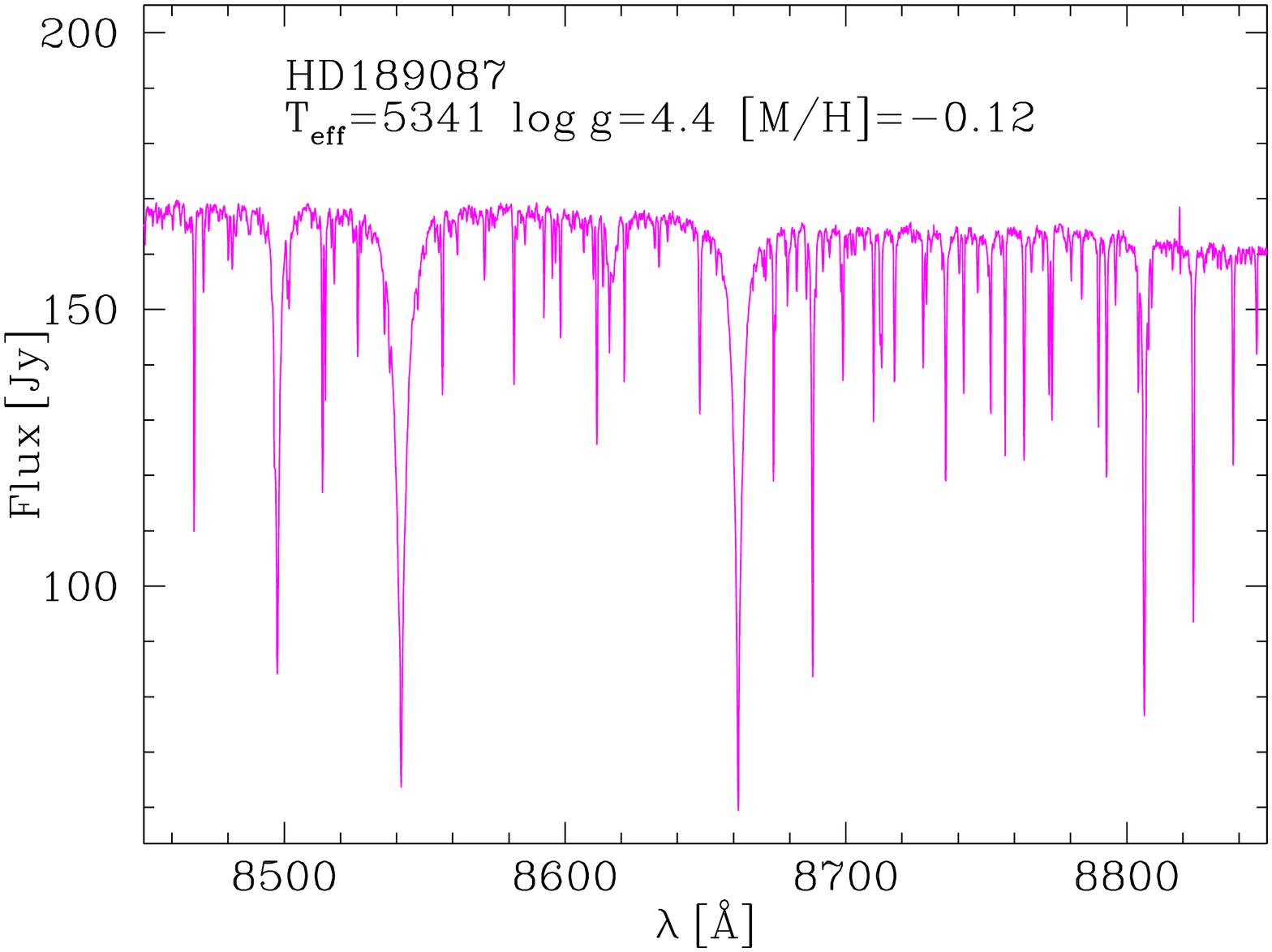}
\includegraphics[width=0.18\textwidth,angle=-90]{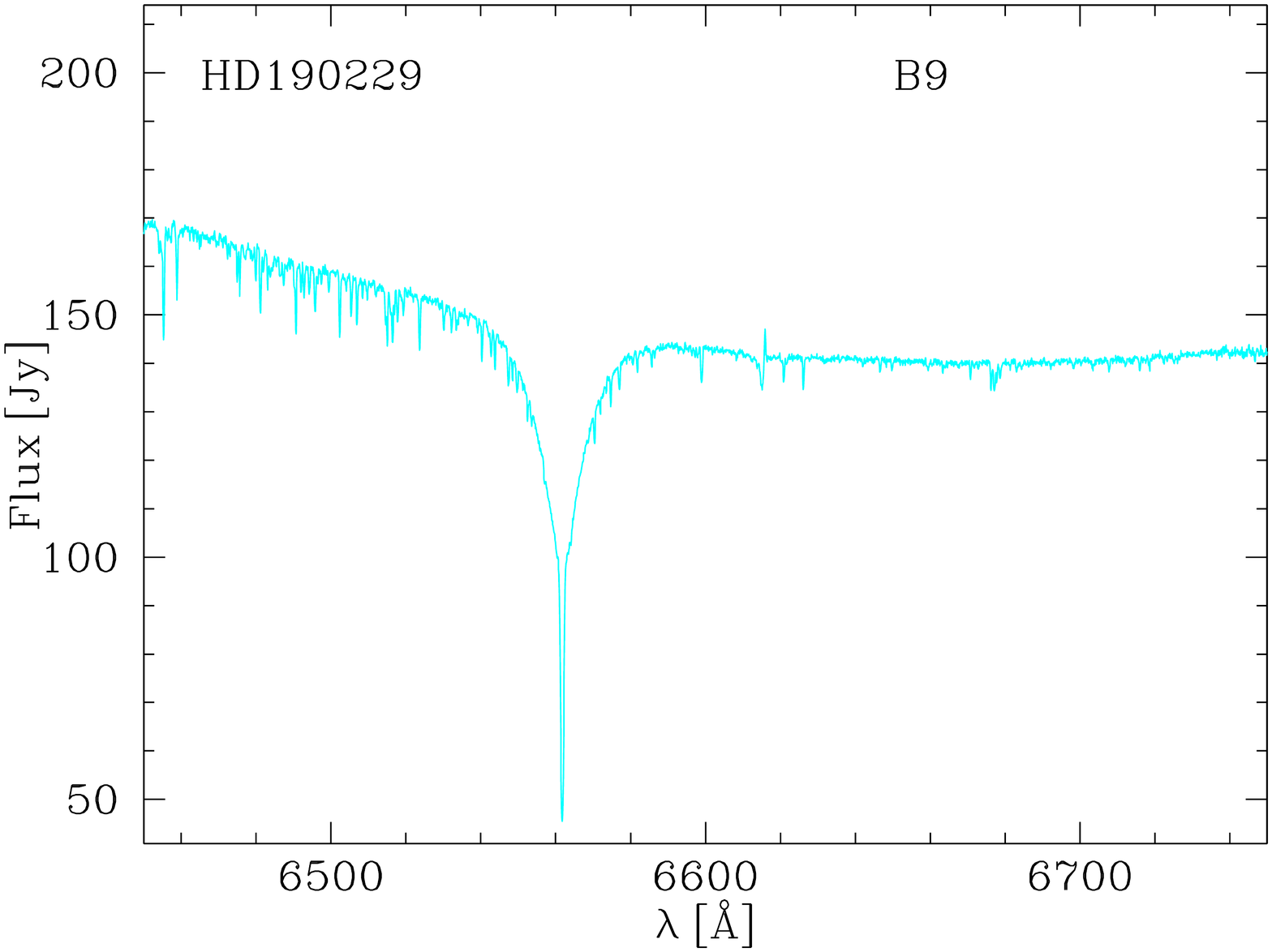}
\includegraphics[width=0.18\textwidth,angle=-90]{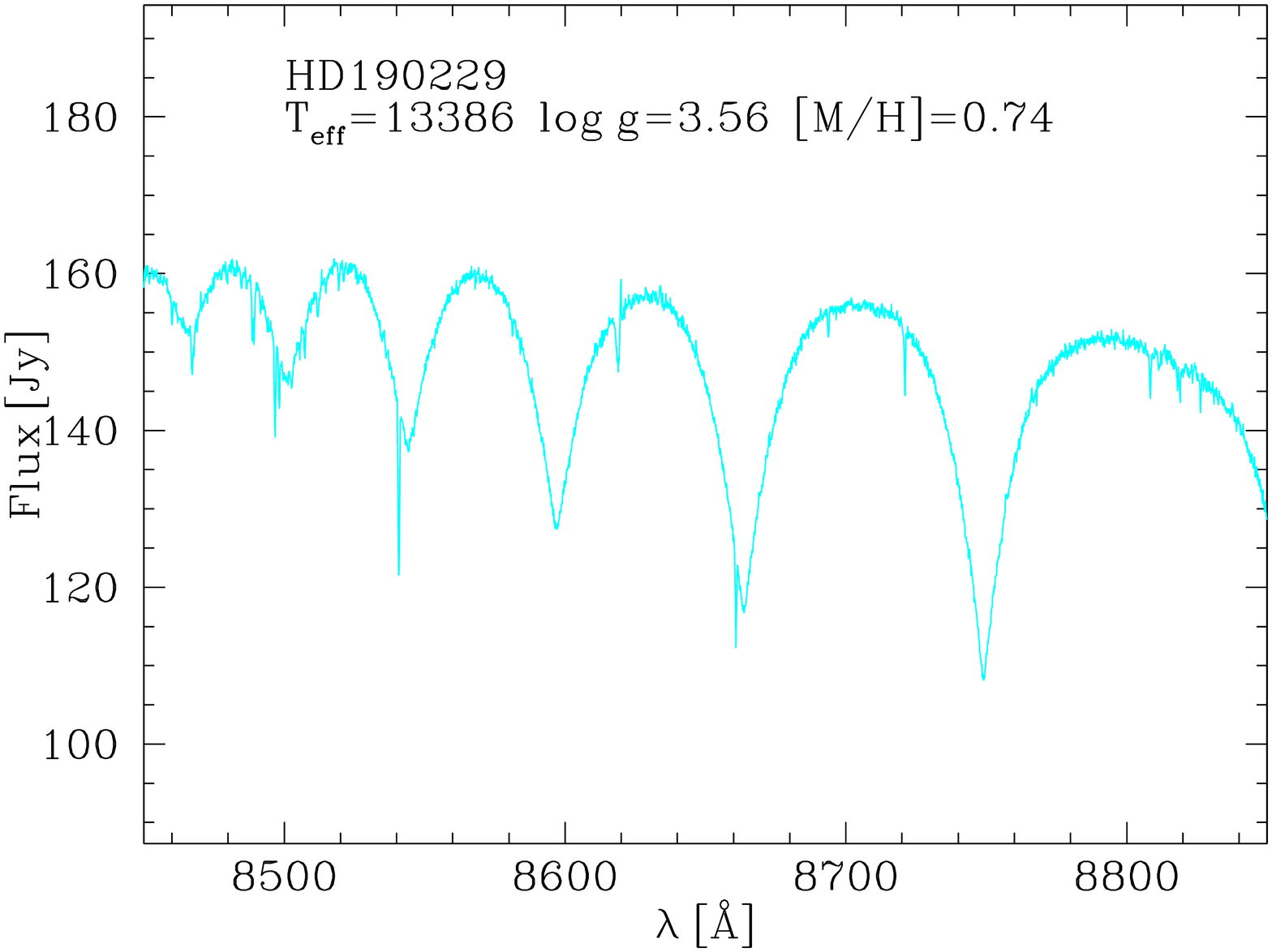}
\includegraphics[width=0.18\textwidth,angle=-90]{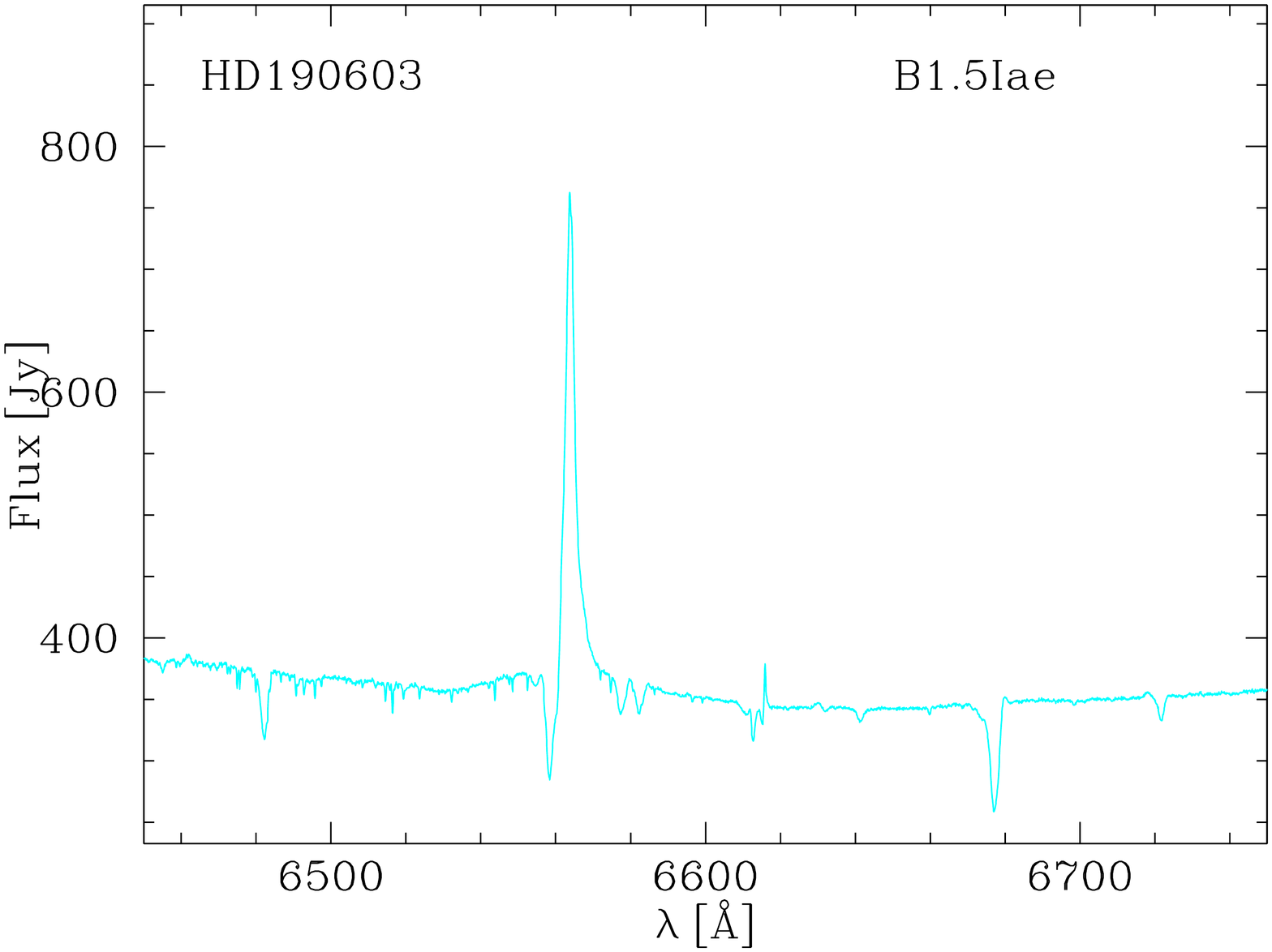}
\includegraphics[width=0.18\textwidth,angle=-90]{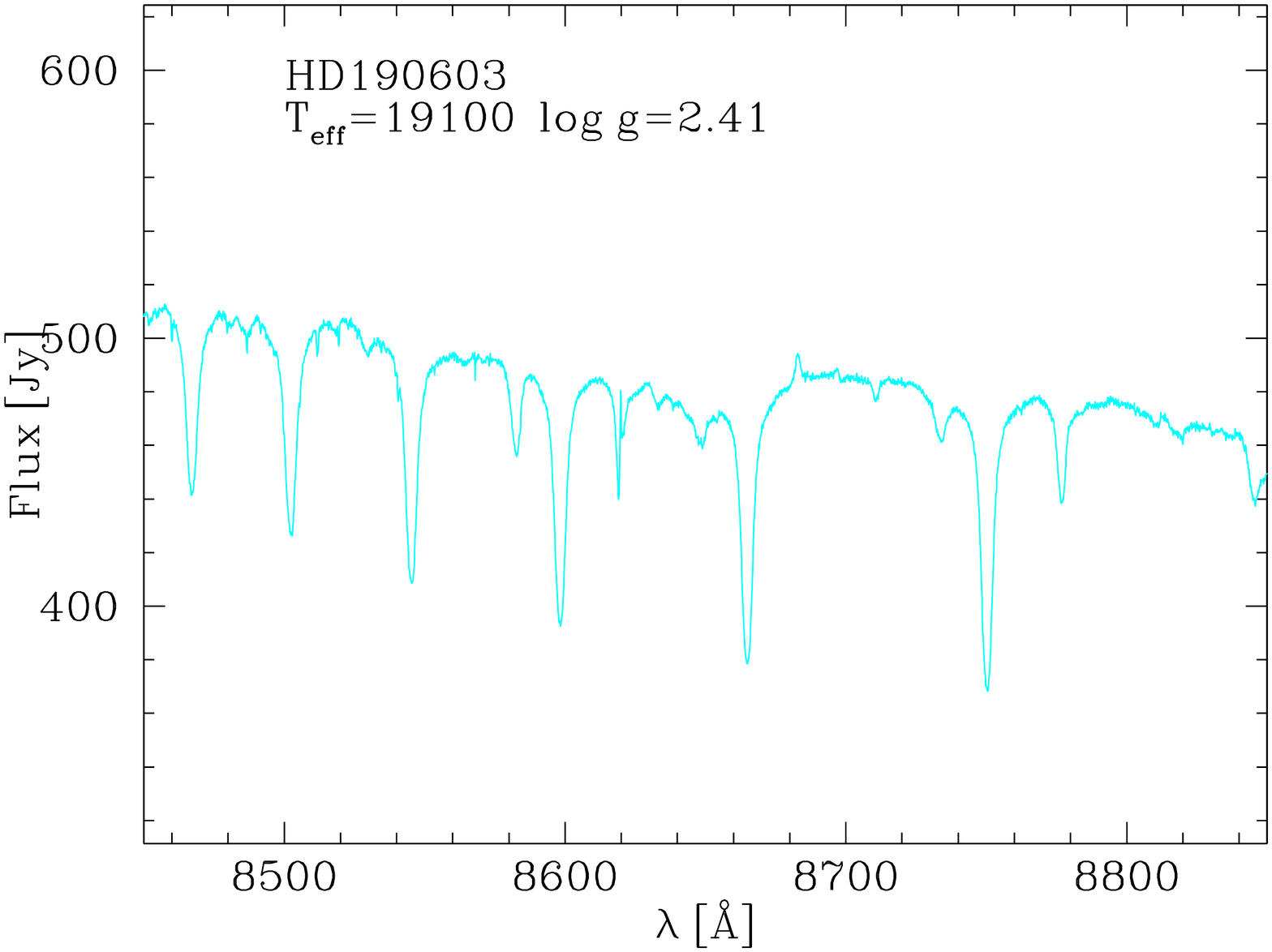}
\includegraphics[width=0.18\textwidth,angle=-90]{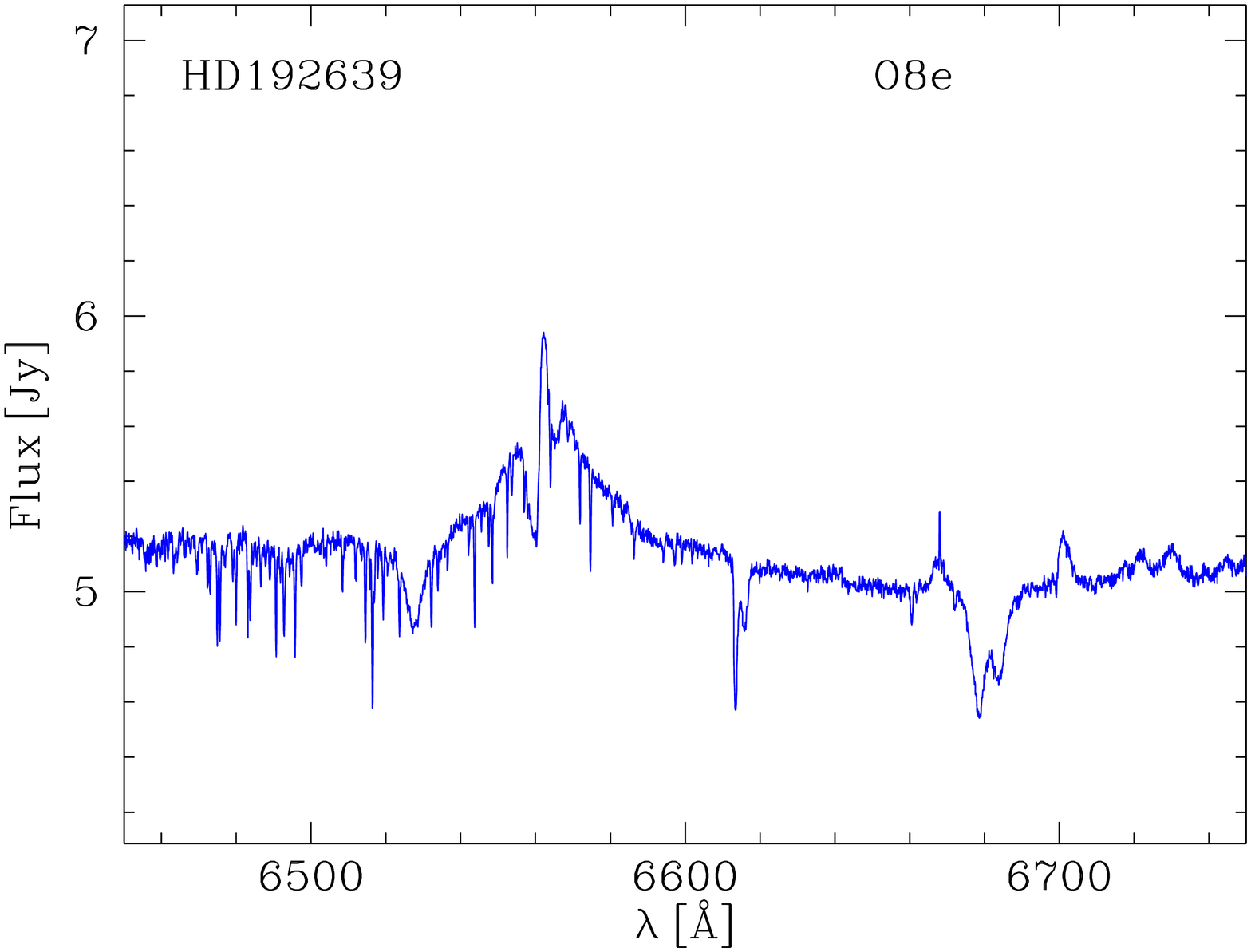}
\includegraphics[width=0.18\textwidth,angle=-90]{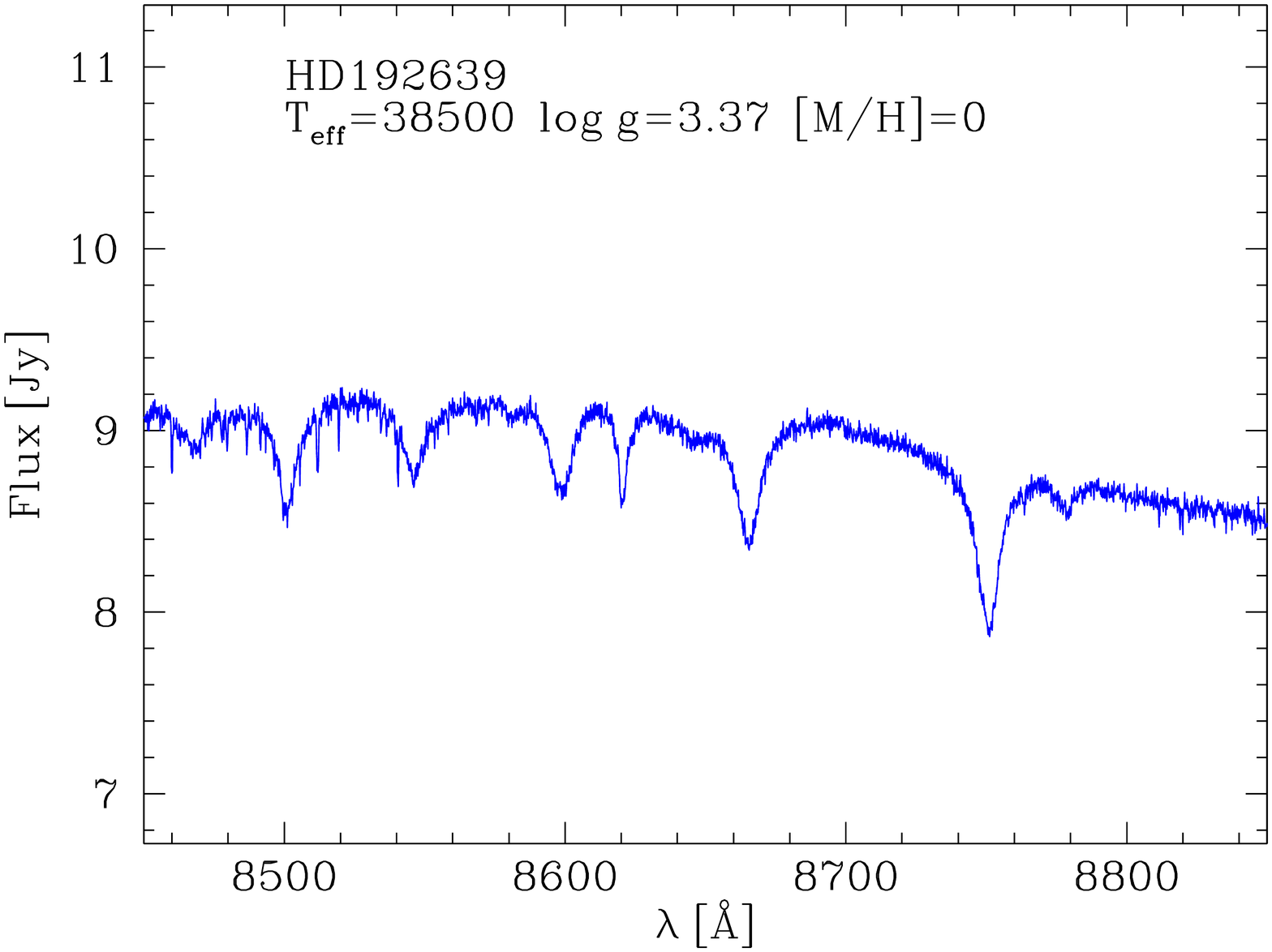}
\includegraphics[width=0.18\textwidth,angle=-90]{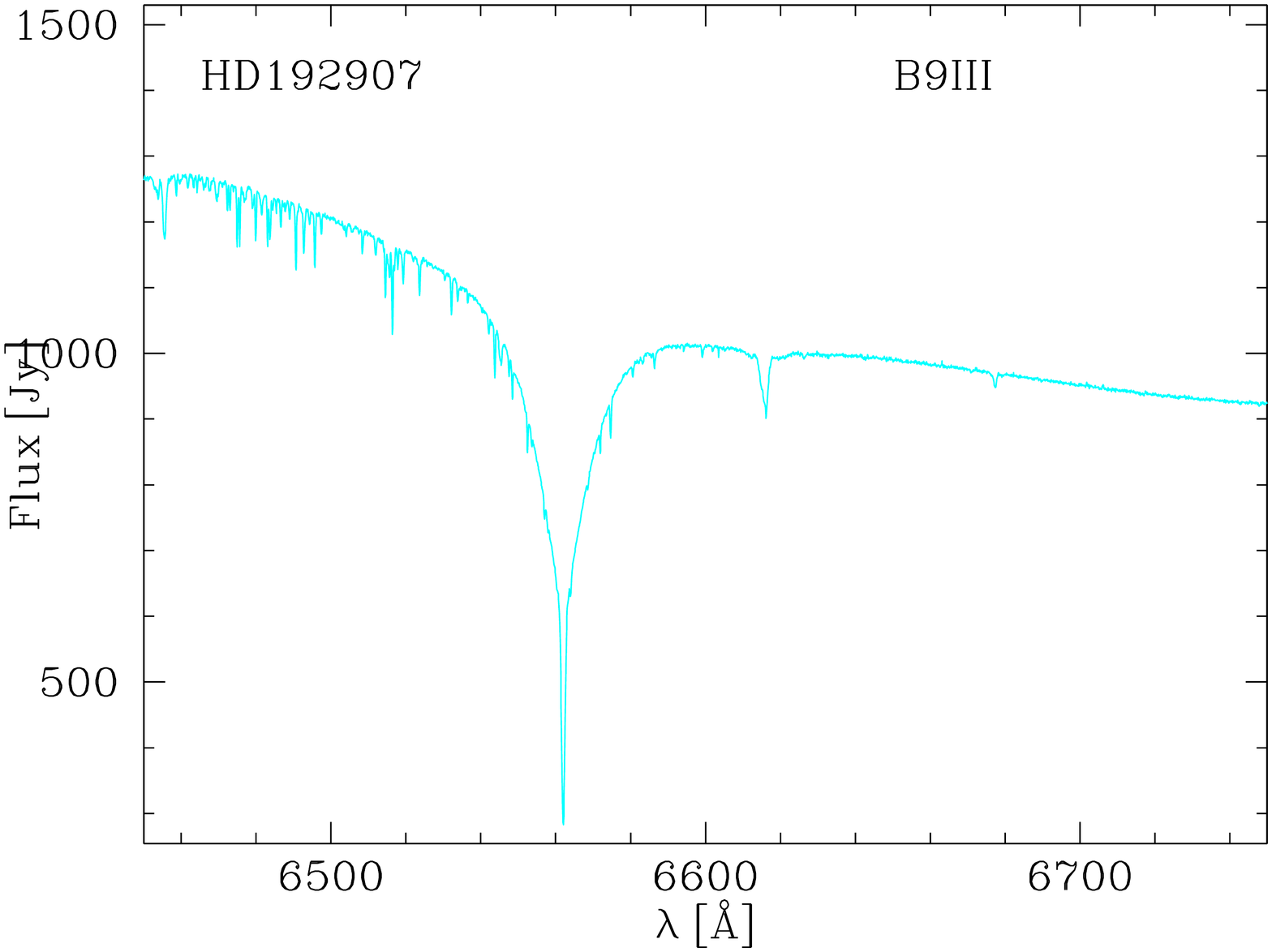}
\includegraphics[width=0.18\textwidth,angle=-90]{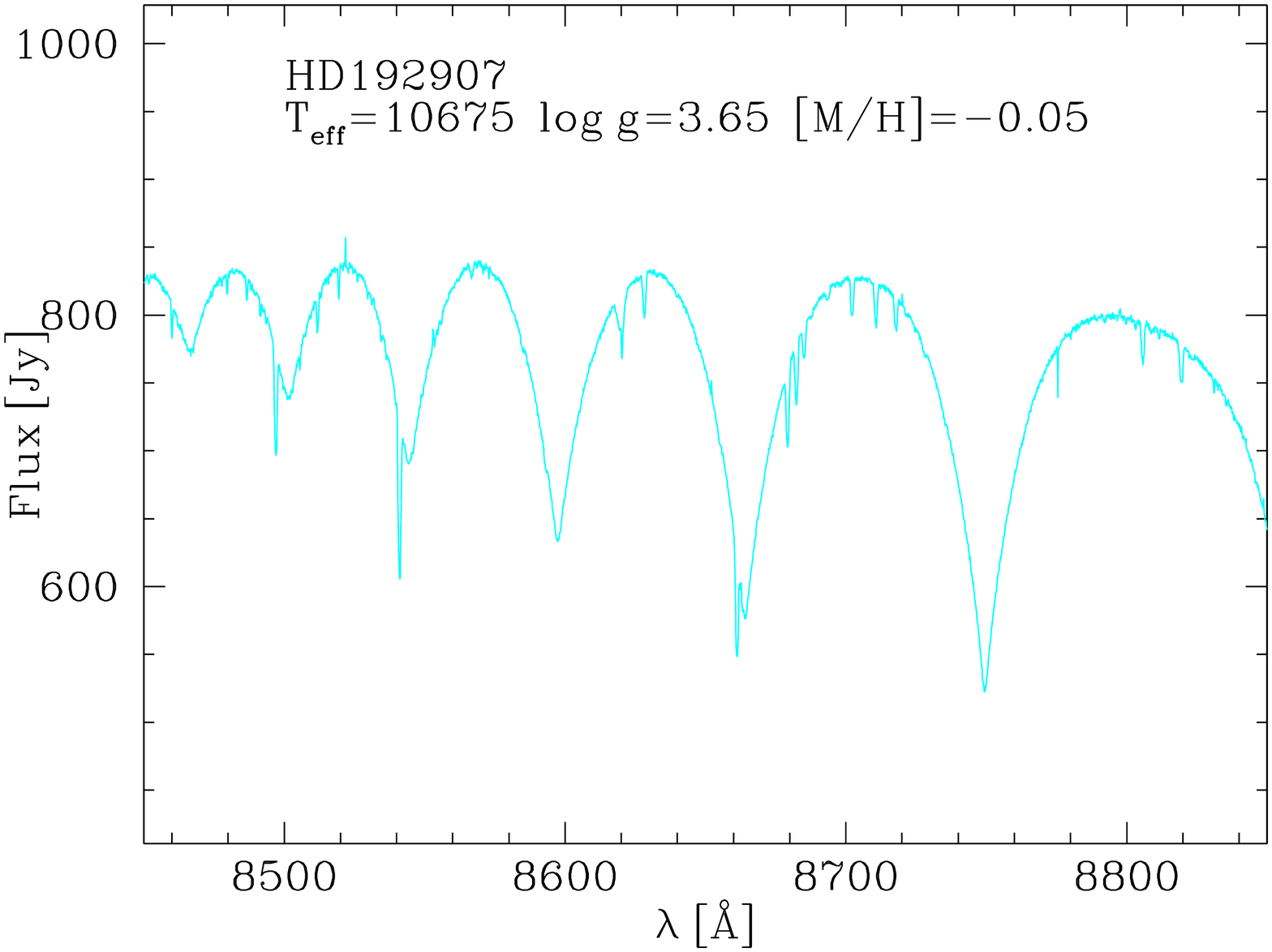}
\includegraphics[width=0.18\textwidth,angle=-90]{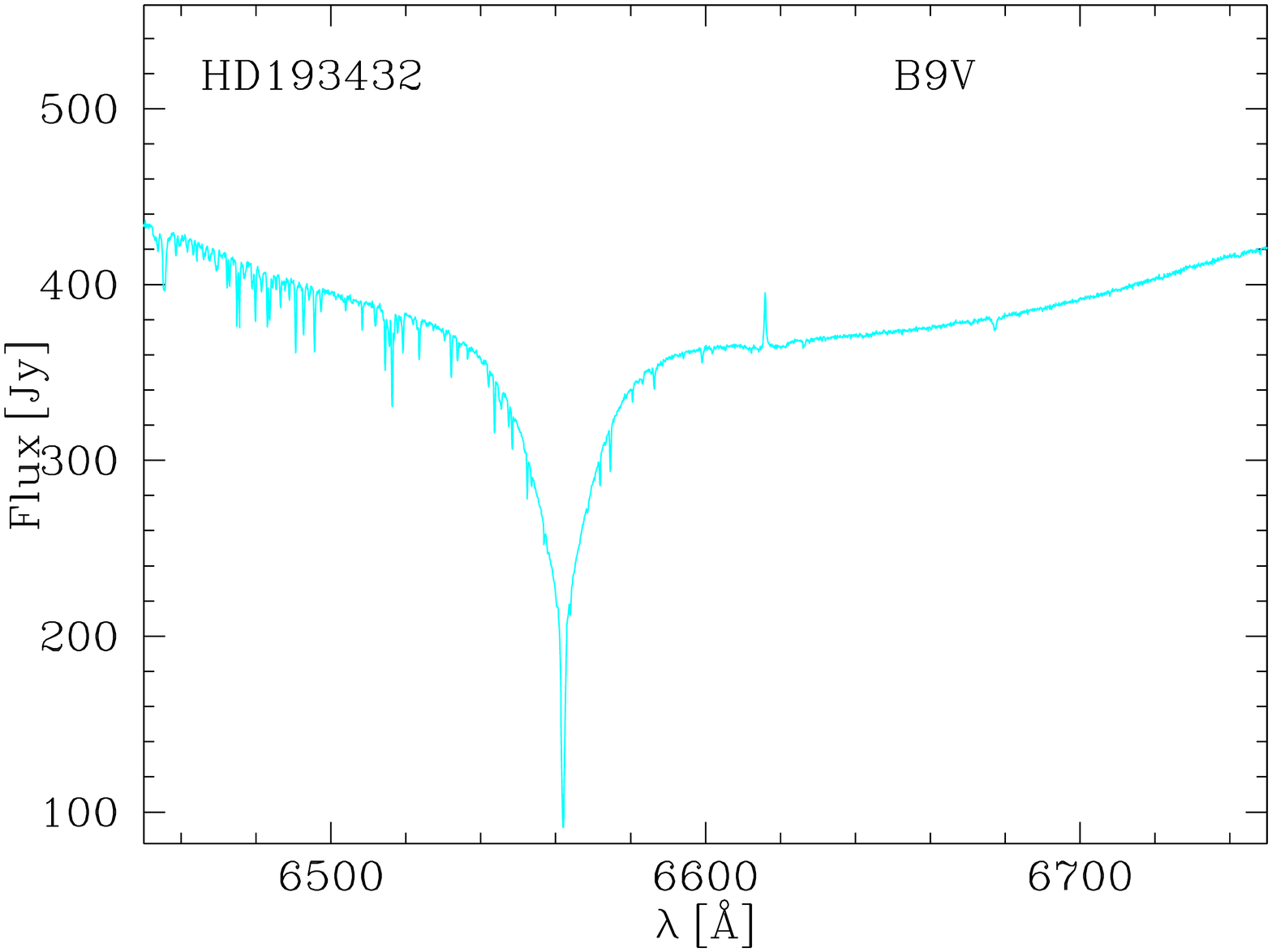}
\includegraphics[width=0.18\textwidth,angle=-90]{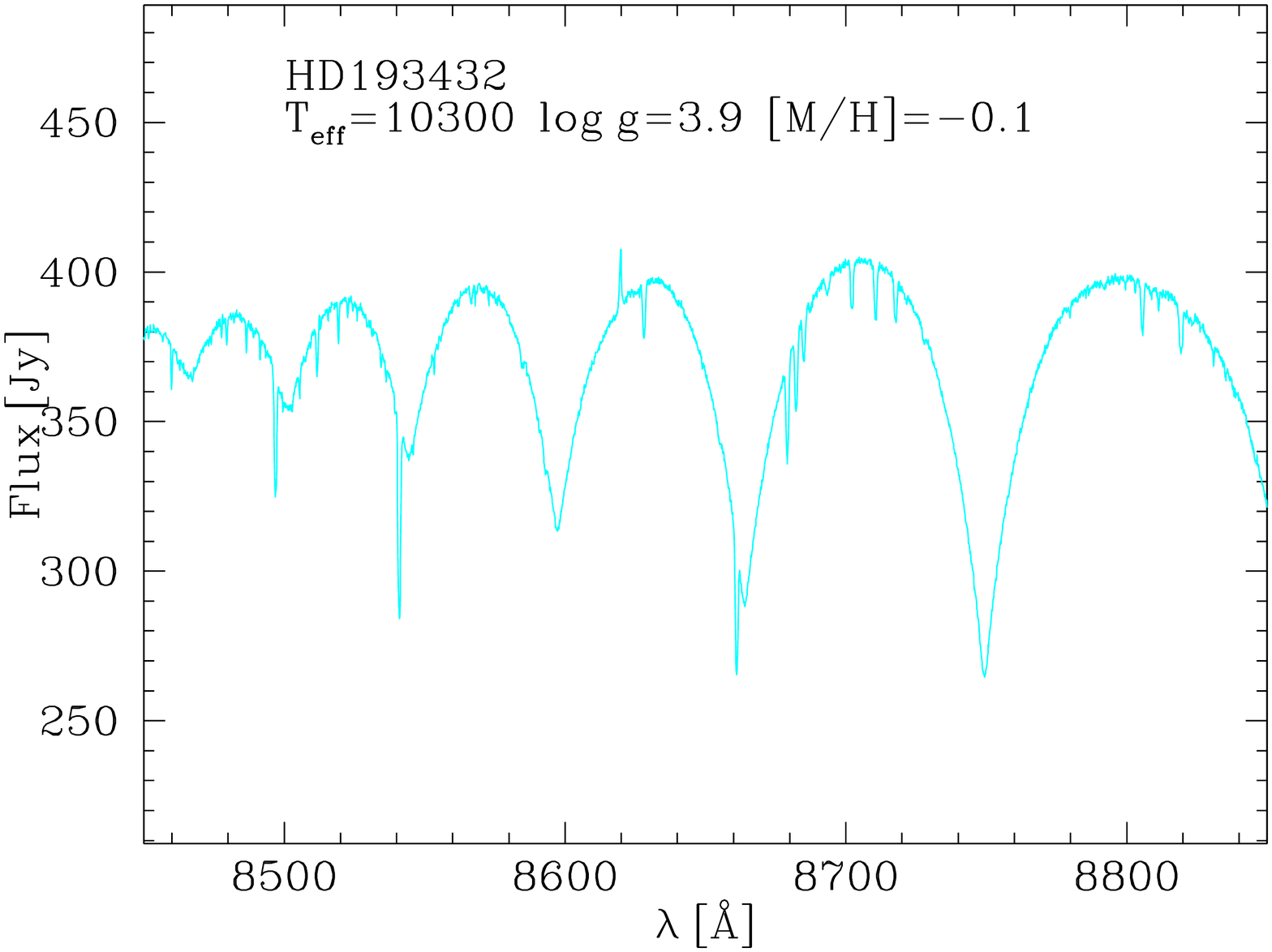}

\contcaption{26. Stars shown in this page are:HD176437, HD180554, HD183144, HD185936, HD187123, HD187879, HD188001, HD188209, HD189087, HD190229, HD190603, HD192639, HD192907 and HD193432.}
\end{figure*}

\begin{figure*}
\includegraphics[width=0.18\textwidth,angle=-90]{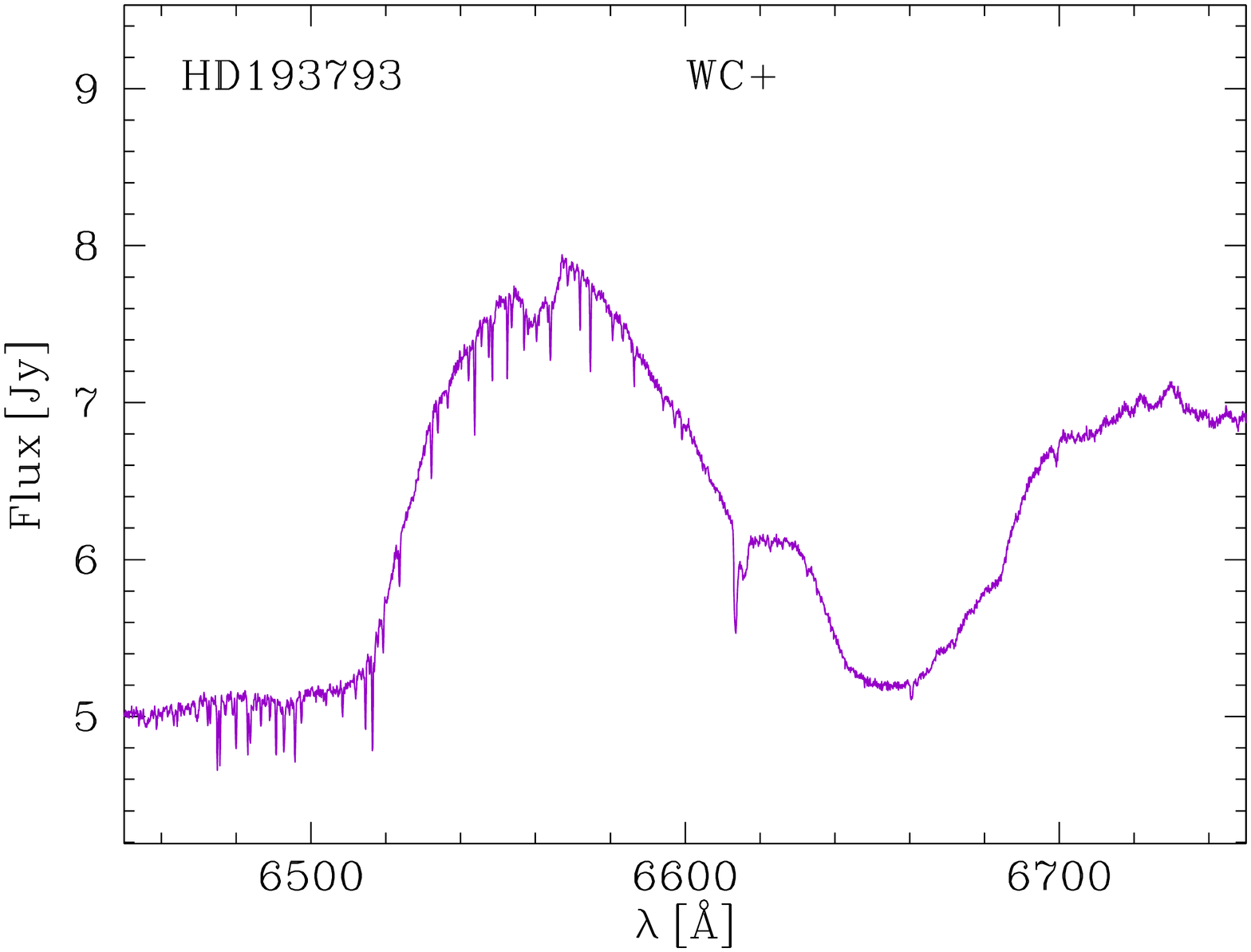}
\includegraphics[width=0.18\textwidth,angle=-90]{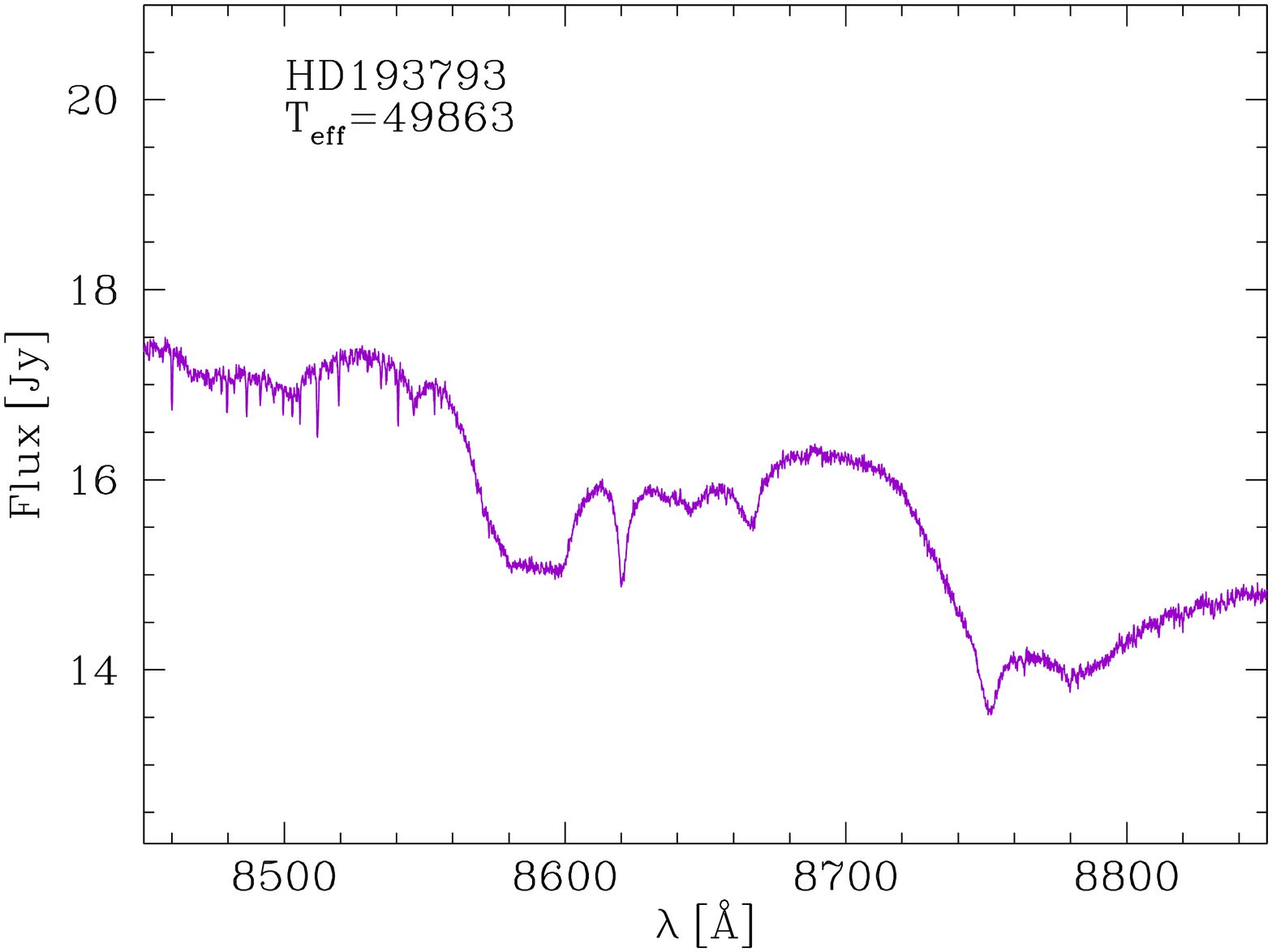}
\includegraphics[width=0.18\textwidth,angle=-90]{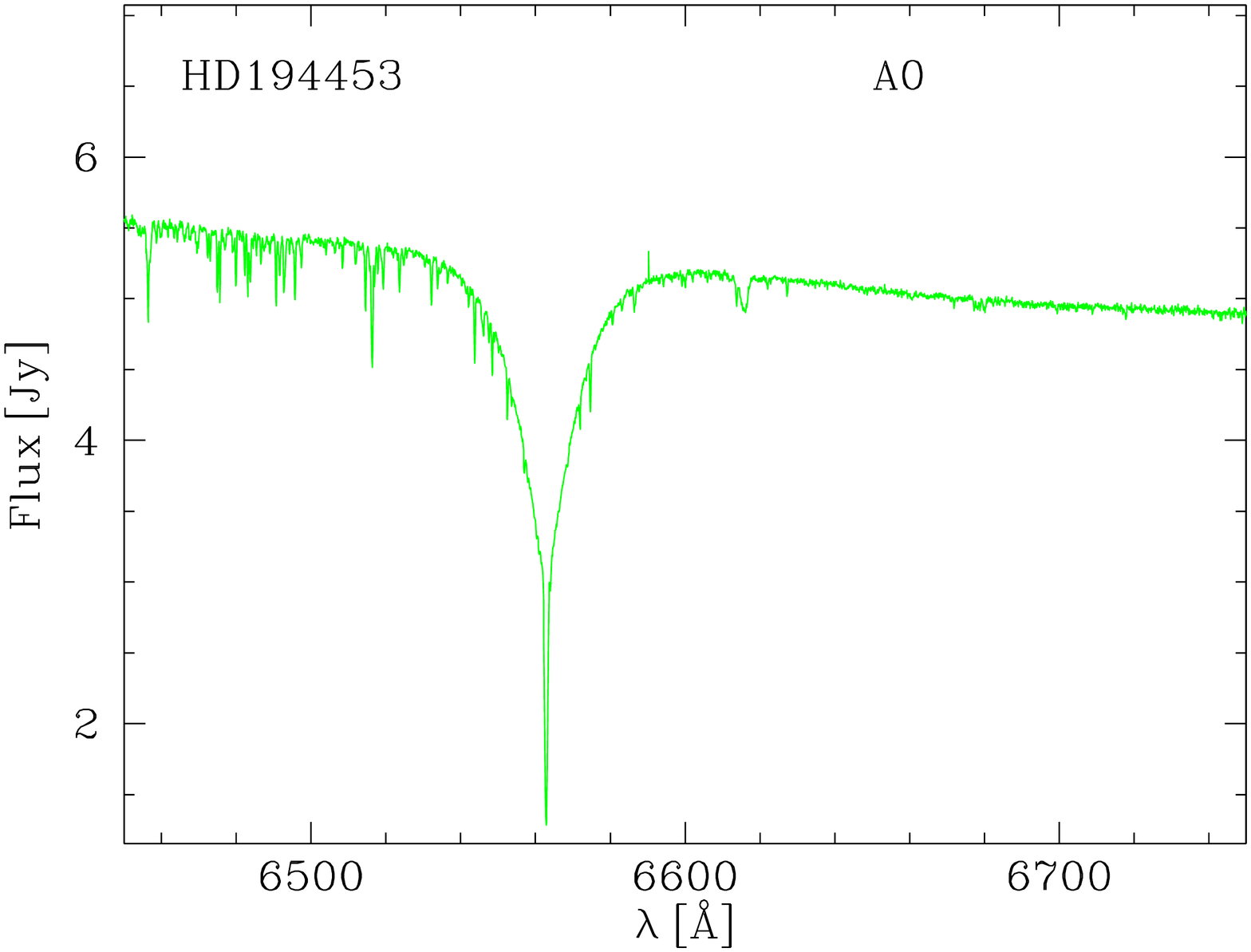}
\includegraphics[width=0.18\textwidth,angle=-90]{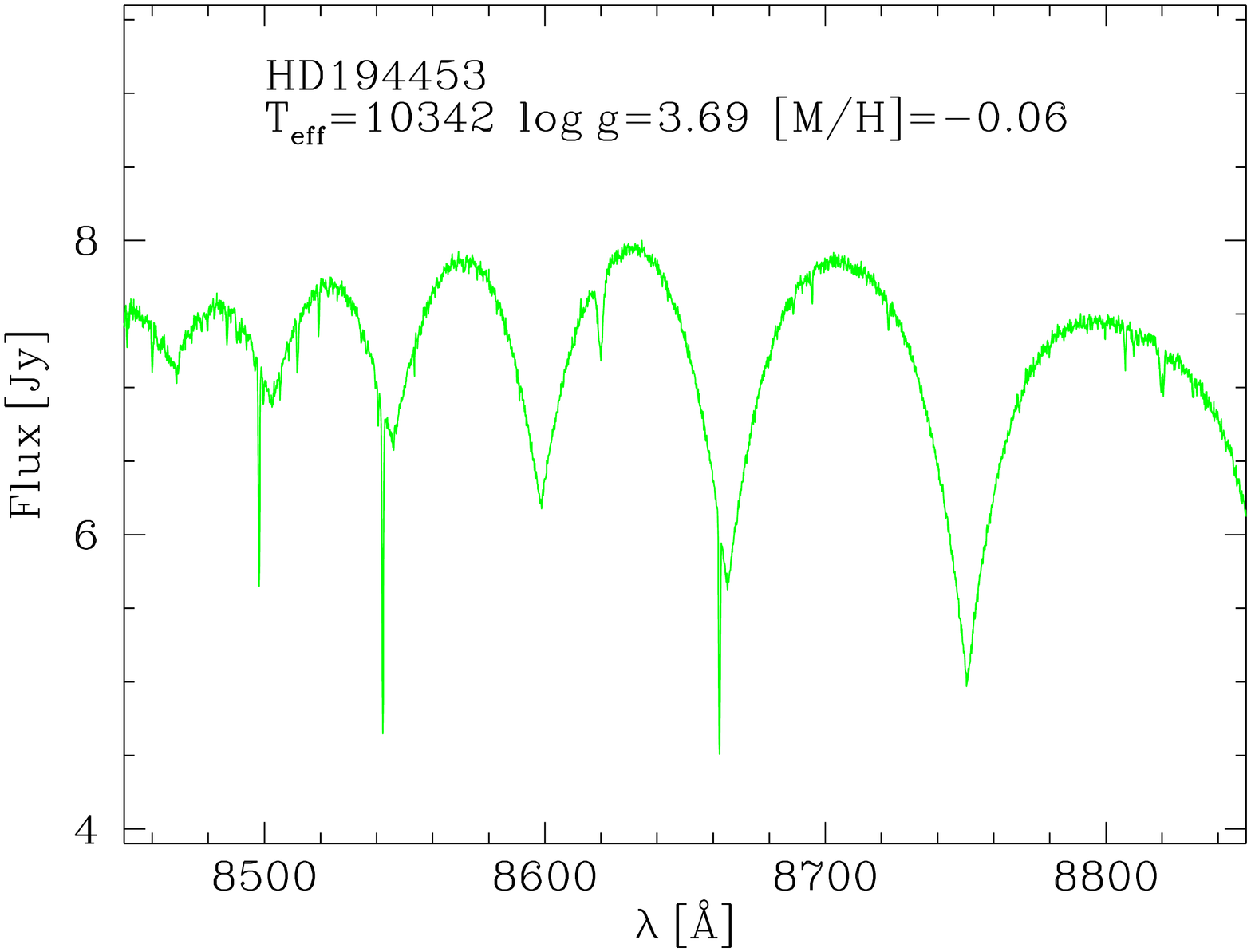}
\includegraphics[width=0.18\textwidth,angle=-90]{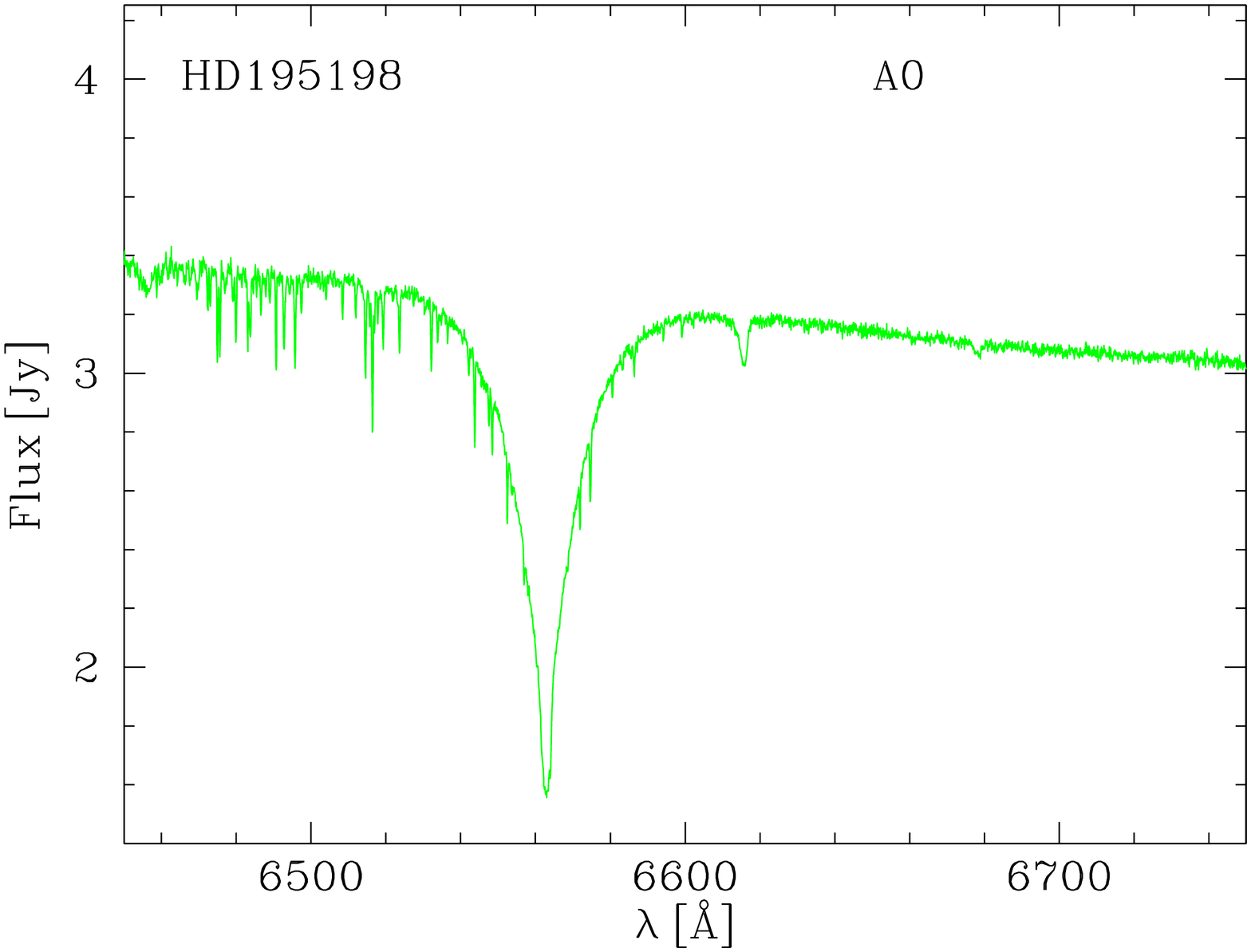}
\includegraphics[width=0.18\textwidth,angle=-90]{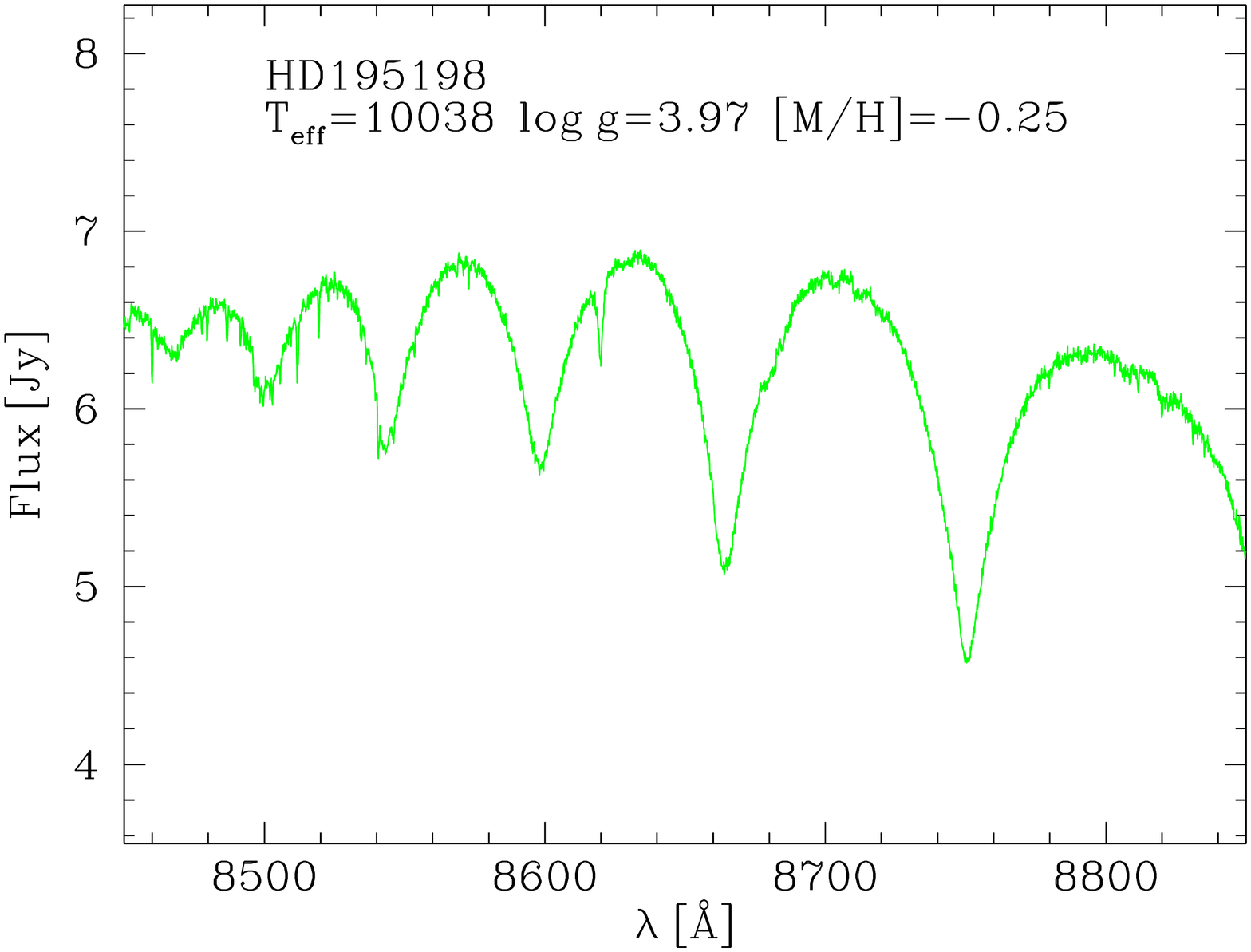}
\includegraphics[width=0.18\textwidth,angle=-90]{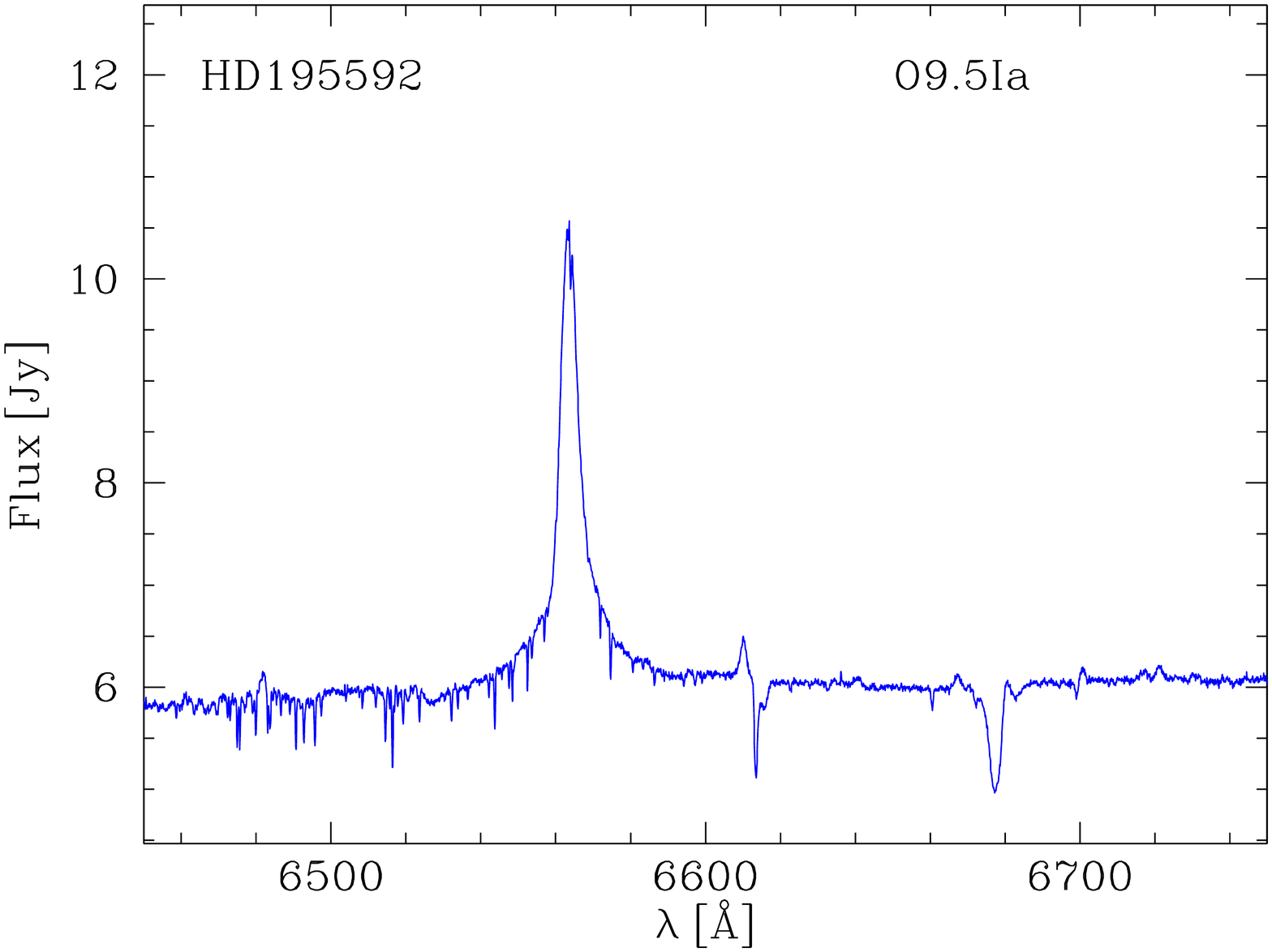}
\includegraphics[width=0.18\textwidth,angle=-90]{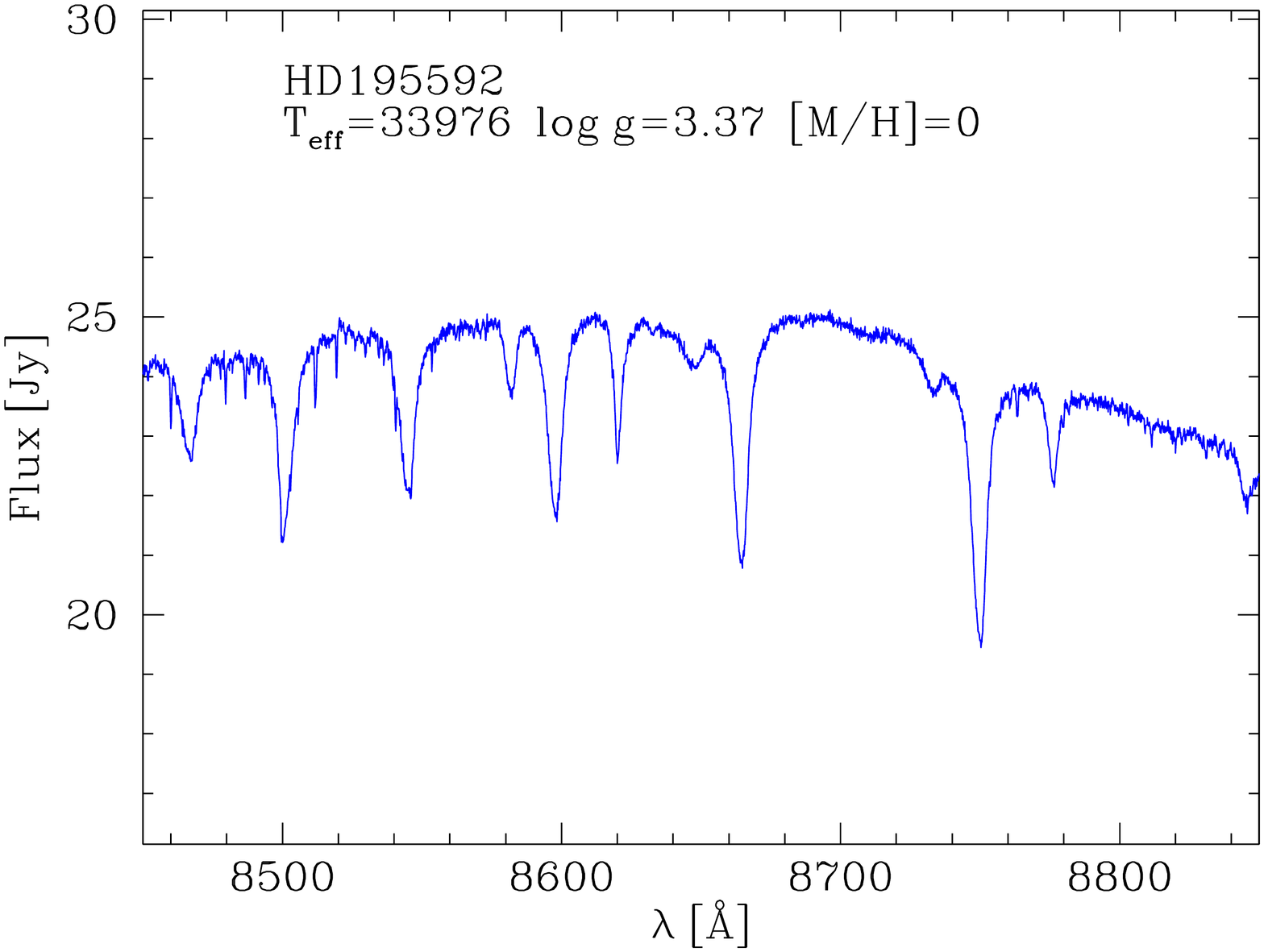}
\includegraphics[width=0.18\textwidth,angle=-90]{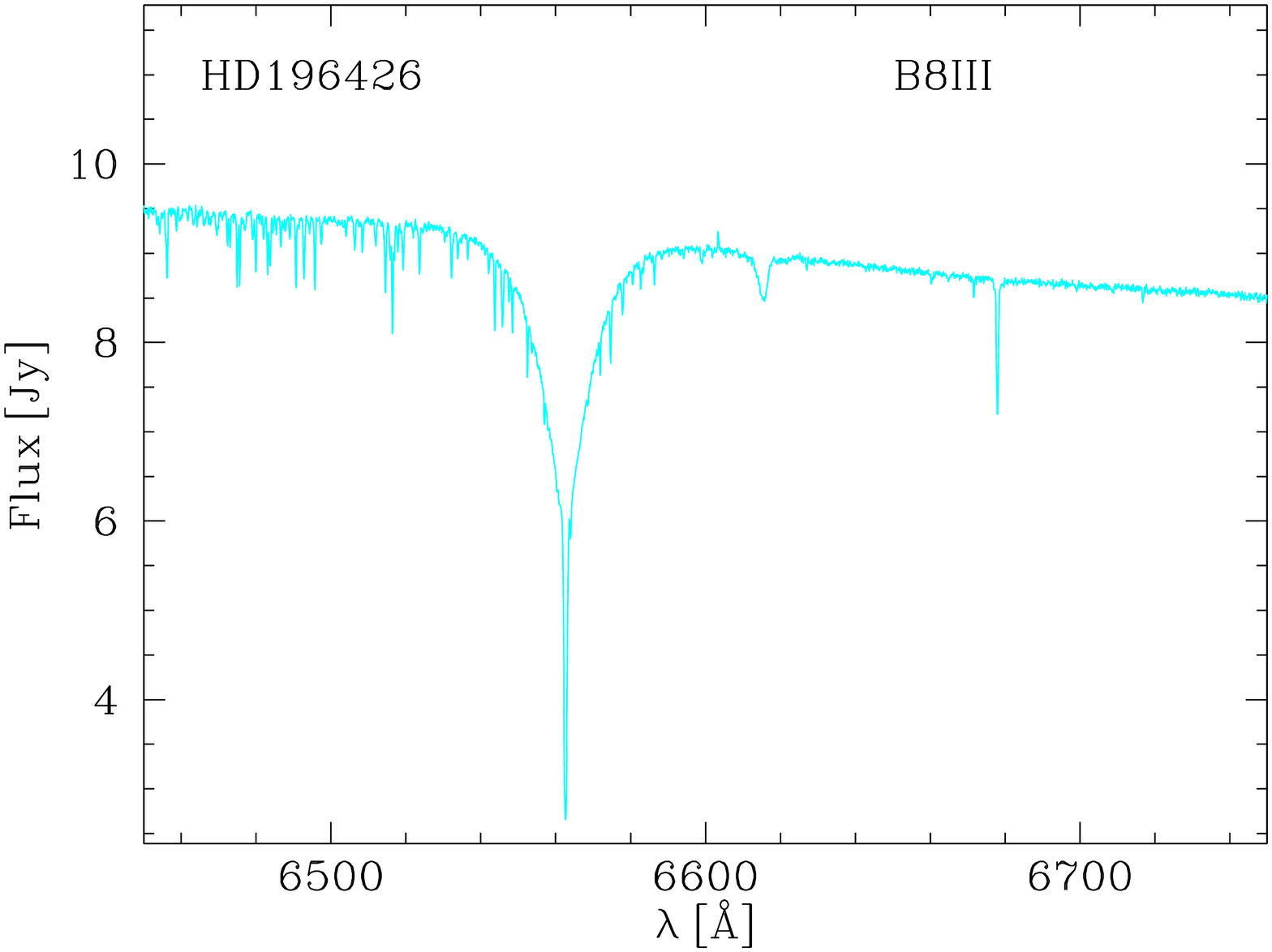}
\includegraphics[width=0.18\textwidth,angle=-90]{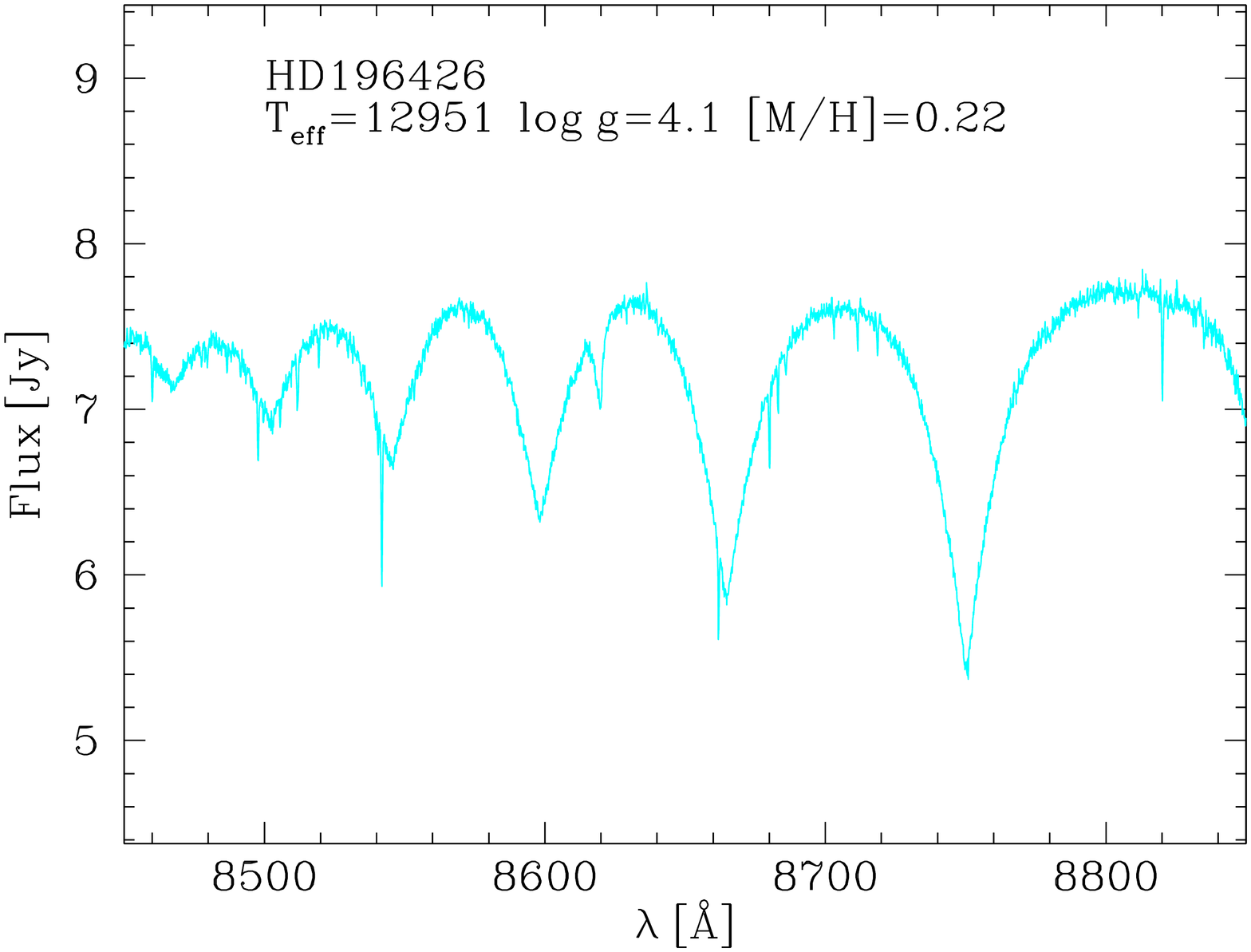}
\includegraphics[width=0.18\textwidth,angle=-90]{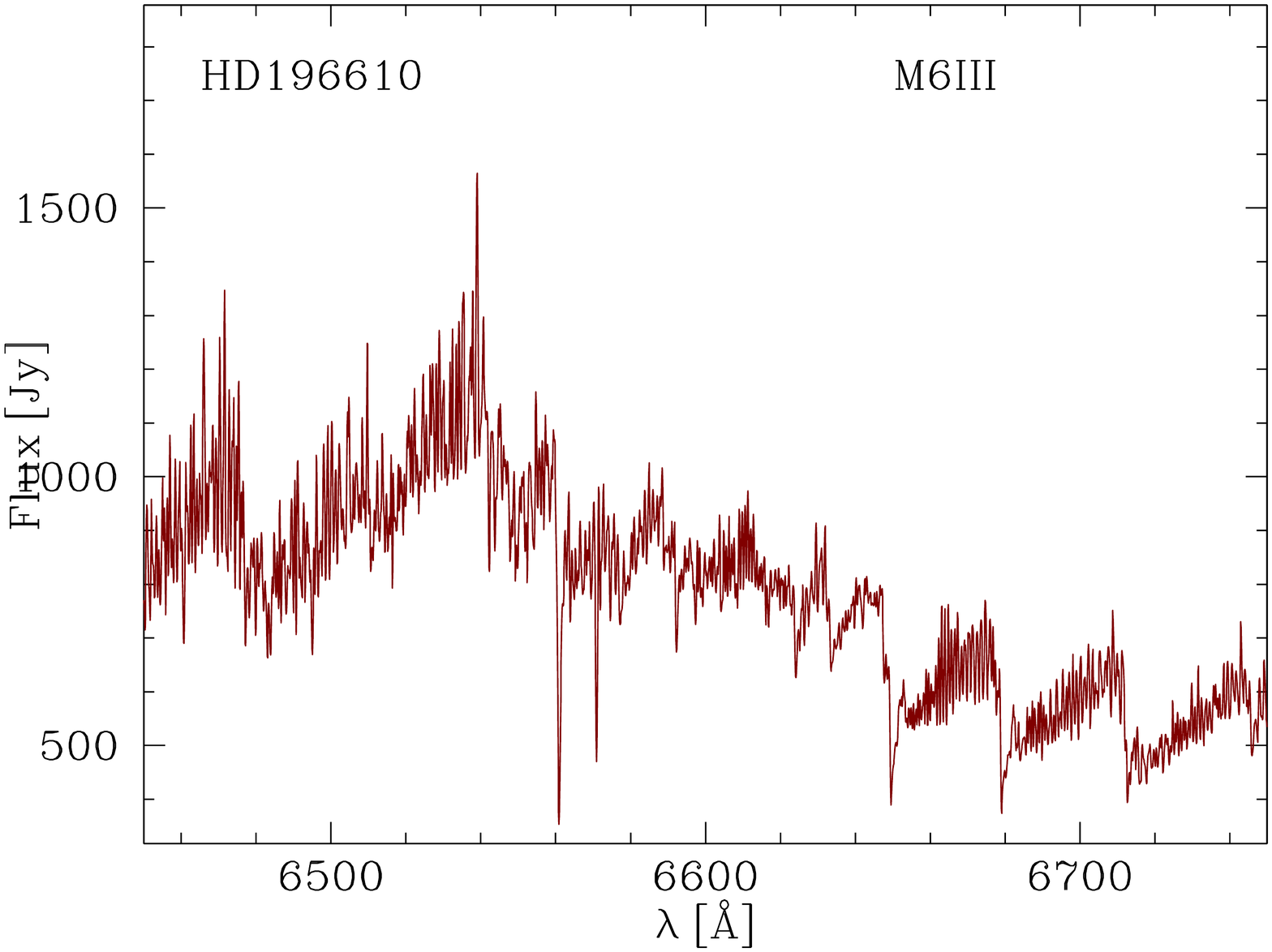}
\includegraphics[width=0.18\textwidth,angle=-90]{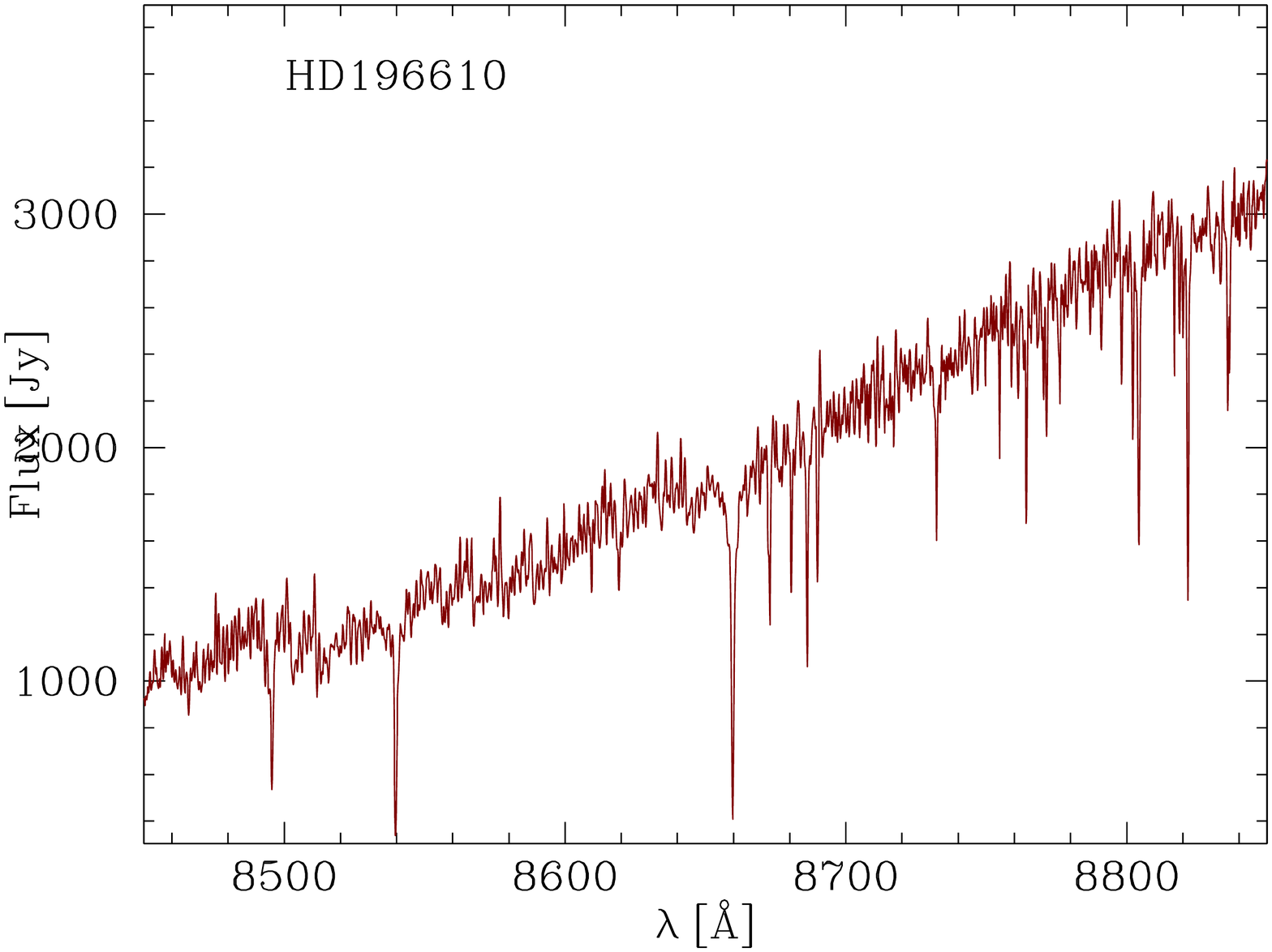}
\includegraphics[width=0.18\textwidth,angle=-90]{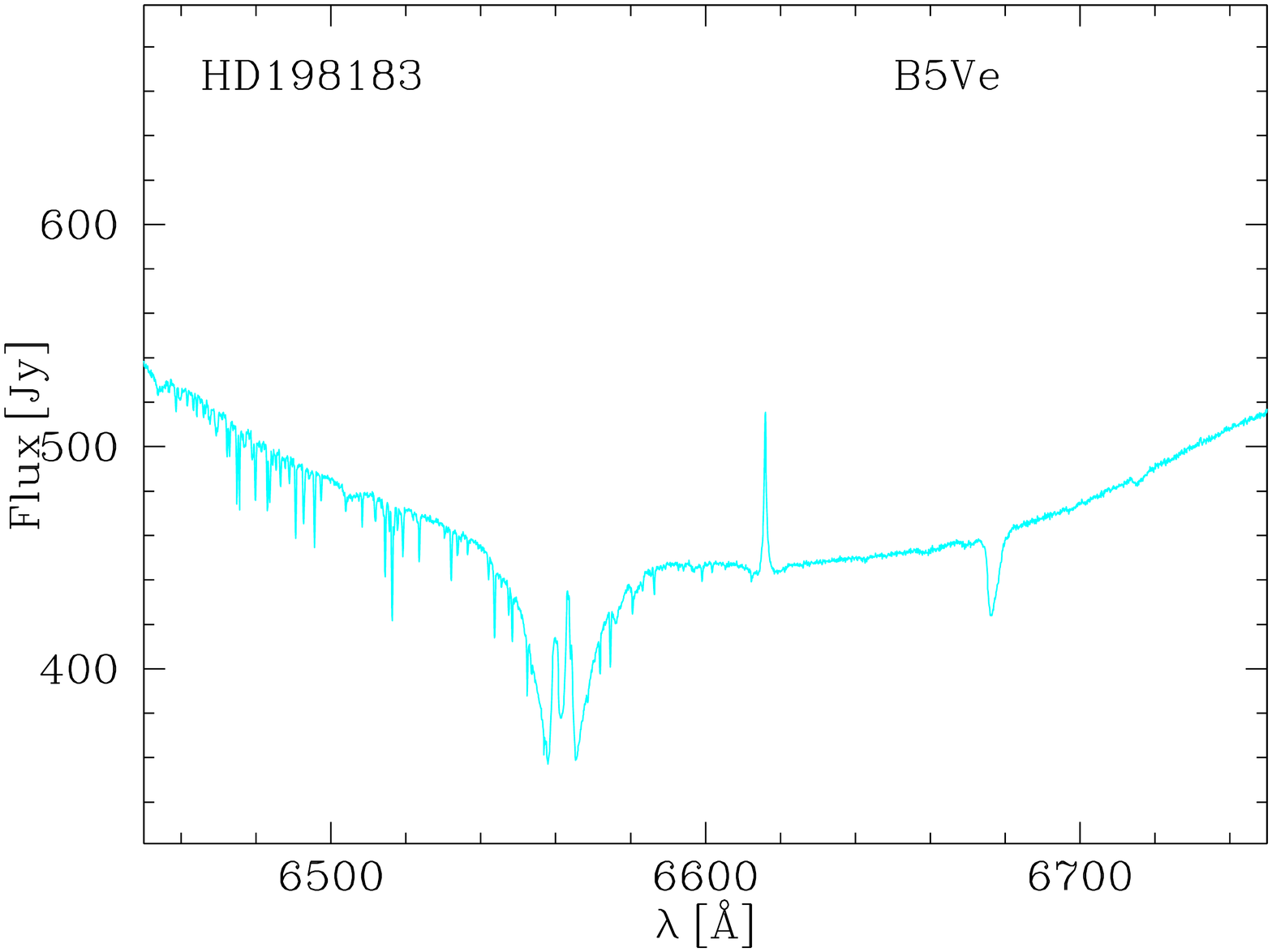}
\includegraphics[width=0.18\textwidth,angle=-90]{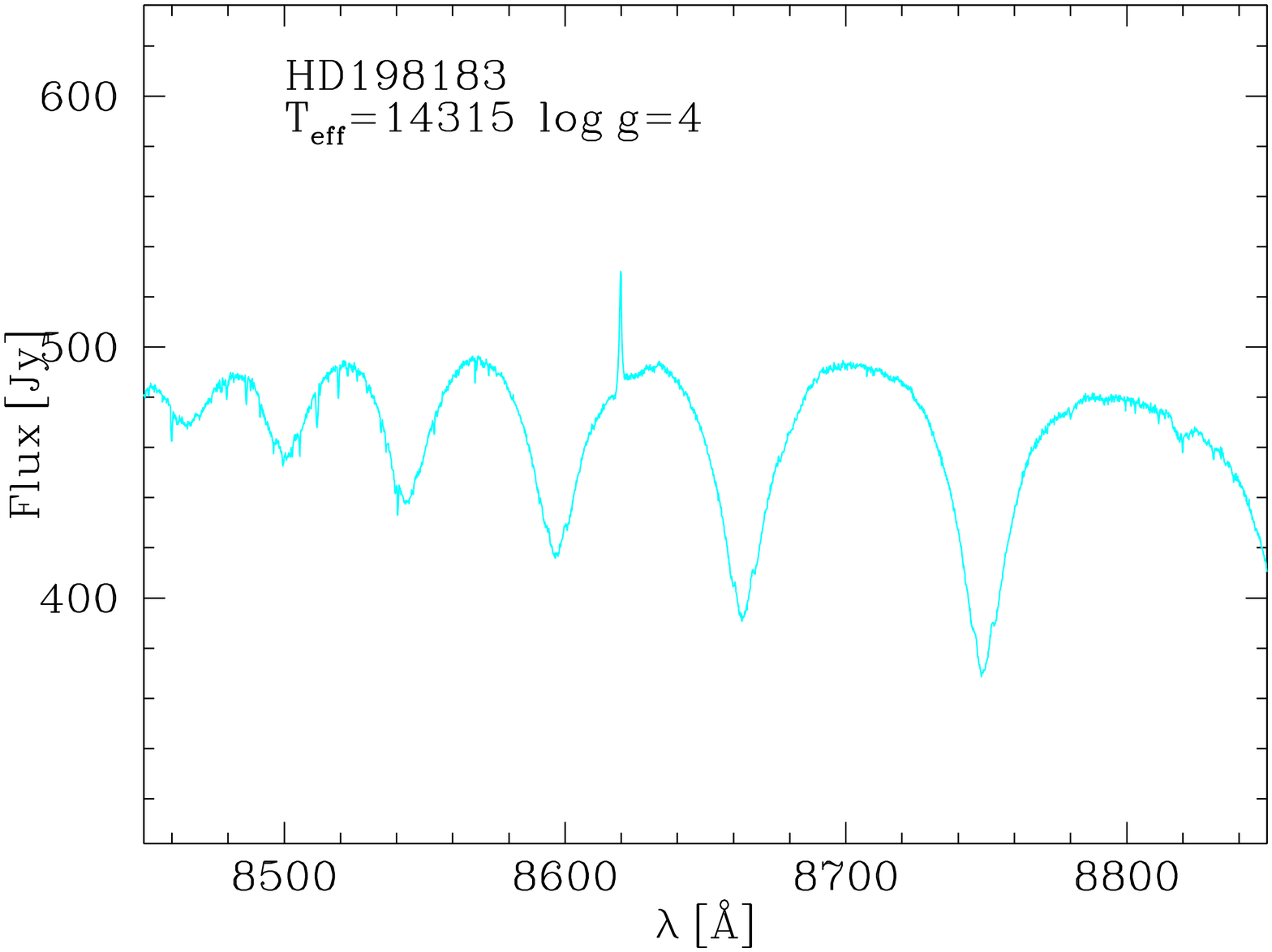}
\includegraphics[width=0.18\textwidth,angle=-90]{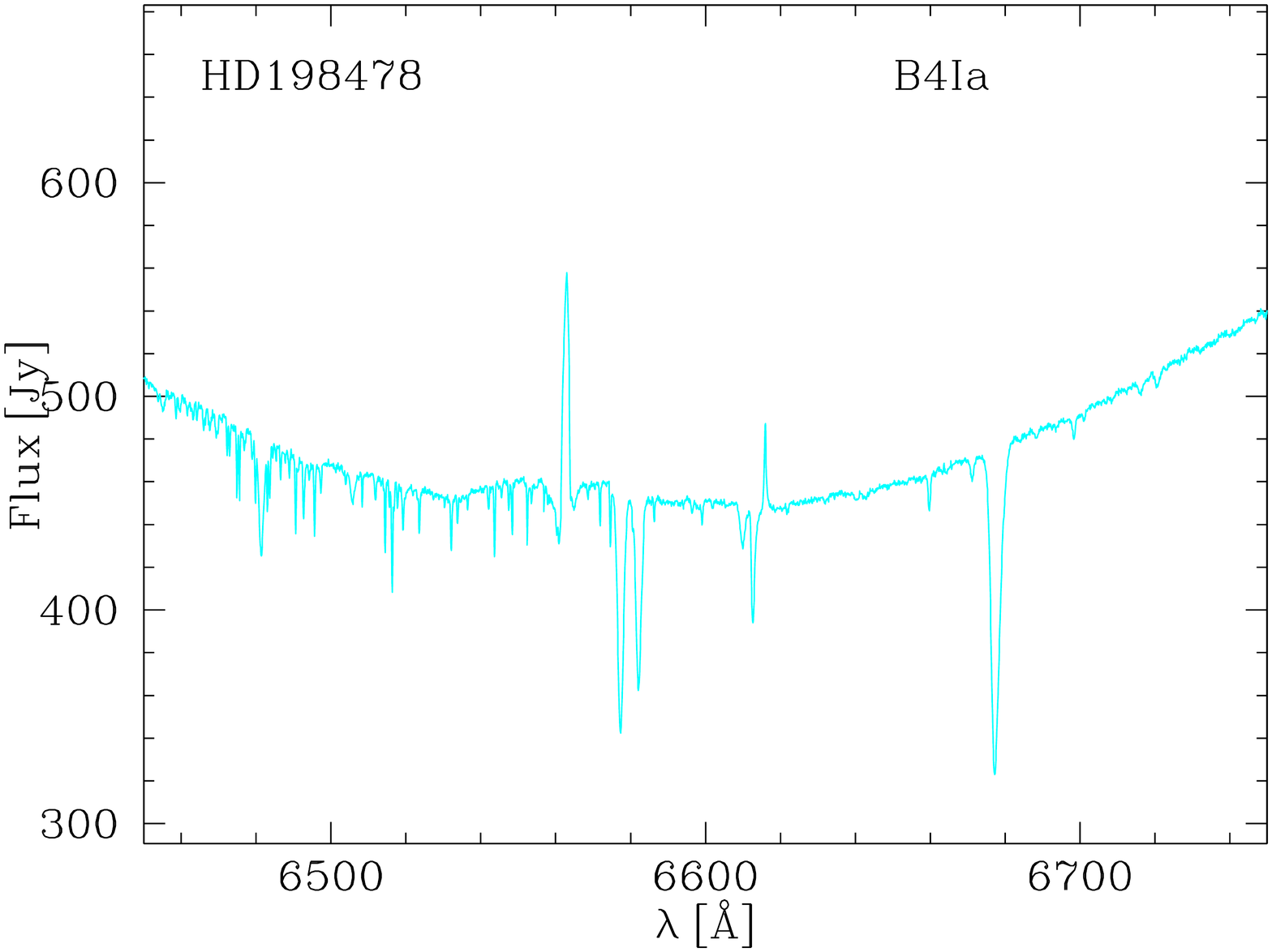}
\includegraphics[width=0.18\textwidth,angle=-90]{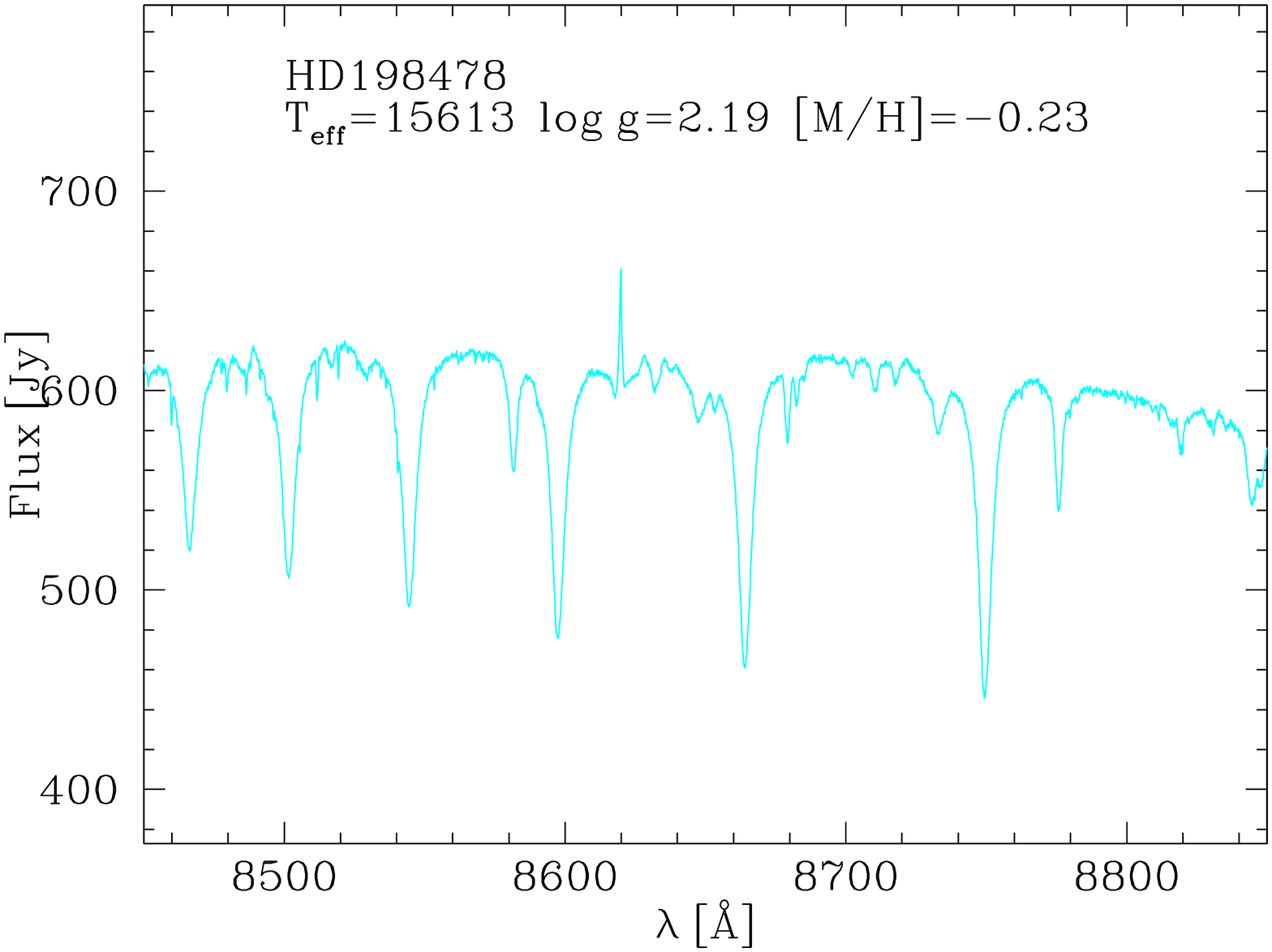}
\includegraphics[width=0.18\textwidth,angle=-90]{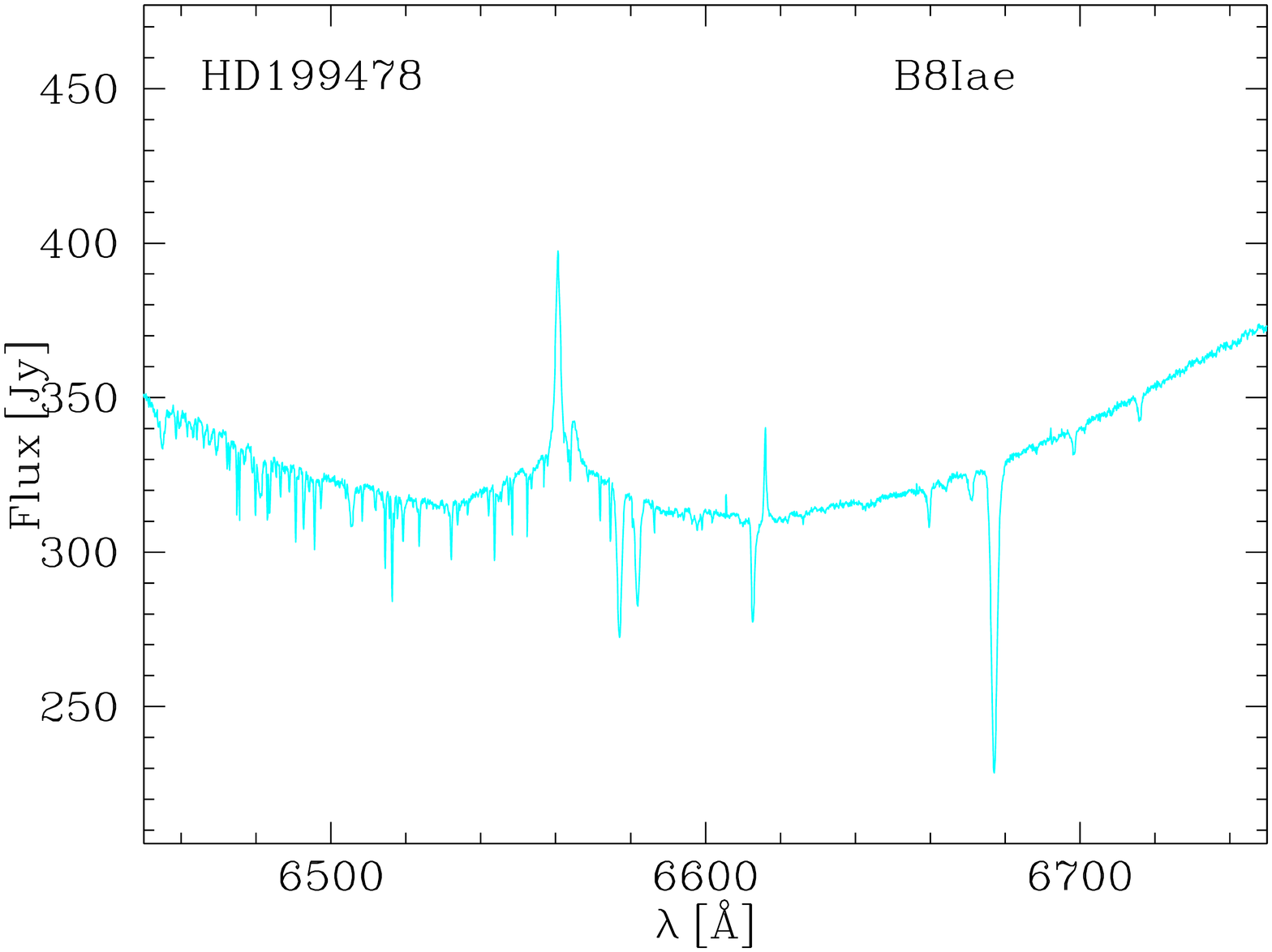}
\includegraphics[width=0.18\textwidth,angle=-90]{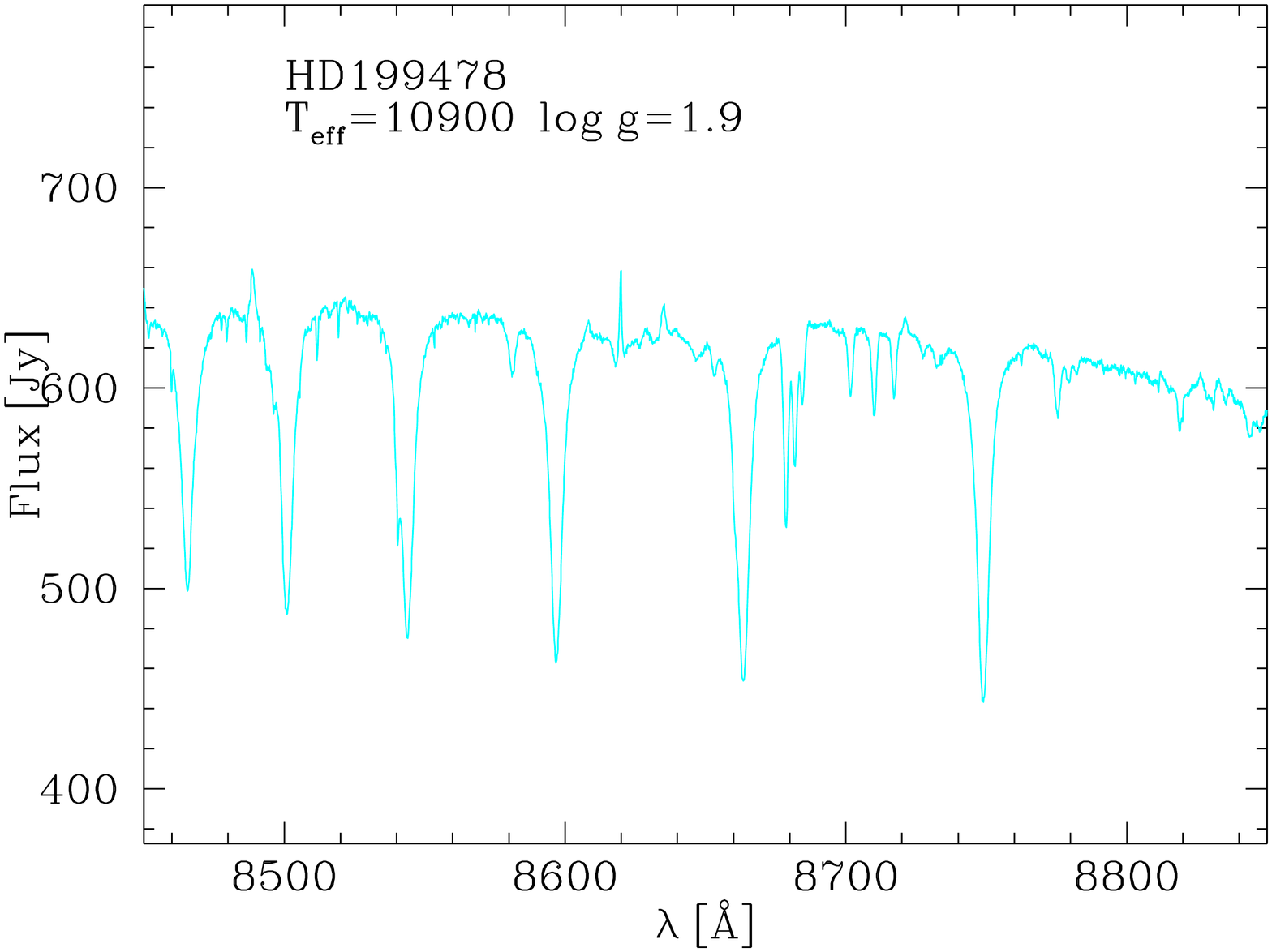}
\includegraphics[width=0.18\textwidth,angle=-90]{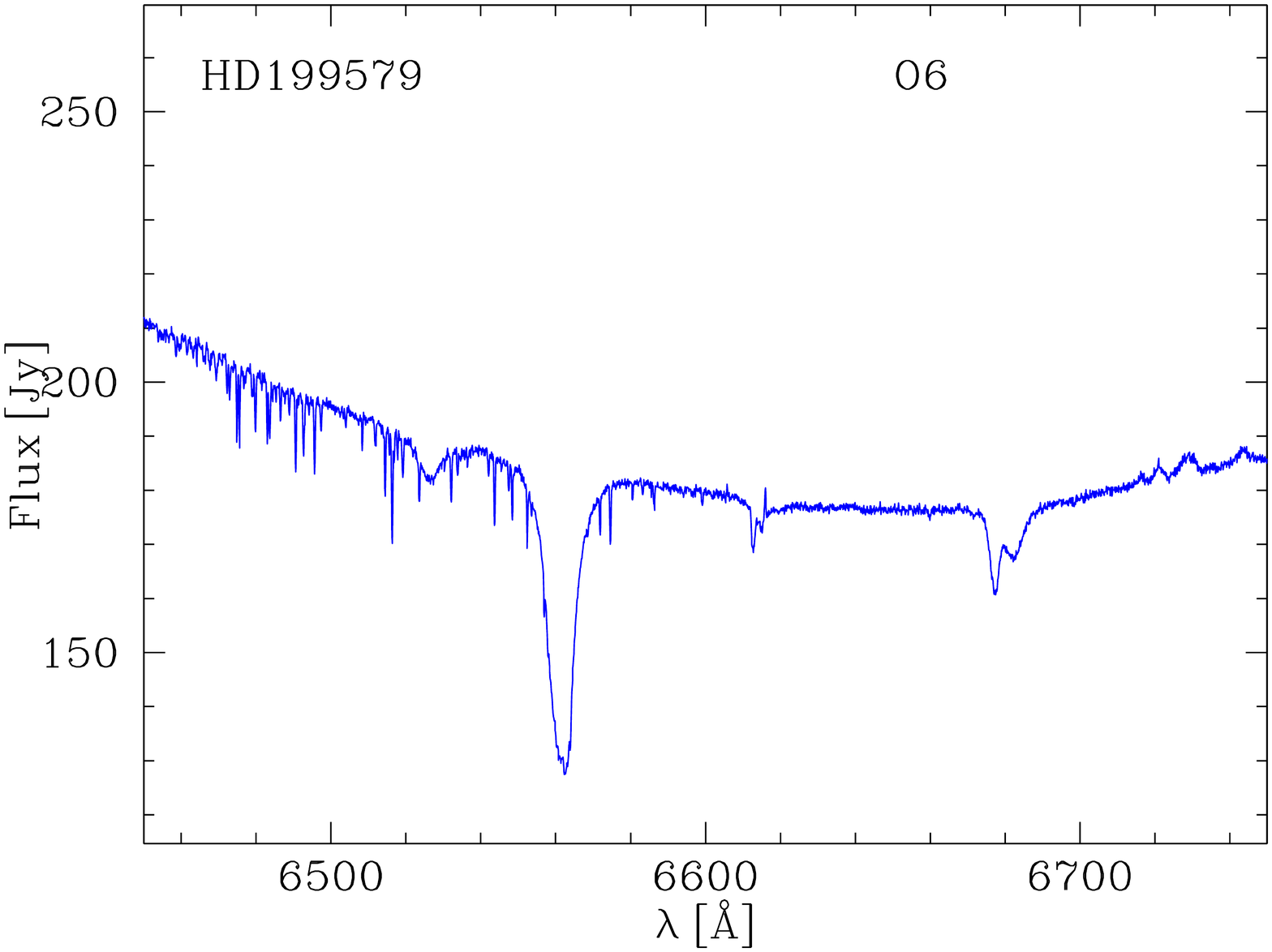}
\includegraphics[width=0.18\textwidth,angle=-90]{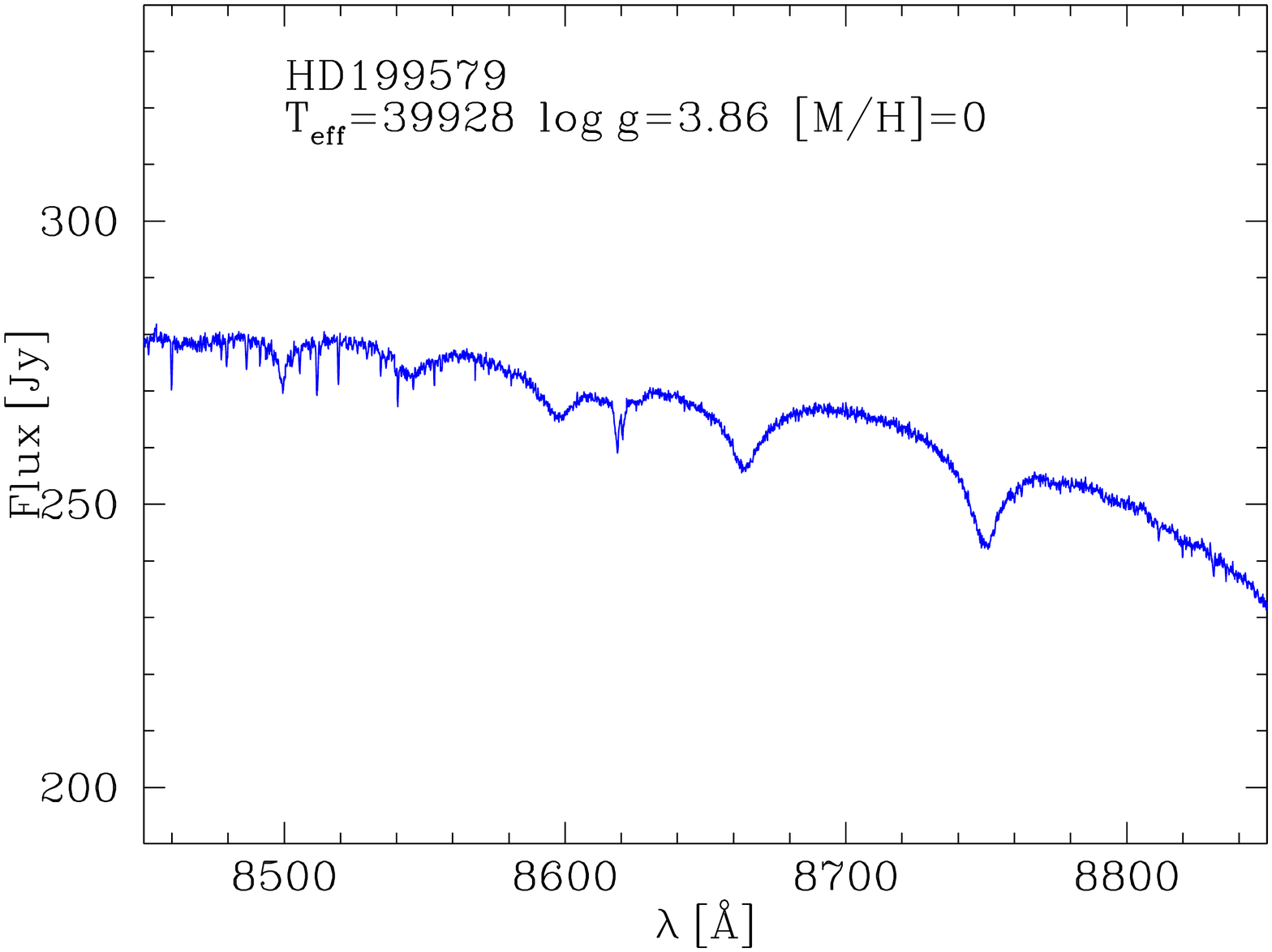}
\includegraphics[width=0.18\textwidth,angle=-90]{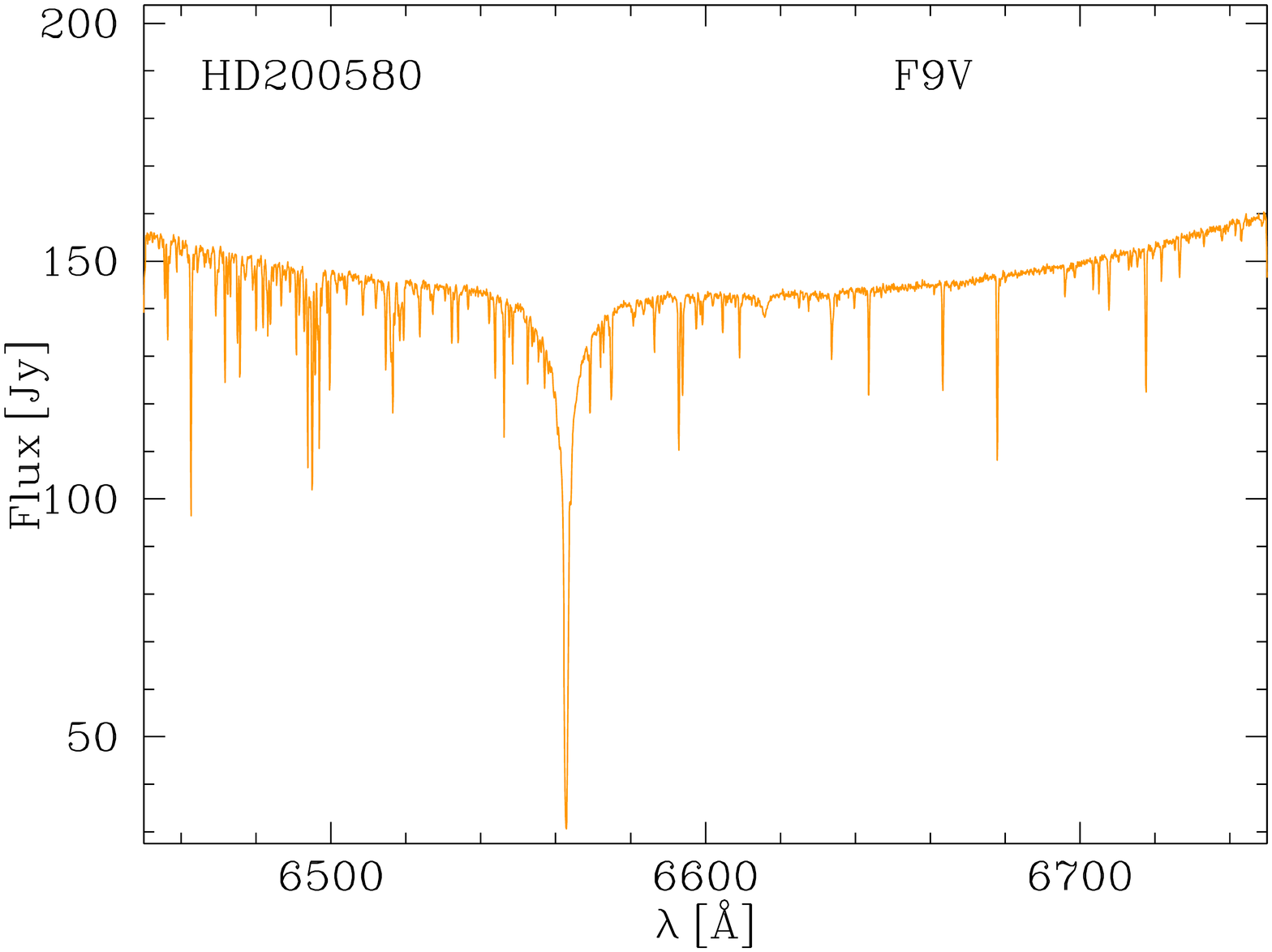}
\includegraphics[width=0.18\textwidth,angle=-90]{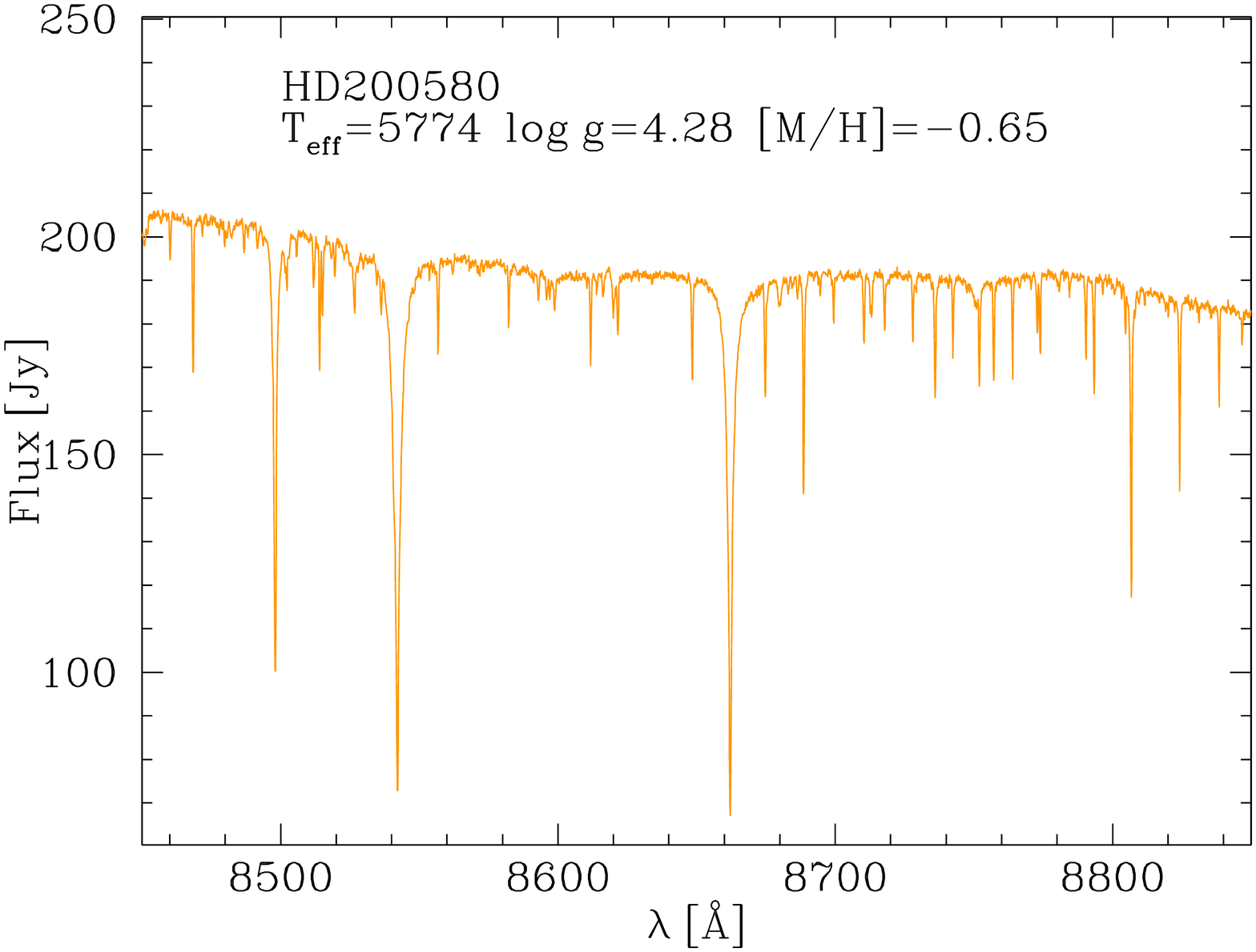}
\includegraphics[width=0.18\textwidth,angle=-90]{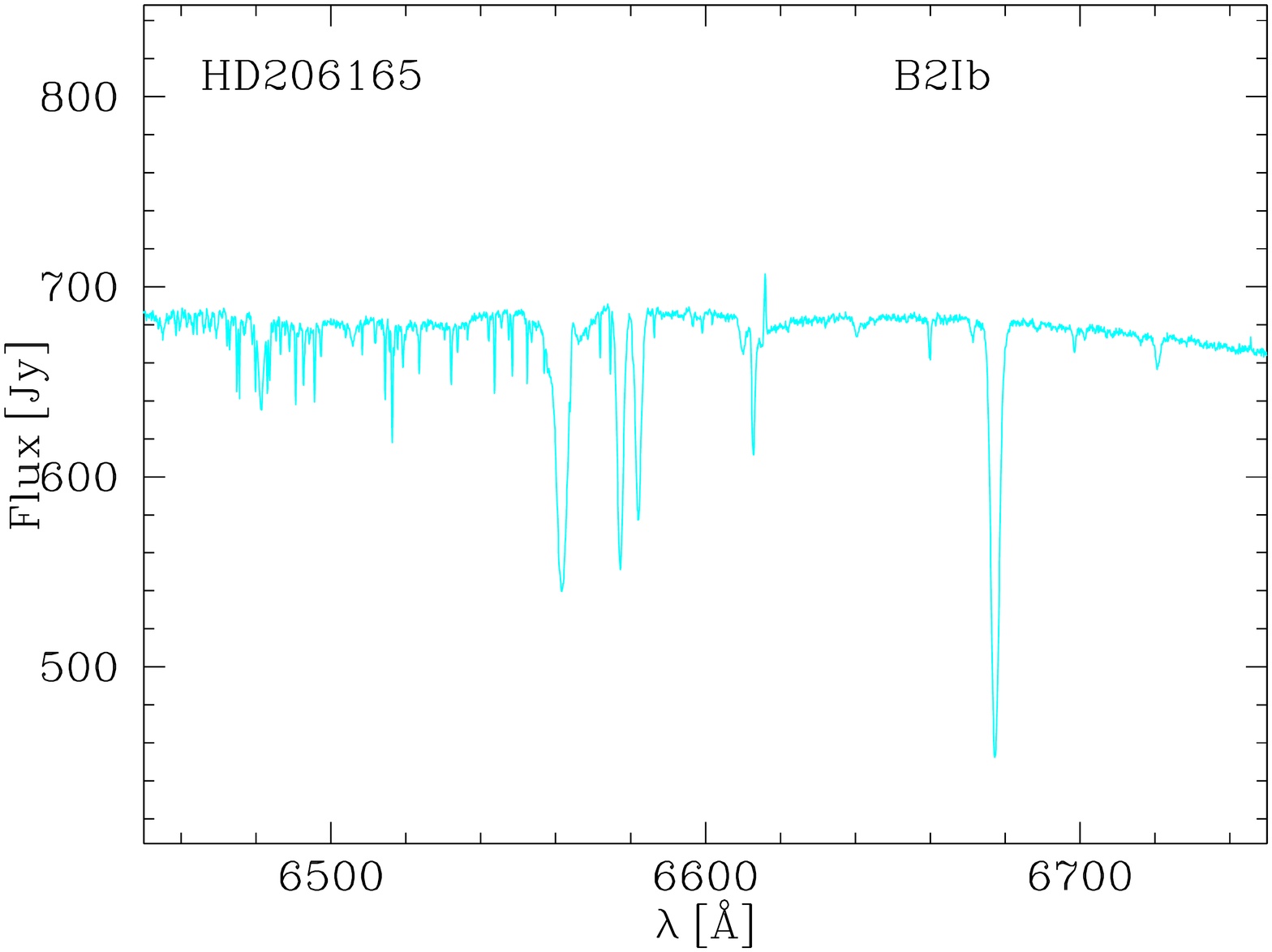}
\includegraphics[width=0.18\textwidth,angle=-90]{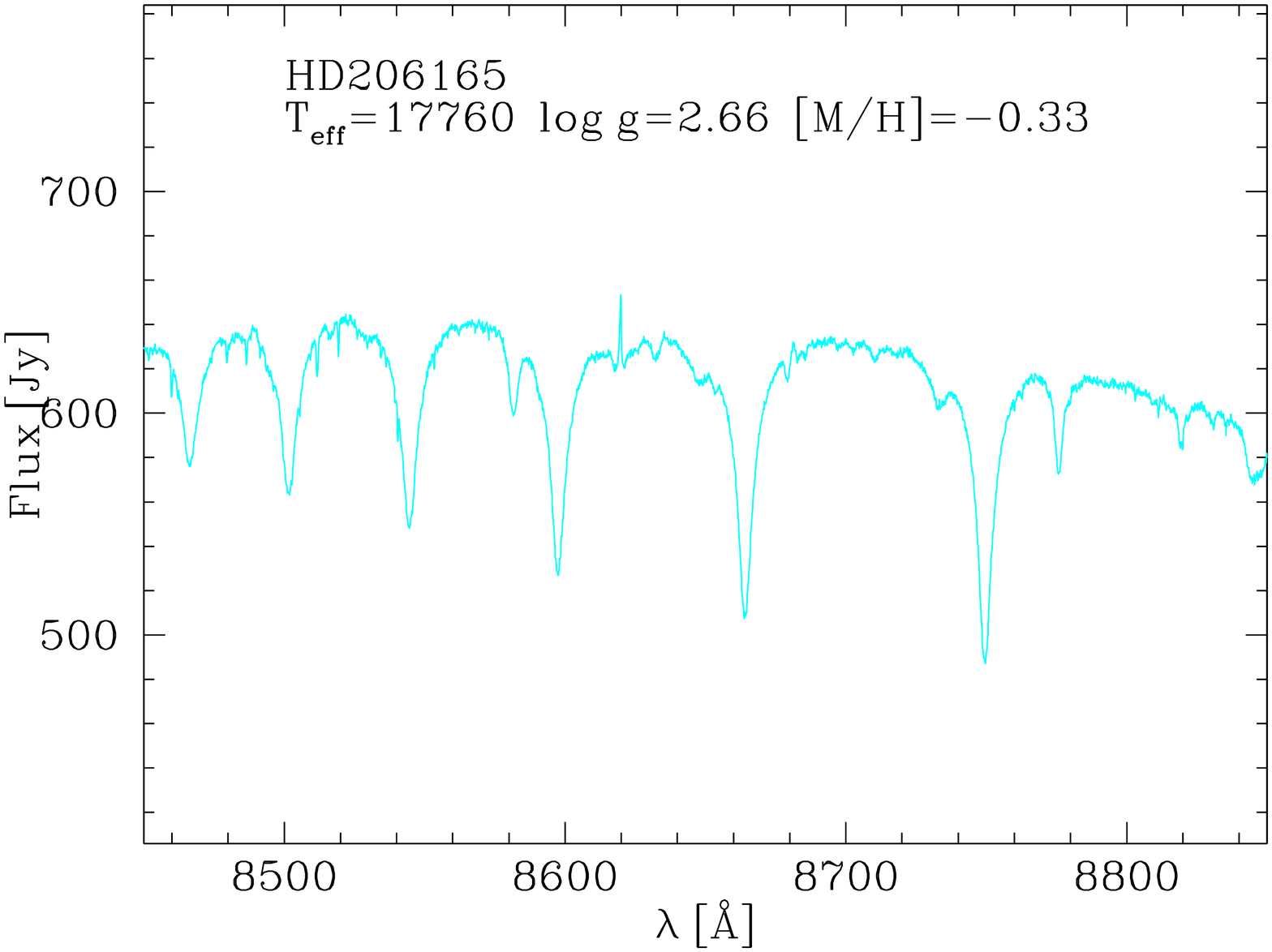}
\includegraphics[width=0.18\textwidth,angle=-90]{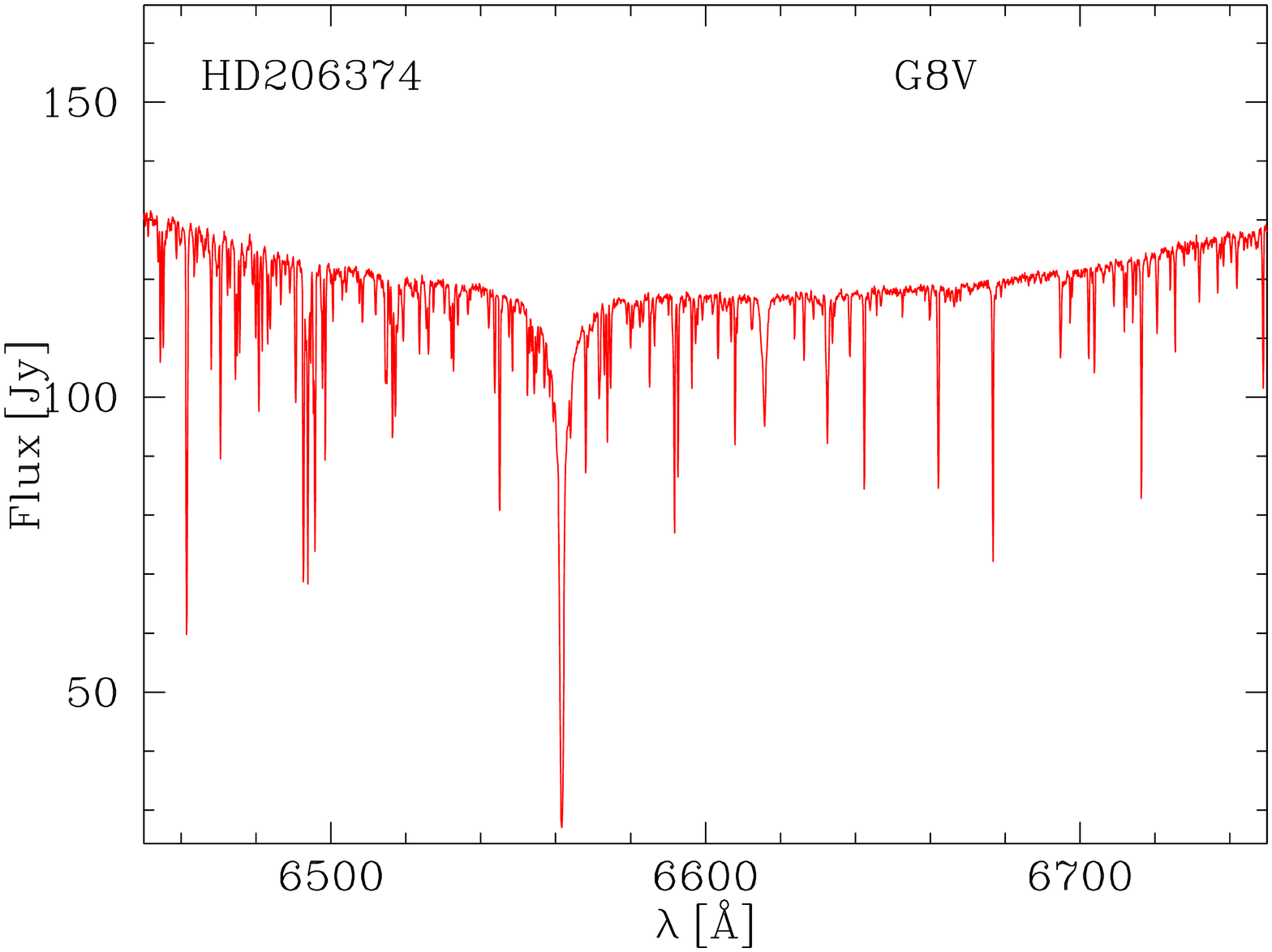}
\includegraphics[width=0.18\textwidth,angle=-90]{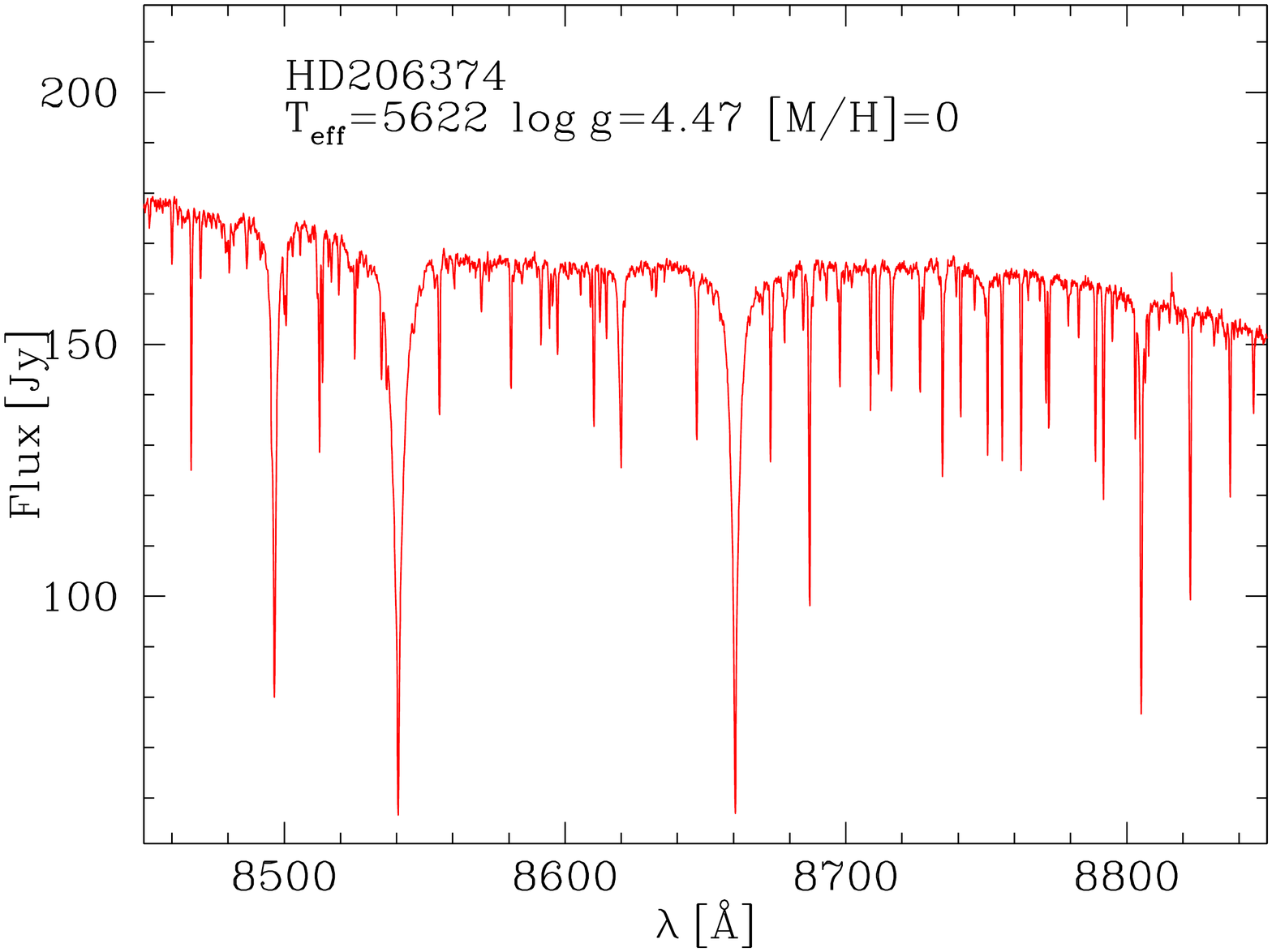}
\includegraphics[width=0.18\textwidth,angle=-90]{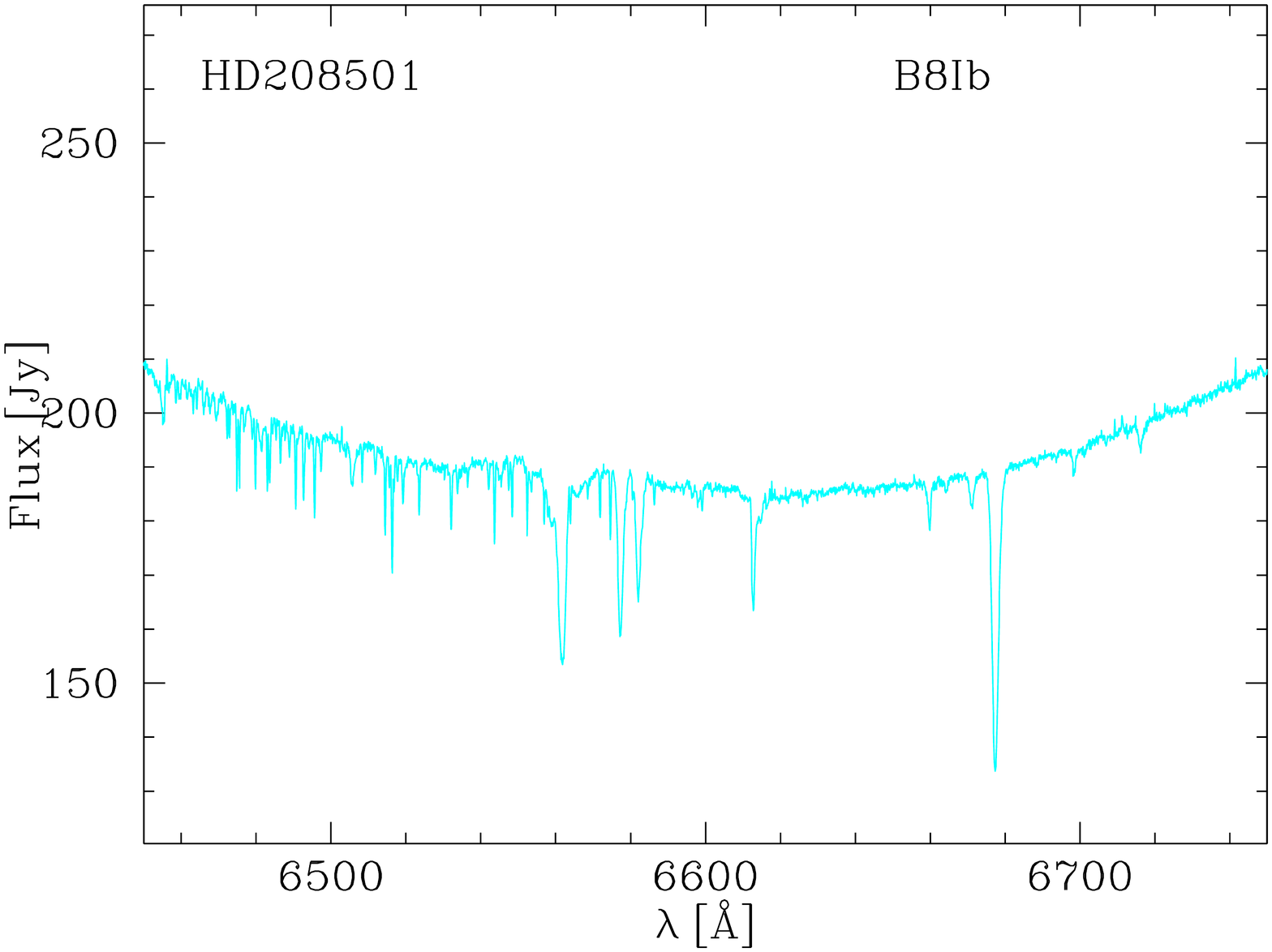}
\includegraphics[width=0.18\textwidth,angle=-90]{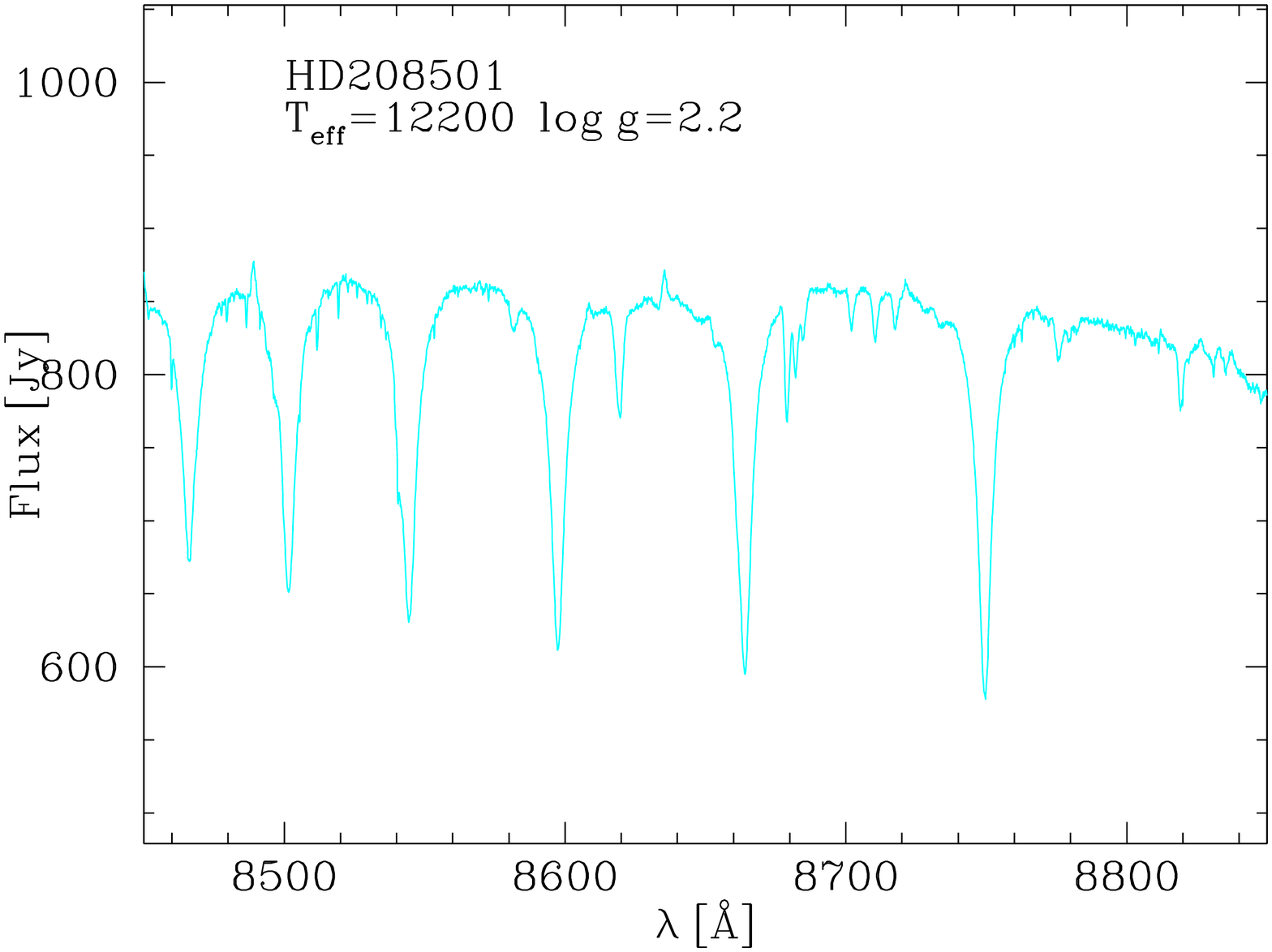}

\contcaption{27. Stars shown in this page are: HD193793, HD194453, HD195198, HD195592, HD196426, HD196610, HD198183, HD198478, HD199478, HD199579, HD200580, HD206165, HD206374 and HD208501.}
\end{figure*}

\begin{figure*}
\includegraphics[width=0.18\textwidth,angle=-90]{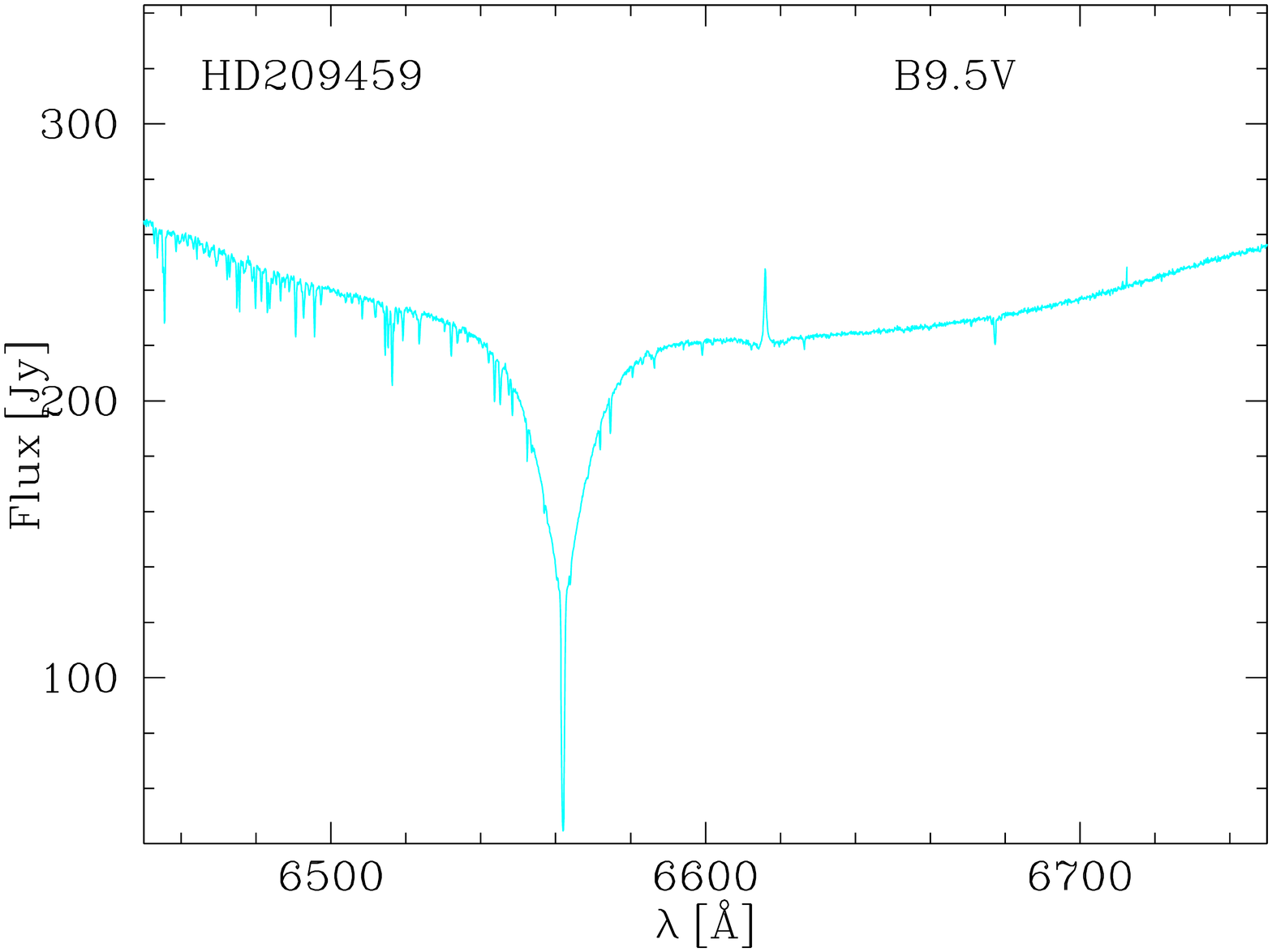}
\includegraphics[width=0.18\textwidth,angle=-90]{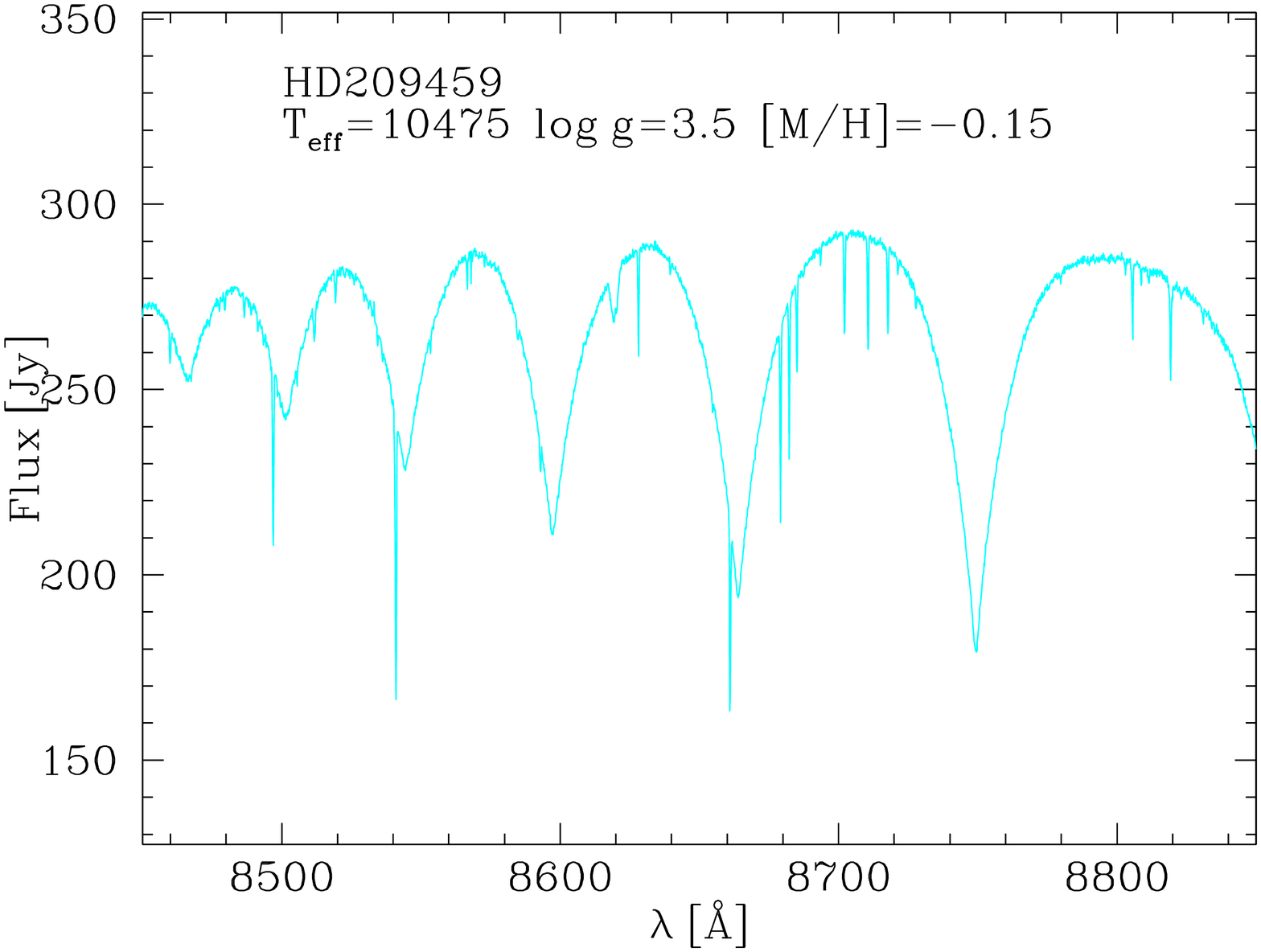}
\includegraphics[width=0.18\textwidth,angle=-90]{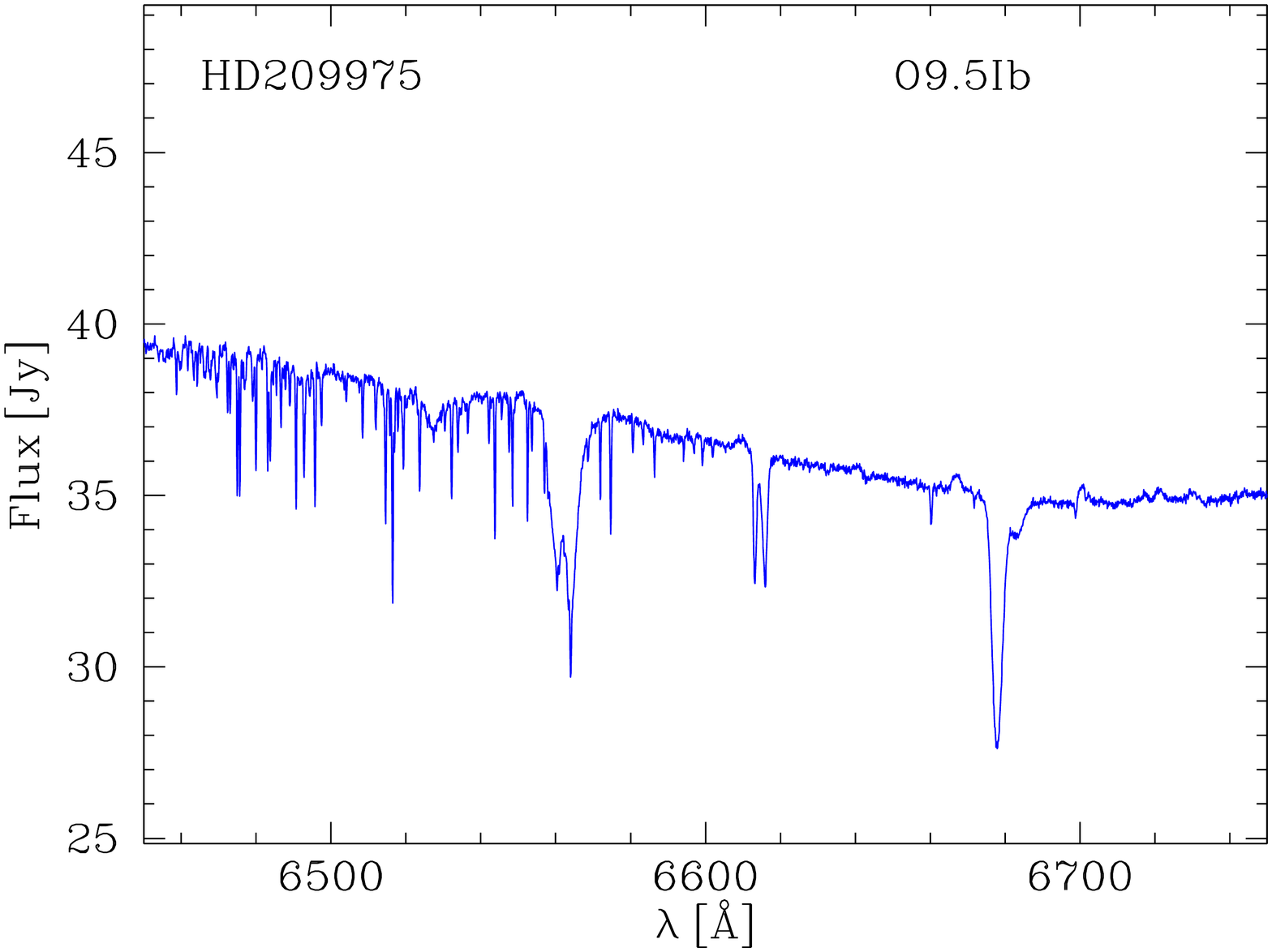}
\includegraphics[width=0.18\textwidth,angle=-90]{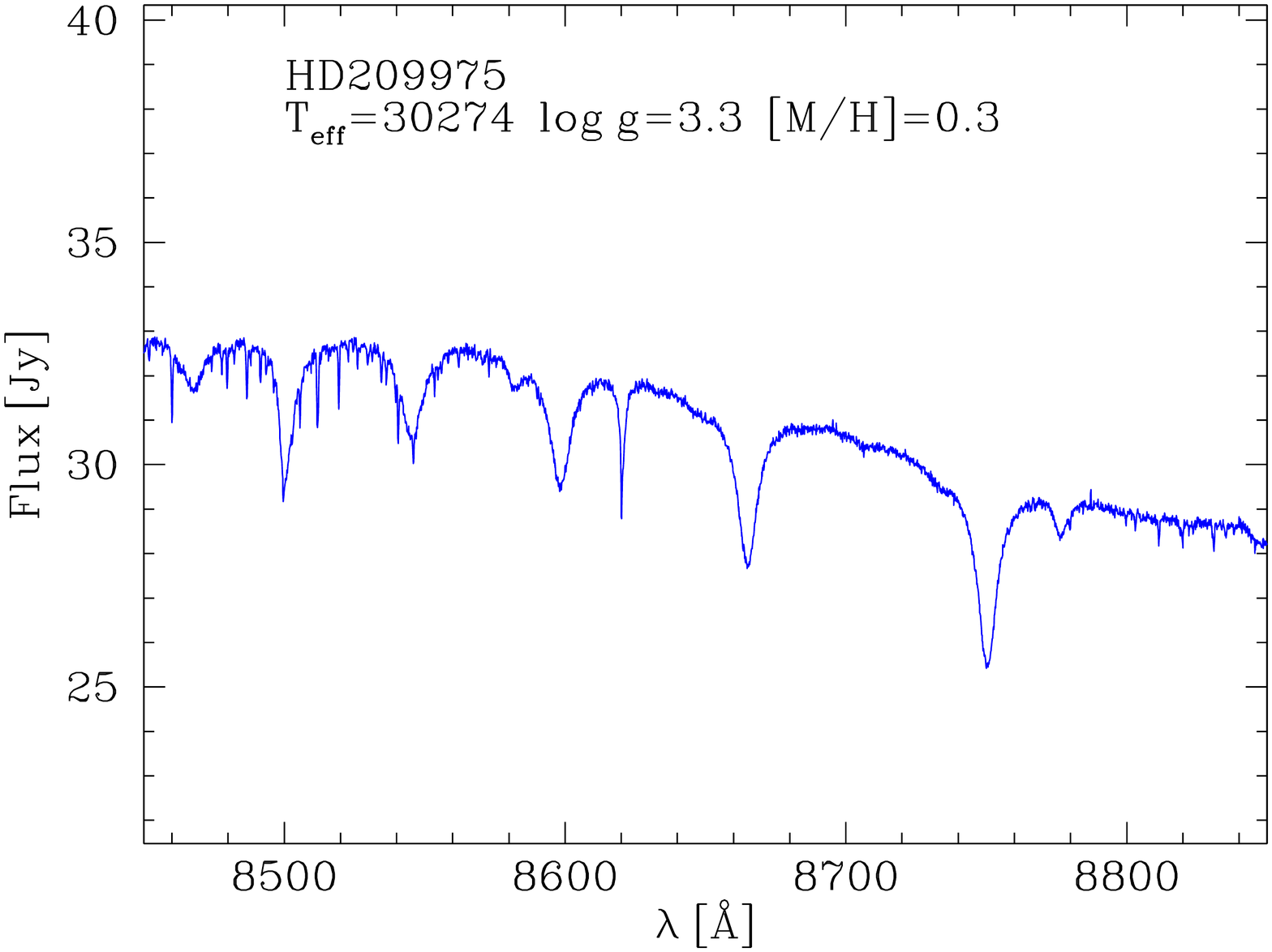}
\includegraphics[width=0.18\textwidth,angle=-90]{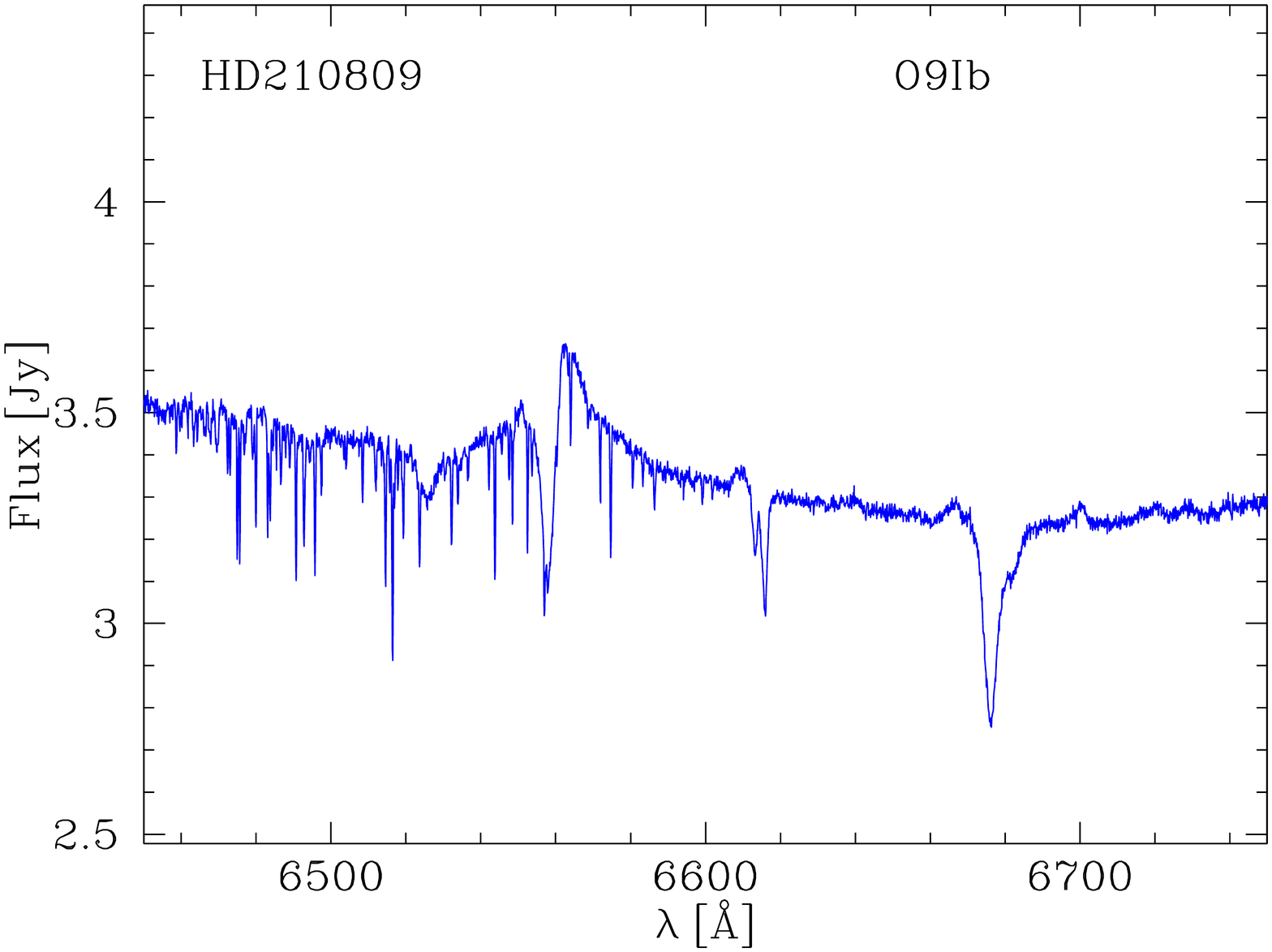}
\includegraphics[width=0.18\textwidth,angle=-90]{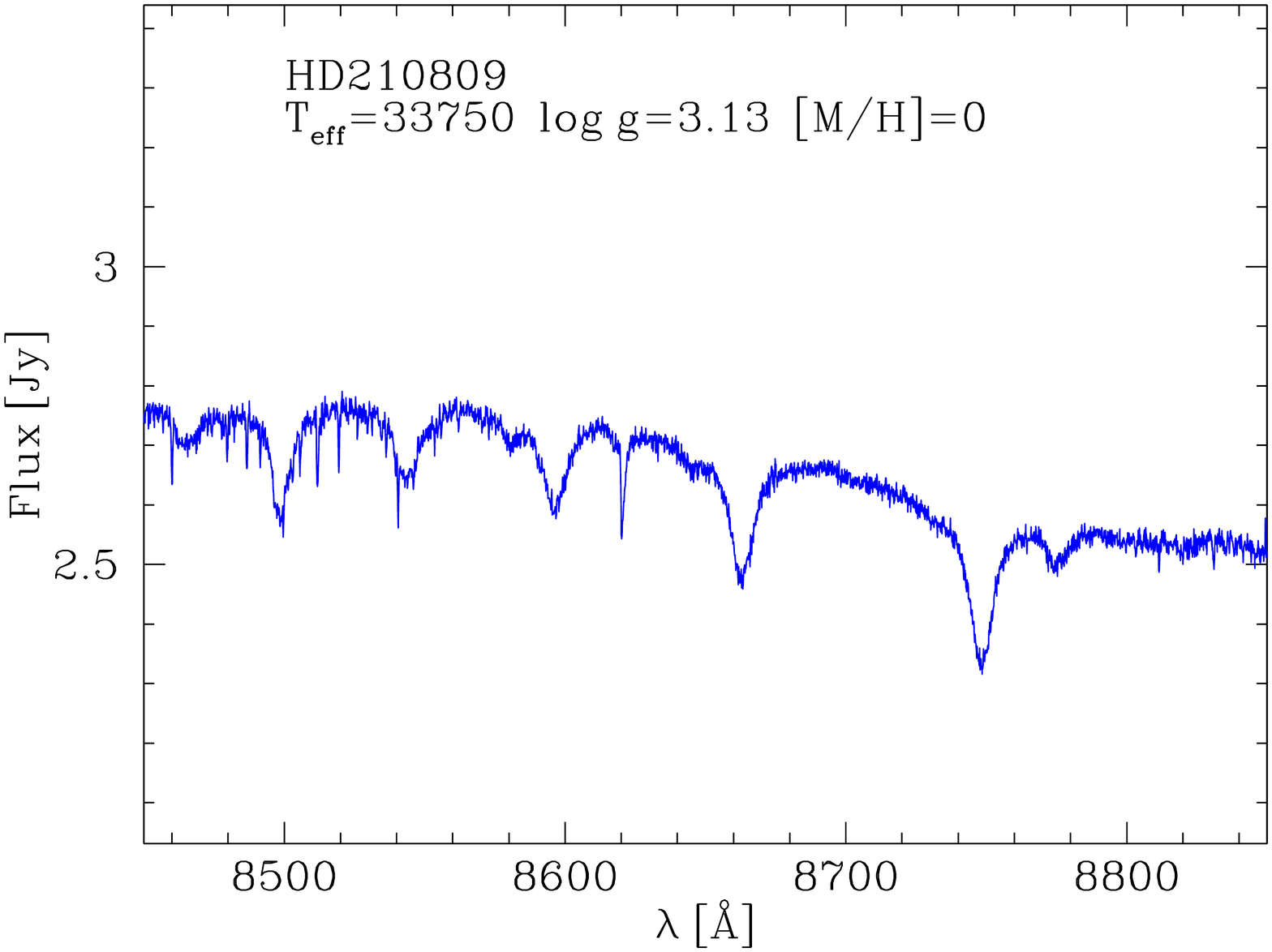}
\includegraphics[width=0.18\textwidth,angle=-90]{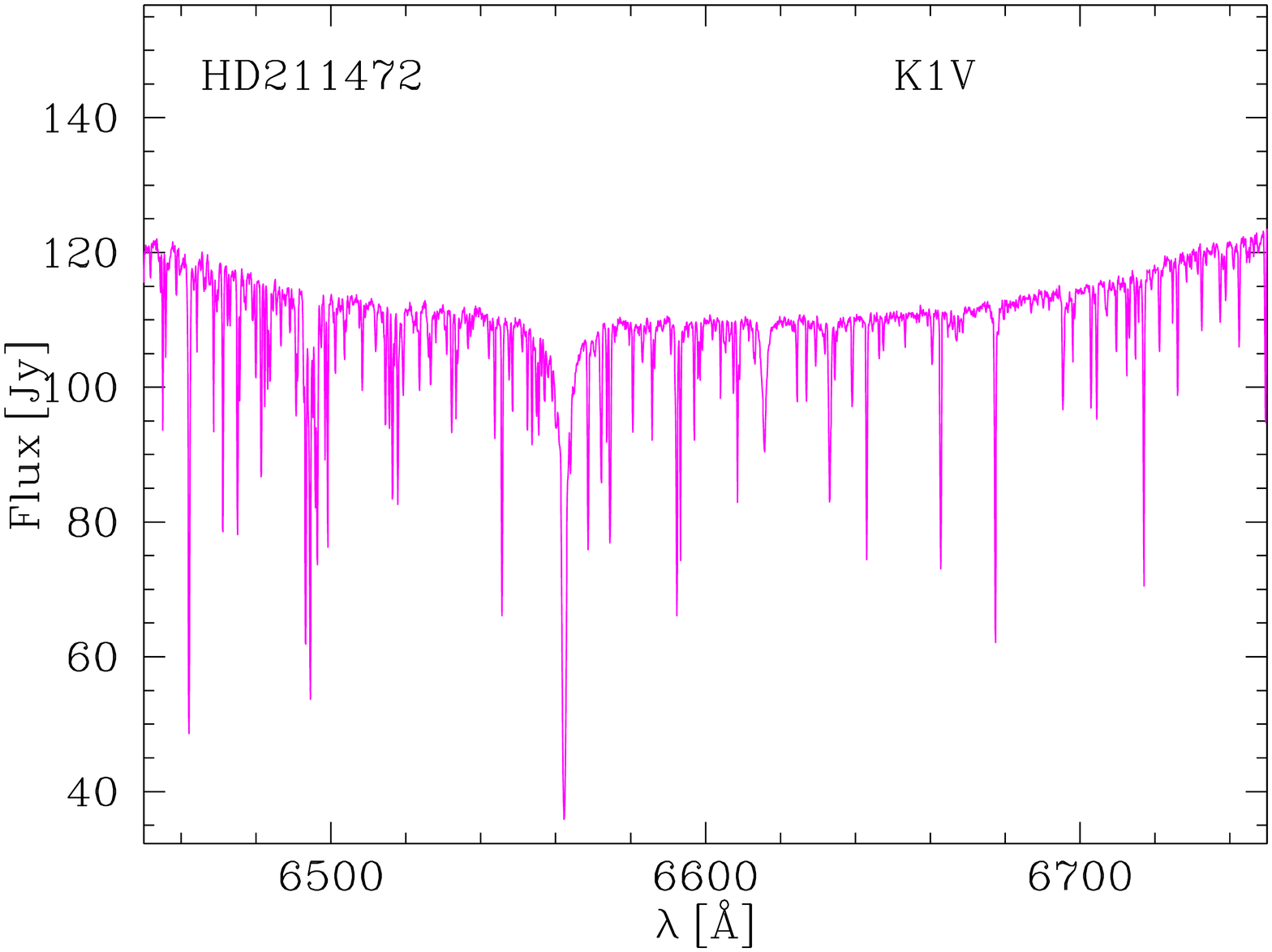}
\includegraphics[width=0.18\textwidth,angle=-90]{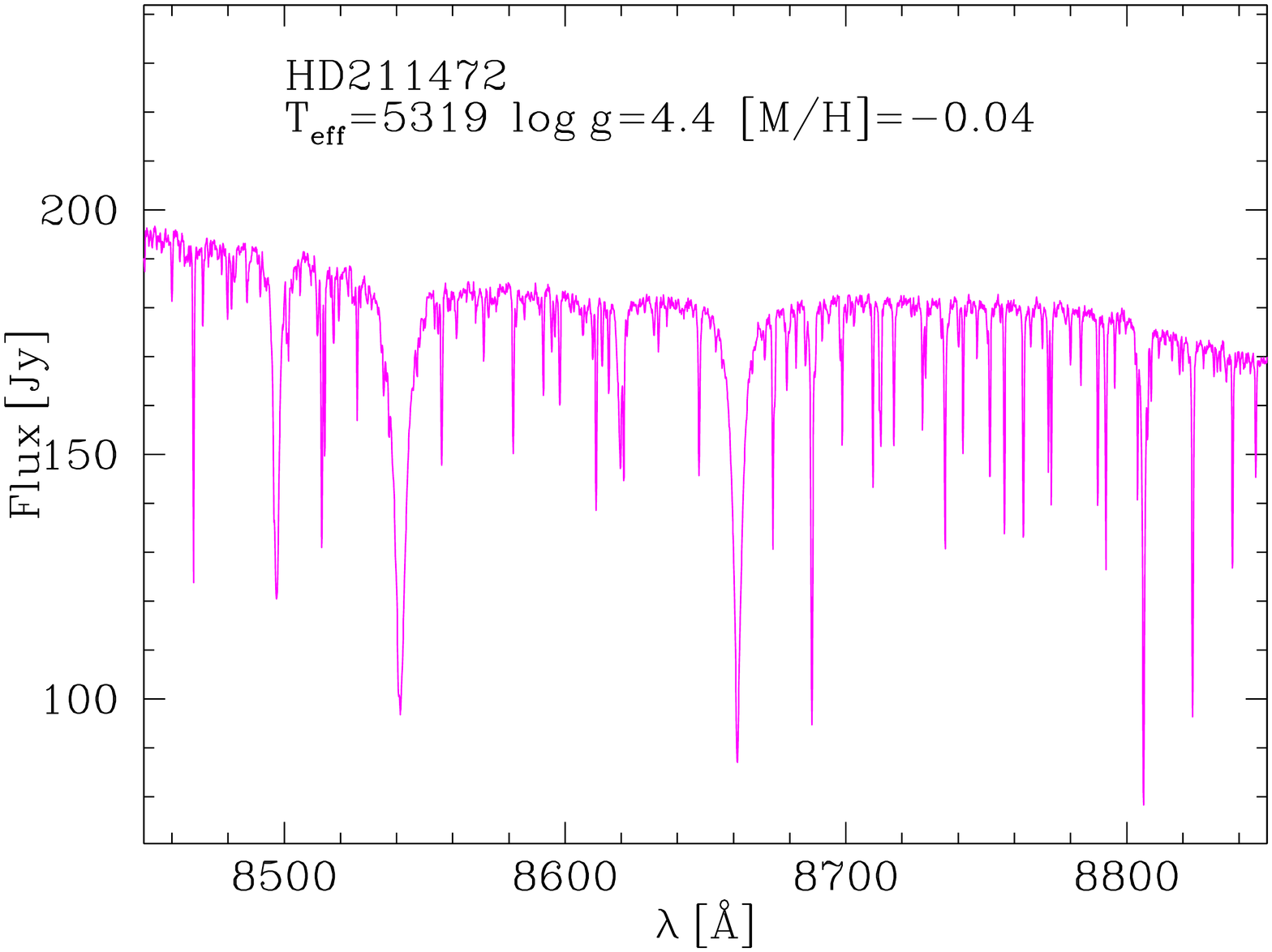}
\includegraphics[width=0.18\textwidth,angle=-90]{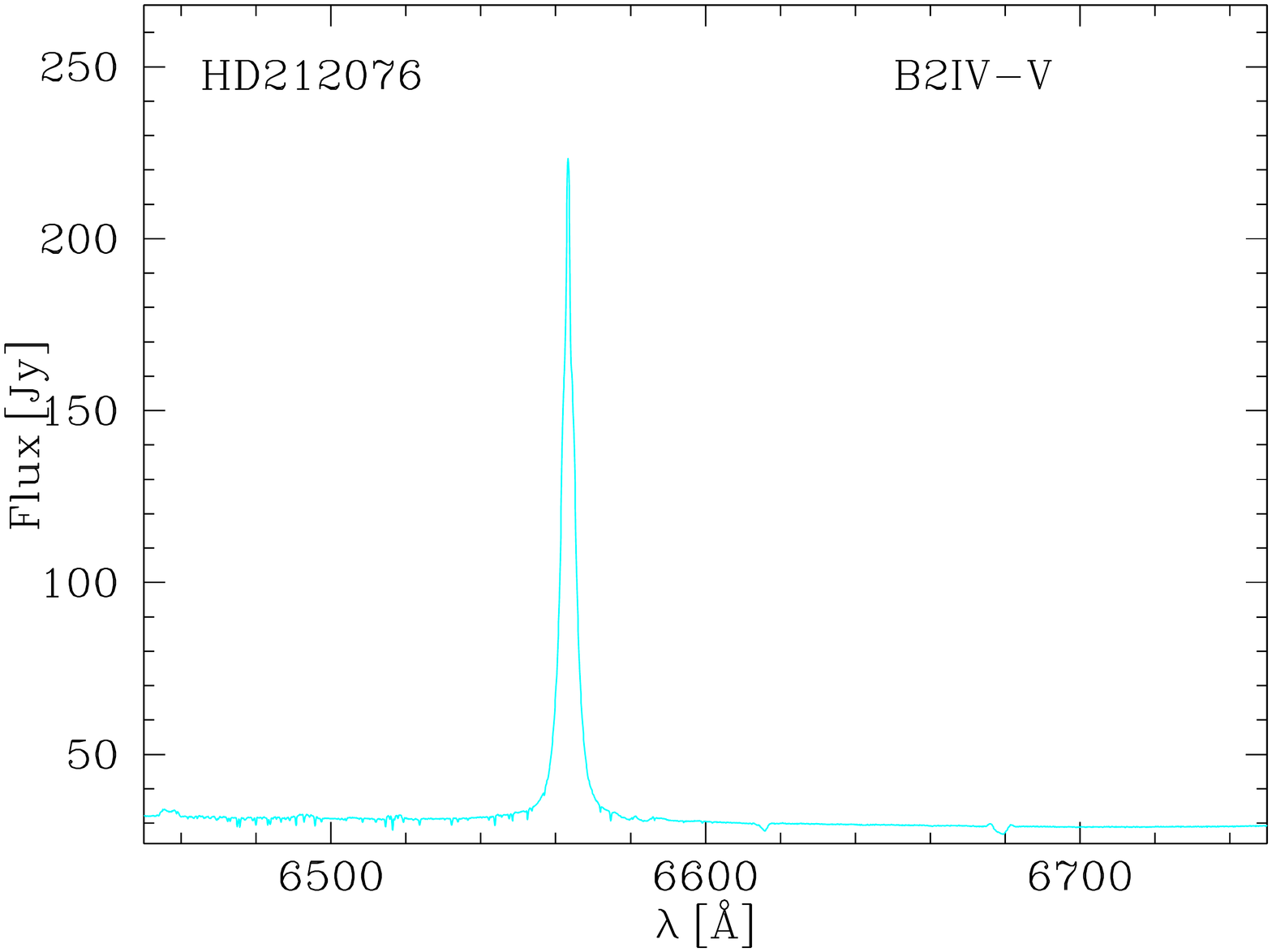}
\includegraphics[width=0.18\textwidth,angle=-90]{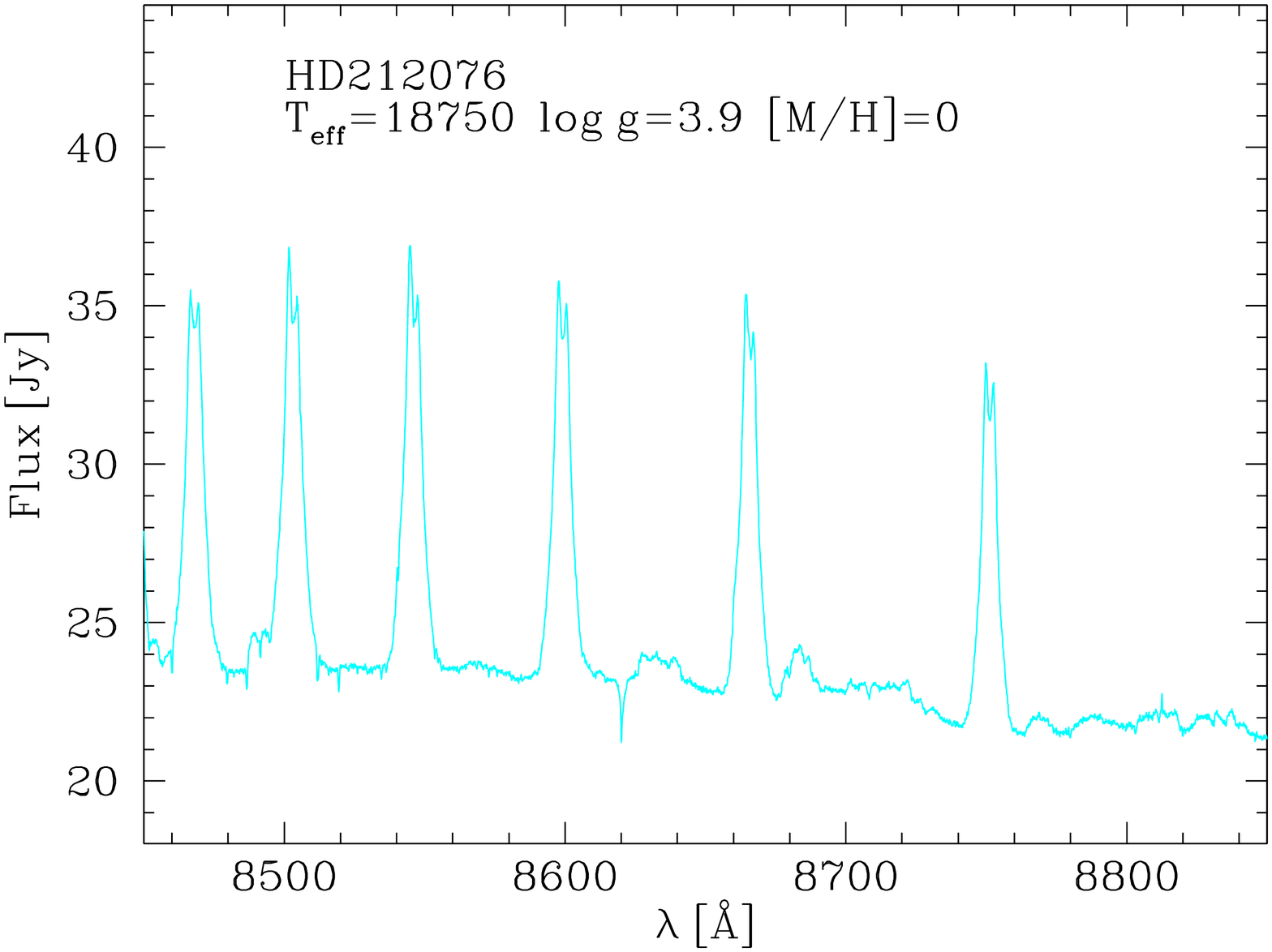}
\includegraphics[width=0.18\textwidth,angle=-90]{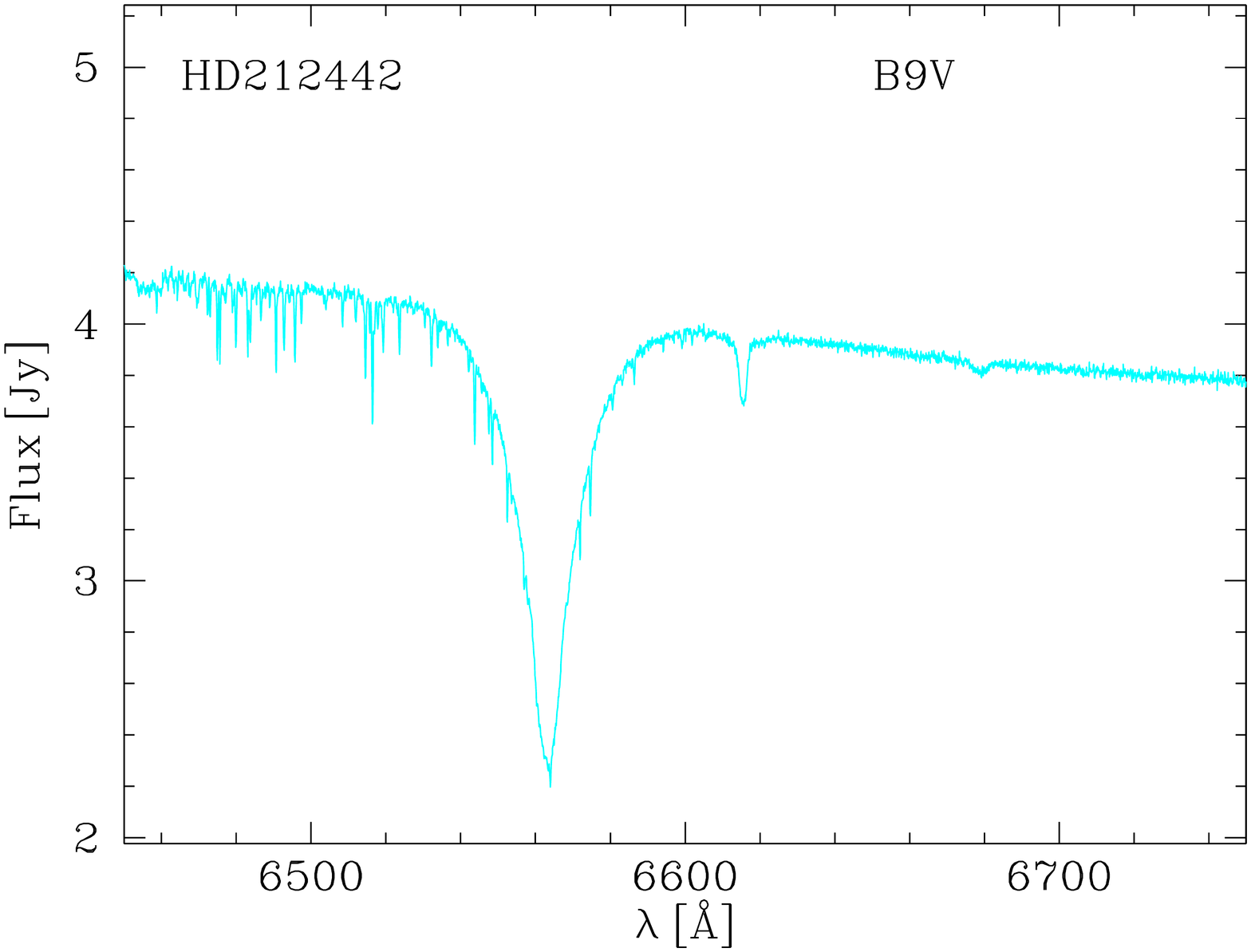}
\includegraphics[width=0.18\textwidth,angle=-90]{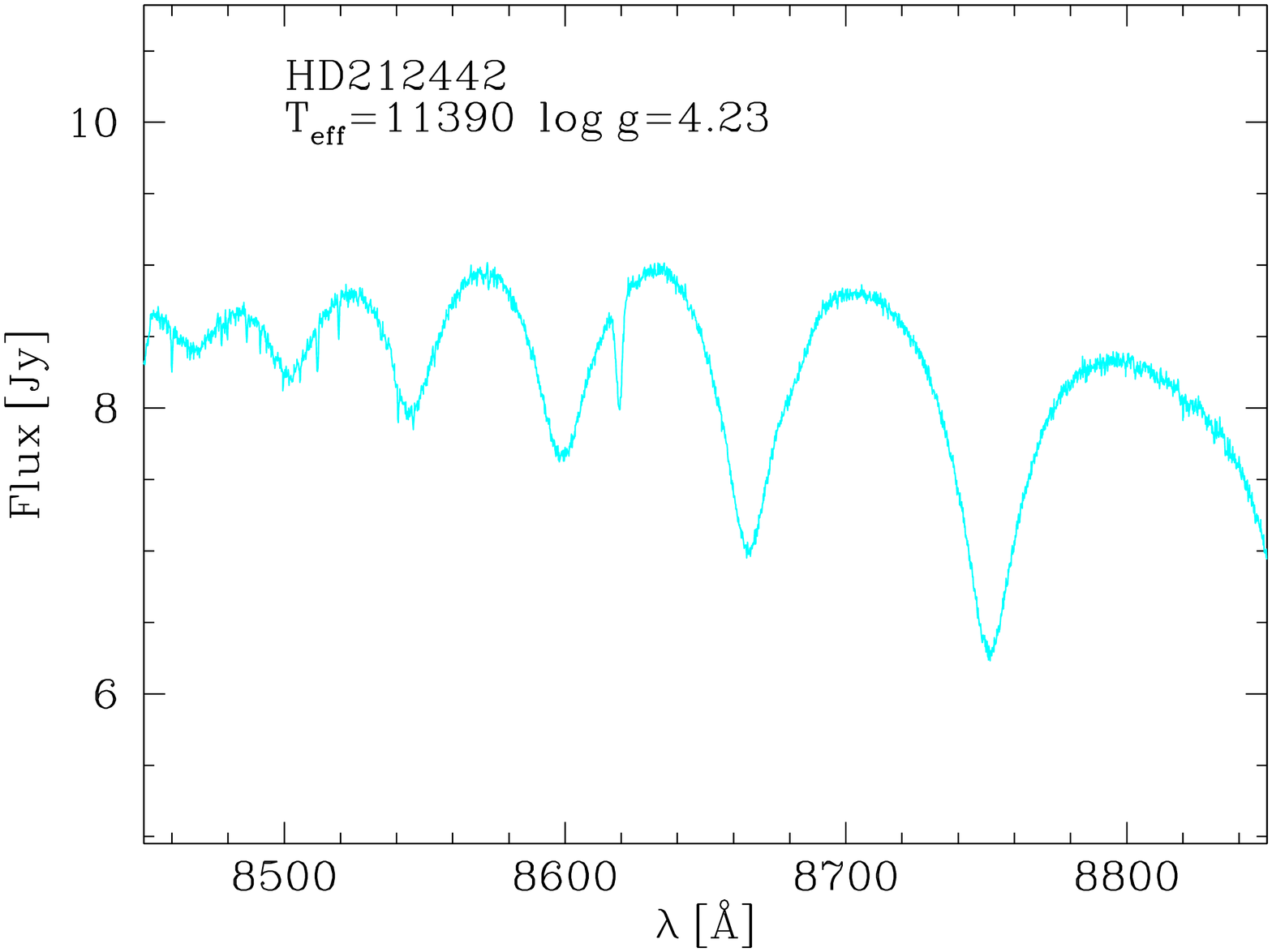}
\includegraphics[width=0.18\textwidth,angle=-90]{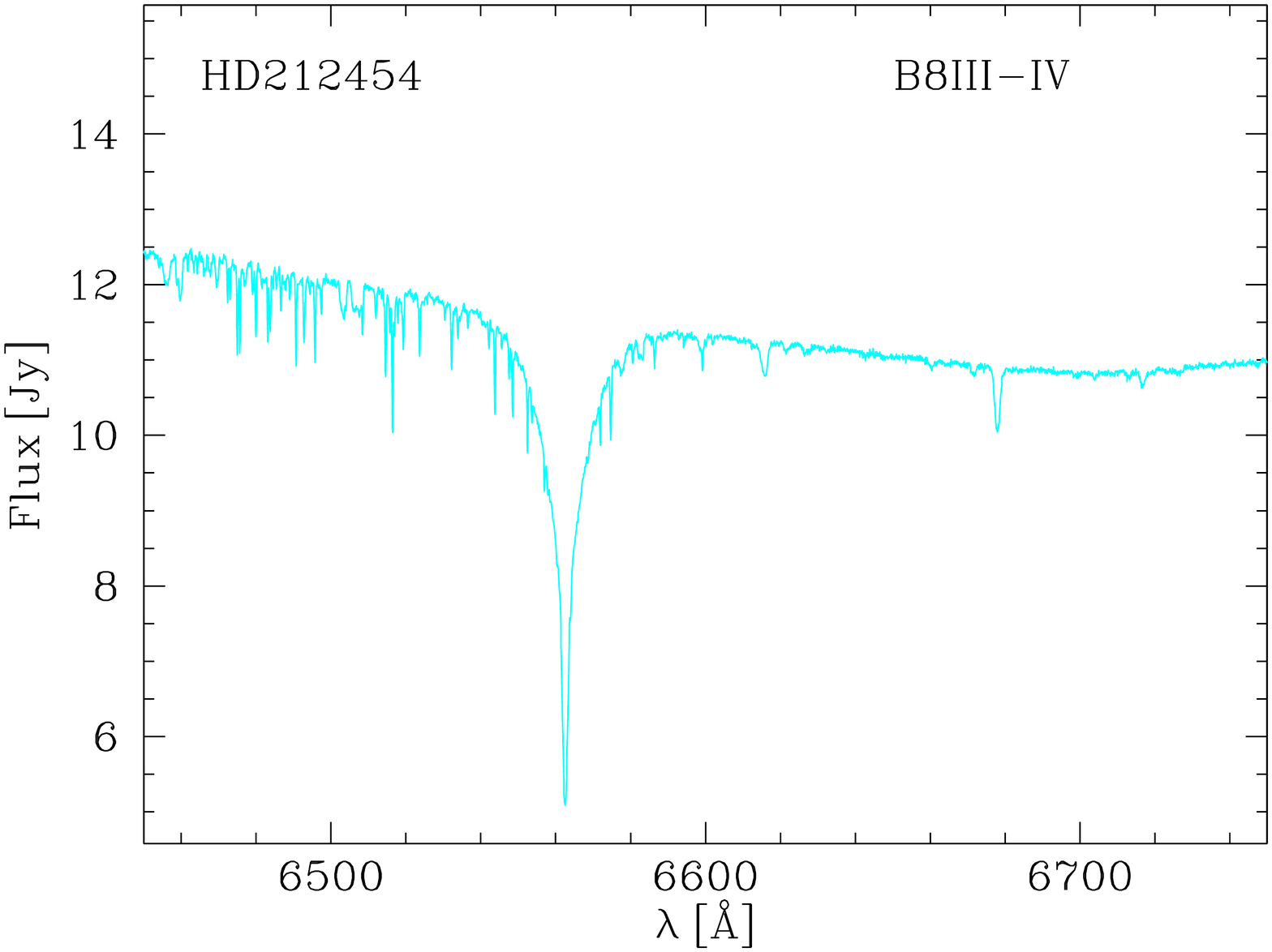}
\includegraphics[width=0.18\textwidth,angle=-90]{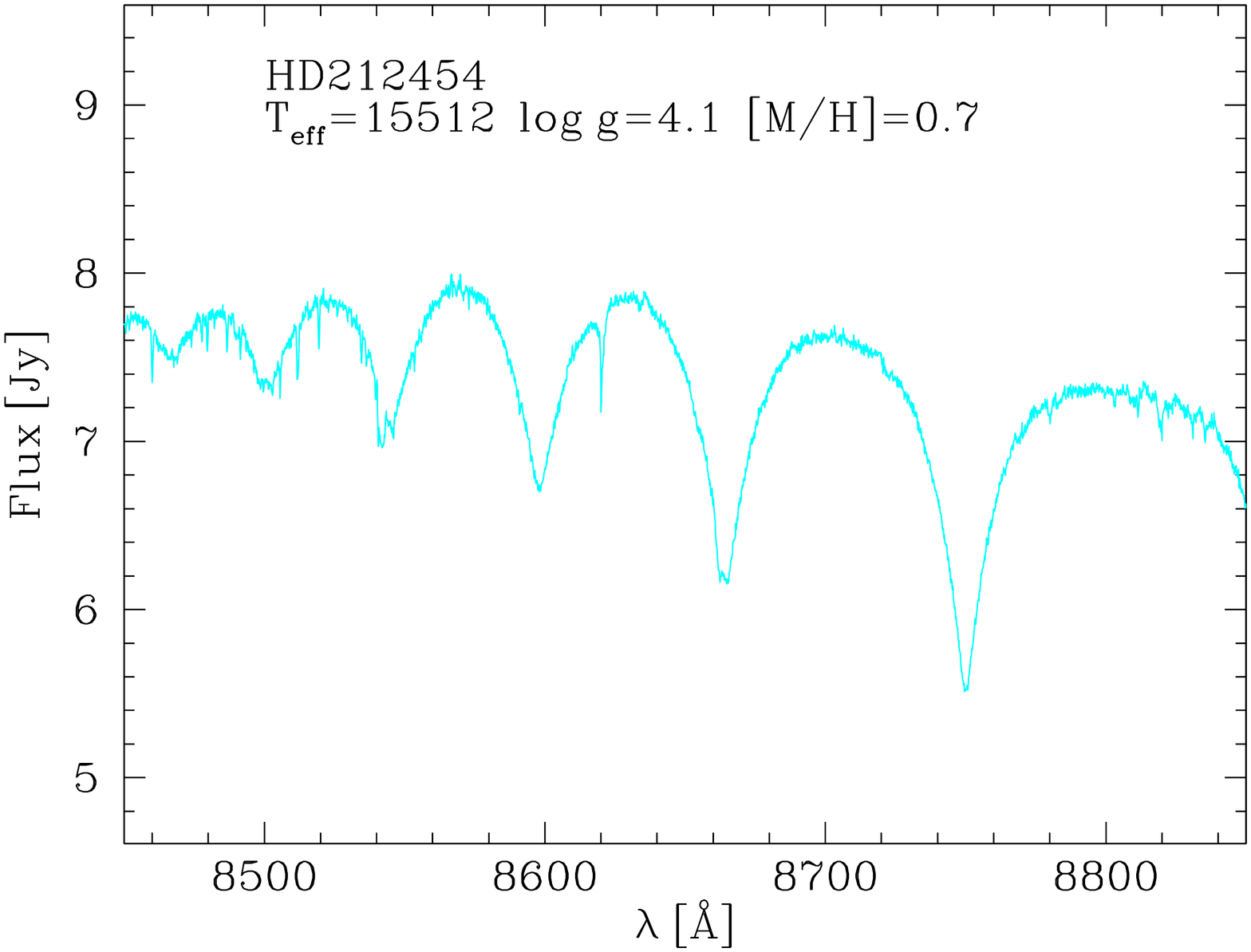}
\includegraphics[width=0.18\textwidth,angle=-90]{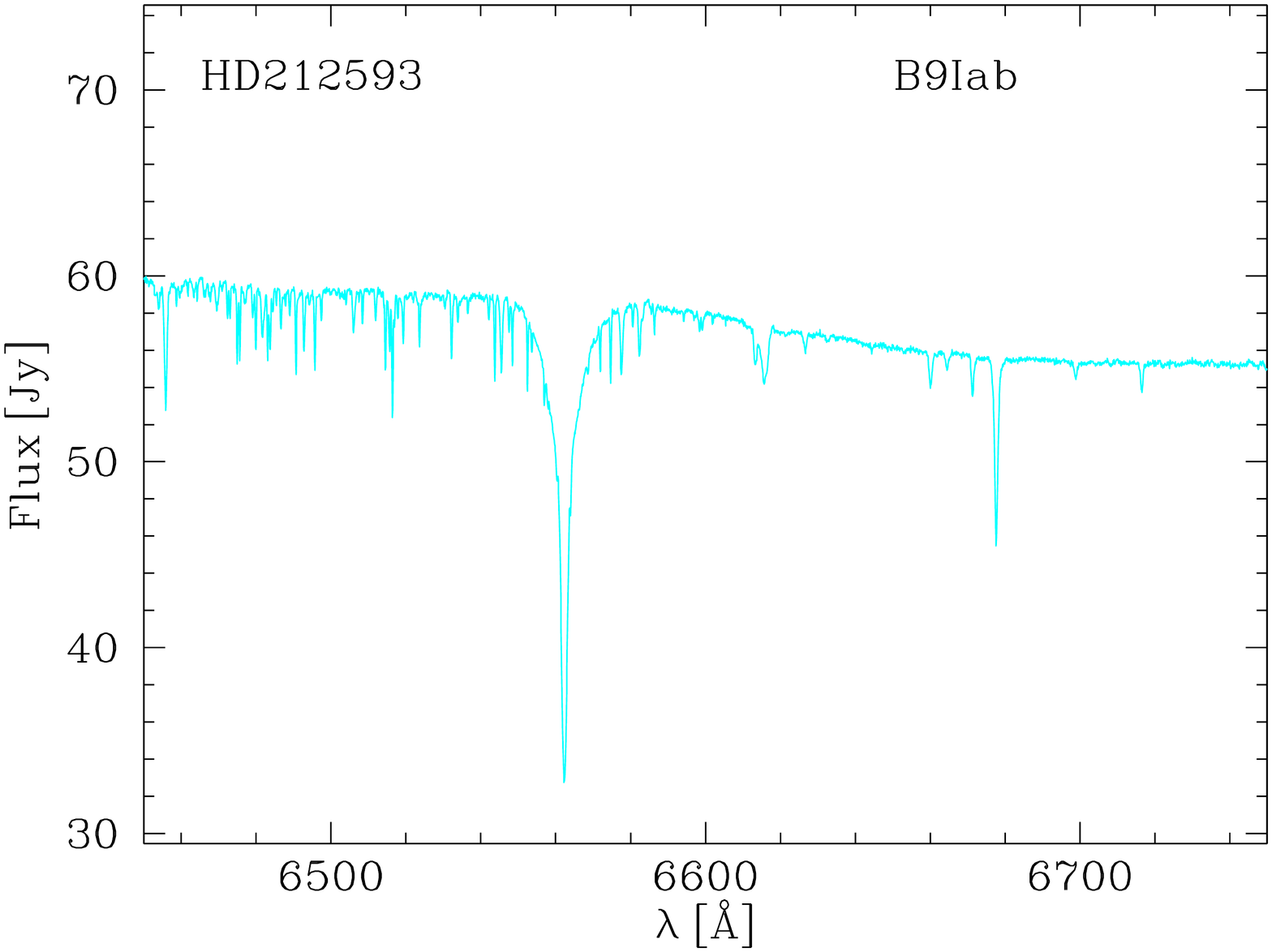}
\includegraphics[width=0.18\textwidth,angle=-90]{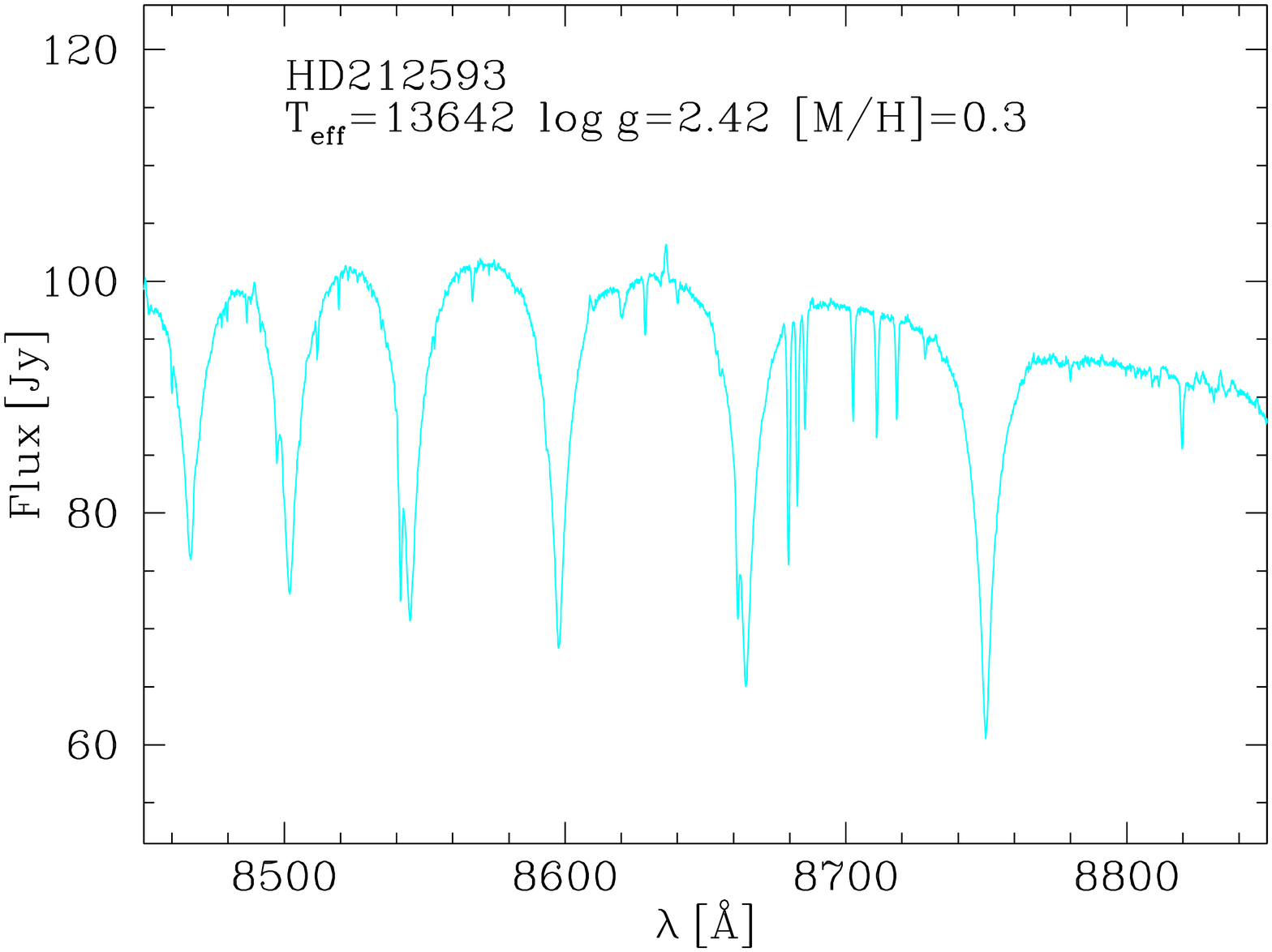}
\includegraphics[width=0.18\textwidth,angle=-90]{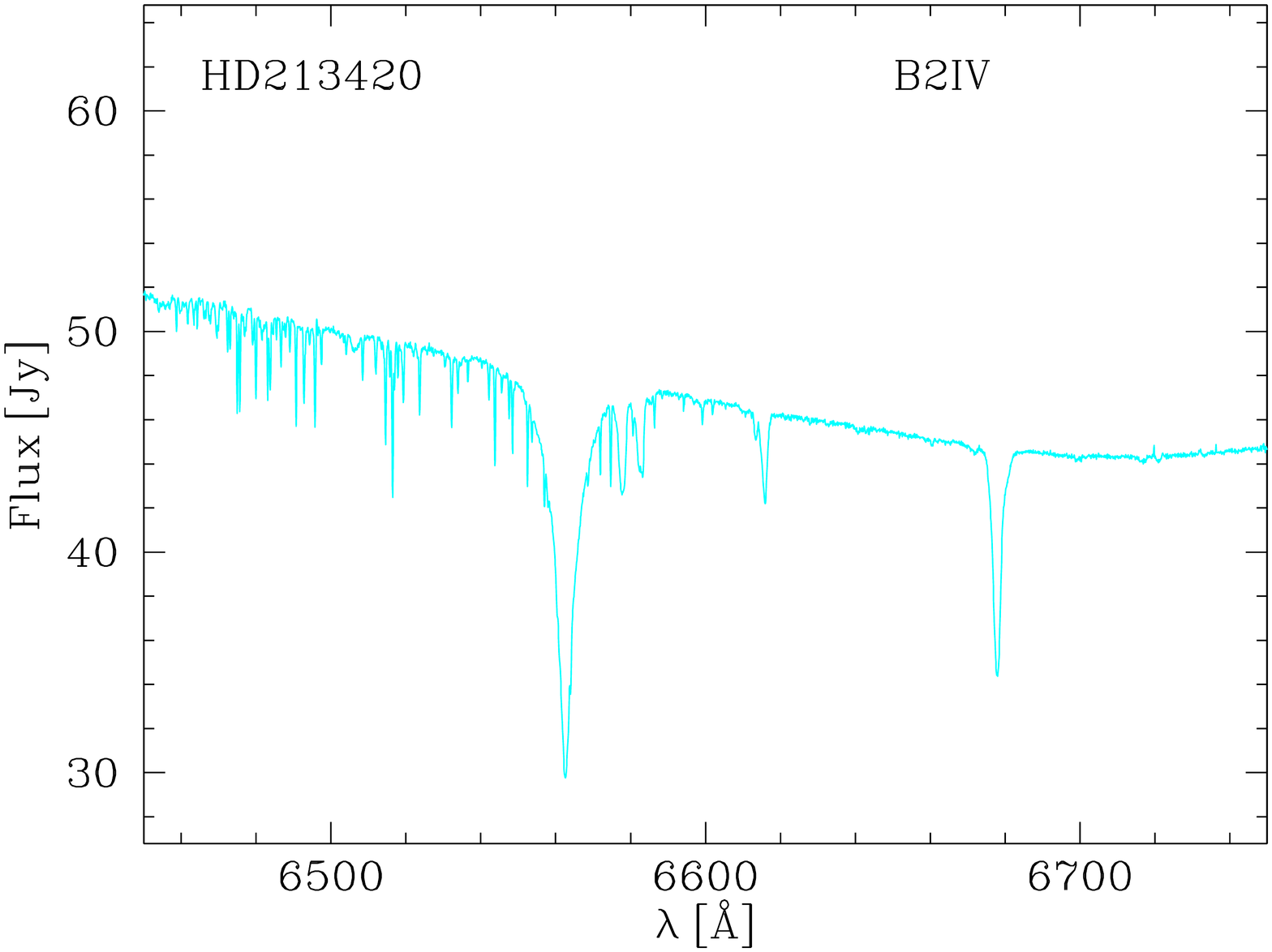}
\includegraphics[width=0.18\textwidth,angle=-90]{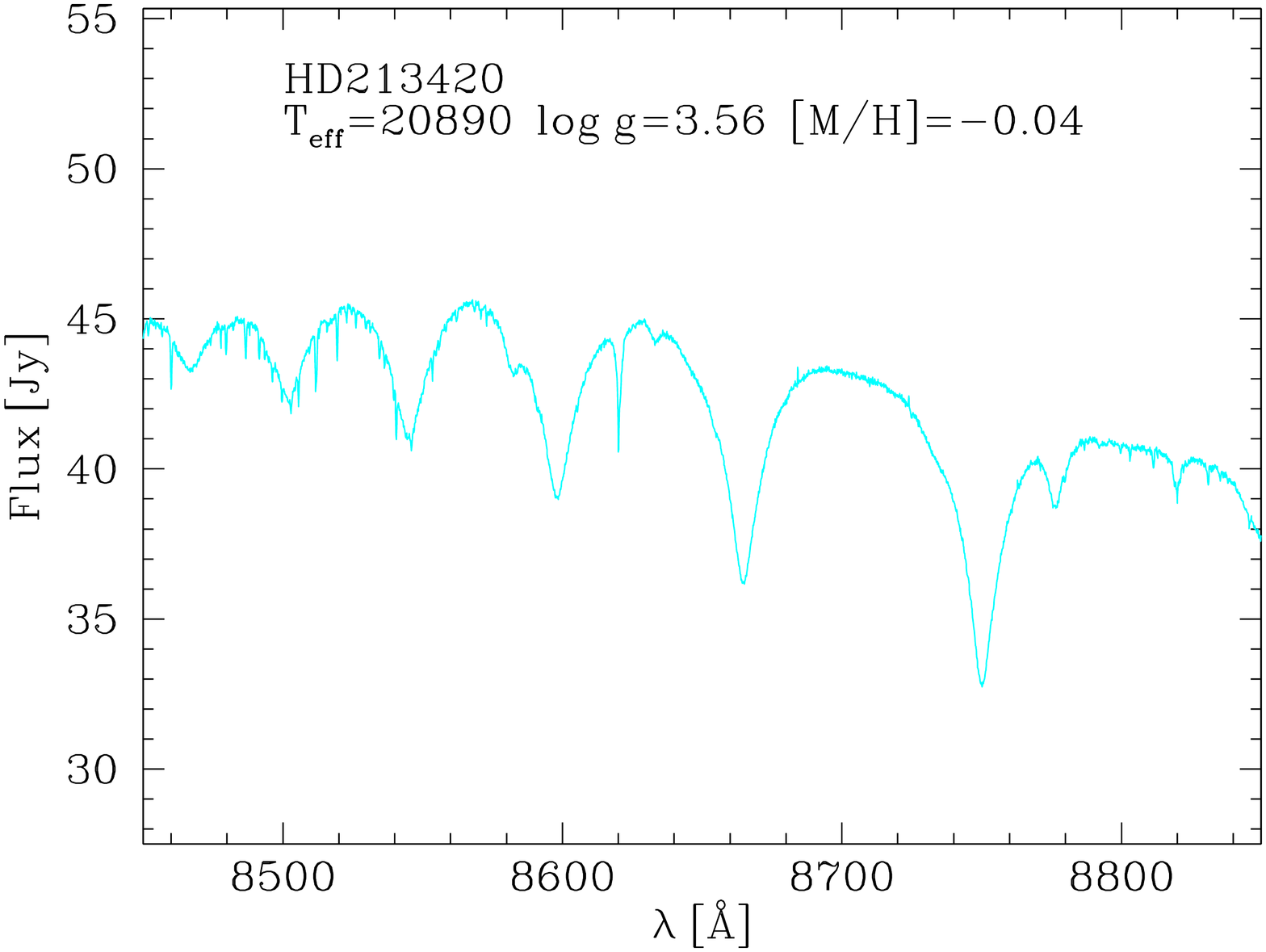}
\includegraphics[width=0.18\textwidth,angle=-90]{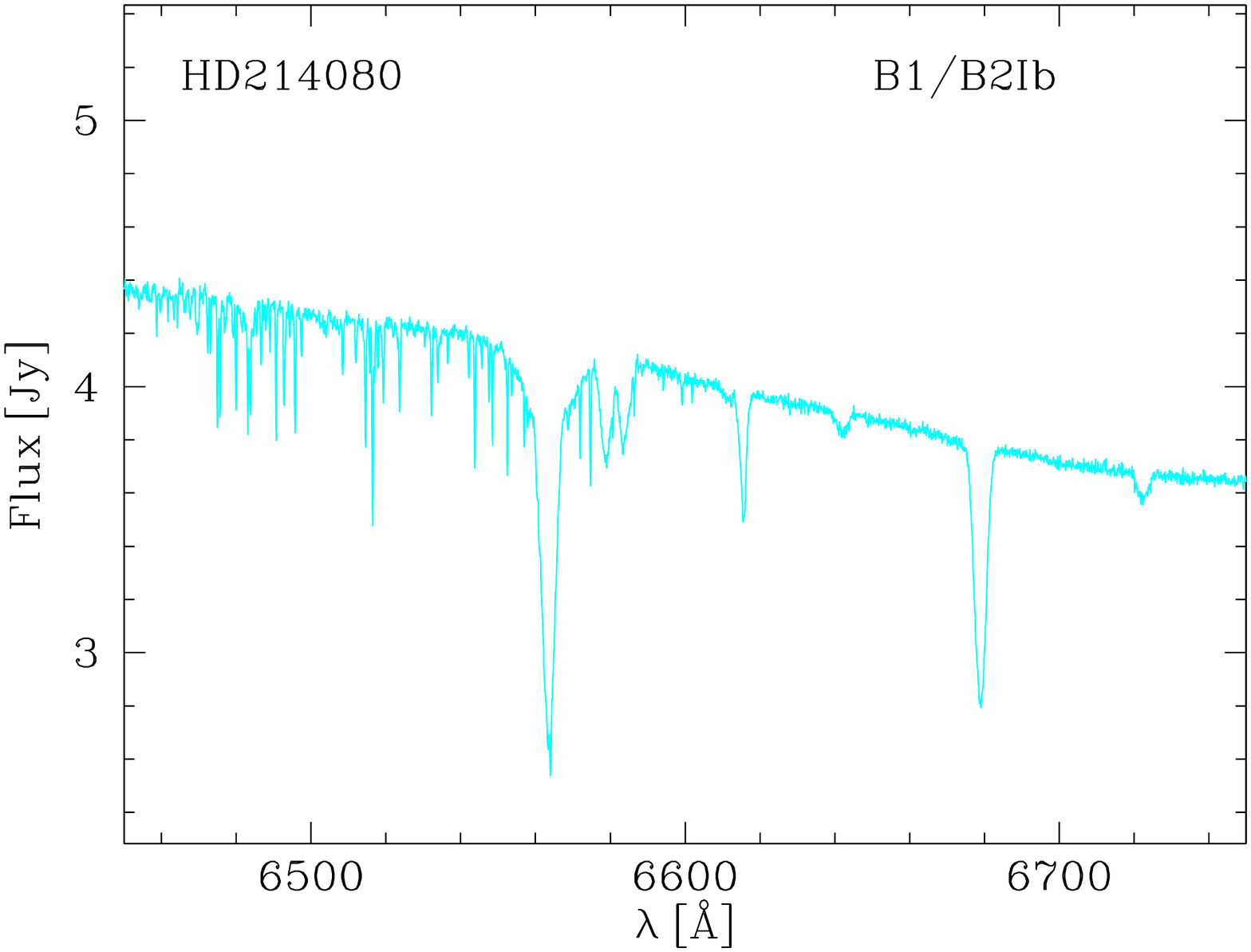}
\includegraphics[width=0.18\textwidth,angle=-90]{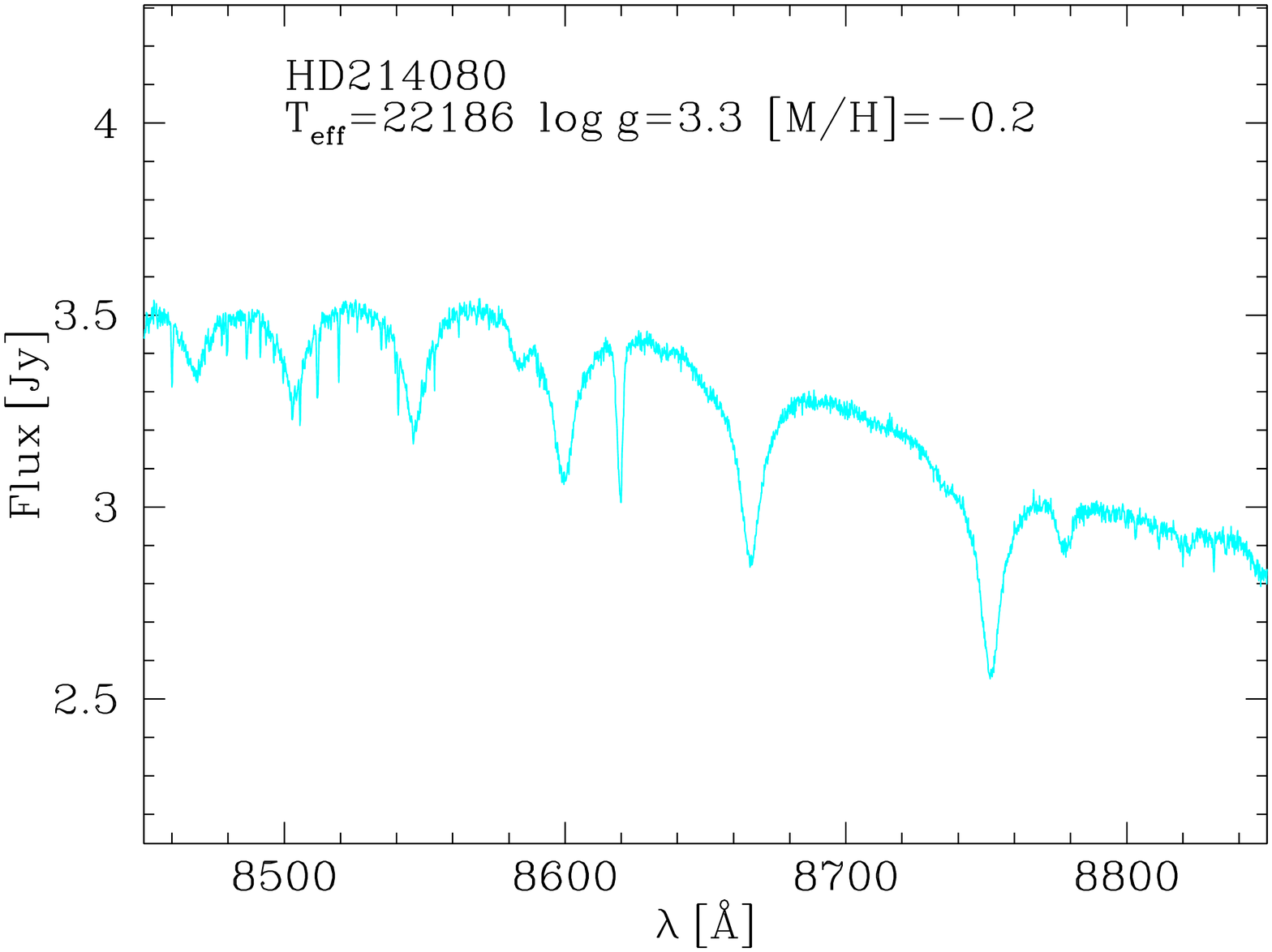}
\includegraphics[width=0.18\textwidth,angle=-90]{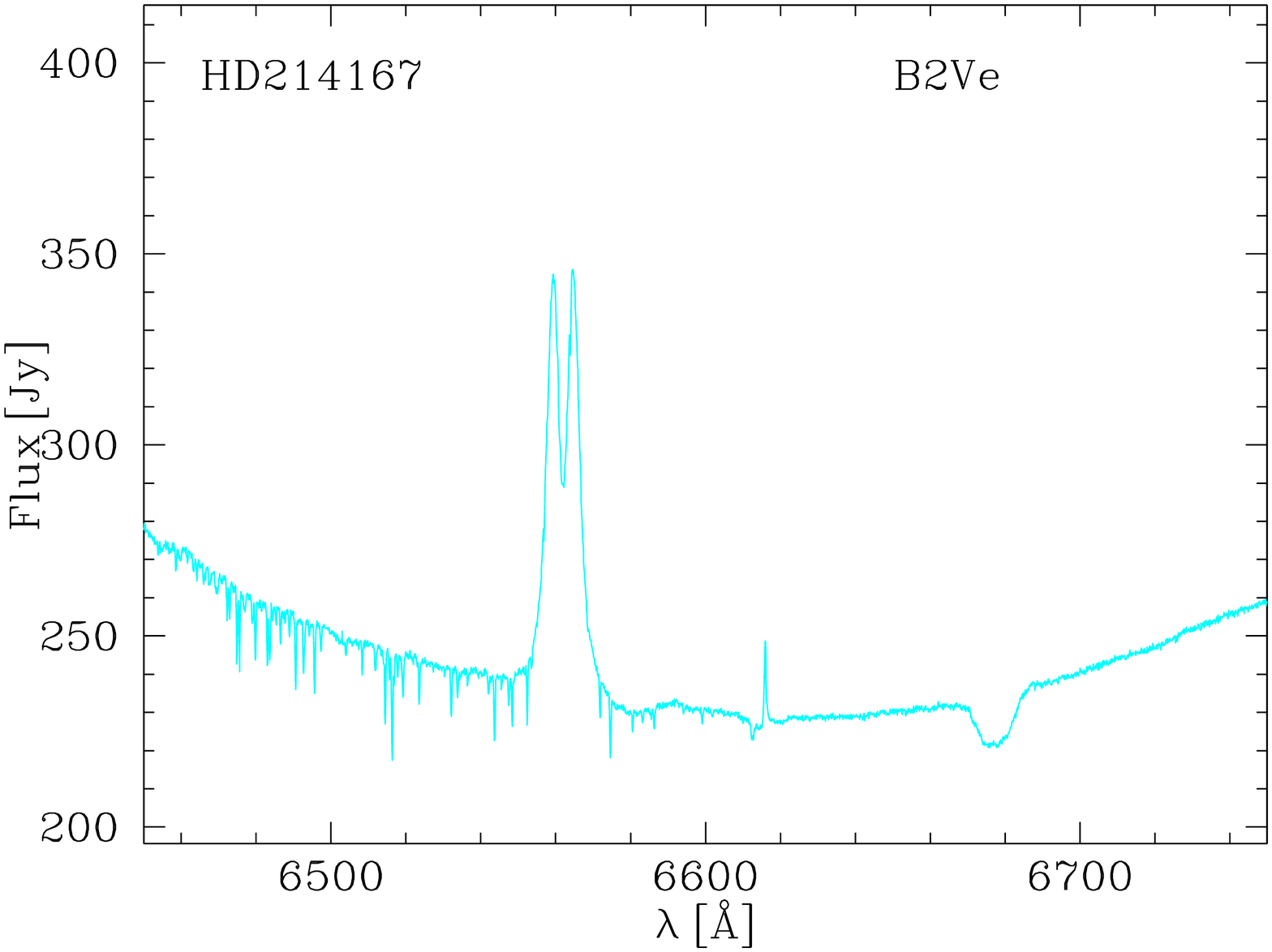}
\includegraphics[width=0.18\textwidth,angle=-90]{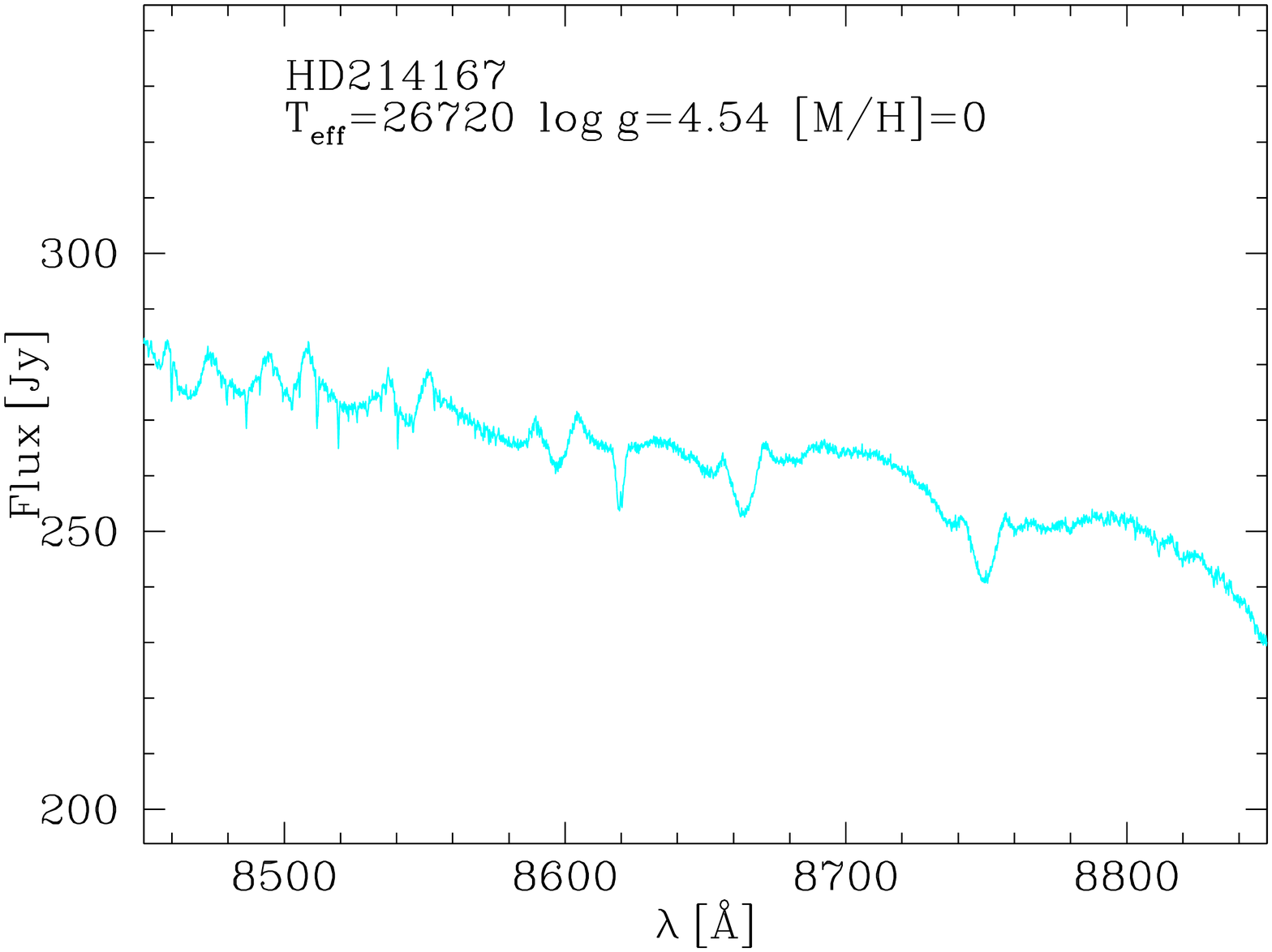}
\includegraphics[width=0.18\textwidth,angle=-90]{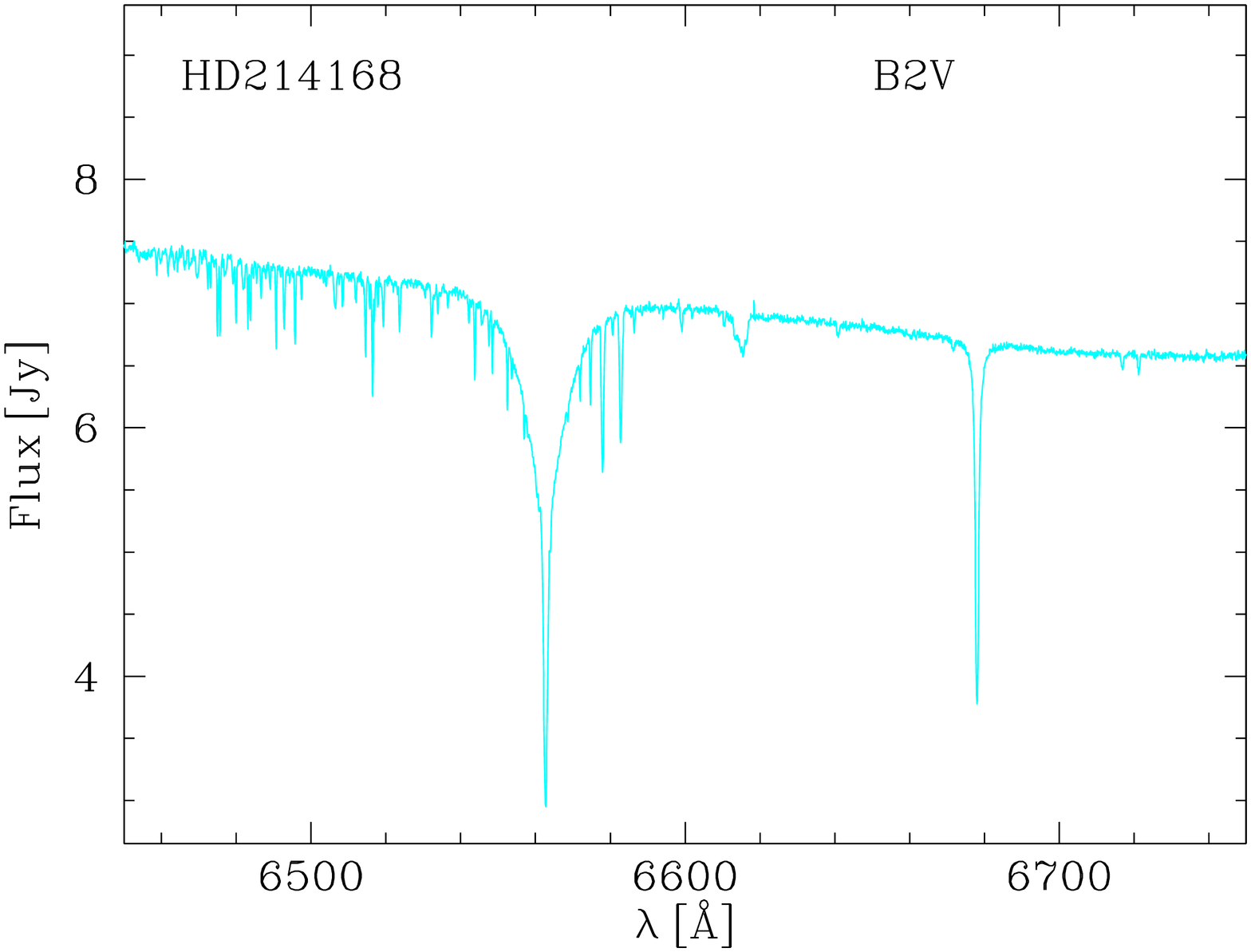}
\includegraphics[width=0.18\textwidth,angle=-90]{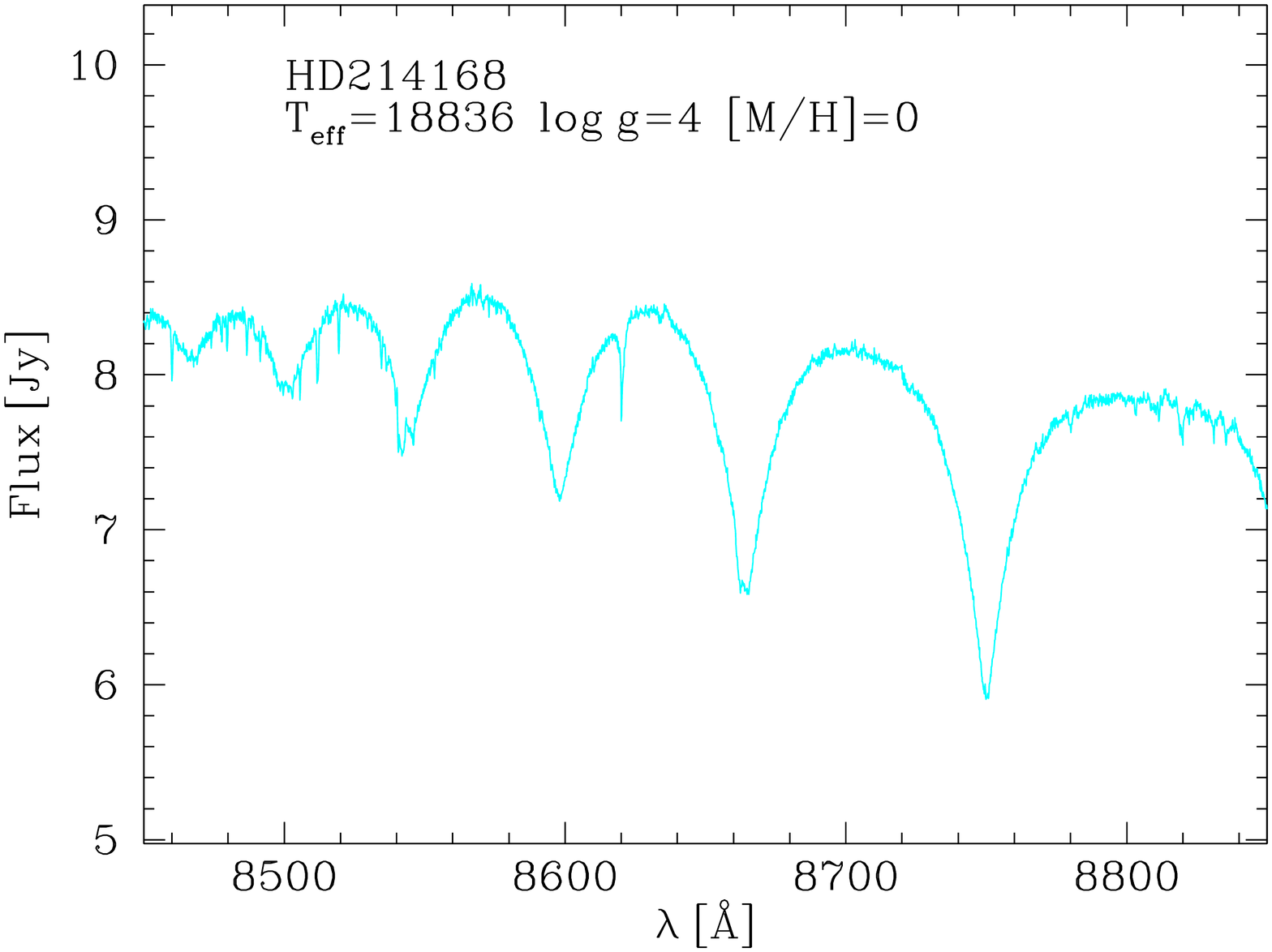}
\includegraphics[width=0.18\textwidth,angle=-90]{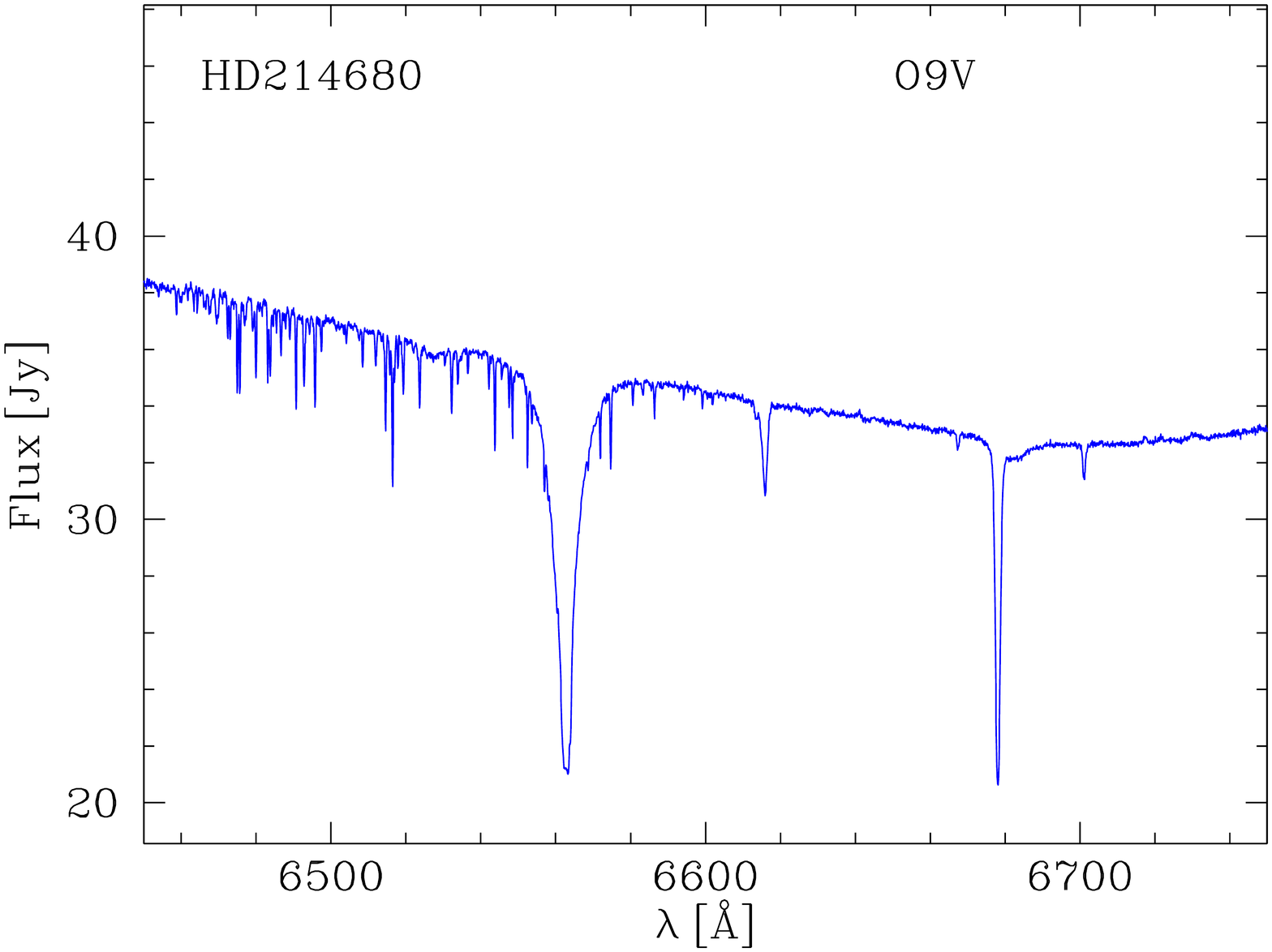}
\includegraphics[width=0.18\textwidth,angle=-90]{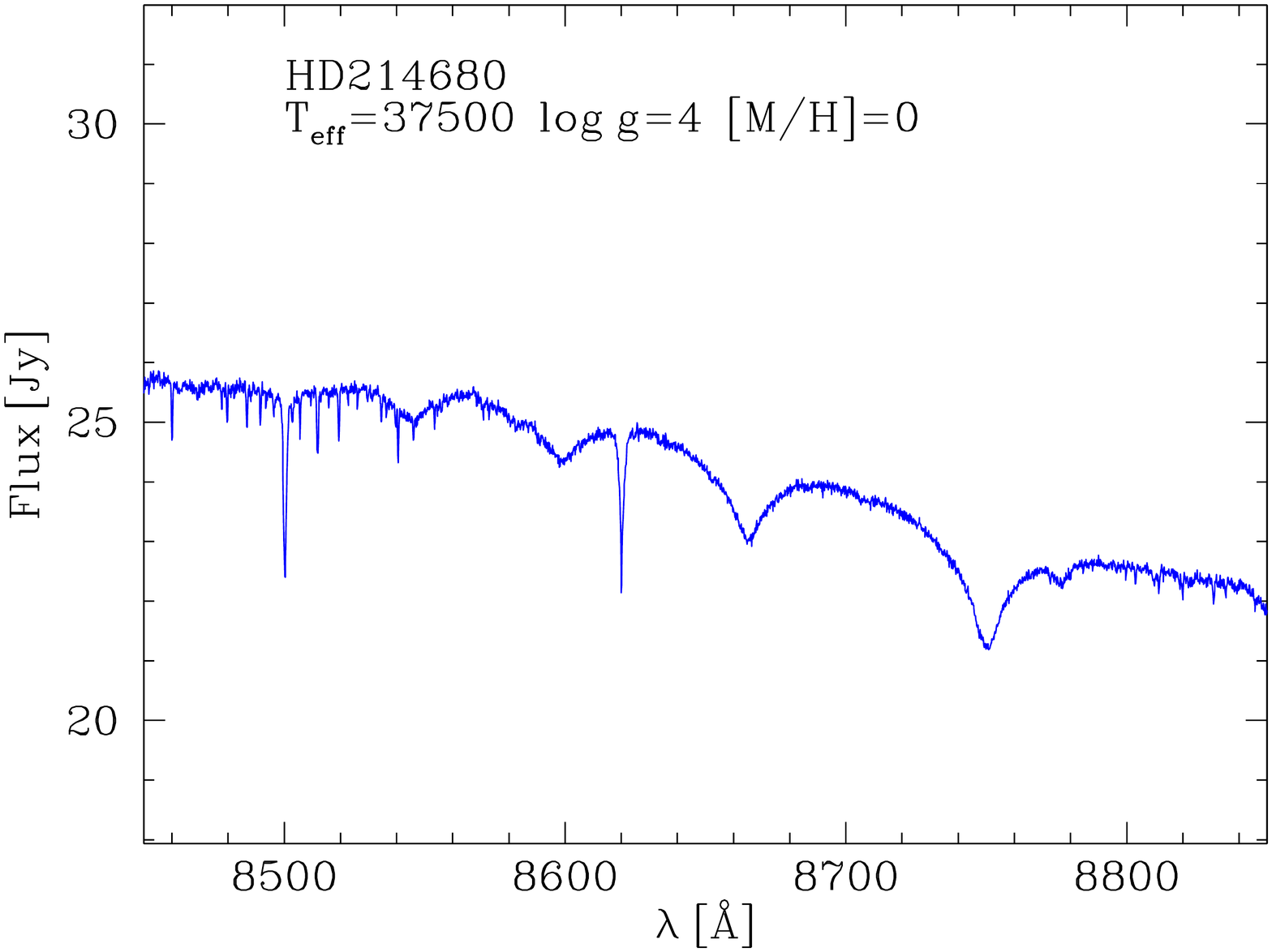}
\includegraphics[width=0.18\textwidth,angle=-90]{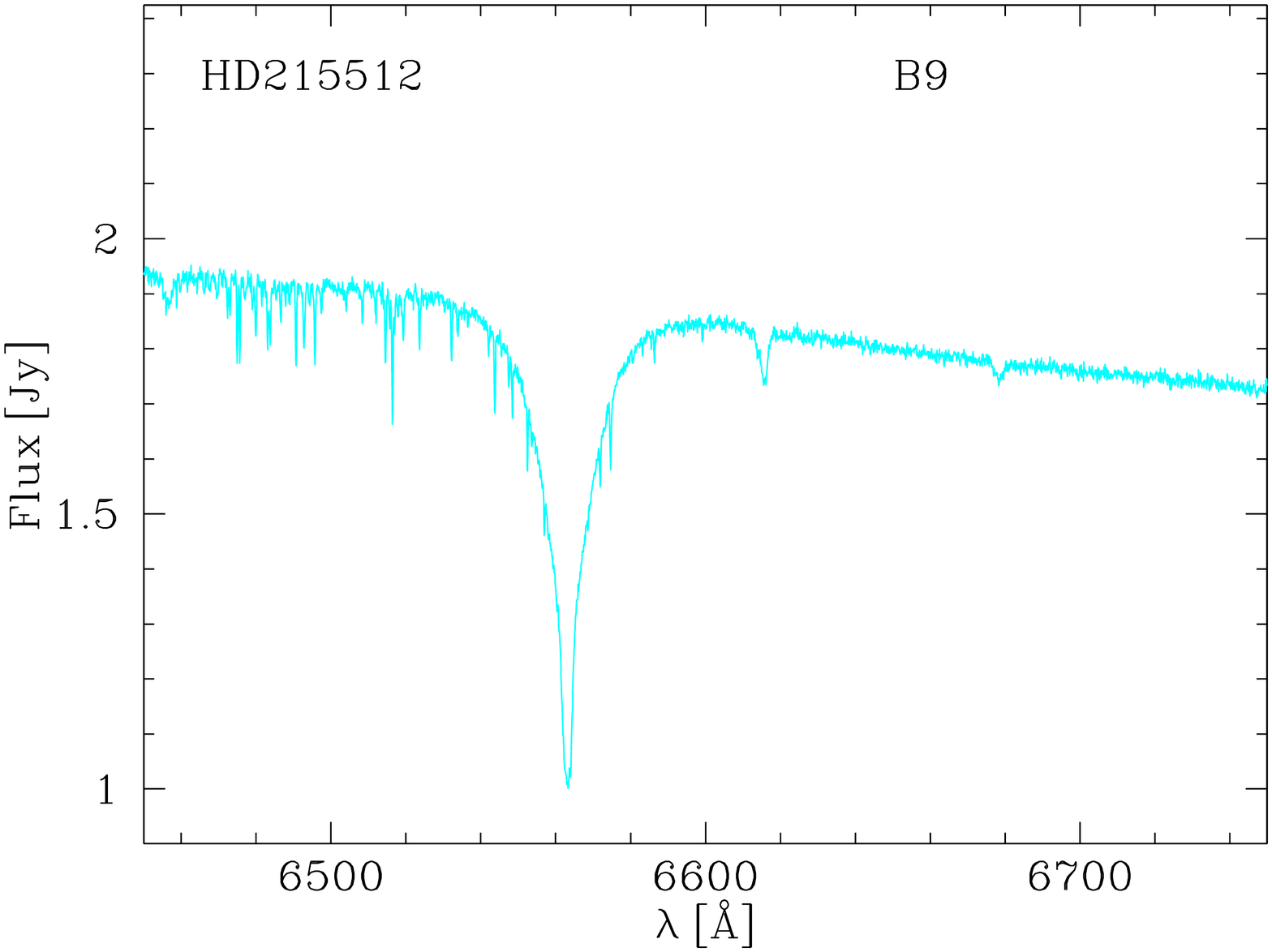}
\includegraphics[width=0.18\textwidth,angle=-90]{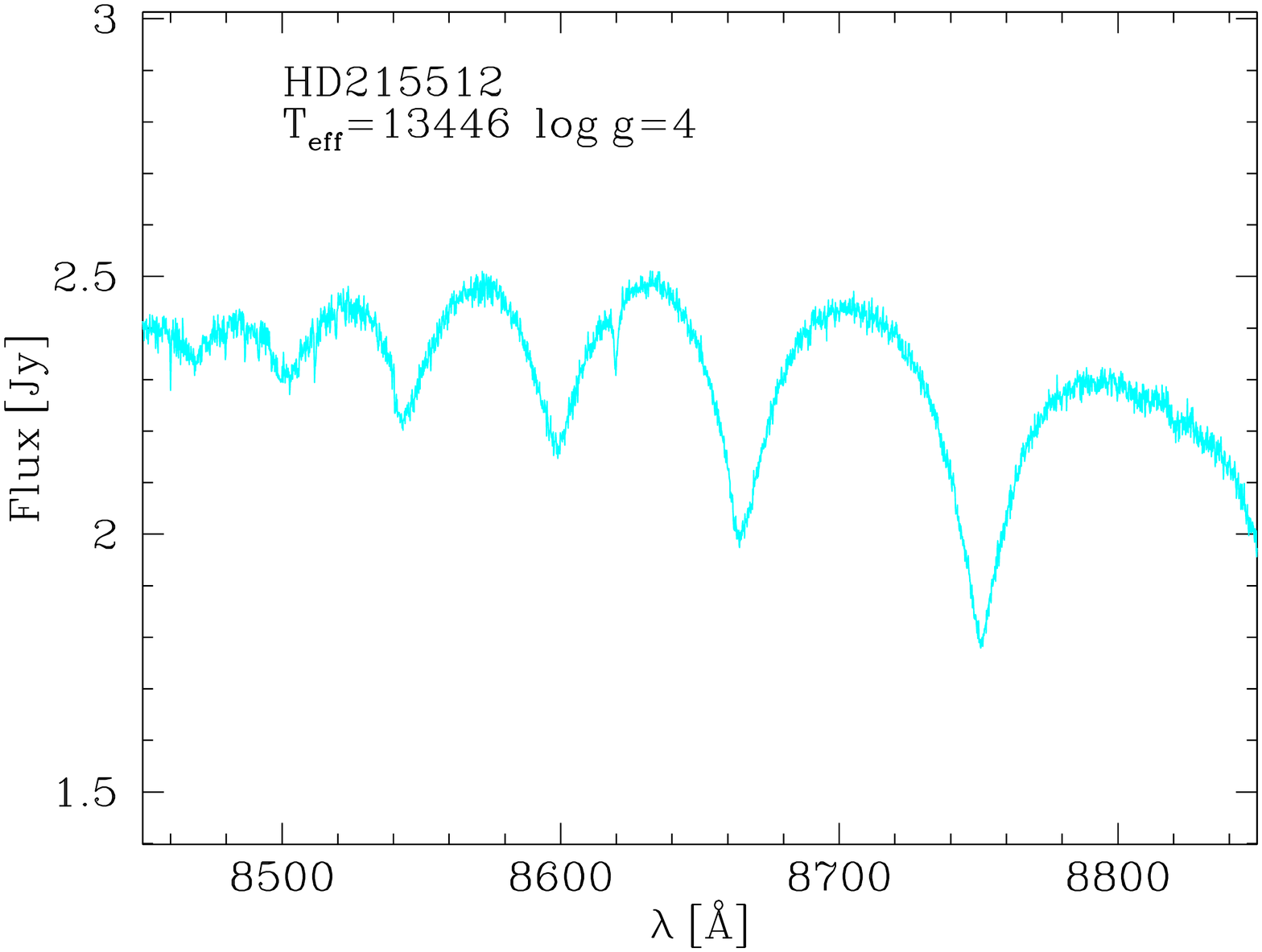}

\contcaption{28. Stars shown in this page are: HD209459, HD209975, HD210809, HD211472, HD212076, HD212442, HD212454, HD212593, HD213420, HD214080, HD214167, HD214168, HD214680 and HD215512.}
\end{figure*}

\begin{figure*}
\includegraphics[width=0.18\textwidth,angle=-90]{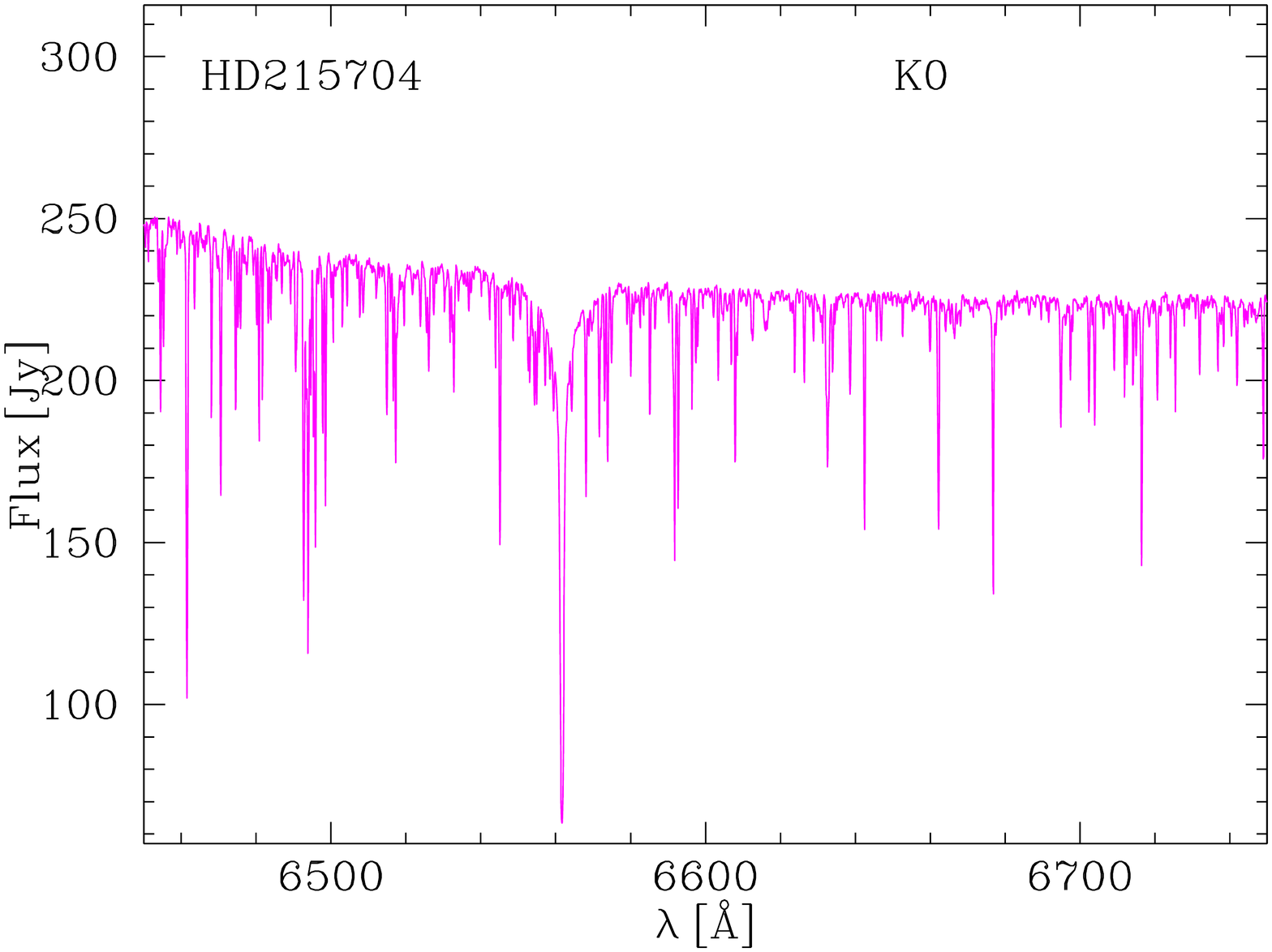}
\includegraphics[width=0.18\textwidth,angle=-90]{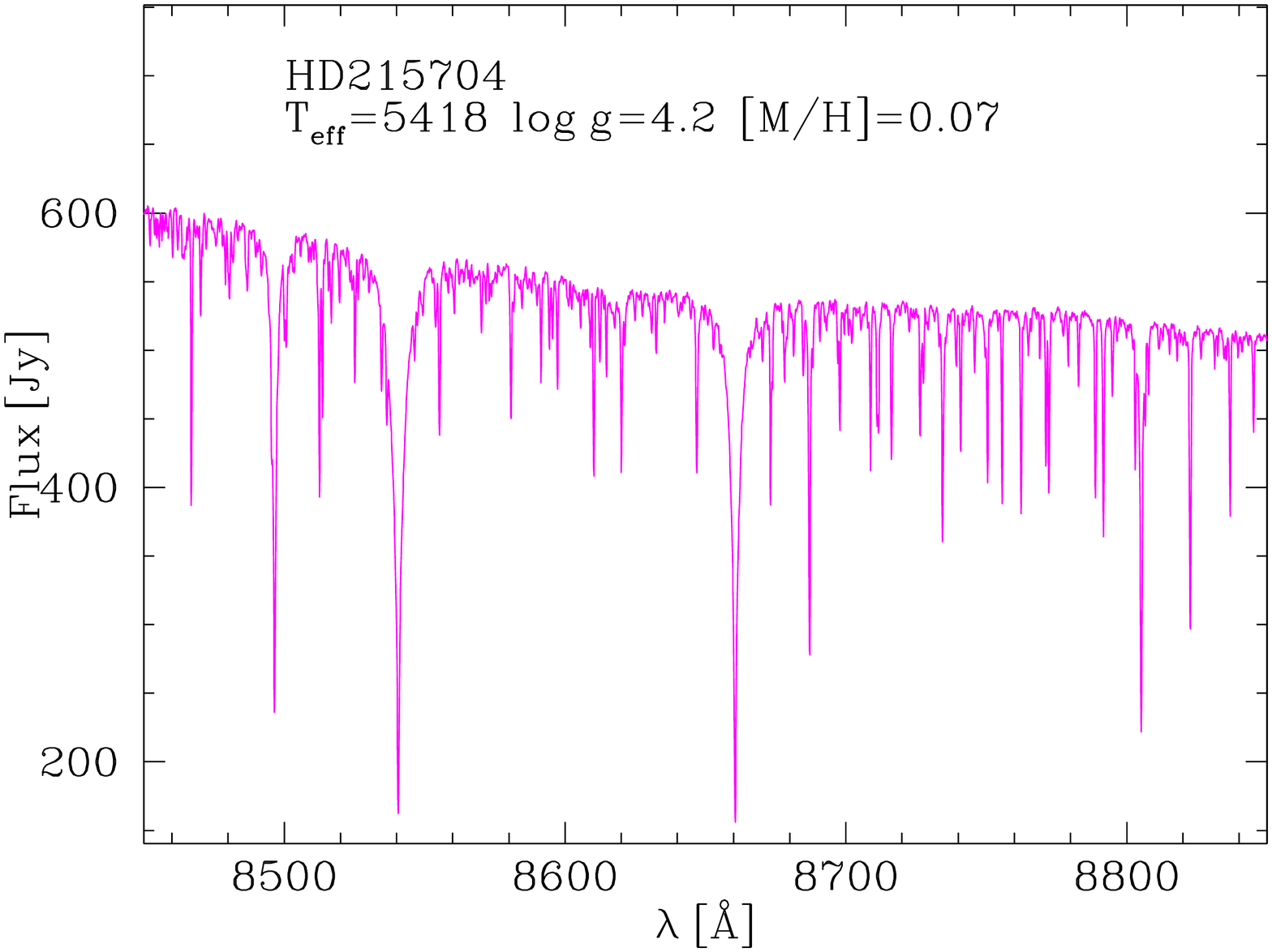}
\includegraphics[width=0.18\textwidth,angle=-90]{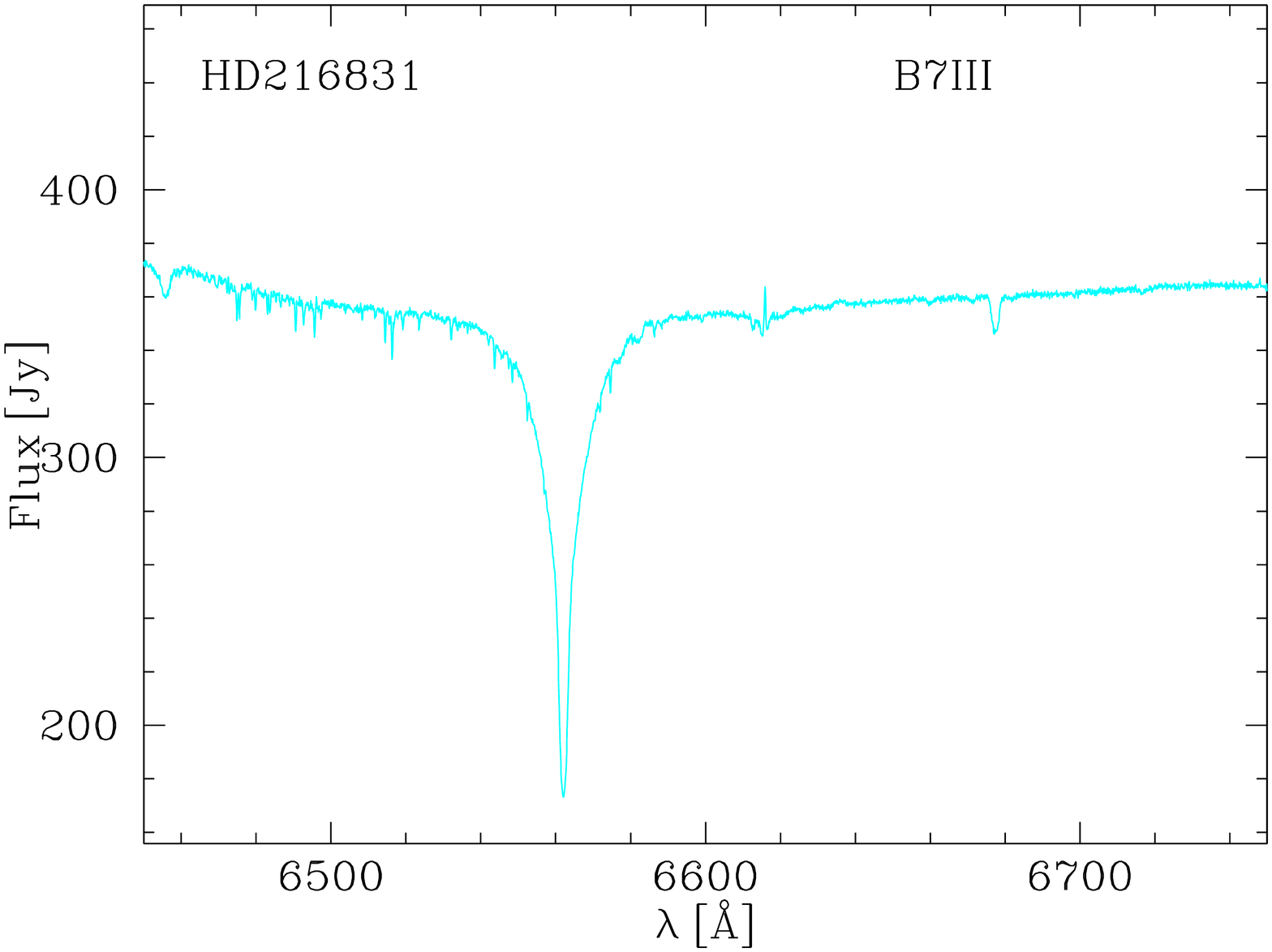}
\includegraphics[width=0.18\textwidth,angle=-90]{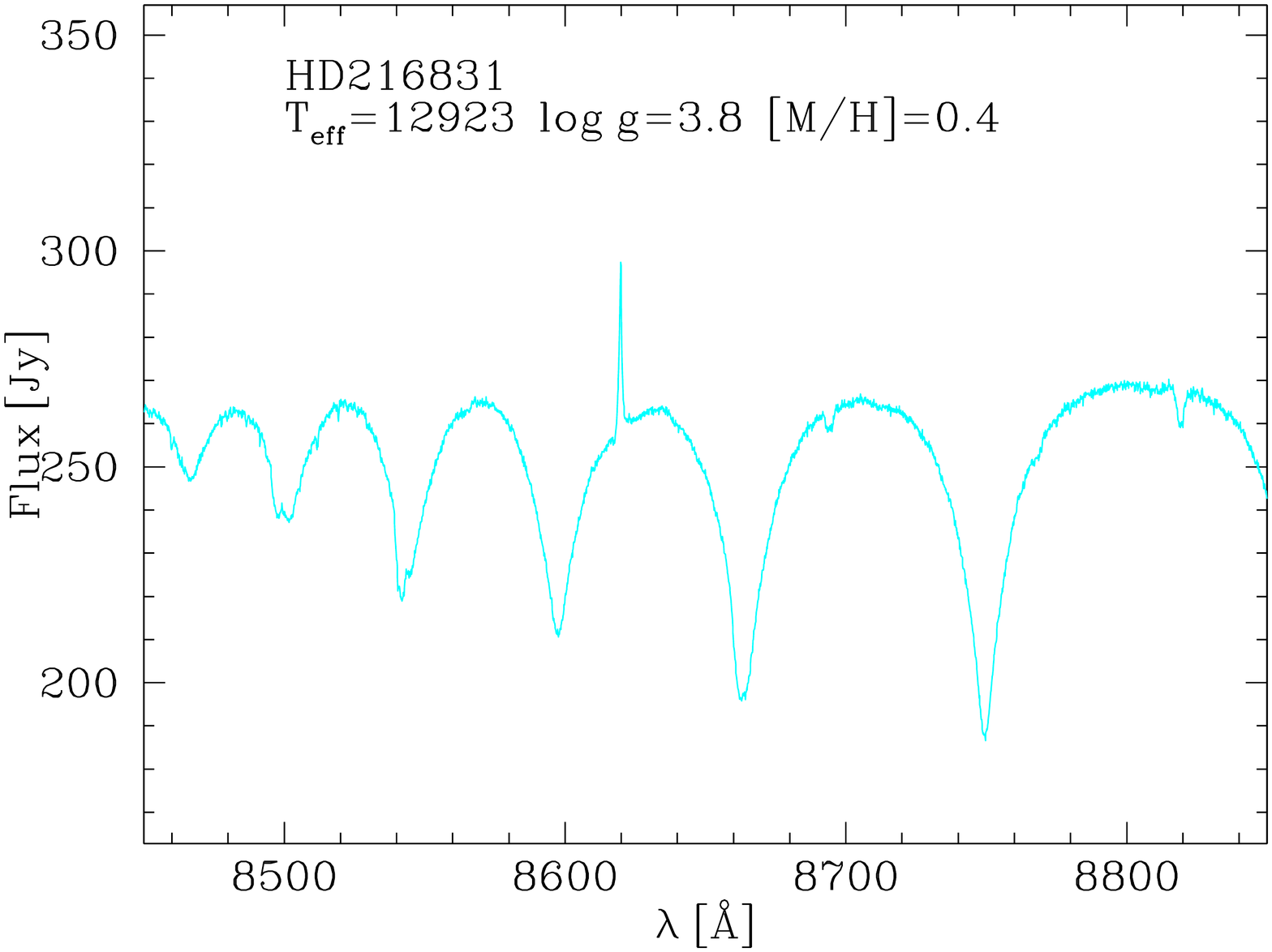}
\includegraphics[width=0.18\textwidth,angle=-90]{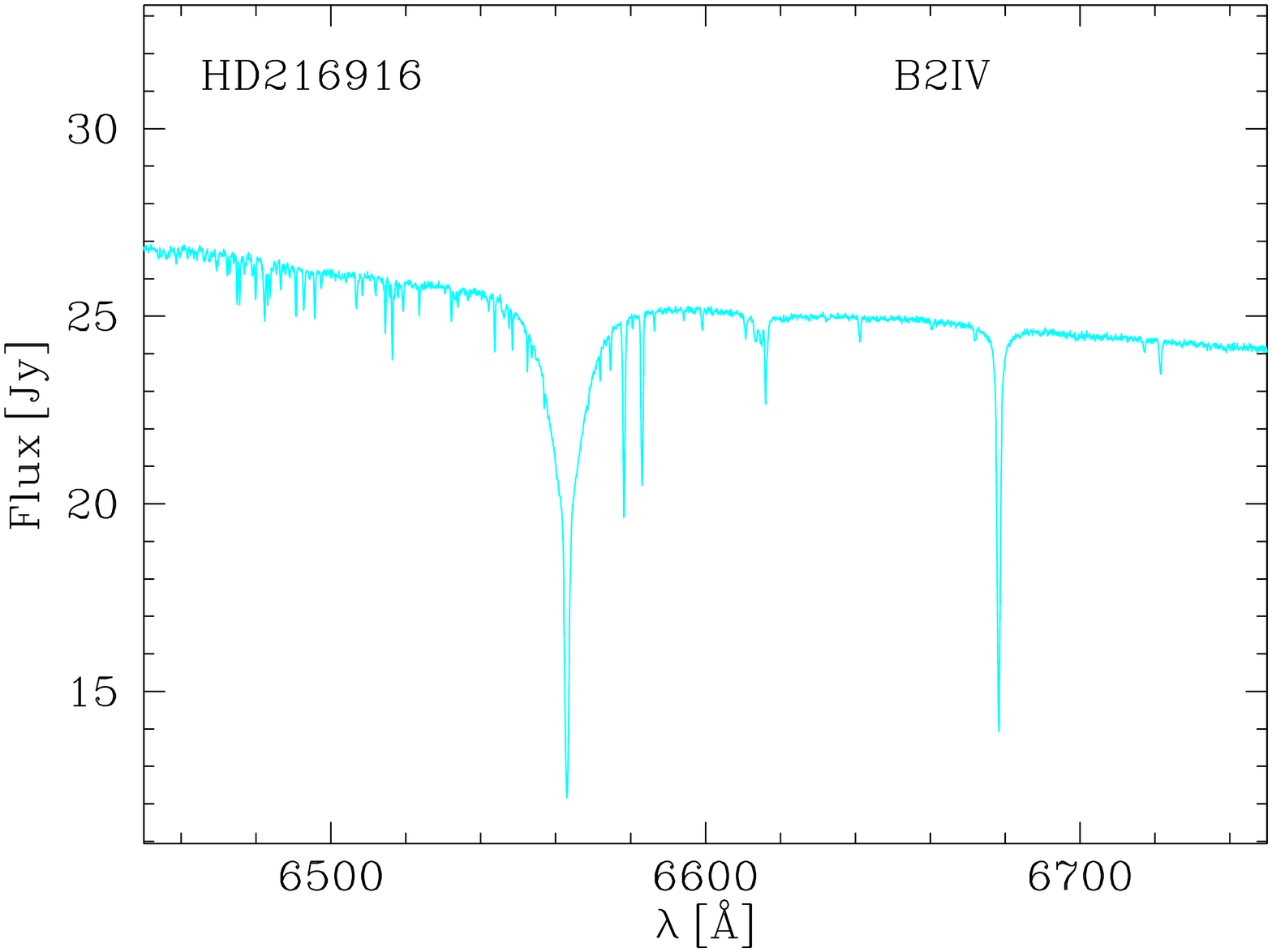}
\includegraphics[width=0.18\textwidth,angle=-90]{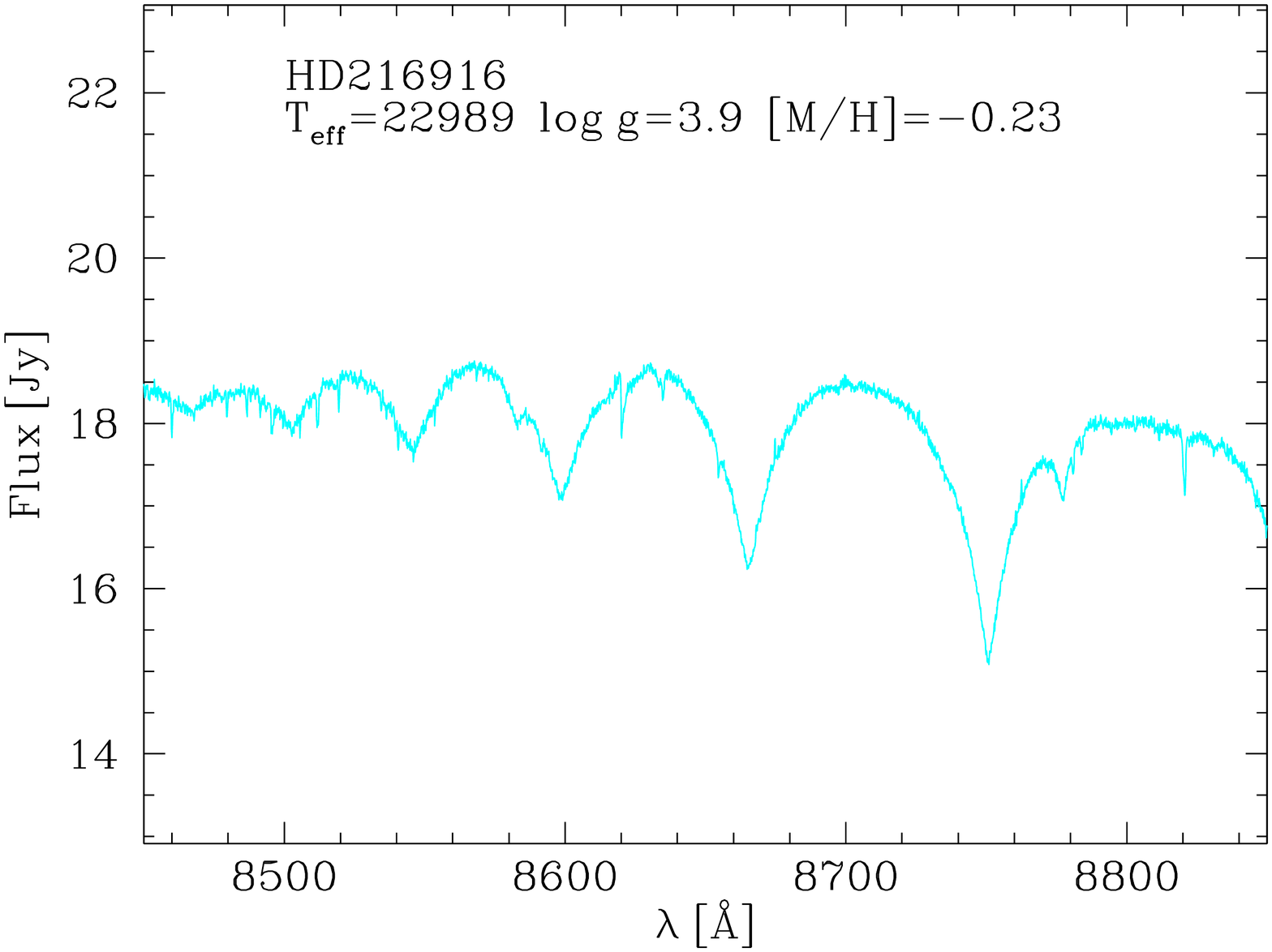}
\includegraphics[width=0.18\textwidth,angle=-90]{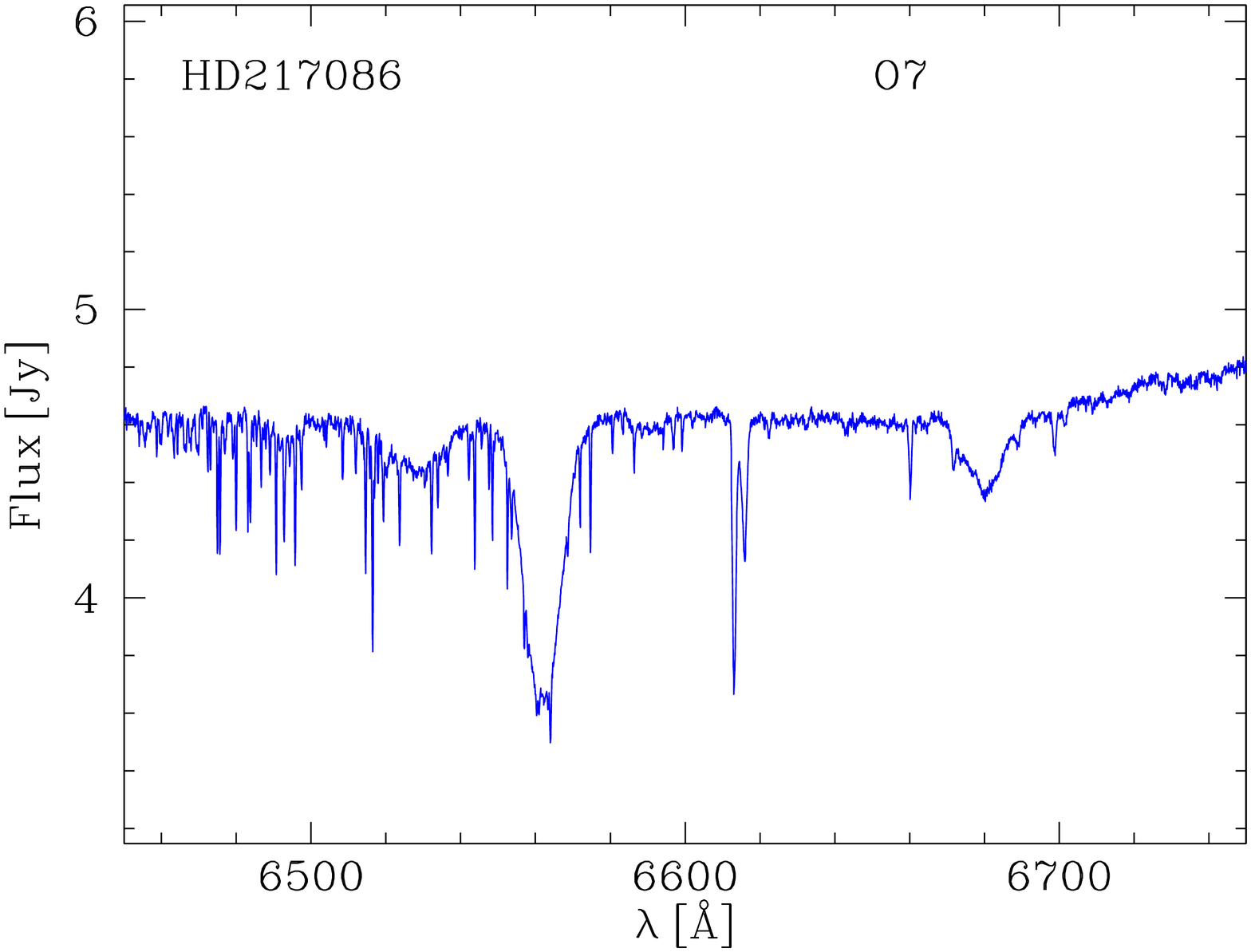}
\includegraphics[width=0.18\textwidth,angle=-90]{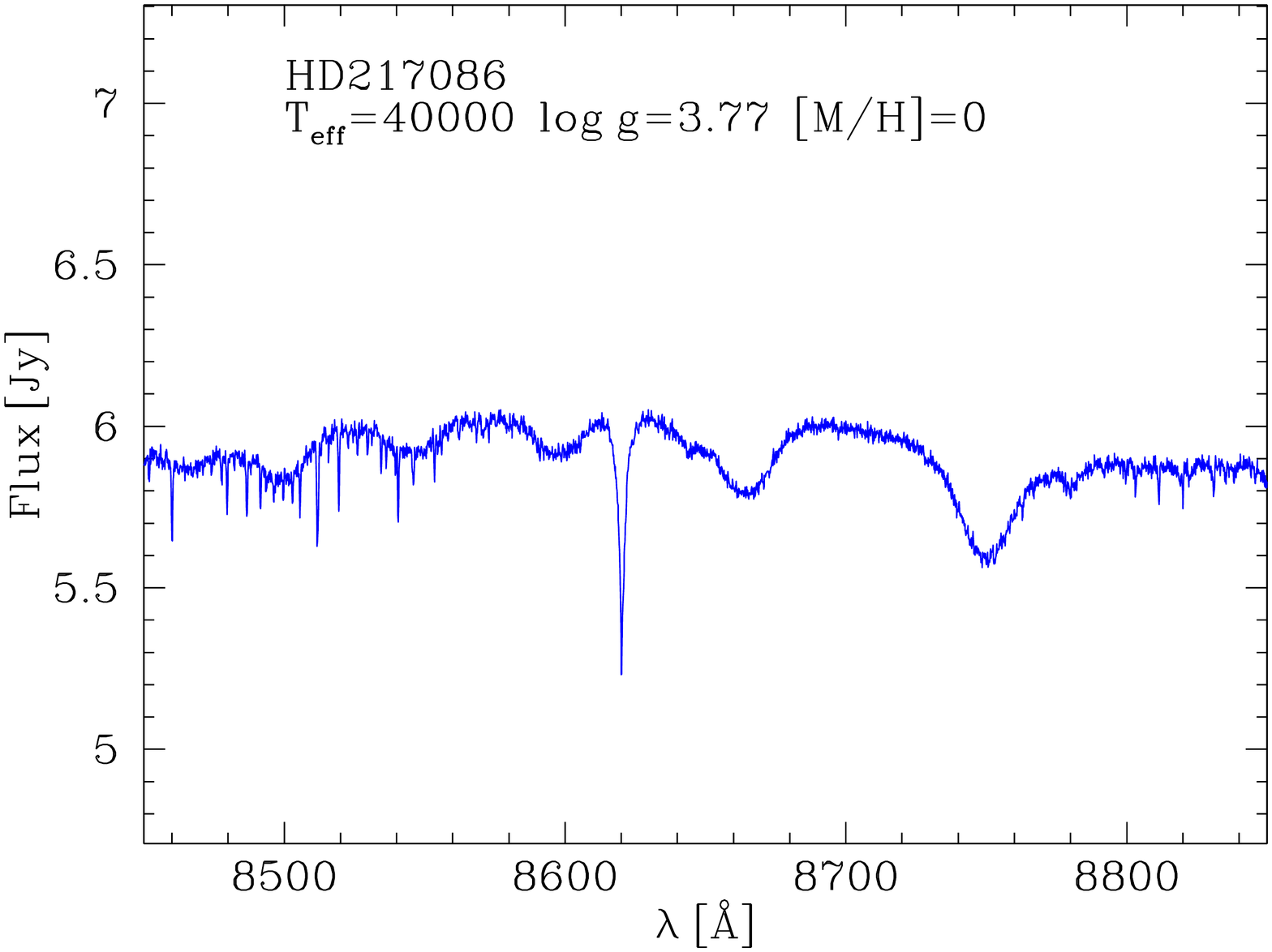}
\includegraphics[width=0.18\textwidth,angle=-90]{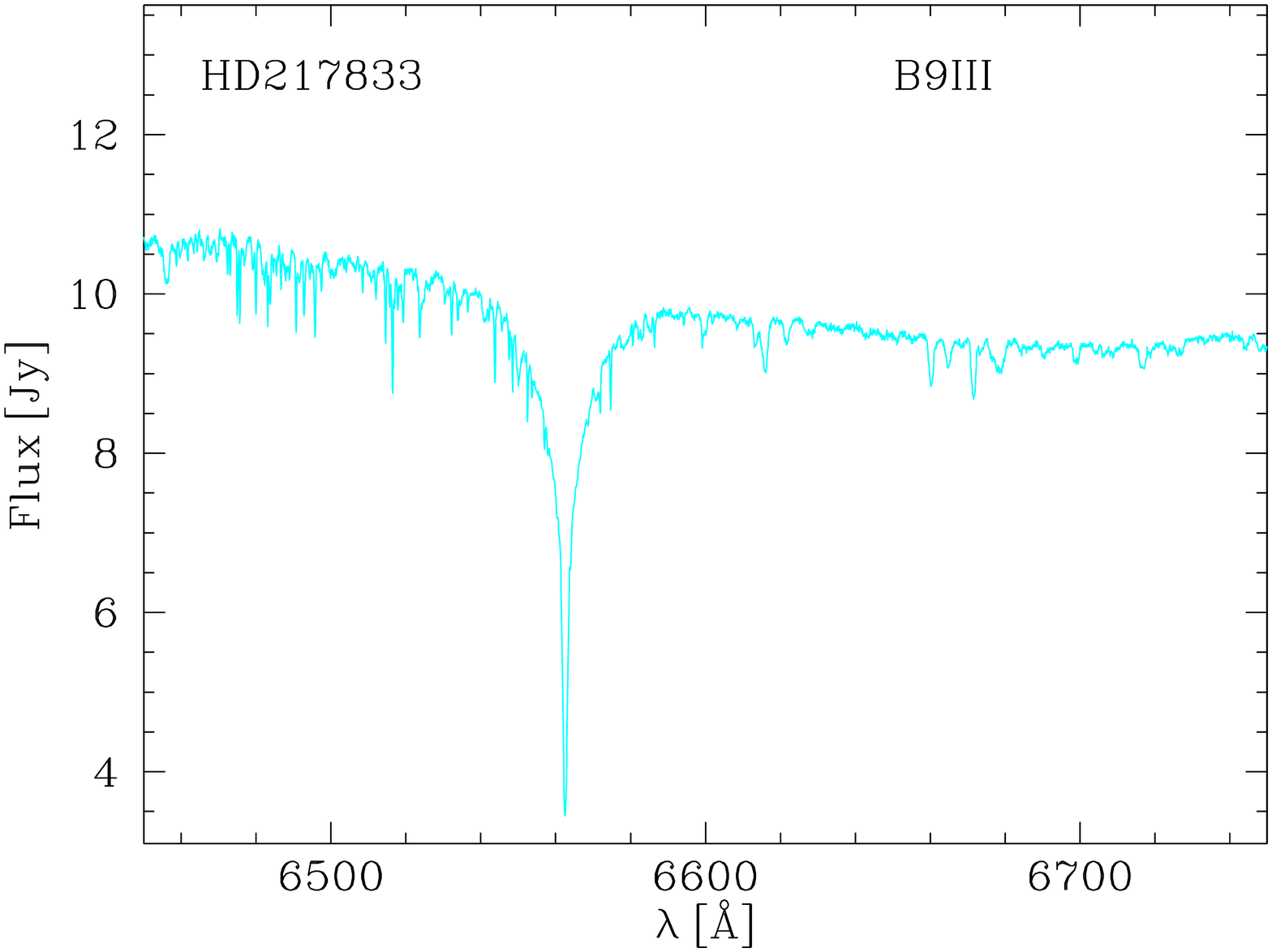}
\includegraphics[width=0.18\textwidth,angle=-90]{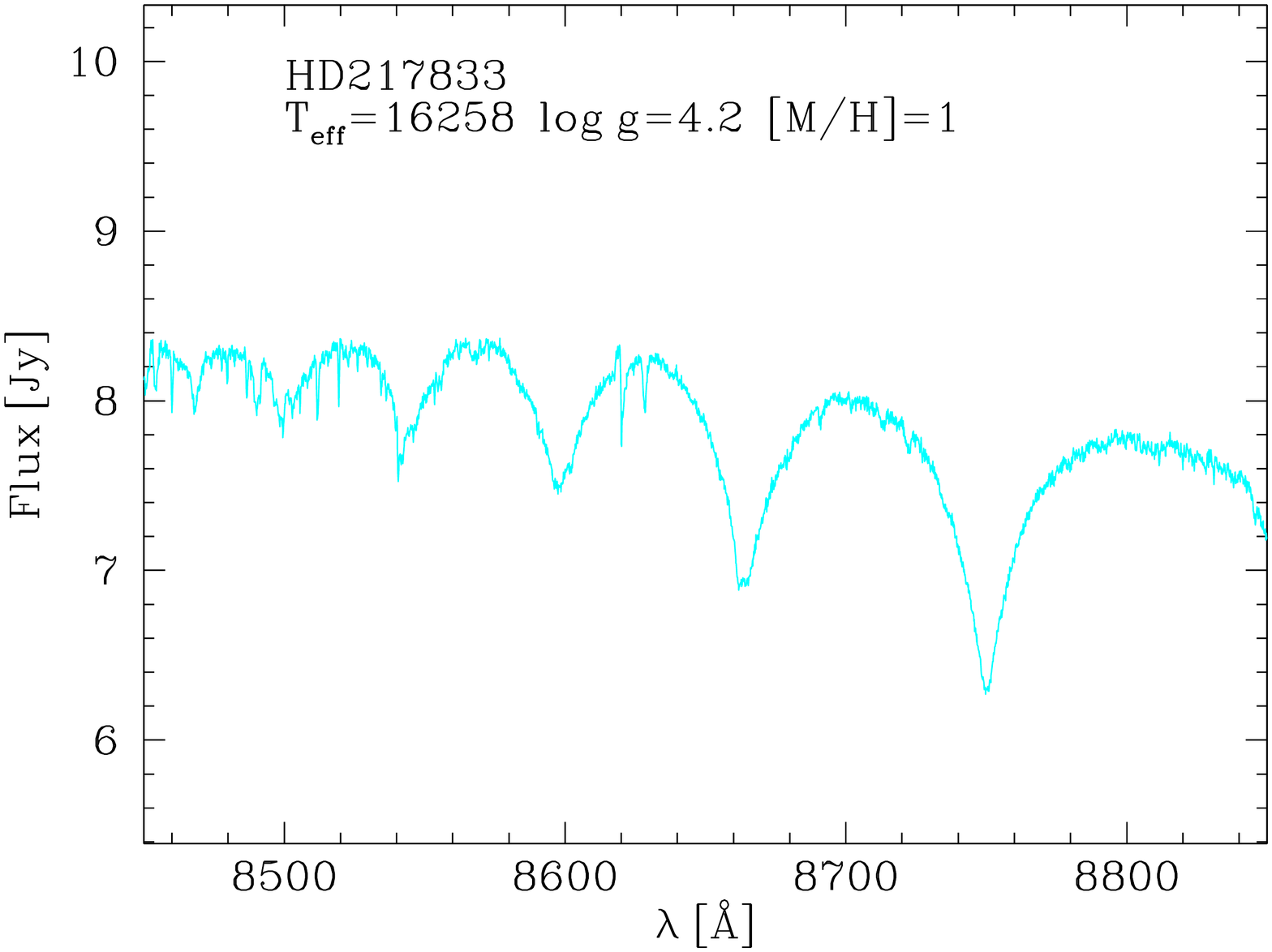}
\includegraphics[width=0.18\textwidth,angle=-90]{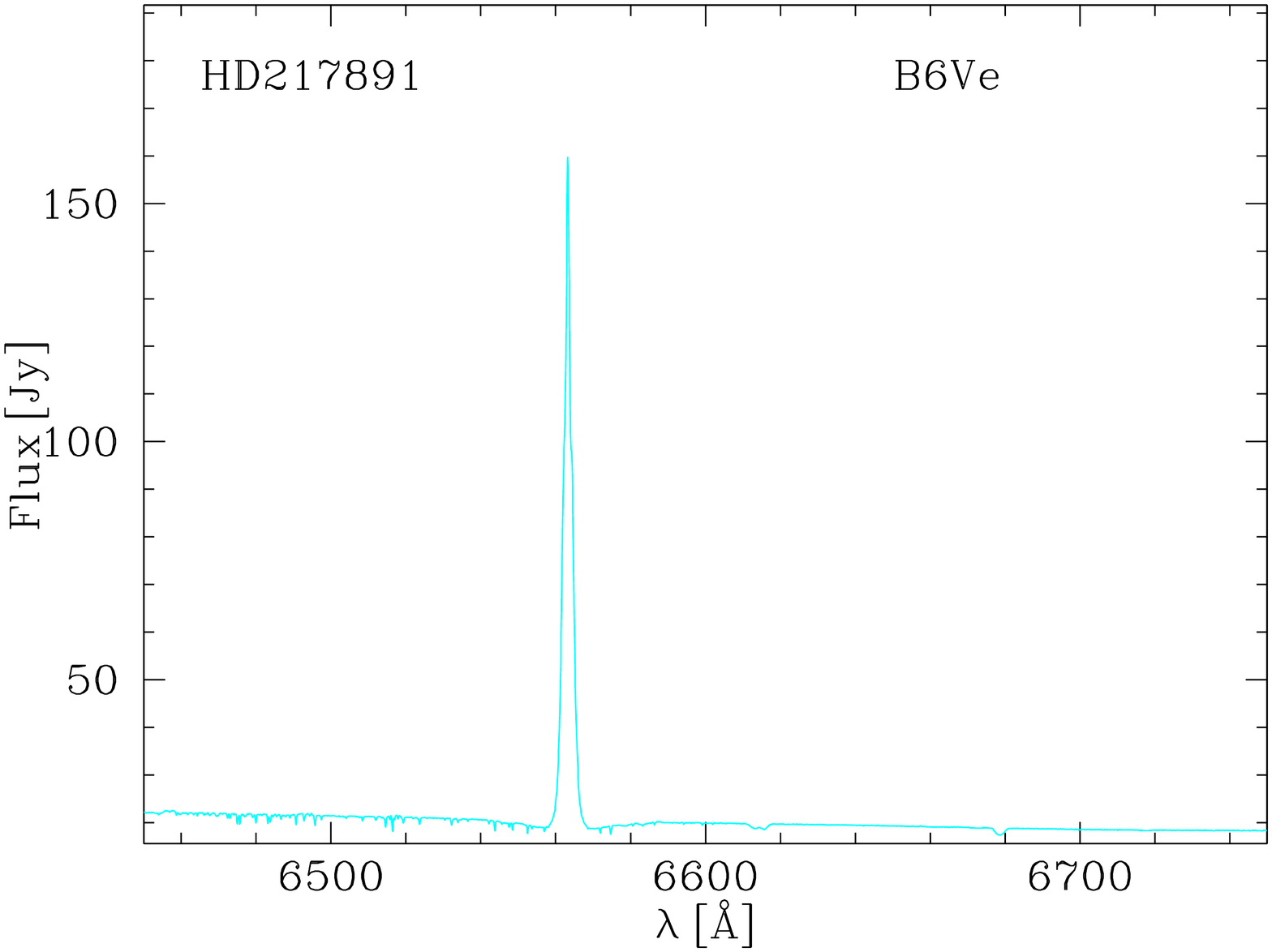}
\includegraphics[width=0.18\textwidth,angle=-90]{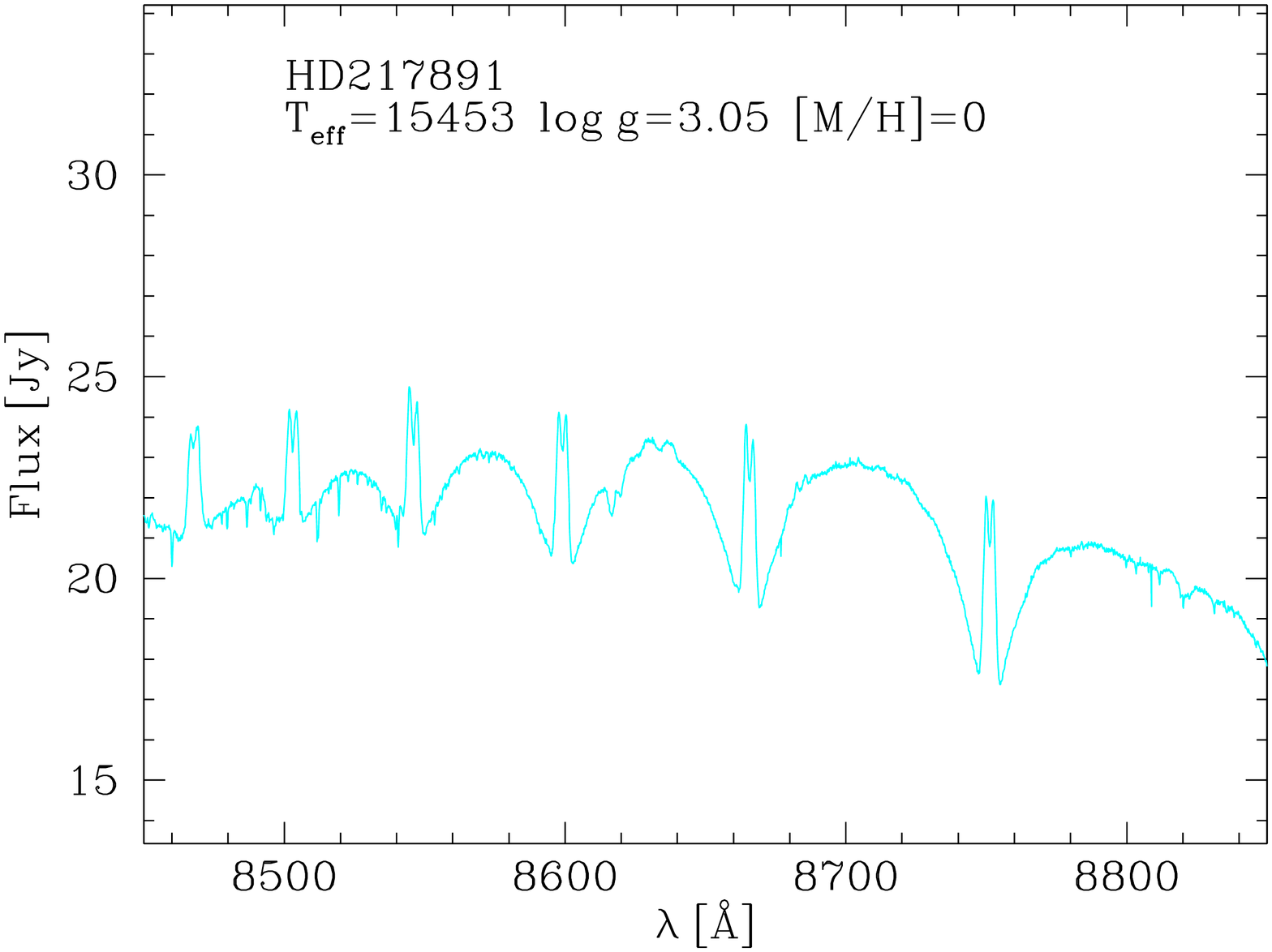}
\includegraphics[width=0.18\textwidth,angle=-90]{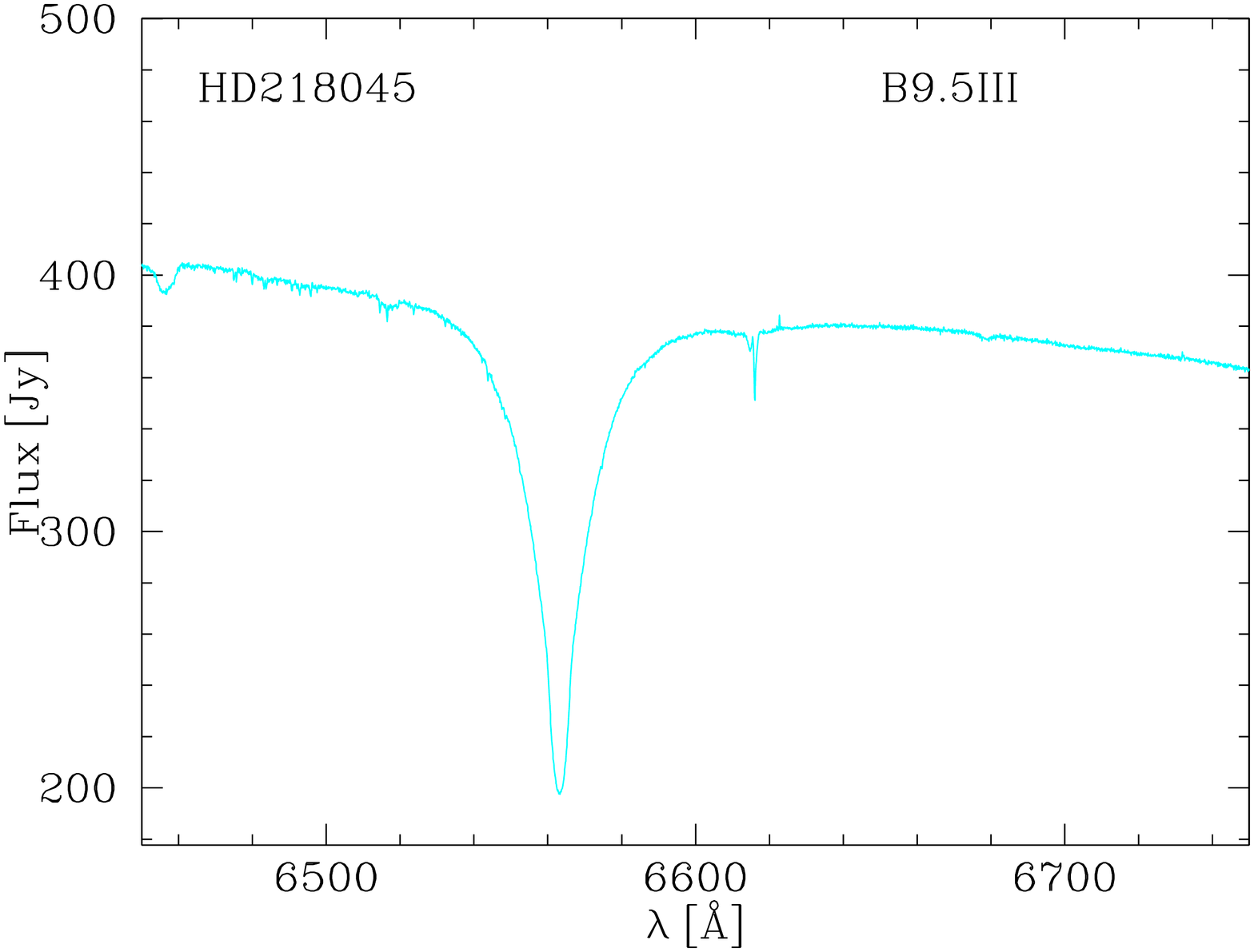}
\includegraphics[width=0.18\textwidth,angle=-90]{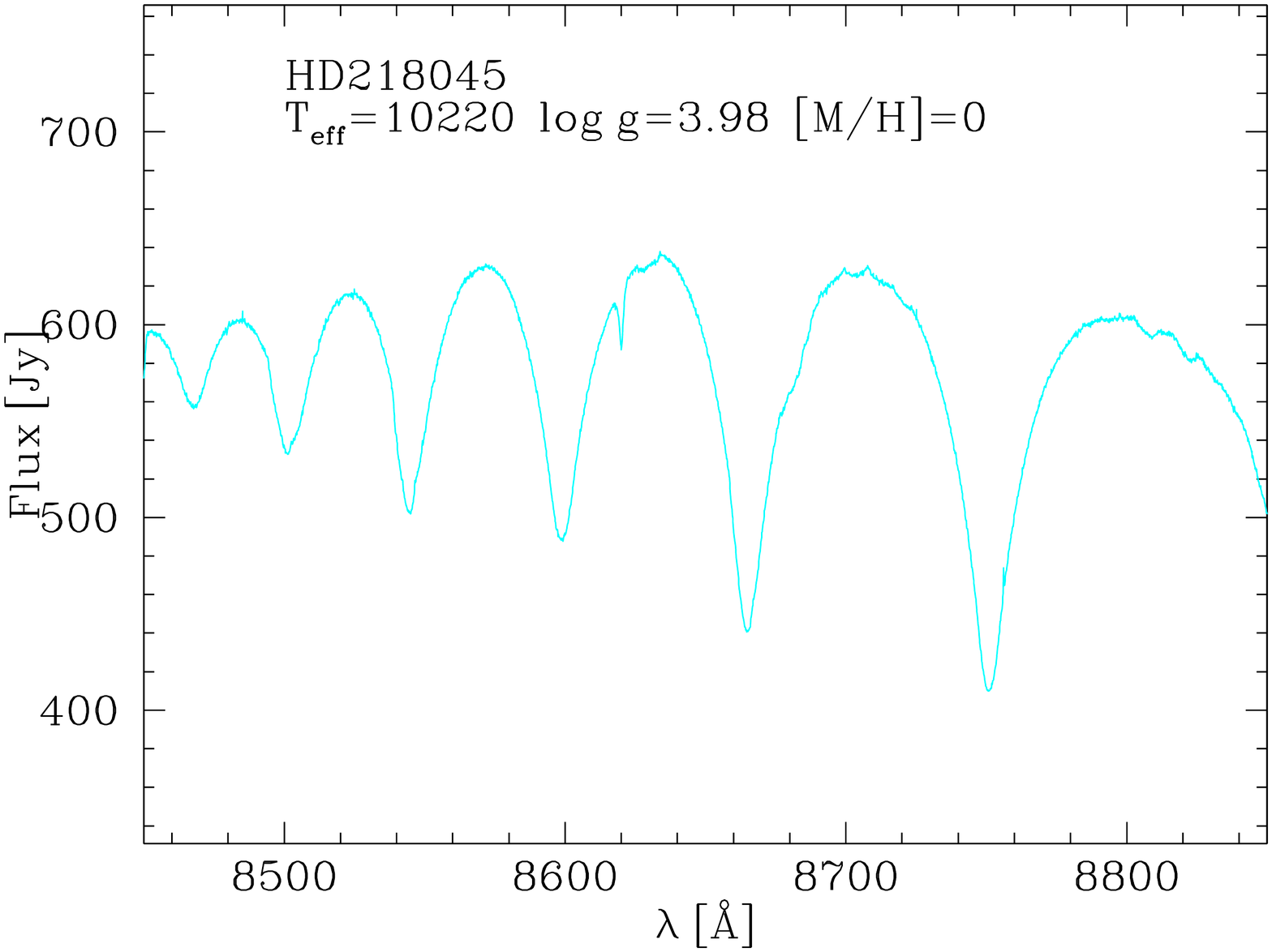}
\includegraphics[width=0.18\textwidth,angle=-90]{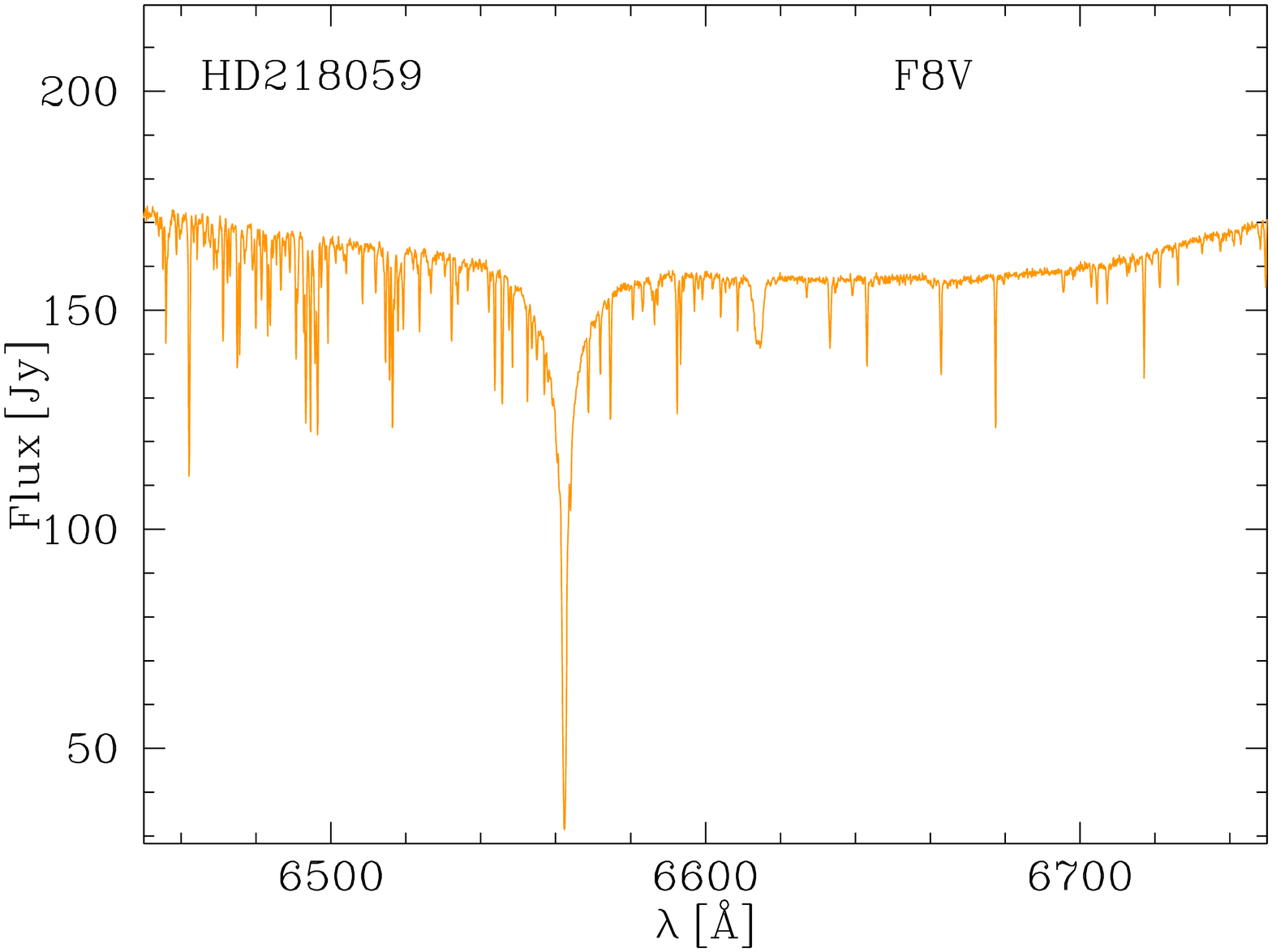}
\includegraphics[width=0.18\textwidth,angle=-90]{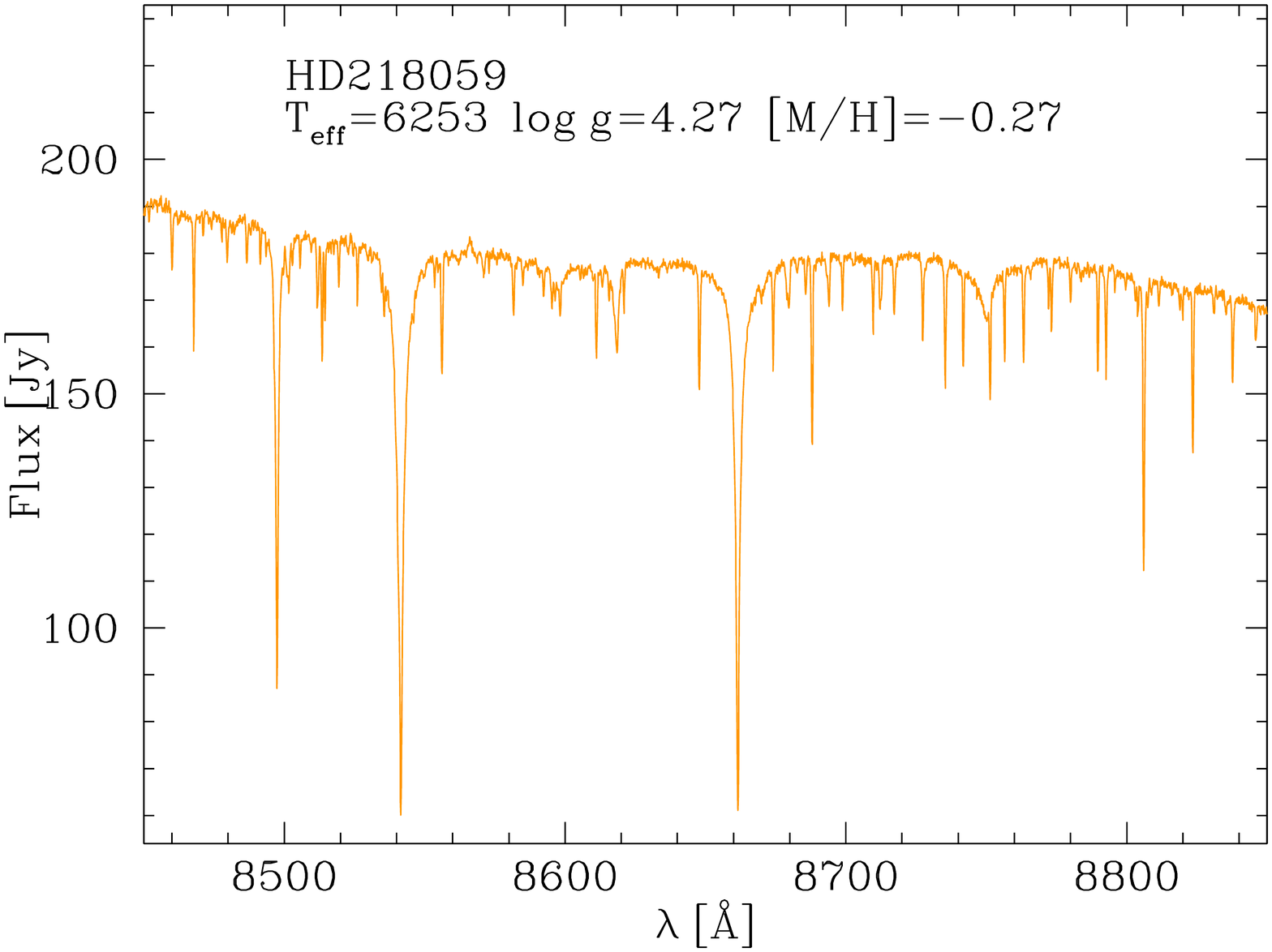}
\includegraphics[width=0.18\textwidth,angle=-90]{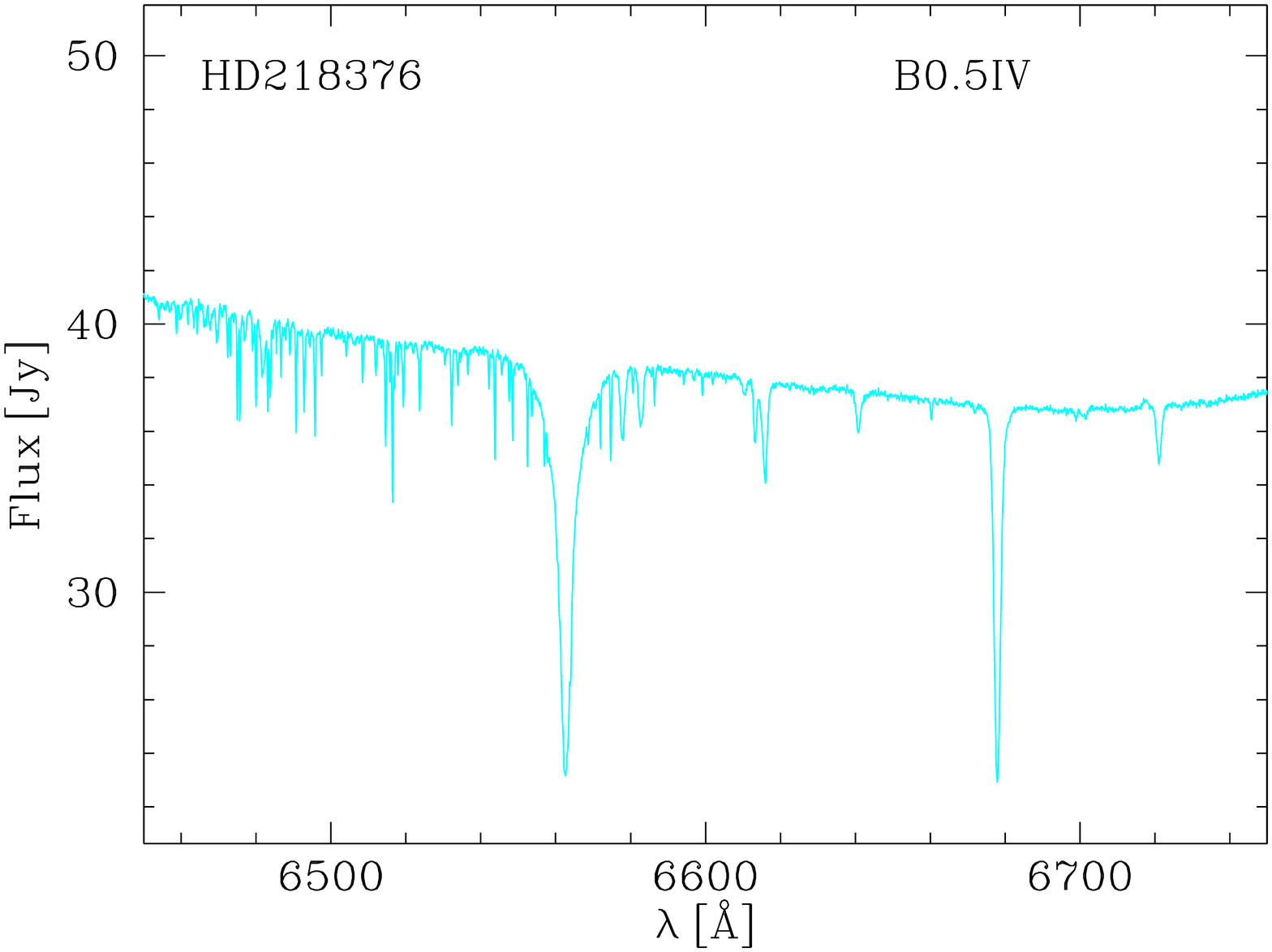}
\includegraphics[width=0.18\textwidth,angle=-90]{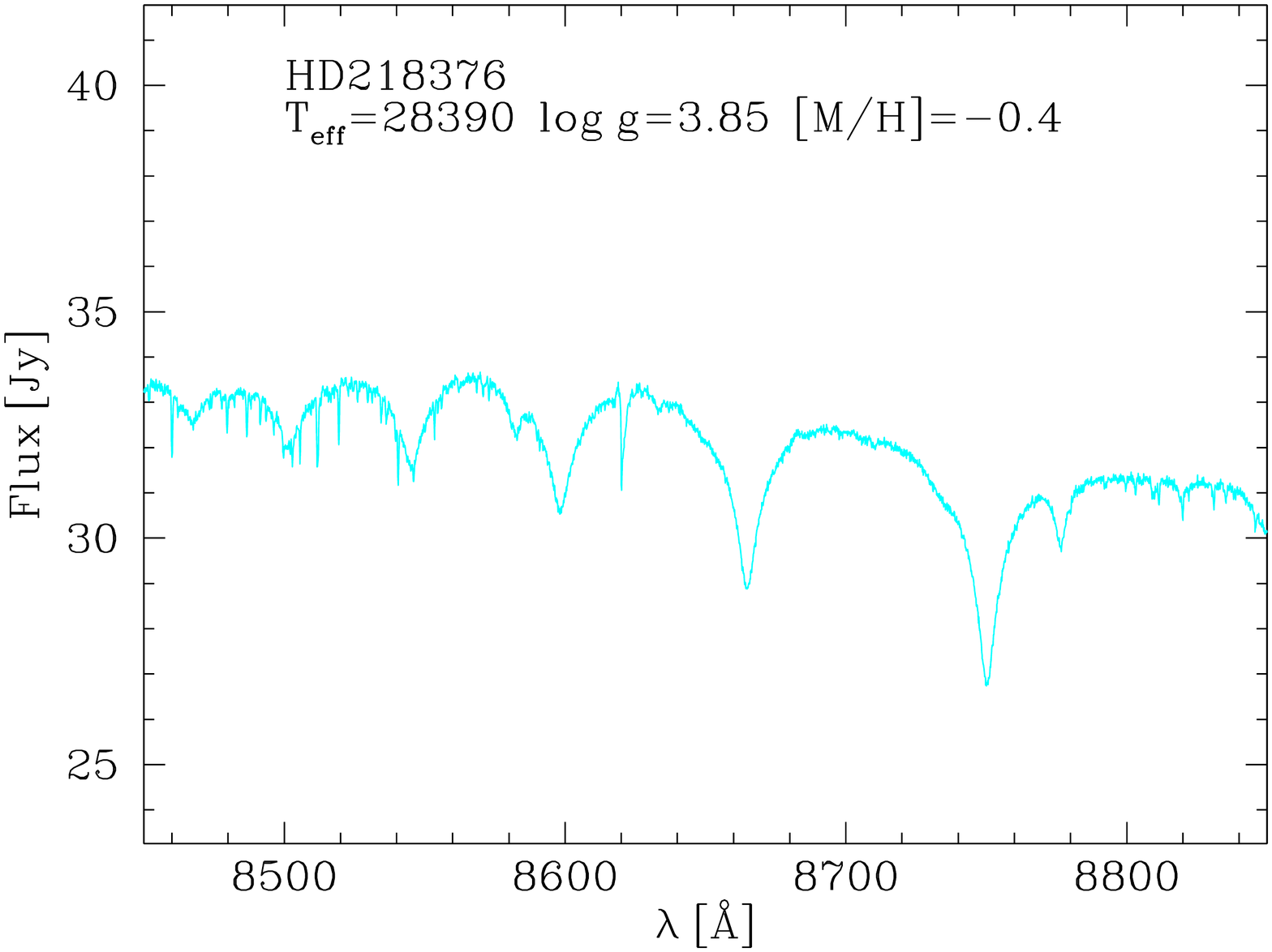}
\includegraphics[width=0.18\textwidth,angle=-90]{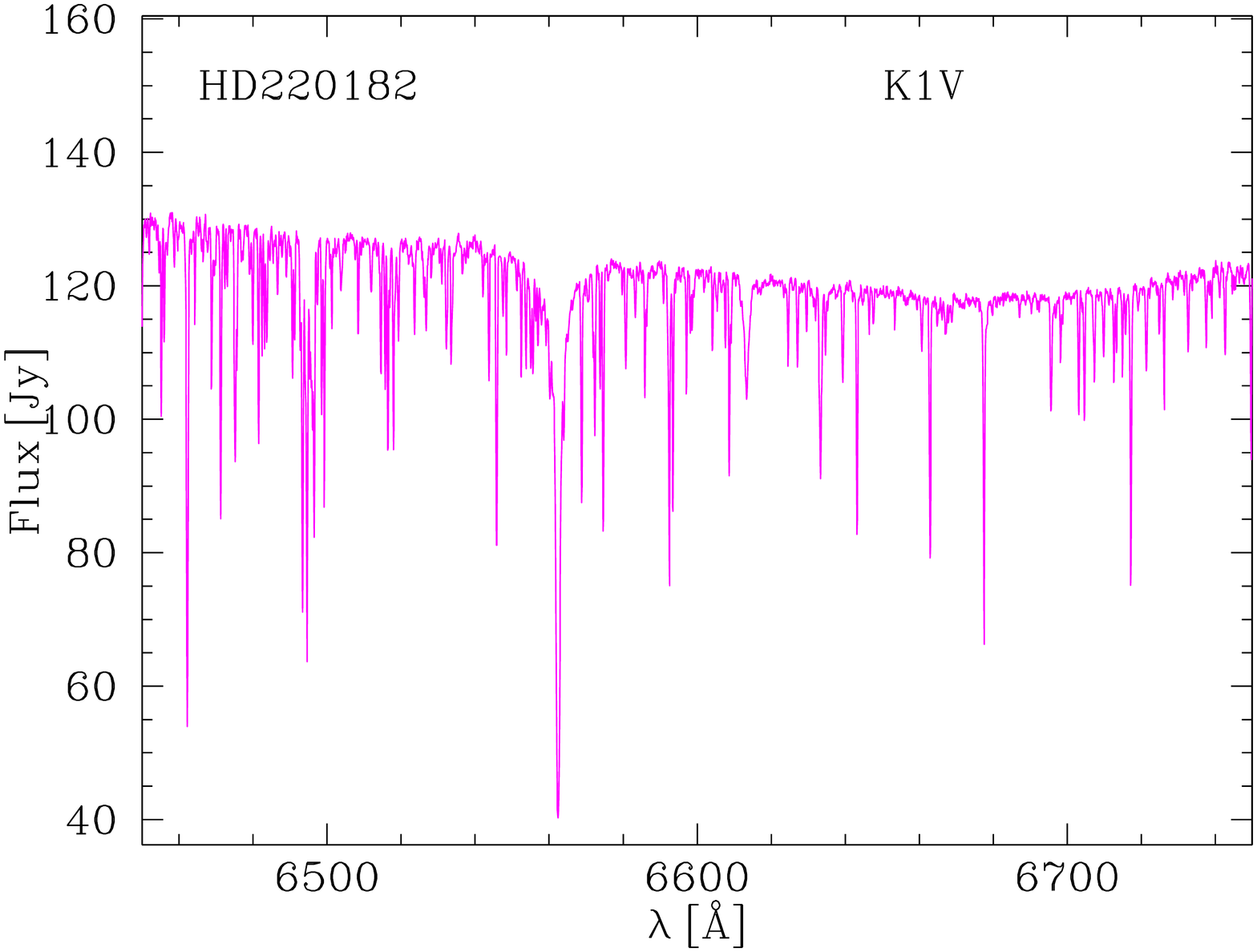}
\includegraphics[width=0.18\textwidth,angle=-90]{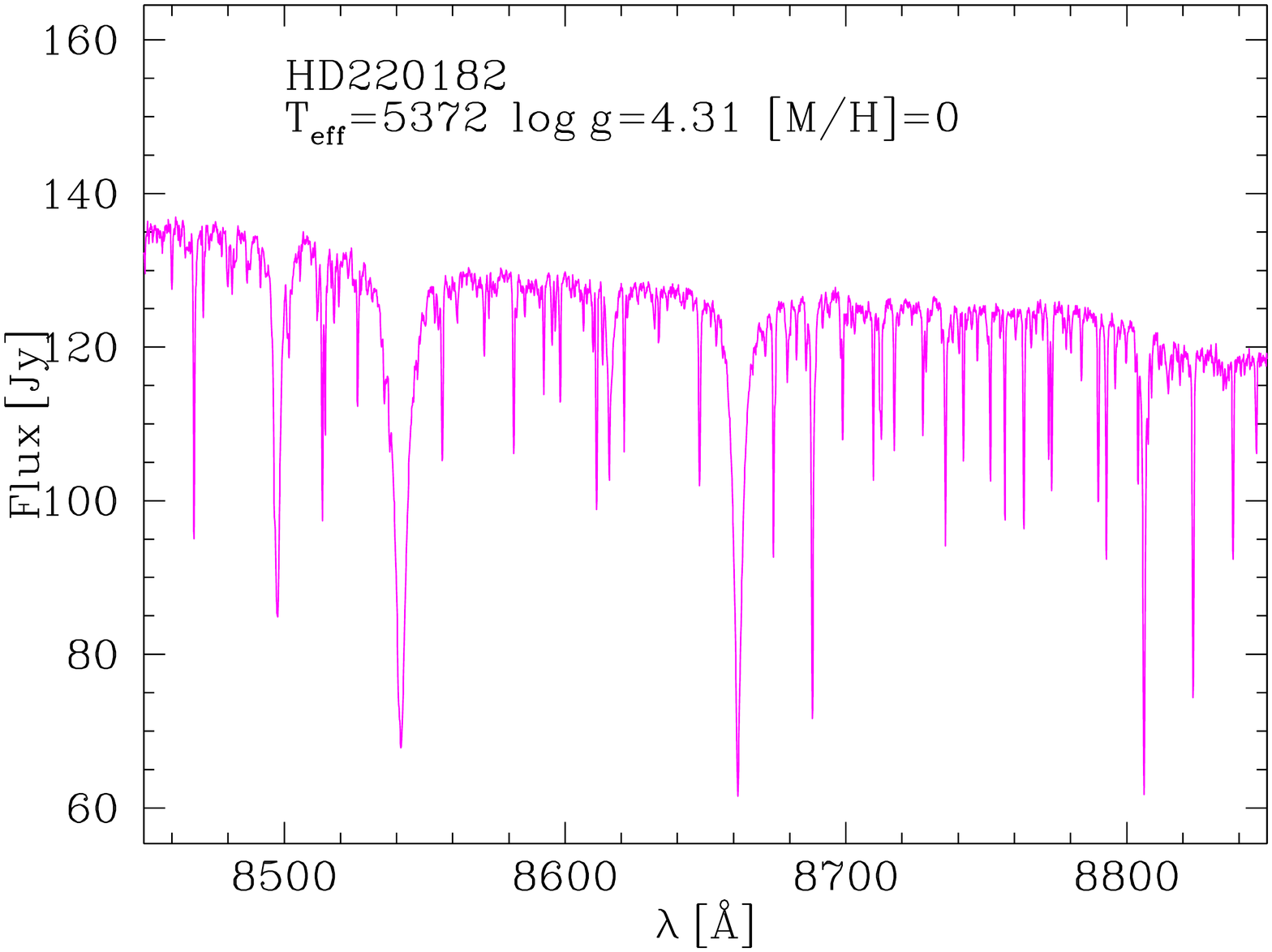}
\includegraphics[width=0.18\textwidth,angle=-90]{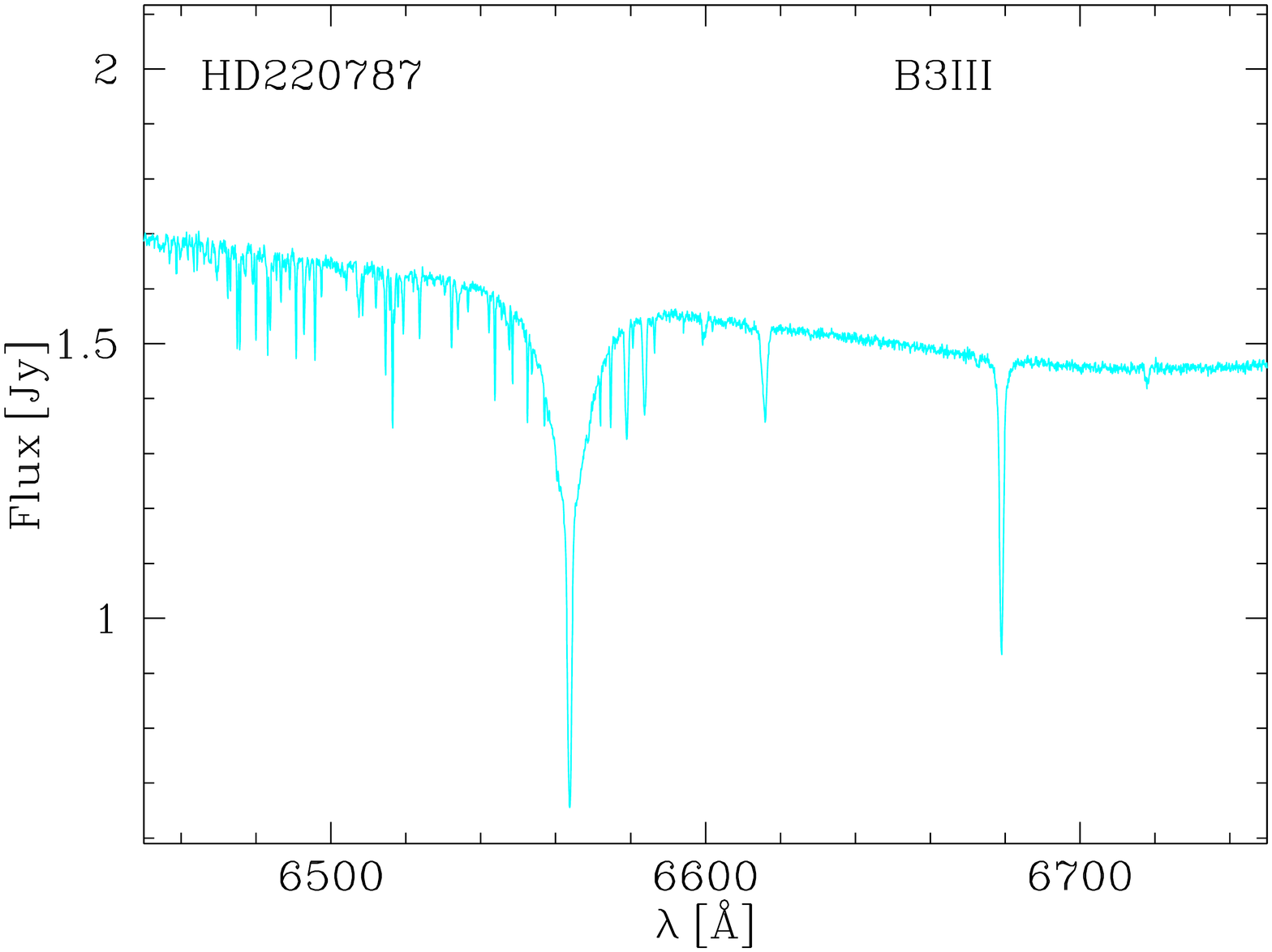}
\includegraphics[width=0.18\textwidth,angle=-90]{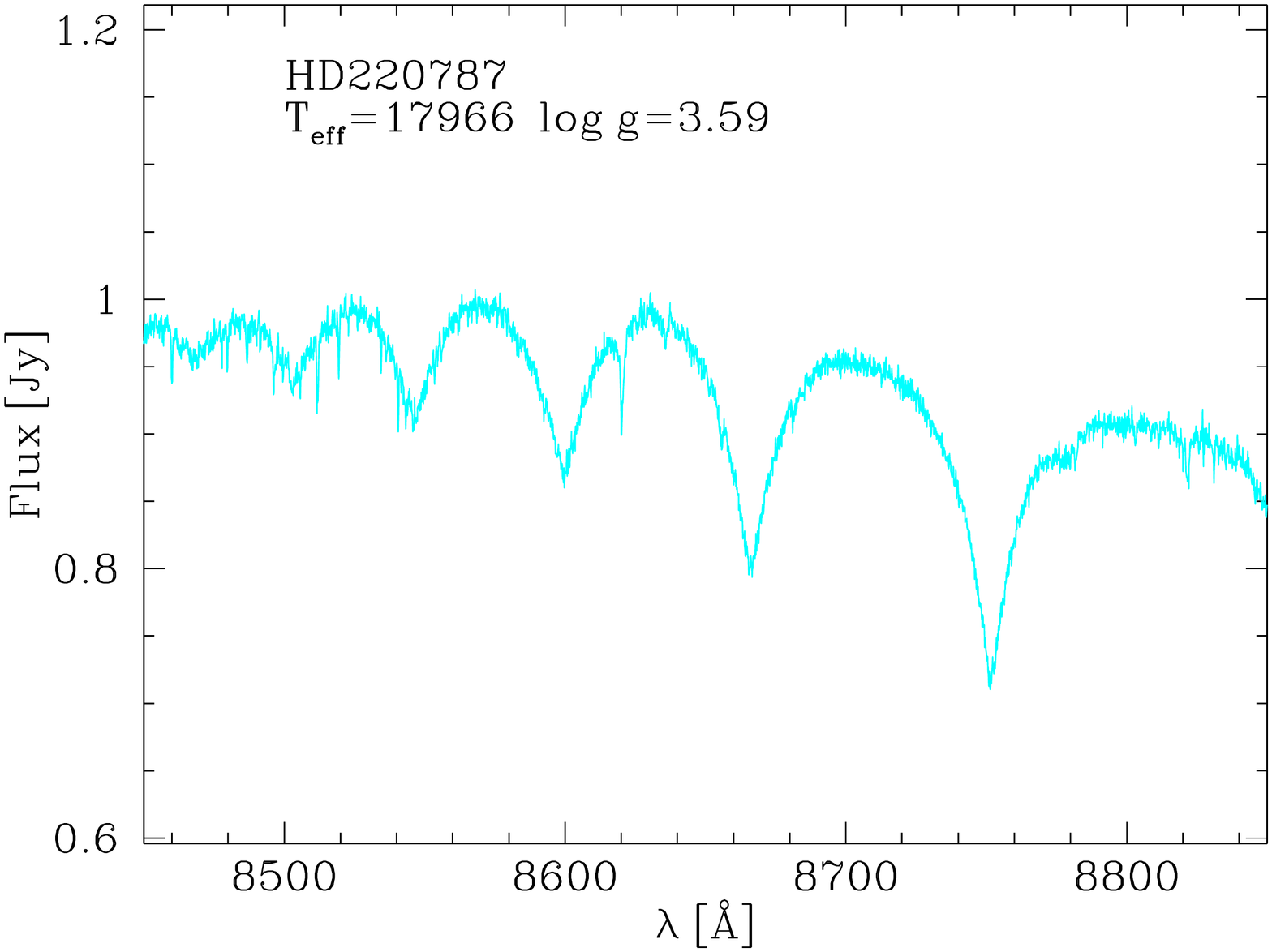}
\includegraphics[width=0.18\textwidth,angle=-90]{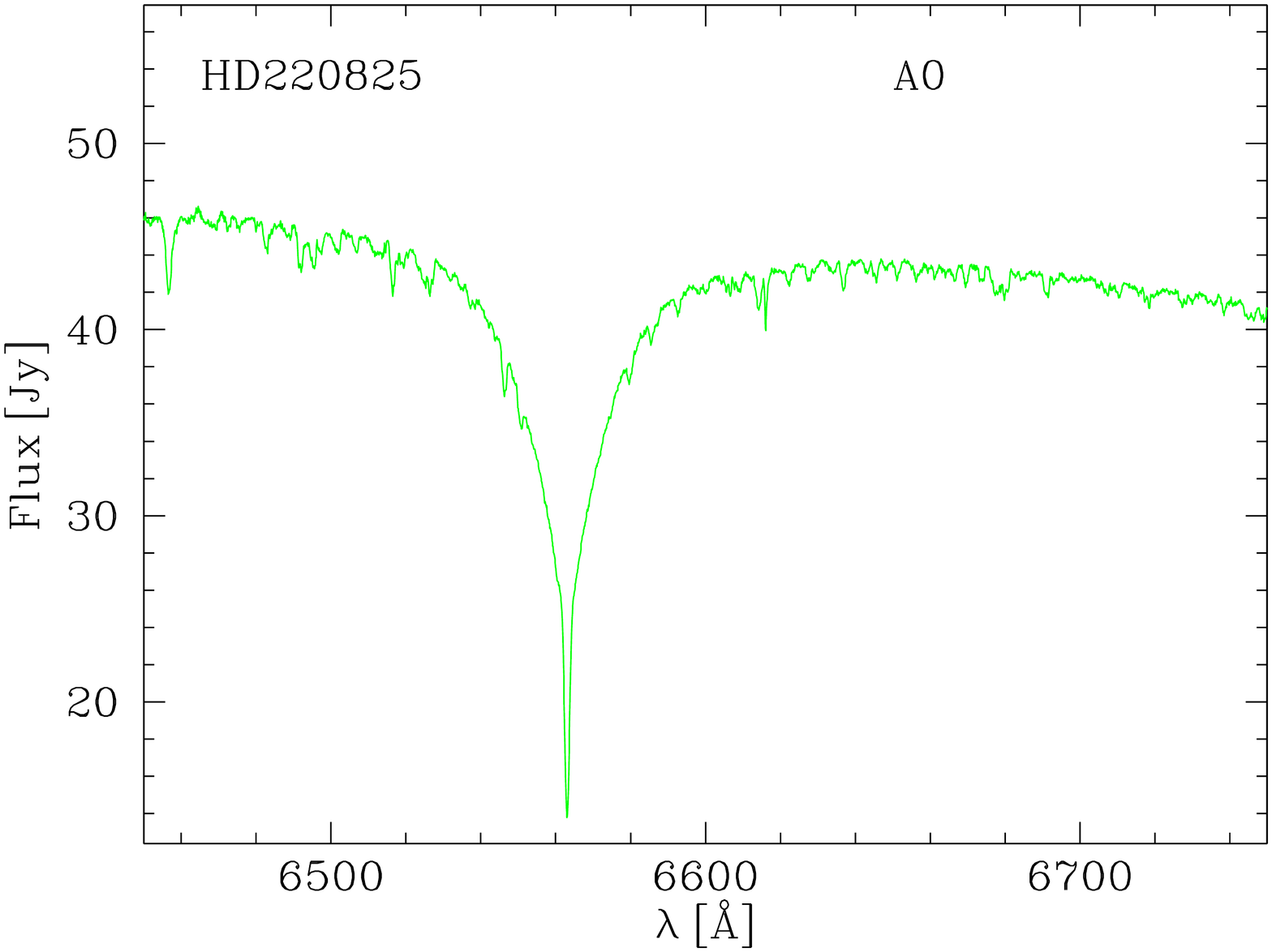}
\includegraphics[width=0.18\textwidth,angle=-90]{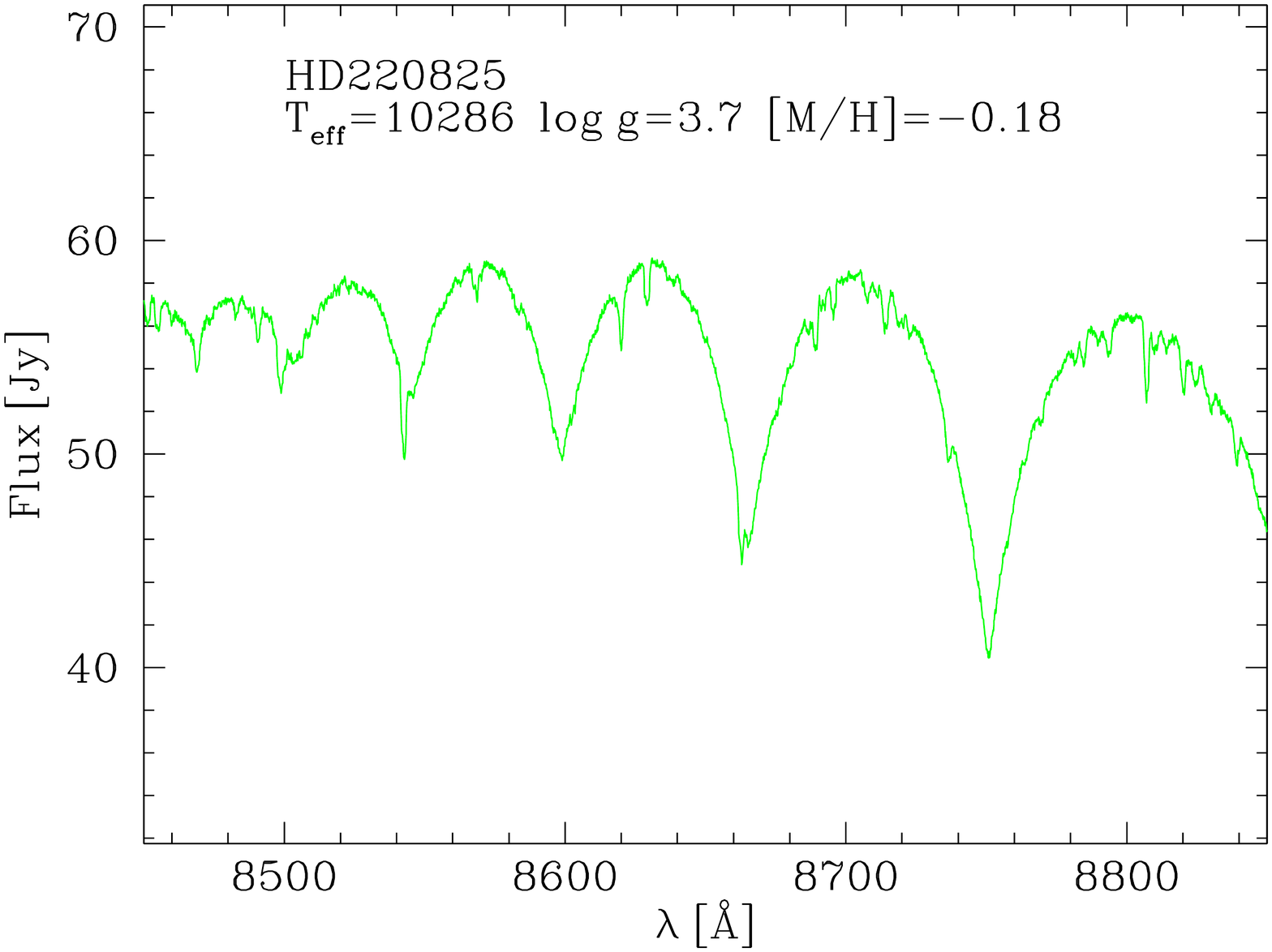}
\includegraphics[width=0.18\textwidth,angle=-90]{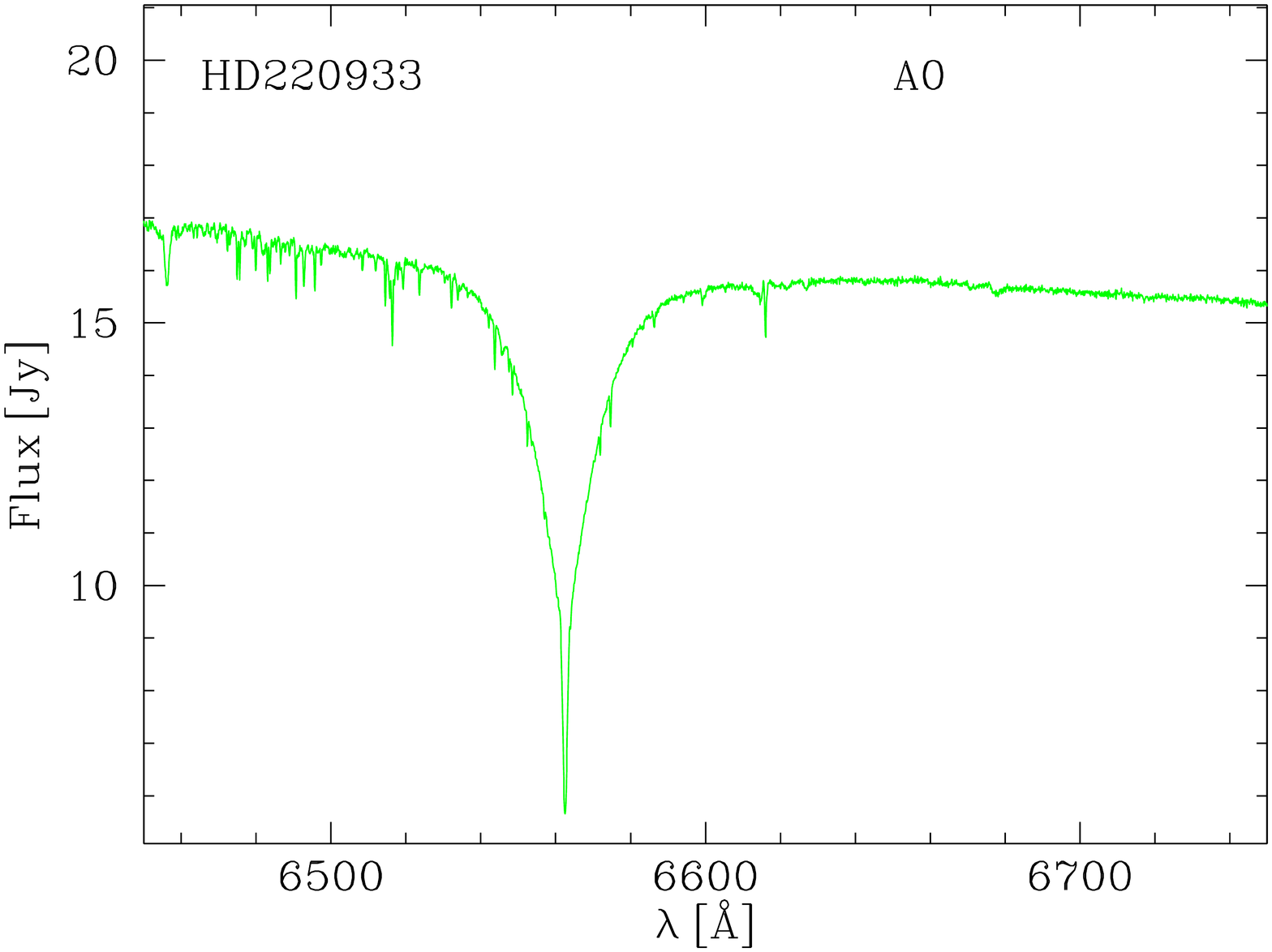}
\includegraphics[width=0.18\textwidth,angle=-90]{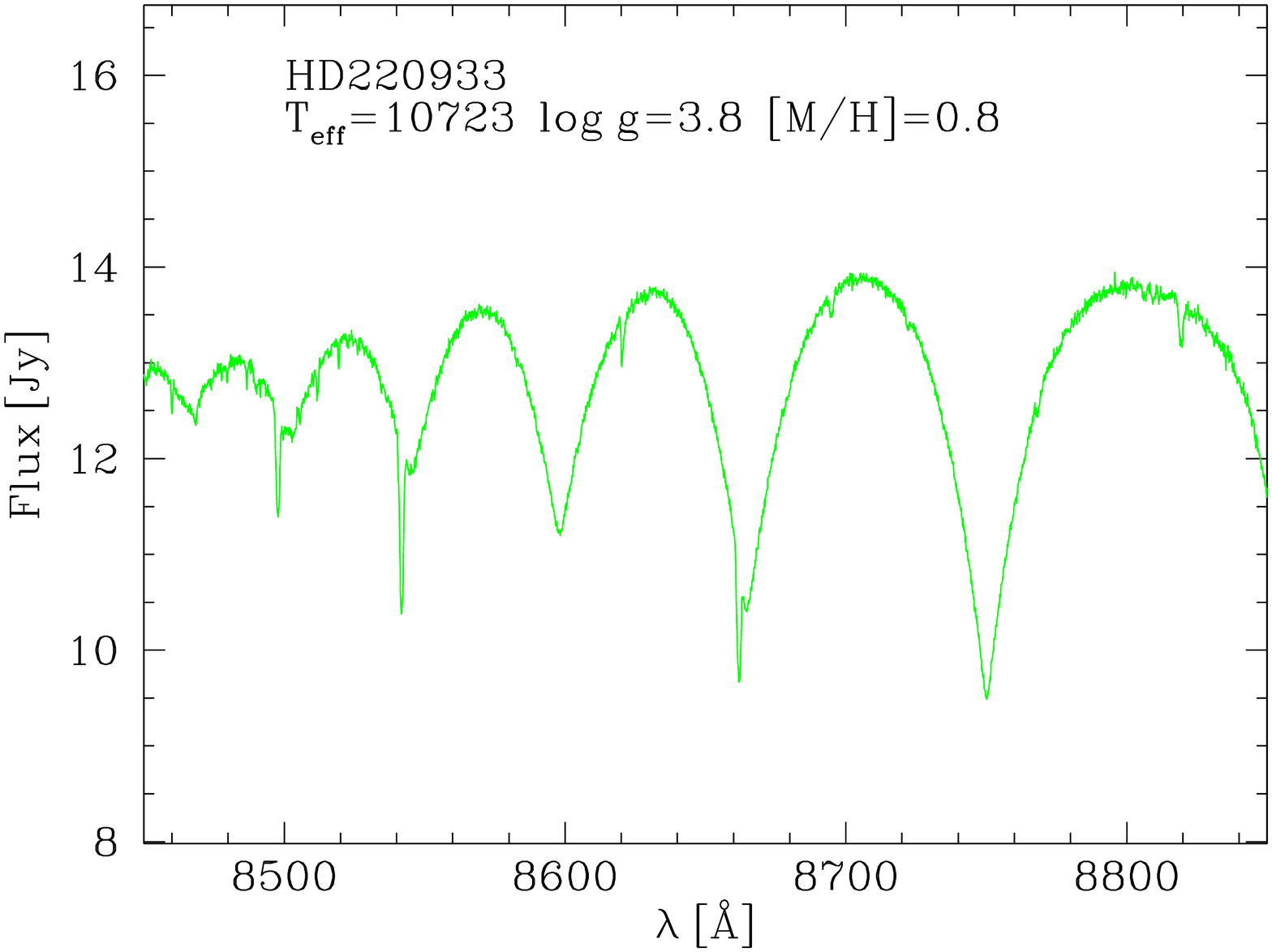}
\includegraphics[width=0.18\textwidth,angle=-90]{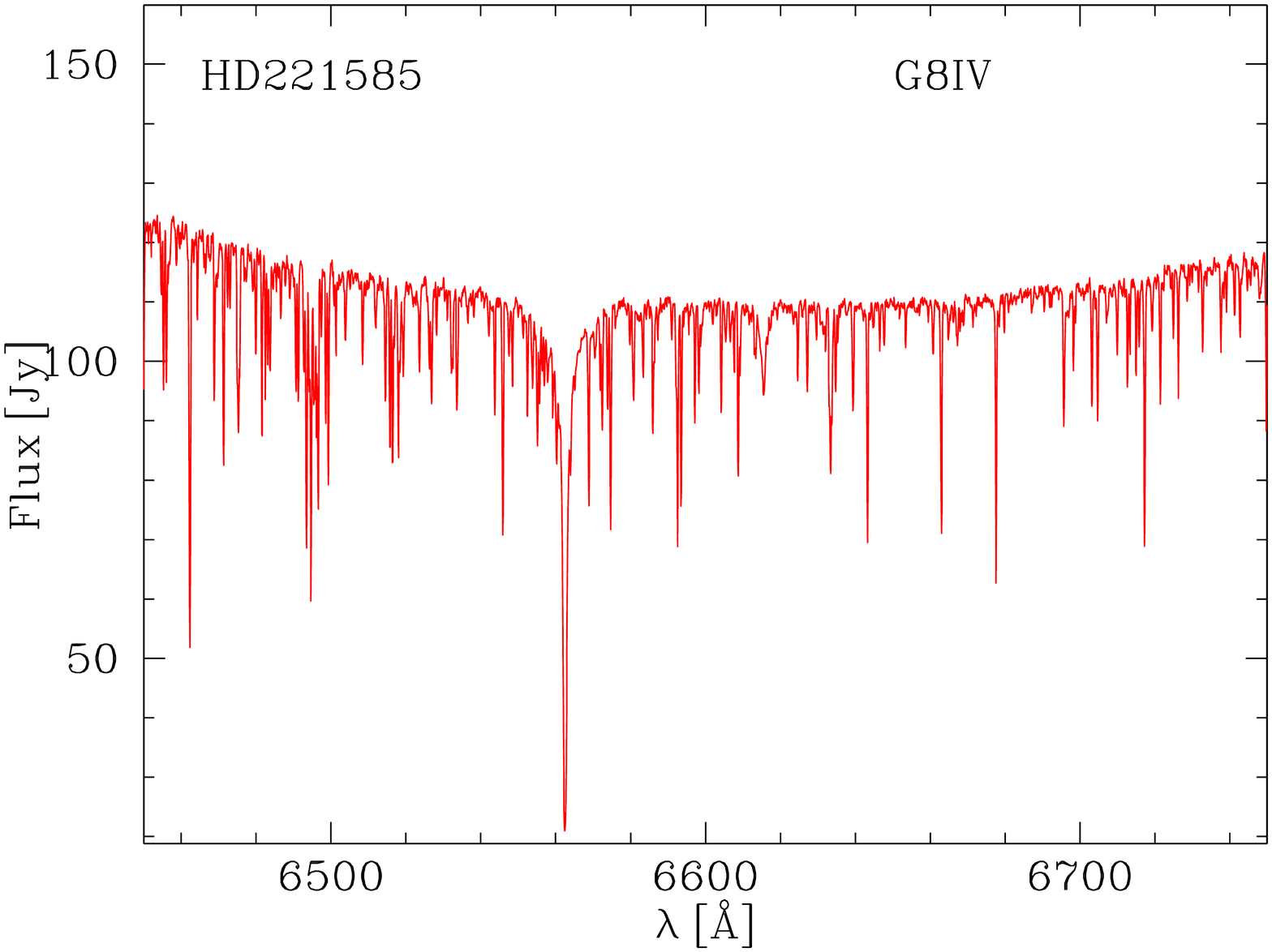}
\includegraphics[width=0.18\textwidth,angle=-90]{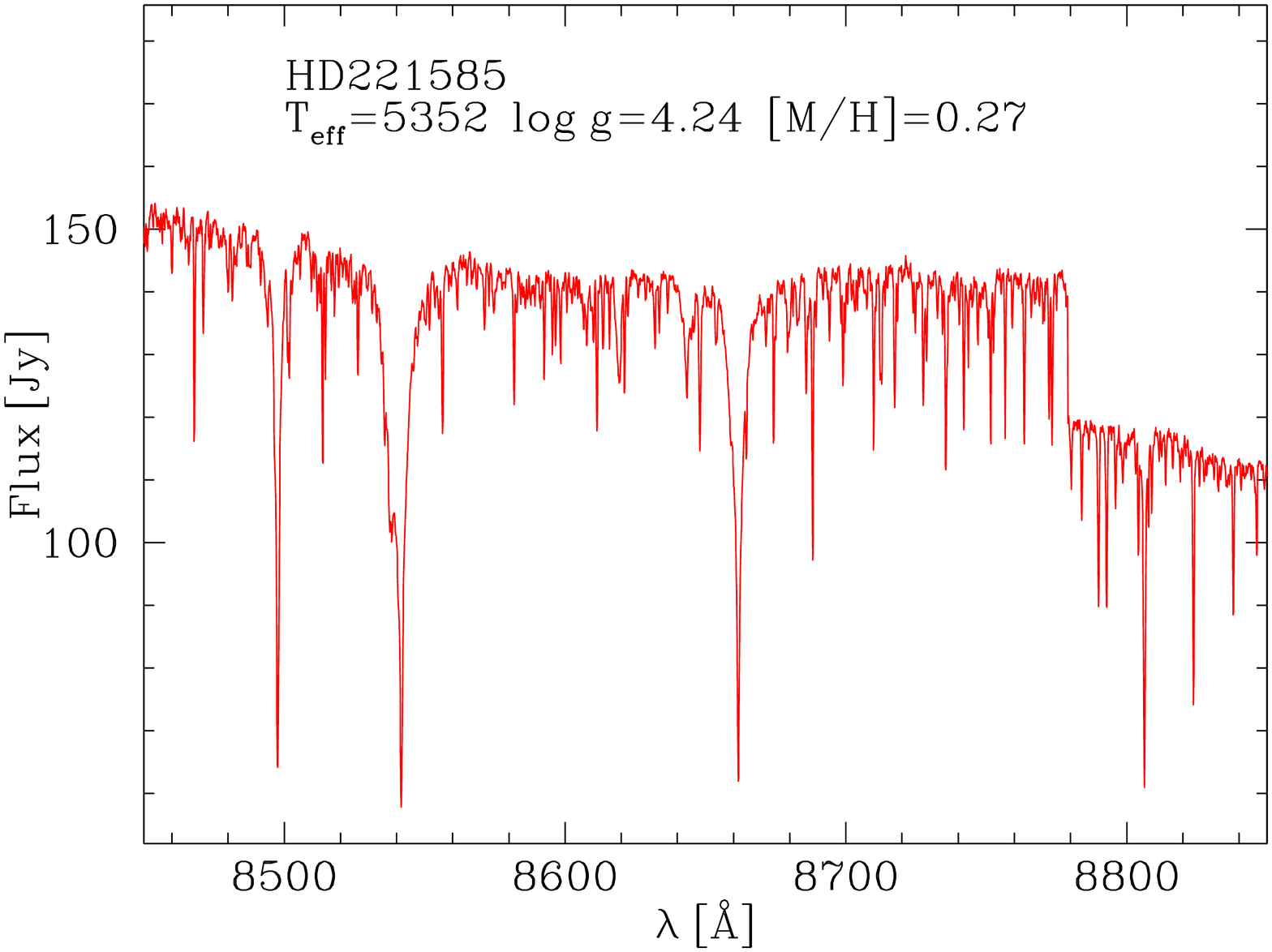}

\contcaption{29. Stars shown in this page are:  HD215704, HD216831, HD216916, HD217086, HD217833, HD217891, HD218045, HD218059, HD218376, HD220182, HD220787, HD220825, HD220933 and HD221585.}
\end{figure*}

\begin{figure*}
\includegraphics[width=0.18\textwidth,angle=-90]{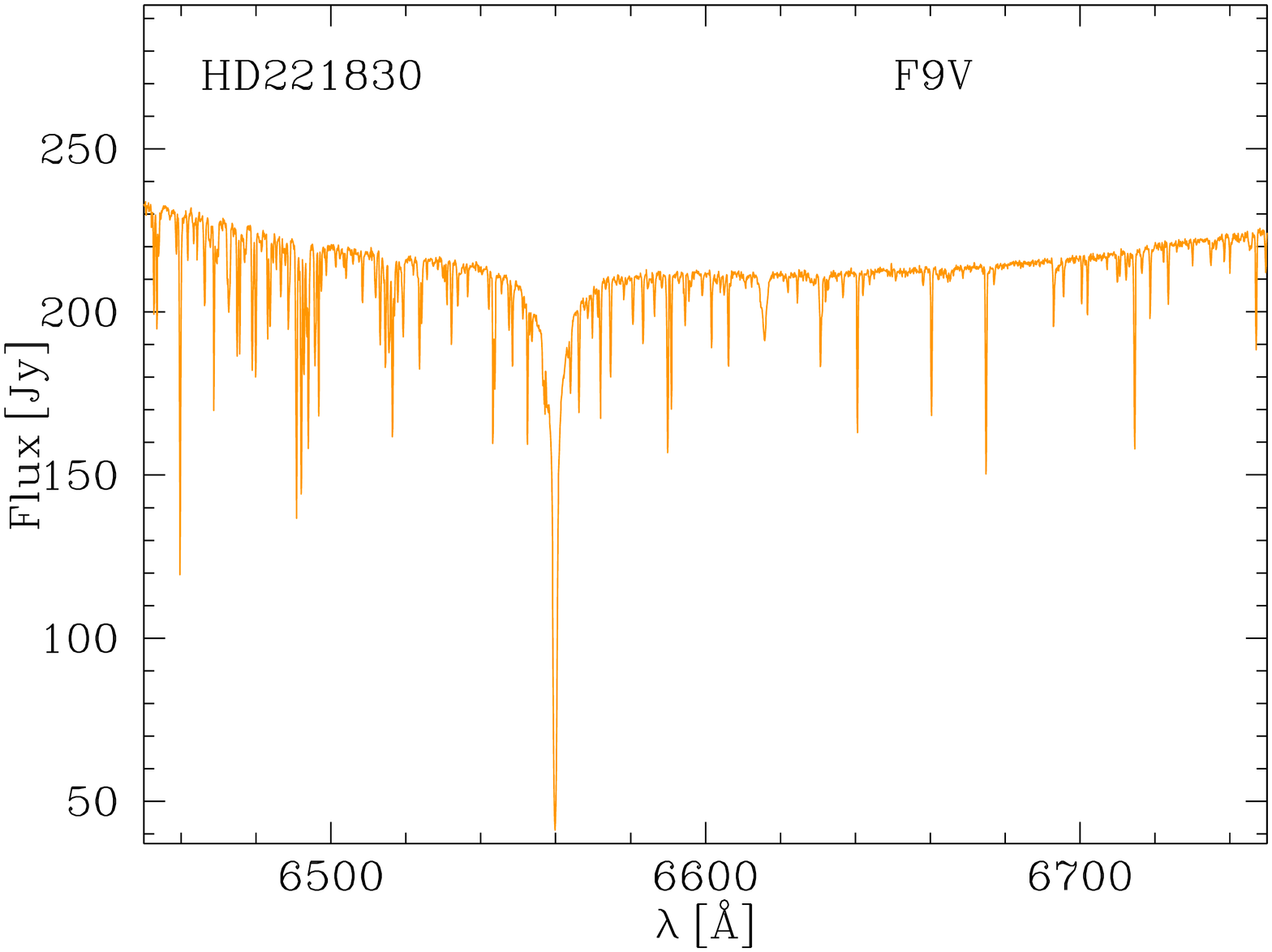}
\includegraphics[width=0.18\textwidth,angle=-90]{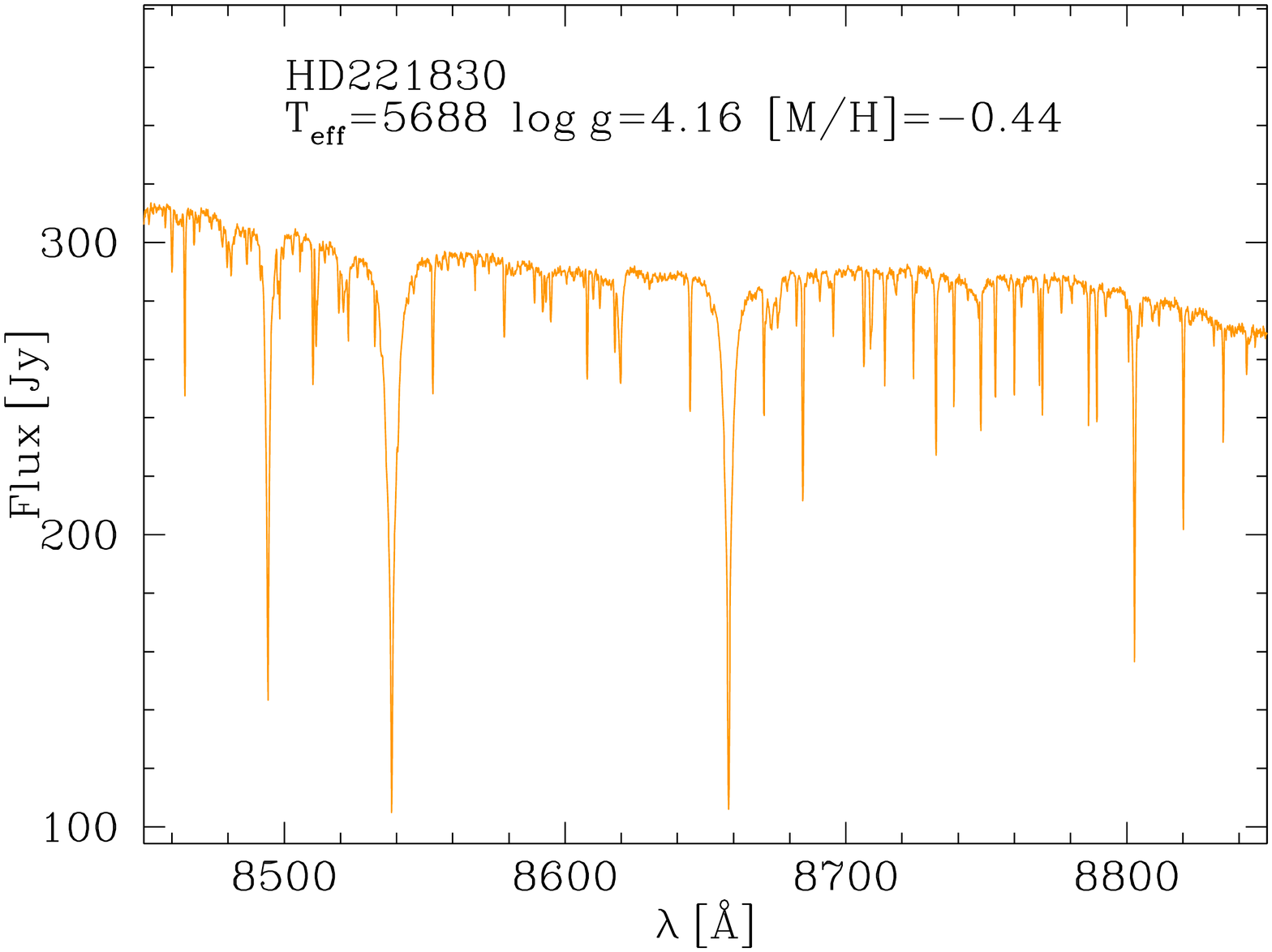}
\includegraphics[width=0.18\textwidth,angle=-90]{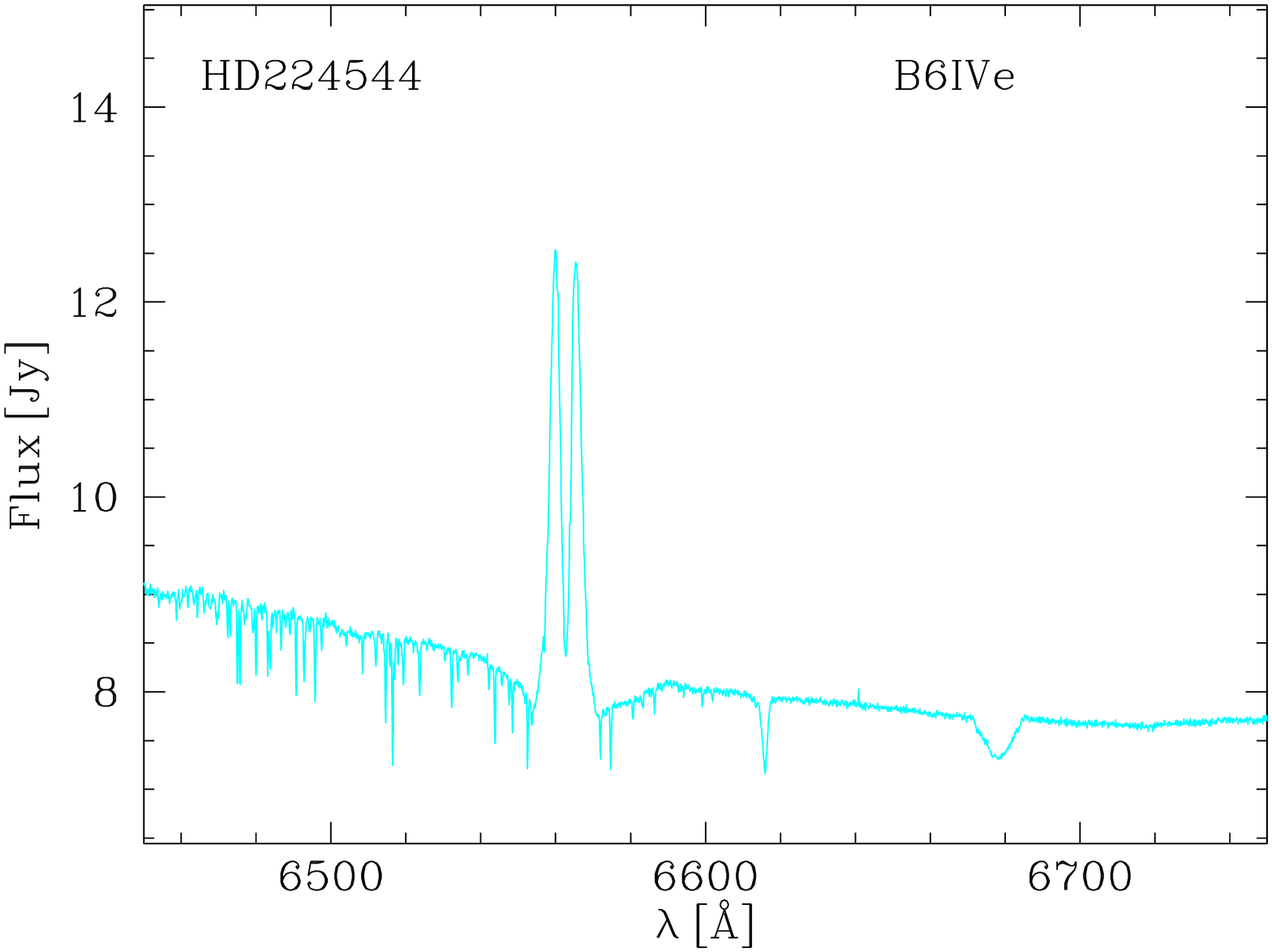}
\includegraphics[width=0.18\textwidth,angle=-90]{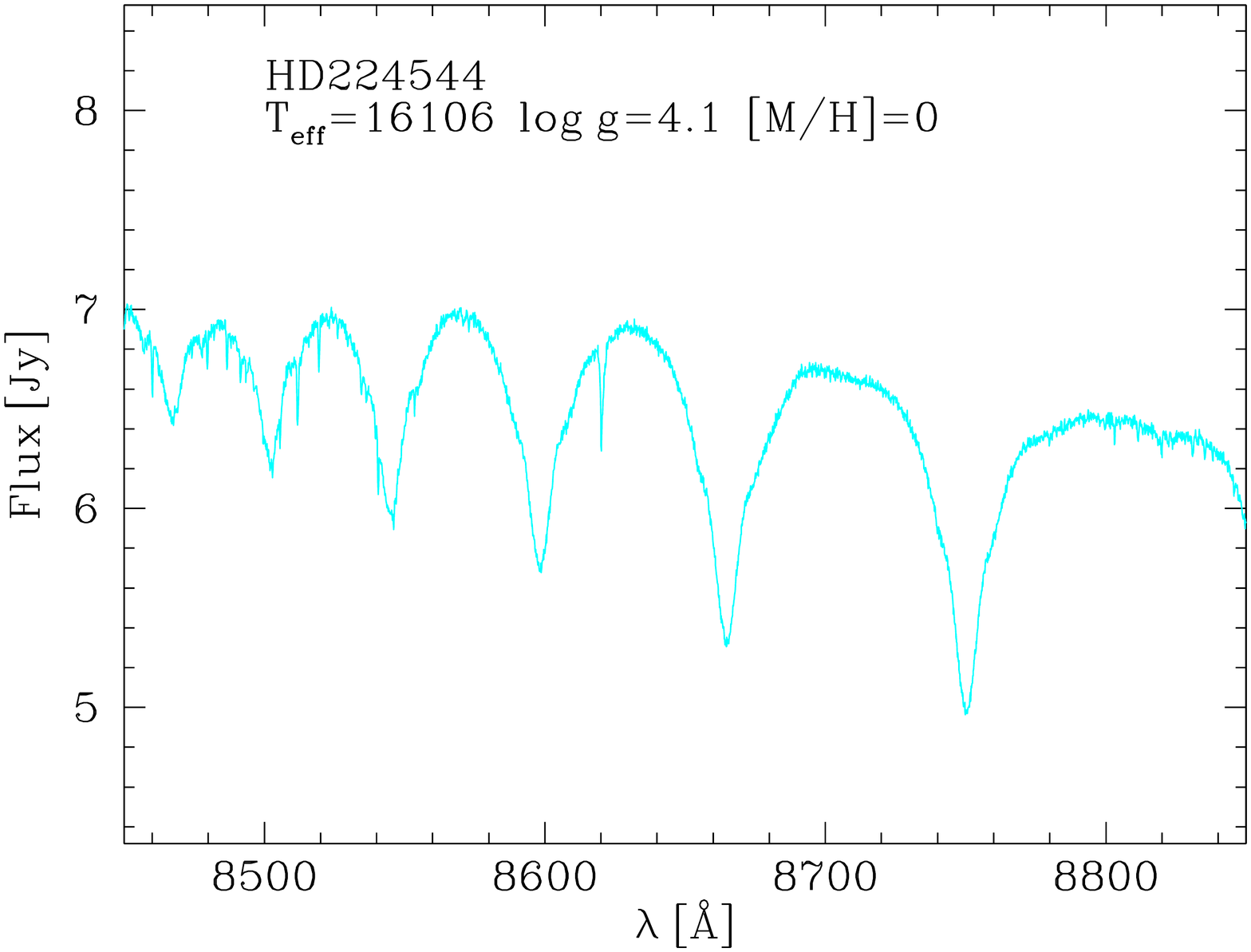}
\includegraphics[width=0.18\textwidth,angle=-90]{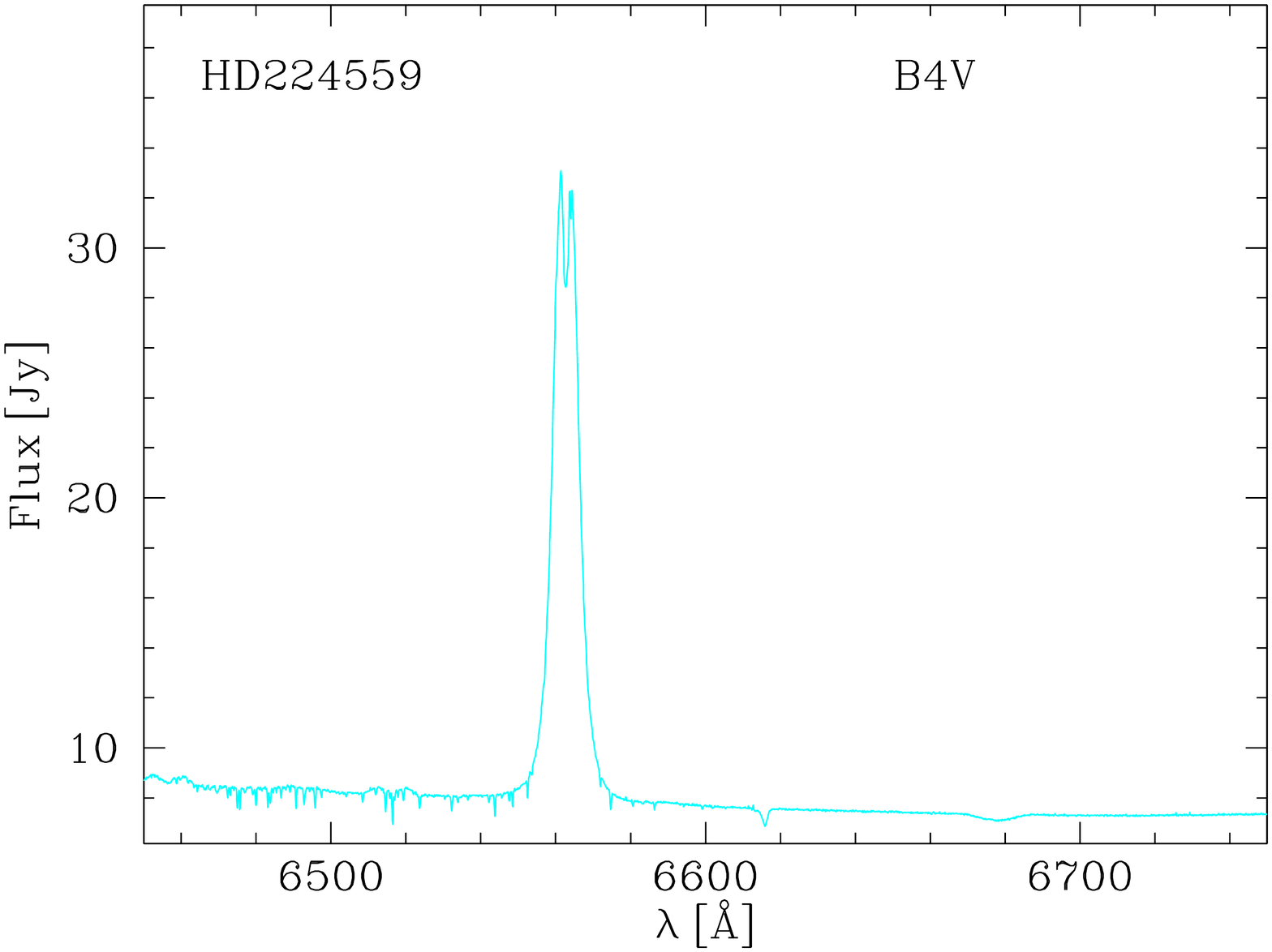}
\includegraphics[width=0.18\textwidth,angle=-90]{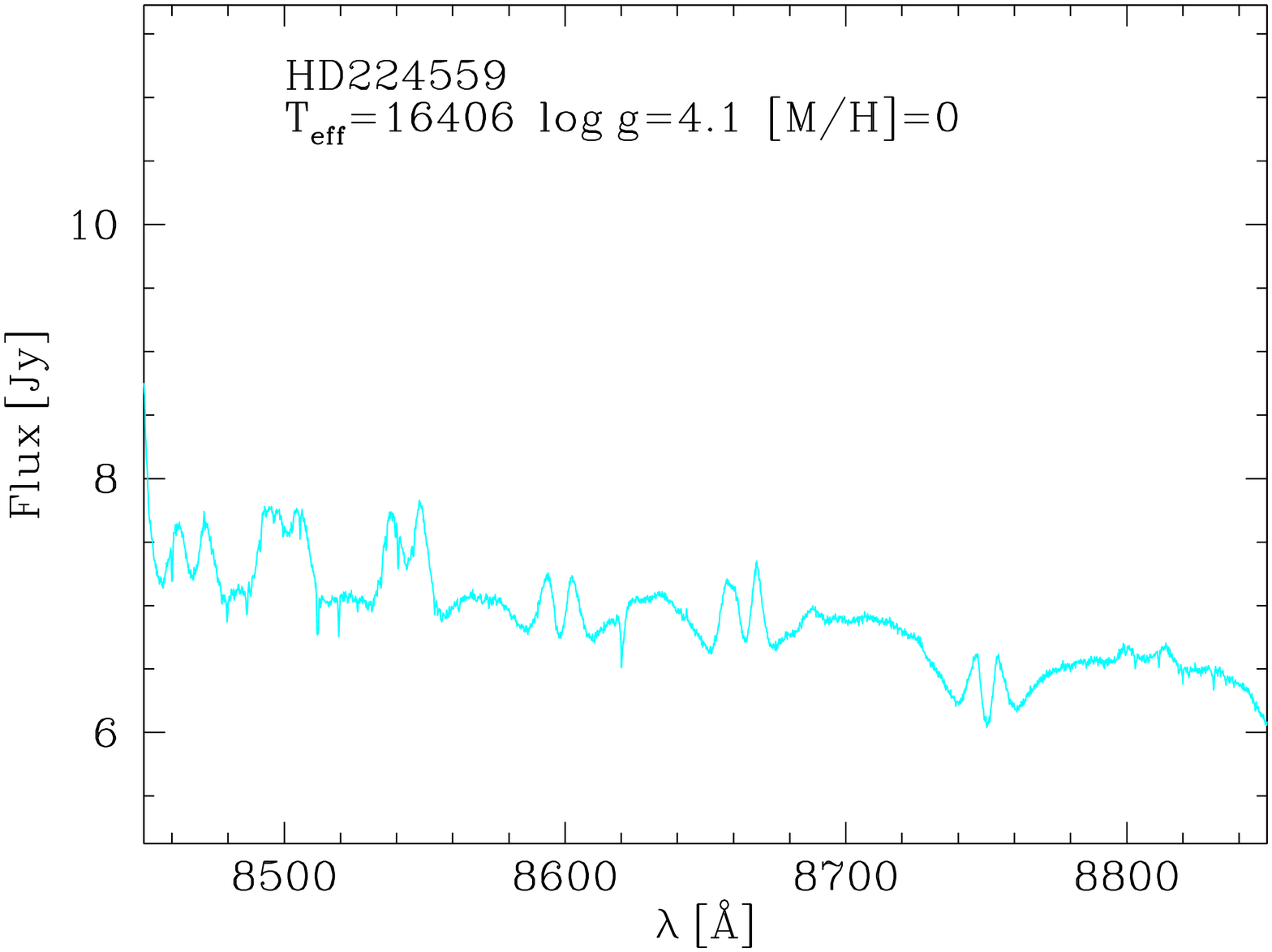}
\includegraphics[width=0.18\textwidth,angle=-90]{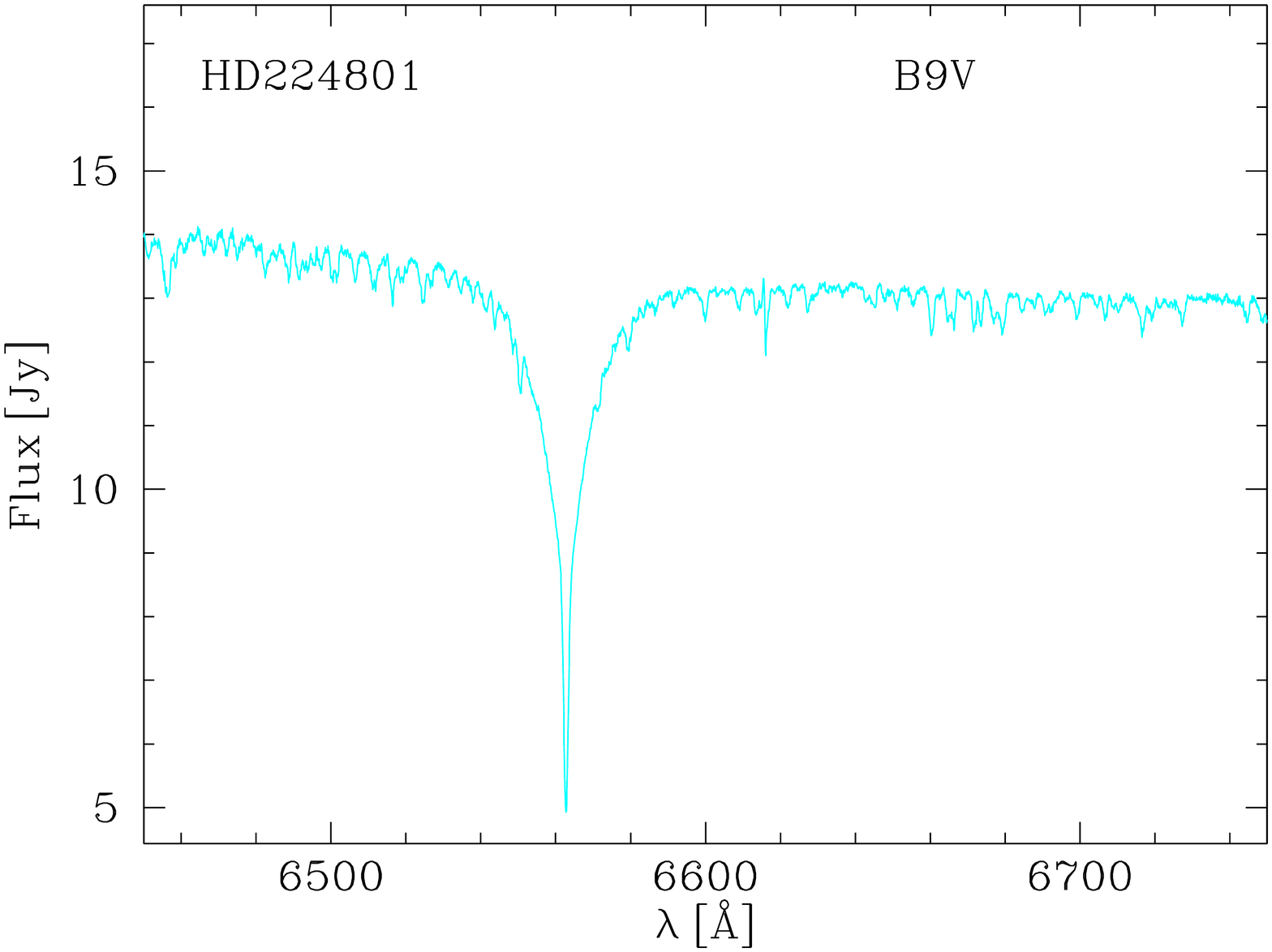}
\includegraphics[width=0.18\textwidth,angle=-90]{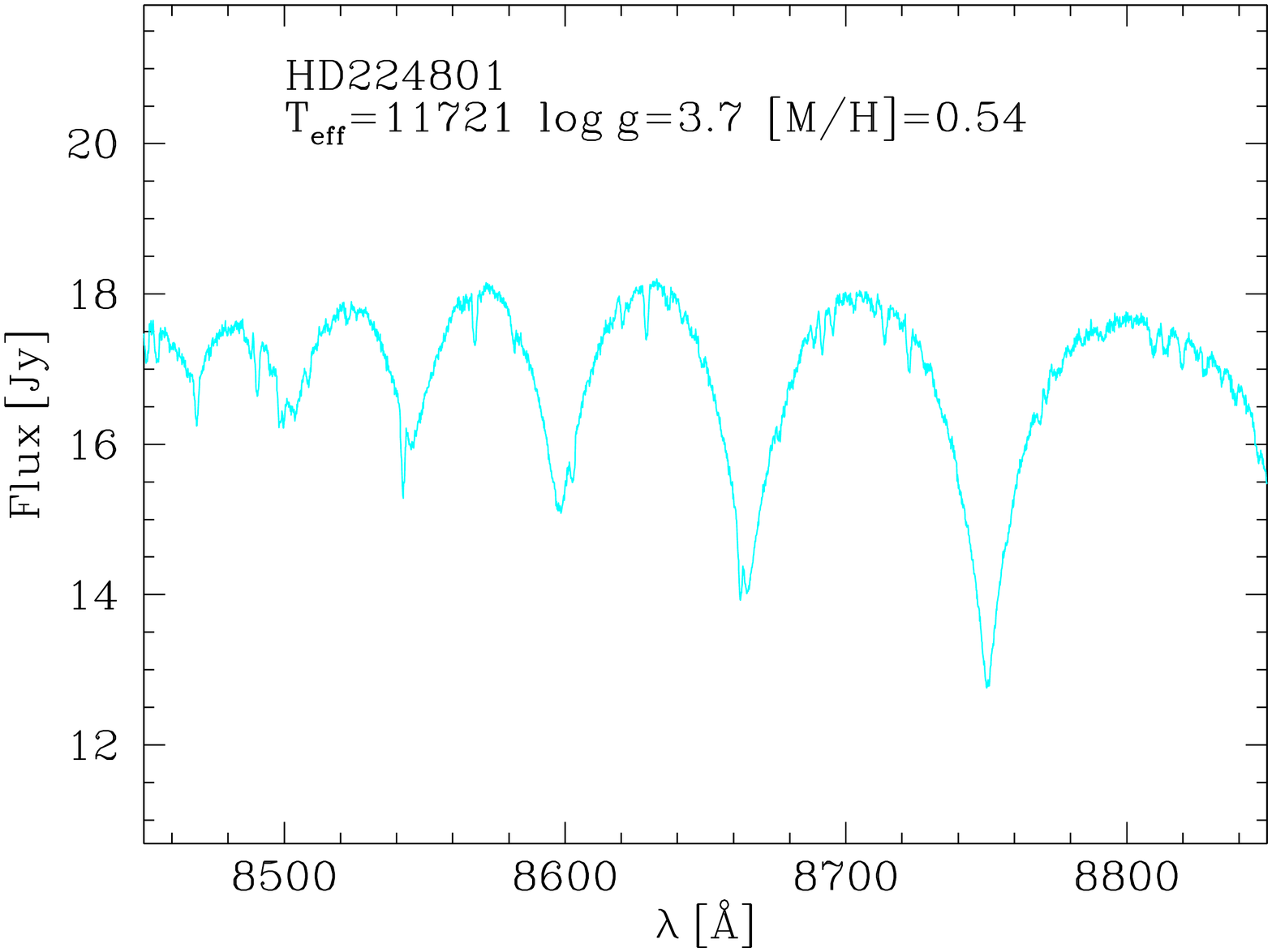}
\includegraphics[width=0.18\textwidth,angle=-90]{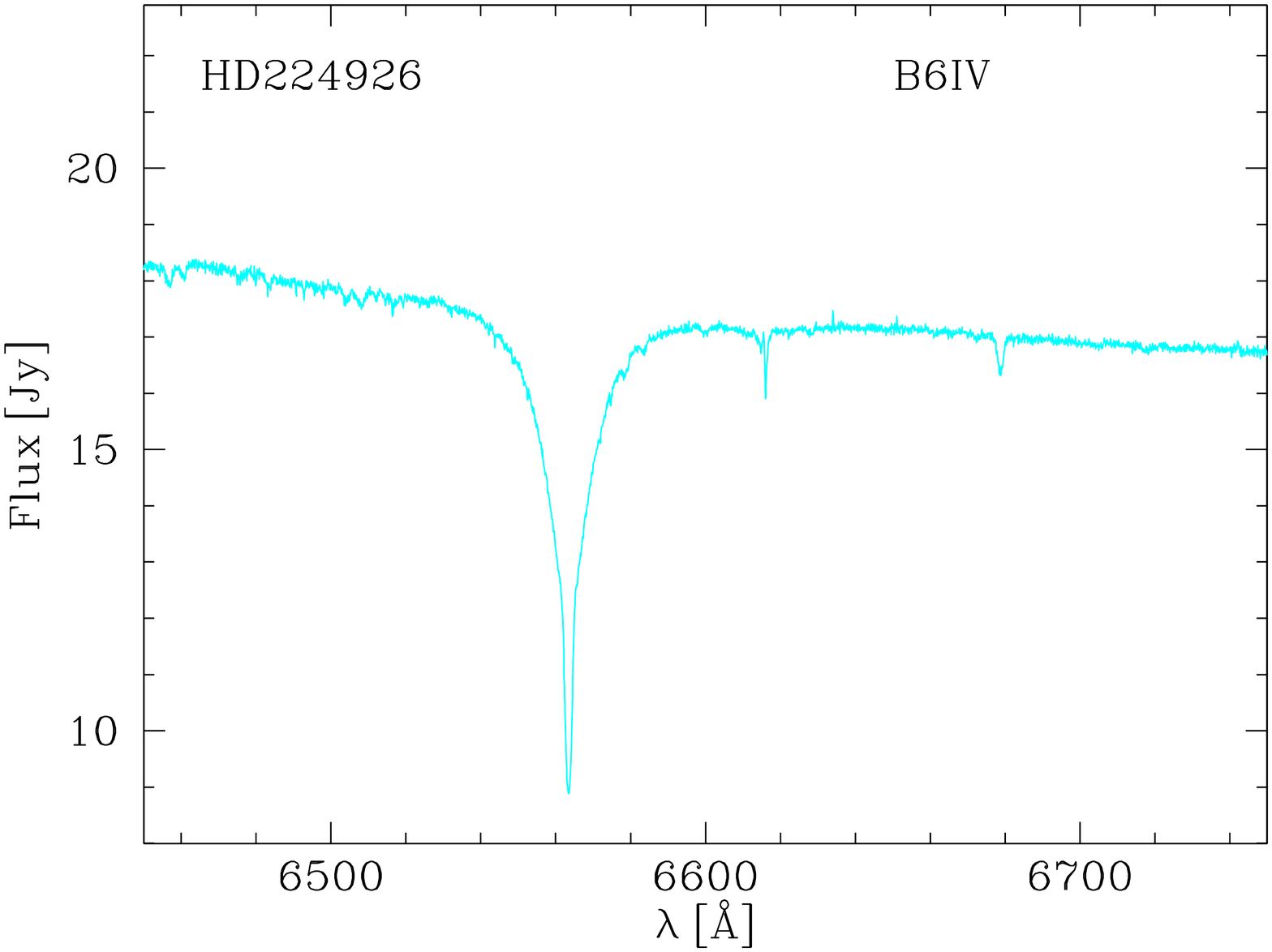}
\includegraphics[width=0.18\textwidth,angle=-90]{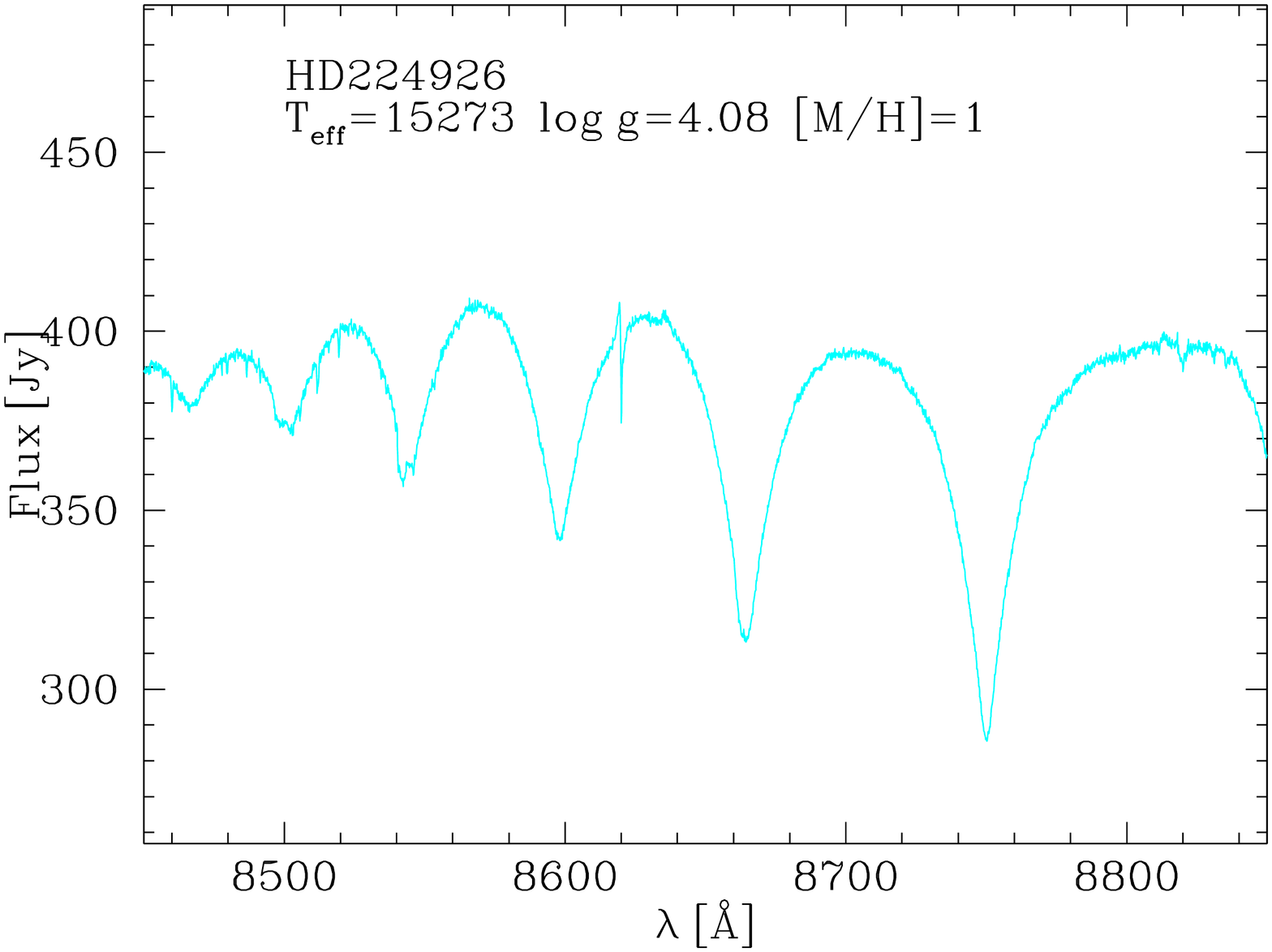}
\includegraphics[width=0.18\textwidth,angle=-90]{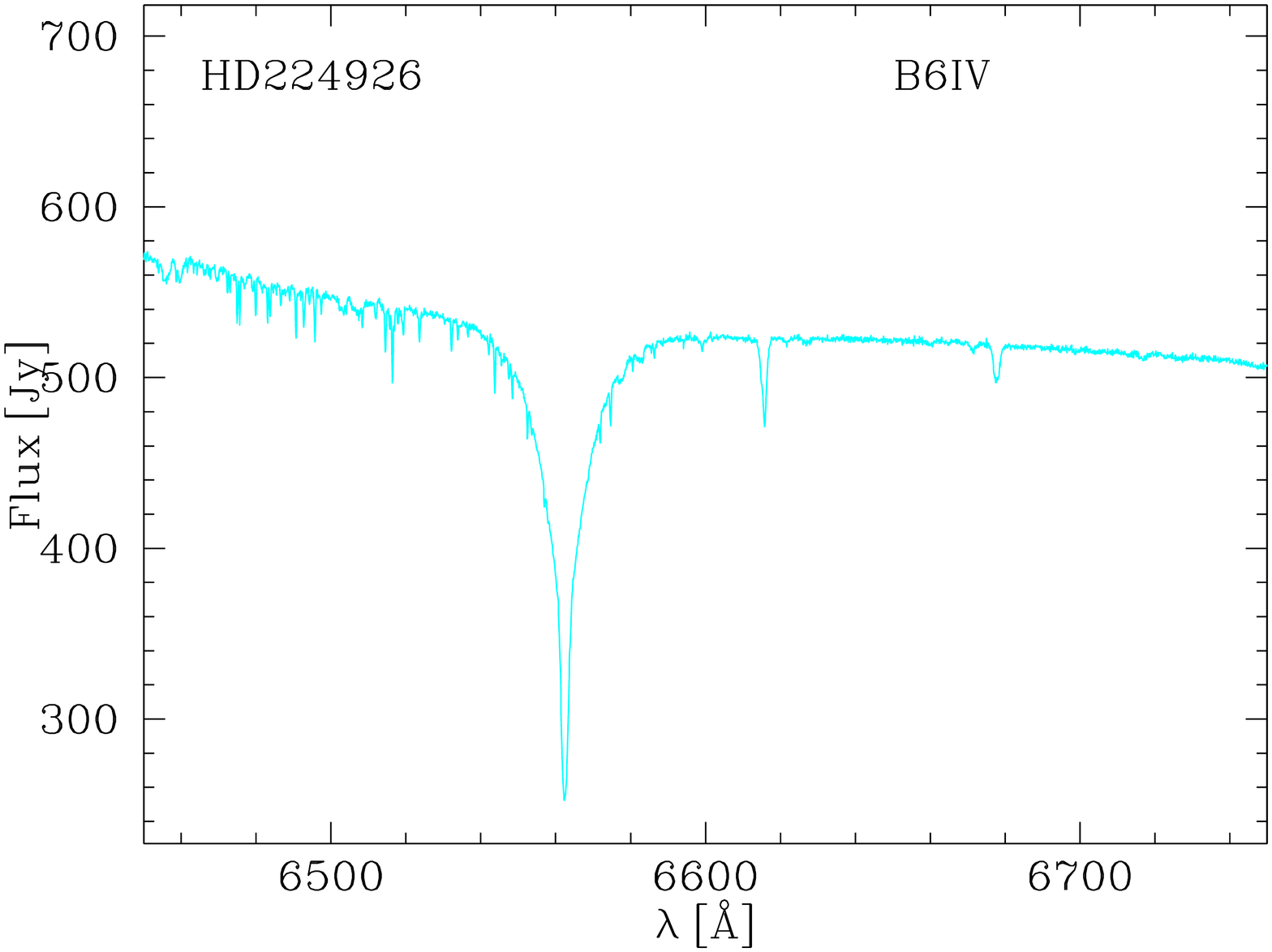}
\includegraphics[width=0.18\textwidth,angle=-90]{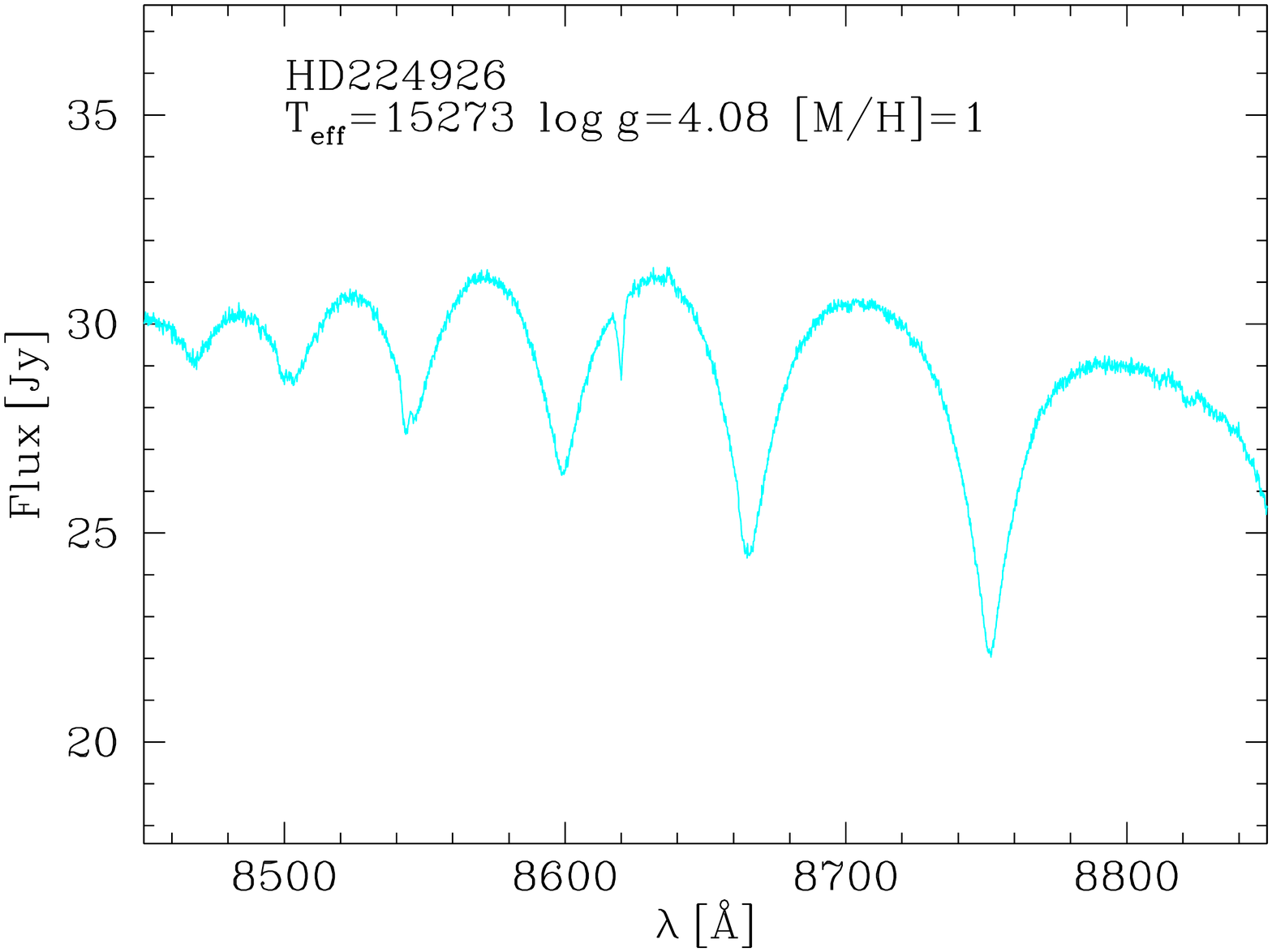}
\includegraphics[width=0.18\textwidth,angle=-90]{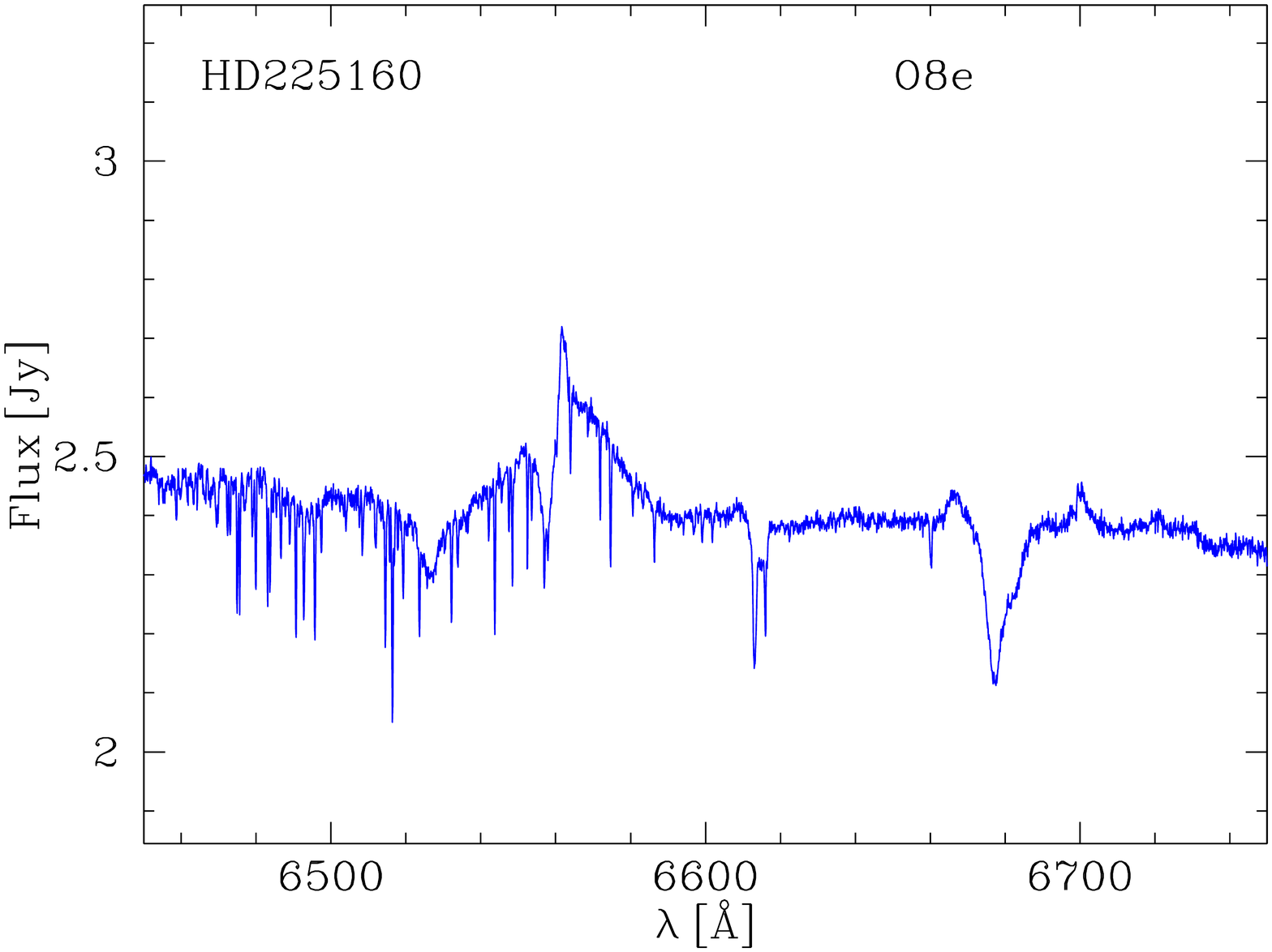}
\includegraphics[width=0.18\textwidth,angle=-90]{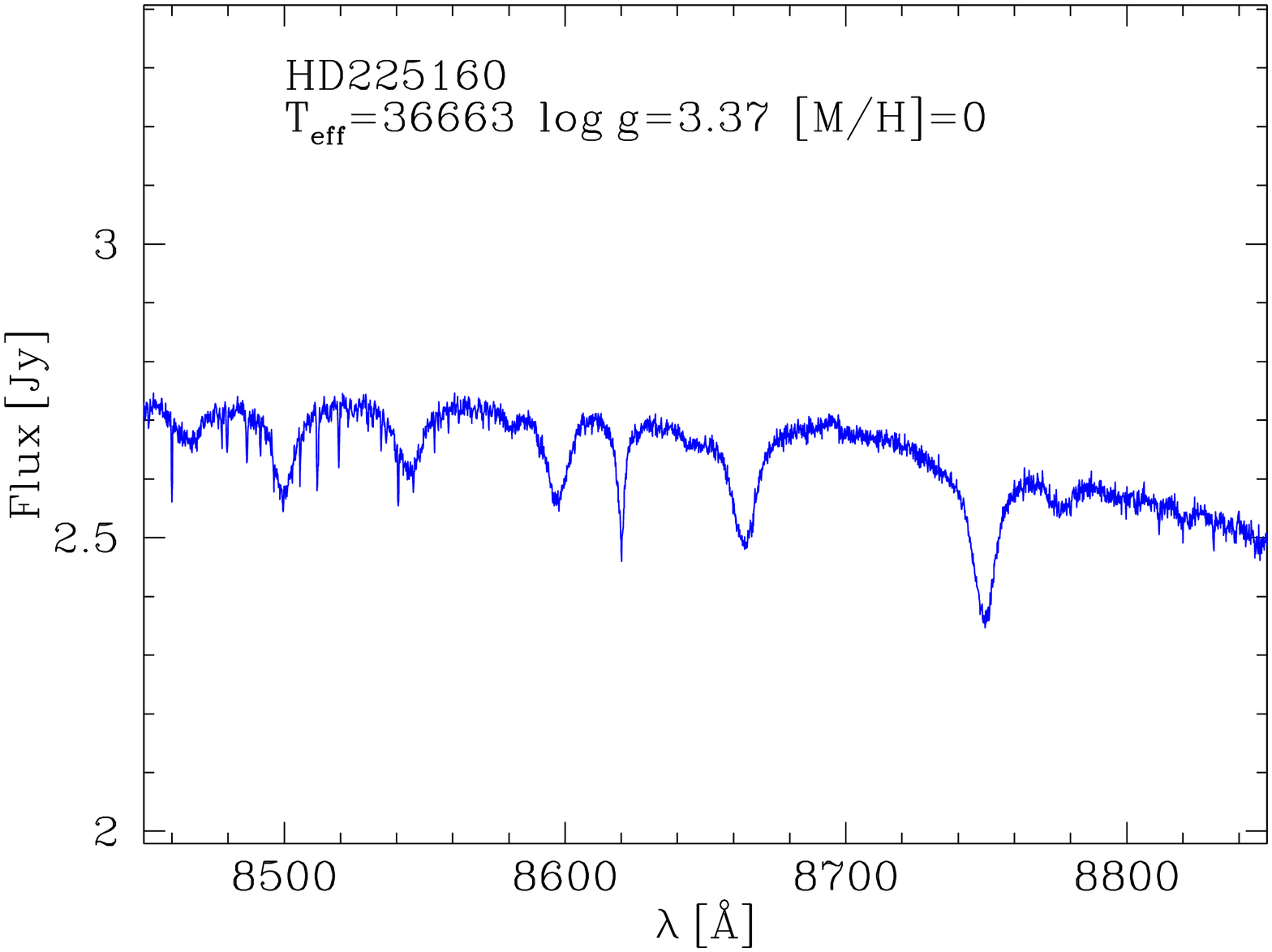}
\includegraphics[width=0.18\textwidth,angle=-90]{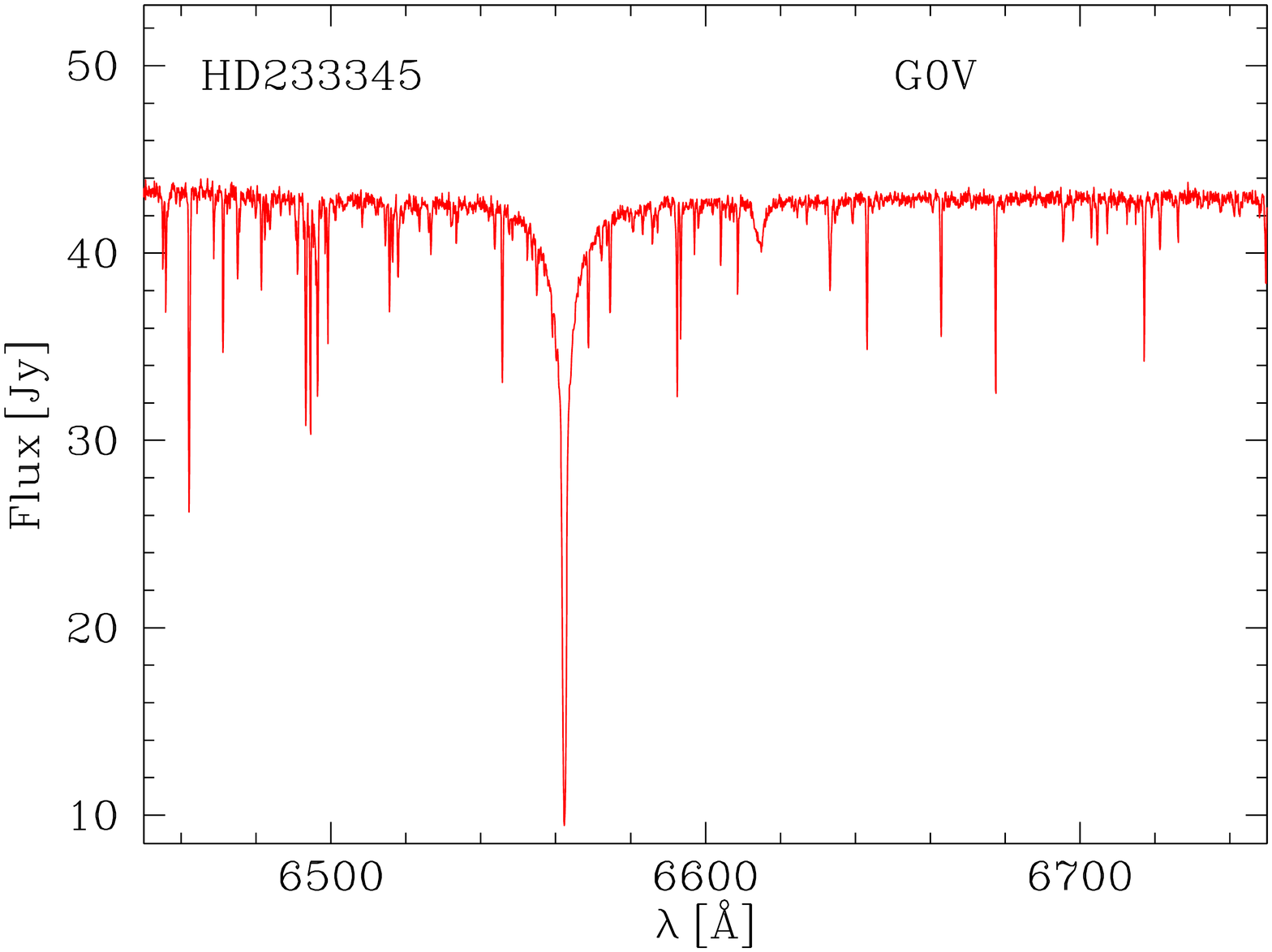}
\includegraphics[width=0.18\textwidth,angle=-90]{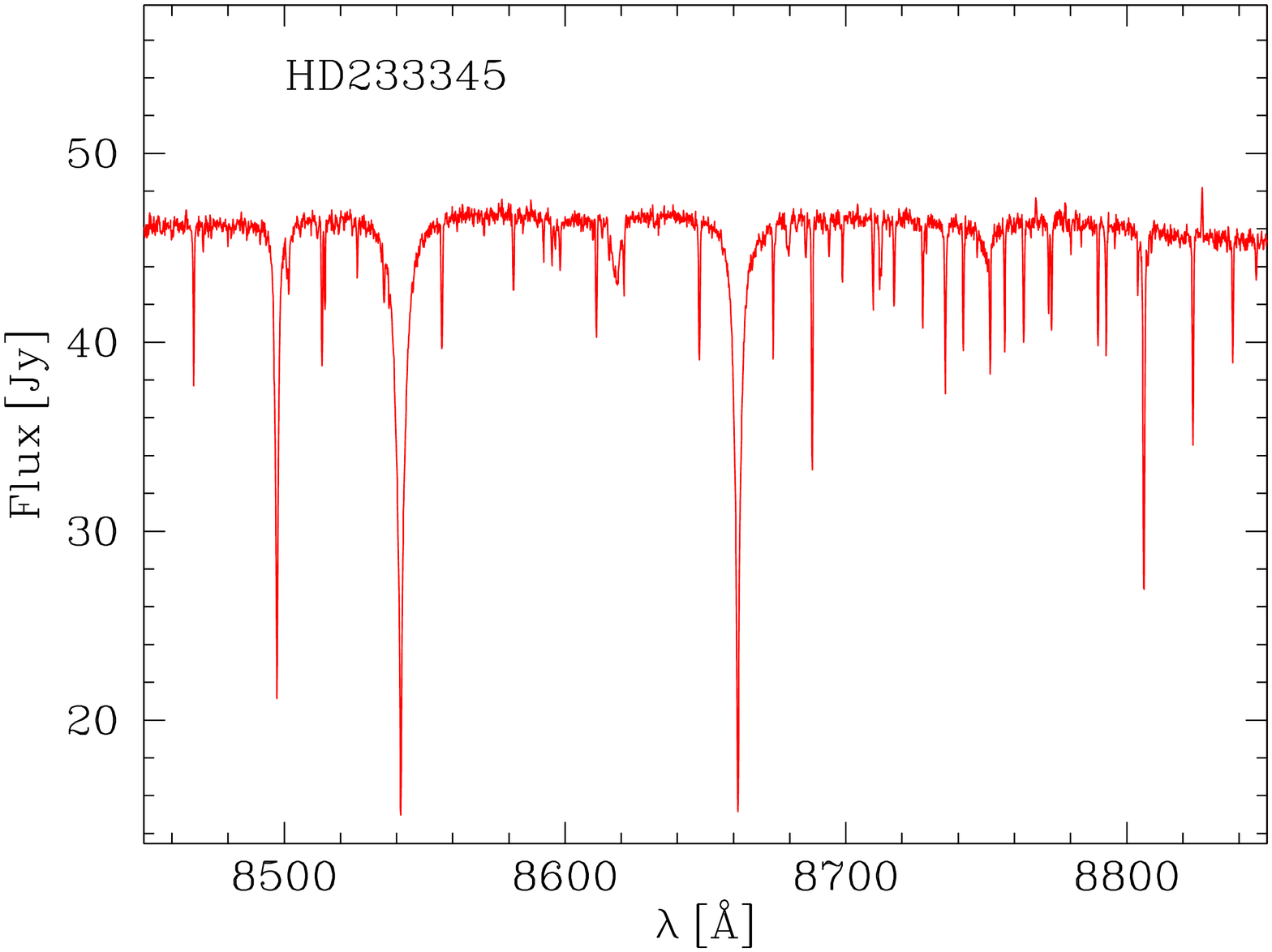}
\includegraphics[width=0.18\textwidth,angle=-90]{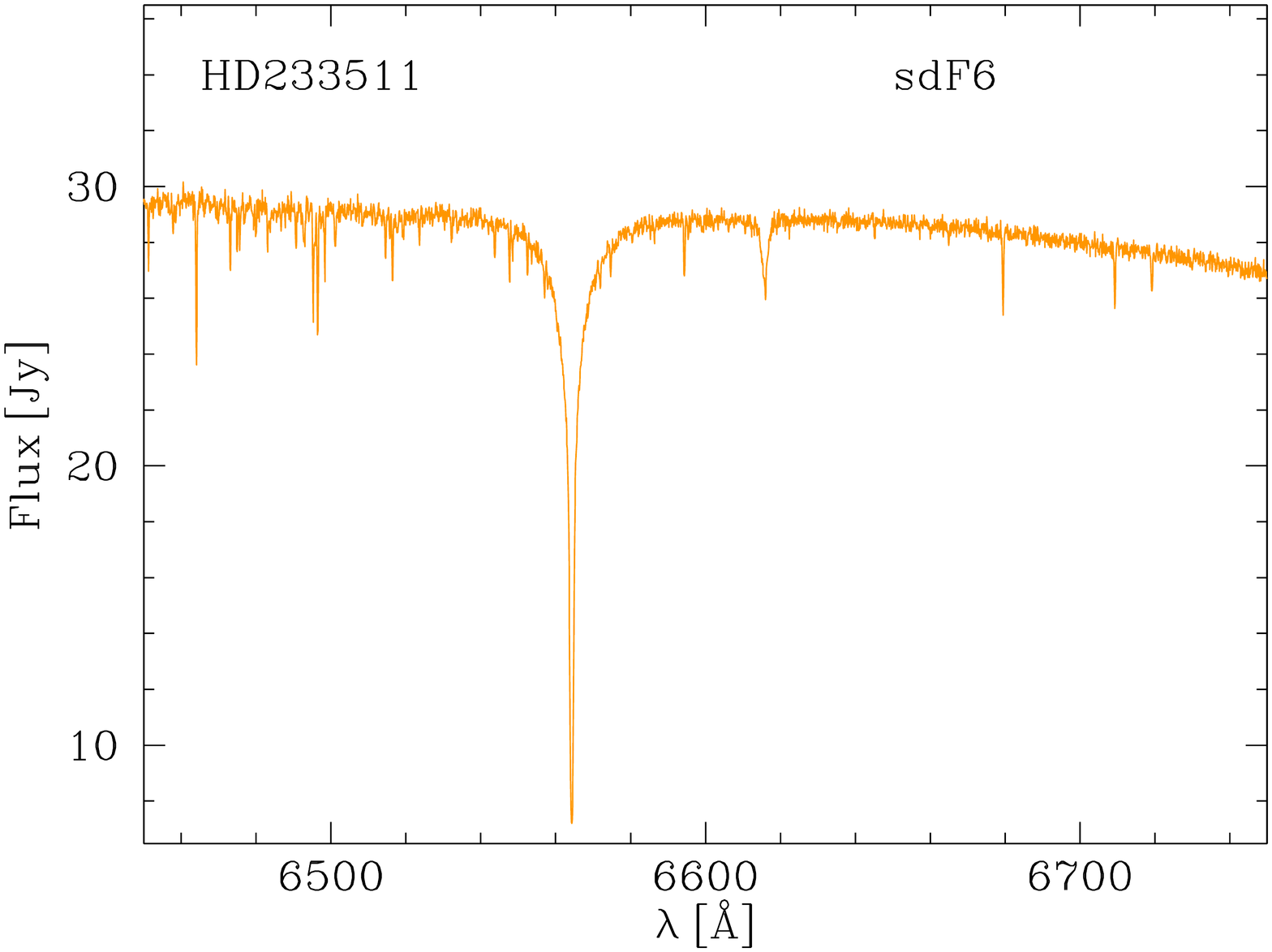}
\includegraphics[width=0.18\textwidth,angle=-90]{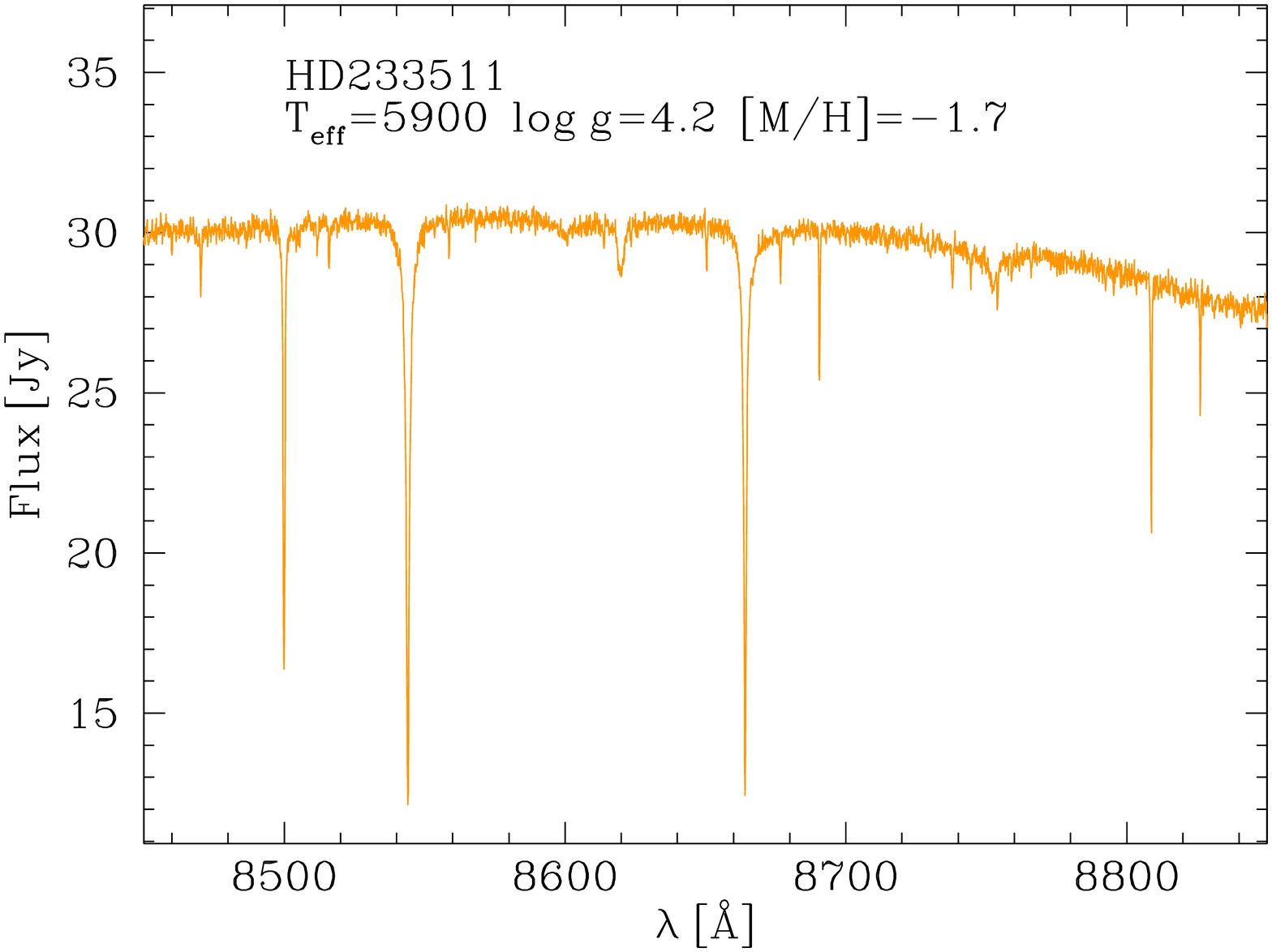}
\includegraphics[width=0.18\textwidth,angle=-90]{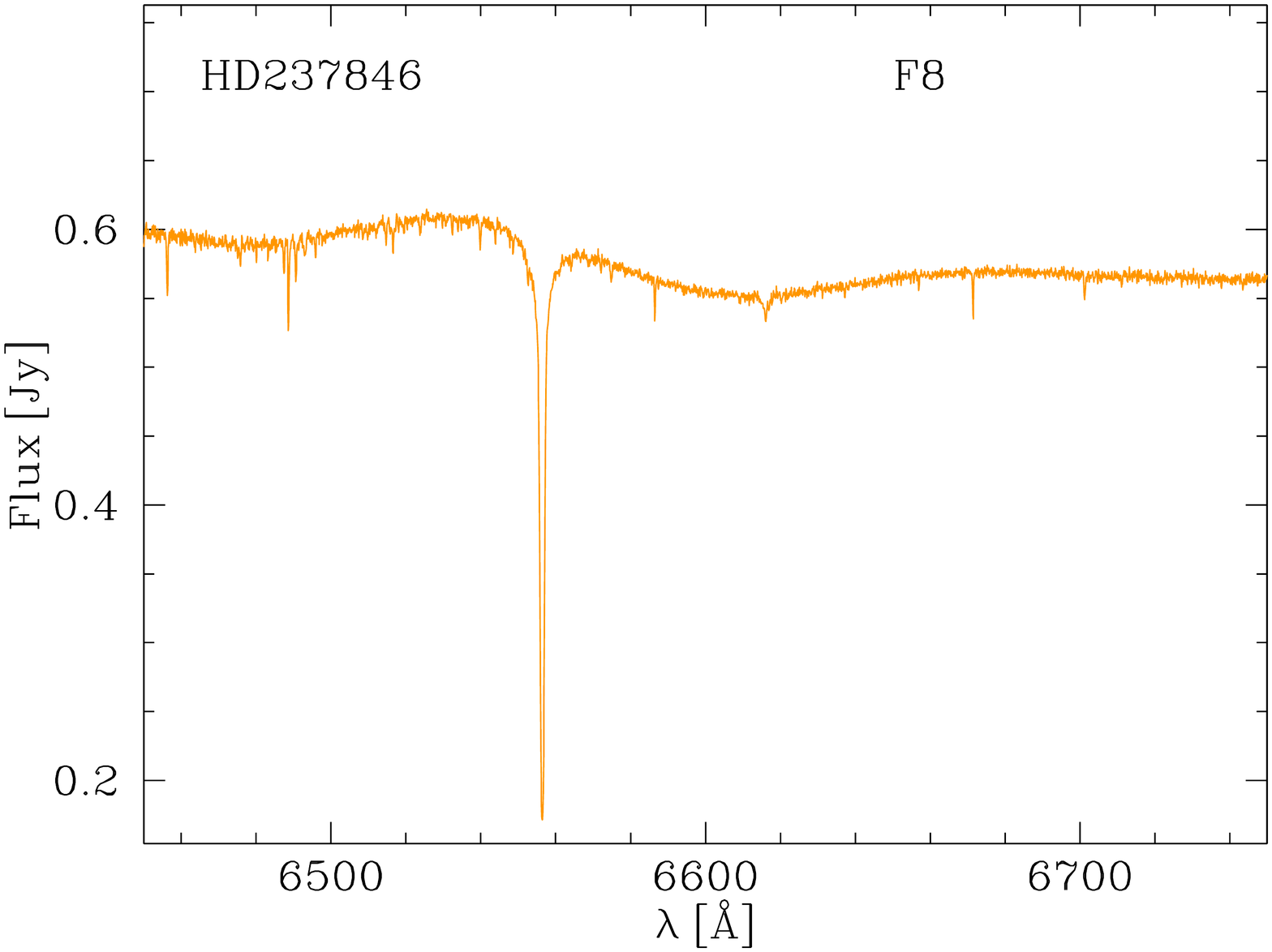}
\includegraphics[width=0.18\textwidth,angle=-90]{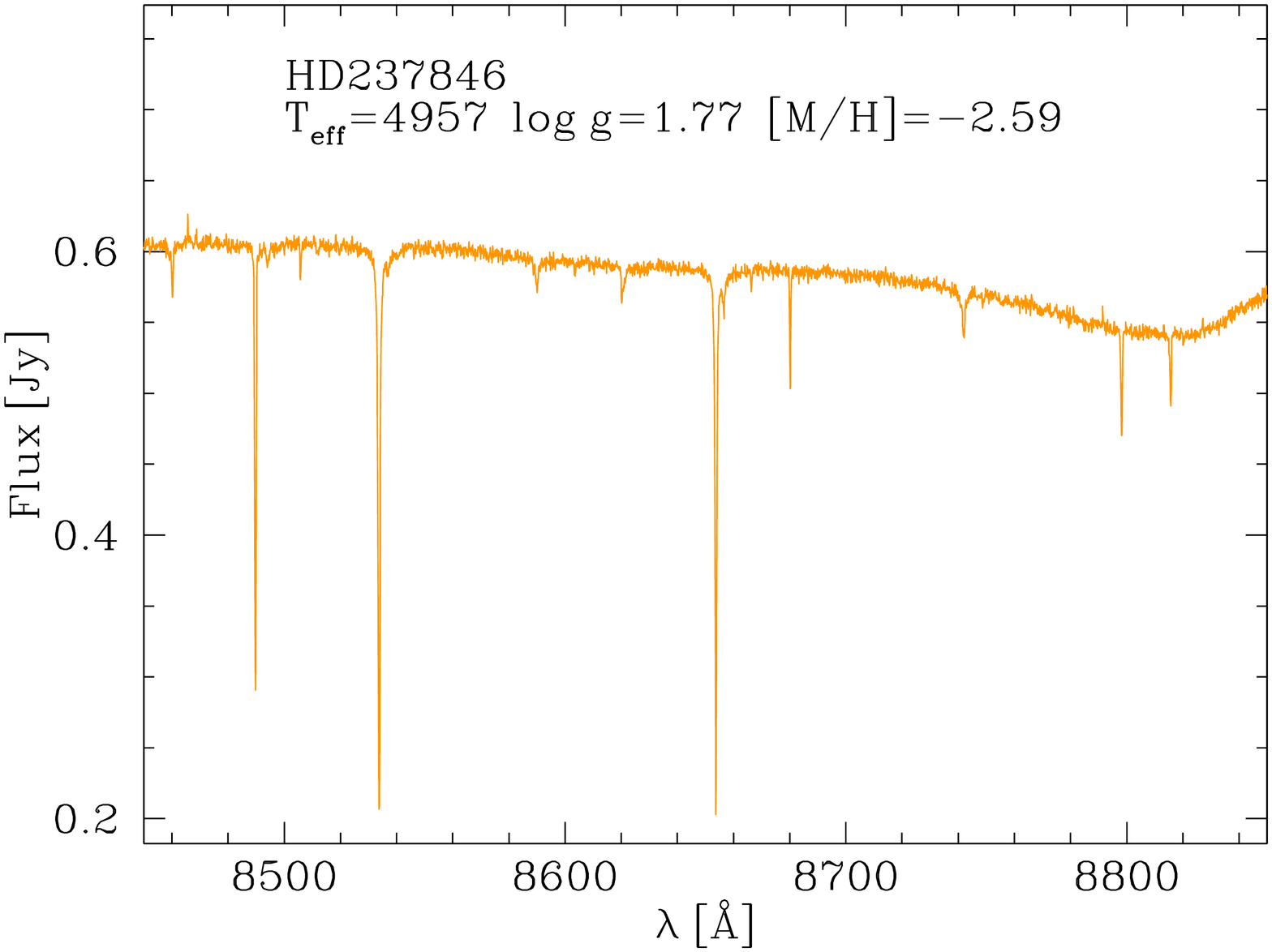}
\includegraphics[width=0.18\textwidth,angle=-90]{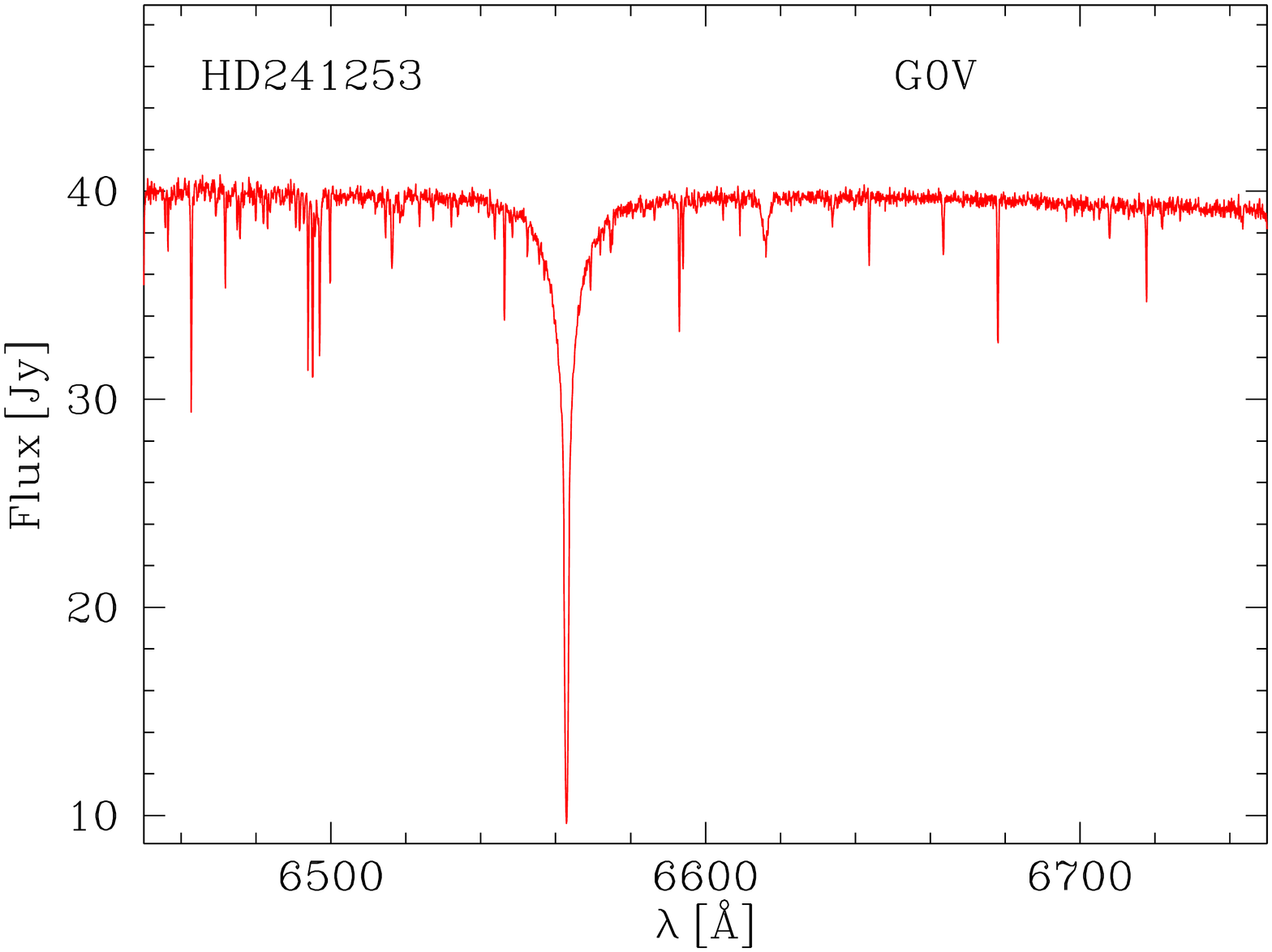}
\includegraphics[width=0.18\textwidth,angle=-90]{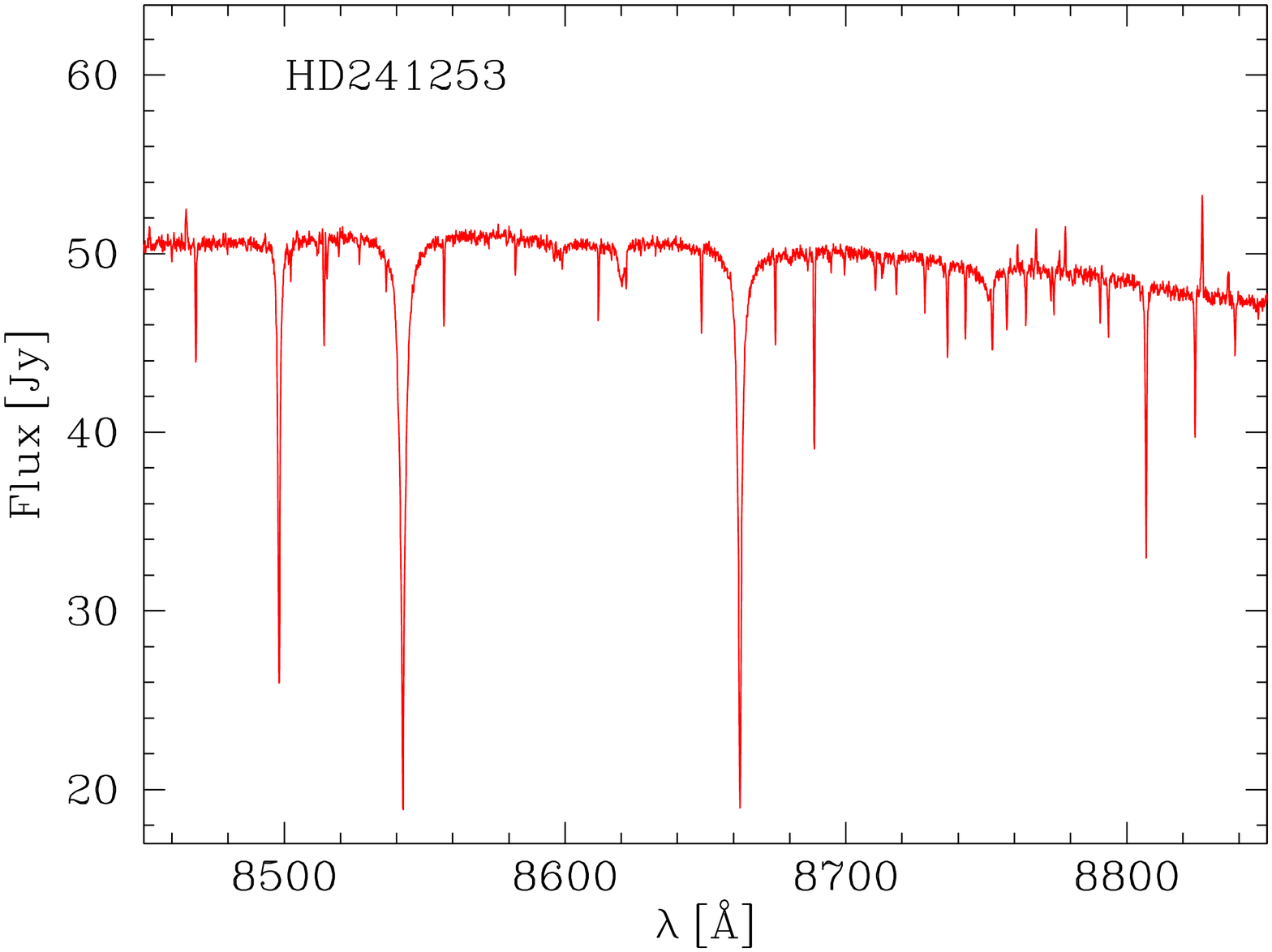}
\includegraphics[width=0.18\textwidth,angle=-90]{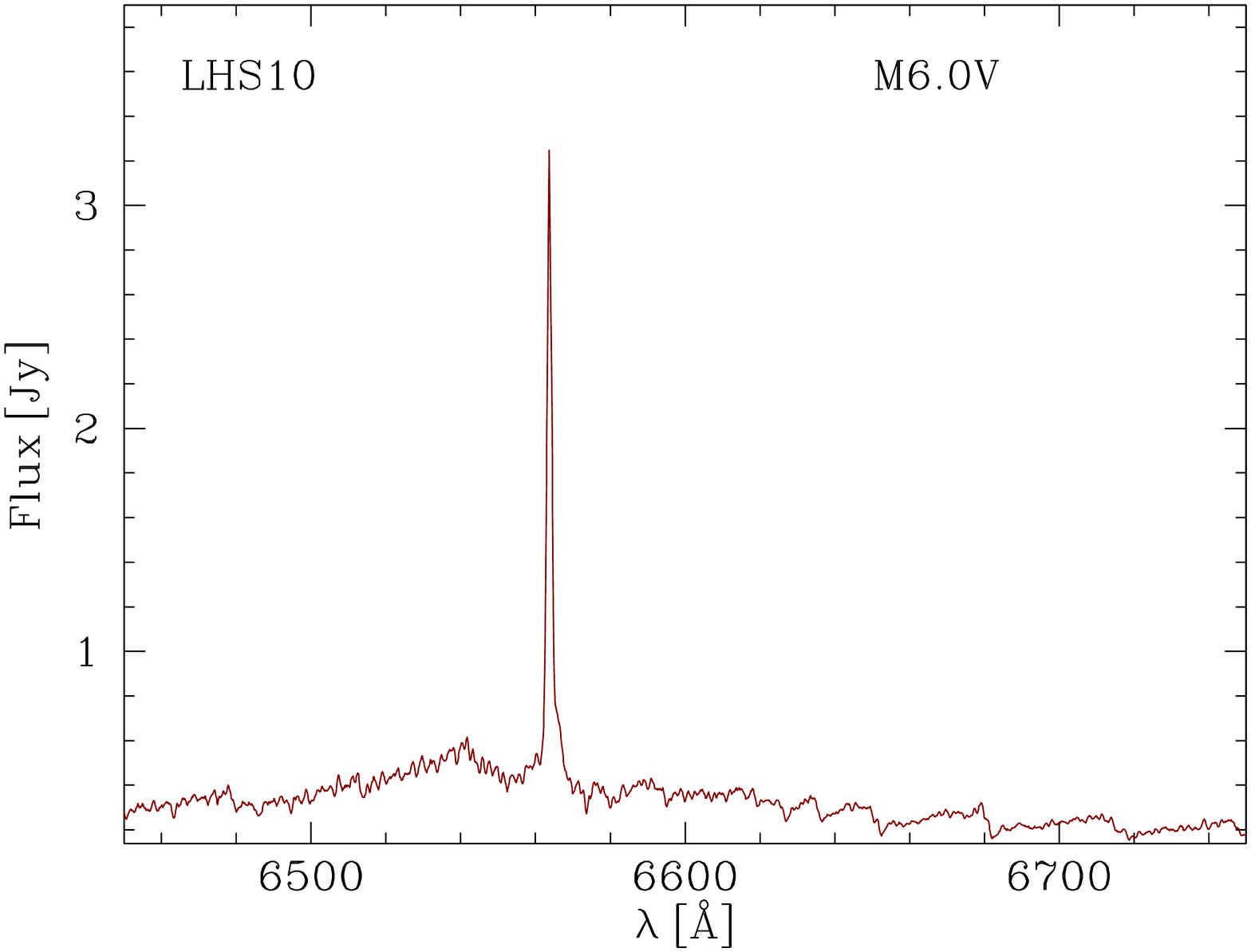}
\includegraphics[width=0.18\textwidth,angle=-90]{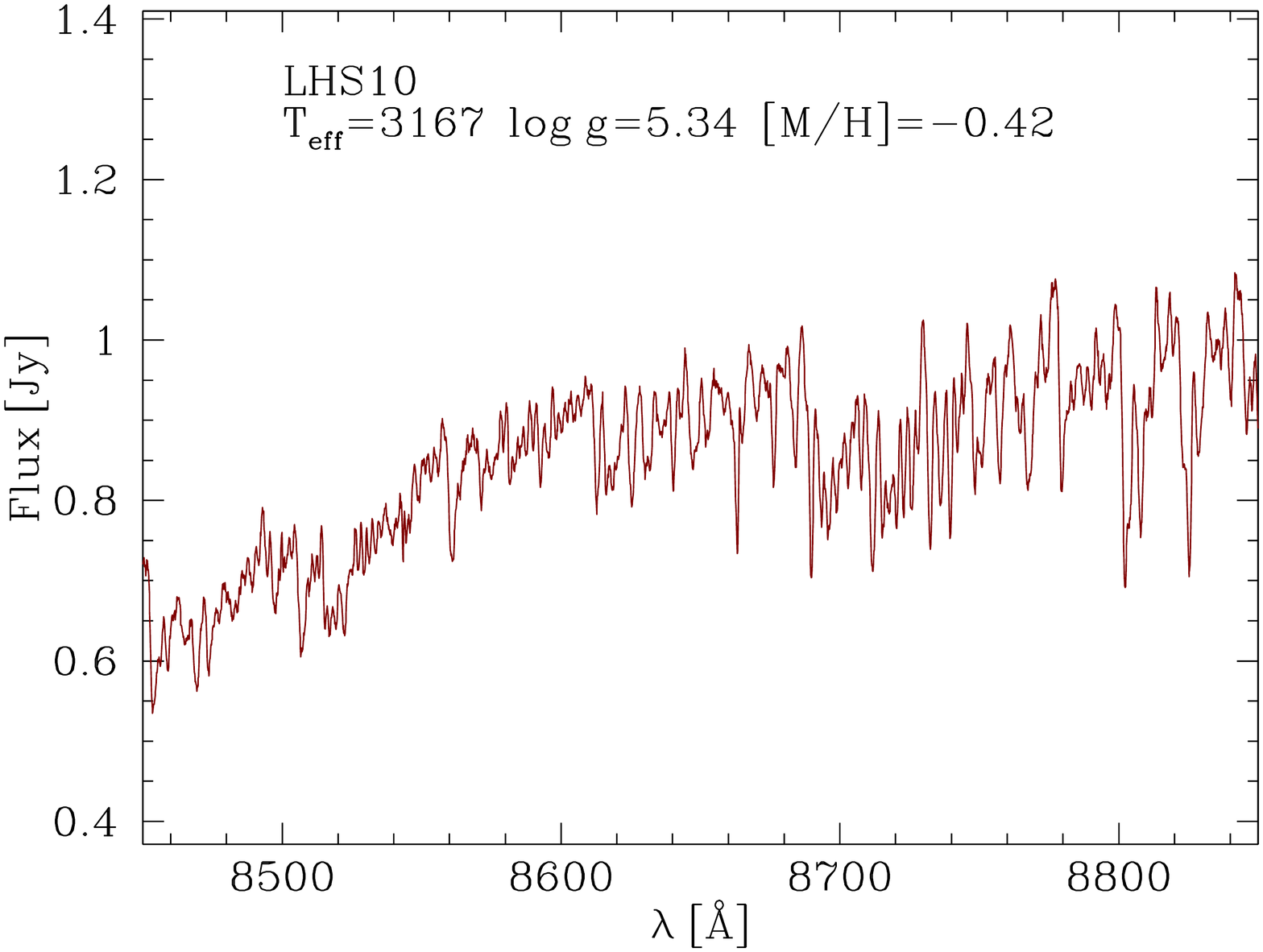}
\includegraphics[width=0.18\textwidth,angle=-90]{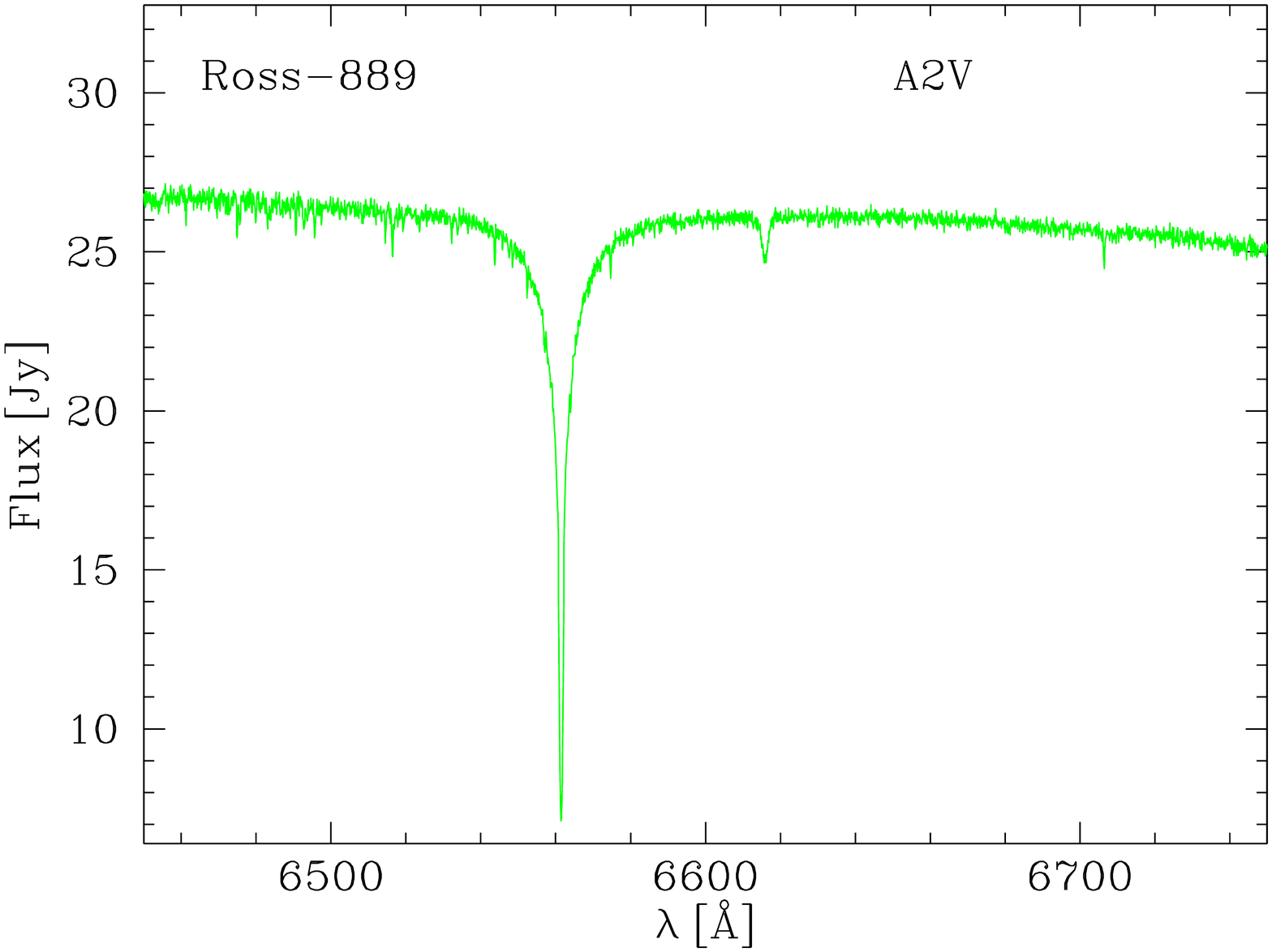}
\includegraphics[width=0.18\textwidth,angle=-90]{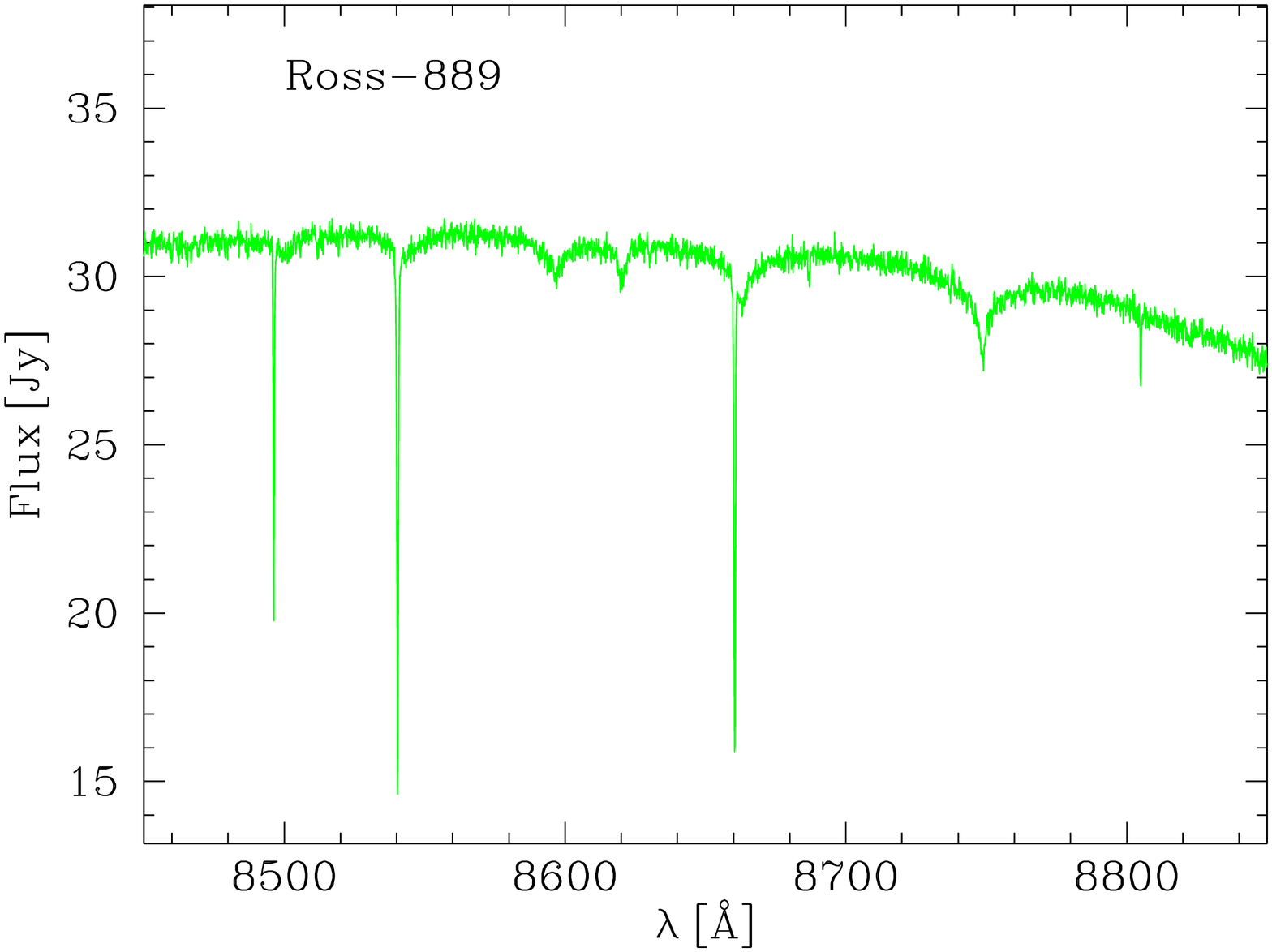}
\includegraphics[width=0.18\textwidth,angle=-90]{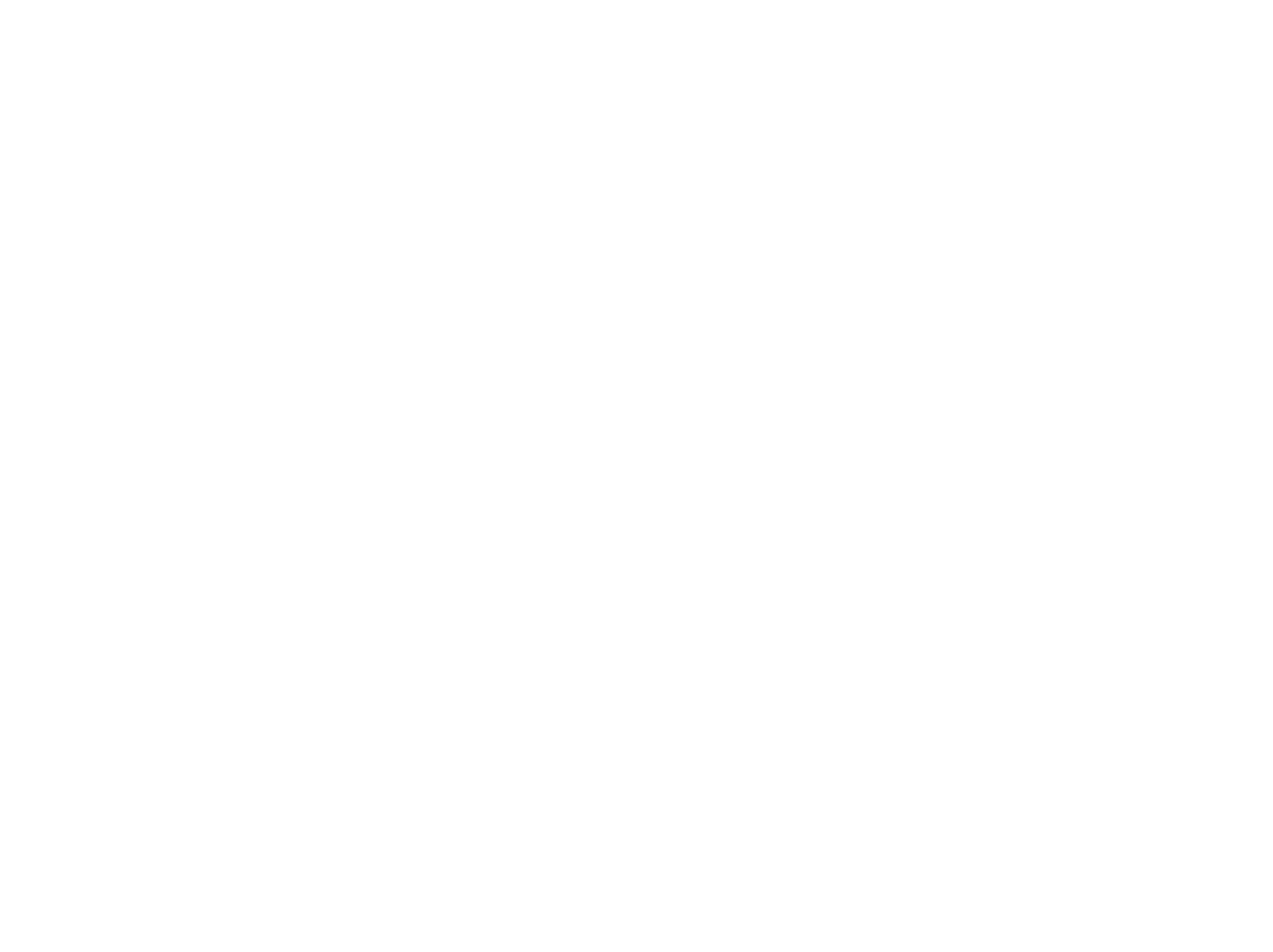}
\includegraphics[width=0.18\textwidth,angle=-90]{void.pdf}

\contcaption{30. Stars shown in this page are: HD221830, HD224544, HD224559, HD224801, HD224926, HD224926, HD225160, HD233345, HD233511, HD237846, HD241253, LHS10 and Ross-889.}
\end{figure*}

\bsp	
\label{lastpage}